\documentclass[11pt,a4paper]{scrartcl}

\usepackage{ECFAreport}
\usepackage{bm}
\usepackage{comment}
\usepackage{ECFAreport_definitions}

\usepackage{xpatch}
\usepackage{makecell}
\usepackage{multicol}
\usepackage{numprint}

\makeatletter
\xpatchcmd\@HepConStyle
 {\edef\@upcode{\updefault}}
 {\ifdefined\shapedefault\edef\@upcode{\shapedefault}\else\edef\@upcode{\updefault}\fi}
 {}{}
\makeatother

\usepackage{arydshln}

\RedeclareSectionCommand[
  beforeskip=1.\baselineskip,
  afterskip=.5\baselineskip]{subsection}
\RedeclareSectionCommand[
  beforeskip=1.\baselineskip,
  afterskip=.5\baselineskip]{subsubsection}

\usepackage{todonotes}
\usepackage{xpatch}
\makeatletter
\xpretocmd{\todo}{\@bsphack}{}{}
\xapptocmd{\todo}{\@esphack}{}{}
\makeatother

\usepackage{booktabs}
\usepackage{csquotes}

\usepackage{SRCH}

\usepackage[italic]{mathastext}

\crefname{equation}{Eq.}{Eqs.}
\Crefname{equation}{Equation}{Equations}
\crefname{figure}{Fig.}{Figs.}
\Crefname{figure}{Figure}{Figures}

\usepackage{pdfpages}
\usepackage{changepage}

\DeclareCiteCommand{\citejournal}[\mkbibbrackets]
  {\usebibmacro{prenote}}
  {\usebibmacro{citeindex}%
   \printtext[bibhyperref]{\printfield{journaltitle}}%
   \iffieldundef{volume}
     {}%
     {\setunit{\addspace}%
     \printtext[bibhyperref]{\printfield{volume}}}%
   \setunit{\addspace}%
   \printtext[bibhyperref]{(\printdate)}%
   \iffieldundef{pages}
     {}
     {\setunit{\addspace}%
     \printtext[bibhyperref]{\printfield{pages}}%
     }%
     }
  {\multicitedelim}
  {\usebibmacro{postnote}}
  
\DeclareCiteCommand{\citesubmit}[\mkbibbrackets]
  {\usebibmacro{prenote}}
  {\usebibmacro{citeindex}%
   \printtext[bibhyperref]{\printfield{journaltitle}}%
   \setunit{\addspace}%
   \printtext[bibhyperref]{(\printdate)}}
  {\multicitedelim}
  {\usebibmacro{postnote}}
  
  \DeclareCiteCommand{\citeconf}[\mkbibbrackets]
  {\usebibmacro{prenote}}
  {\usebibmacro{citeindex}%
   \printtext[bibhyperref]{\printfield{howpublished}}%
   \setunit{\addspace}%
   \printtext[bibhyperref]{(\printdate)}}
  {\multicitedelim}
  {\usebibmacro{postnote}}

\title{}

\nolicence 

\notitlestamp 

\date{\today}

\addauthor{
J.~Altmann,
P.~Skands\\
\emph{Monash University, Wellington Road, Clayton VIC-3800, Australia}
}

\addauthor{
A.~Desai$^{0}$\\
\emph{The University of Adelaide, Adelaide, SA 5005, Australia}
}

\addauthor{
W.~Mitaroff\\
\emph{Institute of High Energy Physics, Austrian Academy of Sciences, 1050 Vienna, Austria}
}

\addauthor{
S.~Pl\"atzer$^{1}$\\
\emph{Institute of Physics, NAWI Graz, University of Graz, Universit\"atsplatz 5, 8010 Graz, Austria}
}

\addauthor{
D.~Dobur,
K.~Skovpen\\
\emph{Ghent University, Proeftuinstraat 86, 9000 Gent, Belgium}
}

\addauthor{
M. ~Drewes,
G.~Durieux,
Y.~Georis,
Y.~Ma\\
\emph{Centre for Cosmology, Particle Physics and Phenomenology (CP3), Universit\'e catholique de Louvain, 1348 Louvain-la-Neuve, Belgium}
}

\addauthor{
A.~De Moor,
K.~Gautam$^{2}$,
E.~Ploerer$^{2}$,
M.~Tytgat\\
\emph{Inter-University Institute for High Energies, Vrije Universiteit Brussel,
Pleinlaan 2, 1050 Ixelles, Belgium}
}

\addauthor{
K.~Mota~Amarilo\\
\emph{Universidade do Estado do Rio de Janeiro, Rio de Janeiro, Brazil}
}

\addauthor{
M.J.~Basso$^{3}$\\
\emph{TRIUMF, Wesbrook Mall, Vancouver, Canada}
}

\addauthor{
J.~Gu\\
\emph{Department of Physics and Center for Field Theory and Particle Physics, Fudan University,
Shanghai 200438, China}
}

\addauthor{
Y.~Zhang\\
\emph{School of Physics, Hefei University of Technology, Hefei 230601, China}
}

\addauthor{
X.-H.~Jiang$^{4}$,
M.~Ruan,
B.~Yan\\
\emph{Institute of High Energy Physics, Chinese Academy of Sciences, Beijing 100049, China}
}

\addauthor{
X.-K.~Wen\\
\emph{School of Physics, Peking University, Beijing 100871, China}
}

\addauthor{
J.~Liao\\
\emph{School of Physics, Sun Yat-sen University, Guangzhou, 510275, China}
}

\addauthor{
C.~Li\\
\emph{School of Science, Sun Yat-sen University (SYSU), Shenzhen, 518107, China}
}

\addauthor{
M.~Ouchemhou, 
T.~Robens\\
\emph{Rudjer Boskovic Institute, Bijenicka cesta 54, 10000 Zagreb, Croatia}
}

\addauthor{
M.~Dam\\
\emph{Niels Bohr Institute, Copenhagen University, Jagtvej 155, 2200 Copenhagen, Denmark}
}

\addauthor{
G.~Bernardi,
K.~Dewyspelaere,
A.~Maloizel,
G.~Marchiori\\
\emph{APC, Universit\'e Paris Cit\'e, CNRS/IN2P3, Paris, France}
}

\addauthor{
S.~Gascon-Shotkin,
E.~Jourd'huy,
C.~Verollet,
J.~Xiao\\
\emph{Institut de Physique des 2 Infinis de Lyon IN2P3-CNRS, rue Enrico Fermi
69622 Villeurbanne, France}
}

\addauthor{
R.~Perez-Ramos$^{5}$\\
\emph{DRII-IPSA, 94200 Ivry-sur-Seine, France}
}

\addauthor{
R.~Madar,
T.~Miralles,
S.~Monteil\\
\emph{LPCA, Universit\'e Clermont Auvergne, CNRS/IN2P3, Clermont-Ferrand, France}
}

\addauthor{
Y.~Amhis,
I.~Combes,
N.~Morange,
R.~P\"oschl,
M.~Reboud,
F.~Richard,
O.~Sumensari\\
\emph{IJCLab, Universit\'e Paris-Saclay, CNRS/IN2P3, 91405 Orsay, France}
}

\addauthor{
G.~Cacciapaglia\\
\emph{Laboratoire de Physique Th\'eorique et Hautes Energies (LPTHE), Sorbonne Universit\'e et CNRS, 75252 Paris, France}
}

\addauthor{
V.~Balagura,
V.~Boudry,
J.-C.~Brient,
C.~Charlot,
C.~Ochando,
R.~Salerno,
Y.~Shi,
Y.~Sirois\\
\emph{Laboratoire Leprince-Ringuet, Ecole Polytechnique, 91128 Palaiseau, France}
}

\addauthor{
O.~Arnaez,
M.~Delmastro,
H.~Taibi,
Z.~Wu\\
\emph{LAPP, 9 Chem. de Bellevue, 74940 Annecy, France}
}

\addauthor{
S.D.~Dittmaier,
M.~Gabelmann\\
\emph{Albert-Ludwigs-Universit{\"a}t, Freiburg, Germany}
}

\addauthor{
D.~Zerwas\\
\emph{DMLab, Deutsches Elektronen-Synchrotron DESY, CNRS/IN2P3, Hamburg, Germany}
}

\addauthor{
J.~Alimena,
M.~Berggren,
F.~Blekman$^{6}$,
B.~Bliewert,
J.~Braathen,
A.~Filipe Silva,
F.~Gaede,
C.~Grojean,
J.~List,
T.~Madlener$^{7}$,
A.~B.~Meyer,
M.T.~N\'u\~nez Pardo de Vera,
K.~Radchenko Serdula,
J.~Reuter,
F.~Sefkow,
J.~Torndal,
A.~Verduras Schaeidt,
G.~Weiglein\\
\emph{DESY, Hamburg, Germany}
}

\addauthor{
K.~K\"oneke\\
\emph{II. Physikalisches Institut, Georg-August-Universit\"{a}t G\"ottingen, G\"ottingen, Friedrich-Hund-Platz 1, 37077 G\"ottingen, Germany}
}

\addauthor{
B.~Dudar,
F.~Yu\\
\emph{PRISMA+ Cluster of Excellence and Mainz Institute for Theoretical Physics, Johannes Gutenberg University, 55099 Mainz, Germany}
}

\addauthor{
S.~Alshamaily,
G.~Brodbek,
U.~Einhaus,
S.~Giappichini,
P.~Goldenzweig,
S.~Keilbach,
J.~Kieseler,
M.~Klute,
M.~Presilla,
F.~Simon,
A.~Wiedl,
X.~Zuo\\
\emph{Karlsruhe Institute of Technology, Karlsruhe, Germany}
}

\addauthor{
P.~Lo Chiatto$^{8}$\\
\emph{Max-Planck-Institut f{\"u}r Physik, Boltzmannstra{\ss}e 8, 85748 Garching bei M{\"u}nchen, Germany}
}

\addauthor{
P.~Bechtle,
M.~Vellasco\\
\emph{University of Bonn, Physikalisches Institut, Nussallee 12, 53115 Bonn, Germany}
}

\addauthor{
K.~Kr\"oninger,
L.~R\"ohrig$^{9}$\\
\emph{TU Dortmund University, Department of Physics, Otto-Hahn-Str. 4a, 44227 Dortmund, Germany}
}

\addauthor{
H.~Bahl,
M.D.~Tat\\
\emph{Universit\"at Heidelberg, Heidelberg, Germany}
}

\addauthor{
A.~Grohsjean,
J~Lahiri,
G.~Moortgat-Pick\\
\emph{University of Hamburg, Hamburg, Germany}
}

\addauthor{
W.~Kilian\\
\emph{Center for Particle Physics Siegen, Universit\"at Siegen, 57068 Siegen, Germany}
}

\addauthor{
T.~Ohl\\
\emph{University of W\"urzburg, W\"urzburg, Germany}
}

\addauthor{
P.~Poulose\\
\emph{Indian Institute of Technology Guwahati, Assam 781039, India}
}

\addauthor{
J.~Dutta\\
\emph{The Institute of Mathematical Sciences, CIT Campus, Tharamani, Chennai, Tamil Nadu, 600113, India}
}

\addauthor{
M.~Mohammadi Najafabadi,
M.~Nourbakhsh,
S.~Tizchang$^{10}$\\
\emph{School of Particles and Accelerators, Institute for Research in Fundamental Sciences (IPM), Artesh Highway, Tehran 1956836681, Iran}
}

\addauthor{
Z.~Baharyioon,
A.~Jafari\\
\emph{Department of Physics, Isfahan University of Technology, Isfahan, 84156-83111, Iran}
}

\addauthor{
A.~Amiri$^{11}$,
K.~Azizi$^{12}$,
H.~Fatehi,
R.~Jafariseyedabad$^{11}$,
M.~Malekhosseini,
S.~Rostami\\
\emph{Department of Physics, University of Tehran, North Karegar Avenue, Tehran 14395-547, Iran}
}

\addauthor{
V.~Keus\\
\emph{Dublin Institute of Advanced Studies (DIAS), Dublin D04 C932, Ireland}
}

\addauthor{
Y.~Soreq\\
\emph{Physics Department, Technion - Israel Institute of Technology, Haifa 3200003, Israel}
}

\addauthor{
M.~Cobal$^{13}$,
G.~Panizzo$^{13}$,
M.~Pinamonti$^{13}$,
L.~Pintucci$^{13}$,
L.~Toffolin$^{13}$\\
\emph{INFN Gruppo Collegato di Udine, Sezione di Trieste, Udine, Via delle Scienze 206, Udine, Italy}
}

\addauthor{
E.~Bagnaschi,
F.~Mescia\\
\emph{INFN, Laboratori Nazionali di Frascati, Via E. Fermi 40, 00044 Frascati, Italy}
}

\addauthor{
W.~Elmetenawee\\
\emph{INFN Sezione di Bari, Via Giovanni Amendola 173, 70126 Bari, Italy}
}

\addauthor{
G.L.~Alberghi,
L.~Bellagamba\\
\emph{INFN Sezione di Bologna, Viale Carlo Berti Pichat 6/2, 40127 Bologna, Italy}
}

\addauthor{
M.~Tammaro\\
\emph{INFN Sezione di Firenze, Via G. Sansone 1, 50019 Sesto Fiorentino, Italy}
}

\addauthor{
E.G.~Gorini,
F.G.~Grancagnolo,
M.P.~Primavera\\
\emph{INFN Sezione di Lecce, Lecce, Italy}
}

\addauthor{
A.~Andreazza$^{14}$,
C.~Meroni$^{14}$,
F.~Sabatini,
R.~Zanzottera$^{14}$,
M.~Zaro$^{15}$\\
\emph{INFN Sezione di Milano, Via Celoria 16, 20133 Milano, Italy}
}

\addauthor{
P.~Azzi,
A.~De Vita,
N.~Selimovi\'c\\
\emph{INFN Sezione di Padova, Via Marzolo 8, 35131 Padova, Italy}
}

\addauthor{
C.~Carloni Calame,
M.~Chiesa,
O.~Nicrosinii,
F.~Piccinini,
G.~Polesello,
N.~Valle\\
\emph{INFN Sezione di Pavia, via A. Bassi 6, 27100 Pavia, Italy}
}

\addauthor{
S.~Ajmal,
O.~Panella\\
\emph{INFN Sezione di Perugia, Via A. Pascoli, 06123, Perugia, Italy}
}

\addauthor{
P.~Azzurri,
D.~Barducci$^{16}$,
D.~Buttazzo,
A.~Dondarini$^{16}$,
A.~Korajac,
G.~Marino$^{16}$,
F.~Palla,
P.~Panci$^{16}$\\
\emph{INFN Sezione di Pisa, Largo Bruno Pontecorvo 3, 56127, Pisa, Italy}
}

\addauthor{
D.~Marzocca\\
\emph{INFN Trieste, Via Valerio 2, 34127 Trieste, Italy}
}

\addauthor{
A.~Vicini\\
\emph{TIFLab, Universit\`a degli Studi di Milano, Via Celoria 16, 20133 Milano, Italy}
}

\addauthor{
N.~De~Filippis$^{17}$,
F.M.~Procacci$^{17}$\\
\emph{Politecnico di Bari, Via Orabona 4, 70126 Bari, Italy}
}

\addauthor{
Alberto Lusiani$^{18}$\\
\emph{Scuola Normale Superiore, Piazza dei Cavalieri 7, 56126 Pisa, Italy}
}

\addauthor{
B.A.~Erdelyi$^{19}$,
R.~Gr\"ober$^{19}$,
A.~Rossia$^{19}$\\
\emph{Dipartimento di Fisica e Astronomia "G. Galilei", Universit\`a degli Studi di Padova, 35131 Padova, Italy}
}

\addauthor{
M.~Bulliri\\
\emph{Universit\`a degli Studi di Roma la Sapienza, Piazzale Aldo Moro 5, 00185 Rome, Italy}
}

\addauthor{
M.~Al-Thakeel$^{20}$,
F.~Maltoni$^{21}$\\
\emph{Universit\`a di Bologna, 40126 Bologna, Italy}
}

\addauthor{
G.~Montagna$^{22}$,
F.~Ucci$^{22}$\\
\emph{Universit\`a di Pavia, 27100 Pavia, Italy}
}

\addauthor{
R.~Franceschini$^{23}$\\
\emph{Universit\`a degli Studi Roma Tre, Via della Vasca Navale 84, 00146 Rome, Italy}
}

\addauthor{
K.~Fujii,
D.~Jeans,
J.~Nakajima\\
\emph{KEK, Tsukuba, Japan}
}

\addauthor{
A.~Das\\
\emph{Hokkaido University, Sapporo 060-0810, Japan}
}

\addauthor{
S.~Narita\\
\emph{Iwate University, Morioka, Iwate 020-8551, Japan}
}

\addauthor{
T.~Takeshita\\
\emph{Shinshu University, Matsumoto, Nagano, Japan}
}

\addauthor{
M.~Ishino,
T.~Mori,
T.~Murata,
W.~Ootani,
T.~Seino,
T.~Suehara,
R.~Tagami,
J.~Tian\\
\emph{The University of Tokyo, 7-3-1 Hongo, Bunkyo-ku, 113-0033 Tokyo, Japan}
}

\addauthor{
K.~Hidaka\\
\emph{Tokyo Gakugei University, Koganei, Tokyo 184-8501, Japan}
}

\addauthor{
H.~Khanpour\\
\emph{AGH University, Faculty of Physics and Applied Computer Science, Al. Mickiewicza 30, 30-055 Krak\'ow, Poland}
}

\addauthor{
J.~Bhom,
P.~Bruckman,
M.~Chrzaszcz,
A.~Kaczmarska,
M.~Kucharczyk,
T.~Lesiak,
P.~Malecki,
M.~Skrzypek,
Z.~Was\\
\emph{The Henryk Niewodnicza\'nski
Institute of Nuclear Physics
Polish Academy of Sciences
ul. Radzikowskiego 152,
31-342 Krak\'ow, Poland}
}

\addauthor{
A.~Price,
W.~P{\l}aczek,
A.~Si\'odmok$^{24}$\\
\emph{Faculty of Physics, Astronomy and Applied Computer Science, ul. Lojasiewicza 11, 30-348 Krak\'ow, Poland}
}

\addauthor{
B.~Brudnowski$^{25}$,
J.~Kalinowski,
J.~Klamka$^{25}$,
K.~Mekala$^{26}$,
K.~Sakurai,
A.F.~\.Zarnecki$^{25}$,
K.~Zembaczy\'nski$^{25}$\\
\emph{Faculty of Physics, University of Warsaw, Warsaw, Poland}
}

\addauthor{
J.~Hajer,
B.~Oliveira\\
\emph{Centro de F\'isica Te\'orica de Part\'iculas (CFTP), Instituto Superior T\'ecnico (IST), Universidade de Lisboa, 1049-001 Lisboa, Portugal}
}

\addauthor{
N.F.~Castro$^{27}$,
G.~Chachamis$^{28}$\\
\emph{LIP -- Laborat\'orio de Instrumenta\c{c}\~ao e F\'isica Experimental de Part\'iculas, Escola de Ci\`encias, Campus de Gualtar, Universidade do Minho, 4701-057 Braga, Portugal}
}

\addauthor{
I.~Bozovic$^{29}$,
G.~Kacarevic,
G.~Milutinovic-Dumbelovic,
I.~Smiljanic,
I.~Vidakovic,
N.~Vukasinovic\\
\emph{Vin\v{c}a Institute of Nuclear Sciences, University of Belgrade, Belgrade, Serbia}
}

\addauthor{
J.~Alcaraz Maestre$^{30}$,
M.~Cepeda$^{31}$,
M.C.~Fouz$^{30}$\\
\emph{Centro de Investigaciones  Energ\'eticas, Medioambientales y Tecnol\'ogicas (CIEMAT)
Madrid, Spain}
}

\addauthor{
J.~Fuster,
A.~Irles,
L.~Mantani,
J.P.~M\'arquez,
V.A.~Mitsou,
M.~Moreno~Ll\'acer,
E.~Musumeci,
B.~Pattnaik,
A.~Saibel,
M.A.~Sanchis,
M.~Vos,
J.~Zurita\\
\emph{IFIC, CSIC-Universitat de Val\`encia, Valencia, Spain}
}

\addauthor{
A.~Ruiz~Jimeno\\
\emph{Instituto de F\'isica de Cantabria, Av de los Castros, 39005 Santander, Spain}
}

\addauthor{
J.A.~Aguilar-Saavedra,
S.~Heinemeyer,
A.~Sabio Vera$^{32}$\\
\emph{Instituto de F\'isica Te\'orica, CSIC-Universidad Aut\'onoma de Madrid, Madrid, Spain}
}

\addauthor{
I.~Corredoira\\
\emph{Instituto Galego de F\'isica de Altas Enerx\'ias (IGFAE), Universidade de Santiago de Compostela
Santiago de Compostela, Spain}
}

\addauthor{
F.~G.~Celiberto$^{33}$\\
\emph{Alcal\'a de Henares, Madrid, E-28805, Spain}
}

\addauthor{
F.~Cornet-Gomez\\
\emph{Campus Universitario de Rabanales, Ctra. N-IVa, Km. 396. 14071, Cordoba, Spain}
}

\addauthor{
M.~Chala,
J.~de Blas$^{34}$\\
\emph{Departamento de F\'isica Te\'orica y del Cosmos, Universidad de Granada,
Campus de Fuentenueva, 18071 Granada, Spain}
}

\addauthor{
M.~Vande Voorde\\
\emph{Department of Physics, Royal Institute of Technology, Stockholm; Sweden}
}

\addauthor{
T.~Sj\"ostrand\\
\emph{Department of Physics, Lund University, Lund, Sweden}
}

\addauthor{
A.~Gall\'en,
R.~Gonzalez Suarez,
G.~Ripellino\\
\emph{Department of Physics and Astronomy, Uppsala University, L{\"a}gerhyggsv{\"a}gen 1, Uppsala, 752 37, Sweden}
}

\addauthor{
S.~Aumiller,
V.~Cairo,
J.~M.~Carceller,
C.~Cornella,
D.~d'Enterria,
J.~Davighi,
M.~M.~Defranchis,
M.~Di Carlo,
F.~Erben,
B.~Francois,
G.~Ganis,
D.~Garcia,
G.~Guerrieri,
A.~J\"uttner,
S.~Kuberski,
A.~Mehta,
L.~Reichenbach$^{35}$,
A.~Sailer$^{7}$,
D.~Schulte,
S.~Sasikumar,
M.~Selvaggi,
J.~Smiesko,
A.~Tolosa-Delgado,
J.~T.~Tsang\\
\emph{CERN, Geneva, Switzerland}
}

\addauthor{
S.~Thor\\
\emph{ETH Z\"{u}rich, 8092 Z\"{u}rich, Switzerland}
}

\addauthor{
L.~Biermann,
M.~Spira\\
\emph{Paul Scherrer Institute
Particle Physics Theory Group LTP
Forschungsstrasse 111
5232 Villigen PSI, Switzerland}
}

\addauthor{
T.~Critchley,
P.~Kontaxakis,
A.~Sfyrla\\
\emph{D\'epartement de Physique Nucl\'eaire et Corpusculaire (DPNC), Universit\'e de Gen\`eve, 1205 Gen\`eve, Switzerland}
}

\addauthor{
S.~Antusch,
A.~Greljo\\
\emph{Department of Physics, University of Basel, 4056 Basel, Switzerland}
}

\addauthor{
M.~Bordone$^{36}$,
F.~Canelli,
A.~Ilg,
G.~Isidori,
B.~Kilminster,
T.H.~Kwok,
V.~Lukashenko,
A.~Macchiolo,
Z.~Polonsky\\
\emph{Physik-Institut, University of Zurich, 8057 Z\"urich, Switzerland}
}

\addauthor{
P.~Koppenburg,
J.~Rojo$^{37}$\\
\emph{Nikhef, Science Park 105, 1098 XG Amsterdam, The Netherlands}
}

\addauthor{
J.~Klari\'c$^{38}$,
V.~Plakkot$^{39}$\\
\emph{University of Amsterdam, Science Park 904, 1098 XH Amsterdam, The Netherlands}
}

\addauthor{
H.~Denizli,
A.~Senol\\
\emph{Department of Physics, Bolu Abant Izzet Baysal University, 14280 Bolu, T\"urkiye}
}

\addauthor{
H.~Duran Yildiz\\
\emph{Institute of Accelerator Technologies, Ankara University, Ankara, T\"urkiye}
}

\addauthor{
E.~Bulyak\\
\emph{NSC KIPT, 1 Academichna Str. Kharkiv UA-61108, Ukraine}
}

\addauthor{
E.~Curtis,
A.-M.~Magnan,
N.~Wardle\\
\emph{The Blackett Laboratory, Imperial College, London SW7 2BW, United Kingdom}
}

\addauthor{
J.~Hayward,
M.~Kenzie,
S.~Williams\\
\emph{Cavendish Laboratory, University of Cambridge, Cambridge CB3 0HE, United Kingdom}
}

\addauthor{
S.~Farrington,
C.~Leonidopoulos$^*$,
J.~Silva,
J.~ter Hoeve\\
\emph{University of Edinburgh, Edinburgh EH9 3FD, United Kingdom}
}

\addauthor{
C.~Englert,
M.~Miralles L\'opez,
J.~Nesbitt,
D.~Protopopescu,
S.~Renner,
A.~Robson$^*$,
Y.~Zhang\\
\emph{University of Glasgow, Glasgow G12 8QQ, Scotland, United Kingdom}
}

\addauthor{
J.M.~Burridge,
E.~Celada,
A.J.~Costa,
V.~Miralles,
A.~Pilkington,
M.~Thomas,
E.~Vryonidou,
A.R.~Wiederhold\\
\emph{University of Manchester, Manchester M13 9PL, United Kingdom}
}

\addauthor{
P.N.~Burrows,
C.~Hays,
G.~Wilkinson\\
\emph{Department of Physics, University of Oxford, Oxford OX1 5NZ, United Kingdom}
}

\addauthor{
K.~Mimasu\\
\emph{University of Southampton, Southampton S017 1BJ, United Kingdom}
}

\addauthor{
B.F.L.~Ward\\
\emph{Baylor University, Waco, TX 76798-7316, USA}
}

\addauthor{
H.~Abidi,
K.A.~Assamagan,
D.~Boye,
E.~Brost,
V.~Cavaliere,
A.E.~Connelly,
~C. L. Del Pio,
G.~Iakovidis,
A.~Li,
M.A.~Pleier,
A.~Sciandra,
S.~Snyder,
R.H.~Szafron,
A.~Tishelman-Charny,
I.~Veli\v{s}\v{c}ek\\
\emph{Brookhaven National Laboratory, Upton, NY 11973-5000, USA}
}

\addauthor{
L.~Gouskos,
L.~Li\\
\emph{Brown University, 182 Hope Street, Providence, RI 02912, USA}
}

\addauthor{
L.~Gray\\
\emph{Fermilab, PO Box 500, Batavia, IL 60510, USA}
}

\addauthor{
Z.~Yu\\
\emph{Theory Center, Jefferson Lab, Newport News, Virginia 23606, USA}
}

\addauthor{
N.~Pinto$^{40}$\\
\emph{Johns Hopkins University, Baltimore, Maryland 21218, USA}
}

\addauthor{
J.~Eysermans\\
\emph{Massachusetts Institute of Technology, Cambridge, MA 02139, USA}
}

\addauthor{
R.~Schwienhorst,
C.-P.~Yuan$^{41}$\\
\emph{Department of Physics and Astronomy, Michigan State University, East Lansing, Michigan 48824, USA}
}

\addauthor{
M.~Larson,
L.~Skinnari\\
\emph{Northeastern University, Boston, Massachusetts 02115, USA}
}

\addauthor{
W.~Chung,
L. S.~Mandacar\'{u} Guerra\\
\emph{Princeton University, Princeton, New Jersey 08544, USA}
}

\addauthor{
D.~Ntounis,
A.~Schwartzman,
C.~Vernieri\\
\emph{SLAC, 2575 Sand Hill Rd, Menlo Park, CA 94025, USA}
}

\addauthor{
V.~Dao\\
\emph{Stony Brook University, 100 Nicolls Rd, Stony Brook, NY 11794, USA}
}

\addauthor{
V.~Slokenbergs\\
\emph{Texas Tech University, Lubbock, Texas, 79409, USA}
}

\addauthor{
N.~Okada\\
\emph{University of Alabama, Tuscaloosa, Alabama 35487, USA}
}

\addauthor{
H.-C.~Cheng\\
\emph{University of California, One Shields Avenue, Davis, California, USA}
}

\addauthor{
W.~Altmannshofer\\
\emph{University of California, Santa Cruz, CA 95064, USA}
}

\addauthor{
M.~Szewc$^{42}$\\
\emph{Department of Physics, University of Cincinnati, Cincinnati, Ohio 45221, USA}
}

\addauthor{
G.W.~Wilson\\
\emph{University of Kansas, Lawrence, KS 66045, USA}
}

\addauthor{
S.~Airen\\
\emph{University of Maryland-College Park,
College Park, Maryland 20740, USA}
}

\addauthor{
A.~Freitas\\
\emph{University of Pittsburgh, Pittsburgh, PA 15260, USA}
}

\addauthor{
S.~Dasu,
A.~Mohammadi,
C.-H.~Nee\\
\emph{University of Wisconsin-Madison, Madison, WI 53706-1390, USA}
}

\addauthor{
T.~Lagouri\\
\emph{Yale University, New Haven CT, USA}
}

\addauthor{
A.~Nesterenko{$^{43}$},
V.A.~Okorokov{$^{44}$}\\
\emph{Affiliated with an institute covered by a cooperation agreement with CERN}
}

\addauthor{
A.~V.~Gritsan{$^{45}$},\\
\emph{Affiliated with an institute covered by a cooperation agreement between the US DOE, US NSF, and CERN}
}

\addauthor{
\begin{flushleft}
\small{
{$^{0}$}Also at ARC Centre of Excellence for Dark Matter Particle Physics, Australia\\
{$^{1}$}Also at Particle Physics, Faculty of Physics, University of Vienna, Boltzmanngasse 5, A-1090 Wien, Austria\\
{$^{2}$}Also at Physik-Institut, University of Zurich, 8057 Z\"urich, Switzerland\\
{$^{3}$}Also at Department of Physics, Simon Fraser University, University Drive W, Burnaby, Canada\\
{$^{4}$}Also at China Center of Advanced Science and Technology, Beijing, China; supported in part by the National Natural Science Foundation of China under grant No. 12342502.\\
{$^{5}$}Also at Laboratoire de Physique Th\'eorique et Hautes Energies (LPTHE), Sorbonne Universit\'e et CNRS, Paris\\
{$^{6}$}Also at University of Hamburg, Hamburg, Germany\\
{$^{7}$}Supported by the EU's Horizon 2020 Research and Innovation programme under Grant Agreement No 101004761\\
{$^{8}$}Also at Weizmann Institute of Science, 234 Herzl Street, POB 26, Rehovot 7610001, Israel\\
{$^{9}$}Also at LPCA, Universit\'e Clermont Auvergne, CNRS/IN2P3, Clermont-Ferrand, France\\
{$^{10}$}Also at Physics Department, Faculty of Sciences, Arak University, Arak, Iran\\
{$^{11}$}Also at Department of Physics, Faculty of Science, Ferdowsi University of Mashhad, Mashhad, Iran\\
{$^{12}$}Also at Department of Physics, Do\v{g}u\c{s} University, Dudullu-\"{U}mraniye, 34775 Istanbul, T\"urkiye\\
{$^{13}$}Also at Dipartimento Politecnico di Ingegneria e Architettura, Universit\`a di Udine, Udine\\
{$^{14}$}Also at Dipartimento di Fisica, Università degli Studi di Milano, 20133 Milano, Italy\\
{$^{15}$}Also at TIFLab, Università degli Studi di Milano, 20133 Milano, Italy\\
{$^{16}$}Also at Universit\`a di Pisa, Lungarno Pacinotti 43, 56126 Pisa, Italy\\
{$^{17}$}Also at INFN Sezione di Bari, Via Orbona 4, 70126 Bari, Italy\\
{$^{18}$}Also at INFN Sezione di Pisa, Largo Bruno Pontecorvo 3, 56127, Pisa, Italy\\
{$^{19}$}Also at INFN Sezione di Padova, Via Marzolo 8, 35131 Padova, Italy\\
{$^{20}$}Also at INFN Sezione di Bologna, Viale Carlo Berti Pichat 6/2, 40127 Bologna, Italy\\
{$^{21}$}Also at Universit\'e catholique de Louvain, 1348 Louvain-la-Neuve, Belgium\\
{$^{22}$}Also at INFN Sezione di Pavia, via A. Bassi 6, 27100 Pavia, Italy\\
{$^{23}$}Also at INFN Sezione di Roma Tre, Via della Vasca Navale 84, 00146 Rome, Italy\\
{$^{24}$}Supported by the National Science Centre, Poland, under awards 2023/50/A/ST2/00224 and 2019/34/E/ST2/00457\\
{$^{25}$}Supported by the National Science Centre, Poland, under the OPUS research project no. 2021/43/B/ST2/01778\\
{$^{26}$}Also at DESY Hamburg.  Supported by the National Science Centre, Poland, under the OPUS research project no. 2021/43/B/ST2/01778\\
{$^{27}$}Also at Departamento de F\'{i}sica, Escola de Ci\^{e}ncias, Universidade do Minho, 4710-057 Braga, Portugal\\
{$^{28}$}Also at LIP -- Laborat\'orio de Instrumenta\c{c}\~ao e F\'isica Experimental de Part\'iculas, Lisboa, Portugal\\
{$^{29}$}Supported by Fund of Science RS, Grant IDEJE HIGHTONE-P\\
{$^{30}$}Supported by Grant PID2021-122134NB-C22 funded by MICIU/AEI/10.13039/501100011033\\
{$^{31}$}Supported by Grants PID2021-122134NB-C22 and CNS2023-144781, funded by MCIU/AEI/10.13039/501100011033 and by European Union NextGenerationEU/PRTR\\
{$^{32}$}Also at Universidad Autonoma de Madrid, E-28049 Madrid, Spain\\
{$^{33}$}Supported by the Atracci\'on de Talento Grant n. 2022-T1/TIC-24176 of the Comunidad Aut\'onoma de Madrid, Spain\\
{$^{34}$}Partially funded by MICIU/AEI/10.13039/501100011033 and FEDER/UE (grant PID2022-139466NB-C21)\\
{$^{35}$}Also at University of Bonn, Physikalisches Institut, Nussallee 12, 53115 Bonn, Germany.  Supported by the Wolfgang Gentner Programme of the German Federal Ministry of Education and Research (grant no. 13E18CHA)\\
{$^{36}$}Also at CERN, Geneva, Switzerland\\
{$^{37}$}Also at Department of Physics and Astronomy, Vrije Universiteit, NL-1081 HV Amsterdam, The Netherlands\\
{$^{38}$}Also at University of Zagreb, Bijeni\v{c}ka c. 32, 10000 Zagreb, Croatia\\
{$^{39}$}Also at Nikhef, Science Park 105, 1098 XG Amsterdam, The Netherlands\\
{$^{40}$}Supported in part by the US NSF grant PHY-2310072\\
{$^{41}$}Supported by the U.S. National Science Foundation under Grant No.~PHY-2310291\\
{$^{42}$}Supported in part by NSF grants OAC-2103889, OAC-2411215 and OAC-2417682 and by DOE grant DE-SC101977; also at International Center for Advanced Studies (ICAS), ICIFI and ECyT-UNSAM, 25 de Mayo y Francia, 1650 San Mart\'{i}n, Buenos Aires, Argentina\\
{$^{43}$} https://orcid.org/0000-0003-4747-699X\\
{$^{44}$} https://orcid.org/0000-0002-7162-5345\\
{$^{45}$} https://orcid.org/0000-0002-3545-7970\\
{$^*$}Editors
}
\end{flushleft}
}

\graphicspath{ {./logos/}{./figures/} }

\DeclareTOCStyleEntry[beforeskip=.2cm]{section}{section}

\begin{document}

\includepdf{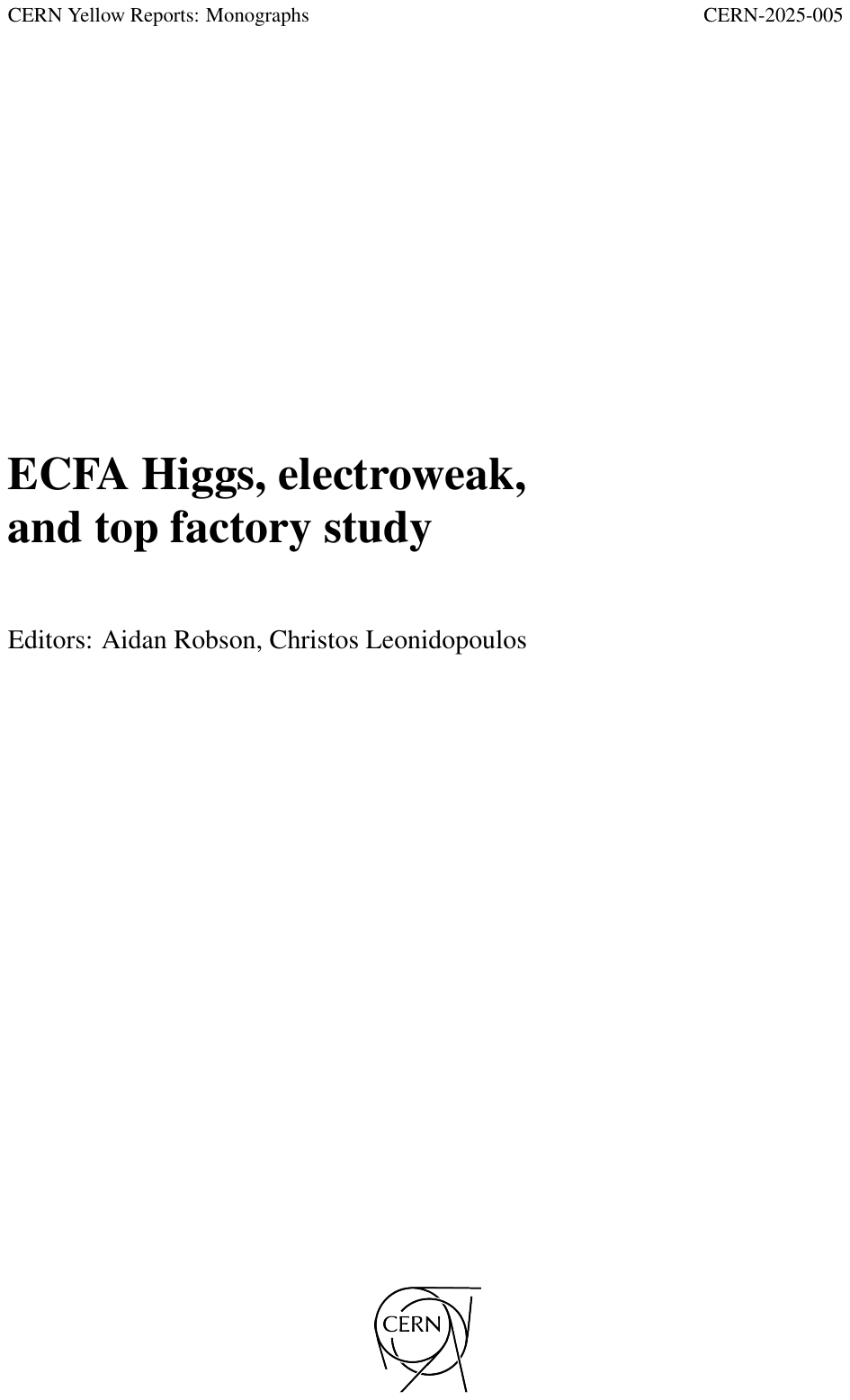}
\includepdf{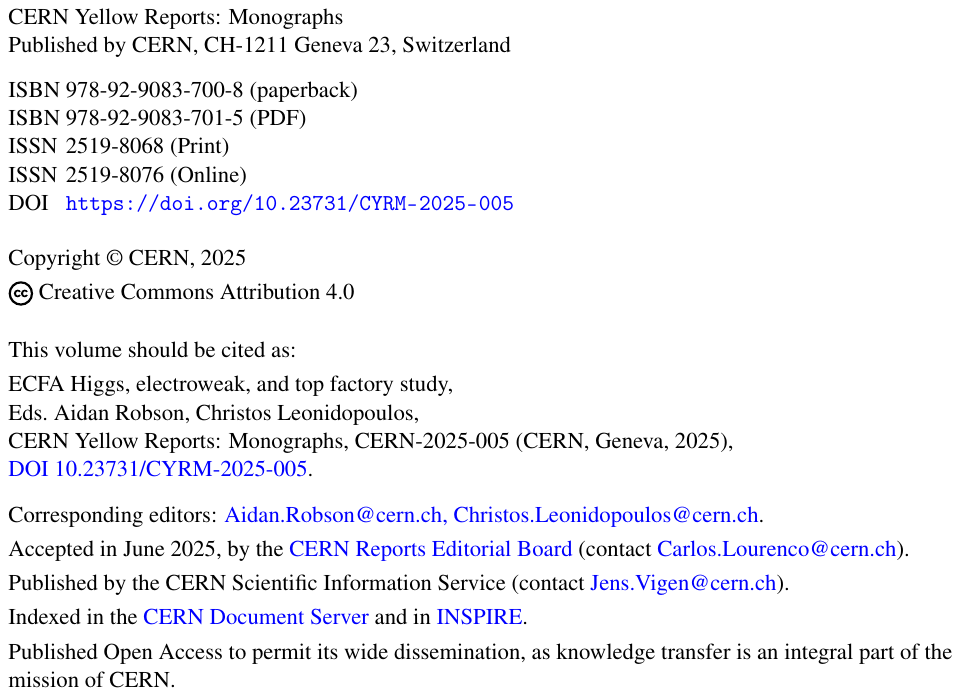}

\titlepage

\tableofcontents

\clearpage

\section{Introduction}
\editors{Aidan, Christos}

\subsection{The ECFA study}

The ECFA Higgs, electroweak, and top Factory Study was set up in response to the recommendation of the European Strategy for Particle Physics Update in 2020 that an electron positron Higgs factory should be the highest-priority next collider.
The initiative recognized the need for the experimental and theoretical communities involved in physics studies, experimental design, and detector technologies towards future Higgs factories to gather, in order to share challenges and expertise, and to explore synergies in the various efforts.
The aim was therefore to bring together the entire \epem Higgs factory effort, fostering cooperation across projects, seeding collaborative research programmes, and increasing the size of the active community.
The work of the study developed through an extensive series of topical meetings, seminars, and mini-workshops organised by working group coordinators and topical conveners; and over the course of three main workshops held at DESY, in Paestum, and in Paris, in October 2022, 2023, and 2024, respectively \cite{ECFAworkshop1,ECFAworkshop2,ECFAworkshop3}.
Furthermore, in order to concentrate efforts and develop critical mass working on areas of particular interest spanning the landscape of the physics potential of a future Higgs/electroweak/top factory, a series of ``Focus Topics'' was defined by expert teams, and the community was invited to contribute in particular to those.

This report provides a snapshot of the work of the study, as input to the next European Strategy Update.

\Cref{sec:detector_r_and_d} discusses the experimental conditions at a future Higgs/electroweak/top factory, the resulting detector requirements and concepts that have been proposed, and progress in the detector development landscape.
Next, \cref{sec:commontools} describes the state of the art and recent progress in the physics analysis tools that form the basis of the studies undertaken, and that are increasingly being developed cross-project.  Particular initiatives coming out of the ECFA study include a common library for luminosity spectra (\cref{sec:lumispectra}), and a new technical benchmarking framework for comparing Monte Carlo generators (\cref{sec:techbench}).  The first Focus Topic, on luminosity measurements, discusses the techniques and precision required for a future Higgs factory (\cref{sec:LUMI}).
This chapter also discusses some of the recent developments in reconstruction algorithms that are used by the physics sensitivity studies that follow.

Higgs physics is addressed directly in \cref{sec:higgs}.
Among the recent work covered, 
a Focus Topic on Higgs-strahlung measurements including angular distributions brings together a number of studies that probe the CP structure of the $\PH\PZ\PZ$ vertex (\cref{sec:ZHang});
recent progress towards sensitivity to the Higgs strange coupling is discussed in a further Focus Topic (\cref{sec:HtoSS}); and
a final Focus Topic brings together important recent developments towards the Higgs self-coupling, both in direct determination via double-Higgs production, and in its extraction from single-Higgs measurements (\cref{sec:Hself}).
Other aspects of electroweak (EW) hysics and quantum chronodynamics (QCD) are discussed in \cref{sec:ewk_qcd}.
Here, dedicated sections discuss the prospects for a $\PW$-boson mass measurement at a future \epem facility (\cref{sec:Wmass}), and the precisions achievable from differential measurements in two-fermion and $\PW\PW$ final states (\cref{sec:TwoF,sec:WWdiff}), as well as what could be learned on fragmentation and hadronisation (\cref{sec:frag-had}).
\Cref{sec:top-quark} covers recent developments on top-quark properties from a threshold scan, top-quark couplings, and exotic top decays; while \cref{sec:globalinterpretations} puts together many of the results from the previous topics into new physics interpretations via global fits.
Direct searches for new physics are covered in \cref{sec:searches},
taking a signature-based approach for Focus Topics on exotic scalars and long-lived particles, and then exploring some particular models.
Finally, aspects of flavour physics are discussed in \cref{sec:flavour}, including prospects for CKM matrix-element measurements and rare decays.

\subsection{Higgs factories \label{sec:runplans}}


Several projects are in an advanced state of development towards realisation as a Higgs/electroweak/top factory.
Each option has different capabilities and strengths, corresponding to the accelerator technology used, and its geometry.
The different accelerator projects have been designed for different centre-of-mass energies.

Among the linear colliders, 
the International Linear Collider (ILC)~\cite{adolphsen2013internationallinearcollidertechnical, barklow2015ilcoperatingscenarios, ILCInternationalDevelopmentTeam:2022izu} targets energies of 250 ($\approx 20$~km long) and 500~GeV ($\approx 30$~km long), with a potential upgrade leading to $\sqrt{s}=1$~TeV ($\approx 50$~km long); while the Compact Linear Collider (CLIC)~\cite{CLICCDR_vol1, Aicheler:2018arh} aims to cover a higher-energy range, with runs foreseen at $\sqrt{s}=380$~GeV, 1.5~TeV, and 3~TeV (with lengths $\approx 11$, 29, and 50\,km, respectively).

The \epem Future Circular Collider (FCC-ee)~\cite{FCC:2018evy} and the Circular Electron Positron Collider (CEPC)~\cite{CEPCStudyGroup:2023quu},
with a circumference of about 90--100~km, are designed for operation between the $\PZ$ resonance pole ($\sqrt{s}=91.2$~GeV) and the $\PQt\PAQt$ threshold ($\sqrt{s} \approx 365$~GeV), with runs at intermediate energy points such as the $\PW\PW$ threshold (160 GeV) and near the peak of the $\PZ\PH$ production cross-section (240 GeV).

Recently, an alternative linear collider project, the Cool Copper Collider (C$^3$)~\cite{Vernieri_2023} has been proposed; it targets energies of 250 and 550~GeV, with copper RF cavities cooled to cryogenic temperatures providing high field gradients, leading to a significantly smaller footprint than the ILC ($\approx 8$~km).  Other proposals such as the Hybrid Asymmetric Linear Higgs Factory (HALHF), using plasma wakefield accelerator technology, are in an earlier stage of development \cite{Foster_2023}.

For reference, the current run-plans for FCC-ee, CLIC, and ILC are given in \cref{table:lumi:FCC,table:lumi:CLIC,table:lumi:ILC}, respectively.  The run-plan for C$^3$ is similar to that for ILC. 
These run-plans and integrated luminosities inform the physics studies and sensitivities that are given later in this report.  In the case of FCC-ee and CLIC, the expected baseline luminosities achievable have recently been updated and so some physics studies quote previously-nominal (lower) integrated luminosities.

\begin{table}[hbt]
\centering{
\begin{tabular}{ l|c|c|c|c|c } 
  \hline
                 &  $\PZ$ pole  &  $\PW\PW$ thresh. & $\PZ\PH$  &  \multicolumn{2}{c}{$\PQt\PAQt$} \\
  $\sqrts$ [GeV] &  88, 91, 94  &  157, 163         & 240       &  340--350  & 365 \\
  \hline
  Luminosity/IP [$10^{34}$\,cm$^{-2}$s$^{-1}$] &
                    140         &  20               & 7.5       &  1.8       & 1.4 \\
  Integrated luminosity/year [$\abinv$] & 68       &  9.6              & 3.6       &  0.83      & 0.67 \\
  Run time [years]   & 4        &  2                & 3         &  1         & 4 \\
  Integrated luminosity [$\abinv$] & 205      &  19.2             & 10.8      &  0.42      & 2.70\\
  \hline
\end{tabular}
}
\caption{The updated baseline FCC-ee operation model with four interaction points (IPs) (December 2024).  The integrated luminosity values correspond to 185 days of physics per year and 75\% operational efficiency, i.e.\ $1.2\times10^7$ seconds per year.  A ramp-up of 50\% in delivered luminosity is assumed over the first two years at the $\PZ$-pole and over the first year at the $\PQt\PAQt$ threshold.  Table from \cite{janot_2024_nfs96-89q08}.
\label{table:lumi:FCC}}
\end{table}

\begin{table}[hbt]
\centering{
\begin{tabular}{ l|c|c|c } 
  \hline
  $\sqrts$ [GeV] &  380  &  1500    & 3000 \\
  \hline
  Repetition frequency [Hz]  & 50 (100)     &  50      & 50   \\
  Luminosity [$10^{34}$\,cm$^{-2}$s$^{-1}$] &
                      2.25 (4.5)            & 3.7   & 5.9  \\
  Run time [years]   & 10                   & 10    & 8  \\
  Integrated luminosity [$\abinv$] & 2.2 (4.3) & 4     & 5 \\
  \hline
\end{tabular}
}
\caption{The updated baseline CLIC operation model (December 2024) \cite{Adli:ESU25RDR}. Two options for 380\,GeV-running are given, with 50 and 100\,Hz repetition rates, respectively.  The integrated luminosity values correspond to 185 days of physics per year and 75\% operational efficiency, i.e.\ $1.2\times10^7$ seconds per year.  A ramp-up of 10\%, 30\%, and 60\% in delivered luminosity is assumed over the first three years at 380\,\GeV.  A ramp-up of 25\% and 75\% is assumed over the first two years at 1.5\,TeV.  Running at $\sqrts=91$\,GeV is an option.
\label{table:lumi:CLIC}}
\end{table}

\begin{table}[hbt]
\centering{
\begin{tabular}{ l|c|c|c } 
  \hline
  $\sqrts$ [GeV] &  250  &  350  & 500 \\
  \hline
  Luminosity [$10^{34}$\,cm$^{-2}$s$^{-1}$] (initial/upgrade) &
                  1.35/2.7  &    &  1.8/3.6 \\
  Run time [years]   & 11   &   0.75  & 9  \\
  Integrated luminosity [$\abinv$] & 2 &  0.2  & 4 \\
  \hline
\end{tabular}
}
\caption{The ILC operation model. The model assumes running for 4.5 years at the initial luminosity to collect 500\,fb$^{-1}$ at $\sqrts=250$\,GeV, upgrading for 1.5 years and then running for a further 5 years with the upgraded luminosity for a total of 2\,\abinv.  The integrated luminosity values correspond to $1.6\times10^7$ seconds per year. A ramp-up of 10\%, 30\%, and 60\% in delivered luminosity is assumed over the first three years of the first phase.  A ramp-up of 25\% and 75\% is assumed over the first two years of subsequent phases. Running at $\sqrts=91$\,GeV, 161\,GeV, and 1\,TeV are options.  C$^3$ follows a similar run plan but with a stage at 550\,GeV instead of 500\,GeV. 
\label{table:lumi:ILC}}
\end{table}

\clearpage
\section{Detector Development}
\label{sec:detector_r_and_d}

Higgs, electroweak, and top factories present a formidable challenge to detectors, in order to match and fully exploit their precision physics potential. 
The developments to tackle this challenge are characterised by long timescales and a need for substantial resources in terms of expertise and infrastructure, and therefore require long-term planning and a stable framework. 

In this section we discuss the experimental conditions future $\epem$ detectors will have to cope with, compiling the most relevant parameters and highlighting differences and commonalities among the various proposed realisations of Higgs/electroweak/top factories. 
We give a short overview of the different detector concepts proposed for large experiments, which have been used in the physics and performance studies presented in the following sections. 
Using a particular example, we show how these concepts need to evolve when porting developments originally targeted at linear colliders to the more demanding conditions at circular machines, where for example the higher rates and tighter alignment requirements present extra challenges. 
We finally present a general picture of the detector developments, briefly summarising the creation and status of implementation of the ECFA Roadmap for Detector R\&D, and highlighting the needs for the near and mid-term future for developing technologies and integrated concepts for the precision physics at the Higgs, top and electroweak scale. 

A more in-depth discussion of the ECFA Detector Roadmap implementation is prepared by the ECFA detector panel. 
Concrete plans for R\&D in the near-term future, proposed to be carried out in the framework of the Detector R\&D (DRD) collaborations and targeted at detectors for Higgs/electroweak/top factories, are presented in the submissions to the European Strategy Process on behalf of the different future collider studies.

\subsection{Experimental conditions}
\label{ssec:detector_r_and_d_condition}
The implementation of the physics programme of future $\epem$ Higgs/Electroweak/Top factories requires accelerators with unprecedentedly high luminosities and detectors with unmatched acceptance and performance in particle reconstruction and identification.
To obtain such performance, a number of alternative accelerator and detector technologies are actively being investigated.


The main beam parameters for linear and circular collider projects are listed in Tables~\ref{tab:linear_collider_params} and~\ref{tab:circular_collider_params}.
Such numbers represent typical values of the different proposals; exact numbers might depend on other parameters such as total power or on further optimisation of the accelerator optics.
\begin{table}[!hbtp]
    \centering
    \begin{tabular}{l|cc|ccc}
    \toprule
    \multirow{2}{*}{Parameter} & \multicolumn{2}{c}{ILC} & \multicolumn{3}{c}{CLIC} \\
              & ILC250  & ILC500 & CLIC380 & CLIC1500 & CLIC3000 \\
         \midrule
         $\sqrt{s}$ [GeV] & 250 & 500 & 380 & 1500 & 3000 \\ 
         Luminosity [$10^{34}/$cm$^2$s] & 1.35 & 1.8 & 2.25 & 3.7 & 5.9 \\
         Train collision frequency [Hz] & 5 & 5 & 50 & 50 & 50 \\
         Bunches/train & 1312 & 1312 & 352 & 312 & 312 \\
         Bunch separation [ns] & 554 & 554 & 0.5 & 0.5 & 0.5 \\
         Train length [$\upmu$s] & 730 & 730 & 0.176 & 0.156 & 0.156 \\
         Beam size at IP $\sigma_x/\sigma_y$ [nm] & 515/7.7 & 474/5.9 & 150/2.9 & 60/1.5 & 40/1 \\
         Crossing angle [mrad] & 14 & 14 & 16.5 & 20 & 20 \\
         \bottomrule
    \end{tabular}
    \caption{Main parameters for linear colliders (typical values).}
    \label{tab:linear_collider_params}
\end{table}
\begin{table}[!hbtp]
    \centering
    \begin{tabular}{l|ccc|ccc}
    \toprule
    \multirow{2}{*}{Parameter} & \multicolumn{3}{c}{FCC-ee} & \multicolumn{3}{c}{CEPC} \\
              & Z  & H & top & Z & H & top \\
         \midrule
         $\sqrt{s}$ [GeV] & 91.2 & 240 & 365 & 91.0 & 240 & 360 \\ 
         Luminosity/IP [$10^{34}/$cm$^2$s] & 182 & 7.3 & 1.33 & 192 & 8.3 & 0.8 \\
         Bunches/beam & 10000 & 248 & 36 & 19918 & 415 & 58 \\
         Bunch separation [ns] & 30 & 1200 & 8400 & 15 & 385 & 2640 \\
         Beam size at IP $\sigma_x/\sigma_y$ [nm] & 8/34 & 14/36 & 39/69 & 6/35 & 15/36 & 39/113 \\
         Crossing angle [mrad] & 30 & 30 & 30 & 33 & 33 & 33 \\
         \bottomrule
    \end{tabular}
    \caption{Main parameters for circular colliders (typical values).}
    \label{tab:circular_collider_params}
\end{table}

An obvious difference between the two classes of accelerators is that linear colliders can only easily accommodate one interaction point (IP), while  circular colliders can accommodate multiple ones (for instance, the baseline for FCC-ee is four IPs). 
A second detector
can still be housed by a linear collider, either (i) using a push-pull scheme to alternate between data-taking phases with one or the other (this scheme is considered for the ILD and SiD detectors at ILC, while it is not included in the CLIC baseline); or (ii) via a second beam-delivery system providing a second interaction point; this latter option has recently been studied for a linear collider facility at CERN (either CLIC or ILC-like).

Another salient difference between the linear and circular machines is the collision frequency: linear accelerators are foreseen to operate with ``trains'' of bunches with a very low duty cycle, with a typical train separation of 20 ms at CLIC and 200 ms at ILC, allowing for power pulsing (i.e.\ turning off between one train and the following) the detector electronics, thus reducing its cooling needs.
Circular machines instead operate with ``continuous'' beams, where the particle bunches are separated in time by amounts as small as 20~ns at the Z-pole, 
making power-pulsing unfeasible.
Both linear and circular colliders require a large crossing angle to avoid parasitic interactions away from the IP; this constraint is more important for circular ones, due to the smaller separation between the bunches, leading to larger crossing angles.

Despite the large instantaneous luminosities of the proposed Higgs factories, the cross sections of signal processes of interest are in general sufficiently small ($<1$ nb for $\sqrt{s} \gtrsim 120$~GeV) to translate into physics rates in the central detectors that are usually easy to manage ($<100$ Hz). However, the resonant enhancement of the cross-section at the $\PZ$ pole leads
to physics rates as high as O(100 kHz) in circular machines running at $\sqrt{s}\approx 91$~GeV, posing constraints on the detectors in terms of response time (to minimise dead time), adoption of zero-suppression techniques (to reduce the data transfer rate), and challenges in implementation of a trigger-less read-out scheme that would in general be preferred in order to minimise systematic uncertainties related to the acceptance of the trigger conditions. 
In addition, the large Bhabha cross section at small angles in circular colliders can lead to O(100 kHz) rates in the very forward regions occupied by the luminosity calorimeters for the instantaneous luminosity measurement, which creates a small but not irrelevant pile-up in $\PZ$ pole running.

All projects would reach high instantaneous luminosities by significantly squeezing the beams colliding at the IP.
As a consequence, particles from one colliding bunch generate intense electromagnetic fields that bend the trajectories of the particles in the other colliding bunch. This leads to radiation off the incoming particles, which can produce backgrounds in the detectors in various forms:
\begin{itemize}
    \item beamstrahlung photons: real photons emitted by one of the incoming particles;
    \item incoherent $\epem$ pairs from the $\PGg\PGg \to \epem$ process initiated by real or virtual photons from two colliding particles. Unlike coherent pair production, where the outgoing particles are produced along the outgoing beam direction and remain in the beampipe, particles from incoherent pair production (IPC) have a flatter polar angle spectrum and can reach the detector, though their energy is typically small;
    \item hadron jets from the high tail of the momentum distribution of $\PGg\PGg \to \PQq\PAQq$ events. They are mostly produced in the central region of the detector, with low momentum.
\end{itemize}
In addition, for circular colliders, bending of the charged beam particles by the dipole magnets near the IP leads to the emission of photons via synchrotron radiation.

Such backgrounds are typically mitigated through careful design of the machine-detector interface region (MDI) including proper shielding, but residual backgrounds have to be properly estimated through detailed simulations and detectors and electronics capable of operating in such an environment have to be designed. 
Beamstrahlung photons do not constitute a major concern for the detectors: their main effect is to increase the size of the IP and the centre-of-mass energy spread, without consequences on the detector performance.

On the other hand, incoherent electron-positron pairs and hadrons can lead to large occupancy and energy deposited in the detectors, calling for excellent space granularity to reduce detector occupancy and potentially also good time resolution for use of timing in offline reconstruction. 
The very large $\Pepm$ rate at low transverse momenta from incoherent pair production impacts the design of the inner and forward regions. As particles with low momenta spiral along helical trajectories around the beam-line under the magnetic field of the detector's solenoid, with smaller radii for lower momentum, lower energy circular colliders can generally afford beam-pipes with smaller radii, down to $\approx 1$~cm, such that the innermost layer of the barrel section of a vertex detector could be located as close as $\approx 1.2$~cm to the beam-line, if rates can be tolerated.

Synchrotron radiation (SR) can be effectively mitigated by proper design of the beam optics around the IP (to limit the critical energy of these photons to 100 keV or lower) and shielding (to absorb them). For instance, at FCC-ee the use of SR mask tips upstream of the interaction region, a beam pipe with copper and high-$Z$ (e.g.\ W) shields outside the central region, and a \SI{5}{\micron} gold coating in the central region absorbs most of the photons. The impact of the remaining ones on the occupancy in the tracking detectors is expected to be small for an all-silicon tracker (typically O(0.1\%), at most $\approx 1\%$ in the end-caps), but can vary significantly in a gaseous-based main tracker, depending on the signal integration time and use of timing information to suppress hits from these photons.

Non-negligible synchrotron radiation can also be emitted by colliding particles moving in the detector solenoidal field along directions that are rotated with respect to the field by large crossing angles. This limits the maximum magnetic field of the detector solenoid, to minimise the impact on the bunch size at the IP and on the luminosity; the effect is larger for circular machines with larger crossing angles, typically limiting the maximum field to 2--3~T.

The different colliding schemes and beam parameters lead to further constraints on the interaction regions and MDI.
For instance, the very small beam size at the interaction point needed to reach the desired instantaneous luminosities in circular colliders requires a very compact MDI, with 
a small distance $L^*$ between the IP and the first focusing quadrupole, which has then to be placed within the detector itself. This also has consequences on the position of the luminosity calorimeter and the tracker acceptance.

Some of the main typical parameters foreseen for the different proposals as used as working hypotheses to assess backgrounds levels  in the detectors and determine their geometrical envelopes are summarised in \cref{tab:interaction_regions}.
\begin{table}[!hbtp]
    \centering
    \begin{tabular}{l|cccc}
    \toprule
    Parameter & FCC-ee & CEPC & ILC & CLIC \\
         \midrule
         $L^*$ [m] & 2.2 & 2.2 & 4.1 & 6 \\ 
Position of final quadrupole wrt detector & inside & inside & outside & outside \\
\multirow{2}{*}{Luminosity calorimeter position} & $z\approx 1$~m, & $z\approx 1$~m, & $z\approx 2.5$~m, & $z\approx 2.5$~m, \\
& 50-100 mrad & 26-105 mrad & 33-80 mrad & 39-134 mrad \\

         Crossing angle [mrad] & 30 & 33 & 14 & 20 \\
         Main solenoid $B$ field [T] & 2 & 3 (2 at Z-pole) & 3.5-5 & 4 \\
         \bottomrule
    \end{tabular}
    \caption{Main parameters for the interaction regions of the linear and circular $\epem$ colliders (typical values).}
    \label{tab:interaction_regions}
\end{table}

Detailed studies of radiation levels have been performed for ILC and CLIC detectors in the past with \texttt{Fluka}. Preliminary estimates, based on the same tools, of expected radiation levels in FCC-ee detectors have been obtained recently.
The main conclusions from these studies for the expected fluence (from non-ionising energy loss, NIEL) and total ionising dose (TID) in the central detectors are the following:
\begin{itemize}
\item ILC: expected NIEL fluence $\leq 10^{11}$~n$_\mathrm{eq}/$cm$^2/$yr, TID~$\leq$1~kGy/yr.
\item CLIC: the highest radiation levels are for $\sqrt{s}=3$~TeV, where one expects a fluence $\leq 2 \times 10^{11}$~n$_\mathrm{eq}/$cm$^2/$yr, and TID~$\leq$ 300~Gy/yr. At lower energies, the lower instantaneous luminosity and energy of the particles lead to reduced fluence and ionising dose delivered to the detector.
\item FCC-ee: the highest radiation levels are at the Z pole ($\sqrt{s}=91.2$~GeV) due to the much larger instantaneous luminosity than for all other centre-of-mass energies. Simulations performed for one of the detector concepts proposed for FCC-ee (IDEA) lead to expected fluence $\leq 2 \times 10^{13}$~n$_\mathrm{eq}/$cm$^2/$yr and TID~$\leq$ few tens of kGy/yr for the Z-pole run.
\end{itemize}
In general these numbers -- even including conservative safety margins -- are a few orders of magnitude below the levels expected in the corresponding regions of the ATLAS and CMS detectors at the high-luminosity phase of the LHC.

\subsection{Detector concepts}
\label{ssec:detector_r_and_d_concepts}

To match the physics goals of the Higgs factories and cope with the constraints imposed by the accelerators, a number of detector concepts are under development (see also 
\cref{ssec:detector_models} for a discussion on the simulation of these detector models).
They mainly differ in the conceptual approach to electromagnetic and hadronic calorimetry, and consequently in calorimeter technology choices.
This leads to different concepts for the overall architecture, e.g.\ where to place the solenoid and how to design the barrel end-cap transition region. 
Different technologies are also proposed for charged particle reconstruction in the main tracking volume, with choices including silicon-based tracking systems and gaseous technologies, e.g. drift chambers, time projection chambers, and straw tubes. 

Two detector concepts have been proposed for ILC: ILD and SiD;
a single detector concept, CLICdet, is being studied for CLIC;
three detector concepts have been proposed for the FCC-ee: CLD (evolved from CLICdet), IDEA, and ALLEGRO. The ILD community is also studying the modifications to the original ILD concept needed in order to use it as a detector for FCC-ee.

The various detector concepts are designed for energy regimes that are not always the same, but they target performance numbers that are broadly similar:
\begin{itemize}
\item a few microns of impact parameter resolution, requiring vertex detectors with single-hit resolution of a few microns, as close as possible ($\lesssim 2$~cm) to the interaction point, with small material budget (order of $0.1\%\; X_0$ per layer);
\item track momentum resolution $\sigma_p/p$ below few per-mille for tracks with momenta below 100 GeV;
\item jet energy resolution around 3--4\% or better for typical jet energies  $30<E_\mathrm{jet}<100$~GeV for proper reconstruction of hadronic decays of $\PW$, $\PZ$ and Higgs bosons. 
\end{itemize}

Essentially, all concepts rely on thin, monolithic pixel sensors for the vertex detector, with small pitch for high spatial resolution and small thickness to limit the effect of multiple scattering.

Charged particle reconstruction in the main volume could be obtained with either a small number of high-resolution space points reconstructed by a silicon detector, such as in SiD, CLICdet or CLD, or by a large number of lower-resolution space points reconstructed by a gaseous device, such as a TPC (ILD), 
a drift chamber (ALLEGRO, IDEA),
or a straw tube tracker. 
A silicon ``wrapping'' layer outside the gaseous devices could provide a high-resolution point for augmenting the precision of momentum measurement and extrapolation of the tracks to the calorimeter, as well as timing information --- potentially with a few tens of ps resolution --- for time-of-flight measurement, without compromising material budget and granularity for the inner detector.

Both ILC detectors (ILD, SiD), as well as CLICdet and CLD employ highly granular ``sandwich'' calorimeters designed for particle-flow reconstruction, located inside the detector's solenoid. Electromagnetic particles are reconstructed in a silicon-tungsten or scintillator-tungsten calorimeter, while hadrons are absorbed and measured by scintillator-steel or RPC-steel calorimeters.
IDEA uses a dual-readout technique to measure the position and energy of electromagnetic and hadronic showers. A crystal calorimeter devised for dual read-out is located inside the solenoid, and a dual-readout fibre calorimeter, outside the solenoid, has steel or copper as absorber and fibres, also detecting both scintillation and Cherenkov signals.
ALLEGRO uses a highly granular noble-liquid (Ar, Kr) sampling calorimeter with Pb or W passive elements for the electromagnetic
section, and scintillating tiles alternated to steel absorbers for the hadronic compartment. The coil is located between the ECAL and the HCAL.

These three different types of calorimeters can in principle achieve jet energy resolutions of 3--4\% for $30<E_\mathrm{jet}<100$~GeV and electromagnetic energy resolution of $8-15\%/\sqrt{E}$
(or better if crystals are used) by either leveraging particle-flow algorithms to determine the type of particles hitting the calorimeters and using either information from the tracker, EM calorimeter or hadron calorimeter for charged particles, photons, and neutral hadrons respectively; or by using a dual readout approach to simultaneously measure the scintillation and Cherenkov components of hadronic showers, to determine event-by-event the electromagnetic fraction and use it to correct the energy response.

Muons traversing the whole detector are tagged by the outermost component of the detectors. 
Most of the detectors rely on instrumenting the iron return yoke of the detector outside of the coil, with either scintillator strips and SiPMs (ILD, SiD), RPCs (ILD, SiD), or $\mu$RWELLs (IDEA).
For ALLEGRO, several options are still under consideration ranging from drift tubes or chambers to resistive plate chambers or micromegas (MM). The muon system will act as a muon tagger but will not provide a full stand-alone muon momentum measurement.

Charged particle identification could be provided by complementary information such as ionisation loss (\dedx) or primary ionisation cluster-counting (\dndx) in gaseous devices, or with time-of-flight measurements using silicon timing layers 
or the front section of the calorimeter.
Dedicated Cherenkov detectors based on ring-imaging Cherenkov light detection are proposed for CLD. 

Significant effort has been put in the past years on 
(i) implementation of the detector geometry in a fully detailed simulation of the detector response based on Geant4; (ii) development of reconstruction algorithms; (iii) optimisation of the detector layout.
A large part of these activities has been shared between the detector concepts for circular machines, for example in terms of simulations for sub-detectors considered by several concepts, 
and a lot of tools and knowledge has been transferred from the linear collider community which has been undertaking these studies for a long time. 
An instructive example is described in the following section.

Important and rapid progress has been made on (i) and (ii), largely due also to the availability of a generic framework for detector geometry description and simulation (\keyhep, described in \cref{sec:softwareeco}) and to various algorithms initially inspired and/or developed by the linear collider communities. 
All detector concepts have now a full detector model based on DD4hep implemented and a working Geant4 simulation (see \cref{sec:simulation}), while further work is required in some cases on the full event reconstruction. Studies are ongoing on (iii) detector layout and its impact on e.g.\ background rates, as well as --- with simplified parametric simulations of expected performance --- on the physics potential.
Now that detailed full simulations become available for all projects, more realistic physics studies and detector optimisations can and should be done, using performance figures of merit to characterise the detector performance for full event quantities such as flavour tagging or heavy boson di-jet mass resolution. 

\subsection{\texorpdfstring{Evolution of detector concepts from linear to circular $\epem$ colliders}{Evolution of detector concepts from linear to circular e+e- colliders}}
\label{ssec:detector_r_and_d_evolution_of_concepts}

The design of a detector is driven not only by the precision required to meet the physics objectives but also by the specific characteristics and operational conditions of the accelerator where it will operate. There are significant differences among the proposed $\epem$ colliders, arising not only from their structural designs --- whether circular or linear --- but also from the different energy ranges that they are intended to cover. 

The first detector concepts proposed were those designed for the ILC: SiD and ILD. Both designs are optimized for Particle Flow Algorithms (PFA) with highly granular calorimeters. The main conceptual difference lies in the tracker, SiD employs only silicon detectors, while ILD incorporates a Time Projection Chamber (TPC) combined with a silicon layer. They served as inspiration for some of the new detector concepts based also on PFA for CLIC, CEPC and FCC-ee, tailored to the specific characteristics of each accelerator. 

Initially, two detector models were defined by adapting the ILC detectors for the higher centre-of-mass energies at CLIC which finally led to the present CLICdet concept. 
In addition to the broader energy range, the CLIC machine environment differs significantly from that of the ILC.  

The small beam sizes and the high energy at CLIC will result in a high background, which will be integrated over multiple bunch crossings due to the beam structure, with bunches separated by 0.5 ns. This restricts the minimum vertex radius to $\sim30$ mm, in comparison with the 16 mm used at the ILD. The background conditions are not well suited for the use of a TPC at 3 TeV leading CLICdet to exclude this option from its design. The CLIC tracking detectors also require readout systems with precise time-stamping capabilities and multi-hit readout functionality for some of the inner strip layers. Compared to SiD, CLICdet incorporates a significantly larger tracking system, particularly with an extended forward region acceptance to better deal with the increase of events with forward boost. The total energy deposited on the calorimeters is very large, imposing strict requirements on their timing resolution to separate the background from the signal of the relevant physics process. Furthermore, the occupancies are too high at the inner part of the end-cap. Solutions such as improved shielding, using alternative active material less sensitive to neutrons or increasing transverse segmentation are envisaged. 
The jet energy range at CLIC extends up to 1 TeV, which complicates the correct assignment of energy deposits among the different particles within the jet. This makes the fine segmentation even more important than for ILC detectors. Furthermore, a deeper calorimeter system is required ($\sim8.5\;\lambda_{I}$ in total) to reduce leakage.

\begin{figure}
    \centering
    \includegraphics[width=0.5\linewidth,trim={0 0.4cm 0 0},clip]{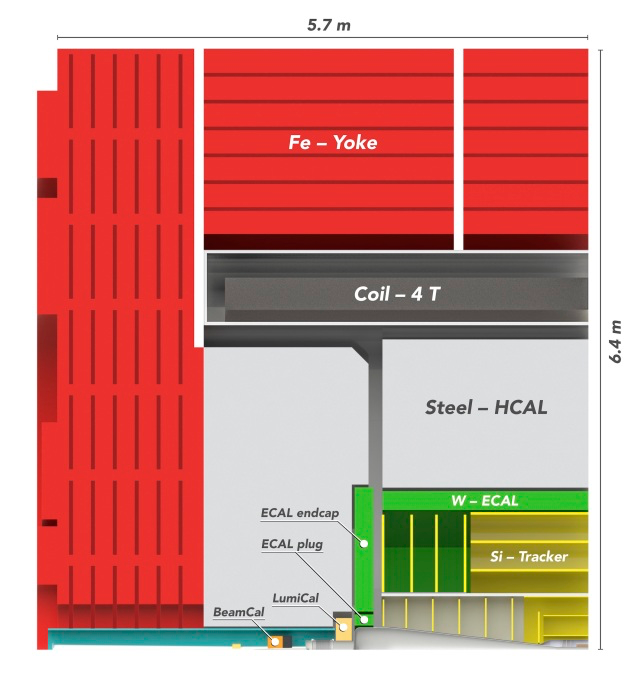}
    \includegraphics[width=0.46\linewidth]{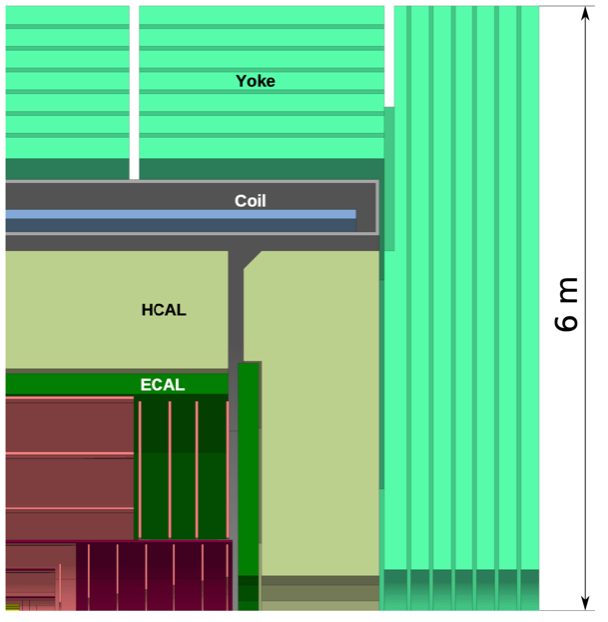}
    \caption{A quadrant of (left) the CLIC detector, and (right) CLD.}
    \label{fig:CLICvsCLD}
\end{figure}

The operating conditions of circular machines differ significantly from those of the linear colliders, and particularly due to operation at the Z-pole, introducing additional challenges for the detectors.  
Detectors must address the complexities of a highly constrained inner region dictated by the MDI. The crossing angle for FCC-ee is 30 mrad, a factor 2 larger with respect to ILC. The last focusing quadrupole is located at just 2.2\,m from the IP, within the detector volume, and the Beamcal is positioned $\sim 1$\,m away, limiting the detector angular acceptance to 100 mrad. Under these conditions the magnetic field of the detector is limited to a maximum of 2\,T, to preserve the beam emittance (at Z pole, a field of 3\,T would reduce the luminosity by a factor 2). To compensate for the lower magnetic field the tracking radius must be increased to preserve the resolution. At the CLD, an adaptation of the CLICdet for the FCC-ee, the tracking region has been extended from 1.5 to 2.5\,m.  Due to the lower centre of mass energy of the FCC-ee, limited to 365 GeV for the top studies, the depth of the CLD hadronic calorimeter has been reduced with respect to the CLICdet, from $7.5$ to $5.5\; \lambda_I$, similar to the one used by the ILD. A quadrant of CLICdet and CLD are shown together in \cref{fig:CLICvsCLD}.

Linear colliders use bunch trains separated by tens or hundreds of milliseconds, allowing the electronics to operate in power pulsing mode. This reduces significantly the power consumption and, consequently, the cooling requirements. In contrast, circular colliders operate with continuous beams of bunches, making power pulsing operation unfeasible. 
With continuous operation, higher data rates need to be processed and more bandwidth is needed for the transfer from the front end, which increases the power consumption further, beyond the power-cycling factor. 
This requires a re-optimization of electronic power consumption and cooling system with a minimized material budget to limit the impact on the detector performance, in both tracking and calorimeter measurements. 

The continuous beam in circular colliders has also implications on the readout scheme, particularly at the Z-pole, where event rates are expected to reach 100 kHz. This rate is comparable to the current maximum rate at Large Hadron Collider (LHC) experiments after the L1 trigger selection, which is further reduced by two orders of magnitude by the High Level Trigger (HLT). For the High-Luminosity LHC (HL-LHC) operation, the final rate will be below 10 kHz. While the DAQ operation is not compromised at the Z-pole, the throughput to disk will probably need to be reduced. To exploit the extremely high luminosities providing high statistical precision, the systematics must be controlled at the 10$^{-5}$ level, which is very challenging for a trigger system. The total data volume, which depends on the detector design, the detectors used and the total number of channels, needs to be evaluated. The detector information being transferred must be optimized, the implementation of real-time algorithms in FPGAs, capable of filtering noise and background hits should be evaluated. 
The feasibility of adopting a trigger-less approach, as employed by LHCb, where software algorithms handle event selection using the full detector readout, should also be investigated.

\subsection{Development of detector technologies}
\label{ssec:detector_r_and_d_DRDs}

The experimental conditions and conceptual considerations presented above show clearly that detectors, which match the precision physics requirements of Higgs/electroweak/top factories, require technologies beyond the present state of the art.
While radiation tolerance, a major theme for the development of the detectors at LHC and HL-LHC, is in general not an issue, performance goals are far more ambitious in terms of momentum and impact parameter resolution for charged particles, energy resolution  for photons and jets, as well as particle identification and luminosity measurement.
Whereas the case has been made that the performance  goals are becoming achievable for experiments at linear colliders, the event rates at circular machines, being much higher and closer to those at the LHC, re-open the challenge, and the demands for highest systematic precision posed by the extremely large datasets at the Z-pole even accentuate it. 
The detector performance assumptions used in the physics studies are realistic, informed extrapolations, but they will only be achieved through a vigorous and targeted R\&D program.

In the course of preparing for the previous update of the European Strategy for Particle Physics, the community had taken stock of the lessons learned from the development of detectors for the LHC, and recognised the long timelines involved and the key importance of a targeted and sustained R\&D program. 
As a consequence, ECFA was mandated 
to set up a process to establish a roadmap for the development of the detectors needed for the future facilities prioritised by the European Strategy.
Following that recommendation, the detector community has undertaken a tremendous effort to establish the needs for strategic R\&D, identify key R\&D themes and to organise the activities on a global level in newly founded DRD collaborations. 

Extensive consultations with the community at large were a core element of the roadmap process. 
In a first round requirements from the various future facilities were collected. 
In a second round, organised in task forces grouped by detector technologies, world-leading experts presented in open symposia~\cite{link-symposia} their visions on detector development for a period covering the following 20--30 years.
The outcome was presented in the ECFA Detector Roadmap document~\cite{Detector:2784893}, which captures the main directions for each technology area in the form of 3 to 5 detector R\&D themes, exposing their relevance for the proposed future facilities and highlighting synergies between concurrent developments as well as between consecutive ones, the earlier serving as stepping stones for the latter. 
Within the facilities driving the roadmap, Higgs/electroweak/top factories are one category among others, but a particularly important one given the high priority of a future $\epem$ collider in the present European Strategy. 
This roadmap continues to serve as a valid guidance for detector R\&D; an incremental update is being proposed only after the current European Strategy Process is concluded and a future flagship project for CERN is identified. 

As a consequence of the very positive reception of the Roadmap by the CERN Council, ECFA was asked to proceed towards a plan for implementation, taking into account the positive experience collected in the past with collaborative detector R\&D. 
This plan~\cite{implementation-plan}, approved in 2022, resulted in the creation of CERN-anchored world-wide detector R\&D (DRD) collaborations, organised along the same technology topics as the Roadmap. 
Proposal preparation teams (PPTs) had been formed which emerged from the roadmap task forces but were considerably extended to ensure broad topical and also geographical representation; in particular, non-European membership was enhanced. 
The PPTs arranged for another round of broad community consultation, with open meetings and calls for input proposals. From this input the proposals were synthesised
and then submitted to the new Detector R\&D committee (DRDC) hosted by CERN and assisted by the ECFA Detector Panel. 
The DRD proposals focus on the next three-year period, with an outlook also given on the subsequent three years, and present concrete projects to be undertaken with available or realistically expected resources. 
In total, eight DRD proposals have meanwhile been successfully reviewed and the collaborations been approved. 
The collaborations have set up their internal structure and have started the MoU process. 

The ECFA Detector Roadmap, with its long-term vision, and the published DRD proposals~\cite{DRD-proposals}, with concrete near-term plans, are essential and integral parts of the detector community's input to the present European Strategy process. In both cases this encompasses the full spectrum of future facilities, but the requirements of Higgs/electroweak/top factories are strong --- if not the strongest --- drivers in the relevant DRDs. 
These are DRD~1 (Gas detectors), DRD~3 (Solid state detectors), DRD~4 (Photodetectors and particle identification) and DRD~6 (Calorimetry), plus the ``horizontal'' collaborations  DRD~7 (Electronics and DAQ) and DRD~8 (Mechanics and cooling). 
The other two, DRD~2 (Liquid detectors for rare event searches and neutrino physics)
and DRD~5 (Quantum and emerging technologies) are mainly addressing the needs of non-collider experiments. 

As explained above, the requirements at circular $\epem$ colliders are the most demanding when compared with those at the ILC or even CLIC. 
Therefore, the  on-going and proposed near-future Higgs/electroweak/top factory-targeted detector R\&D is addressing the challenges posed by FCC-ee (and thus implicitly covering most of the linear collider needs).
At the 3rd workshop of the ECFA Higgs/Electroweak/Top Factory Study, all parallel detector session presentations (except one on simulation) focussed on FCC-ee related
developments. 
The detailed community aspirations for Higgs/electroweak/top factory-motivated R\&D will thus be presented to the European Strategy Group via the input submitted by the FCC study. 
A call for Expressions of Interest to work on FCC sub-detectors has received strong resonance, and the  compiled response will be submitted in a combined document. 
Here we only list the main directions.  

The development of silicon vertex detectors for FCC-ee is strongly inspired by the developments for the ALICE detector upgrade, which represents the forefront in ultra-light thinned and self-supporting silicon sensors and serves as a stepping stone for the Higgs/electroweak/top factory detectors. 
The cooling and support structures are minimized. The silicon is almost the only component in the active area, the sensors are curved forming cylindrical detector layers held in place with carbon foam, without further mechanical support and circuits located at the two edges. Air flow is used as cooling.
Simultaneously achieving required performance goals for FCC-ee in terms of space and time resolution as well as material budget comes into reach only now. 
Realising a trigger-less read-out with realistic beam-related background-induced occupancies needs further advances in microelectronics. 

For silicon-based tracking systems, the light-weight large-scale support structures and services in a $4\pi$ geometry are the main challenge, since the momenta to be measured, from exclusive heavy-meson final states to the Higgs recoil are entirely dominated by multiple-scattering, and innovative solutions are needed to achieve a transparency comparable to that of gas detectors. 
Large area silicon systems are also developed as envelopes for gaseous main trackers, to add precision space points to augment momentum resolution, or time-of-flight functionality for particle identification.
To limit multiple scattering, the material budget should be kept to a minimum; for all-silicon detectors this would likely require thin sensors, monolithic and air-cooled. 
Timing information, if available, could provide an additional handle for improved performance such as for bunch identification and background rejection (for O(ns) resolution) or particle-identification with time-of-flight (for O(tens of ps) resolution); however, excellent timing resolutions would likely increase the power consumption of the front-end electronics, thus making air-flow cooling more challenging.

Gaseous technologies under active development for the main tracker are drift chambers, time projection chambers (TPCs) and straw tubes. 
In addition to lowest-possible material budgets, they offer particle identification capabilities via specific energy loss or ionisation cluster counting. 
The operability of a TPC under background conditions expected for the FCC-ee is still a subject of study. 
A particular challenge for all gas detectors is the concentration of read-out electronics (or cable density) at the end plates, with high data transfer rates, in particular in the case of wave-form sampling for cluster-counting techniques, and corresponding powering and cooling demands being at variance with the need to minimise the material budget in front of high-resolution electromagnetic calorimeters. 

Pion-kaon separation via the specific energy loss is particularly difficult in the relativistic rise region, where the relative difference between pions and kaons is only O(10\%), too small for efficient separation with conventional methods, despite considerable optimisation efforts in the past decades. 
Cluster counting in the time or space domain
has the potential to provide two times better resolution ($\approx 2\%$) and thus better pion/kaon separation and reduced sensitivity to gain fluctuations. 
As an example, to complement silicon-based main tracking detectors, a dedicated system for particle identification based on novel ring-imaging Cherenkov detectors has been conceived. 

All proposed calorimeter technologies aim at 3-dimensional segmentation for topological shower separation, albeit with different emphasis on energy resolution vs.\  imaging performance. 
Finest segmentation is achieved with the technologies using read-out electronics embedded in the active layers, as they have been developed by the CALICE collaboration, originally for application at linear colliders. 
They therefore rely on power-pulsing and have so far not foreseen active cooling. 
The silicon and SiPM-on-Tile technology have also been adopted for the CMS high-granularity end-cap calorimeter upgrade (HGCAL), where they are now being applied at large scale for the first time. Cooling and high-rate read-out solutions have been developed, but with comparatively relaxed requirements on performance and compactness.
Future development must tackle the challenges of FCC-ee whilst maintaining the demonstrated performance  for linear collider conditions.
The other technologies --- crystal, fibre and liquefied noble gas based --- have only more recently stepped up in 3D granularity. Their electronics is mainly located outside the active volume; here the main challenges are in demonstrating the performance of finely segmented prototypes and addressing the questions of scalability to a large $4\pi$ system. 
An intermediate situation is given in the case of a liquefied noble gas calorimeter with the very front-end electronics placed inside the cold volume, capitalising on developments for large-scale cryogenic neutrino detectors.   


Muon detectors will be located at several metres from the interaction point, thus covering a large surface. One of the main challenges is thus to simplify their design and reduce their cost.
An additional point of consideration, which also applies to gaseous detector for charged particle reconstruction in the main tracking volume, is the development of eco-friendly gas mixtures to replace those with potential large greenhouse effects in use in current experiments.

All sub-detector developments are also connected to, and will benefit from ``horizontal'' activities covering multiple fronts, like the one mentioned in the context of DRD 7 and 8. 
Next-generation micro-electronics, as well as advanced light-weight materials and cooling techniques, are relevant for almost all developments mentioned here. 

In addition, the advent of new technologies opens up new perspectives for detectors that are mandatory to be explored. 
One example is fast timing electronics and sensors, opening exciting possibilities not conceivable in the past.  
However, for each sub-system and technology, the balance between performance augmentation and power budget increase must be found.
This is a show-case for underlining the importance of maintaining a strong relation between technological R\&D and studies at the detector concept level. 
Only in the context of a full detector can one strike the right balance between performance gains and implications of system integration.
It is to be emphasised that also here a unified software framework as provided by \keyhep is pivotal not only in establishing coherence and synergy between detector concepts, but also for ensuring an efficient transfer of knowledge generated at the technology frontier, for example of the understanding generated in the analysis of beam test results and implemented in  prototype simulations, to the full detector world. 

Another ``horizontal'' topic is machine learning (ML) and artificial intelligence (AI). 
While particle physics instrumentation is not driving these fields, the potential for applications is ubiquitous, ranging from deep learning based reconstruction methods to generative networks trained for shower simulation, from neural networks implemented in embedded on-detector electronics for data reduction to AI-assisted detector optimisation. 
Also here, maintaining the link between the technology frontier and the ultimate physics performance assigns an essential role to the detector concepts, which are seen at this stage  more as an engineering-informed software model than as a community-building framework which they will no-doubt become in a future phase.

\subsection{Concluding remarks}
\label{ssec:detector_r_and_d_remarks}

The ECFA Detector Roadmap does not only condense the way forward in terms of detector R\&D themes forming the substance of the now founded DRD collaborations, 
but also contains General Strategic Recommendations (GSR) that emphasise the importance of the soil from which this substance grows. 
It should be recalled here how central and indispensable infrastructure such as --- at rank~1 --- test beam facilities are for earthing component R\&D and system simulations with prototypes under realistic conditions. 
Irradiation facilities are indispensable, too, for further advances in our fundamental understanding of radiation damage, without which no bridge to future hadron facilities can be built following FCC-ee.  

Industrial relations have to be strengthened if two or even (as in the FCC-ee baseline) four large detector systems, each pushing technologies beyond the state of the art, are to become a reality. 
A recent example is given by the decline and gradual resurrection of expertise and industrial capabilities to produce the superconductors needed for the experimental solenoids. 
This has to stay in focus and must receive further support, with CERN playing a central role. 

While the creation of the detector roadmap was received with applause 
and its implementation progressed, despite some hiccups on the final stretch, beyond what one originally could have hoped for, 
its implementation continues to need full and unanimous support.
The establishment of DRD collaborations needs to be consolidated, 
since their funding situation is still in a precarious state. 
It is to be recalled that when the roadmap implementation plan was first presented, the main concern expressed by the particle physics community was that the effort could grow prematurely and too fast, in view of the remaining challenges of the ATLAS and CMS upgrades and the emerging opportunities appearing at ALICE and LHCb; not to speak of the rapidly-growing portfolio of ideas directed beyond colliders. 
In order to avoid too much strain on primarily technical human resources, a slow and gradual ramp-up was planned. 
With the recent and unexpected further delays of detector preparations for the HL-LHC, the slow ramp-up has had to be slowed down even further. 
Given the significant investment by the community, in terms of time, effort and intellectual energy, the opportunity for turning this enthusiasm into momentum towards the future must not be missed. 

The detector field is at a critical turning point, facing a challenging transition from a strong focus on the tremendous efforts  for the HL-LHC upgrades, confronting budget-wise, human resources and schedule constraints, towards the open, strongly but differently demanding field of $\epem$ detectors.
This turning point is appearing at different times and speeds for different kinds of expertise; while sensor developers are already seeking and finding new challenges within or outside the field, and ASIC designers will soon leave the HL-LHC stage, system integration and commissioning experts will remain bound for some time to come. 
The clearer the directions from the European Strategy Update under preparation, and the more swiftly they are implemented and reflected in the allocation of resources, the more fruitful and productive this transition will be --- and the more attractive to the bright young researchers planning their careers in particle physics.

\clearpage
\section{Common Developments\label{sec:commontools}}
\editors{Fulvio Piccinini, Patrizia Azzi, Dirk Zerwas}

Physics Analysis Tools cover a broad range of topics ranging from the underlying
software ecosystem through the Monte Carlo generators to the simulation and reconstruction
of physics events. In the last four years the state of the art and new developments were
discussed across the different proposed electron-positron Higgs factories in the ECFA Study workshops as well as in 
five Topical Meetings separately covering Generators~\cite{MCTopical1,MCTopical2}, 
Simulation~\cite{SimTopical1} and Reconstruction~\cite{RecoTopical1,RecoTopical2}, and in 
several Focus Meetings dedicated to luminosity spectra~\cite{Focus1} and Monte Carlo Technical Benchmarks~\cite{Focus2,Focus3}. 

The software ecosystem, \keyhep, contains everything necessary to simulate and reconstruct events. It has 
detector descriptions and the algorithms developed over the years in a common framework and a common Event Data 
Model. \keyhep is supported by and supports all Higgs factory options. It also provides interfaces to many 
generators. \keyhep is described in \cref{sec:softwareeco}.

Monte Carlo generators come in different variants. Some are specialized, i.e.\ are dedicated to a limited set of processes, but at
high precision. Another category includes general multi-purpose Monte Carlo generators providing a broad range of Standard Model
and Beyond the Standard Model processes. The development over the past years has been very rich. Monte Carlo generators have accompanied
the Higgs factories for many years. Generators from the Large Electron Positron (LEP) collider era have been updated. Modern generators developed for the LHC have
extended their capabilities to electron-positron collisions. The current state of the art is described in \cref{sec:gen}.

Beam-beam interactions are specific to each accelerator. A consistent set of the resulting luminosity 
spectra for FCC, CEPC, ILC, CLIC and C$^3$ has been calculated. The results are summarized in \cref{sec:lumispectra}.

The high statistics expected at Higgs factories enable high precision cross section measurements. These
depend crucially on detectors suitable for precision measurements and standard candles, e.g. Bhabha scattering and two photon production, precisely 
predicted/calculated to transform the 
counting rates into integrated luminosity. For this reason an ECFA ``Focus topic'' on luminosity 
was defined. The findings of the group is summarized in \cref{sec:LUMI}.

The multiple generators described in \cref{sec:gen} provide generated events at the same level of precision using 
different techniques. Comparisons are necessary to check their agreement and accompany future
developments. The activity on developing a framework in \keyhep and its results is discussed in \cref{sec:techbench}.

Two aspects of the simulation of detectors are described in \cref{sec:simulation}. The main packages used for fast and full
simulation are described. Over the past years a tremendous effort has been made on implementing detectors in \keyhep (\ddhep).
Thanks to the common framework, full detector concepts are developed from commonly defined sub-detectors. As these contain
detailed implementations of the detectors, including inactive zones, support structures etc, this lays the ground-work
for realistic and robust estimations of the performance in reconstructed simulated physics events. 

The reconstruction algorithms and their performance are described in \cref{sec:reconstruction}. The algorithms
range from tracking via Particle Flow to particle identification and flavour tagging. Some of the algorithms come 
from the LHC experiments. In many cases Machine Learning techniques are leading to improved performance.  
The workshops and \keyhep have increased the cross-project work, i.e., developments 
for one detector at a certain collider based on fast simulation lead to their application at a different collider with full simulation. 
This collaborative work and the comparison of several algorithms, as 
they are available, gives confidence in 
the robustness of the expected performance at the Higgs factories. 

\subsection{The software ecosystem: the \keyhep turnkey software stack}
\label{sec:softwareeco}
\editors{Andre Sailer, Frank Gaede, Gerardo Ganis}

The software system used for these studies is known as \keyhep. The \keyhep
turnkey software stack emerged in 2019~\cite{towardsKey4hep,key4hep_2024} from
an initiative of the linear collider and the FCC communities to combine
their efforts and develop a common software framework to perform 
detector optimisation and physics studies. This software stack provides all the
tools to perform Monte Carlo (MC) generation, parametrised or full simulation,
reconstruction, and analysis for any future~\footnote{Some or all parts of
  \keyhep have also been adopted by concrete experiments such as LUXE or EIC.}
detector concept.

\subsubsection{Core building blocks}

\begin{figure}
  \centering
  \includegraphics[width=0.585\linewidth,trim={0 3cm 0 0},clip]{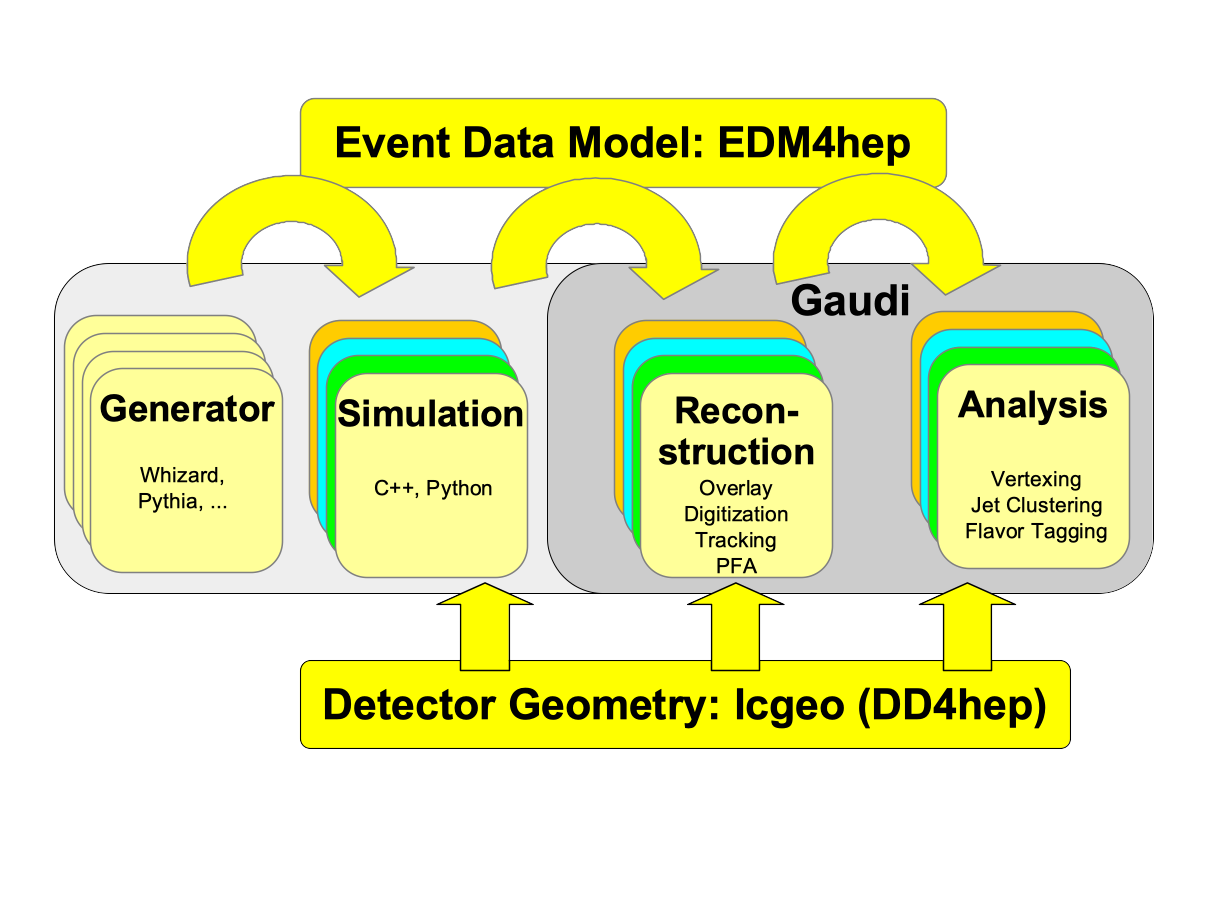}
  \caption{Schematic view of \keyhep core showing the core building blocks and their interaction.}
  \label{fig:key4hep_blocks}
\end{figure}

To benefit from state-of-the art developments, \keyhep aims to integrate
existing tools wherever it is feasible. Therefore, the core building blocks of
\keyhep are \gaudi~\cite{Barrand:2001ny} as the processing framework,
DD4HEP~\cite{Frank:2014zya} for the geometry description, and
\edmhep~\cite{Gaede:2022leb} as the event data model. Only \edmhep is
completely developed within \keyhep. \Cref{fig:key4hep_blocks} shows the
connection between these core building blocks. \edmhep is used to exchange
persistent data between the execution of any processing program. \gaudi
provides geometry information for all steps. \gaudi is used to process in
most steps, except when stand alone MC generators or the \ddsim\
simulation program is used (see \cref{sec:ddsim}).

\begin{figure}
  \centering
  \includegraphics[width=0.99\linewidth]{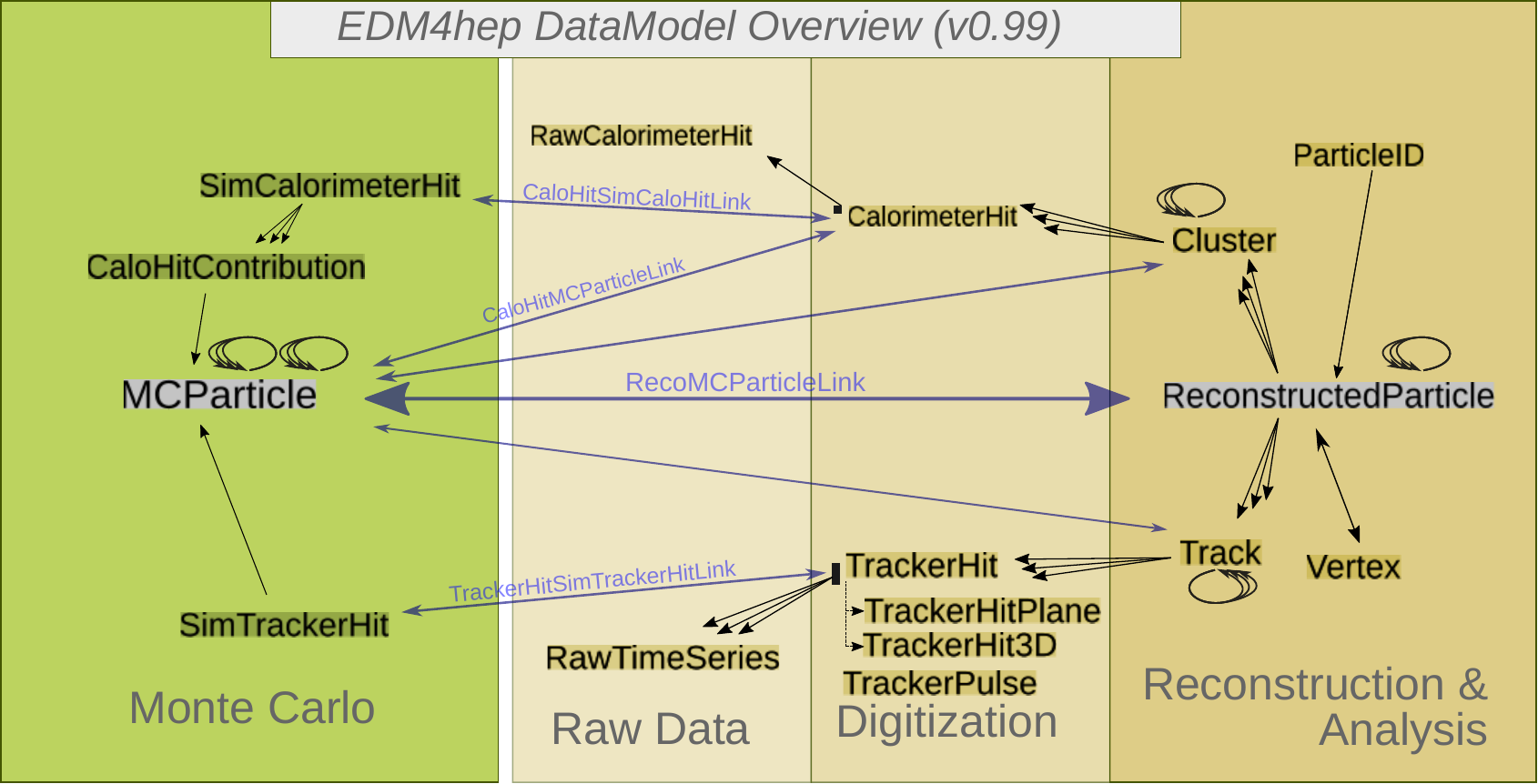}
  \caption{Schematic relationship between entities in the \edmhep event data model.}
  \label{fig:edm4hep_schema}
\end{figure}

\edmhep itself comes from the developments of the \lcio~\cite{Gaede:2003ip} event data model and
FCC-EDM and uses \podio\cite{podio1.0} to output the actual event data model based
on a simple YAML description. \Cref{fig:edm4hep_schema} shows the relationship
between the different entities in the event data model. The Monte Carlo
information (MCParticles, SimTrackerHits, SimCalorimeterHits) is separated from
the data, digitisation, reconstruction and analysis layers, and only available
via links (e.g., CaloHitSimCaloHitLink for the link between CalorimeterHits and
SimCalorimeterHits) from the other objects, but still available for studying
performance and efficiencies. The main reconstructed object is the
reconstructed particle, which is directly connected to potentially
reconstructed objects such as clusters in the calorimeters, tracks or
vertices. Reconstructed particles can also consist of other reconstructed
particles, for example to represent jets. ParticleID objects contain pointers to
the reconstructed particle they potentially identify. The tracks and clusters
contain the lists of calorimeter or tracker hits of which they are made up.

\subsubsection{Infrastructure, testing, and validation}

All development of the \keyhep packages takes place on GitHub, so that everyone
regardless of affiliation can contribute, report issues, or ask questions. 
Discussion also takes place in weekly meetings.
The packages exist in the \href{https://github.com/key4hep}{\keyhep organisation}. 
All contributions to the packages are compiled and
tested (if tests are available) in the so called continuous integration
solution provided via GitHub actions, and once all issues and comments
addressed merged into the mainline repositories. The builds of the \keyhep
software stack are done with spack~\cite{Gamblin2015:spack,Volkl:2021yvp}, and
every night builds for multiple operating systems are deployed to
CVMFS~\cite{blomer_2020_4114078} (\texttt{/cvmfs/sw-nightlies.hsf.org/key4hep})
so that users and automatic tools can use them immediately. After certain
developments are finished, or on user request \emph{releases} are also
permanently deployed to CVMFS (\texttt{/cvmfs/sw.hsf.org/key4hep}). The
software stacks that are built every night are also exercised with simulation
and reconstruction workflows, to ensure that the changes introduced do not
affect anything negatively.

\subsection{Generators\label{sec:gen}}
\editors{Carlo Carloni Calame, Juergen Reuter, Marco Zaro}

Monte Carlo (MC) event generators are the workhorses of the simulation chain for collider experiments. They have profited from decades of developments for LEP, flavour factories, TeVatron, and the LHC, which has tremendously enhanced the level of automation and sophistication. Like at the LHC, for $\Pep\Pem$ Higgs factories, they describe hard scattering processes in the highest technically possible order of perturbation theory, dominated by QED and EW instead of QCD corrections. Hence, theoretical uncertainties are orders of magnitude smaller. Parton distribution
function (PDF) uncertainties for electrons and positrons can be
perturbatively calculated. Underlying event and multiple interactions are not present (however, there are beam-induced backgrounds and overlays, especially for drive-beam driven accelerators like CLIC). 
QCD parton showers and hadronisation proceed in the same way as at LHC, while the exclusive simulations of photons has much higher importance.

In addition to the topical meetings~\cite{MCTopical1,MCTopical2} and ECFA workshops, a two-week CERN workshop on precision calculations and tools~\cite{CERNPrec22} has been important for the work presented in this section. The focus for Higgs/electroweak/top factories has been on the following generators: the three multi-purpose multi-leg generators
\mgfive~\cite{Alwall:2014hca,Frederix:2018nkq},\sherpa~\cite{Gleisberg:2003xi,Sherpa:2019gpd} and 
\whizard~\cite{Kilian:2007gr,Moretti:2001zz}, the two 
multi-purpose shower/hadronisation programs
\herwig~\cite{Bahr:2008pv,Bellm:2015jjp}
and
\pythiasix/\pythiaeight~\cite{Sjostrand:2006za,Sjostrand:2014zea,Bierlich:2022pfr},
and the specialized tools \babayaga~\cite{CarloniCalame:2000pz,CarloniCalame:2001ny,Balossini:2006wc,Balossini:2008xr,CarloniCalame:2019dom},
\bhlumi/\bhwide~\cite{Jadach:1996is,Jadach:1995nk}, \textsc{YFSZ/KoralZ}/\textsc{YFSWW/KoralWW}~\cite{Jadach:1999tr,Skrzypek:1995wd,Jadach:1998gi,Jadach:2001mp}, \textsc{RacoonWW}~\cite{Denner:1999kn}, and \kkmcee~\cite{Jadach:2022mbe}. A detailed overview is given in the input document on MC generators for the US Snowmass Community
Study~\cite{Campbell:2022qmc}. Many interesting details on the
simulation of the complete Standard Model (SM) processes for ILC can be found
in Ref.~\cite{Berggren:2021sju}. Details on the MC simulation can also be
found in the conceptual or technical design reports of the
specific electron-positron collider proposals~\cite{CEPCStudyGroup:2023quu,ILC:2013jhg,Behnke:2013lya,FCC:2018evy,Aicheler:2012bya,Linssen:2012hp} (and also muon colliders~\cite{Accettura:2023ked}, where the demands on MC generators are very similar). As the first part of the simulation chain, the generator section has connections to the simulation~(\cref{sec:simulation},) the simulation of beamstrahlung~(\cref{sec:lumispectra}), and is embedded in the general software ecosystem~(\cref{sec:softwareeco}). During the workshops and meetings listed above, a technical benchmarking between different generators embedded in the Higgs/electroweak/top factory software ecosystem has also been initiated, which is discussed in detail in~\cref{sec:techbench}. It allows comparison of different generators in an automated way, and first results of simple processes show excellent agreement.

This section is structured as follows: we first discuss
the simulation and interfacing of beam spectra in~\cref{sec:beamspectra}, then
hard matrix elements and the treatment of initial-state radiation in~\cref{sec:hard_initial}.  Parton showers, matching and hadronisation are reviewed in~\cref{sec:shower_hadro}. \Cref{sec:specialtools} details specific processes that need a specialized treatment and sometimes specialized tools; in~\cref{sec:bsmsim} we comment on the simulation of Beyond Standard Model (BSM) processes, while in~\cref{sec:sustainablegenerators} we discuss performance and sustainability (e.g.\ phase space sampling and parallelization) of Higgs/electroweak/top factory MC generators, before a short outlook to the future.

\subsubsection{Interfacing to beam spectrum simulations}
\label{sec:beamspectra}

Beam spectra of high-luminosity lepton colliders are not monochromatic, but are rather
Gaussian-shaped for synchrotrons and exhibit long beamstrahlung tails for linear colliders. This is described in detail in~\cref{sec:lumispectra}. For precise simulations, especially detector acceptance studies and assessment of systematic uncertainties, these beam spectra need to be convolved with the MC simulation. 
Gaussian beam spreads are available in many MC generators like e.g.\ \kkmcee, \pythiaeight, and \whizard. A more sophisticated treatment, assuming that the two beam spectra factorize and feature a sharp delta-like peak and power-like tail allows an approximation using a functional parametrisation with six or seven independent parameters. This approach has been codified in the tool
\textsc{CIRCE1}~\cite{Ohl:1996fi} that was used for the simulations of the
TESLA linear collider project. Incarnations of this approach with slight variations are available in \kkmcee, \mgfive~\cite{Frixione:2021zdp}, and \whizard. However, it has been shown that for linear colliders, especially drive-beam accelerators like CLIC, plasma accelerators like HALHF as well as photon colliders, such a parametrisation is
insufficient, both because the peak structure has a very complicated shape, and also the tails  do not exhibit simply a power-like behaviour. As detailed in~\cref{sec:lumispectra}, tools that simulate accelerator spectra like \textsc{GuineaPig} or \textsc{CAIN} produce too low statistics for large event generation samples. To generate a high number of events fast enough, an extrapolation based on 2D-histogrammed fits is used, where a smoothing algorithm is used to avoid too strong fluctuations. This is realized in the \textsc{CIRCE2} algorithm in the \whizard generator. Note that besides the $\Pep\Pem$ components of the spectra, also the
$\Pep\PGg$, $\PGg \Pem$ and $\PGg\PGg$ components are very
important, especially for linear colliders ($\PGg\PGg$-induced
backgrounds).

\subsubsection{Hard matrix elements and resummation of initial-state radiation}
\label{sec:hard_initial}

To match the expected experimental precision, theory uncertainties should be as small 
as permille or sub-permille level. This necessitates going to next-to-leading (NLO) and next-to-next-to-leading (NNLO) order in the electroweak (EW) couplings (and $\alpha_s$ for hadronic final states), and resumming large logarithms that would eventually invalidate fixed-order perturbation theory. The first task is the realm of higher-order matrix elements that we discuss in the first part of this subsection (though clearly the phenomenology at Higgs/electroweak/top factories will be different from LHC as for most processes EW corrections are more important than QCD corrections due
to processes dominated by EW resonance production); while the second deals with the resummation of QED radiation, especially in the initial state. 

\paragraph{Hard matrix elements at fixed order}

At the heart of cross section computations at colliders lies the ability to compute scattering matrix elements in perturbation theory. This is something for which LHC physics leaves a vast legacy. 

In particular, modern matrix-element generators have completely automated the computation of tree-level and one-loop matrix elements based on recursive algorithms (see also~e.g.\ Ref.~\cite{Ohl:2023bvv}), which, in turn, can be combined to deliver predictions accurate at NLO, both in QCD and in EW interactions. Indeed, the problem of computing the essential ingredients, IR-subtracted (tree-level) real-emission matrix elements and one-loop amplitudes has been solved since the beginning of the millennium. In the case of the former, subtraction schemes have been developed~\cite{Catani:1996vz,Catani:2002hc,Frixione:1997np,Nagy:2003qn}, and two of them, Catani--Seymour dipoles and sector subtraction by Frixione, Kunszt and Signer are most widespread, while for the latter different methods exists~\cite{Passarino:1978jh,Davydychev:1991va,Denner:2005nn,Ossola:2006us,Mastrolia:2012bu} implemented in software libraries~\cite{Ossola:2007ax,Hirschi:2011pa,Cascioli:2011va,Peraro:2014cba,GoSam:2014iqq,Denner:2014gla,Hirschi:2016mdz,Denner:2016kdg,Actis:2016mpe,Buccioni:2019sur}. These one-loop providers can be interfaced to phase-space integrators via the Binoth-Les-Houches-Accord
interface~\cite{Binoth:2010xt,Alioli:2013nda}.
Thanks to this, different software packages exist
which deliver NLO-accurate predictions, such as 
\mgfive~\cite{Alwall:2014hca,Frederix:2018nkq},
\sherpa~\cite{Sherpa:2019gpd,Biedermann:2017yoi,Bothmann:2024wqs}, 
\whizard~\cite{Moretti:2001zz,Kilian:2007gr,Bredt:2022dmm}. An almost exhaustive list of NLO QCD corrections for many relevant strongly-interacting final states at Higgs/electroweak/top factories can be found, including up to six jets in the final state,  in Refs.~\cite{Alwall:2014hca,Rothe:2021sml,Bredt:2022nkq}.
Also, predictions at the ``complete-NLO'' accuracy, meaning that they also include contributions formally subleading in the power-counting, have also been made possible, see e.g.\ Refs.~\cite{Frederix:2016ost,Biedermann:2017bss,Reyer:2019obz,Denner:2023eti,Stremmer:2024ecl}. Finally, concerning EW corrections, these can be computed either exactly or in the high-energy (also known as Sudakov~\cite{Sudakov:1954sw}) approximation. Such an
approximation, pioneered by Denner and Pozzorini~\cite{Denner:2000jv,Denner:2001gw} when the automation of EW corrections was still beyond the horizon, has recently regained interest from different
groups~\cite{Bothmann:2020sxm,Pagani:2021vyk, Lindert:2023fcu} with some applications~\cite{Chiesa:2013yma, Bothmann:2021led, Pagani:2023wgc}, including all-order resummation~\cite{Denner:2024yut}. Such an approximation is especially useful at a high-energy lepton collider (CLIC or muon collider, see also Refs.~\cite{Bredt:2022dmm,Ma:2024ayr}).

Going one order further, at NNLO, complete automation is still to be achieved. However, great strides have been made towards it in recent years, and it is not unforeseeable that it will be finally attained on a timescale of 5--10 years. While challenges are conceptually
similar to the ones at NLO, their solution is much more intricate. In this case, several general methods have been developed for the subtraction of IR-singularities~\cite{Catani:2007vq,Gehrmann-DeRidder:2005btv,Currie:2013vh,Braun-White:2023sgd,Braun-White:2023zwd,Fox:2023bma,Czakon:2010td,Czakon:2011ve,Cacciari:2015jma,Boughezal:2011jf,Boughezal:2015eha,Boughezal:2015dva,Gaunt:2015pea,DelDuca:2015zqa,Caola:2017dug,Caola:2018pxp,Delto:2019asp,Devoto:2023rpv,Magnea:2018hab, Magnea:2018ebr,Magnea:2020trj,Bertolotti:2022aih,Bertolotti:2022ohq}, while two-loop amplitudes in QCD are available for up to 5 external particles ($2\to 3$ processes)~\cite{Hartanto:2019uvl,Chawdhry:2020for,Chawdhry:2021mkw,Badger:2021ega,Badger:2021nhg,Badger:2022ncb,Badger:2024sqv,}. In this case, the extension to EW corrections represents a serious challenge due to the
various mass scales that would appear in the loops. Some recent results in this direction, notably for Drell--Yan production at order $\alpha^3 \alpha_s$~\cite{Heller:2020owb,Bonciani:2021zzf,Buccioni:2022kgy,Armadillo:2022bgm,Armadillo:2024nwk}, are encouraging. 
More specifically to the case of electron colliders, first baby steps towards tools for simple
processes like $\Pep\Pem \to \PZ\PH$ (on-shell) have started~\cite{Freitas:2022hyp,Chen:2022dow} as well as $\Pep\Pem \to \PW\PW$~\cite{Actis:2008rb}; note also the progress that has been made within the \textsc{McMule} framework for fully massive fermions geared towards low-energy experiments~\cite{Banerjee:2020rww}. But it will still take many years until that full NNLO automation is nearly as mature as for NLO. 

For both linear and circular variants of Higgs/electroweak/top factories, it is of paramount importance to simulate polarised processes~\cite{Moortgat-Pick:2005jsx}. This applies to the initial state taking into account beam polarisation, but also the final state to produce polarised event samples. Helicity eigenstates for fully polarised initial states are available in the multi-purpose tools \mgfive, \sherpa and \whizard, but also in \kkmcee.  
The most general formalism here is based on spin-density
matrices which allow arbitrary polarisations in the initial state, e.g.\ both longitudinal and transversal polarisation of the lepton beams. This is realized in \whizard. Polarisation fractions can be accounted for in the spin-density matrix; however, also fully polarised samples can be trivially mixed and reweighted. The same spin-density formalism can be used to account for spin correlations when factorizing processes into production of resonances and decays, where also projections on to specific intermediate resonances can be made in all three multi-purpose MCs. 

\paragraph{Resummation of initial-state radiation}

In order to reach (sub) permille-level precision on total cross sections and
differential distributions, besides fixed-order corrections, soft and collinear
photon radiation in the initial state needs to be resummed (see also the details in the~\cref{sec:ewk_qcd}). There are two basic frameworks for resumming QED photon emissions: collinear factorization within a parton-distribution function (PDF) picture and soft or eikonal resummation. The two regimes overlap in the soft-collinear region where double counting has to be avoided. The all-order soft-collinear resummation was spearheaded in Refs.~\cite{Gribov:1972ri, Gribov:1972rt}. Later on, hard-collinear initial-state radiation up to second order and third order~\cite{Kuraev:1985hb, Nicrosini:1986sm, Arbuzov:1999cq, Skrzypek:1990qs,Cacciari:1992pz} have been added to these. Soft or eikonal factorization has been studied in the seminal paper~\cite{Yennie:1961ad}, resumming soft logarithms (also from large-angle emissions) using Yennie--Frautschi--Suura exponentiation. This has also turned out to be a very useful scheme for the resummation of final-state soft emissions for hadronic decays of charge particles, where the YFS algorithm can be also used to generate exclusive soft photon emissions in the complete phase space. YFS exponentiation has been implemented later for LEP MCs~\cite{Jadach:2000ir}, where for simple two-fermion processes a coherent resummation between initial- and final-state radiation can also be performed (CEEX scheme compared to the incoherent sum in the so-called EEX scheme). Recently, the EEX scheme has been automatized within the \sherpa generator~\cite{Krauss:2022ajk}.

On the other hand, collinear factorization resums collinear logarithms instead of soft logarithms and leads to a PDF picture of initial-state ``dressed'' leptons. This has been initiated at LEP for the resummation of leading logarithms (LL)~\cite{Nicrosini:1986sm, Arbuzov:1999cq, Cacciari:1992pz,Skrzypek:1990qs} and recently been carried over to next-to-leading logarithmic (NLL) accuracy~\cite{Frixione:2019lga,Bertone:2019hks,Frixione:2012wtz,Bertone:2022ktl}. NLL QED PDFs are implemented in \mgfive and \whizard. 

Ideally, the resummation of both types of logarithms has to be combined to achieve the highest possible precision, and ideally also in a process-independent way. While there is an algorithmic procedure to improve YFS exponentation (which also can generate
exclusive, though mostly soft, photons) by hard-collinear corrections,
collinear factorization seems to be better suited for combination with
higher fixed-order calculations of the hard process. Each
of the methods is likely to have its own benefits and which is better suited for
what class of processes needs to be studied in the future. For
summaries of these topics, see also~\cite{Heinemeyer:2021rgq,Frixione:2022ofv}.

\subsubsection{Parton showers, matching, and hadronisation}
\label{sec:shower_hadro}

\paragraph{QCD showers and hadronisation}

Parton showers resum large logarithms and provide exclusive multi-jet
events; again, parton showers have benefitted tremendously from two
decades of developments for the LHC. This has driven the development of
(final-state) showers that are accurate at next-to-leading logarithmic
(NLL) order, and hence include effects from colour and spin
correlations beyond the quasi-classical approximation of independent
emissions. Examples of these
are Refs.~\cite{Dasgupta:2020fwr,Herren:2022jej,Nagy:2020rmk,Forshaw:2020wrq}. These
showers are available for jet distributions at Higgs/electroweak/top factories and
have been e.g.\ applied to hadronic Higgs events at future lepton
colliders~\cite{Knobbe:2023njd} or reapplied to LEP data~\cite{Kilian:2011ka}. There are two very active fields of research: (i) to extend NLL shower accuracy to NNLL accuracy~\cite{Hoche:2017iem,Dulat:2018vuy,FerrarioRavasio:2023kyg}, and (ii) to combine parton showers with (automated) analytic resummations~\cite{Banfi:2004yd,Baberuxki:2019ifp}. Going to NNLL accuracy necessitates the inclusion of higher-multiplicity splittings, higher-order QCD corrections in the splitting kernels, taking into account colour and spin correlations and distributing recoil kinematics in a consistent way. Choosing a meaningful value for the scale of $\alpha_s$ in the different steps of the shower is highly non-trivial. While (analytic) resummation allows analytic control over the logarithmic accuracy of the calculation to be kept, it can usually only be applied to very inclusive observables. On the other hand, parton showers allow being fully exclusive in the events, while the proof of a certain logarithmic accuracy is not straightforward. Hence, the interest to combine both approaches and to improve the accuracy of parton showers by analytic resummations.

Another important issue is to keep track of the resonance structure of intermediate heavy particles like $\PZ, \PW, \PQt, \PH$ when transferring the hard process to the parton shower, as the shower needs to know what kinematic features to preserve. This is known e.g.\ from top decays at the LHC~\cite{Jezo:2016ujg}, but the effects are much more visible at a Higgs/electroweak/top factory, e.g.\ the study for $\Pep\Pem\to\PWp\PWm \to jjjj$ in Ref.~\cite{Kilian:2018onl}: resonance peaks could get artificially shifted when not taking care of the resonance history, and also hadronic observables like the total number of photons or neutral hadrons come out wrong. Algorithmically, cross sections of underlying on-shell resonant processes are taken to assign probabilities between different resonance histories and continuum events. 

Before turning to QED showers or matching and merging, we briefly comment on hadronisation. Hadronisation is still based mostly on phenomenological models, either
on Lund string fragmentation or the cluster hadronisation
paradigm. Machine-learning methods aim at simulating hadronisation from trained real data event samples, which might result in realistic hadronic event generation, but, however, do not enhance our
knowledge about the underlying QCD physics. There are estimates that samples of up to \SI{200}{\per\atto\barn} of very clean jet samples from
e.g.\ FCC-ee might necessitate the development of new formalisms for our understanding of hadronisation in order to understand these data. An interesting question is the realistic modelling of colour reconnections in hadronic $\PW\PW$ events. Hadronic decays are simulated within the hadronisation packages \pythiaeight, \herwig and \sherpa. The simulation of $\PGt$ decays is very important for lepton colliders, especially polarised $\PGt$ decays: this is done in \textsc{tauola}~\cite{Jadach:1993hs} or the generic hadronisation tools. 

\paragraph{QED (and electroweak) showers}

As Higgs/electroweak/top factories are lepton colliders, QED showers and the simulation of exclusive photons are of paramount importance: examples are given by the experimental assessment of systematic uncertainties, selection efficiencies, simulation of backgrounds for mono-photon searches etc. These QED showers are mostly realized as interleaved
showers where the QED/EW suppression of emissions from quarks by a
factor of $\alpha/\alpha_s \simeq 1/15$ is dealt with by a combination of
rejection and reweighting algorithms. In such interleaved showers, the QED equivalents of the QCD splitting kernels are being used, such that many generic details of the showers are inherited from the corresponding type of QCD shower. One of the tunable parameters is whether $\PGg \to \Pl\Pl$ are enabled in the QED shower and which charged lepton flavours are active. On the other hand, showers based on soft or eikonal exponentiation instead of collinear splitting kernels are also used, mostly for radiation patterns that do not exhibit QCD radiation features (initial states, decaying hadronic and electroweak resonances etc.). These showers allow electroweak radiative corrections to be combined in a relatively genuine way with the QED shower emissions; examples are tools like \photos~\cite{Barberio:1990ms, Barberio:1993qi, Davidson:2010ew}. 

For (very) high-energy linear lepton colliders like ILC and CLIC, electroweak showers~\cite{Kleiss:2020rcg,Brooks:2021kji,Masouminia:2021kne} also become important. This is connected with the problem of EW fragmentation of inclusive W/Z/H radiation into especially hadronic jets with jet masses around the EW resonances: for EW bosons in the (multi-)TeV region, ``soft'' or quasi-collinear EW radiation can no longer be resolved experimentally, and a more inclusive event description has to be used. This carries over to the initial state, where QED lepton PDFs are generalized to EW (or SM) collinear lepton PDFs including the full (EW) SM content of ``beam'' leptons. 

\paragraph{Matching and merging}

Techniques and algorithms to match fixed-order calculations to
exclusive all-order showers have been developed for hadron colliders
for LO (matrix element corrections in parton showers) to NLO --- MC@NLO~\cite{Frixione:2002ik,Torrielli:2010aw,Frixione:2010ra,Frederix:2020trv},
POWHEG~\cite{Frixione:2007vw}, KrkNLO~\cite{Jadach:2015mza}, CKKW(-L)~\cite{Lavesson:2008ah}, as well as hybrid methods~\cite{Nason:2021xke} 
--- to NNLO (MiNNLOPS \cite{Monni:2019whf,Monni:2020nks}, UN$^2$LOPS~\cite{Hoche:2014uhw}
etc.) and even NNNLO~\cite{Bertone:2022hig}, while consistently merging samples of exclusive
jet multiplicities at NLO into inclusive multi-jet merged samples have been
developed for the LHC~\cite{Lonnblad:2012ix,Frederix:2012ps}. These techniques can be straightforwardly carried over to lepton colliders like a Higgs/electroweak/top factory. While the corresponding matching of
photon emission (and lepton pairs) with (N)NLO EW corrections is in
principle straightforward (e.g.\ in the framework of \babayaga\cite{Balossini:2006wc, Balossini:2008xr}), the complete automation is still work in
progress and will only be available in the future. It is worthnoticing that
some approximate prescriptions for the matching of QED emissions with NLO matrix elements exist
~\cite{Kallweit:2015dum, Bothmann:2021led,Pagani:2023wgc}, whose accuracy however
is not (formally) NLO across the full phase-space. A matching scheme
at LO for matrix element and shower photons can be found
in Ref.~\cite{Kalinowski:2020lhp}. 

A relevant bottleneck that arises when considering higher-order predictions matched to parton-shower is the occurrence of negative weights, which is mostly relevant for MC@NLO-like matching
schemes. The presence of negative weights hampers the
statistical quality of a data sample, requiring 
a larger number of events to be generated than in the case of positive-weighted events in order to attain the same
statistical properties (e.g.\ variance of the cross section). For example, even a moderate
fraction of negative weights of $~\sim 15\%$ translates
to the necessity of generating at least twice the number of events. This fact is particularly problematic
when each event must undergo not only parton-shower and hadronisation, but also detector simulation, which is particularly expensive from the computational point of view. For example, for Higgs/electroweak/top factories it has been shown that the exclusive simulation of the top threshold NRQCD effects matched to continuum NLO QCD is especially plagued by negative event weights.

Recently, various methods have been developed to reduce the fraction of negative weights in Monte-Carlo simulations~\cite{Frederix:2020trv, Andersen:2020sjs, Nachman:2020fff,Andersen:2021mvw, Danziger:2021xvr,Frederix:2023hom,Andersen:2023cku}. The ways to attain such a reduction are different, from suitably altering the subtraction terms, to resampling the events or to spreading the $K$ factor differently across the phase-space. It is not unreasonable to expect even more progress in this direction in the coming years.

Note that there have been attempts of matching initial-state shower with fixed-order EW corrections (using phase-space slicing techniques) that do not produce negative event weights at all~\cite{Kilian:2006cj,Robens:2008sa}. 

\subsubsection{Special processes and dedicated tools}
\label{sec:specialtools}

There are several processes where the standard MC treatment is either not sufficient in their description for the targeted precision, and they need to be treated specially, or they are available in special-purpose MC tools or in dedicated implementations of multi-purpose tools. Here we briefly consider the following processes: (i) two-fermion production $\Pep\Pem \to f\bar{f}$, (ii) Bhabha scattering $\Pep\Pem \to \Pep\Pem$ and $\Pep\Pem \to \PGg\PGg$, (iii) photoproduction of low-$\pT$ hadrons, (iv) the
$\PW\PW$ threshold, (v) the $\ttbar$ threshold. 

\paragraph{Two-fermion processes, Bhabha scattering, hadroproduction}

The simplest physics processes are two-fermion production processes, $\Pep\Pem \to f\bar{f}$. They are very important physics signatures, cf.~\cref{sec:TwoF}, and are background for all other processes. They dominate runs at the $\PZ$ pole. In the LEP era there were several dedicated MCs for these processes, out of which \kkmcee~\cite{Jadach:2022mbe} is one of the most sophisticated. It includes several classes of electroweak virtual corrections and combines them with a coherent scheme (CEEX) of soft-photon exponentiation. Recently, \sherpa~\cite{Krauss:2022ajk} and \whizard have been benchmarked against \kkmcee for this process class. For more details on which higher-order corrections are necessary to match the experimental precision at Giga-Z or Tera-Z; see also \cref{sec:ewk_qcd}.

Bhabha scattering, $\Pep\Pem \to \Pep\Pem$, is a special two-fermion process as it proceeds not only through an annihilation (s-)channel but also through a t-channel contribution, which leads to a very strong forward peak of the differential cross section. This makes Bhabha scattering (together with  two-photon production $\Pep\Pem \to \PGg\PGg$) a very important standard candle for high-precision luminometry (see \cref{sec:LUMI}), as the process is fully dominated by QED and any contamination from new physics is eventually tiny. Matching the projected precision of $10^{-4}-10^{-5}$ on the theoretical side is very challenging and needs a dedicated implementation of EW corrections  together with resummation of all available soft and collinear logarithms; again, see also \cref{sec:ewk_qcd}.

Dedicated tools developed for the high-precision simulation of Bhabha scattering include \bhlumi, \bhwide (for low-angle and wide-angle Bhabha scattering, respectively) and \babayaga. For the latter (which was developed for flavour factories and is based on a QED parton shower matched to exact NLO corrections), developments are planned to simulate Bhabha scattering, $\Pep\Pem\to\PGg\PGg$ \cite{CarloniCalame:2019dom} and $\Pep\Pem\to\mpmm$ at future Higgs/electroweak/top factory energies.

The photoproduction of low-$\pT$ hadrons is one of the largest background processes especially in the central region and for high-energy versions like CLIC. Its simulation is based on fits to total cross
section measurements from the CrystalBall experiment, while the hadronisation for low effective centre-of-mass energies is dominated by pion and kaon production. A special implementation exists e.g.\ in
\whizard. Recently, a comprehensive study of these processes for low-energy colliders has been finalized~\cite{Aliberti:2024fpq}.

\paragraph{\texorpdfstring{$\PW\PW$}{WW} threshold}

For the $\PW\PW$ threshold a very precise determination of the $\PW$-boson mass from the threshold scan necessitates the resummation of QED logarithms~\cite{Beneke:2007zg} which can be added to a fixed-order calculation~\cite{Actis:2008rb}. The fully off-shell process $\Pep\Pem \to \PW\PW \to f\bar{f}f'\bar{f'}$ in \textsc{RacoonWW}~\cite{Denner:1999dt, Denner:1999kn, Denner:1999gp, Denner:2000bj, Denner:2001vr, Denner:2002cg}, was one of the first available for more complicated processes at NLO for $\Pep\Pem$ colliders. \textsc{RacoonWW} treats NLO EW radiative corrections in the double-pole approximation and includes higher-order initial-state radiation (ISR) corrections by a convolution with the collinear LL electron PDFs. On a similar basis, the ``Krak\'ow'' MCs \textsc{KoralW} and \textsc{YFSWW3} \cite{Jadach:2001mp, Jadach:2001uu} feature the resummation of QED corrections in the YFS formalism.

\paragraph{Top threshold}

The top threshold scan is the method which would provide the most precise determination of the top mass and width as the rise of the cross section at threshold depends crucially on these parameters, cf.~\cref{sec:top-quark}. Compared to (relativistic) NLO or NNLO QCD $\Pep\Pem \to \PQt\PAQt$, the threshold production cross section is enhanced by an order of magnitude due to non-perturbative and non-relativistic bound-state effects. These can be described fully inclusively in an analytic calculation that is available at NNNLO NRQCD, together with the resummation of top-velocity logarithms. For the study of experimental
acceptances and systematic uncertainties, this has to be included in an exclusive MC, which has been achieved at the level of NLL NRQCD matched to NLO QCD~\cite{Bach:2017ggt,ChokoufeNejad:2016qux}. For a more precise description, one should go one order further, matching the existing NNLL NRQCD to the (not yet existing) NNLO QCD (for $\Pep\Pem \to \PWp \PQb \PWm \PAQb$), including NLO EW corrections and achieving a matching to soft-gluon emissions (at threshold there is practically no phase space for the emission of hard jets). All of these seem feasible by the time data taking at the Higgs/electroweak/top factory starts. 

\subsubsection{Simulation of physics beyond the Standard Model}
\label{sec:bsmsim}

The automation of simulation of BSM processes, indispensable for searches~(\cref{sec:searches}), builds upon 15 years of development for LHC BSM simulations, developed during several Les Houches workshops and the series of MC4BSM workshops during the years 2006--2018.  Lagrangian-level tools like SARAH~\cite{Staub:2013tta}, LANHep~\cite{Semenov:2008jy} and
FeynRules~\cite{Alloul:2013bka} allow a new physics model to be entered as in a quantum field theory textbook. These tools export physics vertices and coupling constants to MC generators via dedicated interfaces like e.g.\ in Ref.~\cite{Christensen:2010wz}. These
tailor-made interfaces have to be engineered for every pair of
these tools with each MC generator. They have become redundant through
the introduction of a python-syntax based meta-layer within the
Universal Feynman Output (UFO)
interface~\cite{Degrande:2011ua,Darme:2023jdn}. This concentrates the burden of validation on one single interface, both for maintainers of Lagrangian-level tools as well as for MC generators. All different kinds of BSM models relevant for Higgs/electroweak/top factories can be and have been parametrised in terms of UFO models accessible in MC generators, especially dark matter (DM) models, models with complete DM sectors like Higgs portal models, extended Higgs sectors and models with extra light scalars or pseudoscalars (axion-like particles, ALPs) (see also \cref{sec:exotic-scalars-searches}), as well as models with heavy neutral leptons (HNLs), (see \cref{sec:SRCH-NHL}). For HNLs (or some similar new light neutral fermions), oscillation can occur on length scales inside the detectors; these need to be treated with special plugins; see also~e.g.\ Ref.~\cite{Antusch:2022ceb}. Supersymmetric models like the MSSM or its extensions are also available as UFO models and are also hardcoded in \mgfive, \sherpa and \whizard. In addition to specific models, deviations
of the SM are parametrised in terms of effective field theories (EFT) like
SMEFT or HEFT, which are available as well for such simulations in different variants. 
There are many attempts to carry these simulations over from LO to NLO; for NLO QCD
this is possible for most of these models triggered by the needed precision for LHC predictions. NLO EW calculations for BSM models are much more difficult, as consistent renormalisation schemes have to be available which can very likely never be fully
automatized. Also colourful exotics are available in MC generators in an automated way, either for direct production processes for the LHC or by means of EFT operators for Higgs/electroweak/top factories as they appear in e.g.\ Higgs portal and dark sector models. Details on their
implementation and automation in MC generators based on the colour-flow formalism can be found e.g.\ in Refs.~\cite{Ohl:2024fpq,Kilian:2012pz}.

\subsubsection{Sustainable generators: efficiency and performance}
\label{sec:sustainablegenerators}

Sustainability plays a major role in the planning of future colliders. Though the construction and operation of the accelerator (and the detectors) has the largest impact, computing has also a major share (especially the full detector simulation and analysis; see ~\cref{sec:simulation} and~\cref{sec:reconstruction}). Also MC simulations are computationally costly, especially at NLO/NNLO and NNLL. Huge data samples are needed, e.g.\ at the $\PZ$ pole due to the immense instantaneous luminosities, while at the highest energies processes with high particle multiplicities at the hard process level become very tedious, both for the matrix elements as well as the phase space sampling. At lepton colliders like Higgs/electroweak/top factories, for processes that are dominated by EW (resonance) production, simulation usually is classified by the number of the fermions in the final
state: $\Pep\Pem/\Pe\PGg/\PGg\PGg \to 2f, 3f, 4f, 5f, 6f, 7f, 8f, \ldots$ (see also Ref.~\cite{Berggren:2021sju}). With tens to hundreds of thousands of phase space channels, these processes can become very costly even at LO, but much more so at NLO. This is because phase spaces become much more complicated at lepton
colliders with many intertwined electroweak production channels connected by (EW) gauge invariance compared to phase spaces dominated by soft-collinear QCD radiation phenomena at LHC. Such processes need to be sampled in parallel and event generation needs to be
parallelized as well. The latter is always trivially achievable as bundles of events can be generated on different nodes of large computer farms. There are straightforward algorithms for parallelised
matrix element evaluation, e.g.\ parallelising over the helicities of external particles ($2^n$ for $n$ external fermions), the parallelisation of colour flows over different threads in hyper-threading or the usage of virtual machines for matrix elements~\cite{ChokoufeNejad:2014skp}. Phase space adaptation is much more complicated: not
only needing random number chains to be independent on different nodes (as for event generation), but also correlations between different parts of
the phase-space adaptation need to be taken into account. 

The usage of modern and vectorized data structures in order to effectively multi-thread on CPUs or even on GPUs,
and thus reduce the computational cost of simulations has been exploited in different directions.
Examples include the parallelization of the integrator
which shows speed-ups of 30-50 using a vectorized architecture~\cite{Brass:2018xbv,Carrazza:2020rdn}, 
or the parallel evaluation of parton-distribution functions~\cite{Carrazza:2020qwu}. Also, the usage of GPUs for matrix-element evaluation as well as phase-space sampling shows interesting speed-ups, and there are attempts in all three multi-purpose generators (\mgfive, \sherpa and 
\whizard)~\cite{Hagiwara:2013oka,Valassi:2021ljk,Bothmann:2021nch,Carrazza:2021zug,Carrazza:2021gpx,Reuter:2023vei}. A further interesting development is phase-space sampling using machine-learning techniques like invertible neural networks, normalising flows or autoencoders, which could prove a serious alternative to classic phase space adaptation at the time of a next-generation $\epem$ collider.

\subsubsection{Generator summary and outlook}
\label{sec:gensummary}

In summary, MC generators for a Higgs/electroweak/top factory build upon two decades of development for the LHC: the automation of NLO QCD and EW corrections can be applied in the same way at a future facility, but is technically complicated to the numerically challenging LL and NLL QED electron PDFs. This will be reliably available when data taking will start. Future studies will show which processes and energies are in the regime of collinear radiation or of soft radiation which is simulated in the YFS
formalism. Another future task is a universally applicable matching formalism between fixed-order and resummed calculations with exclusive radiation simulations and QED showers in order to achieve per-mil precision for both inclusive and exclusive predictions. 
Parton showers, driven by the LHC physics program, have been pushed to NLL accuracy, given access to non-trivial colour and spin correlations, and will very likely be carried to full NNLL in the future. Hadronisation models
might have to be rethought in an era with gigantic samples of very clean hadronic data from the Z pole.
The inclusion of beam spectra is important for the full simulation of physics processes for the future experiments at lepton colliders. Dedicated simulation frameworks are interfaced to MC generators. Future developments will comprise machine-learning techniques for beamspectrum fitting and simulation. The MC BSM signal simulation has benefitted as well from the developments of LHC, and are
available for multi-purpose MCs for almost arbitrary BSM models using Lagrangian-level tools connected to MCs via a universal python layer interface. A challenge will be EW NLO corrections in arbitrary BSM models, as they
always depend on the availability of appropriate renormalisation schemes. Several important physics processes demand special treatment either because
they are needed at a much higher theoretical precision than the rest or they exhibit very delicate perturbative or nonperturbative corrections. These processes are very often described by dedicated tools, while sometimes multi-purpose generators contain specialized treatments of such processes. Examples are processes for luminometry, the $\PW\PW$ and top threshold, and photon-induced production of low-$\pT$ hadrons.
Past experience has shown that several independent MC tools are needed to allow for an independent assessment of uncertainties from the theoretical modelling. The validation tasks and the assessment of theoretical uncertainties will benefit from automated benchmarking frameworks as have been developed during this ECFA study (\cref{sec:techbench}). MC simulations have been understood as one of the computational bottlenecks of the data analysis at LHC, and this still remains true
at lepton colliders, especially at Tera-Z. Active areas of research are phase-space sampling improvements based on either adaptive multi-channel versions of VEGAS or machine-learning methods based on e.g.\ normalising flows. Another very active field is the usage of
heterogeneous computing where parts of the simulation remains on traditional CPU farms and parts are off-loaded to GPUs. 

The main multi-purpose generators support HepMC~\cite{Buckley:2019xhk} and LHE~\cite{Alwall:2006yp}.  \whizard also supports direct output to \lcio. A converter
to the event data model \edmhep (\cref{sec:softwareeco}) is available in the \keyhep eco-system (\cref{sec:techbench}). 
This ensures the connection to fast and full detector simulation discussed in~\cref{sec:simulation}. For the 
future the definition and the availability of spin information is an important topic. 
\subsection{Beamstrahlung \& luminosity spectra\label{sec:lumispectra}}
\editors{Thorsten Ohl, Daniel Schulte}

In lepton collisions, the initial state consists, at leading order, of
weakly interacting fundamental particles and can be described fully by
their energies, momenta and polarisations.  A priori, this allows for
much more precise theoretical predictions compared to hadron
collisions, where the energies and longitudinal momenta of quarks and
gluons are described by parton distribution functions and can not
be reconstructed on an event-by-event basis.

In practice, this knowledge of the initial state is limited in the
leptonic case as well.  Three sources of uncertainties must be taken
into account: Initial State Radiation (ISR) from radiative
corrections, Beam Energy Spread (BES) from the transport of the beams
through the accelerator optics, and Beamstrahlung (BS) from collective
bunch-bunch interactions.  Here ISR differs from BES and BS, since it
depends on the scale~$Q^2$ of the hard scattering event, while the
latter depend on the beam optics and bunch shapes only.  The treatment
of ISR lies
therefore exclusively in the realm of the event generators discussed in~\cref{sec:gen}.  
These use process-dependent perturbative
calculations, based on matrix element corrections, radiator
functions~$D(x,Q^2)$, parton showers or a careful combination of
those.

Since BES and BS are both independent of the scale of the hard
scattering event, it is possible to handle them universally for all
scattering processes.  In addition, the quantitative description of
BES and BS from first principles involves physics that is outside of
the scope of the event generators in~\cref{sec:gen}.

\subsubsection{Collective bunch-bunch interactions}
Linear colliders require very high luminosities per bunch crossing,
since each bunch passes the interaction region only once.  This leads
to designs using dense bunches with very high space charges and
corresponding strong collective electromagnetic field.  This deflects
the particles in the opposing bunch, causing bremsstrahlung.  In order
to distinguish this process-independent radiation from the process-dependent ISR, it has been called
beamstrahlung~\cite{Blankenbecler:1987rg,Bell:1987rw,Jacob:1987ua,Chen:1988ec}.
Depending on the density and shape of the bunches, particles typically
emit of the order of one BS photon.  These
photons carry typically a few percent of the beam energy; even more
for multi-TeV colliders. The resulting stochastic energy spread leads
to the development of a luminosity spectrum.  The beam parameters are
chosen such that the effect of the degradation of the luminosity spectrum due to BS
is at most similar in size to the degradation that results from ISR.
In addition, the beamstrahlung photons can themselves collide
 with the opposing bunch and contribute to backgrounds.

At circular colliders, the luminosity per bunch crossing can be
smaller.  Thus the bunches are not required to be as dense and
the magnitude of BS is smaller, though not always negligible.

Detailed microscopic simulations of bunch crossings can be performed
with long-established programs such as
\guineapig~\cite{Schulte:1998au} and \cain~\cite{Chen:1994jt}.  Starting from
a given bunch shape and energy distribution of the colliding beams,
these programs track the colliding particles through the interaction
and generate a sample~$\{(x_{i,1},x_{i,2})\}_{i}$ of pairs of energy
fractions $x_i=E_i/E_{\text{design}}$ entering the hard scatterings of
leptons and beamstrahlung photons. The joint probability
density~$D(x_1,x_2)$ describing such a sample can not usually be
factorized into a product of individual densities for the beams.

\subsubsection{Beam transport}
The beams entering the interaction zone are not monochromatic either
and already have a finite energy spread.  In the simplest cases, this
BES can be specified as the width of a Gaussian distribution.
In general, however, the energy distribution can be more complicated.
For example, the acceleration by wakefields in the CLIC~\cite{CLICCDR_vol1}
collider distorts the bunches in such a way that the energy distribution
becomes bimodal.

In the case of circular colliders, the emission of a hard photon can
cause a few particles to lose enough energy that they are lost in
the coming turn. The resulting reduction of the beam lifetime must be
mitigated by choosing the collision parameters appropriately.  It is
necessary to track the bunch after its distortion by one interaction
through the accelerator optics to the next interaction region.  Only
after repeating this for many revolutions around the ring, one can
estimate the stable equilibrium energy distribution and bunch shape.
The increase of the bunch energy spread from the collisions also
leads to a lengthening of the bunches on subsequent turns, that in the
equilibrium typically settles at a factor in the range of just above one
and a few.

\subsubsection{Input to Monte Carlo event generators}
\guineapig~\cite{Schulte:1998au} has been developed to simulate the
emission of beamstrahlung and the generation of secondary beam-beam
background. The code uses a cloud-in-cells model for the two beams and
fast-fourier transformation to calculate the fields. It allows
either for the basic parameters of the colliding beams to be defined, or for
detailed input distributions to be read from files to include any correlation
that might be important.  It produces the beamstrahlung photons and
saves them together with the beam after the collision. It also determines
the luminosity spectra for the original particles as well as the
photons. The distribution of collision energies of the particles and
their location can also be stored in a file. In addition the important
incoherent production of electron-positron pairs and of hadronic
events can be calculated. The particle pairs are tracked through the
fields of the beams that can deflect them strongly.  In addition to
the original version of \guineapig, which has been written in C, a
version that has been translated to C++ is
available, called \guineapigpp. Both agree in all tests that
we performed.

For each collider proposal, with specific beam optics and bunch shapes
at the interaction point, the simulated bunch crossings produce
samples distributed according to a joint luminosity
spectra~$D(x_1,x_2)$ that will serve as input for the particle physics
simulations.  However, particle physics simulations for precision
physics require much larger samples as input.  It is therefore
necessary to model the joint luminosity spectra~$D(x_1,x_2)$, instead
of using the $(x_1,x_2)$-pairs as a fixed size stream of random
numbers.  In addition to the probability density
functions~$D(x_1,x_2)$, efficient generators of non-uniform pairs of
random numbers that are distributed according to~$D(x_1,x_2)$ should
be provided.

In the analysis of experimental data, these simulation results will be
replaced by measurements of reference cross sections that depend on
the beam energy spectra.

\subsubsection{Parametrisations}

\paragraph{\textmd{\circe}}
The simplest approach is a factorized ansatz,
\begin{subequations}
\label{eq:circe1-parametrisation}
\begin{equation}
  D_{p_1p_2}(x_1,x_2) = d_{p_1} (x_1)  d_{p_2} (x_2)\,,
\end{equation}
where the~$p_i$ denote the flavour and polarisation of the colliding
particles.  Since the theoretical treatment of
beamstrahlung~\cite{Blankenbecler:1987rg,Bell:1987rw,Jacob:1987ua,Chen:1988ec}
suggests a power law for the leading order of the energy distributions,
the library \circe~\cite{Ohl:1996fi} used a sum of an unperturbed
$\delta$-peak and a $\beta$-distribution with integrable singularities:
\begin{align}
  d_{e^\pm} (x) &= a_0 \delta(1-x) + a_1 x^{a_2} (1-x)^{a_3} \\
  d_\gamma (x) &= a_4 x^{a_5} (1-x)^{a_6}
\end{align}
\end{subequations}
as ansatz.  In the absence of polarisation dependence, each collider
design was therefore parametrised by seven real
numbers~$a_{0,\ldots,6}$. These numbers were derived with a
least-squares fit from histogrammed results of \guineapig. This simple
parametrisation was extensively used in the simulations that formed
the basis of the TESLA TDR~\cite{Andruszkow:2001hgu} and contemporary designs.
\circe is available as a part of \whizard, but can be compiled and
used independently.

\paragraph{\textmd{\circetwo}}
A parametrisation
like~\cref{eq:circe1-parametrisation} proved inadequate for
describing the luminosity spectra resulting from the wakefield
acceleration at CLIC~\cite{CLICCDR_vol1}, or at laser
backscattering~$\gamma\gamma$ and $\gamma e$
colliders~\cite{ECFADESYPhotonColliderWorkingGroup:2001ikq},
that can also not be
approximated by the factorized~\cref{eq:circe1-parametrisation}.  In order to
address this shortcoming, a successor was developed.  \circetwo
generalizes~\cref{eq:circe1-parametrisation} by dividing the two
dimensional space of $(x_1,x_2)$-pairs into a non-uniform rectangular
grid.  In each collection of cells, a power law or resonance
distribution can optionally be specified.  The weight~$C(x_1,x_2)$ of
each cell is determined from the distribution of $(x_1,x_2)$-pairs in
the simulation results.  The boundaries
of the cells are simultaneously adapted numerically by the \vegas
algorithm~\cite{Lepage:1977sw,Ohl:1998jn} to minimize the variation of
these weights~$C(x_1,x_2)$.  An optional Gaussian smoothing can improve
the parametrisation in regions where the simulation produced just a
few $(x_1,x_2)$-pairs.

The two-dimensional weight distribution~$C(x_1,x_2)$ allows arbitrary correlations to be described, while the adaptable cell boundaries
and power law mappings retain the ability to describe luminosity
spectra~$D(x_1,x_2)$ that typically vary over many orders of
magnitude.  In order to be able to describe the BES, values
of~$x_{1,2}$ above~1 are supported.  The grids that have been adapted
to the simulation results are stored as ASCII files that can be read
by particle physics event generators. \circetwo is again available
both as part of \whizard and as a separate \fortran package, which can
also be linked with \cxx programs and has been integrated into
the \keyhep framework~\cite{key4hep_2024}.

\paragraph{\textmd{\madgraphee}}
The Monte Carlo event generator \madgraphee~\cite{Frixione:2021zdp}
combines the effects of BS and ISR with a convolution of a analytic
parametrisation of BS with ISR structure functions.  For this purpose,
a parametrisation that generalizes and improves
on~\cref{eq:circe1-parametrisation} has been developed.  It
reproduces the results of \guineapig simulations of single bunch
crossings well. For details, see Ref.~\cite{Frixione:2021zdp}.

\paragraph{\textmd{\kkmcee}}
The Monte Carlo Event generator \kkmcee~\cite{Arbuzov:2020coe} uses
the \foam algorithm~\cite{Jadach:1999sf,Jadach:2002kn} to describe the
output of \guineapig and combines this with a Gaussian BES.

\subsubsection{Collider designs}

In addition to the \circe descriptions of
\textmd{ILC}~\cite{Adolphsen:2013jya,Adolphsen:2013kya} beam spectra
distributed with \whizard, \circetwo descriptions are available from
the \whizard page at hepforge
\url{https://whizard.hepforge.org/circe_files/}.  \circetwo
descriptions of the beam spectra at
\textmd{CEPC}~\cite{CEPCStudyGroup:2023quu} are available from the
same place.

The more complex structure of the beam spectra at
\textmd{CLIC}~\cite{CLICCDR_vol1} was one of the motivations to go
beyond~\circe.  Therefore only \circetwo descriptions of the beam
spectra have been prepared by members of the \textmd{CLIC}
collaboration and are available from the \whizard page at hepforge.

The very moderate BS at the proposed Cool Copper Collider
\textmd{C${}^3$}~\cite{Vernieri:2022fae,Ntounis:2024gjw} has revealed
an insufficient masking of almost empty regions in the original design
of \circetwo.  Current versions of \circetwo support a \texttt{null}
map that allows to mask regions of phase space that do not contain
enough events in the \guineapig output and should be discarded.

The beam spectra descriptions for \textmd{FCC-ee}~\cite{FCC:2018evy}
benefit from the addition of the \texttt{null} map to \circetwo, since
the BS is even softer.  The width of the BS distribution and the BES are
comparable and their convolution can be fitted in one step in \circetwo. Other approaches can treat BS and BES seperately.

Work on beam spectra for the proposed hybrid linear Higgs factory
based on plasma-wakefield and RF acceleration
\textmd{HALHF}~\cite{Foster:2023bmq} is ongoing.

\subsection{\focustopic  Luminosity}
\label{sec:LUMI}
\editors{Ayres Freitas, PA}

Precision measurements of the luminosity are important for all cross section and line-shape measurements, in particular the $\PZ$ peak cross section, $\sigma_{\PZ}^0$; the total \PZ width from the line-shape of $\epem \to f\bar{f}$; the \PW boson mass and width from the line-shape of the cross section for $\epem \to \ww $ near threshold; and the total cross section for $\epem \to \PH\PZ$ (used for extracting the effective $\PH\PZ\PZ$ coupling, the total Higgs boson width, and the Higgs self-coupling). At LEP, an absolute calibration of the luminosity with a relative experimental uncertainty of $3.4 \times 10^{-4}$ was achieved~\cite{OPAL:1999clt} using small-angle Bhabha scattering. For future $\epem$ colliders, the luminosity uncertainty is likely to be the limiting factor for several of the measurements listed above, in particular on the \PZ pole, so it is crucial to robustly understand and reduce the uncertainty compared to LEP.

A realistic target for the overall luminosity calibration at the \PZ pole is $10^{-4}$ or better, whereas for the point-to-point luminosity control, i.e.\ the relative uncertainty between two nearby centre-of-mass energies or two beam polarisation settings, one would like to reach $\mathcal{O}(10^{-5})$ precision (see also e.g.\ Ref.~\cite{FCC:2018evy}). 
At intermediate energies, $\ww$ and two-fermion production are the 
highest cross section processes of interest leading 
to anticipated data sets of $\mathcal{O}(10^{7} - 10^{8})$ such events, thus 
motivating a similar $\mathcal{O} (10^{-4})$ experimental target for luminosity precision.
In particular, to obtain a precision of 300~keV on the $\PW$ boson mass from the \PW-pair-production line-shape at threshold, a control of the luminosity uncertainty at the level of a few $10^{-4}$  will be needed~\cite{Azzi:2017iih,Azzurri:2021yvl}.
At higher energies, $\sqrt{s}\, \gtrsim 400$~GeV, $\mathcal{O}(10^{-3})$ precision for the overall luminosity calibration may suffice for the physics goals \cite{BozovicJelisavcic:2013lni}.

\medskip\noindent
The physics processes used for luminosity calibration need to meet certain requirements:
\begin{itemize}
    \item ideally large rate, so as not to be statistics-limited even in small time intervals, and low backgrounds;
    \item good control of the experimental systematic uncertainties (particle ID, acceptance);
    \item reliable, high-precision theory predictions and MC tools, with negligible room for possible new physics contributions.
\end{itemize}

In the following, we summarise the state-of-the-art concerning two processes fitting this bill, namely small-angle Bhabha scattering (SABS) and di-photon production, before summarising the most important open questions.

\subsection*{Small-angle Bhabha scattering (SABS)}
The cross section for Bhabha scattering is very strongly peaked in the forward region, so that the best rate measurement can be performed with a special detector (LumiCal) at $< 100$\,mrad. 
Given the large rate of SABS, systematic uncertainties drive the precision of the integrated luminosity measurement%
, such as detector-related uncertainties, beam-related uncertainties, and uncertainties originating from physics theory and modeling and machine-related interactions. They should be preferably quantified in a full detector simulation including backgrounds in the very forward region. This calls for novel and revised studies, 
in line with the evolving design of the machine-detector interface (MDI) region.

\subsubsection*{Experimental challenges:}
\begin{itemize}
    \item Evaluating the LumiCal acceptance requires control of the detector aperture, position and alignment. Some of the requirements for mechanical precision are more stringent than others (i.e.\ inner radius), posing technological challenges for detector realisation. The uncertainties at the ILC are discussed in Refs.~\cite{stahl2005luminosity,Smiljanic:2024twn}, where each individual effect is evaluated with respect to the target of $10^{-4}$ and $10^{-3}$ precision at the \PZ-pole and higher energies. For the highest precision at the \PZ-pole, the main challenge is metrology requirements for the inner aperture of the LumiCal. The impact of relative position uncertainties between the left and right arm of the luminometer can be reduced using the method of asymmetric counting \cite{OPAL:1999clt}, where the inner and outer radii of the fiducial volume of one side of the luminometer are narrowed by few~\unit{\milli\metre} (which side is chosen on an event-by-event basis). Asymmetric counting may help to relax some alignment challenges,  yet detailed simulations with a concrete luminometer design are needed to establish the definition of the fiducial (counting) volume and impact of mechanical uncertainties on it \cite{Smiljanic:2024twn}.
    
    At FCC-ee and CEPC, the design of the MDI region requires the luminosity monitor to be placed closer to the interaction point (IP) compared to LEP or ILC, which puts even tighter requirements on the geometrical precision for the same angular acceptance uncertainty \cite{Dam:2021sdj,Smiljanic:2024koz}. Different options for the transverse placement of the LumiCal have also been studied \cite{Smiljanic:2024koz}. The following table summarises the requirements for how well the position and dimensions of the luminosity detector need to be controlled. The cited uncertainties include changes between offline metrology and operations, which was a dominant uncertainty for the inner radius acceptance at OPAL~\cite{OPAL:1999clt}. The numbers in the table below 
    correspond to the permissible tolerances to keep each individual source of uncertainty below the level of $10^{-4}$.

    \smallskip
    \begin{tabular}{|l|c|c|c|c|}
    \hline
         & LEP \cite{OPAL:1999clt} & FCC-ee \cite{FCC:2018evy} & CEPC \cite{Smiljanic:2024koz} & ILC  \cite{Smiljanic:2024twn} \\ 
         \hline
         LumiCal distance from IP [m] & 2.5 & 1.1 & 0.95 & 2.48 \\
         \hline
         Precision target & $3.4 \times 10^{-4}$ & \multicolumn{2}{c|}{$10^{-4}$ (\PZ pole)} & $10^{-4}$ (\PZ pole) \\
         &&\multicolumn{2}{c|}{}& $10^{-3}$ (0.25--1 TeV) \\
         \hline
         Tolerance for &&&& \\
         ~~~inner radius [\SI{}{\micron}] & 4.4 & 1 & 1 & 4 \\
         ~~~outer radius [\SI{}{\micron}] &  & 3 & 2 & 55 \\
         ~~~distance btw.\ two LumiCals [$\mu$m] & $\mathcal{O}(100)$ & $<100$ & $<100$ & 200 \\
         \hline
    \end{tabular}
    \medskip

    Any eventual change in the luminometer design or positioning towards smaller polar angles will require control of the inner radius even below micrometre precision.


    \item The luminosity measurement is affected by multiple beam properties and delivery to the IP, including beam-energy asymmetry, energy calibration, IP displacements due to finite transverse beam size and beam synchronisation, and beam-spread effects. These issues have been discussed in Ref.~\cite{Smiljanic:2020wvt} for circular colliders and Refs.~\cite{Rimbault:2007zz,BozovicJelisavcic:2013lni,Smiljanic:2024twn} for linear colliders.
    Energy calibration is important because the selection of Bhabha events over background (e.g.\ from two-photon processes) requires accurate calibration of the LumiCal energy scale. In addition, the Bhabha scattering rate depends on the beam energy, and the luminosity spectrum uncertainty propagates to the integrated luminosity uncertainty (as a potentially limiting factor for linear colliders due to beamstrahlung it is discussed in Ref.~\cite{Lukic:2013fw}). 

    \item SABS receives backgrounds both from physics processes and machine effects. For the latter, incoherent photon conversion to $\epem$ pairs is of importance at linear colliders where beamstrahlung is a relevant source of photons influencing the luminometer occupancy, in particular at higher centre-of-mass energies. The two-photon (Landau--Lifshitz) process as a source of physics background should also be considered. At linear colliders this is discussed in Refs.~\cite{BozovicJelisavcic:2013lni,Abramowicz:2010bg}, showing that this background contributes only at the per-mille level or less.
    
    \item Beam-beam interactions have an impact on SABS through (i) beamstrahlung effects that modify the collision energy spectrum, and (ii) electromagnetic deflection of the outgoing $\Pepm$, the latter being pronounced at lower centre-of-mass energies (\PZ pole). The effects have been studied at linear colliders~\cite{BozovicJelisavcic:2013lni,Lukic:2013fw} and at FCC-ee~\cite{Voutsinas:2019hwu}. The final-state particle deflection leads to a $\mathcal{O}(10^{-3})$ correction~\cite{Voutsinas:2019hwu}, and additional complications arise 
    due to the finite beam-crossing angle at Higgs factories, but at the same time it opens the opportunity of measuring the focusing effect through the acollinearity distribution of Bhabha events.
    
\end{itemize}

\subsubsection*{Theoretical challenges:}

To obtain the luminosity from a measurement of the SABS rate, one needs precise theoretical calculations of the differential Bhabha cross section. The challenges for achieving a theory precision below $10^{-4}$ have been discussed in detail in Refs.~\cite{Jadach:2018jjo,Jadach:2021ayv,Skrzypek:2024gku}.
Bhabha scattering is mostly a QED process, i.e.\ higher order corrections can be reliably calculated. Implementation of these corrections in MC tools is complicated but not a fundamental obstacle.

Production of additional fermions has a significant impact on the simulated Bhabha rates in the 
LumiCal fiducial region. The technology for computing 4-fermion processes at NLO (see \mbox{e.g.} Ref.~\cite{Denner:2005fg}) and 6-fermion processes at LO exists, but these still need to merged in a coherent MC simulation. Inclusion of these contributions should reduce the uncertainty from fermion pair production below $10^{-4}$.

NLO electroweak corrections are missing in existing Bhabha MC tools, but they are straightforward to implement.
Corrections from collinear photon emission and EW higher orders are enhanced at higher energies, thus increasing the theory uncertainty for the luminosity determination there. However, they stay safely below the $10^{-3}$ level for $\sqrt{s}$ up to 1~TeV~\cite{Jadach:2021ayv}.

\begin{figure}
\begin{subfigure}{0.395\textwidth}
\begin{center}
\includegraphics[width=0.45\hsize]{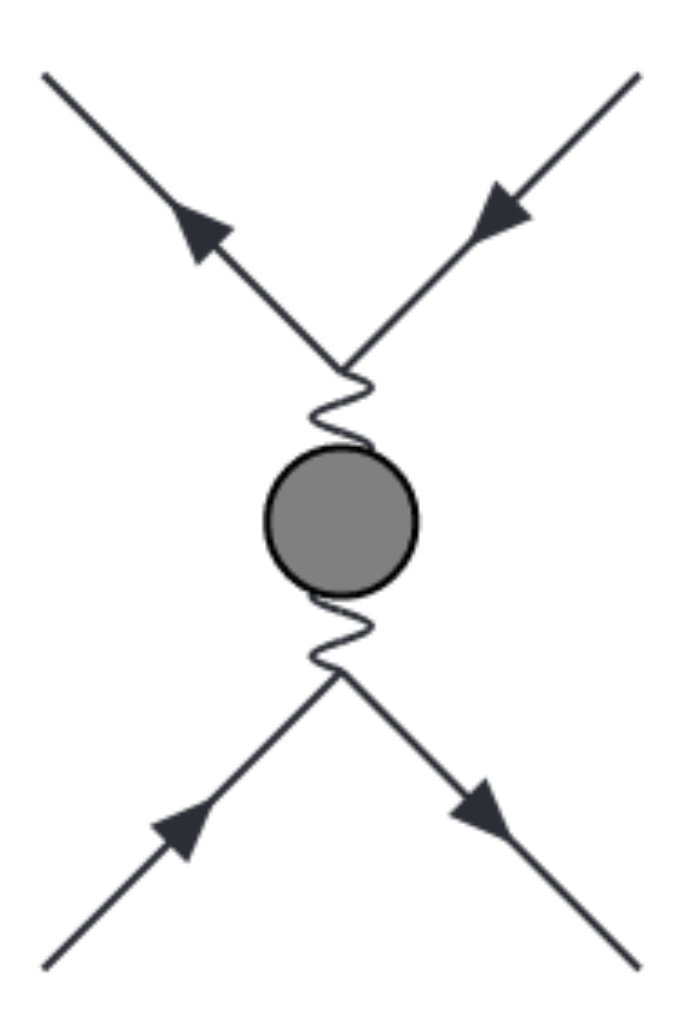}
\end{center}
\caption{\label{fig:lumi:vacpol}}
\end{subfigure}
\begin{subfigure}{0.595\textwidth}
\begin{center}
\includegraphics[width=0.45\hsize]{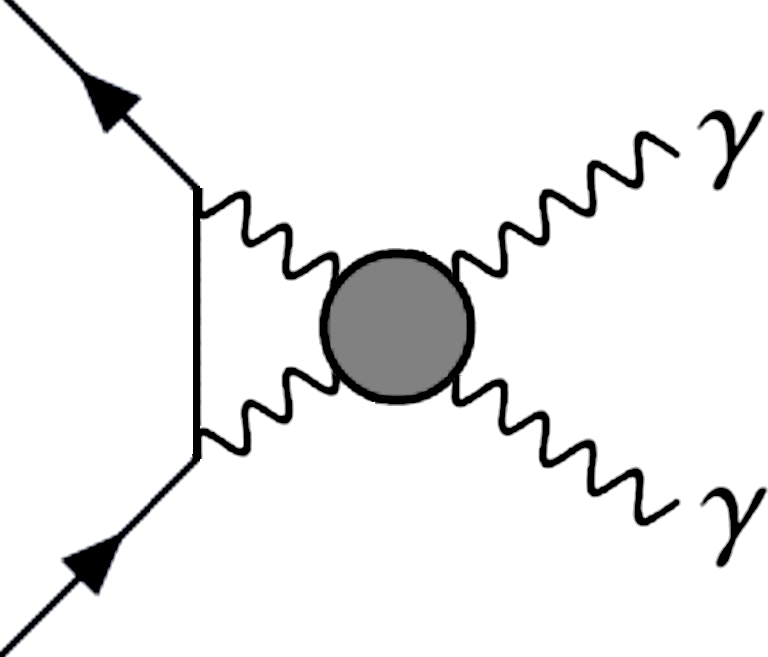}
\end{center}
\caption{\label{fig:lumi:lbl}}
\end{subfigure}
\caption{(a) Vacuum polarisation contribution to Bhabha scattering (from Ref.~\cite{Janot:2019oyi}). (b) Light-by-light correction to di-photon production.}
\label{fig:lumi}
\end{figure}

A significant source of uncertainty arises from hadronic vacuum polarisation in t-channel photon exchange (Fig.~\ref{fig:lumi:vacpol}). This contribution needs to be extracted from data for $\epem \to \text{hadrons}$ or from lattice QCD. With future data (Belle~II, BES~III, CMD-3, SND) it is expected that the resulting uncertainty for the SABS rate can be reduced below the $10^{-4}$ level~\cite{Jadach:2018jjo}, but it may be a limiting factor in the achievable precision.

Thus in summary one obtains the forecast given in \cref{theorSABS} for the theoretical precision for SABS, shown there for the example of FCC-ee at $\sqrt{s}=m_{\PZ}$ \cite{Skrzypek:2024gku}.

    \begin{table}[hb]
    \begin{center}
    \begin{tabular}{|l|c|}
    \hline
    Type of correction / error & Estimate \\ 
    \hline
    ${\cal O}(L_e^2\alpha_\text{QED}^3)$ \quad [$L_e = \ln(|t|/m_{\Pe})$] & $0.1\phantom{0}\times 10^{-4}$ \\
    ${\cal O}(L_e^4\alpha_\text{QED}^4)$ & $0.06\times 10^{-4}$ \\
    ${\cal O}(L_e^0\alpha_\text{QED}^2)$ & $0.17\times 10^{-4}$ \\
    Hadronic vacuum polarisation & $0.6\phantom{0}\times 10^{-4}$ \\
    Extra fermion pairs & $0.27\times 10^{-4}$ \\
    ${\cal O}(\alpha_\text{EW}^2)$ & $0.3\phantom{0}\times 10^{-4}$ \\
    Interference of $\PGg$ radiation from $\Pepm$ & $0.1\phantom{0}\times 10^{-4}$ \\
    \hline
    Total & $0.76 \times 10^{-4}$ \\
    \hline
    \end{tabular}
    \caption{Forecasted theoretical precision for SABS, shown here for the example of FCC-ee at $\sqrt{s}=m_{\PZ}$ \cite{Skrzypek:2024gku}. \label{theorSABS}}
    \end{center}
    \end{table}

There is a variety of available MC tools for the simulation of SABS \cite{Jadach:1996is,Balossini:2006wc,Balossini:2006wc,Arbuzov:2005pt,Banerjee:2020rww,Banerjee:2021mty} that all include NLO QED corrections and various subsets of higher-order corrections (see Ref.~\cite{deBlas:2024bmz} for more information).

While SABS is strongly dominated by QED contributions, it is worth checking whether any contamination of new physics could be present at the level of precision for future $\epem$ colliders. Light d.o.f. far below the electroweak scale are not expected to contaminate the luminosity at the foreseen precision. Therefore, assuming that the new particles are heavier than the centre-of-mass energy, their contribution can be described in terms of dimension-6 operators that modify the $\PZ\Pe\Pe$ couplings or that lead to new 4\Pe interactions. The numerical impact of these operators on SABS has been evaluated using the packages FeynArts \cite{Hahn:2000kx}, SmeftFR \cite{Dedes:2023zws} and BabaYaga \cite{Balossini:2006wc}. When varying the operators' Wilson coefficients within their currently allowed ranges \cite{Falkowski:2017pss}, their relative contribution was found to be as shown in \cref{contribSABS} \cite{Chiesa:2025snt}.
    
    \begin{table}[ht]
    \begin{center}
    \begin{tabular}{|l|c|r|}
    \hline
    Luminometer polar angular coverage & & max.\ rel.\ change of $\sigma_\text{SABS}$ \\
    $[\theta_\text{min},\theta_\text{max}]\;[\unit{\milli\radian}]$ & $\sqrt{s} \; [\unit{\giga\electronvolt}]$ &  [95\% CL] \\[.5ex]
    \hline
    [25,60] \ (LEP, similar to ILC) & 91 & $2 \times 10^{-5}$ \\
    \hline
    [64,86] \ (FCC-ee, similar to CEPC) & 91 &  $7.5 \times 10^{-5}$ \\
    & 240 & $5 \times 10^{-4}$ \\
    \hline
    \end{tabular}
    \caption{Impact of potential new-physics dimension-6 operators on SABS \cite{Chiesa:2025snt}.\label{contribSABS}}
    \end{center}
    \end{table}

As can be seen from the table, the impact of these new-physics operators is not negligible compared to the luminosity precision targets ($10^{-4}$ at the \PZ-pole, better than $10^{-3}$ at \SI{240}{\giga\electronvolt}), with the largest contribution stemming from 4\Pe operators. On the other hand, measurements at the future \epem colliders themselves should allow one to constrain these operators much more tightly. 
A possible solution to this issue was proposed in Ref.~\cite{Chiesa:2025snt}, where luminosity-independent observables are studied to reduce the uncertainties from heavy NP. In particular, for a run around the \PZ-peak, the forward-backward asymmetry $A_\text{FB}(\sqrt{s})$ is found to have a constraining power of around one order of magnitude on contact operators WCs, representing the bulk of the NP uncertainty. In the scenario of a high-energy run at $\sqrt{s}=250 \rm{GeV}$, a new asymmetry was proposed that can constrain the uncertainty of WCs at the level of $10^{-3}$, thus reducing their effect to a negligible level.


\subsection*{Di-photon production}
Di-photon production ($\epem \to \PGg\PGg$) is also a QED-dominated process that avoids some of the challenges of SABS, in particular the severe metrology requirements, final-state deflection issues, and the significant impact of the hadronic vacuum polarisation.
Use of this channel is encouraged by the very high luminosity of circular \epem colliders in their low energy range, in particular at the Z pole.

\subsubsection*{Experimental challenges:}
\begin{figure}
\centering
\includegraphics[width=0.55\textwidth]{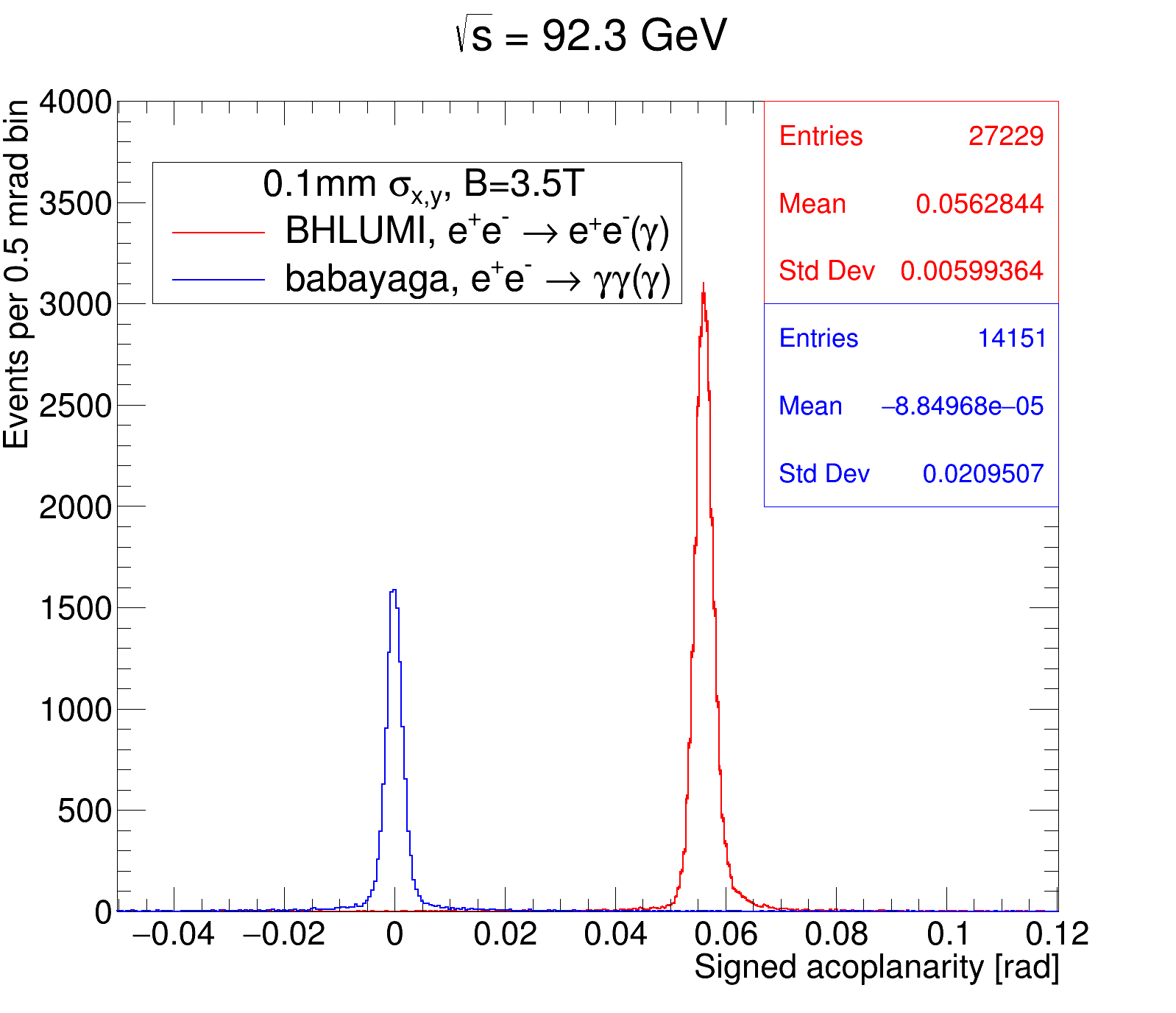}
\caption{Signed acoplanarity distributions at $\sqrt{s}=92.3$~GeV for SABS (generated with BHLUMI \cite{Jadach:1996is}) and $\epem\to\PGg\PGg$ (generated with BabaYaga \cite{Balossini:2008xr}). The relative normalisation is arbitrary. Detector resolution effects are represented by smearing the position of the electromagnetic showers by 0.1~mm in the $x$/$y$-directions. The detector solenoid field is 3.5~T.}
\label{fig:ggaco}
\end{figure}

\begin{itemize}
\item Statistical precision: The cross section for di-photon production is smaller than Bhabha scattering by about two orders of magnitude. To isolate a pure signal sample, it is necessary to separate photons from electrons/positrons. Projected detectors have tracker coverage down to very low forward angles ($|\cos\theta | \gtrsim 0.995$) which allows the cut on the minimum acceptance angle to be defined freely in order to optimize the combined statistical and systematic precision. For projected running scenarios for the Z pole (45 fb$^{-1}$) and the \PH\PZ run (10 fb$^{-1}$), statistical precisions of $2\times 10^{-5}$ and $1\times 10^{-4}$, respectively, can be envisaged. The following discussions of systematics assume a target luminosity precision of $10^{-4}$, but further reductions to the $2 \times 10^{-5}$ level may be achievable with more refined detector design.
\item Backgrounds: Bhabha scattering is an important background to di-photon production, with a cross section that is ${\cal O}(100)$ times larger. A detailed study of this background and the di-photon event selection is currently not available. However, assuming conservatively that the Bhabha background can be controlled with 1\% precision, it would need to be reduced by a factor $10^4$ for an overall luminosity target precision of $10^{-4}$ from di-photon production. This translates to a factor $10^2$ per track, which should be fairly easily doable in the central tracking region. Backgrounds from neutral pion production are expected to be very small.
\item Acceptance/metrology: For $10^{-4}$ precision, the angular acceptance needs to be determined to 30--50~$\upmu$rad \cite{Dam:2020fccws,BlondelDamLumiDet} which is much looser than the acceptance precision required for SABS. However, here this is needed for the central detector, which consists of several components, with potential cracks, etc. 
One possibility under study for FCC-ee is to use diphoton events to measure the calorimeter acceptance to the required level in situ, exploiting the large crossing angle of 30 mrad.  
\item To increase the usable event rate, it would be desirable to extend the fiducial acceptance of $\epem \to \PGg\PGg$ to the angular region covered by the luminosity detector (few degrees). This requires a far-forward detector that is capable of discriminating $\epem \to \PGg\PGg$ events from SABS ($\epem \to \epem)$ events. One concept for this purpose is a high-granularity calorimeter with high angular, position and energy resolution \cite{wilsonecfa}. The bending of $\Pepm$ tracks in the magnetic field of the detector solenoid will lead to an azimuthal acoplanarity of several tens of mrad for forward scattering Bhabha events (e.g.\ 57 mrad for $\sqrt{s}=m_{\PZ}$ for the ILD solenoid and luminosity detector geometry), where there is no acoplanarity for $\PGg\PGg$ events. A luminosity detector with \SI{0.1}{\milli\metre} positional resolution would lead to excellent separation of $\PGg\PGg$ and SABS events, see Fig.~\ref{fig:ggaco}. Initial design studies \cite{wilsonecfa} indicate that such resolution is achievable, together with excellent energy resolution of $<4\%/\sqrt{E/\unit{\giga\electronvolt}}$. Such a detector will also be useful for the analysis of the standard SABS process itself, in particular for controlling beam-induced deflection of the outgoing $\Pepm$ particles.
\end{itemize}

Detailed studies should be carried out, especially considering the issue of the separation of the \epem and $\gamma\gamma$ processes, aiming at optimizing selection cuts towards a detailed design of the detector.

\subsubsection*{Theoretical challenges:}

The photon vacuum polarisation only appears one order higher than in SABS, i.e.\ at NNLO, and thus its uncertainty is negligible~\cite{CarloniCalame:2019dom}. However, there are also hadronic light-by-light (lbl) scattering corrections (Fig.~\ref{fig:lumi:lbl}), which also enter at NNLO, but which is more difficulty to evaluate precisely, i.e.\ it is likely to have a much larger relative uncertainty. This contribution and its impact on $\PGg\PGg$ production is currently unknown, but a rough estimate could be achieved by observing that it is expected to be dominated by tensor states in the s-channel and single- and double-pion states in the t-channel \cite{Aliberti:2024fpq}. Future lattice QCD studies could also be used to evaluate the lbl contribution.

With a large-angle requirement ($|\cos\theta| \lesssim 0.95$), EW corrections are relatively more important than for SABS~\cite{CarloniCalame:2019dom}. By including NLO and leading NNLO EW corrections, the uncertainty from higher-order EW corrections can likely be rendered sub-dominant, but this requires further investigation.

There are only few MC tools currently available for the simulation of $\epem \to \PGg\PGg$ with higher-order corrections, including NLO QED \cite{Berends:1980px}, and NLO QED+electroweak and higher-order leading-log effects through parton showering  \cite{Balossini:2008xr,CarloniCalame:2019dom}. Efforts are underway to calculate and implement NNLO corrections \cite{Banerjee:2020rww}.

The possible contamination from new physics has been studied in Ref.~\cite{Maestre:2022cvs}. Under the assumption that the new particles have masses $\Lambda \gg \sqrt{s}$, it was found that the leading contribution stems from dimension-8 $\Pe\Pe\PGg\PGg$ operators suppressed by $\Lambda^{-4}$. The contributions of operators producing anomalous $\PZ\PGg\PGg$ couplings are suppressed due to the Landau--Yang theorem. The $\Pe\Pe\PGg\PGg$ operators enter into the $\epem \to \PGg\PGg$ cross section with a characteristic angular dependence,
\begin{align}
    \frac{\text{d}\sigma_\text{BSM}}{\text{d}\cos\theta} = 
    \frac{\text{d}\sigma_\text{SM}}{\text{d}\cos\theta} \times \biggl[ 1+ \frac{c_8 s^2}{\Lambda^4}\sin^2\theta \biggr],
\end{align}
which can be used to constrain their size \emph{in situ} from data. Measurements at higher collider energies can achieve even more stringent bounds. In this way it will be possible to limit any new physics safely below the level of $10^{-4}$ at $\sqrt{s}\sim m_{\PZ}$.

\subsection*{Other processes and discussion}

Besides $\epem \to \epem$ and $\epem \to \PGg\PGg$, one could also consider $\epem \to \epem\PGmp\PGmm$ for luminosity measurements. Theoretically, this process could be described through emission of nearly on-shell photons from the $e^\pm$ beams followed by a $\PGg^{(*)}\PGg^{(*)} \to \PGmp\PGmm$ hard scattering process.

The measured cross section at $\sqrt{s}=m_{\PZ}$ amounts to about \SI{2.5}{\nano\barn} for 
events with a  dimuon invariant mass in the  $0.5<m_{\PGmp\PGmm} <45$~GeV range~\cite{L3:1997acq}.
This could in principle allow a luminosity measurement at the $10^{-5}$ level with an integrated luminosity at the \PZ pole of \SI{10}{\per\atto\barn} or larger.
However, this method requires more study to understand both the experimental and theoretical limitations to the achievable precision.

Overall, SABS is advantageous for the point-to-point luminosity control, due its higher yields. However, $\epem \to \PGg\PGg$ and $\epem \to \epem\PGmp\PGmm$ could be promising options for the determination of the total integrated luminosity, with more robust control of systematic uncertainties. Further studies are needed, with detailed simulations and full designs of the luminosity monitors for different collider setups and different detector concepts. For the point-to-point luminosity control, correlations of systematic uncertainties between different nearby centre-of-mass energies need to be investigated.


On the theory side, improved MC tools are crucial, in particular including the production of additional fermion pairs, which are important for understanding the experimental acceptance and signal selection. For beam-induced effects, including beamstrahlung, it will be important to understand how much one has to rely on simulations, or whether these effects can be determined from in-situ measurements.

\subsection{Generator benchmarks\label{sec:techbench}}
\editor{Alan Price}

Monte Carlo (MC) event generators play a crucial role in modern particle physics and will undoubtedly be a cornerstone of any future collider program, as discussed in~\cref{sec:gen}. Unlike the generators developed during the LEP era, which mostly relied on the explicit implementation of processes, contemporary MC tools rely heavily on automated algorithms. This shift has enabled them to handle a broader spectrum of processes with greater flexibility and sophistication.

While MC generators can differ in aspects there are certain foundational areas where consensus is expected. These shared aspects form the basis of the comparisons described in the following sections. The objective of these comparisons is not to evaluate the overall physics potential of the various MCs, but rather to assess their performance and consistency in domains where agreement is both anticipated and necessary. 

By focusing on areas where discrepancies are unlikely, this study aims to provide valuable feedback to help identify technical issues, validate underlying algorithms, and ensure that these tools meet the high precision requirements of future experimental programs. The comparisons are performed at LO accuracy. The tools developed for this purpose allowthis endeavour to be pursued and extended in the future  to higher precision including ISR and higher order corrections. This will be extremely important in the long term perspective, in view of estimating the uncertainties of theoretical predictions at the fully differential level, which can be obtained by critical comparisons between different generators. A first example of the reliability of this program has been obtained with the exhaustive tuned and unleashed comparisons between generators performed at LEP for $\epem \to \Pf \bar{\Pf}$~\cite{TwoFermionWorkingGroup:2000nks,,Bardin:1997xq,Bardin:1999gt},  $\epem \to 4\Pf$~\cite{Beenakker:1996kt,Bardin:1997gc,Boudjema:1996qg,Grunewald:2000ju} and small angle Bhabha scattering~\cite{Jadach:1996gu,Arbuzov:1996eq}.

\subsubsection{Benchmarking in \keyhep}

A key aspect of the benchmarking is the reproducibility of results. To ensure that the results are robust and repeatable, the Python package, k4GeneratorsConfig, which automates the benchmarking process for MC event generation, has been developed. The central philosophy behind this package is that the physics tests should remain independent of the specific MC generator being used. With this goal in mind, the tool accepts universal physics inputs, which are then systematically translated into input formats required for simulation as described in~\cref{sec:softwareeco,sec:simulation}.

The physics information provided to the tool can range from simple details like the collision energy to more complex specifications, such as beam types, particle distributions, and the particular processes to simulate. This abstraction ensures that users can define the physics of their simulation straightforwardly and consistently without needing to account for the idiosyncrasies of different MC generators. Once the input is provided, the package generates runcards or configuration files tailored to the syntax, parameters, and requirements of the specific generators.

k4GeneratorsConfig supports a variety of MC generators relevant for $\epem$ colliders, making it a versatile solution for diverse simulation needs. It not only handles the physics parameters but also configures essential simulation settings, such as random seed values, number of events, and technical details required by the generators. This automation ensures compatibility across generators and drastically reduces the likelihood of human error, providing a reliable way to achieve reproducible results. 

The tool's modular architecture also enables seamless extensibility. New generators can be added by defining their respective templates or translation rules, ensuring that the package remains adaptable as new MC generators or versions are introduced.
Lightweight inheritance is used for tasks common to all generators. Furthermore, k4GeneratorsConfig supports batch processing, allowing users to configure and benchmark multiple generators or physics scenarios efficiently.


The package k4GeneratorsConfig\footnote{https://github.com/key4hep/k4GeneratorsConfig} can be used both standalone and as an integrated component of the \keyhep framework. As a standalone tool, it focuses on generating the required runcards for various MC generators based on user-provided physics inputs. Within the \keyhep framework, the tool goes a step further by generating scripts to automatically execute the corresponding generators, streamlining the simulation process. Additionally, for users working within the framework, k4GeneratorsConfig offers the capability to convert all generated MC events into \edmhep described in~\cref{sec:softwareeco} and other formats by providing a format converter based on the Writer class in the HepMC package~\cite{Buckley:2019xhk}. This ensures compatibility with downstream analysis workflows and promotes consistency across simulations.

The tests for the k4GeneratorsConfig package are run automatically as part of the \keyhep Continuous Integration (CI) pipeline. This integration ensures that the software is consistently validated with every update or change. By leveraging the CI infrastructure, the package undergoes a series of automated tests that check its functionality, compatibility, and performance across different environments. These tests include validating the generation of runcards, checking the conversion of physics information into generator-specific inputs, and ensuring that the results are reproducible and consistent across different MC generators. The CI system runs these tests on multiple platforms, guaranteeing that any issues are detected early in the development process. 


\subsubsection{Benchmark processes}

The results of the cross section calculations generated by the k4GeneratorsConfig tool provide a comprehensive comparison of simulation outputs across various MC generators. At LO parton level results all generators are expected to agree.
The so-called $G_\mu$ scheme was used for with the following electroweak input parameters,

\begin{center}
  \begin{tabular}{rclrcl}
    $m_{\PW}$ & = & 80.419\,\text{GeV}&
        $\Gamma_{\PW}$ & = & 2.085\,\text{GeV} \\
    $m_{\PZ}$ & = & 91.1876\,\text{GeV}&
    $\Gamma_{\PZ}$ & = & 2.4952\,\text{GeV} \\
    $m_{\PH}$ & = & 125.0\,\text{GeV}&
    $\Gamma_{\PH}$ & = & 0.00407\,\text{GeV} \\
    $m_{\Pe}$ & = & 0.000511\,\text{GeV}&
    \\
  $G_{\PGm}$ & = & $1.1663787\times 10^{-5}\,\text{GeV}^{-2}$ & $\alpha_{G_{\PGm}}^{-1}$ & $=$ & 132.511 \\
 \end{tabular}
\end{center}
For the case of $\PGg\PGg$ production with explicit photons in the final state, the $\alpha(0)$ scheme~\cite{Denner:2019vbn} was used.

The simplest two-fermion final state consists of a muon pair $(\PGmp\PGmm)$ at different energies. For this simple setup, the full phase-space is used without any fiducial cuts. The total cross section across the four generators agrees at both energies
as shown in \cref{table:crossSectionsmumu}:
\begin{table}[t!]
	\centering
	\begin{tabular}{c | c  | c }
		 $\epem \rightarrow \PGmp\PGmm$ & 91.2 GeV & 250 GeV\\
		\hline
    KKMC        & 2010.6(2)  & 1.767(8) \\
    Madgraph    & 2010.5(3)  & 1.7665(2) \\
    Sherpa      & 2010.3(1)  & 1.7662(2) \\
    Whizard     & 2010.3(2) & 1.7661(2) \\
	\end{tabular}
	\caption{
     Cross sections for $\epem\rightarrow \PGmp\PGmm$ at different \roots in pb.
      \label{table:crossSectionsmumu}
    }
\end{table}
\begin{figure}
    \centering
    \includegraphics[width=0.45\linewidth]{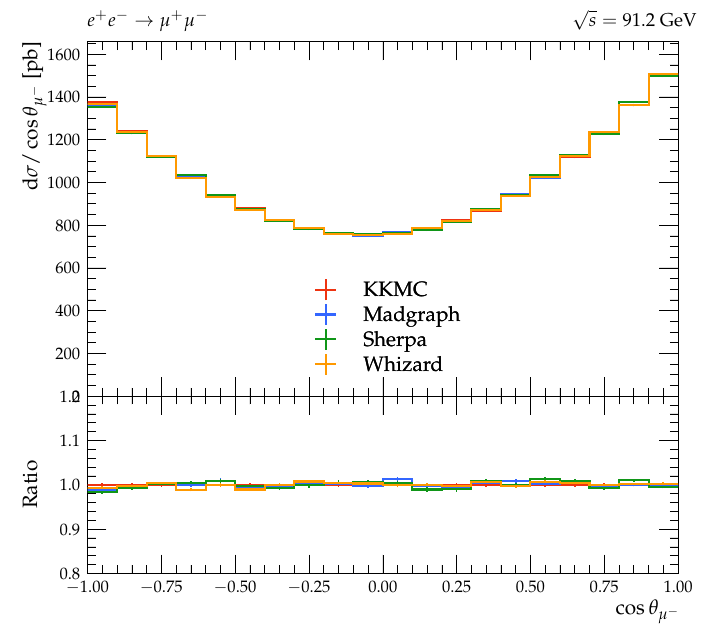}
    \includegraphics[width=0.45\linewidth]{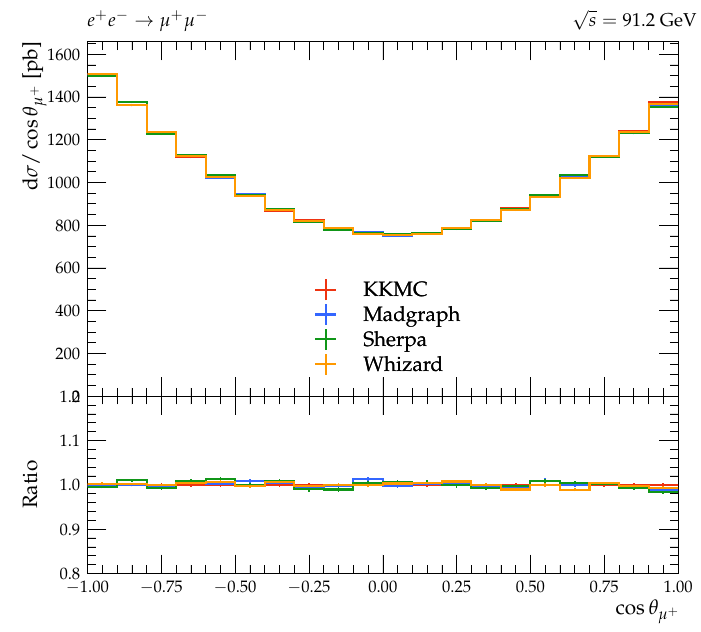}
    \includegraphics[width=0.45\linewidth]{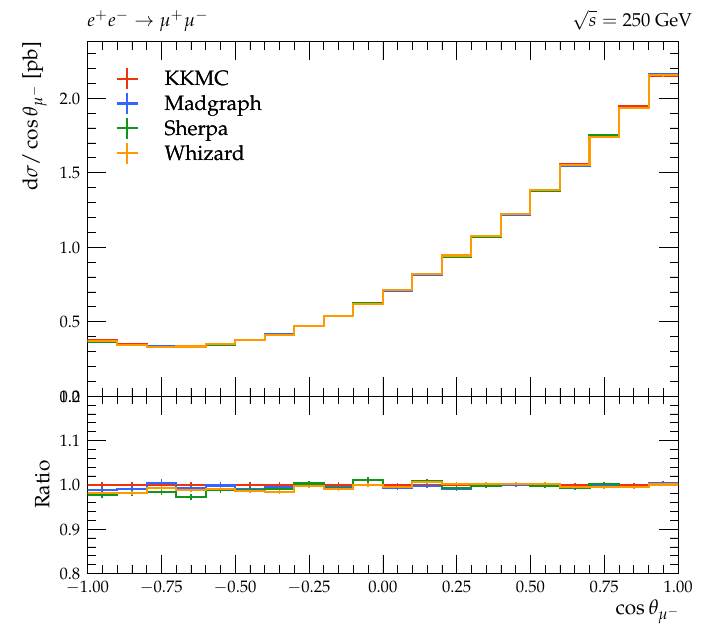}
    \includegraphics[width=0.45\linewidth]{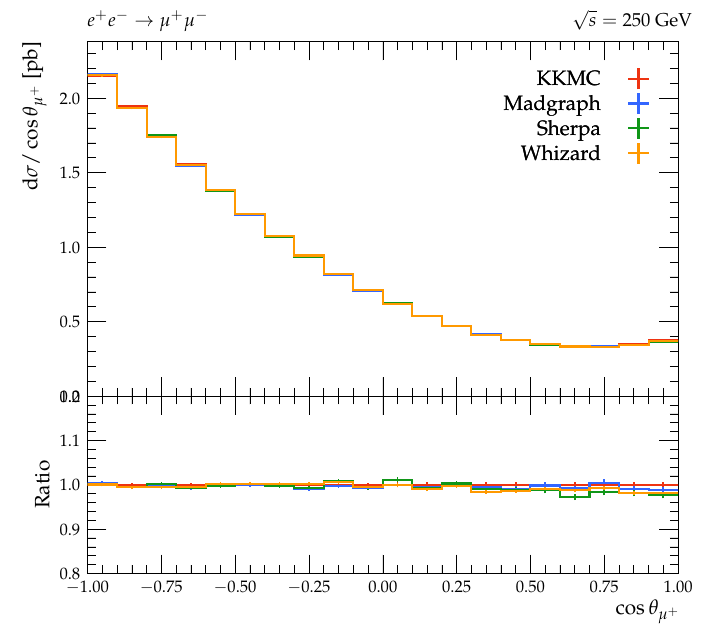}
    \caption{Angular distribution of final-state leptons in $\epem\rightarrow \PGmp\PGmm$.}
    \label{fig:AngularMuMu}
\end{figure}

At the differential level excellent agreement in the angular distributions as show in \cref{fig:AngularMuMu} is observed
as well. On the top row, a somewhat symmetrical distribution while at higher
energies is observed while the effect of the interference become more noticeable. In terms of benchmarking, this also ensures that all generator setups are consistent with respect to the beam axis.

\begin{table}[t!]
	\centering
	\begin{tabular}{c | c | c }
		$\epem \rightarrow \PGg \PGg$ & 91.2 GeV & 250 GeV\\
		\hline
    Babayaga    & 48.39(4) & 6.44(2)  \\
    Madgraph    & 48.38(2) & 6.43(1) \\
    Sherpa      & 48.38(4) & 6.43(1) \\
    Whizard     & 48.31(5) & 6.43(1) \\
	\end{tabular}
	\caption{
     Cross sections for $\epem \rightarrow \PGg \PGg$ at different \roots in pb.
      \label{table:crossSectionsGammaGamma}
    }
\end{table}

The next example is the comparison of $\epem\rightarrow \PGg\PGg$ between the general purpose generators and the dedicated \babayaga \cite{CarloniCalame:2019dom} code. As shown in~\cref{sec:LUMI} this process is important for measuring luminosity at future lepton colliders. In this example, because of the external photons the $\alpha(0)$ input scheme is used. This change of parameters is completely supported by k4GeneratorsConfig, so consistent parameter changes across all the generators are ensured. 

\begin{table}[t!]
	\centering
	\begin{tabular}{c | c  }
		Generator & 250 GeV\\
		\hline
    Madgraph    & 7.72(3)  \\
    Sherpa      & 7.71(2)  \\
    Whizard     & 7.69(4)  \\
	\end{tabular}
	\caption{
     Cross sections for $\epem \rightarrow \PH \PGmp\PGmm$ in fb.
      \label{table:crossSectionsHMuMu}
    }
\end{table}

\begin{figure}
    \centering
    \includegraphics[width=0.5\linewidth]{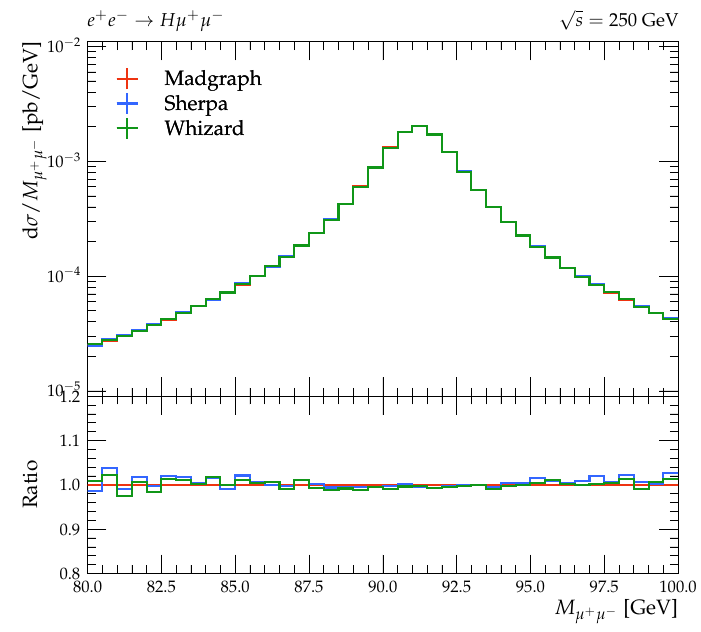}
    \caption{Invariant mass distribution of final state leptons.}
    \label{fig:InvariantMassHmumu}
\end{figure}

The final process is $\epem\rightarrow \PH \PGmp\PGmm$ which is one of the so-called ``golden channels'' for Higgs production at \epem colliders. The input cards generated by k4GeneratorsConfig have produced consistent results across the three general-purpose MC generators. In~\cref{table:crossSectionsHMuMu} the total inclusive cross section for this process at $\roots = \SI{250}{GeV}$ is presented where all the generators agree with each other within the statistical uncertainty. In~\cref{fig:InvariantMassHmumu} the invariant mass distribution of the final-state lepton pair is shown; an observable that will be crucial to the extraction of the Higgs mass for which excellent agreement between the generators is observed. 

\subsubsection{Outlook}

The k4GeneratorsConfig helps to provide a pivotal role in the \keyhep toolchain by automating and streamlining the benchmarking of MC generators into one simple tool. By leveraging this tool, benchmark processes can be seamlessly integrated into the broader \keyhep ecosystem, enabling consistent, reproducible, and comprehensive validation of high-energy physics simulations. This integration will allow the community to quickly and accurately evaluate generator performance for diverse processes, ranging from simple cross section to differential distributions.

Looking ahead, k4GeneratorsConfig is designed with extensibility in mind, allowing it to be easily adapted to accommodate new MC generators as they are developed or updated. By providing a universal interface for physics input, the tool can quickly incorporate additional generators into the benchmarking framework without requiring significant modifications to the existing system. As the tool develops, it will foster a standardised approach to generator benchmarking. This standardisation will not only enhance the comparability of results across different generators but also provide a robust foundation for the development of next-generation simulation workflows in high-energy physics. 
\subsection{Simulation\label{sec:simulation}}
\editors{Andre Sailer, Brieuc Francois, Daniel Jeans}


To estimate realistic expectations for the physics potential of future collider experiments, Monte Carlo simulations taking detector effects into account are essential. To estimate detector efficiencies, reconstruction performance, and to account for background effects, detailed descriptions of the detector geometries are required. This section will report on the tools that are available to perform simulations, both detailed as well as in parametrised form, and include an overview of the detector models that were studied as part of this initiative. The section highlights that common geometry and simulation tools allow sharing of developments and reduced efforts for the community.

\subsubsection{Simulation tools}
\paragraph{\delphes}

\delphes~\cite{delphes} is a fast simulation framework that models detector responses using parametrised descriptions, bypassing the need for full \geant\ simulations. It provides modular components for tracking, calorimetry, and particle identification, enabling rapid prototyping, performance studies, and large-scale event simulations.

The \delphes\ simulation includes a track propagation system embedded in a magnetic field, electromagnetic and hadron calorimeters, and a muon identification system. From these simulated detector responses, physics objects are reconstructed for data analysis. These include basic elements such as tracks and calorimeter deposits, as well as higher-level objects like isolated leptons, jets, and missing momentum. Additionally, \delphes\ includes a basic particle-flow reconstruction algorithm, which optimally combines track and calorimeter information to form particle candidates. These candidates serve as inputs for jet clustering, missing energy calculation, and isolation variable determination.

The performance of individual detectors is parametrised through the so-called \delphes\ cards, which define the sequence of modules to execute and specify particle-dependent efficiencies and resolutions as functions of variables like momentum and rapidity. A wide selection of \delphes\ cards is available in the central \delphes\ GitHub repository~\cite{delphesCards}, including parameterisations for most future HTE factory detectors.

Recent developments in the context of the FCC Feasibility Study Report \cite{FCC-FSR-Vol1, FCC-FSR-Vol2, FCC-FSR-Vol3} have extended \delphes\ with several new features:
\begin{itemize}
	\item  A track reconstruction algorithm that calculates track parameters and their covariance matrices for arbitrary tracking geometries, accounting for effects like multiple scattering.
    \item A module for deriving time-of-flight measurements.
    \item A tool to parametrise the number of ionisation clusters produced by charged particles in gaseous detectors, based on detailed simulations using \textsc{Garfield}~\cite{garfield}.
\end{itemize}

Integration into the \keyhep framework is achieved via the \texttt{k4SimDelphes}~\cite{k4simdelphes} package. This package wraps core \delphes\ functionalities while providing additional capabilities to convert outputs into the \edmhep data model. Standalone tools (\textsc{Delphes*\_EDM4hep}) simplify usage of these extensions, and the \texttt{k4SimDelphesAlg} Gaudi algorithm enables seamless integration of \delphes\ simulations into a complete \keyhep\ workflow.

\paragraph{SGV}

SGV \textit{La Simulation \`a Grande Vitesse} \cite{Berggren:2012ar,SGV} is a
covariance matrix machine,
where the full covariance matrix is constructed from
the generated particles and the detector layout.
As such, for charged particles, it is not a parametrised simulation, but rather a
full simulation, but at a higher level of abstraction than
the microscopic-level simulation used in \geant.
The response of calorimeters, on the other hand, is a parametrised simulation.
The detector geometry is distilled to a simplified, but still accurate, format,
as concentric cylinders in the barrel part of the detector, and as planes in the
endcaps.
To each of the layers, measurement and material properties are attached.
The latter comprises all relevant physical properties of the layer,
such as the geometry, the radiation- and interaction-length, the atomic number, and,
for gaseous detectors, diffusion and mobility constants.

Generated events are either read in from files in \lcio, \stdhep, \hepmc, or \guineapig\ format,
or internally generated.
If needed, the events are boosted to the detector frame.
For charged particles,
the four-momenta, production vertex and detector magnetic field, are used to calculate
the five helix parameters fully representing each particle.
The intersections of the track-helix with all detector layers
up to the end of the tracking system
are calculated analytically.
The local coordinates of the helix are stored in order of
track-length.
From the list of intersected surfaces, the covariance matrix at
the perigee is calculated~\cite{Kalman:1960mft,Billoir:1989mh,Billoir:1983mz} according to the following procedure.
The helix is followed from the outside, starting at the outer-most tracking-detector 
surface. At each recorded intersection, the measurements
the surface contributes are added in quadrature to the relevant elements of the
covariance matrix.
The effect of multiple-scattering~\cite{Lynch:1990sq} at the surface is added to the relevant
elements of the weight matrix (the inverse of the covariance matrix).
The matrix is then once again inverted, and translated
along the helix (in five-dimensional helix-space) to the next
intersected surface, and the procedure is repeated.
This continues until the mathematical
surface representing the point of closest approach is reached.
As each track is followed through the detector, the information on hit-pattern 
is automatically obtained, and is made accessible to later analysis. 

\begin{figure}[t]
  \centering
  \includegraphics[width=0.49\textwidth]{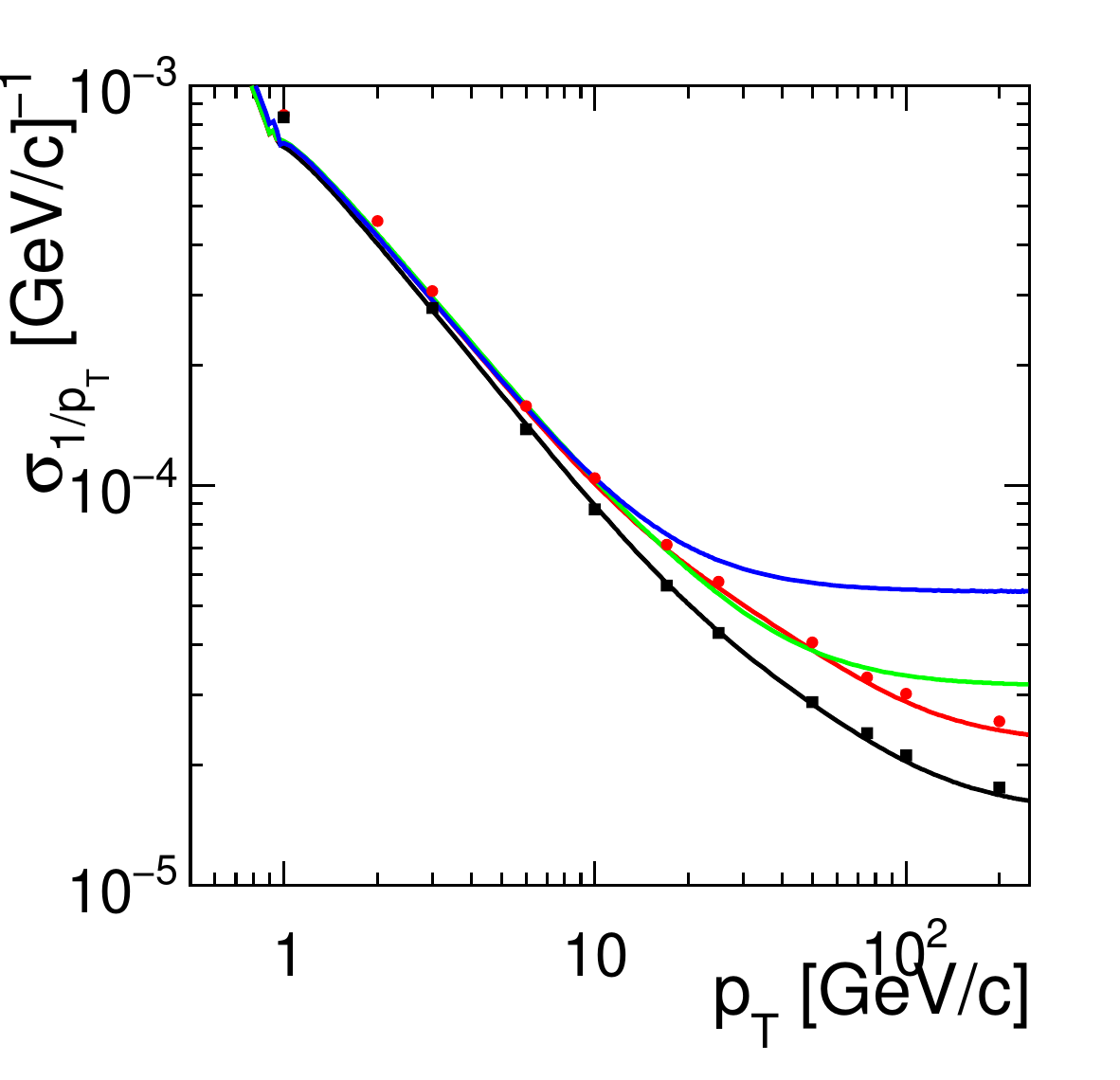}\includegraphics[width=0.49\textwidth,clip,trim=0 20 0 0]{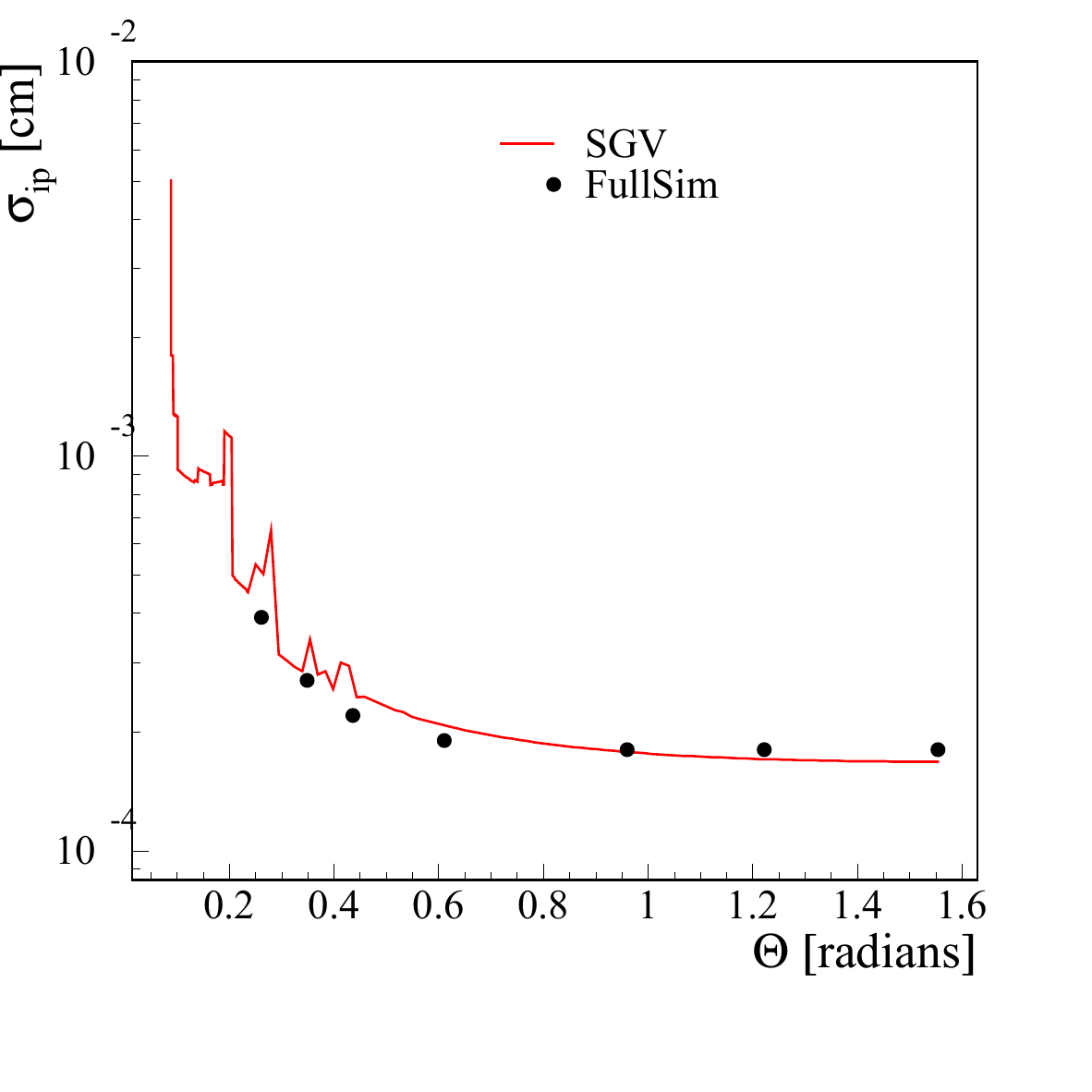}
  \caption{Left: The momentum error $\sigma_{1/p_\mathrm{T}}$ vs. $p_\mathrm{T}$,
for a number of different detector configurations.
Right:  The impact-parameter error $\sigma_{\mathrm{ip}}$ vs. $p_\mathrm{T}$.
The  lines show the SGV result, and the dots show the full simulation
and reconstruction result.}\label{Fig:dp_vs_p_dipvsth}
\end{figure}

The perigee parameters are then smeared according to the calculated 
covariance matrix,
by means of the Cholesky decomposition procedure,
which yields random numbers with correlations between them that are
indeed those of the calculated covariance matrix.
\Cref{Fig:dp_vs_p_dipvsth}
shows a few examples of the excellent agreement between the SGV result and that
obtained by the full simulation and
reconstruction for the same detector configuration.

To simulate calorimeters,
the charged or neutral particle is extrapolated to the intersections with the 
various calorimeters.
From the nature, momentum and position of the particle,
the detector response is
simulated from parameters,
given in the geometry description input-file.
The realism can be increased by simulating confusion between
calorimetric clusters:
clusters might merge, might split, or may get wrongly associated to tracks.
In SGV, information is already 
available about where the particle hits the calorimeters,
and parametrisations of such confusion-errors in the ILD detector are shipped with the
code, and can be used if wanted.
%
%

SGV has no default output format, but comes with plug-ins to create \lcio DST format
output and root trees. The \lcio format is such that it can be directly used by
code designed to analyse full simulation or real data DST files.
In addition, SGV can be run in filtering mode, where the user can decide how a given
event should be processed: It can be ignored already at generation, after detector
simulation, or after detailed analysis. At any stage, it can be decided to
output the generated events passing the selection criteria,
and only proceed with full \geant simulation of these selected events.


%

\paragraph{\ddsim}
\label{sec:ddsim}
The \ddsim~\cite{ddsim} program from \ddhep~\cite{Frank:2014zya,frank15:ddg4} enables users to run
\geant~\cite{GEANT4:2002zbu} simulations of their \ddhep-based detector models. The program focuses on ease of use
while allowing full configurability for the underlying simulation. All aspects of the simulation can be configured, for
example the physics lists, the range cuts, field stepper properties, input sources, and output formats. Most options can
be configured via the command line allowing for fast iteration for interactive investigations. All options can also be
configured via a steering file, which can be mixed with setting some options via the command line.  For further convenience 
all options are extensively documented directly in \ddsim\ via the \texttt{-{}-help} or \texttt{-{}-dumpSteeringFile}
flags. Using the \ddhep\ plugin mechanism, users can further extend the functionality by providing their own plugins,
for example for sensitive detectors, which further extends the flexibility offered by \ddsim. This flexibility
includes the option to plug-in fast-simulation replacement of detailed \geant\ physics. All aspects of the simulation can
be configured without changes to the XML files of the detector models. 

Some of the notable options available in \ddsim\ are the input readers for \edmhep, \lcio, \stdhep, \hepevt, \hepmc, and
\guineapig\ pair files. The treatment of primary particles to be simulated by \geant is highly configurable allowing also
for the realistic simulation of pre-assigned decays and their secondary vertices accounting for material and magnetic
field effects.  For further realism, the Lorentz boost needed to account for crossing-angles, and the option to smear
the interaction point according to some distribution are also available. During the simulation, \ddsim\ keeps track of the
Monte Carlo particle history, allowing one to associate tracker and calorimeter hits with their corresponding particles,
which enables detailed efficiency and performance studies. \edmhep  output files are available by default, which also
include meta data about how the simulation was performed.

The ease-of-use and flexibility offered by \ddsim\ makes it the ideal tool for individual users doing interactive
work to centralised production workflows.


\subsubsection{Detector models\label{ssec:detector_models}}

\paragraph{ALLEGRO}
ALLEGRO (A Lepton Lepton Collider Experiment with Granular Read-Out) is a general-purpose detector concept proposed for FCC-ee, centred around an innovative high-granularity noble liquid ECAL. While still under active development, the baseline design includes a silicon-based vertex sub-detector, a gaseous main tracker, a silicon wrapper, an ECAL with inclined lead absorbers and liquid argon sensitive gaps, an ultra-thin solenoid sharing the same cryostat, an HCAL with iron absorbers and scintillators, and a simple muon tagger.

The software related to this concept is fully integrated into \keyhep: detector geometries are implemented with the \ddhep toolkit~\cite{Frank:2014zya} and are hosted in the \texttt{k4geo} package~\cite{k4geo} (which also hosts the detector geometries of several other concepts for future colliders) while the digitization and reconstruction Gaudi algorithms rely on the \edmhep~\cite{Gaede:2022leb} data model and live in the \texttt{k4RecTracker}~\cite{k4RecTracker} and \texttt{k4RecCalorimeter}~\cite{k4RecCalorimeter} packages.

The ALLEGRO detector model largely exploits the plug-and-play approach provided by \ddhep. For instance, the entire tracking system is directly taken from the IDEA detector implementation described in \cref{sec:idea}. This approach ensures a reasonable material budget before the calorimeters and enables studies requiring tracking information while a dedicated tracking system for ALLEGRO is under development.

Both the ECAL and HCAL benefit from detailed, flexible geometry implementations designed for optimization with particle flow performance. Key parameters such as the shape and material of the absorbers or the dimensions of the readout cells have been made easily configurable. This flexibility is achieved through \ddhep \texttt{Segmentation} classes combined with the possibility to gather multiple physical cells into one readout channel at the digitisation step.

\begin{figure}[ht]
    \centering
    \includegraphics[width=0.85\linewidth]{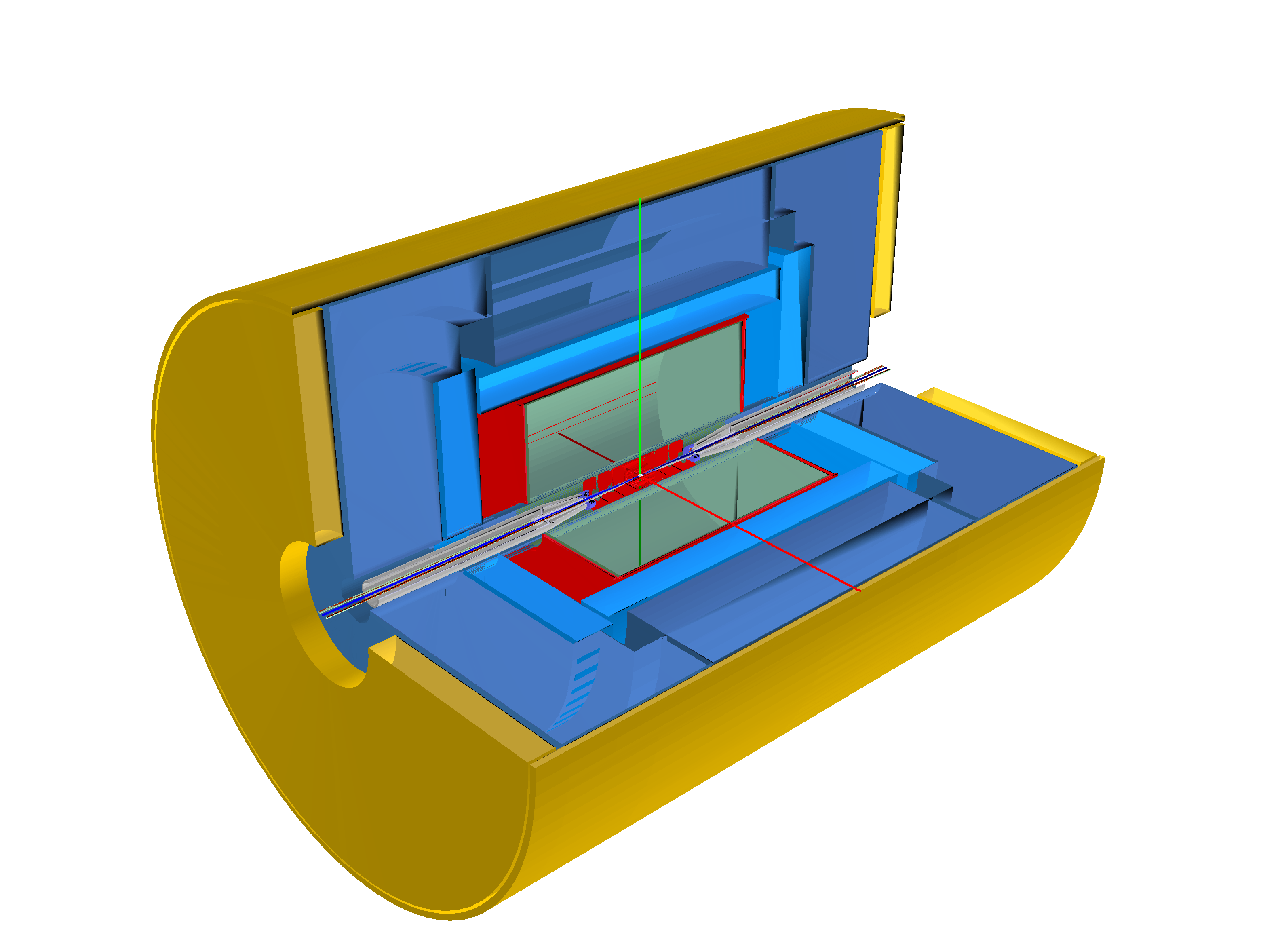}
    \caption{3D visualisation of the ALLEGRO detector model implemented with \ddhep.}
    \label{fig:allegro}
\end{figure}

The muon system is currently a place-holder made of two layers of sensitive material, enabling already first studies relying on truth information. A visualisation of the ALLEGRO \ddhep detector implementation is shown in \cref{fig:allegro}. Additional details on the ALLEGRO detector simulation are available in related documentation~\cite{fccSoftwareNote}.

\begin{figure}[ht]
    \centering
    \includegraphics[width=0.5\linewidth]{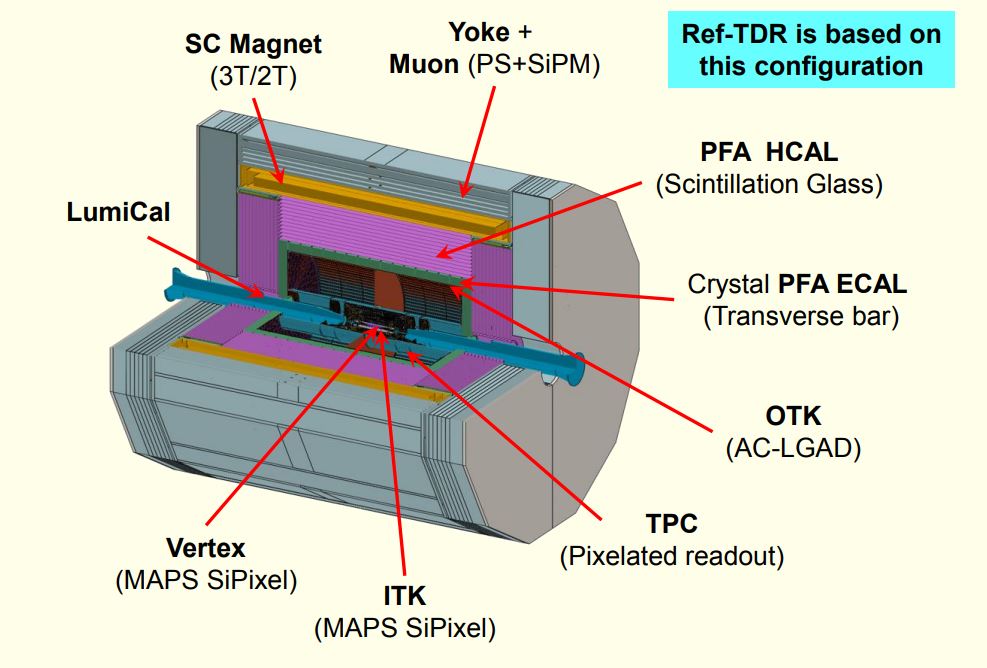}
    \caption{CEPC reference detector.}
    \label{fig:CEPC-det}
\end{figure}

\paragraph{CEPC Reference Detector}
The Circular Electron Positron Collider (CEPC)~\cite{CEPCStudyGroup:2018ghi}
is designed to facilitate electron-positron collisions at a central mass energy of approximately 240 GeV for Higgs boson studies via the $\Pep \Pem \to \PZ \PH$ process. 
Additionally, it will enable collisions at the $\PZ$-boson peak for precise measurements in electroweak physics. 
The reference technical design report 
for the detector is currently being prepared based on the design illustrated in \cref{fig:CEPC-det}. 

The detector consists of seven sub-detectors arranged from the innermost to the outermost radius: Vertex Detector (VTX), Inner Silicon Tracker (ITK), Time Projection Chamber (TPC), Outer Silicon Tracker (OTK), Transverse Crystal-bar Electromagnetic Calorimeter (ECAL), and Scintillation Glass Hadronic Calorimeter (HCAL). 
Surrounding the HCAL is a coil of a \SIrange{2}{3}{\tesla} superconducting magnet, and a flux return yoke is integrated with the Muon Detector (MUON).

The CEPC software (CEPCSW)~\cite{CEPC-software} is being developed using the Gaudi framework as its foundation. By integrating with \keyhep~\cite{Key4hep:2022jnk}, it becomes straightforward to use \edmhep~\cite{Gaede:2022leb} as the event data model and \ddhep~\cite{Frank:2014zya} for detector description. A complete simulation chain has been developed based on a simulation framework in CEPCSW. The simulation chain consists of physics event generation, detector simulation and digitization simulation. They are not executed in the same job option: the event generation and detector simulation are in one job option file, while the digitization and reconstruction are in another job option file. 

Physics event generation is implemented in an algorithm configured with several GenTool instances, while HepMC is used as the intermediate objects. Particle gun, HepEvt reader, StdHep reader, \lcio reader and beam background reader have been implemented as GenTool. A particle gun allows users to shoot multiple particles of certain energies at certain positions in the same event. Different formats readers allow users to read the existing samples, which are already generated before. A beam background reader allows user to specify the input of beam backgrounds.

The detector simulation relies on the \geant~\cite{GEANT4:2002zbu} simulation toolkit, which provides physics models and facilitates particle transport through the defined geometry. In CEPCSW, the \geant run manager is encapsulated within a Gaudi service, allowing the Gaudi framework to control the event loop of the simulation. This service is responsible for initializing the geometry, physics lists, and user actions in \geant, and it also provides standard user interfaces for interaction with the toolkit. A precise description of the detector geometry in \ddhep includes detailed information about the position, shape, dimensions, and material composition of each detector component. DDG4 is used to convert the geometry model from \ddhep to \geant. Sensitive detectors are used to generate hit objects in SimTrackerHit or SimCalorimeterHit type, and the hit objects are inserted into corresponding hit collections according to the sub-detector types. The identifier of each hit is provided from \ddhep, according to the corresponding definition in the \ddhep compact files.

The digitization algorithm is triggered when the energy deposited in the sensitive areas of the detector exceeds a pre-configured threshold. During digitization, the effects of various types of background events are considered. Hits from a signal event are overlaid with those from additional background events before calculating the detector's response.

\paragraph{CLD}

\begin{figure}
    \centering
    \includegraphics[width=0.485\linewidth]{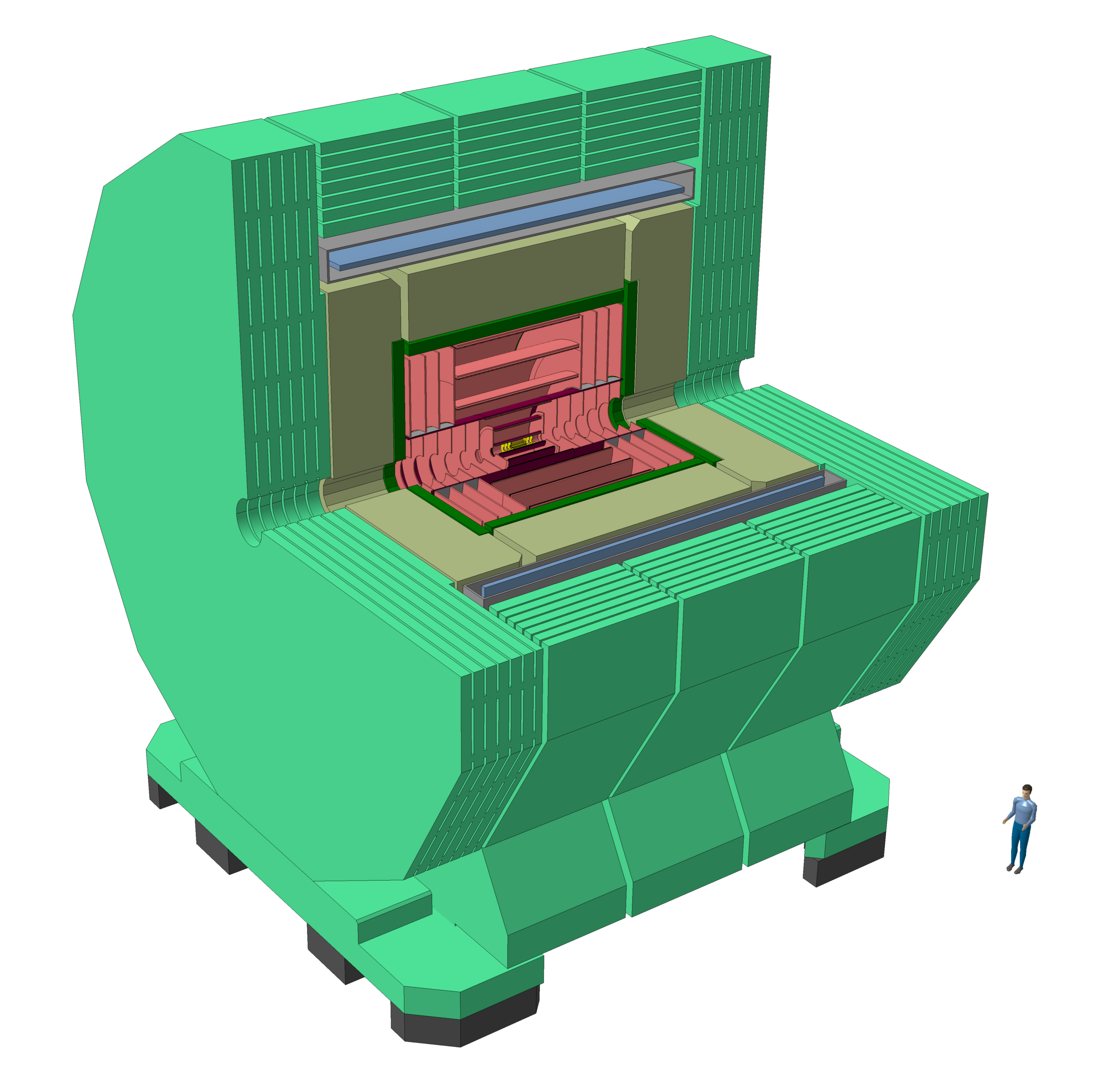}
    \includegraphics[width=0.485\linewidth]{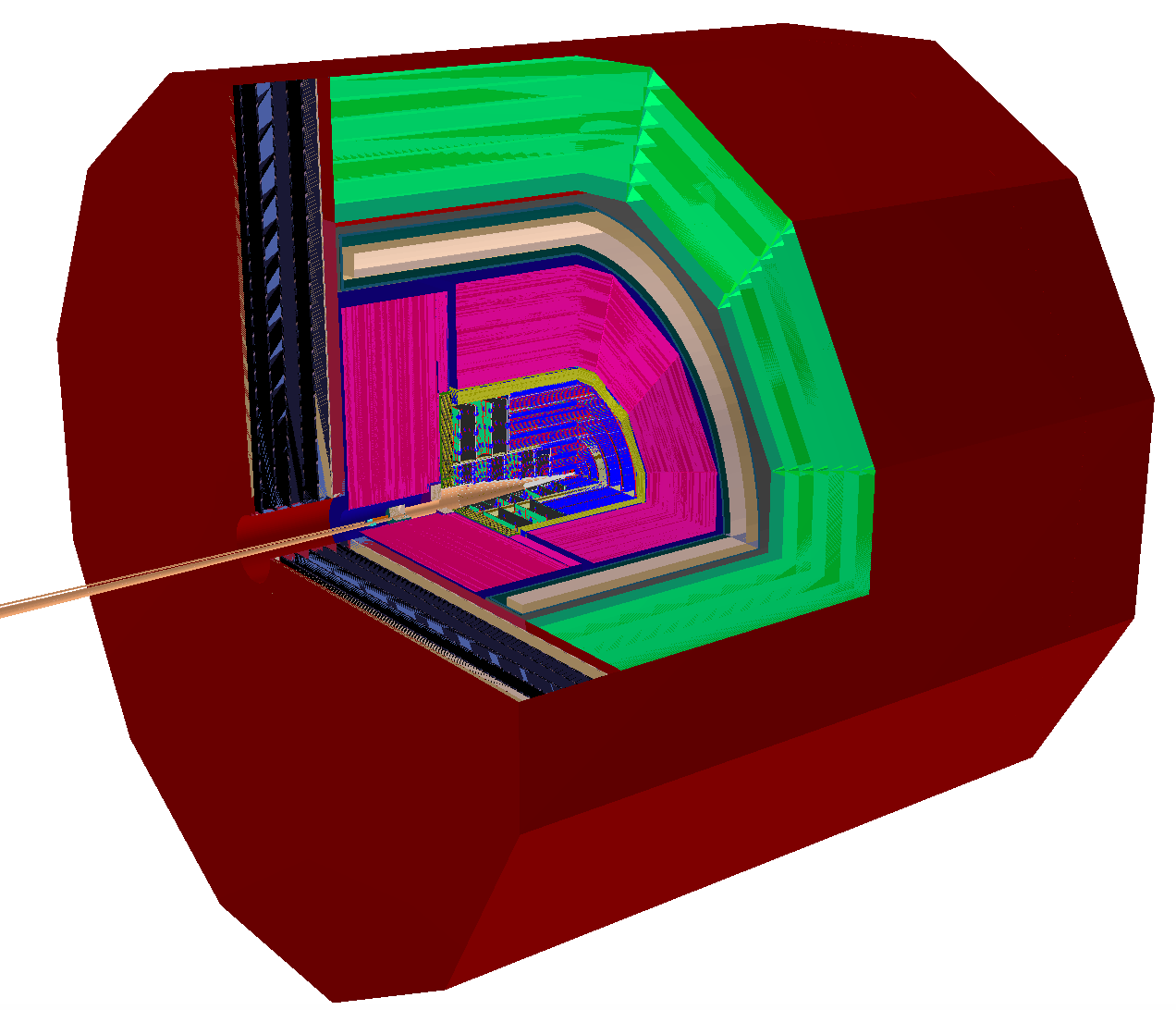}
    \caption{Left: The CLD detector model as envisioned by an engineering drawing. Right: The CLIC detector model as
      implemented in the \ddhep simulation.}
    \label{fig:sim_cld}
\end{figure}

The CLD detector is a detector concept for the FCC-ee~\cite{Bacchetta19}.  It evolved out of the CLICdet detector through adaptations to a \SI{2}{\tesla} solenoid field, reduced vertex detector radius, an increased tracker radius, a thinner hadronic calorimeter, and a modified MDI region. Otherwise the detector is equivalent to CLICdet as described in the next paragraph. Some further optimisation that was done for CLD, was an optimisation of the electromagnetic calorimeter~\cite{Viazlo_2019,diracCalib}. \Cref{fig:sim_cld}(left) shows a CAD drawing of the CLD detector model compared to the \ddhep implementation of CLICdet.

The CLD detector has also been used to host other sub-detectors for which no full simulation model was otherwise available. One variant integrates the ARC (Array of RICH Cells) detector. For this model the silicon tracker was reduced in radius and the ARC detector placed in the freed space to provided particle identification capabilities. For a different variant, the ECal was replaced with a noble-liquid calorimeter and the HCal, Coil and Muon system displaced accordingly, to study particle flow clustering with a fine-grained noble-liquid calorimeter~\cite{Sasikumar:2918988}.

\paragraph{CLICdet}

The CLICdet detector was optimised for electron--positron collisions at CLIC in the energy range of
\SIrange{350}{3000}{\giga\electronvolt}~\cite{AlipourTehrani17,detperf}. The concept evolved out of the lessons learned
from the CLIC conceptual design report~\cite{CLIC_PhysDet_CDR}. The CLICdet concept is based on Particle Flow clustering
for optimal jet reconstruction and beam-background particle
rejection~\cite{Marshall:2012ryPandoraPFA}.

The CLICdet concept consists of a full silicon tracking system, starting with a vertex detector in barrel and endcap
configuration. The endcap consists of helically placed modules to allow for air-cooling of the whole vertex
detector. Surrounding the vertex detector is a silicon tracker. Together at least 10 hits are provided across the full
angular acceptance. The tracker is followed by high-granularity electromagnetic and hadronic calorimeters contained in a
\SI{4}{\tesla} superconducting coil, which itself is surrounded by a return yoke that uses RPCs for muon identification.
The forward region further contains a LumiCal and a BeamCal to measure absolute luminosity via Bhabha events, and to
complement the electromagnetic coverage down to \SI{10}{\milli\radian}. The full detector is implemented in \ddhep with
generic detector drivers that have been re-used for other detector models, in particular the CLD detector that was
derived from CLICdet. The silicon tracking detectors have been designed in view of a realistic material budget, taking
sensor, electronics, mechanical support and cooling requirements into account. \Cref{fig:sim_cld}(right) shows the
detector model as implemented in \ddhep. 

Nominally the same detector would be used for the full energy range covered by the CLIC program; only the vertex
detector and beampipe would be different at the lowest energies, where beam-conditions from beam-induced backgrounds are
less severe than at the highest energies. The tracker and calorimeters could be used at all energies.

The CLICdet model is implemented in \ddhep and available in the k4geo repository. It shares many drivers, such as for
the beampipe, masks, or the LumiCal with other detectors described in this report.

\paragraph{IDEA}
\label{sec:idea}
An ``Innovative Detector for e$^+$e$^-$ Accelerator'' (IDEA) is a detector concept featuring a silicon based vertex detector, an ultra-light, full-stereo drift chamber, a silicon wrapper, dual-readout calorimeters, and a micro-RWELL-based muon system. Two detector configurations are being considered: one with a fibre-based dual-readout sampling calorimeter surrounding the magnet coil, serving as both ECAL and HCAL, and another with an additional crystal-based dual-readout ECAL placed in front of the coil. This section outlines the \ddhep implementation of these sub-detectors.

The vertex sub-detector implementation uses single-hit layers of MAPS and features an accurate representation of the first engineered vertex detector for FCC-ee~\cite{Ilg:2024+2}. The detector driver offers two modes: one constructing standard straight sensors and another employing nearly support-free curved MAPS sensors inspired by the ALICE ITS3~\cite{its3}.

The drift chamber geometry is modeled using hyperboloids that define radial layers, within which twisted tubes emulate the cell volumes hosting the sense and field wires. To enhance \geant navigation performance, an alternative approach using \ddhep segmentation to define cells is under development. The detector driver follows engineering prescriptions to build the full geometry based on a handful of user provided parameters.

The Segmented Crystal EM Precision Calorimeter (SCEPCal) features a projective geometry of longitudinally segmented homogeneous crystals with dual-readout capabilities and a precision crystal timing layer, placed in front. Its geometry construction is fully parametrised, governed by eight input parameters for global dimensions and segmentation. For computational efficiency, a custom \ddhep sensitive detector action manages optical photons.

Two simulation models exist for the fibre dual-readout calorimeter: one with fibres embedded in monolithic metallic towers and another with metallic capillary tubes housing the fibres. This description focuses on the latter version which is more realistic from a manufacturing point of view. The geometry is structured hierarchically as follows. Optical fibres are embedded within metallic tubes, which are arranged in a hexagonal pattern to form towers. These towers are then projectively aligned in $\theta$ to construct a complete stave. The stave is then replicated over $\phi$ for full hermetic coverage. A custom sensitive action optimizes the simulation of optical photons, applying Birk's law and a Poissonian distribution for scintillation photons, while \geant handles Cherenkov photons. Photon propagation to the silicon photomultipliers is parametrised, accounting for reflection and attenuation. This method has been validated with test beam data~\cite{drcaloSensitiveAction}.

\begin{figure}
    \centering
    \includegraphics[width=1\textwidth]{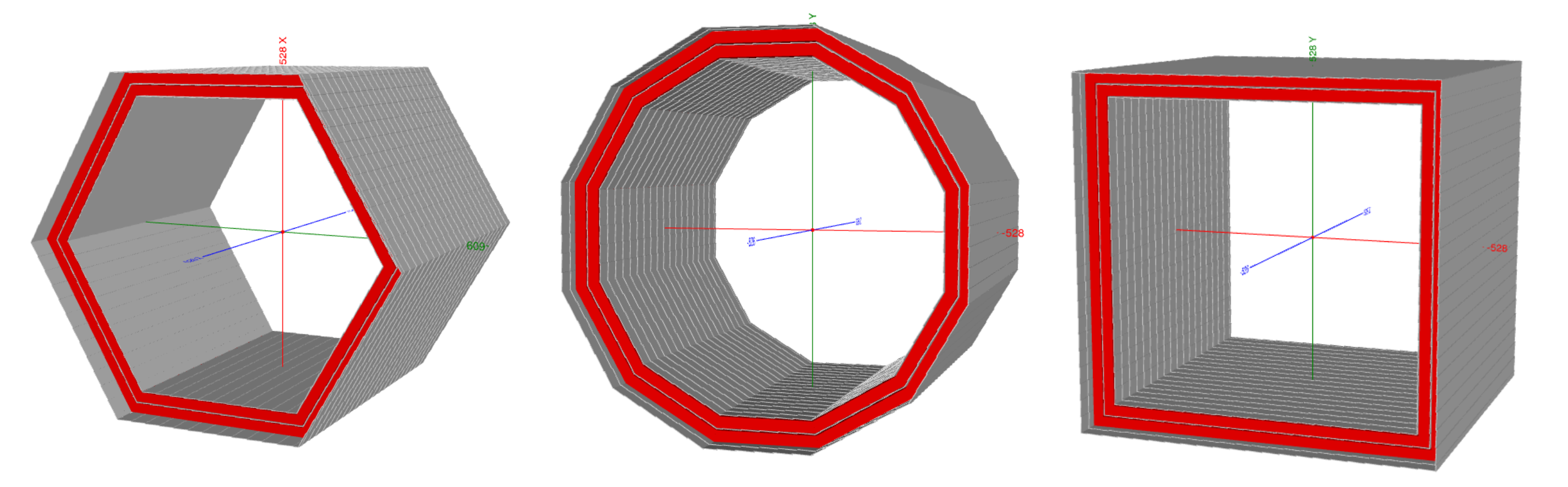}
    \caption{Illustration of the flexibility provided by the detector driver building the IDEA $\mu$-RWELL-based muon systems. From left to right, different barrel implementations with 6, 12 and 4 sides, respectively, are shown. The gray layers represent the sensitive detector components while the return yokes are shown in red.}
    \label{fig:muRWELL}
\end{figure}

The muon system is implemented as a sequence of layers with customisable thickness and material, enabling the modeling of both the yoke and $\mu$-RWELL PCBs with the same code. The muon system forms a layered polyhedron with a user-defined number of sides, as shown in \cref{fig:muRWELL}. The driver automatically places the $\mu$-RWELL PCB ``tiles'' to hermetically fill each side of the polyhedron and ensures hermetic coverage by overlapping PCB tiles at edges to address the insensitive regions of the $\mu$-RWELL. modules.

The two different versions of the IDEA detector concept are obtained by shrinking and pushing the fibre dual readout calorimeter away from the interaction point, updating the muon system radial positions and inserting the crystal dual readout in the created empty space. All of this is done by simple manipulations of \texttt{xml} parameters, thanks to the modularity of the \ddhep implementations. More details on the IDEA detector simulation can be found elsewhere~\cite{fccSoftwareNote}.

\paragraph{ILD}

The ILD concept~\cite{Behnke:2013lya}, developed since around 2008, has been optimised for Particle Flow event reconstruction in electron--positron collisions at the ILC in the energy range of \SIrange{250}{1000}{\giga\electronvolt}. \ddhep is used to describe the detectors and interface to the \geant simulation libraries. All ILD model descriptions are available in the \texttt{k4geo} package.

The ILD has a hybrid tracking system built around a high precision and lightweight vertex detector and a large Time Projection Chamber (TPC), with additional silicon tracking layers bridging between the vertex
detector and TPC, forward tracking disks extending tracking coverage in the forward region, and an envelope tracker providing a
precise track measurement immediately after the TPC to provide excellent momentum resolution.

The highly-granular ``imaging'' calorimeters are optimised for 
Particle Flow performance.
Since several technologies are currently being considered for the calorimeter readout, a ``multi-technology'' calorimeter simulation was developed, in which two technologies are simultaneously simulated in both the ECAL 
(silicon tiles, scintillator strips) and HCAL (scintillator tiles, resistive plate chambers) sections.
This makes use of the fact that the material budget and thickness of active sensor (e.g.\ scintillator tiles) and readout infrastructure (PCB hosting readout electronics)
within calorimeter layers are rather similar, to replace the readout infrastructure with an alternative sensitive technology.

Simulated subdetector performances have been extensively benchmarked against detector prototypes' performance in test beams. Significant effort has been dedicated to
reasonably simulating inactive materials (supports, cables), cracks in detector geometry, and similar features which can affect real detector performance.
The machine-detector interface (MDI) foreseen for ILC is modelled in some detail, including the beam pipe, forward calorimetry (LumiCal, LHCal, BeamCal) as well as supporting systems and final focus quadrupoles. The solenoid provides a 3.5~T central field, and the iron flux return (which also serves to self-shield the detector) is instrumented for muon detection.

\begin{figure}
    \centering
    \includegraphics[width=\linewidth]{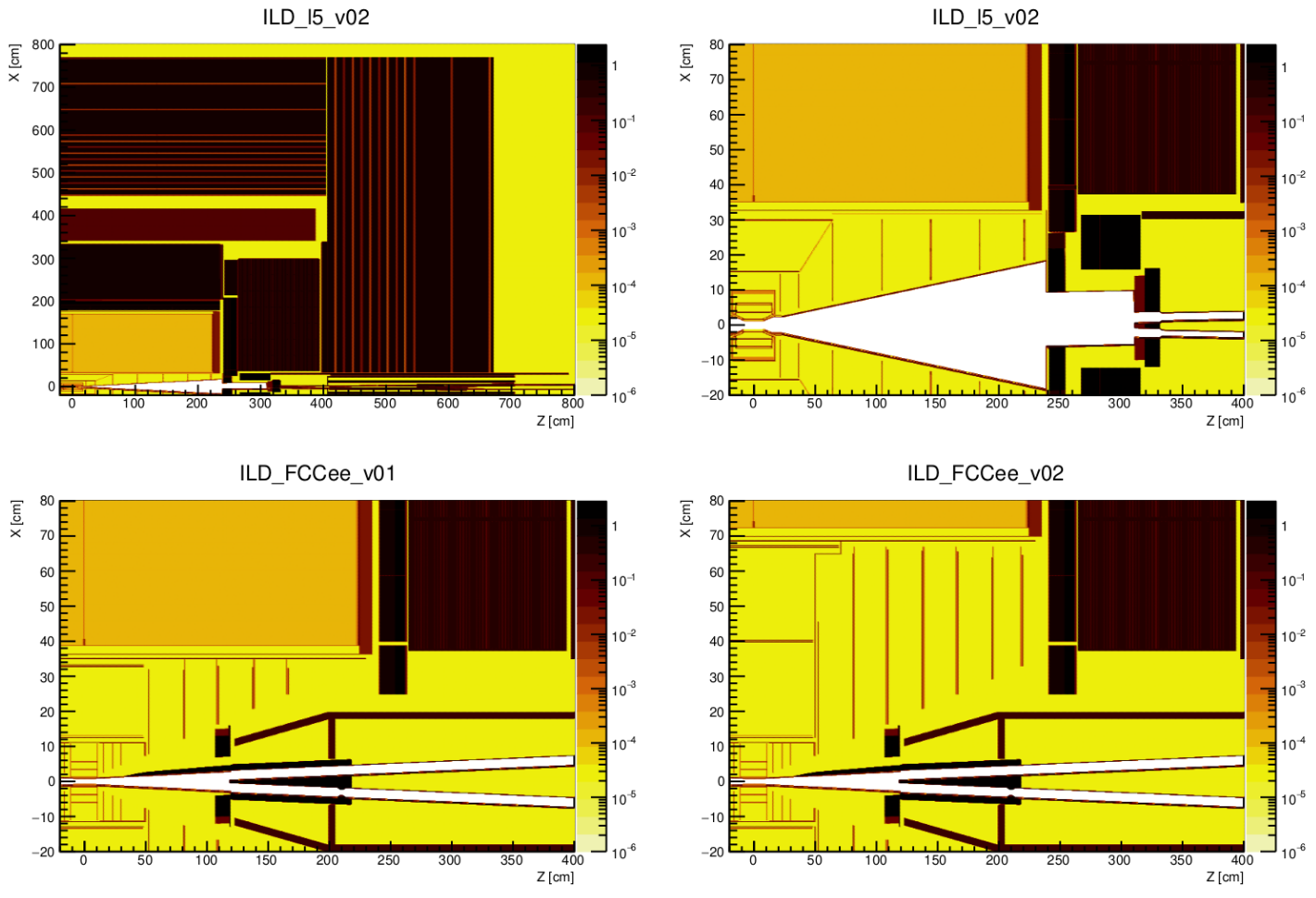}
    \caption{Views of the ILD simulations models, showing material density (colour, in units of $1/X_0$) in a horizontal slice through the detector at $y=0$. Top row: ILD\_l5\_v02, the standard model of ILC (overall view (left) and MDI-centric view (right)); bottom row: MDI-centric views of the two experimental models for FCC-ee, showing the different radii of the TPC (shown in orange).}
    \label{fig:sim_ILD}
\end{figure}

Two variants of the ILD with a larger or smaller ECAL inner radius were simulated and extensively studied for studies towards the Interim Design Report~\cite{ILDConceptGroup:2020sfq}. Large-scale MC samples used for physics studies extensively presented elsewhere in this report use mostly the larger
ILD model, known as ILD\_l5\_v02, which is illustrated in \cref{fig:sim_ILD}.

\paragraph{ILD@FCCee}
Two additional ILD variants have recently been developed, based on ILD\_l5\_v02, but tailored for studies at circular colliders such as the FCC-ee. These models involved minimal changes
to the ILD design necessary to adapt to the Machine Detector Interface of the FCC-ee, in order to minimise the differences with respect to the standard ILD model to enable simpler comparisons. The two options differ in the inner radius of the TPC, with one close to
that of the original ILD, and the other increasing the inner radius by
around 30~cm, as shown in \cref{fig:sim_ILD}.
These models will be used to investigate the effect of the MDI at FCC-ee and to compare the performance of two different tracking detector layouts, with a view to proposing a version of ILD for use at the FCC-ee or other circular Higgs Factory collider.

The implementation of these models was greatly simplified by the plug-and-play nature of 
detector description in \ddhep,
and the use of a common detector geometry repository, \texttt{k4geo}.
The luminosity calorimeter and the new inner tracking region (inside the TPC) have been copied and adapted from that used in CLD,
while the MDI is taken from the common description available in \texttt{k4geo}, including the beampipe, machine solenoids, and shielding.

\paragraph{SiD}

The SiD Detector~\cite{Breidenbach:2021sdo} is a cost-constrained general purpose particle physics experiment optimized for electron-positron linear colliders like ILC. It is also optimized to deliver a well performing particle flow reconstruction~\cite{Thomson:2009rp} for excellent reconstructed jet resolution and all tracking and calorimetry are immersed in a 5~T magnetic field. Its design focuses on leveraging silicon sensing technologies for the main tracking detectors and electromagnetic calorimetry. For the ILC, and accelerators with a similar experimental environment like C3, at 250 GeV centre-of-mass SiD's MAPS-based vertex detector has its innermost layer placed at 1.4\,cm in radius from the beamline offering clean tracking and superb heavy flavour tagging. The vertex detector comprises five layers in total, with outermost radius at 6\,cm. The remainder of the tracking system, to radius 1.25\,m, is a lightweight strip tracker. The electromagnetic calorimeter uses digital MAPS to achieve millimeter-scale two-particle separation, further aiding the particle flow reconstruction in high-energy jets. The hadronic calorimetry and instrumented return yoke are similar to other detector designs like ILD.

The SiD detector is implemented in a full \geant simulation and distributed in the \texttt{k4geo} package~\cite{k4geo} and all studies performed for this document used \ddhep~\cite{Frank:2014zya} and \keyhep~\cite{Key4hep:2022jnk} to generate data. The description of the detector is mostly realistic and meets descriptions of expected average material budgets. However, it does not include a detailed description of material budget or acceptance gaps since final implementation designs are not chosen and the details of the detector geometry are not known. Given the small material budget expected for the vertex detector and tracker, these imprecisions will not affect derived physics performance results. A detailed accounting of SiD detector physics performance benchmarks at ILC can be found in the ILC TDR~\cite{Behnke:2013lya}.

SiD would need to be completely re-optimized for use at a synchrotron, as many of its design decisions depend critically on the use of a 5~T magnetic field. It has been thoroughly validated for a variety of linear collider environments. Beyond this, the detector design remains flexible enough to handle high rate environments, but would need a careful accounting of data throughput and heat load once a readout scheme is identified. The studies to understand the viability of the SiD design at synchrotrons have not yet been performed.

\subsubsection{Outlook}
As outlined above, there has been significant progress in the definition of detectors for future Higgs factory colliders. 
A number of complete detector models are available for use in studies
of the physics potential of future facilities, both at the fast- and full-simulation levels.
These models are based on a common suite of software tools, which has greatly streamlined their development.
Much work remains, for example to improve the description of currently implemented technologies, to develop descriptions of newly-proposed systems, to increase computational efficiency, and to optimise overall detector layouts.
\subsection{Reconstruction\label{sec:reconstruction}}
\editors{Loukas Gouskos, Taikan Suehara, Ulrich Einhaus}


Over recent years, the landscape of reconstruction algorithms for future collider experiments has evolved significantly, driven by advances in detector precision, granularity, and innovative algorithmic techniques. In future electron-positron colliders, the high luminosity and clean event environment create a unique opportunity to achieve unprecedented precision in measurements of the SM parameters, as well as direct searches for new physics. 
Achieving this precision demands that reconstruction algorithms fully leverage the advanced detector designs.

The advent of highly granular detectors enables particle flow algorithms (PFAs) to reconstruct individual particles with exceptional resolution, a necessity for accurate jet reconstruction and energy-momentum measurements. This capability is further enhanced by using sophisticated machine learning (ML) techniques, particularly graph neural networks (GNNs), which optimise the processing of low-level detector signals into higher-level physics objects. In particular, ML-driven approaches are proving instrumental in flavour tagging, jet clustering, and global event reconstruction, crucial for identifying and distinguishing decay products in decays of SM particles and potential new physics signals.

As the experimental requirements expand, so do the applications of these reconstruction algorithms. For example, advanced calorimetry and tracking techniques are not only central to jet reconstruction but also critical for improving particle identification (PID), e.g.\ enhancing the separation of charged particles. The incorporation of timing information, dE/dx, and/or Cherenkov radiation measurements further refines PID capabilities; a notable advantage for clean event environments such as those at future electron-positron colliders. These advancements allow us to unlock the full physics potential by facilitating precise event reconstructions, background suppression, and access to rare or novel decay modes, thereby broadening the physics scope.

All of the reconstruction algorithms are either directly available in the \keyhep ecosystem (e.g., tracking and particle flow) described in \cref{sec:softwareeco} or rely on output generated by \keyhep. In particular the increasing number of neural network approaches utilise established packages outside the framework, which makes it desirable to integrate these or provide coherent interfaces to \keyhep.

In the following sections, we discuss the current state of each major reconstruction steps, highlighting recent advancements and performance results, which are based on the detector simulations described in \cref{sec:simulation}, as well as their relevance to the analyses covered later in this report.

 \subsubsection{Tracking}
Future \epem colliders are designed to explore particle physics with high precision. Tracking systems are crucial in these experiments for reconstructing charged particle trajectories, enabling precise measurements of momentum and impact parameters essential for the physics program. 
Here, we will discuss the tracking reconstruction techniques under development for these concepts and cross-reference their approaches when relevant. The similarities between FCC-ee and CEPC are particularly pronounced, as both designs leverage closely related methodologies and technologies for their tracking reconstruction.

\textbf{Tracking at ILC.} The ILC proposes two primary detector concepts: ILD and SiD as described in \cref{sec:simulation}. 

The ILD~\cite{Behnke:2013lya} concept  is 
designed to deliver high-precision tracking performance through a tracking system consisting of a Time Projection Chamber (TPC), a silicon vertex detector (VTX), Silicon Inner Tracker (SIT), and Forward Tracking Detector (FTD).

ILD's track reconstruction algorithm~\cite{Gaede:2014aza} is developed within the iLCSoft framework, which integrates \ddhep for detector simulation, Marlin as the main event-processing application, and \lcio for standardized data storage. \lcio's Track class records multiple TrackStates (fitted track parameters) at important points along each track, including the interaction point, first and last tracker hits, and calorimeter interface. ILD uses a perigee track parametrisation, specifying curvature ($\Omega$), 
impact parameters ($d_0$ and $z_0$), and directional angles ($\phi_0$ and $\tan\lambda$), allowing compact, efficient track description.

Track reconstruction in ILD involves several stages: clustering, seeding, pattern recognition, and track fitting. In the TPC, charge clusters are identified with high spatial density to provide continuous 3D tracking, achieving a radial resolution of \SI{100}{\micron} per hit. Track seeding is performed by the Clupatra algorithm in the TPC, using an outside-in approach, where track candidates are initiated from the outer layers and extended inward, minimizing ambiguities. In the silicon detectors, the SiliconTracking algorithm generates seeds by combining triplets of hits within the VTX, SIT, and FTD layers. These seeds are refined through a Kalman filter that dynamically updates track parameters with each additional hit, compensating for uncertainties due to scattering and material interactions.

Pattern recognition is handled by the Clupatra and SiliconTracking algorithms, with Clupatra connecting adjacent clusters based on spatial alignment, and SiliconTracking extending seeds through helical paths. The Kalman filter plays a dual role, smoothing tracks during pattern recognition and performing final track fitting. This recursive filtering optimises track parameters iteratively, achieving the ILD's stringent requirements of the asymptotic momentum resolution of $\displaystyle{\delta(1/\pT) = 2 \times 10^{-5}~\text{GeV}^{-1}}$  
and impact parameter resolution of $\displaystyle{\sigma_d = 5~\upmu\text{m} \oplus \frac{10~\upmu\text{m}}{p \sin^{3/2}\theta}}$.

Post-fitting, ILD applies quality control to remove duplicates and refine track purity, achieving tracking efficiency exceeding 99.5\% and a fake rate below 0.1\% above a $\pT$ of about \SI{300}{MeV}. The ILD's low material budget -- around 5\% radiation length for the TPC and 0.15\% per layer for silicon trackers -- minimises scattering effects, preserving track accuracy across the detector. This layered, systematic approach makes ILD's tracking system a robust and efficient framework for high-precision studies at the ILC.

In the silicon tracker, pattern recognition is performed by extending seeds along helical trajectories using the SiliconTracking algorithm. Both systems are synchronized to ensure that tracks identified in the TPC are matched with their corresponding hits in the silicon detectors, providing an integrated view of the trajectory.

SiD~\cite{Aihara:2009ad,Behnke:2013lya} at the ILC adopts an all-silicon tracking system, providing a compact, design optimized for precision tracking and fast timing.

The reconstruction process begins with clustering, where hit positions in the finely segmented silicon detectors are transformed into a unified format. The vertex detector, with a resolution of $\lesssim 5~\upmu\text{m}$, provides precise 3D spatial information for identifying tracks originating near the interaction point. 
These initial hits are used to seed track candidates, which are formed by combining hit clusters across multiple layers.
A helical trajectory model is applied to estimate parameters such as curvature ($\omega$)
and impact parameters ($d_0$ and $z_0$), providing an initial description of the particle's path through the detector.

Global pattern recognition algorithms then extend these tracks into the outer tracking layers, where the helical model is used to predict and match hits across the silicon strip layers. 
In regions transitioning between the barrel and endcap geometries, stereo strip configurations provide 3D space points that enhance the accuracy of pattern recognition and allow seamless reconstruction across the detector's fiducial volume. 
Tracks from decays or interactions outside the vertex detector are also reconstructed using calorimeter-assisted tracking algorithms. 
Here, entry points from the electromagnetic calorimeter are used to seed track candidates, which are then extrapolated back into the tracker.

Once initial track candidates are formed, the reconstruction enters a refinement phase. A Kalman filter is applied to iteratively refine the trajectory, correcting for material effects such as multiple scattering and energy loss. This method ensures high precision in momentum and vertex parameter estimates while maintaining computational efficiency. The track fitting algorithms are augmented with alignment corrections to mitigate distortions arising from mechanical tolerances or material deformations, further enhancing accuracy.

SiD achieves an asymptotic momentum resolution of $\displaystyle{\delta(1/\pT) = 5 \times 10^{-5}\text{GeV}^{-1}}$ for tracks with $\pT > 1\text{GeV}$, comparable to other high-precision trackers. The vertex detector provides an impact parameter resolution of $\displaystyle{\sigma_r\phi, \sigma_rz = 5 \oplus \frac{10}{p \sin^{3/2} \theta}~\SI{}{\micron}}$, enabling precise reconstruction of secondary vertices, which is critical for flavour tagging and rare decay studies. Tracking efficiency exceeds 99\%, with robust performance maintained even in high-multiplicity environments. The silicon-based design minimizes material, with a contribution of $\sim 0.8\%$ radiation length per layer, ensuring minimal scattering and preserving high track purity.

Compared to other ILC and FCC designs (discussed later in the section), SiD's fully silicon-based architecture prioritizes compactness and fast reconstruction. This contrasts with the hybrid designs of ILD and FCC-ee IDEA, which incorporate TPCs or drift chambers. SiD's reliance on discrete silicon layers 
simplifies the reconstruction algorithms while maintaining high precision, whereas TPC-based or drift chamber systems offer continuous tracking but require additional computational complexity to resolve ionization trails or timing ambiguities.
This design makes SiD particularly effective for environments with low to moderate track multiplicities.

\textbf{Tracking at the FCC-ee.} The Future Circular Collider (FCC-ee) is currently studying two main detector concepts to address its high-precision tracking requirements: CLD and IDEA as described in \cref{sec:simulation}.

CLD~\cite{Bacchetta19,CLICdp:2018vnx} at FCC-ee includes pixel and microstrip detectors layered across the vertex, inner, and outer tracker regions, offering high spatial resolution while minimizing material interaction. These features allow the CLD tracker to achieve exceptional momentum and spatial resolution, optimized for handling the high hit densities expected in FCC-ee's high-luminosity environment. The overall concept of CLD's tracking system shares significant similarities with the SiD detector at the ILC, both emphasizing compact all-silicon architectures with high spatial resolution and precise tracking performance. However, key differences in detector design and reconstruction algorithms distinguish CLD from SiD.

The silicon vertex detector in the CLD comprises three double layers in the barrel region and three double-layer disks per side in the forward region. With a pixel size of $25 \times 25~\SI{}{\micron}^2$ and a silicon sensor thickness of \SI{50}{\micron}, the vertex detector achieves a spatial resolution of approximately \SI{3}{\micron}, critical for precise impact parameter measurements. The tracker extends to an outer radius of $2.15$~m, surpassing the $1.5$~m radius of SiD, to compensate for the lower magnetic field of 2~T at FCC-ee compared to 3.5~T at the ILC. This adjustment ensures robust curvature measurements in the transverse momentum region. SiD, on the other hand, adopts a more compact geometry tailored for moderate track multiplicities, with a slightly higher magnetic field supporting momentum resolution.

The track reconstruction was originally developed for CLICdet, from which CLD has evolved.  It begins with clustering, where hits in the silicon layers are grouped based on spatial proximity. The high granularity of the pixel detectors supports precise seed formation, where combinations of triplets or quadruplets of hits across adjacent layers initialise track candidates. For pattern recognition, conformal tracking (CT) is employed, which maps the curved trajectories of charged particles in the magnetic field into straight lines in the $u-v$ coordinate space~\cite{Brondolin:2019awm}. 
This transformation simplifies the pattern recognition problem by linearising particle paths, effectively reducing the complexity of hit connections. The SiD algorithm, relies on direct helical trajectory models for pattern recognition, which are computationally simpler but less suited for the high-luminosity conditions and larger event sizes of FCC-ee.

The conformal mapping in CLD is coupled with a cellular automaton (CA) algorithm to identify and propagate track segments. The CA links hit pairs (or ``cells'') based on directional alignment and spatial consistency, iteratively extending track segments to form complete candidates. This method is particularly effective in reconstructing displaced tracks from secondary vertices, critical for flavour tagging and heavy-flavour physics. 
While SiD employs a global pattern recognition approach to achieve similar goals, the lack of conformal mapping and CA-based techniques makes SiD's approach more straightforward but less scalable for densely populated events.

Following pattern recognition, a Kalman filter is applied to iteratively refine the track parameters. The Kalman filter dynamically updates the trajectory by incorporating measurement uncertainties, multiple scattering, and energy loss due to material effects. However, CLD's implementation integrates enhanced material corrections to handle the denser silicon detector layout and the more challenging beam-induced background conditions of FCC-ee. The SiD algorithm benefits from its compact design and fewer layers, simplifying the propagation of track parameters.

Performance benchmarks for the CLD tracking system demonstrate its suitability for FCC-ee's physics goals. It achieves a transverse asymptotic momentum resolution of:
$\displaystyle{\frac{\sigma(\Delta \pT)}{\pT} = 3 \times 10^{-3}~\mathrm{GeV}^{-1}}$
for muons with $\pT > 1~\mathrm{GeV}$ and large polar angles, and an impact parameter resolution of: $\displaystyle{\sigma(\Delta d_0) = \SI{5}{\micron} \oplus \frac{\SI{15}{\micron}}{p \sin^{3/2} \theta}}$
\cite{Bacchetta:2019fmz}. The tracking efficiency exceeds 97\%, with a fake rate below 4.4\%, also for tracks with $\pT > 1\mathrm{GeV}$. These benchmarks match the performance of the SiD detector at the ILC while addressing the additional challenges posed by FCC-ee's higher track multiplicities.

IDEA features a design balancing the low-material budget and continuous tracking capability of the drift chamber with the high spatial precision of the silicon layers. IDEA's tracking system is optimized for high-precision reconstruction of particle trajectories, particularly for low-momentum particles, while maintaining robust performance in high-multiplicity environments~\cite{Barchetta:2021ibt}.

The track reconstruction process~\cite{Barchetta:2021ibt,DeFilippis:2022} begins with a local pattern recognition method specifically designed for the stereo geometry and high granularity of the drift chamber. The drift chamber, which spans a radial range of 800--\SI{1800}{mm}, uses 55 stereo wire layers filled with a helium-isobutane gas mixture (90:10). 
Each layer provides timing and spatial information with a transverse spatial resolution of approximately \SI{100}{\micron} per layer. In the initial step, hit clusters are formed by grouping ionisation signals based on spatial proximity and timing measurements. This process uses advanced cluster timing and cluster counting techniques, which improve both particle identification and spatial resolution by leveraging the precise timing of ionization signals along the wires.

Then, a combinatorial method where pairs of hits in consecutive stereo wire layers of the drift chamber are combined to form initial track candidates. The stereo wire geometry, consisting of 55 layers, enhances the reconstruction of the longitudinal trajectory. At this stage, the left-right ambiguity inherent in drift chamber measurements is resolved using timing offsets and the specific configuration of the stereo wires. These timing offsets, derived from the drift times of ionization clusters, allow precise determination of the track positions.

Once the initial seeds are formed, the tracks are propagated outward using a trajectory propagation algorithm implemented via the genFit2 framework~\cite{genFit2,DeFilippis:2022}. 
This algorithm iteratively incorporates new hits into the track candidates while correcting for material interactions, multiple scattering, and energy loss. The genFit2 framework uses the Runge-Kutta method for precise numerical propagation of the track parameters, ensuring high accuracy even in complex detector geometries 

For the formation of preliminary track seeds, a CA algorithm is employed for track linking. The CA algorithm matches and propagates track segments by identifying pairs of hits (cells) that are spatially and directionally consistent. These track segments are iteratively extended by adding compatible cells, resulting in full track candidates. This method is particularly effective in reconstructing displaced tracks from secondary vertices, which are crucial for flavour tagging and heavy-flavour physics analyses. The CA's localised approach ensures efficient reconstruction even in the high-multiplicity events typical of FCC-ee, maintaining both computational efficiency and high accuracy.

Following this stage, the Kalman filter is applied to refine the tracks. The filter uses the trajectory and hit data to iteratively adjust the track parameters, accounting for effects such as energy loss, multiple scattering, and drift chamber-specific uncertainties like timing variations. The Kalman filter also integrates hits from the silicon vertex detector, which provides high-resolution measurements near the interaction point with a spatial resolution of $5~\mu \text{m}$. 
This hybrid approach improves the overall precision of track reconstruction, particularly for displaced vertices and secondary decay products. The final fitting process relies on a $\chi^{2}$-based optimization implemented in genFit2, ensuring that track parameters are consistent across all detector subsystems

Timing information from the drift chamber's ionisation signals is used to resolve ambiguities and filter noise hits effectively. This timing data, combined with stereo measurements, enhances the accuracy of trajectory reconstruction in high-occupancy events. The genFit2 interface ensures the integration of magnetic field and material effects, optimizing track fitting for both linear and helical trajectories.

The IDEA detector achieves excellent tracking performance across a wide momentum range. The transverse asymptotic momentum resolution is approximately $\displaystyle{\frac{\sigma(\Delta \pT)}{\pT} = 6 \times 10^{-4}\mathrm{GeV}^{-1}}$
for charged particles with $\pT \sim 1\mathrm{GeV}$ and large polar angles. The impact parameter resolution is
$\displaystyle{\sigma_d = \SI{2.4}{\micron} \oplus \frac{\SI{20}{\micron}}{p \sin^{1/2} \theta}}$,
which ensures precise reconstruction of displaced vertices, a critical feature for heavy-flavour tagging and searches for new physics. Tracking efficiency exceeds 99.5\% for single particles, with a fake rate maintained below 5\%, demonstrating the robustness of the reconstruction algorithms even in high-luminosity environments typical of FCC-ee.

In addition to rule-based methods, both CLD and IDEA detectors are exploring machine learning (ML) techniques to further enhance tracking efficiency, particularly for complex events at high luminosity. Graph Neural Networks (GNNs) are a primary ML tool under consideration, as they treat hits as nodes and potential connections between them as edges, enabling a flexible and highly adaptable pattern recognition method. The GNN model can dynamically learn the relationships among hits, improving track reconstruction by effectively handling the combinatorial complexity that arises in densely populated events.

Preliminary studies~\cite{Garcia:2024} indicate that GNN-based pattern recognition achieves a tracking efficiency significantly larger than the CA-based approaches in CLD, maintaining efficiencies close to 100\% with similar fake rate. 
In IDEA, GNNs assist in resolving ambiguities in drift-based measurements, improving track purity and reducing fake rates by accurately associating hits under high occupancy and background conditions. Additionally, ML-based tracking reduces computational demands in event reconstruction, providing scalability that could potentially support real-time data processing.

\textbf{Tracking at the CEPC.}
The CEPC reference detector has undergone detailed studies. Its drift chamber, integrated with silicon trackers, introduces unique features for both tracking and particle identification (PID). This hybrid configuration leverages the complementary strengths of silicon detectors for precision and drift chambers for low-material continuous tracking, making it distinct from the SiD/CLD detectors at the ILC and FCC-ee, respectively. The silicon-based CEPC concepts align more closely with the SiD/CLD detectors in terms of their emphasis on conformal tracking and high spatial resolution with minimal material impact~\cite{Liu:2024pah}.

The CEPC drift chamber employs a two-stage track reconstruction process consisting of track finding and fitting. In track finding, silicon-seeded tracks are propagated into the drift chamber using the Combinatorial Kalman Filter (CKF) algorithm, adapted from its implementation in Belle II. This iterative algorithm projects initial track estimates into the drift chamber volume, identifying candidate hits by considering the spatial proximity and alignment with the extrapolated trajectory. 
A left-right ambiguity inherent in drift chambers is resolved at this stage using timing offsets and the stereo wire configuration. The trajectory propagation uses the Runge-Kutta method to iteratively integrate material effects such as energy loss and multiple scattering.

To address limitations in CKF, particularly in resolving ambiguous or missing hits, a hit salvaging algorithm is implemented. This approach examines the residual distance between the extrapolated trajectory and candidate hits to recover lost measurements. These salvaged hits are integrated into the track using GenFit-based fitting, refining the track's quality and improving hit efficiency. As an additional enhancement, hits from the external silicon tracker (SET) are appended by projecting the drift chamber tracks outward. Any SET hit falling within a $3\sigma$ window of the extrapolated position is added to the candidate tracks, improving reconstruction accuracy.

Track fitting is performed using the GenFit framework~\cite{Rauch:2015}, which optimizes track parameters through a least-squares approach, minimizing the residuals between measured and extrapolated positions. GenFit accounts for various detector effects, including inhomogeneous magnetic fields and material-induced scattering, providing globally consistent track parameters. This method ensures robust momentum and impact parameter resolutions, even in the presence of challenging background conditions.

The GenFit package has been extended within CEPC software to include custom interfaces for geometry and material interaction. These include the CEPCMaterialProvider and CEPCMagneticFieldProvider, which extract relevant parameters from the DD4hep detector description for precise modeling of the detector environment.

For single muon tracks with $\pT = 10~\mathrm{GeV}$, the transverse momentum resolution is $\Delta \pT/\pT < 0.14\%$. 
The spatial resolution, defined as the residual between the drift distances and fitted positions, is approximately $\SI{106}{\micron}$, slightly better than the design specification of $\SI{110}{\micron}$. These results demonstrate the high spatial accuracy and fitting reliability achieved through the use of GenFit and CKF.

The tracking efficiency for single particles, including electrons, muons, and pions, remains above $99.8\%$ in clean conditions, and it remains robust under $20\%$ background noise levels. 

While full integration of ML techniques such as graph neural networks is not yet implemented, exploratory work is underway to assess their potential in enhancing tracking performance~\cite{Liu:2018mqb}.

\textbf{Conclusion.}
The tracking systems for future $\epem$ colliders exhibit diverse approaches tailored to their physics goals. Fully silicon-based designs, like SiD at the ILC and CLD at FCC-ee, achieve exceptional precision, with a transverse asymptotic momentum resolutions of  $\delta(1/\pT) = 5 \times 10^{-5} \mathrm{GeV}^{-1}$  and  $7 \times 10^{-5} \mathrm{GeV}^{-1}$ , respectively, and impact parameter resolutions optimized for high spatial accuracy. 
Hybrid systems, such as IDEA and CEPC's drift chamber design, combine low-material gaseous trackers with high-resolution silicon detectors, achieving resolutions of  $\sim \SI{100}{\micron}$  per layer and tracking efficiencies exceeding 99.5\%. 
Advanced reconstruction techniques, including Kalman filters, combinatorial methods, and emerging machine learning tools, ensure robust performance across all designs, positioning these detectors to meet the stringent demands of next-generation particle physics experiments.

\subsubsection{Calorimetry and particle flow} 

\textbf{Introduction and PandoraPFA.}
Many calorimeter designs for HTE factories are based on Particle Flow calorimetry.
Particle Flow is a concept to separate particles inside jets by highly-granular calorimeters.
The jet energy resolution can be improved by assigning tracks to calorimeter clusters
and use momenta of tracks instead of cluster energies.
PandoraPFA \cite{Thomson:2009rp} is a long-used Particle Flow algorithm for LC studies, which demonstrated around 3\% jet energy resolution at 100 GeV jets with ILD SiW-ECAL + AHCAL configuration.
PandoraPFA is implemented as a modular framework with multiple sub-algorithms such as
cone clustering of hits, topological association of clusters, track-cluster matching and reclustering, and thus possible to adapt with various new ideas and detector configurations.
Applications to the ILD/SiD/CLICdet/CLD detectors have been already established and adaptation to Allegro is ongoing.

\textbf{Arbor and related studies on CEPC.}
Arbor is the PFA algorithm that has been used for the CEPC physics and detector studies. As its name suggests, the idea of Arbor is to reproduce the tree topology of particle shower (especially the hadronic ones) at high granularity calorimeter, by create oriented connectors between nearby hits. The reconstructed tree topology then enhances the separation of nearby particles, which is critical for the PFA reconstruction. 

Conventionally, the performance of PFA at CEPC is quantified by the Boson Mass Resolution (BMR), which refers to the relative mass resolution of massive, hadronically decayed bosons, for example Higgs boson decays into a pair of gluons. Though optimization studies, it became clear that the detector of electron positron Higgs factory need to deliver a BMR better than 4\%, to separate the qqH signal from qqZ background using the recoil mass to the di-jet system. At the CEPC CDR studies, the combination of Arbor and CDR baseline detector reached a BMR of 3.7\% \cite{Ruan:2018yrh}; while though detector optimization --- usage of a Glass-Scintillator HCAL --- the BMR reaches 3.4\% \cite{Hu:2023dbm}. 

The concept of 1-1 correspondence reconstruction has been developed by combining Arbor and Artificial Intelligence. The latter is used to identify and control various of confusions raised in the Arbor reconstruction. Quantitative analysis shows that 95\% of visible energy could actually be reconstructed into PFOs that preserve the 1-1 correspondence mapping with the visible particles. These reconstructed particles are then identified into 10 different species, corresponding to 5 types of charged particles, photons, and 4 types of neutral hadrons. The charged particle and photon have inclusive simultaneous identification efficiencies around 98\% to 100\%, while the neutral hadrons could be simultaneously identified with typically efficiency of 80\%. 1-1 correspondence reconstruction suppresses significantly the confusion of PFA, and the BMR has been consequently improved to 2.8\% \cite{Wang:2024eji}. 1-1 correspondence reconstruction describes the physics event holistically, and provide new methodologies for both physics analysis and control of systematic effects. 

\textbf{GNN-based Particle Flow Algorithm.}
The complicated nature of the calorimeter clustering and particle flow gives
opportunity to improve the performance by utilising machine-learning algorithms.
Among many studies, there is an ongoing study on fully implementing particle flow
with single network including graph structure. It utilises GravNet-based network,
which uses distance-based convolution in a virtual coordinate, and Object Condensation
loss function which attracts hits of the same cluster in another virtual coordinate.
A preliminary study \cite{Suehara:2024qoc} shows better cluster-level performance over PandoraPFA in ILD simulation and further studies on energy regression is ongoing.


\textbf{Dual-readout calorimetry.}
The dual-readout calorimeter approach is motivated by its ability to achieve superior energy resolution compared to traditional calorimetry, particularly due to the non-compensation in hadronic showers. 
The method leverages two distinct signals: scintillation light sensitive to all charged particles and Cherenkov light, which predominantly detects relativistic electromagnetic shower components. By measuring the ratio of these signals, the electromagnetic fraction of a shower can be corrected on an event-by-event basis. This technique significantly improves energy resolution, critical for precise measurements of hadronic decays of W, Z, and H bosons at future colliders such as the FCC-ee and CEPC. 
Dual-readout technology has also the potential to minimize systematic uncertainties due to fluctuations in nuclear interactions and provides linear responses across a wide energy range, making it ideal for jet energy measurements.

There are different options for a dual-readout calorimeter~\cite{Aleksa:2021ztd}, such as a fully fibre-based electromagnetic (EM) and hadronic (HAD) calorimeter~\cite{INFNRD-FA:2020gzh}, or a hybrid design consisting of a crystal electromagnetic calorimeter (ECAL) followed by a fibre-based hadronic calorimeter (HCAL)~\cite{Lucchini:2020bac}. In this section, we focus on the latter; however, the general principles of reconstruction remain similar across both approaches.

The dual-readout particle flow algorithm (DR-PFA), e.g. Ref.~\cite{Lucchini:2022vss},  is specifically tailored to exploit the dual-readout corrections to compensate for hadronic shower non-linearity and delivering a near-linear response for neutral hadrons.

The detector simulation utilizes a cylindrical ECAL composed of lead tungstate (PWO) crystals with a transverse segmentation of $1 \times 1 \, \text{cm}^2$, providing an electromagnetic energy resolution of $3\%/\sqrt{E}$. The HCAL section consists of longitudinally unsegmented scintillating and Cherenkov fibres embedded in copper, achieving hadronic energy resolutions of $(25-30\%)/\sqrt{E}$. 
The simulation is currently ported to \geant~\cite{GEANT4:2002zbu}, and effects related to instrumental response, such as photo-statistics fluctuations and SiPM-based noise, are incorporated into the signal digitization to ensure a realistic representation of the detector performance.

The DR-PFA operates in sequential stages. Photon identification is performed by clustering ECAL hits into neutral seeds with energies above 100 MeV, where no associated charged track exists within a radius of $\Delta R = 0.013$. 
A transverse shower shape parameter $R_{\text{transverse}} = E_{\text{seed}} / \sum_i E_{\text{hit},i} (\Delta R_i < 0.013)$ is used to clean up photon candidates. 
Track matching follows by associating hits in the ECAL and HCAL to charged tracks, iteratively minimizing the residual between the expected and measured energy response. The unmatched hits, primarily from neutral hadrons, are corrected using the following rule: $E_{\text{hit}} = (S - \chi C)/(1 - \chi)$,
where $S$ and $C$ are the scintillation and Cherenkov signals, respectively, and $\chi$ is the calibration constant for the calorimeter section. Jet clustering is performed using the FastJet package, combining hits from photons, charged tracks, and dual-readout-corrected neutral hadron contributions.

Performance studies demonstrate the capability of the DR-PFA to achieve jet energy resolutions of 4.5\% at 45 GeV, improving to 3\% for higher jet energies. Compared to calorimeter-only reconstruction, the DR-PFA reduces the resolution by nearly 25\%~\cite{Lucchini:2022vss}. 
The dual-readout corrections provide linear responses for hadrons and Gaussian-like energy distributions, mitigating the intrinsic limitations of non-compensating calorimeters. Angular resolutions below $0.01$~mrad for 125 GeV jets further highlight the potential of this approach for the FCC/CEPC physics goals.

\textbf{Conclusion.}
Particle Flow Algorithms and dual-readout techniques hold significant promise for advancing particle energy measurements at future colliders such as FCC-ee and CEPC. While PandoraPFA has established benchmarks for high-granularity detectors, the Dual-Readout Particle Flow Algorithm demonstrates competitive performance by leveraging scintillation and Cherenkov signals to mitigate hadronic shower non-linearity, achieving state-of-the-art jet energy resolutions of 4.5\% at 45 GeV and 3\% at higher energies, with angular resolutions below 0.01 mrad. Emerging approaches, such as GNNs, show potential to further improve performance by optimizing the reconstruction process and exploring more of the detector's true potential.
 
\subsubsection{Particle identification \label{sec:com:reco:PID}}
PID in its most general meaning refers to the identification (ID) of the species of the individual reconstructed particles, \ie photons, leptons and hadrons sufficiently long-lived to directly interact with the detector and leave an identifiable imprint.
It is, however, usually used to refer to the identification of detector-stable charged particles, and dedicated PID subdetectors often specifically aim to provide hadron identification.
Here, we will discuss advances in the simulation of PID systems by technology, covering the ID of electrons, muons, pion, kaons, protons, photons and neutrons, with a focus on pion-kaon separation where applicable.

\textbf{Cluster shapes.} Practically all recent and future detector concepts have adopted an approach to basic PID via cluster topology, \ie the use of tracking to identify the charge of a particle and then a dedicated calorimeter system, which allows to differentiate between early, \ie ECal, showers of electrons and photons and late, \ie HCal, showers of charged and neutral hadrons.
Further tracking layers outside of the hadronic calorimeter identify muons, hence their name muon chambers.
This very general approach is common to all future collider detector concepts and is usually implemented as a module in the corresponding ParticleFlow algorithm chain.
It provides very efficient tagging of electrons, muons, photons and neutral hadrons above an energy of a few GeV.
For ILD it was found that Pandora ParticleFlow had some inefficiencies, in particular with regard to electron and muon ID, which could be enhanced.
Hence a dedicated LeptonID tool \cite{CommonReco_LeptonID} was added to improve the electron and muon ID efficiency and purity, in particular at lower momenta.

\textbf{Calorimeter hadron ID.} In the approach described above, there is no hadron ID provided, considering the showers of pions, kaons and protons have very similar topologies.
However, a new study \cite{CommonReco_CaloPID} showed that with sufficient (spatial and/or temporal) granularity the differences in shower topology can in fact be exploited to provide a modest level of hadron ID.
Heavier particles tend to shower later and with a larger radius, which can be used to enhance separation.

\textbf{Gaseous tracker d$E$/d$x$ / d$N$/d$x$.}  For dedicated charged particle ID gaseous trackers have proven effective in the past, so ILD is proposing a TPC and IDEA a drift chamber (DC) as central tracking devices.
These already provide good PID via the conventional approach of correlating the momentum of charged particles and their specific energy loss \dedx. The average energy loss is strongly tilted due to the Landau distribution governing the individual ionising interactions, which is typically mitigated by using a truncated mean method, as described \eg\ in Ref.~\cite{CommonReco_Aoki2022}. This way, the TPC achieves a \dedx\ resolution of about 4.5\% for \SI{1.3}{m} tracks, the DC also of about 4.5\% for \SI{2}{m} tracks.
This can be significantly enhanced via the so-called cluster counting. This method aims to instead determine the number of ionising interactions, which does not suffer from the Landau tails.
It was already proposed decades ago but needs a high sample granularity along a track, either spatially or temporally, which is now made possible by new readout technologies.
For both the TPC and the DC algorithms have been developed to enable this cluster counting approach and were applied to test beam measurements of prototypes.

For a highly segmented TPC readout that provides granularity in space, Ref.~\cite{CommonReco_Aoki2022} implemented an algorithm for source extraction used in astronomical images to enable cluster counting in the 2D projection of the detected energy depositions showing the dependence of effective resolution on granularity.
With this approach the effective \dedx\ resolution of the ILD TPC in simulation can be improved to about 3.5\% with a granularity of O(\SI{100}{\micron}).
Using a fully pixelised TPC with a granularity of \SI{55}{\micron}, Ref.~\cite{CommonReco_PixelTPC} implemented a cluster reconstruction scheme via a template fit with an inverse-distance weighting of individual active pixels and applied it to beam test measurements, which results in an effective \dedx\ resolution for ILD of 2.5\% (\SI{1.3}{m} tracks).

A DC readout with ps-timing per sensitive wire provides granularity in time and several methods have been developed to extract the number of clusters from the wire signal.
Reference~\cite{CommonReco_DCdNdx} implemented a simple peak-counting method as well as a one using a template fit with a single-electron signal as template.
This enabled an effective \dndx\ resolution of 3\% for the IDEA DC (\SI{2}{m} tracks) when applied to beam test measurements in an initial analysis \cite{CommonReco_DCdNdxTalk}.
With more refinement, there are prospects to improve this further and get closer to the analytical expectation of about 2\%.
Reference~\cite{CommonReco_DCdNdxTalkNN} implemented a cluster counting approach using two neural networks, one for peak finding and one for cluster number extraction.
While this has not been applied to data yet, it shows higher efficiency and purity in direct comparison to the aforementioned template fit method, which also raises prospects to get closer to the analytical expectation.

Overall, the base \dedx\ resolutions correspond to effective pion-kaon separation over a large momentum range from a few \SI{100}{\MeV} to about 20 to \SI{30}{\GeV}, with a gap around \SI{1}{\GeV}, the so-called blind spot.
With cluster counting, the separation power is significantly increased, leading to an increase of the upper bound of the usable momentum range to 50 to \SI{100}{\GeV} and a minimisation of the blind spot.

\textbf{Time of flight.}
A ps-timing capability can also be used to measure the time of flight (TOF) of charged particles.
Combined with their momentum and track length their mass can be directly calculated.
The small beam spot in e+e- colliders allows the starting point of a TOF measurement to be determined by the machine clock rather than a dedicated reference measurement in the vertex detector.
The corresponding end point is chosen as late as possible to enable a lever arm as long as possible.
This can be realised via a dedicated timing layer between tracker and calorimeter or by utilising subdetectors which are foreseen anyway, \ie the outermost part of the tracker
or the inner part of the calorimeter, equipped with corresponding timing capabilities.
Since the calorimeter offers a multitude of measured energy depositions per particle, it would require a less stringent timing requirement per hit compared to a single dedicated timing layer,
but may still pose a significant hardware implementation challenge if applied to a large fraction of the calorimeter system.
Hardware developments for ps-timing in recent years, in particular for pile-up rejection at the HL-LHC, can provide between 30 and \SI{100}{ps} precision per detector hit, with newer fundamental R\&D pointing closer to \SI{10}{ps} for future detectors.

Based on this, a number of TOF assessments have been implemented in large detector reconstruction.
The basic approach in \delphes simulation is simple: the mass is directly calculated from the track length, momentum and TOF smeared with an assumed timing resolution given at a suitable point, \eg\ the entry point in the calorimeter.

In full-sim, momentum and track length need to be reconstructed, as must be the measured time if taken from the calorimeter system.
This has been implemented by Ref.~\cite{CommonReco_ThesisDudar} for the ILD detector and led to several interesting findings.
The track length goes into the reconstructed mass with the same exponent as the timing, so it needs to be known with at least the same precision, \ie typically at or below the per-cent level. In a gaseous trackers, such as ILD's TPC, this can be achieved through a harmonic mean of the track length along the quasi-continuous tracking \cite{Mitaroff:2021yzp,CommonReco_ThesisDudar}. Furthermore, TOF is an end-to-end measurement, as opposed to the continuous measurement of \eg\ \dedx. There, thanks to truncation minor reconstruction errors are usually cut out and compensated, while any misreconstruction in a TOF measurement typically leads to an unusable result. This makes a highly efficient track reconstruction a necessity for high efficiency and purity TOF, \ie for a high separation power.

Combining hit times by an arithmetic mean for hits inside a cylinder in the 10 ECal layers and within a certain time window as well as assuming a per-hit resolution of \SI{50}{ps}, Ref.~\cite{CommonReco_ThesisDudar} achieved a combined per-particle resolution of about \SI{17}{ps}. The application of a neural network approach to the same data by Ref.~\cite{CommonReco_ThesisHelms} resulted in a per-particle resolution of \SI{15.5}{ps}.

TOF can provide pion-kaon separation over the range from a few 100 MeV to between 3 and 10 GeV, depending on the effective timing resolution per particle.
This range also covers in particular the \dedx\ / \dndx blind spot around 1 GeV.

\textbf{Cherenkov radiation.}
Another long-used PID technology is the ring imaging Cherenkov detector (RICH).
Since it requires a separate subdetector which takes up space and adds material before the calorimeter, it was not considered for a Higgs factory detector optimised for ParticleFlow for a long time.
With the growing understanding of the importance of PID as well as the the ability to minimise the impact on material budget through the utilisation of Silicon photomultipliers (SiPMs), however, RICH systems have been proposed more recently for detectors that use full-Si trackers and have thus no PID above 10 GeV via \dedx\ / \dndx.
For SiD, a single-phase (gaseous) RICH \cite{albert2022strangequarkprobenew} has been proposed, providing pion-kaon separation from 5 to 30--40 GeV.
For CLD, a dual-phase RICH, with an aerogel and a gaseous phase called ARC \cite{CommonReco_ARCTalk}, has been proposed and (partially) implemented in an adapted CLD detector model. It provides pion-kaon separation up to 45 GeV, depending on the used gas. The aerogel part will provide additional coverage of the low-momentum range once implemented in simulation. 
The used gas is a major concern: traditional gas mixtures are based on fluoro-carbo-hydrates, strongly climate-active gases which are expected to be banned in some years. Noble-gas based alternatives are being studied and can likely provide similar performances.

\textbf{Combination and assessment.}
The multitude of PID observables per detector concept and between detectors introduces the need for a coherent combination and comparison of the PID evaluation.
This is addressed by the Comprehensive Particle Identification (CPID) tool \cite{CommonReco_CPID}.
Its modular structure allows for different observables as well as different combination algorithms to be utilised at run time, which provides the flexibility desired for comparative studies of detector designs.
CPID also makes use of the ``p-value'' method for PID performance assessment, rather than a simple separation power.
While the separation power is reasonably well defined for sufficiently Gaussian underlying distributions, such as the track dE/dx distribution (using truncation), it is overestimating the performance for non-Gaussian distribution, such as the reconstructed mass using TOF which typically has long tails that overlap between species.
The p-value characterises this overlap and can therefore be used with arbitrary distributions -- including Gaussians, which allows for a re-translation to a well-defined separation power.

\textbf{Conclusion.}
Overall, PID has been identified as a desirable detector capability. It appears realistic to have detectors at a future Higgs factory that cover large momentum ranges in (hadron) PID, with TOF for low and dE/dx / dN/dx or RICH for high momenta.
The reconstruction algorithms of these systems are still being developed and refined and may allow for better performances than currently envisioned. Full-simulation is crucial to understand the full capabilities and limitations of each technology in the context of a complete detector concept.

 \subsubsection{Isolated lepton and photon tagging}
 
\textbf{$e$/$\mu$ tagging.}
Isolated lepton tagging comprises of two algorithmic parts: identifying the lepton, and determining its level of isolation.
The identification of electrons and muons can be done with high efficiency using the basic cluster shapes information described in the corresponding PID section above. The LeptonID tool is a new development that improves this aspect.
The level of isolation generally means isolation from jets, since leptons can be jet constituents, while in many analyses one looks for leptons that come directly from the initial hard scattering, e.g.\ as daughters of heavy bosons.
Cuts on the angular distance to neighbouring particles in combination with the particles' momentum allow for a highly efficient tagging of isolated electrons and muons in all detector concepts envisioned for a future Higgs factory.

\textbf{Photon tagging.}
Photon tagging is implemented with similar conditions to electron tagging. The current implementation for ILD relies on particle ID provided by PandoraPFA, and a cone selection is applied for the isolation condition. There are two cones, to cluster activity related to final state radiation or material interaction in an inner cone, and identify neighbouring activities in an outer cone to separate jet photons. Further optimization is desired for studies for which the performance of the photon identification is critical.

\textbf{$\tau$ tagging.}
Identifying isolated tau leptons and distinguishing them from hadronic jets is an essential task for many analyses.
A standardised algorithm would help physics analyses using taus
mainly for event selection. For more specific analyses heavily depending on tau characteristics (\eg Higgs to tau pairs), detailed analyses such as identifying decay modes can be implemented after the standard tau tagging algorithm.
Leptonic decay of taus can be identified as isolated leptons, so here we only consider hadronic decays.

There is an implementation (TauJetClustering) originally developed for ILD analyses \cite{Kawada:2015wea}.
The algorithm consists of a clustering and a selection part.
For the clustering, it selects the track with highest energy as a seed, and clusters surrounding particles based on invariant mass and opening angle. The number of charged particles is limited as most of the taus have one, three or five tracks. An optional selection algorithm distinguishes taus from hadronic jets, which is necessary for analysis of partially hadronic final states. It was originally optimized to separate two-tau events such as Higgs decaying to two taus with a hadronic $\PZ$ decay, from semileptonic four-fermion backgrounds such as $\PQq\PQq\PGt\PGn$. 
The selection consists of an isolation condition with two cones, whose inner cone specifies the tau decay products while the outer cone ensures isolation from other hadronic activity. The energy fraction of each cone can be set as parameters. Other selections based on impact parameters and secondary lepton identification are available,
but currently only cone conditions are used for analyses.

There is an alternative approach, mainly adopted by FCC-ee studies, that hadronic taus are treated as a flavour of jet. In this approach, the clustering of taus is done with ordinary jet clustering algorithms and the tagging of taus is possible via flavour tagging algorithms described in the later sections. The advantage of this approach is that it is easier to use modern ML-based flavour identification algorithms, for which high performance is expected. In contrast, misclustering of taus combined into hadronic jets can degrade the overall tau-tagging performance.
Comparison of these methods should be done as future work.
 
 \subsubsection{Jet clustering}



\textbf{Jet clustering with LCFIPlus.}
Jet clustering is one of the most important reconstruction tasks for analyses including jets. In LCFIPlus \cite{Suehara:2015ura}, jet clustering is implemented to maximise performance of jet flavour tagging for events including multiple heavy-flavour jets, which often suffers from jet mis-clustering causing mis-counting of the number of heavy-flavour jets. In LCFIPlus, jet clustering is done after vertex finding so that vertex information is used to avoid reconstructed jets including multiple heavy mesons.
The basic algorithm is pair-wise with a choice of pairing functions: either Durham \cite{Catani:1991hj}, $\kT$, or Valencia \cite{Boronat:2014hva}. 
Each reconstructed vertex is treated as a single particle at clustering, and any pairs including two vertices optionally yield a higher value on the pairing functions to avoid clustering multiple vertices to single jets. To avoid separating cascade decays, a threshold on the opening angle can be applied and angles lower than the threshold forcibly clustered. Secondary leptons can also be treated similarly to the vertices.
Another characteristic of the jet clustering in LCFIPlus is pileup separation. Algorithms like $\kT$ are intrinsically implemented with such a feature, for clustering particles near the beam axis that are to be removed. In LCFIPlus all algorithms including Durham are capable of pileup rejection by clustering particles with the beam direction. Several tunable parameters are available to adjust the strength of the pile-up rejection.

\textbf{FastJet.} 
FastJet~\cite{Cacciari:2011ma} is a standalone jet clustering package widely used in studies in hadron colliders as well as Higgs factories. It covers a wide variety of jet algorithms used for \epem colliders such as Durham, Cambridge/Aachen as well as generalized $\kT$ algorithms. It is common to use the anti-$\kT$ algorithm to reject low-energy pileup at the forward region.
For the ILD implementation, LCFIPlus allows these jets to be treated the same way as those produced by its internal algorithm.

\textbf{Study on colour singlet identification.}
Colour singlet identification (CSI) refers to the procedure to identify the colour singlet origin of each final state particle in the full hadronic events, for example, to distinguish the final state particles originating from the Higgs boson or $\PZ$ boson in $\PZ\PH$ events, or, from which Higgs bosons in $\PGn\PGn\PH\PH$ events. Given that the majority of ZH events decay into fully hadronic final states, while the Higgs self-coupling is critical for understanding the Higgs potential, CSI is critical to the success of Higgs factories. 

CSI could be achieved through the conventional method using Jet Clustering and matching. However the jet clustering could group particles from different colour Singlet origins into the same jet. Quantitative studies show that for typical 4-jet events, e.g.\ fully hadronic WW and ZZ events at 240 GeV, the performance of jet clustering plus matching methods result in roughly 50\% of the events with a successfully reconstructed colour singlet, while the other half is close to a random combination of final state particles \cite{Zhu:2019uve}.

\subsubsection{Flavour tagging \label{sec:com:flavourtagging}}
In recent years, a number of new, neural-network based tools for flavour tagging have been devised, which all strive to use the available information in jets in an optimal way to determine its originating quark's flavour. Several implementations of the ParticleNet \cite{Qu:2019gqs} tagger and of the ParticleTransformer \cite{Qu:2022mxj} are described in the following, after a reference to the established tool LCFIPlus \cite{Suehara:2015ura}.
Due to the rapid developments and their variety we can not cover all approaches.
Instead it seems desirable to set up ways for a coherent comparison of the various algorithms to be able to down-select in the future and continue with the most successful and promising tagger implementations.


\textbf{Flavour tagging with LCFIPlus.}
LCFIPlus is a long-used software tool (released in 2013) which consists of vertex finder, jet clustering, additional treatment of vertices after jet clustering (called JetVertexRefiner) and Boosted Decision Tree (BDT)-based flavour tagging. The vertex finder is implemented with a tear-down method for the primary vertex and a build-up method for secondary vertices. Using kinematic selections it is possible to reconstruct vertices before running the jet clustering resulting in improved performance of heavy-flavour jets. The JetVertexRefiner associates jets and vertices (in case an external jet clustering method is used) and finding single-track vertices with secondary tracks around jet axis. It also constrains the number of vertices in each track to two by refitting vertices if there are more than two vertices. After this process, a BDT-based flavour tagging algorithm is applied, with input variables from displaced tracks and vertices. Only high-level variables like vertex position, multiplicity and mass are used unlike modern DNN-based methods which usually utilise low-level information of all tracks and neutral particles. This results in a moderate performance of 6.3\% c-jet and 0.79\% uds-jet efficiency at b-tagging of 80\% efficiency in ILD full simulation of 2-jet events (\SI{250}{\GeV} $\PGn\PGn\PQq\PQq$) \cite{Suehara:2024qoc}. A significant number of analyses have been done with LCFIPlus, while an important effort has been made with a DNN-based algorithm expected to give significant improvement of performance.

\textbf{ParticleNet-based flavour tagging at ILD.}
One flavour tagger based on ParticleNet has been implemented for the ILD detector concept.
It uses three consecutive edge convolutions and then two layers of a fully connected net.
The number of features is 14 from secondary vertices and 38 from the individual constituents of the jets to be tagged and the features are largely chosen analogously to LCFIPlus to provide comparability.
The training data are dedicated 6-jet and 4-jet samples at 500 and 250 GeV in ILD full-simulation.
The initial implementation \cite{CommonReco_FT_Meyer} aims for a direct comparison with LCFIPlus, using similar inputs and generating the same tag types (b, c and ``other''). An update, mentioned in Ref.~\cite{CommonReco_FT_Einhaus}, has added PID information (ILD's TPC \dedx\ and TOF) and a dedicated s-tag.
The mistag rates for c- and uds-jets at 80\% b-tag efficiency are improved by about a factor 1.5 compared to LCFIPlus. The additional s-tag vs.\ uds-background has a moderate performance with an AUC of about 0.7 and a uds-mistag rate of 27\% at an s-tag efficiency of 60\%.

\textbf{ParticleNet-based flavour tagging at IDEA.}
The ParticleNet flavour tagging algorithm for the IDEA detector  concept (ParticleNetIDEA)~\cite{Bedeschi:2022rnj} utilizes input features derived from reconstructed jet constituents, focusing on four-vectors and track-related information. These features provide detailed kinematic and positional data essential for jet substructure and flavour discrimination. 
In this context a  fast-tracking algorithm was developed for the IDEA detector, which delivers precise measurements of track parameters and computes the full track covariance matrix. This enables detailed propagation of uncertainties into the tagging features, significantly enhancing the reconstruction of jet substructure and improving particle identification capabilities, particularly for challenging cases like strange jet tagging.
Additionally, it incorporates techniques to improve particle identification (PID), leveraging advanced track-based PID methods, which are particularly relevant for enhancing strange jet tagging performance.

Unlike the ILD implementation, which uses full simulation, ParticleNetIDEA tagger is trained using FastSimulation to speed-up data preparation, and detector and algorithm optimization.

ParticleNetIDEA is able to reach 90\% b-tagging efficiency, with a c-jet mistag rate of approximately 2\% and a uds-jet mistag rate around 0.2\%. For strange jet tagging at 60\% efficiency, the uds-mistag rate is less than 1\%.
Further development explores fully connected graph architectures, particularly Transformer-based models. Preliminary results indicate up to a $\approx50\%$ improvement in background rejection at equivalent signal efficiency, showcasing the transformative potential of these advanced techniques for flavour tagging~\cite{FlavourTagging_IDEA}.

Systematic studies~\cite{IDEApixelThicknessStudies} show the importance of the single-point
resolution of the silicon pixel layers of the IDEA vertex detector
in c-tagging as well as the rejection of strange-jets. The impact of
the silicon tracker thickness was
studied where, after retraining, a degradation of the light jet
rejection and c-tagging efficiency was observed.
The performance is sensitive to the number of sensor layers and their
distance from the beam pipe, as expected. Further studies quantify the
impact
of PID information such as cluster counting and time of flight
measurements on ud-, b-, and c-tagging.

\textbf{Flavour Tagging with Particle Transformer at ILD.}
Particle Transformer (ParT) is a transformer-based algorithm which gives currently one of the best performance on heavy flavour tagging in LHC experiments.
Its application to ILD full simulation 
is being studied \cite{Suehara:2024qoc}\cite{Tagami:2024gtc}. In this study, impact parameters, error matrices and kinematic variables of all tracks and kinematic variables of all reconstructed neutral particles are used as input variables. Separate embedding layers for tracks and neutral particles are prepared before connecting to the self-attention layers. A sample with 1 million jets from $\PGn\PGn\PQq\PQq$ events at $\sqrts=250$~GeV with ILD full simulation are used in this study, and 80\% of the sample is used as training.
Optimization in a 3-category (b, c, and uds) classification scheme results in a b-tagging performance of 0.48\% and 0.14\% efficiencies on c and uds jets at 80\% efficiency b selection, and a c-tagging performance of 0.86\% and 0.34\% on b and uds jets at 50\% efficiency c selection. The rejection is about 10 times better compared to LCFIPlus. 
Further optimization is ongoing.

\textbf{Flavour Tagging with Particle Transformer at FCC-ee.}
The application of ParT is also studied in fast simulation using the IDEA concept and a modified IDEA configuration with a Silicon Tracker (IDEA-ST), as well as in CLD full simulation~\cite{aumiller_2024_e537w-n6886}. 
The jet-tagger
uses seven categories (u, d, s, c, b, g, and $\tau$) and takes the same
input features as ParticleNet. With this model, the IDEA tagger
performance improves to an 80\% c-tagging efficiency with a ud mistag rate of 0.2\% (compared to 1\% with ParticleNet). With particle
identification, IDEA achieves an s-tagging efficiency of 60\% with a ud mistag rate of approximately 10\%. In contrast, the IDEA-ST detector (without particle identification) can only reach a ud mistag rate of 30\% for the same tagging efficiency. The CLD full simulation tagging performance is observed to be comparable to the ILD full simulation described above.

\textbf{Transformer-based Neural Network: DeepJetTransformer} 
DeepJetTransformer \cite{Blekman:2024wyf} is a jet flavour-tagging algorithm exploiting a transformer-based neural network that is substantially faster to train.
The DeepJetTransformer network uses information from particle flow-style objects and secondary vertex reconstruction as is standard for b- and c-jet identification supplemented by additional information, such as reconstructed $V_0$s and $\PKpm/\PGppm$ discrimination, typically not included in tagging algorithms at the LHC. The model is trained as a multiclassifier to identify all quark flavours separately and performs excellently in identifying b- and c-jets. An s-tagging efficiency of 40\% can be achieved with a 10\% ud-jet background efficiency. The impact of including $V_0$s and $\PKpm/\PGppm$ discrimination is presented.
The network is applied on exclusive $\PZ \to \PQq\PAQq$ samples to examine the physics potential and is shown to isolate $\PZ \to \PQs\PAQs$ events. 
Assuming all other backgrounds can be efficiently rejected, a $5\sigma$ discovery significance for $\PZ \to \PQs\PAQs$ can be achieved with an integrated luminosity of 60 nb$^{-1}$, corresponding to less than a second of the FCC-ee run plan at the Z resonance.

\textbf{Jet origin identification with ParticleNet and ParticleTransformer.}
Jet origin identification (JoI) \cite{Liang:2023wpt} is a newly developed idea for high energy frontier that aims to distinguish jets originating from different quarks and gluons. It classifies jets originating from 11 different coloured particles simultaneously, namely five species of quark (except top quark), their anti-particles, and gluon. The concept of JOI combines jet flavour tagging, jet charge measurement, strange jet identification, gluon-quark jet differentiation, etc., at the same time.  

JoI has been realized in CEPC studies, by applying the ParticleNet or ParticleTransformer framework to the fully reconstructed di-jet events of the CEPC reference detector. In this set up, JoI delivers simultaneous jet flavour tagging efficiencies of 90\%/80\%/70\% for b/c/s jet, and nearly 40\% efficiency for u and d jets. The jet charge flip rate is under control typically at 7--20\%. 

JoI has a strong impact at the physics reach of a Higgs factory, enabling to obtain limits on the Higgs rare and exotic hadronic decay branching ratios (uu, dd, ss; uc, ds, db, sb) at a level of $10^{-3}$ to $10^{-4}$ level with only $\PGn\PGn\PH$ and $\Pl\Pl\PH$ events at CEPC. For the $\PH\rightarrow\PQs\PQs$ decay, this upper limit corresponds to three times the SM prediction. Compared to the current analysis, JoI improves the precision of H$\rightarrow$cc measurements by roughly a factor of 2. Jet origin identification also boosts the precision of the weak mixing angle measurements near the Z pole, the CKM matrix elements measurements using W boson decays, as well as flavour physics measurements, e.g.\ Bs oscillations. 

\textbf{Conclusion.}
Driven by machine learning developments, a plethora of neural network approaches have been adapted for flavour tagging.
They easily outperform more traditional methods, in particular BDT-based ones.
With increasing complexity, the networks are capable of utilising an increasing amount of the information provided by the detector systems, which is why absolute performances depend strongly on the simulated technology and its level of detail.
The ``final'' performance numbers remain open and under constant improvement. In addition, tags have been expanded to include in particular strange jets, but are now also targeting u- and d-jets, as well as the jet charge sign.
Studies also cover the performance dependence on individual observables, which can ultimately inform the detector hardware development.
With the increasing number of implementations, it also becomes increasingly desirable to have direct comparisons between them on defined data sets and a coherent way of assessing their results.

\subsection{Summary}
\label{sec:WG2summary}
\editors{Patrizia Azzi, Fulvio Piccinini, Dirk Zerwas}

Over the course of the last years, the development of the Physics Analysis Tools for Higgs factories has made 
great strides. The software ecosystem, \keyhep is supported by and supports all Higgs factory variants providing 
interfaces to generators and the full life cycle from event generation through simulation to reconstruction.

Multiple Monte Carlo event generators from the LEP era, LHC and those developed specifically for Higgs factories
are now available for multiple processes. A consistent set of luminosity spectra has been provided
for all Higgs factories.
All of this enables detailed comparisons and the estimation of uncertainties. Automated
comparisons are being developed. 

In addition to fast simulations, detailed simulations for detectors have been developed.  Residing in
\keyhep the assembly of full concepts from this basis is facilitated, irrespective of the underlying
collider for which they were developed.

The common use of \keyhep enables also the improvement of reconstruction algorithms such as tracking,
particle flow, particle identification and flavour tagging. Modern techniques such as Machine Learning
improve the expected performance, but maybe more important the cross checks performed in the common
framework render robust the estimation of the expected performance.

Summa summarum, the Physics Analysis Tools suite for Higgs factories is solid, leading to a robust basis 
for physics analyses. This work makes the high precision physics goals for Higgs factories reachable.

One last word:
MC generators are based on collaborative efforts with time-consuming service tasks like user support, validation and portability. Especially young MC developers need continuous support from the community to ensure active MC collaborations at the time of data taking at a Higgs factory. The same is true for the software framework and the basic algorithms where a well-supported core group of software developers is essential to help achieve these aims. The Higgs factory program 
of high precision analyses can only succeed if the work on MC generators, the software ecosystem \keyhep and the algorithms
is supported with grants and permanent positions.

\clearpage
 \section{Developments in Higgs Physics}\label{sec:higgs}
\editors{Chris Hays, Karsten K\"oneke, Fabio Maltoni}

The Higgs boson that was first observed in 2012 by the ATLAS and CMS Collaborations with a mass of 125~\GeV~\cite{HIGG-2012-27,CMS-HIG-12-028} is the newest SM particle to be measured by experiments. It has been studied in detail at the LHC, and the upgraded experiments at the HL-LHC will provide further insight. This includes the LHCb experiment, which can provide additional constraints on Higgs-boson decays to a pair of bottom or charm quarks~\cite{LHCb:2013gxz}. 

\begin{figure}[h!]  
    \begin{subfigure} {.45\linewidth}
        \centering
        \includegraphics[width=\linewidth]{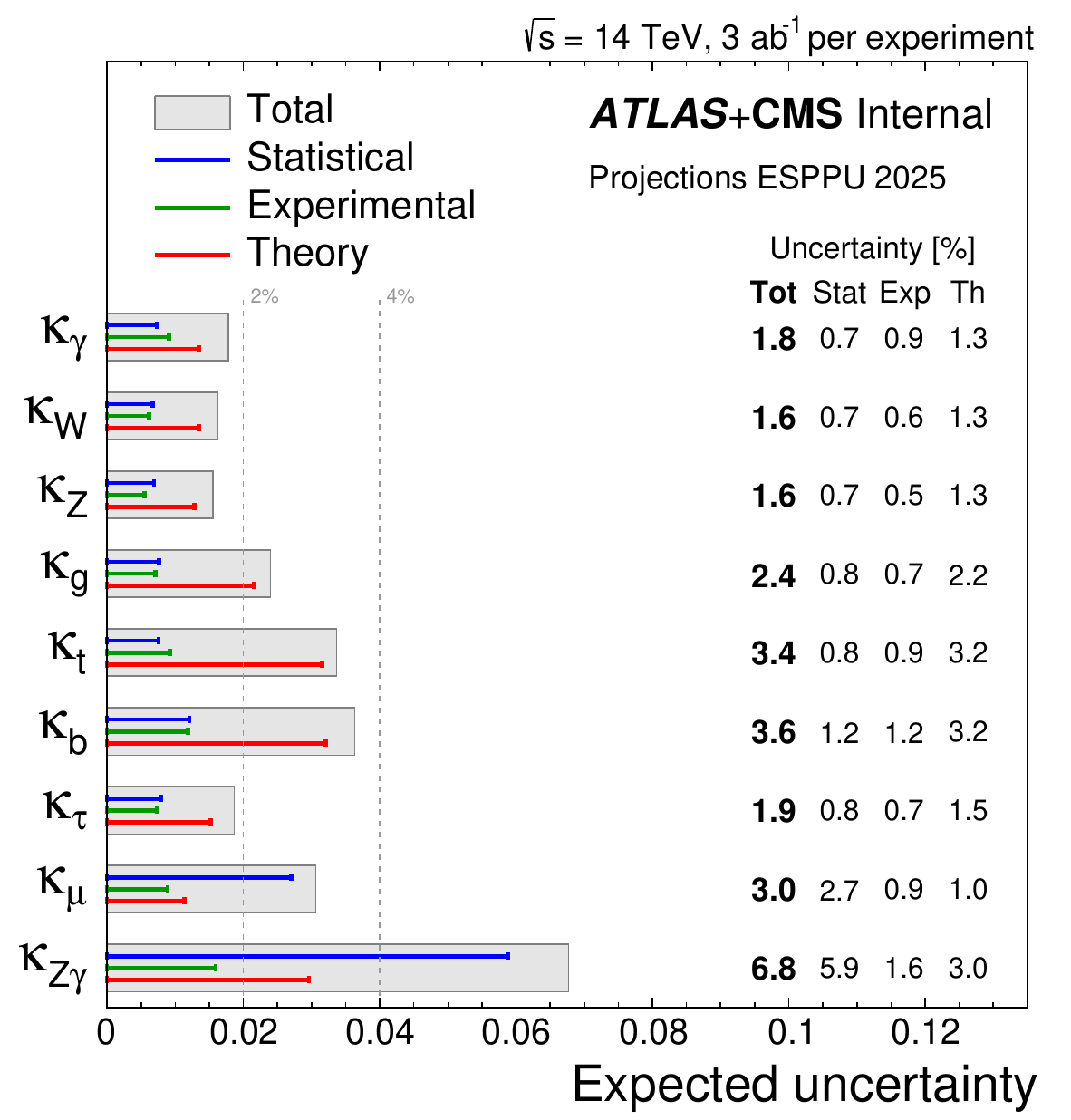}
        \caption{}
        \label{fig:higgs:hllhccouplings}
    \end{subfigure}
    \quad
    \begin{subfigure}{.55\linewidth}  
        \centering
        \includegraphics[width=\linewidth]{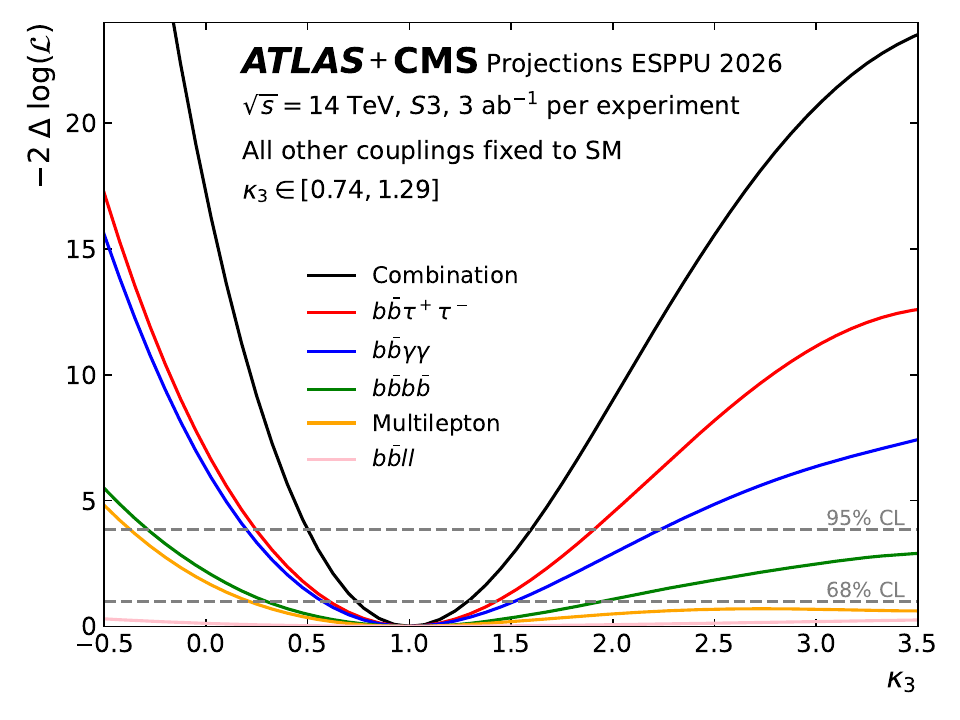}
        \caption{}
        \label{fig:higgs:hllhclambda}
        \end{subfigure}
    \caption{Expected precision of Higgs boson coupling measurements at the HL-LHC.  The expected precision of the Higgs boson coupling modifiers to other SM particles is shown on the left, while the Higgs boson self-coupling measurement projection is shown on the right. 
    }
    \end{figure}

The expected HL-LHC precision on the coupling between the Higgs boson and $\PZ$ bosons is 1.6\%, and goes up to about 7\% for the $\PH\PZ\PGg$ coupling (see \cref{fig:higgs:hllhccouplings}). However, these and other couplings are extracted under an assumption about the Higgs boson width, as the measurements combine production and decay processes.  Experiments at an $\epem$ collider can provide an absolute $\PH\PZ\PZ$ coupling determination with an inclusive measurement of the production cross section. This feature allows precise absolute determinations of many Higgs couplings at an $\epem$ collider, given  loose assumptions about the $\epem \to \PZ\PH$ process.  

There have been extensive studies on the sensitivity to many aspects of Higgs physics
for all of the major \epem\ Higgs factory options \cite{ILCInternationalDevelopmentTeam:2022izu,Abramowicz_2017,robson2018updatedclicluminositystaging,FCC:2018evy,CEPCPhysicsStudyGroup:2022uwl},
which have also been brought together in global fits,
for example those of Refs. \cite{DeBlas:2019qco,deBlas:2022ofj}. 
This report does not attempt to survey all of this previous work, 
but rather to summarize additional recent results that have been
developed in the context of the ECFA study, and submitted to it. 

\Cref{sec:ZHang} discusses studies of $\PZ\PH$ production and various interpretations, including \CP properties of the Higgs couplings.
An $\epem$ collider also allows measurements of the couplings of the Higgs boson to second-generation quarks, which are unattainable at the LHC if their values are of the order predicted by the SM. Even upper limits close to the SM value can be reached for the electron Yukawa 
at circular $\epem$ colliders running at $\sqrt{s}=125$~\GeV, via direct $s$-channel production $\epem\to\PH$, as described in \cref{sec:eYukawa} in the context of FCC-ee. The determination of the Higgs boson coupling to strange quarks will be challenging, and the extensive ongoing investigations into the prospects are described in \cref{sec:HtoSS}.

A key Higgs parameter is the strength of the trilinear Higgs boson coupling, $\lambda_{\PH\PH\PH}$, the measurement of which is frequently presented as the ratio of the observed value to the SM prediction, $\kappa_{\lambda}\equiv \lambda_{\PH\PH\PH}/\lambda_{\PH\PH\PH}^{\rm SM}$. At the HL-LHC, the latest combined extrapolation by the ATLAS and CMS Collaborations yielded an expected uncertainty on $\kappa_{\lambda}$ of  [--26\%,+29\%]. 
At $\epem$ centre-of-mass energies below about 500~\GeV, precise $\epem \to \PZ\PH$ cross section measurements allow the self-coupling to be determined through loop effects, 
while from around 500~\GeV the self-coupling can be determined 
via Higgs pair production.  Details of the possible constraints are provided in \cref{sec:Hself}.


\subsection{\texorpdfstring{\focustopic $\PZ\PH$ production and angular studies}{\focustopic ZH production and angular studies}}
\label{sec:ZHang}
\editors{Ivanka Bozovic, Chris Hays, Markus Klute, Sandra Kortner, Cheng Li, Ken Mimasu, Gudrid Moortgat-Pick}

$\PZ\PH$ production and decay provides sensitivity to Higgs-boson properties and couplings, and can be used to perform a range of measurements at the EW scale.  The $\PZ\PH$ cross-section is affected by modifications to the Higgs-boson mass, width, and couplings (strength and structure).  In addition, given the expected precision, sensitivity to, for example, the Higgs-self or Yukawa couplings can enter through loop effects.  Prospects for differential cross-section measurements have been investigated, and sensitivity to Higgs-boson properties have been determined.  Recent results obtained in the context of FCC-ee are presented in Sections ~\ref{sec:HiggsMass} and~\ref{sec:HiggsWidth}.  

The $\PH\PZ\PZ$ coupling strength and structure can be studied through a wide range of methods and observables. Examples are given in Section~\ref{sec:HZZ}. First, the sensitivity of a dedicated measurement of this coupling using the $\PH\to \PZ\PZ$ decay is presented for several centre-of-mass energies.  Second, a series of studies on the \CP properties of the HZZ coupling is presented using the angular distributions of the final state leptons, a matrix-element-based observable, and neural-network-based observables.  Next, the potential for improving \CP sensitivity using polarised beams is discussed, followed by the potential for probing the \CP properties of HVV couplings at higher $\sqrt{s}$. Finally,  we present a perspective study on quantum entanglement in $\PH \to \PV\PV$ decays.

Among other precision studies, the proposals to improve estimates on  the strength and structure of the $\PH\tau\tau$ coupling sensitivity have received particular attention.  Here we present three studies: the first on the coupling strength in the context of the FCC-ee; the second on the \CP properties using quantum observables; and the third on quantum entanglement.
  
\subsubsection{\texorpdfstring{Higgs boson mass and model-independent $\PZ\PH$ cross-section}{Higgs boson mass and model-independent ZH cross-section}}
\label{sec:HiggsMass}
\editor{Jan Eysermans - abstract 67}

Electron--positron collisions close to the $\PZ\PH$ production cross section maximum a unique opportunity to precisely measure the properties of the Higgs boson.  A per-mille level measurement of the total $\PZ\PH$ cross-section is essential for determining the absolute HZZ coupling, which sets the absolute scale of the coupling in a model-independent way that is not possible at hadron colliders.
Determining the Higgs boson mass with a precision smaller than its width (4.1 MeV in the Standard Model) is necessary to either constrain or measure the electron Yukawa coupling via direct $\epem \rightarrow \PH$ production at \roots~=~125~\GeV. 

As an example, a recent study in the context of FCC-ee investigates the $\PZ\PH$ production cross-section at both \roots~=~240~\GeV and \roots~=~365~\GeV, focusing on leptonic decays of the associated Z boson (electrons and muons), as well as the Higgs mass analysis obtained from the recoil mass distribution~\cite{eysermans_2024_a68b8-3mt57}. Over $2\times 10^6$ events are expected at \roots~=~240~\GeV~and an additional $4\times 10^5$ events at \roots~=~365~\GeV. Results quoted are obtained from Monte Carlo events using the \whizard~+~\pythia event generator, interfaced with the fast simulation package \delphes. The IDEA detector with a crystal electromagnetic calorimeter is used as a reference detector. Furthermore, results are normalised to an integrated luminosity of \SI{10.8}{\per\atto\barn} and  \SI{3}{\per\atto\barn} for \roots~=~240 and 365~\GeV~respectively. 


\begin{figure}
  \begin{center}
    \includegraphics[width=0.5\linewidth]{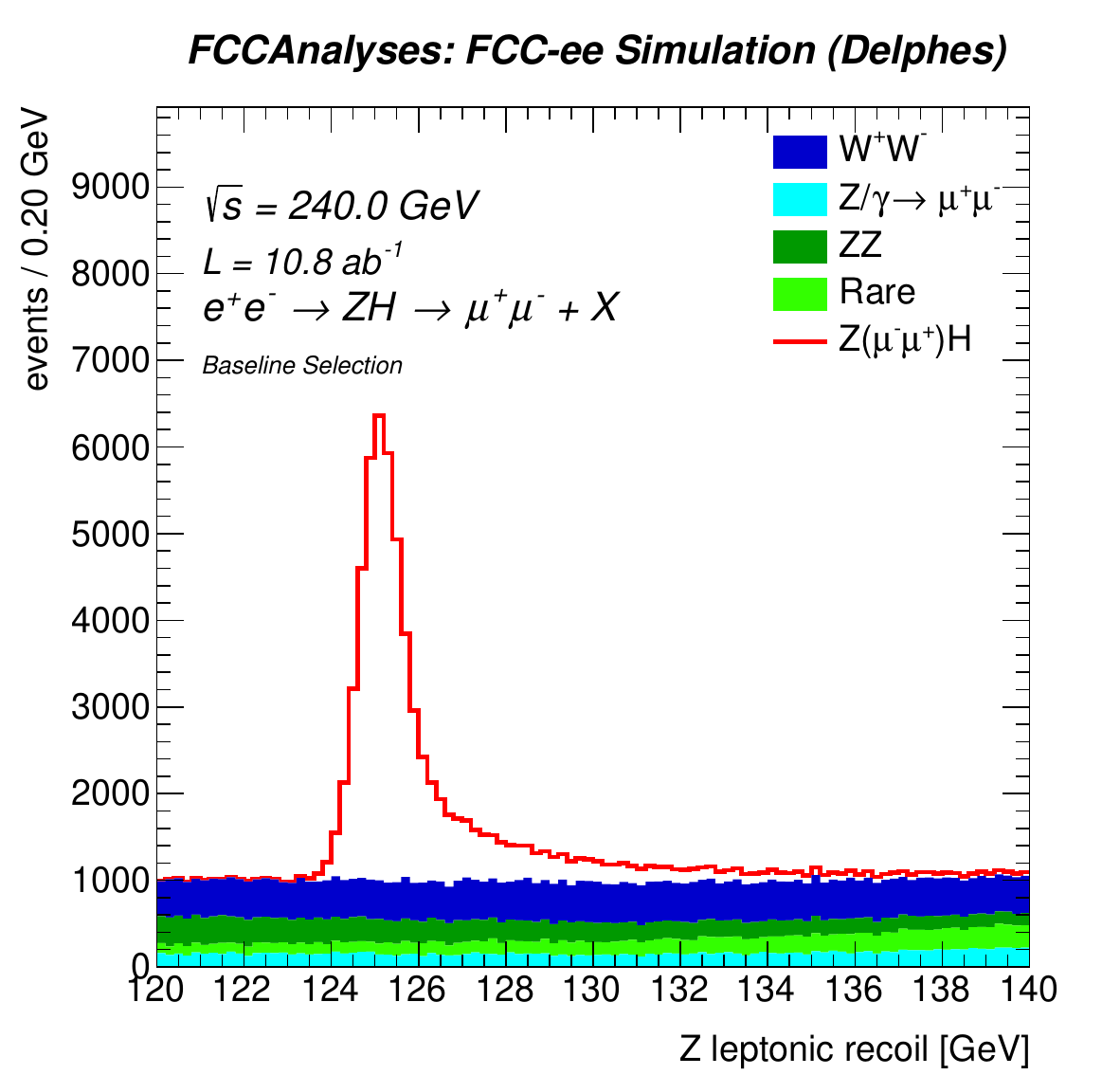}
    \caption{Recoil mass distribution after all selections at \roots~=~240~\GeV.}
    \label{fig:recoil}
  \end{center}
\end{figure}

The $\PZ\PH$ cross-section analysis, performed in both $\PZ(\mumu)\PH$ and $\PZ(\epem)\PH$ final states, relies heavily on precise event selection to effectively suppress backgrounds from \ww, $\PZ\PZ$, and $\PZ\PGg$ processes. To ensure independence from specific Higgs boson decay modes, the selection uses only the kinematic properties of the leptons. Events are selected by first identifying a pair of opposite-sign leptons, requiring at least one of the leptons to be isolated. The invariant mass of this lepton pair must be compatible with the Z boson mass ($86 < m_{\ell\ell} < 96~\GeV$), alongside specific requirements on the dilepton momentum [ $20(50) < p_{\Pl\Pl} < 70(150)~\GeV$ at \roots~=~240(365)~\GeV,~respectively]. Lastly, the recoil mass of the dilepton system must be within the Higgs boson mass window of  $120 < m_{\text{rec}} < 140~\GeV$. The resulting recoil mass distribution after all selections is shown in Figure~\ref{fig:recoil} for the centre-of-mass energy of \roots~=~240~\GeV.

To further enhance sensitivity to the $\PZ\PH$ signal, the selected events are subjected to a BDT, trained on the lepton-based input variables that maintain independence from specific Higgs boson decay modes. Additional variables based on the lepton angles are also included. The training is conducted separately for electron and muon events, as well as for both centre-of-mass energies. To retain the full statistical power of all selected events, the BDT output distribution is employed to extract the $\PZ\PH$ cross-section uncertainty through a binned maximum likelihood fit. At \roots~=~240~\GeV, the results show uncertainties of 0.76\% for the muon final state and 0.92\% for the electron final state, yielding a combined uncertainty of 0.58\%. At \roots~=~365~\GeV, the uncertainties are 1.91\% for the muon final state and 2.13\% for the electron final state, with a combined uncertainty of 1.42\%. Systematic uncertainties are found to be small compared to the statistical ones, and the total uncertainties are 0.59\% and 1.48\% at 240 and 365~GeV, respectively.


The Higgs mass is determined by reconstructing the recoil mass in both $\PZ(\mumu)\PH$ and $\PZ(\epem)\PH$ events, combined with precise knowledge of the centre-of-mass energy \roots. This analysis is performed exclusively at \roots~=~240~\GeV, as the sensitivity at \roots~=~365~\GeV~is negligible due to lower statistics and poorer recoil mass resolution at the higher energy. The latter is primarily affected by the increased beam energy spread and larger initial-state radiation.

Starting from the baseline selection criteria for the cross-section analysis, an additional selection is applied using $\text{cos}(\theta_{\text{miss}}) < 0.98 $, with $\theta_{\text{miss}}$ the polar angle of the missing energy vector, to further reduce the $\PZ/\PGg$ background. The remaining events are categorized based on the polar angles of the leptons: both central, one central and one forward, and both forward. This allows differentiation of the recoil mass resolutions, accounting for the varying material budget in different regions of the tracker. For each category, the recoil mass distribution is fitted using an analytic function comprising two Crystal Ball functions and a Gaussian. This procedure is repeated for off-peak samples ($\pm~50$~\MeV) relative to the nominal Higgs mass of 125~\GeV, with the backgrounds merged and modeled using a polynomial function.

A maximum likelihood fit is used to determine the Higgs mass sensitivity based on analytic shape models, yielding uncertainties of 3.92~\MeV~for the muon final state and 4.95~\MeV~for the electron final state, with a combined uncertainty of 3.07~\MeV. The electron channel shows slightly lower performance due to its poorer resolution and the presence of a small VBF contribution in the t-channel, which broadens the recoil mass distribution.

The primary systematic uncertainty arises from the centre-of-mass energy, conservatively estimated to be 2~\MeV~based on $\PZ\rightarrow \Pl\Pl$ radiative return events. When this uncertainty, along with other less significant uncertainties, is included in the likelihood fit, the Higgs mass uncertainties are 4.74~\MeV~for the muon final state and 5.68~\MeV~for the electron final state, with a combined uncertainty of 3.97~\MeV. The inclusion of systematic effects increases the overall uncertainty by 30\% compared to the statistical-only result.

\begin{table}[ht]
\centering
\begin{tabular}{llll}
\toprule
\textbf{Fit configuration}              & \textbf{\mumu} channel    & \textbf{\ee} channel  & combination  \\
\midrule
Nominal IDEA + crystal ECAL (2T field)  & 3.92 (4.74)               & 4.95 (5.68)           & 3.07 (3.97) \\ 
\midrule
IDEA + crystal ECAL with 3T field       & 3.22 (4.14)               & 4.11 (4.83)           & 2.54 (3.52) \\ 
\midrule
IDEA with silicon tracker (2T)          & 5.11 (5.73)               & 5.89 (6.42)           & 3.86 (4.55) \\ 
\midrule
Ideal resolution                        & 3.12 (3.95)               & 3.58 (4.52)           & 2.42 (3.40) \\ 
\bottomrule
\end{tabular}
\caption{Statistical uncertainty on the Higgs mass (MeV) for various detector configurations. The values in brackets represent the statistical+systematic uncertainties.}
\label{tab:mass:fit_results}
\end{table}

Table~\ref{tab:mass:fit_results} summarizes the results for various detector configurations. When the reconstructed muons and their uncertainties are substituted with generator-level kinematics (``ideal resolution''), the total mass uncertainty --- dominated by selection requirements and backgrounds --- improves by 21\% for the statistical-only case and 14\% for the statistical plus systematic case. Increasing the magnetic field from 2T to 3T yields an improvement of 17\% (statistical only) and 11\% (statistical + systematic), although this improvement is ultimately limited by the beam energy spread, which becomes the dominant factor. Conversely, replacing the IDEA drift chamber with a heavier silicon tracker degrades the uncertainty by 26\% (statistical only) and 15\% (statistical + systematic).


In summary, it is expected that at FCC-ee, using di-muon and di-electron $\PZ\PH$ events, the $\PZ\PH$ cross-section can be measured in a model-independent way with a relative precision of 0.59\% at \roots~=~240 and 1.48\% at \roots~=~365~\GeV. Using a similar event selection, the Higgs boson mass can be determined with an uncertainty of 3.97~MeV, including the dominant systematic uncertainty from the centre-of-mass energy determination.

\subsubsection{\texorpdfstring{Higgs boson width from the $\PZ\PH, \PH \to \PZ\PZ^*$ cross section measurement}{Higgs boson width from the ZH, H -> ZZ* cross section measurement}}
\label{sec:HiggsWidth}
\editor{Nicolas Morange - abstract 95}

The measurement of the Higgs boson's natural width, \GH, holds significant importance as any deviation from the SM prediction could indicate the presence of new physics, such as the decay of the Higgs boson into previously unknown particles. According to the Standard Model, the Higgs boson's natural width is predicted to be approximately 4.1 MeV; a value too small for direct measurement. As a result, indirect methods have been employed at the LHC, utilizing the differing dependencies of on-shell and off-shell Higgs production on the particle's width \cite{ATLAS:2023dnm,CMS:2022ley}. Such techniques often make significant theoretical assumptions, including the universality of the Higgs boson's on-shell and off-shell couplings to vector bosons.

In contrast, the ability to measure the inclusive production cross-section of the $\PZ\PH$ process  in \epem collisions without assuming specific Higgs boson decay modes enables the determination of the absolute coupling between the Higgs and $\PZ$ bosons. Combining this measurement with data on the $\PZ\PH, \PH\to \PZ\PZ^*$ process, along with the $\PH\to \PZ\PZ^*$ branching ratio, allows for the calculation of the total Higgs boson width without any additional theoretical assumptions:
\begin{equation}
    \GH \propto \frac{\sigma_{\PZ\PH}^{2}}{\sigma_{\PZ\PH,\PH(\PZ\PZ^*)}}.
\end{equation}
Assuming that the $\PZ\PH$ production cross-section and the $\PZ\PH$, $\PH\to \PZ\PZ^*$ measurements uncertainties have close-to-no correlation, the uncertainty on the Higgs width measurement can be decomposed as:
\begin{equation}
    \frac{\delta\GH}{\GH} \propto \sqrt{\left(2\frac{\delta\sigma_{\PZ\PH}}{\sigma_{\PZ\PH}}\right)^{2} + \left(\frac{\delta\sigma_{\PZ\PH,\PH(\PZ\PZ^*)}}{\sigma_{\PZ\PH,\PH(\PZ\PZ^*)}}\right)^{2} }.
\end{equation}
Considering only statistical uncertainties, a 0.3\% relative uncertainty on the inclusive $\PZ\PH$ production cross-section can be achieved at the FCC-ee using leptonic $\PZ$ tagging. As will be shown below, the relative uncertainties on the measurement of \GH\ will therefore be dominated by the precision on the determination of the $\PZ\PH/\PH\to \PZ\PZ^*$ cross-section.

A sensitivity study for the $\PZ\PH/\PH\to \PZ\PZ^*$ process measurement at the FCC-ee, with a centre-of-mass energy of 240 GeV, has been conducted \cite{morange_2023_ysabc-wm427,HindTaibi2024,}. 
As the signature originates from the decay of three \PZ bosons, $\PZ\PZ\PZ^*$, this study considered various final states, including those with either two or four leptons resulting from the decay of $\PZ$ bosons. Targeted event selections were focused on $\PZ\PH$ events where $\PZ$ bosons decay into electron--positron, muon--antimuon, neutrino--antineutrino, or quark--antiquark pairs. Post-selection, the contribution from vector-boson-fusion are minor and categorized as background. The study utilized fast simulation techniques for particle propagation and interaction with detector materials, incorporating parametrised efficiencies and resolutions. An assumed integrated luminosity of \SI{10.8}{\per\atto\barn}, equivalent to three years of operation with four detectors at distinct interaction points, was used for this analysis.

The provided information represents only a concise overview of the analysis. Detailed findings and methodologies will be outlined in the forthcoming FCC-ee feasibility study \cite{FCC-FSR-Vol1} and a dedicated article currently under preparation.


The studies are based on Monte Carlo event samples of the main signal and background processes, generated at \epem centre-of-mass energies of 240 GeV and normalised using their theoretical cross-sections to the target integrated luminosity of \SI{10.8}{\per\atto\barn}. The beam energy and the size of the interaction point are smeared according to the expected beam energy spread and size of the FCC-ee luminous region.

The \whizard generator \cite{Kilian:2007gr,Moretti:2001zz} has been used to produce signal events, where a Higgs boson is produced together with an \epem, \mm, \vv, $\PGt^+\PGt^-$ or \qq\ pair. The \pythiaeight generator \cite{Sjostrand:2014zea} has been used for the production of the main background event samples considered in this analysis: $\epem \to \PWp\PWm$ and $\epem \to \PZ\PZ$.

The generated events are passed through a parametric simulation of the detector response implemented in \delphes\ \cite{deFavereau:2013fsa}. The reconstruction efficiencies and resolutions for various particle hypotheses are those expected for the IDEA detector with a crystal electromagnetic calorimeter upstream of a dual-readout fibre calorimeter. Lepton misidentification is assumed to be negligible. The kinematic properties of the jets are assumed to be determined from those of the constituents via a particle-flow algorithm.


Two analyses are performed in parallel: one considers final states with two leptons\footnote{In the following, the term ``lepton'' will be used to identify electrons or muons, while $\PGt$ leptons are excluded from the definition.}, along with either two quarks generating hadronic jets and two neutrinos, or four hadronic jets; and another considers final states with four leptons and either two quarks generating hadronic jets or two neutrinos. The presence of on-shell $\PZ$ bosons in the final state is sought by selecting events with an opposite-sign, same-flavour lepton pair with invariant mass close to that of the $\PZ$ boson, or using a looser selection when a leptonic off-shell $\PZ$ is considered in the four-lepton case. Only the most sensitive combinations are considered. All analyses cluster the reconstructed particles, excluding the $\PZ$ boson decay products, using the exclusive $\kT$ Durham algorithm \cite{CATANI1991432,Cacciari:2011ma}. The particles are clustered into two or four jets depending on the considered topology. A selection on the event missing energy characterises events with a $\PZ$ boson decaying into neutrinos.

The signal is contaminated by non-resonant ($\epem \to \PWp\PWm$ and $\epem \to \PZ\PZ$) and resonant (mostly $\PH \to \PWp\PWm$, $\PH\to\PGt^+\PGt^-$) backgrounds. Dedicated selections are deployed to reduce specific backgrounds affecting some of the considered topologies, such as a selection on the recoil mass to distinguish between $\PZ$ bosons from $\PZ\PH$ production and those from Higgs decay, or a selection on the properties of a photon in the event to reduce $\PH \to \PZ \PGg$ contamination. The selections are designed to allow categorization of events targeting a given final state into orthogonal categories.

Dedicated BDT discriminants, trained on the kinematical properties of the objects in the final states for signal and backgrounds, are additionally used to increase the sensitivity of the two-lepton analysis. Some of the final states, when relevant, split the event classes into subcategories with different signal-to-background ratios to enhance the sensitivity of the measurement and better constrain the backgrounds. Each of the studied analysis region selects tens to hundreds of $\PZ\PH,\HZZ$ signal events; this translates into a large final statistical component to the expected total uncertainty. A 10\% systematic uncertainty on the main non-resonant backgrounds is considered when performing the maximum-likelihood fits to the selected candidates, as well as a 5\% uncertainty on some resonant backgrounds (e.g.\ $\PH \to \PWp\PWm$) when relevant.


The results for each analysis are determined by a binned, maximum-likelihood fit to the distribution of a quantity that discriminates between the signals and the background. The chosen quantity for the fit in each considered channel is tabulated in Table~\ref{tab:ZHHZZ_results}, alongside the channel signal-to-background ratio and measurement sensitivity. Each channel is labelled according to the target decay of each $\PZ$ boson in the final state: $\PZ_1$ refers to the $\PZ$ boson produced in association with the Higgs boson, $\PZ_2$ is the on-shell $\PZ$ boson from the \HZZ\ decay, and $\PZ_3$ the off-shell $\PZ^*$ boson.

\begin{table}[!t]
  \centering
  \begin{tabular}{ccccccc}
    \hline \hline
    \multicolumn{3}{c}{Channel} & \multicolumn{2}{c}{Sensitivity} \\    
    $Z_1$ & $Z_2$ & $Z_3$ & fit quantity & $\mu_{ZH, H\to ZZ^*}$ \\
    \hline
    $\LL$ & $\vv$ & $\jj$ & BDT discriminant         & 1 $\pm$ 0.050 \\
    $\LL$ & $\jj$ & $\vv$ & BDT discriminant         & 1 $\pm$ 0.073 \\
    $\vv$ & $\LL$ & $\jj$ & BDT discriminant         & 1 $\pm$ 0.047 \\
    $\LL$ & $\jj$ & $\jj$ & BDT discriminant         & 1 $\pm$ 0.084 \\
    \hline
    $\LL$ & $\LL$ & $\jj$ & $m^{\mbox{rec}}_{\LL_1}$ & 1 $\pm$ 0.131 \\
    $\LL$ & $\jj$ & $\LL$ & $m_{\jj+\LL_3}$          & 1 $\pm$ 0.131 \\
    $\jj$ & $\LL$ & $\LL$ & $m_{\LL_2+\LL_3}$        & 1 $\pm$ 0.127 \\           
    $\vv$ & $\LL$ & $\LL$ & $m_{\LL_2+\LL_3}$        & 1 $\pm$ 0.268 \\
    \hline
    \hline
  \end{tabular}
  \caption{Quantity used in the maximum likelihood fit and expected sensitivity to the $\epem \to \PZ\PH, \HZZ$ cross-section of the channels discussed in this study. $\PZ_1$ refers to the \PZ boson produced in association with the Higgs boson, $\PZ_2$ is the on-shell \PZ boson from the \HZZ\ decay, $\PZ_3$ the off-shell $\PZ^*$ boson. The details of the variable entering the BDT discriminants of the 2-lepton channels are found in Ref.~\cite{morange_2023_ysabc-wm427}. $m^{\mbox{rec}}_{\LL_1}$ is the recoil mass against the dilepton system of $\PZ_1$; all other variables refer to the invariant mass of system formed by objects listed in the subscript.}
  \label{tab:ZHHZZ_results}
\end{table}

The combination of the three most sensitive two-lepton channels ($\LL\vv\jj$, $\LL\jj\vv$, $\vv\LL\jj$) leads to a precision of 3.1\%. The inclusion of the $\LL\jj\jj$ channel improves the overall sensitivity to 2.9\%. The combination of the four-lepton channels considered in the second analysis leads to a precision of about 7.2\%; their naive combination with the two-lepton channels results in an overall sensitivity of about 2.7\%.

When considering an integrated luminosity of \SI{10.8}{\per\atto\barn} in \epem collisions at $\sqrt{s} = 240$ GeV, FCC-ee could measure the natural width of the Higgs boson with a precision of approximately 2.7\%. 
This value should be considered an upper limit, as further improvements are expected with more advanced analysis techniques and the inclusion of additional final states. In the $\LL\vv\jj$ channel, only half of the possible final states have been studied so far, and background contributions from misidentified hadronic taus could be significantly reduced with dedicated reconstruction algorithms. The $\LL\jj\jj$ channel is expected to have potential in certain subcategories, while the four-lepton channel could benefit from the inclusion of tau final states and further optimization, given that current results are less performant than earlier full-simulation studies. Additionally, previous studies suggest that the determination of the Higgs width via VBF production at $\sqrt{s} =365$\,GeV could achieve a precision comparable to that of the $\PZ\PH$ production mode. With these improvements, a precision better than 1\% may be achievable, though further studies would be required for a more precise quantitative estimate.

\subsubsection{\texorpdfstring{$\PH\PZ\PZ$ coupling}{HZZ coupling}}
\label{sec:HZZ}
\editor{Fabio == Merging various contributions}
\newcommand{\OHBtil}{\ensuremath{\widetilde{\mathcal{O}}_{\Phi\widetilde{B}}}\xspace}
\newcommand{\OHWtil}{\ensuremath{\widetilde{\mathcal{O}}_{\Phi\widetilde{W}}}\xspace}
\newcommand{\OHWtilB}{\ensuremath{\widetilde{\mathcal{O}}_{\Phi\widetilde{W}B}}\xspace}
\newcommand{\cHBtil}{\ensuremath{c_{\Phi\widetilde{B}}}\xspace}
\newcommand{\cHWtil}{\ensuremath{c_{\Phi\widetilde{W}}}\xspace}
\newcommand{\cHWtilB}{\ensuremath{c_{\Phi\widetilde{W}B}}\xspace}
\newcommand{\Zee}{\ensuremath{\PZ\to\Pe\Pe}\xspace}
\newcommand{\Zmm}{\ensuremath{\PZ\to\PGm\PGm}\xspace}

In this section we focus on the extraction of the properties, in particular strength and CP-parity, of the  $\PH\PZ\PZ$ coupling from studying rates and distributions in $\epem \to \PZ\PH$ production and/or $\PH\to \PZ\PZ$ decay. We start by reporting a sensitivity study  on the coupling strength. CP properties are then studied for a circular collider and linear colliders. We conclude with the exploration of observables to measure quantum entanglement in the $\PH\to \PZ\PZ$ decay. 

\epem colliders running above the $\PZ\PH$ threshold provide access to the strength and structure of the $\PH\PZ\PZ$ coupling. The simplest possible BSM scenario, i.e.\ a simple arbitrary rescaling of the SM coupling, can be constrained through the measurement of $\epem \to \PZ\PH$ cross section considering  $\PZ\to \ell^+\ell^-$  and being inclusive of all possible \PH boson decays. A global fit in a $\kappa$ framework can also be employed providing sensitivities at the subpercent level, i.e.\ at about $0.2\%$~\cite{deBlas:2019rxi,robson2020clichiggscouplingprospects}. 

More generally, the structure of the HZZ coupling can be modified by the presence of higher dimensional operators, giving rise to new CP-even and CP-odd structures, see e.g.\ Ref.~\cite{deBlas:2019rxi}. 
In the SM the only source of CP violation stems from fermion mixing in the charged currents, while the Higgs boson is predicted to have CP-even, flavour-diagonal interactions. As it stands, it is insufficient to account for the matter dominance observed in the universe, therefore searches for additional sources of \CP violation are well-motivated to connect the SM with paradigms of baryogenesis through new interactions. Detecting non-zero CP-odd components in the Higgs interactions with the SM particles, would therefore clearly point to physics beyond the Standard Model. Departures from the SM can be efficiently parametrised in terms of a limited set of (flavour conserving) dimension-6 operators~\cite{Grzadkowski:2010es}.  Assuming that new physics at some high energy scale, $\Lambda$, provides additional sources of \CP violation, the effects at low energy scales can be expressed as operators in an effective Lagrangian:
$$\mathcal{L_{\text{SMEFT}}}= \mathcal{L}_{\text{SM}}^{(4)} + \sum_i\frac{c_i^{(6)}}{\Lambda^2}\mathcal{O}_i^{(6)}
+ \sum_i\frac{c_j^{(8)}}{\Lambda^4}\mathcal{O}_j^{(8)} + \dots,
$$
\noindent where $\mathcal{O}_i^{(6)}$ and $\mathcal{O}_j^{(8)}$ parametrise generic extensions of the SM at low energy. 

At Higgs factories, both production $\epem \to \PZ\PH$ distributions as well as Higgs decays $\PH\to \PZ\PZ$ can provide observables that are sensitive to \CP-violation in the $\PH\PV\PV$ vertex. The subset of operators that are \CP-odd and that affect $\PH\PV\PV$ interactions is: 
\begin{equation}
\label{eq:ops}
\OHBtil = \Phi^{\dagger}\Phi B^{\mu\nu}\widetilde{B}_{\mu\nu} \, , \;\; 
\OHWtil = \Phi^{\dagger}\Phi W^{i\,\mu\nu}\widetilde{W}^i_{\mu\nu} \, , \;\;
\text{and} \;\;
\OHWtilB = \Phi^{\dagger}\sigma^i \widetilde{W}^{i\,\mu\nu}B_{\mu\nu} \; ,
\end{equation}
leading to CP-violating $\PV\PV\PH, \PV\PV\PH\PH$ and $\PV\PV\PV\PH\PH$ couplings. 
Focusing only on the neutral EW $\PV\PV\PH$ couplings, after EWSB 
we obtain 
\begin{eqnarray}
\delta{\cal L}^{HVV}_\text{CPV} =  \frac{h v}{\Lambda^2} \Big[
{\tilde c_{aa}} A_{\mu\nu}\tilde{A}_{\mu\nu}  
+{\tilde c_{za}} Z_{\mu\nu}\tilde{A}_{\mu\nu} 
+ {\tilde c_{zz}}  Z_{\mu\nu}\tilde{Z}_{\mu\nu} \Big]\,,
\label{eqn:hVV-CPV}
\end{eqnarray}  
where $\tilde V_{\mu\nu}= \frac12 \epsilon^{\mu \nu \rho\sigma} V_{\rho\sigma}$ and the three $\tilde c_{aa,za,zz}$ couplings are linear combinations of the three Wilson couplings  $c_{\Phi\widetilde{B},\Phi\widetilde{W},\Phi\widetilde{WB}}$. \footnote{Also the operator $W^+_{\mu\nu} \tilde{W}^-_{\mu\nu}$ is present, whose coefficient $c_{ww}$ is not independent from $\tilde c_{aa,za,zz}$.}. 

Focusing only on HZZ interactions and adding the CP-even terms at dim=4 and dim=6 leads to a  phenomenological Lagrangian 
\begin{equation}
         \mathcal{L}^{HZZ} =  \frac{h}{v} \left[\cos \xi_{CP}  
         \left( k_{SM} m_Z^2  Z_{\mu}{Z}^{\mu} +  {k}_{zz} Z_{\mu\nu}{Z}^{\mu\nu} \right)   +  \sin \xi_{CP} \tilde{k}_{zz}    Z_{\mu\nu}\tilde{Z}^{\mu\nu}\right], \label{eq:effl2}
\end{equation}
where we have introduced a dimensional rescaling factors $k$~\cite{Artoisenet:2013puc}.

Considering only dimension-6 operators, the scattering amplitude is:
$$|\mathcal{M}|^2 = |\mathcal{M}_{\text{SM}}|^2 + 2Re(\mathcal{M}^*_{\text{SM}}\mathcal{M}_{\text{d6}}) + |\mathcal{M}_{\text{d6}}|^2 \; .$$

The interference term $\mathcal{M}^*_{\text{SM}}\mathcal{M}_{\text{d6}}$ is the leading correction to the SM, proportional to $1/\Lambda^2$, while other contributions are proportional to $1/\Lambda^4$.  The interference term is \CP-odd and produces asymmetries in \CP-odd observables, but integrates to zero for \CP-even observables. ``Squared'' dimension-6 interactions compete with dimension-8 interactions, but for the \CP-odd interactions listed above lead to \CP-even modifications. \CP-odd observables targeting the interference term can therefore be used to set genuinely \CP-sensitivity limits on the Wilson coefficients associated with these operators. 

\subsubsection*{HZZ coupling strength sensitivity}

\editor{Ivanka Bozovic - abstract 6 == Edited by Fabio}

As an example of the sensitivity that can be achieved by a single observable measurement at different energies, let us consider $\text{BR}(\PH\to \PZ\PZ^*)$ as a proxy for the HZZ coupling, at centre-of-mass energies of 350 GeV, 1.4 TeV and 3 TeV \cite{Abramowicz_2017,Vukasinovic:2022}. Using full simulations based on integrated luminosities of 1 ab$^{-1}$, 1.5 ab$^{-1}$ and 5 ab$^{-1}$, respectively, rates are measured corresponding to the product of the Higgs production cross-section $\sigma$ and $\text{BR}(\PH\to \PZ\PZ^*)$. 

The signal chosen is the $\PZ\PZ^*$ semileptonic final state, owing to the smaller irreducible background than in the hadronic final state. In Higgsstrahlung at 350 GeV CLIC, the primary $\PZ$ is assumed to decay hadronically.  The analysis is based on  the identification of exactly two isolated leptons per event while the remaining particle flow objects \cite{Thomson:2013} are clustered into four (two) jets. Isolation criteria are based on kinematic observables in the event: energy and impact parameter of a particle, ratio of particle's energy deposition in calorimeters, and track vs. cone energy around the isolated lepton candidate. A multivariate analysis \cite{hoecker:2009tmvatoolkitmultivariate} is applied to further suppress background on the basis of kinematic and event-shape variables. The relative statistical precision $\delta$ on $\text{BR}(\PH\to \PZ\PZ^*)$ from the fully simulated measurements is obtained as the inverse of the statistical significance $S$, where $S = N_{S}/ \sqrt{N_{S}+N_{B}}$ for the selected numbers of signal $N_{S}$ and background $N_{B}$ events. 
The corresponding statistical precision is further scaled to $\delta_{\mathrm{projected}}$ assuming the integrated luminosities foreseen in the current CLIC staging scenario with an electron beam polarisation of 80\%. 

We find 
$\delta = 20\%,~ 5.6\%$, and $3.0\%$ for $\sqrts=350~\GeV,~ 1.4~\TeV$, and $3~\TeV$, respectively; 
and $\delta_{\mathrm{projected}} = 10\%,~ 2.8\%$, and $2.4\%$ corresponding to 4.3 ab$^{-1}$, 4 ab$^{-1}$, and 5 ab$^{-1}$ (with 80\% beam polarisation at $\sqrts=3~\TeV$).


These results show 
that  high-energy operation of an \epem collider allows a percent level statistical precision already with a single observable measurement. As mentioned above,   optimized global fits can then improve the precision of the HZZ coupling strength to the per-mille level \cite{Barklow:2018}. 


\subsubsection*{CP sensitivity}
\editors{Andrei Gritsan, Nicholas Pinto, Valdis Slokenbergs - abstract 84}

\begin{figure}[b!]
    \centering
    \includegraphics[width=0.32\linewidth]{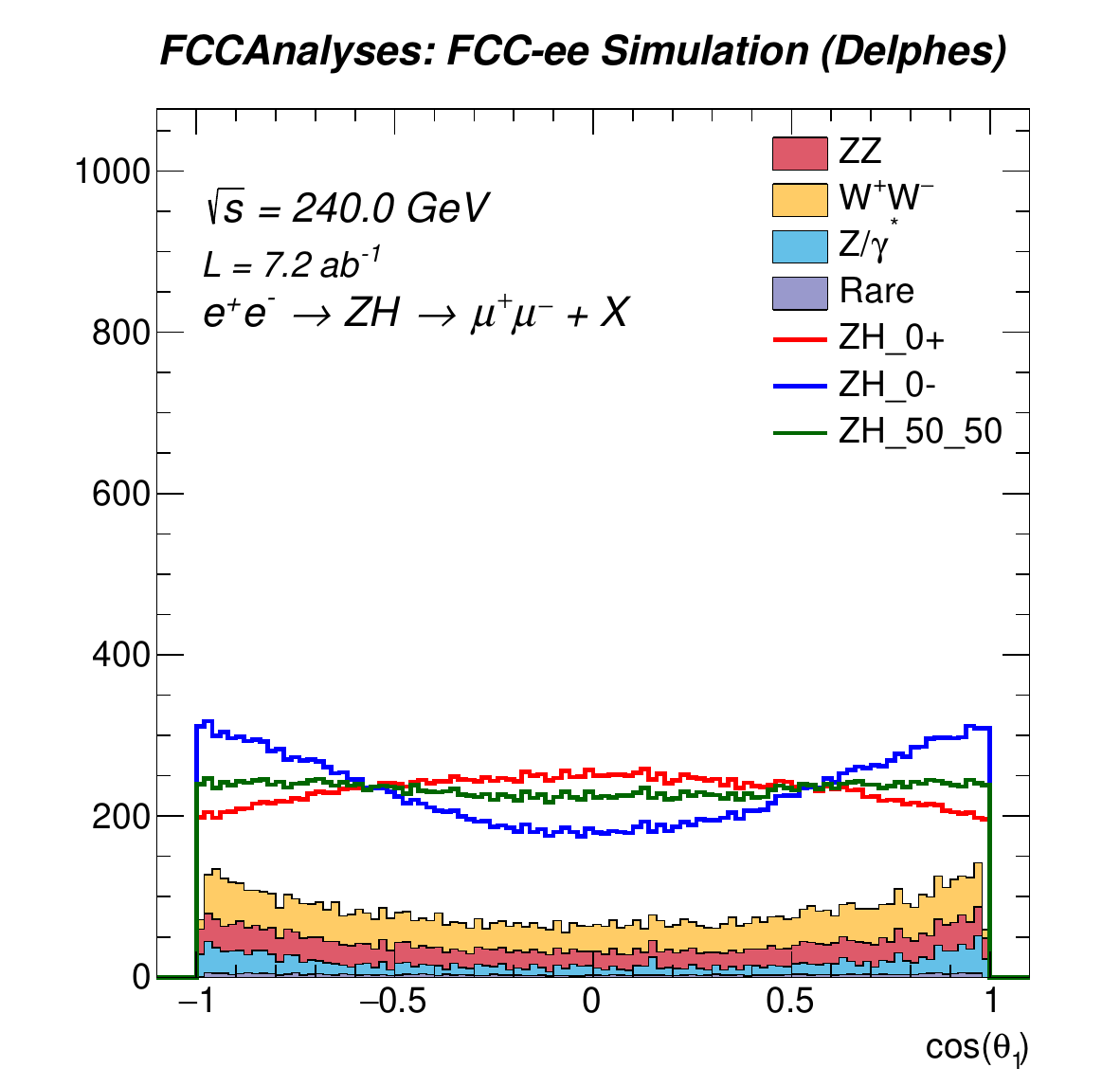}
    \includegraphics[width=0.32\linewidth]{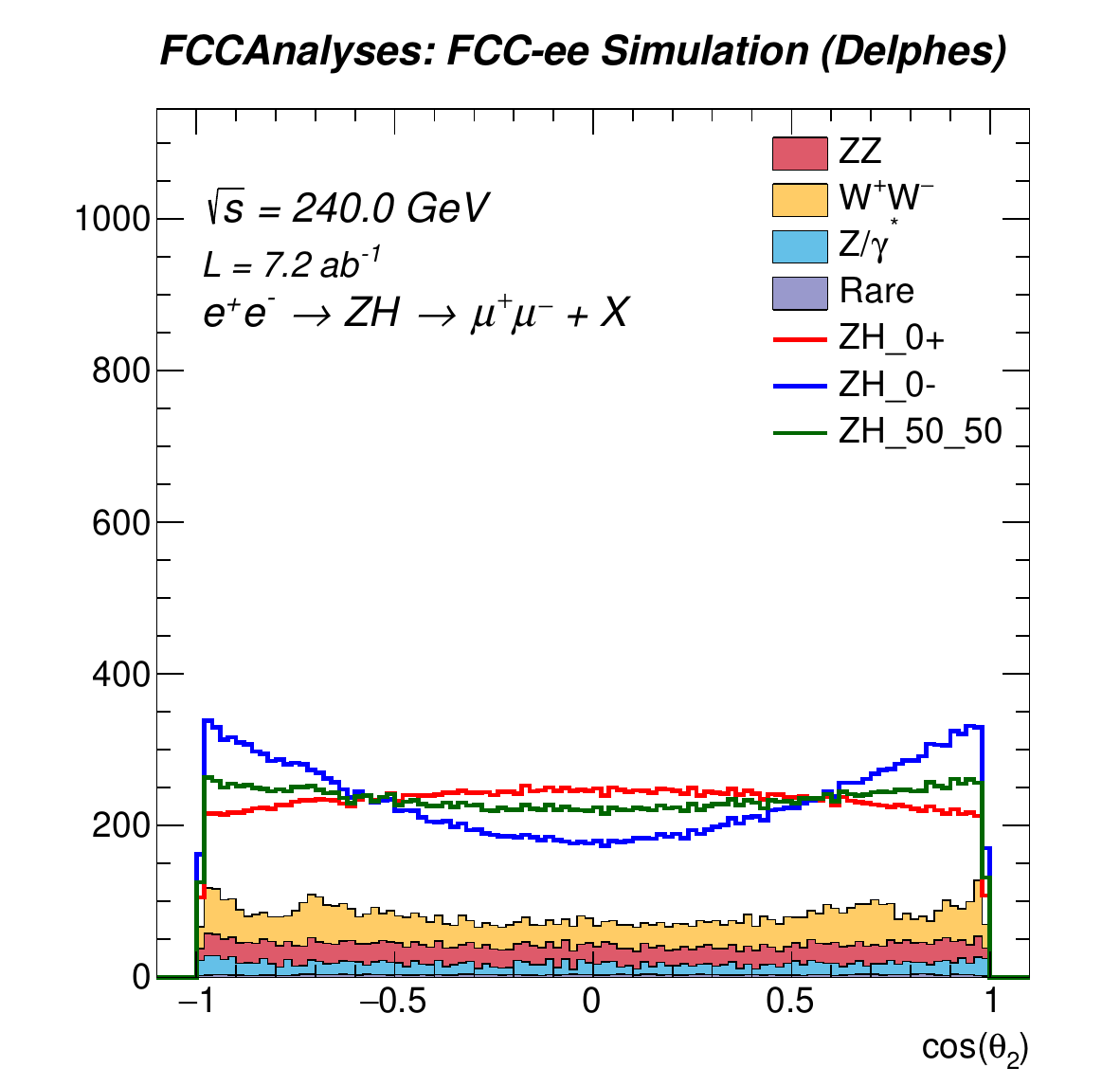}
    \includegraphics[width=0.32\linewidth]{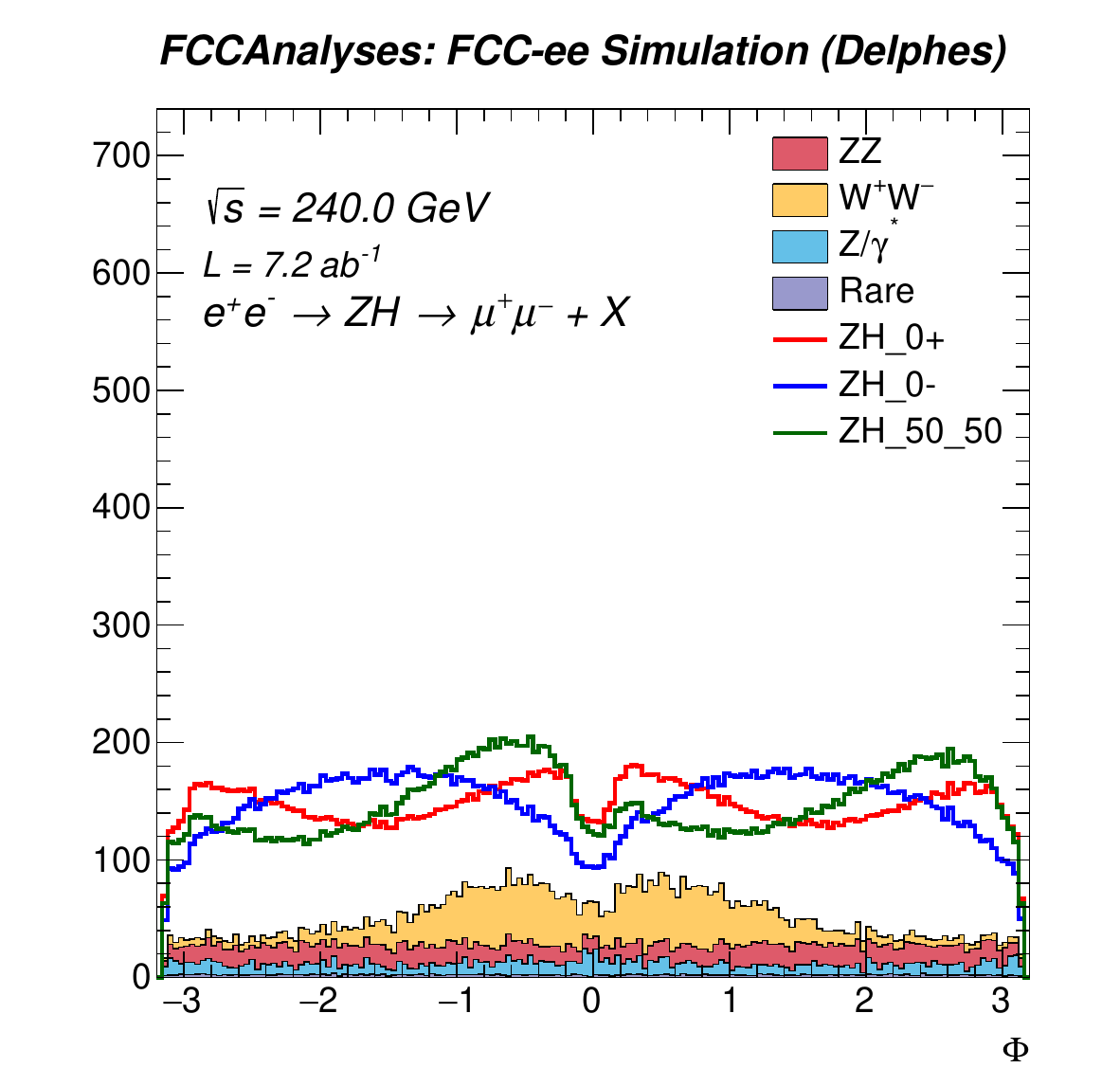}
    \caption{Reconstructed angular distributions in the process $\epem \rightarrow  \PZ\PH \rightarrow  \mumu \PX$ and its backgrounds.
    $\theta_1$ and $\theta_2$ are the polar angles of the two muons, while $\Phi$ is the angle between the plane of the muons, and the plane defined by the incoming electron beam and the $\PZ$ boson direction.
    The maximum \CP violation is apparent in the asymmetry produced by a 50\% mixture of \CP-odd and \CP-even contributions (green).}
    \label{fig:angles}
\end{figure}

At a centre-of-mass energy of \SI{240}{\giga\electronvolt}, the Future Circular Collider is expected to produce of the order of $10^6$ events of the Higgsstrahlung  process $\Pep\Pem\to\PZ\PH$. Relevant kinematic distributions  for probing \CP-violation at the $\PH\PZ\PZ$ vertex can be easily identified; see \cref{fig:angles}.
In the Snowmass-2013 and Snowmass-2022 studies~\cite{Gritsan:2022php}, a benchmark parameter 
was established for the \CP-odd contribution based on production-invariant decay process:
\begin{eqnarray}
f_{\CP}^{\PH\PX}\equiv
\frac{\Gamma^{\CP\mathrm{-odd}}_{\PH\to \PX}
}{\Gamma^{\CP\mathrm{-odd}}_{\PH\to \PX}
+\Gamma^{\CP\mathrm{-even}}_{\PH\to \PX}}
\,,
\label{eq:fCP} 
\end{eqnarray}
where the partial decay $\PH\to \PX$ width is calculated with either the $\CP$-odd or $\CP$-even part of the amplitude.  In the case of the $\PH\PZ\PZ$ vertex, the reference decay is $\PH\rightarrow \PZ\PZ\rightarrow 2\Pe2\PGm$.
%
%

The targeted signal process in this analysis is $\Pep\Pem\to\PZ\PH$. 
Four final states for the $\PZ$ are considered: $\PZ\to\Pep\Pem, \PGmp\PGmm$ (leptonic channels), 
and $\PQq\PAQq, \PQb\PAQb$ (hadronic channels), where 
$\PQq\PAQq$ refers to $\PQu\PAQu, \PQd\PAQd, \PQs\PAQs, \PQc\PAQc$. 
The $\PH$ boson is allowed to decay inclusively and is not reconstructed. 
Because the centre-of-mass energy ($\sqrt{s}$) of each event is precisely known, 
the four-momentum of the $\PH$ boson can be reconstructed using total four-momentum 
conservation with the knowledge of the reconstructed four-momentum of the $\PZ$. 
In the leptonic channels, the dominant backgrounds considered are 
$\Pep\Pem\to\PWp\PWm,\, \Pep\Pem\to\PZ\PZ,\,\Pep\Pem\to\PGmp\PGmm$ and $\Pep\Pem\to\PGtp\PGtm$. 
In the hadronic channels, the dominant backgrounds are $\Pep\Pem\to\PZ\PZ$ and $\Pep\Pem\to\PQq\PAQq$ 
(in the case of the latter background, $\PQq = \PQu,\,\PQd,\,\PQs,\,\PQc,\,\PQb$). 
The $\Pep\Pem\to\PWp\PWm$ background dominates the $\PZ\to\PQq\PAQq$ channel, 
but is suppressed in the $\PZ\to\PQb\PAQb$ channel.

Signal Monte Carlo (MC) samples are generated using \whizard\ 
with parton showering done in \pythiasix. 
Background samples are generated and showered using \pythiaeight. 
Reconstruction and detector simulation are performed by the fast simulation program \delphes. 
The concept detector simulated is the Innovative Detector for Electron-positron Accelerators (IDEA). 
Matrix-element re-weighting is performed by the MELA package.
Transition probabilities calculated from generator-level kinematic observables are used to re-weight the simulated, 
SM distributions to ones where the interaction at the $\PH\PZ\PZ$ vertex is fully \CP-odd 
or a mixture of \CP-even and \CP-odd. 

A set of preselection requirements is applied for each channel. In the leptonic channels, only events containing 
at least two leptons of opposite charge are selected. Both of these leptons are required to have momentum with 
a magnitude greater than \SI{20}{\giga\electronvolt}, which minimizes the noise from soft radiative processes. 
At least one lepton is required to be relatively isolated from the rest of the products of the event. 
In the hadronic channels, preselection requires that events have no more than two charged leptons. 
Any charged lepton entering the analysis is required to have momentum < \SI{20}{\giga\electronvolt}. 
Following these requirements, all reconstructed particles in the event are exclusively clustered to four 
jets by the Durham-$\kT$ algorithm
used in the \fastjet\ clustering program.

The $\PZ\to ff$ difermion system that minimizes the expression $\chi^2 = 0.6(m_{\PZ} - m_{ff})^2 - 0.4(m_{\PH} - m_{recoil})^2$ 
is selected as the candidate for the~$\PZ$.



A set of kinematic  requirements is applied in the leptonic channels:
on the invariant mass of the selected dilepton system $86 < m_\text{dilepton} < 96$\,GeV;
on the three-momentum of the dilepton system $45 < p_\text{dilepton} < 55$\,GeV;
on the recoil mass $124 < m_\text{recoil} < 127$\,GeV;
on the cosine of the missing momentum $|\cos{\theta_\text{miss}}| < 0.98$;
and on the angle between the momentum of the $\PZ$ and the momentum of one of its decay products $|\cos{\theta_{2}}| < 0.99$.
 A looser set of requirements is applied on the same observables in the hadronic analysis as well as a cut on jets from the Z tagged as gluons to suppress background.
 An additional requirement on a visible energy $>$ \SI{150}{\giga\electronvolt} is applied. 

\begin{figure}[t!]
    \centering
    \includegraphics[width=0.5\linewidth]{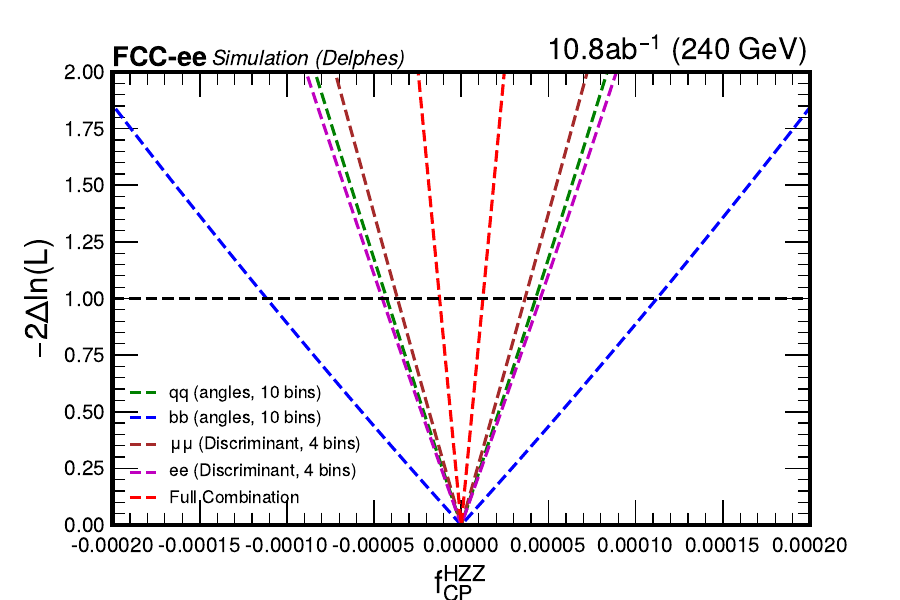}
    \caption{The expected likelihood scans of $f^{\PH\PZ\PZ}_{CP}$ with $\PZ\to \PQb\PAQb, \PQq\PAQq, \Pep\Pem,$ and $\PGmp\PGmm$ channels at FCC-ee, based on 10.8\,ab$^{-1}$ of data collected at $E_{\epem}=240$\,GeV.}
    \label{fig:scan}
\end{figure}

In the leptonic channel, two optimal observables are computed using the matrix element technique:
$D_{0^-}$, which is sensitive to the pure \CP-odd contribution, and 
$D_{\CP}$, which corresponds to the interference between the \CP-even and \CP-odd components. 
Three templates characterise the signal parameterization using MC simulation that,
together with the background, is incorporated into the likelihood model with a free parameter $f^{\PH\PZ\PZ}_{\CP}$.
In the hadronic channel, the kinematic observables used are $\cos{\theta_1}, \cos{\theta_2},$ and $\Phi$ as shown in \cref{fig:angles}.
The expected rates of events are normalised to the 10.8\,ab$^{-1}$ of data collected at $E_{\epem}=240$\,GeV.
\Cref{fig:scan} shows the expected constraints on $f^{\PH\PZ\PZ}_{\CP}$ at FCC-ee,
based on 10.8\,ab$^{-1}$ of data collected at $E_{\epem}=240$\,GeV.
The combined constraint is $f^{\PH\PZ\PZ}_{CP}=\pm1.2\times10^{-5}$, which is $\pm 4.5\times10^{-5}$ with $\PZ\rightarrow\Pep\Pem$, $\pm 3.7\times10^{-5}$ with $\PZ\rightarrow\PGmp\PGmm$, $\pm 4.3\times10^{-5}$ with $\PZ\rightarrow\PQq\PAQq,$ and $\pm 1.1\times10^{-4}$ with $\PZ\rightarrow\PQb\PAQb$.

In conclusion, we have estimated the constraints on \CP-violating contributions 
in the Higgsstrahlung process at FCC-ee, based on 10.8\,ab$^{-1}$ of data collected at $E_{\epem}=240$\,GeV.
In the combined leptonic and hadronic $\PZ$ boson channels, an expected sensitivity of $f^{\PH\PZ\PZ}_{CP}=\pm1.2\times10^{-5}$ is projected.
Our results are similar to those obtained in the Snowmass-2022 studies~\cite{Gritsan:2022php}, 
when adjusted for the same luminosity, but they are more dependable due to the comprehensive detector studies conducted.

\subsubsection*{Improving $\PH\PZ\PZ$ \CP constraints with neural-network-based observables}
\editor{Aidan Robson - abstract 74}

Further studies explore the extra sensitivity to \OHBtil, \OHWtil, and \OHWtilB, as defined in \cref{eq:ops}, that can be obtained using neural-net (NN) based observables over direct angular observables.

Studies are carried out using fast simulation.  Events are generated using \textsc{MadGraph5\_aMC@NLO}. 
\textsc{SMEFTSim 3.0} is used to include the anomalous interactions from the EFT operators.
Detector effects are modeled via \delphes\ parameterizations.
SM events are simulated and normalisation factors applied to cover missing higher-order effects.  
Cross-checks on event yields are made with relevant LHC and future-collider studies.

In the first instance $\epem\to\PZ\PH$ events are studied with \Zee or \Zmm, and \Hbb, at $\sqrt{s}=240$\,GeV using the \textsc{IDEA} concept \delphes\ card and at $\sqrt{s}=250$\,GeV using the \textsc{ILCgen} \delphes\ card.  A simple kinematic selection is made following Ref.~\cite{azzi2012prospectivestudieslep3cms}, plus two b-tags at the 85\% efficiency working point.  A data sample corresponding to 5\,\abinv of integrated luminosity is assumed, and compared with 2\,\abinv of integrated luminosity using polarised beams under the baseline ILC polarisation scheme where  67.5\% of the data is taken with the electron beam --80\% polarised and the positron beam +30\% polarised (--80,+30), 22.5\% with the electron beam +80\% polarised and the positron beam --30\% polarised (+80,--30), and the remaining 10\% shared equally between (--80,--30) and (+80,+30).

Traditional angular observables are those such as angles between decay planes, and the rapidity-ordered azimuthal angle between two objects, for example $\Delta\phi_{\ell\ell} = \phi_1 - \phi_2$ between the $\PZ$ decay leptons, where $y_1 > y_2$.
The NN-based analyses follow the approach of Refs.~\cite{Bhardwaj:2021ujv} and \cite{PhysRevD.107.016008}.  
As the \CP asymmetries arise from the interference between SM and \CP-odd amplitudes, 
interference-only events are generated for each EFT operator.
These interference samples are split into constructive and destructive interference 
and the NN is trained to distinguish between the two samples.
A single observable is constructed from the NN classifications, i.e.\ 
$O_{NN}=P_+ - P_{-}$, 
such that a dedicated observable is optimised for each EFT operator.

It is enlightening to consider the performance of different genuinely \CP-sensitive observables such $\Delta\phi_{\ell\ell}$ in direct comparison against neural-net improved observables. We can expect good statistical control at future facilities, hence, there is the possibility of augmenting $\Delta\phi_{\ell\ell}$ with measures of momentum transfer to gain sensitivity. In particular, the invariant mass $m_{12}$ is motivated as an additional observable as the interference term swaps sign around the $\PZ$ pole in the electron final state. 

\begin{figure}
  \centering
  \includegraphics[width=0.4\textwidth]{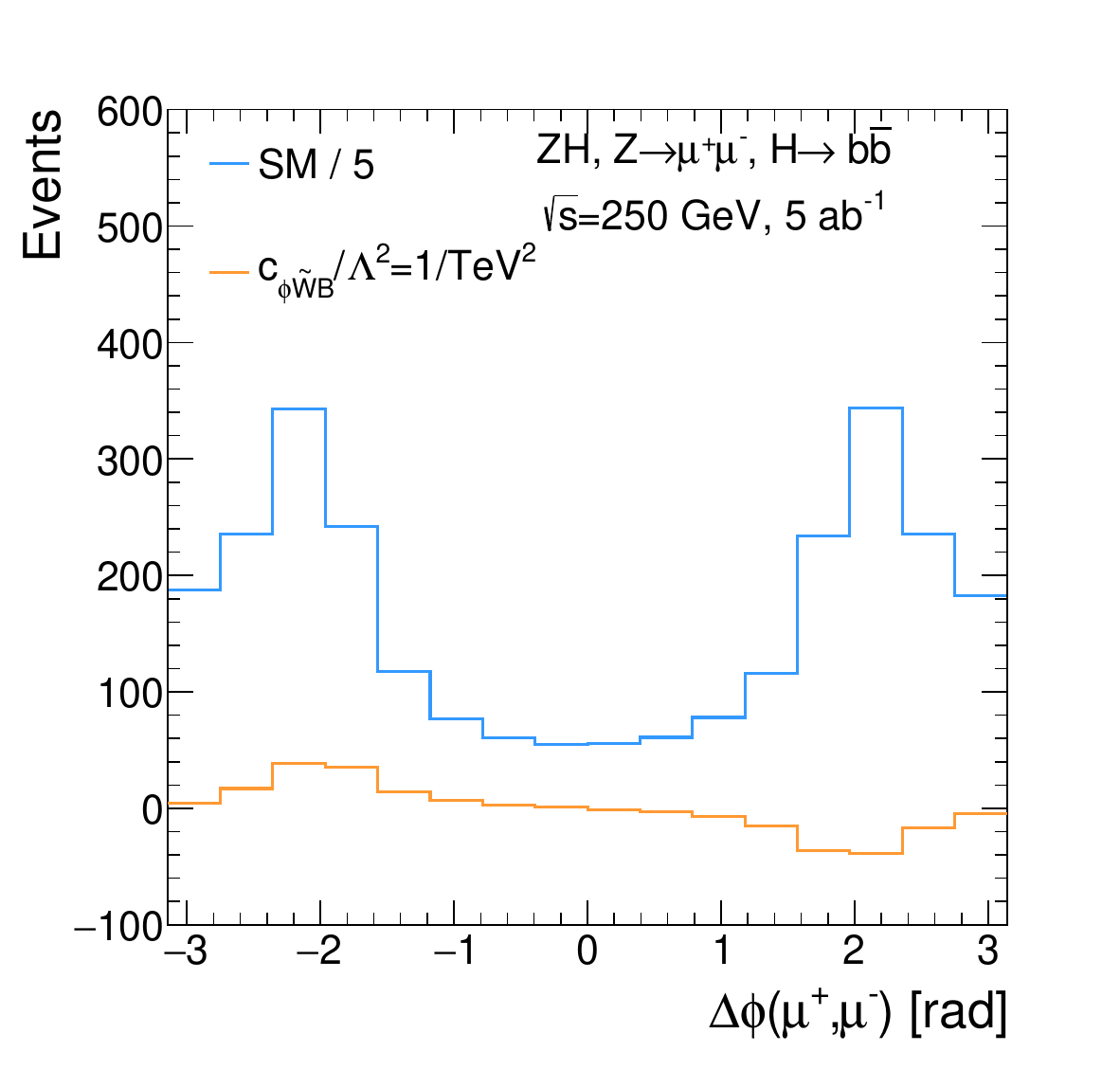}
  \includegraphics[width=0.4\textwidth]{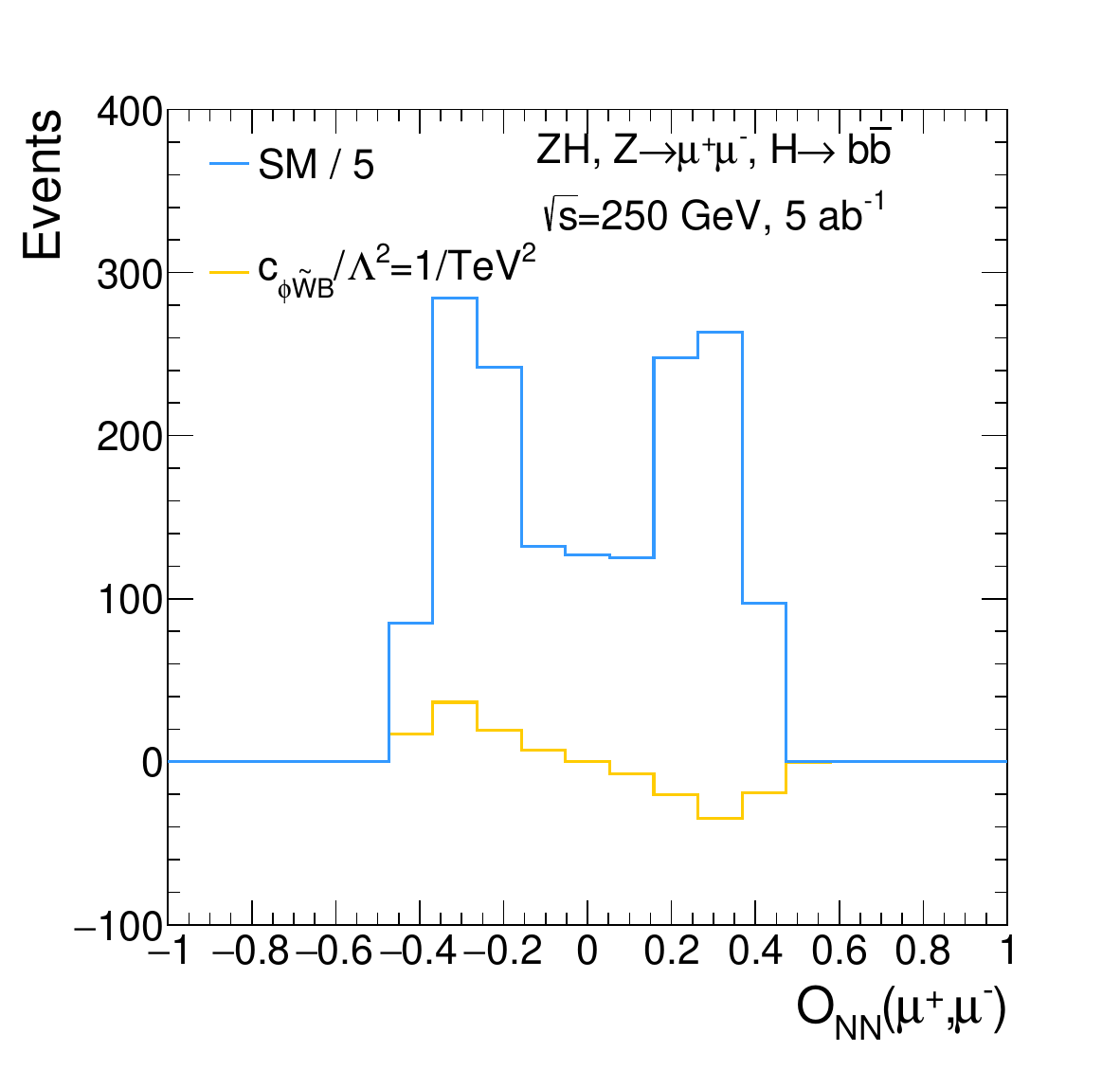}
  \caption{Sensitivity of \CP-sensitive observables for the $\OHWtilB$ operator, in the $\PGmp\PGmm\PQb\PQb$ final state: (left) event yield as a function of $\Delta\phi_{\ell\ell}$; (right) event yield as a function of $O_{NN}$.
  \label{fig:cpInHZZplots1}}
\end{figure}

\begin{figure}
  \centering
  \includegraphics[width=0.4\textwidth]{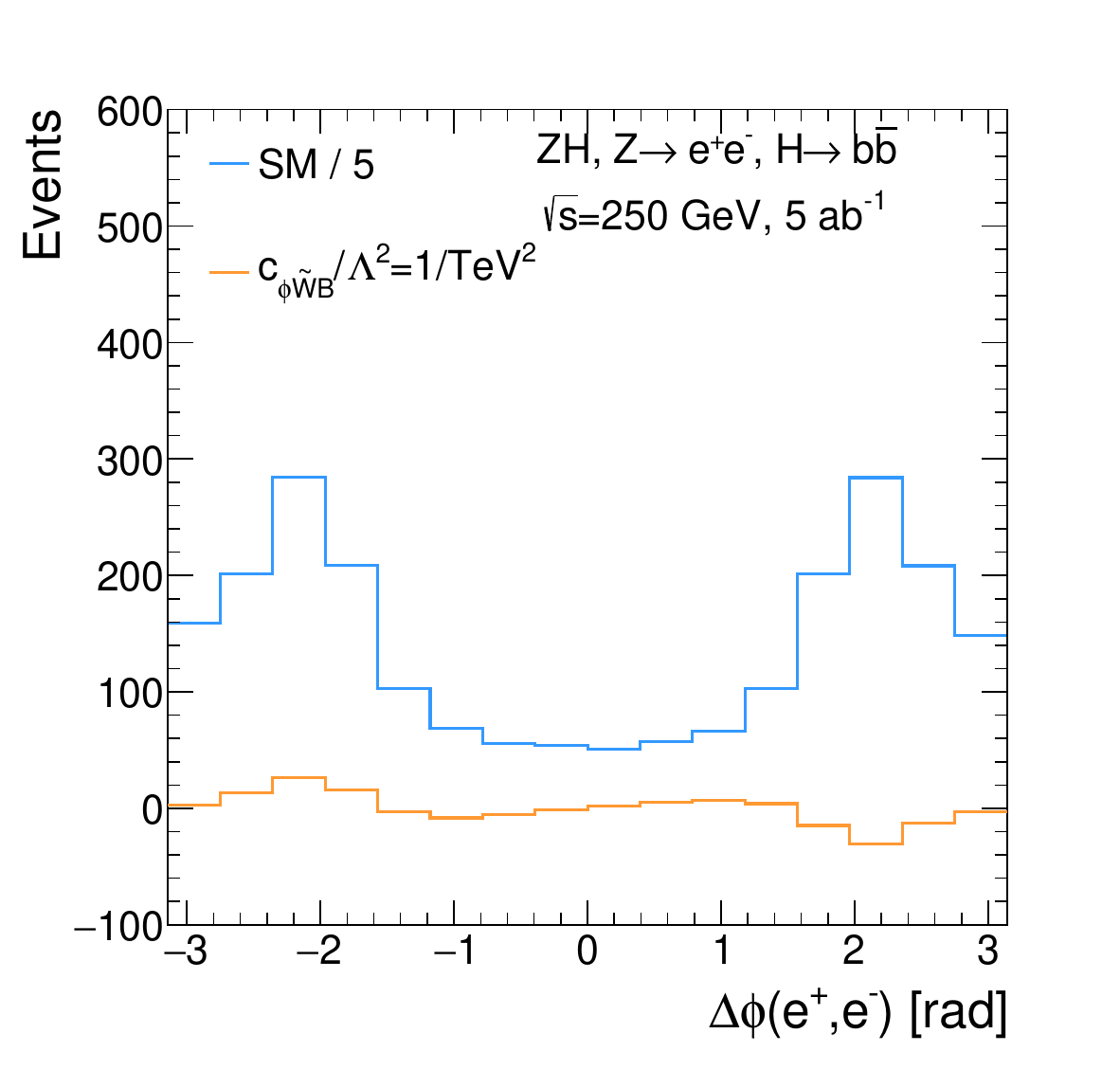}
  \includegraphics[width=0.4\textwidth]{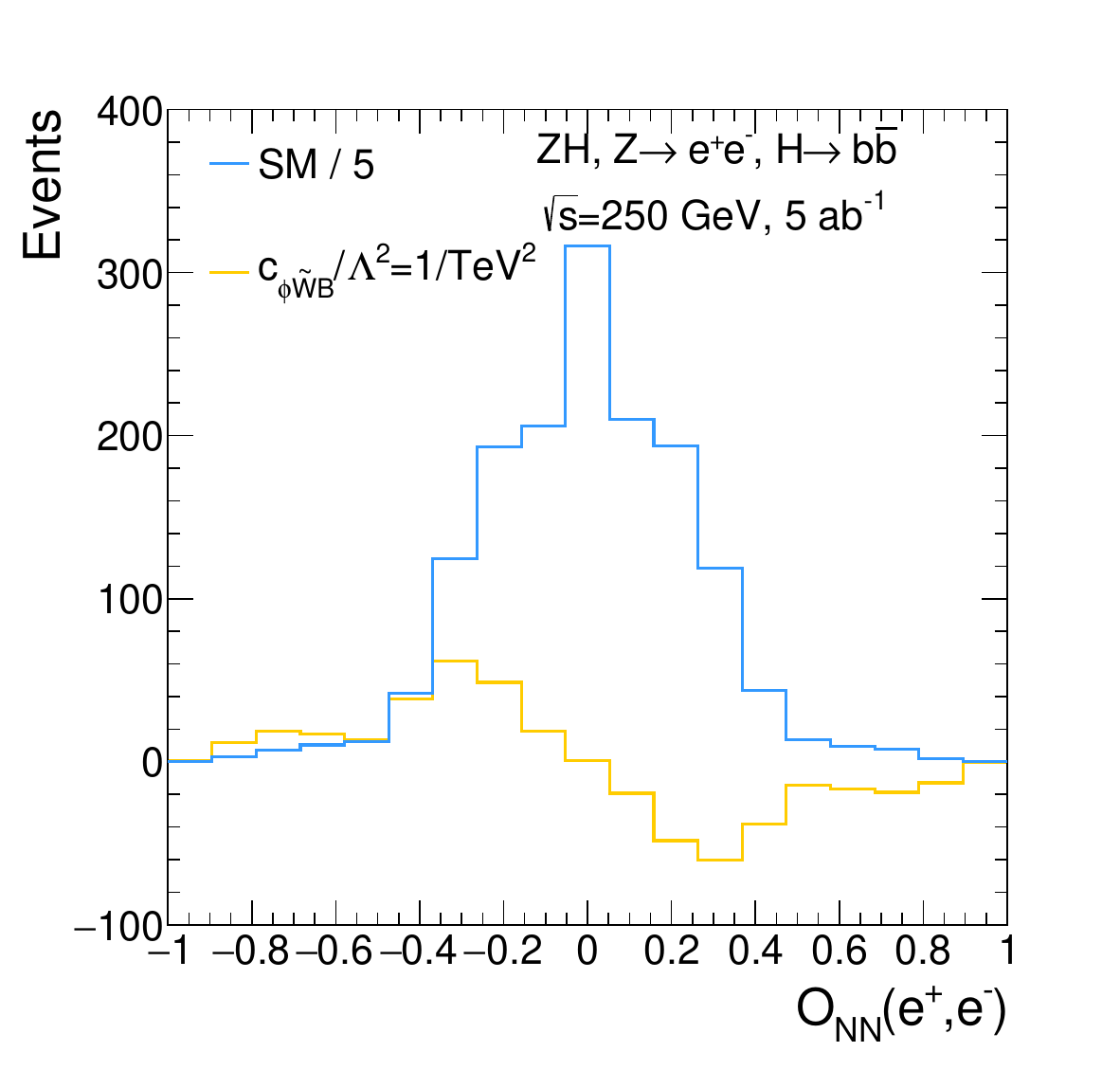}
  \caption{Sensitivity of \CP-sensitive observables for the $\OHWtilB$ operator, in the $\Pem\Pep$ inclusive final state: (left) event yield as a function of $\Delta\phi_{\ell\ell}$; (right) event yield as a function of $O_{NN}$.
  \label{fig:cpInHZZplots2}}
\end{figure}

\Cref{fig:cpInHZZplots1} shows the expected event yields in the di-muon final state in events with two b-tags, as a function of $\Delta\phi_{\ell\ell}$ and $O_{NN}$, respectively.
Similarly, \cref{fig:cpInHZZplots2} show the expected event yields in the di-electron final state. 
The interference effects are enhanced relative to the SM prediction for the neural-net-based observable. Permutation feature importance techniques show that this enhancement is caused by a sign-flip in the interference spectrum around the $\PZ$ pole.
This feature is not observed in dimuon final states, indicating that interference between VH and VBF diagrams is the likely origin.

Constraints on the Wilson coefficients are derived using a profile likelihood test and are shown in \cref{table:cpInHZZresults1}. 
The procedure accounts for statistical fluctuations in the observable. Experimental systematic effects are not included, but these are expected to have a very small effect on the measured asymmetries, as discussed in Ref.~\cite{PhysRevD.107.016008}.
For the $\Delta\phi_{\ell\ell}$ observable, the constraints on the $\OHWtilB$ operator are weaker than the constraints obtained for the other operators. $\OHWtilB$ is known to be notoriously difficult to constrain (see e.g.\ Ref.~\cite{Bernlochner:2018opw}), whereas the hierarchy between $\OHWtil$ and $\OHBtil$ reflects the usual $U(1)_Y$ vs. $SU(2)_L$ coupling hierarchy. Additional kinematic information provides substantially tighter constraints. Going beyond simply exploiting 2d correlations, the neural network observable $O_{NN}$ distils the available \CP information into a highly sensitive measure of \CP violation. For $\OHWtil$, which is the only operator to affect the Higgs boson's interactions with the $\PW$ boson, the 2d $\Delta\phi_{\ell\ell}-m_{12}$ correlations formidably reflect the amplitude's \CP information.
While the sensitivity is dominated by the statistics of the dataset, \cref{table:cpInHZZresults1} shows that longitudinal beam polarisation can compensate for a smaller dataset, particularly for $\OHWtil$.

The initial studies indicate that all of the sensitivity comes from analysis of the $\PZ\to\ell\ell$ topology, hence the analysis is being extended to remove the requirement on a particular Higgs final state and treat it inclusively; a factor of ~2-3 improvement in sensitivity is likely when considering the yield increase from BR and tagging-efficiency, after accounting for the larger backgrounds.

\begin{table}[t!]
\centering{
\begin{tabular}{ l|c|c|c||c|c|c } 
 & \multicolumn{3}{c||}{ 5\,\abinv unpolarised} & \multicolumn{3}{c}{ 2\,\abinv polarised}\\
\hline
 & \multicolumn{3}{c||}{ 95\% confidence intervals [TeV$^{-2}$]} & \multicolumn{3}{c}{ 95\% confidence intervals [TeV$^{-2}$]} \\
\hline
  Observable & \cHWtil/$\Lambda^2$ & \cHBtil/$\Lambda^2$ & \cHWtilB/$\Lambda^2$  & \cHWtil/$\Lambda^2$ & \cHBtil/$\Lambda^2$ & \cHWtilB/$\Lambda^2$ \\
\hline
$\Delta\phi_{\ell\ell}$ & [-0.40, 0.40] & [-0.70, 0.70] & [-0.92, 0.92] & [-0.44, 0.44] & [-0.85, 0.85] & [-1.11, 1.11] \\
$\Delta\phi_{\ell\ell}$ vs $m_{12}$ & [-0.30, 0.30] & [-0.15, 0.15] & [-0.26, 0.26] & [-0.35, 0.35] & [-0.28, 0.28] & [-0.34, 0.34] \\
$O_{NN}$ & [-0.22, 0.22] & [-0.12, 0.12] & [-0.20,0.20] & [-0.22, 0.22] & [-0.18, 0.18] & [-0.26,0.26] \\
\end{tabular}
}
\caption{Sensitivity to the interactions of \cref{eq:ops} from $\epem\to\PZ\PH$ events with \Zee or \Zmm, and \Hbb, assuming 5\,\abinv of unpolarised data or 2\,\abinv of polarised data (preliminary results).
\label{table:cpInHZZresults1}}
\end{table}

\subsubsection*{\CP tests with polarised beams}
\editor{Cheng Li - abstract 66}
We now consider the additional leverage that initial state polarisation for the electron-positron beams can provide to the determination of CPV HZZ interactions, by identifying two CP-odd observables \cite{Li:2025ouv} and employing the parametrisation of Eq.~\ref{eq:effl2}.

To study HZZ interactions, we consider Higgs-strahlung $\epem \rightarrow \PZ\PH$. By taking the polarisation of the initial beams into account, the spin-density matrix of the Higgs-strahlung $\rho_{\lambda_r \lambda_u}$ can be derived by applying the Bouchiat-Michel formula \cite{Bouchiat:1958yui}, obtaining
\begin{equation}
                {\rho}=
                (1-P_-^3P_+^3)A+(P_-^3-P_+^3)B +\sum_{mn}^{1,2}P_-^m P_+^n C_{mn}.
\end{equation}

Using the interactions obtained by Eq.~\eqref{eq:effl2},  the total amplitude squared can be cast in the following form
\begin{equation}
            \begin{split}
                    |\mathcal{M}|^2 &=(1-P_-^3P_+^3)( \cos^2\xi_{\CP}\,\mathcal{A}_\text{\CP-even} + {\sin2\xi_{\CP}\,\mathcal{A}_\text{\CP-odd}} + \sin^2\xi_{\CP}\,\widetilde{\mathcal{A}}_\text{\CP-even})\\
                    &+(P_-^3-P_+^3)(\cos^2\xi_{\CP}\,\mathcal{B}_\text{\CP-even} + {\sin2\xi_{\CP}\,\mathcal{B}_\text{\CP-odd}} + \sin^2\xi_{\CP}\,\widetilde{\mathcal{B}}_\text{\CP-even})\\
                    &+\sum_{mn}^{1,2}P_-^m P_+^n\left(\cos^2\xi_{\CP}\,\mathcal{C}^{mn}_\text{\CP-even} + {\sin2\xi_{\CP}\,\mathcal{C}^{mn}_\text{\CP-odd}} + \sin^2\xi_{\CP}\,\widetilde{\mathcal{C}}^{mn}_\text{\CP-even} \right)\, , 
            \end{split}
            \label{eq:ampsqbsm}
            \end{equation}
where the \CP-invariant parts with $\cos^2\xi_{\CP}$ and $\sin^2\xi_{\CP}$ are both \CP-conserving, while the \CP-mixing terms with $\sin2\xi_{\CP}$ violate the \CP symmetry. 

\begin{figure}[h]
    \centering
    \tikzstyle{every node}=[scale=2]
\begin{tikzpicture}[line width=1pt,scale=2]
\draw [-stealth](-1,0) -- (0,0);
\draw [-stealth](0,0) -- (0.7,0.7);
\draw [stealth-](0,0) -- (1,0);
\draw [-stealth](0,0) -- (-0.7,-0.7);
\draw[thin, ] (0.2,0) arc (0:-135:0.2);
\node [scale=0.3] at (0.1, -0.3) {$\theta_{\PH}$};
\node [scale=0.3] at (-0.8, -0.8) {$\PH$};
\node [scale=0.3] at (-1, -0.1) {$\Pem$};
\node [scale=0.3] at (1, -0.1) {$\Pep$};
\node [scale=0.3] at (0.3, 0.4) {$\PZ$};
\draw [-stealth](0.7,0.7) -- (0.9,1.1);
\draw [-stealth](0.7,0.7) -- (1.1,0.9);
\draw [dotted] (0.7, 0.7) -- (0.35, 0);
\draw[thin, ] (0.5,0) arc (0:63.4:0.15);
\node [scale=0.3] at (0.6, 0.1) {$\theta_{\PGmm}$};
\node [scale=0.3] at (1, 1.2) {$\PGmm$};
\node [scale=0.3] at (1.2, 1) {$\PGmp$};

\draw[help lines, ->] (0, -1.5) --(0,1.5);
\draw[help lines, ->] (-1.5, 0) --(1.5, 0);
\node[scale = 0.5] at (1.4, -0.2) {$\frac{\Vec{p}_{e^-}}{|\Vec{p}_{e^-}|}$};
\node[scale = 0.5] at (0, 1.8) {$\frac{\vec{s}_{e^-}}{|\vec{s}_{e^-}|}$};

\end{tikzpicture}~~~~~~\begin{tikzpicture}[line width=1pt,scale=2]
\draw [-stealth](0,0) -- (-0.7,0.7);
\draw [-stealth](0,0) -- (0.7,-0.7);
\draw[thin, ] (0.2,0) arc (0:315:0.2);
\node [scale=0.3] at (0.2, 0.2) {$\phi_{\PH}$};
\node [scale=0.3] at (0.8, -0.8) {$\PH$};
\node [scale=0.3] at (-0.3, 0.4) {$\PZ$};
\draw [-stealth](-0.7,0.7) -- (-0.6,0.9);
\draw [-stealth](-0.7,0.7) -- (-0.8,0.5);
\draw [dotted] (-0.7, 0.7) -- (-1.05, 0);
\draw[thin, ] (-0.9,0) arc (0:63.4:0.15);
\node [scale=0.3] at (-0.8, 0.1) {$\phi_{\PGmm}$};
\node [scale=0.3] at (-0.5, 1.0) {$\PGmm$};
\node [scale=0.3] at (-0.9, 0.5) {$\PGmp$};

\draw [dotted] (-0.7, 0.7) -- (-1.0,1.0);
\draw[thin, ] (-0.67, 0.8) arc (63.4:135:0.1);
\node [scale=0.3] at (-0.75, 0.95) {$\Delta\phi$};

\draw[help lines, ->] (0, -1.5) --(0,1.5);
\draw[help lines, ->] (-1.5, 0) --(1.5, 0);
\node[scale = 0.5] at (1.4, -0.2) {$\frac{\Vec{s}_{e^-}\times\Vec{p}_{e^-}}{|\Vec{s}_{e^-}\times\Vec{p}_{e^-}|}$};
\node[scale = 0.5] at (0, 1.8) {$\frac{\vec{s}_{e^-}}{|\vec{s}_{e^-}|}$};

\end{tikzpicture}
    \caption{The coordinate system in the laboratory frame for the $\epem \rightarrow \PH \PGmp \PGmm$ process with initial beams polarisation.}
    \label{fig:coord}
\end{figure}
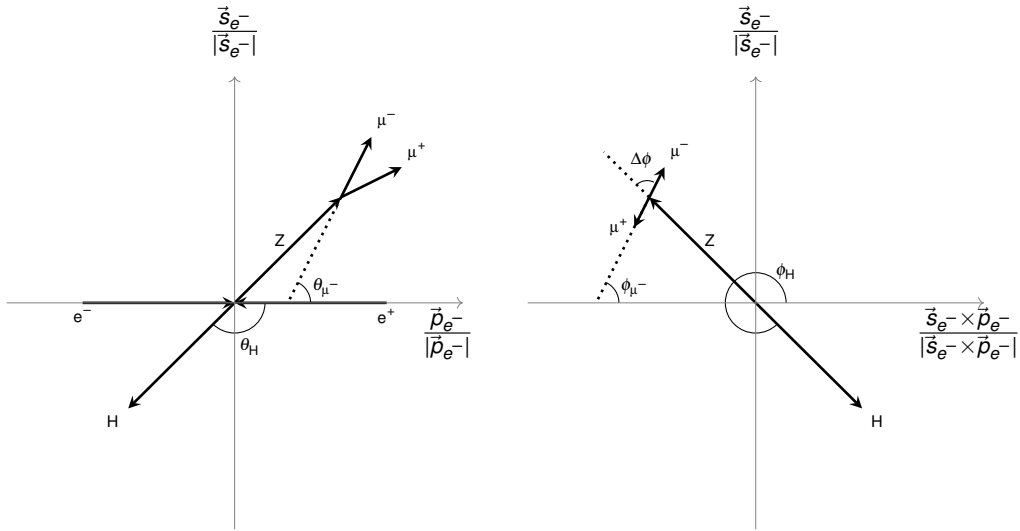

The \CP-violating terms $\mathcal{A}_\text{\CP-odd}$, $\mathcal{B}_\text{\CP-odd}$ and  $\mathcal{C}^{mn}_\text{\CP-odd}$ can be extracted by the triple-products, which are given by
            \begin{align}
                &\mathcal{A}_\text{\CP-odd}, \mathcal{B}_\text{\CP-odd}\propto \epsilon_{\mu\nu\alpha\beta} [p_{e^-}^\mu p^\nu_{e^+} p^\alpha_{\PGmp} p^\beta_{\PGmm}]\propto (\vec{p}_{\PGmp}\times \vec{p}_{\PGmm})\cdot \vec{p}_{e^-}, \\
                &\mathcal{C}^{mn}_\text{\CP-odd} \propto \epsilon_{\mu\nu\rho\sigma}[({p}_{e^-} +{p}_{e^+})^\mu p_{\PGmp}^\nu p_{\PGmm}^\rho s_{e^-}^\sigma ]\propto(\vec{p}_{\PGmp}\times\vec{p}_{\PGmm})\cdot \vec{s}_{e^-}.
            \end{align}
The latter triple-product corresponds to  the azimuthal-angle difference between the $\PGmp\PGmm$ plane and the spin of the electron. Therefore, defining the orientation of the azimuthal plane by fixing the direction of electron transverse polarisation, the $C^{\text{\CP-mix}}_{mn}$ depends directly on the azimuthal angle of $\phi_{\PGmm}$ (see Fig.~\ref{fig:coord}).

Starting from triple products in the \CP-odd amplitude terms, one can define two \CP-odd observables
\begin{equation}
    \mathcal{O}^T_{\CP} = \eta_{\PH} \sin 2 \phi_{\PGmm}\,,\end{equation}
and
\begin{equation}
   \mathcal{O}^{UL}_{\CP} = \cos\theta_{\PGm}  \sin (\phi_{\PGm}-\phi_{\PH})\,.
\end{equation}
through which asymmetries can be built:
\begin{equation}
        \mathcal{A}_{\CP} = \frac{1}{\sigma_\text{tot}} \int\operatorname{sgn}(\mathcal{O}_{\CP}){d\sigma} = \frac{N(\mathcal{O}_{\CP}<0)-N(\mathcal{O}_{\CP}>0)}{N(\mathcal{O}_{\CP}<0)+N(\mathcal{O}_{\CP}>0)},
            \label{eq:asyN}
\end{equation}
where $N$ denotes the corresponding number of events. 
Since the SM is \CP conserving for the neutral current, the SM background for this asymmetry is negligible. Statistical fluctuations lead to an uncertainty of this asymmetry that can be estimated through a binomial distribution and given by $\Delta\mathcal{A}_{\CP} = \sqrt{\frac{1-\mathcal{A}_{\CP}^2}{N_\mathrm{tot}}}$.

In principle, both observables $\mathcal{O}^T_{\CP}$ and $\mathcal{O}^{UL}_{\CP}$ are sensitive to the \CP-violation. The observable $\mathcal{O}^{UL}_{\CP}$ can be always measured for any initial polarisation states. However, the $\mathcal{O}^T_{\CP}$ can only be measured when initial beams are transversely polarised. Consequently, one can combine the observables $\mathcal{O}^T_{\CP}$ and $\mathcal{O}^{UL}_{\CP}$ to improve the sensitivity to \CP-violation, when the transverse polarisation of electron-positron beams are imposed. 

Therefore, we can propose to combine the two observables by 
\begin{equation}
    \chi^2 = \left(\frac{\mathcal{A}^T_{\CP}}{\Delta\mathcal{A}^T_{\CP}}\right)^2 +\left(\frac{\mathcal{A}^{UL}_{\CP}}{\Delta\mathcal{A}^{UL}_{\CP}}\right)^2.
\end{equation}
If only longitudinally polarised beams are available  $\mathcal{O}^{UL}_{\CP}$ can be used. 

In summary, the beam polarisation can be helpful to improve the sensitivity to \CP-violation. The longitudinally polarisation can enhance the total cross-section and suppress the statistical uncertainties. The transverse polarisation can offer an additional observable, and combining the two observables would also improve the precision of \CP-violating coupling. 

\subsubsection*{$\PH\PZ\PZ$ \CP studies at $\sqrt{s}=1$~TeV}
\label{sec:HVVCPILC}
\editor{Ivanka Bozovic - abstract 3}
 
We now report on the reach for constraining  \CP-odd interactions in the HZZ vertex, by studying Higgs boson production in $ZZ$-fusion, at 1 TeV ILC, with unpolarised beams~\cite{vukasinovic:2024}. 

Using a full reconstruction of SM background and fast simulation/reconstruction of the signal, we have simulated 8 ab$^{-1}$ of data collected with the ILD detector \cite{ILDConceptGroup:2020sfq}. 
As a result we find that the \CP mixing angle between scalar and pseudoscalar interactions can be measured with the statistical uncertainty of 3.8 mrad at 68\% CL, corresponding to $f_{\CP} = 1.44 \times 10^{-5}$, for the pure scalar state. This is the first result on sensitivity of an \epem collider to measure $f_{\CP}$ in the Higgs production vertex in vector boson fusion (VBF). Since the study does not assume any particular polarisation scheme, it can be used as a reference for any \epem collider capable of operating at around 1 TeV centre-of-mass energy.

\begin{figure}[h!]  
    \begin{subfigure} {.5\linewidth}
        \centering
        \includegraphics[width=\linewidth]{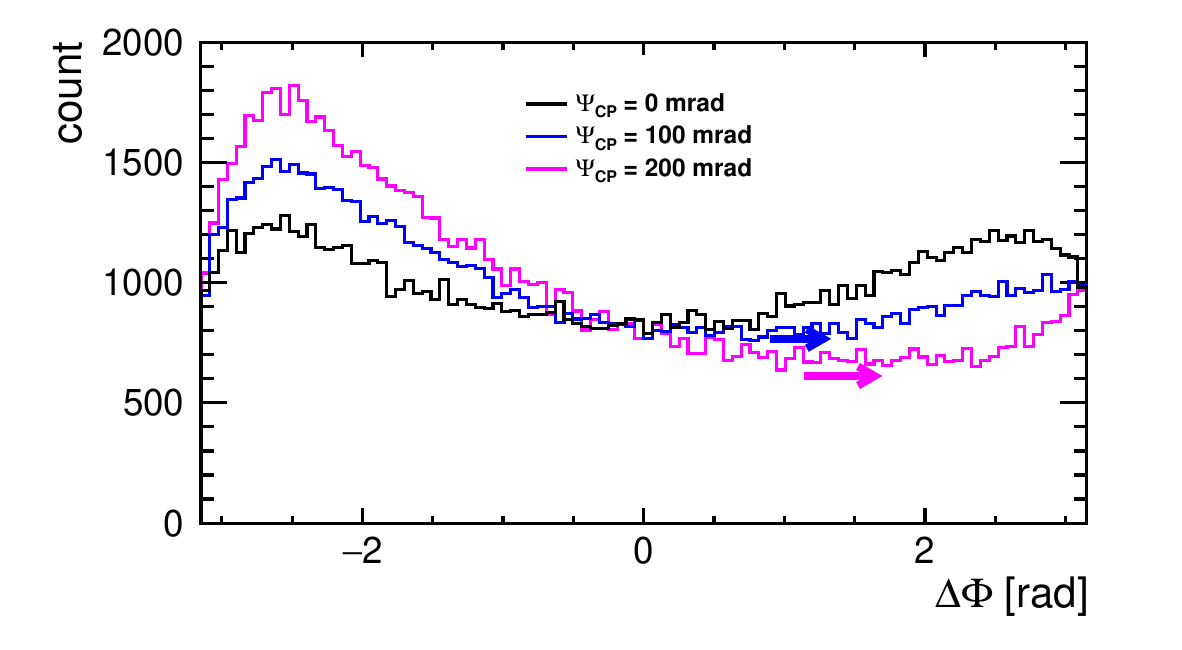}
        \caption{}
        \label{fig:hzz_angles}
    \end{subfigure}
    \quad
    \begin{subfigure}{.5\linewidth}  
        \includegraphics[width=\linewidth]{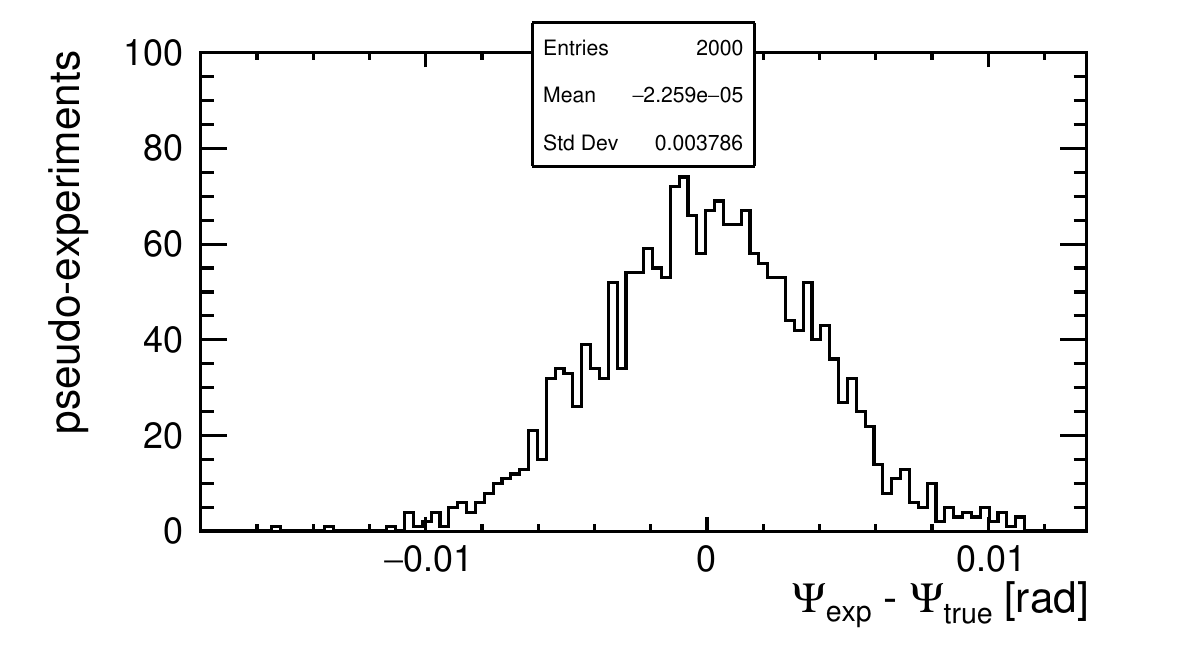}
        \caption{}
        \label{fig:hzz_pseudoexp}
    \end{subfigure}
    \caption{ (a) $\Delta\Phi$ distribution for different mixing angles ($\Psi_{CP}=\xi_{CP}$ in the text) illustrating the shift of the $\Delta\Phi$ minimum for non-zero values of $\xi_{CP}$. (b) Statistical dispersion of measured $\xi_{\CP}$ values ($\Psi_{\mathrm{exp}}$) w.r.t. the true ones ($\Psi_{\mathrm{true}}$).}
\end{figure}

To interpret the obtained precision of measurement of the mixing angle in terms of sensitivity to the \CP-odd amplitude $f_{\CP}$, the framework described in Ref.~\cite{gritsan:2022snowmass} is assumed, that is that $f_{\CP}$ will vary from zero as $\sin^{2}(\Delta(\xi_{\CP}))$ for the pure scalar state, where $\Delta(\xi_{\CP})$ is the absolute statistical uncertainty of the mixing angle measurement. The sensitivity of $\xi_{\CP}$ measurement to $f_{\CP} \sim 10^{-5}$ illustrates the potential of a TeV linear collider to probe Higgs \CP-odd component in VBF Higgs production channels.

\subsubsection*{Quantum entanglement in $\PH\to\PV\PV$}
\editor{Juan Antonio Aguilar Saavedra}
The precision with which Higgs production at an $\epem$ collider will be accessible opens also the way to studies of the quantum properties of 4 fermion final states arising from the $\PH \to \PV_1 \PV_2$ decays. 

The angular momentum in the Higgs boson decay into two weak bosons (which we consider as distinguishable) is fully determined by the tripartite density operator $\rho_{L S_1 S_2}$ involving the orbital angular momentum and the  spins of the two bosons. In particular, this operator can be used to determine the entanglement between any two parties of the $L S_1 S_2$ system. The SM predicts a large entanglement in most cases. Using as entanglement measure an observable $N$~\cite{Plenio:2007zz} called negativity, which takes values between 0 and 1, with 1 indicating largest possible entanglement. For $\PH \to \PZ \PZ$ decays one predicts
\begin{align}
& L-(S_1 S_2) :\quad N = 0.757 \,, && L-S_1 :\quad N = 0.105 \,, \notag \\
& S_1-(L S_2) :\quad N = 0.998 \,, && L-S_2 :\quad N = 0.105 \,, \notag \\
& S_2-(L S_1) :\quad N = 0.998 \,, && S_1 - S_2 :\quad N = 0.843 \,,
\end{align}
with similar values for $\PH \to \PW \PW$~\cite{Aguilar-Saavedra:2024whi}. The numbers on the left side quantify the entanglement between one subsystem and the rest; with all of them being non-zero, the system is said to have genuine tripartite entanglement. The numbers on the right side correspond to the entanglement between two parties when the degrees of freedom of the third one are integrated out. The level of entanglement of the tripartite state, as well as possible violation of Bell inequalities, can be reconstructed from angular distributions of the decay products.

Angular distributions in the decay $\PH \to \PV_1 \PV_2$  can be parametrised either using a fixed reference system $(\hat x, \hat y, \hat z)$, or a moving one $(\hat r, \hat n, \hat k)$, where $\hat k$ is taken in the direction of one of the decay products in the Higgs rest frame. The former, referred to as ``canonical basis'', is the convenient one to express $\rho_{L S_1 S_2}$. The latter, also known as ``helicity basis'', is very useful for measurements, because the orbital angular momentum component in the direction of motion vanishes, which greatly simplifies the description. For example, there are 18 non-zero decay amplitudes in the canonical basis, while in the helicity basis the number is reduced to only three quantities $a_{\lambda_1 \lambda_2}$, with $(\lambda_1,\lambda_2) = (1,1)$, $(-1,-1)$ or $(0,0)$. These amplitudes in general depend on the off-shell boson mass $m_{V^*}$.

In the helicity basis the four-dimensional angular distribution is~\cite{Aguilar-Saavedra:2022wam,Aguilar-Saavedra:2022mpg}
\begin{equation}
\frac{1}{\sigma}\frac{d\sigma}{d\Omega_1d\Omega_2} = \frac{1}{(4\pi)^2}\left[ 1 +a_{lm}^1 Y_l^m(\theta_1, \phi_1) + a_{lm}^2 Y_l^m(\theta_2, \phi_2)    + c_{l_1 m_1 l_2 m_2}  Y_{l_1}^{m_1}(\theta_1, \phi_1)Y_{l_2}^{m_2}(\theta_2, \phi_2)  \right] \,,
\label{ec:dist4D}
\end{equation}
where $\Omega_1 = (\theta_1, \phi_1)$, $\Omega_2 = (\theta_2,\phi_2)$ are the angles that determine the orientation of the spin analysers $\Pf_1$, $\Pf_2$ for the two bosons $\PV_1$, $\PV_2$ in their respective rest frame, and $Y_l^m$ are the usual spherical harmonics. 
This angular distribution can be considered either differentially as a function of $m_{V^*}$, or integrated over this variable. In either case, its determination requires a full reconstruction of the final state, because the momenta of $\Pf_{1,2}$ have to be boosted to the rest frame of the decaying bosons. 

At the tree level, the coefficients $a_{lm}^{1,2}$ and $c_{l_1 m_1 l_2 m_2}$ are related to the elements of the spin-density operator $\rho_{S_1 S_2}$ in the helicity basis, which can be expanded in a basis of irreducible tensor operators $T^l_m$ as
\begin{eqnarray}
\rho_{S_1 S_2} & = & \frac{1}{9}\left[
\mathbf{1}_3 \otimes \mathbf{1}_3 + A^1_{lm} \, T^l_{m} \otimes \mathbf{1}_3 + A^2_{lm} \, \mathbf{1}_3 \otimes T^l_{m}  + C_{l_1 m_1 l_2 m_2} \, T^{l_1}_{m_1} \otimes T^{l_2}_{m_2}
\right] \,.
\label{ec:rhoAC}
\end{eqnarray}
The relation is $a_{lm}^1 = A_{lm}^1 B_l^1$, $a_{lm}^2 = A_{lm}^2 B_l^2$, $c_{l_1 m_1 l_2 m_2} = B_{l_1}^1 B_{l_2}^2 C_{l_1 m_1 l_2 m_2}$, with constants $B_l^i$ that represent the spin analysing power of the fermions $\Pf_i$, $i = 1,2$. For decay into charged leptons one has $B_1 = - \sqrt{2 \pi} \eta_\ell$, $B_2 = \sqrt{2\pi/5}$, with $\eta_\ell = \pm 1 $ for $\PW^\mp$ (taking as spin analyser the charged lepton) and $\eta_\ell \simeq 0.13$ for $\PZ$ (taking as spin analyser the negative lepton). Next-to-leading order (NLO) effects have been studied for $\PH \to \PZ \PZ$~\cite{Grossi:2024jae}. For terms with $l = 2$ the corrections with respect to the leading order are quite small, whereas for spin correlation terms $c_{1 m_1 1 m_2}$, which are suppressed by a small $\eta_\ell^2 \simeq 0.01$ factor, NLO corrections are larger. 

The statistics of Higgs bosons expected at a future $\Pep \Pem$ collider is one order of magnitude smaller than at LHC Run 3, and two orders of magnitude smaller than at the HL-LHC. Still, lepton colliders can explore decay modes that have large backgrounds in hadron collisions. Among them, the semileptonic decay modes
\begin{itemize}
\item $\PH \to \PW \PW \to \Pl \PGn \PQq \PAQq$
\item $\PH \to \PZ \PZ \to \Plp \Plm \PQq \PAQq$
\end{itemize}
are especially promising. The discrimination between quarks and anti-quarks, necessary for the direct measurement of the parameters of the distribution (\ref{ec:dist4D}), may be circumvented by assuming \CP conservation in the decay. In such a scenario, the three helicity amplitudes for $\PH \to \PV_1 \PV_2$ (which depend on $m_{V^*}$) can be determined from a binned measurement of the parameters $a_{20}^{1,2}$ in Eq.~(\ref{ec:dist4D}), namely
\begin{equation}
a_{11} = a_{-1-1} = \left[ \frac{1}{4} + \frac{1}{2 \sqrt 2} \frac{a_{20}^i}{B_2^i} \right]^{1/2}  \qquad
a_{00} = - \left[ \frac{1}{2} - \frac{1}{\sqrt 2} \frac{a_{20}^i}{B_2^i} \right]^{1/2} \,.
\end{equation}
The parameters $a_{20}^i$ multiply spherical harmonics $Y_2^0(\theta,\phi)$ that are even in $\cos \theta$, therefore their experimental determination does not rely on the identification of quarks versus anti-quarks. Once the helicity amplitudes are measured, the density operator $\rho_{L S_1 S_2}$ can be determined in a model-independent fashion. A study for the $\PZ \PZ \to 4\Pl$ channel has shown that even despite the need of a (coarse) binning on $m_{V^*}$, there is a great statistical gain when determining the parameters of the density operators by using this method~\cite{Aguilar-Saavedra:2024whi}. 

Concerning the rest of decay modes, the fully-hadronic ones seem difficult due to the combinatorics required to properly identify the quarks resulting from each boson decay. The fully-leptonic final state $\PW^+ \PW^- \to \Plp \PGn \, \Plm \PGn$ is more favourable than in hadron collisions~\cite{Barr:2021zcp} but still the number of kinematical constraints (five, for the missing energy and on-shell conditions for $\PH$ and one $\PW$ boson) is smaller then the number of unknowns (six, for the two neutrino momenta) and the final state cannot be uniquely reconstructed. The possibility of a kinematical reconstruction via a likelihood fit or a multivariate method remains to be explored.

In conclusion, $\epem$ colliders will provide a unique opportunity to study quantum correlations in H to vector bosons decays.

\subsubsection{\texorpdfstring{$\PH\PGt\PGt$ coupling}{Htautau-coupling}}
\label{sec:Htautau}
\editor{Fabio}
In this section we focus on the extraction of the properties, in particular strength and CP-parity, of the  $\PH\PGt\PGt$ coupling from studying rates and distributions in the $\PH\to \PGt\PGt$ decay. We consider first a study performed in the context of the FCC-ee on the strength of the coupling and then on the CP properties. Then we report a phenomenological study on the CP properties and finally an exploration of the quantum properties of the spins of the $\PGt$ leptons.  

In general, in an effective field theory at dimension 6 one can consider the modifications induced in the $\PH\PGt\PGt$ coupling, by operators of the form $\frac{c_{\tau \Phi}(\Phi^{\dagger}\Phi) }{\Lambda^2} \bar L \Phi \tau_R+h.c.$. This type of operator gives rise to possibly (P-violating/C-violating) CP-violating (yet flavour-diagonal) interactions of the Higgs boson with fermions that in the EW broken phase can be parametrised as   
\begin{equation}
{\cal L}^{h\tau \tau}_\text{CPV}   = -\bar \kappa_\tau m_\tau \frac{h}{v} 
\bar{\psi}_\tau (\cos \delta + i  \gamma_5 \sin \delta) \psi_\tau\,,
\label{eqn:hff-CPV}
\end{equation}
where the angle $\delta$ parametrises the departure from the CP-even case. Another, equivalent parametrisation employs $\kappa_\tau = \bar \kappa_\tau \cos \delta$ and $\tilde\kappa_\tau = \bar \kappa_\tau \sin \delta$, where $\kappa_\tau=1+\delta y_\tau$ in the notation used for the CP conserving cases in the $\kappa$-framework (with $\kappa>0$). The pure scalar coupling corresponds to $\delta=0$ ($\tilde\kappa_\tau=0$),  a pure pseudoscalar coupling to $\delta=90^\circ$ ($\kappa_\tau=0$), while CP violation occurs in all other intermediate cases.

In the literature, the $\PH \PGt \PGt$ \CP phase measurement at ILC using different tau decay modes
have also been explored \cite{Bower:2002zx,Desch:2003rw}.
The studies in Refs.~\cite{Harnik:2013aja, Jeans:2018anq}
exploit the $\PGt^\pm \to \rho^\pm \PGn$ channel 
and suggest that the sensitivity can reach $\sim 4 ^{\circ}$,\footnote{Note, however, that these analyses do not include the effect of energy mismeasurement.}
which is in line with the expectations of first theoretical studies, i.e.,  $2.8^{\circ}$ \cite{Berge:2013jra}, including also other decay modes, e.g.\ $\PGt^\pm \to a^\pm \PGn$ and $\PGt^\pm \to \Plpm \PGn \PAGn$.
A recent study \cite{Chen:2017bff} using a likelihood analysis based on the Matrix-Element
claims that the \CP-phase can be measured in the accuracy of $2.9^{\circ}$ at the ILC.

At HL-LHC, the resolution of the $\PH \PGt\PGt$ \CP-phase measurement is expected to reach  
$\sim 11^{\circ}$ using the tau decay modes
$\PGt^\pm \to \PGppm \PGn$ \cite{Hagiwara:2016zqz}
and $\PGt^\pm \to \rho^\pm \PGn$ \cite{Harnik:2013aja}.
Combining comprehensive decay modes with the Matrix-Element likelihood method, the resolution may reach
$5.2^{\circ}$ \cite{Chen:2017nxp}.
On the other hand, however, Ref.~\cite{Askew:2015mda}
claims that the detector effect severely impacts on 
the performance of \CP measurements at HL-LHC
and the \CP phase hypothesis $\delta = 0$ can be distinguished from $\delta = 90^{\circ}$ 
only at 95\,\% CL with $\PGt^\pm \to \PGppm \PGn$ channel.

\subsubsection*{$\PH\PGt\PGt$ coupling sensitivity}
\editor{Sofia Giappichini - abstract 46}

The large Higgs boson statistics together with the clean environment at an \epem collider can bring spectacular gains in our knowledge of the $\PH\to\PGtp\PGtm$ interaction. 
In this study we explore the prospective sensitivity of the $\Pep\Pem\to\zhsm$, $\PH\to\PGtp\PGtm$ cross section measurement at FCC-ee. 

Both signal and background processes have been simulated in the Monte Carlo event generator \whizard v3.0.3 \cite{Kilian:2007gr,Moretti:2001zz} including a parametrisation of initial state radiation. Subsequently, \textsc{Pythia6} \cite{Sjostrand:2006za} was used to handle showering and hadronisation as well as tau decays. \textsc{Delphes} v3.5.1pre05 \cite{deFavereau:2013fsa} simulated the response of IDEA detector~\cite{IDEA1}.
We tested two jet algorithms: the \epem inclusive generalized $\kT$ algorithm with the jet-radius parameter equal to $R=0.5$ and a cut on the merging scale of $p_{\mathrm{T},j}>$\SI{2}{\giga\electronvolt}, and the \epem exclusive Durham algorithm with $n_{\textrm{jets}}$ equal to the expected number of jets.

To reconstruct hadronic tau decays, we have compared two methods based on the initial reconstruction of jets in the events. In the explicit reconstruction method, the four-vector of the taus, the charge and mass, are based on the jet constituents spatially close to the leading charged pion for jets not containing leptons. Reconstructed taus need to have a mass smaller than \SI{3}{\giga\electronvolt}. The number of charged pions and photons in the candidates is then used to infer the decay mode. The second method is based on applying a ParticleNet tagger, trained on di-jet Higgs events for FCC-ee \cite{Bedeschi:2022rnj}, where the taus are defined by a tau score above 0.5, while quark jets are defined as jets that did not meet this criterion. Leptonic taus decays are identified by looking directly at the isolated leptons in the event. The visible part of the four-momentum of the Higgs boson can then be reconstructed from the taus' four-momenta.

We organise our analysis based on the decays of the \PZ ($\PZ\to\Pl\Pl$ with $\Pl=\Pe,\PGm$, $\PZ\to\PQq\PQq$ with $\PQq=\PQu, \PQd, \PQs, \PQc, \PQb$, and $\PZ\to\PGn\PGn$) and Higgs boson ($\PH\to\PGt_{\Pl}\PGt_{\Pl}$, $\PH\to\PGt_{\Pl}\PGt_h$, and $\PH\to\PGt_h\PGt_h$ where $\PGt_{\Pl}$ and $\PGt_h$ denote leptonic and hadronic tau decays). After the basic event selection to define the categories, we have optimised cuts on the reconstructed objects and event observables in each category to enhance the signal-to-background ratio, separately for each category of $\PZ$ decay mode (leptonic, hadronic, neutrino). These include requirements on the reconstructed $\PZ$ mass, collinear mass, and recoil mass (for leptonic and hadronic $\PZ$ decays) and on the missing energy (for $\PZ\to\PGn\PGn$).  In all cases the $\PGt$ candidates are required to be separated via $\Delta R_{\PGt\PGt}>2, \cos\theta_{\PGt\PGt}<0$ (where $\theta_{\PGt\PGt}$ is the angle between the two \PGt momenta), and $|\cos{\theta^\text{miss}}| < 0.98$.

Parallel to the cut-based event selection, we have also trained a boosted decision tree (BDT) classifier in each of the categories except for the leptonic \PZ decays.


\begin{figure}[h]
    \begin{subfigure}[h]{0.33\textwidth}
        \centering
        \includegraphics[trim={5mm 0 7mm 0},clip,width=\textwidth]{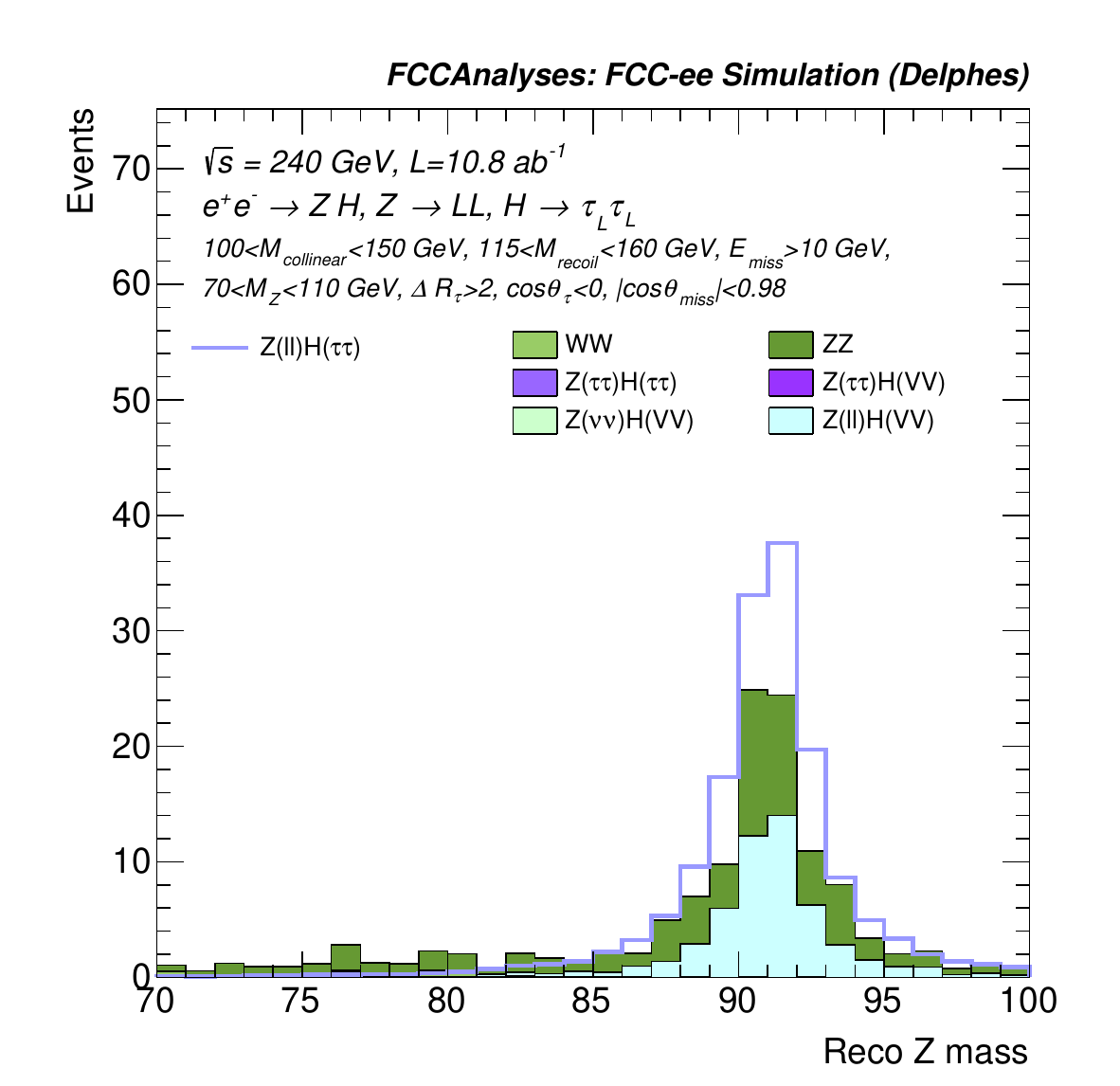}
        \subcaption{$\PZ\to\Pl\Pl$, $\PH\to\PGt_{\Pl}\PGt_{\Pl}$.}
    \end{subfigure}
    \begin{subfigure}[h]{0.33\textwidth}
        \centering
        \includegraphics[trim={5mm 0 7mm 0},clip,width=\textwidth]{figs/Higgs/Recoil.pdf}
        \subcaption{$\PZ\to\Pl\Pl$, $\PH\to\PGt_{\Pl}\PGt_h$.}
    \end{subfigure}
    \begin{subfigure}[h]{0.33\textwidth}
        \centering
        \includegraphics[trim={5mm 0 7mm 0},clip,width=\textwidth]{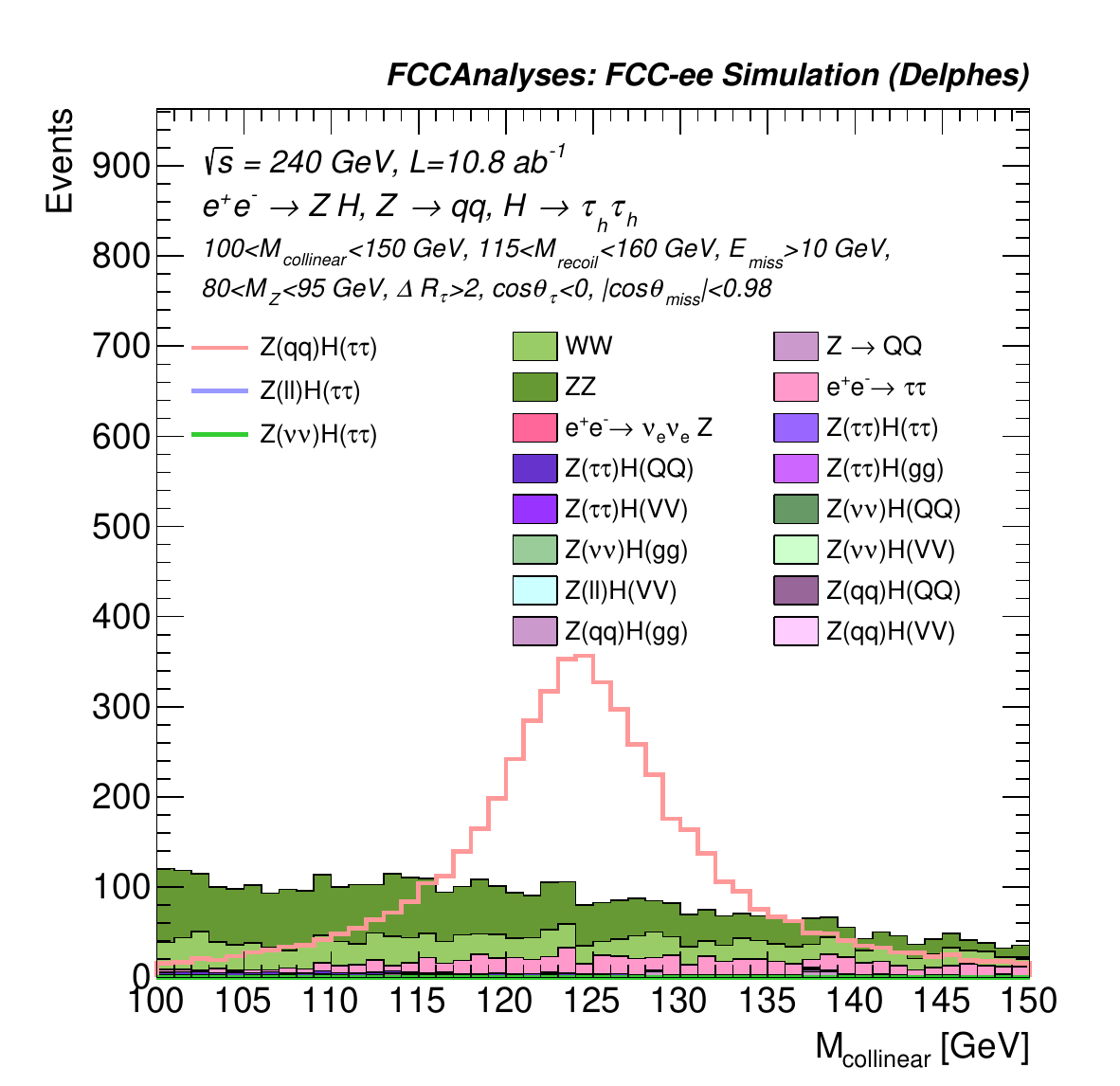}
        \subcaption{$\PZ\to\PQq\PQq$, $\PH\to\PGt_h\PGt_h$.}
    \end{subfigure}
    \caption{Some reconstructed variables for some of the categories studied, using ParticleNet tau reconstruction with inclusive jets. All signal and background samples are shown after passing the cuts indicated in the legends.}
    \label{fig:KIT-Htautau-plots_ll_zmass}
\end{figure}

\begin{figure}[h]
    \centering
    \begin{subfigure}[h]{0.32\textwidth}
        \centering
        \includegraphics[width=\linewidth]{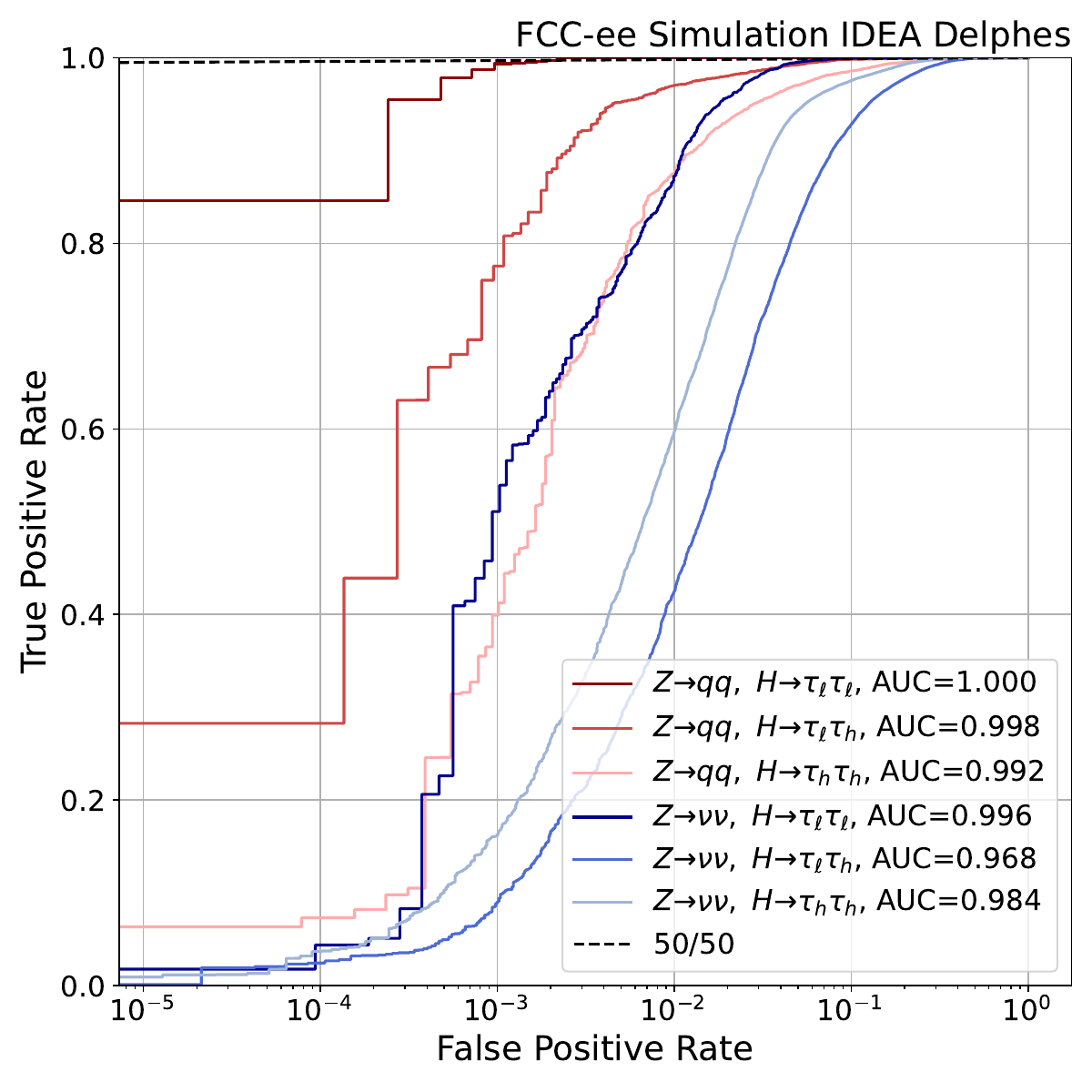}
    \end{subfigure}
    \begin{subfigure}[h]{0.35\textwidth}
        \centering
        \includegraphics[width=\linewidth]{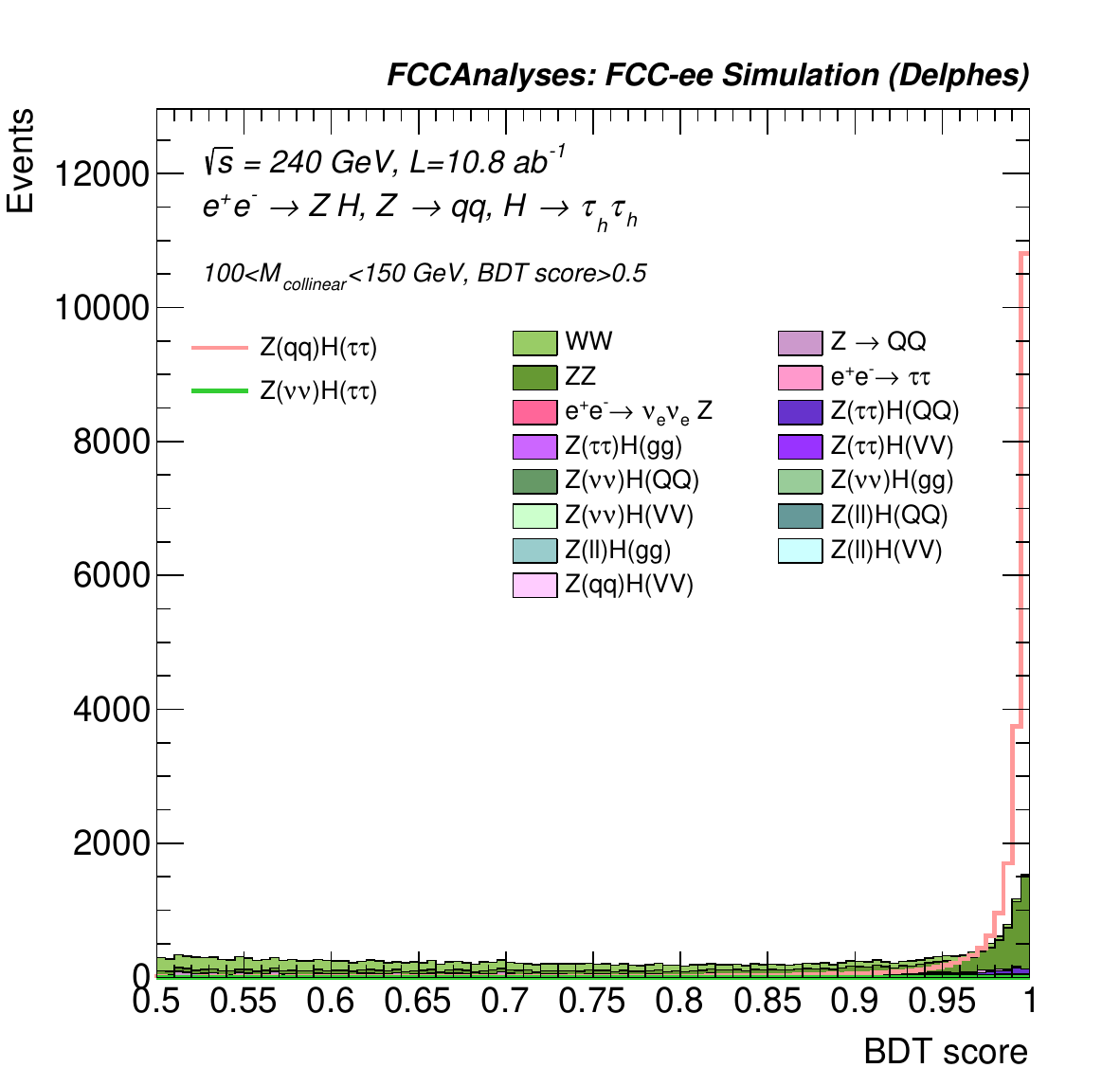}
    \end{subfigure}
    \caption{ROC curves for each of the BDT training and corresponding AUC value in the left plot. On the right, the distribution of the BDT score for $\PZ\to\PQq\PQq$, $\PH\to\PGt_h\PGt_h$. Both plots correspond to ParticleNet tau reconstruction with exclusive jet clustering.}
    \label{fig:KIT-Htautau-ROC}
\end{figure}

The cross section measurement has been extracted with the CMS Combine tool \cite{CMS:2024onh} using binned maximum-likelihood fits. 
With the cut-based analysis, the recoil mass distribution for the $\PZ\to\Pl\Pl$ and $\PZ\to \PQq\PQq$ categories are used in the fit, while for the $\PZ\to\PGn\PGn$ category the visible mass was fitted. 
In categories where the BDT training is available, an alternative fit is also performed directly on the BDT outputs instead of the mass variables.
Combining all categories of \PZ and Higgs decays with the cut-based strategy, the relative uncertainty on $\sigma_{\zhsm}\times\mathcal{B}$($\PH\to\PGtp\PGtm$) is found to be 0.95\% by using the ParticleNet tau reconstruction on inclusively clustered jets. This result is further improved with the BDT selection, where the best performance is found by applying the ParticleNet tau reconstruction on exclusively clustered jets, with the combined relative uncertainty of the cross-section measurement equal to 0.74\%.

In conclusion, the relative uncertainty expected at FCC-ee for $\PH\to\PGtp\PGtm$ in \zhsm events, $\sqrt{s}=$ \SI{240}{\giga\electronvolt} with $\mathcal{L}_{int}=$ \SI{10.8}{\per\atto\barn}, is 0.74\%.


\subsubsection*{Quantum entanglement in $\PH\to\PGt\PGt$}
\editor{Kazuki Sakurai}
The prospects of quantum information measurement with $\PH \to \PGtm\PGtp$ at future lepton colliders have been studied in Refs.~\cite{Altakach:2022ywa, Ma:2023yvd, Fabbrichesi:2024wcd}.
Quantum measurements can detect classes of spin correlations that cannot be realised in classical theories. 
These include quantum entanglement and Bell non-locality. 
When two particles are entangled, the property of one particle cannot be described independently of the other particle within quantum mechanics.  
On the other hand, if a particle pair is in a Bell non-local state, there exists a set of measurements about these particles whose data cannot be described by Local Real Hidden Variable (LRHV) theories.
This class of theories demands that any physical property of a local object, say a particle, is fully described without referring to the information about space-like separated regions (locality) and predetermined prior to the act of measurement (reality). 
Generally, the measurement outcomes of such theories can be described by a set of unknown (hidden) variables, $\lambda$, with the occurrence $P(\lambda)$.
Let us suppose a pair of particles, $A$ and $B$, have interacted locally in the past and then separated far apart. 
Alice measures the spin $a$-component of particle $A$, and Bob measures the spin $b$-component of particle $B$.
Their outcomes are denoted by $x$ for Alice and $y$ for Bob with $x,y \in \{ +1, -1\}$.
In LRHV theories, the probability distribution of the event can be written as 
\begin{equation}
P(x,y|a,b) = \int d \lambda P(\lambda) P_A(x|a,\lambda) P_B(y|b,\lambda)  \,.
\label{prob}
\end{equation}
On the right-hand side, the probability distributions of Alice ($P_A(a|x,\lambda)$) and Bob ($P_B(b|y,\lambda)$) are factorised. 
Equation (\ref{prob}) is often considered to be the defining equation of LRHV theories.

As we will see below, quantum measurements can also detect a type of spin correlation that cannot be attained within quantum theories. 
If the observation of such a correlation is confirmed, quantum mechanics is falsified.  
Since quantum mechanics may be modified at a short distance scale (for example, to be reconciled with gravity), a high-energy test of quantum mechanics is a valuable experimental programme. 

In this section, we briefly describe the analysis and results presented in Ref.~\cite{Altakach:2022ywa}.

The spin state of the two tau leptons produced from the Higgs decay, $\PH \to \PGtm \PGtp$, is given by the maximally entangled 2-qubit \cite{Altakach:2022ywa,Fabbrichesi:2022ovb}
\begin{equation}
| \Psi_{H \to \PGt \PGt}(\delta) \rangle = \frac{1}{\sqrt{2}} \left( | +,- \rangle + e^{i 2 \delta} | -,+ \rangle \right)\,,
\label{eq:pure}
\end{equation}
where $\delta$ is the \CP-phase in the $\PH \PGt \PGt$ interaction.
The spin correlation matrix is then given by
\begin{equation}
C_{ij} \equiv \langle \hat s^-_i \hat s^+_j \rangle,
~~~~
C = \begin{pmatrix}
\cos 2 \delta & \sin 2 \delta & 0  \\
-\sin 2 \delta & \cos 2 \delta  & 0  \\
0 & 0  & -1  \\
\end{pmatrix}\,.
\label{eq:Cmat}
\end{equation}
In this expression, $\hat s^\mp_i$ is the $i$-th component of the $\PGt^\mp$ spin.
The $i = 1,2,3$ component corresponds to the 
directions of the normalised vectors ($\mathbf r$, $\mathbf n$, $\mathbf k$) in the helicity basis.

Using the definitions in Eq.\ \eqref{eq:pure} and 
in Eq.\ \eqref{eq:Cmat}, the \textit{concurrence} \cite{Hill:1997pfa,Wootters:1997id}, a measure of entanglement, is given by
\begin{equation}
{\cal C}[\rho] = \text{max} \left[ 0,\, \frac{D_{+} + C_{kk} - 1}{2},\,
\frac{D_{-} - C_{kk} - 1}{2} 
\right]\,,
\label{eq:crho_tau}
\end{equation}
with $D_{\pm} \equiv \sqrt{(C_{rn} \pm C_{nr})^2 + (C_{rr} \mp C_{nn})^2}$.
The concurrence is a non-negative function of the spin density operator, $\rho$, and vanishes for all separable (nonentangled) states.
${\cal C}[\rho]$ takes the maximum value 1 for maximally entangled states.
In fact, for the spin state of the tau pair in $\PH \to \PGtm\PGtp$, we have ${\cal C}[\rho] = 1$.

Another metric connected to entanglement (but not equivalent in case of mixed states) is related to the so-called CHSH inequality \cite{Clauser:1969ny}, one of the Bell inequalities.
This is a statement about the combination of spin correlations defined by 
\begin{equation}
\langle R_{\text{CHSH}} \rangle =  \frac{1}{2} \langle \big[ 
\hat s^A_a (\hat s^B_b + \hat s^B_{b'}) + 
\hat s^A_{a'} (\hat s^B_{b} - \hat s^B_{b'})
\big]
\rangle
\,,\label{eq:R}
\end{equation}
where $\hat s^I_i$ ($I = A,B$, $i=a,b$) is the spin $i$-component of particle $I$.
Alice, the observer of particle $A$, has two possible spin measurement axes, $a$ and $a'$, while Bob, the observer of particle $B$, has measurement axes $b$ and  $b'$.
The spins are normalised (rescaled) so that the outcomes of Alice and Bob are either $+1$ or $-1$.
In LRHV theories, the spin components of $A$ and $B$ are predetermined prior to the act of measurement. 
In that case, one of $(\hat s^B_b + \hat s^B_{b'})$ and $(\hat s^B_b - \hat s^B_{b'})$ must vanish and the other takes value two.  
Therefore, $\bra R_{\rm CHSH} \ket$ cannot exceed 1 in LRHV theories. 
In quantum mechanics, on the other hand, we can write $R^2_{\rm CHSH} = 1 - \frac{1}{4} [ \hat s^A_a, \hat s^A_{a'} ][ \hat s^B_b, \hat s^B_{b'} ]$.
Since the commutator $[\hat s^A_a, \hat s^A_{a'}]$ cannot have eigenvalues larger than 2 (c.f.\ $[\sigma_i, \sigma_j] = 2 \epsilon_{ij} \sigma_k$), quantum mechanics cannot give $\bra R_{\rm CHSH} \ket$ larger than $\sqrt{2}$:
\begin{equation}
R_\text{CHSH} \le \left\lbrace
\begin{array}{ll}
1 &  (\mathrm{HLV} \text{~\cite{Clauser:1969ny}})\\
\sqrt{2} &  (\mathrm{QM} \text{~\cite{Tsirelson}})
\end{array}
\right.
\end{equation}
for all sets of measurement axes $a, a', b$ and $b'$.
Observation of $R_\text{CHSH} > 1$ ($\sqrt{2}$) would, therefore, falsify Hidden Local Variable theories (Quantum Mechanical theories).   
The tau pair from $\PH \to \PGtm\PGtp$ saturates the quantum bound: $R_\text{CHSH}^\text{max} = \sqrt{2}$, where the maximum is taken over all sets of measurement axes. 

Suppose that in the rest frame of the $\PGtm$ the tau spin is polarised into $\mathbf{s}$ direction ($|\mathbf{ s}| = 1$).  
The $\PGtm$ decays into a decay mode, $f$, producing a detectable particle $d$.  
The conditional probability that the particle $d$ takes the direction $\mathbf{u}$ ($|\mathbf{ u}| = 1$) when the $\PGtm$ spin is polarised in $\mathbf{ s}$ direction is given by \cite{Bullock:1992yt}
\begin{equation}
P(\mathbf{ u}|\mathbf{ s}) \, = \, 1 + \alpha_{f,d} \,   \mathbf{ s} \cdot \mathbf{ u}\,,
\label{eq:prob}
\end{equation}
with the normalisation $\int \frac{d \Omega}{4 \pi} P(\mathbf{ u}|\mathbf{ s}) = 1$,
where $\alpha_{f,d} \in [-1, 1]$ is called the spin analysing power.
For the CP counterpart, $(f,d) \xleftrightarrow{\text{CP}} (\bar f, \bar d)$, $\alpha_{\bar f, \bar d} = - \alpha_{f, d}$.

We denote the $\PGtp$ polarisation by $\mathbf{ \bar s}$ ($|\mathbf{ \bar s}| = 1$). 
The direction of its decay product, $d'$, measured at the rest frame of the $\PGtp$, is represented by a unit vector $\mathbf{ \bar u}$.
We want to relate the spin correlation $\bra \mathbf{ s} \otimes \mathbf{ \bar s} \ket$
with the angular correlation $\bra \mathbf{ u} \otimes \mathbf{ \bar u} \ket$ since the latter is measurable. 
Using the probability distribution \eqref{eq:prob}, one can derive \cite{Altakach:2022ywa}
\begin{equation}
C_{ab} \,=\, \bra s_a \bar s_b \ket \,=\, \frac{9}{\alpha_{f,d} \alpha_{f',d'}} \bra u_a \bar u
_b \ket  \,,
\label{eq:angle-to-spin}
\end{equation}
where $u_a \equiv \mathbf{ u} \cdot \mathbf{ a}$, $\bar s_b \equiv \mathbf{ \bar s} \cdot \mathbf{ b}$, etc.\ are the components with respect to arbitrary unit vectors $\mathbf{ a}$ and $\mathbf{ b}$.
Using this relation, one can compute both the concurrence ${\cal C}[\rho]$ (Eq.\ \eqref{eq:crho_tau}) and the CHSH variable $R_\text{CHSH}$ (Eq.\ \eqref{eq:R}) from the angular correlation observables $\bra u_a \bar u_b \ket$.  In $\PH \to \PGtm\PGtp$, the set of four measurement axes ($\mathbf{ a}_*$, $\mathbf{ a'}_*$, $\mathbf{ b}_*$, $\mathbf{ b'}_*$) that maximises $R_\text{CHSH}$ can be easily obtained \cite{Altakach:2022ywa}.
%
%
We study the quantum information measurement in $\PH \to \PGtm\PGtp$ at future lepton colliders, assuming two collider settings, ILC and FCC-ee, listed in Table \ref{tb:colliders} in \cref{sec:CPtau}.
In the analysis we focus on the $\PGtpm \to \PGppm \PGn$ channel, where the spin analysing power is maximum, $\alpha_{f,d} = 1$ for $f = \PGtm \to \PGpm \PGn$ and $d = \PGpm$.
At lepton colliders, the Higgs is produced via the associated production $\epem \to \PZ \PH$.
From the total initial momentum, $P_\text{in}^\mu = (\sqrt{s}, 0, 0, 0)$ and the reconstructed $\PZ$ momentum, $p_{\PZ}^\mu$,
the Higgs momentum four-vector, $p_{\PH}^\mu$, can be obtained as $p_{\PH}^\mu = P_\text{in}^\mu - p_{\PZ}^\mu$.
The distribution of the recoil mass, $m_\text{recoil} = \sqrt{(P_\text{in} - p_{\PZ})^2}$,
therefore sharply peaks at the Higgs mass in the signal.
By selecting events that fall within a narrow window, $|m_\text{recoil} - m_{\PH}| < 5$ GeV, one can achieve background/signal $\sim 0.05$ with a signal efficiency of $93\%$ and $96\%$ for the ILC and FCC-ee, respectively.  

We generate signal and background events with {\textsc{MadGraph5\_aMC@NLO}}
\cite{Alwall:2014hca} at leading order in the Standard Model, i.e.~$(\kappa, \delta) = (1,0)$ and we employ the {\textsc{TauDecay}} package for $\tau$ decays~\cite{Hagiwara:2012vz}.

On the right-hand-side of Eq.\ \eqref{eq:angle-to-spin},
the direction $u_a$ ($\bar u_b$) is measured in the rest frame of the respective mother tau lepton $\PGtm$ ($\PGtp$).
This means that in order to measure the spin correlation, it is crucial to reconstruct the $\PGtm$ and $\PGtp$ rest frames accurately.   
This requires the precise reconstruction of two neutrino momenta.
For six unknown neutrino momentum components, there are two mass-shell constraints: $m_{\PGt}^2 = (p_{\PGnGt} + p_{\PGpm})^2$ and $m_{\PGt}^2 = (p_{\PAGnGt} + p_{\PGpp})^2$,
and four conditions from the energy-momentum conservation:
$(P_\text{in} - p_{\PZ})^\mu = (p_{\PGnGt} + p_{\PGpm} + p_{\PAGnGt} + p_{\PGpp})^\mu$.
By solving those six constraints for the six unknowns, an event can be fully reconstructed up to twofold solutions: $i_s = 1,2$. 
The best solution can be selected by constructing and minimising the log-likelihood function incorporating the impact parameter information of the charged pion tracks from the tau decays.
The exact method has been developed in Ref.~\cite{Altakach:2022ywa} by adopting the formalism outlined in Refs.~\cite{Harland-Lang:2011mlc, Harland-Lang:2012zen, Harland-Lang:2013wxa}.

To take into account the energy mismeasurement, 
we smear the energies of all visible particles in the final state as
$E^\text{true} \,\to\, E^\text{obs} = (1 + \sigma_E \cdot \omega) \cdot E^\text{true}$
with the energy resolution $\sigma_E = 0.03$ \cite{FCC:2018evy, Behnke:2013lya}
for both ILC and FCC-ee,
where $\omega$ is a random number drawn from the normal distribution. 

The biggest experimental challenge in the $\PH \to \PGtm\PGtp$ channel is that ${\PGt}$s are heavily boosted because $m_{\PGt} \ll m_{\PH}$.
This means that going back to the rest frame of $\PGtm$ and $\PGtp$, one has to apply large boosts.  
Small reconstruction errors in the lab frame are amplified by the large boosts in the tau rest frame, which often spoil the measurement.  
In fact, it has been demonstrated that entanglement and CHSH violation cannot be observed in the simple kinematical method described above \cite{Altakach:2022ywa}. 

To overcome this problem and refine the accuracy of the reconstruction, we include the impact parameter information. 
Since tau leptons are marginally long-lived, $c \tau = 87.11$ $\mu$m \cite{Workman:2022ynf}, and highly boosted, one can observe a mismatch between the interaction point and the origin of the $\PGppm$ in $\PGtpm \to \PGn\PGppm$.
The impact parameter $\vec{b}_\pm$ is the minimal displacement of the extrapolated $\PGppm$ trajectory from the interaction point. 
The magnitude of the impact parameter $|\vec b_{\pm}|$ follows an exponentially falling distribution with the mean $|\vec b_{\pm}| \sim \SI{100}{\micron}$ for $E_{\PGtpm} \sim m_{\PH}/2$,
which is significantly larger than the experimental resolutions \cite{Behnke:2013lya}.
In our numerical simulation, 
we take constant values 
$\sigma_{b_T} = \SI{2}{\micron}$ (transverse)
and $\sigma_{b_z} = \SI{5}{\micron}$ (longitudinal) 
for the impact parameter resolutions,
although the actual resolutions are 
functions of the track momentum and
the polar angle $\theta^*$ from the beam direction.
The above modeling with the constant parameters gives a reasonable approximation for
the track momentum $\sim 100$ GeV
and $\theta^* \gtrsim 20^\circ$ 
as can be seen in Figure II-3.10 in Ref.\ 
\cite{Behnke:2013lya}.

If all quantities are accurately measured, the impact parameter, ${\vec{b}_\pm}$, from the $\PGtpm \to \PGn\PGppm$ decay,
is related to the directions of $\taup$ and $\pip$
and their angle $\Theta_\pm$ \cite{Hagiwara:2016zqz}.
We use this information to curb the effects of energy mismeasurement. The details of this procedure are reported in Ref.~\cite{Altakach:2022ywa}. 

\begin{table*}[t!]
\begin{center}
\begin{tabular}{|c || c |} 
 \hline
  & ~~~ILC~~~ \\ 
 \hline \hline
 $C_{ij}$ & 
 $\begin{pmatrix} 
 0.830 \pm 0.176 & 0.020 \pm 0.146 & -0.019 \pm 0.159 \\
 -0.034 \pm 0.160 & 0.981 \pm 0.1527 & -0.029 \pm 0.156 \\
 -0.001\pm 0.158 & -0.021 \pm 0.155 & -0.729\pm 0.140
 \end{pmatrix}$
  \\ 
 \hline
 ${\cal C}[\rho]$  & $0.778\pm 0.126$   \\
 \hline
 $R_\text{CHSH}^*$ & $1.103 \pm 0.163$  \\
 \hline
\end{tabular}
\vspace{3mm}
~\\
~\\
\begin{tabular}{|c || c |} 
 \hline
  & ~~~FCC-ee~~~ \\ 
 \hline \hline
 $C_{ij}$ & 
 $\begin{pmatrix} 
 0.925 \pm 0.109 & -0.011 \pm 0.110 & 0.038 \pm 0.095 \\
 -0.009 \pm 0.110  & 0.929 \pm 0.113  & 0.001 \pm 0.115  \\
 -0.026 \pm 0.122  & -0.019 \pm 0.110  & -0.879 \pm 0.098  
 \end{pmatrix}$
  \\ 
 \hline
 ${\cal C}[\rho]$  & $0.871 \pm 0.084$   \\
 \hline
 $R_\text{CHSH}^*$ & $1.276 \pm 0.094$  \\
 \hline
\end{tabular}
\end{center}
\caption{Result of quantum property measurements with a log-likelihood method incorporating the impact parameter information. The SM predictions read $C_{ij} = {\rm diag}(1,1,-1)$, ${\cal C}[\rho] = 1$
and $R_\text{CHSH}^* = \sqrt{2}$.
\label{tb:tauresult_2}}
\end{table*}

In Table \ref{tb:tauresult_2} we show the result of our quantum information measurements. 
We see that for both ILC and FCC-ee the components of the $C$-matrix are correctly measured.  
The concurrence ${\cal C}[\rho]$
is also measured with good accuracy. 
and the formation of entanglement is observed at
more than 5\,$\sigma$.
Observation of CSHS violation is the most challenging one since it is the strongest quantum correlation.
As can be seen in the last line in Table \ref{tb:tauresult_2},
the violation of the CHSH inequality is confirmed at the FCC-ee at $\sim 3\,\sigma$ level,
while $R^{*}_\text{CHSH} > 1$ is not observed at the ILC beyond the statistical uncertainty.
The superior performance of FCC-ee is attributed to the fact that the beam energy resolution of FCC-ee is much better than ILC.
The precise knowledge of the initial state momentum is crucial to reconstruct the rest frame of $\PGtpm$ accurately.

\subsubsection*{\CP studies in $\PH\to\PGt\PGt$}
\editor{Kazuki Sakurai}
\label{sec:CPtau}


If the $C$-matrix is measured with good accuracy, one can use it to constrain the \CP-phase $\delta$. 
From Eq.\ \eqref{eq:Cmat} we see that
only the $rn$ part (i.e.\ the upper-left $2 \times 2$ part) of the $C$-matrix is sensitive to $\delta$. By comparing the measured $C$-matrix entries in the $rn$ part and the prediction in Eq.\ \eqref{eq:Cmat}, we construct the $\chi^2$ function as \cite{Altakach:2022ywa}
\begin{equation}
\chi^2(\delta) = \frac{ \left( C_{rr} - \cos 2 \delta \right)^2 }{ \sigma^2_{rr} } +  
\frac{ \left( C_{rn} - \sin 2 \delta \right)^2 }{ \sigma^2_{rn} } 
\,+\, \frac{ \left( C_{nn} - \cos 2 \delta \right)^2 }{ \sigma^2_{nn} } 
+ \frac{ \left( C_{nr} + \sin 2 \delta \right)^2 }{ \sigma^2_{nr} } \,,
\label{eq:chi2_delta}
\end{equation}
where $C_{ij}$ and $\sigma_{ij}$ are the central value and the standard deviation, respectively, obtained from the result of the quantum information measurement with the $\PGt^\pm \to \PGppm \PGn$ channel summarised in \cref{tb:tauresult_2} in the previous section.
The goodness of fits is $\chi^2_\mathrm{ min}(\mathrm{ ILC})/\mathrm{ d.o.f.} = 0.93/3$ 
and
$\chi^2_\mathrm{ min}$(FCC-ee)/{d.o.f.} = 0.86/3,
where we used the benchmark collider parameters listed in 
Table \ref{tb:colliders}.

\begin{table}[t!]
\begin{center}
\begin{tabular}{r | c | c} 
  & ~~~ILC~~~ & ~~~FCC-ee~~~ \\ [0.5ex] 
 \hline 
 energy (GeV) & 250 & 240  \\ [0.5ex]
 luminosity (ab$^{-1}$) & 3 & 5  \\[0.5ex]
 beam resolution $e^+$ (\%) & 0.18 & $0.83 \times 10^{-4}$  \\[0.5ex]
beam resolution $e^-$ (\%) & 0.27  & $0.83 \times 10^{-4}$  
\\[0.5ex]
$\sigma(e^+ e^- \to HZ)$ (fb) & 240.1  & 240.3 
\\[0.5ex]
\# of signal ($\sigma \cdot \mathrm{ BR} \cdot L \cdot \epsilon$) & 385 & 663
\\[0.5ex]
\# of background ($\sigma \cdot \mathrm{ BR} \cdot L \cdot \epsilon$) & 20 & 36
\end{tabular}
\caption{Parameters for benchmark lepton colliders \cite{Baer:2013cma, FCC:2018evy}. Only the main background, $\Pep\Pem \to \PZ \PGtp \PGtm$, is considered, where $\PGtp \PGtm$ are produced from off-shell $\PZ/\gamma$. The numbers of signal and background reported here include the decay branching ratios and the efficiency of the event selection, $|m_{\mathrm{ recoil}} - m_{\PSH}| < 5$ GeV. \label{tb:colliders} }
\end{center}
\end{table}

\begin{figure}[t!]
    \centering
	\includegraphics[width=0.5\linewidth]{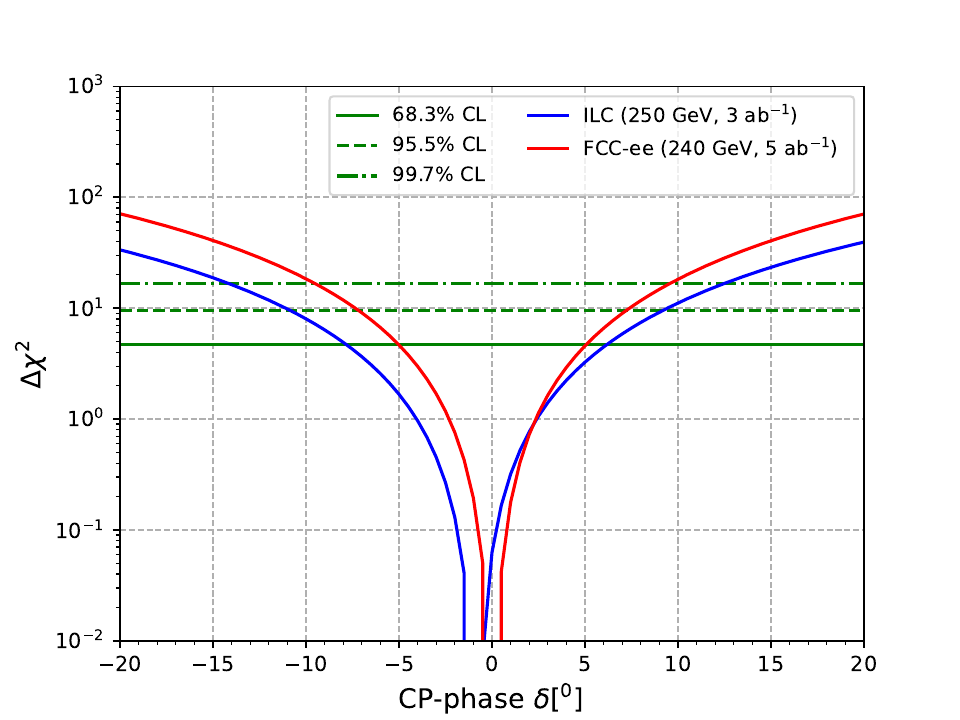}\,
	\caption{$\Delta \chi^2$ as a function of the \CP phase $\delta$ \cite{Altakach:2022ywa}.
	}
	\label{fig:chi2}
\end{figure}

The minimum of $\chi^2$ appears at the vicinity of three \CP-conserving points: $\delta = 0$, $\pm 180^{\circ}$ (\CP-even) and $\pm 90^{\circ}$ (\CP-odd).
Focusing on the minimum around $\delta = 0$,
the 1, 2 and 3\,$\sigma$ regions of $\delta$
obtained from this analysis are listed in Table \ref{tb:cp}.
\begin{table}[t!]
\begin{center}
\begin{tabular}{ c | c | c } 
~~~~CL~~~~ & ILC & FCC-ee \\ [0.5ex] 
\hline 
 68.3\,\% & $[-7.94^{\circ},6.20^{\circ}]$  & $[-5.17^{\circ}, 5.11^{\circ}]$ \\
 95.5\,\% & $[-10.89^{\circ},9.21^{\circ}]$  & $[-7.36^{\circ}, 7.31^{\circ}]$ \\
 99.7\,\% & $[-13.84^{\circ},12.10^{\circ}]$  & $[-9.21^{\circ}, 9.21^{\circ}]$   \\
\end{tabular}
\end{center}
\caption{Expected sensitivities on the \CP phase $\delta$ \cite{Altakach:2022ywa}.
\label{tb:cp}}
\end{table}
The analysis is based on $\Delta \chi^2(\delta) \equiv \chi^2(\delta) - \chi^2_\mathrm{ min}$, whose values around $\delta = 0$ are plotted in Fig.~\ref{fig:chi2}.
We note that the allowed windows are asymmetric. This is due to the statistical uncertainty of the 100 pseudo-experiments.

We see that the resolution of $\delta$ obtained from this analysis is roughly $\sim 7.5^{\circ}$ (ILC) and $\sim 5^{\circ}$ (FCC-ee) at 1\,$\sigma$ level.
These results should be compared with the resolutions 
obtained in the standard approach for the $\PH \to \PGtp \PGtm, \PGt^\pm \to \PGppm \PGn$ channel,
which exploits the angle $\varphi^\star$ 
\cite{Kramer:1993jn,Was:2002gv}
defined between the two planes, each spanned by the pair of momentum 3-vectors $(\vec{p}_{\PGpp}, \vec{p}_{\PGtp})$ and
$(\vec{p}_{\PGpm}, \vec{p}_{\PGtm})$ in the Higgs rest frame.
Using the same analysis described in  the previous section 
and the statistical method based on 100 pseudo-experiments,
we find the resolution of the \CP-phase 
with the $\varphi^\star$ method
is $6.4^{\circ}$ for FCC-ee.
This shows that the proposed method based on Eq.\ \eqref{eq:chi2_delta} is at least as good as the standard method with $\varphi^\star$.

Under the \CP conjugation, the $C$-matrix transforms as 
$C \xrightarrow{\CP} C^T$.
This fact can be used for a model-independent test for \CP violation.
To measure the asymmetry in the $C$-matrix, 
we define \cite{Altakach:2022ywa}
\begin{equation}
A = (C_{rn} - C_{nr})^2 + (C_{nk} - C_{kn})^2 + (C_{kr} - C_{rk})^2 \geq 0 \,.
\end{equation}
An experimental verification of $A \neq 0$ immediately confirms the violation of \CP. 

From the result summarised in \cref{tb:tauresult_2}, $A$ is obtained as 
\begin{equation}
A \,=\,
\left\{
\begin{array}{ll}
0.168 \pm 0.131 &~~~\mathrm{ (ILC)}
\\
0.081 \pm 0.060 &~~~(\mathrm{ FCC}\text{-}\mathrm{ ee})
\end{array}
\right.
\end{equation}
Here, the error corresponds to a 1\,$\sigma$ statistical uncertainty obtained from 100 pseudo-experiments. 
The result is consistent with 
the Standard Model (i.e.\ absence of \CP violation)
at $\sim 1\,\sigma$ level.

In the explicit model leading to \eqref{eq:Cmat},
we obtain $A = 4 \sin^2(2 \delta)$.
One can interpret the above model-independent result 
within this model and derive bounds on $\delta$.
In the domain around $\delta = 0$, the following limits are obtained at $1\,\sigma$ \cite{Altakach:2022ywa}
\begin{equation}
|\delta| <
\left\{
\begin{array}{l}
7.9^{\circ} ~~~\mathrm{ (ILC)}
\\
5.4^{\circ}~~~(\mathrm{ FCC}\text{-}\mathrm{ ee})
\end{array}
\right.\,,
\end{equation}
consistent with the limits obtained in the $\chi^2$ analysis (see Table \ref{tb:cp}).

\subsubsection{Higgs hadronic couplings}
\label{sec:higgsHadronicFCC}


Fully hadronic final states are produced in the vast 
majority of Higgs boson decays in the SM:
61\% of Higgs boson decays produce quark-antiquark pairs,
8\% yield gluon-gluon pairs, and about 14\% lead to hadrons
via intermediate production of \ww, \zz or \tptm.
Reconstructing and precisely measuring the branching fraction for
these decays is crucial for measuring the Yukawa couplings
of the Higgs boson to quarks, the effective coupling to
gluons, and potentially can also provide additional constraints
on the couplings to gauge weak bosons and to the \PGt lepton.

The sensitivity of the measurement of the signal strengths
of Higgs bosons decaying to hadronic final states
has previously been studied for ILC, CLIC, and CEPC \cite{Ono:2012oyw,ILCInternationalDevelopmentTeam:2022izu,Abramowicz_2017,Bai:2019qwd}.
In this section, a recent study performed in the context of FCC-ee is reported. 
A much deeper discussion of the strange-quark coupling is given in \cref{sec:HtoSS}.

The event selection targets \ZH events with \PZ bosons decaying either
to $\epem, \mpmm, \nunubar$, or \qqbar. The contribution from vector-boson fusion is small after the selection
and considered as part of the signal.

The \whizard generator~\cite{Kilian:2007gr,Moretti:2001zz} has been used to produce signal events, where a Higgs
boson is produced together with a \ee, \mumu, \nunubar, or \qqbar pair.
The same generator has also been used to produce background samples such as
\ee$\to$~\ee, ee$\to$~\mumu, and ee$\to$~\nuenuebar\PZ.
The \pythiaeight generator~\cite{Sjostrand:2014zea} has been used for the production of the other 
background event samples (inclusive \ww, \zz, ee$\to$~\qqbar production).
Samples are generated for centre-of-mass energies of 240 and 365~\GeV, assuming 
integrated luminosities of \SI{10.8}{\per\atto\barn} and \SI{3.0}{\per\atto\barn}, respectively.

The generated events are passed through a parametric simulation of the detector
response implemented in \delphes~\cite{deFavereau:2013fsa}.
The reconstruction efficiencies and resolutions for various particle hypotheses
are those expected for the IDEA detector with a crystal electromagnetic
calorimeter upstream of a dual-readout fibre calorimeter.
Lepton misidentification is assumed to be negligible.
The kinematic properties of the jets are assumed to be determined from those
of the constituents via a particle-flow algorithm.
Jet flavour tagging is performed by a dedicated algorithm~\cite{Bedeschi:2022rnj} based on a graph
neural network exploiting several low-level features of the input constituents.
The network provides seven output scores, corresponding to the probabilities
that a jet originates from the hadronisation of a b, c, s, u, d
quark or a gluon, or from the hadronic decay of a \PGt lepton.

Three analyses are performed in parallel, targeting three different final states:
a \textit{leptonic} analysis targeting \zll (\Pl=\Pe,\PGm) decays;
a \textit{invisible} analysis targeting \zvv decays; and
a \textit{hadronic} analysis targeting \zqq decays.
The leptonic analysis selects events with an opposite-sign, same-flavour lepton
pair with invariant mass close to that of the \PZ boson, vetoing events with additional
high-momentum leptons. The other two analyses apply a veto on the presence of
high-momentum leptons, yielding datasets orthogonal to the first one.
The invisible (hadronic) analysis requires (vetoes) events with a large ``missing'' mass,
calculated as the invariant mass of the system recoiling against all reconstructed
objects.

All the analyses cluster the reconstructed particles (excluding the
two \PZ boson decay products in the leptonic analysis) into jets.
The exclusive \kT \textit{Durham} algorithm~\cite{CATANI1991432,Cacciari:2011ma} is used. The particles are
clustered into two (four) jets in the leptonic and invisible (hadronic) analyses.
In the leptonic and invisible analysis, the two jets are assumed to originate from
the decays of the Higgs boson; in the hadronic analysis, the pairing of the four
jets and the association to either a \PZ or Higgs boson decay is performed
based on the output scores of the jet flavour tagger for each jet and on the
invariant masses of the jet pairs.
A few additional requirements suppress the main backgrounds, while maintaining
a large efficiency for the signal.
Loose preselections are applied to the quantities that are used for
signal-vs-background discrimination in maximum-likelihood fits to the
selected candidates: the mass of the system recoiling against the
two leptons in the leptonic analysis; the missing mass and the invariant 
mass of the dijet pair in the invisible analysis;
the invariant mass of the dijet pair from the Higgs boson decay and
the invariant mass of the system recoiling against it in the hadronic analysis.
The kinematic selection criteria are different at 240 and 365~\GeV\ to take into
account the different Lorentz boost of the Higgs and \PZ bosons in the two
scenarios.

At $\sqrt{s}=240$~\GeV, the typical signal selection efficiencies of the
three analyses for events with Higgs bosons decaying to quarks
or gluon pairs are: 60\% for the leptonic analysis, 80\% for the
invisible analysis, and 85\% for the hadronic analysis.
The expected signal yield for \Hbb events is about 50k/230k/690k
for the leptonic/invisible/hadronic analyses;
for other decays it scales approximately with the ratios of branching
fractions. For \Hss, the expected yield is 23/100/300 for the three analyses.
At $\sqrt{s}=365$~\GeV, the efficiencies are around 45\%/65\%/70\% for the
three channels, about 15--20\% relatively smaller than at $\sqrt{s}=240$~\GeV;
combined with the 70\% lower integrated luminosity and the 40\% lower \ZH
production cross-section, they lead to expected signal yields which are almost
an order of magnitude smaller than at 240~\GeV.

After the event selection, events are classified in orthogonal categories enriched
in dijet pairs of a given flavour or in events from a given Higgs boson decay.
The leptonic and invisible analyses use a multi-output neural network to classify events
in \bb, \cc, gg, \ssbar, \ww, \zz, \tptm-enriched categories. The network uses as inputs the seven
output scores of the flavour tagger for each of the two jets
and the minimum distance between the particles that are
clustered together by the jet clustering algorithm in the $4\to 3$ and $3\to 2$
clustering steps (2, 3 and 4 indicate the number of jets and particle candidates
in the event at a given step).
An alternative version of the invisible analysis and the hadronic analysis uses the
highest sum of the output scores of the taggers for a given flavour hypothesis
for the jet candidates from the Higgs boson decay to classify the events in
\bb, \cc, gg, \ssbar, \uu, \dd, \tptm-like categories.
All the analyses further split the event classes into two or three subcategories with
different signal-to-background ratio to enhance the sensitivity of the measurement.

The results for each analysis are determined by a binned, maximum-likelihood fit to the
distribution of one or two quantities that discriminate between the signals and
the background. These are the recoil mass in the leptonic analysis, the visible
mass and the recoil mass in the invisible one, and the invariant mass of the dijet
candidate from the Higgs boson decay and the invariant mass of the system
recoiling against it in the hadronic analysis.
Templates for the various signal and background processes are determined
from the simulation. The normalisation of the signal processes is floating,
expressed as the product of a signal strength $\mu_i$ ($i$=\bb, \cc, ...)
times the SM expected yield for the corresponding Higgs boson decay in
the targeted \PZ boson decay channel. The background normalisations are
constrained to the expectations within a 5\% uncertainty. Statistical
uncertainties in the templates from the limited amount of simulated events
are included.

The results for a given centre-of-mass energy are combined
by performing a simultaneous fit to the categories of the three different final states, correlating the signal
strengths and the normalisation parameters of the same background processes across the categories. For
the analysis of $\sqrt{s}=365$~\GeV\ simulations in the $\PGn\PGn\PH$ final state, different templates are used
for $\PZ\PH$, $\PZ\to\PGn\PGn$, and for the W-boson fusion (WBF).
The systematic uncertainties on the background normalisations are constrained from the sidebands,
and as such scale with the integrated luminosity. The results of the analysis at $\sqrt{s}=240$~\GeV\
are given in \cref{tab:signal_strengths_240}, and 
the preliminary results of the analysis at $\sqrt{s}=240$~\GeV\ are given in \cref{tab:signal_strengths_365}.
For the $\PWp\PWm$ fully hadronic final state we obtain with the 240~\GeV\ dataset a precision on the signal strength of 1.1\%
that requires to be combined at a later stage with dedicated analyses targeting the semi- and fully-leptonic final states.

In summary, with an integrated luminosity of 10.8\,\abinv at $\sqrt{s}=240$~\GeV, the signal strength of Higgs bosons decaying to
\bb, gg, and \cc could be measured with precisions close to 0.2\%, 0.8\%, and 1.6\%, respectively, and
a precision close to 1\% could also be obtained for the $\PH\to\PW\PW$ fully-hadronic decay;
while data collected at $\sqrt{s}=365$~\GeV\ could provide measurements with uncertainties around 2--3 times larger.
A deeper discussion of the $\PH\to\ssbar$ is given in the next section.

\begin{table}[!htbp]
  \centering
  \begin{tabular}{lrrrr}
    \toprule
    final state & \Hbb & \Hcc & \Hgg & \Hss \\
    \midrule
    \zll      & 0.68 & 4.02 & 2.18 & 234  \\
    \zqq      & 0.32 & 3.52 & 3.07 & 409  \\
    \zvv      & 0.33 & 2.27 & 0.94 & 137  \\
    \midrule                              
    comb      & 0.21 & 1.65 & 0.80 & 105  \\
    \bottomrule
  \end{tabular}
  \caption{Uncertainty (in \%) at 68\% confidence level on the signal strengths
    in the various Higgs boson decay channels at $\sqrt{s}=240$~\GeV,
    for the three analyses and their combination, corresponding to an
    integrated luminosity of $10.8\,\abinv$.}
  \label{tab:signal_strengths_240}
\end{table}  

\begin{table}[!htbp]
  \centering
  \begin{tabular}{lrrrr}
    \toprule
    final state & \Hbb & \Hcc & \Hgg & \Hss \\
    \midrule
    \zll      & 1.74 & 11.3 & 5.74 & 1170  \\
    \zqq      & 0.51 & 3.87 & 3.05 & 564  \\
    \zvv      & 0.69 & 4.24 & 2.82 & 414  \\
    $\PGn\PGn\PH$ (WBF)  & 0.68 & 3.99 & 2.64 & 295  \\
    \midrule                              
    comb ($\PZ\PH$)     & 0.41 & 3.13 & 2.21 & 356  \\
    \bottomrule
  \end{tabular}
  \caption{Uncertainty (in \%) at 68\% confidence level on the signal strengths
    in the various Higgs boson decay channels at $\sqrt{s}=365$~\GeV,
    for the three analyses and their combination.}
  \label{tab:signal_strengths_365}
\end{table}


\subsection{\focustopic \texorpdfstring{$\PH\to\PQs\PAQs$}{H->ss}}
\label{sec:HtoSS}
\editors{The whole HtoSS expert team}
%
\subsubsection{Introduction}
\label{sec:HtoSS:Intro}
One of the crucial aspects of the physics program at future Higgs factories is the precise measurement of the Higgs boson couplings to fermions. This has been studied thoroughly within the last European Strategy Update~\cite{deBlas:2019rxi, Cepeda:2019klc, Cerri:2018ypt} and more recently in the context of the 2021 Snowmass community exercise~\cite{Dawson:2022zbb, Narain:2022qud}.
The study of the couplings of the Higgs boson to the light quarks (\ie, up, down, and strange quarks) was considered nearly impossible due to the small branching ratio when assuming SM couplings as well as the difficulty in identifying the flavour of quark-initiated jets (flavour tagging). Nevertheless, the exploration of the Higgs coupling to the strange quark $y_s$ has emerged with increasing interest, given also its tight connections with detector technologies and layout optimisation. Enabling sensitivity to inclusive $\PH\to\PQs\PAQs$ production would allow for a complete exploration of the second-generation Yukawa couplings, and go beyond the current LHC's limited reach for $y_s$ via $\PH\to\upvarphi\PGg$ decays~\cite{Kagan:2014ila, ATLAS:2017gko}.

At the LHC, in addition to the small branching fraction, the rare $\PH\to\PQs\PAQs$ decay mode is inaccessible with the current detector capabilities. In fact, one of the most powerful handles to identify a strange-quark-initiated jet (strange-tagging) is the possibility to distinguish between kaons and pions up to tens of GeV in momentum. This relies on dedicated detector subsystems which are not included in the LHC multi-purpose detectors, as will be detailed in \cref{sec:HtoSS:Detector_performance}. Furthermore, the overwhelming multi-jet production rate at the LHC inhibits the study of strange, up, and down quark couplings with inclusive $\PH\to\PQq\PAQq$ decays, in addition to the dominant $\PH\to\bb$ decay mode. Therefore, future Higgs factories present a unique opportunity to probe new physics frontiers with the strange quark, but this requires the design of detectors and algorithms that enable such studies. Along with the SM scenario, this signature provides the possibility of testing several BSM theories that allow for extended Higgs sectors, as discussed in \cref{sec:HtoSS:Theory}.

In the past years, preliminary proof-of-concept investigations at future colliders have been performed. Some of them~\cite{Bedeschi:2022rnj} focus primarily on strange-tagging algorithms and some others~\cite{Duarte-Campderros:2018ouv, Albert:2022mpk} include also their application to $\PH\to\PQs\PAQs$ searches, interpretations in BSM frameworks, and potential detector designs. The assumptions and the detector concepts used in these studies differ, as summarised in \cref{sec:HtoSS:Recap}. All the results show promising avenues and motivate more in-depth explorations and future harmonisation. However, modelling the fragmentation of strange quarks is a complex matter. While this was not specifically addressed in the above studies, a discussion is now presented in \cref{sec:HtoSS:Fragmentation} of this document, highlighting the synergies and complementarity between the Higgs and SM measurement programs at future Higgs factories.

Analyses developed since Snowmass~2021 are summarized in \cref{sec:HtoSS:Recap}.
Target detector performance aspects relating to strangeness tagging, and current tagging performance are described in \cref {sec:HtoSS:Detector_performance}.
Algorithmic developments for strangeness-tagging specifically in the context of $\PH\to\PQs\PAQs$ are reviewed in  \cref{sec:HtoSS:Tagger}.
Finally, \cref{sec:HtoSS:Future} presents exciting opportunities for new studies to be carried out in the future.

\subsubsection{Theoretical motivation and phenomenological landscape}
\label{sec:HtoSS:Theory}

The current measurements of the Higgs coupling to $\PQb$ and $\PQt$ quarks confirm the SM predictions of $y_{\PQb,\PQt}\propto m_{\PQb,\PQt}$ ~\cite{ATLAS:2018kot,CMS:2018nsn,CMS:2018uxb,ATLAS:2018mme,ATLAS:2024yzu,CMS:2023vzh,CMS:2020zge}.
Therefore, one can conclude that the mass of the heavy quarks is dominated by the SM Higgs mechanism and then constrain deviations from the SM prediction, for example, in the form of dimension-six operator $\bar{q}q H H^\dagger H$.
The $c$-quark Yukawa is probed via inclusive~\cite{Delaunay:2013pja,Perez:2015aoa,CMS:2022psv,ATLAS:2022ers,ATLAS:2024yzu} and exclusive rates to Higgs and a vector meson~\cite{Bodwin:2013gca,ATLAS:2018xfc,CMS:2018gcm,CMS:2022fsq,ATLAS:2022rej} and kinematical effects~\cite{Bishara:2016jga,ATLAS:2022qef,CMS:2023gjz}, where the tightest bound is $\sim10$ times its SM value~\cite{CMS:2022psv,ATLAS:2024yzu}.
Moreover, the non-universality of the Higgs Yukawa couplings was established both in the lepton and quark sectors~\cite{Perez:2015aoa}.
However, the situation is different for light quarks, $\PQu$, $\PQd$, and $\PQs$, which are very challenging to probe due to the tiny expected rates and the large QCD backgrounds.
Currently, these can be searched for via exclusive decays~\cite{Kagan:2014ila,Konig:2015qat,CMS:2024tgj,ATLAS:2023alf}, or indirectly via global fits~\cite{Kagan:2014ila} or kinematical effects~\cite{Soreq:2016rae,Yu:2016rvv,Falkowski:2020znk}, resulting in rather weak bounds.
However, correlated observables in models can extend this reach further~\cite{Egana-Ugrinovic:2021uew}.

Probing the Higgs-strange Yukawa coupling is motivated for a number of reasons.
First, it will shed light on how the strange quark receives its mass: is it from the SM Higgs, or from another mechanism related to the first two generations (see \eg\ Refs.~\cite{Ghosh:2015gpa,Altmannshofer:2015esa})?
Second, it can be utilised to probe BSM models and in particular flavour models (see~Ref.~\cite{Giudice:2008uua} and Refs.~\cite{Cepeda:2019klc,Cerri:2018ypt} and references within as well as more recent work in Refs.~\cite{Egana-Ugrinovic:2019dqu, Egana-Ugrinovic:2021uew,Nir:2024oor, Giannakopoulou:2024unn, Erdelyi:2024sls}).
Finally, there would be particularly strong motivation to probe the Higgs--strange coupling directly, if there were to be a sign of BSM physics in the Higgs couplings that could be attributed to the light quark Yukawa couplings or the gluon.

A summary of models that can modify the Higgs couplings to fermions including the $\PQs$ Yukawa can be found in Refs.~\cite{Cepeda:2019klc,Cerri:2018ypt}.
Assuming that the new physics is associated with a heavy scale, $\Lambda$, the modification for the Yukawa coupling can be captured by the following irrelevant operator
\begin{align}
    \label{eqn:hssEFT}
    \frac{1}{\Lambda^2}\bar{s}_L s_R H H^\dagger H \, .
\end{align}
The resulting deviation from the SM can be written as
\begin{align}
    \kappa_{\PQs}
    \equiv
    \frac{y_{\PQs}}{y_{\PQs}^{ \mathrm{SM}}}
    = 1 + \frac{1}{y_{\PQs}^{\mathrm{SM}}}\frac{v^2 }{\Lambda^2 } \, .
\end{align}
It is translated to the following scale
\begin{align}
    \Lambda
    \approx
    \frac{14\,{\mathrm{TeV}}}{\sqrt{|\kappa_{\PQs}-1|}}
    \approx
    \frac{2\,\mathrm{TeV}}{\sqrt{\kappa_{\PQs}/40}}
    \, .
\end{align}
where the last step is inspired by the bound on $\kappa_{\PQs} < 40$ from the global fit of the Higgs data~\cite{Kagan:2014ila,Perez:2015aoa}.

\begin{figure}[htbp]
    \centering
    \includegraphics[width=0.7\textwidth]{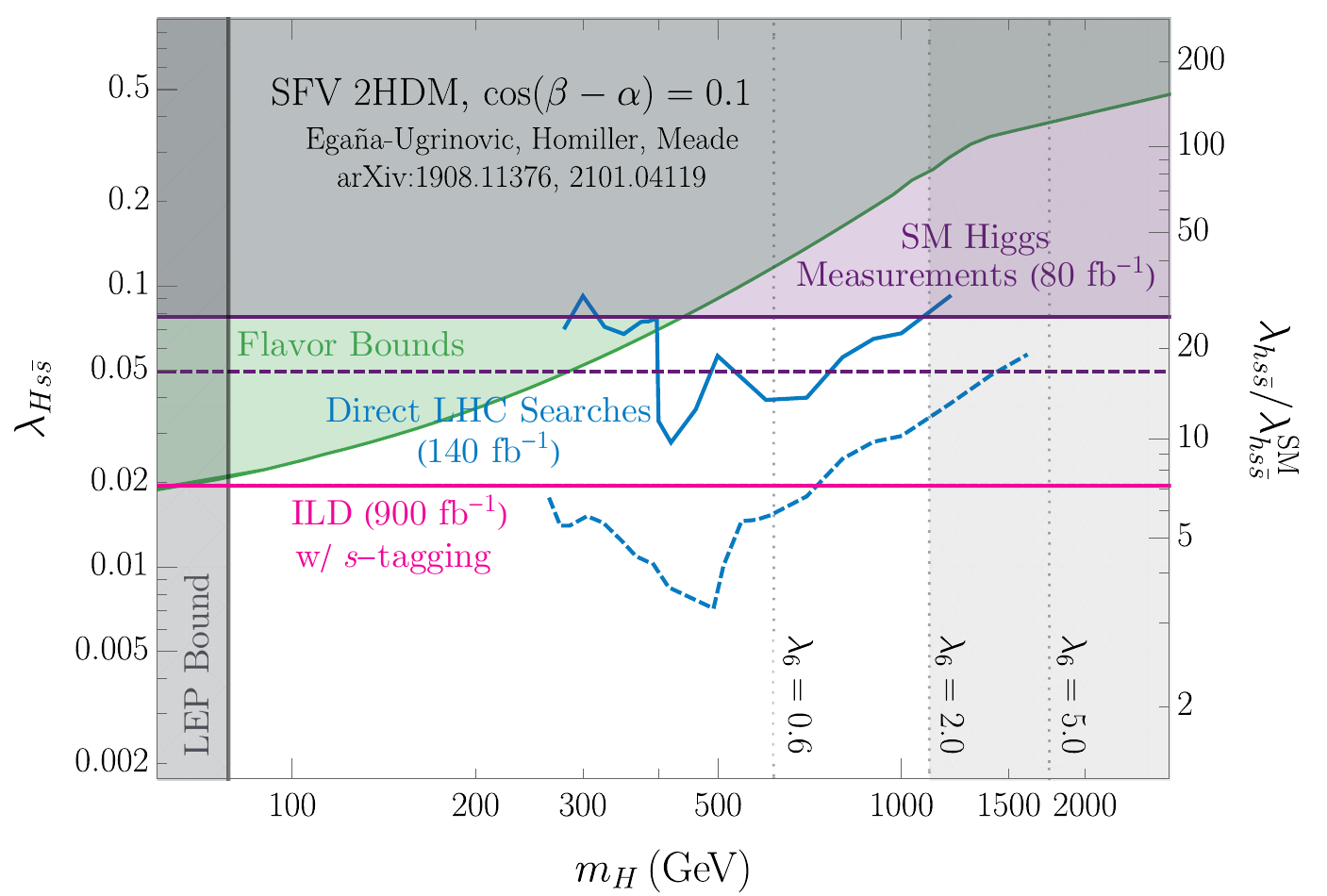}
    \caption{95\% CL bounds on the CP-even Higgs-strange Yukawa coupling, $\lambda_{\PH\PQs\PAQs}$, as well as on \qty{125}{GeV} SM Higgs-strange Yukawa coupling, $\kappa_{\PQs} \equiv \lambda_{\PH\PQs\PAQs}/\lambda_{\PH\PQs\PAQs}^\textrm{SM}$, for the 2HDM model described in Refs.~\cite{Egana-Ugrinovic:2019dqu, Egana-Ugrinovic:2021uew}. See Ref.~\cite{Albert:2022mpk} for more information.}
    \label{fig:limits_ild}
\end{figure}

Nevertheless, it is important to note that while an EFT description can be used to parametrise the effects at low energy, unless there is a power counting that explains why this is the leading effect while also suppressing flavour-changing neutral currents~(FCNC), it is not particularly insightful.
In particular, to generate a large enough effect in Eq. (\ref{eqn:hssEFT}) that it is relevant for measurements requires a large tree-level coupling between new physics states that couple to both the strange quark and the Higgs.
This can be accomplished in one of two ways: either through a new scalar two-Higgs-doublet model (2HDM) state or vector-like quarks~(VLQ).
However, in either case that means there are also correlated observables that are not understood from the bottom up EFT perspective.
Recently there has been work that tries to analyse these cases for a 2HDM where a symmetry can be implemented to be consistent with flavour data~\cite{Egana-Ugrinovic:2019dqu,Egana-Ugrinovic:2021uew} and for VLQ where FCNC suppression is put in by hand~\cite{Erdelyi:2024sls}.
As can be seen in Figure~\ref{fig:limits_ild}, in the case of a 2HDM strange flavour tagging at a future Higgs factory as discussed here can extend beyond the LHC reach when comparing with correlated observables.

\subsubsection{Interpretation as Higgs-strange Yukawa coupling}
\label{sec:HtoSS:Yukawa}
Since the Higgs couplings to the third generation fermions have been
measured at the 10\,\% level at the LHC \cite{ATLAS:2024fkg,CMS:2024gzs} and will be pinned down to the per-cent or sub-per-cent level at future \epem colliders \cite{Narain:2022qud}, the
next step will be to obtain information about the Higgs couplings to the second generation.
A starting point is provided by the sensitivity to the muon and charm Yukawa couplings at the LHC, but more accurate measurements are expected at \epem colliders.
In this context a relevant question arising is related to the Yukawa coupling to strange quarks.
First studies for FCC-ee are quite promising.
However, the sophisticated treatment of signal and background processes is of high relevance.
This section contributes to the present state of the art on the theoretical side.

\begin{figure}[t]
    \centering
    \vspace*{-2.5cm}
    \includegraphics[width = 1.20\textwidth]{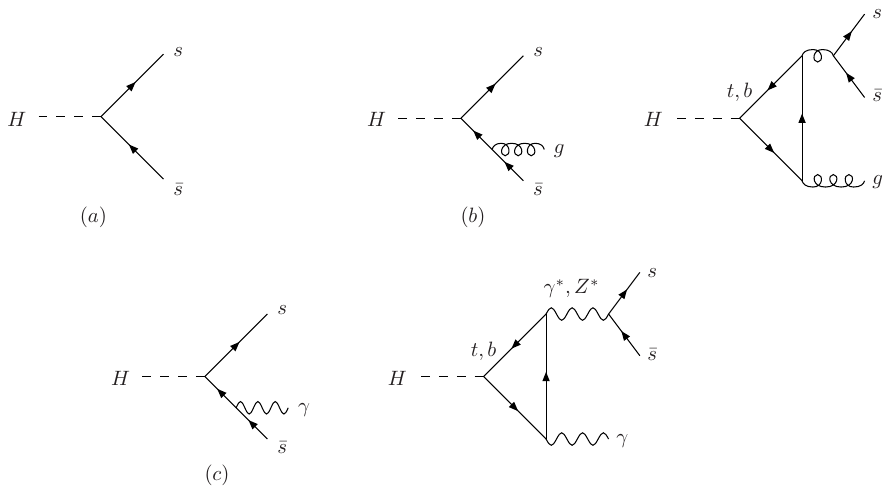}
    \vspace*{-19.0cm}
    \caption{Typical diagrams contributing to (a) $\PH\to\PQs\PAQs$, (b) $\PH\to\PQs\PAQs\Pg$ and (c) $\PH\to\PQs\PAQs\PGg$.}
    \label{fg:h2ss_dia}
\end{figure}

Higgs decays into strange quarks are induced by the strange Yukawa
coupling, see \cref{fg:h2ss_dia}(a). QCD and electroweak corrections to the LO process are of moderate size (for a review see
\eg\ Ref.~\cite{Spira:2016ztx}), if the strange Yukawa coupling is
expressed by the $\overline{MS}$ strange mass $\overline{m}_{\PQs}(M_{\PH})$ at the scale of the Higgs mass $M_{\PH}$. The
inclusive strange Yukawa-induced branching ratio of $\PH\to\PQs\PAQs$ amounts to $0.022\%$, thus being very small and below the per-mille level. This immediately poses the question of competing Higgs decays with strange quarks in the final state that are not induced by the strange Yukawa coupling. The leading processes of this kind are the loop-induced strong and weak Dalitz decays of the Higgs boson, $\PH\to\PQs\PAQs + \Pg/\PGg$, see 
\cref{fg:h2ss_dia}(b,c). At the inclusive level they contribute at the per-cent level to the total branching ratios, \ie\ are
more than an order of magnitude larger than the strange-Yukawa-induced branching ratio of $\PH\to\PQs\PAQs$. However, in order to clear up the situation of the measurement of the strange-Yukawa coupling, the corresponding distributions need to be investigated in more detail, since appropriate cuts on the invariant mass of the $\PQs\PAQs$ pair in the final state will affect the loop-induced and Yukawa-induced contributions in different ways.

\begin{figure}[h!]  
    \vspace*{-4.0cm}
    \begin{subfigure} {.5\linewidth}
        \centering
        \includegraphics[width=\linewidth]{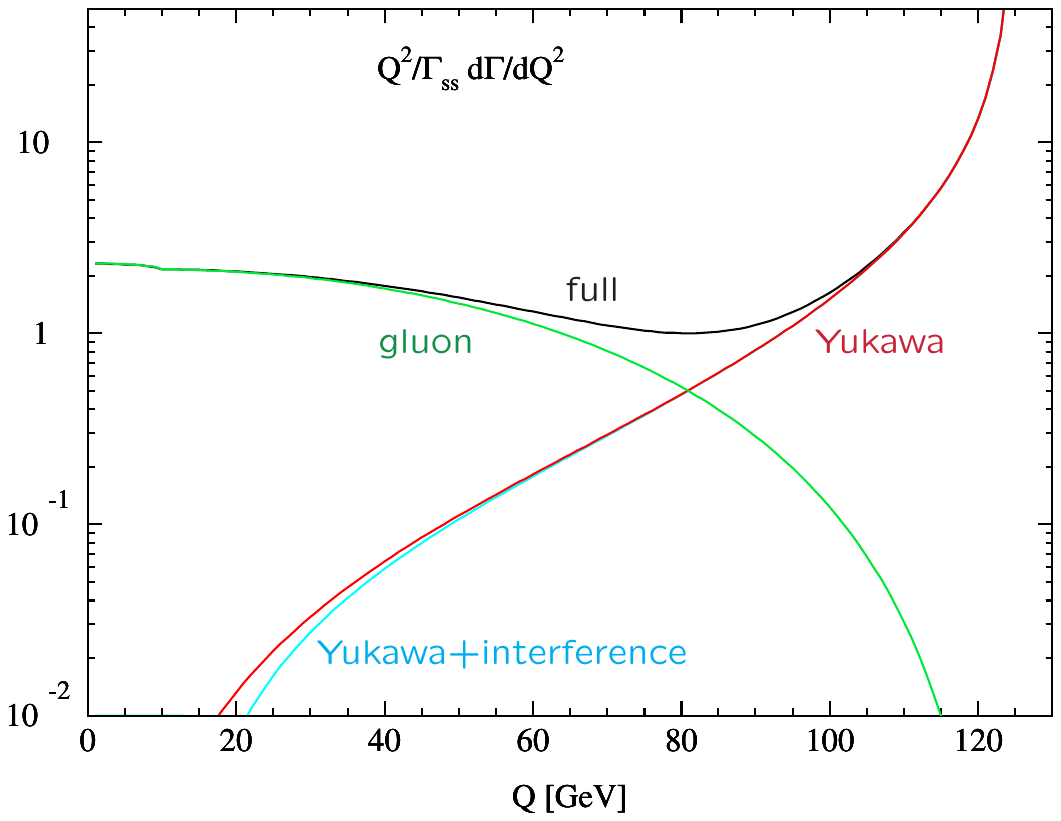}
        \vspace*{-3.3cm}
        \caption{Full distribution as well as individual contributions of the Yukawa- and gluon-induced part and their interference term. No resummation of soft gluon effects has been performed at the upper end of the spectrum.}
        \label{fg:strongdalitz} 
    \end{subfigure}
    \quad
    \begin{subfigure}{.5\linewidth}
        \centering
        \includegraphics[width=\linewidth]{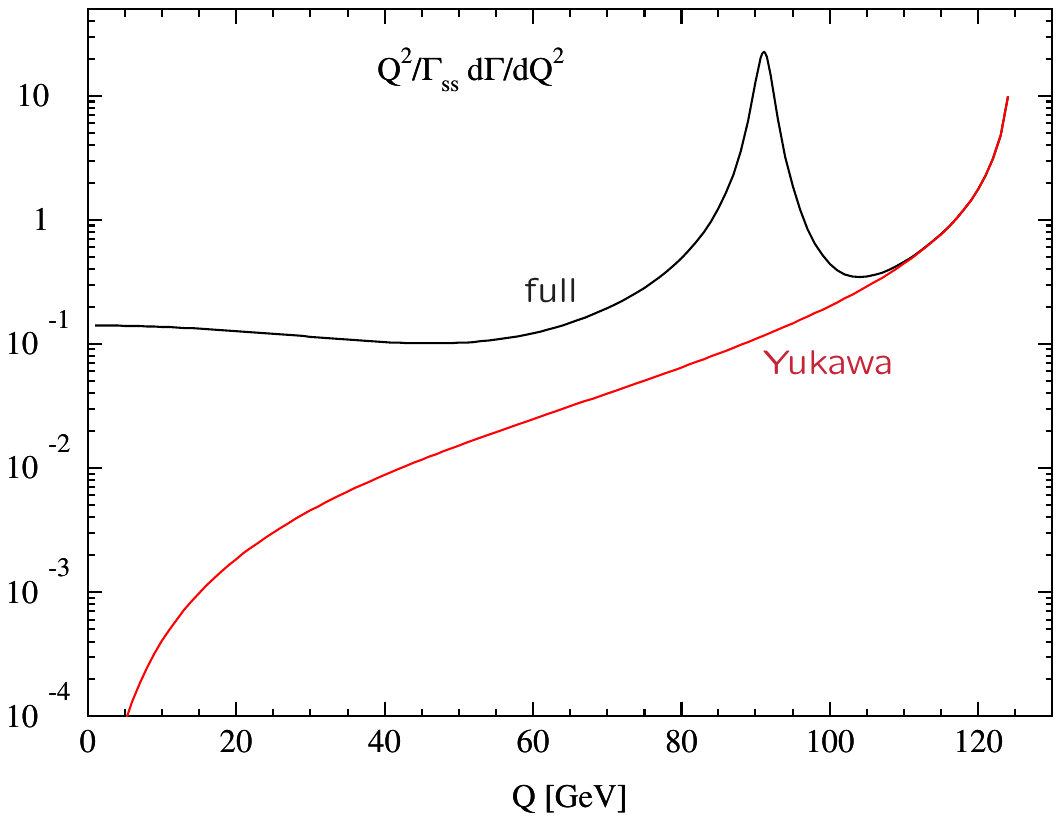} 
        \vspace*{-3.0cm}
        \caption{Full distribution as well as individual contribution of the Yukawa-induced part. No resummation of soft photon effects has been performed at the upper end of the spectrum.}
        \label{fg:weakdalitz}
    \end{subfigure}
    \caption{(a) The differential partial width of the strong Dalitz decay $\PH\to\PQs\PAQs\Pg$, and (b) the weak Dalitz decay $\PH\to\PQs\PAQs\PGg$ as a function of the invariant $\PQs\PAQs$ mass $Q$ normalised to the inclusive partial decay width $\Gamma(\PH\to \PQs\PAQs)$. The kinematical strange-mass has been identified with the $\overline{MS}$ mass.}
\end{figure}


We have calculated the full strong and weak Dalitz decays $\PH\to\PQs\PAQs + \Pg/\PGg$ taking into account the interference with the Yukawa-induced part of the calculation. The final result for the strong Dalitz decay $\PH\to\PQs\PAQs\Pg$ is shown in \cref{fg:strongdalitz} as a function of the invariant $\PQs\PAQs$ mass $Q=M_{\PQs\PAQs}$ with the individual contributions of the loop-induced (gluon) contribution, the Yukawa-induced one and the interference added to the latter. Provided there will be a sufficient mass resolution of the strange jet pair, the region of larger $Q$ values could be singled out so that the strange Yukawa coupling could be extracted. The distribution of \cref{fg:strongdalitz} is normalised to the total inclusive partial width $\Gamma(\PH\to\PQs\PAQs)$. In addition, the omission of the bottom loops or taking the top loops only in the heavy-top limit (HTL) does not modify the picture significantly.

The analogous calculation of the weak Dalitz decay $\PH\to \PQs\PAQs + \PGg$ \cite{Abbasabadi:1996ze,Abbasabadi:2004wq,Abbasabadi:2006dd,Dicus:2013ycd,Chen:2012ju,Passarino:2013nka,Sun:2013rqa,Kachanovich:2020xyg,Kachanovich:2021pvx} is presented in \cref{fg:weakdalitz} where the full contribution including all ingredients and the Yukawa-induced one is shown. Thanks to the $\PZ$-boson resonance, the loop-induced weak Dalitz decay contribution is of similar magnitude as the strong Dalitz decay. However, the upper end of the distribution in $Q$ is again dominated by the Yukawa-induced contribution so that the resolution of the invariant mass of the strange jet pair is of crucial relevance for the sensitivity to the strange Yukawa coupling.


Open questions concerning this sensitivity are related to the detailed definition of the strange mass at the jet level and its relation to the strange mass involved in the Yukawa coupling, and fragmentation effects of the more prominent Higgs decays into bottom and charm quarks into strange jets. 
The fragmentation is further discussed in Section~\ref{sec:HtoSS:Fragmentation}.

\subsubsection{\texorpdfstring{Fragmentation modelling for $\PH\to\PQs\PAQs$: state of the art and challenges}{Fragmentation modelling for H->SS: state of the art and challenges}}
\label{sec:HtoSS:Fragmentation}

\subsubsection*{Introduction}

Assuming that strangeness taggers will be trained on a combination of data-driven and MC simulation samples~\cite{Albert:2022mpk}, their performance -- and uncertainties -- will depend at least in part on the fidelity of the MC modelling. This, in turn, will depend both on the level of theoretical accuracy that MC simulation will be able to reach at that time, as well as on the quality of the constraints that can be placed on their free parameters.

From a MC point of view, the fragmentation of partons into hadrons is modelled in four main steps: 1)~fixed-order matrix elements, 2)~parton showers, 3)~hadronisation, and 4)~hadron (and $\tau$) decays. 
The state of the art and active fields of research in parton showers, matching, and hadronisation is discussed in \cref{sec:shower_hadro}.  Fragmentation and hadronisation is also discussed in \cref{sec:frag-had}.
%
For parton showers, it does seem realistic at this point to assume that MC treatments capable of reaching perturbative accuracies of at least NNLO+NNLL will be available for Higgs decays and their backgrounds within a time scale of a few years \cite{Butler:2023glv,Begel:2022kwp,Boughezal:2022cbl,Campbell:2022qmc}. This should be sufficient to guarantee a perturbative fidelity at or better than the one-percent level.

New measurements from the LHC have restarted some old debates -- \eg\ on colour reconnections (CR) -- and raised new ones -- \eg\ on ``collective effects'' in small systems \cite{ALICE:2016fzo,CMS:2010ifv,Li:2012hc,Schlichting:2016sqo,ALICE:2022wpn,Nagle:2018nvi,Citron:2018lsq,Bierlich:2024odg,Altmann:2024icx}. The extent to which lessons learned from these studies will impinge on precision modelling of hadronic final states in \epem annihilation is not yet clear.

Here, we summarise some of the aspects of (string) fragmentation modelling that may be relevant to strangeness tagging in $\PH\to \PQs\PAQs$ and remark on open questions. We also emphasise that the ultimate test for hadronisation models can be delivered by a future \epem collider with excellent particle identification capabilities, complemented by hadron-collider data.

\subsubsection*{String Fragmentation}

\begin{figure}[t]
\centering
    \includegraphics[width = 0.58\textwidth]{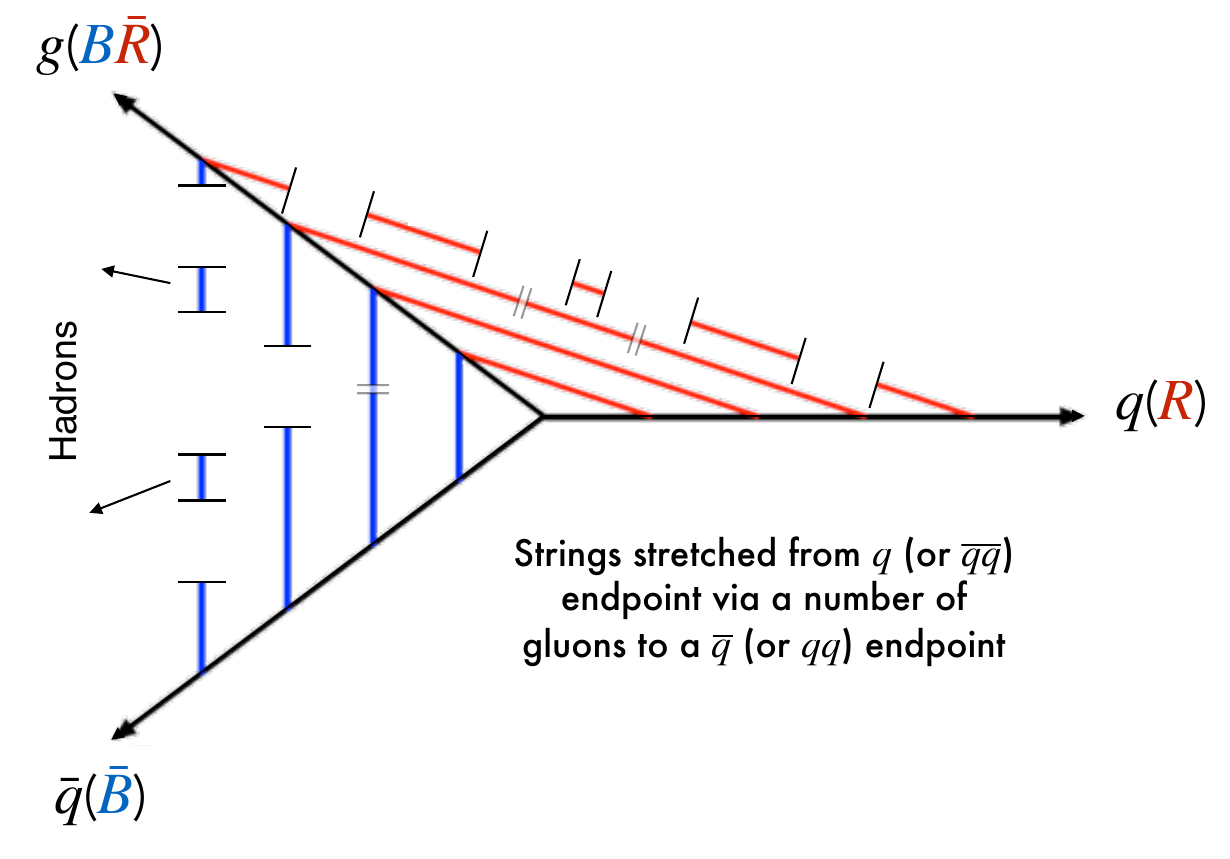}
    \caption{Illustration of a string stretched between a red quark (right endpoint), via a blue-antired gluon (kink), to an antiblue antiquark (bottom left endpoint). The sketch shows the string position at different times, as the three partons move away from their common origin in the centre of the figure. Overlaid on the string worldsheet towards late times are $\PQq\PAQq$ fluctuations that become string breaks, eventually resulting in hadrons.
	\label{fig:stringFrag}}
\end{figure}

Hadronisation in \pythia~\cite{Bierlich:2022pfr} is modelled using Lund string fragmentation~\cite{Andersson:1983ia,Andersson:1997xwk}, in which the confining field between coloured partons is represented as a 1+1 dimensional string characterised by a constant tension $\kappa\sim 1$~GeV/fm, with dynamical evolution governed by the string world sheet.
As such, colour (anti)triplets (\ie, quarks and antiquarks) correspond to string endpoints, and octets (\ie, gluons) form transverse kinks on strings.
An example of a $qg\bar{q}$ string configuration fragmenting into hadrons is shown in \cref{fig:stringFrag}.

In high-energy collision processes, partons move apart at high energies, which in turn stretches the confining potentials and leads to string breaking, \ie, string fragmentation. In the baseline Lund model, string breaks are assumed to proceed by a tunnelling mechanism that is modelled on the QED Schwinger mechanism~\cite{Schwinger:1951nm}. This predicts that quark-antiquark pairs should be created with Gaussian $p_\perp$ distributions and Gaussian suppression of higher-mass quarks,
\begin{equation}
     {\cal P}(m_q^2, p^2_{\perp q}) ~\propto~ \exp\left( \frac{-\pi m^2_q}{\kappa} \right)\exp\left( \frac{-\pi p^2_{\perp q}}{\kappa} \right) ~\equiv~
    \exp\left( \frac{-\pi m^2_{\perp q}}{\kappa} \right), \label{eqn:schwinger}
\end{equation}
where $m_{\perp q}^2=m_q^2 + p_{\perp q}^2$ is called the transverse mass (squared) of the produced quark and $p_{\perp q}$ is measured relative to the local string axis.

\begin{figure}[t]
    \centering
    \includegraphics[width = 0.7\textwidth]{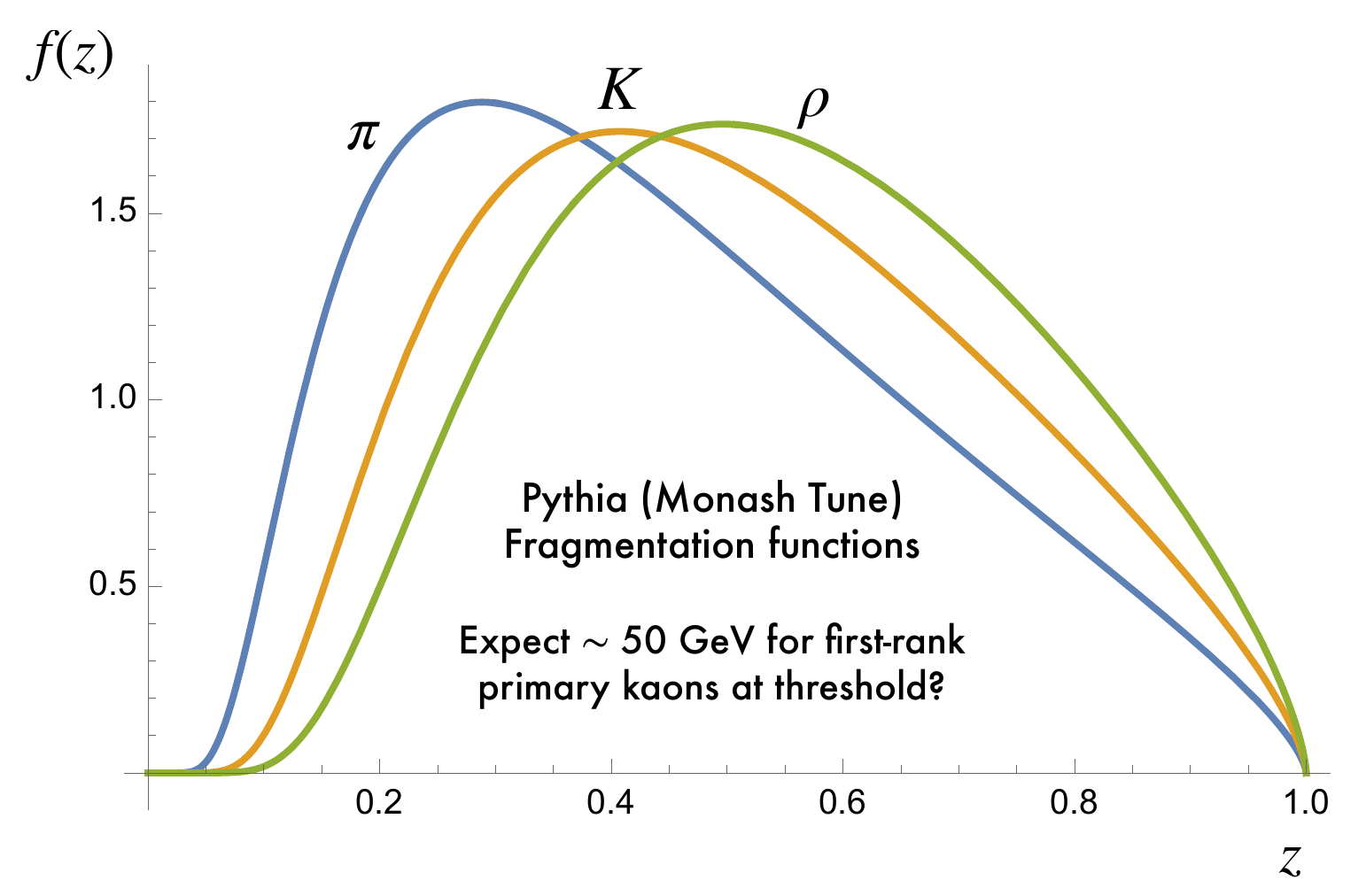}
	\caption{The Lund symmetric fragmentation function $f(z)$ for default (Monash) values of $a$, $b$, and $\left<p_\perp\right>$, for pions, kaons, and rho mesons, illustrating that heavier particles tend to have harder spectra.
	\label{fig:ffs}}
\end{figure}

Another relevant aspect of the Lund model is that string breaks are causally disconnected. This means that the time ordering of the breaks has no physical significance. Hence in order to simplify hadronic mass constraints, the fragmentation procedure is carried out iteratively, picking a random string endpoint and producing one on-shell hadron at a time. The fraction of the endpoint momentum taken by the produced hadron, $z$, in each break is then probabilistically chosen by the ``Lund symmetric fragmentation function'' which takes the form
\begin{equation}
	f(z) = N \frac{1}{z}(1-z)^a \exp\left(\frac{-b m_{\perp h}^2}{z}\right),
	\label{eqn:lundFragFn}
\end{equation}
where $m_{\perp h}$ is the transverse mass of the produced hadron, $N$ is a normalisation constant, and $a$ and $b$ are free model parameters with default (Monash 2013~\cite{Skands:2014pea}) values $a=0.68$ and $b=0.98\,\mathrm{GeV}^{-2}$. Comparisons of $f(z)$ for pions, kaons, and rho mesons are shown in \cref{fig:ffs}, for an average $\left<p_{\perp h}\right>$ value $\sim 0.335\,\mathrm{GeV}$.

Recent discussions of constraints on the $a$, $b$, and $p_\perp$-broadening parameters can be found in Refs.~\cite{Skands:2010ak,Skands:2014pea,Christiansen:2015yqa,Amoroso:2018qga,Jueid:2023vrb}. We note in particular that a convenient reparametrisation of $f(z)$ was introduced in Ref.~\cite{Amoroso:2018qga} which allows to keep the average value $\left<f\right>$ (which is very well constrained) fixed while varying its tails, cf. \\~\texttt{StringZ:deriveBLund} in the \pythia documentation.

Given the Lund string model as described above, we now explore some of the implications for strangeness tagging in $\PH\to \PQs\PAQs$ events and their backgrounds. Let it be emphasised though that this is a \emph{model}, not written in stone, and hence experimental cross checks and constraints are crucial to validate the aspects of the modelling that taggers might be sensitive to.

\subsubsection*{Implications for Strangeness Tagging}

In $\PH\to \PQs\PAQs$ events the produced strange quarks necessarily become string endpoints.
As a consequence of the symmetric treatment of the string-fragmentation procedure, the resultant strange hadrons formed will be labelled as so-called rank-1 primary hadrons, with ``rank'' referring to the order hadrons are produced from the string endpoint, and ``primary'' referring to the fact that the hadron is not produced as a ``secondary'' from the decay of another hadron. Below, we shall also use the word ``leading'' by which we mean a simple ordering in energy of hadrons within a jet.

\begin{figure}[t]
\centering
    \vspace*{1mm}\includegraphics[width = 0.87\textwidth]{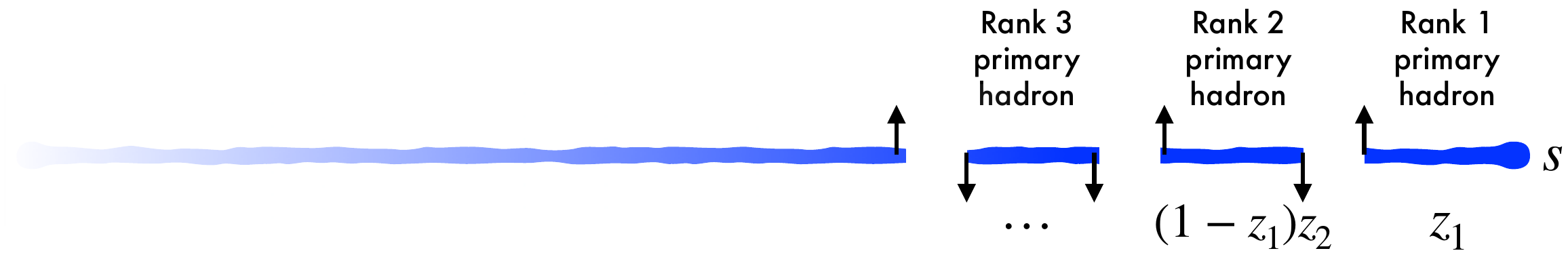}\vspace*{-2mm}
	\caption{Illustration of longitudinal momentum fractions, $z$, and (correlated) $p_\perp$ kicks in string fragmentation.
	\label{fig:zFrac}}
\end{figure}

The fact that $\PH\to \PQs\PAQs$ produces rank-1 strange hadrons has a few key ramifications. First, one should note that rank 1 does not necessarily equate to the given hadron being the leading one in the respective jet.
Given the iterative fragmentation procedure from string endpoints, the rank-$i$ hadron will take a fraction $z_i$ of the remaining endpoint momentum, according to the fragmentation function, Eq.~\eqref{eqn:lundFragFn}.
For $\PH\to \PQs\PAQs$ this means that the first-rank strange hadron takes a fraction $z_1$ of the strange-quark momentum, and the second-rank hadron will take a fraction $z_2$ of the \emph{remaining} momentum, \ie, $(1-z_1)z_2$ of the initial strange-quark momentum, as illustrated in \cref{fig:zFrac}.
Accordingly, one would in general expect the first-rank hadron -- and thus in the case of $\PH\to \PQs\PAQs$, the strange hadron -- to be the hardest hadron formed.
This however is of course not necessarily the case given these $z$-fractions are sampled probabilistically, and there remains a finite probability that the first-rank hadron is softer than the second-rank one (or, with successively smaller probability, a higher-rank one). This would happen if the first-rank hadron takes a small $z$~value and the second- (or higher-)rank one takes a larger total $z$~fraction. How often this occurs depends on the width of the fragmentation function(s) used in the model, and although their averages are well constrained, their widths (and hence the fluctuations from break to break) are much less so.

\begin{figure}[t]
\centering
    \vspace*{1mm}\includegraphics[width = \textwidth]{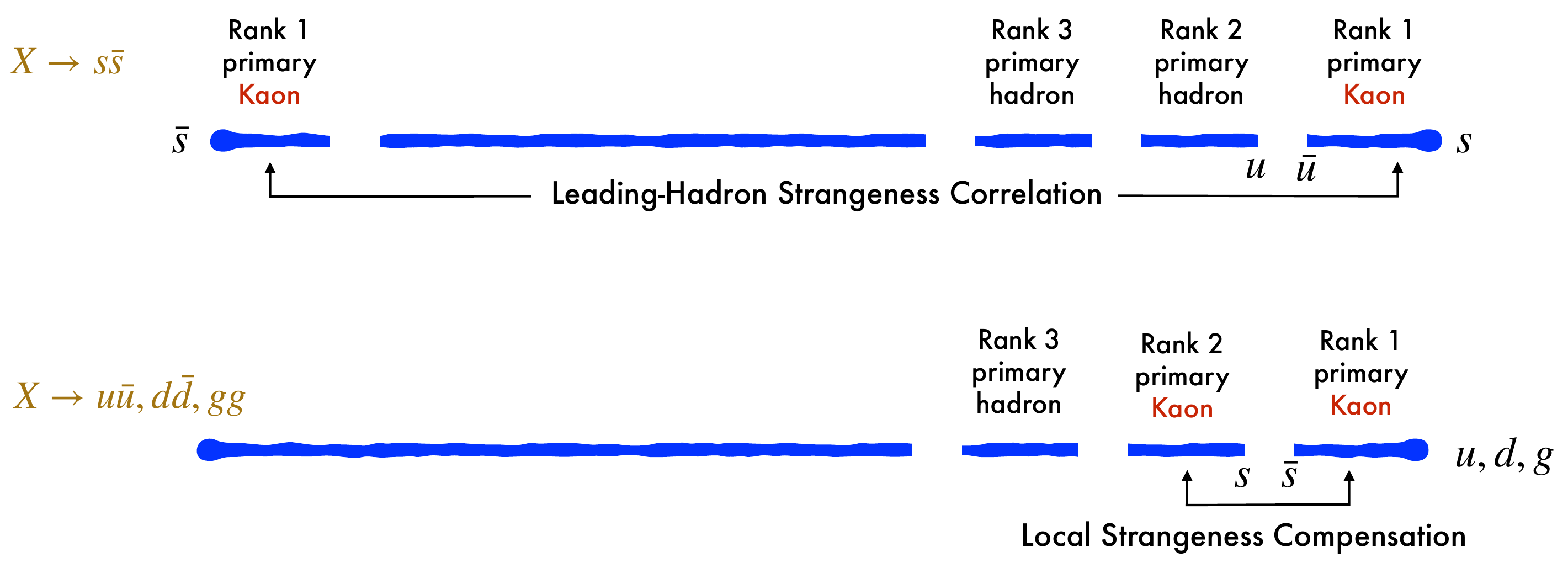}\vspace*{-2mm}
	\caption{Strangeness correlations in the decay of a heavy colourless particle, $X$, into $\PQs\PAQs$ (signal, upper panel) vs.\ decay into $\Pg\Pg$ and lighter quark flavours (background, lower panel). The signal is characterised by global (long-distance) strangeness conservation, between the first-rank hadrons in each of the two jets, while the backgrounds are characterised by local (short-distance) strangeness conservation, within each jet, and suppression (though not absence) of strangeness in the first-rank hadrons in each jet.
	\label{fig:strangeHad}}
\end{figure}

Another key ramification of the strange quarks in $\PH\to \PQs\PAQs$ contributing to rank-1 hadrons is that the resultant hadron is formed from the contribution from only a single string break, meaning it only receives a single $p_\perp$~kick. Contrastingly, higher rank hadrons have all constituents sourced from string breaks and thus receive two $p_\perp$~kicks. Such a $p_\perp$ contribution to both rank-1 and higher rank hadrons is illustrated in \cref{fig:strangeHad}.
Given this is the case, first rank hadrons may be able to be tagged by having a comparatively smaller $\langle p_\perp\rangle$ than other higher rank hadrons, which for $\PH\to \PQs\PAQs$ would manifest as strange hadrons with lower $\langle p_\perp\rangle$.

A further consequence of the strange quarks forming string endpoints in $\PH\to \PQs\PAQs$ is that the resultant strange hadrons lead to global strangeness conservation, without necessarily being highly correlated.
In contrast, higher rank strange hadrons expected to be more highly correlated as the strangeness is sourced from $\PQs\PAQs$ pair creation via string breaks, whereby the strange and antistrange quarks form adjacent to one another as illustrated in the bottom panel of \cref{fig:strangeHad}. This leads to the two resulting strange hadrons also forming adjacent to each other, and a local strangeness conservation.

\subsubsection*{Modelling Uncertainties and Variations}

In addition to the obvious possibility of cross checking string-based hadronisation models with ones based on clusters, such as those of HERWIG~\cite{Bewick:2023tfi} or SHERPA~\cite{Chahal:2022rid,Knobbe:2023njd}, parametric variations within the context of a single modelling paradigm are also of interest. Here we comment on a set of such variations, for strangeness tagging in the context of \pythia's string model, with the caveat that there is no certain guarantee that these variations are exhaustive.

In the following, we refer to \pythia parameters by their names in the code. We assume that readers unfamiliar with these will look up the corresponding definitions in the program's manual and documentation.

To gauge the effect of a lower-rank hadron acquiring higher energy than the first-rank one, one could vary the width of the fragmentation function $f(z)$. Since the mean of $f(z)$ is quite well constrained, this variation could be done using \texttt{StringZ:deriveBLund = on}~\cite{Amoroso:2018qga}, to keep the mean constant, and the varying the \texttt{StringZ:aLund} ($a$) parameter in a conservative range constrained by the high-$x$ tail of hadron momentum spectra and multiplicity distributions at LEP. One could also consider the effect of allowing a different $a$ parameter for strange quarks, using \texttt{StringZ:aExtraSQuark}.

Varying the degree of locality of the correlations in flavour and $p_\perp$ is currently only possible in the baryon sector, via the so-called popcorn mechanism. However it may be possible to at least mimic the effect also in the meson sector by considering very wide fragmentation functions, corresponding to low values of the \texttt{StringZ:aLund} parameter. This should ``shuffle'' the hadrons around more, effectively smearing out their correlations.

One can also investigate changes in the balance between transverse and longitudinal fragmentation, by adjusting the $p_\perp$-broadening parameter \texttt{StringPT:sigma}, noting its anticorrelation with $\left<f\right>$.
One may also want to investigate the effect of different shower models, including VINCIA~\cite{Brooks:2020upa} (which is being developed towards NNLO accuracy for $\PH\to \PQs\PAQs$~\cite{Campbell:2021svd}) and/or the effect of variations of perturbative parameters in the shower including not only the standard renormalisation-scale variations but also, \eg, variations of nonsingular terms in the shower kernels~\cite{Mrenna:2016sih}. If one sees large effects of the latter, strategies to increase the matching and merging with matrix elements should be pursued.

We note that it should be possible to carry out many of the variations mentioned so far, using the frameworks for variational weights in showers~\cite{Giele:2011cb,Mrenna:2016sih} and hadronisation~\cite{Bierlich:2023fmh}. Work is in progress to extend the latter to be able to handle changes to the hadrochemistry of jets as well.

\begin{figure}[t]
    \centering
    \includegraphics[width = 0.95\textwidth]{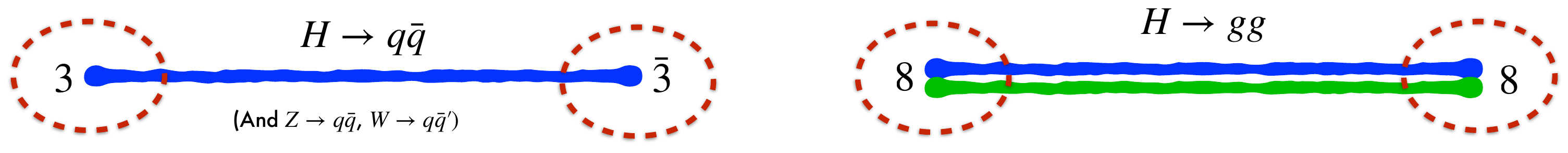}
    \caption{Illustration of $\PH\to \PQq\PAQq$ and $\PH \to gg$ string configurations. In baseline \pythia, the only difference between quark and gluon fragmentation is that the latter are connected to two string pieces instead of one. Thus, no new or separate parameters are introduced to describe gluon jets. This accounts for some of the key differences between quark and gluon jets, such as multiplicity ratios $\propto 2 \sim C_A/C_F$. But this is also obviously a simplified picture, the reliability of which depends upon experimental validation. The dashed red circles are intended to highlight the fragmentation regions associated with the ``leading'' components of the respective jets.
    \label{fig:Hgg}}
\end{figure}

Moving on to quark vs.\ gluon jet properties, cf. \cref{fig:Hgg} (see also the studies in Refs.~\cite{Gras:2017jty,Knobbe:2023njd}),
it is difficult to vary these independently in \pythia. We here mention QCD colour reconnections~\cite{Christiansen:2015yqa}, colour ropes~\cite{Bierlich:2014xba}, and/or closepacking effects~\cite{strangepacking} as options that may affect the relative properties of octet vs.\ triplet fragmentation. This question is also under intense investigation at LHC where one exploits a combination of infrared safe and infrared-sensitive observables to constrain it~\cite{Gras:2017jty}.
Further qualitative modelling variations
could include thermodynamical string fragmentation~\cite{Fischer:2016zzs}, explicit modelling of hyperfine splitting effects~\cite{Bierlich:2022vdf}, effects of a fluctuating or time-dependent string tension~\cite{Pirner:2018ccp,Hunt-Smith:2020lul}, and/or effects of string excitations~\cite{excitedstrings}. At least the first of these options is possible to investigate with current versions of \pythia, while we expect more of the latter to be available in the future.

Finally, we emphasise that useful in-situ constraints on many of these variations can presumably be made via samples of $\PWp\to \PQc\PAQs$ events, which would give access to leading-strangeness jets of a similar hardness as those expected from Higgs decays, and where the opposite side could be tagged by the presence of the charm quark. And, for the background modelling, gluon jets in \epem contexts can be relatively cleanly identified, \eg, in $\PZ\to \PQb\PAQb\Pg$ events with good $\PQb$-tagging. Given a sufficiently large statistical sample, the limiting behaviour of the latter for ``hairpin'' configurations (in which the $\PQb\PAQb$ are very close together, i.e.\ the closest one can get to a ``$\PZ\to \Pg\Pg$''-like event configuration) can also be studied. Several other less clean or direct cross checks could also be envisioned, some of which could be done in $\Pp\Pp$ collisions at the LHC. We remark that the constraining power of any such studies will presumably depend quite sensitively on particle-identification capabilities, without which one is effectively blind to the details of the hadronisation process.


\subsubsection{Overview of existing experimental studies}
\label{sec:HtoSS:Recap}

Experimental studies to explore the possibilities of testing the Higgs interactions with the strange quark have been growing over the last years. In the Snowmass~2021 Energy Frontier report~\cite{Narain:2022qud}, the work presented in Ref.~\cite{Albert:2022mpk} is singled out as a new proposed measurement.
Since then, this research area has gained a prominent role in the study of Higgs couplings at future Higgs factories and several simulation-based analyses have been performed using different flavour tagging algorithms and detector/accelerator scenarios. \cref{tab:analyses_table} summarises the available results.

\begin{table}[h]
    \centering
    \renewcommand{\arraystretch}{1.5} 
    \resizebox{\textwidth}{!}{ 
    \begin{tabular}{|>{\centering\arraybackslash}p{2.5cm}| 
                    >{\centering\arraybackslash}p{1.5cm}|
                    >{\centering\arraybackslash}p{2.5cm}|
                    >{\centering\arraybackslash}p{3cm}|
                    >{\centering\arraybackslash}p{2.5cm}|
                    >{\centering\arraybackslash}p{2.0cm}|
                    >{\centering\arraybackslash}p{2.0cm}|
                    >{\centering\arraybackslash}p{2.5cm}|}
        \hline
        \textbf{Accelerator} & \textbf{Detector Concept} & \textbf{Dedicated Tagger} & \textbf{Analysis Strategy} & \textbf{Results (BR)} & \textbf{Results ($k_{\PQs}$)} & \textbf{References} & \textbf{Additional Notes}\\
        \hline
        ILC @ 250~GeV, 2~ab$^{-1}$ expected, 0.9~ab$^{-1}$ considered & ILD & Yes. RNN with jet and track-level information. Uses truth-based PID and also LCFIPLus~\cite{Suehara:2015ura} scores for $\PQb$-, $\PQc$-, other-jets & $\PZ\PH$ production, $\PZ\to\Pl\Pl$ and $\PZ\to\PGn\PAGn$. Cut-based. Sum of leading and sub-leading strange-jet score used to apply final cut (optimised for maximal significance) & Not available & $k_{\PQs} < 7 \times \textrm{SM}$ at 95\% CL with 900~fb$^{-1}$ & Ref.~\cite{Albert:2022mpk} & Full Simulation samples. General interpretability due to truth-based PID. \\
        \hline
        FCC-ee @ 240~GeV, 10.8~ab$^{-1}$, @ 365~GeV, 3.0~ab$^{-1}$ & IDEA & Yes, ParticleNet. Uses \dndx and TOF for PID information & $\PZ\PH$ production, $\PZ\to\Pl\Pl$, $\PZ\to\PGn\PGn$ and $\PZ\to\PQq\PAQq$. Use shape information of discriminant variable, fit all couplings simultaneously & $\sigma(\PZ\PH)\times\textrm{BR}(\PH\to\PQs\PAQs)$ O(100\%) at 68\% CL with 10.8~ab$^{-1}$ at 240~GeV. O(460\%) at 68\% CL with 3.0~ab$^{-1}$ at 365~GeV. & Not available & Ref.~\cite{Bedeschi:2022rnj}. 
        \cref{sec:higgsHadronicFCC} & Fast Simulation based on \delphes. \\
        \hline
        CEPC @ 240~GeV, 20~ab$^{-1}$ & CEPC Baseline Detector & Yes, ParticleNet. Emulates scenarios with perfect lepton ID, perfect charged hadron ID and perfect $K_0$ ID & $\PZ\PH$ production, $\PZ\to\Pl\Pl$ and $\PZ\to\PGn\PGn$. BDT separating signal from background using the 11 tagger scores. & Upper limit on $\textrm{BR}(\PH\to{\PQs\PAQs}) = 0.75 \times 10^{-3}$ with 20~ab$^{-1}$ & $k_{\PQs} < 1.7 \times \textrm{SM}$ at 95\% CL with 20~ab$^{-1}$ & Ref.~\cite{Liang:2023wpt} & Full Simulation samples. \\
        \hline
        ILC @ 250~GeV, 2~ab$^{-1}$& SiD & Yes, ParticleNet with improved calorimeter granularity & $\PZ\to\Pl\Pl$, $\PZ\to\PGn\PGn$ and $\PZ\to{\PQq\PQq}$ & $\sigma(\PZ\PH)\times\textrm{BR}(\PH\to{\PQs\PAQs})$ O(300\%). & To evaluate and add higher energy run & Based on IDEA & Analysis sensitivity estimated by extrapolation. \\
        \hline
    \end{tabular}
    }
    \caption{Summary of existing $\PH\to\PQs\PAQs$ simulation-based analyses.}
    \label{tab:analyses_table}
\end{table}

\subsubsection{\texorpdfstring{Target detector performance aspects for $\PH\to\PQs\PAQs$}{Target detector performance aspects for H->ss}}
\label{sec:HtoSS:Detector_performance}

\begin{figure}[t]
\centering
    \vspace*{1mm}\includegraphics[width = 0.6\textwidth]{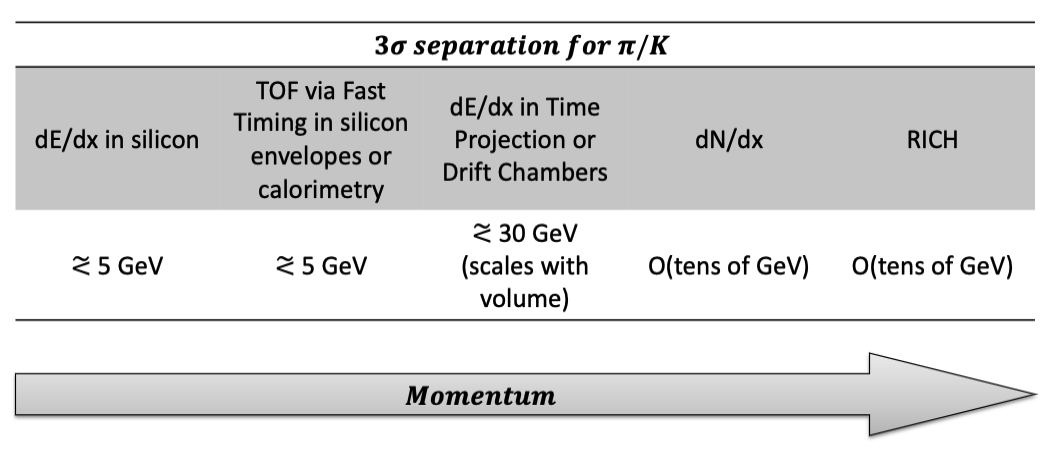}\vspace*{-2mm}
	\caption{Comparison among particle identification capabilities as a function of the momentum ranges covered by different detector technologies.
	\label{fig:PiD_detectors_momentumRange}}
\end{figure}

Particle identification (PID) capabilities, primarily for kaons and pions, at high momenta ($\gtrsim\qty{10}{GeV}$) are a key feature for future detector designs in order to enable the experimental identification of strange-quark-initiated jets. Hadrons are identified by their mass, by combining momentum and velocity information. Assuming that the particle momentum is inferred from the trajectory curvature in a magnetic field, the remaining challenge is to measure its velocity. Various technologies allow the identification of hadrons in different momentum ranges, as summarised in \cref{fig:PiD_detectors_momentumRange}.

Several studies are underway both to design detector concepts and to assess the impact that a newly introduced PID-capable sub-detector would have on the reconstruction of physics objects, as introduced in \cref{ssec:detector_models} and \cref{sec:com:reco:PID}. In this section, we summarise 
the progress. 
The technologies themselves are broadly applicable to detectors at any future lepton collider.

At ILC~\cite{Behnke:2013xla, ILCInternationalDevelopmentTeam:2022izu} and the proposed ILD~\cite{ILD:2019kmq, ILDConceptGroup:2020sfq}, a combination of ionisation energy loss and time-of-flight (TOF) has been studied~\cite{Einhaus:2021abe, Einhaus:2023amy}. Measurements of ionisation energy loss are realised using ILD's time projection chamber (TPC). Instead of measuring the total ionised charge, ILD counts the number of ionising interactions, so-called ``cluster counting'' (\dndx), resulting in an improved correlation between the measured energy and the momentum of the incident particle and PID separation power, $\mathcal{S}$ (measured in units of standard deviations)~\cite{Einhaus:2019jvg}. Measurements of \dndx\ are complemented by time-of-flight (TOF) measurements, instrumented either via a silicon envelope of the TPC or layers of the electromagnetic calorimeter. The timing values, when combined with a known track length, yield particle velocity and mass. Altogether, the measurements of \dedx and TOF (assuming a timing resolution of \qty{30}{ps}) result in a $\mathcal{S} > 2$ kaon-pion separation up to $\sim$\qty{50}{GeV}.

Gaseous ring-imaging Cherenkov (RICH) detectors~\cite{Seguinot:1976yp} have also been studied as instruments for PID at future lepton colliders. RICH detectors rely on the angle of an emitted Cherenkov cone to calculate a particle's velocity and therefore mass; examples of RICH detectors include SLD's CRID~\cite{Vavra:1999lzh} and DELPHI's gaseous RICH detector~\cite{Albrecht:1999ri}. For ILD or the Silicon Detector (SiD)~\cite{Aihara:2009ad} at the ILC, this includes the RICH detector proposed in Ref.~\cite{Basso:2023zuq}; for the CLIC-Like Detector (CLD)~\cite{Bacchetta:2019fmz} at the FCC-ee, this includes the Array of RICH Cells (ARC) detector~\cite{FortyFCCWeek, TatECFAWS}. Both detector concepts employ a slim design, $\sim$\qty{20}{cm} in radial depth, with a low-material budget, $<10\%$ of a radiation length, to limit their impact on tracking and calorimetry. As a gaseous radiator, C$_4$F$_{10}$ at \qty{1}{bar} is the preferred choice for both detector concepts, and each relies on arrays of spherical coated mirrors (\eg, made of low-mass beryllium) to focus the Cherenkov light onto the sensitive detector elements. The preferred choice for these detector elements are silicon photomultipliers (SiPMs), owing to their excellent efficiency, granularity, and compact readout. Some of the most critical parameters affecting performance include the SiPM pixel size and the angular resolution of the incident tracks. SiPM noise can be eliminated by cooling ($+2$--3\,$^\circ$C for the RICH detector in Ref.~\cite{Basso:2023zuq}; $-40$\,$^\circ$C for the ARC detector) as well as by using timing cuts on the difference between the time of a particle's signal and a Cherenkov photon's signal (\eg, $\pm\qty{200}{ps}$ would be sufficient). Depending on the contributions to the total uncertainty on the Cherenkov angle, $\mathcal{S} > 3$ kaon-pion separation up to $\sim$\qty{50}{GeV} may be achieved. For the ARC detector, there are ongoing studies to simulate the detector in \geant and reconstruct Cherenkov angles~\cite{TolosaDelgadoFCCWS, PezzuloECFAWS}. In order to build these RICH detectors, the efficiency, size, and cost of SiPMs must further improve (as is expected to happen over the next decade) and low-mass support structures and mirrors must be used.

In addition to detector capabilities enabling kaon-pion separation, vertexing as well as calorimetry remain of utmost importance for flavour tagging. Further information on the PID technologies which have been studied for some of the proposed detectors is presented in \cref{sec:HtoSS:Tagger}.

\subsubsection{\texorpdfstring{Algorithm R\&D: jet flavour tagger for $\PH\to\PQs\PAQs$}{Algorithm R\&D: jet flavour tagger for H->ss}}
\label{sec:HtoSS:Tagger}

Flavour tagging is discussed in general in \cref{sec:com:flavourtagging}.
Strange tagging, especially strange-gluon separation is essential for Higgs to strange analysis. The strange jet can be tagged mainly by identifying high-momentum kaons (of which strange jets have higher fraction than others) while the difference on kinematics (jet multiplicity, jet mass, \etc) can be used to separate gluon jets.

In ILD full simulation, developments are ongoing to combine a new MVA-based particle ID framework, Comprehensive PID (CPID)~\cite{Einhaus:2023oqy}, and a new flavour tagger based on Particle Transformer (ParT)~\cite{Qu:2022mxj}. CPID utilizes \dedx obtained by ILD Time Projection Chamber and time-of-flight information (with 100 picosecond resolution assumed) from 10 front layers of the electromagnetic calorimeter to efficiently separate kaons from protons and pions for a wide momentum range. The output PID probabilities for all tracks inside jets are provided to the input variables of ParT-based tagger, which classify jets as $\PQb$, $\PQc$, $\PQs$, $\PQu$, $\PQd$ and $\Pg$. ParT takes features of all tracks and neutral particles, with separate embedding layers, to calculate attention weight among particles. The specific feature of ParT from standard transformer is ``interaction'' input, which are variables calculated from 4-momenta of each pair of the particles to bias attention weight of the Transformer. The network has been trained with $\PGn\PGn\PH \to \PGn\PGn{}\PQq\PQq$ final states, which give 2-jet events with the same flavour and variable energies, as an ideal training sample for the flavour tagging. The current network results in background efficiency of 25.7\% for gluon jets, and 42.7\% for $d$-jets, when a selection efficiency of 80\% for strange jets is applied, obtained with full detector simulation~\cite{Suehara:2024qoc}.

The \textsc{DeepJetTransformer}~\cite{Blekman:2024wyf} is an alternative DL-based algorithm for strange jet tagging that utilises transformer models, similar to other tagging algorithms such as ParticleNet. This algorithm leverages the capabilities of transformer models to analyse the complex structure of jets and identify subtle features indicative of strange quark origin. Of particular importance in this work is the role of V$^0$ vertex reconstruction in enhancing the performance of strange jet tagging, especially for jets with low charged kaon multiplicities. V$^0$s, specifically $\PKzS$ and $\PGL^0$ , are particles that decay into a pair of oppositely charged tracks. The reconstruction and identification of these V$^0$ vertices provide additional handles for distinguishing strange jets from light quark jets, as strange jets typically exhibit a higher multiplicity of kaons.
Including V$^0$ variables in the \textsc{DeepJetTransformer} algorithm significantly
improves strange tagging efficiency, by up to $\approx20\%$ in cases without $\PKpm$ identification and by $\approx5\%$ with perfect $\PKpm$ identification, both with a background efficiency of 10\% for strange quark versus up or down quark discrimination. This improvement underscores the value of precise V$^0$ reconstruction, supported by a high-performance vertex
detector with excellent spatial resolution and a lightweight tracker with numerous measurement points. V$^0$ vertex information thus enhances the
algorithm's ability to effectively identify strange jets, especially when charged kaon identification is limited or unavailable.

\subsubsection*{Strange Tagging Performance at SID and IDEA}

A comparative study \cite{Ntounis:2025itg} between different detector concepts, specifically IDEA and SiD, has revealed important insights into $s$-tagging capabilities. The investigation centred on calorimeter and PID, which would enable distinguishing between kaons, pions, and protons with momentum up to tens of \GeV  and calorimeter performance.  While certain detector concepts, such as IDEA and ILD, are expected to have PID  through \dndx\ or \dedx\ measurements, as well as TOF information, other detector concepts, such as SiD and CLD, could benefit from the addition of a dedicated RICH subdetector for PID. However, it remains to be evaluated how adding such a subdetector would impact the performance of other detector systems, particularly the calorimeters.

The study utilized the \textsc{ParticleNet} tagger with \delphes\ fast simulation, analysing $\PZ(\rightarrow \PGn\PGn)\PH(\rightarrow jj)$ events. Three detector configurations were examined:
\begin{itemize}[noitemsep]
    \item Standard SiD configuration matching the TDR performance (all-silicon vertex and tracking, no PID)
    \item IDEA detector configuration (silicon vertex detector, and drift chamber surrounded by silicon wrapper; PID through \dndx\ and TOF) 
    \item And IDEA-like detector with its tracking system replaced by that of CLD (silicon tracker, no PID)
\end{itemize}

The calorimeter resolutions for both detector concepts are characterised by the values shown in \cref{tab:HssCalorimeterRes}.

\begin{table}[h!]
    \centering
    
    \begin{tabular}{|c|c|c|}
    \hline
        $\mathbf{{\sigma(E)}/{E}}$ & \textbf{SiD} & \textbf{IDEA} \\
        \hline
        \textbf{ECAL} & $\tfrac{17\%}{\sqrt{E}} \oplus 1\%$ & $\tfrac{3\%}{\sqrt{E}} \oplus \tfrac{0.2\%}{E} \oplus 0.5\%$ \\
        \hline
        \textbf{HCAL} & $\tfrac{55.9\%}{\sqrt{E}} \oplus 9.4\%$ & $\tfrac{30\%}{\sqrt{E}} \oplus \tfrac{5\%}{E} \oplus 1\%$ \\ \hline
    \end{tabular}
    \caption{Calorimeter resolutions for SiD and IDEA. \label{tab:HssCalorimeterRes}}
\end{table}

The analysis revealed that PID capabilities significantly drive improvements in $s$-tagging performance, particularly in discriminating against light-quark and gluon jets. Enhanced calorimeter resolution showed strong benefits primarily for strange versus charm and bottom quark discrimination. For an 80\% signal efficiency working point, the IDEA detector configuration achieved notably lower mistag rates compared to the nominal SiD design: light quark mistag rates improved from 54\% to 25\%, gluon mistag rates from 24\% to 12\%, and charm mistag rates from 6.0\% to 3.3\%. Signal efficiency working points are given in the \textsc{ParticleNet} rows of \cref{tab:HtoSS:TaggerSummary}.


\subsubsection*{IDEA Detector Studies as a function of tagger performance}

A comprehensive study of the IDEA detector's inner components reveals significant impacts on jet-flavour identification capabilities at the FCC-ee. The official \textsc{Delphes} fast simulation framework with
key additions, such as the \textsc{TrackCovariance}~\cite{TrackCovariance}, \textsc{TimeOfFlight}~\cite{TimeOfFlight} and \textsc{ClusterCounting}~\cite{ClusterCounting} packages discussed in Ref.~\cite{FCCFlavTag}, is exploited to simulate the detector response. 
The baseline configuration features a vertex detector with $25\times25\,\SI{}{\micron}^2$ pitch and \SI{3}{\micron} single-point resolution, with the innermost layer at 1.2\,cm from the beamline. Several detector variations were investigated:

\begin{itemize}
    \item Single-point resolution variations of $\pm65\%$ around the nominal value of $3\,\mu\text{m}$ showed substantial effects on $c$-tagging, with factor $\sim$2 changes in $s$-rejection efficiency;
    \item Material budget variations of $\pm50\%$ demonstrated limited benefits from reduction but significant degradation ($\sim$1.5x) in $c$-tagging with increased material;
    \item Displacement of all silicon layers by 0.5 cm away from the beampipe resulted in 2-3x worse rejection performance for both $\PQb$- and $c$-tagging;
    \item PID capabilities were evaluated through \dndx\ and ToF variations.
\end{itemize}

These detector variations' impact on Higgs coupling measurements in $\PZ\PH$ hadronic final states was quantified through comprehensive analysis retraining. The study revealed that there was robust Higgs decay tagging efficiency across different detector configurations. The impact from ToF removal, silicon layer displacement, and material budget changes was limited. However, the study found that there was significant degradation in Higgs coupling measurements without cluster-counting information. A summary of the relative changes are shown in \cref{tab:HtoSS:coupling-impact}.

\begin{table}[h!]
    \centering
     \begin{tabular}{|ccccc|}
        \hline
        \multicolumn{1}{|l|}{\multirow{2}{*}{}} & \multicolumn{4}{c|}{68\% CL  precision} \\ \cline{2-5}
        \multicolumn{1}{|l|}{} & \multicolumn{1}{c|}{$\mu_{Hbb}$} & \multicolumn{1}{c|}{$\mu_{Hcc}$} & \multicolumn{1}{c|}{$\mu_{Hgg}$} & $\mu_{Hss}$ \\ \hline
        \multicolumn{1}{|c|}{Baseline} & \multicolumn{1}{c|}{$\pm$ 0.3\%} & \multicolumn{1}{c|}{$\pm$ 4.2\%} & \multicolumn{1}{c|}{$\pm$ 2.8\%} & \begin{tabular}[c]{@{}c@{}}+674\%\\ -669\%\end{tabular} \\ \hline
        \multicolumn{5}{|c|}{Relative change compared to baseline ($\mu_{variation}$/$\mu_{baseline}$)} \\ \hline
        \multicolumn{1}{|c|}{No TOF} & \multicolumn{1}{c|}{x1.3} & \multicolumn{1}{c|}{x1.02 (upper limit only)} & \multicolumn{1}{c|}{x1} & x1.03 \\ \hline
        \multicolumn{1}{|c|}{No \dndx} & \multicolumn{1}{c|}{x1.3} & \multicolumn{1}{c|}{x1.07} & \multicolumn{1}{c|}{x1.07} & x1.6 \\ \hline
        \end{tabular}
   \caption{Impact of detector variations on Higgs coupling measurements at 68\% CL. Baseline uncertainties are shown for $\mu_{\PH\rightarrow \PQb\PAQb}$, $\mu_{\PH\rightarrow \PQc\PAQc}$, $\mu_{\PH\rightarrow \Pg\Pg}$ and $\mu_{\PH\rightarrow \PQs\PAQs}$, followed by relative changes when removing ToF and \dndx.}
    \label{tab:HtoSS:coupling-impact}
\end{table}

\subsubsection*{Jet flavour tagging at FCC-ee 91~GeV with a transformer-based neural network}

This study presents advances in jet flavour identification through \textsc{DeepJetTransformer}, a transformer-based neural network capable of distinguishing between $\PQb$, $\PQc$, $\PQs$, $\PQu$, and $\PQd$ quarks.

The research utilizes $\epem \rightarrow \PZ \rightarrow \PQq\PAQq$ processes at $\sqrt{s} = 91.2$ GeV, employing \textsc{Delphes} for reconstruction and \fastjet-3.3.4 for jet clustering with the exclusive $\epem~k_{\text{T}}$ algorithm. The \textsc{DeepJetTransformer} model incorporates comprehensive input features including particle flow information, particle identification, secondary vertex data, and V$^{0}$ variables.

The tagger achieves significant discrimination capabilities:
\begin{itemize}
    \item $\PQb$-jet identification reaches $99\%$ efficiency against light quarks ($\PQs$,$\PQu$,$\PQd$) and $86\%$ against $\PQc$-jets both at a $0.1\%$ background rate;
    \item $\PQc$-jet tagging achieves $90\%$ efficiency at $10\%$ $\PQb$-jet background ($70\%$ at $1\%$ background);
    \item $\PQs$-jet discrimination shows $40\%$ efficiency at $10\%$ $\PQu\PQd$-jet background rate, see also the \textsc{DeepJetTransformer} rows of \cref{tab:HtoSS:TaggerSummary};
    \item Limited but measurable discrimination between $\PQu$- and $\PQd$-jets is also achieved.
\end{itemize}

The model demonstrates particular strength in isolating pure samples of strange quark decays from $\PZ$ bosons, achieving $5\sigma$ significance with $\SI{60}{\per\nano\barn}$ of luminosity. This would offer a substantial validation dataset for future applications at 250 GeV centre-of-mass energy collisions. 

The transformer-based architecture offers advantages in training speed compared to graph neural networks, making it particularly suitable for detector concept studies. Further details are provided in Ref.~\cite{Blekman:2024wyf}.

\subsubsection*{Summary and overview of flavour tagger performances}

The different strange-tagging algorithms, their strange-jet tagging efficiencies and mistag rates, are summarized in \cref{tab:HtoSS:TaggerSummary}.

\begin{table}[h!]
    \centering
     \footnotesize
    \begin{tabular}{|c|c|c|c|c|c|c|c|c|c|}
        \hline
        \multirow{2}{*}{Tagger}  & \multirow{2}{*}{Detector} & \multirow{2}{*}{Simulation}      & \multirow{2}{*}{Sample} & \multirow{1}{*}{$\mathbf{\epsilon{(s)}}$}  & \multicolumn{4}{c|}{\textbf{Mistag rates [\%]} } & \multirow{2}{*}{Comment} \\
                                              &              &                                   &                                    &           [\%]        &  $g$ & $u/d$ & $c$  & $\PQb$    &  \\ \hline \hline
        \multirow{6}{*}{\textsc{ParticleNet}} & SiD          & \multirow{6}{*}{\textsc{Delphes}} & \multirow{6}{*}{$Z(\nu \nu)H(jj)$} & \multirow{3}{*}{90}   & 45   & 70 & 9.1  & 1.3  &  \\ 
                                              & IDEA         &                                   &                                    &                       & 26   & 38 & 5.4 & 0.59 &  \\ 
                                              & IDEA w/ CLD tracker     &                                   &                         &                       & 29   & 41 & 6.4 & 0.76 &  \\ 
                                              \cline{2-2}\cline{5-9}
                                              & SiD   &                                   &                                    & \multirow{3}{*}{80}   & 24   & 54 & 6.0 & 0.68  &  \\ 
                                              & IDEA         &                                   &                                    &                       & 12   & 25 & 3.3 & 0.27  &  \\
                                              & IDEA w/ CLD tracker   &                                   &                           &                       & 13   & 28 & 4.0 & 0.36  &  \\
                                              \hline \hline

        \multirow{3}{*}{\textsc{ParT}} & ILD    & \scriptsize Full w/ CPID~\cite{Einhaus:2023oqy} & \multirow{3}{*}{$Z(\nu \nu)H(jj)$} & \multirow{3}{*}{80}  & 25.7 & 42.7 &  &   & \tiny $\dedx$ in TPC and  \\ 
                                       & ILD    & \scriptsize Full w/ truth PID                   &                                    &                      & 23.2 & 38.0 &   &   & \tiny ToF in calo \\ \cline{6-10}
                                       & IDEA & \textsc{Delphes}                                &                                    &                      & 20.3 & 29.6 &   &   &  \\ \hline \hline

        \multirow{2}{*}{\textsc{\tiny DeepJetTransformer}} & \multirow{2}{*}{IDEA} & \multirow{2}{*}{\textsc{Delphes}} & \multirow{2}{*}{$Z(jj)$} & 80        &      & 37.8 & 7.4 & 0.10 & \multirow{2}{*}{\parbox{2,3cm}{\tiny Fast training, $V^0$ reco, emulated uniform PID with  $\epsilon_{K}=90\%$, $\epsilon_{\pi}=10\%$}}  \\ \cline{5-9}
                                       &                              &                         &                                    &                     40 &      & 10.1 &  1.3 &  <0.01 &  \\ \hline 

    \end{tabular}
   \caption{Overview over studies of flavour taggers for strange-tagging and resulting mistag rates for different working points, collider options, and detector concepts. Note that a direct comparison between different flavour taggers performed with different data sets or at different detector designs is difficult and should not be taken at face value.  ParticleNet values taken from Ref.~\cite{Ntounis:2025itg}. }
    \label{tab:HtoSS:TaggerSummary}
\end{table}


\subsubsection{Ideas for future studies}
\label{sec:HtoSS:Future}

In future Higgs factories, advanced studies of fragmentation and hadronisation will be crucial for enhancing strangeness tagging capabilities, directly impacting the sensitivity of measurements of the Higgs coupling to second-generation quarks. As discussed in \cref{sec:HtoSS:Fragmentation}, modeling the fragmentation and hadronisation of quarks is particularly delicate. Current uncertainties, especially those arising from gluon splitting into quark pairs and heavy-flavour fragmentation, can introduce significant uncertainties in tagging performance. To address these challenges, research is focusing on systematically exploring variations in fragmentation models and deepening our understanding of non-perturbative QCD effects.

Two main avenues of action are currently being pursued to tackle these challenges. The first one is focused on  studies using simulated events produced with different configurations of event generators. By simulating conditions expected at future colliders, such as the Tera-Z runs anticipated at the FCC-ee, which would produce $6\times 10^{12}$ $\PZ$ bosons, we aim to understand how such vast statistics can be exploited to improve our understanding of strange-quark fragmentation and enhance strangeness tagging performance. These simulations enable detailed investigations of fragmentation functions, hadronisation parameters, and particle correlations within jets initiated by strange quarks. By refining both theoretical models and experimental analysis strategies through these simulations, we can proactively reduce systematic uncertainties associated with fragmentation processes before the actual data becomes available. This work helps in assessing the potential impact of various detector designs and informs the development of advanced tagging algorithms.

Another critical effort involves the re-analysis of existing data, particularly from LEP. There is ongoing work to reproduce and translate these historical datasets into modern analysis frameworks such as Key4HEP~\cite{Ganis:2021vgv}, which provides a universal format across proposed future colliders. By applying the latest developments in theoretical calculations and experimental tools to these datasets, the aim is to re-express LEP measurements in terms of variables and observables most relevant for current and future studies. This re-analysis allows for a better understanding and improvement of the modelling of fragmentation and hadronisation processes, potentially providing important  inputs to tune and validate the Monte Carlo event generators. Such efforts bridge the gap between past data and future experimental needs, enhancing the overall robustness of strangeness tagging methodologies.

\subsubsection*{Physics motivated ML for improved strange Yukawa measurements}
\label{sec:HtoSS:Futureml}

Deep neural network-based jet taggers such as particle-net~\cite{Qu:2019gqs}, particle-transformer~\cite{Qu:2022mxj}, and GN1~\cite{Shlomi:2020ufi, Duperrin:2023elp} are state of the art neural networks, deployed by the CMS and ATLAS collaborations for their latest physics studies. Similar architectures have been studied in the context of strange tagging~\cite{Kats:2024eaq,Blekman:2024wyf} as well. All these networks are trained through supervised learning methods, in which a particular jet flavour is set as a target label. Assigning flavour to a jet in fixed order perturbation theory is a subtle issue due to flavoured quark-antiquark pair production at very low energies. In recent times several  improved ``flavoured'' jet algorithms~\cite{Czakon:2022wam,Gauld:2022lem,Caola:2023wpj} have been proposed addressing this issue. A thorough investigation of the impact of low energetic flavour splitting is worth venturing in the context of the $\PH \rightarrow \PQs\PAQs$ study. In the boosted scenario, where the emitted $\PQs\PAQs$ pairs are within geometric vicinity, an accurate flavour assignment will be an intricate issue, having impact on the determination of the strange Yukawa measurement.

While graph representation of a hadronic jet seems to be more preferred over image representation, due to its flexibility in handling irregular detector geometry and sparse shower profiles as input, a higher order data representation like hypergraphs may be worth investigating. Hypergraphs are designed to capture higher-order correlations among elements of a set and thus may be capable of capturing the correlations between two strange jets originating from the $\PH$ decay, which are colour connected. In recent times hypergraphs have been proved to perform better in the context of particle flow event reconstructions~\cite{DiBello:2022iwf} and thus it may be worth investigating their impact on $\PH \rightarrow \PQs\PAQs$ reconstruction.

\subsubsection*{Exotics Higgs decays}
\label{sec:HtoSS:ExoHiggs}
Strange tagging can provide a new experimental handle to extend the searches for exotic Higgs decays to scalars which decay to pairs of strange quarks. This is a final state that is inaccessible at the HL-LHC because of the lack of $\PQs$-tagging. When the mass of the scalar $a$ is smaller than 3~GeV, only decays to gluons, $\PQu$, $\PQd$, and $\PQs$ quarks are allowed. While the HL-LHC will search for gluons and $\PQu$/$\PQd$/$\PQs$ quark final states, it is not able to distinguish among the three light flavour topologies. An electron-positron Higgs factory with particle ID capabilities up to 30~GeV can extend the discovery reach to exotic Higgs decays to scalars and, in particular, enable the $\PH \to \Pa\Pa \to \PQs\PAQs\PQs\PAQs$ channel for full coverage of all hadronic final states. Identifying $a \to s \bar{s}$ decays presents various experimental challenges that need to be studied. First, given the very light mass of the scalars, the strange quark anti-quark pairs will have a very small opening angle, creating a boosted jet topology even if the jets have relatively low momentum. This means that PID detectors will need to be able to provide efficient pion-kaon separation for tracks produced at very small opening angles. This is likely not to be an issue for a RICH detector placed at a large radius, but detailed studies based on full simulations of cluster counting in a drift chamber tracker will be critical to establish its performance in this challenging final state. The plans for this study consist of i) determining the physics sensitivity of $\PH \to \Pa\Pa \to \PQs\PAQs\PQs\PAQs$ under different assumptions of ideal particle ID, and ii) carrying out systematic studies using full simulation to evaluate the performance of cluster counting PID and RICH as a function of the properties of $\Pa \to \PQs\PAQs$ jets. Such studies are expected to provide valuable information for the design of Higgs Factory subdetectors. Beyond particle ID, highly boosted jets from exotic Higgs decays to light ($\leq 5$~GeV) decays provide a challenging benchmark for particle flow reconstruction studies.


\subsubsection{Outlook}
\label{sec:HtoSS:Conclusion}

The unique potential of \epem colliders to detect Higgs decays into strange-quark-initiated jets offers unprecedented opportunities for probing the properties of the Higgs boson, particularly in relation to its interactions with strange quarks, which will remain inaccessible at the High-Luminosity LHC. The ambitious studies performed in the last years highlight that, pursuing this research line has a promising potential to enable discoveries in the Higgs sector if a combination of advanced theoretical models, refined jet-tagging algorithms, and specialised detector technologies are developed in the next decades.
Strange-tagging capabilities rely on detector features, such as kaon-pion separation at high momentum, and advanced machine learning techniques which have emerged as essential tools for isolating strange jets from background. 
The impact of various detector configurations on tagging performance has been evaluated. In addition to the key particle identification capabilities enabling kaon-pion separation, vertexing and calorimetry are also crucial for flavour tagging, as any compromise in their performance would hinder the physics reach. The time-of-flight subsystem offers only marginal improvement for strange-tagging.
Z-pole calibration remains a foundational assumption, targeting per mille-level precision in particle tagging. This calibration underpins the precise measurements required for Higgs studies and aids in refining strange-tagging algorithms essential for the high granularity expected from future detectors.
Moreover, the likelihood of observing decays of the Standard Model Higgs boson into strange quarks depends on the integrated luminosity that will be available at the next $\epem$ collider, and this in turn depends on the choice of accelerator technology and the prioritisation in running schedules at different centre-of-mass energies. These preliminary studies rely on the study of the $\PZ\PH$ associated production at 240/250~GeV centre-of-mass energy. A detailed analysis of Higgs boson production (including $\PGn\PGn \PH$ production) at 365/500~GeV centre-of-mass energy is not available at this time.  
Along with probing the Standard Model, strange tagging and particle identification can provide new probes for physics Beyond the Standard Model via, for example, exotics Higgs decays into scalar particles. With continued innovation in these areas, future Higgs factories offer the possibility of unlocking new insights into the Higgs mechanism, deepening our understanding of fundamental particle interactions, and potentially uncovering BSM physics.


\subsection{Other rare Higgs couplings}

\subsubsection{Impact of vector-like quarks on Yukawa couplings}
\editor{Nudžeim Selimović - abstract 17}
\subsubsection*{Introduction}
The Yukawa couplings of the first- and second-generation quarks are notoriously difficult to measure due to their smallness and the difficulty to tag the light quarks. Nevertheless, various proposals have shown that the first generation quark Yukawa couplings can be constrained to $\mathcal{O}(\text{few}\; 100)$ times their SM value, the strange Yukawa coupling to $\mathcal{O}(\text{few} \; 10)$ times the SM value and to $\mathcal{O}(1)$ for the charm Yukawa coupling at the HL-LHC either measuring exclusive radiative final states with vector mesons~\cite{Bodwin:2013gca, Konig:2015qat, Perez:2015lra, Perez:2015aoa, Alte:2016yuw, deBlas:2019rxi, dEnterria:2023wjq}, reconstructing charm jets from Higgs decays or associated production~\cite{Brivio:2015fxa, Aguilar-Saavedra:2020rgo}, or exploiting kinematic distributions of onshell or offshell Higgs production~\cite{Delaunay:2013pja, Zhou:2015wra, Yu:2016rvv, Bishara:2016jga, Soreq:2016rae,  Bonner:2016sdg, Carpenter:2016mwd, Yu:2017vul, deBlas:2019rxi, Alasfar:2019pmn, Alasfar:2022vqw, Vignaroli:2022fqh, Balzani:2023jas, Yan:2023xsd}. Instead, at $\Pep \Pem$ colliders tagging of the light quark flavours might be possible \cite{Duarte-Campderros:2018ouv, Kamenik:2023hvi, Liang:2023wpt}, improving the bounds that can be set on the light Yukawa couplings. 
\par
Here, we will discuss \textit{all} the simplified models that lead to modifications of the light quark Yukawa couplings when matched at tree-level to Standard Model effective field theory (SMEFT) without involving any
$s$-channel resonances decaying into di-jets.\footnote{The latter is for instance the case in the two-Higgs doublet model discussed in the context of enhanced light quark Yukawa couplings in Ref.~\cite{Egana-Ugrinovic:2019dqu, Giannakopoulou:2024unn} and models with an extra scalar and a vector-like quark representation. } The models we consider include two states of vector-like quarks (VLQs) that couple to the quarks of the light generations. In addition to causing deviations in the light quark Yukawa couplings, the models also generate tree-level modifications to the fermion couplings with the $\PZ$ and $\PW$ bosons. This implies that they can be constrained by electroweak precision tests and hence potentially discovered at the FCC-ee. In this note, we discuss the implications of the FCC measurements on the question of how large the light quark Yukawa couplings can become within the context of these models. This extends the discussion of Ref.~\cite{Erdelyi:2024sls}.  

\subsubsection*{UV models}
\begin{table}[t]
    \centering
        \renewcommand{\arraystretch}{1.3}
    \begin{tabular}{|c|c||c|c||c|c|}
    \hline
    Model & VLQs &   Model & VLQs & Model & VLQs \\
    \hline
    1     & $(3,1)_{2/3}+(3,2)_{1/6}$    &  4   & $(3,1)_{-1/3}+(3,2)_{-5/6}$    & 7    & $(3,2)_{1/6}+(3,3)_{2/3}$ \\
    2     & $(3,1)_{-1/3}+(3,2)_{1/6}$ &  5   & $(3,2)_{1/6}+(3,3)_{-1/3}$       & 8    & $(3,2)_{7/6}+(3,3)_{2/3}$   \\
    3     & $(3,1)_{2/3}+(3,2)_{7/6}$    &  6   & $(3,2)_{-5/6}+(3,3)_{-1/3}$    &      &                                           \\
    \hline
    \end{tabular}
    \caption{UV models containing VLQs, with charges under the SM gauge group (SU(3)$_c$, SU(2)$_L$)$_{\text{U}(1)_Y}$.}
    \label{tab:VLQpairsmodels}
\end{table}
The leading contribution to the deviations of the up (down) quark Yukawa couplings with respect to their SM value is provided by the $\mathcal{O}_{\PQu\PH}$ ($\mathcal{O}_{\PQd\PH}$) dimension-six operators in the Warsaw basis \cite{Grzadkowski:2010es}. Such operators can be generated at tree-level by integrating out vector-like quarks (VLQs). Starting from the VLQ interaction Lagrangian reported in Ref.~\cite{deBlas:2017xtg}, we construct eight models containing two representations of VLQs with the requirement that there exists a VLQ-VLQ-Higgs interaction term. To this end, the dimension-6 contribution to $\mathcal{O}_{\PQu\PH}$ or respectively $\mathcal{O}_{\PQd\PH}$ does not get suppressed by the small SM quark Yukawa couplings. \Cref{tab:VLQpairsmodels} details the particle content of each of the UV models considered.

In order to study these models they were matched at the one-loop level to the Warsaw basis using \texttt{SOLD}~\cite{Guedes:2023azv} and \texttt{Matchete} \cite{Fuentes-Martin:2022jrf}. At tree-level, apart from $\mathcal{O}_{\PQu\PH}$  and $\mathcal{O}_{\PQd\PH}$, they match to operators of the class $\psi^2 \PH^2 D$ that lead to modifications in the fermion couplings to the massive gauge bosons. Furthermore, we get one-loop contributions to operators of type $\PH^4 D^2$ and $X^2 \PH^2$, which can be constrained by electroweak precision probes and Higgs physics. Including the full one-loop matching allows us in particular not only to bound the coupling of the VLQ with the Higgs doublet and a light quark but also the VLQ-VLQ-Higgs coupling. In this way, we become sensitive to all new physics couplings that enter the Yukawa modifications.

Apart from Higgs physics and electroweak precision tests, the models receive constraints from direct searches currently excluding masses up to $\SI{1.6}{\tera\electronvolt}$~\cite{ATLAS:2024zlo,Erdelyi:2024sls}. At the HL-LHC one can expect probing mass scales up to $\SI{2.4}{\tera\electronvolt}$ with direct searches for VLQs decaying to $\PW$ bosons and a jet \cite{Freitas:2022cno}.
Furthermore, the models can be constrained by flavour physics. Assuming that the VLQs couple only to one generation, bounds on SU(2)$_L$ singlets and triplets stem from unitarity constraints on the CKM matrix, while the case of the triplets is in addition constrained by $\Delta F=2$ transitions~\cite{Bona:2024bue, ParticleDataGroup:2024cfk}. 
Currently, those bounds range from $\SI{1.6}{}$ to $\SI{3.2}{\tera\electronvolt}$ times the coupling of the singlet/triplet to the light SM quark and the Higgs field \cite{Erdelyi:2024sls}. 

\subsubsection*{Results}
\begin{figure}
    \centering
    \includegraphics[width=0.48\linewidth]{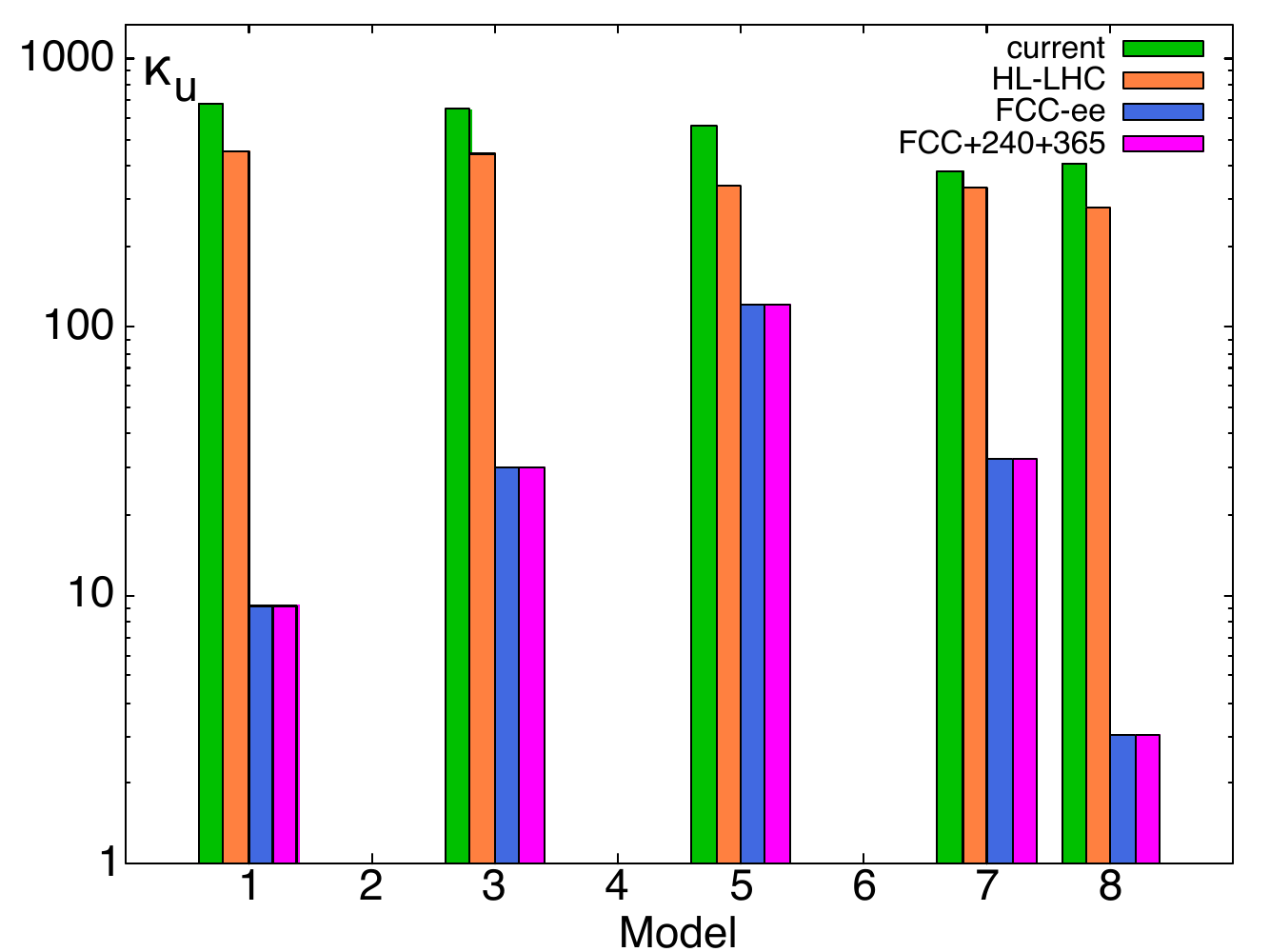}
    \,\,
    \includegraphics[width=0.48\linewidth]{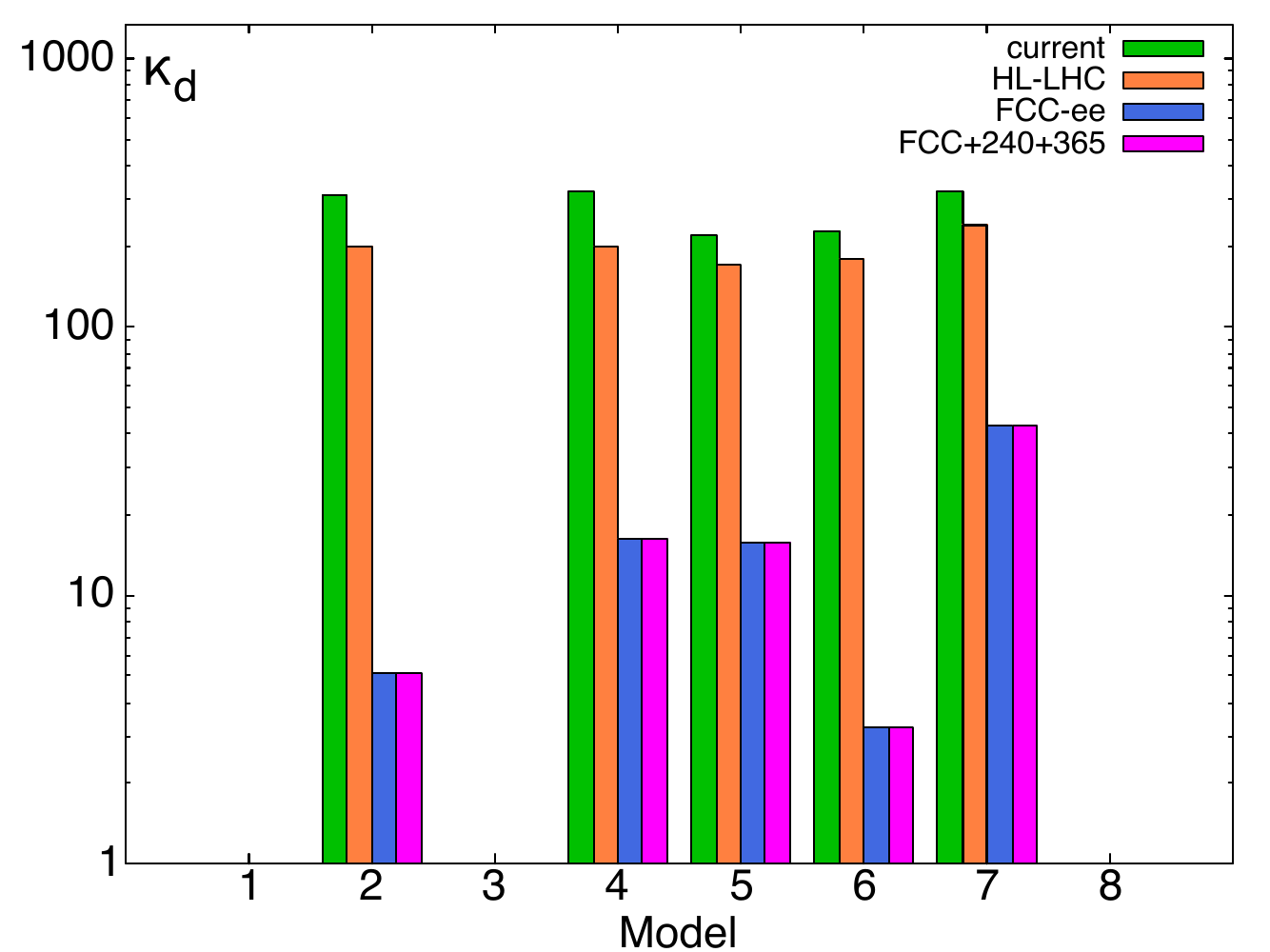}\\
    \includegraphics[width=0.48\linewidth]{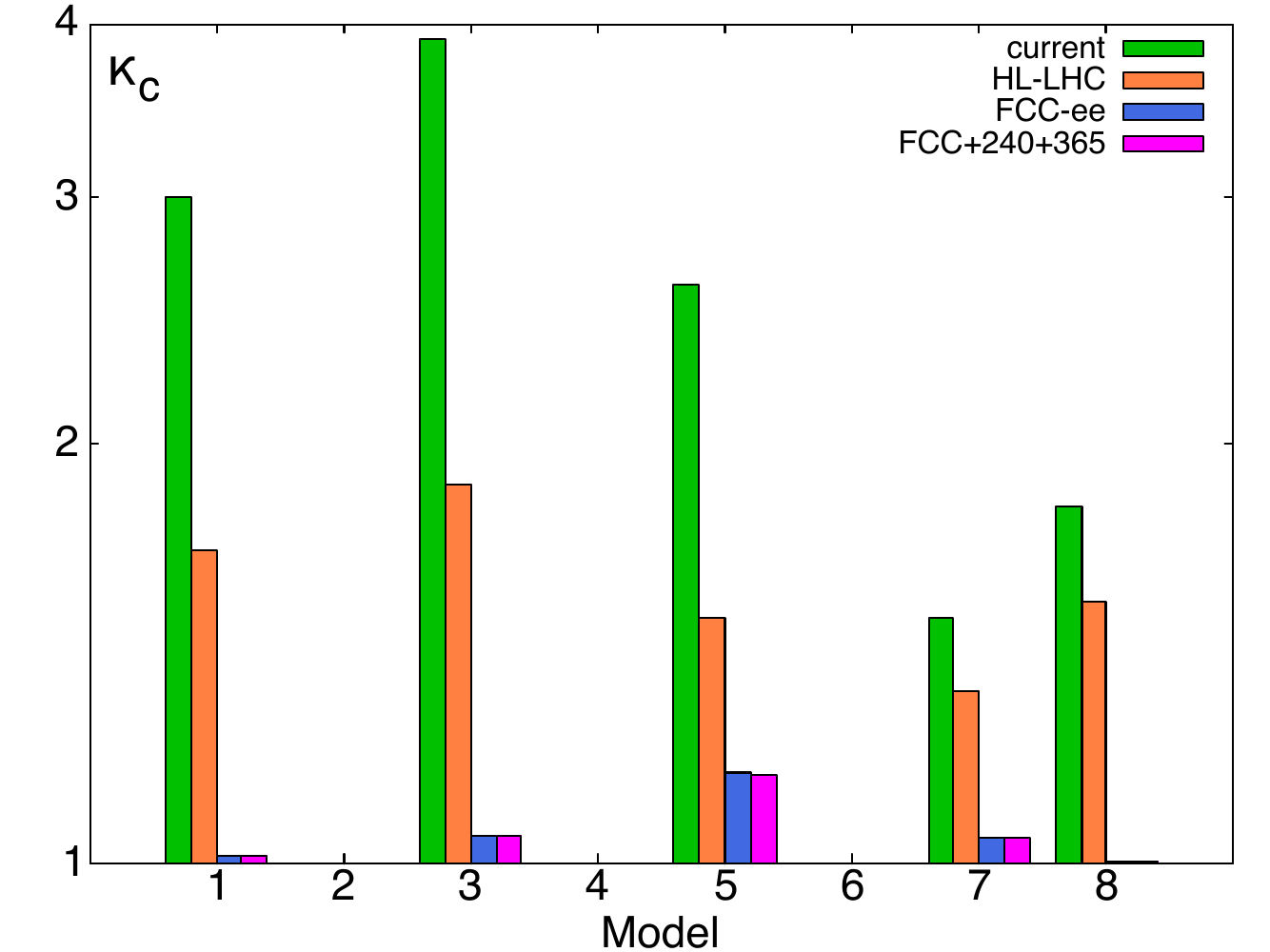}
    \,\,
    \includegraphics[width=0.48\linewidth]{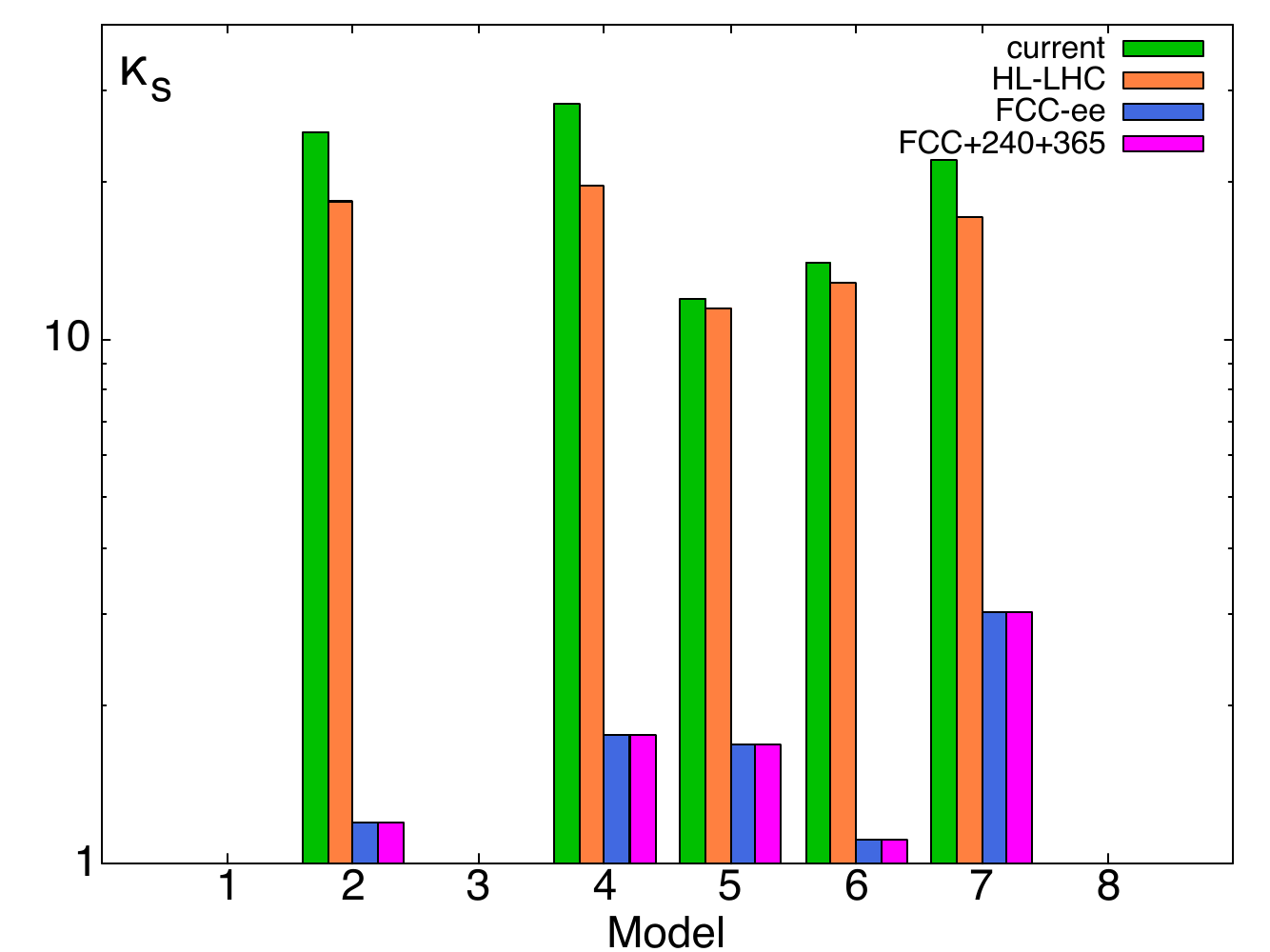}
    \caption{Maximum allowed coupling enhancements at a $95\%$ CL for the considered models.}
    \label{fig:kappau}
\end{figure}

To find the largest allowed values for the coupling enhancements $\kappa_{\PQq}=g_{\PH \PQq \PQq}/g_{\PH\PQq \PQq}^{\text{SM}}$ ($\PQq=\PQu$, $\PQd$, $\PQc$ and $\PQs$), four different fits were performed and are summarised in Fig. \ref{fig:kappau}, where one can observe that not all models generate both up- and down-type coupling enhancements: Models 1, 3, and 8 can only enhance the up-type couplings, Models 2, 4, and 6 affect the down-type couplings and, finally, Models 5 and 7 can modify both.

Going from left to right, the values labeled as ``current'' were obtained in Ref.~\cite{Erdelyi:2024sls}, considering the current Higgs, electroweak, and flavour physics data. To pass direct detection constraints, the mass scale of the VLQs was set to $\SI{1.6}{\tera\electronvolt}$. The remaining three sets of $\kappa_{\PQq}$-values were constructed considering projections for future colliders. To this end, the mass of the VLQs was taken to be $\SI{2.4}{\tera\electronvolt}$ \cite{Freitas:2022cno}. The next set of $\kappa_{\PQq}$-values, ``HL-LHC'', is again the result of a combined fit, however, the current Higgs data was replaced by the HL-LHC CMS projections of Ref.~\cite{Cepeda:2019klc}, using only total signal strengths. Differential measurements might improve on this \cite{Bishara:2016jga, Soreq:2016rae, Balzani:2023jas}. 

The ``FCC-ee'' set of coupling enhancements arises from the electroweak fit based on Refs.~\cite{DeBlas:2019qco, Bernardi:2022hny} taking into account the Z-pole run. We also investigated a combined electroweak and Higgs fit using additional projections for the 240 GeV and 365 GeV runs, ``FCC+240+365'', using the precision on the Higgs couplings reported in Ref.~\cite{FCC:2018evy}. The absence of significant changes in the largest allowed values of $\kappa_{\PQq}$ indicates the dominance of the Z-pole run in probing the light quark Yukawas in these scenarios.

Overall, \Cref{fig:kappau} summarises the exceptional potential of the FCC-ee to uniquely probe models featuring pairs of vector-like quarks and giving rise to light quark Yukawa enhancement.

Finally, we note that, in the same spirit, new physics scenarios involving vector-like leptons or new scalar fields interacting with the SM leptons have the potential to significantly enhance the electron Yukawa coupling. A systematic classification of simplified models representing such new physics frameworks, along with their associated phenomenological implications, has been presented in Ref.~\cite{Erdelyi:2025axy}. Unlike the case of quarks, these models are subject to stringent constraints from the anomalous magnetic moment of the electron which can provide constraints comparable to those imposed by the FCC-ee. However, certain classes of models allow for significant deviations in the electron-Yukawa coupling, which can only be effectively probed in a dedicated run at the Higgs pole mass.

\subsubsection{\texorpdfstring{Higgs-electron Yukawa coupling}{Electron Yukawa}}
\label{sec:eYukawa}
\subsubsection*{Physics potential}
\editors{David D'Enterria}

Confirming the mechanism of mass generation for the stable visible elementary particles of the universe, composed of $\PQu$ and $\PQd$ quarks plus the electron (and neutrinos), is experimentally very challenging because of the low masses of the first-generation fermions and thereby their small Yukawa couplings to the Higgs field (the neutrino mass generation remains a BSM problem in itself). In the SM, the Yukawa coupling of the electron is $y_{\Pe} = \sqrt{2} m_{\Pe}/v = 2.8\times 10^{-6}$ for $m_{\Pe}(m_{\PH}) = 0.486\times 10^{-3}$\,GeV and Higgs vacuum expectation value $v = (\sqrt{2}\mathrm{G_F})^{-1/2} = 246.22$\,GeV, and measuring it appears hopeless at hadron colliders because the $\PH \to \epem$ decay has a tiny partial width due to its dependence on the ${\Pe}^\pm$ mass squared:
\begin{equation}
\Gamma(\PH\to\epem) = 
\frac{\mathrm{G_F}m_{\PH}m_{\Pe}^2}{4\sqrt{2}\,\pi}\left(1-\frac{4\,m^2_{\Pe}}{m^2_{\PH}}\right)^{3/2} = 2.14\times 10^{-11} \GeV\,,
\label{eq:Gamma_H_ee}
\end{equation}
which corresponds to a $\mathcal{B}(\PH\to\epem)\approx 5\times 10^{-9}$ branching fraction for the SM Higgs boson with $m_{\PH}=125$\,GeV mass and $\Gamma_{\PH} = 4.1$\,MeV total width. At the LHC, such a final state is completely swamped by the Drell--Yan $\epem$ continuum whose cross section is many orders of magnitude larger. The first LHC searches with about 20~fb$^{-1}$ of pp collisions at 8~TeV, assuming the SM Higgs production cross section, lead to an upper bound on the branching fraction of $\mathcal{B}(\PH\to\epem) < 1.9\times 10^{-3}$ at 95\% CL, corresponding to an upper limit on the Yukawa coupling $y_{\Pe}\propto \mathcal{B}(\PH\to\epem)^{1/2}$ of 600 times the SM value~\cite{CMS:2014dqm}. Such results were further updated in~\cite{ATLAS:2019old,CMS:2022urr}, exploiting about 140~fb$^{-1}$ of pp data at $\sqrts = 13$~TeV and reaching an observed upper limit of $\mathcal{B}(\PH\to\epem) < 3.0 \times 10^{-4}$ at 95\% CL. This latter value translates into a current upper bound on the Higgs boson effective coupling modifier to electrons of $|\kappa_{\Pe}| < 240$. Assuming that the sensitivity to the $\PH \to \epem$ decay scales simply with the square root of the integrated luminosity, the HL-LHC phase with a $\LumiInt = 2\times 3\,\mathrm{ab}^{-1}$ data sample (combining ATLAS and CMS results) will result in $y_{\Pe}\lesssim 100 y^\mathrm{\textsc{sm}}_{\Pe}$. Based on searches for the similar $\PH \to \PGm\PGm$ channel, one can expect upper limits on $\mathcal{B}(\PH\to\epem)$ to be further improved by factors of about four by adding more Higgs production categories and using advanced multivariate analysis methods, eventually reaching $y_{\Pe}\lesssim 50 y^\mathrm{\textsc{sm}}_{\Pe}$ at the end of the HL-LHC.

About ten years ago, it was first noticed that the unparalleled integrated luminosities of $\LumiInt \approx 10$\,ab$^{-1}$/year expected at $\sqrts = 125$\,GeV at the FCC-ee would make it possible to attempt an observation of the direct production of the scalar boson and thereby directly measure the electron Yukawa coupling~\cite{dEnterria:2014,dEnterria:2017dac}. Subsequently, various theoretical~\cite{Jadach:2015cwa,Greco:2016izi,Dery:2017axi,Davoudiasl:2023huk,Boughezal:2024yjk}, simulated data analysis~\cite{dEnterria:2021xij}, and accelerator~\cite{Zimmermann:2017tjv,telnov2020monochromatizationeecolliderslarge,Faus-Golfe:2021udx} works discussed different aspects of the $\epem\to\PH$ measurement. The Feynman diagrams for $s$-channel Higgs production (and its most statistically significant decay, see below) and dominant backgrounds are shown in Fig.~\ref{fig:eeH_diags} (left). 
The resonant Higgs cross section in $\epem$ collisions at a given centre-of-mass (CM) energy $\sqrts$ is theoretically given by the relativistic Breit--Wigner (BW) expression:
\begin{equation}
\sigma_\mathrm{ee\to H} = \frac{4\pi\Gamma_{\PH}\Gamma(\PH\to\epem)}{(s-m_{\PH}^2)^2 + m_{\PH}^2\Gamma_{\PH}^2}.
\label{eq:sigma_H_ee}
\end{equation}
From this expression, it is first clear that an accurate knowledge of the $m_{\PH}$ value is critical to maximize the resonant cross section. Combining three $\epem\to \PH\mathrm{Z}$ measurements at FCC-ee (recoil mass, peak cross section, and threshold scan), a $\mathcal{O}$(2\,MeV) mass precision is achievable~\cite{Azzurri:2021nmy} (\cref{sec:HiggsMass}) before any dedicated $\epem\to\PH$ run. In addition,  the FCC-ee beam energies will be monitored with a relative precision of $10^{-6}$~\cite{Blondel:2021zix}, warranting a sub-MeV accuracy of the exact point in the Higgs lineshape being probed at any moment. For $m_{\PH} = 125$\,GeV, Eq.~(\ref{eq:sigma_H_ee}) gives $\sigma_\mathrm{ee\to H} = 4\pi\mathcal{B}(\PH\to\epem)/m_{\PH}^2 = 1.64$~fb as peak cross section. Two effects, however, lead to a significant reduction of the Born-level result: (i) initial-state $\gamma$ radiation (ISR) depletes the cross section and generates an asymmetry of the Higgs lineshape, and (ii) the actual beams are never perfectly monoenergetic, \ie\ the collision $\sqrts$ has a spread $\delta_{\sqrts}$ around its central value, 
further leading to a smearing of the BW peak. The reduction of the BW cross section due to initial-state photon emission(s) is of factor of 0.35 and leads to $\sigma_\mathrm{ee\to H} = 0.57$~fb~\cite{Jadach:2015cwa}. The additional impact of a given CM energy spread on the Higgs BW shape can be quantified through the convolution of BW and Gaussian distributions, \ie\ a relativistic Voigtian function. Figure~\ref{fig:eeH_diags} (right) shows the Higgs lineshape for various $\delta_{\sqrts}$ values. The combination of ISR plus $\delta_{\sqrts} = \Gamma_{\PH} = 4.1$\,MeV reduces the peak Higgs cross section by a total factor of 0.17, down to $\sigma_\mathrm{ee\to H} = 0.28$~fb. Though tiny, the cross section for any other $\epem\to\PH$ production process, through W and Z loops, is further suppressed by the electron mass for on-shell external fermions (chirality flip) and has negligible cross sections~\cite{Altmannshofer:2015qra}. 

\begin{figure}[htpb!]
\centering
\includegraphics[width=0.27\columnwidth]{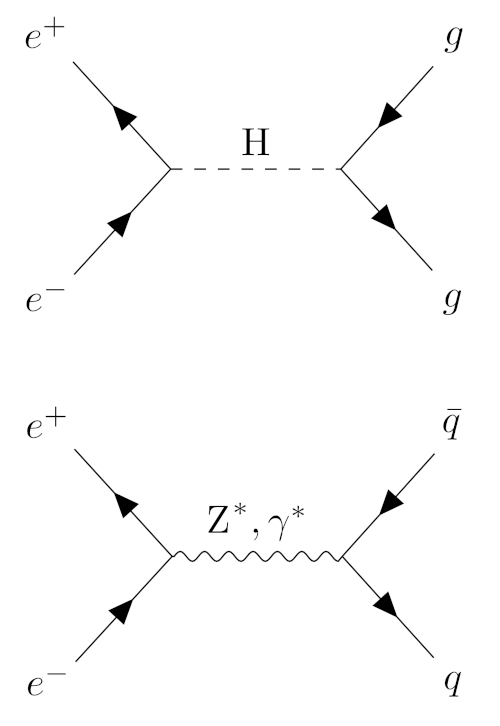}\hspace{0.7cm}
\includegraphics[width=0.53\columnwidth]{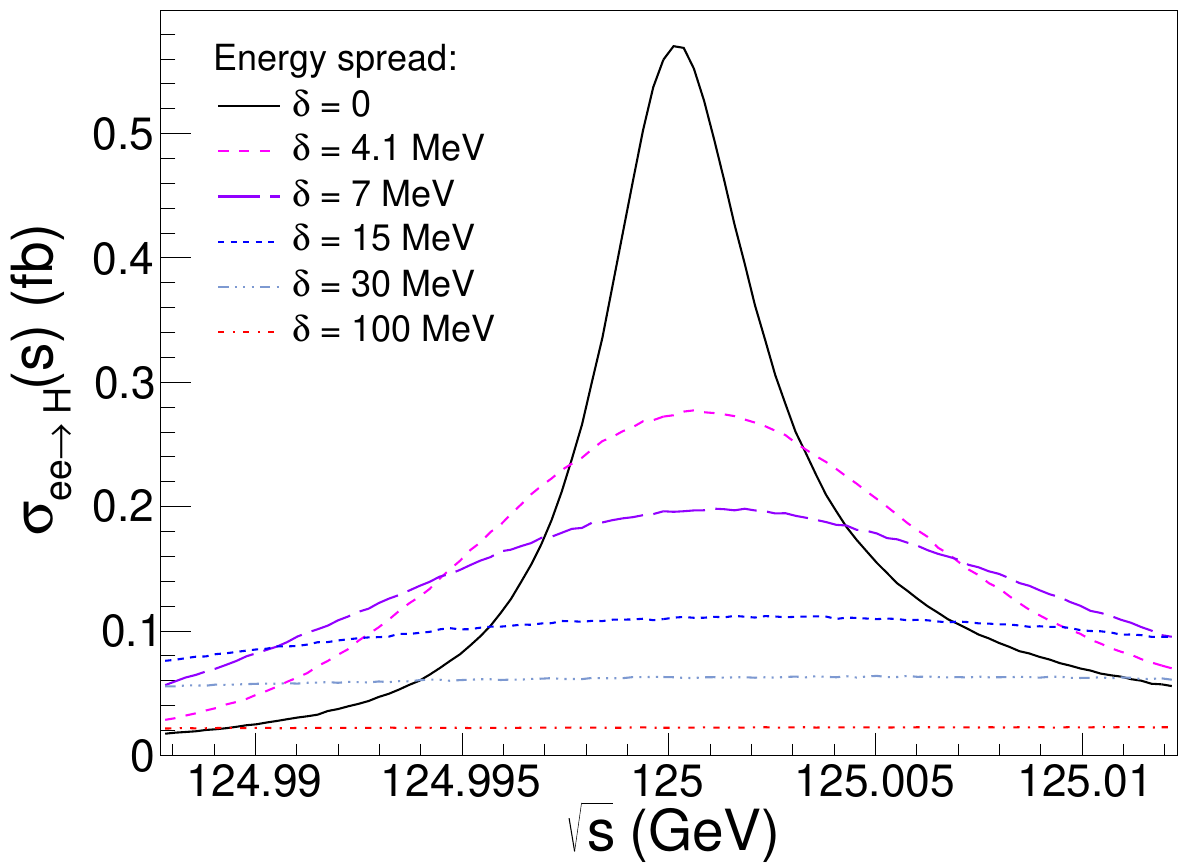}
\caption{Left: Diagrams for the $s$-channel production of the Higgs boson decaying into two gluon jets (upper) and reducible $\PZ^*$ quark dijet backgrounds (lower) in  $\epem$ at $\sqrts = 125$~GeV. 
Right: Resonant Higgs production cross section at $\sqrts = 125$~GeV, including ISR effects, for several $\epem$ CM energy spread values: $\delta_{\sqrts}$ = 0, 4.1, 7, 15, 30, and 100\,MeV~\cite{Jadach:2015cwa}.
\label{fig:eeH_diags}}
\end{figure}

The three main challenges of the $\epem\to \PH$ measurement have been discussed in Ref.~\cite{dEnterria:2021xij}: (i) the need to accurately know (within MeV's) beforehand the value of the Higgs boson mass where to operate the collider, (ii) the smallness of the resonant cross section (few hundred ab) due to ISR and beam  energy spread ($\delta_{\sqrts}$) that requires to monochromatize the beams, \ie\ to reduce $\delta_{\sqrts}$ at the few MeV scale, while still delivering large (few ab$^{-1}$) integrated luminosities $\LumiInt$, and (iii) the existence of multiple backgrounds with orders-of-magnitude larger cross section than the Higgs signal decay channels themselves.  As mentioned above, the knowledge of $m_{\PH}$ with a few MeV accuracy seems feasible at FCC-ee as per dedicated studies reported in \cref{sec:HiggsMass}. The latest developments of the monochromatization schemes at FCC-ee, point (ii), are summarised below. The challenge (iii) has been addressed in detail in Ref.~\cite{dEnterria:2021xij} where a generator-level study was performed choosing a benchmark monochromatization point leading to $(\delta_{\sqrts},\LumiInt)=(4.1\,\mathrm{MeV},10\,\mathrm{ab}^{-1})$, corresponding to a peak $s$-channel cross section of $\sigma_{\epem\to\PH} = 280$\,ab, and 2\,800 Higgs bosons produced. The strategy to observe the resonant production of the Higgs boson is based on identifying final states consistent with any of the H decay modes, that lead to a small excess (but, hopefully, statistically significant when combined together) of the measured cross sections with respect to the theoretical expectation for their occurrence via background processes alone, involving $\PZ^*, \gamma^*$, or $t$-channel exchanges. For this purpose, large simulated event samples of signal and associated backgrounds have been generated with the \pythia~8 MC code~\cite{Sjostrand:2014zea} for 11 Higgs boson decay channels. A simplified description of the expected experimental performances has been assumed for the reconstruction and (mis)tagging of heavy-quark ($\PQc$, $\PQb$) and light-quark and gluons ($\PQu\PQd\PQs\Pg$) jets, photons, electrons, and hadronically decaying tau leptons. Generic preselection criteria have been defined targeting the 11 Higgs boson channels, suppressing reducible backgrounds while keeping the largest fraction of the signal events. A subsequent multivariate analysis of $\mathcal{O}(50)$ kinematic and global topological variables, defined for each event, has been carried out. BDT classifiers have been trained on signal and background events, to maximize the signal significances for each individual channel. The most significant Higgs decay channels are found to be $\PH\to \Pg\Pg$ (for a gluon efficiency of 70\% and a $\PQu\PQd\PQs$-for-$\Pg$ jet mistagging rate of 1\%), and $\PH\to\PW\PW^*\to\ell\nu\,jj$. The digluon final state is the most sensitive channel to search for the resonant Higgs boson production (Fig.~\ref{fig:eeH_diags} left, upper) because it features a moderately large branching fraction ($\mathcal{B}\approx 8\%$) while the irreducible $\PZ^*\to\Pg\Pg$ background is forbidden by the Landau--Yang theorem. The most important experimental challenge is to reduce the light-quark for gluon mistagging rate to the 1\% level (while keeping the efficiency for $\PH\to\Pg\Pg$ channel at 70\%) to keep the overwhelming $\PZ^*\to \uubar, \ddbar, \ssbar$ backgrounds (Fig.~\ref{fig:eeH_diags} left, lower) under control. Such a mistagging rate is a factor of about 7 times better than the current state-of-the-art for jet-flavour tagging algorithms~\cite{Bedeschi:2022rnj}, but it is not unthinkable after all the experimental and theoretical improvements in our understanding of parton radiation and hadronisation expected after 15 years of studying light and heavy quarks and gluon jets in the clean \epem environment at the FCC-ee~\cite{Proceedings:2017ocd}.

\begin{figure}[htbp!]
\centering
\includegraphics[width=0.49\textwidth]{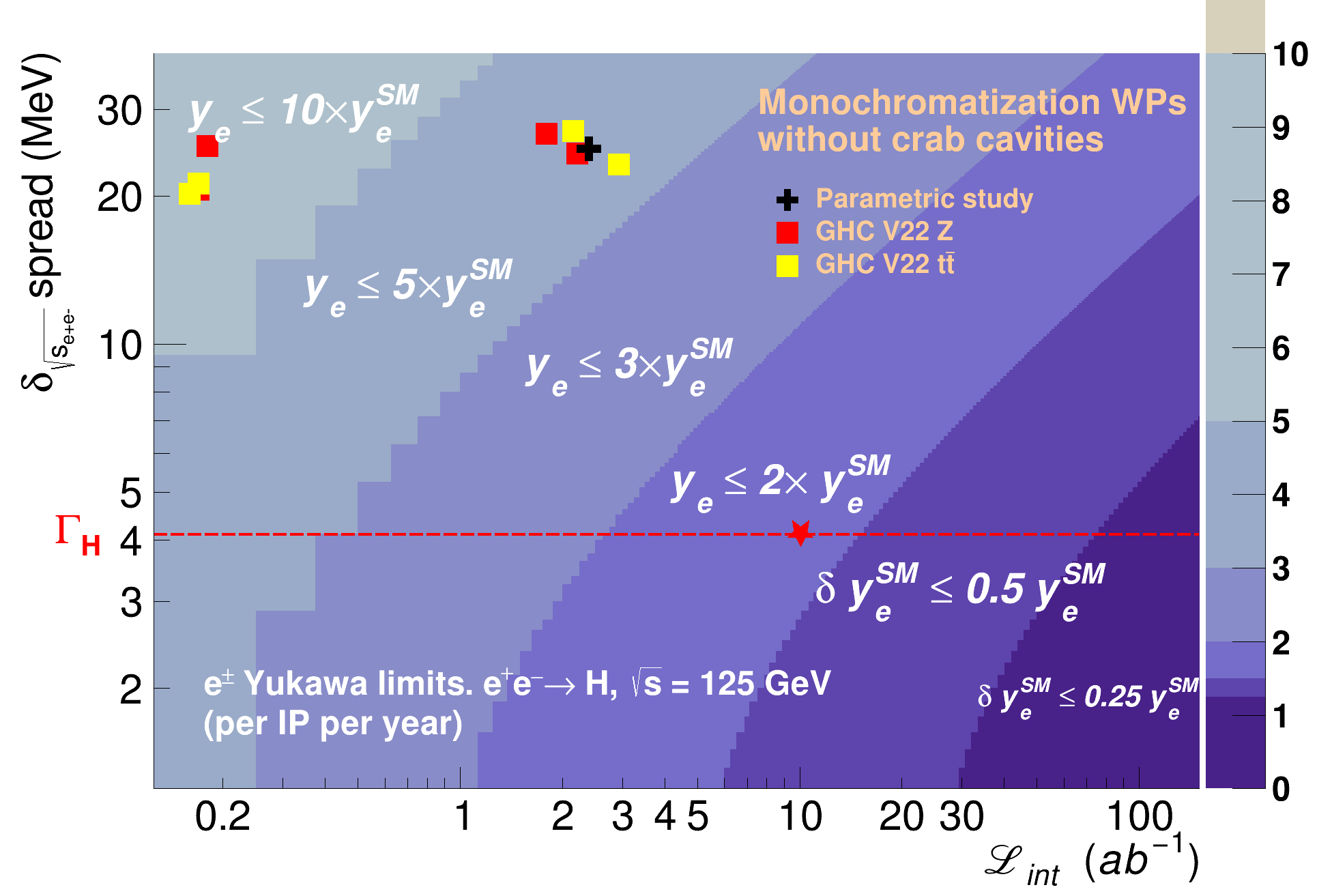}
\includegraphics[width=0.49\textwidth]{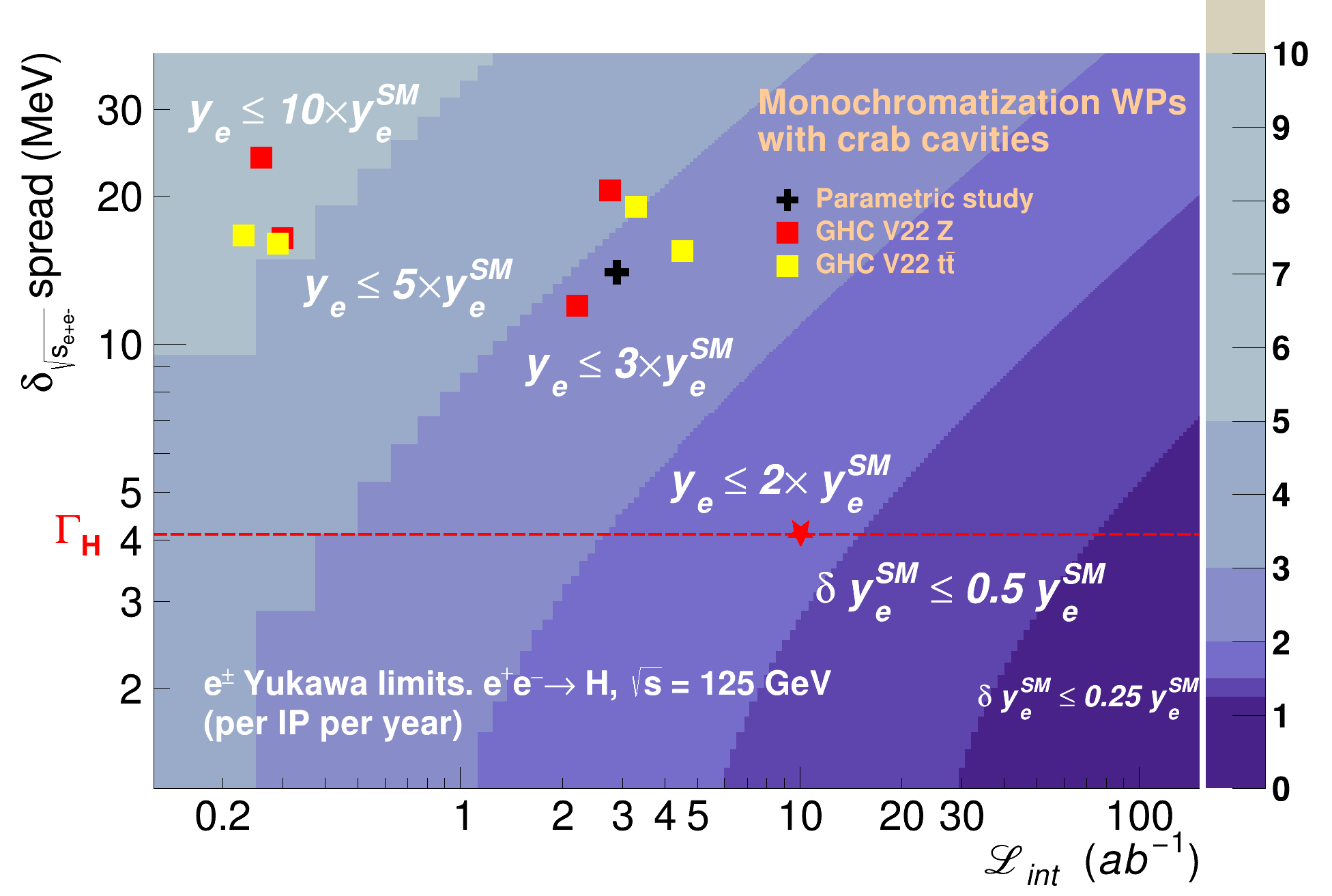}
\caption{Upper limits contours (95\% CL) on the electron Yukawa $y_{\Pe}$ in the CM-energy-spread $\delta_{\sqrt{s}}$ vs.\ integrated-luminosity $\mathcal{L}_{\text{int}}$ plane, without (left) and with (right) crab cavities. The red star over the $\delta_{\sqrts} = \Gamma_{\PH} =4.1$\,MeV red-dashed line, indicates the reference point assumed in the physics simulation analysis~\cite{dEnterria:2021xij}.  The black cross indicates the previously achieved working point with self-consistent parametric monochromatization~\cite{ValdiviaGarcia:2022nks,Faus-Golfe:2021udx}. The red and yellow squares indicate the monochromatization points based on simulations of the ``GHC V22 Z'' and ``GHC V22 $\ttbar$'' optics, respectively~\cite{Zhang:2024sao}.
\label{fig:eeH_signif}}
\end{figure}

Combining all results for an accelerator operating at $(\delta_{\sqrts},\LumiInt)=(4.1\,\mathrm{MeV},10\,\mathrm{ab}^{-1})$, a $1.3\sigma$ signal significance can be reached for the direct production of the Higgs boson, corresponding to an upper limit on the electron Yukawa coupling at 1.6 times the SM value: $|y_{\Pe}|<1.6|y^\mathrm{\textsc{sm}}_{\Pe}|$ at 95\% confidence level (CL), per FCC-ee interaction point (IP) and per year. Based on this benchmark result and the parametrised dependence of the resonant Higgs cross section on $\delta_{\sqrts}$ (Fig.~\ref{fig:eeH_diags}, right), bidimensional maps of $\epem\to\PH$ significances and electron-Yukawa sensitivities have been determined in the $(\delta_{\sqrts},\LumiInt)$ plane. Figure~\ref{fig:eeH_signif} shows the 95\% CL upper limit contours on the electron Yukawa coupling strength as a function of the energy spread and integrated luminosity with the red star (on the red-dashed line corresponding to a reference monochromatized CM energy spread equal to the Higgs boson width) indicating the result of this benchmark study.

\subsubsection*{Electron Yukawa results at FCC-ee with monochromatization}

\begin{figure}[htbp!]
\centering
\includegraphics[width=0.9\textwidth]{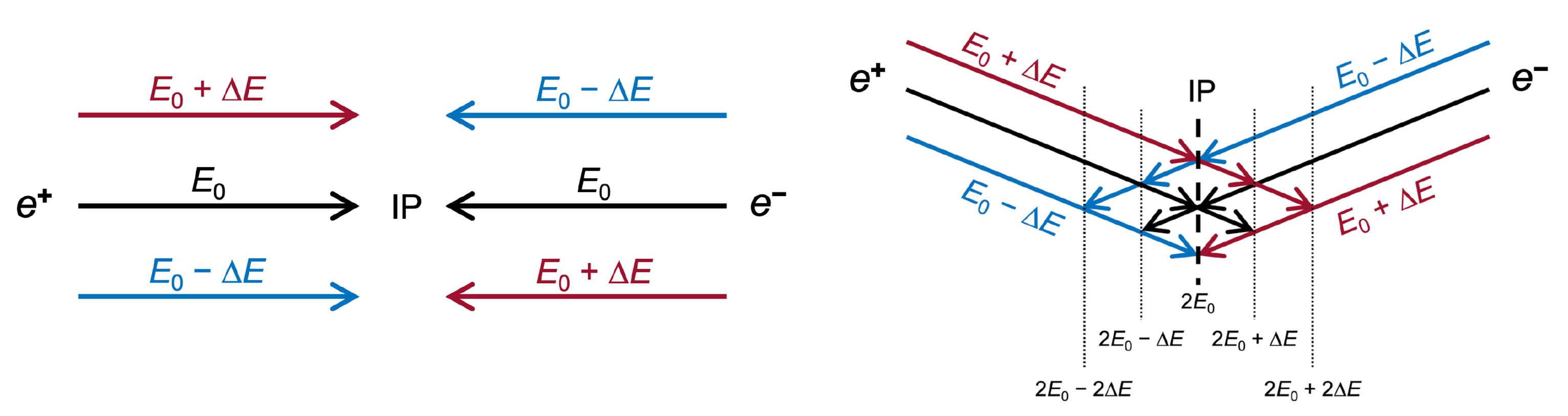}
\caption{Schematic of the principle of monochromatization shown for head-on collisions (left) and for collisions with a crossing angle (right).  In both cases opposite-sign correlations between the transverse position in the beam and energy lead to a reduction in the spread of collision energy compared with the uncorrelated case.}
\label{fig:monochromschem}
\end{figure}

A key requirement of the $\epem\to\PH$ run to actually produce a minimum amount of Higgs boson is to have a CM energy spread much smaller than the $\delta_{\sqrts}\approx 50$~MeV value of the conventional collision scheme at 125~GeV, caused by synchrotron radiation. Reducing $\delta_{\sqrts}$ to the few-MeV level of the natural SM Higgs width requires beam ``monochromatization''~\cite{Renieri:1975wt}, a technique that relies on creating opposite correlations between spatial position and energy deviations within the colliding beams with nominal beam energy ($E_0$). In such a configuration, the CM energy $\sqrts = 2E_0 + \mathcal{O}(\Delta E)^2$ is reduced without necessarily decreasing the inherent energy spread of the two individual beams. Figure~\ref{fig:monochromschem} shows a schematic of the principle of monochromatization for beams that collide head on and for those that collide with a crossing angle ($\theta_c$).  The current baseline design of FCC-ee corresponds to the crossing-angle configuration, as it is not foreseen to deploy crab cavities.  In both configurations, the correlations between transverse (either horizontal or vertical) position in the beam and energy lead to a spread in collision energy that is lower than in the uncorrelated case.

In terms of beam optics, monochromatization can be achieved by generating a nonzero dispersion function with opposite signs for the two beams at the IP, by adding the necessary components within the interaction region (IR) of the collider. A nonzero dispersion function at the IP in the horizontal and/or vertical directions ($D_{x,y}^{\ast} \neq 0$) enlarges the IP transverse beam size ($\sigma_{x,y}^{\ast}$) which in turn affects the luminosity, $\mathcal{L} \propto 1/(\sigma_{x}^{\ast} \sigma_{y}^{\ast})$. The monochromatization factor ($\lambda$) is defined as:
\begin{equation}
    \lambda = \sqrt{1 + \sigma_{\delta}^2 \left( \frac{D_x^{\ast 2}}{\varepsilon_{x} \beta_{x}^{\ast}} + \frac{D_y^{\ast 2}}{\varepsilon_{y} \beta_{y}^{\ast}} \right) }
\label{eq:lambda_monochrom}
\end{equation}
with $\sigma_{\delta}$ the relative energy spread, $\varepsilon_{x,y}$ the transverse emittances and $\beta_{x,y}^{\ast}$ the betatron functions at the IP. For any value of $\lambda$ achieved, the $\delta_{\sqrts}$ and the $\mathcal{L}$ in the monochromatization operation mode are given by:
\begin{equation}
    \delta_{\sqrts} = \frac{\sqrt{2} E_0 \sigma_\delta}{\lambda} \: \text{ and } \: \mathcal{L} = \frac{\mathcal{L}_0}{\lambda}\:,
\label{eq:delta_sqrts}
\end{equation}
where $\mathcal{L}_0$ represents the luminosity for the same values of $\beta_{x,y}^{\ast}$ but without $D_{x,y}^{\ast}$. Consequently, the design of a monochromatization scheme requires considering both the IR beam optics and the optimization of other collider parameters to maintain the highest possible luminosity. The implementation of monochromatization at the FCC-ee has been continuously improved, starting from self-consistent parametric studies~\cite{ValdiviaGarcia:2016rdg,Zimmermann:2017tjv,Faus-Golfe:2021udx,ValdiviaGarcia:2022nks}. The latest developments~\cite{Zhang:2024sao,Zhang:2024phd} include a detailed study of the IP region optics design exploring different potential configurations and their implementation in the FCC-ee global lattice, along with beam dynamics simulations and performance evaluations including the impact of beamstrahlung (BS). These studies indicate that generating either nonzero $D_{x}^{\ast}$, $D_{y}^{\ast}$, or a $D_{x}^{\ast}$/$D_{y}^{\ast}$ combination are all viable approaches, offering the potential to significantly reduce $\delta_{\sqrts}$ and facilitate the resonant Higgs boson production.

The baseline FCC-ee standard lattice design is the so-called ``Global Hybrid Correction'' (GHC) optics~\cite{Oide:2016mkm,Keintzel:2023myn,vanRiesen-Haupt:2024ore}, designed to operate at four different beam energies of 45.6, 80, 120 and \SI{182.5}{\giga\electronvolt}, allowing physics precision experiments at the Z-pole (Z mode), the W-pair-threshold (WW mode), the ZH-maximum (ZH mode) and above the top-pair-threshold ($\ttbar$ mode), respectively. It allows for four experimental IRs where the $\Pep$ and $\Pem$ beams are brought to collision from the inside outwards with a $\theta_{c}= \SI{30}{\milli\radian}$ angle in the horizontal plane, as well as a potential vertical crab-waist scheme. The most natural way to implement monochromatization in such FCC-ee lattice type is reconfiguring the local chromaticity correction horizontal dipoles to generate a nonzero $D_{x}^{\ast}$, while maintaining the same $\theta_{c}$. A wide $\sigma_{x}^{\ast}$ helps to mitigate the BS impact on $\sigma_{\delta}$, while preserving a small $\sigma_{y}^{\ast}$ is crucial for attaining a high $\mathcal{L}$. Taking into account the reference parameter table~\cite{Keintzel:2023myn} for the FCC-ee GHC lattices with horizontal betatron sizes ($\sigma_{x\beta}^{\ast} = \sqrt{\varepsilon_{x} \beta_{x}^{\ast}}$) at the IP of the order of \si{\milli\meter} and a $\sigma_{\delta,\text{SR}}$ of $\sim$0.05\% at \textit{s}-channel Higgs production energy ($\sim$\SI{125}{\giga\electronvolt}), a $D_{x}^{\ast}$ of around \SI{10}{\centi\meter} is required to achieve a monochromatization factor of 
$\lambda\approx$ 5 to 8 
from Eq.~(\ref{eq:lambda_monochrom}). Guinea-Pig simulations~\cite{guinea-pig} were performed to accurately assess the performance of the FCC-ee GHC monochromatization IR optics, taking into account the BS impact. The particle distribution at the IP was simulated as an ideal Gaussian distribution, comprising 40000 particles, and defined by the following global optical performance parameters: $E_0$, $\sigma_{\delta}$, $\varepsilon_{x,y}$, $\beta_{x,y}^{\ast}$, $D_{x,y}^{\ast}$, $\sigma_{z}$, and $\theta_{c}$. For each configuration, the $\delta_{\sqrts}$ (from the distribution of the CM energy) and $\mathcal{L}$ were determined. Tables~\ref{tab:monochromZ} and~\ref{tab:monochromtt} indicate the values of CM energy spread and luminosity obtained for the best two operation modes: ``GHC V22 Z'' (where the lattice is optimised for operation at the Z pole) and ``GHC V22 $\ttbar$'' (optimised for operation above the $t\bar{t}$ threshold), respectively (in both `V22' designates the 2022 configuration).

\begin{table}[htbp!]
\centering
\begin{tabular}{lccccc}\hline
    Parameter  [Unit]  & Std.\ ZES & ZH4IP & ZH2IP & ZV & ZHV \\\hline
    CM energy spread $\delta_{\sqrts}$ [\si{\mega\electronvolt}] & 69.52 & 26.80 & 24.40 & 25.25 & 20.58 \\
    Luminosity / IP $\mathcal{L}$ [$10^{34}$ \si{\per\centi\meter\squared\per\second}] & 44.8 & 15.0 & 18.4 & 1.46 & 1.42 \\
    Integrated Luminosity / IP / year $\mathcal{L}_{\text{int}}$ [\si{\ab}$^{-1}$] & 5.38 & 1.80 & 2.21 & 0.18 & 0.17 \\\hline
\end{tabular}
\caption{Values of $\delta_{\sqrts}$, $\mathcal{L}$, and $\mathcal{L}_{\text{int}}$ for various setups of the ``FCC-ee GHC V22 Z''  monochromatization IR optics~\cite{Zhang:2024sao}.\label{tab:monochromZ}}
\end{table}

\begin{table}[htpb!]
\centering
\begin{tabular}{lccccc}\hline
    Parameter [Unit]  & Std.\ TES & TH4IP & TH2IP & TV & THV \\\hline
    CM energy spread $\delta_{\sqrts}$ [\si{\mega\electronvolt}] & 67.20 & 27.10 & 23.16 & 20.23 & 21.24 \\
    Luminosity / IP $\mathcal{L}$ [$10^{34}$ \si{\per\centi\meter\squared\per\second}] & 71.2 & 17.9 & 24.5 & 1.37 & 1.42 \\
    Integrated Luminosity / IP / year $\mathcal{L}_{\text{int}}$ [\si{\ab}$^{-1}$] & 8.54 & 2.15 & 2.94 & 0.16 & 0.17 \\
    \hline
\end{tabular}
\caption{Values of $\delta_{\sqrts}$, $\mathcal{L}$, and $\mathcal{L}_{\text{int}}$ for various setups of the ``FCC-ee GHC V22 $\ttbar$'' monochromatization IR optics~\cite{Zhang:2024sao}.\label{tab:monochromtt}}
\end{table}

The physics performances of the different FCC-ee GHC monochromatization IR optics considered  (Tables~\ref{tab:monochromZ} and~\ref{tab:monochromtt}) are plotted as red (yellow) squares for the ``GHC V22 Z'' (``GHC V22 $\ttbar$'') setups in the $(\delta_{\sqrt{s}},\LumiInt)$ plane shown in Fig.~\ref{fig:eeH_signif}, from which the corresponding 95\% CL upper limits contours for the $y_{\Pe}$ coupling can be read off. 
The results without crab cavities are shown in Fig.~\ref{fig:eeH_signif} (left).
The physics performances of all designed monochromatization IR optics with nonzero $D_{x}^{\ast}$ are comparable to or even exceed those of the previous FCC-ee self-consistent parameters (black cross). The ``MonochroM TH2IP'' optics achieves the best $\delta_{\sqrt{s}}$-$\mathcal{L}_{\text{int}}$ benchmark, with $\delta_{\sqrts} = \SI{23.16}{\mega\electronvolt}$ and $\mathcal{L}_{\text{int}} = \SI{2.94}{\ab}$. This corresponds to an upper limit (95\% CL) of $\lvert y_{\Pe} \rvert < 3.2\,\lvert y_{\Pe}^{SM} \rvert$ for the Higgs-electron coupling, per IP per year. For the monochromatization IR optics with nonzero $D_{y}^{\ast}$, the best physics performance benchmark is achieved with the ``MonochroM TV'' optics, yielding an upper limit (95\% CL) of $\lvert y_{\Pe} \rvert < 7.1\,\lvert y_{\Pe}^{SM} \rvert$. The ``MonochroM ZHV'' and ``MonochroM THV'' optics, which incorporate combined IP dispersion, do not provide improved physics performance compared to the optics with only nonzero $D_{x}^{\ast}$. This is primarily due to the increased luminosity loss caused by the $\varepsilon_{y}$ blow-up when accounting for the impact of BS.
With the same analysis under the head-on collision configuration including crab cavities, the physics performances of all proposed monochromatization IR optics are further improved~\cite{Zhang:2024phd}, as shown in Fig.~\ref{fig:eeH_signif} (right). The best $\delta_{\sqrt{s}}$ vs.\ $\mathcal{L}_{\text{int}}$, achieved with the ``MonochroM TH2IP'' optics, yielding $\delta_{\sqrts} = \SI{15.46}{\mega\electronvolt}$ and $\mathcal{L}_{\text{int}} = \SI{4.51}{\ab^{-1}}$, indicates an upper limit of $\lvert y_{\Pe} \rvert < 2.6\,\lvert y_{\Pe}^{SM} \rvert$. It should be noted that the use of crab cavities is not currently part of the baseline collision scheme for FCC-ee, as the design assumes a large  Piwinski angle (given by: $\varphi = (\sigma_{z}/\sigma_{x}^{\ast})\tan (\theta_{c}/2)$, where $\sigma_{z}$ is the bunch length)~\cite{Raimondi:2007vi}, which does not need, or even contradicts, crab cavities.

\subsubsection*{Possibilities with an energy recovery \epem collider}
\label{sec:eYukawa:ReLiC}
\editors{Karsten K\"oneke}

The very preliminary results in this section are a result of the Higgs/top/electroweak subgroup conveners asking what a possible future energy recovery \epem collider \`a la \textit{Recycling Linear Collider} (ReLiC) could potentially achieve in measuring the electron Yukawa coupling strength to the Higgs boson when operated at an \epem centre-of-mass energy of \SI{125}{\giga\electronvolt}, i.e., the mass of the observed Higgs boson. The very preliminary results were presented at a Higgs/top/electroweak mini-workshop by Vladimir Litvinenko~\cite{Litvinenko:2023feb}, and are summarised briefly in this section, in places using exact phrases from that presentation. Further studies will be required. 

A possible future ReLiC would have several key advantages over other \epem colliders. The recycling of the particles would mitigate the need for the high-intensity positron source necessary for a linear collider. The beam energy would be recovered. It could reach extremely high instantaneous luminosity and, if built as a very long linear machine, very high energy. Furthermore, it allows for high degrees of polarisation of both electron and positron beams, which could allow the $s$-channel Higgs boson production cross section to be enhanced over certain background processes. 

Previous ReLiC studies focused on reaching high energies. However, a beam energy of \SI{62.5}{\giga\eV} warrants a completely different approach to the design of the interaction region (IR). For example, a very modest 2--3 times bunch compression and decompression is sufficient for lossless particle recovery. Less than 1\% of particles radiate beamstrahlung photons, which reduces mono-energetic collisions by 2\%. Since energy-recovery-based colliders use fresh beams, the necessary dispersion can be introduced in the IR for monoenergetic collisions without adverse effect on the beam emittance. Since electron and positrons beams propagate through different (left and right) accelerator structures, dispersion with opposite signs of electrons and positrons $D_{e^+}=-D_{e^+}$ can be created using magnets -- no electrostatic elements are needed. Using $D_x=\SI{12}{\centi\m}$ in the IR with a $\beta^*_x=\SI{5}{\centi\m}$ will provide for an energy spread in the \epem collision of less than \SI{1}{\mega\eV}. This mode can be achieved in ReLiC without loss of luminosity. 

The studied machine parameters are summarised in \cref{tab:eYukawa:ReLiC}. The resulting very small centre-of-mass beam energy spread $\delta_{\sqrt{s}}$ is determined as
\begin{equation}
\delta_{\sqrt{s}} = E \cdot \frac{\sqrt{2\cdot \epsilon_x\cdot \beta^*_x}}{D_x} = \SI{1.4}{\mega\eV}.
\label{eq:ReLiC:breamSpread}
\end{equation}
One year, corresponding to $\SI{2e7}{\s}$, of running such a ReLiC$_{125}$ collider, operating at a centre-of-mass energy of \SI{125}{\giga\electronvolt}, with a tiny beam energy spread of $\delta_{\sqrt{s}} = \SI{1.4}{\mega\eV}$, would result in an integrated luminosity of \SI{90}{\per\atto\barn}. This could potentially allow for achieving a $5\sigma$ discovery of the electron Yukawa coupling to the observed Higgs boson, if such a coupling behaves according to the Standard Model of particle physics.

\begin{table}[h!]
    \centering
    \begin{tabular}{l|r}
        \hline
        Parameter                          &  Assumed value \\ \hline
        Centre-of-mass energy              &  \SI{125}{\giga\electronvolt} \\
        Length of accelerator              &  \SI{5}{\km} \\
        Number of particles per bunch      &  $1.0\times 10^{11}$ \\
        Beam current                       &  \SI{38}{\milli\ampere} \\
        Normalised $\epsilon_x$            &  \SI{4.0}{\milli\metre \milli\rad} \\
        Normalised $\epsilon_y$            &  \SI{1.0}{\micro\metre \milli\rad} \\
        Relative beam spread in interaction region ($\sigma_{\mathrm{E}}/E$)&  $1.6\times 10^{-4}$  \\
        $\beta_x$                          &  \SI{0.05}{\metre} \\
        $\beta_y$, matched                 &  \SI{0.2}{\milli\metre} \\
        $D_x$                              &  \SI{0.08}{\metre} \\
        $\sigma_y$                         &  \SI{1}{\milli\metre} \\
        Disruption $D_x$                   &  0.0 \\
        $D_y$                              &  109 \\  
        \hline \\[-2.3ex] 
        Total instantaneous luminosity     & \SI{4.5e36}{\per\square\centi\metre\per\s} \\
        Integrated luminosity per year ($=\SI{2e7}{\s}$)  &  \SI{90}{\per\atto\barn} \\
        \hline
    \end{tabular}
        \caption{Possible parameters for a ReLiC$_{125}$ \epem collider operating at a centre-of-mass energy of \SI{125}{\giga\electronvolt}. }
    \label{tab:eYukawa:ReLiC}
\end{table}



\subsubsection{Flavour-violating Higgs decays}
\subsubsection*{SUSY predictions}
\editor{Keisho Hidaka}

%
Here we discuss the future \epem collider sensitivity for Quark Flavour Violating 
(QFV) SUSY in Higgs decays.                                                
This addresses the nature of the SM-like Higgs boson discovered at LHC.
Here we study a possibility that it is the lightest Higgs boson \PShz of the MSSM,   
focusing on the decays $\PShz \to \cc, \bb, \PQb\PAQs, \PGg \PGg, \Pg \Pg$           
with special emphasis on SUSY QFV. This work is based on Refs. \cite{Bartl:2014hcc, 
Eberl:2016hbb, Eberl:2018hgg, Hidaka:2022pso, Hidaka:2023pst, Hidaka:2024lcw}. 
We focus on the $\PSQc_{L/R} - \PSQt_{L/R}$ and $\PSQs_{L/R} - \PSQb_{L/R}$ 
mixing, which are described by the QFV parameters $M^2_{Q23}$, $M^2_{U23}$, 
$T_{U23}$, $T_{U32}$ and $M^2_{Q23}$, $M^2_{D23}$, $T_{D23}$, $T_{D32}$, 
respectively. The parameters $T_{U33}$ and $T_{D33}$ which induce the 
$\PSQtL - \PSQtR$ and $\PSQbL - \PSQbR$ mixing, respectively, are 
also involved. We assume that R-parity is conserved and that the lightest 
neutralino \PSGczDo is the lightest SUSY particle (LSP). 

%
%
We compute the decay widths $\Gamma(\PShz \to \cc)$, 
$\Gamma(\PShz \to \bb / \PQb \PAQs)$ at full 1-loop level and 
the loop-induced decay widths $\Gamma(\PShz \to \PGg \PGg)$, 
$\Gamma(\PShz \to \Pg \Pg)$ at NLO QCD level in the MSSM with QFV 
\cite{Bartl:2014hcc,Eberl:2016hbb,Eberl:2018hgg}. 
For the first time, we perform a systematic  
MSSM-parameter scan for these decay widths, respecting all the relevant 
theoretical and experimental constraints described in Ref.~\cite{Eberl:2021csv}.
We generate the MSSM-parameter points randomly in the ranges shown in 
Table 1 of Ref. \cite{Eberl:2021csv}. No GUT relation 
for the gaugino masses $M_1$, $M_2$, $M_3$ is assumed. 

%
%
We find that the large Higgs-squark-squark trilinear couplings $T_{U23,32,33}$, 
$T_{D23,32,33}$, large $M^2_{Q23}$, $M^2_{U23}$, and $M^2_{D23}$ can lead 
to large MSSM 1-loop corrections to these decay widths, resulting in large 
deviations of these MSSM widths from their SM values. 

The main MSSM 1-loop corrections to $\Gamma(\PShz \to \cc)$ stem 
from the lighter up-type squarks ($\PSQu_i$) - gluino ($\PSg$) loops 
at the decay vertex, where $\PSQu_i$ are strong $\PSQc_{L,R}$ - $\PSQt_{L,R}$ 
mixtures (see Fig. 1(a) in Ref.~\cite{Hidaka:2022pso}). The large Higgs-squark-squark 
trilinear couplings $T_{U23,32,33}$ can enhance the $\PShz-\PSQu_i-\PSQu_j$ couplings, 
resulting in enhancement of the $\PSQu_i$-$\PSg$ loop corrections to 
$\Gamma(\PShz \to \cc)$. 

The main MSSM 1-loop corrections to $\Gamma(\PShz \to \bb)$ and 
$\Gamma(\PShz \to \PQb \PAQs)$ stem from 
(i) $\PSQu_i$ - chargino ($\HepParticle{\PSGc}{k}{\pm}\Xspace$) loops at the decay vertex which 
have $\PShz-\PSQu_i-\PSQu_j$ couplings enhanced by large $T_{U23,32,33}$ 
(see Fig. 1(b) in Ref.~\cite{Hidaka:2022pso}) and 
(ii) $\PSQd_i$ - $\PSg$ loops at the decay vertex, where $\PSQd_i$ are 
strong $\PSQs_{L,R}$ - $\PSQb_{L,R}$ mixtures, which have $\PShz-\PSQd_i-\PSQd_j$ 
couplings to be enhanced by the large Higgs-squark-squark trilinear 
couplings $T_{D23,32,33}$ (see Fig. 1(c) in Ref.~\cite{Hidaka:2022pso}). 
Hence, the large $T_{U23,32,33}$ and $T_{D23,32,33}$ can enhance the 
$\PSQu_i$ - $\HepParticle{\PSGc}{k}{\pm}\Xspace$ and $\PSQd_i$ - $\PSg$ loop 
corrections to $\Gamma(\PShz \to \bb)$, $\Gamma(\PShz \to \PQb \PAQs)$, respectively. 

%
%
From our MSSM parameter scan we find that the QFV decay branching 
ratio $\text{B}(\PShz \to \PQb \PQs) \equiv \text{B}(\PShz \to \PQb \PAQs) + \text{B}(\PShz \to \PAQb \PQs)$ 
can be as large as $\sim 0.15\%$ \cite{Hidaka:2024lcw} (see also \cite{Gomez:2015hbs}) while it 
is almost zero (< \SI{1.0d-7}) in the SM. The ILC250+500+1000 sensitivity to 
this branching ratio could be $\sim 0.1\%$ at 4$\sigma$ signal significance~\cite{Tian:2023hbs} (see also Ref.~\cite{Barducci:2017hbs}).
Note that the expected upper bound on $\text{B}(\PShz \to \PQb \PQs)$ at FCC-ee 
is \cite{Selvaggi:2024hbs}:
$\text{B}(\PShz \to \PQb \PQs)$ < \SI{4.5d-4}~(95\%~CL).
The expected upper bound on $\text{B}(\PShz \to \PQb \PQs)$ at CEPC 
is \cite{Liang:2023hbs}:
$\text{B}(\PShz \to \PQb \PQs)$ < \SI{2.2d-4}~(95\%~CL).
Note that the LHC and HL-LHC sensitivity should not be so 
good due to the huge QCD background \cite{Barducci:2017hbs}. 

\begin{figure}[htpb!]
\hspace{-0.7cm}
\centering
\includegraphics[width=0.49\columnwidth]{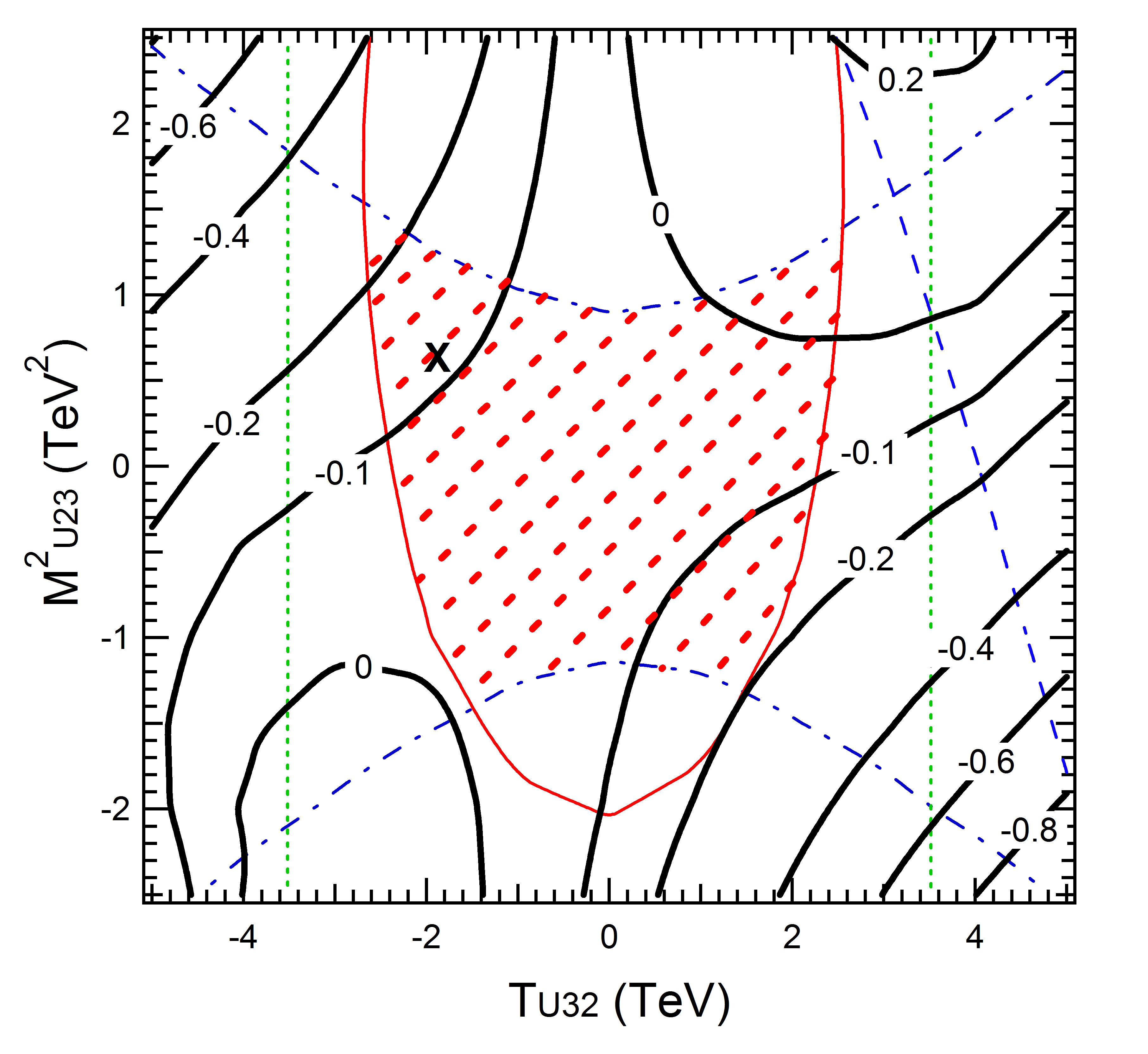}
\includegraphics[width=0.49\columnwidth]{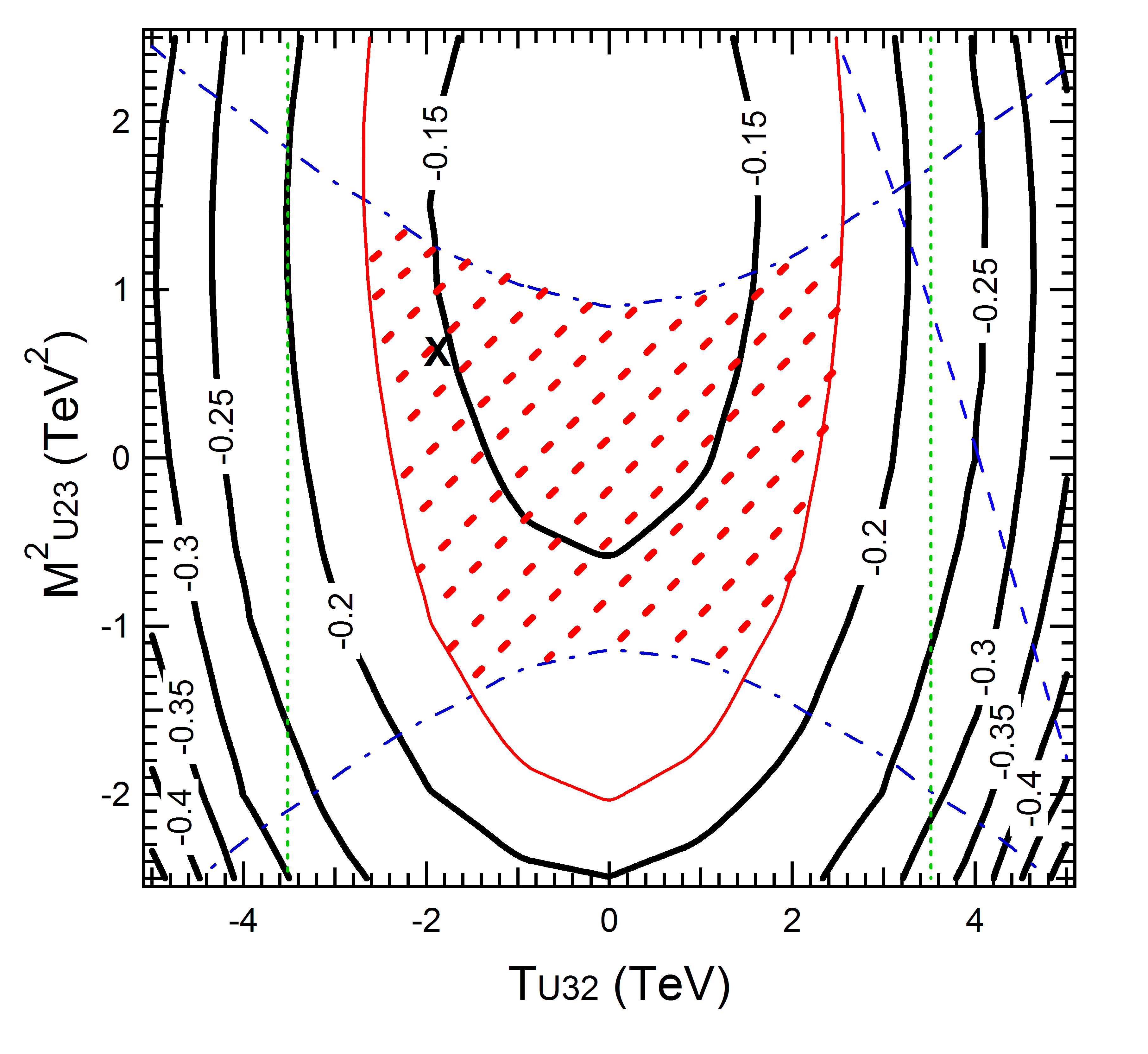}
\caption{Contour plots of DEV(c) (Left) and DEV(b) (Right) around P1 
in the $T_{U32}$-$M^2_{U23}$ plane. ``X" marks P1. The red hatched region 
is allowed by all the constraints including the expected sparticle mass 
limits from HL-LHC.
\label{fig:DEVcb_Cont}}
\end{figure}

%
%
We define the relative deviation of the MSSM width from the SM width as 
$\text{DEV(X)} \equiv \Gamma(\PShz \to X \bar X)_{\text{MSSM}}/\Gamma(\PShz \to X \bar X)_{\text{SM}} - 1$, (X=c,b).
DEV(X) is related to the coupling modifier 
$\kappa_X \equiv \text{C}(\PShz X \bar X)_{\text{MSSM}}/\text{C}(\PShz X \bar X)_{\text{SM}}$ via $\text{DEV(X)}=\kappa_X^2 -1$. 
In Fig. \ref{fig:DEVcb_Cont} we show the contour plot of DEV(c) and DEV(b) in the 
$T_{U32}$-$M^2_{U23}$ plane around the benchmark scenario P1 described in 
Ref.~\cite{Hidaka:2022pso}, which satisfies all the relevant theoretical and experimental 
constraints including all the expected SUSY-particle (sparticle) mass limits 
(including ($\mass{\PSA/\PSHp}$, $\tanbeta$) limits) from negative sparticle-search 
in the HL-LHC experiment \cite{CidVidal:2018cyr, Bose:2022snm}.
We find that DEV(c) can be very large 
(about --30\% to 10\%) in a sizable region allowed by all the constraints 
including the expected sparticle mass limits from the future HL-LHC experiment, 
in strong contrast with the prediction in the MSSM with Minimal 
Flavour Violation (MFV). 
We see also that DEV(b) can be very large 
(about --10\% to --18\%) in a sizable region allowed by all the constraints 
including the expected sparticle mass limits from the future HL-LHC experiment.

According to Refs.~\cite{Blas:2022snm, Hidaka:2024lcw}, the expected absolute 1$\sigma$ errors 
of DEV(c) and DEV(b) measured at ILC are given by 
$\Delta \text{DEV(c)}$ = (3.6\%, 2.4\%, 1.8\%) and $\Delta \text{DEV(b)}$ = (1.7\%, 1.1\%, 0.9\%) 
at (ILC250, ILC250+500, ILC250+500+ 1000) together with HL-LHC. 
Similar results are obtained for the future lepton colliders other than ILC 
(i.e.\ CLIC, FCC-ee, CEPC, muon collider) in Refs.~\cite{Blas:2022snm, Hidaka:2024lcw}.
Hence, future lepton colliders can observe such large deviations 
DEV(c) and DEV(b) at very high significance. 

%
%
From our MSSM parameter scan and contour plot analysis, 
we find DEV(\PGg) and DEV(\Pg) can be sizable simultaneously: 
DEV(\PGg) can be as sizable as $\sim \pm1\%$, 
and DEV(\Pg) can be as large as $\sim +4\%$ and $\sim -7\%$. 
The expected absolute 1$\sigma$ errors of DEV(\PGg) and DEV(\Pg) 
measured at ILC are given by $\Delta \text{DEV}(\PGg)$ = (2.4\%, 2.2\%, 2.0\%) 
and $\Delta \text{DEV}(\Pg)$ = (1.8\%, 1.4\%, 1.1\%) at (ILC250, ILC250+500, 
ILC250+500+1000) together with HL-LHC \cite{Blas:2022snm, Hidaka:2024lcw}. Similar results 
are obtained for the future lepton colliders other than ILC in Refs.~\cite{Blas:2022snm, Hidaka:2024lcw}.
Hence, future colliders 
together with HL-LHC can observe 
such a sizable deviation DEV(\Pg) at high significance while they can not 
observe such a moderate deviation DEV(\PGg) significantly.

%
In summary, we have found that the deviations of these MSSM decay 
widths from the SM values can be quite sizable in the MSSM with general 
QFV, in strong contrast to the usual studies in the MSSM with MFV. 
Here, we remark that SUSY QFV parameters (such as $T_{U23}$, $T_{U32}$) 
can have an important impact on \textit{not only} QFV observables 
(such as $\text{B}(\PShz \to \PQb \PQs)$), but also quark flavour conserving (QFC) 
observables (such as DEV(c) and DEV(b)). 
Future lepton colliders such as ILC, CLIC, FCC-ee, CEPC and the muon collider 
can observe such sizable deviations from the SM at high signal significance 
even after the potential non-discovery of SUSY particles  at the HL-LHC.
In case the deviation pattern shown here is really observed at the lepton 
colliders, then it would strongly suggest the discovery of QFV SUSY (the 
MSSM with general QFV). 

\subsubsection*{Projected sensitivity}\label{subsubsec:projSensitivityHiggs}
\editors{Arman Korajac, Manuel Szewc, Michele Tammaro}

The SM predicts the branching ratio of $\PH\to \PQb\PQs,\PQb\PQd,\PQc\PQu$ to be strongly suppressed by the GIM mechanism, while the $\PH\to \Pe\PGm,\Pe\PGt,\PGt\PGm$ processes are strictly forbidden. These decay channels are thus very sensitive to possible New Physics contributions. The effective FCNC couplings of quarks to the Higgs boson can be parametrised as
\begin{equation}\label{eq:HbsLagrangianPheno}
	\mathcal{L} \supset \, y_{ij}(\PAQq_{i,L} \PQq_{j,R})\PH + y_{ji}(\PAQq_{j,L} \PQq_{i,R})\PH + \mathrm{h.c.} \,, 
\end{equation}
where $ij=\PQb\PQs,\PQb\PQd,\PQc\PQu$, and similarly for leptons by replacing $\PQq\to\Pl$. This Lagrangian can be obtained as the low energy realization of various extensions of the SM, e.g., the addition of vector-like quarks~\cite{Fajfer:2013wca,Barducci:2017ioq}, or in the Two-Higgs-Doublet Model (2HDM)~\cite{Branco:2011iw,Crivellin:2013wna}.

At future lepton colliders, such as FCC-ee and CEPC, about $10^6$ $\PH$ will be produced on-shell at the $\PZ\PH$ threshold run, $\sqrt{s} = \SI{240}{\giga\electronvolt}$. Due to the smallness of the SM flavour changing branching ratios, reported below, for these relatively small statistics the expected sensitivity is well above the SM values, and we can only set upper bounds on the decay branching ratios.

\paragraph{Quarks:} Updated computations of $\PH\to\PQq_i\PQq_j$ yield~\cite{Benitez-Guzman:2015ana, Aranda:2020tqw,Kamenik:2023hvi}
\begin{equation}
{\cal B}(\PH\to \PQb\PQs) = \left( 8.9 \pm 1.5 \right)\times10^{-8}\,, \quad {\cal B}(\PH\to \PQb\PQd) = \left( 3.8 \pm 0.6 \right)\times10^{-9}\,, \quad {\cal B}(\PH\to \PQc\PQu) = \left( 2.7 \pm 0.5 \right) \times 10^{-20}\,,
\end{equation}
where ${\cal B}(\PH\to \PQq_i \PQq_j) = {\cal B}(\PH\to \PAQq_i \PQq_j) + {\cal B}(\PH\to \PQq_i \PAQq_j)$. The reported uncertainties contain both the theoretical uncertainties from numerical inputs and the leading two-loop corrections to the amplitudes. 
 
Present direct bounds on flavour changing Higgs boson decays come from searches for decays to undetermined final states, $\PH\to\text{undet.}$, at LHC. The ATLAS and CMS analyses lead to the rather poor limit ${\cal B}(\PH\to\text{undet.})<16\%$~\cite{ATLAS:2021vrm,CMS:2022dwd}, which translates to $|y_{ij},y_{ji}| \lesssim 7\times 10^{-3}$. 

Indirect bounds on flavour changing Higgs couplings can be extracted from measurements of meson mixing parameters, e.g.\ the mass splitting of $B_s-\bar B_s$. Assuming no large cancellations in the effective couplings, these bounds are
\begin{equation}
{\cal B}(\PH\to \PQb\PQs) < 2\times10^{-3}\,, \quad {\cal B}(\PH\to \PQb\PQd) < 10^{-3}\,, \quad {\cal B}(\PH\to \PQc\PQu) < 2\times10^{-2}\,,
\end{equation}
which translate to $|y_{ij},y_{ji}|\lesssim 10^{-3}$. Additional bounds can be extracted by Higgs decays to a photon and flavourful vector mesons, $\PH\to\PGg + VM^0$~\cite{dEnterria:2023wjq,ATLAS:2024dpw,CMS:2024tgj}; these exclusive channels are either poorly or not at all constrained, thus provide much weaker bounds.

For the FCC-ee, the total detector acceptance for $\PH$ hadronic decays is expected to be $\sim70\%$. 
In order to reduce the background and enhance the signal, it is convenient to select events where the $\PZ$ boson decays to charged leptons even at the expense of further reduction in statistics. In that case, the main background arises from $\PH\to\PQq\PQq$ decays~\cite{Kamenik:2023hvi}, which mimics the signal when one of the jets is mistagged.

To project the sensitivity of future colliders, we exploit the method described in Refs.~\cite{ATLAS:2022ers,Faroughy:2022dyq,CMS:2020vac,Kamenik:2023hvi}, leveraging jet-flavour taggers to create when possible a binned two-dimensional distribution from which to extract the signal. As a working case, we choose the Particle-Net tagger performances reported for the FCC-ee, utilizing the IDEA detector concept~\cite{Bedeschi:2022rnj,Bedeschi:2022rnjtaggerWP}. Indicating by $\epsilon_\beta^q$ the efficiency of the $q$-tagger to tag the $\beta$-flavoured jet, with $\beta=\{g,\PQs,\PQc,\PQb\}$, we have
\begin{equation}\label{eq:taggerperform:Higgs}
	\begin{split}
		\epsilon_\beta^{\PQb} &= \{0.02,\,0.001,\,0.02,\,0.90\}\,, \\
		\epsilon_\beta^{\PQs} &= \{0.09,\,0.80,\, 0.06,\, 0.004\}\,, \\
		\epsilon_\beta^{\PQc} &= \{0.02,\,0.008,\,0.80,\,0.02\}\,.
	\end{split}
\end{equation}  
We do not consider $\PQd$ and $\PQu$ taggers, given that they are still currently in development. Because of this, we consider two taggers for the $\PQb\PQs$ channel and only one tagger for the $\PQb\PQd$ and $\PQc\PQu$ channels.  Notice that, thanks to the excellent capabilities of state-of-the-art flavour taggers, backgrounds that require two mistagged jets are greatly suppressed, e.g.\ $\PH\to\PQc\PQc$ backgrounds for searches of $\PH\to\PQb\PQs$. Systematic uncertainties on the tagger efficiencies are expected to be at or even below the $1\%$ level; however the reach on $\PH$ flavour changing decays will be limited by the low statistics.

The expected reaches are
\begin{equation}
{\cal B}(\PH\to \PQb\PQs)<9.6\times10^{-4}\,, \quad {\cal B}(\PH\to \PQc\PQu)<2.5\times10^{-3}\,, 
\end{equation}
which translate to the bounds $|y_{\PQb\PQs},y_{\PQs\PQb}|\lesssim5\times10^{-4}$ and $|y_{\PQc\PQu},y_{\PQu\PQc}|\lesssim 8\times10^{-4}$, respectively. The $\PH\to \PQb\PQd$ suffers from a lower statistics and the absence of a $\PQd$ tagger, leading to the weaker constraint ${\cal B}(\PH\to \PQb\PQd)<5\times10^{-3}$.

Thus, future colliders can improve the strongest indirect constraints on flavour-changing Higgs couplings by a factor of a few for $\PH\to \PQb\PQs$ and $\PH\to \PQc\PQu$, and in general exclude regions where the New Physics signals could be hidden by large cancellations between coupling.

\paragraph{Leptons:} Lepton flavour changing decays of the Higgs are forbidden in the SM, thus any signal would be a smoking gun for New Physics. Presently, direct bounds have been obtained by exclusive searches at LHC from both ATLAS and CMS~\cite{ATLAS:2023mvd, CMS:2021rsq, CMS:2023pte}, leading to
\begin{equation}
	{\cal B}(\PH\to \Pe\PGt) < 2.0\times10^{-3}\,, \quad {\cal B}(\PH\to \PGm\PGt) < 1.5\times10^{-3} \,, \quad {\cal B}(\PH\to \Pe\PGm) < 4.4\times10^{-5}\,.
\end{equation}
Flavour violating decays of charged leptons will lead to indirect bounds on Higgs off-diagonal couplings~\cite{Harnik:2012pb}. While $\PH$ decays involving the $\PGt$ are poorly constrained, with ${\cal B}(\PH\to \Pe\PGt/\PGm\PGt) \lesssim 0.1$, flavour changing decays of the muon, as $\PGm\to3\Pe$ and $\PGm\to\Pe\PGg$, lead to limits as strong as ${\cal B}(\PH\to \Pe\PGm) \lesssim 10^{-9}$. 

Oppositely to the previous case, here we can select $\PZ$ boson hadronic decays; the total width of the latter channel is $\Gamma(\PZ\to\text{had.})\simeq67\%$, with a detector acceptance of nearly $100\%$~\cite{Agapov:2022bhm}. The main suppression in the signal acceptance will thus come from reconstruction and identification of the final state leptons. In particular, decays involving the $\PGt$ will suffer from low signal efficiency, as the $\PGt$ has to be reconstructed either from the decay to a $\PGm$ and missing energy from neutrinos, $\PGt\to\PGm\PGn\PAGn$, or from the ``3-prong" decay to pions, $\PGt\to3\PGp+\PGn$. Following the detailed analysis of Ref.~\cite{Qin:2017aju}, the efficiency for the $\Pe\PGm$ channel is $\sim40\%$, while it reduces to $\sim5\%$ for the $\Pe\PGt$ and $\PGm\PGt$ cases.

The projected bounds on branching ratios at future circular lepton colliders are
\begin{equation}
	{\cal B}(\PH\to \Pe\PGt) \lesssim 1.6\times10^{-4}\,, \quad {\cal B}(\PH\to \PGm\PGt) \lesssim 1.4\times10^{-4} \,, \quad {\cal B}(\PH\to \Pe\PGm) \lesssim 1.2\times10^{-5}\,,
\end{equation}
which translate to the limits on the respective couplings
\begin{equation}
	|y_{\Pe\PGt},y_{\PGt\Pe}|\lesssim3.6\times10^{-4}\,, \quad |y_{\PGm\PGt},y_{\PGt\PGm}|\lesssim3.4\times10^{-4}\,, \quad|y_{\Pe\PGm},y_{\PGm\Pe}|\lesssim1.0\times10^{-4}\,.
\end{equation}
Apart from the $\Pe\PGm$ case, these bounds show a significant improvement over present limits.


\vspace*{.1cm}
\subsection{\focustopic Higgs self-coupling}
\label{sec:Hself}
\editor{Junping Tian, Jorge de Blas}

\subsubsection{Introduction}

Many open questions of particle physics are related to the Higgs sector and, in particular, to the Higgs potential. 
In the SM, the form of the potential follows directly from the symmetries of the theory and the requirement of renormalisability,
\begin{equation}
V_{\rm SM}(\phi)=-\mu^2 \phi^\dagger \phi + \lambda (\phi^\dagger \phi)^2=\frac 12 m_{\PH}^2 \PH^2+\lambda_{\PH\PH\PH}\PH^3 + \frac 14\lambda_{\PH\PH\PH\PH} \PH^4 + \ldots~~~(\mu^2,~\lambda >0).
\label{eq:VSM}
\end{equation}
However, its key structural properties, determined by the signs of the quartic and quadratic terms, are decided ad-hoc to trigger Electroweak Symmetry Breaking (EWSB) and to have a potential that is bounded from below, respectively. 
%
This leaves open the question of whether there is underlying new physics that explains the origin and true dynamics of EWSB.
The Higgs potential also plays a role in determining how 
the electroweak phase transition (EWPT) in the early universe took place, and could be linked to possible explanations of the observed asymmetry between matter and anti-matter in the universe in terms of electroweak baryogenesis. 

After the Higgs boson discovery at the LHC, the values of all the terms in \cref{eq:VSM} are unambiguously predicted in the SM in terms of the vacuum expectation value, $v$, (or, alternatively, the Fermi constant) and the Higgs boson mass: 
\begin{equation}
\mu^2=\frac 12 m_{\PH}^2,~~~~\lambda=\frac{\lambda_{\PH\PH\PH}}{v}=\lambda_{\PH\PH\PH\PH}=\frac{m_{\PH}^2}{2 v^2}.
\end{equation}
As for any other SM prediction, e.g.\ all the other Higgs couplings and EW observables, the observation of a modification of the Higgs boson self-interactions with respect to these predictions 
would give evidence of the presence of new physics.
%

The actual form of the Higgs potential that is realised in nature and its physical origin are, however, largely unknown up to now. 
While the bilinear term proportional to $\PH^2$ is related to the measured value of $m_{\PH}$, and tells us about the potential around the electroweak (EW) minimum, 
understanding the shape of the potential away from that point requires an experimental determination of the  
Higgs boson three- and four- (and possibly higher) point self-couplings. 
The observable effects of these interactions have only been loosely constrained experimentally so far. 
In this regard, it is important to clarify that the Higgs boson self-coupling(s), like any other coupling, are not themselves physical observables, hence their determination is tied to the interpretation of an observable (e.g.\ the single Higgs boson or Higgs pair production cross sections) within a given theory framework. Therefore, a fully model-independent measurement of a Higgs coupling is not possible as a matter of principle at any collider, and the choice of both an appropriate theory framework and the observables to be interpreted play a role in the sensitivity to the strength of the Higgs (self-)interactions. 
In this section, we focus our attention on the challenges surrounding the experimental determination, and theory interpretation, of the Higgs trilinear coupling $\lambda_{\PH\PH\PH}$, proportional to $\PH^3$, whose measurement will provide the next step in understanding the shape of the Higgs potential. The main probes 
used currently in LHC experiments as well as those possible at future $\epem$ colliders are briefly described in what follows, with the progress made in the latter being the core focus of the section.

\begin{figure}[ht!]
\centering
  \includegraphics[width=0.9\textwidth]{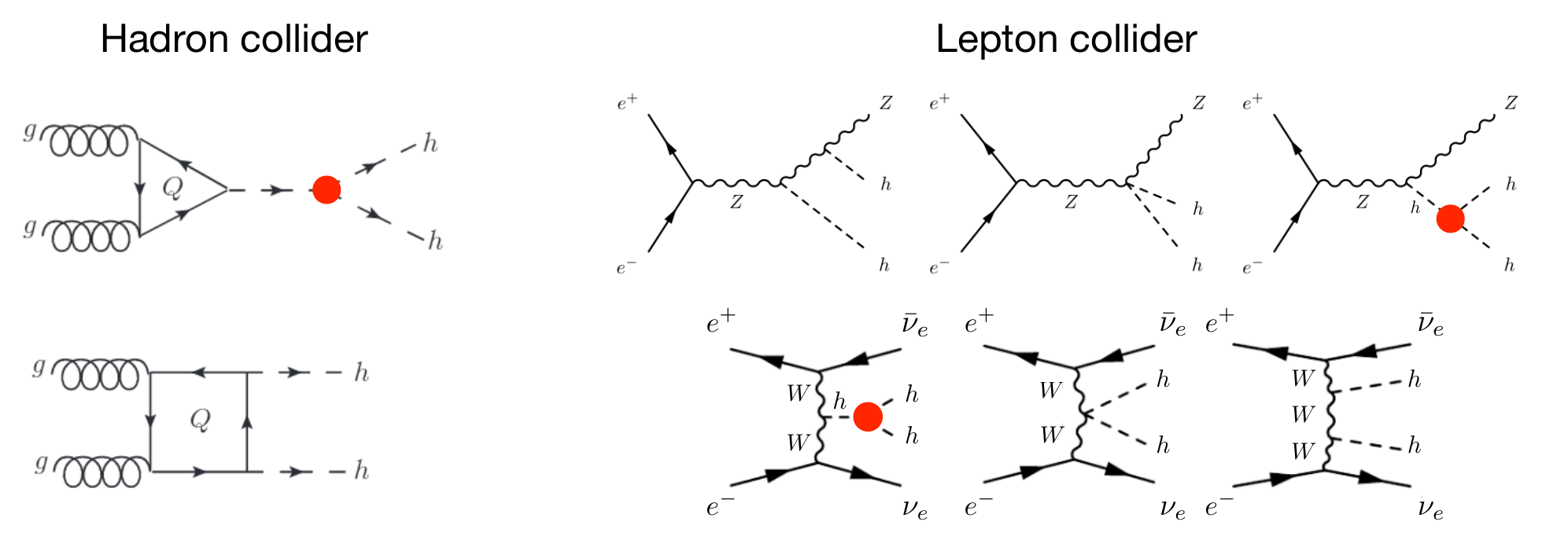}
\caption{Diagrams contributing to the main Higgs pair-production production mechanism at hadron and $\epem$ colliders.\label{fig:HHxs_diagrams}}
\end{figure}

\begin{figure}[ht!]
\centering
  \includegraphics[width=0.45\textwidth]{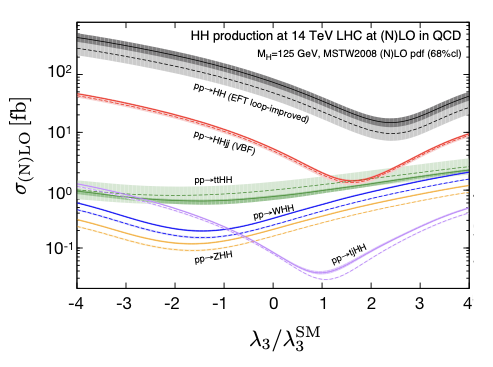}
~
  \includegraphics[width=0.5\textwidth]{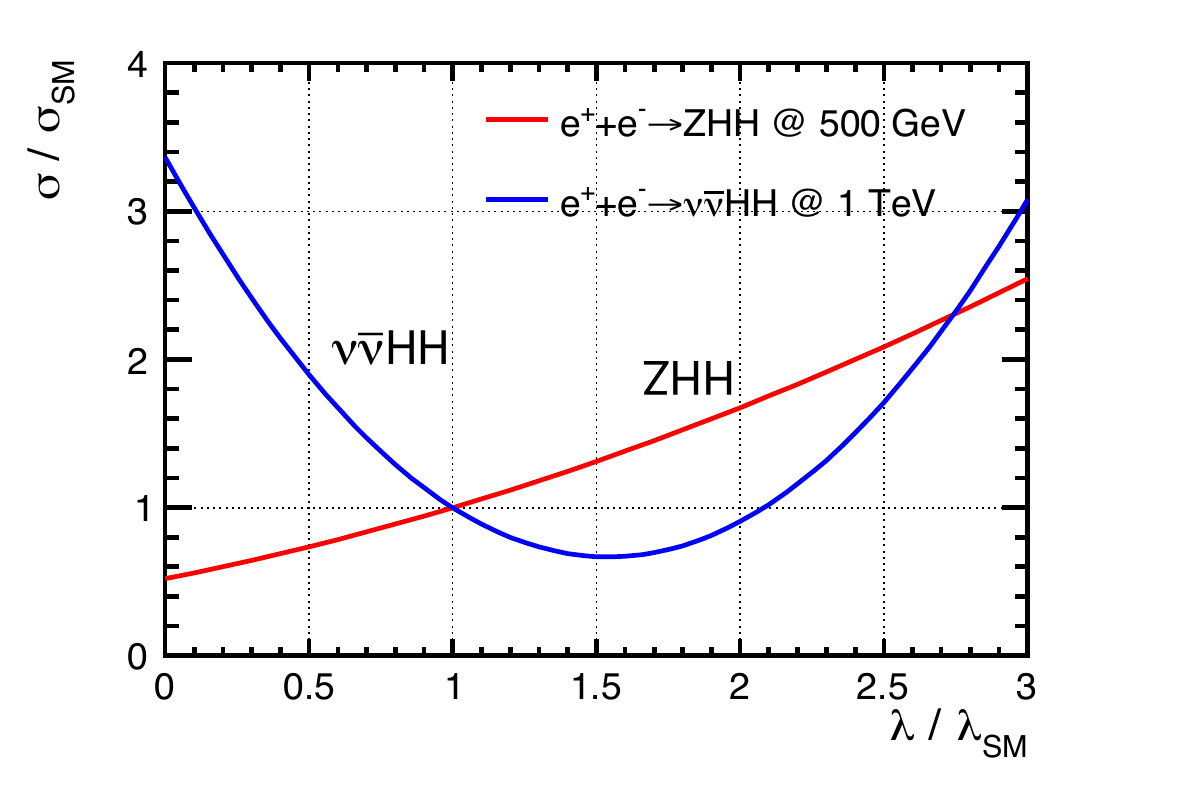}
~
  \includegraphics[width=0.5\textwidth]{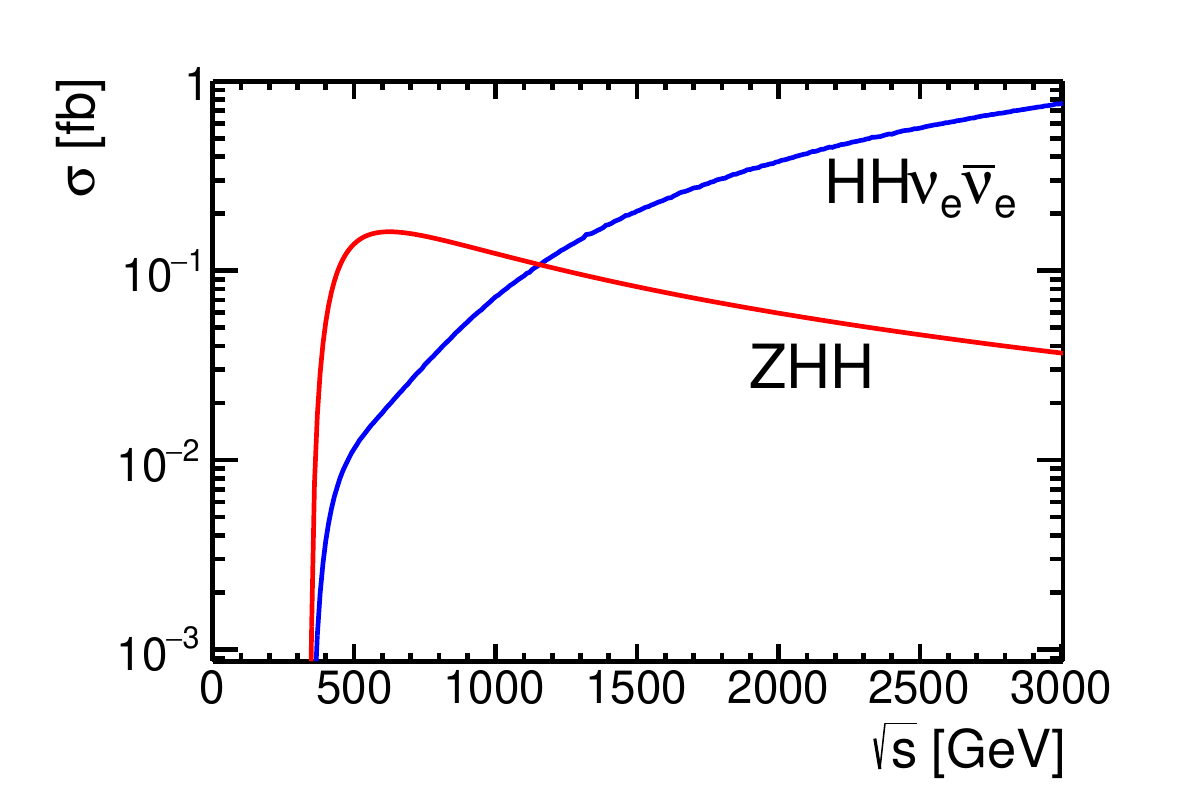}
\caption{Higgs pair-production cross section, as a function of $\lambda_{\PH\PH\PH}$, at the LHC~\cite{Frederix:2014hta} (top left) and $\epem$ colliders~\cite{DiMicco:2019ngk} (top right). 
All other interactions contributing to the processes are assumed to take their SM values. 
(Bottom) $\epem\to \PH\PH + X$ SM production cross section as a function of $\sqrt{s}$~\cite{DiMicco:2019ngk}.}
\label{fig:HHxs_kappalambda}
\end{figure}

The existing constraints on $\lambda_{\PH\PH\PH}$ from the LHC have mainly been obtained from the searches for the Higgs pair production process, dominated by the gluon-fusion channel. 
Assuming that all other Higgs boson couplings besides $\lambda_{\PH\PH\PH}$ are fixed to their SM values, the current bounds on the di-Higgs production cross section from ATLAS and CMS can be translated into limits on the Higgs self-coupling, $-1\lesssim \kappa_\lambda \lesssim 7 $ at 95$\%$ C.L., with $\kappa_\lambda\equiv \lambda_{\PH\PH\PH}/\lambda_{\PH\PH\PH}^\textrm{SM}$~\cite{ATLAS:2024ish,CMS:2022dwd}. 
These results will be significantly improved  with the high-luminosity upgrade of the LHC.  Assuming the SM value of $\lambda_{\PH\PH\PH}$ is realised in nature, current HL-LHC projections~\cite{ATL-PHYS-PUB-2025-018} give an expected 68$\%$ C.L. uncertainty for $\kappa_\lambda$ of $1 ^{+0.29} _{-0.26}$ (see also \cref{fig:higgs:hllhclambda}). Further improvements are  expected when all (single- and double-Higgs) production and decay channels are included, and when the extended acceptance of the HL-LHC detectors is taken into account.
 It should be noted that, in the gluon fusion channel, the leading-order one-loop diagram containing $\lambda_{\PH\PH\PH}$ and the top Yukawa coupling $y_{\PQt}$ enters together with a box diagram involving the coupling factor $y_{\PQt}^2$, see \cref{fig:HHxs_diagrams}. Because of the destructive interference between the contributions from these two diagrams, the total cross section for di-Higgs production changes very substantially, if $\lambda_{\PH\PH\PH}$ is varied around the SM value. This is illustrated in the top-left panel of \cref{fig:HHxs_kappalambda}, where the Higgs pair production cross section is shown as a function of $\lambda_{\PH\PH\PH}$ for the different production modes. In this regard, it should also be noted that 
 a change in $\lambda_{\PH\PH\PH}$ would not only affect the total cross section, but also the kinematic properties of the Higgs pairs, e.g.\ their invariant mass, and hence the signal acceptance and efficiency. Both the modified kinematic properties and the single Higgs production (which benefits from a constructive interference) at the HL-LHC can be harnessed to mitigate the loss in statistical sensitivity due to the smaller cross section for moderately large values of $\lambda_{\PH\PH\PH}$, around the minimum in \cref{fig:HHxs_kappalambda}.


While the existing bounds on $\lambda_{\PH\PH\PH}$ are rather weak, 
they can still probe so-far untested parameter regions of physics beyond the SM (BSM), in those scenarios where corrections to 
$\lambda_{\PH\PH\PH}$ are much larger than those to the interactions between
$\PH$ and the gauge bosons and fermions (noting that the interaction between the Higgs boson and gauge bosons and fermions, and EW observables, can be measured much more precisely). 
Power counting arguments, both from an effective field theory perspective~\cite{Chang:2019vez,Falkowski:2019tft,Durieux:2022hbu} and in particular classes of models~\cite{Kanemura:2002vm,Durieux:2022hbu,Bahl:2022jnx,Bahl:2023eau},
suggest that such a hierarchy of corrections can indeed be generated by new physics at the tree,
1-loop and even 2-loop level. 
This is discussed and illustrated with some examples in \cref{sec:largeH3}. 
%
Furthermore, an $\mathcal{O}(1)$ upward shift in $\lambda_{\PH\PH\PH}$ is also motivated in many scenarios giving rise to a strong first-order EWPT which is for instance required for electroweak baryogenesis~\cite{Sakharov:1967dj}. (A large value of $\lambda_{\PH\PH\PH}$ could therefore also have cosmological consequences, detectable as a gravitational wave stochastic background at the future space-based observatory LISA~\cite{Grojean:2006bp, Caprini:2015zlo}.)
A first-order electroweak phase transition can occur in many BSM theories proposed for several decades. These include singlet scalar extensions, triplet scalar extensions, two-Higgs doublet models, supersymmetric models, left-right symmetric models, Pati-Salam model, Georgi--Machacek model, Zee--Babu model, extended composite Higgs models, neutrino mass models and hidden sector models involving scalars, and others (see the extended list of references compiled in Ref.~\cite{Ramsey-Musolf:2024zex}). 
For a recent example in the case of the 2HDM see e.g.\ Ref.~\cite{Biekotter:2022kgf} , where the parameter region featuring a strong first-order EWPT and a potentially detectable gravitational wave signal at the future space-based observatory LISA is correlated with an enhancement of $\lambda_{\PH\PH\PH}$ compared to the SM value by about a factor of 2.
On the other hand, it is also important to note that large $\mathcal{O}(1)$ corrections to $\lambda_{\PH\PH\PH}$ are not expected to be obtained in the more natural models explaining electroweak symmetry breaking, like the minimal supersymmetry or composite Higgs models, without some tuning, and some special dynamical features are needed (discussed in \cref{sec:largeH3}) to generate large deviations of $\lambda_{\PH\PH\PH}$ and to remain compatible with all other experimental constraints (e.g.\ single-Higgs measurements or electroweak precision data). 

At future colliders, information about $\lambda_{\PH\PH\PH}$ could be obtained in different and complementary ways. 
At $\epem$ centre-of-mass energies below about 500~\GeV, constraints can be obtained using precise $\epem \to \PZ\PH$ cross section measurements, while above this energy the self-coupling can be determined via Higgs pair production. Both of these methods are accessible at the HL-LHC.

\begin{figure}[ht!]
\centering
   \includegraphics[width=0.495\textwidth]{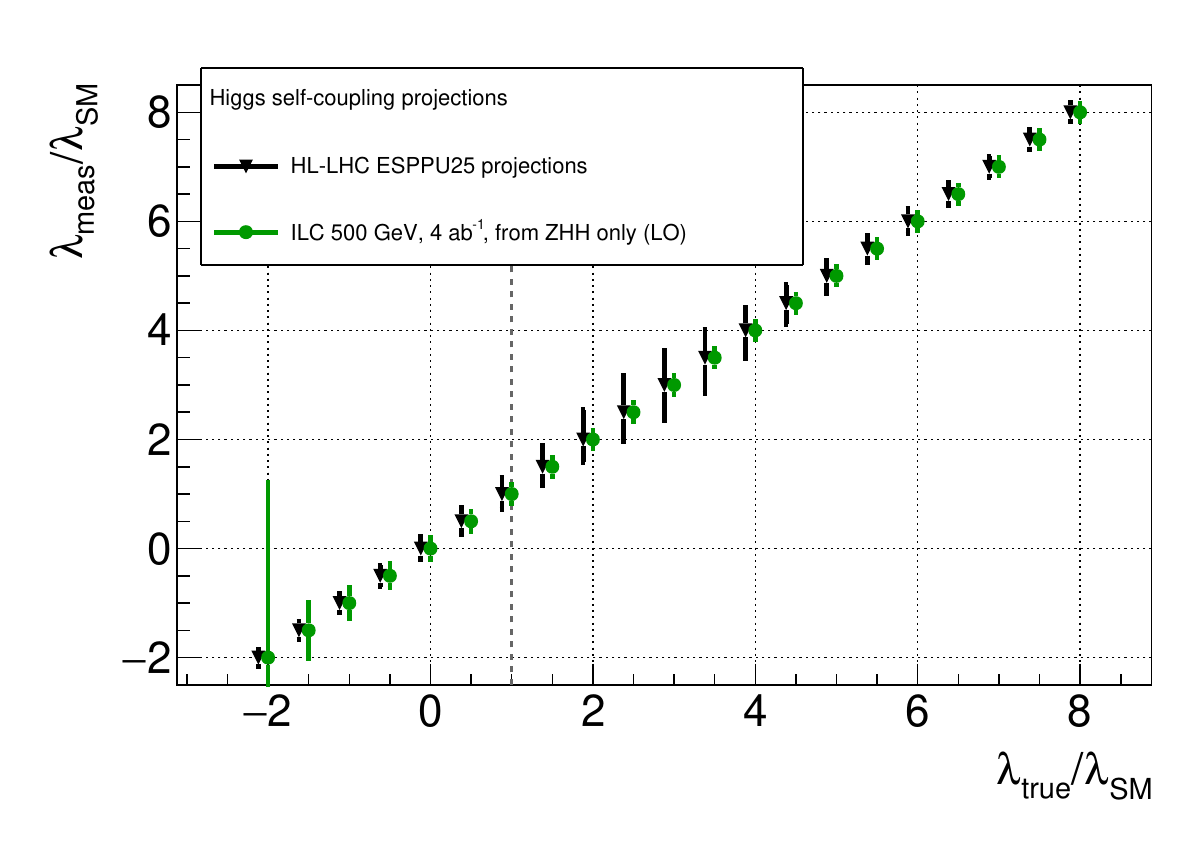}
   \includegraphics[width=0.495\textwidth]{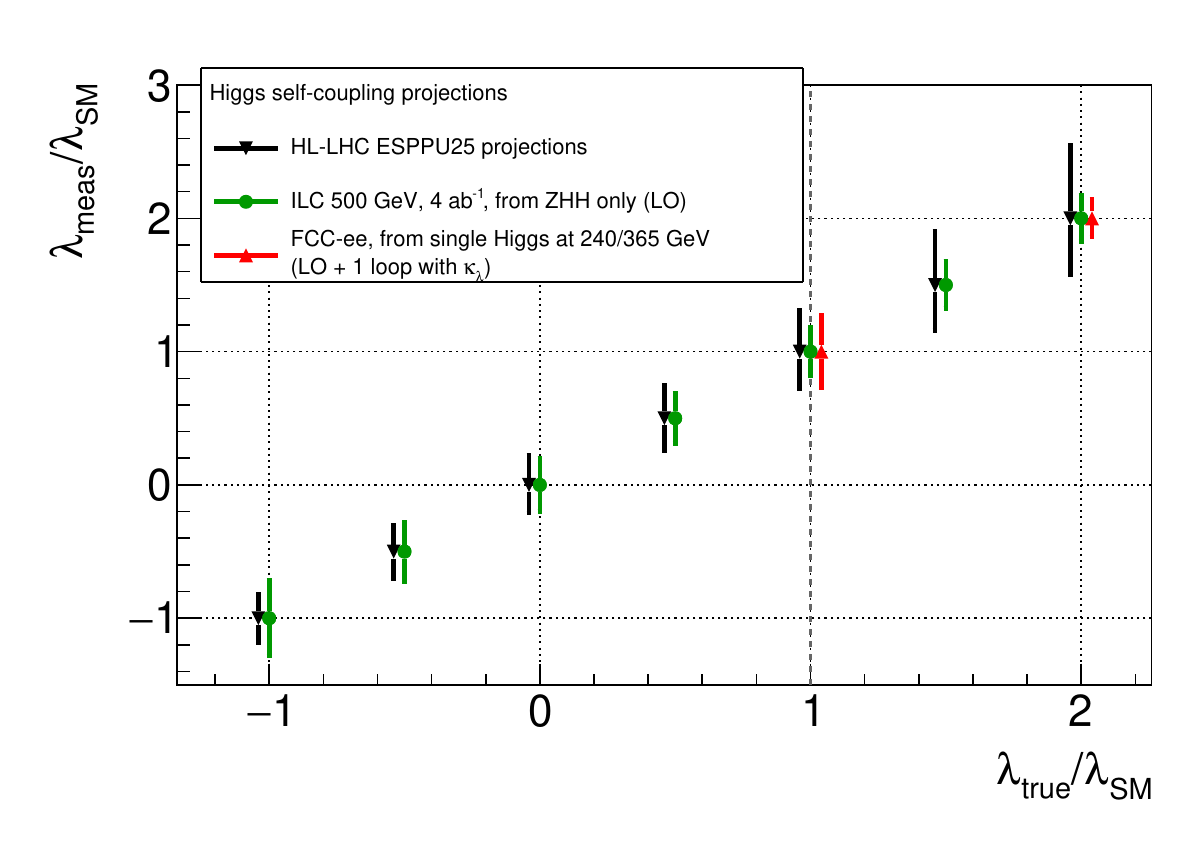}
\caption{Projected accuracies for $\lambda_{\PH\PH\PH}$ from Higgs pair production as a function of the actual value of $\lambda_{\PH\PH\PH}$ that is realised in nature. The right-hand panel zooms in on the region close to the SM value.  ILC sensitivities are from the study reported in \cref{sec:HHprod}, HL-LHC sensitivities from \cite{ATL-PHYS-PUB-2025-018}, and FCC-ee sensitivities from \cite{FCC-FSR-Vol1}.
\cref{fig:KlambdaKi} gives further results from single-Higgs measurements.
}
\label{fig:kappalambda}
\end{figure}

Starting with Higgs pair-production: at $\epem$ colliders a CM energy of at least 500~GeV would be needed to measure the $\epem \to \PZ\PH\PH$ process 
(this is also the case for the weak-boson fusion process $\epem \to \PGne\PAGne\PH\PH$, though this would benefit from much larger CM energies); see bottom panel in \cref{fig:HHxs_kappalambda}.
As will be discussed in \cref{sec:HHprod}, assuming the SM value of $\lambda_{\PH\PH\PH}$ is realised in nature,
a 500 GeV (4\abinv, $|P(\Pem,\Pep)|=(80\%,30\%)$) linear $\epem$ collider would obtain a determination of 
$\kappa_\lambda = 1 \pm 0.18$ from $\PZ\PH\PH$ production alone and using current tools.  
Other energies and potential improvements are discussed in \cref{sec:HHprod}. 
For $\kappa_\lambda>1$, 
assuming the $\PH\PH\PZ\PZ$ interaction is SM-like, the interference terms involving the $\lambda_{\PH\PH\PH}$ interaction contribute  
constructively to $\PZ\PH\PH$ production at an $\epem$ collider operating around 500 GeV, leading to a growth in the cross section as shown in the top-right panel of \cref{fig:HHxs_kappalambda}. 
This is unlike the leading process at hadron colliders, gluon fusion, where the interference is destructive. 
Despite the destructive interference, as can be seen in \cref{fig:kappalambda}, 
the relative precision of the HL-LHC determination improves somewhat with increasing $\kappa_\lambda$,
going from $\kappa_\lambda = 1^{+0.29}_{-0.26}$ to $\kappa_\lambda = 2^{+0.55}_{-0.42}$ ($^{+27\%}_{-21\%}$) \cite{ATL-PHYS-PUB-2025-018}.
In this same case, the determination of $\lambda_{\PH\PH\PH}$ at 500 GeV with 4\,\abinv would improve to $\kappa_\lambda = 2 \pm 0.18$ (9\%). 
%
As mentioned above, such large values of $\kappa_\lambda$ would be favoured in scenarios giving rise to a strong first-order EWPT.\footnote{
Related to the possibility of measuring the Higgs pair production process and, aside from detecting possible enhancements in the value of $\kappa_\lambda$, we note here that in BSM scenarios with extra scalars, one typically finds additional triple Higgs couplings (THCs) involving also heavy Higgses that could be probed in $\epem \to \PH\PH +X$. Moreover, if the new scalars are sufficiently light, such couplings could be tested via a resonant contribution. The effects that such scenarios could have on $\epem \to \PZ\PH\PH$ is explored in \cref{sec:BSMHhh}, with special emphasis on the impact of loop corrections to these THCs. 
}
Conversely, from the same \cref{fig:kappalambda} one observes that the improvement with respect to HL-LHC would be milder if $\lambda_{\PH\PH\PH}$ turns out to be smaller than 1. In particular, for $\lambda_{\PH\PH\PH}< 0$, the HL-LHC and future $\epem$ colliders at around 500 GeV would obtain similar precision.

Below the threshold for Higgs pair production processes, 
future $\epem$ Higgs factories
are also sensitive to the effects of the Higgs self-coupling via its virtual loop 
effects in precision observables. 
Specifically, single Higgs observables, like the production cross section and branching ratios, receive one-loop contributions that depend on $\lambda_{\PH\PH\PH}$, while in the predictions for the electroweak precision observables at the Z~pole and for the W-boson mass $\lambda_{\PH\PH\PH}$ enters at the two-loop level. In particular, special attention has been paid to the determination
of the Higgs self-coupling from measurements of single-Higgs processes, where sub-percent precision would bring sensitivity to quantum effects of $\lambda_{\PH\PH\PH}$~\cite{McCullough:2013rea}. 
As will be discussed in \cref{sec:eeZH_H3}, FCC-ee would obtain a determination of $\kappa_\lambda = 1 \pm 0.27$ (27\%) or $\kappa_\lambda = 2\pm 0.28$ (14\%) using this approach \cite{FCC-FSR-Vol1}.  In combination with HL-LHC the precision obtainable would be 18\% for $\kappa_\lambda = 1$ and 11.5\% for $\kappa_\lambda = 2$.
Thus, $\epem$ precision single-Higgs measurements open the possibility of testing $\lambda_{\PH\PH\PH}$ in a way that is complementary to Higgs pair production. 
For example, in single-Higgs processes $\lambda_{\PH\PH\PH}$ enters in diagrams with two off-shell and one on-shell Higgses, 
while in pair production, both at hadron and lepton colliders, $\lambda_{\PH\PH\PH}$ contributes in a diagram
where the situation is reversed, with one off-shell and two on-shell Higgses.
%
%
%
%
In general, the loop contributions involving $\lambda_{\PH\PH\PH}$ compete with larger lowest-order contributions, and with other loop contributions (e.g.\ a top-quark loop). The sensitivity to $\lambda_{\PH\PH\PH}$ via loop effects may also be more affected by the theoretical uncertainties induced by unknown higher-order contributions and by the experimental errors of the SM input parameters. 
The latter, however, are known or will be measured with a precision that suggests these parametric uncertainties are under control~\cite{Freitas:2019bre}, as will be discussed in \cref{sec:eeZH_H3}.

\begin{figure}[h]\centering
\includegraphics[width=.5\textwidth]{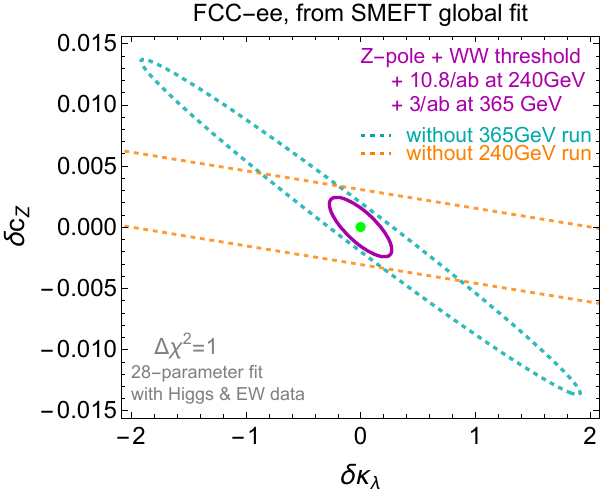}
\caption{$1\sigma$ sensitivity achievable on the trilinear self-coupling of the Higgs boson ($\delta \kappa_\lambda=\kappa_{\lambda}-1$, shown on the $x$ axis) with 240 and \SI{365}{\giga\electronvolt} runs at the FCC-ee.
Aside from the single Higgs coupling to the Z boson ($\delta c_{\PZ}$, shown on the $y$ axis), 
26 additional Higgs and EW coupling modifications have been marginalised over to obtained the ellipses shown.
}
\label{fig:sensitivityH3CC}
\end{figure}

For the study of sensitivity to $\lambda_{\PH\PH\PH}$ in single Higgs processes, the possible deviations in the Higgs self-coupling with respect to the SM have been typically parametrised within the so-called SMEFT formalism, see \cref{sec:globalSMEFT}, truncated to dimension six.\footnote{This includes only effects of ${\cal O}(1/\Lambda^2)$ with $\Lambda$ the heavy scale acting as cutoff of the EFT. As is also shown in the results in \cref{sec:globalSMEFT}, the inclusion of effects of ${\cal O}(1/\Lambda^4)$ with the precision achievable at future colliders would seem to have only a small impact on the dimension-six results, suggesting the EFT expansion is well behaved.}   
The studies presented in Ref.~\cite{deBlas:2019rxi} (see Figure 11 in that reference) reported the sensitivity to $\lambda_{\PH\PH\PH}$ for different future colliders, assuming the SM value is realized in nature, i.e.\ $\kappa_{\lambda}=1$. Although these studies take into account the effect of projected SM theory uncertainties mentioned above and also illustrate some illuminating points about the interplay of different EFT contributions, they were restricted to the EFT calculations available at that time and, as we discuss below, recent EFT calculations of additional NLO effects could affect the $\lambda_{\PH\PH\PH}$ interpretation. 
For instance, in
an EFT analysis including only the 1-loop contributions from $\lambda_{\PH\PH\PH}$ as well as any possible contribution at leading-order, measurements at a single centre-of-mass energy show an approximate degeneracy, which can efficiently be lifted by performing measurements at a second $\sqrt{s}$~\cite{DiVita:2017vrr}.
With runs at \SI{240}{} and \SI{365}{\giga\electronvolt}, the FCC-ee would achieve the sensitivity pictured in \cref{fig:sensitivityH3CC}. 
 However, without any extra model assumptions, a consistent analysis in the EFT approach at next-to-leading order (NLO) requires the introduction of contributions from 
 all the operators that are relevant at that order. 
 This introduces 
 a large number of additional EFT contributions, in particular from $\epem\PQt\PAQt$ operators, which are currently poorly constrained. This opens the possibility of testing also these new effects, though at the expense of potentially reducing the precision in $\kappa_{\lambda}$.
 For the main production process, $\epem \to \PZ\PH$, such NLO contributions have only been recently identified in Refs.~\cite{Asteriadis:2024qim,Asteriadis:2024xts}. 
Details on this calculation and its implications are discussed in \cref{sec:eeZH_NLO}. 
The study of the effect of these contributions in the determination of $\lambda_{\PH\PH\PH}$ and of the observables needed to constrain them 
will be the topic of future work.

While the previous paragraphs and the different parts of this chapter focus on introducing/discussing aspects of the use of single- and double-Higgs measurements as probes of the Higgs boson self-coupling separately, it is important to remember that these are not to be considered as alternative probes but, as mentioned above, rather complementary ones. Indeed, since single- and double-Higgs processes are affected differently by new physics, 
a disagreement between a determination of $\kappa_\lambda$ from both methods with similar precision would evidence the presence of additional new effects, making clear the synergetic aspects of the two approaches. Therefore, as in any other of situation where a sign of new physics may be discovered in the data, 
having as many different types of observables as possible where one can test the new effect would help in characterising its origin.  
It should also be remembered that a future hadron collider FCChh could reach 3\% or better precision on $\kappa_\lambda$.


\subsubsection{\texorpdfstring{Generating large corrections to the Higgs self-coupling in new physics models}{Generating large corrections to lambda(HHH) in BSM models}}\label{sec:largeH3}

As mentioned above, the dynamics and nature of the EWPT are closely connected to the Higgs self-coupling value. 
For instance, if a strong first-order phase transition is needed to explain the observed matter/anti-matter asymmetry, this is something that cannot be realised in the SM but can be obtained in many BSM theories, typically in correlation with a large (order one) deviation of $\lambda_{\PSh\PSh\PSh}$ from its SM value~\cite{Grojean:2004xa,Kanemura:2004ch,Noble:2007kk, Katz:2014bha, Huang:2016cjm}.

\begin{figure} 
{\includegraphics[width=.47\textwidth]{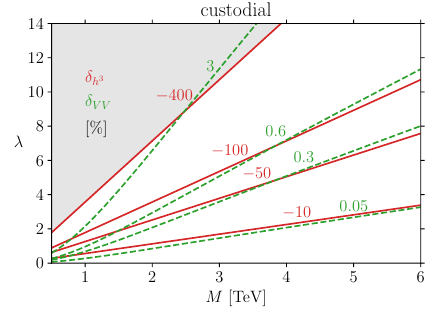}}
~~~~~{\includegraphics[width=.5\textwidth]{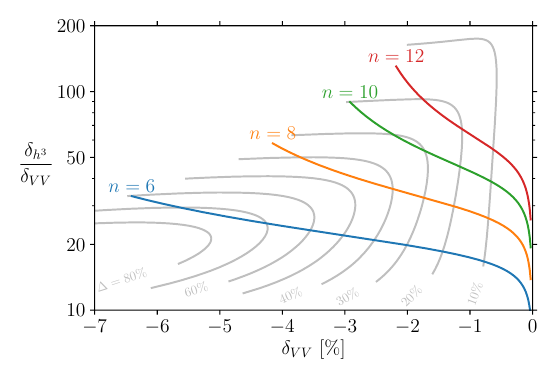}}
{\caption{(Left) Single-Higgs coupling and self-coupling deviations expected for a custodial electroweak quadruplet model, as a function of the heavy state coupling, $\lambda$, and mass, $M$. From Ref.~\cite{Durieux:2022hbu}. 
In the grey region, vacuum stability requires sizeable contributions from potential terms not included in the analysis. 
(Right) Ratio of self-coupling to $HVV$ corrections in the Gegenbauer's Twin model for several values of the charge of global symmetry breaking spurion. In the empty upper-left corner of the figure, the SM vacuum is no longer the global minimum of the potential and therefore becomes unstable. From Ref.~\cite{Durieux:2022hbu}.
In both figures, $\delta_{h^3}$ and $\delta_{VV}$ correspond to the quantities $\delta \kappa_\lambda$ and $\delta \kappa_V$ used in this report, respectively.
}\label{fig:ParametricLargek3}}
\end{figure}

Given the already tight constraints on the Higgs couplings to gauge bosons and heavy fermions, and their foreseen improvements at the (HL-)LHC, one can wonder if there is still room for possible large deviations in the Higgs self-couplings.  
In generic models of heavy decoupling new physics, the deviations in all Higgs couplings are expected to be roughly of the same order. For instance, in models that follow the {\it Strongly-Interacting Light Higgs} (SILH) power counting~\cite{Giudice:2007fh}, like composite Higgs or decoupling supersymmetric models, we expect 
\begin{equation}
\label{eq:k3SILH}
\delta \kappa_\lambda \sim \delta \kappa_V \sim \frac{v^2}{f^2},
\end{equation}
where the $f$ parameter is related to the typical coupling $g_\star$ and mass scale $m_\star$ of the new dynamics by $f \sim m_\star/g_\star$. (Similarly to $\kappa_\lambda$, we have introduced $\kappa_V$ as the Higgs coupling to vector bosons, normalised to the SM value.)
There are two main ways to depart from this relation: 
\begin{enumerate}
{\item enforcing a particular structure of the heavy new physics;}
{\item having non-decoupling/relatively light particles.}
\end{enumerate}
Even in the first class, the relation in \cref{eq:k3SILH} can be avoided in many ways and perturbativity arguments only give an upper bound on the ratio of the Higgs coupling deviations \cite{Durieux:2022hbu}:
\begin{equation}
\label{eq:ratio}
\left|  \frac{\delta \kappa_\lambda}{\delta \kappa_V}\right| \lesssim \mathrm{min} \left( \left(\frac{4\pi v}{m_{\PH}}\right)^2 , \frac{m_\star^2}{m_{\PH}^2} \right).
\end{equation}
However, among UV models with single particle extensions~\cite{deBlas:2017xtg}, only the one with a custodial electroweak quadruplet explicitly realises a parametric enhancement of the Higgs self-coupling deviation and gets close to saturate the bound in \cref{eq:ratio}, see \cref{fig:ParametricLargek3} (left). 
Another way to get parametrically large $\delta \kappa_\lambda$ is to rely on Goldstone bosons resulting from large charged spurion symmetry breaking, as in the Gegenbauer's Twin models~\cite{Durieux:2022sgm,Durieux:2022hbu}, as illustrated in the right panel of \cref{fig:ParametricLargek3}. Other dynamics that lead to large $\delta \kappa_\lambda$ include (i)  scenarios in which the Higgs is a generic bound state of a strongly coupled dynamics (i.e.\ not a Goldstone boson), (ii)  bosonic technicolour scenarios,  (iii)  Higgs-portal models.

Examples of the second class of models are models with extended scalar sectors because of large higher-order radiative corrections from the additional BSM scalars (which we generically denote $\Phi$)~\cite{Kanemura:2002vm,Kanemura:2004mg,Aoki:2012jj,Kanemura:2015fra,Kanemura:2015mxa,Arhrib:2015hoa,Kanemura:2016sos,Kanemura:2016lkz,He:2016sqr,Kanemura:2017wtm,Kanemura:2017wtm,Kanemura:2017gbi,Chiang:2018xpl,Basler:2018cwe,Senaha:2018xek,Braathen:2019pxr,Braathen:2019zoh,Kanemura:2019slf,Basler:2020nrq,Braathen:2020vwo,Bahl:2022jnx,Bahl:2022gqg,Bahl:2023eau,Aiko:2023xui,Basler:2024aaf}. 
These corrections are controlled by couplings of the general form $g_{\PH\PH\Phi\Phi}\propto(m_\Phi^2-\mathcal{M}^2)/v^2$, where $m_\Phi$ is the physical mass of the scalar $\Phi$, $\mathcal{M}$ is some mass scale controlling the decoupling of the BSM states, and $v$ is the SM vacuum expectation value. In scenarios where a splitting occurs between $\mathcal{M}$ and $m_\Phi$, the $g_{\PH\PH\Phi\Phi}$ couplings can grow rapidly and produce large BSM contributions to $\kappa_\lambda$. It should be emphasised that these loop effects involving BSM scalars and $g_{\PH\PH\Phi\Phi}$ couplings are not a perturbation of the tree-level trilinear Higgs coupling, but rather a new class of contributions only entering at the loop level. They can therefore become larger than the tree-level contribution, without being associated with a violation of perturbativity (this situation is analogous to that of loop-induced processes like e.g.\ $\PH\to\PGg\PGg$ or $\PH\to\PGg\PZ$). 

\begin{figure} 
{ \includegraphics[width=.47\textwidth]{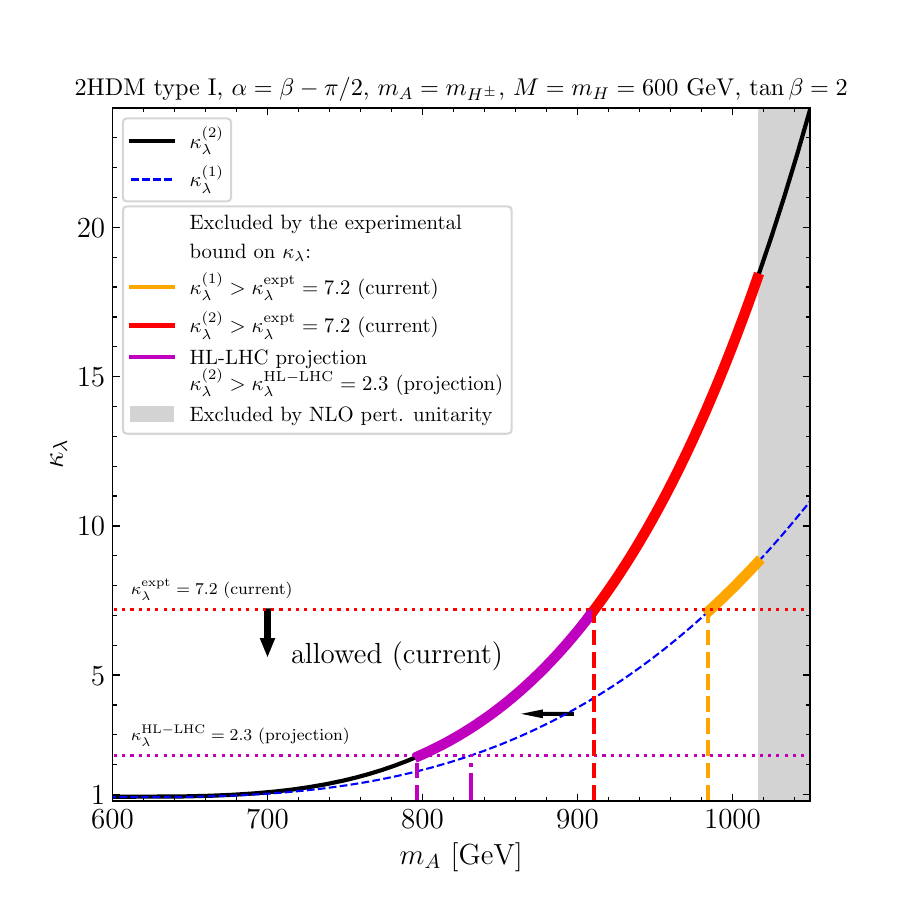}}
~~~{ \includegraphics[width=.45\textwidth]{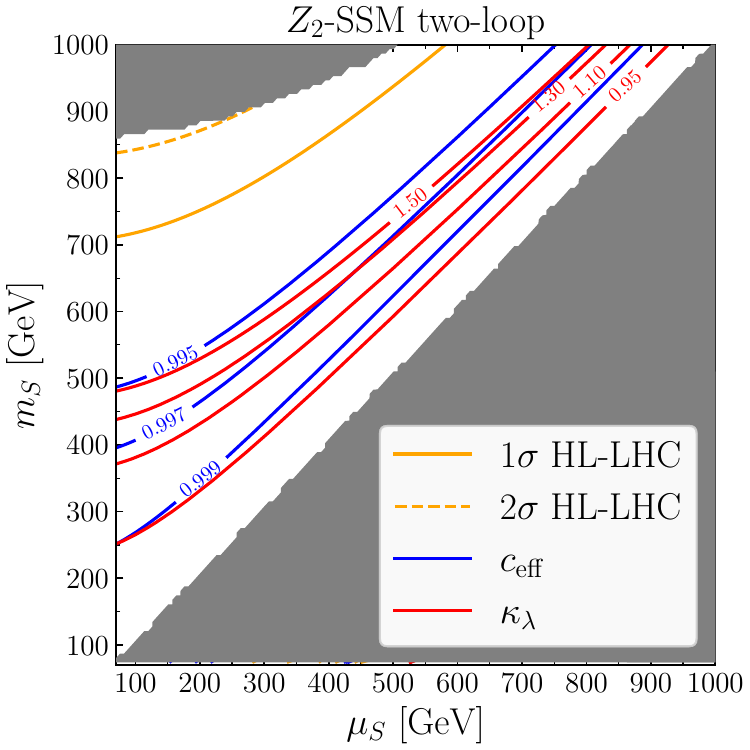}} 
{\caption{(Left) One- and two-loop predictions for $\kappa_\lambda$ as a function of $m_{\PSA}=m_{\PSHpm}$ in an aligned 2HDM scenario of Yukawa type I (see footnote~\ref{footnote:alignmentlimit}). The other BSM input parameters are chosen to be $M=m_{\PH}=\SI{600}{\giga\electronvolt}$ and $t_\beta=2$. 
(Right) Contour lines of $\kappa_\lambda$ (red) and $c_\text{eff}$ (blue), computed at two loops, in the $\{\mu_S,m_S\}$ parameter plane of the $Z_2$-SSM (with $\lambda_S=0$). The orange solid and dashed lines indicate the regions of parameter space probed by single-Higgs measurements at the HL-LHC (assuming SM-like central values) at the $1\sigma$ and $2\sigma$ levels respectively. 
}\label{fig:benchmark_masssplittingeffects}}
\end{figure}

The left panel of \cref{fig:benchmark_masssplittingeffects} presents an example of mass-splitting effects in $\kappa_\lambda$ up to the two-loop level in an aligned 2HDM~\cite{Bahl:2022jnx}. 
(In what follows, and for the rest of this subsection, we will slightly change the notation, using $\PSh$ to refer to the SM-like Higgs boson, to distinguish it from the other possible heavy Higgs bosons that appear, e.g.\ in 2HDM models, for which the notation $\PH, \PH^\pm$ is used.)
For the scenario considered here,\footnote{We refer the reader to Ref.~\cite{Braathen:2019zoh,Bahl:2022jnx} for details on our choices of conventions and notations for the CP-conserving 2HDM. $M$ is the BSM mass scale of the 2HDM (corresponding to the generic $\mathcal{M}$), $m_{\PH},\ m_{\PSA},\ m_{\PSHpm}$ are the masses of the BSM scalars, and $\alpha$ and $\beta$ are respectively the mixing angles in the CP-even sector and in the CP-odd and charged sectors. The alignment limit~\cite{Gunion:2002zf} means that the EW vacuum expectation value is aligned in field space with the detected Higgs boson $\PH$ of mass 125 GeV, and corresponds to $\alpha=\beta-\pi/2$ in terms of mixing angles. As a consequence of this limit, which has to be enforced in absence of any structural dynamics,
all couplings of $\PSh$ are SM-like at the tree level. \label{footnote:alignmentlimit}} with $M=m_{\PH}=\SI{600}{\giga\electronvolt}$, $m_{\PSA}=m_{\PSHpm}$ and $\tan\beta=2$, all relevant theoretical and experimental constraints have been verified. The comparison of the two-loop prediction for $\kappa_\lambda$ with recent experimental bounds on this coupling modifier~\cite{ATLAS:2022jtk} at the LHC allows excluding significant parts of the otherwise unconstrained parameter space of the model. For instance, in this specific scenario, the mass range $m_{\PSA}=m_{\PSHpm}\gtrsim \SI{900}{\giga\electronvolt}$ (or equivalently mass splittings $m_{\PSA}-m_{\PSH}\gtrsim\SI{300}{\giga\electronvolt}$) would lead to the two-loop prediction for $\kappa_\lambda$ to be larger than 7.2~\cite{ATLAS:2024ish} 
and is therefore excluded.  
Additionally, the mass range $m_{\PSA}=m_{\PSHpm}\gtrsim\SI{800}{\giga\electronvolt}$ of this scenario would be probed at the HL-LHC, as indicated in purple in the left panel of \cref{fig:benchmark_masssplittingeffects}. 
%
Importantly, the mass-splitting effects shown in that same figure 
and the exclusions obtained with current LHC data, or expected with future colliders, are not restricted to the 2HDM, but have been found to occur in various BSM theories with additional scalars -- as can now be studied with the public tool \texttt{anyH3}~\cite{Bahl:2023eau}. 

The occurrence of large radiative corrections in $\lambda_{\PSh\PSh\PSh}$ raises the questions of how sizeable higher-order corrections to other Higgs properties involving $g_{\PSh\PSh\Phi\Phi}$ couplings can become, and where one would first observe a deviation from the SM predictions in these types of scenarios. Power counting arguments~\cite{Bahl:2024xyz} show that, in the limit of large $g_{\PSh\PSh\Phi\Phi}$ couplings, the leading one-loop BSM contributions to the trilinear Higgs coupling are of $\mathcal{O}(g_{\PSh\PSh\Phi\Phi}^2)$, while those in single-Higgs couplings grow at most linearly with $g_{\PSh\PSh\Phi\Phi}$. 

As a simple but illustrative example, we consider the $Z_2$-SSM,\footnote{We refer the reader to e.g.\ Ref.~\cite{Braathen:2020vwo} for an overview of the notations and conventions we employ in this work (we consider here the $N=1$ case of the $O(N)$-symmetric SSM of Ref.~\cite{Braathen:2020vwo}).} i.e.\ a real-singlet extension of the SM with an unbroken global $Z_2$ symmetry. Due to this symmetry, the BSM scalar $S$ does not mix with the detected Higgs boson at 125 GeV. Its mass takes the form $m_S^2=\mu_S^2+\lambda_{\PH\Phi}v^2$ where $\mu_S$ is the singlet Lagrangian mass term and $\lambda_{\PH\Phi}$ the portal quartic coupling between the singlet and the (SM-like) doublet. The coupling $\lambda_{\PH\Phi}$ plays in the $Z_2$-SSM the exact role of the generic $g_{\PSh\PSh\Phi\Phi}$ coupling in the discussion above. In this model, single-Higgs couplings $g_{\PSh XX}$, where $X$ can be a gauge boson or fermion, only receive BSM contributions via external-leg corrections --- there are no mixing effects at the tree level, and moreover vertex-type corrections do not appear because of the singlet nature of the BSM scalar and the unbroken $Z_2$ symmetry.  
This allows obtaining compact expressions for the single-Higgs coupling modifier $c_\text{eff}\equiv g_{\PSh XX}/g_{\PSh XX}^\text{SM}$ at one and two loops~\cite{Bahl:2024xyz}. 
In the right panel of \cref{fig:benchmark_masssplittingeffects}, we present contour lines for $\kappa_\lambda$(red) as well as the quantity $c_\text{eff}$ (blue), both computed at the two-loop order, in the $\{\mu_S,m_S\}$ parameter plane of the $Z_2$-SSM. As a conservative choice, we fix $\lambda_S=0$ in the plane; this cancels the effects from two-loop terms involving $\lambda_S$ in $\kappa_\lambda$ and $c_\text{eff}$ (we note that for these contributions, like for those involving only $g_{\PSh\PSh\Phi\Phi}$ couplings, the effects in $\kappa_\lambda$ have a higher scaling in powers of $\lambda_{\PH\Phi}$ than those in $c_\text{eff}$). We implemented the expected levels of accuracy on single-Higgs couplings at the HL-LHC as ``measurements'' (assuming SM-like central values) in \texttt{HiggsSignals}~\cite{Bahl:2022igd}, and orange lines in the left panel of  \cref{fig:benchmark_masssplittingeffects} correspond respectively to the $1\sigma$ (solid) and $2\sigma$ (dashed) exclusion ranges from HL-LHC measurements. 
The values indicated in each contour line for $\kappa_\lambda$ ($c_\text{eff}$) indicate the maximum (minimum) value of the parameter 
predicted in the region below the line.
Therefore, these lines provide information about the regions of the $Z_2$-SSM parameter space for which large corrections to $\kappa_\lambda$ can be obtained while those to $c_\text{eff}$ are small. 
As can be observed, comparing for instance the uppermost red and blue contour lines in the right panel of \cref{fig:benchmark_masssplittingeffects}, values of $\kappa_\lambda$ up to 1.5 could be obtained while corrections 
to $c_\text{eff}$ are kept below $0.5\%$.

Other examples where this hierarchy of corrections can be obtained from large radiative corrections include, for instance, the Inert Doublet Model (IDM), a variant of the 2HDM in which the second doublet is charged under an unbroken $\mathbb{Z}_2$ symmetry, see e.g.\ Refs.~\cite{Deshpande:1977rw,Barbieri:2006dq,Braathen:2019zoh,Aiko:2023nqj} for an overview of the model.
As in the above example for the $Z_2$-SSM, using the two-loop predictions for $\kappa_\lambda$ (from Refs.~\cite{Braathen:2019pxr,Braathen:2019zoh,Aiko:2023nqj}) and the one-loop BSM deviation in the $g_{\PSh\PZ\PZ}$ coupling, one can find regions of the IDM parameter space where large ${\cal O}(1)$ corrections of $\kappa_\lambda$ are predicted, while keeping the modifications of the Higgs couplings to vector bosons well below $1\%$. 

Of course, future Higgs/electroweak/top factories will be making precise measurements of $c_\text{eff}$ and $\kappa_\lambda$ that will increasingly constrain the available parameter space.

As we discuss next, the large values of $\lambda_{\PSh\PSh\PSh}$ discussed in this section
could be probed via double-Higgs production 
at future \epem colliders operating at CM energies around 500 GeV. 
Similarly, while corrections to single Higgs coupling can remain small in these models, 
deviations at the level of a few tens of percent in the value of $\kappa_\lambda$ would also 
induce effects at one-loop level in single-Higgs observables that could be tested 
at future \epem Higgs factories operating at more than one centre-of-mass energy such as FCC-ee, as is explained in \cref{sec:eeZH_H3}.


\subsubsection{\texorpdfstring{Progress in Higgs-pair production at $\epem$ colliders}{Progress in Higgs-pair production at e+ e- colliders}\label{sec:HHprod}}

Double Higgs production at $\Pep\Pem$ colliders with centre-of-mass energies $\sqrt{s} \geq$ \SI{500}{\giga\electronvolt} comprises both di-Higgs production from $\PW\PW$ fusion, $\PGn\PAGn\PH\PH$, dominant at centre-of-mass energies above \SI{1.2}{\tera\electronvolt} (see \cref{fig:HHxs_kappalambda}), as well as double Higgs-strahlung, $\PZ\PH\PH$, dominant at lower centre-of-mass energies. Both processes have been studied about $10$ years ago in detailed Geant4-based simulation of the ILD detector concept, considering the $\PH\PH \rightarrow  4\PQb$~\cite{Durig:2016jrs, Tian:2013} and $\PH\PH \rightarrow \PQb\PAQb\PW\PW^*$~\cite{Kurata:2013} final states.
For the SM value of the tri-linear Higgs self-coupling $\lambda_{\PH\PH\PH}$, the combination of the two channels yielded precisions of \SI{27}{\%} at \SI{500}{\giga\electronvolt} and \SI{10}{\%} when combining both centre-of-mass energies. A few years later, CLICdp studied the prospects at $\sqrt{s}=$\SI{1.4}{\tera\electronvolt} and \SI{3}{\tera\electronvolt}~\cite{Roloff:2019crr}, also in detailed Geant4-based simulation, reaching precisions down to $[-\SI{8}{\%},+\SI{11}{\%}]$, again assuming the SM value of $\lambda_{\PH\PH\PH}$. It should be noted that in particular for the $\PZ\PH\PH$  process, i.e.\ the measurements at around \SI{500}{\giga\electronvolt}, it has been shown explicitly that, for decoupling new physics, all other parameters entering the interpretation of the $\PZ\PH\PH$ cross-section will be determined with sufficient precision at a Higgs factory so that their impact on the extraction of $\lambda_{\PH\PH\PH}$ is negligible~\cite{Barklow:2017awn}. 
On the other hand, having relatively light new particle exchanges, e.g.\ in the last diagram in \cref{fig:HHxs_diagrams}, would still affect the interpretation of a possible effect in the data in terms of $\kappa_\lambda$. Being a tree-level exchange limits the type of possible new contributions, but additional observables/assumptions would still be needed to determine $\lambda_{\PH\PH\PH}$ separately from the 
additional new physics effects the process gives access to. (See, e.g.\ \cref{sec:BSMHhh} for the case of trilinear interactions of two Higgs bosons with an extra scalar.)

\begin{figure}[htbp]
    \centering
    \begin{subfigure}{.5\textwidth}
    \centering
        \includegraphics[width=0.95\textwidth]{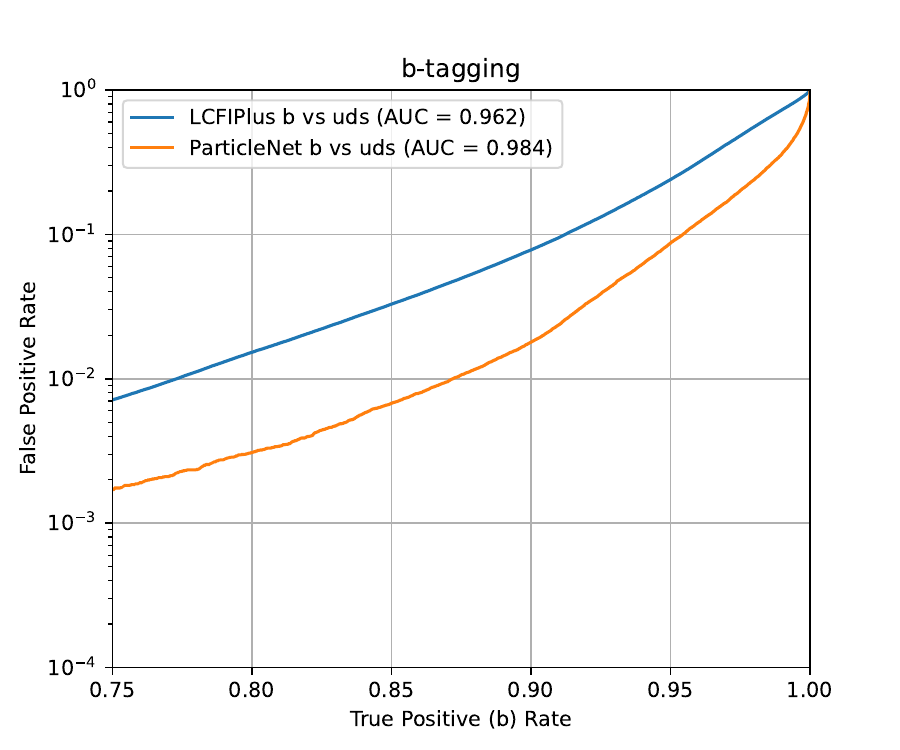}
        \caption{}
        \label{fig:anaimprove:flav}    
    \end{subfigure}\hfill%
    \begin{subfigure}{.5\textwidth}
    \centering
        \includegraphics[width=0.95\textwidth]{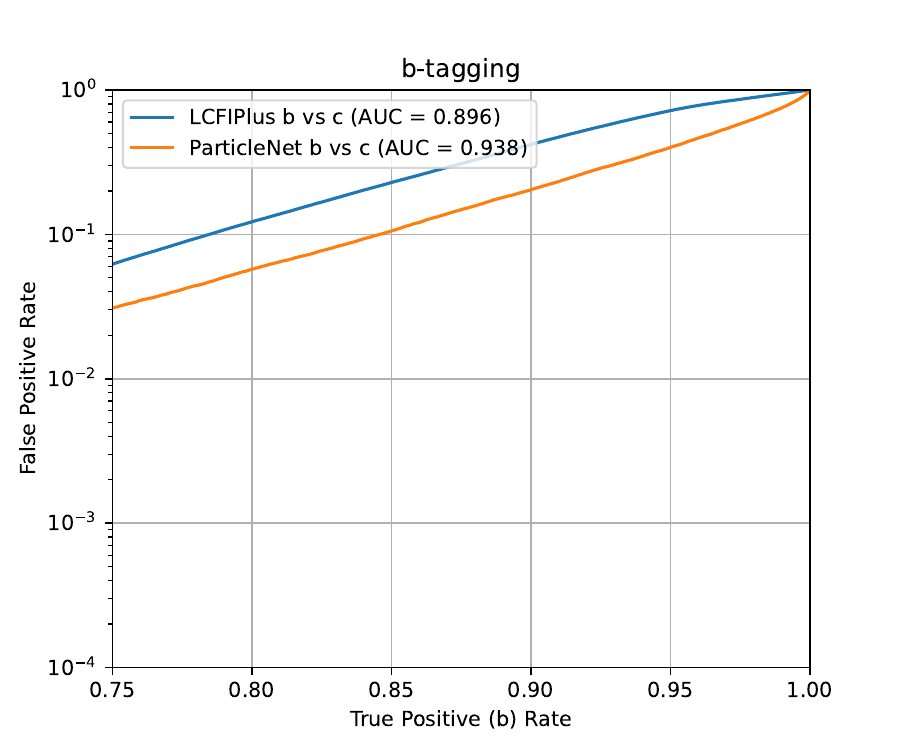}
        \caption{}
    \end{subfigure}\hfill%
    \begin{subfigure}{.5\textwidth}
        \centering
        \includegraphics[width=0.95\textwidth]{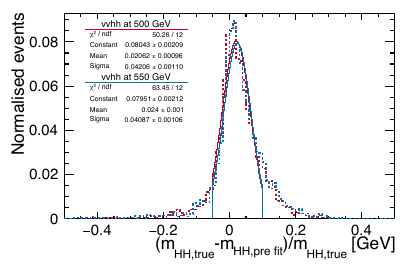}
        \caption{}
        \label{fig:anaimprove:kine}    
    \end{subfigure}%
    \begin{subfigure}{.5\textwidth}
        \centering
        \includegraphics[width=0.95\textwidth]{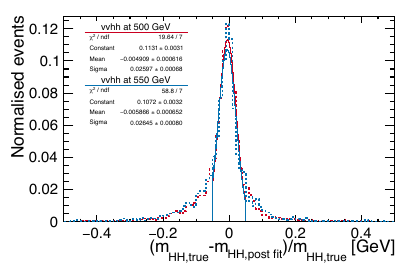}
        \caption{}
    \end{subfigure}%
    \caption{Reconstruction improvements considered in the updated $\PZ\PH\PH$ projection: (a), (b) \PQb -tagging efficiency (true positive rate) increases by about 10\% (relative) for the same background efficiency (false positive rate) at typical working points of the previous analysis (80\% \PQb -tagging efficiency with LCFIPlus) (c), (d) the resolution on the di-Higgs invariant mass improves by about a factor of two when the new semi-leptonic decay correction and kinematic fit are applied.}
    \label{fig:anaimprove}
\end{figure}

The  analysis of $\PZ\PH\PH$ production at \SI{500}{\giga\electronvolt} is being updated in detailed simulations of the ILD detector concept, covering the $\PH\PH \rightarrow \PQb\PAQb \PQb\PAQb$ and $\PZ \rightarrow$ $\PQq\PAQq$ / $\Pep\Pem$ / $\PGmp\PGmm$ / $\PGn\PAGn$ channels. While a full new analysis is ongoing, several key ingredients of the analysis have been updated to more modern analysis tools based on the MC production for the ILD Interim Design Report~\cite{ILDConceptGroup:2020sfq}. This concerns in particular the jet-flavour tagging, for which by an update from LCFIPlus~\cite{Suehara:2015ura} to a ParticleNet-based approach~\cite{Qu:2019gqs} so far a relative increase of the \PQb -tagging efficiency of 10\% at the same level of background rejection could be achieved on 6-jet events as shown in Figure~\ref{fig:anaimprove:flav}. Since the analysis requires at least three jets to be \PQb -tagged to reject $\PQt\PAQt$ and $\PZ\PZ\PH$ (with $\PZ \rightarrow$ $\PQc\PAQc$ /  $\PQs\PAQs$ / $\PQu\PAQu$ / $\PQd\PAQd$ ) backgrounds, the reduced inefficiency enters to the third power. The most signal-like $\PZ\PZ\PH$ background with at least one of the $\PZ \rightarrow \PQb\PAQb$ can only be rejected based on event kinematics. In this respect, an improvement of at least 10\% higher efficiency at same background level is expected from recent advances in correcting for semi-leptonic \PQb -decays and kinematic fitting~\cite{Einhaus:2022bnv}, as well as in using the full matrix elements for $\PZ\PH\PH$  and $\PZ\PZ\PH$. As example, Fig.~\ref{fig:anaimprove:kine} shows the improvement in the di-Higgs invariant mass reconstruction in the $\PZ \rightarrow \PGn\PAGn$ channel. For now, these two improvements will be propagated through the previous analysis~\cite{Durig:2016jrs}. In view of even significantly better flavour tagging tools reported in \cref{sec:com:flavourtagging} (see also e.g.\ Refs.~\cite{Bedeschi:2022rnj, Qu:2022mxj, Suehara:2024qoc}),  
as well as the potential expected from using machine-learning (ML) through-out the event selection, this is a quite conservative assumption. All numbers below have been scaled to the standard ILC running scenario~\cite{Barklow:2015tja}, i.e.\ \SI{4}{\per\atto\barn} with $P(\Pem,\Pep) = (\mp 80\%, \pm 30\%)$.

From this re-analysis of the channels studied in Ref.~\cite{Durig:2016jrs}, the obtained precision on the $\PZ\PH\PH$ production cross-section improves from the \SI{21.1}{\%} obtained in Ref.~\cite{Durig:2016jrs} to now \SI{16.0}{\%} 
for the SM case. 

In combination with the $\PH\PH \rightarrow \PQb\PAQb \PW\PW^*$ channel, the precision on the cross-section improves from the \SI{16.8}{\%} obtained in Ref.~\cite{Durig:2016jrs} to now \SI{12.8}{\%}, again for the SM case, while the corresponding precision on the self-coupling improves from \SI{26.6}{\%} to \SI{20.8}{\%}. By including $\PZ \rightarrow \PGtp\PGtm$, as well as $\PH\PH \rightarrow \PQb\PAQb \PGtp\PGtm$ and other modes well reconstructible in the $\Pep\Pem$ environment, the expected precision on the $\PZ\PH\PH$ cross-section improves further to \SI{11.2}{\%}, corresponding to a precision on a SM-like self-coupling value $\uplambda_{\mathrm{SM}}$ of \SI{18.1}{\%}.
This sensitivity of 18\% on $\uplambda_{\mathrm{SM}}$ corresponds to \SI{4}{\per\atto\barn} collected at $\sqrt{s}=500$~\GeV, analysed using currently-available tools.

One of the main limiting factors not yet addressed by novel algorithms is the jet clustering. Assuming that future developments, e.g.\ based on ML, will improve the di-jet mass resolution, we estimate that $\uplambda_{\mathrm{SM}}$ could be determined with a precision of \SI{15}{\%} from $\PZ\PH\PH$ production alone -- this just illustrates some of the further room for improvement. 

The analysis of the $\PW\PW$ fusion production has been investigated in Refs.~\cite{Tian:2013, Kurata:2013} at a centre-of-mass energy of \SI{1}{TeV}. At \SI{500}{GeV}, the separation of the $\PW\PW$ fusion contribution is still ongoing work and will be covered in future updates. If the $\PW\PW$ fusion cross-section at \SI{500}{GeV} could be constrained to the level of \SI{120}{\%} from the $\PH\PH \rightarrow \PQb\PAQb \PQb\PAQb$, translating to \SI{86}{\%} in combination with the other $\PH\PH$ decay modes, then combining the $\PW\PW$ fusion and $\PZ\PH\PH$ measurement would allow to constrain the SM value of $\uplambda_{\mathrm{SM}}$ to a precision of \SI{17}{\%} with current tools and \SI{14}{\%} when assuming improved jet clustering. 

The results summarised above have been extrapolated to other centre-of-mass energies. In particular, partial MC productions have been performed in full, Geant4-based simulation of ILD as well as using SGV~\cite{Berggren:2012ar} and centre-of-mass energies of \SI{550}{\giga\electronvolt} and \SI{600}{\giga\electronvolt}, showing that beyond the expected scaling with cross-section, both the flavour tag as well as the kinematic reconstruction profit from the higher boost of the produced bosons. This is true in particular for the jet clustering, which benefits from less confusion towards higher COM energies \cite{Torndal:2023mmr}.

\begin{figure}[htbp]
    \centering
    \begin{subfigure}{.5\textwidth}
    \centering
        \includegraphics[width=0.95\textwidth]{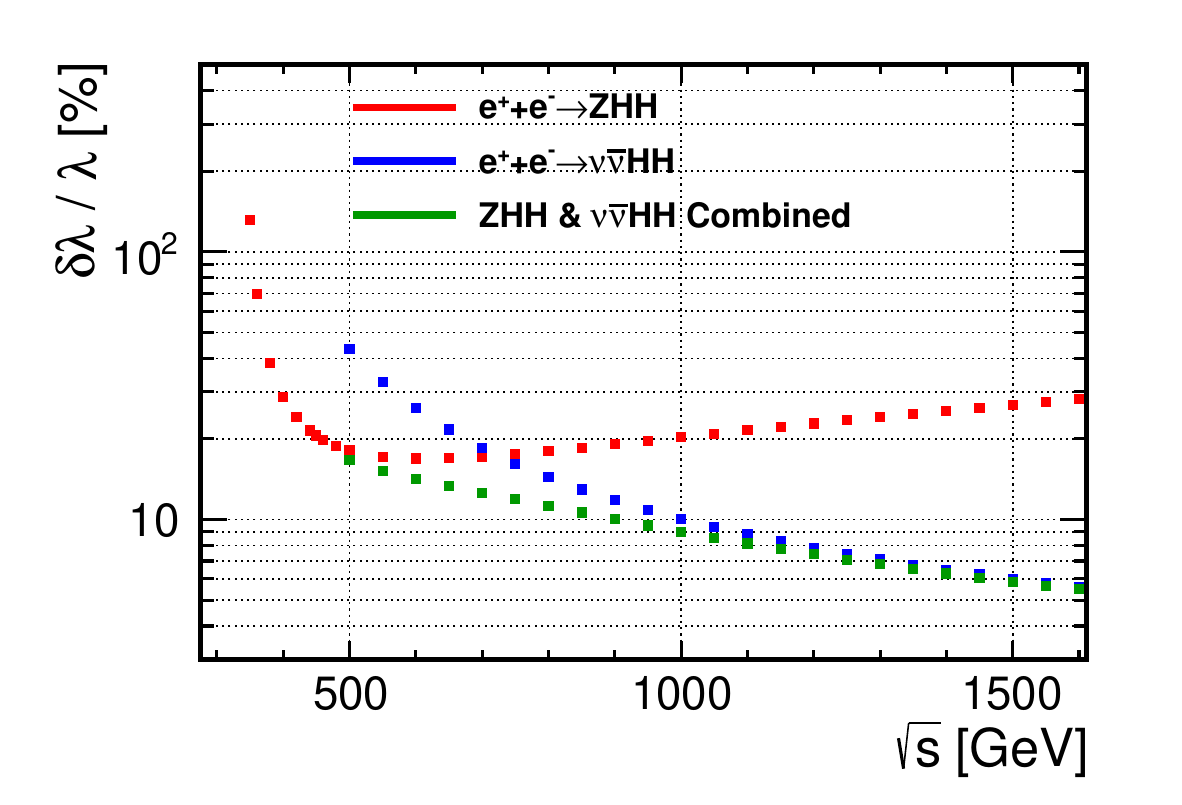}
        \caption{}
        \label{fig:lambda:ZHH_WWfusion}    
    \end{subfigure}\hfill%
    \begin{subfigure}{.5\textwidth}
        \centering
        \includegraphics[width=0.95\textwidth]{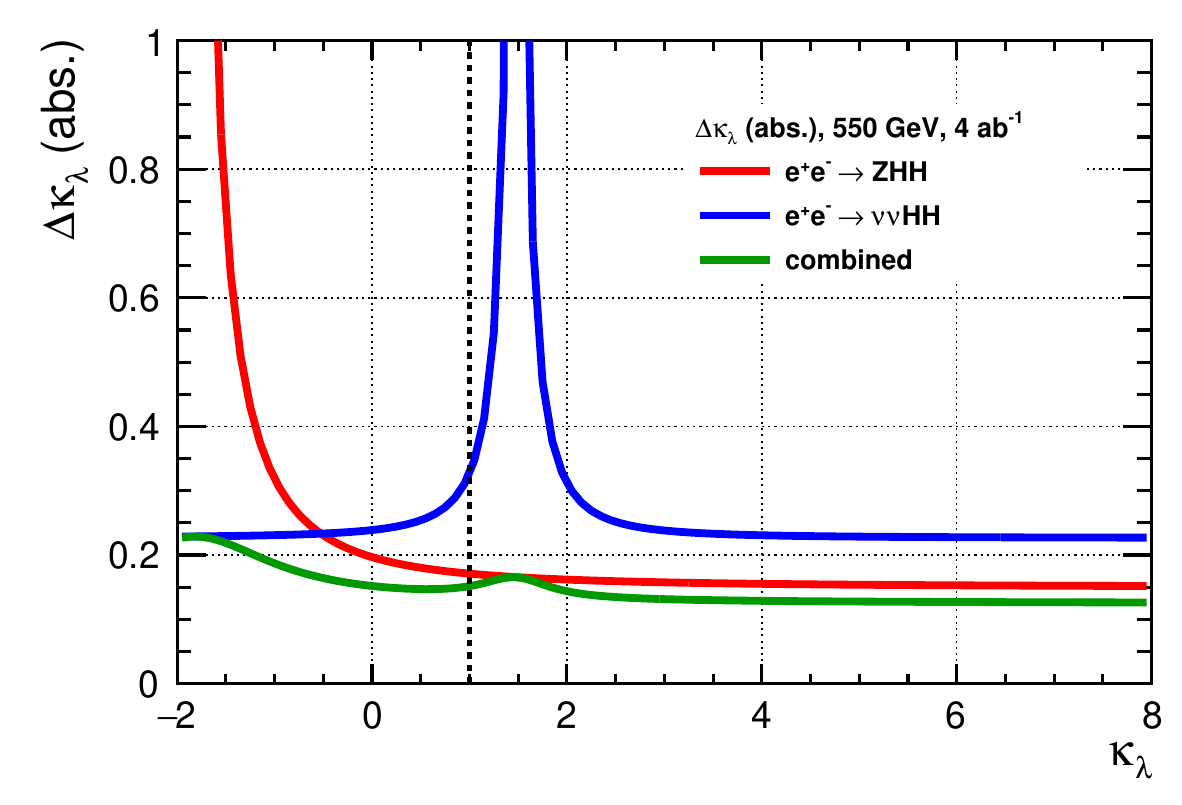}
        \caption{}
        \label{fig:lambda:BSM}    
    \end{subfigure}%
    \caption{Projected precisions on the tri-linear Higgs self-coupling $\lambda_{\PH\PH\PH}$ for double Higgs-strahlung, di-Higgs production from $\PW\PW$ fusion, and their combination: (a) relative precision as a function of the centre-of-mass energy and (b) absolute uncertainty on the determination $\kappa_{\lambda}$ as a function of the value assumed to be realized in nature. }
    \label{fig:lambda}
\end{figure}

\Cref{fig:lambda:ZHH_WWfusion} shows the resulting evolution of the expected precision on $\uplambda$ with the centre-of-mass energy, for both production modes separately and for their combination. For instance at $\sqrt{s}=$\SI{550}{\giga\electronvolt}, which is the envisioned energy for C$^3$ as well as for a Linear Collider Facility at CERN, the expected combined precision on $\uplambda_{\mathrm{SM}}$ for 4.4\abinv\ is \SI{15}{\%} using current tools; a substantial improvement over the \SI{26.6}{\%} of Ref.~\cite{Durig:2016jrs}.

Enlarging the scope to the BSM case, it is important to note that the two production modes play very complementary roles due to their very different interference patterns. This is illustrated in \cref{fig:lambda:BSM}, which shows the behaviour of the expected precision as a function of $\lambda_{\PH\PH\PH}/\lambda_{\PH\PH\PH}^{\mathrm{SM}}$ for the two production modes at $\sqrt{s}=$\SI{550}{\giga\electronvolt}. This shows that measurements of $\PZ\PH\PH$  production at centre-of-mass energies around \SI{500}{\giga\electronvolt} play a crucial role in pinning down the BSM behaviour of the tri-linear Higgs self-coupling.


\subsubsection{\texorpdfstring{Probing the Higgs self-coupling via single-Higgs processes}{Probing the Higgs self-coupling via single-Higgs processes}}\label{sec:eeZH_H3}

The measurements of single-Higgs processes that could be taken at any
of the future \epem Higgs factories, namely the inclusive cross section and rates for different decay channels of the Higgs boson, 
are typically interpreted in terms of single-Higgs couplings to the SM fermions and vector bosons, quantities that in the SM contribute
to these processes at tree level. But, clearly, owing to the quantum nature of these observables, such interpretation needs to 
be improved beyond this leading order approximation, as what these experiments measure are not single-Higgs couplings, but
combinations of such couplings and all the other interactions that enter in the process at the quantum level.

In the SM, the 1-loop corrections coming from the Higgs self-coupling to, e.g.\ the $\epem \to \PZ\PH$ cross section at 240 GeV,  
are relatively small, increasing the LO cross section by $\sim1.5\%$. In the full set of NLO electroweak contributions, bosonic and fermionic corrections
partially cancel, so the total NLO correction amount to $\sim 3\%$. 
Furthermore, the NNLO from fermion loops are currently known to increase only the NLO result by $0.7\%$~\cite{Freitas:2023iyx}, and the uncertainty from missing bosonic electroweak corrections has been estimated to be less than $0.3\%$~\cite{Freitas:2023iyx}. Therefore, it is reasonable to expect 
that, with some moderate progress in the SM calculations, by the time of a future Higgs factory the SM uncertainty due to higher-order corrections will be further reduced and become much smaller than the leading $\lambda_{\PH\PH\PH}$ effects. Similarly, benefiting from future EW and Higgs measurements, parametric uncertainties from the EW input parameters are also expected to be very small, well below the experimental precision.~\footnote{
An estimate using the SM part of the calculation of \cite{Asteriadis:2024xts} gives an expected parametric uncertainty at FCC-ee well below $0.1\%$.
It is of particular relevance to improve the precision of the Higgs boson mass to around or below 20 MeV. See Ref.~\cite{Freitas:2019bre} for estimates on the corresponding SM uncertainties on Higgs decay rates.}  

Thus, with an experimental precision at the per mille level in $\epem \to \PZ\PH$, it is clear that a future \epem Higgs factory would provide a clean test of the SM prediction for $\lambda_{\PH\PH\PH}$. 
The precision of such test, or the one on the determination of $\lambda_{\PH\PH\PH}$ if everything but this parameter is assumed to be SM-like, could be obtained parametrising the modifications of single-Higgs observables in terms of the following contribution to the Higgs potential
\begin{equation}
\Delta V(\PH)=(\kappa_\lambda -1) \frac 12 \frac{m_{\PH}^2}{v} \PH^3 .
\label{eq:VHgen}
\end{equation}
%
%
In general, however,  
estimating the actual precision of the determination of $\lambda_{\PH\PH\PH}$ that would be possible from this (or any other) 
type of measurements is not something that can be addressed in a completely model-independent way. For instance,
different new particles could be exchanged virtually, e.g.\ in the EW loops, opening the possibility of testing 
many other new physics effects, but also requiring a commensurate number of other observables to identify the 
origin of a possible non-SM effect in the data. 
As mentioned in the previous section, this can be addressed in a fairly general way for the case of decoupling new physics. 
The use of additional observables needed to constrain all the possible additional loop effects and extract $\lambda_{\PH\PH\PH}$ in this case is briefly discussed at the end of this chapter and is part of future work.
%

\begin{figure}[h]\centering
\includegraphics[width=.5\textwidth]{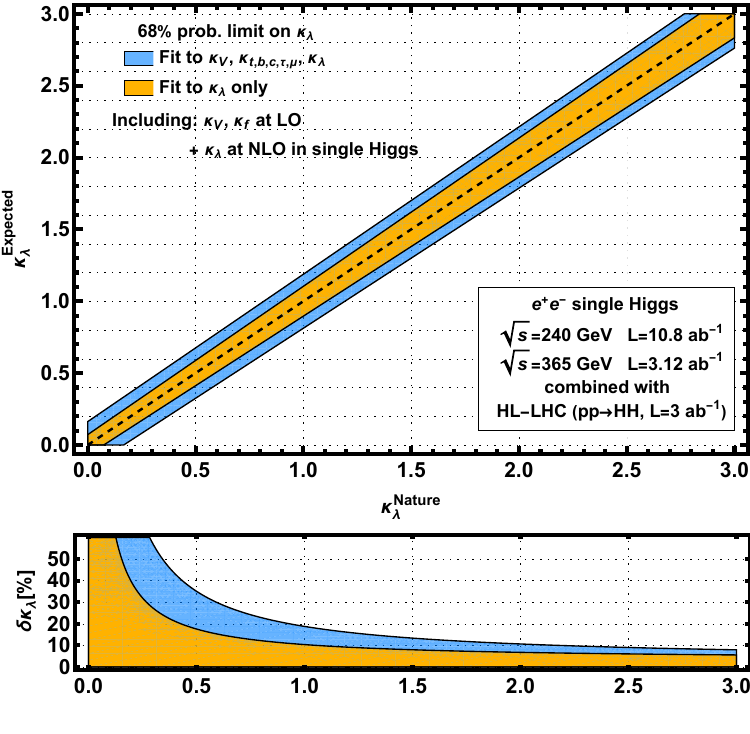}
\caption{$1\sigma$ sensitivity on the trilinear self-coupling of the Higgs boson
achievable with the single-Higgs measurements at 240 and 365 GeV from FCC-ee,
in combination with the HL-LHC results from double-Higgs production. 
The results are obtained assuming different values of $\kappa_\lambda$ are realized in nature, and presented under two assumptions in the Lagrangian in \cref{eq:Lnonlinear}: (orange) all parameters are set to their SM values except for $\kappa_\lambda$; (blue) all couplings are considered as floating parameters and marginalized over to obtain the sensitivity to $\kappa_\lambda$. The lower panel shows the relative uncertainty. 
}
\label{fig:KlambdaKi}
\end{figure}

\Cref{eq:VHgen} could be used to extend the simplest Lagrangian description of 
the scenarios of the so-called $\kappa$ framework~\cite{LHCHiggsCrossSectionWorkingGroup:2012nn} 
considering only modifications of SM-like interactions to include corrections to the Higgs self-coupling,
%
\begin{equation}
\Delta {\cal L}=\Delta V(\PH) +  2 m_Z^2 \kappa_V \frac{\PH}{v}Z_\mu Z^\mu +  2 m_W^2 \kappa_V \frac{\PH}{v}W_\mu W^\mu - \sum_\psi m_\psi \overline{\psi}_L  \kappa_\psi \frac{\PH}{v} \psi_R +{\rm h.c.} ,
\label{eq:Lnonlinear}
\end{equation}
where we have assumed custodial symmetry. 
A simple exercise performing a frequentist fit of \cref{eq:VHgen} and \cref{eq:Lnonlinear} to the expected precision of all the single-Higgs measurements at \epem colliders using two centre-of-mass energies, taking as an example FCC-ee at 240 and 365 GeV (from Ref.~\cite{deBlas:2022ofj}, updated to the FCC-ee baseline with 4 interaction points~\cite{FCC-FSR-Vol2}, see \cref{table:lumi:FCC} for the actual luminosities), would give the constraints on $\kappa_\lambda$ shown in \cref{fig:KlambdaKi}. 
The results are obtained including also the information that will be available from the HL-LHC
studies of Higgs pair production. 
In the figure we consider not only the possibility of $\kappa_\lambda=1$ but also use Asimov datasets generated with \cref{eq:VHgen} 
for different values of $\kappa_\lambda$. We use the theoretical expressions of Refs.~\cite{Degrassi:2016wml,DiVita:2017vrr} to describe the one-loop dependence of $\kappa_\lambda$, while the effects of the modifications of other Higgs couplings are kept at leading order. 
One thing to note is that, with $\kappa_\lambda$ entering at the loop level in single-Higgs observables, for moderate deviations from the SM value 
the change in the single Higgs cross sections is rather small and the absolute precision on $\kappa_\lambda$ at \epem does not change substantially. 
This leads to a determination of $\kappa_\lambda = 1 \pm 0.27$ (27\%) or $\kappa_\lambda = 2\pm 0.28$ (14\%) for FCC-ee.  In combination with HL-LHC the precision obtainable would be 18\% for $\kappa_\lambda = 1$ and 11\% for $\kappa_\lambda = 2$.
It is important to note that a disagreement between FCC-ee and HL-LHC on a similar-precision $\kappa_\lambda$ extraction would be a sign of new physics in the $\PZ\PH$ loops, making the measurement from single-Higgs production very much synergistic with that of pair-production at HL-LHC.

To give precise meaning to the modifications in \cref{eq:Lnonlinear}, this must be embedded in a concrete model or in
a well-defined effective field theory. 
The projections available for $\kappa_\lambda$ from single-Higgs measurements in Refs.~\cite{DiVita:2017vrr,deBlas:2019rxi} were prepared in the so-called Standard Model Effective Field Theory (SMEFT).
We refer to \cref{sec:globalinterpretations} for a small description of the assumptions defining this framework
to describe the low-energy effects of decoupling new physics. 
Following the power-counting of the SMEFT, at the leading order in the EFT expansion --- a power series in $1/\Lambda$, with $\Lambda$ the cut-off of the EFT --- there are many more types of possible sources of BSM effects than those in \cref{eq:Lnonlinear}, including not only modifications of the SM Higgs properties, but also of the EW interactions and types of operators absent in the SM. 
While the large number of BSM effects provides a more general description of new physics scenarios, global analyses including different types of observables are required if one wants to individually constrain all the different directions in the EFT parameter space.
This is particularly relevant when going beyond tree level in the description of new physics effects, 
as needed for the determination of $\kappa_\lambda$ from single-Higgs couplings, 
as this increases the number of SMEFT operators that can enter in the analysis via their loop contributions. 
The status of the NLO calculations required for this and steps needed to extend the results in Refs.~\cite{DiVita:2017vrr,deBlas:2019rxi} are described in the next section.


It is important to note that, while these global analyses of the sensitivity to $\lambda_{\PH\PH\PH}$ in single-Higgs measurements performed within the EFT formalism depend on a large number of parameters that need to be constrained from data, a consequence of this is that the resulting $\lambda_{\PH\PH\PH}$-sensitivity is expected to be a conservative estimate of that which would be possible in actual models described by the EFT. 
Indeed, interpretations within specific models, which match into lower-dimensional slices of the EFT parameter space, add extra theory correlations between the different operators, reducing the 
uncertainty that would be present should all these operators be independent degrees of freedom in the study. For a trivial example, a model predicting only a deviation of the Higgs self-coupling would match into the results of a SMEFT fit with only one operator to dimension six, $(\phi^\dagger \phi)^3$. For our purposes, this is equivalent to considering only the effect given by \cref{eq:VHgen}. As illustrated in the example in \cref{fig:KlambdaKi} with the improvement in the $\kappa_\lambda$ determination when this is the only degree of freedom in the fit (compare the orange band with the blue one, obtained including additional parameters), this model would be better constrained than in \cref{fig:sensitivityH3CC}, where one marginalizes over many other SMEFT parameters that would be zero in this particular scenario.
Finally, as in any other theory calculation, the level of agreement of the EFT fit results when matched to a model and that in the particular scenario will depend on the precision of the EFT theory calculations, both in the computation of observables as well as in the matching.\footnote{For example, it is well known that in some types of scenarios with extra scalar doublets it is important to include terms of dimension eight in the EFT expansion~\cite{Dawson:2022cmu}. See also \cite{Contino:2016jqw} for a more general discussion on the assessment of accuracy in the interpretation of the EFT description.} 
Continuing the effort of improving these interpretations to higher orders both in the EFT expansion as well as in perturbation theory, like the calculation described next, is therefore important to extend the range of applicability of current EFT results.
Conversely, this also highlights the need to, in the meantime, augment these interpretations with studies directly in terms of well-motivated models in those cases where they may not be described accurately in terms of current SMEFT calculations. 

\subsubsection{\texorpdfstring{$\epem \to \PZ\PH$ at NLO in the SMEFT}{e+ e- -> ZH at NLO in the SMEFT}}\label{sec:eeZH_NLO}

\begin{figure}
    \centering
    \includegraphics[height=69pt]{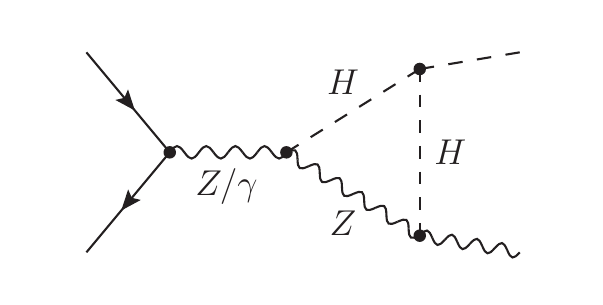}
    \includegraphics[height=69pt]{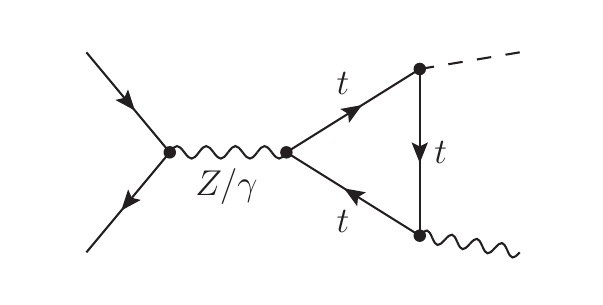}
    \includegraphics[height=69pt]{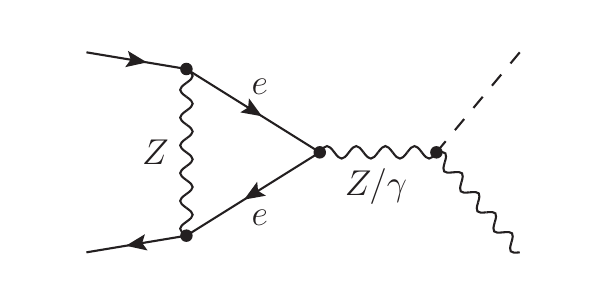}\\[8pt]
    \includegraphics[height=69pt]{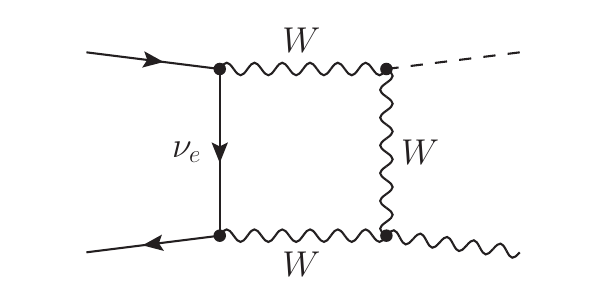}
    \includegraphics[height=69pt]{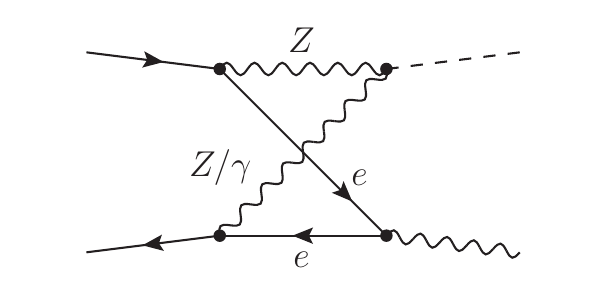}
    \includegraphics[height=69pt]{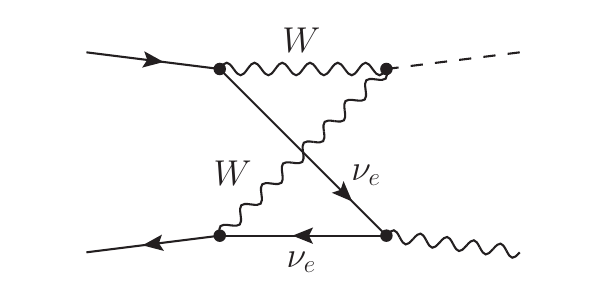}\\[8pt]
    \includegraphics[height=69pt]{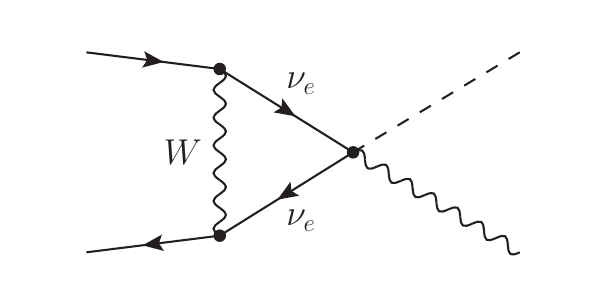}
    \includegraphics[height=69pt]{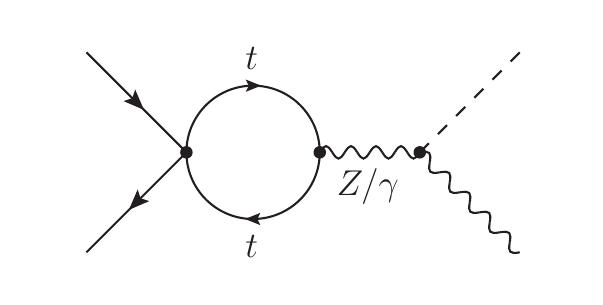}
    \includegraphics[height=69pt]{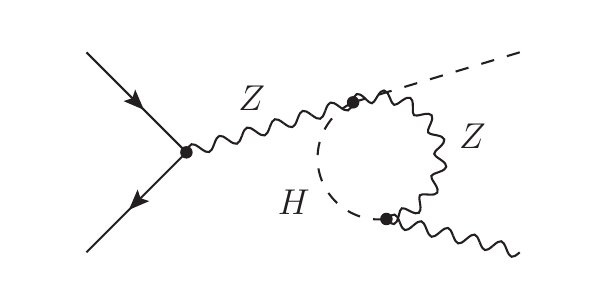}
    \caption{
    }
    \caption{Example of 1-loop diagrams for contributions to the $\epem \to \PZ\PH$ cross section in the SMEFT.}\label{fig:1loop}
\end{figure}

We conclude this section discussing the progress on the theory side regarding the determination of $\lambda_{\PH\PH\PH}$ from single Higgs measurements, where the sensitivity to the Higgs self-coupling in single Higgs processes comes from one-loop level contributions, in the context of the SMEFT.  
Within the SMEFT framework, the full set of NLO contributions from dimension-six operators 
to $\epem \to \PZ\PH$ has recently been computed, first for operators involving the top quark~\cite{Asteriadis:2024qim} and then to the whole dimension-six basis~\cite{Asteriadis:2024xts}. A sample of such contributions is shown in \cref{fig:1loop}. From the point of view of the $\lambda_{\PH\PH\PH}$ SMEFT interpretation, 
these calculations open the possibility of extending previous studies in the direction of a model-independent interpretation within the assumptions of this EFT, see \cref{sec:globalSMEFT}.~\footnote{To achieve this, calculations of the NLO effects in other processes relevant for Higgs physics at $\epem$ are still missing, e.g.\ $\epem\to \nu\bar{\nu} \PH$ (via $W$ boson fusion).}

\begin{figure}[t]\centering
\includegraphics[width=0.35\textwidth]{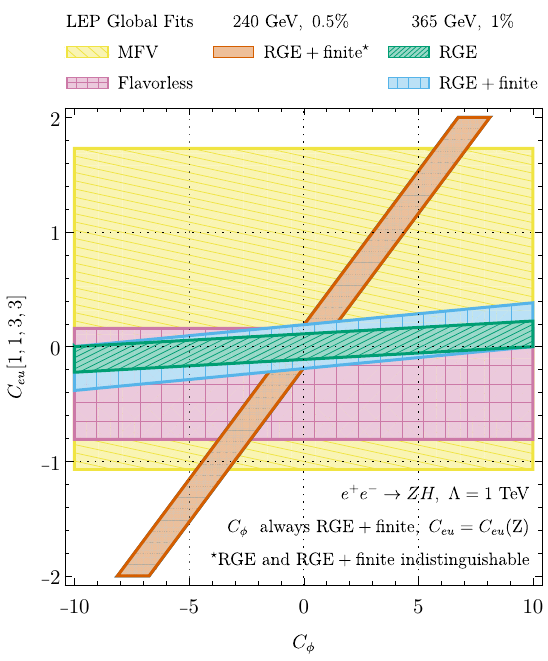}%
~~~~~~~~~~~~~~~~~~\includegraphics[width=0.35\textwidth]{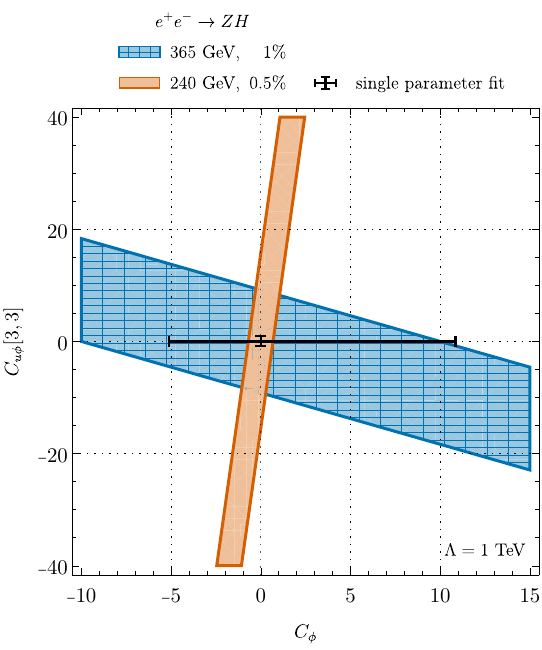}%
\caption{Left: Joint constraints on the trilinear Higgs self-coupling ($x$ axis) and on right-handed $\bar{\Pe}\Pe\PAQt\PQt$ operator ($y$ axis) from LEP, as well as \SI{240}\, and \SI{365}{\giga\electronvolt} runs.
Right: Similarly for the top-quark Yukawa operator ($y$ axis). The bands show the sensitivity that a 0.5\% measurement of $\sigma(\PZ\PH)$ at $\sqrt{s}=240$\,GeV, and 1\% at $\sqrt{s}=365$\,GeV, give on these parameters. 
Figures reproduced from Ref.~\cite{Asteriadis:2024xts}.
}
\label{fig:top-operators}
\end{figure}

\begin{figure}[h]\centering
\includegraphics[width=0.35\textwidth]{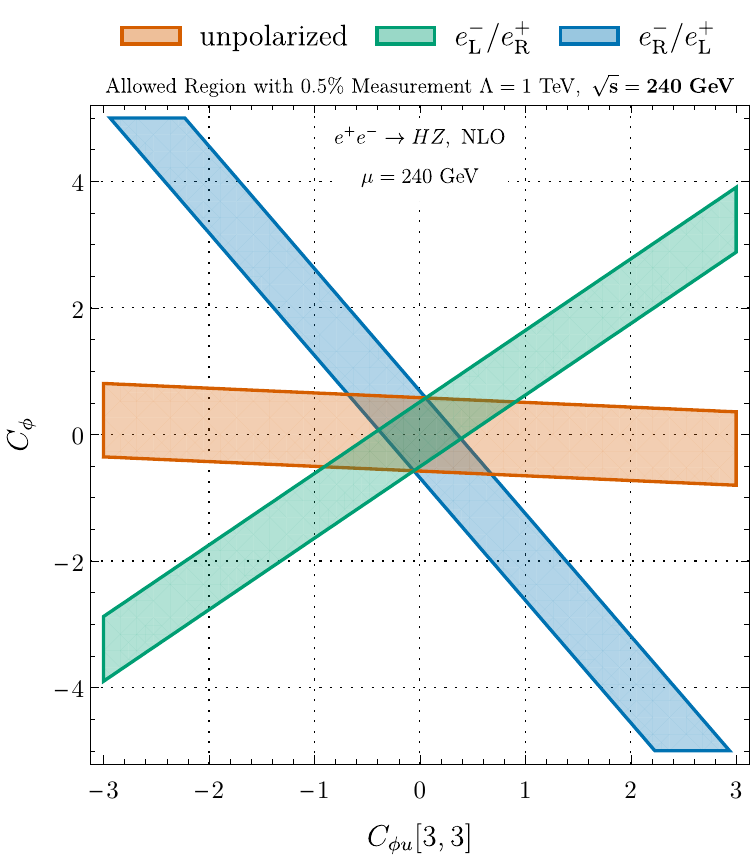}%
~~~~~~~~~~~~~~~~~~\includegraphics[width=0.35\textwidth]{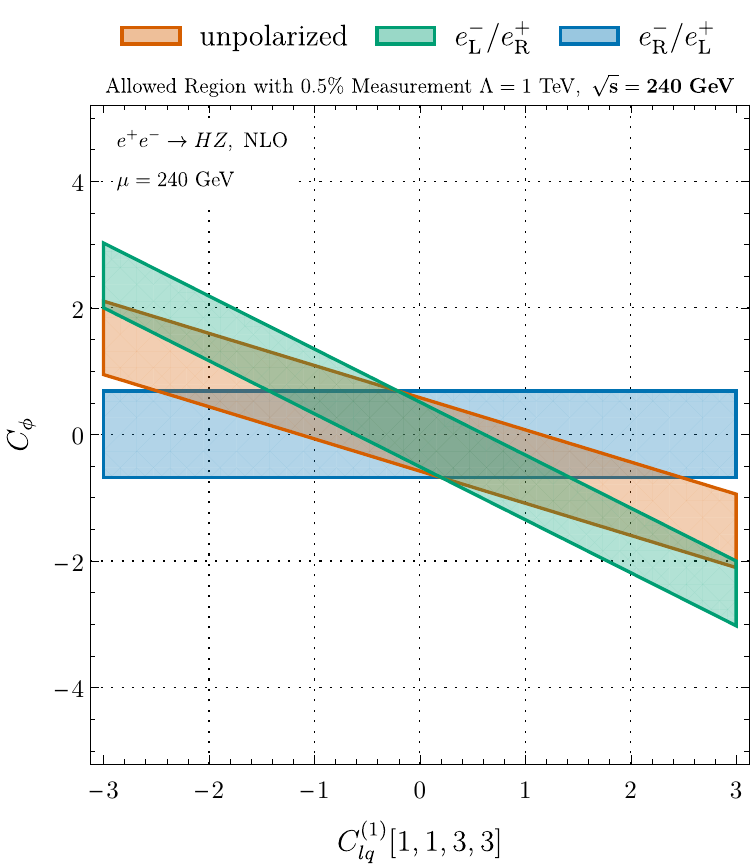}%
\caption{Left: contributions from modifications of the Higgs tri-linear coupling operator ($x$ axis) on the cross section $\epem \to \PZ\PH$ at \SI{240}{\giga\electronvolt}, correlated with those from $C_{\phi u}[3,3]$, which modify the $\PZ\PQt\PAQt$ vertex, and depending on the polarisation of the $\epem$ system. Right: the same for the $\epem \PQt\PAQt$ operator with coefficient $C_{lq}^{(1)}[1,1,3,3]$ ($y$ axis). The bands show the sensitivity that a 0.5\% measurement of $\sigma(\PZ\PH)$ at $\sqrt{s}=240$\,GeV gives on these parameters. 
Figures reproduced from Ref.~\cite{Asteriadis:2024qim}.
}
\label{fig:top-operators-2}
\end{figure}

Aside from the operators entering at LO, there are a total of 6 bosonic (4 of them CP violating), 9 two-fermion and 14 four-fermion dimension-six operators that contribute to $\epem \to \PZ\PH$ at NLO.
Of particular importance are the contributions of the top-quark operator mentioned before, e.g.\ $\epem \PQt\PAQt$ contact interactions, which are poorly constrained by current data but, as we will discuss below, will be measured at future \epem colliders. 
In this regard, and as in the case of the contributions from the Higgs self-coupling, 
considering modifications to top-quark interactions entering in other loop contributions to $\epem\to \PZ\PH$, runs at two centre-of-mass energies also provide complementary sensitivity and tight two-parameter constraints~\cite{Asteriadis:2024qim}, as shown in \cref{fig:top-operators}. 
Similarly, polarised beams can provide complementary information on different corrections, and help to separate the contributions of the Higgs self-coupling operator from those from top operators, as shown, e.g.\ in \cref{fig:top-operators-2}.


The large number of different interactions entering at NLO would suggest that in the single-Higgs analysis at NLO there 
will be a considerable amount of degeneracy. 
With more operators simultaneously included, more observables need to be combined to constrain all directions 
in a global fit
and single out the potential new contributions to $\lambda_{\PH\PH\PH}$. 
In Higgsstrahlung production itself, angular distributions and radiative return to lower effective centre-of-mass energies could for instance be considered~\cite{talkecfa}.
Outside Higgs physics, the above-mentioned top interactions could be tested at the tree level in $\epem \to \PQt\PAQt$ at \SI{350}-\SI{365}{\giga\electronvolt}. Finally, some of the same top operators enter in Z-pole observables also at the loop-level, and a combined analysis exploiting the extreme high precision of the Tera-Z run could also help in a global analysis. 
The determination 
of $\lambda_{\PH\PH\PH}$ thus relies on 
having precision measurements of a large enough number of observables to constrain any possible BSM effects in other contributions that enter in the calculations of the observables to the same order in perturbation theory as the leading contributions of $\lambda_{\PH\PH\PH}$.
%

There is currently an ongoing effort to extend the $\lambda_{\PH\PH\PH}$ EFT determination from single-Higgs measurements, including these recent NLO effects, and other aspects related to this EFT interpretation. 








\clearpage
\section{Developments in Electroweak Physics \& QCD}
\label{sec:ewk_qcd}

The various centre-of-mass energy stages proposed by future \epem collider projects provide opportunities for a wealth of measurements in the electroweak and QCD sectors.  Prospects for a precision $\PW$-boson mass measurement and $\PW$-boson couplings are discussed in \cref{sec:Wmass} and \cref{sec:Wcouplings}.  The electroweak precision measurements achievable from the very large statistics of two-fermion final states both at and above the $\PZ$-pole are discussed in \cref{sec:TwoF}.  Other $\PZ$-boson and neutrino measurements are addressed in \cref{sec:otherZ}.  \Cref{sec:WWdiff} presents new results on the sensitivity to triple-gauge couplings that could be achieved through differential $\PW\PW$ measurements.  \Cref{sec:frag-had} addresses fragmentation and hadronisation, both in terms of what could be learned from future \epem colliders, and where this could be a limiting factor on other measurements.  Finally \cref{sec:otherParams} briefly considers other SM parameters that enter the interpretation of electroweak precision measurements.

\subsection{\focustopic W boson mass measurement \label{sec:Wmass}}
\label{sec:ewqcd_wmass}
\editors{Paolo Azzurri, }
The mass of the $\PW$ boson ($m_{\PW}$) is a fundamental parameter of the standard model (SM). 
A precise experimental $m_{\PW}$  determination is highly desirable to test the SM consistency and as a means to reveal possible new physics effects~\cite{Haller:2018nnx}.
The level of precision to which  $m_{\PW}$ could be determined at a future $\epem$ collider with a scan of cross sections at the pair-production  threshold, and from measurements of decay kinematics is greatly superior to current measurements~\cite{Workman:2022ynf}, and expected to by far exceed  measurements at the LHC which  yield uncertainties around 10 MeV~\cite{Dainese:2703572,CMS:2024nau}.

To obtain an estimate of the achievable theory and experimental  systematic uncertainties at am $\Pep\Pem$ Higgs factory, the potential of analysis, calibration and event reconstruction methods, currently in development, should be assessed. Several complementary methods will be used to measure $m_{\PW}$. Ideally their systematic uncertainty should match or be better than the statistical precision of the data.
While $\PW$-pair threshold cross section scans are expected to be less impacted by systematic uncertainties, detector performance will play a crucial role in $m_{\PW}$ measurements that make use of kinematic information obtained from event reconstruction.
In the following we outline the main methods envisaged.

\subsubsection*{\boldmath $\PW$-pair production at threshold}

The rapid rise of the $\PW$-pair production cross section near its kinematic threshold can be exploited to obtain a precise and direct determination of the $\PW$ boson mass.
The threshold method is clean and simple as it just involves classifying and counting events.
A small amount of LEP $\Pep\Pem$ collision data taken in 1996 at a single energy point near 161\,GeV allowed the $\PW$ mass to be determined with a precision of 200\,MeV~\cite{Barate:1997mn,Abreu:1997sn,Acciarri:1997xc,Ackerstaff:1996nk}.

Collecting over 1\,ab$^{-1}$ of integrated luminosity at threshold energies will enable a determination of the $\PW$ mass with a statistical uncertainty below 1\,MeV~\cite{Abada:2019lih}.
Measuring the threshold cross section at more than one energy point will allow measurement of both the $\PW$ mass and width, with similar precision~\cite{Azzi:2017iih,Azzurri:2021yvl}.
Polarised collisions can enhance or suppress $t$-channel signal production, enabling further control of the background~\cite{Wilson:2016hne,LCCPhysicsWorkingGroup:2019fvj}.

\subsubsection*{\boldmath $\PW$-pair decay kinematics}
The primary method to measure the $\PW$ mass and width at LEP2 was with the kinematic reconstruction of semileptonic ($\qqbar\ell\nu$) and fully hadronic ($\qqqq$) 
$\PW$-pair decays~\cite{ALEPH:2006cdc,DELPHI:2008avl,L3:2005fft,OPAL:2002hhr,ALEPH:2013dgf}.
These results were obtained by imposing the constraint that the total four-momentum in the event should be equal to the known initial centre-of-mass energy and zero momentum.

It is foreseen that performing similar measurements with future collider data would yield a statistical precision of a few MeV or less~\cite{LCCPhysicsWorkingGroup:2019fvj,Azzurri:2021yvl},
but the impact of systematic uncertainties is more difficult to predict; in particular, those arising from the modelling of non-perturbative QCD effects in the $\PW$-boson hadronic decays, that stood out in the LEP2 measurements~\cite{ALEPH:2013dgf}.

In the standard (LEP2-style) $\PW$-mass kinematic reconstruction, non-perturbative QCD uncertainties  arise on two fronts. 
The first is from overall uncertainties on the modelling of core jet properties, in particular on
the jets boost ($\beta_\text{jet} = p_\text{jet}/E_\text{jet}$) that is a key ingredient of the  kinematic fit.  
Uncertainties on $\beta_\text{jet}$ will affect similarly both the $\qqbar\ell\nu$ and $\qqqq$ channels.
The second source is from colour reconnection (CR) effects, that can 
lead to an important impact and uncertainty on $m_{\PW}$, but only in the $\qqqq$ channel. 

The energy spectrum endpoints in leptonic decays,  that are free from QCD uncertainties, can also be used to measure the $\PW$ mass~\cite{LCCPhysicsWorkingGroup:2019fvj}.
In the case of fully leptonic decays, a pseudomass~\cite{OPAL:2002hhr} can be computed and employed for a $\PW$-mass measurement.
Using only the lepton decay kinematics degrades the statistical sensitivity of the measurement, that 
is finally expected to achieve a precision 
of 3--5\,MeV with this approach~\cite{LCCPhysicsWorkingGroup:2019fvj}.

All the above methods to determine the W mass, including the threshold lineshape, necessitate a precise knowledge of the collision energy 
as the beam energy uncertainty propagates directly to the 
W -ass systematic uncertainties with 
$\Delta m_{\PW}(E_\text{beam}) \simeq \Delta E_\text{beam}$.
A future circular collider would provide a calibration of the beam energy by resonant depolarisation with a precision well below 1\,MeV, up to the W-pair threshold~\cite{Abada:2019lih}.
Finally a direct measurement of the hadronic mass, without kinematic constraints, can be performed~\cite{Anguiano:2020qpk} where the uncertainties will be dominated by the hadronic energy scale calibration.

\subsection*{Theoretical and phenomenological targets and challenges }

\subsubsection*{Precise predictions of total and differential $\PW$-pair production}

Accounting for the high experimental precision in the $m_{\PW}$ determination
at $\epem$ colliders in predictions 
requires a proper inclusion of radiative corrections
to the $\epem\to$~4f processes with intermediate $\PW$ pairs.
Depending on the $\epem$ CM energy $E_\text{CM}$, 
different physical effects and corrections 
dominate, and the target precisions are not the same.
Customised treatments are required for CM energies near the 
W-pair threshold ($|E_\text{CM}-2m_{\PW}|<n\Gamma_{\PW}$, $n\sim2{-}3$),
at intermediate energies of a typical future circular collider
($2m_{\PW}+n\Gamma_{\PW} < E_\text{CM} < $250\,GeV),
or at higher energies.

``Improved-Born approximations''
based on leading universal corrections (initial-state radiation, 
Coulomb singularity, effective couplings)
can be formulated uniformly for the whole energy range, but
are good only to $\sim2\%$ for integrated cross sections
up to intermediate energies (and deliver only $\sim10\%$
accuracy for $E_\text{CM}\sim$ 1\,TeV).
For intermediate energies, a pair of on-shell $\PW$ bosons 
dominates the cross section, so that NLO corrections can be included
in the ``double-pole approximation (DPA)''~\cite{Denner:2000bj,Jadach:2001mp}, which is based on
the leading term of an expansion of amplitudes about the $\PW$ resonance poles. 
The quality of the DPA was estimated to $0.5{-}1\%$ for integrated
cross sections and later confirmed in a comparison
to NLO predictions for the full $\epem\to 4\text{f}$ cross sections~\cite{Denner:2005fg} which should be good within
a few $0.1\%$ even in the $\PW$-pair threshold region, where the DPA breaks down
(see Section\,6.5.5 of Ref.~\cite{Denner:2019vbn} for more details and 
original references).
At threshold, the currently best calculation is based on complete 
NLO results for $\epem\to$~4f~\cite{Denner:2005fg}
and partial higher-order effects 
for the total cross section from an effective field theory (EFT) framework~\cite{Beneke:2007zg,Actis:2008rb}, suggesting a theory uncertainty on $m_{\PW}$ of about 3\,MeV for $m_{\PW}$ extracted from a threshold measurement~\cite{Actis:2008rb}, excluding the uncertainty from QED initial-state radiation.

The quality of differential predictions in dominant phase-space regions 
widely follows the precision estimates for integrated cross sections,
but the precision significantly deteriorates in regions where 
resonant $\PW$-pair production becomes less dominant, e.g.\ in the
backward direction of $\PW$ production or at very large CM energies
(see also  Section\,8 of Ref.~\cite{Frixione:2022ofv}).

To account for the leap in precision from LEP to a future $\epem$ collider, 
predictions for $\PW$-pair production have to be improved considerably,
most notably the treatment of initial-state radiation~\cite{Jadach:2019bye,Jadach:2019wol}.
At threshold, NNLO corrections to the ``hard'' $\PW$-pair production and $\PW$ decay
processes as well as leading corrections beyond NNLO can be calculated within
the EFT approach, and a precision of 0.5\,MeV in $m_{\PW}$ from an
energy scan near threshold seems feasible~\cite{Freitas:2019bre}. 
For intermediate and high energies, a full NNLO calculation for $\PW$-pair 
production in DPA would be most desirable, a task that should be
achievable in the next decade, anticipating further progress at
the frontier of loop calculations.

\subsubsection*{Estimates and control of QCD effects} 
Accurate jet evolution and prediction of radiation patterns are important ingredients for the $\PW$-mass measurement.

Hadronisation models might impact jet properties such as the boost, as well as the energy scale due to differences in the baryon ratios. For the jet energy calibrations from $\PZ$-peak events, beam energy, jet mass and jet momentum are affected by baryon and strange ratios.
Hadronisation models can also affect the measured jet energy, resolution and jet mass via the charged/neutral ratio fraction of very low and high  $\pT$ tracks and the shower shape. $\PZ$-peak events have a different flavour composition from $\PW$-boson events. Studies of $\PQb$ and $\PQc$ tags exist but $\PQs$ quarks must also be considered.

Colour reconnection (CR) in all-hadronic events affects particles with large $\Delta R$ to the jet (parton) direction by pulling them towards or away from the jet, since they change the colour topology underpinning the hadronisation process. These have a large lever arm on the jet direction. At LEP2, the effect was reduced with cuts on $\Delta R$ or cuts on soft particles or weights. Theoretically well-behaved algorithms, e.g.\ Cambridge/Aachen with freeze-out, are needed to address this issue. Experimentally, CR can be accessed using particle flow in the region between jets. Different models with similar measured flow can have significantly different effects on the reconstructed jet direction, and thus the $\PW$ mass. All ``realistic'' CR models should be considered to estimate the possible impact in lack of further constraints. Constraints on CR can be obtained from change of jet direction (or $\PW$ mass) by changing parameters of jet algorithms (e.g.\ jet freeze out). Jet algorithms that do not include soft particles with large $\Delta R$ in the jet associated with the partons from the $\PW$-boson decay might have larger modelling uncertainties. $\PZ$-boson peak events with two jets can be used to calibrate the direction resolution.

Newly developed colour reconnection and hadronisation models \cite{Christiansen:2015yqa,Gieseke:2017clv,Gieseke:2018gff,Platzer:2022jny} should be considered to put further constraints on the predictivity for such final states.

\subsection*{Detector performance and analysis methods}
Detector and analysis performances are particularly relevant for measurements with the reconstruction of the $\PW$-boson decay kinematics.
In the case of the reconstruction of fully leptonic or semileptonic $\PW$-pair decays, the determination of lepton energy scale is a key ingredient. 
Previous studies concluded that the evaluation of $\PJGy \to \mumu$ events could provide the most precise constraints of the track momenta, as the $\PJGy$ mass is currently known to the level of 1.9 ppm, corresponding to about 6\,keV~\cite{Workman:2022ynf}. However, at $\epem$ colliders the production of $\PJGy$ is expected to be statistics limited. More recent studies reported in the ECFA workshop~\cite{ecfa22_wilson}, based on Refs.~\cite{Madison:2022spc,Rodriguez:2020qhf}, have explored how a combination of mass information from several well measured particles, in particular $\PKzS\to \PGpp\PGpm$ and $\PGL \to \Pp \PGpm$ can be used an absolute calibration of track momenta at a future high-energy $\epem$ collider. With sufficient statistics it is estimated that a centre-of-mass energy uncertainty of $2 \times 10^{-6}$ could be achieved, competitive with the uncertainty in the centre-of-mass energy expected at FCC-ee.

In contrast, the jet energy scale is not expected to be known better than the level of $10^{-4}$, with dominant uncertainties arising from modelling uncertainties such as flavour dependence, hadronisation, colour reconnection, as outlined above.
Nevertheless, measurements of the $\PW$-boson mass in the hadronic channel, without kinematic constraints, will  provide complementary results with uncorrelated uncertainties, useful for cross-checking and combination with the measurements in other channels.

Particle reconstruction and flavour identification in jets are important for the event selection and to calibrate and control flavour-dependent effects. In current experiments, novel machine-learning techniques have proven to facilitate unprecedented performance in signal-to-background separation and flavour identification.

If external information about the beam energy is used in the reconstruction of hadronic ($\qqqq$) or semileptonic ($\qqbar\ell\\PGn$) $\PW$-pair decays, the hadronic jet energy scale can be constrained by the four momentum conservation in the event, removing the importance of the  experimental jet energy determinations. 
Still, a precise modelling and control of the hadronic jets boost ($\beta = p/E$) will be essential for precise $m_{\PW}$ measurements from $\qqqq$ and  $\qq\ell\\PGn$ events. 

In view of the need for extreme precision on jet internal properties such as the boost, it will be highly desirable for detector capabilities to be highly efficient and precise in identifying and measuring all particles inside hadronic jets.  
Excellent particle reconstruction capabilities should extend to low momentum particles away from the jet axis, and in the inter-jet regions, where colour reconnection (CR) effects in the $\qqqq$ channel can be determined and constrained {\em in situ}.

An ultimate aim should be to fit simultaneously $\PW\PW$, $\PZ\PZ$ and $\PZ\gamma$ leptonic and hadronic  decay modes, in order to extract a determination of the $m_{\PW}/m_{\PZ}$ ratio with potentially large cancellations of most systematic uncertainties, from theory, experiment, and beam energy. 


\subsection{Precision W-boson coupling measurements \label{sec:Wcouplings}}
In a future $\epem$ physics program a total of 40-150~$10^6$ W-pair events can be produced, mostly at  $\sqrt{s}$= 240--250~GeV and at $\sqrt{s}$= 162~GeV energies.

As the LEP2 program has demonstrated, a very large fraction of the produced W-pair events 
can be collected (85-95\% efficiency), with low background levels (90--95\% purity), 
in all decay final states, including fully hadronic $\qqqq$, semi-leptonic $\Pl\PGn\qqbar$, 
and fully leptonic $\Pl\PGn\Pl\PGn$ channels, with $\Pl=\Pe$, $\PGm$ and $\PGt$. 

The event yields in the $\qqqq$ channel, the three  $\Pl\PGn$qq channels, and the six 
$\Pl\PGn\Pl\PGn$ channels can be combined, taking into account their cross-contaminations and correlations, in order to fit the W decay branching ratios, and the total W-pair cross sections. 
A fit that does not assume W-lepton coupling universality can be performed to extract the leptonic decay couplings $B_{\Pe}$, $B_{\PGm}$ and $B_{\PGt}$, while a fit that assumes lepton universality can be performed to extract the 
hadronic decay coupling $B_{\PQq}$. In both fits the sum of leptonic and hadronic branching fractions is constrained to unity. 

\begin{table}[htbp]
\begin{center}
\begin{tabular}{|c|c|c|c|c|} \hline
Decay mode relative precision & $B(\PW\rightarrow\Pe\PGn)$ &$B(\PW\rightarrow\PGm\PGn)$ &$B(\PW\rightarrow\PGt\PGn)$ &$B(\PW\rightarrow\qqbar)$ \\ \hline
LEP2 & 1.5\% & 1.4\% & 1.8\% & 0.4\% \\ 
LHC & 1.0\% & 0.8\% & 2.1\% & 0.3\% \\ 
future $\epem$ & 3$\times 10^{-4}$  & 3$\times 10^{-4}$  & 4$\times 10^{-4}$  & 1$\times 10^{-4}$ \\  \hline
\end{tabular}
\caption{Relative precision on the determination of the W decay branching ratios.
Final combined results with LEP2 data~\cite{ALEPH:2013dgf} are compared with recent results obtained with LHC data~\cite{CMS:2022mhs},
and the projected precision obtainable with future $\epem$ data.}
\label{tab:WBR}
\end{center}
\end{table}%

Projected precisions on the W boson decays to hadrons, $\Pe$ $\PGm$ and $\PGt$ leptons achievable with a future $\epem$ program are shown in Table~\ref{tab:WBR}, and compared with LEP2~\cite{ALEPH:2013dgf} 
and LHC~\cite{CMS:2022mhs} precisions. In these projections the impact of systematic uncertainties on the future $\epem$ precisions will be comparable but not much larger than the statistical uncertainty.  This will be achievable by using data-driven methods on independent data, to constrain the leading systematics, e.g.\ using tag and probe methods to measure the selection performances of jet reconstruction and lepton identification.
The estimated improvement of the future $\epem$ precision with respect to existing LEP2 and LHC results ranges from $\sim 30$ for the hadronic decays, to $\sim 50$ for muon decays.

Within the Standard Model the W-boson hadronic branching ratio $B_{\PQq}$ is related to the 
Cabibbo-Kobayashi-Maskawa (CKM) quark mixing matrix, and to the strong coupling constant $\alpha_S$ through the relation

\begin{equation}
R_{\PW} = \frac{B_{\PQq}}{1-B_{\PQq}} =  \left( 1+ \frac{\alpha_S(m^2_{\PW})}{\pi}\right) 
\sum_{i=\PQu,\PQc; j=\PQd,\PQs,\PQb} |V_{ij}|^2 .
\label{eq:WBR}
\end{equation}

Assuming CKM unitarity with 
$S_{\PW} = \sum_{i=\PQu,\PQc; j=\PQd,\PQs,\PQb} |V_{ij}|^2 = 2$, the $B_{\PQq}$ determination can be used 
to extract the value of $\alpha_S(m^2_{\PW})$. Focusing on fitting directly the $R_{\PW}$ 
ratio of hadronic to leptonic decay rates, the projected achievable 
$\alpha_S(m^2_{\PW})$ relative precision is 0.2\%.

If the CKM unitarity is not assumed in the sum, and $\alpha_S(m^2_{\PW})$ is taken
form other independent precision determinations, the $B_{\PQq}$ and $R_{\PW}$ measurements 
can be used in turn to provide a stringent test of CKM unitarity for the five lightest quarks $S_{\PW}$ at the precision level of few $10^{-4}$.
The determination of $S_{\PW}$ can in turn be employed to derive the value of the CKM element $|V_{\PQc\PQs}|$, with a precision bounded by the uncertainty on the sum of the other five elements in $S_{\PW}$. 

The flavour tagging of jets from W decays can be exploited to perform more direct measurements of $|V_{\PQc\PQs}|$ and of the overall fraction of W boson decays to charm quarks $R_{\PQc}$~\cite{ALEPH:1999djo,OPAL:2000jsa,DELPHI:1998hlc}. The final achievable statistical precision  on $R_{\PQc}$ should be at the $10^{-3}$ level, or better. 

Future $\epem$ data will also allow exploration of the more rare  $\PW\rightarrow\PQb\PQc$ and $\PQb\PQu$ events, where respectively $\sim 10^5$ and $\sim 10^3$  decays are expected from a total of $\sim200$M \PW boson decays. Also in the case of \PW-boson jet-flavour determinations we expect that dominant systematic uncertainties, related to jet-tagging performances, 
will be data-driven and constrained by the collected luminosity.  
These measurements will lead to direct determinations of the $|V_{\PQc\PQb}|$ and $|V_{\PQu\PQb}|$  CKM matrix elements 
with precisions better than 1\% for $|V_{\PQc\PQb}|$ and around 5\% for $|V_{\PQu\PQb}|$, therefore improving  the knowledge of the quark-mixing matrix. 
More details on these possible measurements are given in \cref{sec:flav-Vcb}.
Such measurements were not  exploitable at LEP2 given the low statistics of collected W decays. 

Any other rare decay of the W boson can be explored with future $\epem$  data with a sensitivity that can probe the level of $10^{-7}$ decay probabilities. In this context particular interest will be in measuring exclusive radiative decays~\cite{Grossman:2015cak} that will provide stringent tests of the QCD factorisation formalism and enable novel searches for new physics.
Other exclusive rare hadronic decays~\cite{Mangano:2014xta,Melia:2016knk} will also be accessible with future $\epem$  data.

\subsection{\focustopic 2-fermion final states \label{sec:TwoF}}
\editors{Adrian Irles -- EXP: Adrian Irles, Daniel Jeans, Manqi Ruan, THEORY: Emanuele Bagnaschi, Alessandro Vicini, Juergen Reuter, Ayres Freitas, Bernnie Ward}
\newcommand{\AFBq}{\ensuremath{A_{\text{FB}}^{\PQq}}\xspace}
\newcommand{\AFBb}{\ensuremath{A_{\text{FB}}^{\PQb}}\xspace}
\newcommand{\AFBc}{\ensuremath{A_{\text{FB}}^{\PQc}}\xspace}

\subsubsection{Introduction}

The precision of the determination of the EW couplings of gauge bosons to fermions is expected to improve by several orders of magnitude at future \epem colliders \cite{Belloni:2022due} with respect to the legacy measurements from LEP and SLC \cite{ALEPH:2005ab}.
Such precision will be achievable thanks to the higher luminosities, longitudinally polarised beams (in the case of linear colliders), a wider range of collider energies, precise modern detectors with improved reconstruction, and improved theoretical modelling.

The unprecedented statistical power provided by future colliders will require a great effort on the control and understanding of systematic uncertainties from theory and experiment. Indeed, a $\PZ$-pole run foreseen by FCC-ee/CEPC will offer more than two orders of magnitude smaller statistical uncertainties than those of previous measurements~\cite{FCC:2018evy,CEPCPhysicsStudyGroup:2022uwl}. 
A significant improvement in precision could also be reached 
at the ILC~\cite{ILCInternationalDevelopmentTeam:2022izu}.
This requires remarkably stable operation of the detectors and accelerators.

The LEP and SLC colliders probed the gauge structure of the SM at the quantum level, finding an overall good agreement with theory predictions. However, some tensions in the determination of the weak effective mixing angle for different flavours are still unresolved \cite{ParticleDataGroup:2024cfk}. Future colliders will be key in clarifying these issues and probing BSM physics in other observables.

Furthermore, for the investigation of the Higgs sector and for searches for new physics at higher energies, more precise determinations of the EW couplings to fermions are required \cite{DeBlas:2019qco,Ellis:2020unq}. 

Projections for the determination of the electroweak couplings of the $\PZ$ boson to fermions from measurements at a future \epem collider running at the $\PZ$ pole have been reviewed as part of the Snowmass 2021/22 Study \cite{Belloni:2022due}, see Tab.~\ref{tab:ewpocomp}.
More work is required to exploit final states involving light quark families, for instance, using strange-tagging techniques. Related studies are discussed in \cref{sec:HtoSS}.

\begin{table}[tb]
\centering
  \begin{tabular}{|c|c|c|c|c|c|c|}
    \hline
    Quantity       & current &    ILC250      & ILC-GigaZ   &      FCC-ee            &  CEPC      &  CLIC380     \\
    \hline
    $\Delta\alpha(m_{\PZ})^{-1}\;(\times 10^3)$ &    18$^*$  & 18$^*$  &             &    3.8 (1.2)  & 18$^*$ &           \\
    $\Delta m_{\PZ}$ (MeV)    & 2.1$^*$      &    0.7 (0.2)  &  0.2           &    0.004 (0.1)   & 0.005 (0.1)   &    2.1$^*$       \\
    $\Delta\Gamma_{\PZ}$ (MeV)     & 2.3$^*$ &    1.5 (0.2)  &  0.12    &    0.004 (0.025)   & 0.005 (0.025)   &    2.3$^*$       \\ 
    \hdashline
    $\Delta A_{\Pe}\;(\times 10^5)$   & 190$^*$ &    14 (4.5)  &   1.5 (8)  &    0.7 (2)   & 1.5 (2)   &    60 (15)       \\ 
    $\Delta A_{\PGm}\;(\times10^5)$  & 1500$^*$ &    82 (4.5)  &   3 (8)  &    2.3 (2.2)   & 3.0 (1.8)   &    390 (14)      \\    
    $\Delta A_{\PGt}\;(\times10^5)$ & 400$^*$ &   86 (4.5)  &   3 (8)  &    0.5 (20)   & 1.2 (20)   &    550 (14)      \\        
    $\Delta A_{\PQb}\;(\times10^5)$   & 2000$^*$ &   53 (35)  &   9 (50)  &    2.4 (21)   & 3 (21)  &    360     (92)  \\
    $\Delta A_{\PQc}\;(\times10^5)$  & 2700$^*$  &  140 (25)  &  20 (37)  &    20 (15)   & 6 (30)   &    190     (67)  \\ 
    \hdashline
    $\Delta \sigma_\text{had}^0$ (pb)  & 37$^*$ &           &          &    0.035 (4)      & 0.05 (2)     &    37$^{*}$       \\     
    $\delta R_{\Pe}\;(\times10^3)$    & 2.4$^*$ &    0.5 (1.0)    &   0.2 (0.5)  &    0.004 (0.3)   & 0.003 (0.2)   &    2.5 (1.0)       \\
    $\delta R_{\PGm}\;(\times10^3)$  &  1.6$^*$  & 0.5 (1.0)    &   0.2 (0.2)  &    0.003 (0.05)  & 0.003 (0.1)   &    2.5 (1.0)       \\
    $\delta R_{\PGt}\;(\times10^3)$ &  2.2$^*$ &  0.6 (1.0)    &   0.2 (0.4)  &    0.003 (0.1)   & 0.003 (0.1)   &    3.3 (5.0)       \\
    $\delta R_{\PQb}\;(\times10^3)$    & 3.1$^*$ &   0.4 (1.0)    &   0.04 (0.7)  &   0.0014 ($<0.3$)   & 0.005 (0.2)   &    1.5 (1.0)       \\
    $\delta R_{\PQc}(\times10^3)$    &  17$^*$ &  0.6 (5.0)    &   0.2 (3.0)  &    0.015 (1.5)   & 0.02 (1)   &    2.4 (5.0)       \\
    \hline    
  \end{tabular}
\caption{Electroweak precision observables extracted from two-fermion processes at future \epem colliders: statistical error (estimated experimental systematic error). $\Delta$ ($\delta$) stands for absolute (relative) uncertainty, while * indicates inputs taken from current data \cite{ParticleDataGroup:2024cfk}. Table adapted from Ref.~\cite{Belloni:2022due}.}
\label{tab:ewpocomp}
\end{table}

Final states with two fermions will also be studied at higher-energy runs.
These data will be more challenging to interpret in terms of electroweak couplings, but they can be used to put constraints on higher-dimensional four-fermion contact operators (see next subsection). Moreover, the mechanism of radiative return allows to study the invariant mass distributions of the fermion pair,
including the $\PZ$-pole region, also at beam energies much larger than half the $\PZ$ mass. To exploit this opportunity, the development of new experimental and theoretical techniques has been started~\cite{Mizuno:2022xuk}.

\subsubsection{Theoretical and phenomenological aspects}
\editors{Emanuele Bagnaschi, Alessandro Vicini, Juergen Reuter, Ayres Freitas, Bennie Ward}

The cross section for fermion pair production at the $\PZ$ peak will be measured with a $\mathcal{O}(10^{-4})$ relative precision at future \epem colliders. Following the approach that was chosen at the time of LEP and SLC colliders it is possible to parametrise the $\PZ$ resonance in terms of pseudo-observables, like \PZ branching ratios $R_f$ or asymmetry parameters $A_f$. To match the experimental precision, these have to be computed including three-loop and leading higher-order electroweak corrections~\cite{Blondel:2019qlh,Freitas:2019bre}, which are not fully available today. The relative statistical precision a few GeV away from the $\PZ$ peak and at higher energies will be better than $\mathcal{O}(10^{-3})$, which requires NNLO electroweak theory predictions. Second order corrections can indeed be as large as a few percent, with a non-trivial energy behavior expressed by the running of the electromagnetic coupling \cite{Chiesa:2024qzd} or by EW Sudakov logarithms \cite{Jantzen:2005az}. Such effects are in several cases of opposite sign, featuring non-trivial cancellations. In addition, these fixed-order calculations will need to be integrated with Monte Carlo generators for the simulation of QED/QCD showering. The high luminosity expected for the \PZ-pole runs, particularly for FCC-ee and CEPC, raises the question about the precision attainable in the deconvolution of QED and QCD corrections, to extract the values of the pseudo-observables.

Achieving these goals calls for a multi-year dedicated effort by the theory community, but some work is already progressing. For example, NNLO mixed QCD-EW corrections to $\epem \to \PQq\PAQq$, with exact energy dependence at any value of $\sqrt{s}$, could be studied using the results of Refs.~\cite{Bonciani:2021zzf,Buccioni:2022kgy,Armadillo:2022bgm,Armadillo:2024nwk}. Techniques for the calculation of NNLO electroweak corrections to $2\to 2$ scattering processes and N$^3$LO corrections to electroweak precision pseudo-observables have been developed \cite{Song:2021vru,Armadillo:2022ugh,Liu:2022chg,Chaubey:2022jfi,Freitas:2022hyp,Chen:2022mre,Bi:2023bnq,Niggetiedt:2023uyk}, which will enable more phenomenological results in the near future.
Besides electroweak corrections, QCD contributions are important for quark final states. Recent work has improved the understanding of QCD radiation impacts on flavour identification of hadronic jets \cite{Czakon:2022wam,Gauld:2022lem,Caola:2023wpj}.

It is desirable to combine results for electroweak loop calculations with predictions for real photon emission in a Monte Carlo framework that allows for the generation of exclusive events. A promising approach is based on the Yennie-Frautschi-Suura (YFS) method for soft photon resummation. In the YFS method, the $n$-photon phase-space is treated exactly, and it can, in principle, be extended systematically to any given order. Implementations of exclusive YFS exponentiation are available in the Monte Carlo programs KKMC \cite{Jadach:1999vf,Jadach:2022mbe} and SHERPA \cite{Krauss:2022ajk}. While soft-photon emission is exponentiated to all orders in the YFS approach, collinear logarithms would a priori need to be incorporated order by order. However, it was recently shown that the leading collinear logarithms can also be resummed in the YFS methodology \cite{Jadach:2023pka}, and this improvement will be important for the development of high-precison simulation tools.
An alternative approach for the matching of fixed-order EW results with the resummation of QED effects, considering those terms enhanced by an initial state collinear logarithmic factor, can be obtained by treating the incoming leptons as composite systems and describe their partonic content in terms of PDFs \cite{Frixione:2019lga}. The latter, in contrast with the proton case, can be computed from first principles and allow the exploitation of the results developed in QCD to describe scattering processes at hadron colliders \cite{Frixione:2021zdp,Bertone:2013vaa}.
Also see \cref{sec:shower_hadro,sec:specialtools} for a discussion of Monte Carlo methods and tools.

Besides accurate SM calculations, one also needs a parametrisation of possible BSM effects. Near the \PZ resonance, the leading BSM effects can be captured by pseudo-observables, as already mentioned above.
One caveat, however, is the possibility of a $\PZ^{\prime}$ boson quasidegenerate with the \PZ boson, where a two-particle parametrisation of the lineshape will be required, as discussed in Ref.~\cite{LoChiatto:2024guj}.

 Another approach would be to directly fit the SM effective leptonic weak mixing angle, used as a Lagrangian input, to the physical observables~\cite{Chiesa:2019nqb}, performing a consistency test of the model.

Rather than complementary physics programs, the study of the production of a fermion pair at the $\PZ$ resonance and above should be considered in a single approach: the large statistics of the cross sections in the proximity of the resonance might be sufficient to identify a deviation from the SM behaviour. The combination of several points at different energies and (for linear collider facilities) different initial state longitudinal beam polarisations might confirm the presence of a pattern, relying on the increasing size of the effects, which compensates for the larger statistical errors at high energy runs.

The parametrisation of any deviation from the best available SM prediction, assuming that new physics appears at scales sufficiently larger than the ones of the measurements, can be expressed with the introduction of gauge-invariant higher-dimension operators in the SMEFT framework.
A recent study~\cite{Allwicher:2024sso} has shown that the measurements obtained with the Tera-Z run will already be able to probe new physics up to mass scales of a few to tens of TeVs for models that generate dimension-6 operators at tree level.
The additional inclusion in SMEFT fits of two fermion observables obtained at higher energies may be able to further extend the reach of this kind of analyses. This approach has already been used, for instance, in Refs.~\cite{deBlas:2022ofj,Celada:2024mcf,Greljo:2024ytg}.
In particular, in Ref.~\cite{Greljo:2024ytg} the authors present an optimized analysis using machine learning-based flavour tagging, achieving up to two orders of magnitude improvement in the precision of electroweak observables defined by hadronic cross-section ratios $ R_{\PQb}, R_{\PQc}, $ and $ R_{\PQs} $ at the $ \PW\PW $, $ \PZ\PH $, and $ \PQt\PAQt $ FCC-ee runs, offering an unprecedented test of the SM. They also demonstrate significant potential in probing flavour-conserving four-fermion interactions within the SMEFT framework, far exceeding current LEP-II and LHC bounds, while highlighting complementarity with $ \PZ $-pole observables, as illustrated by the $ W $ and $ Y $ oblique parameters in Fig.~3 of Ref.~\cite{Greljo:2024ytg}.

\subsubsection{Experimental aspects}
\editors{Adrian Irles, Daniel Jeans, Manqi Ruan}

Below are listed some critical detector design targets, which include algorithm development that should be addressed to maximize the potential of precision physics in the two-fermion final state for all proposed collider options. To address these issues, full-simulation studies with realistic detector designs and detailed Monte Carlo simulations are mandatory, including detailed beam dynamics modelling.

\begin{itemize}
    \item Design of inner tracking detector systems and their influence on vertex finding, including the optimization of the forward region.
    \item Charged hadron identification detectors, especially of kaons for b, c, s tagging and charge measurement using \dedx, \dndx, time-of-flight detectors, RICH detectors, etc.
    \item Impact of ECAL design (e.g.\ granularity and energy resolution) on tau decay mode identification.
    \item Novel flavour-tagging algorithms profiting from the excellent granularity and vertexing capabilities expected from the detectors. The algorithms should be able to fully reconstruct single heavy hadrons originating in heavy quark decays.
    \item Impact of high-energy photon identification. This will be needed for measurements in the continuum where the return to the $\PZ$ pole contributes to the background but not to the signal. These photons tend to be found in the forward regions. 
    \item Hermeticity of detectors: to study $\PZ$ couplings at 240/250 GeV with radiative return events, we will have to look at events where the ISR photon has escaped the detector through the beam pipe. Proper estimations of the missing energy (via angular measurements, for example) are required, and these depend on the detector hermeticity.
  
\end{itemize}

The optimisation of the detector concepts should also address the challenges associated with the different operation scenarios: high-energy vs low-energy scenarios, high-rates ($\PZ$ pole) or lower rates ($\PH\PZ$ threshold and above), discussed in \cref{sec:twof_Zpole} and \cref{sec_twof:aboveZ}, respectively. Special attention has been given to the charged-hadron identification capabilities and flavour tagging, with parallel efforts in detector R$\&$D, algorithm development and physics potential studies. The identification and tagging algorithm aspects are discussed in Chapter 2 of this report, while the physics potential aspects are discussed in this section.

Future Higgs Factories running at the $\PZ$ pole, especially the FCC with the \TeraZ run with expected event statistics of the order \num{6e12} $\PZ$ decays, offer unprecedented precision in the field of electroweak precision observables (EWPOs). 
However, the systematic uncertainties from heavy-quark EWPOs, which are expected to show sensitivity to New Physics effects in loops, including the top quark, have to be commensurate with the statistical ones. 
From the world's most precise measurements of these observables in the beauty sector from LEP and SLD times~\cite{ALEPH_Rb_measurement_lifetime_mass, OPAL_Rb_measurement, ALEPH_AFB_measurement, OPAL_AFB_measurement, SLD_Rb_Rc}, the leading source of systematic uncertainties has been the misidentification of the quark flavour. 
Several efforts are being conducted to develop more robust and efficient flavour tagging and quark charge measurement algorithms. Most of them are based on modern machine learning algorithms~\cite{Liang:2023wpt,Tagami:2024gtc,Blekman:2024wyf} and are reported extensively in Section 2. In addition, profiting from the superb vertexing and charged hadron identification capabilities expected for future Higgs Factories detectors, the development of algorithms reconstructing hadrons produced on quark-decays and tagging single or double hemispheres in the detector allow for a minimization of mistagging rates and maximal control of systematic uncertainties. These techniques were not fully exploited at LEP or at SLC in the past due to moderated flavour tagging capabilities and/or integrated luminosities (compared with current expectations for all proposed collider options). Studies at the \PZ pole \cite{Rb_AFBb_exclusive} and above \cite{Irles:2024ipg} show the potential of this method to maximally control systematic uncertainties. Studies at the \PZ pole and above are documented in this report.

\subsubsection*{Full simulation studies}

Prior to the ECFA Higgs/electroweak/top factory study, several full simulation studies were performed on \epem collisions producing $\PQb\PAQb$, $\PQc \PAQc$, $\PQs \PAQs$, and $\PGt^+ \PGt^-$ final states. These studies have been finalized and discussed within the ECFA-HTE workshop.
They are based on high-energy collision (250 GeV, 500 GeV) simulations performed by the ILD concept group using the ILC beam conditions and ILD model.
Ref.~\cite{Jeans:2019brt} reports the work on the $\PGt^+ \PGt^-$  reconstruction at 500 GeV for different decay modes and the potential of measuring the $\PGt$ polarisation.
Several studies have been conducted using final states $\PQb\PAQb$ and $\PQc \PAQc$, in particular,  Refs.~\cite{Irles:2023nee,Irles:2023ojs,Marquez:2023guo,Irles:2024ipg}, based on single hadron identification and charge reconstruction and performing single vs double tagging measurements.
Recently, a related new work on $\PQs \PAQs$ and light-quark pairs has been performed \cite{Okugawa:2024dks}. These three activities are reported in detail in Section \ref{sec_twof:aboveZ}, emphasizing the main experimental challenges (reconstruction, detector design, etc.), which are shared at all common stages of the different Higgs Factory proposals. These studies are described in detail in Section \ref{sec_twof:aboveZ}.

The ECFA Higgs/electroweak/top factory study has also fostered a set of studies on two-fermion final states at the \PZ pole, with some studies that are summarized in the following sections. This work is performed at different levels of realism in simulation and assuming different colliders' scenarios (the FCC-ee \TeraZ being the predominant and more promising in terms of highest statistical power) and will be continued after this report is released, increasing the level of realism on the simulation.
The prospects of using single hadron identification and hemisphere reconstruction at the \PZ-pole for the first time with full simulation are discussed in Ref.~\cite{Rb_AFBb_exclusive}.
In Ref.~\cite{Cobal:2025shq}, the bottom quark forward-backward asymmetry from semileptonic decays of the bottom quark and jet-charge reconstruction is studied.
The study reported here is performed with full simulation of events at FCC-ee \TeraZ and the IDEA detector involving heavy quark final states. Another study included in this report discusses the prospects of reconstructing the forward-backward asymmetry for strange quarks, profiting from modern flavour tagging algorithms at FCC-ee \TeraZ and the IDEA detector (using fast simulation). 
Another work, started in Ref. \cite{Mekala:2024xal}, proposes a method to determine the flavour of two-quark production investigating the QED final-state-radiation, profiting from the high statistics foreseen at all colliders and the excellent photon-identification capabilities expected for all proposed detectors. This work has been started using fast-simulation tools, and the authors plan to perform full simulation studies in the near future.
These studies and a novel study addressing the potential to study two-tau final states are described in detail in \cref{sec:twof_Zpole}.

Full-simulation studies for comparing the potential of $\PZ$-pole running and radiative return events at e.g.\ 250 GeV CM energy are still missing and should be addressed.

In all these works, flavour tagging, particle identification (kaon identification, photon identification), and precise vertexing in the barrel and the forward region are emphasised.
These studies have triggered the development of novel algorithms and optimized detector designs. In the following, a sample of ongoing physics studies is reported. For detailed reports on reconstruction algorithm development we refer the reader to  \cref{sec:HtoSS} and Chapter 2.

\subsubsection{\texorpdfstring{Ongoing studies at \PZ-pole}{Ongoing studies at Z-pole}}
\label{sec:twof_Zpole}


\subsubsection*{\texorpdfstring{Reconstruction of EW observables for $\PQb$- and $\PQc$-quarks at Tera-Z using hemisphere-tagger algorithms}{Reconstruction of EW observables for \PQb- and \PQc-quarks at Tera-Z using hemisphere-tagger algorithms}} \label{sec:twof_Zpole:hem}

The explicit reconstruction of a selected list of hadron decays in independent detector hemispheres in light of $\mathcal{O}(\num{e12})$ events reaches ultra-pure tagging performances, even for strange quarks \cite{Rb_AFBb_exclusive}. 
In Ref.~\cite{Rb_AFBb_exclusive}, hemisphere tagging is explored using full simulation of \PZ-pole events reconstructed with the IDEA detector. This technique is applied to estimate the potential for partial-decay width and forward-backward asymmetry measurements for \PQc and \PQb-quarks . For the latter, since charge measurements are required, only charged mesons and baryons were included in the study to overcome mixing dilutions.
The concept of explicitly reconstructing $\PQb$ hadrons in $Z\to \PQq\PAQq$ events is expected to produce superb purity on the flavour-tag of the hemisphere, reaching purities that will be considered compatible with \SI{100}{\percent} in the following.

With this, any systematic uncertainties attached to the light-quark misidentification are expected to be drastically reduced if not removed. Furthermore, for the \PQb case, by considering only the charged $\PBp$-meson decays, as well as the $\PGLzb$-baryon decays, charge confusion from neutral $B$-meson mixing is overcome, and the dominant remaining source of systematic uncertainty arises from QCD corrections. These account for high-energy gluon emissions from the initial $\PQb$ quark, which might lead to confusing the hemispheres depending on the energy of the radiated gluon. However, fully reconstructed $\PBp$ and $\PGLzb$ serve as optimal estimators, not only for the $\PQb$-quark direction and charge but also for the amount of radiated gluons, which directly corresponds to the energy of the reconstructed hadrons. Two scenarios have been considered depending on the precisions with which the QCD corrections are known theoretically: a more conservative and a more optimistic scenario with \SI{5}{\percent}, and \SI{1}{\percent}, respectively. In both cases, thresholds on the minimally required $\PQb$-hadron energy have been found, at which $\sigma_\text{stat.}(\AFBb) = \sigma_\text{syst.}(\AFBb)=\num{2.3e-5}$ if a $\SI{1}{\percent}$ correlation correction is needed.

\begin{figure}[t]
    \begin{subfigure}[t]{0.48\textwidth}
        \centering
        \includegraphics[width = \textwidth]{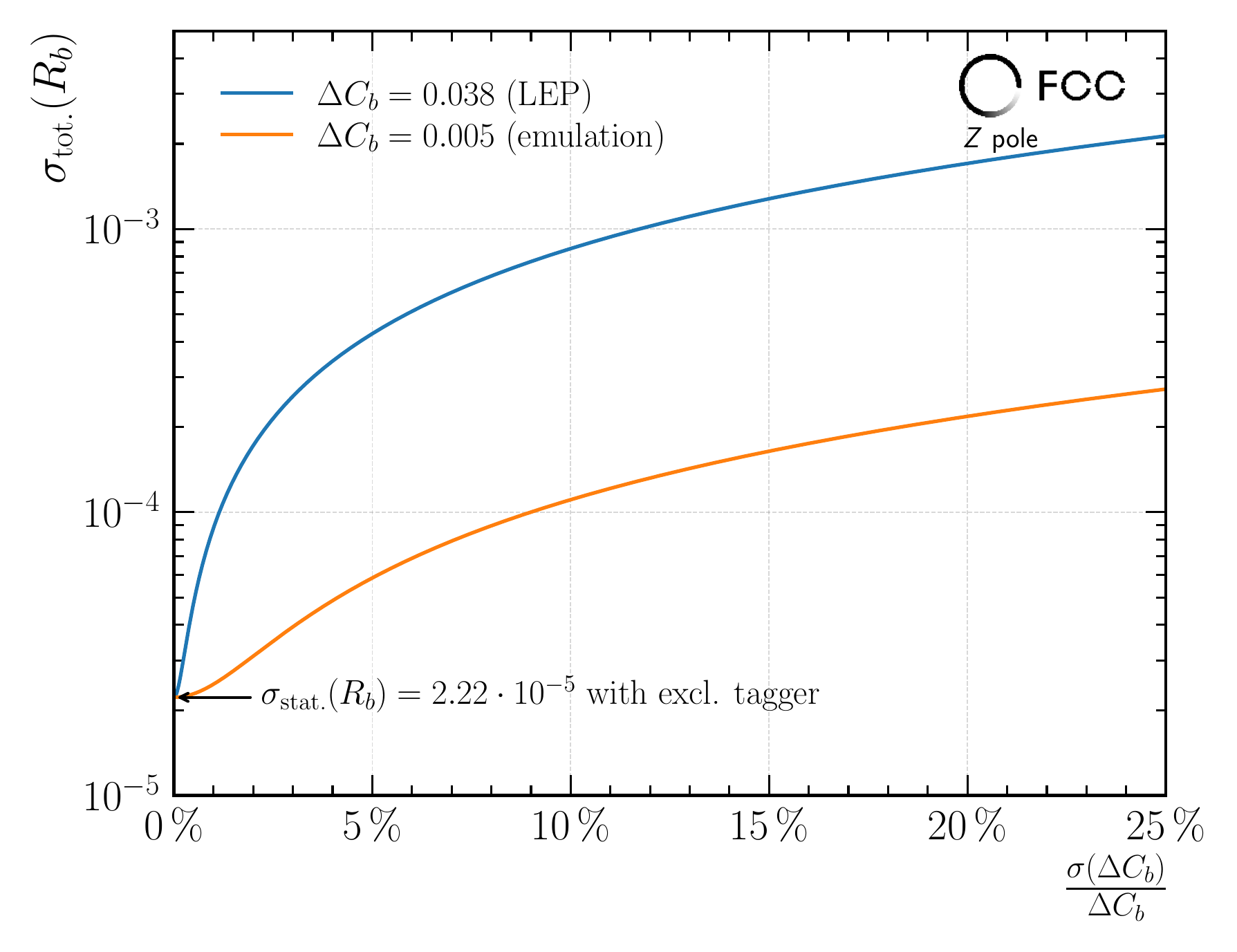}
        \caption{Total uncertainty of $R_{\PQb}$ as a function of the relative precision of $\Delta C_{\PQb}$ for two nominal values.}
    \end{subfigure}\hfill
    \begin{subfigure}[t]{0.48\textwidth}
        \centering
        \includegraphics[width = \textwidth]{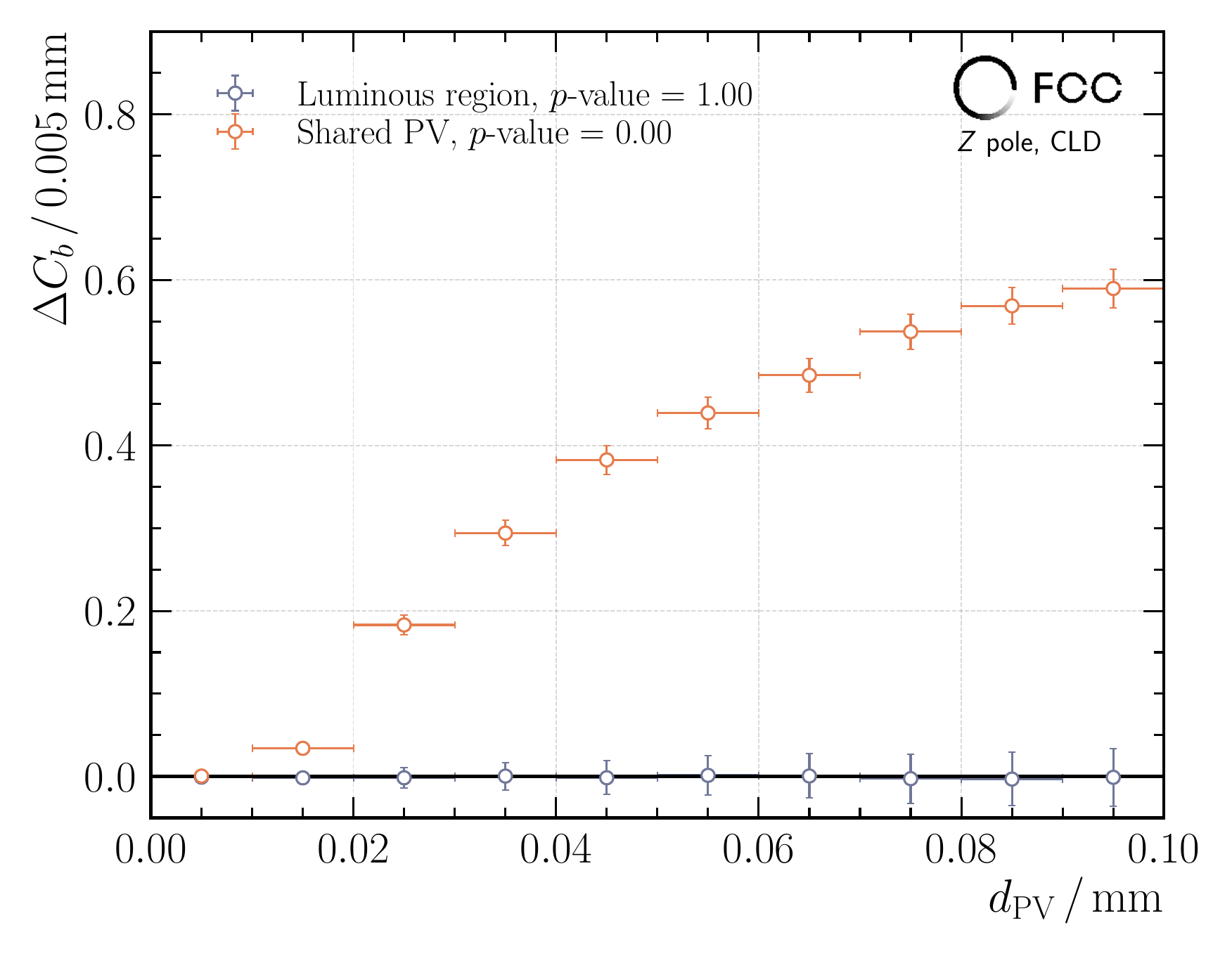}
        \caption{$d_\text{PV}$ is a measure for the spatial displacement of the reconstructed PV of the event from the true one. The nominal value of $\Delta C_{\PQb}$ strongly depends on this favour of one hemisphere, which is introduced with $d_\text{PV}\neq 0$.}
    \end{subfigure}
    \caption{Additional plots, that stress the importance of $\Delta C_{\PQb}$ for the measurement of $R_{\PQb}$. Techniques have been developed to mitigate its effect and achieve a measurement with competitive statistical and systematic uncertainties.}
    \label{fig:Rb_figures}
\end{figure}

The prospects for the \PQb-quark partial-decay width would profit from a flavour-unambiguous tagging procedure. In that case, the set of double-tag equations \cite{ALEPH:2005ab,Irles:2023ojs,Rb_AFBb_exclusive} are simplified since no (or minimal) contributions from mistagging appear. In that case, the tagging efficiency and the partial width will be extracted from the data itself, provided the correlation factor is estimated with simulations and its uncertainty, $\Delta C_{\PQb}$, is small. 
With the statistical precision in reach of $\sigma_\text{stat.}(R_{\PQb}) = \num{2.2e-5}$, any departure from zero of $\Delta C_{\PQb}$ significantly increases the systematic uncertainty attached to $R_{\PQb}$. Two scenarios have been considered in \cref{fig:Rb_figures}: assuming $\Delta C_{\PQb} = 0.038$ found by the ALEPH Collaboration in blue, and an emulated value of \num{0.005}. A gain in precision of about an order of magnitude can be achieved when controlling $\Delta C_{\PQb}$ at the zero-level.
The main source of $\Delta C_{\PQb}\neq 0$ has been found from the LEP Collaborations to originate from the reconstruction of a shared primary vertex (PV) in the event, which biases the tag of the opposite hemisphere, if the one in front has (not) been tagged~\cite{ALEPH_Rb_measurement_lifetime_mass}. This effect has been investigated and reduced up to the level of statistical agreement with zero by considering only tracks for the $\PQb$-hadron reconstruction that are incompatible with the beam spot (luminous region). The impact of the PV determination in terms of displacement of the reconstructed PV from the true one, $d_\text{PV}$, is shown in the right panel of Fig.~\ref{fig:Rb_figures} for the two approaches. While $\Delta C_{\PQb}$ increases significantly as a function of $d_\text{PV}$, it stays constant at zero for the luminous-region approach.
Furthermore, the inclusive value of $\Delta C_{\PQb}$ has been determined to be
\begin{align*}
    \Delta C_{\PQb}^{\text{shared PV}} &= \num{0.035(3)}(\text{stat.})\,,\\
    \Delta C_{\PQb}^{\text{luminous region}} &= \num{0.001(0.003)}(\text{stat.})\,, 
\end{align*}
where the statistical uncertainty corresponds to the limited size of fully simulated events used in this study.  

To not further increase the total uncertainty of $R_{\PQb}$, $\Delta C_{\PQb}$ must be known with a relative precision of at least \SI{10}{\percent} assuming a determination with the luminous-region approach. With this assumption, the statistical and systematic uncertainties read as
\begin{equation}
    \sigma_\text{stat.}(R_{\PQb}) = \sigma_\text{syst.}(R_{\PQb}) = \num{2e-5}\,.
\end{equation}
Further gain in the overall precision is in reach by considering also partially reconstructed $\PQb$-hadron candidates providing excellent purities as well. However, this would require a reevaluation of the hemisphere efficiency correlation. 

In contrast to $R_{\PQb}$ with a flavour-unambiguous tagging assumption, the double-tag equations are extended in the case for $R_{\PQc}$, by adding new factors due to mistagging rates becoming non-negligible.
Since the background contamination from the \PQc-quark flavour tagging is itself a source of systematic uncertainty, a balanced trade-off must be found between the loss in statistical precision and the reduction in systematic uncertainty from high-purity charm tags. At FCC-ee with a projected number of $\PZ$-boson decays of the order of \num{e12}, a pure selection of candidates from about 20 charm-hadron decay modes has the potential to mitigate the leading source of uncertainty arising from the contamination of $Z\to \PQb\PAQb$ events. Due to the high statistics expected at \PZ-pole runs and the high flavour tagging and hadron reconstruction capabilities, it is expected that \PQc-quark EWPOs will be also measured with unprecedented precision. 
However, more sophisticated methods to further reduce the uncertainties need to be employed, such as multivariate analysis techniques. Further application of the charm tagger for the measurement of the $\PQc$-quark forward-backward asymmetry $\AFBc$ is work in progress.

\subsubsection*{Precision Measurements of Strange Quark Forward-Backward
Asymmetry at the FCC-ee} \label{sec:twof_Zpole:afbss}

With the previous era accelerator facilities and detectors, typically the identification of bottom and maybe charm quarks was reliable, but with the increased interest in the origin of the mass and generational structure of the standard model, and with modern jet flavour taggers becoming available that can also tag strange and maybe even up and down quarks~\cite{Qu:2022mxj, Blekman2025}, it becomes possible also to measure the $A_{\text{FB}}$ for other quark flavours than $A_{\text{FB}}^{\PQb}$ and $A_{\text{FB}}^{\PQc}$, something previously not possible with competitive precision~\cite{Cossutti:2625562, SLD:2000jop}. In this work, we assess the sensitivity to measure $A_{\text{FB}}$ for bottom, charm, and strange quarks, with a representative collider and detector scenario (IDEA detector concept \cite{IDEA1} at FCC-ee \cite{Benedikt:2651299}).

The simulated samples consist of the process $\epem \to \PZ \to \qqbar$ (and $\epem \to \PZ \to \lplm$ for background) at $\sqrt{s}$ of $\SI{91.2}{\giga\electronvolt}$. \pythiaeight \cite{Bierlich_2203} is used for event generation, parton showering, and hadronisation. Values of $\sin^{2}\theta_{\PW}$ were unmodified. \textsc{Delphes}\xspace \cite{de_Favereau_2014} is used for event reconstruction assuming the IDEA detector concept. As \textsc{Delphes}\xspace does not simulate fake tracks, all reconstructed tracks correspond to genuine charged particles. Jet clustering is performed with $\fastjet$ \cite{Cacciari_2012} using the exclusive $\epem~\kT$ algorithm \cite{Catani:1991hj} and requiring exactly two jets. The event sample is scaled to correspond to a luminosity of $125 \mathrm{ab^{-1}}$.

Following fiducial cuts on the jet momentum ($|p_\text{jet}|>\SI{20}{\giga\electronvolt}$) and the jet axis $\cos\theta_{\text{jet}} < 0.97$, the following selection criteria were applied on the sample to reduce/remove irriducible backgrounds: the invariant mass of the two-jet system was constrained to reduce background contributions from 2-photon production and the tau decay channel of the Z boson ($ 81.2 <\text{M}_{\text{inv}}<\SI{101.2}{\giga\electronvolt}$); events were required to have more than two reconstructed tracks to reduce the electron and muon channel background ($N_{\text{tr}}>2$); the jet momenta were required to be dominated by hadrons instead of muons by imposing $(\sum|p^{\PGmpm}|/|p_{\text{jet}}| < 0.9)$ and $(\sum|p^{\text{had}}|/|p_{\text{jet}}| > 0.01)$, to further suppress leptonic Z boson decays.

Particle Transformer \cite{Qu:2022mxj}, tuned for the FCC-ee, is used for jet flavour tagging, which is applied sequentially to identify $\PQb$-, $\PQc$-, and $\PQs$-flavour jets. The working points used are chosen to have $0.1\%$, $0.1\%$, and $1\%$ mistag rates, respectively. An event is selected if both jets pass the threshold to be tagged with the same flavour. The flavour mistagging background is of the order of $0.01\%$ for all three flavours.

Momentum-weighted jet charge, defined as
\begin{equation}
    \text{jet charge} = \dfrac{\sum\limits_{i \in \text{jet}} q_i\cdot |p_i|^{0.3}}{|p_{\text{jet}}|^{0.3}},
    \label{eq:jet_charge}
\end{equation} is used to separate the quark-antiquark jets, where $q_i$ ($p_i$) is the charge (momentum) of jet constituents and $p_{\text{jet}}$ is the momentum of the jet. An event is selected when the two jets have opposite jet charge. A loose working point is selected, where the jet charge misidentification leads to the dominant background being of the order of $10\%$. This is a conservative approach; in realistic analyses, the charge misidentification can be optimized much more stringently with simultaneous fits of the same-charge contribution or the use of charge-sensitive event variables.

\begin{figure}
    \centering
    \includegraphics[width=0.48\linewidth]{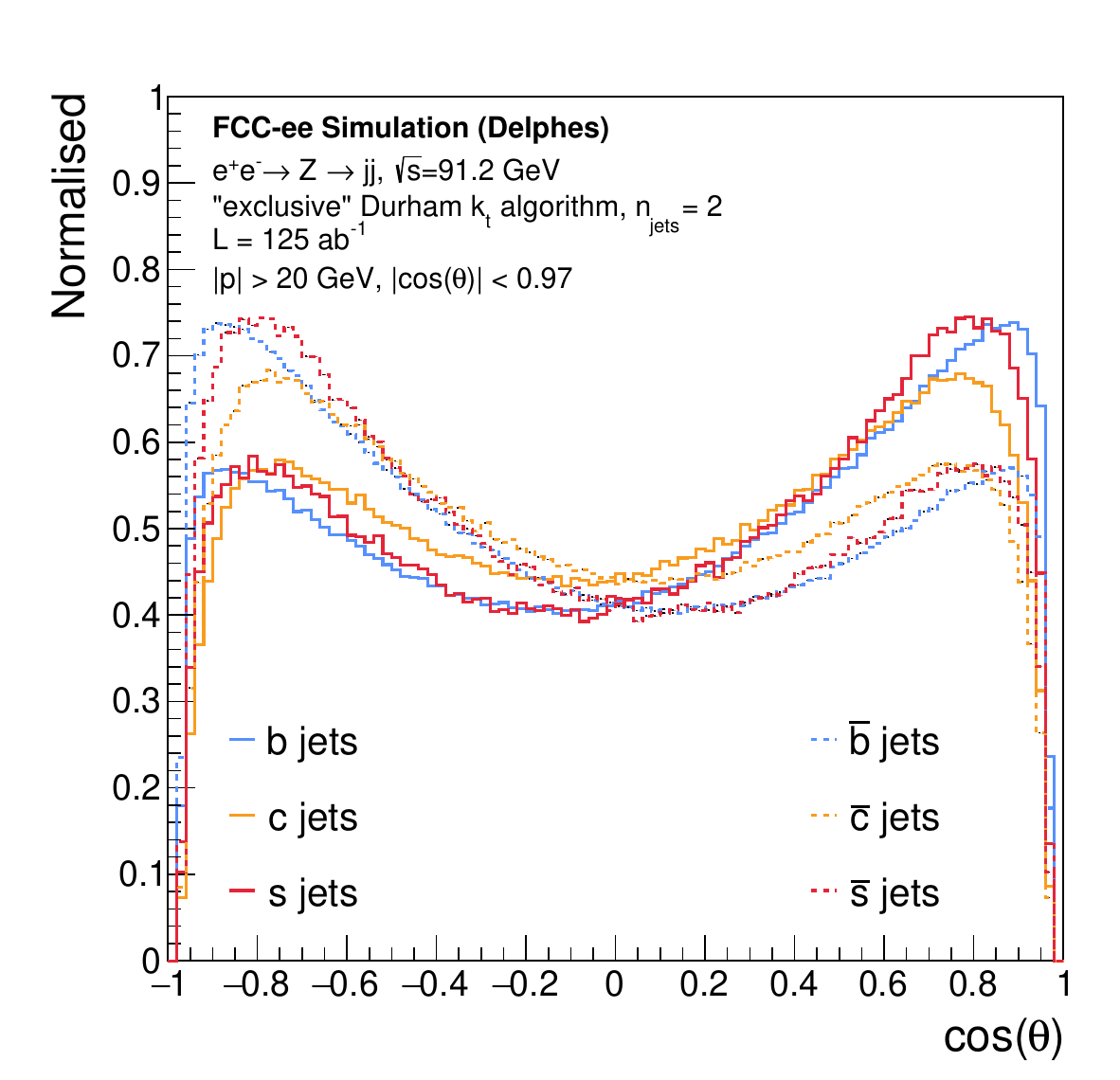}
    \includegraphics[width=0.48\linewidth]{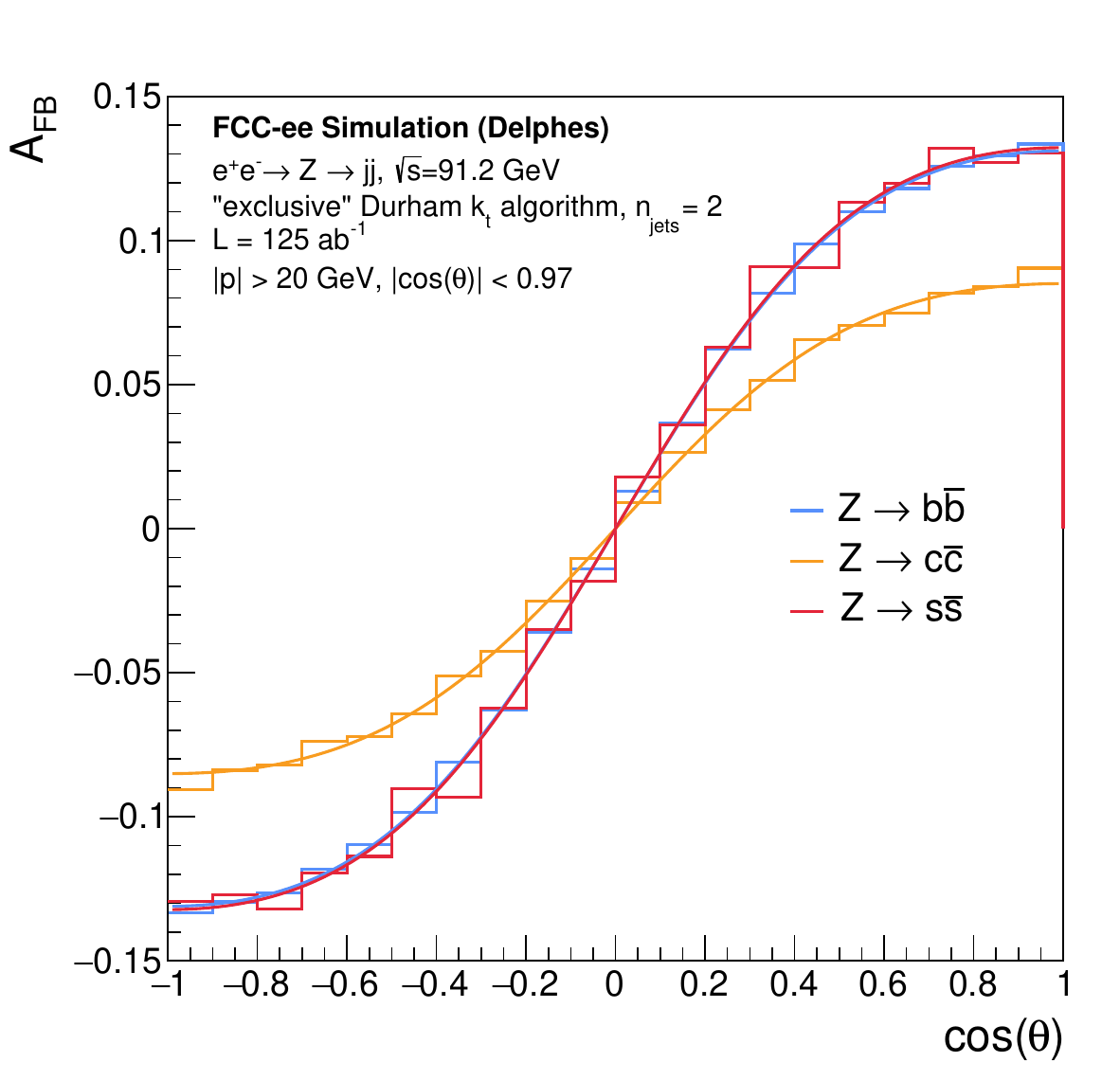}
    \caption{Left: The $\cos\theta$ distribution of the quark-antiquark jets of bottom (in blue), charm (in orange), and strange (in red) flavours. The distribution corresponds to the signal events after applying the preselection criteria, jet flavour tagging, and momentum-weighted jet charge tagging. Right: $A_{\text{FB}}^{\PQq}$ distribution over the entire range of the cosine of the polar angle ($\theta$) for bottom (in blue), charm (in orange), and strange (in red) events after applying the preselection criteria, jet flavour tagging, and momentum-weighted jet charge tagging.}
    \label{fig:asymmetry_fitted_all}
\end{figure}

To perform the $A_{\text{FB}}$ measurement, the signal events are selected after passing the previously listed jet quality criteria and flavour and jet charge tagging. The leptonic and mistagging backgrounds are assumed to be modelled reasonably well (with associated uncertainties) and consequently removed. The $\cos\theta$ distribution of the quark and antiquark jets for each flavour are shown in \cref{fig:asymmetry_fitted_all}. The forward-backward asymmetry for a quark $\PQq$ is measured for each bin $i$ as:
\begin{equation}
    A_{\text{FB}}^{\PQq,i} 
    = \dfrac{\sigma_{\text{F}}^{\PQq} - \sigma_{\text{B}}^{\PQq}}{\sigma_{\text{F}}^{\PQq} + \sigma_{\text{B}}^{\PQq}}
    = \dfrac{N^{\PQq}(i) - N^{\PAQq}(i)}{N^{\PQq}(i) + N^{\PAQq}(i)},
\end{equation}
where $N^{\PQq}(i)$ is the number of entries in bin $i$ of the $\cos\theta$ distribution of the jets containing the quark $\PQq$ and $N^{\PAQq}(i)$ is the same for jets containing the antiquark $\PAQq$. This distribution of $A_{\text{FB}}^{\PQq}$ is re-binned and fitted over the entire range of $\cos\theta$ with the function:
\begin{equation}
    A_{\text{FB}}^{\PQq}(\cos\theta) = 4 \left(\dfrac{1-4|Q_\text{e}|\sin^{2}\theta^{\text{eff}}}{1+8Q^{2}_\text{e}(\sin^{2}\theta^{\text{eff}})^{2}-4|Q_\text{e}|\sin^{2}\theta^{\text{eff}}} \right) \left(\dfrac{1-4|Q_\text{q}|\sin^{2}\theta^{\text{eff}}}{1+8Q^{2}_\text{q}(\sin^{2}\theta^{\text{eff}})^{2}-4|Q_\text{q}|\sin^{2}\theta^{\text{eff}}} \right) \dfrac{\cos\theta}{1+\cos^{2}\theta}
\end{equation}
and $\sin^{2}\theta^{\text{eff}}$ as the fit parameter \cite{ALEPH:2005ab}. The asymmetry distributions, along with the fit, are presented in \cref{fig:asymmetry_fitted_all}.

The resulting forward-backward asymmetry values for the three flavours are:

\begin{align*}
    A_{\text{FB}}^{\PQb} = 0.0983222 \pm 0.0000012~\text{(stat.)} \\
    A_{\text{FB}}^{\PQc} = 0.0637820 \pm 0.0000020~\text{(stat.)} \\
    A_{\text{FB}}^{\PQs} = 0.0991786 \pm 0.0000026~\text{(stat.)}, \\
\end{align*}
where the $A_{\text{FB}}^{\PQb}$ and $A_{\text{FB}}^{\PQc}$ values were used to test the robustness of the $A_{\text{FB}}^{\PQs}$ measurement method but not used in the measurement.

An alternative recent assessment of the systematic and statistical uncertainty on $A_{\text{FB}}^{\PQs}$ gives values of the order of $10^{-5}$ \cite{ablondel_2025}. This estimate was obtained by extrapolating the ALEPH measurements of $A_{\text{FB}}$ in inclusive hadronic decays \cite{DECAMP1991377, ALEPH:1996qlh}, and the $\PQb$-asymmetry measurement~\cite{ALEPH:2001mdb}. That analysis is based on hemisphere charge asymmetries and their first and second moments, and is robust against QCD gluon radiation, a dominant systematic effect. Statistical uncertainties were dominant at LEP, with systematic uncertainties at the $10^{-3}$ level at that time, primarily driven by the experimental statistics on jet charge determination (see e.g.\ Table 5 from Ref.~\cite{ALEPH:2001mdb}).
The uncertainty from the charge separation is expected to decrease by a factor of $300$ at the FCC-ee, while the modern flavour taggers \cite{Bedeschi_2022, Blekman2025} and planned detector improvements are assumed to improve the systematic uncertainties originating from flavour purities by more than a factor of $15$. 
With the anticipated improvements in the measurement and modelling of strange particle and baryon production, and of secondary interactions, together with the extensive internal controls offered by modern analysis methods, the total statistical and systematic uncertainties for this method were optimized to be of similar size by applying a substantially tighter flavour tag requirement, and projected to be around $10^{-5}$. This confirms that systematic uncertainties will likely be dominant over the statistical uncertainty determined earlier in this study. 

This investigation shows that the forward-backward asymmetry measurement in the strange decay channel of the $\PZ$ boson is possible with a statistical precision of $2.6 \cdot10^{-6}$ at the FCC-ee, considering a luminosity of $125~{\mathrm{ab}^{-1}}$ at the $\PZ$ resonance and the current developments in jet flavour tagging. Using an alternative approach, the systematic uncertainty on $A_{\text{FB}}^{\PQs}$ was shown to be possible to be constrained to below $10^{-5}$.

\subsubsection*{Determining \PZ-boson couplings to quarks studying final-state-radiation.}

Future Higgs factories operating at the $\PZ$-pole would produce $10^{9}$--$10^{12}$ $\PZ$ bosons, allowing for precise measurement of the $\PZ$-boson couplings to fermions. Here, we report progress in a recent work that aims to establish the potential of measuring electroweak couplings of quarks, especially highlighting the measurement for light quarks, inspecting the QED final-state-radiation in two-fermion final states.  A similar measurement has already been performed at LEP~\cite{DELPHI:1991utk, L3:1992kcg, L3:1992ukp, DELPHI:1995okj, OPAL:2003gdu}. The main idea relies on the fact that up- and down-type quarks differ in electric charge, and thus their electromagnetic couplings are distinguishable. The coupling strength of the $\PZ$ boson to a given fermion, $c_f$, is conventionally defined as a sum of its squared vector and axial couplings. The total width of the $\PZ$ boson to hadrons, $\Gamma_\text{had}$, is proportional to the sum of the couplings for down- and up-type quarks:
\begin{equation}
    \Gamma_\text{had} \sim (3c_{\PQd} + 2c_{\PQu}).
\end{equation}
Analogously, the total width to radiative hadronic decays for exactly one photon emission, $\Gamma_{had+\gamma}$, scales as
\begin{equation}
    \Gamma_{had+\gamma} \sim \frac{\alpha}{2\pi} f(y_{\text{cut}}) \left(3Q_{\PQd}^2 c_{\PQd} + 2Q_{\PQu}^2c_{\PQu}\right),
\end{equation}
where $\alpha$ is the electromagnetic coupling constant, $f(y_{\text{cut}})$ is a form factor depending on an arbitrary parameter $y_{\text{cut}}$ incorporating the isolation criteria for photons and $Q_{\PQd}$ ($Q_{\PQu}$) is the electric charge of down-type (up-type) quarks. In the following, we will assume $y_{\text{cut}}$ is the photon transverse momentum with respect to the jet direction, $q^{\text{T}}$. Since $Q_{\PQd} \ne Q_{\PQu}$, the expressions for the radiative and the total hadronic widths include different coupling combinations, and the couplings of the up- and down-type quarks can potentially be disentangled.

In the study summarized here and started in Ref. \cite{Mekala:2024xal}, the authors collect two-jet events inclusively and classify them into 10 categories, corresponding to all the combinations of 4 possible jet tags (``light'', $\PQs$, $\PQc$, $\PQb$). Additionally, they tag events with exactly one isolated photon. To estimate the statistical precision, they fit cross sections to minimise the $\chi^2$-test statistic by comparing the numbers of expected and ``measured'' events, where the second number is drawn from the Poisson distribution.

A realistic and precise Monte Carlo simulation is crucial for the measurement. The process of $\Pep \Pem \to \PQq \PAQq$ is often perceived as a benchmark point for Monte Carlo generators, but the reconstruction of isolated photons poses several challenges. Not only does the final-state photon radiation (FSR) from Matrix Elements (ME) have to be included, but also the initial-state radiation (ISR), as well as their matching to parton showers and hadronisation modelling. As a recipe, one can generate data samples using fixed-order ME calculations, with exclusive emissions of hard photons, and match them with ISR structure functions and FSR showers accounting for collinear and soft emissions. Building upon \cite{Kalinowski:2020lhp}, a dedicated matching procedure for simulating photons using both ME calculations (for hard emissions) and ISR structure function and FSR showers (for soft emissions) has been developed in \whizard~\cite{Kilian:2007gr,Moretti:2001zz}.

The difference in quark charges affects only the photons emitted from the final state and hence, experimental selection should be optimised to enhance the numbers of measured photons of this kind and suppress the other contributions (ISR and decays of hadronisation products). To simulate detector effects, \textsc{Delphes 3.5.0}~\cite{deFavereau:2013fsa} was used with the \textit{ILCgen} cards which were modified to cluster all the photons into jets. It allowed for testing photon isolation criteria independently of the current assumptions.

\begin{figure}[t]
    \centering
    \includegraphics[width=0.5\textwidth]{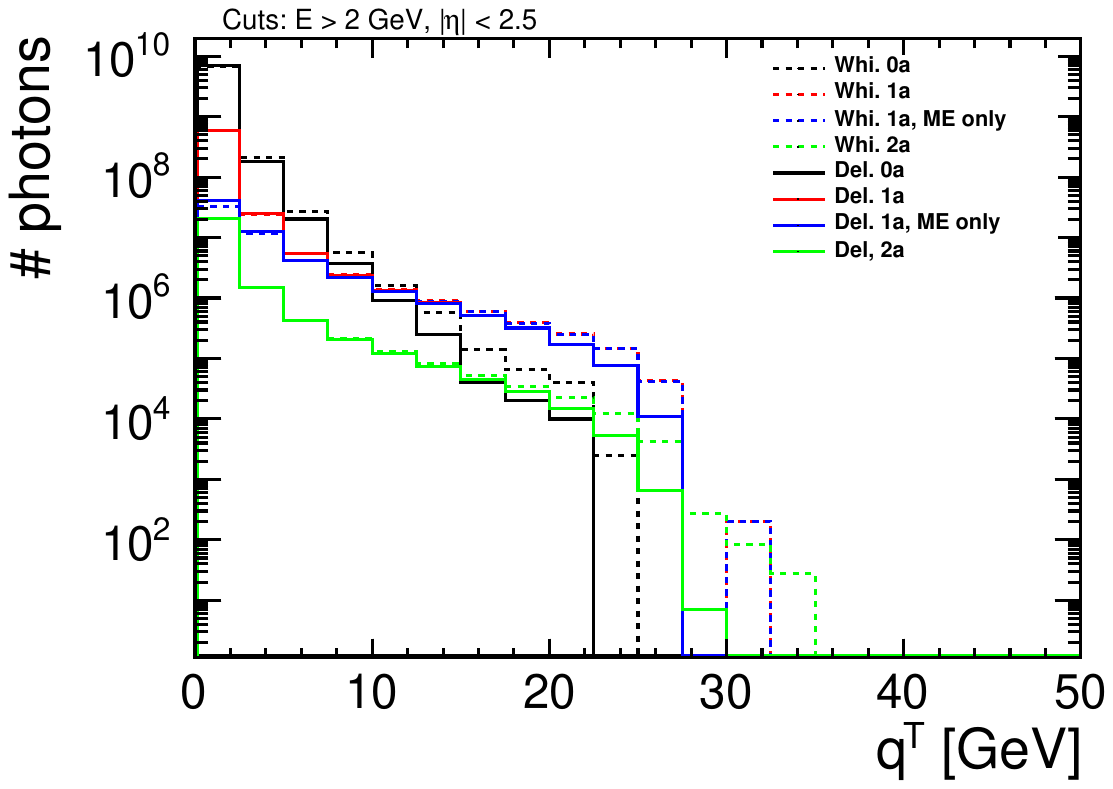}
    \caption{The number of photons visible in the detector as a function of $q^{\text{T}}$. See text for details.}
    \label{mekala-fig2:qt}
\end{figure}

Figure \ref{mekala-fig2:qt} compares the distributions of photons at the generator and detector level for different generated photon multiplicities. Besides the distributions for 0, 1 and 2 ME-photon samples, the distribution of the 1 ME-photon sample is plotted where only ``proper'' photons (those generated in the full ME picture, not coming from showers or hadronisation) are tagged. The plot shows that above 10 GeV the 1-ME-photon sample becomes dominant. 

To optimise the cut, the authors studied seven sources of uncertainties: luminosity of 0.01\%, acceptance of radiative (non-radiative) events of 5\% (50\%), and tagging of $\PQb$ jets of 1\%, $\PQc$ jets of 2\%, $\PQs$ jets of 5\% and light quarks of 10\%. The preliminary assumptions were used as a ``proof of concept'' of the statistical framework. 

Preliminary results suggest that a sub-percent precision will be achievable for light quarks while for heavier quarks, a sub-permille precision is envisioned.

\subsubsection*{Towards a measurement of the tau polarisation.}

The tau lepton is a particularly interesting particle for testing the electroweak sector of the Standard Model, thanks to its heavy mass and rich decay phenomenology. 
In high-energy \epem colliders, tau pairs are produced through the process  $\epem \to \PZ/\PGg^* \to \PGtp\PGtm$.   
At the \PZ-pole energy, assuming no longitudinal beam polarisations in this case, the production cross section of
$\sigma(\epem \to \PZ \to \PGtp\PGtm) = 1476.58 \, \text{pb}^{-1}$  at $\sqrt{s} = \SI{91.188 }{\giga\electronvolt}$ will yield an unprecedented large $\PGtp\PGtm$ sample, with the added bonus of precise momentum reconstruction capabilities and very low-background environment.  
Tau measurements pose demanding detector requirements on momentum resolution, on the knowledge of the vertex detector dimensions, on e/$\mu$/$\pi$ separation over the whole momentum range, and require fine granularity and high efficiency in the tracker and electromagnetic calorimeter.
A comprehensive program of tau studies will be performed on this tau pair sample, aimed at achieving high-precision measurements
including the measurement of the tau polarisation and the extraction of fundamental electroweak properties through its study.  

The tau polarisation, $\mathcal{P}_{\PGt}$, provides a sensitive probe of the couplings of the \PZ and the weak interaction. It can be measured through the angular distributions and energy spectra of the tau decay products.  It can be expressed as 
$\mathcal{P}_{\PGt} \equiv \frac{\sigma_+ - \sigma_-}{\sigma_+ + \sigma_-}$,
where $\sigma_{+}$ and  $\sigma_{-}$ are the cross sections for producing left-handed and right-handed tau leptons, respectively.
The polarisation can be further expressed as a function of the two neutral current asymmetry parameters ($\mathcal{A}_e$ and $\mathcal{A}_{\PGt}$), taking into account as well its dependence on the direction of the tau expressed as the angle between the tau momentum and the electron beam ($\theta$)~\cite{ALEPH:2005ab,Eberhard:201688}:
\begin{equation}
    \mathcal{P}_{\PGt} \left( \cos \theta  \right) = - \frac{\mathcal{A}_{\PGt} (1+\cos^2 \theta ) + 2 \mathcal{A}_e \cos\theta } {1+ \cos^2\theta + 2 \mathcal{A}_e \mathcal{A}_{\PGt} \cos \theta }.
\end{equation}

Measuring $\mathcal{P}_{\PGt}(\cos\theta)$ yields nearly independent determinations of $\mathcal{A}_{\PGt}$ and $\mathcal{A}_e$. Consequently, $\PGt$ polarisation measurements provide not only a determination of $\sin^{2}\theta_\text{eff}$ but also test the hypothesis of the universality of the couplings of the \PZ to the electron and $\PGt$ lepton.  Integrating over $\cos\theta$ we obtain $\mathcal{P}_{\PGt} (total) = -\mathcal{A}_{\PGt}$.  

At LEP, $\mathcal{A}_{\PGt}$ and $\mathcal{A}_{\Pe}$ were measured to be $\mathcal{A}_{\PGt} (\text{LEP}) = 14.39 \pm 0.35 \, (\text{stat}) \pm 0.26 \, (\text{syst}) \%$ and $\mathcal{A}_e (\text{LEP}) = 14.98 \pm 0.48 \, (\text{stat}) \pm 0.09 \, (\text{syst}) \%$~\cite{ALEPH:2005ab}, dominated by the statistical uncertainties. The systematic uncertainty for $\mathcal{A}_{\PGt}$ is found to be significantly larger than the $\mathcal{A}_{\Pe}$ one: this is due to a cancellation of charge and $\cos{\theta}$-independent systematic uncertainties in the latter case, leading to an extremely precise measurement. The most sensitive results are obtained in the decay modes to 
 pions ($\PGt\to\PGp\PGn$) and rho ($\PGt\to\PGr\PGn\to\PGp\PGpz\PGn$). Uncertainties varied largely between LEP experiments. This underlines the importance of the detector concept. A key aspect is having excellent performance in the reconstruction and identification of both charged particles and photons.   At FCC-ee, the large data samples and the excellent performance in charged particle and photon identification promised by future detectors will lead to much smaller
uncertainties overall.  Assuming a factor of 10 improvement in systematics over the LEP measurements leads to an estimated systematic uncertainty for $\mathcal{A}_{\PGt}$ that could be as good as $0.02\%$, and even lower for $\mathcal{A}_{e}$.  

To go beyond this estimation, a full simulation study that starts by developing a tau reconstruction is necessary. The results documented here are based on the reconstruction of $\PGt$ leptons with full simulation of the CLD detector~\cite{Bacchetta:2019fmz} and should be understood as a proof of concept, that will be refined in the future. 
Tau identification relies on reconstructing charged hadrons and photons with precise energy resolution. Around 65$\%$ of tau decays are hadronic, primarily manifesting as one or three charged hadrons accompanied by photons coming from the decay of $\pi^0$. Effective identification of $\pi^0$s is key. 
For rare decays, discrimination between pions and kaons will also be relevant.  
A first algorithm for reconstructing hadronic tau decays at the FCC-ee has been developed based on PandoraPFA~\cite{THOMSON200925} candidates as input. This approach involves identifying tau candidates by clustering charged candidates (assumed to be pions) and nearby photons into tau-lepton candidates. The identification process focuses on reconstructing the main decay modes of the tau: $\PGt \to \pi \nu_{\PGt}$: a single charged pion; $\PGt \to \PGr \nu_{\PGt}$: involving $\PGppm$ and a $\pi^0$; $\PGt \to \Pa_1 \nu_{\PGt}$: with $a_1$ decaying into either $\PGppm 2\PGpz$ or $\PGppm\PGpmp\PGppm$. The first two decay modes ($\PGp$ and $\PGr$) are expected to provide the most sensitive measurements for the polarisation study. Decays of the taus to light leptons are considered in the algorithm and removed based on the identification of these light leptons by PandoraPFA. 
To address limitations in the PandoraPFA approach, a Graph Neural Network (GNN) method is being developed in parallel. The GNN takes the tracks and calorimeter clusters as inputs. This method aims to improve separation between similar decay modes, for example, aiding in the distinction between merged photons and resolved diphotons.

A first study of tau polarisation, including optimal variables, has been done with the PandoraPFA method.  This first analysis is limited to $\PZ\to\PGt\PGt$ events in which one of the taus decays leptonically in one hemisphere and the other tau hadronically. 
This will be expanded in future iterations of the analysis to improve the coverage of the phase-space and to benefit from determining the tau direction possible in the case both taus decay hadronically~\cite{ALEPH:2001uca}. 
Only tau backgrounds from category migration are considered in the analysis for now. The Monte Carlo samples used in the analysis have been generated in \pythia 8.3~\cite{Bierlich:2022pfr,Ilten:2012zb}.
To model the behaviour of samples with anomalous polarisation, a weighting technique~\cite{Alcaraz:293383}, based on the dependence of the polarisation on the kinematics of the tau and, in particular, on the theta of the outgoing meson, is used. This approach can be validated with additional samples with specific settings for $\Pol_{\PGt}=\pm 1$ generated in \pythiaeight. 

\begin{figure}[h!]
    \centering
    \includegraphics[width=0.32\textwidth]{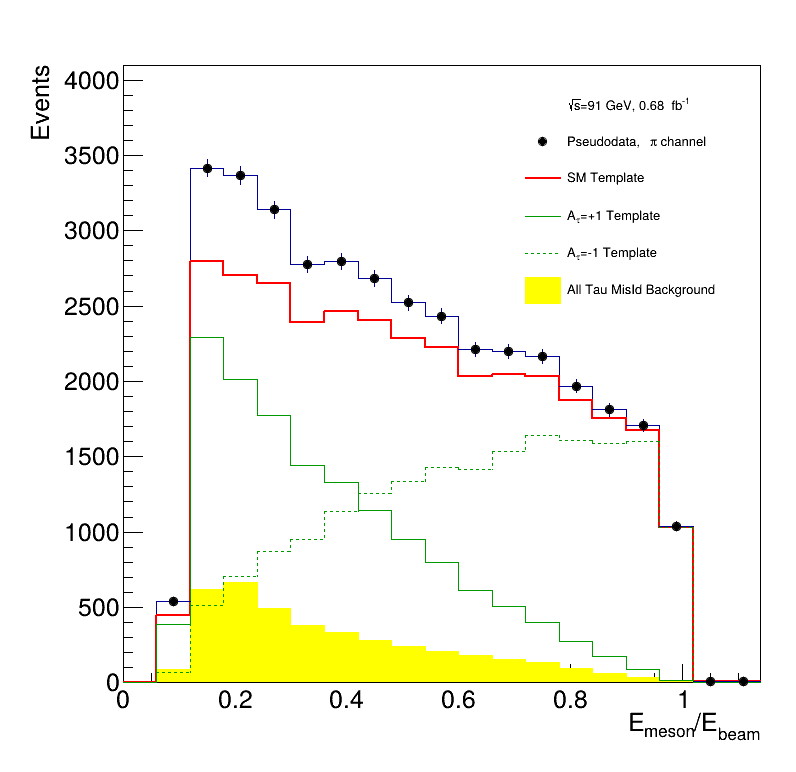} 
    \includegraphics[width=0.32\textwidth]{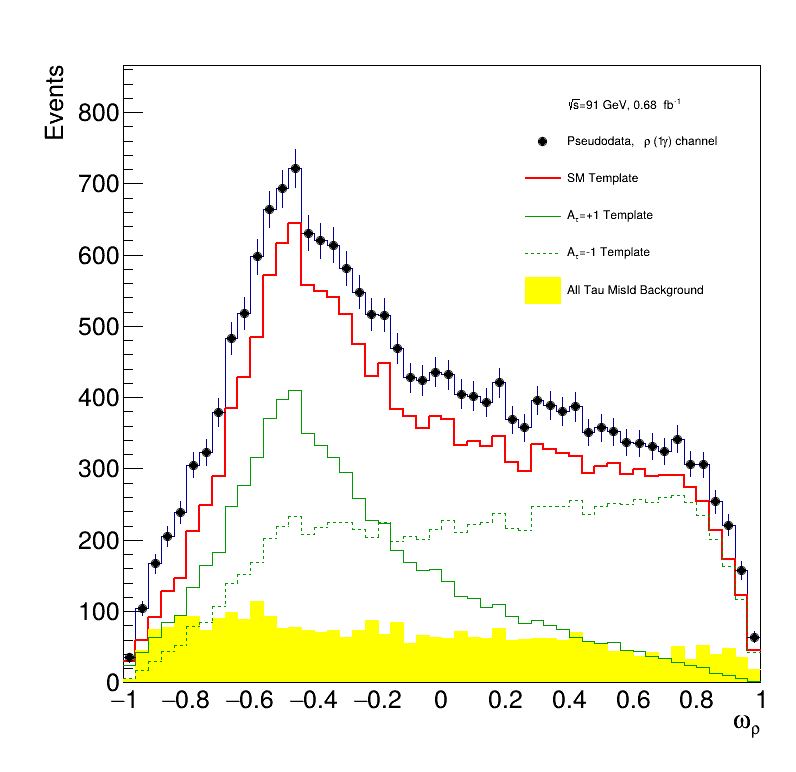}
    \includegraphics[width=0.32\textwidth]{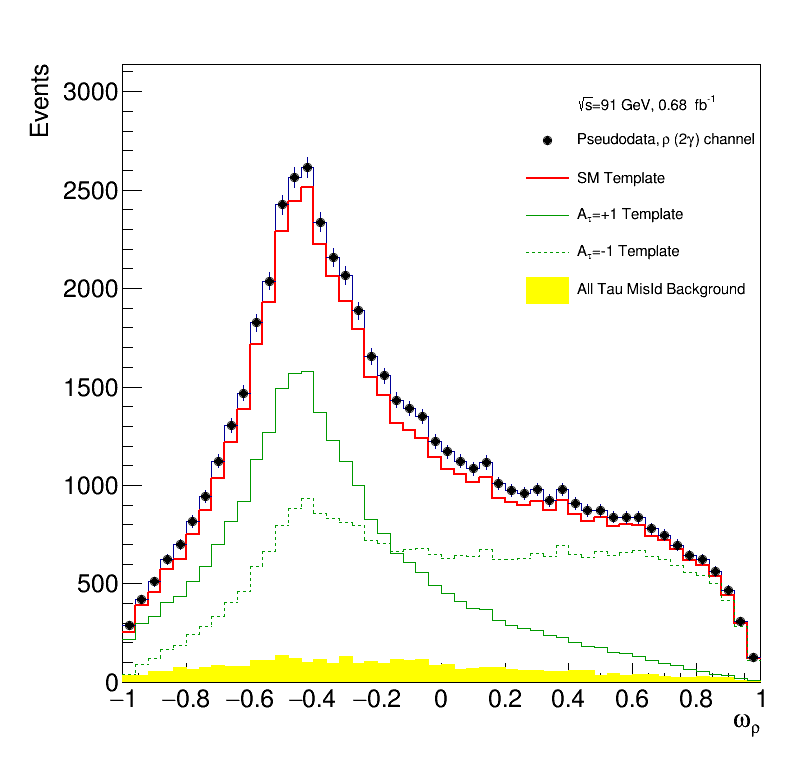}
    \caption{Optimal variables in the $\PGp$ (right),  $\PGr$  with one identified photon, (middle) and $\PGr$ with two identified photons (right) decay modes.}
    \label{fig:XVariabThreeChannels}
\end{figure}

Figure~\ref{fig:XVariabThreeChannels} shows the optimal variables for the study of the polarisation after this very simple selection for the three main configurations considered in the analysis: a single charged pion, a charged pion and a single (merged) photon, and a single pion and two resolved photons. The first mode is designed to select  $\PGt\to\PGp\PGn$ decays. In this case the optimal variable is directly the energy fraction  $x_\pi = \frac{E_\pi}{E_\text{beam}}$. 
The latter two categories correspond to the $\PGt\to\PGr\PGn\to\PGp\PGpz\PGn$ channel with either one or both of the photons resulting from the $\PGpz$ decay reconstructed. Here, the optimal observable in LEP was defined to be
$\PGo_{\PGr} = \frac{W_+(\theta^*, \psi) - W_-(\theta^*, \psi)}{W_+(\theta^*, \psi) + W_-(\theta^*, \psi)}$, 
where $W_+$ and $W_-$ represent the angular distributions of the $\PGr$ decay products for different helicity states, and $\theta^*$ and $\psi$ are angles describing the decay products in the $\PGt$ rest frame\cite{Nikolic:1996nj}.  The weighted $\Pol_{\PGt}=\pm 1$ signal templates are shown in comparison to the SM prediction. Additionally, the backgrounds coming from tau misidentification are also shown. No other backgrounds are considered at this stage. Further studies of the full simulation samples and optimization of the reconstruction algorithm are in progress. 

A comprehensive extraction of the polarisation parameters requires analysing the data in multiple bins of $\cos \theta$. This binned analysis allows for a simultaneous determination of $\mathcal{A}_{\PGt}$ and $\mathcal{A}_e$ and goes beyond this preliminary study. As a first step to gauge the potential uncertainties, we have performed a limited analysis that aims for $\mathcal{A}_{\PGt}$, which can be extracted independently of $\cos \theta$. This has been done through a log-likelihood fit of the optimal variable in the $\PGr$ channel with two resolved photons. 
For a dataset corresponding to an integrated luminosity of $17 \, \text{ab}^{-1}$, which corresponds to the data collected by only one FCC-ee experiment during a single year, the statistical uncertainty of the  polarisation asymmetry extracted with this very simple approach is found to be
$\Delta \mathcal{A}_{\PGt} \approx 0.007\%$.  This corresponds to a very small fraction of the full dataset available at FCC-ee: the final statistical uncertainty can be considered negligible even for the final analysis binned in terms of $\cos{\theta}$.

Future steps in the analysis will assess the impact of systematic variables on the optimal observables based on the performance of different detector configurations. Backgrounds will be expanded to include Bhabha scattering, significant for the $\mathcal{A}_{\Pe}$ measurement. 

\subsubsection{Ongoing studies above the Z pole}
\label{sec_twof:aboveZ}

A full-simulation study of the $\PGtp\PGtm$ final state is reported in Ref.~\cite{Jeans:2019brt}, including discussion of event selection and background rejection, tau decay mode reconstruction, and the decay mode-specific extraction of the tau polarisation estimators. The main experimental causes of sensitivity loss compared to a perfect detector with perfect reconstruction were also investigated. Improvements in the reconstruction of this final state, allowing explicit reconstruction of the tau lepton momenta, were first discussed in Ref.~\cite{Yumino:2022vqt} and are further developed below.

The works for the $\PQq\PAQq$ include the study of data-driven techniques (such as double tagging and double charge measurement) for the determination of $R_{\PQq}$ and $\AFBq$ reducing the usage of Monte Carlo simulations for the modelling of systematic uncertainties such as fragmentation functions or angular correlations due to QCD corrections. These techniques could not be maximally exploited in past experiments due to reduced yields and/or vertexing capabilities. In Refs.~\cite{Irles:2023nee,Irles:2023ojs,Marquez:2023guo,Irles:2024ipg}, the potential of using \dedx or \dndx for charged hadron identification has also been studied and proven to have a large impact on the final uncertainties, especially for the $\PQc$-quark case, where the kaon identification becomes crucial for the charge reconstruction of the jet. These studies prove that with an optimal detector design, allowing high purity and efficient quark identification capabilities and data-driven techniques (single vs double hemisphere tagging), it is possible to evaluate and minimize the uncertainties without relying on Monte Carlo simulations.
These studies have been performed in the context of linear colliders with energy stages above the \PZ-pole and the $\PZ\PH$ threshold but, of course, these studies are also highly relevant to different running scenarios and collider options since the experimental challenges share many commonalities.

\subsubsection*{Reconstruction and use of \PGt spin information}

The \PGt-pair final state provides unique possibilities for
testing the Standard Model and discovering the effects of new physics, thanks to having access to the spin
orientation of the final state fermions by consideration of the
distribution of the \PGt decay products.
Both the longitudinal polarisation, which distinguishes between
positive and negative helicity \PGt, and transverse spin components
can be probed, as well as spin correlations between the two
final state taus.
The basic detector-level observables which one would like to reconstruct
are the \PGt-pair invariant mass and centre-of-mass scattering angle, along
with estimators of the tau spin orientation (which we call ``polarimeters'').

\begin{figure}[h]
\begin{center}
  \includegraphics[width=0.5\textwidth]{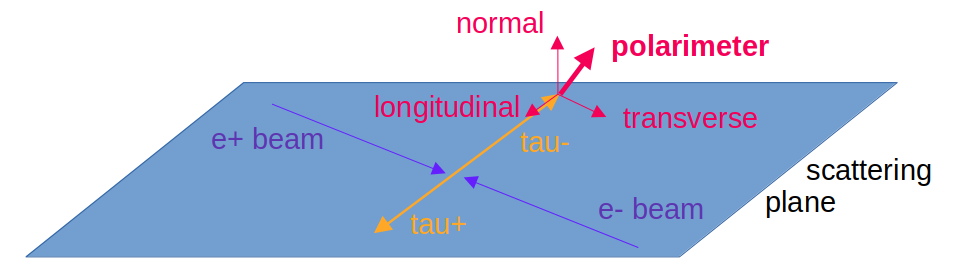}
  \hspace{0.05\textwidth}
  \includegraphics[width=0.35\textwidth]{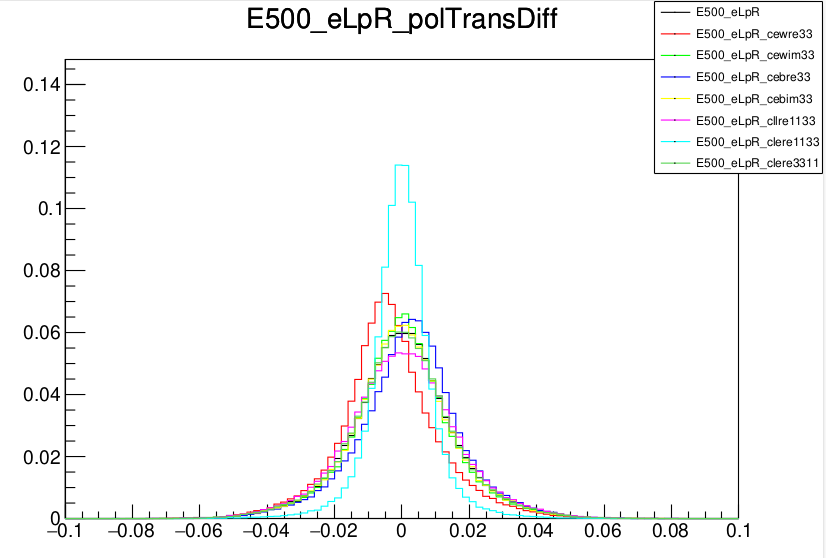} 
  \caption{Left: definition of polarimeter components.
    Right: illustration of the variation of a particular combination of polarimeter components
    (the difference of the two taus' transverse components),
    for di-\PGt production at 500~GeV centre-of-mass with
    100\% polarised left-handed electron and right-handed positron beams.
    Different colours represent different SMEFT operators.}
  \label{fig:ewk_qcd_tauPolHiM:eft}
\end{center}
\end{figure}

In Ref.~\cite{Jeans:2019brt} the authors used the \madgraph event generator~\cite{madgraph} interfaced with the
TauDecay library~\cite{taudecay} to
investigate the sensitivity of \PGt spin observables to physics beyond
the Standard Model in the context of Standard Model Effective Field Theory (SMEFT). A number
of dimension-6 operators which affect the \PGt-\PGt-\PW and \PGt-\PGt-\PB vertices
are being considered, as well as 4-fermion coupling between the electron and
\PGt. Figure~\ref{fig:ewk_qcd_tauPolHiM:eft} shows the definition of polarimeter components, and an example of
variations in a polarimeter observable with some SMEFT coefficients.

The reconstruction of \PGt leptons is challenging because of the neutrinos
produced in their decay, which reduces the power of kinematic constraints.
If the invariant mass and rest-frame of the tau-pair are known,
as is the case at the \PZ pole where ISR is limited,
then the ``cone-method'' can be used to reconstruct the tau momenta~\cite{Yumino:2022vqt}.

At higher energies, ISR and beamstrahlung induce significant variations
in the \PGt-pair invariant mass and rest frame, so an alternative approach is required.
A method is proposed by the authors which can reconstruct \PGt momenta in this case of reduced kinematic
constraints, which uses, in addition, the detailed trajectories of charged decay products of the \PGt in the
vicinity of the interaction point.
The method assumes that a single ISR photon has escaped detection, and scans over possible momenta of this photon.
At each candidate momentum, a solution is searched for which is consistent with
4-momentum conservation, the \PGt invariant mass,
and with both \PGt's being produced at the same point along the nominal beamline and decaying on the
charged pion trajectories. 
This method can identify zero, one, or multiple solutions per event.
In the case of multiple solutions, there is no obvious way to decide which is correct,
so all solutions are considered, producing a per-event distribution of estimated event quantities.

\begin{figure}[h]
\begin{center}
  \includegraphics[width=0.45\textwidth]{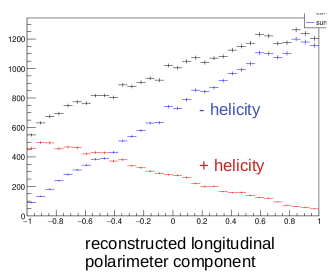}
  \caption{
    Example of the reconstructed longitudinal polarimeter distribution, using as input
    true values of pion momenta and trajectories. Extracting the average \PGt polarisation
    involves fitting the total (black) distribution as the sum of positive/negative
    (red/blue) helicity contributions.
  }
\label{fig:ewk_qcd_tauPolHiM:pol}
\end{center}
\end{figure}

As benchmark, the authors consider the tau decay modes with one or two pions, representing over a third of
\PGt decays, which are experimentally relatively easy to reconstruct and for which
optimal polarimeter observables can easily be extracted.
Once the full kinematics of the \PGt decay are reconstructed,
the well-known forms of optimal polarimeters in the one- and two-\PGp decay
can be used to estimate the \PGt spin orientation~\cite{Yumino:2022vqt}.
An example of the reconstructed polarimeters' longitudinal component
is shown in Fig.~\ref{fig:ewk_qcd_tauPolHiM:pol}.

This method has been tested using as input the true momenta of pions taken from the Monte Carlo, as well as
estimates of these quantities after detector reconstruction using the {\em Simulation \`{a} Grande Vitesse} (SGV)
package~\cite{Berggren:2012ar}, whose parameters are modelled on the expected performance of the ILD concept.
Further work is planned to examine the sensitivity of this analysis to detector performance.



\subsubsection*{Experimental challenges and prospects for BSM discovery in heavy-quark pair production above the Z-pole energy}


In Ref.~\cite{Irles:2024ipg} and here, the prospects for discovering benchmark GHU models {\cite{Agashe:2004rs,Medina:2007hz,Hosotani:2008tx,Funatsu:2014fda,Funatsu:2019xwr,Funatsu:2020haj,Funatsu:2021yrh,Funatsu:2023jng,Yamatsu:2023bde} using the forward-backwards asymmetry, \AFBq, and observables combining different running energy scenarios at future colliders are discussed. In a nutshell, these benchmark models are characterised by the following:

The $A$ models ($A_1$ and $A_2$) and $B$ models ($B_j^\pm$
with $j=1,2,3$ indicating the sign of the lepton bulk
masses) as described in Ref.~\cite{Irles:2024ipg}, are adopted as benchmark points: 
\newline
$A_1:\to m_{\PZ^{\prime}}=7.19$ \TeV;
$A_2:\to m_{\PZ^{\prime}}=8.52$ \TeV,\newline
$B_1^\pm:\to m_{\PZ^{\prime}}=10.2$ \TeV;
$B_2^\pm\to m_{\PZ^{\prime}}=14.9$ \TeV;
$B_3^\pm\to m_{\PZ^{\prime}}=19.6$ \TeV;\newline
where $\PZ^\prime$ is the first KK $\PZ$ boson. The masses of the first KK 
$\PGg$ and $\PZ_R$ bosons are similar to those of the first
KK $\PZ$ boson.

The precise reconstruction of the \AFBq with \PQq being \PQc- or \PQb-quarks poses several challenges. The natural way to address these challenges is using highly-performant flavour tagging algorithms and the usage of single-vs-double hemisphere tagging as discussed in \cref{sec:twof_Zpole:hem}, with the caveat of having in addition large contributions from other SM backgrounds, such as the \PZ radiative return events. The experimental input is based on detailed simulations of the ILD at centre-of-mass energies of 250 and 500 \GeV~ and extrapolated to 1000 \GeV. In contrast with other studies, the fully differential cross section $\text{d}\sigma/\text{d} \cos\theta$ reconstruction is studied. With the current layout of the ILD detector for ILC, the $\text{d}\sigma/\text{d} \cos\theta$ can be well reconstructed, with homogeneous reconstruction efficiency in $-0.9<\cos\theta<0.9$. Novel geometries or reconstruction algorithms should be considered to increase the acceptance. Similar detector performances and design challenges are expected for circular collider experiments, requiring careful optimization of the forward-regions to maximize the reconstruction of ISR photons and heavy quarks at low angles.

The studies demonstrate that a high level of statistical precision is attainable even using the double-tagging techniques.
In Ref.~\cite{Irles:2024ipg}, state-of-the-art LCFIPlus, which relies on \textit{traditional} Machine Learning (BDT) algorithms, was utilized for flavour tagging. Additionally, ILD, like most detector concepts for future Higgs Factories, offers the crucial capability of providing charged-kaon identification across a broad momentum spectrum using TPC information.
Experimental systematic uncertainties were found to be sub-dominant.

Figure \ref{fig:1} shows the expected uncertainties for the reconstruction of \AFBb and \AFBc using full-simulation of signal and background events in the ILD model of Ref.~\cite{ILD:2020qve}, plus two additional scenarios on the detector model and reconstruction tools. The study was done for the available samples of ILC250 and ILC500.
Studies performed with the ILD model adapted to circular colliders operating up to the $\PQt\PAQt$ threshold are to be performed.

\begin{figure}[!ht]
    \centering
            \includegraphics[width=0.4\textwidth, trim={0 0 0 7.5cm},clip]{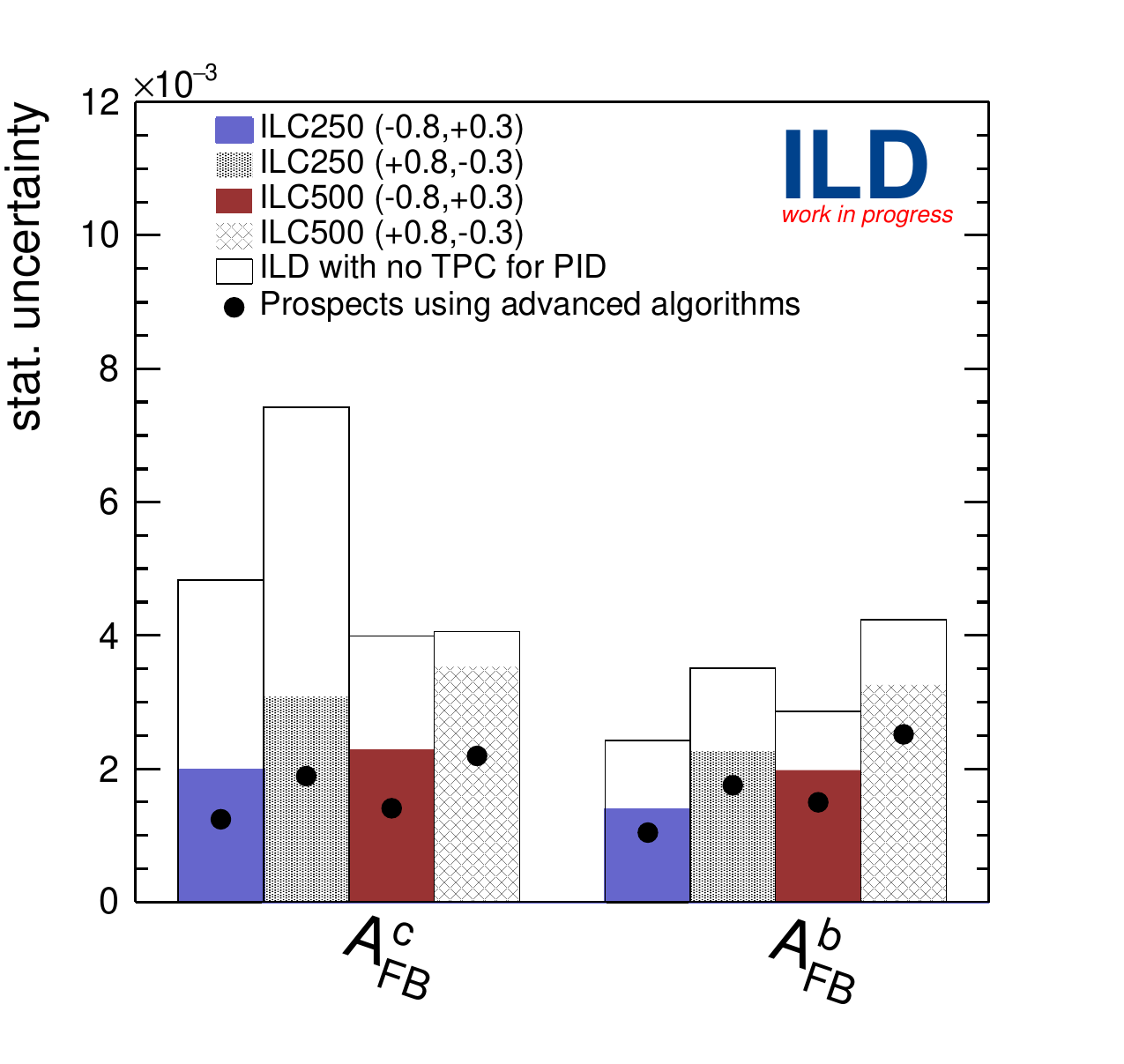} 
    \caption{Estimated statistical uncertainties on \AFBc and \AFBb using ILD full simulation and reconstruction at ILC250 and ILC500. Two alternative scenarios of ILD model and reconstruction are shown.}
    \label{fig:1}
\end{figure} 

Figure \ref{fig:2} discusses the sensitivity to BSM models in different possible running scenarios, including scenarios at the $\PZ\PH$ threshold with no longitudinal beam polarisation. For the latter case, it shows that the gap in sensitivity will be filled with a factor two of integrated luminosity.

 Finally, three scenarios of different reconstruction capabilities are compared in Figs. \ref{fig:1} and \ref{fig:2}: a detector without charged-hadron particle identification capabilities (\textit{i.e.,} a hypothetical version of ILD without a TPC-based central tracker); a state-of-the-art ILD detector with reconstruction tools described in Ref.~\cite{ILD:2020qve} and cluster counting for the charged-hadron particle identification using the TPC; and an improved scenario after applying modern reconstruction techniques based on advanced Artificial Intelligence models. These latter prospects are obtained by simply extrapolating from the \textit{baseline} case considering the expected improvements on flavour-tagging discussed in Ref. \cite{Tagami:2024gtc}. The same improvement is assumed for the charged-hadron particle identification. It is important to remark on the importance of charged-hadron particle identification (PID), especially for the \PQc-quark case at ILC250, although the benchmark models discussed here show small sensitivity for ILC250.

Considering only the scenario of running up to the $\PZ\PH$ threshold, $m_{\PZ}^\prime\simeq10$ TeV can be reached at the evidence level. These limits could of course be improved by including more fermion final states.
This study also highlights the importance of precise vertexing, high acceptance forward tracker detectors, efficient charged-hadron particle identification capabilities, and high-energy reach, above the \PZ pole. The addition of longitudinal beam polarisation would also play a positive impact on the sensitivity.

\begin{figure}[!ht]
    \centering
    \begin{tabular}{cc}
        \includegraphics[width=0.4\textwidth, trim={0 0 0 7.5cm},clip]{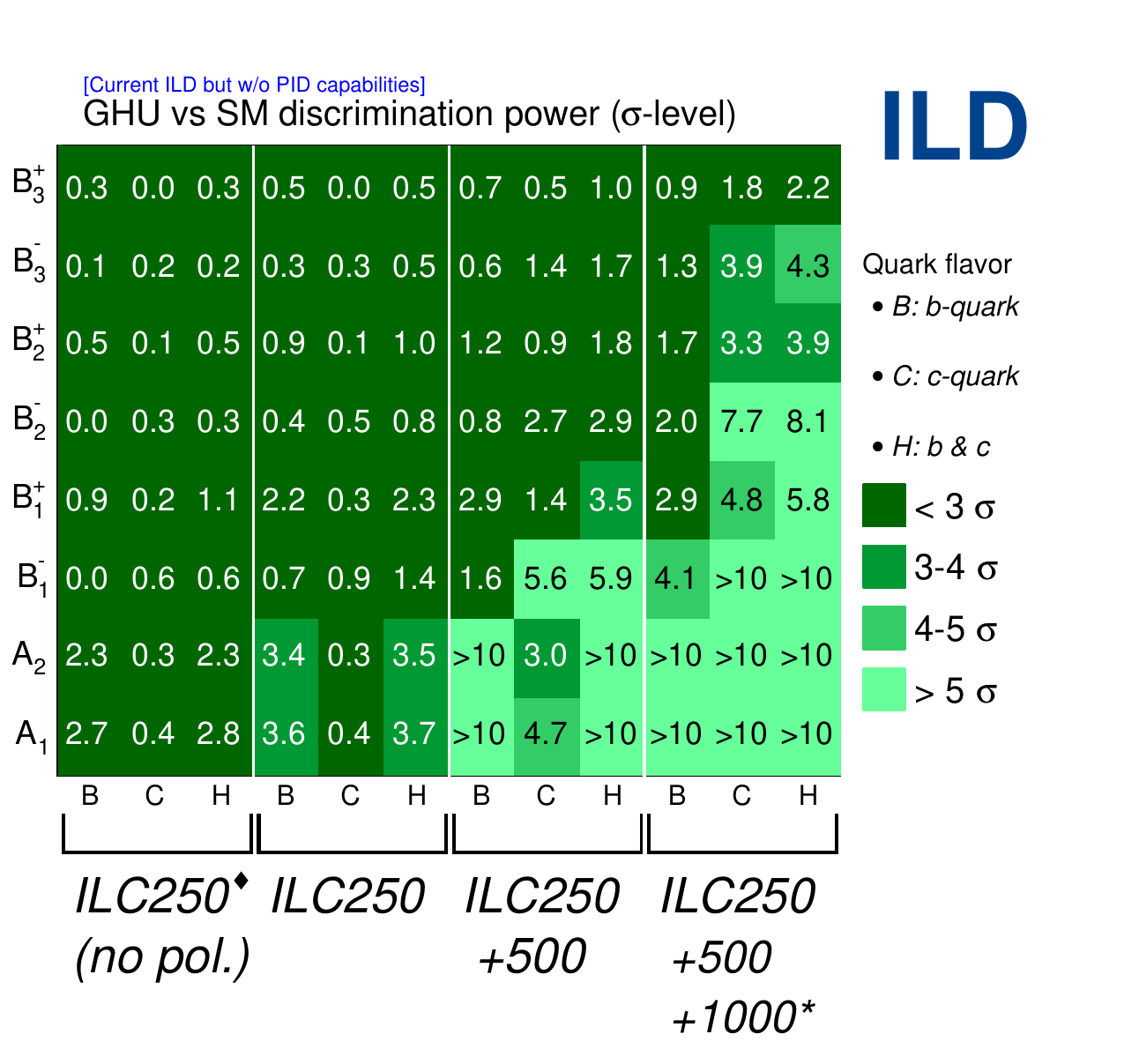} &
        \includegraphics[width=0.4\textwidth, trim={0 0 0 7.5cm},clip]{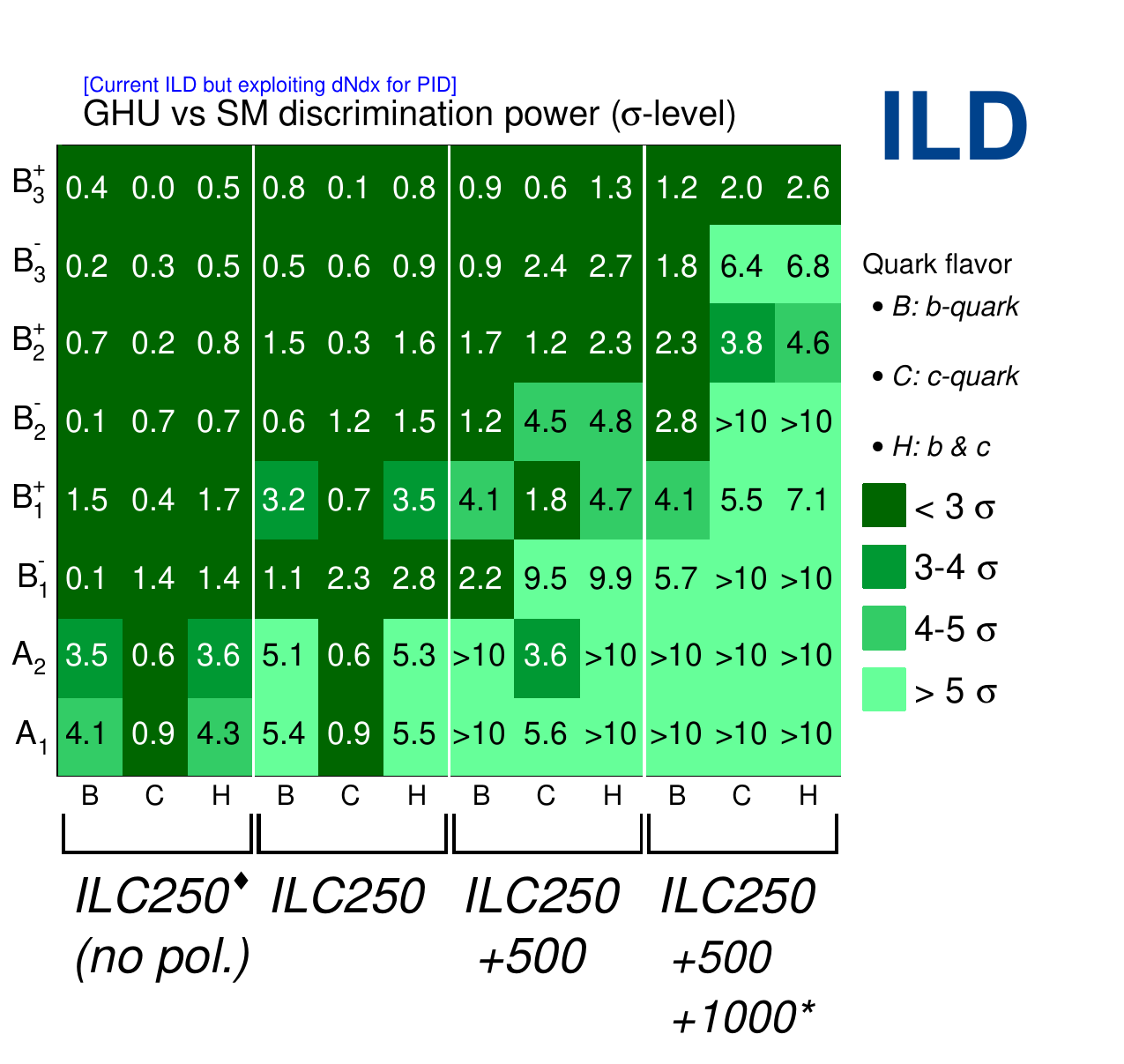} \\
    \end{tabular}    
    \includegraphics[width=0.4\textwidth, trim={0 0 0 7.5cm},clip]{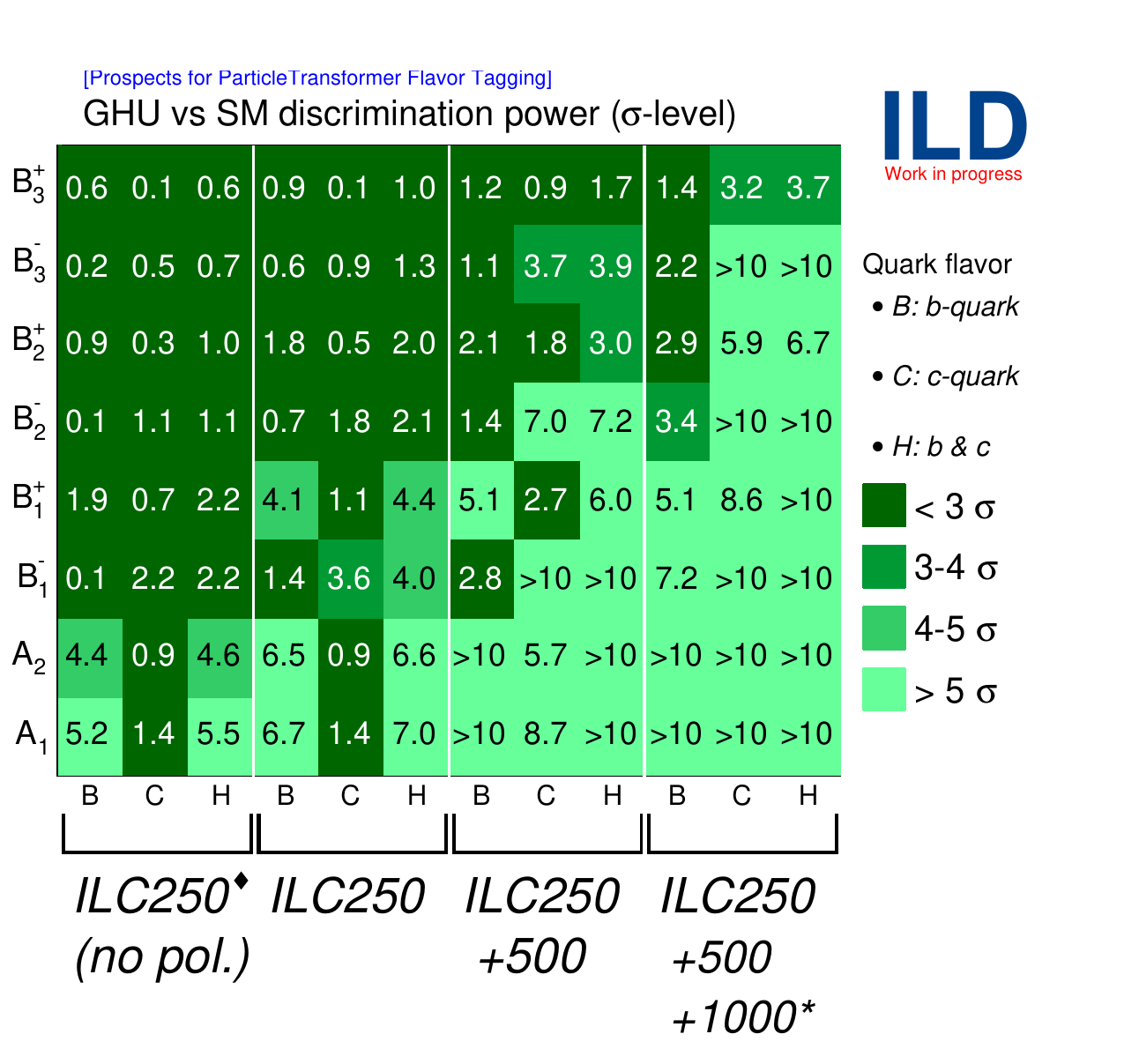} \\

    \caption{Statistical discrimination power between the GHU models described in the text and the SM. Different running scenarios of ILC are compared: ILC250$^{\blacklozenge}(no~pol.)$ (hypothetical case with no beam polarisation and 2000 \fbinv of integrated luminosity), ILC250 (2000 \fbinv), ILC500 (4000 \fbinv), and ILC1000* (8000 \fbinv, not using full simulation studies but extrapolations of uncertainties from ILC500). }
    \label{fig:2}
\end{figure}


\subsection{Other Z-boson and neutrino interactions \label{sec:otherZ}}

\subsubsection{Electron couplings from transversely polarised beams}
\editors{Xin-Kai Wen, Bin Yan, Zhite Yu, C.-P. Yuan}

The investigation of electroweak dipole moments is essential for exploring the internal structure of particles. 
These moments can be effectively described by a subset of dimension-six operators in the Standard Model Effective Field Theory (SMEFT)~\cite{Buchmuller:1985jz,Grzadkowski:2010es}, characterised by flipping fermion chirality. 
This property causes dipole operators to contribute to the cross section either quadratically at $\mathcal{O}(1/\Lambda^4)$ or in association with the fermion mass at $\mathcal{O}(1/\Lambda^2)$, where $\Lambda$ denotes the scale of new physics (NP) beyond the Standard Model (SM). 
Consequently, constraining the Wilson coefficients of light-fermion dipole operators has proven extremely challenging in current global SMEFT analyses.
To address this issue, we propose incorporating polarised observables associated with the transverse spin of light fermions into such analyses. 
As an interference effect between different helicity states, the transverse spin compensates for the helicity-flipping effect from the dipole operators, allowing them to interfere with SM interactions at $\mathcal{O}(1/\Lambda^2)$ without mass suppression or dependence on additional NP effects.
Therefore, this approach will greatly enhance the sensitivity power. 

Although the same approach applies to other generations and can be extended to quark dipole moments~\cite{Boughezal:2023ooo,Wang:2024zns,Wen:2024cfu}, we illustrate this idea for the first-generation leptonic dipole operators in the SMEFT~\cite{Wen:2023xxc}, 
\begin{equation}
	\mathcal{L}_{\text{eff}}
	= -\frac{1}{\sqrt{2}} \bar{\ell}_L \sigma^{\mu\nu}
		\left( g_1 \Gamma_B^{e\P} B_{\mu\nu} + g_2 \Gamma_W^{\Pe} \sigma^a W_{\mu\nu}^a \right) \frac{H}{v^2} e_R 
		+ {\text{h.c.}}
\end{equation}
where we redefine the electron dipole couplings $\Gamma_{\PGg}^{\Pe}=\Gamma_W^e-\Gamma_B^{\Pe}$ and $\Gamma_{\PZ}^{\Pe}=c_W^2\Gamma_W^{\Pe}+s_W^{\P2}\Gamma_B^{\Pe}$ in the mass eigenstate basis.
The imaginary parts of these couplings, if nonzero, violate parity and $CP$ symmetries.

The dipole couplings $\Gamma_{\PZ,\PGg}^{\Pe}$ are most suitably probed at an $\Pem \Pep$ collider with transversely polarised beams, through measurements of the azimuthal distributions in the production of final state $i = \PZ\PH$, $\PZ\PGg$, $\PWp\PWm$, or $\PGmp\PGmm$.
The transverse spin $\bm{s}_T = b_T (\cos\phi_0, \sin\phi_0)$ or $\bar{\bm{s}}_T = \bar{b}_T (\cos\bar{\phi}_0, -\sin\bar{\phi}_0)$ of the electron or positron, respectively, enters the cross section through off-diagonal elements of the density matrices,
\begin{equation}
	\Sigma^i(\phi,\bm{s},\bar{\bm{s}})
	= 	\rho_{\alpha_1\alpha_1^\prime}({\bm{s}}) 
	\rho_{\alpha_2\alpha_2^\prime}(\bar{{\bm{s}}})
	\mathcal{M}^i_{\alpha_1\alpha_2}(\phi)
	\mathcal{M}_{\alpha_1^\prime\alpha_2^\prime}^{i *}(\phi)
	=\Sigma^i_{UU} + b_T\Sigma^i_{TU}(\phi) + \bar{b}_T\Sigma^i_{UT}(\phi) + b_T \bar{b}_T \Sigma^i_{TT}(\phi), 
\end{equation}
where $\rho(\bm{s})$ is the fermion spin density matrix, and $b_T$~$(\bar{b}_T) > 0$ denotes the magnitude of the electron's (positron's) transverse spin.
$\mathcal{M}^i_{\alpha_1\alpha_2}(\phi)$ is the helicity amplitude of $\Pem_{\alpha_1} \Pep_{\alpha_2} \to i$, with the azimuthal angle $\phi$ defined for either particle in $i$.
Using rotation symmetry, one has $\mathcal{M}_{\alpha_1\alpha_2}(\phi) = e^{i(\alpha_1-\alpha_2)\phi} \mathcal{A}_{\alpha_1\alpha_2}$, where $\mathcal{A}_{\alpha_1\alpha_2} = \mathcal{M}_{\alpha_1\alpha_2}(0)$, independent of $\phi$.
Then the azimuthal dependence of the single-spin quantities $\Sigma_{TU, UT}$ is, 
\begin{equation}
	\Sigma_{TU}
		= \frac{1}{2} {\operatorname{Re}} \left[ e^{i (\phi-\phi_0) } \left(\mathcal{A}_{+-} \mathcal{A}_{--}^* + \mathcal{A}_{++} \mathcal{A}_{-+}^* \right)\right], 
	\qquad
	\Sigma_{UT} 
		= \frac{1}{2} {\operatorname{Re}} \left[ e^{i (\phi-\bar{\phi}_0) } \left(\mathcal{A}_{+-} \mathcal{A}_{++}^* + \mathcal{A}_{--} \mathcal{A}_{-+}^*\right) \right],
\end{equation}
both yielding characteristic $\cos\phi$ and $\sin\phi$ distributions that can be observed to probe the dipole couplings $\Gamma_{\PZ,\PGg}^{\Pe}$, which govern the chirality-flipping amplitudes $\mathcal{A}_{\pm\pm}$.

For simplicity, we integrate over the polar angle and consider only the $\phi$-dependent differential cross section,
\begin{equation}
	\frac{2\pi \, d\sigma^i}{\sigma^i \, d\phi}
	= 1 + A_R^i(b_T, \bar{b}_T; \operatorname{Re} \Gamma_{\PZ,\PGg}^{\Pe}) \cos\phi 
		+ A_I^i(b_T, \bar{b}_T; \operatorname{Im} \Gamma_{\PZ,\PGg}^{\Pe}) \sin\phi + b_T \, \bar{b}_T \, B^i \cos2\phi
	+ \mathcal{O}(1/\Lambda^4),
\end{equation}
where the coefficients $A_{R,I}^i$ depend linearly on $b_T$, $\bar{b}_T$ and the real ($R$) or imaginary ($I$) parts of the dipole couplings $\Gamma_{\PZ,\PGg}^{\Pe}$, while $B^i$ is given only by the SM.
It is convenient to extract them by constructing the azimuthal asymmetry observables, dubbed as single transverse spin asymmetries (SSAs), 
\begin{equation}
	A_{LR}^i
	\,= \frac{\sigma^i(\cos\phi>0) - \sigma^i(\cos\phi<0)}{\sigma^i(\cos\phi>0) + \sigma^i(\cos\phi<0)}
	= \frac{2}{\pi} A_R^i,
	\quad
	A_{UD}^i
	\,= \frac{\sigma^i(\sin\phi>0) - \sigma^i(\sin\phi<0)}{\sigma^i(\sin\phi>0) + \sigma^i(\sin\phi<0)}
	= \frac{2}{\pi} A_I^i.
\end{equation}
By measuring these observables for the various channels of $i$, one can constrain both the real and imaginary parts of the dipole couplings with linear sensitivity, respectively, as well as the potential $CP$ violation.
\begin{figure}[h]
	\centering
	\includegraphics[scale=0.32]{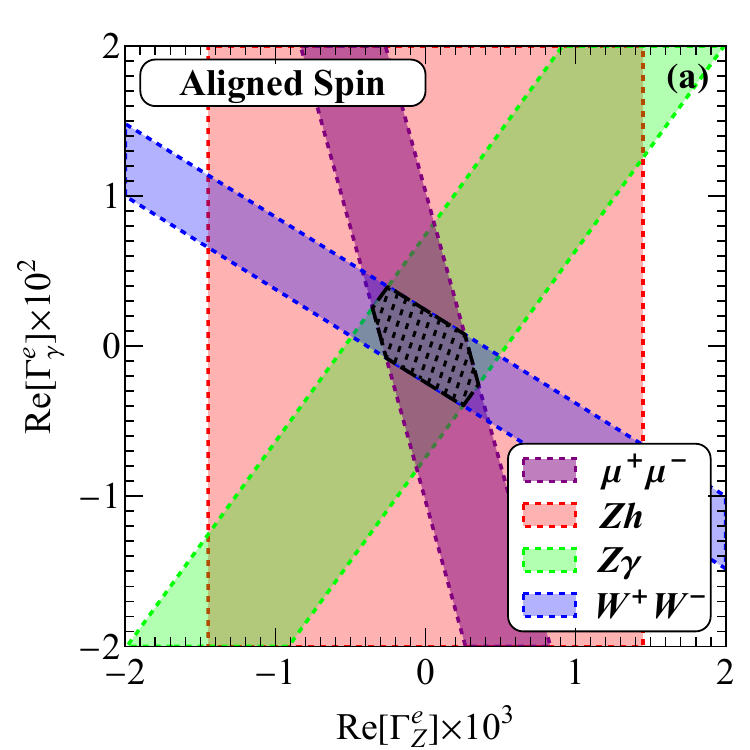}
	\includegraphics[scale=0.32]{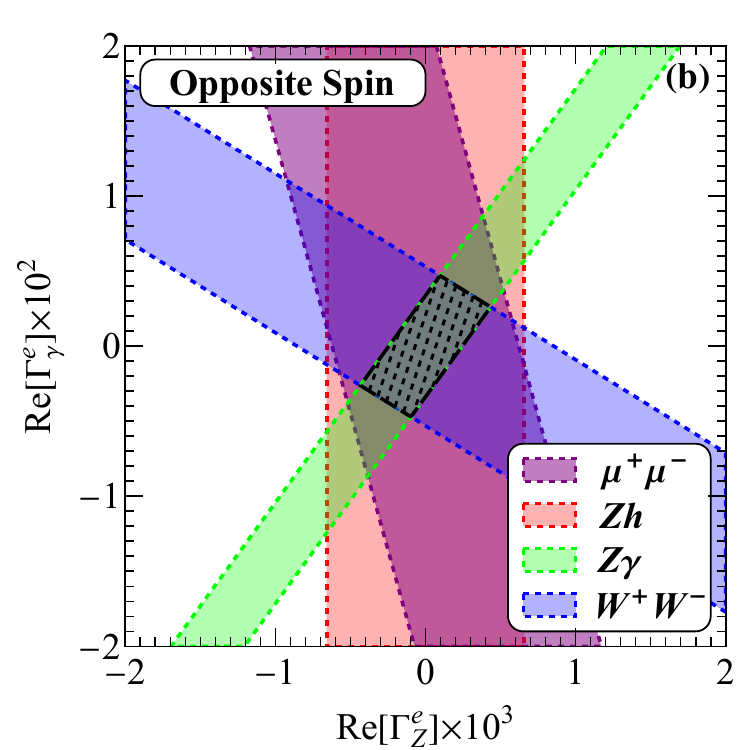}
	\includegraphics[scale=0.32]{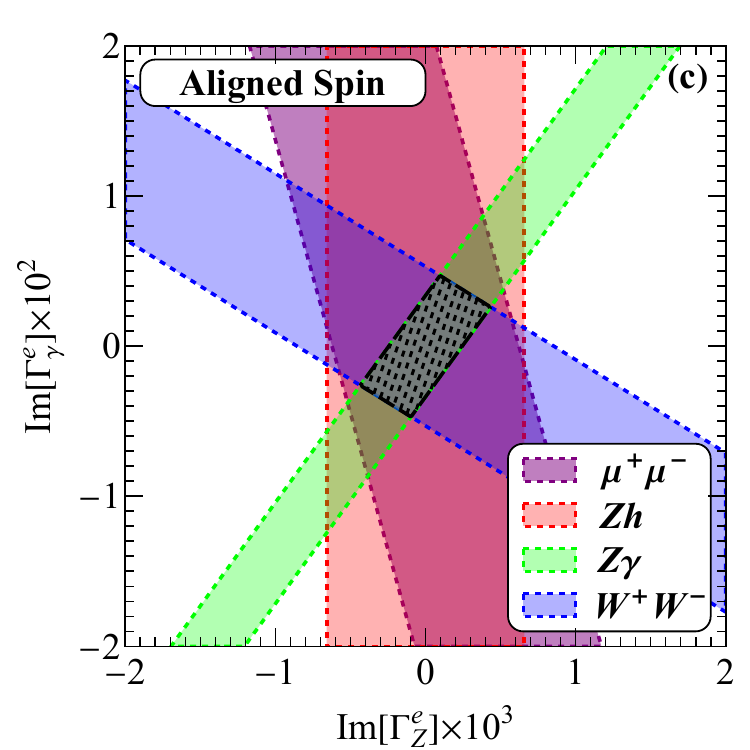}
	\includegraphics[scale=0.32]{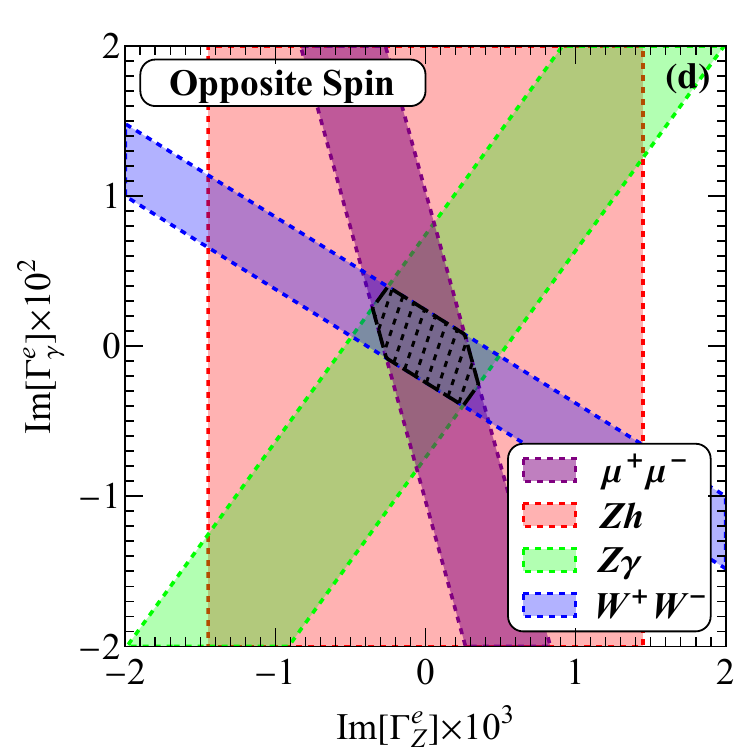}
	\caption{Expected constraints from the spin asymmetry observables on the electron dipole couplings $\Gamma_{\PZ,\PGg}^{\Pe}$.
	}
	\label{Fig:SSALimit}
\end{figure}

We choose the benchmark centre-of-mass energy at $\sqrt{s}=\SI{250}{\giga\electronvolt}$ with $(b_T, \, \bar{b}_T) = (0.8,\, 0.3)$ and integrated luminosity of $\mathcal{L}=\SI{5}{\per\atto\barn}$.
The beam spin directions can be optimized to maximize the sensitivity, for which we study two scenarios: the aligned spin setup $(\phi_0,\bar{\phi}_0)=(0,0)$ and the opposite spin setup $(\phi_0,\bar{\phi}_0)=(0,\pi)$.
Assuming the experimental measurements centre around the SM prediction, i.e., $\langle A_{LR, \, UD}^i \rangle = 0$, we present in \cref{Fig:SSALimit} the expected constraints on the dipole couplings at 68\% confidence level.
Combining all the four processes allows $\Gamma_{\PZ}^{\Pe}$ and $\Gamma_{\PGg}^{\Pe}$ to be separated, and yields the same constraints on their real and imaginary parts, with typical upper limits of $\mathcal{O}(0.01\%)$ for $\Gamma_{\PZ}^{\Pe}$ and $\mathcal{O}(0.1\%)$ for $\Gamma_{\PGg}^{\Pe}$. 
These limits are significantly stronger compared to other methods applied in Drell--Yan and $\PZ$-pole processes, which usually constrain $|\Gamma_{\PZ,\PGg}^{\Pe}|$ to around $\mathcal{O}(1\%)$ even when considering one operator at a time.

\subsubsection{Flavour changing 
Z-boson decays}
\editors{Arman Korajac, Manuel Szewc, Michele Tammaro}

The SM predicts the branching ratio of $\PZ\to \PQb\PQs,\PQb\PQd,\PQc\PQu$ to be strongly suppressed by the GIM mechanism, while the lepton flavour violating processes, $\PZ\to \Pe\PGm,\Pe\PGt,\PGm\PGt$, are forbidden. Precise measurements of these decay channels are then highly sensitive to possible New Physics contributions. We parametrise the latter in terms of effective FCNC couplings to the $\PZ$ boson,
\begin{equation}\label{eq:ZbsLagrangianPheno}
\mathcal{L} \supset \, g_{ij}^L(\PAQq_{i,L} \gamma_\mu \PQq_{j,L})\PZ^\mu + g_{ij}^R(\PAQq_{i,R} \gamma_\mu \PQq_{j,R})\PZ^\mu \,, 
\end{equation}
where $ij=\PQb\PQs,\PQb\PQd,\PQc\PQu$, with analogous definitions for lepton flavour-changing couplings. This Lagrangian can be obtained as the low-energy realization of various extensions of the SM, e.g.\ the addition of vector-like quarks~\cite{Fajfer:2013wca,Barducci:2017ioq}.

In their operational phase at the centre-of-mass energy $\sqrt{s} =  \SI{91}{\giga\electronvolt}$, corresponding to the $\PZ$ pole mass, both CEPC and FCC-ee are expecteded to produce $ N_{\PZ} = \SI{5d12} \, \PZ$ bosons. Considering this impressive statistical yield, the reach for quark decays will be mostly limited by the systematic uncertainties on the tagger performances, while for lepton decays it will depend on the particle identification capability of the detector.

\paragraph{Quarks:} Updated computations yield~\cite{PhysRevD.22.214, PhysRevD.27.570, PhysRevD.27.579,Kamenik:2023hvi}
\begin{equation}
{\cal B}(\PZ\to \PQb\PQs) = \left( 4.2 \pm 0.7 \right)\times10^{-8}\,, \quad {\cal B}(\PZ\to \PQb\PQd) = \left( 1.8 \pm 0.3 \right)\times10^{-9}\,, \quad {\cal B}(\PZ\to \PQc\PQu) = \left( 1.4 \pm 0.2 \right) \times 10^{-18}\,,
\end{equation}
where ${\cal B}(\PZ\to \PQq_i \PQq_j) = {\cal B}(\PZ\to \PAQq_i \PQq_j) + {\cal B}(\PZ\to \PQq_i \PAQq_j)$. The reported uncertainties contain both the theoretical uncertainties from numerical inputs and the leading two-loop corrections to the amplitudes. 

While existing direct limits on the non-standard hadronic decays of the $\PZ$ follow from the agreement of the measurement and the SM prediction for the $\PZ$ hadronic width~\cite{OPAL:2000ufp}, giving $\mathcal{B} (\PZ \to \PQq\PQq') < 2.9 \times 10^{-3}$ at $95 \, \% \,\mathrm{CL}$, the low-energy constraints are much more powerful. The effective couplings in \cref{eq:ZbsLagrangianPheno} will contribute to a wide range of observables, as D and B meson mixings, leptonic and semi-leptonic decays. In particular, for $\PZ\to \PQb\PQd$ and $\PZ\to \PQb\PQs$, these transitions essentially constrain the New Physics effect in flavour changing processes to be well below the SM value; thus the limit on direct observations of these branching ratios is at the moment the SM value itself, within uncertainties. For the $\PZ\to \PQu\PQc$ decay instead, we obtain ${\cal B}(\PZ\to \PQu\PQc)< 4 \times 10^{-7}$ from the flavour global fit. Additional bounds, although much weaker, can be obtained by rare $\PZ$ decays to a photon and a vector meson, $\mathrm{\PZ \to VM^0+\PGg}$~\cite{dEnterria:2023wjq,ATLAS:2024dpw,CMS:2024tgj}.

At the FCC-ee, the total detector acceptance for hadronic $\PZ$ decays is estimated to be $\sim0.994$~\cite{Agapov:2022bhm}.   Coupled with the accurate mass and energy reconstruction capabilities, this allows most of the background to be removed such that only other SM hadronic decays, $\PZ \to \PQq\PQq$, need to be considered as possible noise signal, where one of the two jets is mistagged. We therefore employ an identical analysis to the case of Higgs flavour-changing decays~\cite{ATLAS:2022ers,Faroughy:2022dyq,CMS:2020vac,Kamenik:2023hvi}, including the same jet-flavour tagger performances (see \cref{eq:taggerperform:Higgs} and the discussion in \cref{subsubsec:projSensitivityHiggs}). Analogously, backgrounds from processes where a double mistag is needed to provide a fake signal can also be safely disregarded.
We find that in order to probe the SM value for ${\cal B}(\PZ\to \PQb\PQs)$, we would need very performant taggers with very low systematic uncertainties (<0.1\%). Similar results hold for $\PZ \to \PQb\PQd$ and $\PZ \to \PQu\PQc$. For $0.1\%$ systematic uncertainties, the expected reaches are
\begin{equation}
{\cal B}(\PZ\to \PQb\PQs)<7.33\times10^{-6}\,, \quad {\cal B}(\PZ\to \PQc\PQu)<4.06\times10^{-4}\,.
\end{equation}
One could in principle apply a similar metholodogy to the $\PQb\PQd$ channel. However, there is currently not a reliable $\PQd$-jet tagger. Employing only a $\PQb$-tagger we obtain the weaker constraint ${\cal B}(\PZ\to \PQb\PQd)<2.45\times10^{-4}$.

Thus we conclude that, for realistic efficiencies and uncertainties, indirect probes set BSM bounds that are much stronger than the projected FCC-ee reach.

\paragraph{Leptons:} As lepton flavour changing decays of the $\PZ$ are forbidden in the SM, any clear signal would point to the existence of New Physics. 
The current direct constraints have been derived from dedicated searches conducted at the LHC by ATLAS~\cite{ATLAS:2021bdj, ATLAS:2022uhq}, resulting in the following limits at 95~\%~CL:
\begin{equation}
	{\cal B}(\PZ\to \Pe\PGt) < \SI{5.0d-6} \,, \quad {\cal B}(\PZ\to \PGm\PGt) < \SI{6.5d-6} \,, \quad {\cal B}(\PZ\to \Pe\PGm) < \SI{2.62d-7}\,.
\end{equation}

Low energy probes of charged lepton flavour violation, as muon and tau exotic decays ($\PGm\to3\Pe$, $\PGm\to\Pe\PGg$, $\PGt\to\PGr\PGm$ etc.), will pose indirect limits on the $\PZ$ boson off-diagonal couplings~\cite{Calibbi:2021pyh}. For the $\PGm\Pe$ case, these bounds can reach ${\cal B}(\PZ\to \Pe\PGm) \lesssim \num{e-13}$, while in the $\PGt$ sector, we have typically ${\cal B}(\PZ\to \PGt\PGm/\PGt\Pe) \lesssim \num{e-7}$.  

The projected bounds on branching ratios at future circular lepton colliders are \cite{Dam:2018rfz}:
\begin{equation}
	{\cal B}(\PZ\to \Pe\PGt) \lesssim (\SI{d-8} - \SI{d-10})\,, \quad {\cal B}(\PZ\to \PGm\PGt) \lesssim \SI{d-9} \,, \quad {\cal B}(\PZ\to \Pe\PGm) \lesssim \SI{d-9}\,,
\end{equation}
where the range of values for $\PZ \to \PGm \Pe$ reflects whether or not particle identification via \dedx will be available. As expected, these will improve the limits on decays involving the $\PGt$ lepton, while for $\PGm\Pe$ the low energy bounds dominate. 
This would lead to the following projected bounds on the lepton flavour-violating $\PZ$ couplings (see \cref{eq:ZbsLagrangianPheno}):

\begin{equation}
	|g^{L,R}_{\Pe\PGt}|\lesssim \SI{3.2d-3}\,, \quad |g^{L,R}_{\PGm\PGt}|\lesssim \SI{3.7d-3}\,, \quad|g^{L,R}_{\Pe\PGm}|\lesssim\SI{7.3d-4}\,,
\end{equation}
assuming only one coupling being turned on at time. More detailed studies are warranted, as the very large samples of $\PGt$ decays at FCC-ee will allow for significantly improved tests of LFU \cite{FCC:2018byv}.

\subsubsection{
Z-boson decays in models with right-handed neutrinos}
\editor{Mikael Chala}
%
The see-saw type I model~\cite{Minkowski:1977sc,Yanagida:1979as,Mohapatra:1979ia} is arguably one of the most appealing explanations for neutrino masses. In its simplest incarnation, it extends the SM with right-handed neutrinos $N$ of very large mass $m_N$, so that a natural mixing between $N$ and the SM neutrinos induces, after EWSB, masses for the latter of order $m_\nu \sim v^2/m_N$, with $v$ the Higgs VEV.
However, less simple (and perhaps more realistic) realizations of this model allow $N$ to be at the EW scale itself~\cite{Pilaftsis:1991ug,Borzumati:2000mc}, and generally involve further heavier particles~\cite{Dev:2009aw,Dev:2016dja}. If both conditions hold, then the physics at current accessible energies is precisely described by the EFT of the SM extended with $N$, also known as NSMEFT~\cite{delAguila:2008ir,Aparici:2009fh,Bhattacharya:2015vja}. The corresponding Lagrangian reads:
\begin{equation}
    L_\text{NSMEFT} = L_{\text{SM}+N} + \frac{1}{\Lambda^2}\sum_i O_i + \mathcal{O}\left(\frac{1}{\Lambda^3}\right)\,,
\end{equation}
where $\Lambda$ stands for the cut-off of the EFT. The operators $O_i$ span a basis of dimension-6 interactions; they can be found in Ref.~\cite{Liao:2016qyd}. This theory is well understood theoretically, including its quantum structure. It has been renormalised at one loop in a series of works~\cite{Chala:2020pbn,Datta:2020ocb,Datta:2021akg,Ardu:2024tzb}, and the matching on to its low-energy version (in which the top quark, the Higgs and the $Z$ and $W$ bosons are integrated out) is known to tree level~\cite{Chala:2020vqp}.

Under some mild assumptions~\cite{Biekotter:2020tbd}, it can be argued that, for $1\lesssim m_N\lesssim100$ \SI{}{\giga\electronvolt}, and not too a large cut-off ($\Lambda\lesssim 10$ \SI{}{\tera\electronvolt}), $N$ decay mostly via loop operators~\footnote{These are of the form $O_{NV} = \overline{L}\sigma_{\mu\nu} N\tilde{\Phi} V^{\mu\nu}$ with $L$ and $\Phi$ representing the lepton and Higgs doublets, respectively, and with $V$ being either a photon or a $Z$ boson.} following $N\to \\PGn\\PGg$  and, to a lesser extent, via tree-level four-fermion interactions into $\Plp\Plm\PGn$ or into quarks. The corresponding widths are of the order $10^{-15} m_N^3/\SI{}{\giga\electronvolt}^2$ and $10^{-20} m_N^5/\SI{}{\giga\electronvolt}^4$; respectively.

Several signatures of the production and subsequent decay of $N$ have been explored at colliders, including rare top decays $\PQt\to N \PQb\Pl, N\to \nu\PGg$~\cite{Alcaide:2019pnf,Biekotter:2020tbd}, rare Higgs decays $\PH\to N\nu, N\to\nu\PGg$ as well as a variety of meson decays, long-lived signals, and others~\cite{Duarte:2016caz,Duarte:2019rzs,Cottin:2021lzz,Beltran:2021hpq,Barducci:2022hll,Barducci:2022gdv,Barducci:2023hzo,Beltran:2023ksw}. We also find rare $\PZ$ decays of the form $\PZ\to N\PGn$ and $\PZ\to NN$. Depending on the $N$ decay mode, they give rise to a variety of signals: $\PZ\to \\PGn\\PGn\PGg$, $\PZ\to \PGn\PGn\PGg\PGg$, $\PZ\to\PGn\PGn\Plp\Plm$, $\PZ\to \PGn\PGn \Plp\Plm\Plp\Plm$ as well as channels with quarks; see Fig.~\ref{fig:rarezdecays} for example diagrams. To the best of our knowledge, none of these modes has been studied either phenomenologically or experimentally. In what follows, we elaborate on the tremendous sensitivity that an \epem collider, operating at the Z pole and producing about $5\times 10^{12}$ Z bosons, has to these processes. To this aim, we provide rough numbers for the first three cases. In all cases, we have that $\Delta\Gamma(\PZ)\sim \frac{m_{\PZ}^3 v^2}{\pi \Lambda^4}$. Assuming $\mathcal{O}(1)$ couplings in the EFT and a $\Lambda = 1$ TeV, we find the expected branching ratio $\Delta\mathcal{B}(\PZ)\sim 10^{-3}$.

\begin{figure}[t]
\begin{center}
\includegraphics[width=0.9\columnwidth]{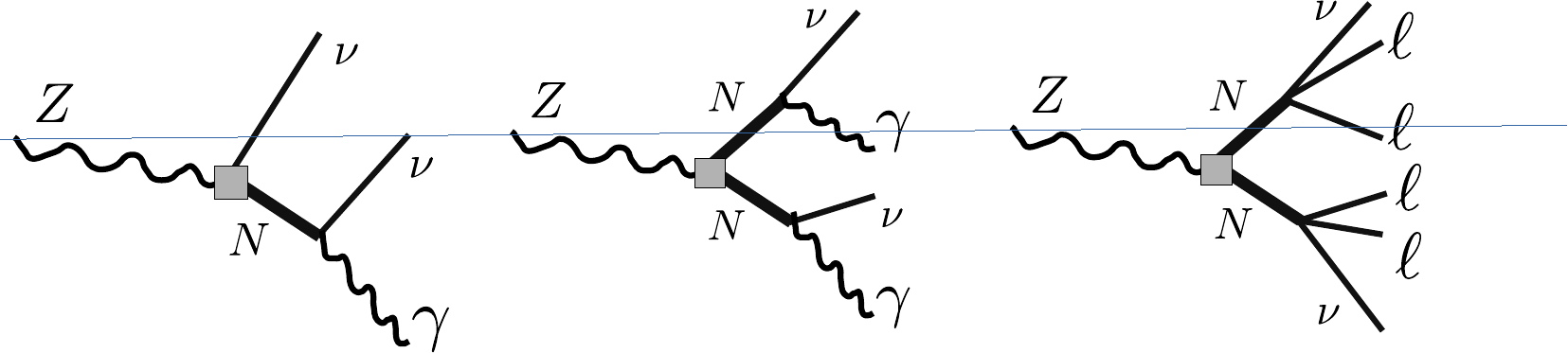}
\end{center}
\caption{Example diagrams of $\PZ$ rare decays mediated by a neutral lepton $N$.}\label{fig:rarezdecays}
\end{figure}

For the first mode, the current experimental bound on the corresponding branching ratio, ensuing from monophoton events in $\Pep\Pem$ collisions at the $\PZ$ pole by the L3 collaboration~\cite{L3:1997exg}, is $\mathcal{B}(\PZ\to\PGn\PGn\PGg)<3.2\times 10^{-6}$. Following this reference, and being rather conservative, we assume an event selection efficiency of $\sim 0.2$ (compared with $\sim 0.85$ in the aforementioned work) and require $10$ observed events. This implies an expected bound at a future $\Pep\Pem$ facility of $\mathcal{B}(\PZ\to\\PGn\PGn\PGg)\lesssim 10^{-11}$, which amounts to testing $\Lambda\sim 100$ TeV.

For the second mode, the current experimental bound on the corresponding branching ratio is $\mathcal{B}(\PZ\to\PGn\PGn\PGg\PGg) < 3.1\times 10^{-6}$. It follows from searches in $\Pep\Pem$ collisions at LEP by the OPAL collaboration in events with $m_{\PGg\PGg}=(60\pm 5)$ GeV~\cite{OPAL:1993ezs}. Following the same reasoning as for the previous channel (in this case though, the event selection efficiency of the original analysis is $\sim 0.56$), we obtain also $\mathcal{B}(\PZ\to\\PGn\PGn\PGg)\lesssim 10^{-11}$ and so again $\Lambda\sim 100$ TeV. We find no reason to think that different diphoton mass windows could affect the bound significantly.

For the third mode, the current experimental bound on the corresponding branching ratio can be inferred to be $\mathcal{B}(\PZ\to\PGn\PGn\Plp\Plm\Plp\Plm)\lesssim 6\times 10^{-7}$ from current measurements of $\mathcal{B}(\PZ\to\Plp\Plm\Plp\Plm) = (4.41\pm0.30)\times 10^{-6}$~\cite{ATLAS:2021kog}. (For comparison, the SM prediction is $(4.50\pm0.01)\times 10^{-6}$~\cite{ATLAS:2014jlg}.) It is hard to extrapolate from this bound, derived from analysis of events produced in a hadron facility, to a future search at a high-energy lepton collider. Still, since the background can be assumed to be negligible to good approximation, it is not unreasonable to expect the corresponding sensitivity to be as strong as in the previous channels.

\subsubsection{Four-fermion interactions with neutrinos}
\editors{Yu Zhang, Jaijun Liao}

 Since the explanation of neutrino oscillations requires the neutrino masses to be nonzero, the observation of neutrino oscillations provides a clear evidence of new physics beyond the Standard Model (SM).
A model-independent way of studying new physics in neutrino experiments was given in the effective field theory framework of nonstandard neutrino interactions (NSI);  for reviews see Refs.~\cite{Ohlsson:2012kf, Miranda:2015dra, Farzan:2017xzy}. In this framework, NSI are typically described by dimension-six four-fermion operators.
The Lagrangian of neutral current (NC) NSI with electrons can be written as, 
\begin{eqnarray}
\label{eq:NSIee}
\mathcal{L}_{\text{NSI}}^{\text NC,e}
&=&-2\sqrt{2}G_F\epsilon_{\alpha\beta}^{eL}(\bar{\nu}_\alpha\gamma^\mu P_L \nu_\beta) (\bar{e}\gamma_\mu P_L e)
-2\sqrt{2}G_F\epsilon_{\alpha\beta}^{eR}(\bar{\nu}_\alpha\gamma^\mu P_L \nu_\beta) (\bar{e}\gamma_\mu P_R e)
\\
&=&-\sqrt{2}G_F\epsilon_{\alpha\beta}^{eV}(\bar{\nu}_\alpha\gamma^\mu P_L \nu_\beta) (\bar{e}\gamma_\mu e)
+\sqrt{2}G_F\epsilon_{\alpha\beta}^{eA}(\bar{\nu}_\alpha\gamma^\mu P_L \nu_\beta) (\bar{e}\gamma_\mu \gamma^5 e)\,,
\end{eqnarray}
where    
\begin{equation}
\label{eq:lr2va}
\epsilon_{\alpha\beta}^{eV} \equiv \epsilon_{\alpha\beta}^{eL} + \epsilon_{\alpha\beta}^{eR}\,, 
\quad \epsilon_{\alpha\beta}^{eA} \equiv \epsilon_{\alpha\beta}^{eL} - \epsilon_{\alpha\beta}^{eR}\,,
\end{equation}
with $\alpha$, $\beta$ labeling  the lepton flavours ($\Pe, \PGm, \PGt$), and
$\epsilon^{eV}_{\alpha\beta}$ ($\epsilon^{eA}_{\alpha\beta}$) describing  the
strength of the new vector (axial-vector) interactions between electrons.

NSI with electrons will lead to new contributions to the monophoton process $\epem \to \PGn \PAGn \PGg$ at electron colliders. An early study of NSI with electrons using monophoton events at LEP has been given in Ref.~\cite{Berezhiani:2001rs}, and constraints on NSI with electrons from neutrino scattering,  solar and reactor neutrino experiments are performed in Refs.~\cite{Davidson:2003ha, Barranco:2005ps, Bolanos:2008km}. A combination of the data from LEP, neutrino scattering, solar and reactor neutrino experiments can be found in Refs.~\cite{Barranco:2007ej, Forero:2011zz}. 
The $\epem$ collider experiment LEP can help in removing degeneracy in the global analysis of constraints on NSI parameters. Moreover, the current constraints on tau-type NSI with electrons are mainly coming from LEP. However, due to cancellation between the left-handed (vector) and right-handed (axial-vector) NSI parameters, the allowed ranges of NSI with electrons are still large if both the left and right-handed NSI parameters exist. Considering that will provide much more data than LEP, it is necessary to study the constraints on NSI with electrons from the proposed future $\epem$ collider experiments.

Since we simulate the data assuming the SM, the point (0,0) will always be included in the allowed regions. Also, the allowed regions from each running mode will lie between two concentric circles. From Fig.~(\ref{fig:cepceelr}-\ref{fig:cepcmmlr}), we see that the direction from the SM point (0,0) to the circle centre with $\sqrt{s}=240\ \mathrm{GeV}$ is almost parallel to that with $\sqrt{s}=160\ \mathrm{GeV}$, and the allowed regions with $\sqrt{s}=160\ \mathrm{GeV}$ can be superseded by $\sqrt{s}=240\ \mathrm{GeV}$ in the neighbourhood of the SM point (0,0). For electron-type NSI, 
the direction from the SM point (0,0) to the circle centre with $\sqrt{s}=91.2\ \mathrm{GeV}$
is approximately perpendicular to that with the other two running modes. Therefore, the allowed parameter space of electron-type NSI is severely constrained by combining the data from three running modes. Even if both $\epsilon_{\Pe\Pe}^{eL}$ and $\epsilon_{\Pe\Pe}^{eR}$ are present, the allowed ranges for $|\epsilon_{\Pe\Pe}^{eL}|$ or $|\epsilon_{\Pe\Pe}^{eR}|$ can be constrained to be smaller than 0.002. 
As for muon-type or tau-type NSI, the direction from the SM point (0,0) to the circle centre with $\sqrt{s}=91.2\ \mathrm{GeV}$ is approximately opposite to that with the other two running modes. Hence, the allowed regions will become long ellipses after combining the data from three running modes. If both $\epsilon_{\PGm\PGm,\PGt\PGt}^{eL}$ and $\epsilon_{\PGm\PGm,\PGt\PGt}^{eR}$ are present, the allowed ranges of muon-type and tau-type NSI are about one order of magnitude weaker than that of electron-type NSI.

In Table \ref{tab:cepc}, we present the 90\% C.L. constraints on only one parameter at a time by fixing the remaining parameters to zero at CEPC. One can find that with three different running modes, CEPC can
lead to a great improvement in constraining each NSI parameter compared to the previous global analysis of the LEP, CHARM II, LSND and reactor data~\cite{Barranco:2007ej}.
Except for $\epsilon_{\PGm\PGm,\PGt\PGt}^{eV}$, one can constrain all NSI parameters to be smaller than 0.002 by combining the data from three running modes. 

\setlength{\tabcolsep}{2mm}{
	\begin{table}
		\renewcommand\arraystretch{1.5}
		\begin{tabular}{c|c|c|c|c|c}
			\hline \hline
			& CEPC-91.2 & CEPC-160 & CEPC-240 & CEPC-combined & Previous Limit\\ 
			&  $L=16$ ab$^{-1}$ &  $L=2.6$ ab$^{-1}$ & $L=5.6$ ab$^{-1}$ & $L=24.2$ ab$^{-1}$ & 90\% Allowed  \cite{Barranco:2007ej} \\ 
			\hline 
			$\epsilon_{\Pe\Pe}^{eL}$  &[-0.0037,0.0037]&[-0.0036,0.0035] &[-0.0010,0.0010] & [-0.00095,0.00095] &[-0.03,0.08]\\
			\hline 
			$\epsilon_{\Pe\Pe}^{eR}$  &[-0.0017,0.0017] &[-0.014,0.015] &[-0.0065,0.0070] &  [-0.0017,0.0017] & [0.004,0.15]\\
			\hline
			$\epsilon_{\Pe\Pe}^{eV}$   &[-0.0024,0.0023] &[-0.0094,0.0093] &[-0.0024,0.0024]  & [-0.0017,0.0017] & --\\ 
			\hline 
			$\epsilon_{\Pe\Pe}^{eA}$  &[-0.0065,0.0066] &[-0.0057,0.0057] &[-0.0018,0.0018]  & [-0.0017,0.0017] & --\\
			\hline 
			$\epsilon_{\PGm\PGm/\PGt\PGt}^{eL}$  &[-0.0014,0.0014] &[-0.012,0.012] &[-0.0055,0.0053] &[-0.0013,0.0013] &[-0.03,0.03]/[-0.5,0.3] \\
			\hline      
			$\epsilon_{\PGm\PGm/\PGt\PGt}^{eR}$  &[-0.0017,0.0017] &[-0.014,0.015] &[-0.0065,0.0070] &[-0.0017,0.0017] &[-0.03,0.03]/[-0.3,0.4]\\
			\hline        	
 			$\epsilon_{\PGm\PGm/\PGt\PGt}^{eV}$  &[-0.013,0.015] &[-0.194,0.074] &[-0.085,0.033]   & [-0.013,0.014] & --\\
 			\hline 	
 			$\epsilon_{\PGm\PGm/\PGt\PGt}^{eA}$  &[-0.0015,0.0015] &[-0.013,0.013] &[-0.0061,0.0060] &[-0.0015,0.0015] & --\\
			\hline 	 \hline 
		\end{tabular} 
		\caption{Constraints of NSI parameters by varying only one parameter
		at a time at CEPC. For comparison, the previous reported results from
	the global analysis with the data of the LEP, CHARM II, LSND, and reactor pieces are shown in the last column.}
		\label{tab:cepc}
	\end{table}
}

\begin{figure}[htbp]
	\begin{centering}
		\includegraphics[width=0.45\columnwidth]{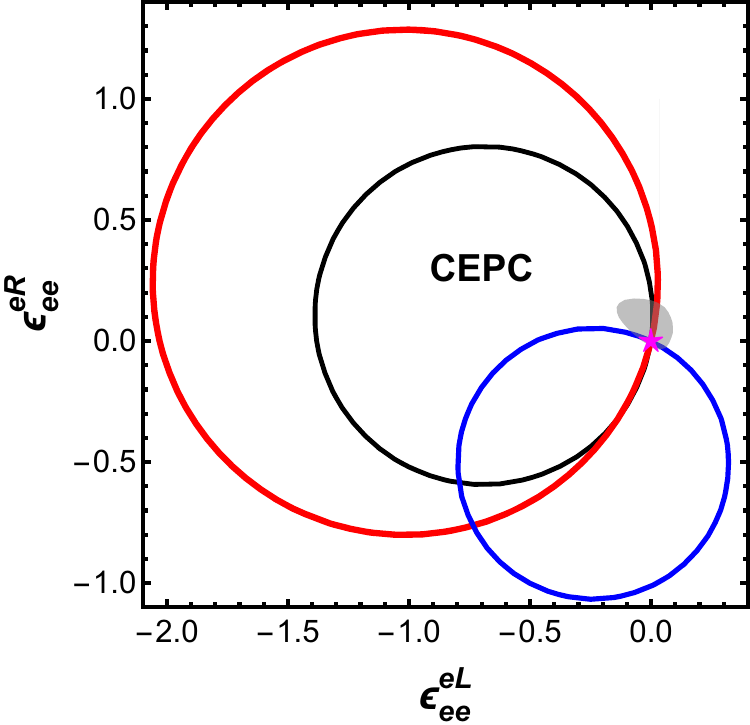}
		\includegraphics[width=0.48\columnwidth]{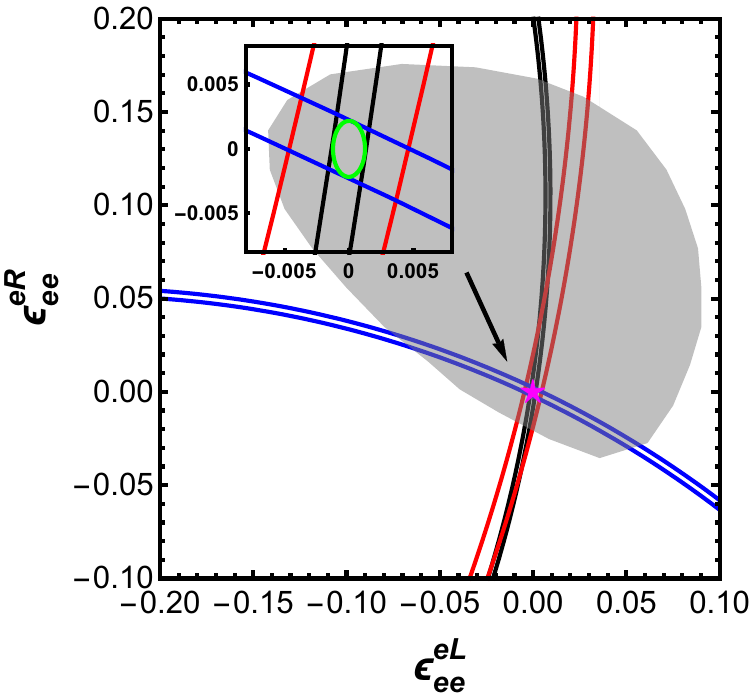}
		\caption{The allowed  90\% C.L. region for electron-type neutrino NSI in the planes of $(\epsilon_{\Pe\Pe}^{eL},\ \epsilon_{\Pe\Pe}^{eR})$ at future CEPC with 5.6 ab$^{-1}$ data of $\sqrt{s}=240$ GeV (Black), with 2.6 ab$^{-1}$ data of $\sqrt{s}=160$ GeV (Red), and with 16 ab$^{-1}$ data of $\sqrt{s}=91.2$ GeV (Blue), respectively.
	   The  allowed 90\% C.L. regions  arising
	from the global analysis of the LEP, CHARM, LSND, and reactor data \cite{Barranco:2007ej},  are shown in the shaded grey regions.
	With all the data collected in all three running modes, the combined result is labelled in green and shown in the right panel.
	}
		\label{fig:cepceelr}
	\end{centering}
\end{figure}

\begin{figure}[htbp]
	\begin{centering}
		\includegraphics[width=0.45\columnwidth]{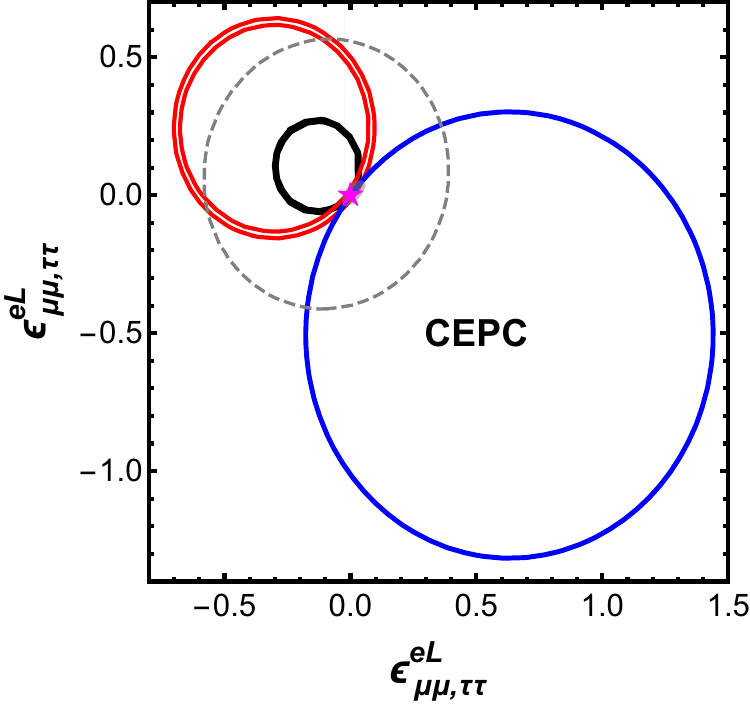}
		\includegraphics[width=0.45\columnwidth]{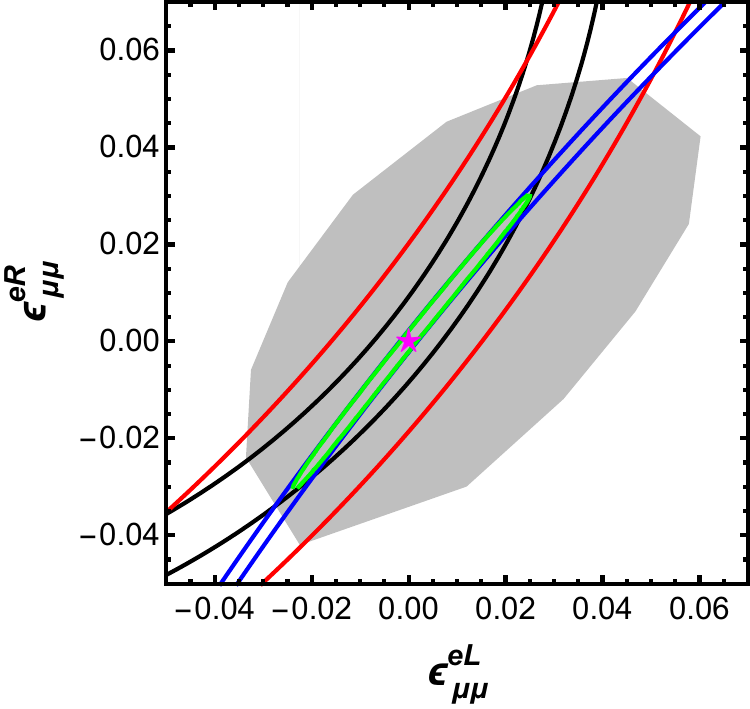}
		\caption{
		The allowed  90\% C.L. region for muon-type and tau-type neutrino NSI in the planes of $(\epsilon_{\PGm\PGm,\PGt\PGt}^{eL},\ \epsilon_{\PGm\PGm,\PGt\PGt}^{eR})$at future CEPC with 5.6 ab$^{-1}$ data of $\sqrt{s}=240$ GeV (Black), with 2.6 ab$^{-1}$ data of $\sqrt{s}=160$ GeV (Red), and with 16 ab$^{-1}$ data of $\sqrt{s}=91.2$ GeV (Blue), respectively. 
			The  allowed 90\% C.L. regions for muon-type arising
			from the global analysis of the LEP, CHARM II, LSND, and reactor data \cite{Barranco:2007ej},  are shown in the shaded grey regions, and the allowed 90\% C.L. regions for tau-type NSI arising from the LEP data are shown in dashed grey lines. 	With all the data collected in all three running modes, the combined result is labelled in green and shown in the right panel.}
		\label{fig:cepcmmlr}
	\end{centering}
\end{figure}

\subsubsection{Neutrino anomalous magnetic moment}
\editors{Emidio Gabrielli and Luca Marzola}

Experimental evidence from solar and atmospheric neutrino oscillation experiments~\cite{SNO:2001kpb, Super-Kamiokande:1998kpq} show that at least two neutrino species have mass. If neutrinos are Dirac particles, this observation then allows for the exploration of the magnetic and electric dipole moments of these particles. The corresponding electromagnetic neutrino current $J_{\mu}(x)$ projected onto momentum eigenstates gives a matrix element parametrised as
\begin{eqnarray}
\langle p^{\prime} |J_{\mu}(x=0)|p\rangle =e~\bar{\nu}(p^{\prime}) \Big(F_1(q^2)\gamma_{\mu}+ i\frac{F_2(q^2)}{2m_{\nu}}
\sigma_{\mu\nu} q^{\nu}\Big) \nu(p)\, ,
\label{eq:J}
\end{eqnarray}
where $e$ is the electromagnetic unit charge, and $F_{1}(q^2)$ and $F_{2}(q^2)$ are the electric and magnetic form factors, respectively. These depend on $q^2\equiv (p-p^{\prime})^2$, with $p$ and $p^{\prime}$ being the neutrino 4-momenta. The symbol $m_{\nu}$ indicates the Dirac neutrino mass and, as usual, $\sigma_{\mu \nu}=i[\gamma_{\mu},\gamma_{\nu}]/2$. Then, in units of the Bohr magneton $\mu_B$, the magnetic moment $k$ of a neutrino $\nu$ is given by
\begin{equation}
k \mu_B =\frac{F_2(0)}{2m_{\nu}}\, .
\end{equation}

The present bounds on the magnetic moment of the $\PGt$-neutrino ($k_{\PGt}$) are several order of magnitudes below the corresponding constraints that hold for electron and muon neutrinos~\cite{ParticleDataGroup:2024cfk}. Experimentally, the neutrino magnetic moment can be directly constrained at collider experiments by analyzing the inclusive production in $\epem$ annihilations of a neutrino pair plus a photon at the $\PZ$-boson resonance
\begin{equation}
\epem\to \PZ\to \PGn \PAGn \PGg\,.
\label{process}
\end{equation}
For the signal that we seek, the photon ($\gamma$) must be emitted from either of the neutrinos in the final state via the magnetic dipole interaction in~\cref{eq:J}, as shown in the Feynman diagrams (a) and (b) of Fig.~\ref{fig:FD}.

\begin{figure}[t]
  \begin{center}
  \includegraphics[width=\linewidth]{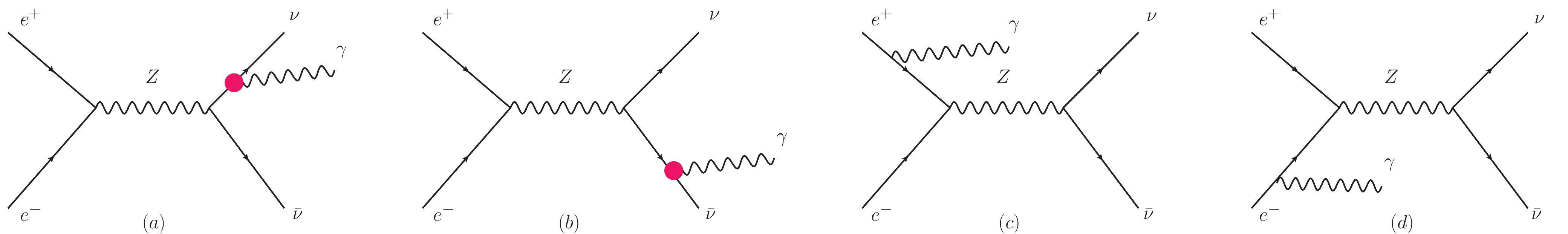}
  \caption{\small Dominant Feynman diagrams for the processes $\epem\to \PZ\to \PGn \PAGn \PGg$ at the $\PZ$-boson resonance. The red dots in the diagrams (a) and (b) indicate the vertex induced by the effective neutrino magnetic dipole moment sourced by~\cref{eq:J}.\label{fig:FD}
  }
  \end{center}
  \end{figure}

The differential cross section for the process in~\cref{process} can be approximated as~\cite{Gould:1994gq}
\begin{eqnarray}
  \frac{\text{d}\sigma}{E_{\gamma} \text{d}E_{\gamma} \text{d}\cos{\theta_{\gamma}}}
  =\frac{\alpha^2k^2}{96\pi}\mu_B^2~R(s^2_W)\left(
\frac{s-2\sqrt{s}E_{\gamma}+\frac{1}{2}E_{\gamma}^2\sin^2{\theta_{\gamma}}}
  {(s-M_Z^2)^2+M_Z^2\Gamma_Z^2}\right)\, ,
\end{eqnarray}
where $\theta_\gamma$ and $\EG$ are the scattering angle and energy of the photon in the $\epem$ centre-of-mass (c.o.m.) frame, respectively, $\sqrt{s}$ is the c.o.m. energy, and $s_W=\sin{\theta_W}$, with $\theta_W$ being the Weinberg angle. The $\PZ$-boson mass and width are respectively indicated with $\MZ$ and $\Gamma_Z$, whereas the function $R$ is given by
\begin{equation}
R(x)=\frac{8x^2-4x+1}{x^2(1-x)^2}\, .
\end{equation}
As the flavour of the produced neutrinos cannot be selected, the process is sensitive to the combination $k^2\equiv \sum_{\alpha=\Pe,\PGm\PGt} k^2_\alpha$, where $k_{\Pe,\PGm,\PGt}$ are the magnetic moments of electron, muon, and $\PGt$ neutrinos.

By integrating over the scattering angle and using the Breit-Wigner approximation, we obtain the linear dependence of the $\PZ$-boson width on the rescaled photon energy, given by
\begin{equation}
\frac{\text{d} \Gamma}{\text{d} x_{\gamma}}=\frac{\alpha~k^2 M_Z^3~ \mu_B^2}{24\pi^2c_W^2s_W^2}F(x_{\gamma})\, 
\label{eq:distr}
\end{equation}
where $c_W\equiv\cos{\theta_W}$, $x_{\gamma}=E_{\gamma}/M_Z$, and $F(x)=x(1-2x+x^2/3)$. The distribution $\frac{\text{d} \Gamma}{\text{d} x_{\gamma}}$, shown in Fig.~\ref{fig:distr}, has a maximum near $E_{\gamma}\simeq 24.4~{\textrm{GeV}}$, which is peculiar to the magnetic dipole interactions.

\begin{figure}[h!]
  \begin{center}
  \includegraphics[width=0.5\linewidth]{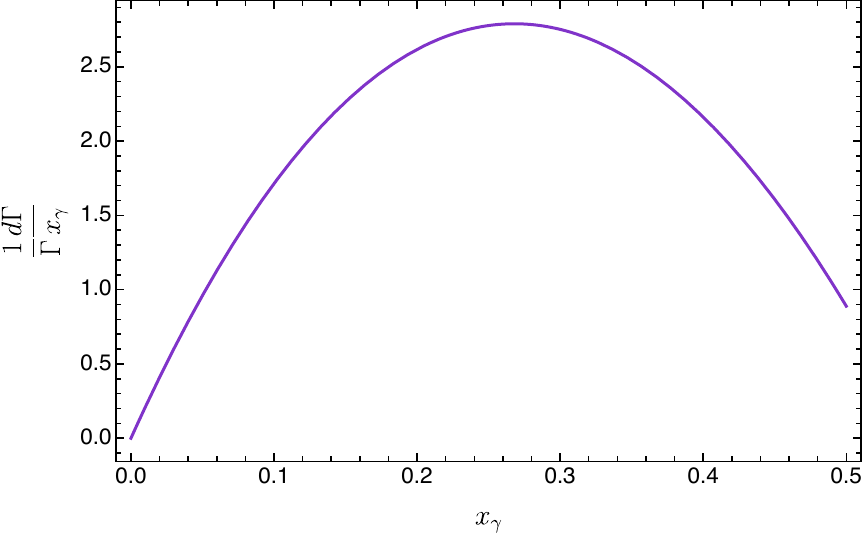}
  
  \caption{\small The normalised distribution in \cref{eq:distr} as a function of the rescaled energy of the photon $x_\gamma=\EG/M_Z$. Its peculiar shape could allow disentangling the origin of possible excesses measured in the number of mono-photon events. \label{fig:distr}
  }
  \end{center}
  \end{figure}

The process in \cref{process} consists of a single final-state photon accompanied by missing energy, carried by the neutrinos. The main Standard Model (SM) background for the energy region of interest is given by the Feynman diagrams (c) and (d) in Fig.~\ref{fig:FD}, in which the emission of a photon is due to the initial $\epem$ states.

The proposal of constraining the magnetic moment of the $\nu_\tau$ 
based on single-photon final states was considered in Ref.~\cite{Gould:1994gq}, for LEP searches~\cite{DELPHI:1996drf,L3:1997exg,OPAL:1994kgw,ALEPH:1993pqw}, and previously in Ref.~\cite{Grotch:1988ac} in the context of $\epem$ collider experiments at lower energies.

At LEP, the DELPHI~\cite{DELPHI:1996drf} and L3~\cite{L3:1997exg} collaborations analyzed this process at the $\PZ$ peak to establish one of the most stringent constraints on $k_{\tau}$. More recently, analogous bounds have been investigated in the context of new physics searches at future $\epem$ colliders such as CLIC and ILC~\cite{Gutierrez-Rodriguez:2006fzr,Aydin:2006nc,Gutierrez-Rodriguez:2011rdm,Binh:2018teg,Billur:2018cnk,Gutierrez-Rodriguez:2018kxt}.

The L3 analysis~\cite{L3:1997exg} used an integrated luminosity of ${\textrm{L}}_{\textrm{LEP}}=137~{\rm pb}^{-1}$ and cuts optimized for a photon energy $E_{\gamma}> 15$\,GeV and scattering angles  $20^{\circ}<\theta_\gamma< 34.5^{\circ}$, $44.5^{\circ}<\theta_\gamma< 135.5^{\circ}$, 
giving a total of 14 events after selections.  The result is consistent with the Monte Carlo simulation of the SM expectation due to production of neutrino pairs accompanied by initial-state radiation, radiative Bhabha events and $\epem\to \PGg \PGg (\PGg)$~\cite{L3:1997exg}, which predicted 14.1 events altogether. Under the assumption of lepton universality in the $Z$ decay to neutrinos, this yields the following upper bounds at 90\% confidence level (CL) on the magnetic moment of $\PGt$-neutrino~\cite{L3:1997exg}
\begin{equation*}
k^{\rm LEP}_{\PGt}< 3.3 \times 10^{-6} ~~ 90\%~{\rm CL}\, .
\label{boundsLEP}
\end{equation*}

We now turn to the prospects for investigating the $\tau$-neutrino magnetic moment through the process in \cref{process} at the future FCC-ee collider. After running for 4 years at the $\PZ$-boson
resonance, the FCC-ee is expected to produce $3\times 10^{12}$ $\PZ$ bosons~\cite{FCC:2018evy,VenturiniDelsolaro:2023wqm}, thereby dramatically improving over the statistics limit of the LEP program. The corresponding expected integrated luminosity collected over this period is ${\rm L}_{\rm FCC}=150~{\rm ab}^{-1}$.
By using the same kinematic cuts mentioned above that were optimized for L3, the reach of the FCC-ee can be estimated by simply rescaling the LEP bound by the fourth root of the luminosity ratio $({\rm L}_{\rm LEP}/{\rm L}_{\rm FCC})^{1/4}\simeq 1/31$, giving
\begin{equation}
k^{\rm FCC}_{\tau}< 1.1 \times 10^{-7} ~~ 90\%~{\rm CL} \, .
\label{boundsFCC}
\end{equation}

Presently, the strongest bound on $k_{\PGt}$ is given by the DONUT collaboration by analyzing the scattering of $\PGt$-neutrinos on electrons $\nu_{\PGt}~ \Pe^- \to \PGn_{\PGt} ~\Pe^- $, yielding~\cite{DONUT:2001zvi,ParticleDataGroup:2024cfk}
\begin{equation}
k^{\rm DONUT}_{\PGt}< 3.9 \times 10^{-7} ~~ 90\% ~{\rm CL} \, .
\label{boundsDON}
\end{equation}
We stress that the limit in \cref{boundsDON} applies exclusively to $k_{\PGt}$, while the one derived from the process in \cref{process} includes all the contributions of the three magnetic moments of neutrinos.

In conclusion, after four years of operation at the $\PZ$ peak, the FCC-ee could offer the possibility of improving the present limit \cref{boundsDON} by a factor of almost 4. This estimate maintains the kinematic cuts used to optimize the L3 analysis at LEP, hence it is plausible that further improvements could be achieved once the FCC-ee detectors and machine will be understood and dedicated kinematic cuts identified.

\subsection{\focustopic WW differential measurements \label{sec:WWdiff}}
\editors{Alexander Grohsjean}

Constraints on gauge boson interactions are essential for global interpretations in both the SMEFT framework and UV-complete models. In scenarios where the electroweak symmetry is linearly realized, new physics contributions to anomalous triple gauge couplings are intrinsically linked to modified Higgs couplings. This connection establishes a profound complementarity between measurements in the gauge and Higgs sectors, enabling a more comprehensive exploration of potential deviations from Standard Model predictions.
In the most general formulation, the vertex between a neutral and two charged electroweak gauge bosons encompasses 14 complex couplings, amounting to 28 real parameters~\cite{Gounaris:293937}. Within the SM, only four of these couplings are expected to have values of one, while all remaining parameters are predicted to be zero. Any deviation from this configuration would serve as a precise indicator of physics beyond the SM. Due to the limited constraining power of LEP data on the full set of trilinear gauge couplings (TGCs), the number of parameters traditionally analyzed is substantially reduced by imposing conservation laws such as C, P, and CP invariance, along with electromagnetic gauge invariance and $SU(2) \times U(1) $ symmetry. This leaves three independent parameters, commonly taken to be $ g_{1,Z} $, $\kappa_\gamma $, and $\lambda_Z$ which are used to define the corresponding Lagrangian
\begin{equation}
\begin{aligned}
\Delta \mathcal{L}^{\text{aTGC}} = & \, i e \delta \kappa_{\PGg} A^{\mu\nu} W^{+}_{\mu} W^{-}_{\nu} \\
& + i g \cos \theta_W \left[ \delta g_{1Z} (W^{+}_{\mu\nu} W^{-\mu} - W^{-}_{\mu\nu} W^{+ \mu}) Z^\nu + (\delta g_{1Z} - \frac{g'^{2}}{g^2} \delta \kappa_{\PGg}) Z^{\mu\nu} W^{+}_{\mu} W^{-}_{\nu} \right] \\
& + \frac{i g \lambda_{\PZ}}{m^2_{\PW}} \left( \sin \theta_W W^{+ \nu}_{\mu} W^{- \rho}_{\nu} A^{\mu}_{\rho} + \cos \theta_W W^{+ \nu}_{\mu} W^{- \rho}_{\nu} Z^{\mu}_{\rho} \right).
\end{aligned}
\end{equation}

The sensitivity of the HL-LHC to anomalous triple gauge couplings (aTGC) has been estimated by extrapolating existing differential measurements of the leading lepton transverse momentum in inclusive diboson production~\cite{Azzi:2019yne}. While the cross-section of inclusive $\PW \PW$ production can be measured with roughly twice the precision compared to $\PW \PW$ electroweak scattering~\cite{ATL-PHYS-PUB-2022-018}, both processes provide complementary information. Their combination is a critical input for global EFT fits, enabling a more comprehensive exploration of deviations from Standard Model predictions~\cite{Bellan:2021dcy}.
By taking existing ATLAS and CMS measurements and assuming a 50\% reduction in systematic uncertainties, the sensitivity achieved with VBS-only fits at the HL-LHC is projected to surpass current constraints from combined diboson and VBS analyses. For certain EFT operators, improvements of at least a factor of 10 over current bounds are anticipated~\cite{Ethier:2021ydt}.

Numerous studies have examined the potential of future colliders to probe aTGCs. The most comprehensive theoretical analysis —- ignoring effects from detector resolution, backgrounds, or systematic uncertainties -— demonstrates that at a high-energy $\epem$ collider operating at $\sqrt{s}=\SI{500}{\GeV}$ with polarised beams, including transverse polarisation, all 28 real coupling parameters can be disentangled and constrained effectively~\cite{Diehl:2002nj, Diehl:2003qz}. More recently, the sensitivity to the three couplings of the LEP parametrisation was examined within a SMEFT fit to the Higgs and electroweak sectors, relying on optimal observables at the theory level~\cite{deBlas:2022ofj}. This study projects an improvement in precision exceeding a factor of 100 relative to the HL-LHC. The effects of detector resolution and residual backgrounds on the optimal observable technique were analyzed in Ref.~\cite{Chai:2024zyl}, showing that while both can introduce significant biases in central values, machine learning techniques can effectively provide the necessary corrections.

Studies incorporating realistic detector simulations have thus far been conducted only for the ILC at $\sqrt{s}=\SI{500}{\GeV}$ and \SI{1}{\TeV}~\cite{Marchesini:2011aka, Rosca:2016hcq, ILC_TDR_Detectors}. Due to constraints on \geant-based Monte Carlo simulations at the time, these studies focused exclusively on $\PW$-pair production. They analyzed three out of five angular observables within a binned analysis, resulting in highly conservative outcomes. These studies served as inputs for fits conducted in, for example, Ref.~\cite{Barklow:2017suo}. So far, projections for lower centre-of-mass energies were only derived by interpolating between the \SI{500}{\GeV}/\SI{1}{\TeV} projections and actual LEP2 results~\cite{Karl:2019hes}, which indicate that at \SI{250}{\GeV}, sensitivity is expected to degrade by a factor of 4-5 relative to \SI{500}{\GeV} for the same final states. The inclusion of single-\PW final states was found to provide a substantial precision gain in the $\Pe\PGn\qqbar$ channel, with improvements by a factor of 2-3 over analyses limited to $\PW$-pair events.

The first comprehensive study addressing experimental uncertainties and their mitigation has been reported in Ref.~\cite{Beyer:2022ofv}. This work demonstrates that by combining data from various 2-fermion and 4-fermion final states, the relevant physics observables (including TGCs as well as forward-backward asymmetries) can be separated from systematic effects—such as those due to luminosity, beam polarisation, energy scale, and detector acceptance—in a combined fit that includes both physics and nuisance parameters. Notably, this approach minimizes the residual impact of systematic uncertainties on the extracted physics observables, particularly when both beams are polarised, thus providing sufficient redundancy across data sets to effectively constrain the fit.

To achieve an improved sensitivity estimate for differential $\PW \PW$ measurements beyond the interpolation approach discussed in Ref.~\cite{Karl:2019hes}, dedicated analyses were carried out for ILD at $\sqrt{s}=\SI{250}{\GeV}$ and CLD at $\sqrt{s}=\SI{240}{\GeV}$. Diboson events were generated using \whizard~\cite{Kilian:2007gr,Moretti:2001zz}, processed through \pythiasix~\cite{Sjostrand:2006za} for parton showering, and subsequently passed through the full \geant simulation of the ILD and CLD detectors, respectively.

For the ILC scenario, event samples were obtained from the official mass production campaign~\cite{Ono:2021opz}, while for the CLD case, dedicated samples were privately simulated using \ddsim. Reconstruction for both experiments was performed within the \marlin framework, with CLD employing the \keyhep \marlin wrapper~\cite{placido_fernandez_declara_2024_13837467}. Event samples were then appropriately mixed and re-weighted to reflect unpolarised data for the FCC-ee case and polarised data for the ILC case.

The analysis examines the sensitivity of five key observables: the production angle of the $\PW^-$, $\cos \theta_{\PW^-}$, defined as the angle between the electron beam and the $\PW^-$, along with the polar and azimuthal angles of the fermion pair in the rest frame of the parent $\PW^-$ or $\PW^+$. These angular observables are measured in the semi-leptonic decay channel, utilizing the charged lepton and the reconstructed neutrino.

Given the importance of measuring the charged lepton and the increased difficulty in the electron channel thereof, the electron acceptance of the detectors was evaluated in a first step as shown in \cref{fig:wwdiff_eff_theta}.

The extraction of constraints on triple gauge couplings is most effectively performed using optimal observables. \Cref{fig:oos} illustrates the expected reconstruction of these optimal observables and compares them to the truth level expectation.
The three observables correspond to the couplings in the LEP parametrisation, which are also used in linear dim-6 SMEFT fits.
As can be seen from the plot, the detector-resolution effects have almost no impact on the size and shape of the ellipses and only introduce a bias to the central values.

The performed studies have successfully set the foundations needed to perform a full optimal observable-based analysis of $\PW\PW$-pair production. In the signal-only comparison, the optimal observable technique has proven itself to be extremely stable under detector resolution effects. However, we expect additional influence from background contamination in the ongoing full analysis which needs to be quantified before actual projections for further global interpretations will be provided.

\begin{figure}
     \centering
     \begin{subfigure}[b]{0.45\textwidth}
         \centering
         \includegraphics[width=\textwidth]{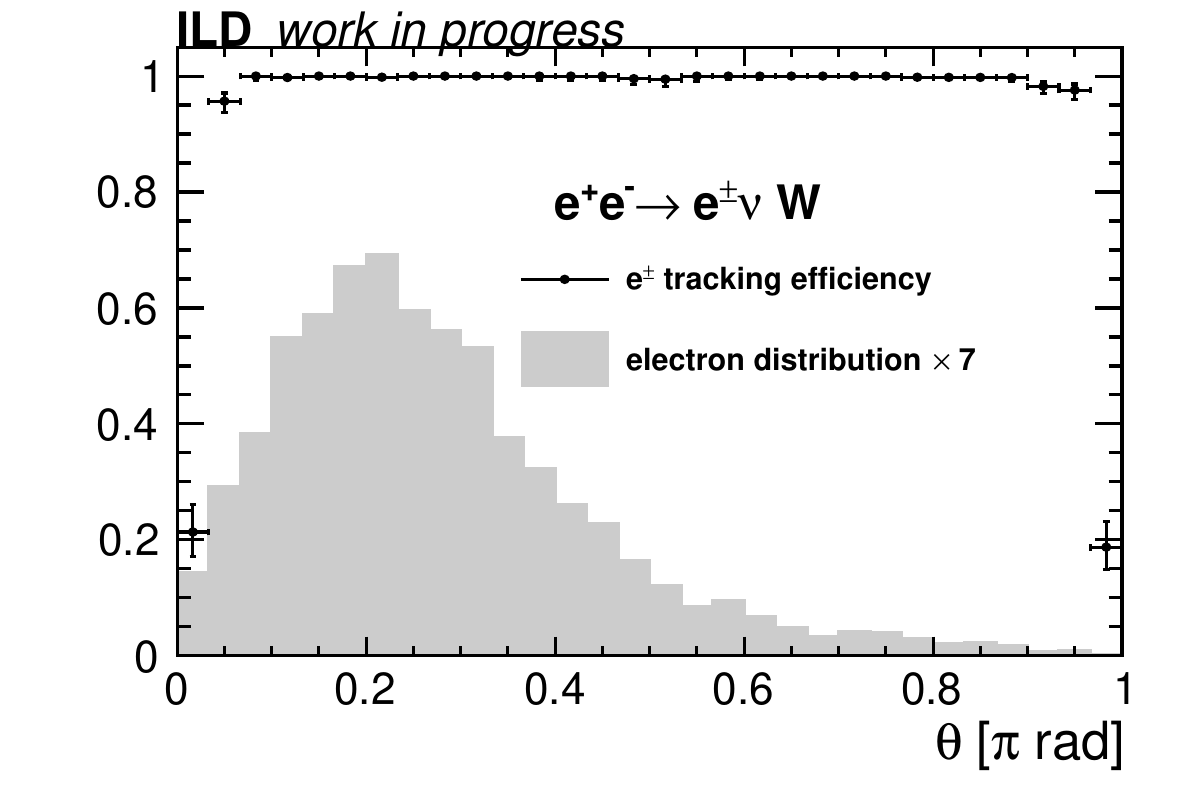}
         \label{fig:wwdiff_eff_theta_ild}
     \end{subfigure}
     \hfill
     \begin{subfigure}[b]{0.45\textwidth}
         \centering
         \includegraphics[width=\textwidth]{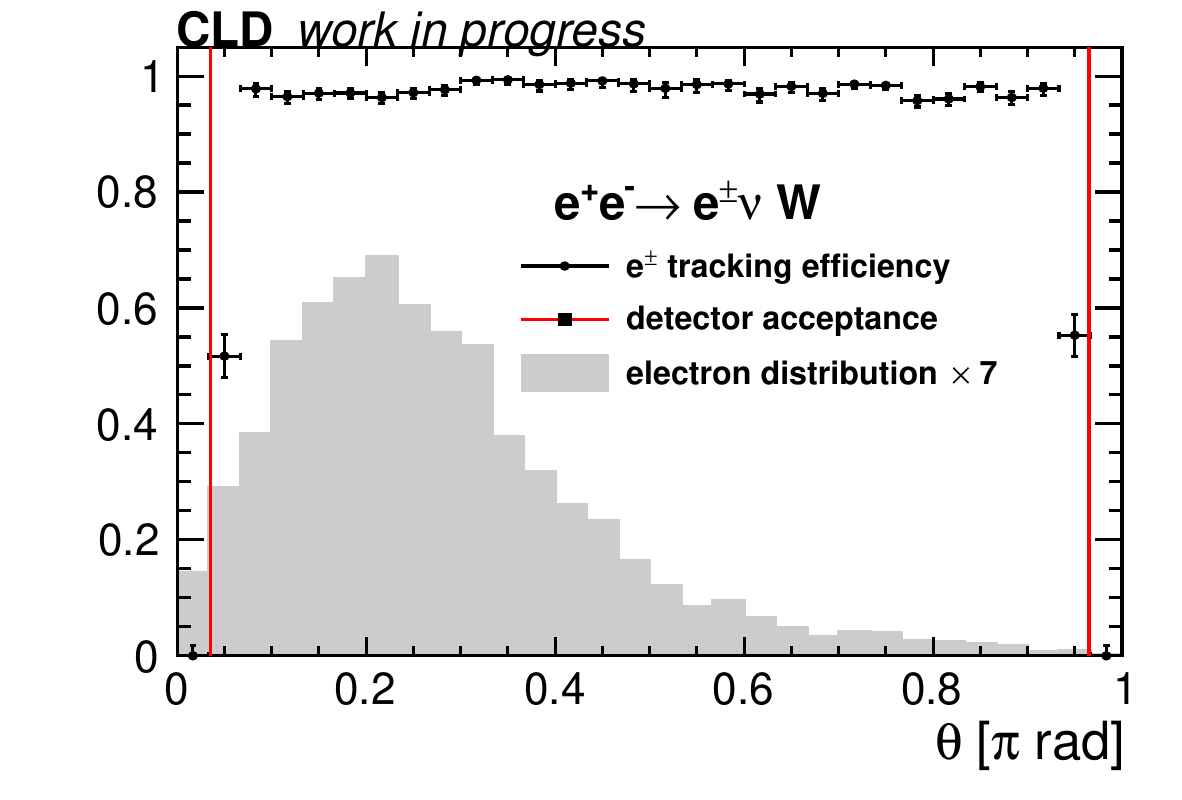}
         \label{fig:wwdiff_eff_theta_cld}
     \end{subfigure}
        \caption{Electron angular tracking efficiency, assessed with the same performance evaluation for CLD and ILD. The angular distribution of the electrons in the final state is shown in grey for comparison.}
        \label{fig:wwdiff_eff_theta}
\end{figure}

\begin{figure}
     \centering
     \begin{subfigure}[b]{0.32\textwidth}
         \centering
         \includegraphics[width=\textwidth]{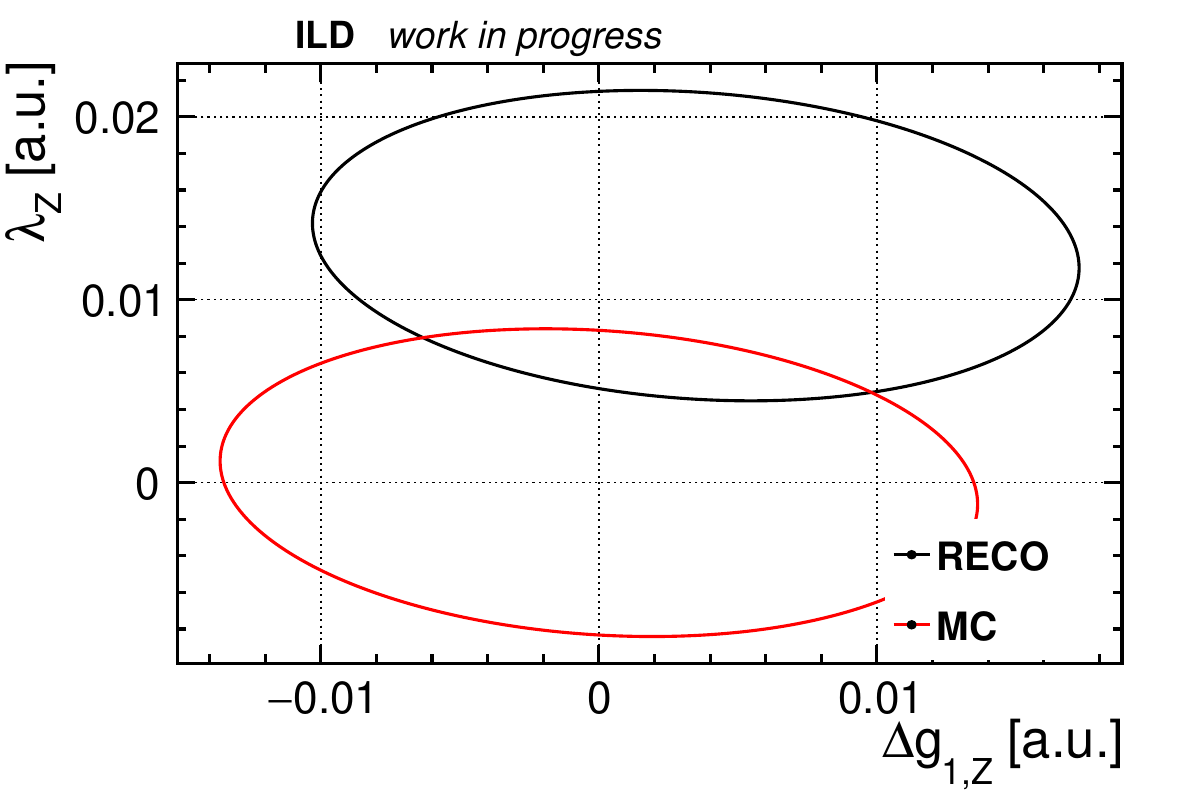}
         \label{fig:oo_dg1z}
     \end{subfigure}
     \hfill
     \begin{subfigure}[b]{0.32\textwidth}
         \centering
         \includegraphics[width=\textwidth]{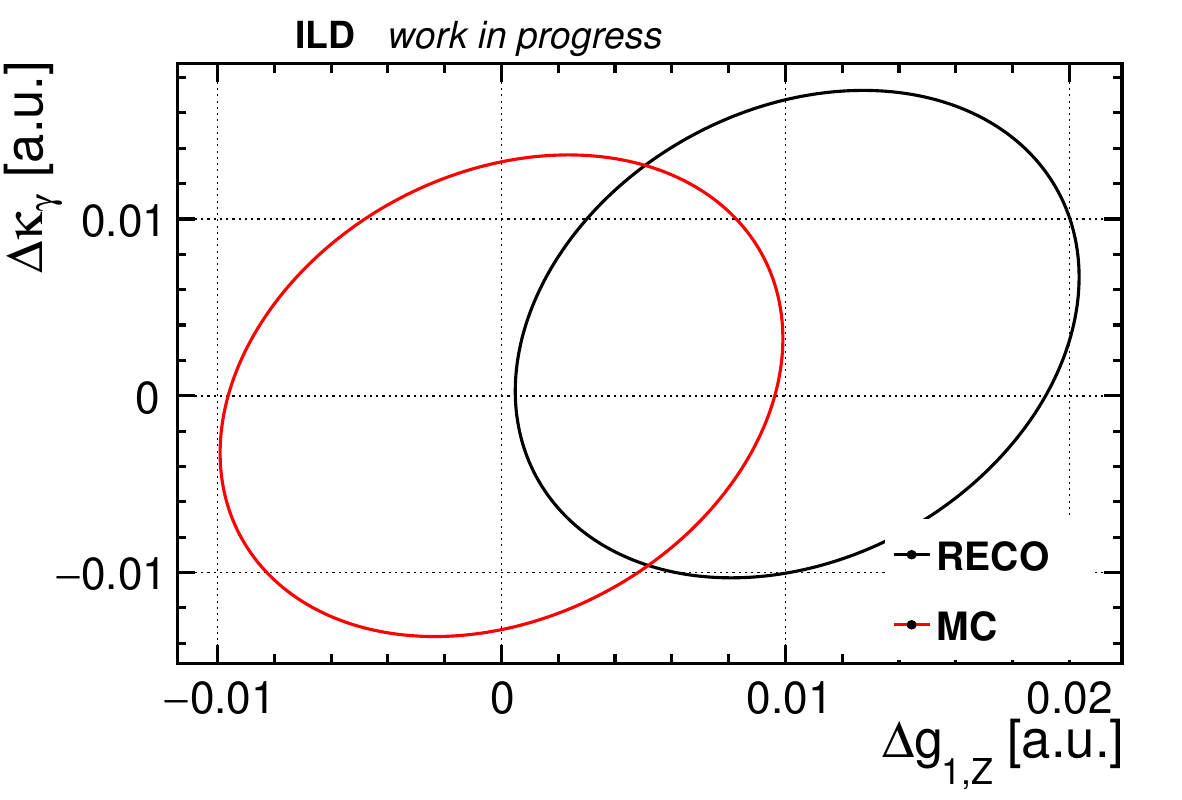}
         \label{fig:oo_dkg}
     \end{subfigure}
     \hfill
     \begin{subfigure}[b]{0.32\textwidth}
         \centering
         \includegraphics[width=\textwidth]{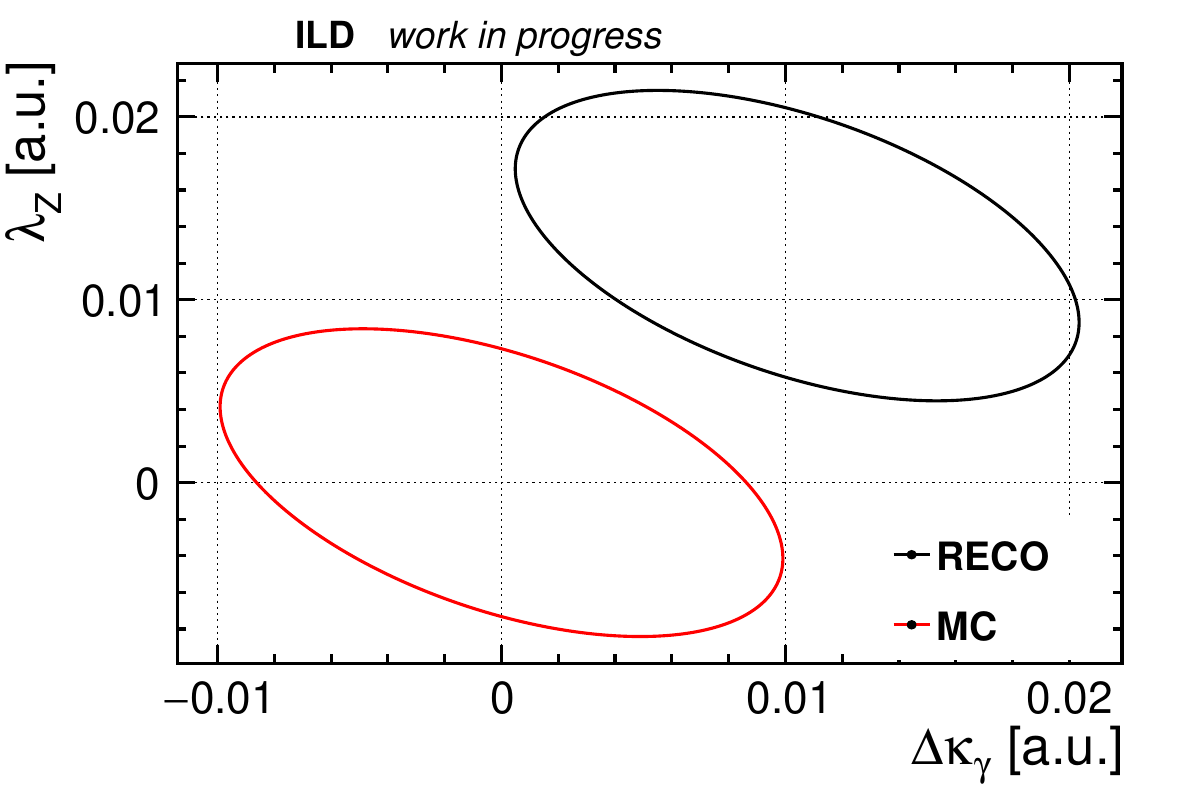}
         \label{fig:oo_lz}
     \end{subfigure}
        \caption{One-standard-deviation-ellipses for the three TGC in the LEP parametrisation. The truth level ellipses (red) is centred at (0, 0) while the reconstruction level one (black) is shifted by the differences of the means of the observables.}
        \label{fig:oos}
\end{figure}



\subsection{\focustopic Fragmentation and hadronisation \label{sec:frag-had}}
\editors{Adrián Irles, Ayres Freitas, Andreas Meyer, Paolo Azzurri, Simon Plätzer, Torbjörn Sjöstrand,... }
  
Fragmentation and hadronisation of quarks and gluons are key ingredients for the precision modeling of physics processes.
The issues are entangled, and their systematic uncertainties are expected to be limiting factors in precision Higgs and electroweak measurements at a future Higgs factory. To assess their impact, it is important to review the state of the art, to project how future theory developments could reduce current uncertainties, and to devise ways to constrain fragmentation and hadronisation from measurements at the Higgs factories. 

The full QCD description of an event can be split into perturbative and
nonperturbative aspects, although the two are often intertwined. 
The former involves matrix elements for the hard process,
parton showers for the subsequent softer emissions, and a consistent
matching and merging of them. Much work is ongoing in extending
matrix elements and showers to higher orders, so improved perturbative
predictions can confidently be foreseen in the years to come, as is 
described in \cref{sec:gen}.

The nonperturbative aspects mainly deal with the hadronisation
of quarks and gluons into primary hadrons, and the subsequent
decays of these primaries into the final-state observable particles.
Hadronisation is typically modelled by string or cluster fragmentation. 
A fair amount of studies are ongoing here, not least stimulated by
unexpected LHC observations. This is likely to further our understanding
also for $\epem$ colliders, but in a less easily quantifiable way.
Decays seldom attract much attention, but is not to be dismissed
as trivial. Notably for B hadrons, the lack of measured reasonably
complete decay tables is a problem, exacerbated by the possibility
for a given final state to arise via different interfering intermediate
states. 

In the following we will discuss a few of the open questions on the
theory side, along with target observables for experimental exploration. 
Notably this includes the target topics of the fragmentation of $\PQb$
quarks into B hadrons and the splitting of gluons into heavy-quark pairs.  A related discussion on fragmentation, with special emphasis on strangeness production, can be found in \cref{sec:HtoSS:Fragmentation}.

\subsubsection{Fragmentation functions}

In the high-precision limit, fragmentation functions will not be universal, i.e.\,they are expected to depend on observables and initial states. It is argued that the factorisation of the perturbative and non-perturbative parts of the problem is not possible without dedicated tuning of free parameters in the required fragmentation model used. There are ongoing developments in disentangling hadronisation and fragmentation, specifically the cross-talk between parton shower and fragmentation. New work is needed for NLL-accurate showers.

There has been a lot of progress recently on the computation of perturbative bottom and charm fragmentation functions (FFs).  Heavy quark FFs can be computed
in perturbation theory in QCD, starting from initial conditions at a reference scale $\mu_0\sim m_\text{Q}$ (with $m_\text{Q}$ the mass of the heavy quark) 
and employing the time-like DGLAP evolution equations to evolve them up to any other scale. 
Initial conditions for the gluon- and heavy-quark-initiated fragmentation into a
heavy quark are known at order $\alpha_{\mathrm{S}}$~\cite{Mele:1990cw,Mele:1990yq} and have been computed at order $\alpha_{\mathrm{S}}^2$~\cite{Melnikov:2004bm,Mitov:2004du}. 

The time-like DGLAP evolution equation is implemented in public codes such as \textsc{QCDnum}~\cite{Botje:2010ay},
\textsc{FFevol}~\cite{Hirai:2011si}, \textsc{APFEL}~\cite{Bertone:2013vaa} or \textsc{MELA}~\cite{Bertone:2015cwa}, up to NNLL logarithmic accuracy.  
In Ref.~\cite{Ridolfi:2019bch}, the role of gluon-initiated fragmentation to heavy quarks has been considered, and the coupled time-like 
evolution of bottom quarks and gluons is considered in detail. This is important as, while at LEP and at Tevatron the 
$\Pg\to \bb$ splitting mechanism was considered subdominant, this is no longer the case at
the LHC. In Ref.~\cite{Maltoni:2022bpy} the perturbative component of the fragmentation function of
the $\PQb$~quark to the best of the present theoretical knowledge was presented. The fixed-order calculation
to order $\alpha_{\mathrm{S}}^2$ of the fragmentation function at the initial scale~\cite{Melnikov:2004bm,Mitov:2004du} is matched with soft-emission
logarithm resummation to next-to-next-to-leading logarithmic accuracy, so that order-$\alpha_{\mathrm{S}}^2$
corrections are accounted for exactly, and logarithmically enhanced contributions from
loops of $\PQb$~quarks are included. In Ref.~\cite{Czakon:2021ohs} the perturbative computation of the $\PQb$-quark fragmentation function at NNLO + NNLL is supplemented 
by the fit of the non-perturbative component $\PQb\to$ B~hadrons. Similarly, the perturbative component of $\PQc$-quark fragmentation and the fit of the non-perturbative component $\PQc\to$ D hadrons are discussed in Ref.~\cite{Bonino:2023vuz}. A key question is how can this progress can be implemented in exclusive Monte Carlo simulations.

While perturbative calculations combined with fitted non-perturbative components are adequate for inclusive fragmentation functions, exclusive predictions (e.g.\ to study the effects of experimental acceptances and cuts) require a different approach:
parton showers are needed 
for the perturbative step, and fragmentation
models for the non-perturbative one. These two steps are intertwined and convolved in a complicated space of momenta and colours, so that fragmentation in general cannot be described as a convolution of a
perturbative state and a fragmentation function $f(z)$ with $0 < z < 1$.
In string fragmentation, the colour strings stretched between partons can be attached so as to pull a heavy hadron either to
a lower or to a higher momentum than the mother quark, depending on the string topology at hand \cite{Norrbin:2000zc}. Thereby
also $z > 1$ becomes possible, which for some observables can have a significant impact relative to naive expectations. 
Similar considerations also exist for cluster fragmentation models.

\begin{figure}[htb]
\includegraphics[width=0.33\textwidth]{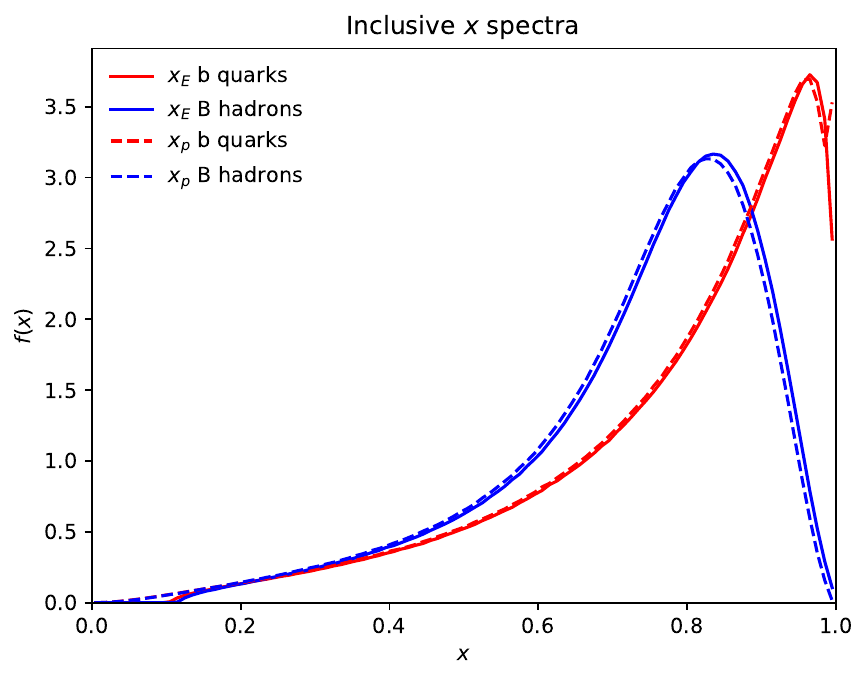}%
\includegraphics[width=0.33\textwidth]{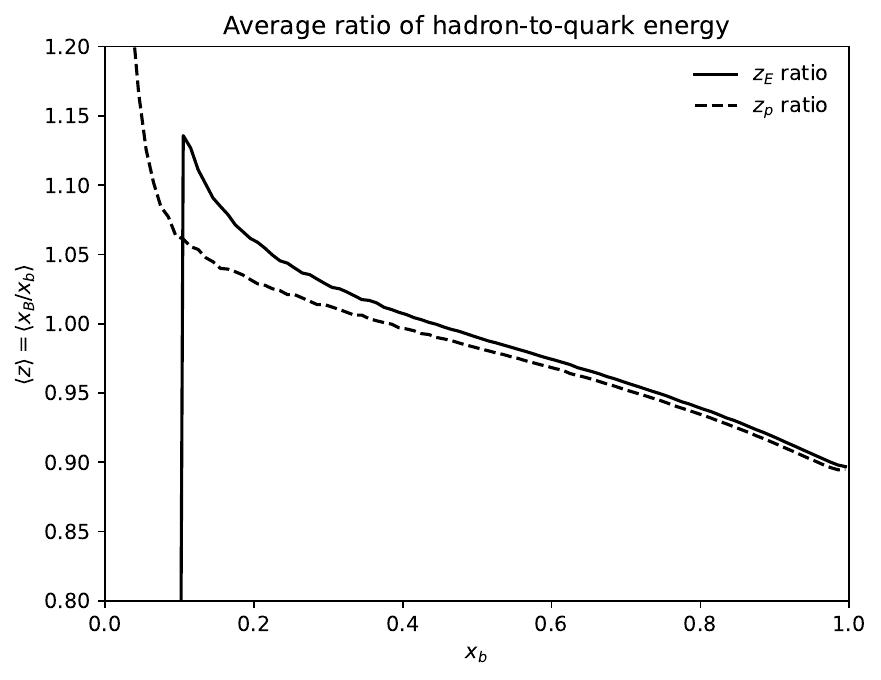}%
\includegraphics[width=0.33\textwidth]{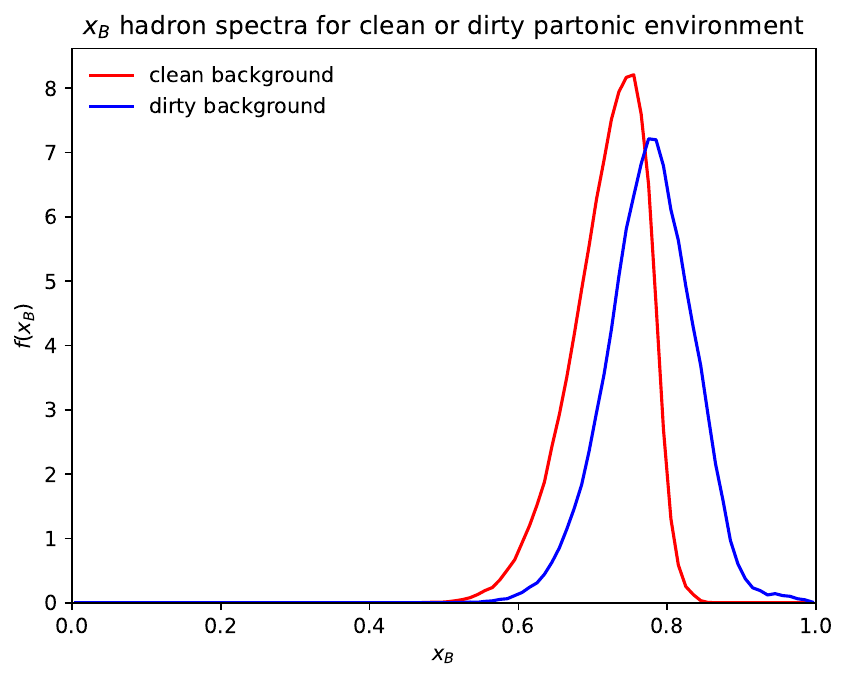}\\
\hspace*{\fill}(a)\hspace*{\fill}\hspace*{\fill}(b)\hspace*{\fill}%
\hspace*{\fill}(c)\hspace*{\fill} 
\caption{(a) $x$ spectra of $\PQb$~quarks and B~hadrons.
(b) The average hadron-to-parton $x$ ratio as a function of 
the parton $x$.
(c) The B~hadron spectrum when $0.78 < x_{\PQb} < 0.80$.}
\label{fig:test487}  
\end{figure}

To illustrate the nontrivial behaviour of String Fragmentation (SF)  and
Cluster Fragmentation (CF), in the context
of fragmentation functions,  a simple study of $\epem \to \PZz \to \PQb \PAQb$
at the $\PZz$ peak, using \textsc{Pythia} 8.312 is shown here. 
Initial-state 
QED radiation is switched off, so that the CM frame is well-defined. 
The few events where there have been $\Pg \to \PQb\PAQb$ branchings
are rejected. In the others, the respective primary daughter B
hadron is identified.

Now study energy fractions $x = x_E = 2 E / E_\text{CM} = 2 E /m_{\PZ}$,
or momentum ditto $x = x_p = 2 |\mathbf{p}| /m_{\PZ}$.
\Cref{fig:test487}(a) shows the $x$ spectra of the $\PQb$ quarks
and the B~hadrons, normalised to unit area. The B~hadrons are 
softer than the $\PQb$~quarks, as expected. But next consider,
for each $\PQb$--B pair, the ratio $z = x_\text{B} / x_{\PQb}$, and plot 
its average value $\langle z \rangle$ as a function of $x_{\PQb}$,
\cref{fig:test487}(b). Already here we see an environment effect, 
that hadronisation ``slows down'' a high-energy $\PQb$, 
$\langle z \rangle < 1$, but ``speeds up'' a low-energy one, 
$\langle z \rangle > 1$. In an Independent Fragmentation (IF) approach one would expect the $z$ 
ratio always to be below unity, and $\langle z \rangle$ to be constant. 

To further zoom in on the role of the environment, now only consider
the $\PQb$~quarks in a narrow bin $0.78 < x_{\PQb} < 0.80$,
near the peak of the $x_{\PQb}$ spectrum. Draw a cone of 
size $\theta = 0.5$ (on the unit sphere) around the $\PQb$~quark. Sum up 
the background energy $E_\text{cone}$, being that of all final
partons inside this cone, except for the $\PQb$ itself, and 
scale it to an $x_\text{cone} = 2 E_\text{cone} / m_{\PZ}$.
Define clean (inside-the-cone) background events as those where
$x_\text{cone} < 0.05$ and dirty background ones as those having
$x_\text{cone}  > 0.15$. Then compare the $x_\text{B}$ spectra for the
clean and dirty cases, \cref{fig:test487}(c). As foretold, the dirty
environment offers a larger pool of nearby energy,  some of which
can be absorbed back into the B~hadron during the hadronisation
process, giving a harder B~hadron than in the clean case.

Unfortunately, \cref{fig:test487}(c) is contingent on knowing the 
partonic state before hadronisation, both that of the $\PQb$~quark and that 
of the surrounding region. It does not directly translate into
observables in the hadronic final state.

\subsubsection{Gluon splitting}

The splitting of gluons into $\bb$ or $\cc$ is only modeled in the perturbative step of the process (not in the string/cluster fragmentation) but still, charm and bottom masses are parameters in the shower. Other relevant differences may be in the treatment of the strong coupling differently for heavy quarks or (massless) gluons in the limit of $\pT=0$. Separation of $h\to$ gluons from $h \to \bb/\cc$ ($\sqrt{s}=m_{\PZ}$ and beyond) is affected.
Details are described in Ref.~\cite{sjostrand_gQQ}.

\subsubsection{Improved hadronisation models}

Hadronisation, the conversion of an entire partonic ensemble of particles with individual colour charges into an observed ensemble of hadrons is a complex process. Within the string or the cluster model, conversion to hadrons takes place after the parton shower and accounts for the local colour connections. colour reconnection can re-distribute momenta in between colour connected subsystems. For the current status of hadronisation and colour reconnection in multi-purpose event generators we refer to the respective release notes and references therein \cite{Bewick:2023tfi,Bierlich:2022pfr,Bothmann:2024wqs}.

The recent development of hadronisation models highlights the interplay with perturbative evolution, in momentum as well as in colour space \cite{Christiansen:2015yqa,Gieseke:2017clv,Gieseke:2018gff,Chahal:2022rid,Altmann:2024odn} and can be confronted with full-$N_c$ parton evolution \cite{AngelesMartinez:2018cfz,DeAngelis:2020rvq}, as well as higher-logarithmic accuracy parton showering \cite{Dasgupta:2020fwr,Forshaw:2020wrq,Nagy:2020rmk,Herren:2022jej,vanBeekveld:2024wws,Preuss:2024vyu} in order to reduce uncertainties of the combined description of infrared sensitive observables. Interesting connections also arise in the context of interpreting the parton shower infrared cutoff as a factorization scale \cite{Platzer:2022jny,Hoang:2024zwl}.

Putting such constraints on hadronisation models raises the question whether they would be flexible enough to describe a wealth of current data, such that these developments should be complemented, and can profit from, machine-learned modeling of the same effect \cite{Ghosh:2022zdz,Ilten:2022jfm,Chan:2023ume,Chan:2023icm,Bierlich:2024xzg}, which will not necessarily provide a direct measure of the theoretical uncertainty but can help to constrain building blocks in newly developed models. A number of additional observables able to constrain hadronisation models have also been discussed at earlier ECFA workshops \cite{ecfa22_krauss} as well as at the recent Parton Shower and Resummation workshop \cite{psr24_kiebacher}. At any rate, hadronisation models predict a richer phenomenology than would be captured by fragmentation functions alone and so should be confronted with existing and novel measurements to further constrain their respective components and the cross talk with modern parton shower algorithms.

\subsubsection{Colour reconnection}

Colour reconnection (CR) is a common name for mechanisms that lead 
to colour flows that are not expected at first sight. For lepton 
colliders the main impact is that seemingly separate colour singlet
systems get to be intertwined. The prime example here is $\PW$ pair
production, $\PWp \PWm \to (\PQq_1\PAQq_2)(\PQq_3\PAQq_4)$, where by 
CR the fragmenting systems instead are $\PQq_1\PAQq_4$ and $\PQq_3\PAQq_2$.
Such effects were observed at LEP 2, favouring a CR rate of approximately 50\%, but unfortunately low statistics did not make for high significance
\cite{ALEPH:2013dgf}. Studies for future $\epem$ colliders can be found 
e.g.\ in Ref.~\cite{Christiansen:2015yca}.

Many different CR models have been proposed. Some common physics 
components used in the modelling, singly or in combination, include:
\begin{itemize}
\item Reduction of string length, cluster masses or some other
``free energy'' measure, whereby also the average hadron multiplicity
is reduced.
\item Physical space--time overlap of the hadronising systems. The further
above threshold a $\PWp\PWm$ pair is, the faster the two move apart, and the 
less the likelihood of overlapping. If nevertheless there is a CR,
the effect on that event could be larger, however.
\item Consistent colour assignments in an $N_C = 3$ setup, whereby
only some reconnections become allowed.
\end{itemize}
Further principles and models may be foreseen in the future, not least
driven by LHC studies.

In addition to primary $\PWp\PWm$ and $\PZz\PZz$ production, 
also Higgs decays to $\PWp\PWm$ and $\PZz\PZz$
pairs can be relevant, e.g.\ to search for any CP-odd admixture in
the Higgs, where CR can affect the angular correlations used to
probe for such effects. The $\PQt\PAQt$ process with hadronic decays
involves three singlets, and thus the possibility of increased effects. 
Effects of CR  with rapidity gaps in hadronic multi-jet events at the $\PZ$ pole have been studied
with LEP data~\cite{OPAL:2003njc,L3:2003ohc,Buschbeck:2002gig,ALEPH:2006jcu},  yielding measurements of CR effects below the predictions of available models.  
Future large $\epem$ data samples at the Z pole  will be very useful to refine such measurements and constrain CR models.

Bose--Einstein effects among final-state hadrons is another area
where colour singlets could get mixed up. No such interconnection
effects were found at LEP 2, but the issue should be kept open.

\subsubsection{Relevance for the physics program of a Higgs/Top/EW Factory}
Jets and in particular heavy-flavour jets play an important role in many of the flagship measurements of Higgs/Top/EW Factories. As examples, we highlight here the connection to other ECFA Focus Topics:

\begin{description}

\item[\boldmath Precise study of $\Ph\rightarrow \Pg\Pg/\bb/\cc$: ] 
Future Higgs Factories will provide sensitivity to these topologies providing capabilities to fully explore the second generation of Yukawa couplings, which is out of reach at the LHC. However, current uncertainties in gluon splitting into heavy quarks would introduce large systematic uncertainties in the measurements. The questions arising are: how to consistently implement gluon splitting in parton shower tools (modeling and free parameters) and how to evaluate the impact of incomplete modeling of the gluon splitting when determining the $\Ph\rightarrow \Pg\Pg/\bb/\cc$  couplings. 
This issue is discussed in  
\cref{sec:HtoSS}.

\item[\boldmath Precise determination of W mass and cross section:] 
$\PW$ mass measurements at future Higgs factories are expected to deliver statistical accuracies at the sub-MeV level exploiting the W-pair threshold cross section lineshape~\cite{LCCPhysicsWorkingGroup:2019fvj, Azzurri:2021yvl}. 
Alternative measurements from the kinematic reconstructuon of \PW-pair decay products 
also have the potential to achieve MeV-level  precision on $m_{\PW}$ , but for this method
the control of systematic uncertainties will be crucial. At LEP2, the modeling of non-perturbative QCD effects in W boson
hadronic decays was a dominant source of systematic uncertainties. Further theoretical and experimental studies are required to estimate the size of such uncertainties at future colliders.
This issue is discussed in  
\cref{sec:Wmass}.

\item[\boldmath \PZ -- b/c couplings:] 
What would be the impact of these uncertainties on the extraction of \PZ--b/c couplings at the \PZ pole?  In Ref.~\cite{AlcarazMaestre:2020fmp} it is demonstrated that hadronisation uncertainties have a significant impact on determinations of the partial widths normalised to total hadronic width ($R_\text{b,c}$), the forward-backward asymmetries ($A_\text{FB}^\text{b,c}$), or left-right asymmetries at $\epem$ colliders, even after application of cuts to reduce their impact. The size of these uncertainties could be a limiting factor when operating at the $\PZ$ pole in the high-luminosity scenarios of FCC-ee. 
This issue is discussed in 
\cref{sec:TwoF}.
\end{description}

\subsubsection{Target physics observables}

A summary of recommended measurements is given in 
Table~\ref{table:bcfrag}. Related observables for $\epem$ and pp are compared.
In practice, the measurements of the observables would be intertwined. 
The following measurements should also be considered:

\renewcommand{\arraystretch}{1.3}
\begin{table}
\rotatebox{-90}{
\small
\begin{tabular}{|p{0.30\textheight}p{0.25\textheight}p{0.35\textheight}|}
\hline
\textbf{Observable} & $\epem$ & pp \\
\hline
\textbf{Event shapes and angular distributions} & & \\
\textbf{Inclusive B/D production cross section} & 
primary production is well known from theory, so any ``excess'' is from gluon splitting &
combines primary production, gluon splitting, and MPI (multiparton interactions) contributions, each with significant theoretical uncertainties \\
\textbf{Flavour composition} as far back in decay chains as can be traced (even equal $D^{*0}$ and $D^{*+}$ rates gives unequal D$^0$ and D$^+$ ones) &
we do not expect sizeable momentum dependence, but it is interesting to contrast mesons and baryons for smaller ones &
significant $\pT$ dependence observed and to be studied further, also high- vs. low-multiplicity events, rapidity, ..., which is important for development/tuning of colour reconnection models \\
\textbf{Particle-antiparticle production asymmetries} &
none expected, except tiny from CP-violation in oscillations &
asymmetries expected and observed from $p$ flavour content, increasing at larger rapidities; relates to how string (and cluster?) fragmentation connects central rapidities to beam remnants \\
\textbf{Momentum spectra} &
$dn/dx_E$ with \mbox{$x_E$ = $2E_\text{had} / E_\text{cm}$}; basic distribution for tuning of ``fragmentation function" &
$dn/d\pT$ and $dn/dy$ give basic production kinematics, but the many production channels give less easy interpretation \\
\textbf{Energy flow around B/D hadrons}, excluding the hadron itself, as a test that dead cone effects are correctly described &
$\text{d}E/\text{d}\theta$ where $\theta$ is the distance from B/D on the sphere &
$\text{d}\pT/\text{d}R$ where $R$ is the distance in ($\eta$, $\phi$) or ($y$, $\phi$) space, only applied for B/D above some $\pT$ threshold \\
\textbf{B/D hadron momentum fraction} of total $E$ or $\pT$ in a jet, with $x = \pT^\text{had} / \pT^\text{jet}$, as a test of the fragmentation function combined with almost collinear radiation, suitably for some slices of $\pT$ (and in addition with a veto that no other B/D should be inside the jet cone, so as to suppress the gluon splitting contribution) &
draw a jet cone in $\theta$ around B/D and measure $x$ &
draw a jet cone in $R$ around B/D and measure $x$ \\
\textbf{B/D~hadron multiplicity}, as a measure of how often several pairs are produced & & \\
\textbf{Separation inside B/D~pairs}, where large separation suggests back-to-back primary production, while small separation suggests gluon splitting &
separation in $\theta$ &
separation both in $\phi$ and in $R$, since for primary production $\phi = \pi$ is hallmark with $\eta/y$ separation less interesting, while gluon splitting means $R$ is small while $\phi$ and $y/\eta$ individually are less interesting \\
\textbf{Hardness difference} within (reasonably hard) pairs, $\Delta = (\pT^\text{max} - \pT^\text{min}) / (\pT^\text{max} + \pT^\text{min})$, where for gluon splitting $x^2 + (1 - x)^2$ translates to $1 + \Delta^2$ &
separately for small or large $\theta$ &
separately for large or small $\phi$ \\[0.5mm]
\hline
\end{tabular}
}
\caption{Target physics observables at $\epem$ and pp.}
\label{table:bcfrag}
\end{table}

\begin{itemize}
\item For a pair with small separation, say $\theta/R < 0.7$, draw a cone around the midpoint of the two, say again $\theta/R = 0.7$, and find the fraction $x = (\pT^\text{had,1} + \pT^\text{had,2}) / \pT^\text{jet}$, to quantify loss to showers and hadronisation. This loss would be reduced by colour reconnection which could combine the $\bb$ or $\cc$ quark pairs into a singlet, rather than the default octet where the two pairs fragment separately.
\item In events with two B/D~pairs, many observables become possible. There are four possible particle-antiparticle pairs (more if B~mixing is considered), each of which can be studied according to the two points above. In addition, a pair with a small separation would suggest a gluon splitting, while one with a large ditto is a primary production. For pp, two back-to-back pairs would suggest MPI. One can try to classify events into most likely history and study the relative composition of (a) two separate hard processes (MPIs, pp only); (b) one hard process and one gluon split; (c) two gluon splits on the same side of the event; and (d) two gluon splits on opposite sides.
\item Even if one B/D is missed in pp collisions, and only three B/D~hadrons are observed, one can study the three pairings and see whether either pair has a small $R$ or a large $\phi$. Again relative rates will provide info on the composition of production mechanisms.
\end{itemize}

\subsubsection{Target detector performance, analysis methods and tools}
\begin{itemize}
    \item Large tracker acceptance as well as very good vertexing and flavour tagging capabilities (including light quarks and gluon quarks)
    \item Jet charge measurements, including charge hadron identification capabilities (see above), and fit a representative set of observables for hadronisation calibration (see above).
    \item Samples for hadronic observables using different hadronisation models and parameters. Full simulation is required to understand flavour tagging capabilities. Existing tools are e.g.\ the generators \textsc{Pythia}~\cite{Bierlich:2022pfr}, \textsc{Herwig}~\cite{Bahr:2008pv,Bellm:2015jjp}, \textsc{Sherpa}~\cite{Sherpa:2019gpd} and the tuning tools \textsc{Professor}~\cite{Buckley:2009bj}, \textsc{Rivet}~\cite{Bierlich:2019rhm}.
    \item Access to LEP Archived Data. LEP data (and simulations) have been partially archived to allow their use for physics analyses after the closure of the collaboration. 
    The use of archived data is authorised to former members of the collaboration and collaborators.  However, the understanding and reprocessing of analysis is still challenging 
    and depends on the safeguarding of the different collaboration's analysis frameworks and mini-data at CERN. 
    Recent efforts \cite{Badea:2019vey} have been driven to re-analyse these data by exporting the data and simulations to more modern and accessible formats, for instance, the MIT Open Data format. A systematic approach for the exportation of such archived data and software tools to the {\textsc Key4hep} environment should be considered by the Higgs Factory community. This would allow the validation of newer calculations and MC tools with existing data.
\end{itemize}    

\subsubsection{Summary and open questions}
\begin{enumerate}
    \item How can the recent progress in the Perturbative FF framework, supplemented by the fits of the non-perturbative component, be implemented and used in practice, e.g.\ in PS Monte Carlo simulations?
    \item What is the quantitative impact of uncertainties from parton-shower, fragmentation and hadronisation on flagship Higgs/Top/EW measurements - and which level of precision will be required? 
    \item Which measurements of particle rates, species, distributions are needed in order to constrain fragmentation and hadronisation models to the required level of precision?
    \item Which detector capabilities are required and to which extent do the proposed detector concepts provide these?
    \item To which extent could LEP data be useful and how could they be made accessible to test new calculations and MC tools? 
\end{enumerate}

\subsection{Determination of fundamental Standard Model parameters \label{sec:otherParams}}
\label{sec:fundamentalparameters}
\editors{Ayres Freitas}
The interpretation of electroweak precision measurements, such as effective Z-bsoon couplings, within the SM or BSM models requires similarly precise measurements of various other parameters: the Fermi constant, $G_\mu$, the top quark and W-boson masses, the strong coupling constant, $\alpha_s$, and the running of the electromagnetic coupling $\Delta\alpha(m_{\PZ}) = 1-\frac{\alpha(0)}{\alpha(m_{\PZ})}$. Improving the precision of the latter two quantities is challenging.

Currently $\alpha_s$ is known with an uncertainty of 0.8\% \cite{ParticleDataGroup:2024cfk}. At future $\epem$ colliders, it can be determined with high statistical precision from measurements of jet variables, tau decays, or \PZ decay branching fractions \cite{dEnterria:2022hzv}. The first two methods are limited by systematic uncertainties from non-perturbative hadronisation effects (see e.g.\ Refs.~\cite{Kardos:2020igb,Pich:2020qna}). On the other hand, the inclusive branching ratio $R_{\Pl} = \Gamma[\PZ\to\text{had.}]/\Gamma[\PZ\to\Pl^+\Pl^-]$, which is sensitive to $\alpha_s$ through higher-order corrections, receives a negligible impact from non-perturbative QCD effects. However, the extraction of $\alpha_s$ from $R_{\Pl}$ assumes the validity of the SM and may be spoiled by possible BSM physics (this would also apply to W-boson branching ratios)~\cite{dEnterria:2020cpv}. High-statistics measurements of $R_{\Pl}$ at $\sqrt{s} \ll m_{\PZ}$ may help to break the degeneracy between $\alpha_s$ and BSM effects. Alternatively, $\alpha_s$ may be obtained from lattice calculations, which are also expected to become significantly more precise in the future \cite{DelDebbio:2021ryq}. Depending on the method used, an uncertainty of 0.3\%--0.1\% for $\alpha_s$ could be achievable in the future, in which case it would not be a limiting factor for electroweak precision tests at HTE factories.

Similar, the uncertainty for $\Delta\alpha$, which is currently about 2\%, would need to be reduced by a factor 3--5 \cite{Belloni:2022due}. Most determinations of $\Delta\alpha$ use data for $R(s) = \frac{\sigma[\epem \to \text{had.}]}{\sigma[\epem \to \mumu]}$ at low energies and perturbative QCD (pQCD) at higher scales. New data from experiments at VEPP, BES and SuperKEKB, 4-loop pQCD and some pQED results, and more precise determinations of $m_{\PQb}$, $m_{\PQc}$ and $\alpha_s$ can help to improve the precision for $\Delta\alpha$ by a factor 2--3 \cite{Jegerlehner:2019lxt}. Note that fourth-order pQCD contributions, including resummation of $\pi^2$-enhanced terms, is already available in the massless quark limit (see e.g.\ Refs.~\cite{Nesterenko:2017wpb,Nesterenko:2019rag}), but only partial results for the quark mass effects are known at this order \cite{Maier:2017ypu}.
Alternatively, $R(Q^2)$ could be determined on the lattice for spacelike $Q^2$ with competitive preicison for $\Delta\alpha$ \cite{Ce:2022eix}, and an improvement by a factor 2--3 is also achievable here in the forseeable future \cite{MeyerWG1PREC}. Finally, one could determine $\alpha(m_{\PZ})$ directly (without using $\Delta\alpha$) from measurements of $\epem \to \mumu$ at values of $\sqrt{s}$ a few GeV below and above the \PZ peak \cite{Janot:2015gjr}. With ${\cal O}$(100 ab$^{-1}$) luminosity, available e.g.\ at FCC-ee, the precision for $\alpha(m_{\PZ})$ can be improved by more than a factor 3 compared to today. This method crucially depends on multi-loop calculations for this process.

\clearpage
\section{Developments in Top Physics \label{sec:top-quark}}
\editors{Marcel Vos, Roberto Franceschini}

All electron-positron collider projects propose to include a rich top physics program, providing a precise characterisation of top-quark properties and interactions. A scan of the centre-of-mass energy through the top pair production threshold at $\sqrt{s} \sim 2 m_{\PQt}$ is foreseen in all projects, with a total integrated luminosity of 100 to \SI{400}{\per\femto\barn}. The circular collider projects moreover envisage a high-luminosity run at somewhat higher energy (\SI{1.5}{\per\femto\barn} at $\sqrt{s} = $ \SI{365}{ \giga\electronvolt} in the FCC-ee program, \SI{1}{\per\femto\barn} at $\sqrt{s} = $ \SI{360}{\giga\electronvolt} for CEPC). CLIC's baseline initial energy stage at $\sqrt{s} = $ \SI{380}{\giga\electronvolt} is above the top pair production threshold, while the linear collider projects all propose runs at higher energies where new associated production processes, such as $\epem \rightarrow \ttbar\PH$ production, open up at around \SI{500}{\giga\electronvolt}. At multi-\SI{}{\tera\electronvolt} energies, electroweak single top production and vector-boson-fusion production of top quarks become relevant.

\paragraph{Motivation for top-quark measurements}

The combination of precision measurements on the $\PZ$-pole at the previous generations of lepton colliders with precise determinations of the $\PW$-boson, Higgs boson and top-quark masses at hadron colliders provide strong constraints on the electroweak sector. This data set tests the relations between the masses of the $\PW$ boson, Higgs boson and the top quark~\cite{deBlas:2022hdk,Haller:2018nnx,Baak:2014ora}. With the current central values of the SM parameters, the electroweak vacuum is meta-stable~\cite{Degrassi:2012ry}, as the Higgs self-coupling is driven negative at high energy scales. 

A new high-energy electron-positron collider will dramatically improve the precision of the key Standard Model parameters. Measurements of top-quark pair production at the pair production threshold, discussed in the next section, yield a top-quark mass with a precision of \SI{50}{\mega\electronvolt}. In combination with precise determinations of the strong coupling (with a precision of 0.1--0.3\%, see \cref{sec:fundamentalparameters}) and $\PW$-boson mass (with a precision of order \SI{1}{\mega\electronvolt}, see \cref{sec:ewqcd_wmass}), the electroweak fit reaches a new level of precision. 

\subsection{\focustopic Top-quark properties from the threshold scan}

A scan of the centre-of-mass energy of an electron-positron collider through the top-quark pair production threshold provides a powerful method of determining top-quark properties~\cite{Gusken:1985nf, Strassler:1990nw, Guth:1991ab}. At $\sqrt{s} \sim 2 m_{\PQt}$ the cross section rises sharply. While the location of the threshold is highly sensitive to the top-quark mass, the shape can be used to determine its width. The cross section in the threshold region is sensitive to the exchange of soft gluons and virtual Higgs bosons, leading to a strong dependence on the strong coupling and the top-quark Yukawa coupling. A measurement of the top-quark pair production cross section --- and potentially other observables~\cite{Martinez:2002st} --- at several centre-of-mass energies around the threshold then gives access to these parameters.

\subsubsection{Predictions for top-quark pair production at threshold}

Precise predictions are available for the threshold region~\cite{Hoang:2000yr, Hoang:2013uda, Beneke:2007zg, Beneke:2015kwa, Beneke:2016kkb}, with non-relativistic QCD (NRQCD) calculations reaching N$^3$LO. While the connection of direct top-quark mass measurements at hadron colliders to the pole mass remains ambiguous~\cite{Hoang:2020iah}, the result obtained from the threshold scan has a clear interpretation in a field-theoretical mass scheme. The calculations are performed in specific short-distance mass schemes, such as the "1S" or "PS" schemes, and the result can be converted to the $\overline{\text{MS}}$ scheme with a loss of precision~\cite{Marquard:2015qpa} that is limited to tens of \SI{}{\mega\electronvolt}. The conversion does require that a precise determination of the strong coupling is available~\cite{Vos:2016til}.

Recent studies of the potential of the threshold~\cite{Seidel:2013sqa,CLICdp:2018esa} employ the \textsc{ QQThreshold} code~\cite{Beneke:2016kkb} to predict the cross section and estimate the theory uncertainty. This code performs calculations at N3LO in non-relativistic QCD (NR-QCD) around the \ttbar threshold, including electroweak, Higgs, and finite-width effects, as well as QED ISR corrections. In the calculation, both resonant and non-resonant diagrams are included. The top-quark mass is defined in the potential-subtracted (PS) scheme~\cite{Beneke:1998rk}.

\begin{figure}
    \centering
  \includegraphics[width=0.7\linewidth]{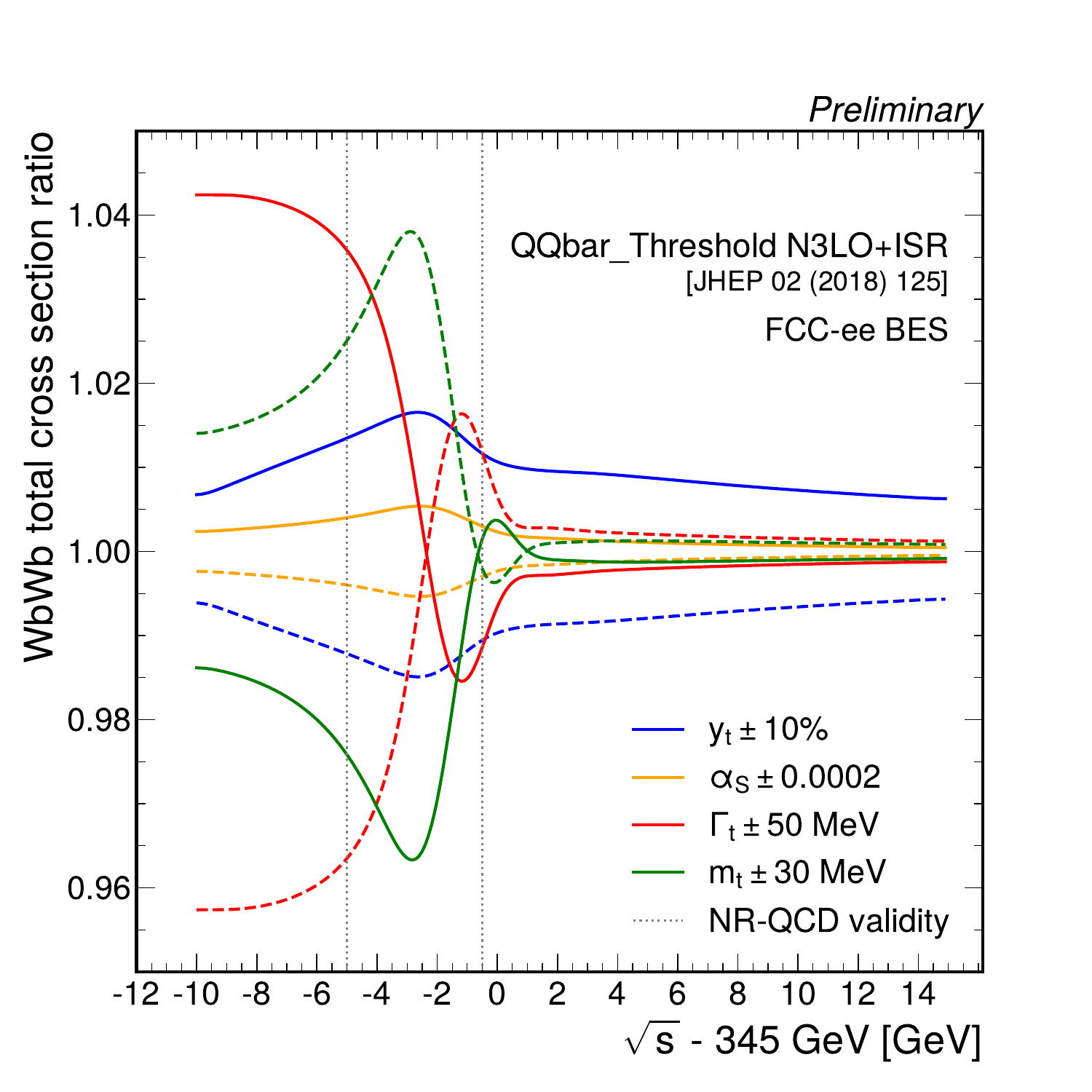}
      \caption{The dependence of the $\epem \rightarrow \PWm\PQb\PWp\PAQb$  cross section on the top-quark mass, width, Yukawa coupling, and $\alpha_\mathrm{s}$ in the region around the top-quark pair production threshold \cite{Defranchis:2025auz}.}
    \label{fig:top_mass_threshold}
\end{figure}

The dependence of the prediction for the $\epem \rightarrow \PWm\PQb\PWp\PAQb$ production cross section on the top-quark mass and width, and the top-quark Yukawa and strong coupling in the threshold region, is shown in Fig.~\ref{fig:top_mass_threshold}. As also seen in previous work~\cite{CLICdp:2018esa,Li:2022iav}, the range between 340 and \SI{345}{\giga\electronvolt} contains the highest sensitivity to $m_{\PQt}$ and $\Gamma_{\PQt}$. In this region, a non-negligible dependence on $\alpha_\mathrm{S}$ and $y_{\PQt}$ is also observed. On the other hand, the cross section at $\sqrt{s} = \SI{365}{\giga\electronvolt}$ is found to depend mildly on $m_{\PQt}$, $\Gamma_{\PQt}$, and $\alpha_\mathrm{S}$, while preserving a sizeable dependence on $y_{\PQt}$ arising from virtual corrections to the $\PZ\ttbar$ vertex. In BSM scenarios, the threshold lineshape could be altered by e.g.\ operators that alter the top-quark electroweak couplings. The sensitivity of the threshold scan to BSM effects remains to be studied.

\subsubsection{Experimental studies}

A new fast-simulation study has been performed on the experimental signal selection \cite{Defranchis:2025auz}. Events for the $\epem \rightarrow \PWm\PQb\PWp\PAQb$ signal are generated with \whizard~\cite{Kilian:2007gr}. The dominant $\epem \rightarrow \PWm\PWp +$~jets background is generated with \pythia. Detector effects are simulated using the \delphes~\cite{deFavereau:2013fsa} parametrisation of the IDEA detector concept at FCC-ee. 

In this study, we consider centre-of-mass energies ranging from 340 to \SI{365}{\giga\electronvolt}. For centre-of-mass energies around the \ttbar threshold an integrated luminosity of \SI{41}{\per\femto\barn} per scan point is assumed, corresponding to 10\% of the total integrated luminosity foreseen for the \ttbar threshold scan. For the \SI{365}{\giga\electronvolt} point, an integrated luminosity of \SI{2.65}{\per\atto\barn} is considered. 

The analysis targets the fully-hadronic and lepton+jets final states decays of the \PW boson pairs excluding hadronic decays of \PGt leptons. This combination of final states corresponds to a total branching fraction of above 80\%. In the lepton+jets final state, electrons and muons with momentum larger than \SI{12}{\giga\electronvolt} are selected. This selection achieves an acceptance of 99.5\% independently of $\sqrt{s}$, limited by the geometrical acceptance of the detector. Leptons are required to fulfil isolation requirements. Events with exactly one isolated lepton are classified as lepton+jets events, events with no isolated leptons as fully-hadronic events. 

An inclusive jet clustering is then performed. The jet multiplicity provides good rejection against the $\PW\PW$ background, in both the fully hadronic and the lepton+jets channel. 
A parametrised simulation of the jet flavour-tagging algorithm is applied to the reconstructed jets in order to identify those arising from \PQb~quarks. The b-tagging efficiency is 90 $\pm$ 1\%. 
In the sample with two \PQb-jets, a BDT is trained to reject the irreducible $\PW\PW\PZ$ background, with $\PZ\rightarrow \PQb\PAQb$ decay. The BDT uses the jet kinematic properties and the invariant mass of the two \PQb-jets. The jet multiplicity distribution for events with one \PQb-tagged jet and the BDT classifier for events with two \PQb-tagged jets are shown for the fully hadronic channel in Fig.~\ref{fig:top_mass_bdt}.

The jet multiplicity distribution and BDT output are used to perform a template-based maximum likelihood fit to extract the signal and background cross sections and the \PQb-tagging efficiency. The fit is performed simultaneously to the two channels. With this configuration, the impact of the $\PWp\PWm$ background on the measured signal cross section rates can be controlled to the per-mille level, well below the statistical uncertainty in the signal. 

Systematic uncertainties corresponding e.g.\ to the \PQb-jet identifications and the background normalisations are controlled in-situ in the fit. Given the high efficiency of the BDT selection, we will assume 100\% efficiency in the signal in the following. It has to be noted that events failing the lepton requirements are included in the hadronic selection, which further justifies the above assumption. Finally, we expect fully-leptonic events and hadronic \PGt decays to be equally well reconstructed at FCC-ee.

\begin{figure}
    \centering
    \includegraphics[width=0.49\linewidth]{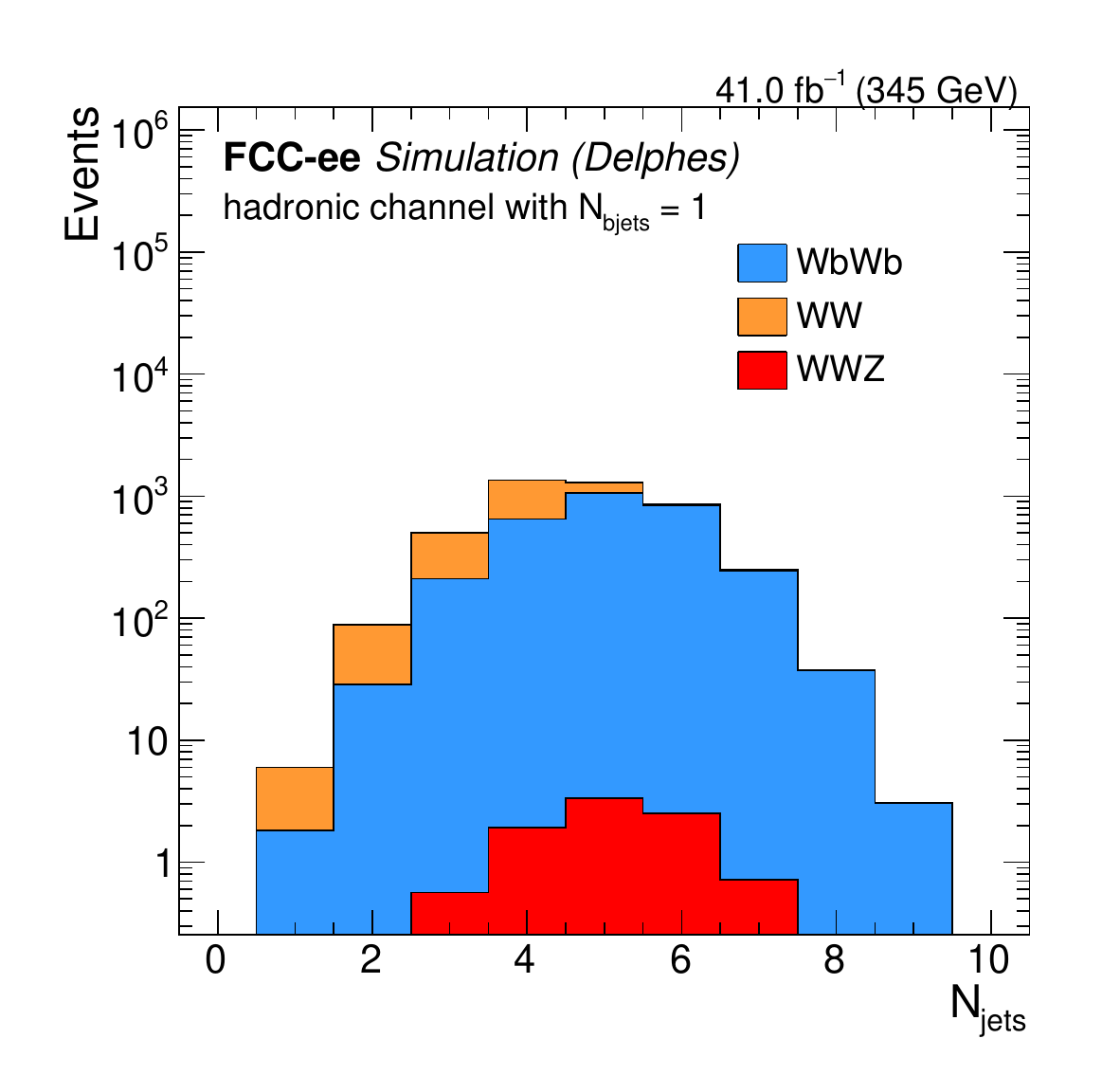}
      \includegraphics[width=0.49\linewidth]{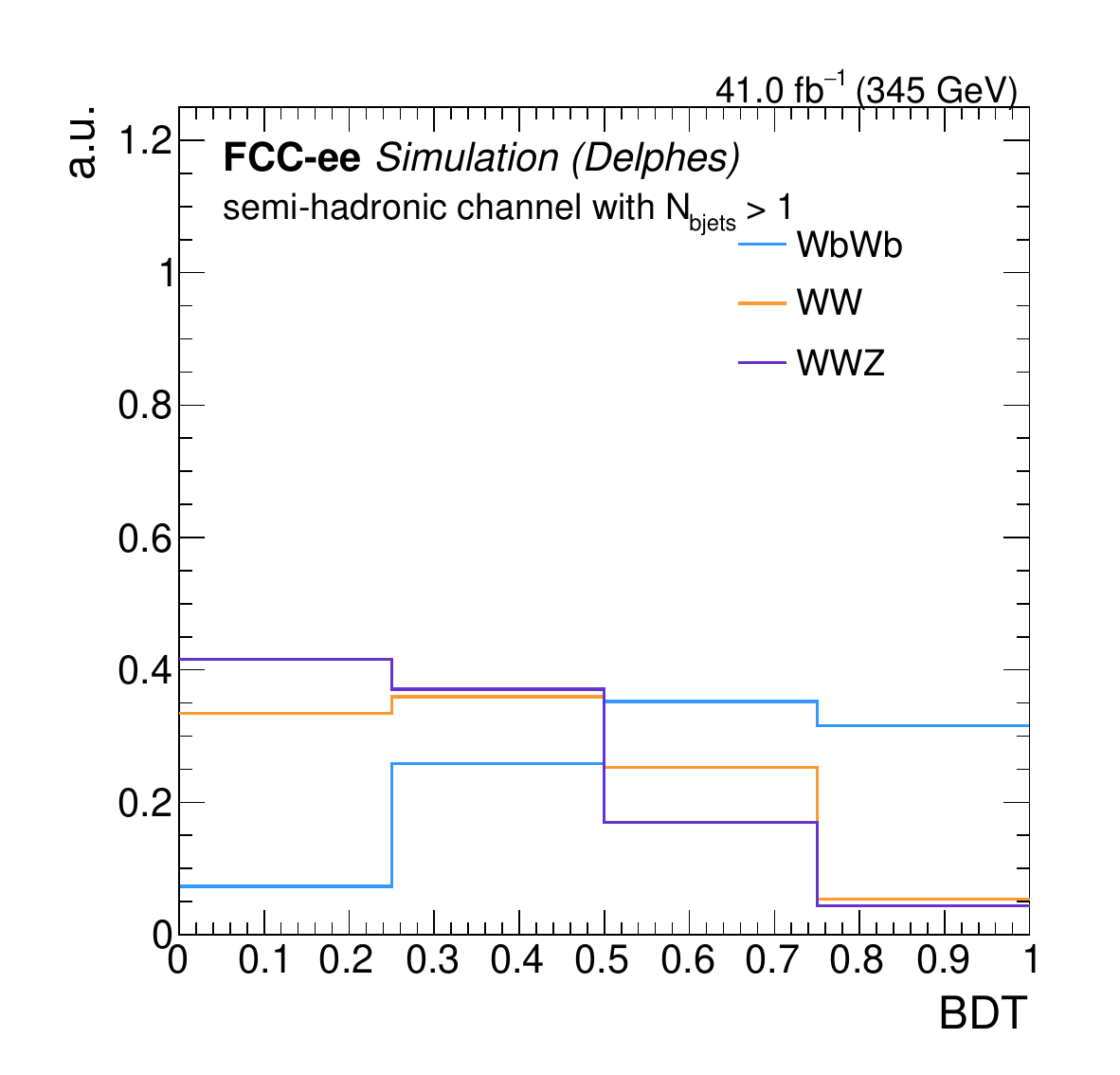}
    \caption{The jet multiplicity distribution for events with one b-tagged jet (left panel) and the BDT output for the events with two b-tags (right panel) that are used to extract the $\epem \rightarrow \PWm\PQb\PWp\PAQb$ cross section at each point of the threshold scan \cite{Defranchis:2025auz}. }
    \label{fig:top_mass_bdt}
\end{figure}

A simultaneous fit of $m_{\PQt}$, $\Gamma_{\PQt}$ is performed on the cross section measurements at the ten centre-of-mass energies. Pseudo-data are generated assuming statistical uncertainties only. The parametric dependence on $\alpha_{\mathrm{S}}$ and $y_{\PQt}$ are included as a nuisance parameter. The expected precision for $\alpha_\mathrm{S}$ at the \PZ boson pole~\cite{dEnterria:2020cpv} is taken as an ancillary measurement that constrains the value of $\alpha_{\mathrm{S}}(m_{\PZ})$ within $\pm 0.0001$. 

Template cross sections are generated as a function of $\sqrt{s}$ for different values of the top-quark PS mass, top-quark width and using the \textsc{QQbar\_Threshold} code~\cite{Beneke:2016kkb}. The calculated cross section is convoluted with the expected beam energy spread (BES) for FCC-ee of 0.23\%. 

The results of the fit are shown in Fig.~\ref{fig:top_machine_systematics}.
The values of $m_{\PQt}$ and $\Gamma_{\PQt}$ are determined with a statistical uncertainty of 4 and \SI{10}{\mega\electronvolt}, respectively. Remarkably, $y_{\PQt}$ is also measured with 1.7\% statistical uncertainty, when the point at $\sqrt{s} = $ 365~\GeV{} is included in the fit and the top EW couplings are assumed to be precisely known. This statistical precision is comparable to that expected for colliders operated above the $\ttbar\PH$ threshold ~\cite{CLICdp:2018esa}. We also observe some additional constraining power on the value of $\alpha_\mathrm{S}$, which is improved by about 15\% compared to the input uncertainty. 

The dependence of the result on the renormalisation scale of the NR-QCD calculation is shown in the second plot on the top row of Fig.~\ref{fig:top_machine_systematics}. The impact of varying the nominal scale by a factor two and a factor one half is found to be of the order of \SI{35}{\mega\electronvolt} (\SI{25}{\mega\electronvolt}) for $m_{\PQt}$ ($\Gamma_{\PQt}$). Parametric uncertainties on the top-quark mass and width from the strong coupling and the top-quark Yukawa coupling, shown on the central row of Fig.~\ref{fig:top_machine_systematics}, range from 2 to 4~\MeV. Considering a precision of $10^{-4}$ on $\alpha_{\textrm S}$ from the $\PZ$ and $\PW$ hadronic branching fraction and a 3\% precision on the top-quark Yukawa coupling from the HL-LHC, the total parametric uncertainty remains below \SI{5}{\mega\electronvolt}. The theoretical uncertainty remains therefore the limiting factor with state-of-the-art theoretical calculations. A dedicated calculation, that \textit{matches} the NRQCD calculation for the threshold region with a fixed-order calculation for the continuum, is necessary to reliably estimate the theoretical uncertainty on $y_{\PQt}$.

\subsubsection{Machine-related systematic uncertainties}

The {\em ideal} threshold scan collects data at a number of sharp values of the centre-of-mass energy. In practice, different types of colliders have different challenges in approximating the ideal threshold scan. In the $\ttbar$ run at circular colliders, beam particles lose a significant fraction of their energy (approximately \SI{9}{\giga\electronvolt} at the FCC-ee) in each turn. The energy is restored by the RF system. As a result of the energy loss in each turn, the energy of the electron and positron beam will differ for the multiple interaction points around the ring. Collisions are slightly asymmetric. The beam energy spread is estimated to be a few per mille (0.23\% at FCC-ee, slightly less at CEPC), approximately Gaussian and independent of the interaction point. At linear colliders that achieve high luminosity by shrinking the beam spot, beamstrahlung is the most important effect. The interaction of the beam particles with the electromagnetic field of the opposing beam causes a long tail towards lower centre-of-mass energy. 

\begin{figure}
    \centering
      \includegraphics[width=0.4\linewidth]{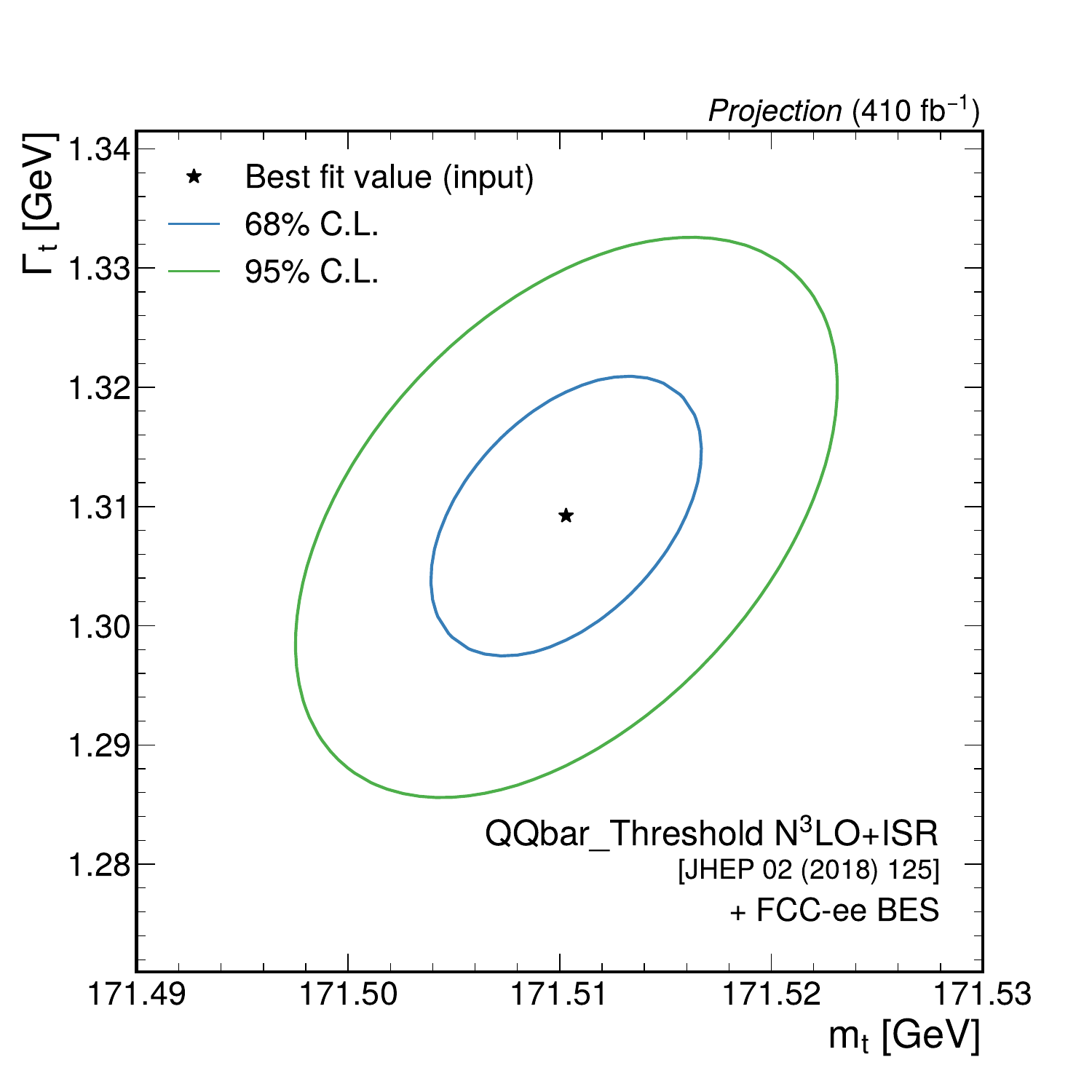}
      \includegraphics[width=0.4\linewidth]{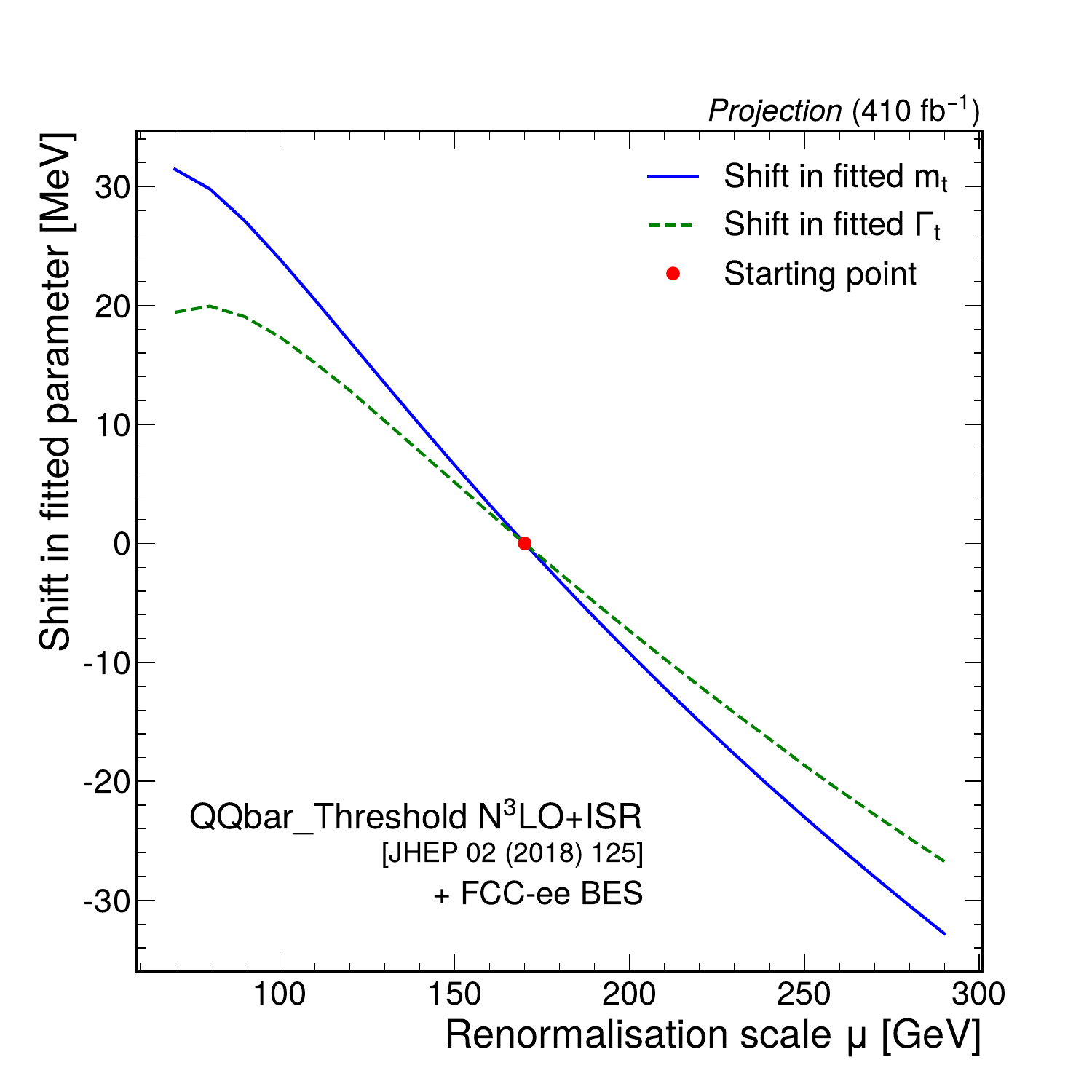}
     \includegraphics[width=0.4\linewidth]{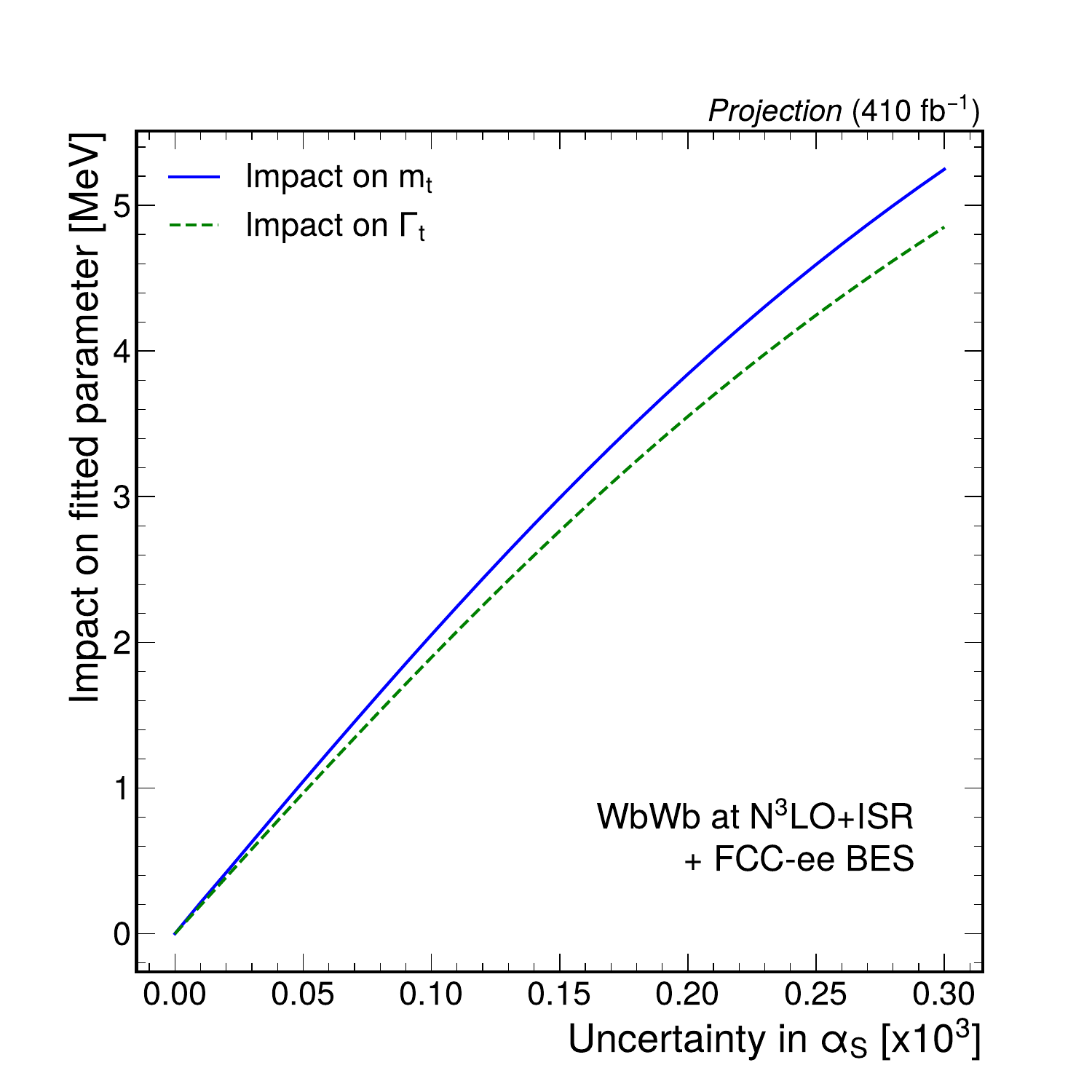}
      \includegraphics[width=0.4\linewidth]{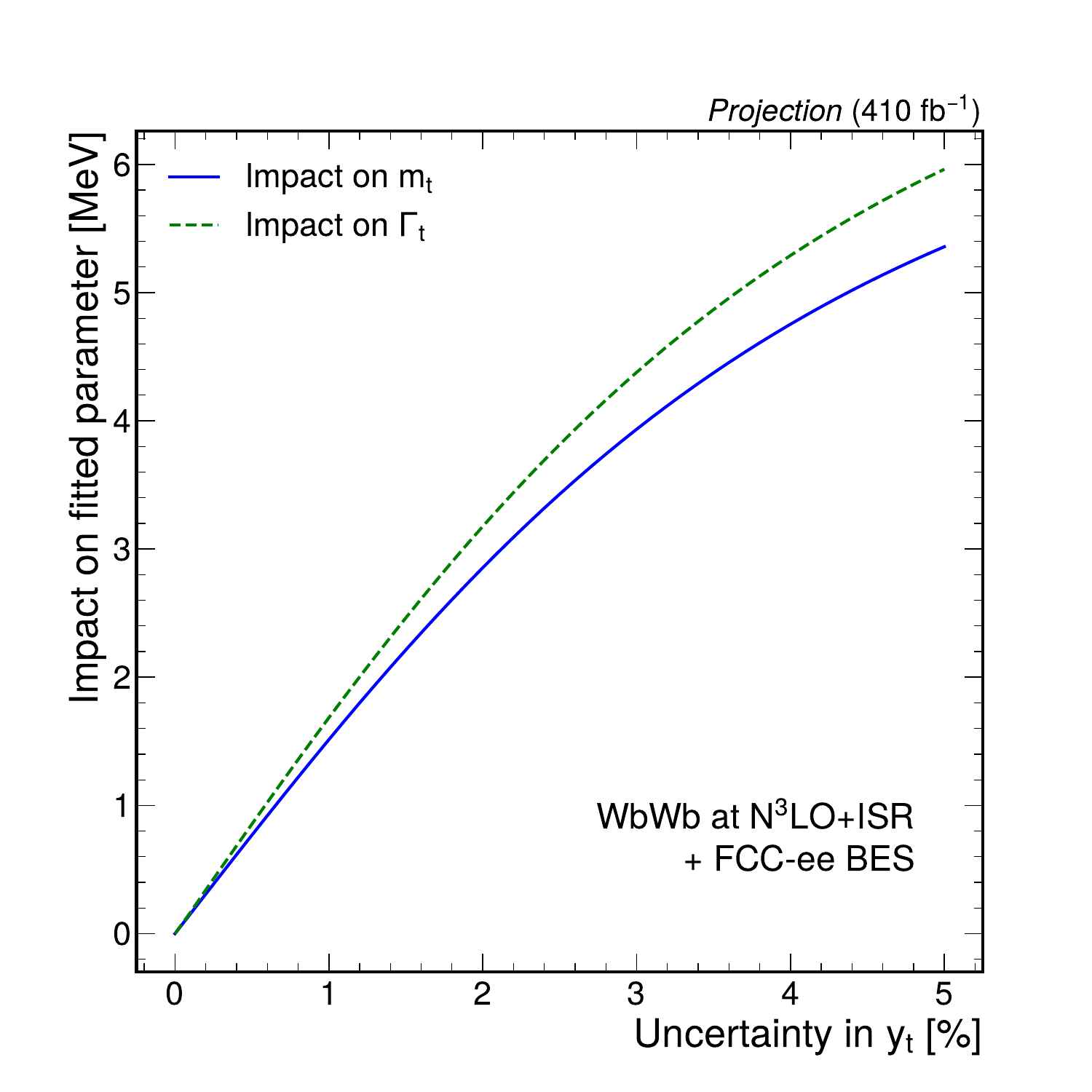}
    \includegraphics[width=0.4\linewidth]{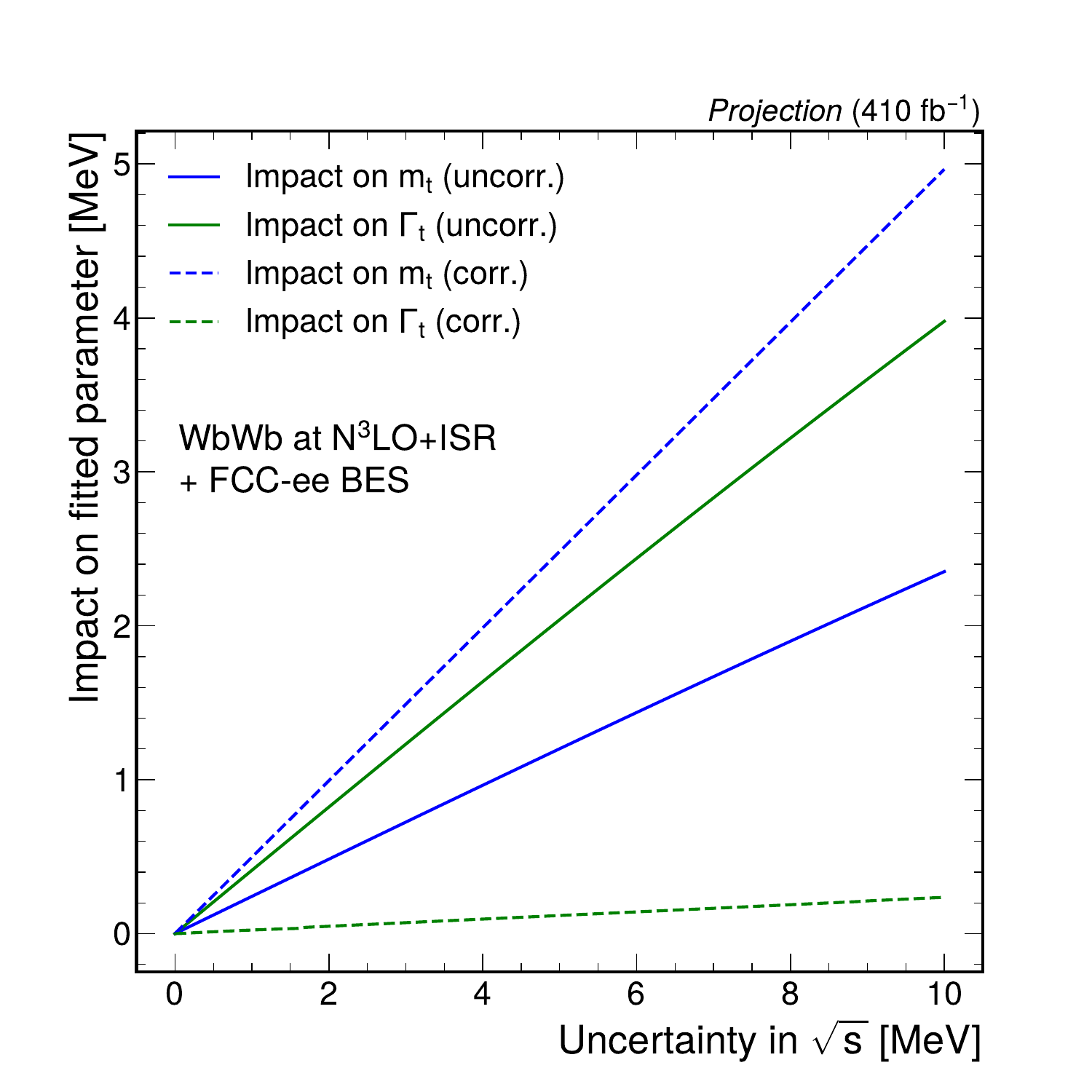}
      \includegraphics[width=0.4\linewidth]{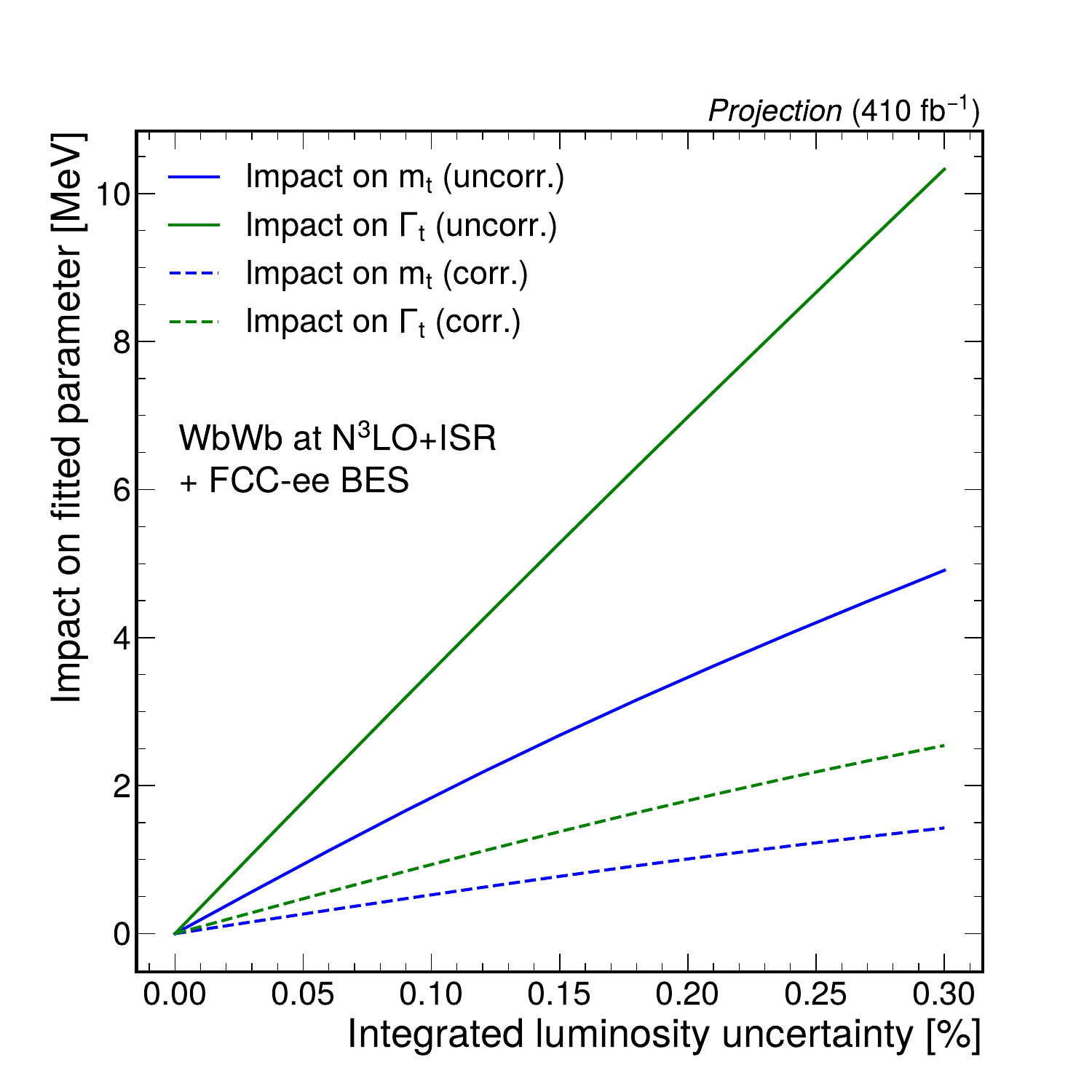}
    \caption{The uncertainties on the top-quark mass and width. The plots on the top row show the statistical uncertainty and the estimate of the uncertainty due to missing high-order corrections from varying the renormalisation scale in the calculation. The central row shows the impact of the external parameters, the strong coupling $\alpha_{\textrm S}$ and the top-quark Yukawa coupling $y_{\PQt}$. The plots on the bottom row present the impact of uncertainties in the beam energy and luminosity calibration \cite{Defranchis:2025auz}.}
    \label{fig:top_machine_systematics}
    \end{figure}
    
For precise top-quark mass and width measurements the luminosity and beam energy at each point of the threshold scan must be precisely known. The impact of the uncertainty of the beam energy and luminosity calibration on the extracted top-quark mass and width are shown in Fig.~\ref{fig:top_machine_systematics}. Uncorrelated uncertainties affect each scan point differently. These uncorrelated sources stem e.g.\ from the limited statistics of the calibration sample at each centre-of-mass energy. Correlated uncertainties are considered that affect the luminosity or beam energy of all threshold scan points in the same way.  These include systematic uncertainties in reference values such as the acceptance for Bhabha scattering or $\PGg\PGg $ events in the luminosity calibration or the $\PW$-boson or $\PZ$-boson mass in the beam energy calibration~\cite{Beguin:2710098, Blondel:2019jmp}. In cases where they have been evaluated, the correlated systematic uncertainties can be controlled to below the uncorrelated, statistical components. 
    
The impact of the beam energy and luminosity calibration is rather small: their contributions of these uncertainties remain sub-dominant up to uncertainties on the beam energy of \SI{15}{\mega\electronvolt} (i.e.\ a relative uncertainty of $1 \times 10^{-4}$) and luminosity uncertainties up to 1 per mille. Both these goals are expected to be achievable.

The {\em luminosity spectrum}, the distribution of centre-of-mass energies of the collisions delivered by the machine, must be known precisely, as well. It can be reconstructed accurately from Bhabha events~\cite{Poss:2013oea} or from $\epem \rightarrow \PGmp\PGmm$ events~\cite{Blondel:2019jmp} and their impact on the mass and width extraction is expected to be sub-dominant~\cite{Simon:2014hna,CLICdp:2018esa}.

\subsubsection{Results for the top-quark mass and width}

\begin{table}[h!]
    \centering
    \begin{tabular}{l|c|c|l}
    uncertainty  & ${m_{\PQt}^{\mathrm{PS}}}$ [MeV] & $\Gamma_{\PQt}$ [MeV] & comment \\ \hline
statistical         & 3.7 & 9.6 & FCC-ee, $410~\mathrm{fb^{-1}}$ \\
b-tagging, background & - & - & controlled in-situ \\
luminosity calibration (uncorr.) &  0.6 & 1.1  & $\delta L/L = 1 \times 10^{-3}$~\cite{Poss:2013oea} \\
luminosity calibration (corr.) &  0.3 & 0.5  & $\delta L/L = 0.5 \times 10^{-3}$ \\
beam energy calibration (uncorr.) & 1.2 &  2.0  & $ \delta \sqrt{s} = 5~\mathrm{MeV}$ \cite{Blondel:2019jmp,Beguin:2710098} \\
beam energy calibration (corr.) & 1.2 &  0.1  & $ \delta \sqrt{s} = 2.5~\mathrm{MeV}$ \\
beam energy spread (uncorr.) &  0.6 & 1.1  & $\delta \Delta E =  1\%$ \cite{Blondel:2019jmp} \\
beam energy spread (corr.) &  $<$ 0.1 & 1.5  & $\delta \Delta E = 0.5\%$ \\
parametric ($\alpha_{\textrm S}$) &  2.0   & 1.9    &  $\delta \alpha_{\textrm S} = 1 \times 10^{-4}$ \\
parametric ($y_\mathrm{t}$) &  3.8   & 4.5    &  $\delta y_{\PQt} = 3\%$~\cite{Azzi:2019yne} \\
\hline
total profiled & 6.2 & 11.3 & \\
theory, unprofiled (scale)       & 35 & 25 & $\mathrm{N^3LO}$ NRQCD~\cite{Beneke:2016kkb} \\ 

    \end{tabular}
     \caption{Uncertainties on the determination of the top-quark threshold mass and width from a scan of the centre-of-mass energy. The uncertainties for the top-quark mass are given in the potential-subtracted scheme implemented in the NRQCD calculation~\cite{Beneke:2016kkb}; the uncertainty on the conversion to the $\overline{\text{MS}}$ mass is not accounted for \cite{Defranchis:2025auz}. }
     \label{tab:top_mass_uncertainties}
\end{table}

The precision for the extraction of the top-quark mass and width from the $\epem \rightarrow \PWm\PQb\PWp\PAQb$ production cross section as a function of the centre-of-mass energy in the region around the top-quark pair production threshold is summarized in Table~\ref{tab:top_mass_uncertainties}. The considered sources include the statistical uncertainty, contributions from theory and uncertainties related to the beam energy and luminosity calibration. For the latter, indicative estimates of the uncorrelated and correlated components have been propagated to the mass and width. Experimental systematic uncertainties from the background normalisation or the b-tagging efficiency are found to be negligible in the approach described above.

The uncertainties on the top-quark mass in Table~\ref{tab:top_mass_uncertainties} are given in the threshold mass scheme that is used in the theory templates. The conversion to the $\overline{\text{MS}}$ mass is not accounted for. With the current state-of-the art calculation to four loops in QCD~\cite{Marquard:2015qpa}, the threshold mass can be converted to the $\overline{\text{MS}}$ mass with a precision of approximately $10$--\SI{20}{\mega\electronvolt}. The contribution to this uncertainty from $\alpha_{\textrm S}$ is \SI{7}{\mega\electronvolt}, given the assumed precision of $\delta \alpha_{\textrm S} = 10^{-4}$.

A \ttbar threshold scan at an electron-positron collider has the potential to measure top-quark properties with unprecedented precision. The final precision is expected to be dominated by the scale uncertainties in the NRQCD prediction. This result underlines once more the need for theoretical advancements to fully profit from the physics potential of future $\Pep\Pem$ colliders.

\subsubsection{Determination of the top-quark Yukawa coupling}
\label{sec:top_mass_yukawa}

The cross section in the threshold region is quite sensitive to the top-quark Yukawa coupling, through diagrams where the top and anti-top quarks exchange a Higgs boson.  Horiguchi et al.~\cite{Horiguchi:2013wra} claimed a statistical precision of 4\% can be achieved on the Yukawa coupling. The effect of the top-quark Yukawa coupling is, however, nearly flat over the threshold region. In a simultaneous fit of multiple parameters, the Yukawa coupling is nearly degenerate with the strong coupling, and both are strongly affected by the theory uncertainty due to missing higher orders.

The addition of a point well above the pair production threshold (\SI{365}{\giga\electronvolt} in this study) may help resolve the degeneracy. The finite electroweak corrections still have a sizable effect, while the $\alpha_{\textrm S}$ dependence due to soft-gluon exchange is expected to be much reduced. A fit that includes a high-precision point well above threshold may then break the degeneracy. A quantitative estimate requires a matched calculation and a careful assessment of the correlations between theory uncertainties. 

Operation of an electron-positron collider around the top-quark pair production threshold offers the opportunity of a competitive measurement of the top-quark Yukawa coupling

\subsubsection{Mass measurements above the top-quark pair production threshold}

Radiative events, where the top-quark pair is produced in association with a photon, allow for a measurement of the top-quark mass~\cite{Boronat:2019cgt} with good precision using the large volume of data collected above the pair production threshold. Measuring the cross section as a function of the energy of the photon, it is possible to isolate events where the top-quark pair has an invariant mass close to twice the top-quark mass, that have excellent mass sensitivity. This measurement has the same unambiguous interpretation as the threshold scan. 

A {\em direct} measurement of the top-quark mass can be performed with good statistical precision~\cite{CLICdp:2018esa}. This measurement is crucial to connect the top-quark mass to the mass parameter of Monte Carlo generators.

At electron-positron colliders operated at a centre-of-mass energy above \SI{1}{\tera\electronvolt} the analysis of boosted top-quark jets enables yet another alternative determination of the top-quark mass~\cite{Fleming:2007qr}. First-principle predictions at particle level are possible for observables like the hemisphere mass that are inclusive in the top-quark decay products~\cite{Bachu:2020nqn}, enabling the extraction of the top-quark mass in a field-theoretical mass scheme with excellent sensitivity.

\subsection{\focustopic Top-quark couplings}

Operation of an electron-positron collider above the top-quark pair production threshold yields direct access to the top-quark couplings to the $\PZ$-boson and the photon~\cite{Amjad:2015mma, Bernreuther:2017cyi,Janot:2015gjr}. The lepton collider measurements complement the top-quark physics program at hadron colliders. The LHC and Tevatron have studied the strong interaction of the top quark in detail in top-quark pair production, and have characterised the charged-current interaction. The couplings to the $\PZ$-boson and photon are accessible only through associated $\Pp\Pp\to \ttbar\PZ$ and $\Pp\Pp\to \ttbar\PGg$ production at hadron colliders and are less precisely measured. The golden channel for a precise and robust determination of the top-quark Yukawa coupling, both at hadron and lepton colliders, is top-quark pair production in association with a Higgs boson ($\Pp\Pp\to \ttbar\PH$ and $\epem \rightarrow \ttbar\PH$). The combination of the HL-LHC and $\epem$ data yields a complete characterisation of the top quark.

The study of Ref.~\cite{Janot:2015gjr} demonstrates the use of energy and angular observables of the charged lepton from top decay to constrain the form factors of the Lagrangian. The final-state information is able to disentangle the photon and $\PZ$-boson components, even when the beam particles are not polarised. This study was revisited recently~\cite{Keilbach2024} with FCC-ee fast simulation, confirming the key finding of the study. Work is ongoing to provide results in the SMEFT basis.

Most recent studies~\cite{Durieux:2018tev, Durieux:2019rbz,Schwienhorst:2022yqu} assess the potential of electron-positron colliders to characterise the top-quark sector of the Standard Model in terms of bounds on the coefficients of dimension-six operators of the Standard Model Effective Field Theory (SMEFT). The values of the Wilson coefficients are constrained in fits to a projected data set, which combines existing ATLAS and CMS measurements extrapolated to the HL-LHC scenario with projections for electron-positron colliders from Ref.~\cite{Durieux:2018tev}. The complementarity of hadron colliders and electron-positron colliders is explicit in this framework. Hadron colliders provide stringent bounds on four-fermion operators with two light quarks and two heavy quarks $\PQq\PAQq\ttbar$. Lepton colliders yield exquisite sensitivity to the two-lepton-two-heavy quark operators ($\epem\ttbar$).    

\subsubsection{Fit to the top sector of the SMEFT}\label{sec:TopSMEFT}


The top physics program at the Tevatron and the LHC yields a precise characterisation of all interactions of the top quark. Its QCD interactions are strongly bounded by high-precision differential measurements in top-quark pair production. The charged-current interactions, through the $\PQt\PW\PQb$ vertex, are studied in top-quark decay and electroweak single top-quark production. Run 2 of the LHC has opened up the study of rare top-quark production processes. Associated production of a top-quark pair or single top quark with a photon or a $\PZ$-boson yields constraints on the top-quark couplings to neutral gauge bosons~\cite{BessidskaiaBylund:2016jvp}, and $\Pp\Pp\to \ttbar\PH$ production is the golden channel for measurements of the top-quark Yukawa coupling~\cite{CMS:2018uxb,ATLAS:2018mme}. In recent years, these analyses have evolved from searches aiming for observation of these SM processes to precise and differential measurements. Finally, four-top-quark production was observed recently~\cite{ATLAS:2023ajo,CMS:2023ftu} and provides constraints on the $\ttbar\ttbar$ vertex. 

The experimental results from the LHC are combined with legacy results from the Tevatron and LEP and SLC results in SMEFT fits of the top sector. Projections by ATLAS and CMS of the HL-LHC program are available for several processes~\cite{Azzi:2019yne}. An extrapolation from run 2 results is made by the IFIT/C~\cite{Durieux:2019rbz} and SMEFiT collaborations~\cite{Celada:2024mcf}. These groups adopt variants of the "S2" scenario of Ref.~\cite{Azzi:2019yne}, where statistical and experimental systematic uncertainties scale with the inverse square root of the luminosity and current theory and modelling uncertainties are scaled by 1/2. The resulting 95\% C.L. bounds on the Wilson coefficients of SMEFT operators involving the top quark are shown in Fig.~\ref{fig:top_smeft_bounds}. A global fit of the Higgs, electroweak and top sectors by the SMEFiT collaboration~\cite{Celada:2024mcf} is discussed in Section~\ref{sec:globalinterpretations}. 

\begin{figure}
    \centering
    \includegraphics[width=0.99\linewidth]{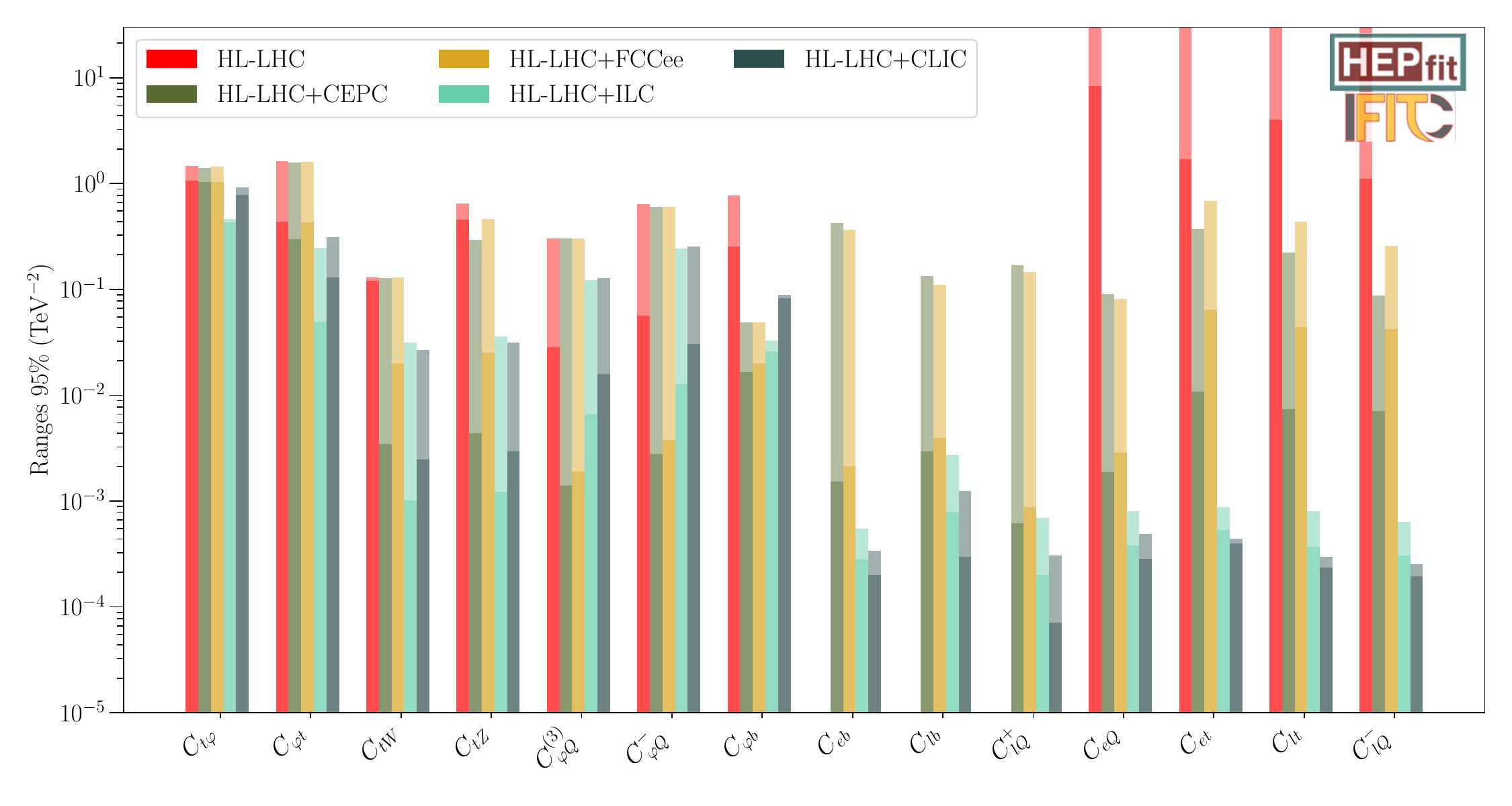}
    \caption{The 95\% C.L. bounds on the Wilson coefficients of SMEFT operators involving top and bottom quarks. The four-fermion operators with two light quarks and two heavy quarks are bounded to $\mathcal{O}(0.2-0.5  \mathrm{TeV}^{-2}) $ by the HL-LHC and are not presented. The FCC-ee and CEPC programmes include runs at the $\PZ$-pole, the Higgs run at $\sqrt{s}=$ 240~GeV, the \ttbar{} threshold scan and runs at \SI{360}{\giga\electronvolt} (CEPC) or \SI{365}{\giga\electronvolt} (FCC-ee). The ILC programme includes runs with polarised beams at \SI{250}{\giga\electronvolt}, \SI{500}{\giga\electronvolt} and \SI{1}{\tera\electronvolt}, while for CLIC runs are foreseen at \SI{380}{\giga\electronvolt}, \SI{1.5}{\tera\electronvolt} and \SI{3}{\tera\electronvolt}. The  integrated luminosities envisaged in each project are given in \cref{sec:runplans}. Figure based on Ref.~\cite{IFITC2024}.  }
    \label{fig:top_smeft_bounds}
\end{figure}
The expected 95\% CL confidence bounds on the Wilson coefficients of dimension-six operators involving top and bottom quarks are shown in Fig.~\ref{fig:top_smeft_bounds} from Ref.~\cite{IFITC2024}. The full length of the vertical bars represent the result of a global fit, where all 29 degrees of freedom are varied simultaneously. The section with a darker shading corresponds to the individual bounds, where all coefficients except one are set to 0. The 14 coefficients of $\PQq\PAQq\PQt\PAQt$ operators do not improve beyond HL-LHC and are not shown here. The remaining 14 coefficients are grouped into four two-fermion operators that exclusively affect top-quark production ($C_{\PQt \varphi}$ through $C_{\PQt\PZ}$ in the leftmost part of the plot), three $\epem\PQb\PAQb$ operators ($C_{\Pe\PQb}$, $C_{\Pl\PQb}, C_{\Pl Q}^+$) and three $\epem\PQt\PAQt$ operators ($C_{\Pe\PQt}$, $C_{\Pl\PQt}, C_{\Pl Q}^-$). Three operators, $C_{\varphi Q}^{(3)}$, $C_{\varphi Q}^-$ and $C_{\Pe Q}$, affect both top-quark pair and bottom quark pair production.

The first, red bar presents the results of the "S2" projection for the HL-LHC with an integrated luminosity of \SI{3}{\per \atto \barn}. Most coefficients improve by a factor two to four in comparison with the current bounds based on LHC run 2 data supplemented by Tevatron and LEP/SLD legacy results. In the "S2" scenario progress at the HL-LHC is limited by the uncertainty on the SM prediction and modelling uncertainties.

The subsequent vertical bars show the impact of adding $\epem$ data to the HL-LHC results, in several electron-positron collider scenarios: dark green for CEPC, orange for FCC-ee, teal for ILC and dark grey for CLIC. 

Top-quark pair production at lepton colliders yields excellent sensitivity to the SMEFT operators that alter the top-quark couplings to the neutral gauge bosons, which are only constrained through associated production ($\Pp\Pp\rightarrow \PQt\PAQt\PX$ and $\Pp\Pp \rightarrow \PQt\PX\PQq$, with $\PX=\PGg,\PZ,\PH$) at hadron colliders. The couplings to the $\PZ$-boson and the photon can be disentangled using polarised beams~\cite{Amjad:2015mma} or the final-state polarisation~\cite{Janot:2015gjr}. The constraints from the $\epem \rightarrow \ttbar$ data are expected to lead to a marked improvement of the bounds on most of the purely bosonic and two-fermion operators entering the SMEFT fit, by up to two orders of magnitude in some cases~\cite{IFITC2024,Celada:2024mcf}.

The four-fermion operators with two charged leptons and two heavy quarks are very poorly constrained by the LHC experiments. Fits based on a parametrisation including $\Lambda^{-4}$ terms can yield bounds of $\mathcal{O}$ (1~TeV$^{-2}$)~\cite{CMS:2020lrr}, these bounds rely heavily on the dimension-six-operator-squared terms. In a fit including only $\Lambda^{-2}$ terms, fits on LHC data cannot currently provide global bounds. The best bounds on these operators are obtained at lepton colliders operated at very high energy -- up to \SI{1}{\tera\electronvolt} with super-conducting RF cavities, up to several \SI{}{\tera\electronvolt} with a drive-beam scheme, and to several tens of \SI{}{\tera\electronvolt} in a muon collider or wakefield facility - provide the most stringent bounds on $\epem\PQt\PAQt$ operators, as the sensitivity increases strongly with energy.

\subsubsection{The top-quark Yukawa coupling}

The Yukawa coupling of the top quark, of order one in the Standard Model, is arguably one of the most interesting parameters of the theory. The "golden" mode to determine its value at a hadron collider is $\Pp\Pp\to \ttbar\PH$ production. The current precision is of the order of 10\%. Projections by ATLAS and CMS envisage a precision of about 3\% at the end of the HL-LHC program~\cite{Azzi:2019yne}. These correspond to the "S2" scenario, that envisages a substantial decrease of statistical uncertainties and experimental systematic uncertainties and a more modest reduction of theory and modelling uncertainties. 

Other LHC analyses take advantage of the dependence on the top-quark Yukawa coupling of EW diagrams in $\Pp\Pp\to \ttbar\ttbar$ production~\cite{ATLAS:2023ajo} and loop level contributions to top-quark production~\cite{CMS:2020djy}. A combination of $\ttbar\PH$ data with these alternative determinations yields a value of the Higgs width~\cite{Cao:2016wib,ATLAS:2024mhs}. The top-quark Yukawa coupling can be determined indirectly in $\ttbar$ production at $\epem$ colliders, too, with the best sensitivity close to threshold, as discussed in Section~\ref{sec:top_mass_yukawa}. 

The Higgs factory stage of a future lepton collider provides sensitivity through Higgs decays that proceed through top-quark loops, as discussed in Section~\ref{sec:globalinterpretations} on global interpretations. Associated production of a top-quark pair and a Higgs boson ($\epem \rightarrow \ttbar\PH$) requires a centre-of-mass energy greater than \SI{500}{\giga\electronvolt}. Full-simulation studies have been performed by Price et al.~\cite{Price:2014oca} and the CLIC detector and physics group~\cite{CLICdp:2018esa}. These results have been extrapolated to updated operating scenarios by e.g.\ Ref.~\cite{deBlas:2022ofj}, yielding the projections of Table~\ref{tab:top_yuk}. A global fit result is not available for FCChh, SPPC and the muon collider. A detailed detector study remains to be performed for these projects. For reference, the table includes the result of phenomenological studies into the power to constrain the top-quark Yukawa coupling of the $\Pp\Pp \rightarrow \PQt\PAQt\PH$ production process at a hadron collider operated at $\sqrt{s} = $ \SI{100}{\tera\electronvolt}~\cite{Mangano:2015aow} and of $VV \rightarrow \PQt\PAQt$ production at a  \SI{10}{\tera\electronvolt} muon collider~\cite{Liu:2023yrb}.

\begin{table}[]
    \centering
    \begin{tabular}{c|c|c|c|c|c|c|c|c|c|}
    \multicolumn{2}{c|}{ Values in \% units }   & LHC  & HL-LHC & ILC500 & ILC550 & ILC1000 & CLIC & FCChh & $\mu$-coll\\\hline
    \multirow{2}{*}{$\delta y_{\PQt}$} & Global fit  & 12\% & 5.1\%   & 3.1\%   & 2.6\%   & 1.5\%   & 3.0\%  & - & - \\
                                  & Indiv. fit  & 10\% & 3.7\%   & 2.8\%   & 2.3\%   & 1.4\%   & 2.5\% & 1\% & 1.5\% \\
    \end{tabular}
    \caption{
    Uncertainties for the top-quark Yukawa coupling at 68\% probability for different scenarios, in percentage. The ILC500, ILC550 and CLIC scenarios also include the HL-LHC. The ILC1000 scenario includes also ILC500 and HL-LHC. Numbers for lepton colliders are based on an extrapolation in Ref.~\cite{deBlas:2022ofj} of detailed studies in Refs.~\cite{Price:2014oca,CLICdp:2018esa}. The FCChh and muon collider projections are based on Refs.~\cite{Mangano:2015aow,Liu:2023yrb}.}
    \label{tab:top_yuk}
\end{table}


\subsection{\focustopic  Exotic top decays \label{sec:extt}}
\editors{Roberto Franceschini}
\paragraph{Predictions for top-quark FCNC rates in BSM models}
\subsubsection{\texorpdfstring{$\PQt \to \PQc \PZ$ in Randall--Sundrum}{t->cZ in Randall--Sundrum} \label{sec:top-quark-RS}}


\paragraph{Flavour violation in Randall--Sundrum}
In the Randall--Sundrum (RS) model, all SM model particles are identified as zero modes of the 5D fields used in this model. This  can be interpreted as the SM fields being the lowest energy solutions of the ``particle in a box'' quantum mechanics textbook problem extended to 5D and with a non-constant potential. Drawing from the same analogy, all SM particles have higher excitations, usually referred to as Kaluza--Klein (KK) modes, which appear around some scale of new physics scale that we denote as $M_{\text{KK}}$. KK modes corresponding to the $\PZ$ boson ($\PZ_{KK}$) in general have flavour violating couplings to the SM fermions in the mass eigenbasis in the RS model. Due to mixing of the SM $\PZ$ boson  with the $\PZ_{KK}$ after EWSB, tree-level flavour violating couplings of the top quark arise in the RS model. This leads to a flavour violating decay of the top quark into a $\PZ$ boson and a charm quark, which was calculated in Ref.~\cite{Agashe:2006wa} and depends on $M_{\text{KK}}$ as follows,
\begin{equation}
    \textrm{Br}( \PQt \to \PZ \PQc) \sim 10^{-5} \left( \frac{3 \, \textrm{T} \electronvolt}{M_{\text{KK}}}\right)^4 \left (\frac{(U_R)_{23}}{0.1} \right)^2, \label{eq:airen-BRt2Zc-RS}
\end{equation}
where $(U_R)_{23}$ parametrises the degree of flavour violation in the couplings of $\PZ_{KK}$. In the minimal model with no flavour symmetries, $(U_R)_{23}$ is completely fixed to an $\mathcal{O}(1)$ factor \cite{Agashe:2004cp}
\begin{equation}
    (U_R)_{23} \sim \frac{m_c}{m_t \lambda_{CKM}^2} \approx 0.1,
\end{equation}
where $m_{\PQc}$ and $m_{\PQt}$ are the masses of the charm and the top, and $\lambda_{CKM}$ is the Wolfenstein parameter for the CKM matrix. Therefore, the prediction for $\textrm{Br}( \PQt \to \PZ \PQc)$ only depends on the scale $M_{\text{KK}}$, more specifically, on the mass of $\PZ_{KK}$, $M_{\PZ_{KK}}$. 

The mixing of SM gauge bosons with gauge KK modes also leads to other consequences in flavour observables and beyond. In the 2013 Snowmass study~\cite{TopQuarkWorkingGroup:2013hxj} the prediction of RS for $\PQt \to \PZ \PQc$ was the largest among the models considered for this final state. This prediction was obtained by setting $M_{KK}$ to $3~\TeV$, which was the lowest value consistent with the electroweak precision data from LEP. Here we consider bounds from direct searches for the KK modes at the LHC with the goal to update the prediction for the BR in \cref{eq:airen-BRt2Zc-RS} in light of the LHC searches for RS. At the end of this section we also comment briefly on the impact of the improved electroweak precision data from the HTE factory.

\paragraph{Randall--Sundrum searches at the LHC}

In the minimal RS model, $M_{\text{KK}}$ corresponds to a common mass scale where  \emph{all} the KK modes of the first level appear. 
Under the assumption of a common mass scale one can put  bounds exploiting the most visible KK mode, e.g.\ the gluon KK $g_{\text{KK}}$, to infer a limit on the KK of the $\PZ$ boson. However, the RS model admits considerable mass differences between the KK modes, and it is possible to separate the scales of various KK modes with some moderate model building~\cite{Agashe:2017wss} without altering the core features of the model. Exploiting this freedom, the KK states that give rise to the most visible signatures, e.g.\ the KK gluon, can be made heavy enough to become irrelevant to gain information on the $\PZ$ KK that generates the FCNC interactions. 
Other states, such as the KK of the $\PW$ boson, are more tightly related to the KK of the $\PZ$ due to these states forming a triplet under the weak $SU(2)$. 
For this reason we pursue bounds on both $\PZ$ and $\PW$ KK modes from CMS and ATLAS searches. 

In RS the $\PW$ and $\PZ$ KK resonances decay mostly into $\PW$ and $\PZ$ bosons and Higgs bosons, giving rise to VV and VH final states, in addition to a usually sub-leading $\PQt\PAQt$ final state. The branching ratio into a specific final state depend on $M_{\text{KK}}$ and other model choices discussed in Refs.~\cite{Agashe:2008jb, Agashe:2007ki}. For this reason, we consider a range of BR values for each decay mode. 
Also, we found that the $\PQt\PAQt$ final state gives very weak bounds; henceforth we only focus on VV and VH final states. 

CMS and ATLAS have performed direct searches for a triplet of vector bosons \cite{CMS:2022pjv, ATLAS:2020fry, ATLAS:2022enb} decaying into VV and VH final states  and have found no indication of new physics. The limits reported do not directly apply to Randall--Sundrum models, as they   use a simplified model, the Heavy Vector Triplet (HVT) model~\cite{Pappadopulo:2014qza}, to report their results. A variant of this model called HVT-B mimics KK electroweak resonances in RS model quite closely. There a few important differences between the two models that are worth noting. 
In HVT-B, the branching fraction for decay into VV and VH are fixed. Each decay mode BR reflects just the number of degrees of freedom its final state contains. On the contrary, in RS the branching fractions can favor a particular final state depending on the model parameters~\cite{Agashe:2008jb, Agashe:2007ki}. Additionally, the couplings of the new vectors to SM fermions, especially to the top quark, is quite model dependent. To assess the impact of these results on the RS model, we recast the results from CMS and ATLAS using full Run2 data, corresponding to about $\SI{138}{\per\femto\barn}$ luminosity,  taking into account these differences between the models.

\Cref{fig:airen-RS_limits} shows results of our recast~\cite{Airen:2025aaa} using the limits from the CMS diboson search Ref.~\cite{CMS:2022pjv}.  ATLAS results are comparable, but slightly less stringent, and we do not show them here. Results from searches of $\ttbar$ resonances put less stringent bounds, and we do not consider them here. We find that the best limits in the CMS analysis are set by production of $\PW_{KK}$ and subsequent decay into $\PW \PH$ final state. The bound we obtain from Run2 results is roughly $\SI{3.7}{\textrm{T}\electronvolt}$. A naive extrapolation based on the improvement with luminosity comparing with partial Run2 data set analysis leads us to expect this bound to improve to $\SI{4.5}{\textrm{T}\electronvolt}$ for a single experiment with the   luminosity $\SI{3000}{\per\femto\barn}$ of HL-LHC, putting the combined limit from two experiments well above $\SI{5}{\textrm{T}\electronvolt}$.  Taking this expected bound on $M_{\text{KK}}$ at HL-LHC into account, the upper bound  from \cref{eq:airen-BRt2Zc-RS} is $\textrm{Br}( \PQt \to \PZ \PQc) \lesssim 10^{-6}$. Thus, in the RS model one should expect to observe less than one event of $\PQt \to \PZ \PQc$ per million $\ttbar$ pairs produced at the HTE factory.

\begin{figure}
    \centering
    \begin{subfigure}{0.4\textwidth}
        \includegraphics[width=\textwidth]{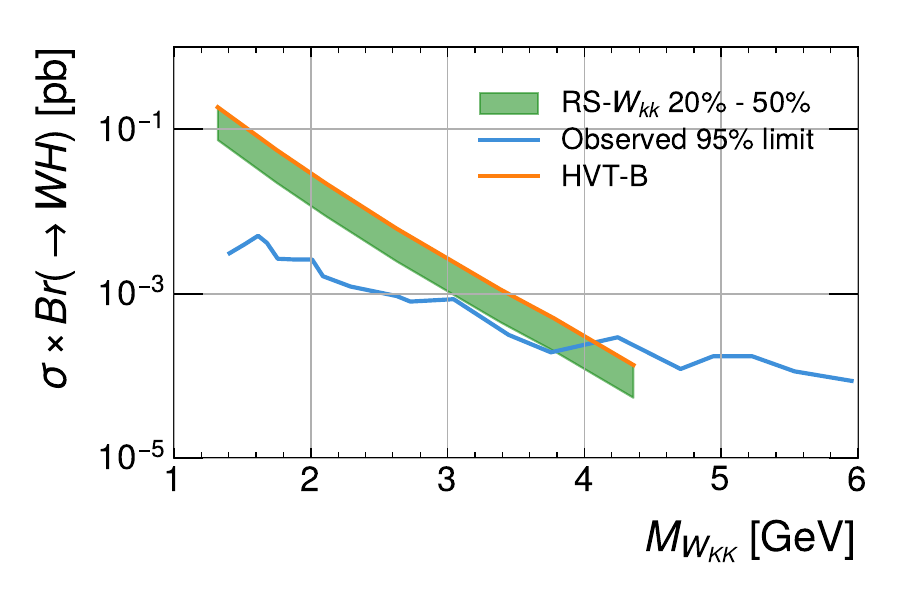}
    \end{subfigure}
    \hspace{1cm}
    \begin{subfigure}{0.4\textwidth}
        \includegraphics[width=\textwidth]{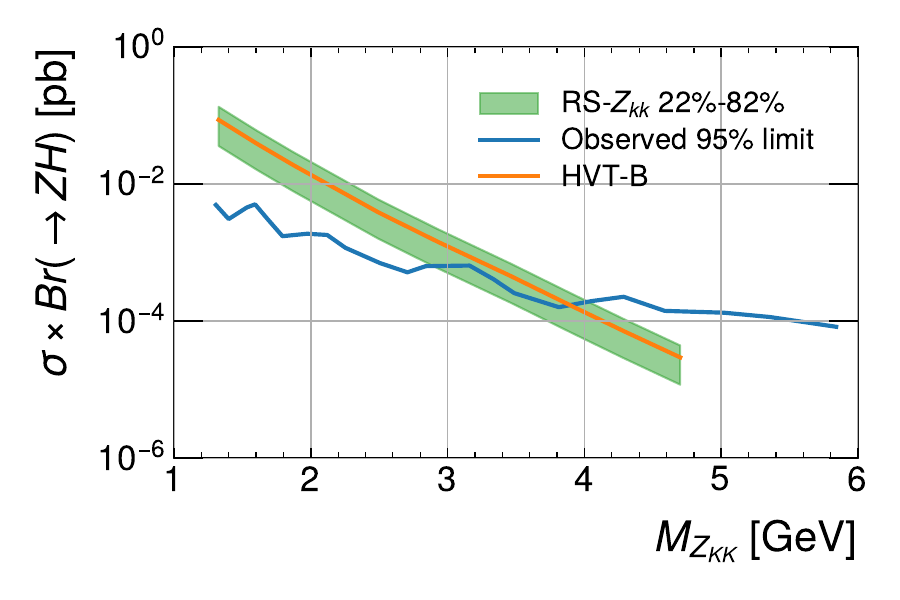}
    \end{subfigure}
    \caption{Recast results \cite{Airen:2025aaa} using $\SI{138}{\per\femto\barn}$ data from the CMS diboson search Ref.~\cite{CMS:2022pjv}. Blue lines indicate the observed cross-section limits at 95\% C.L.\,. Orange lines show the prediction for the cross-section in the HVT-B model. The green shaded regions show the prediction for the cross-section in RS. The width of the band captures the variability in the branching fractions that exists in the RS model. The left panel shows results using the $\PW \PH$ final state. The right panel  shows slightly less stringent bounds from the $\PZ \PH$ final state. 
    }
    \label{fig:airen-RS_limits}
\end{figure}

\paragraph{Before the top factory}

The   $\PZ$ boson coupling responsible for $\PQt \to \PZ \PQc$, can also lead to single top production  
\begin{equation}
    \Pep \Pem \to \PQt \PAQc + c.c.
\end{equation}
 at the Higgs factory stage at \SI{240}{\giga\electronvolt} or \SI{250}{\giga\electronvolt} through a $\PZ$ in the s-channel. 

Expected limits on FCNC couplings of the top quark have been calculated using the SMEFT framework~\cite{Aguilar-Saavedra:2018ksv} at the Circular Electron Positron Collider (CEPC) \cite{Shi:2019epw}. The analysis pursued in that work is expected to be valid also for the Higgs/electroweak/top factory. 
Indeed, we were able to reproduce limits on the relevant combinations of new physics couplings studied in Ref.~\cite{Shi:2019epw}. In place of  a \delphes~\cite{deFavereau:2013fsa} parametrisation of the CEPC detector we used cross-section computed at LO in perturbation theory with \mgfive \cite{Alwall:2014hca} applying a smearing of momentum applied on partonic level events. We broadly confirm results of \cite{Shi:2019epw} and anticipate that partonic level simulations can be used for exploration of the potential of single top production at the $\PZ \PH$ threshold stage of the Higgs factory. In addition we explored the reach at other possible centre of mass energies, finding interesting results, which   will be presented in forthcoming works~\cite{Airen:2025aaa, Agashe:2004cp}. 

In RS models,  out of the several that have been studied in SMEFT in Ref.~\cite{Shi:2019epw}, only the following CP-preserving two-fermion operators are generated at tree level~\cite{Agashe:2003zs}:
\begin{equation}
    O_{\phi q}^{1(ij)} = (\phi i \overset{\text{$\leftrightarrow$}}{D_\mu} \phi)(\bar{q_i}\gamma^\mu q),
\end{equation}
\begin{equation}
    O_{\phi q}^{3(ij)} = (\phi i \overset{\text{$\leftrightarrow$}}{D^I_\mu} \phi)(\bar{q_i}\gamma^\mu \tau^I q),
\end{equation}
\begin{equation}
    O_{\phi u}^{(ij)} = (\phi i \overset{\text{$\leftrightarrow$}}{D_\mu} \phi)(\bar{u_i}\gamma^\mu u).
\end{equation}

The bounds on the coefficients FCNC operators can be translated into a corresponding $\textrm{Br}( \PQt \to \PZ \PQc)$   to make it easier to compare the reach of single top production and the search for FCNC top decays. We find that the $\PZ\PH$ threshold Higgs factory can probe $\textrm{Br}( \PQt \to \PZ \PQc)$  up to $6\times 10^{-6}$
for $\SI{5.6}{\per\atto\barn}$. 

This demonstrates a reach for FCNC $\PQt \PZ \PQc$ interactions at the $\PZ\PH$ threshold at least comparable to what the top factory can probe. Motivated by this finding we have made a preliminary study of the sensitivity of single top production at centre of mass energies below $\sqrt{s} = \SI{240}{\giga \electronvolt}$. Assuming a luminosity that changes as $\mathcal{L}\sim E^{-4}$ we find that sensitivity to single top is retained  all the way to its threshold, showing a sensitivity on-par with the sensitivity at \SI{240}{\giga \electronvolt} for a centre of mass energy around $\SI{200}{\giga \electronvolt}$.

In addition to the sensitivity from single top production, we highlight that significant information  on the sources of the FCNC interactions in the RS model is expected to come from the $\PZ$ pole run, which is expected to significantly improve precision on electroweak observables and could possibly give hints for, or  new  bounds on, the KK mass scale. Currently available prospects~\cite{deBlas:2019rxi} on electroweak observables, e.g.\ on the $\hat{S}$ parameter,  show sensitivity to $M_{\text{KK}}$ around \SI{11}{\tera \electronvolt}, which is equivalent to a branching ratio around $5 \times 10^{-8}$ according to \cref{eq:airen-BRt2Zc-RS}. 

 \paragraph{Discussion and outlook \label{sec:airen-conclusions-RSfcnc}}

 The LHC bounds on vector KK excitations of the weak gauge bosons have surpassed the indirect limits from LEP and are expected to be sensitive to a mass scale for the KK $\PW$ and $\PZ$ in excess of  $\SI{5}{\textrm{T}\electronvolt}$ considering combinations of bounds from ATLAS and CMS experiments at the HL-LHC. The suppression on top quark FCNC that is entailed by these bounds results in an expected rate of less than one $\PQt \to \PZ \PQc$ decay per million $\ttbar$ pairs produced at the HTE factory. The same coupling involved in this decay can be studied in single top-quark production $\Pep \Pem \to \PQt \PAQc + c.c.$ leading to comparable or stronger sensitivity than that of decays. Naturally, the study of decays at the top factory retain sensitivity also to decay modes that cannot be probed by the $\Pep \Pem$ initial state, e.g.\ decays in gluons, studied in \cref{sec:top-quark-twohiggs}, and decay modes into BSM states studied in \cref{sec:fcnc-bsm-search}.

\subsubsection{\texorpdfstring{$\PQt \to \PQc \Pg$ in 2HDM}{t->cg in 2HDM} \label{sec:top-quark-twohiggs}}
\paragraph{2HDM links with flavour}
In a general model with two Higgs doublets it is possible to couple both doublets to all species of fermions of the Standard Model. This in general leads to large deviations in flavour observables. Therefore, \thdm that are most often considered in the literature have some kind of symmetry that singles out one of the two Higgs doublets and make each of them interact specifically with only a subset  of the fermions. Another approach to mitigate the flavour issues of a generic \thdm is to invoke a coupling structure in flavour space for the couplings of the doublets. These structures may be a simple ansatz~\cite{Sher:2022aaa,Babu:2018uik} or have relations to symmetries that can be imposed on the doublets and the SM fermions~\cite{Alves:2018kjr}. 

In most cases the third generation is singled out as possibly having larger flavour violation than the first two. This is in part due to the larger masses of third generation fermions, and in part to the existence of several bounds from the large set of measurements about flavour in the first and second generation.
In addition to flavour constraints on models with an extended scalar sector, the measurements of the properties of the Higgs boson at \SI{125}{\giga\electronvolt} point to specific regions of the large parameter space of these models. 

In general the two Higgs doublets can both contain a fraction of the physical degree of freedom that ends up being eaten by the weak vector bosons in the Higgs mechanism. In other words, neither of the two Higgs doublets  coincide with the state that couples with weak vector bosons and acquire a vacuum expectation value $v=(\sqrt{2} G_{F})^{-1/2}$. Furthermore, the two Higgs doublets in general do not coincide with mass eigenstates either. Therefore, out of these two possible misalignments between gauge and mass and symmetry breaking eigenstates  one has to pick a special condition to hold for the new doublet to not give rise to unacceptably large deviations from the SM coupling picture of the \SI{125}{\giga\electronvolt} Higgs boson. This condition is usually referred to as the ``alignment limit'' and it is given as $\cos(\beta-\alpha)=0$ in the standard notation of two Higgs doublet models, with $\alpha$ being the angle between gauge and mass eigenstates and $\beta$ the angle that defines the symmetry breaking eigenstates (one which takes \vev and one that does not).  

The special alignment condition which makes the 125~GeV Higgs boson be fully Standard Model-like does not necessarily imply anything about the flavour structure of the couplings of the other doublet. Therefore, it is possible to imagine flavour violation to happen in processes in which the second doublet participates, while leaving no trace of flavour violating couplings in the 125~GeV Higgs. This regime of the \thdm is a subset of general models studied in earlier surveys, e.g.\ in Ref.~\cite{Atwood:1996vj} and in recent community exercises such as the Snowmass study 2013~\cite{TopQuarkWorkingGroup:2013hxj}. The relevance of this regime of the \thdm has become more prominent as the Higgs data has settled on couplings that broadly agree with the predictions of the SM. In view of the current situation in the study of Higgs boson properties we assume to be in the alignment limit throughout in the following.

\paragraph{Flavour violation in 2HDM}

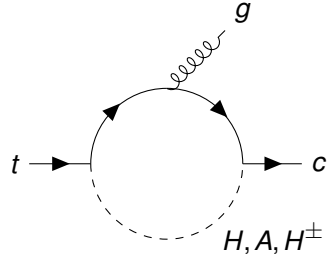
\begin{figure}[th!]
\begin{centering}

\begin{tikzpicture}
  \begin{feynman}
    \vertex (t) at (-1, 0) {$t$};
    \vertex (c) at (0, 0);

    \vertex (a) at (1,1);
     \vertex (q) at (3,0) {$c$};
          \vertex (d) at (2,0);
    \vertex (b) at (1,-1);
    \vertex (f) at (2.4,-1){$H,A,H^{\pm}$}; 

     \vertex (g) at (2,2) {$g$}; 

    \diagram*{
      (t) -- [fermion] (c),
       (a) -- [gluon] (g),
    };

 \draw[fermion] (a) arc [start angle=90, end angle=0, radius=1cm];
  \draw[fermion] (c) arc [start angle=180, end angle=90, radius=1cm];


 \draw[scalar] (c) arc [start angle=180, end angle=360, radius=1cm];
     \diagram*{
      (d) -- [fermion] (q),
    };
  \end{feynman}
\end{tikzpicture}

\end{centering}
\caption{Representative Feynman diagram for the $\topgluoncharm$ decay. \label{fig:t2cgluon2HDM-Bulliri}}
\end{figure}

All the particles contained in the extra doublet can mediate flavour violation and can participate in the $\topgluoncharm$ decay. An example Feynman diagram is shown in \cref{fig:t2cgluon2HDM-Bulliri}, with possible intermediate states being the neutral $\Hneutrals$ and charged states $\Hcharged$. These states can mediate flavour violating transitions through off-diagonal elements of their Yukawa couplings matrices. A commonly adopted parametrisation, that we borrow from Refs.~\cite{Sher:2022aaa,Atwood:1996vj,Kim:2015zla}, is to assume Yukawa matrices with the following structure for their elements 
\begin{equation}
  \xi_{ij} = \lambda_{ij} \sqrt{2\cdot m_i m_j}/v. \label{eq:flavour-ansatz-Bulliri}
\end{equation}
Each of the states of the \thdm is sensitive in principle to different flavour violating couplings, e.g.\ the charged state probes $\PQb\leftrightarrow\PQc $ flavour changing interactions, while the neutral state probes $\PQt\leftrightarrow\PQc$ flavour changing interactions. This implies that bounds on each contribution can arise from a different set of measurements and limits on flavour changing couplings beyond the Standard Model. For example the charged state receives bounds from not giving an excessive change in the observed rate for $\btosgamma$  transitions, while the neutral bosons contribution is constrained  e.g.\ by not giving rise to excessive $\cctott$, or same-sign top production $\cctottSS$.
In a fully general model one could allow each of these flavour changing effects to be fully independent, leading to a very high-dimensional parameter space for the model, possibly including large deviations from the theoretically more motivated case of \cref{eq:flavour-ansatz-Bulliri} with all $\lambda_{ij}\sim O(1)$.

In Ref.~\cite{Kim:2015zla} constraints from $\bmesons$ and $\kmesons$ mesons oscillations, $\btaunu$, $\btosgamma$, $\PZ\to \PQb \PAQb$, the $\rho$ parameter, same sign top-quark production $\protonproton \to \PQt \PAQt$ at LHC, and $R(D)$ and $R(D^{*})$ have been considered to constrain the three masses of the physical states of the \thdm  and seven flavour dependent couplings of these states, that, in the common notation of Refs.~\cite{Sher:2022aaa,Atwood:1996vj,Kim:2015zla}, are 
$\xi_{tt}$,
$\xi_{tc}$,
$\xi_{tu}$,
$\xi_{bb}$,
$\xi_{bs}$, 
$\xi_{bd}$,
$\xi_{\tau\tau}$.

\paragraph{Results}
There are multiple solutions that would satisfy the constraints from all the observables considered in Ref.~\cite{Kim:2015zla}, as the predictions for certain observables may be sensitive to cancellations among couplings. For instance the contribution to $\btosgamma$ may cancel for appropriately chosen $\PQt\PAQt$ and $\PQb\PAQb$ couplings. This makes it difficult to obtain an absolute upper bound for the  branching ratio  $\topgluoncharm$ in this model.   Ref.~\cite{Kim:2015zla} has made several predictions for upper bounds on $\topgluoncharm$ in different regions of the parameter space of the model which satisfy the constraints using different degrees of cancellations in the new physics effects on the measurements used to put constraints on the $\xi$ couplings and the masses of the fields in the extra doublet. 
Upper bounds for $\topgluoncharm$ are shown in \cref{fig:t2cg-upperbound-Bulliri} assuming very little cancellation (green), some degree of cancellation and in particular a reduced new physics contribution to $\PB$ mesons oscillations and $\btosgamma$ compared to is natural size (black), some degree of cancellation and a particularly reduced new physics contribution to $\PB$ mesons oscillations with a large CP-asymmetry phase $\phi_{ccs}$ in $\PQb \to \PQc$ decays (grey).
\begin{figure}[htbp]
\begin{center}
\includegraphics[width=0.45\linewidth]{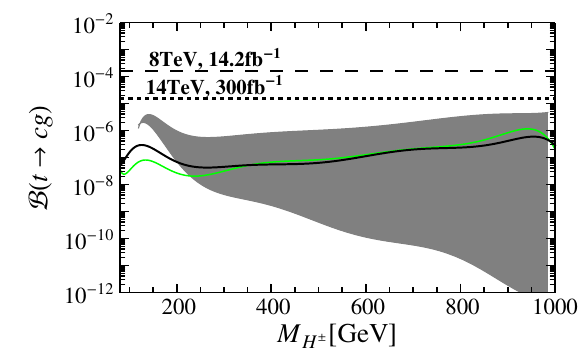} 
\caption{ The green and black lines denote the upper bounds on the  branching ratio  $\topgluoncharm$  as a function of $M_{\PSHpm}$ from two different solutions of the combined flavour constraints described in Ref.~\cite{Kim:2015zla}. The gray region corresponds to a different solution of the same constraints with more fine-tuned choice of couplings. The dashed line denotes the current upper bound at LHC, while the dotted line is for the future sensitivity at \SI{14}{\tera\electronvolt} with $\SI{0.3}{\per\atto\barn}$ luminosity.}
\label{fig:t2cg-upperbound-Bulliri}
\end{center}
\end{figure}

\paragraph{Outlook}
The analysis of Ref.~\cite{Kim:2015zla} shows that  BR$(\topgluoncharm)$ larger than $10^{-6}$ is not compatible with flavour and electroweak measurements  sensitive to a new doublet and its flavour violating couplings. The impact of improvements on the measurements used as experimental inputs in Ref.~\cite{Kim:2015zla} is being evaluated \cite{Bulliri:2025aaa}. 
The improvement of measurements in the electroweak fit will lead to improved values for the $\rho$ parameter. Measurements from Belle~II and LHCb will also provide further constraints. 
In  addition the $\PZ$ pole run and subsequent runs of the \factory factory will improve measurements of $\PZ\to \PQb \PAQb$ and rates for $\PB$ meson decays beyond the results of Belle~II, delivering a sensitivity to $\topgluoncharm$ much below the observable rate at the $\ttbar$ stage of the \factory factory. 

\subsubsection{Direct search for new states in top-quark decays\label{sec:fcnc-bsm-search}}
\paragraph{Motivation}

Searches for top-quark FCNC have traditionally concentrated on SM final states, such as $\PQt \to \PZ \PQc$,  $\PQt \to \Pg \PQc$, $\PQt \to \PGg \PQc$, $\PQt \to \PSH \PQc$. The rates for these processes are exceeding small in the SM, so the observation of top-quark FCNC at any observable level would be a clear sign of new physics. These FCNC transitions are generated in BSM models due to misalignment of flavour and mass eigenstates of the SM matter interactions with new physics states. As the mass of these new states gets probed and pushed to higher values by direct searches at the LHC, the predictions for these FCNC to SM final states get  smaller and smaller. Some models traditionally thought as targets for these searches are starting to be pushed to such a high mass scale that the amount of top-quark FCNC in SM final states is hardly observable, as discussed at the beginning of this \cref{sec:extt}.

While it is worth pursuing FCNC in purely SM final states as a probe of models already considered in previous literature, e.g the Snowmass exercises~\cite{TopQuarkWorkingGroup:2013hxj}, the capabilities of the HTE factory extend much further. In fact, the clean environment of an $\Pep \Pem$ machine allows to tackle scenarios in which the LHC can have serious trouble to have any sensitivity.

Extending the possible final states of the FCNC top-quark decay to encompass new physics states gives countless possibilities. For instance the new physics FCNC can result in a fully hadronic final state, which can prove to be quite challenging in the LHC environment. More generally the new physics signature associated with the top-quark FCNC can be dirty enough for the LHC to not be sensitive, thus leaving behind blind spots in the LHC searches.

To assess the capabilities of the HTE factory we consider a top-quark FCNC decay 
$$ \PQt \to \PQc \phi\;,$$
where $\phi$ is a new physics scalar light enough to be an on-shell decay product in the top-quark decay. It can be a new scalar part of the Higgs sector related to the breaking of the electroweak symmetry~\cite{Banerjee:2018fsx}, e.g.\ in a singlet or doublet extensions of the SM, or may be a new scalar related to the breaking of flavour symmetries of the SM. The details of these models will determine the strength of the flavour changing $\phi \PQt \PQc$ coupling as well as the decay channels available to $\phi$. It is reasonable to assume that the strength of the couplings of $\phi$ tracks the masses of the SM particles it couples to. This is the pattern observed for scalars that inherit their couplings from those of the SM Higgs boson, as well as a generic prediction of classes of models that break the flavour symmetry of the SM and aim to explain the masses and mixing of the SM quarks and leptons using spontaneous symmetry breaking.

For the above reasons, it is reasonable to assume that $\phi\to \PQb \PAQb$ is the main decay channel of $\phi$. In the following we study this possibility in detail, but we keep in mind that other decays could be interesting as well, e.g.\ $\phi \to \PQc \PAQc$~\cite{Carmona:2021seb}, decays into light quarks $\phi \to \PQq \PAQq$, or decays into invisible final states $\phi \to invisible$. We leave these further decay modes for future study, as possible tests of the flavour tagging capabilities of the detectors of the HTE factory and in general to test the kinematic reconstruction performance of the various types of detectors proposed.

\paragraph{Signal and backgrounds}

The possibility of observing exotic \ttbar decays has been studied for the $\sqrt{s} = 365 \; \GeV$ Run of FCC--ee with projected luminosity of  \SI{1.5}{\per\atto\barn}, computing the expected BR limits for the decays $\PQt \to \PQc \phi$ and $\phi \to \bb$.
For this study, the event generator \mgfive~\cite{Alwall:2014hca} was used to create the Monte Carlo (MC) samples considering \Pep\Pem collisions at $\sqrt{s} = \SI{365}{\giga\electronvolt}$. Only the dominant background contributions arising from \ttbar + jets events were included in this study. For the signal events, one of the top quarks is forced to decay via $\PQc\phi$, with $\phi \to \bb$, using the TopFCNC UFO model~\cite{Degrande:2014tta,Durieux:2014xla}, while the second top quark decays via the SM couplings to $\PQt \to \PQb \PWp \to \PQb (\Pl \PGnl)$ the masses of [15, 20, 50, 125] \SI{}{\giga\electronvolt} were simulated for the scalar particle. The generated events were processed through \pythia\ version 8.311~\cite{Bierlich:2022pfr}, to model parton showering, hadronisation, and underlying event. The detector modeling was done by \delphes~\cite{deFavereau:2013fsa}, using the cards for the IDEA detector and the \keyhep\ event data model (\edmhep)~\cite{ganis2021key4hepframeworkfuturehep}. \cref{tab:kmotaama_samples} summarizes the MC samples that were included in the study.

\begin{table}[h!]
    \centering
    \begin{tabular}{c | c} 
    \hline
    Sample & $\sigma$ \\
    \hline
    \ttbar + jets &   \SI{ 0.3432 }{\pico\barn}  \\ 
    $\PQt \to \PQc\phi$, $m_\phi = \SI{15}{\giga\electronvolt}$  &  \SI{1.663}{\femto\barn} \\
    $\PQt \to \PQc\phi$, $m_\phi = \SI{20}{\giga\electronvolt}$ &  \SI{ 1.643}{\femto\barn}\\
    $\PQt \to \PQc\phi$, $m_\phi = \SI{50}{\giga\electronvolt}$  &  \SI{1.415}{\femto\barn} \\
    $\PQt \to \PQc\phi$, $m_\phi = \SI{125}{\giga\electronvolt}$  &  \SI{0.3737}{\femto\barn} \\
    \hline
    \end{tabular}
    \caption{Summary of the samples generated for this analysis.}
    \label{tab:kmotaama_samples}
\end{table}

\paragraph{Reconstruction and Selection}

With the samples in \edmhep\ data format, the events were analyzed using the \textsc{FCCAnalyses} framework~\cite{helsens_2024_13871482}. We apply the Durham exclusive jet clustering algorithm requiring four jets in an event using the \fastjet~\cite{Cacciari:2011ma,Cacciari:2005hq} package, perform a flavour-tagging of the jets using the particle transformer model trained on Higgs samples, and transform the \edmhep\ data format into a flat ntuple including various important variables for the selection of the signal process.

For the event selection, a few preselection requirements were used before applying a Boosted Decision Tree (BDT) multivariate analysis techniques to discriminate signal from background events: 

\begin{itemize}
    \item At least one lepton (\Pe or \PGm), with p >  \SI{5}{\giga\electronvolt}, $|\eta| < 2.9$ and relative isolation $\Delta R/p$ < 0.5;
    \item At least two jets b--tagged (score > 0.5) and at least one c--tagged (score > 0.5) jet.
\end{itemize}

As input, the BDT included:
\begin{itemize}
    \item The invariant mass and $\Delta R$ of the jet pair with closest mass to the scalar;
    \item The invariant mass of the triple jet with the closest mass to the top quark;
    \item The scores of b--tag and c--tag of all jets of the event; 
    \item The invariant mass of the lepton and the missing energy; 
    \item The $\Delta R$ between the lepton and the jet not used to reconstruct the top mass;
    \item The invariant mass of the lepton and missing energy vector;
    \item The missing energy.
\end{itemize}
A working point of 0.97 was chosen for the final selection. \cref{fig:kmotaama_BDToutputs} displays the outputs of the BDTs used for each dataset.

\begin{figure}[!htbp]
    \centering
    \begin{subfigure}{0.24\textwidth}
        \includegraphics[width=\textwidth]{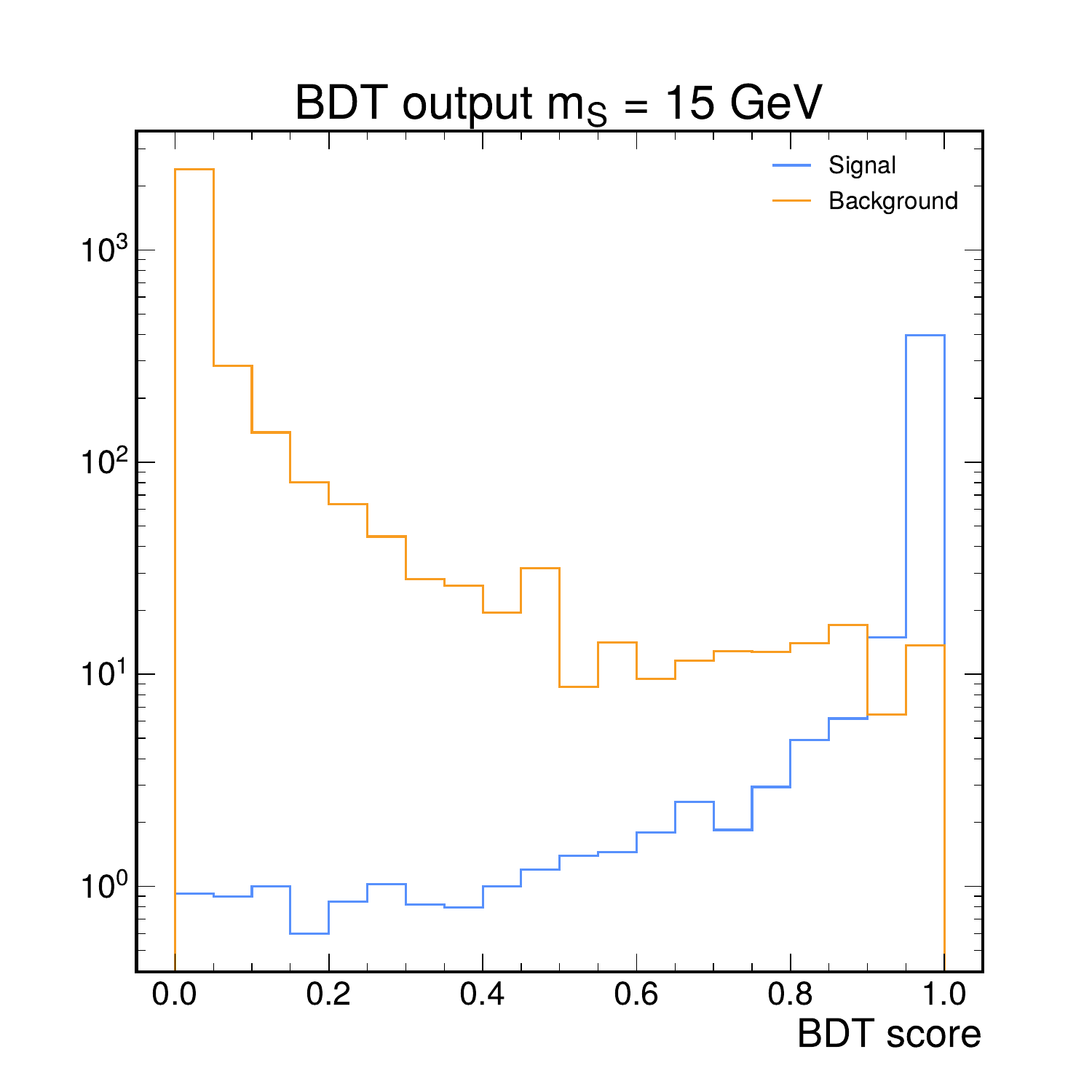}
    \end{subfigure}
    \hfill
    \begin{subfigure}{0.24\textwidth}
        \includegraphics[width=\textwidth]{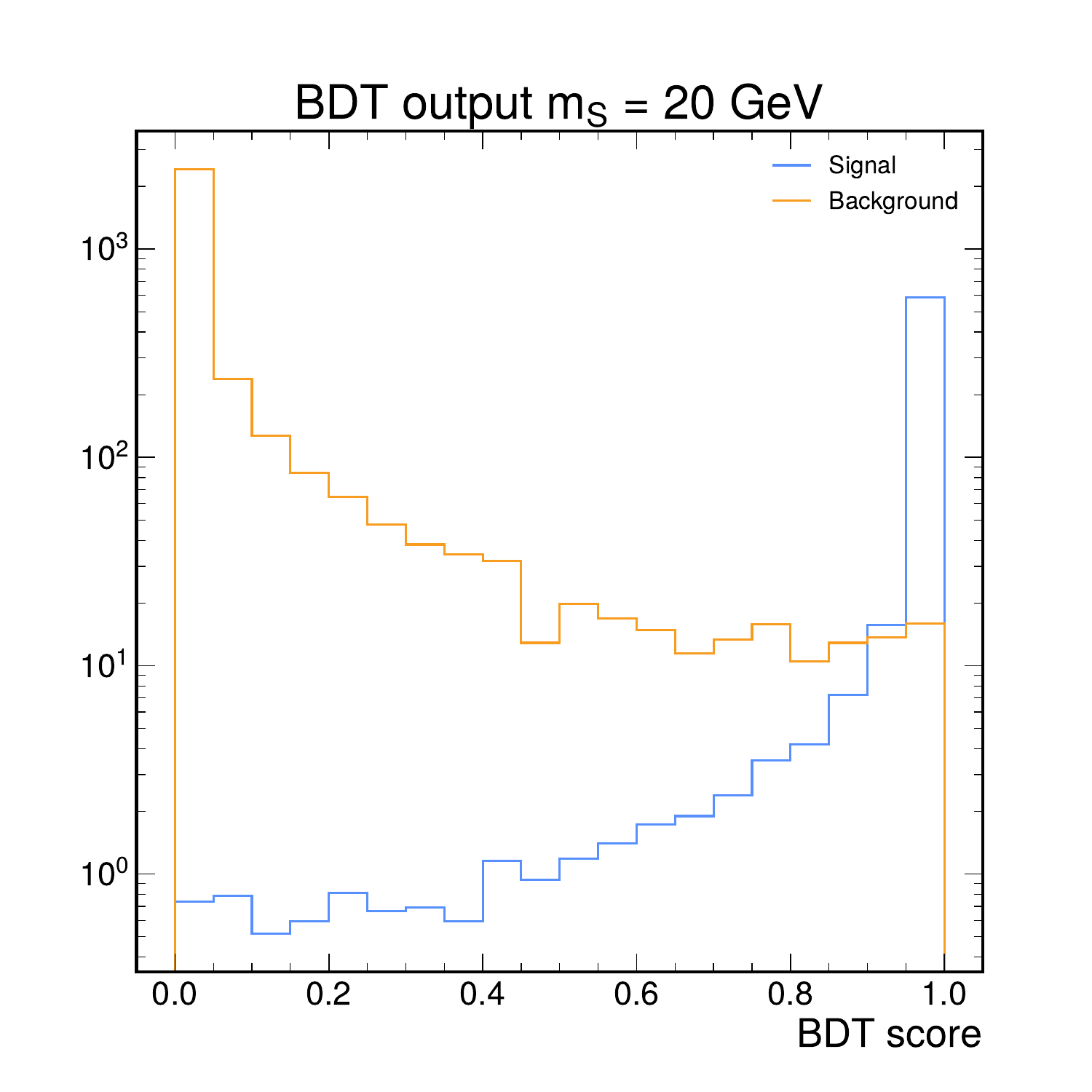}
    \end{subfigure}
    \hfill
    \begin{subfigure}{0.24\textwidth}
        \includegraphics[width=\textwidth]{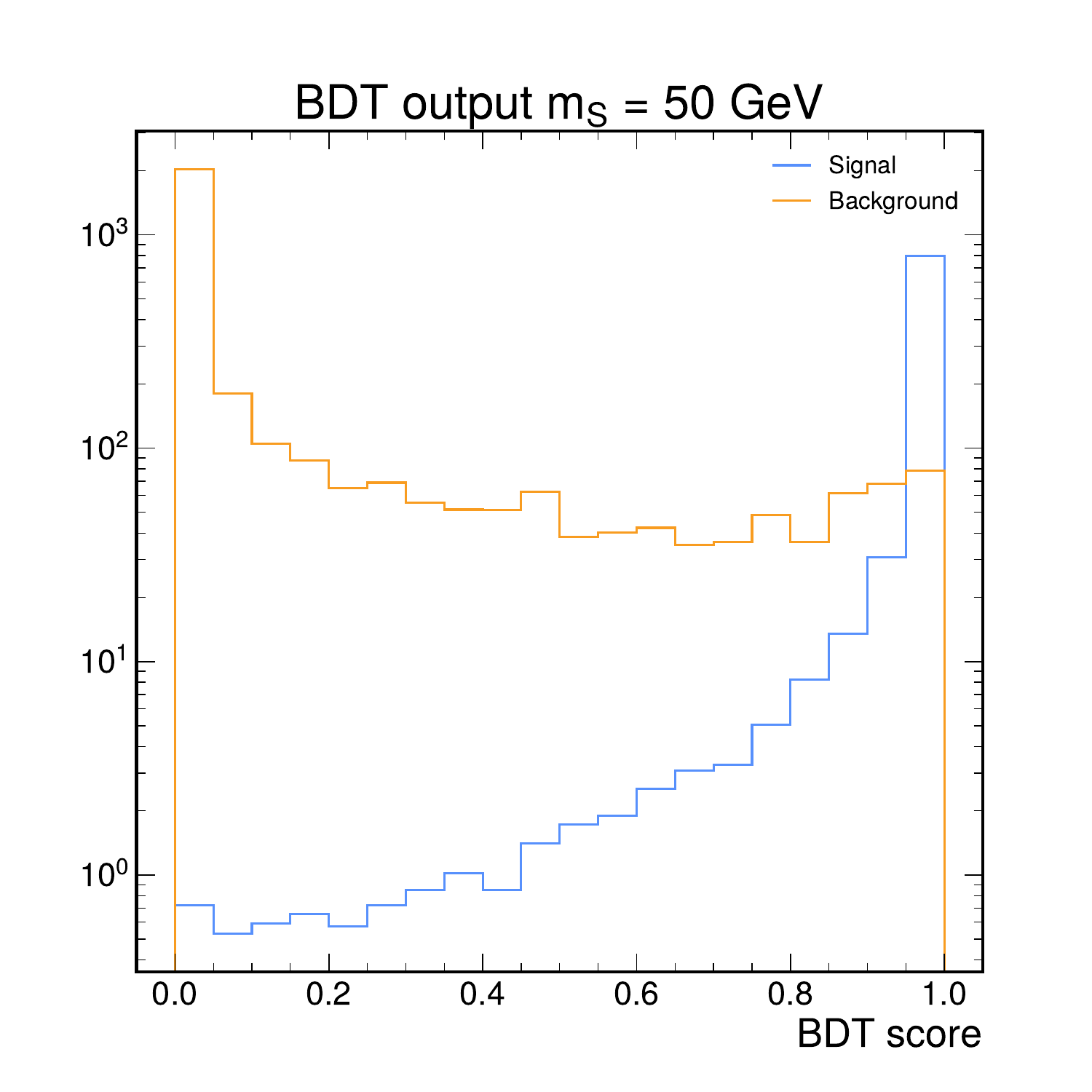}
    \end{subfigure}
    \hfill
    \begin{subfigure}{0.24\textwidth}
        \includegraphics[width=\textwidth]{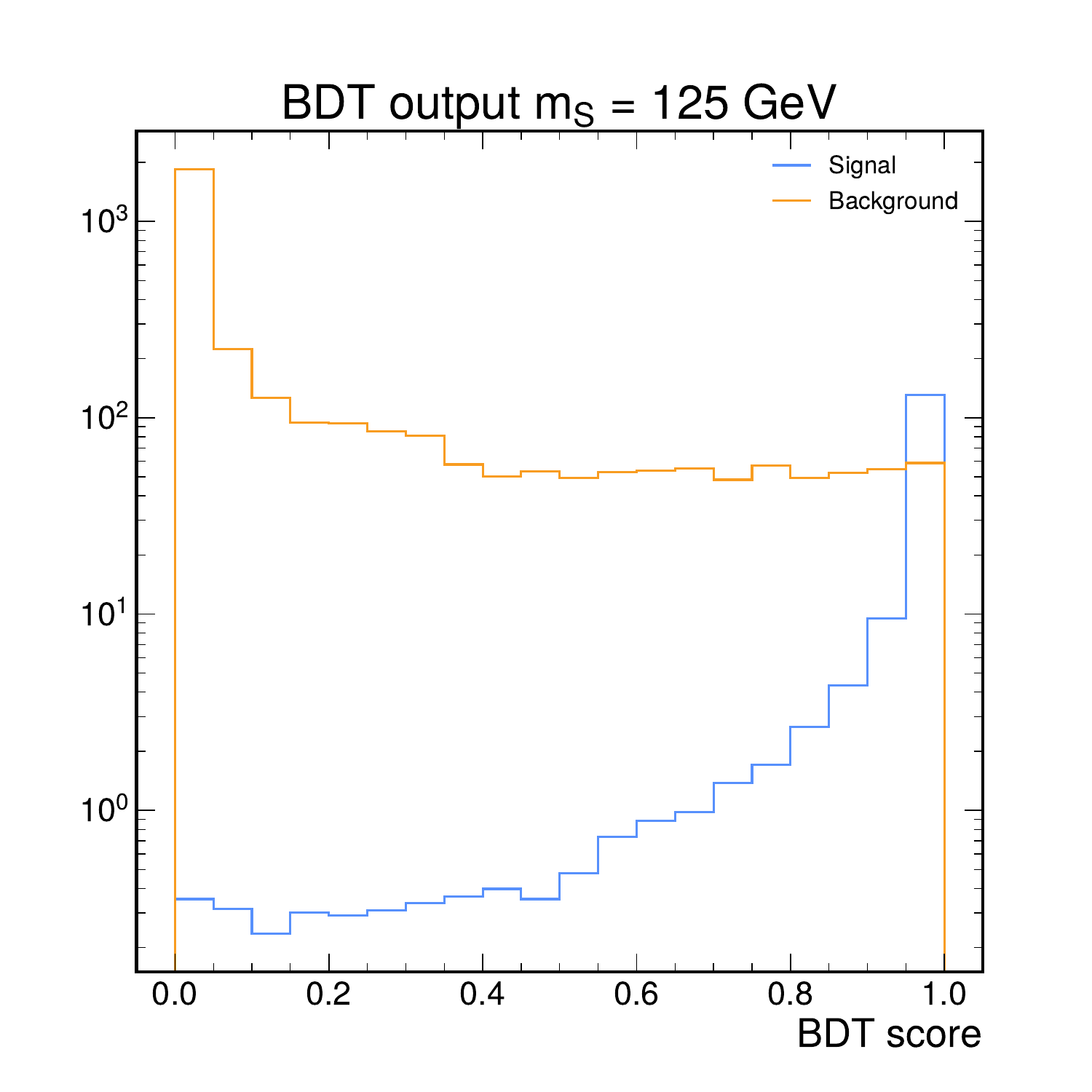}
    \end{subfigure}
            
    \caption{The BDT discriminant corresponding to each hypothesis on the assumed scalar mass.}
    \label{fig:kmotaama_BDToutputs}
\end{figure}

\paragraph{Results}

The \textsc{Combine} statistical analysis tool~\cite{cms2024cms} was used to determine the expected 95 \% confidence level (CL) intervals. In \cref{fig:kmotaama_limit}, the dashed line shows the expected limit for each of the considered scalar masses, together with the $\pm1 \sigma$ and $\pm2 \sigma$ CL intervals (in green and yellow bands), respectively. Over most of the mass range investigated these limits are comparable to what was obtained by the ATLAS Collaboration using Run2 data~\cite{ATLAS:2023mcc}. At a HTE factory we demonstrate that it is possible to obtain a sensitivity for light scalars, below \SI{20}{\giga\electronvolt}, which are difficult to study at hadron colliders due to complications in jet reconstruction arising from significant contributions associated with pileup interactions. We also remark that our study can be further improved by e.g.\ introducing a BDT to perform the top-quark reconstruction. Previous studies have shown that an exclusion limit of $8.8 \times 10^{-5}$ for $\PQt \to \PQc \PH$~\cite{CLICdp:2018esa} can be attained at CLIC.

\begin{figure}[!htbp]
    \centering
    \includegraphics[width=0.5\linewidth]{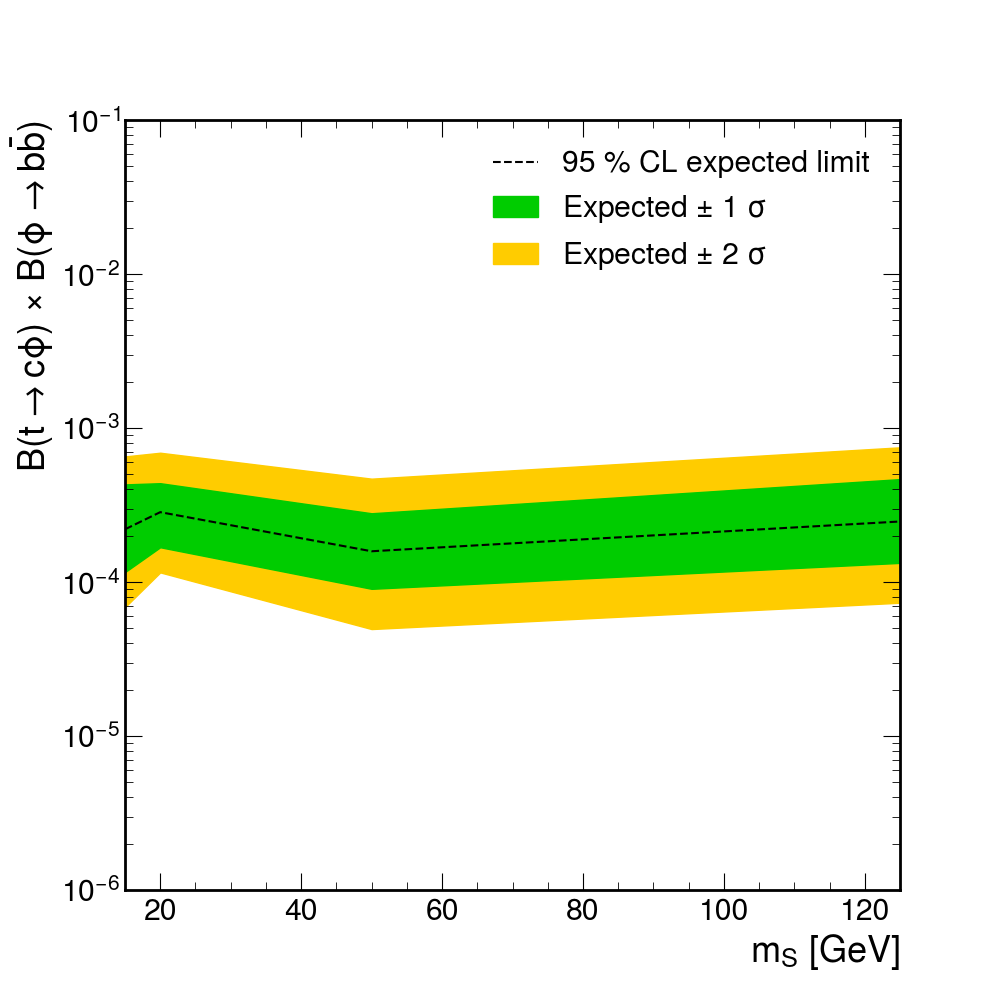}
    \caption{Expected 95\% CL upper limits for the BR$(\PQt \to \PQc \phi)$. The green and yellow bands show the 68\% and 95 \% confidence intervals, respectively.}
    \label{fig:kmotaama_limit}
\end{figure}

\clearpage
\section{Global Interpretations}
\label{sec:globalinterpretations}
\editors{Jorge de Blas}

The purpose of any future collider experiment is to further our knowledge of what new physics explains the open issues of the Standard Model. From the point of view of future $\epem$ Electroweak/Higgs/Top factories, the exquisite precision that will be possible in the measurements of physical observables of different sectors of the SM will bring an unprecedented
indirect sensitivity to possible virtual effects of new particles. In a given UV model, deviations will typically be expected in several of these measurements, and assessing the sensitivity in measuring the model parameters will, in general, require a global study, combining the information from all the processes to which the model can contribute. In previous sections, we described progress in the measurements of different types of observables and, in some cases, the implications these would have on particular new physics scenarios. 
The aim of this section is to synthesize these findings by integrating insights from EW, Higgs, and top quark measurements to provide a comprehensive interpretation of new physics.

In \cref{sec:globalSMEFT} we discuss the combination from the point of view of the so-called Standard Model Effective Field Theory (SMEFT), which has been already used in some of previous chapters, whereas in \cref{sec:BSM_SMEFT} the results are interpreted in terms of particular BSM extensions, by matching single-particle extensions of the SM to the SMEFT and deriving limits on the characteristic mass scale of these simplified models under certain assumptions for their couplings to the SM. Particular attention is devoted to the sensitivity that would be attained at the Tera-Z run, where some of the most precise measurements across all sectors accessible at future \epem colliders will be possible. This is discussed in \cref{sec:SMEFT_TeraZ}. 

\subsection{\texorpdfstring{Global SMEFT fits at future $\epem$ colliders}{Global SMEFT fits at future e+e- colliders}}\label{sec:globalSMEFT}

In the current situation, with no experimental indication of what type of models should be focused on in these global studies of indirect effects of new physics, the techniques of Effective Field Theories (EFT) become very useful. 
In the EFT framework, a consistent description of new physics effects across different observables can be obtained with minimal reference to the details of the UV theory. In the case of the SMEFT, the full EFT is constructed starting from the observation that the SM seems to provide a good description of most processes at the EW scale and, based on the fact that no sign of new particles has been found
yet in experimental direct searches, making the following assumptions about the new physics beyond the SM (BSM): 
(1) its energy scale is higher than the energies explored so far; and 
(2) its effects decouple as the masses of the new particles increase. 
Thus, in the infrared, the SMEFT Lagrangian is written in terms of:
(a) the SM symmetries; (b) the SM fields (with the Higgs belonging to a $SU(2)_L$ doublet\footnote{This extra assumption is not strictly necessary based on experimental observations. An alternative formulation where the Higgs is a singlet is given by the so-called {\it Higgs Effective Field Theory}).}  ); and the decoupling of new physics allows (c) BSM effects to be parametrised in terms of higher-dimensional operators, suppressed in the Lagrangian by inverse powers of the cutoff scale of the EFT, $\Lambda$:
\begin{equation}
{\cal L}_\textrm{SMEFT}={\cal L}_\textrm{SM} + \sum_{d>4} \frac{1}{\Lambda^{d-4}}\sum_{i} C_i^{(d)} {\cal O}_i^{(d)},
\label{eq:SMEFT}
\end{equation}
where ${\cal O}_i^{(d)}$ are dimension-$d$ operators built exclusively with SM fields, respecting Lorentz and SM gauge invariance. 
The information about the details of the UV physics is encoded in the values of the {\it Wilson coefficients} $C_i^{(d)}$, which can be obtained via matching of the EFT with a given model. 
The SMEFT Lagrangian constructed in this way provides a description of infrared effects of any ultraviolet model complying with these very minimal assumptions. It is important to emphasize that only in this sense, this and any other EFT can be understood as {\it model-independent} as, for instance, models predicting light states that somehow could have escaped direct searches could not be mapped into the SMEFT. Despite this, based on current observations, with the apparent (though not conclusive) lack of any new states at the EW scale suggesting a mass gap between this and any new physics scale, the SMEFT formalism is arguably one of the most phenomenologically reasonable agnostic parametrisations of BSM deformations at low energies.

Assuming lepton and baryon number conservation, the leading order new physics effects in \cref{eq:SMEFT} are given by dimension-six operators.
Current SMEFT studies at future colliders are performed to this order in the EFT expansion. In preparation for the {\textit{2020 Update of the European Strategy for Particle Physics}} (ESPP2020)~\cite{deBlas:2019rxi} and later for the {\textit{Snowmass 2021}} process~\cite{deBlas:2022ofj}, several SMEFT studies of the physics potential in terms of the EW and Higgs measurements at the different future collider options were presented. These results, prepared using the {\texttt{HEPfit}} framework~\cite{DeBlas:2019ehy} and expressed in terms of the sensitivity to the modifications of the interactions of the SM particles predicted by the SMEFT, the so-called {\textit{effective couplings}} described in Ref.~\cite{deBlas:2019rxi}, are summarized in~\cref{fig:SM21_smeft_bounds}. This figure has been updated with respect to Ref.~\cite{deBlas:2022ofj} with: (1) the latest run scenarios of the FCC-ee project including four interaction points, with total integrated luminosities of 10.8 ab$^{-1}$ at the 240 GeV run and 3.1 ab$^{-1}$ at 340-365 GeV; (2) the new staging plan for CLIC, increasing the luminosity collected at the 380 and 1500 GeV stages to 4.3 ab$^{-1}$ and 4.0 ab$^{-1}$, respectively~\cite{Adli:ESU25RDR}. Numerical values for the sensitivity to modifications in each coupling at each project can be found in Table 29 of Ref.~\cite{deBlas:2022ofj}. The previously mentioned updates need to be taken into account for the FCC-ee and CLIC. These affect mostly the Higgs coupling sensitivity and, for the most part, can be 
extrapolated by a simple statistical scaling to the new luminosities of each project. 

\begin{figure}
    \centering
    \includegraphics[width=0.99\linewidth]{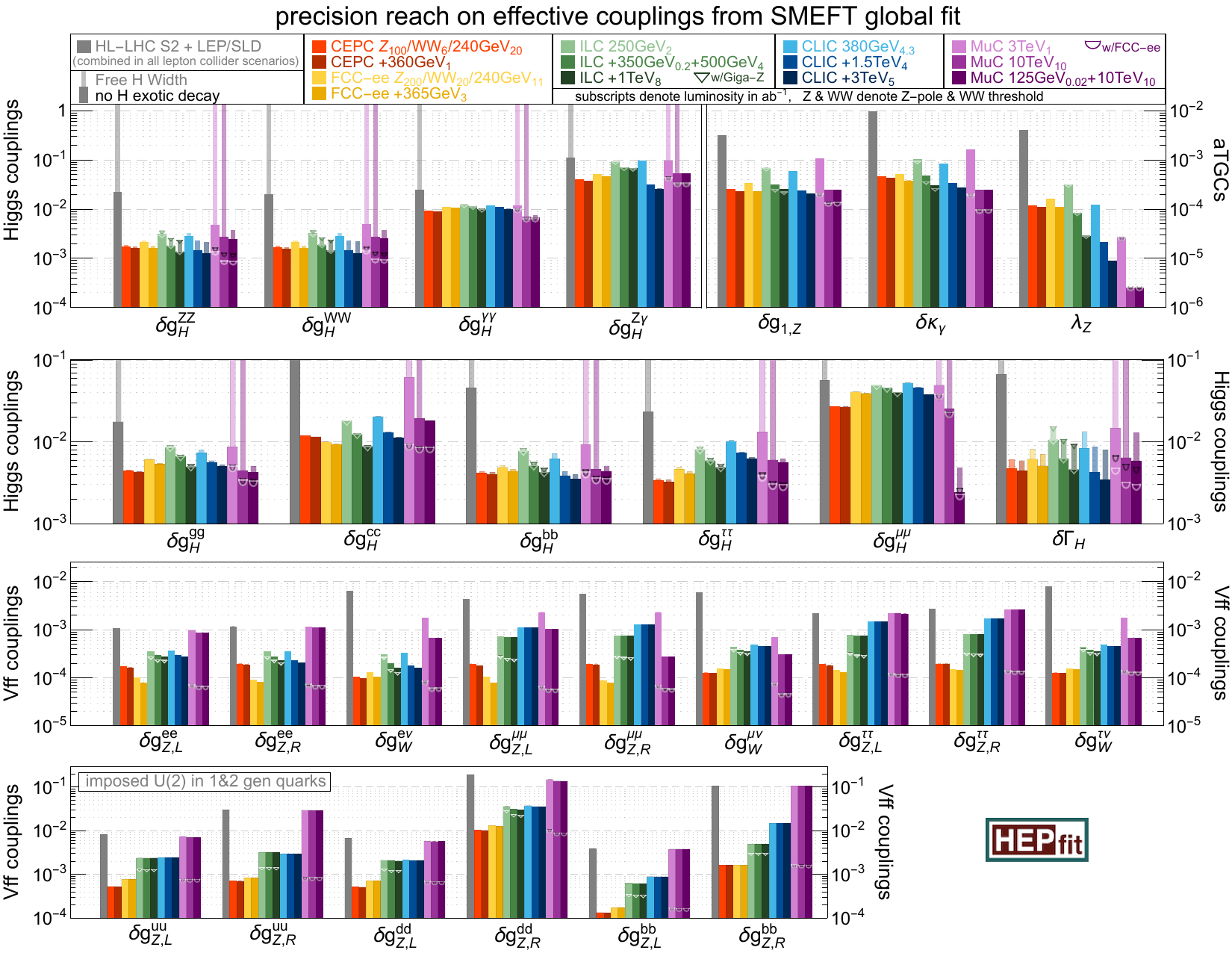}
    \caption{Comparison of the 1$\sigma$ (68$\%$ probability) sensitivity to new physics modifications in the interactions of the Higgs and EW vector bosons, obtained projecting the SMEFT results into the {\it effective couplings} of these particles to the other SM fields. Figure updated with respect to the results presented in~\cite{deBlas:2022ofj}. See text for details. 
    }
    \label{fig:SM21_smeft_bounds}
\end{figure}

The top sector was arguably not considered in the same detail as the electroweak and Higgs ones in the ESPP2020 studies, and this was partially addressed during the {\textit{Snowmass 2021}}, where a global study of new physics in top couplings in the SMEFT was also presented in Ref.~\cite{deBlas:2022ofj}. The update of such studies of the top-quark properties has been discussed in \cref{sec:TopSMEFT}. In this section we present developments in terms of extending these SMEFT studies to a full EW/Higgs/Top fit.

\subsubsection{Combining EW/Higgs/Top measurements in the SMEFT at future colliders}\label{sec:SMEFT_EW_H_Top}

In this section we review the extension of previous efforts to a more global SMEFT study, incorporating the information of a comprehensive set of top processes into the analysis of the electroweak and Higgs sectors. This is based on the study in Ref.~\cite{Celada:2024mcf}, carried out within the \textsc{SMEFiT} framework~\cite{Hartland:2019bjb,Ethier:2021ydt,Ethier:2021bye,Giani:2023gfq,terHoeve:2023pvs},
and focuses specifically on circular colliders, namely the FCC-ee and the CEPC. 

On the experimental side, the study considers the latest scenarios in the baseline run plan of each project, and includes \PZ-pole EWPOs, fermion-pair, Higgs, diboson, and top-quark production, using optimal observables for both the $\PWp\PWm$ and the $\PQt\PAQt$ channels.
HL-LHC projections are also considered, obtained extrapolating from the Run-2 measurements, and not optimised for the increase in luminosity.

On the theory side, the analysis is performed both at linear and at quadratic order in the EFT expansion, includes $n_{op}=50$ independent Wilson coefficients, and accounts for NLO QCD corrections to the EFT cross-sections for the hadron collider observables. The flavour assumption is U$(2)_q\times$U$(3)_d\times$U$(2)_u\times(\text{U}(1)_{\ell}\times \text{U}(1)_{e})^3$ 
and dimension-six operators are defined in the Warsaw basis, with small modifications following Ref.~\cite{Aguilar-Saavedra:2018ksv}. \footnote{Furthermore, the flavour assumption is slightly relaxed to include the operators ${\cal O}_{b\varphi}$ and ${\cal O}_{\tau\varphi}$ that modify the bottom-quark and tau-lepton Yukawa interactions.}
As the baseline of this analysis, the full set of current EWPOs and several recent measurements of Higgs, diboson, and top quark
production data from the LHC Run-2 are considered. A total of $n_{dat}=445$ data points are included in the fit.
From this, in the left panel of Fig.~\ref{fig1} we display the sequential impact of the separate $\sqrt{s}$ runs at the FCC-ee in the global SMEFT fit based on $\mathcal{O}( \Lambda^{-2})$ calculations.
The constrained operators are classified into 2-light-2-heavy four-fermion, two-fermion, and purely bosonic operators. 
The dominant constraints are provided by the measurements at $\sqrt{s}=240$ GeV, with \PZ-pole EWPOs and the runs with $\sqrt{s}=161$ ($\PWp\PWm$ threshold) and 365 ($\PQt\PAQt$ threshold)  necessary to achieve the ultimate
constraining potential of the FCC-ee.
As compared to the post-HL-LHC situation, the FCC-ee measurements can pin down several Wilson coefficients with a precision improved by two orders of magnitude. 

\begin{figure}[h]
    \centering
\includegraphics[width=0.50\linewidth]{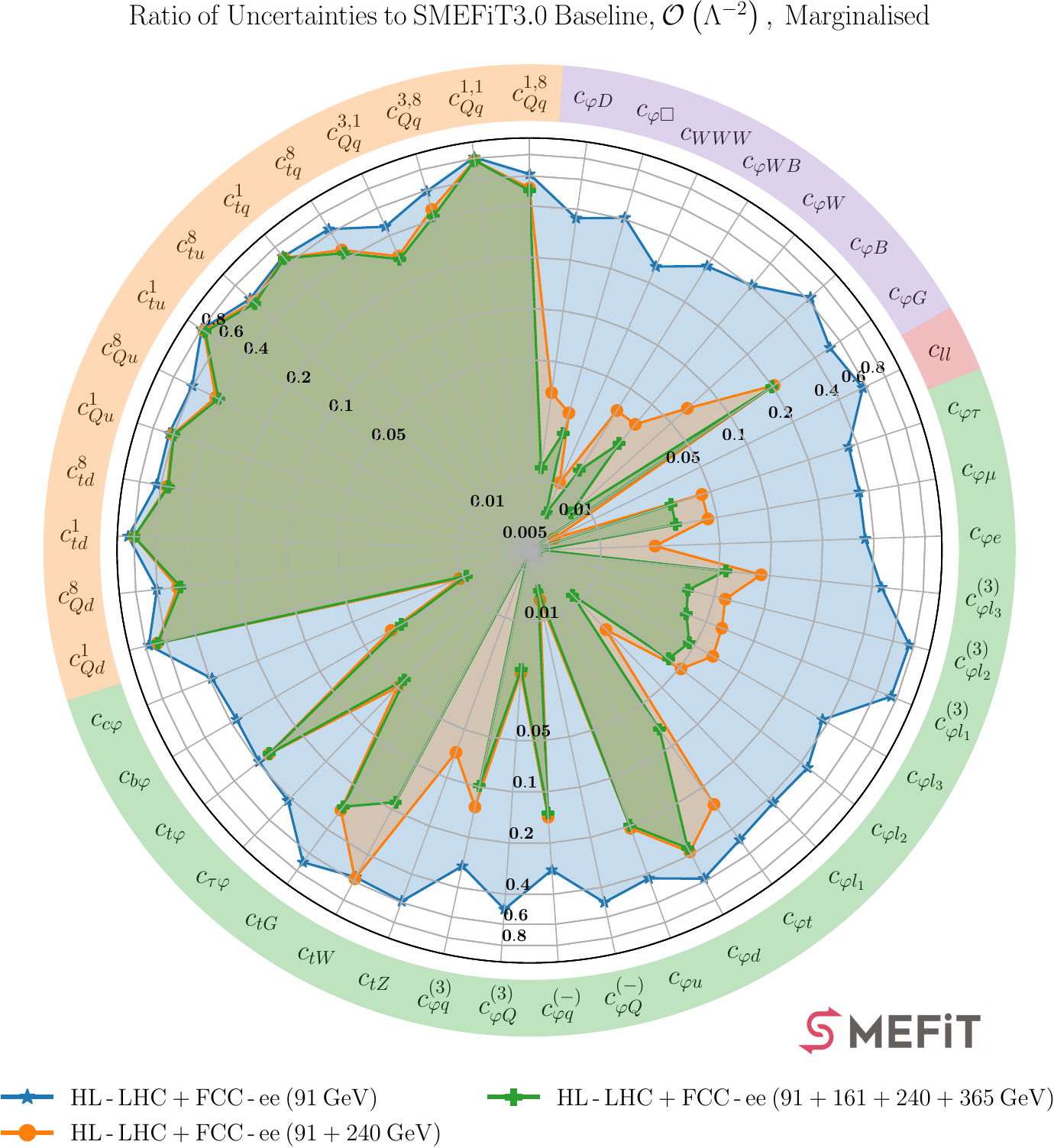}
\includegraphics[width=0.478\linewidth]{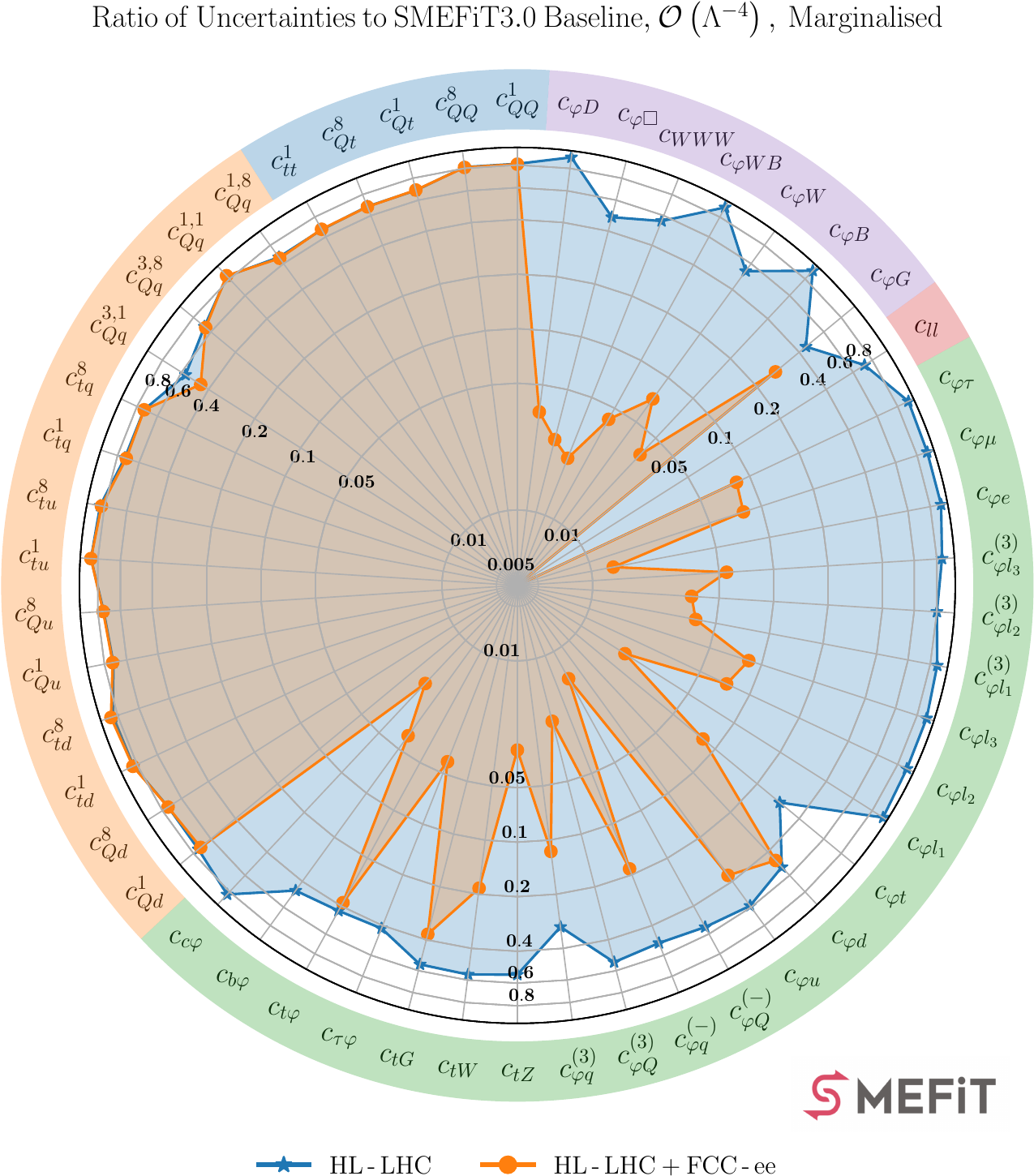}
    \caption{Left: The sequential impact of the separate $\sqrt{s}$ runs at the FCC-ee in the SMEFT fit based on $\mathcal{O}( \Lambda^{-2})$ calculations.
    Right: The impact of the HL-LHC and FCC-ee including  $\mathcal{O}( \Lambda^{-4})$ corrections. 
    }
\label{fig1}
\end{figure}

The right panel of Fig.~\ref{fig1}  displays the impact of the HL-LHC and FCC-ee in the global SMEFT fit once quadratic, $\mathcal{O}( \Lambda^{-4})$, corrections are accounted for.
%
The impact in the SMEFT parameter space of the FCC-ee
is similar in linear and quadratic fits. 
The fact that the results of the linear fit remain unchanged when adding quadratic effects would suggest that, with the precision of future measurements, the uncertainty associated with effects of ${\cal O}(\Lambda^{-4})$ is expected to be small. 
Although it is not shown here, as one would expect the EFT reach for both FCC-ee and CEPC is found to be very similar.

With the main addition compared to previous studies at future colliders being the top dataset, we compare in \cref{fig:Ops_DataSets} the impact of these observables on the bounds from the global fit. This is clear for the top dipole operators $c_{\PQt\PW}$ and $c_{\PQt\PZ}$, whose bounds are completely dominated by top data, whereas for $c_{\PQt G}$, bounds from top and Higgs observables seem to be more comparable ($c_{\PQt G}$ enters in gluon fusion and $\PH\to \Pg\Pg$).\footnote{On a related note, these processes, together with $\PH \rightarrow \PGg \PGg$~\cite{Jung:2020uzh}, also provide the strongest sensitivity to the $O_{\PQt\PGf}$ operator, that shifts the top quark Yukawa coupling.
However, the fit cannot resolve all operator coefficients that enter these decays and the global bounds on the top quark Yukawa coupling remains highly dependent on the tree-level sensitivity in $\ttbar\PH$ production.}  
Looking at the coefficients $c_{\varphi \PQt}$, $c_{\varphi Q}^{-}$, and $c_{\varphi Q}^{(3)}$, which modify the right-handed and left-handed top couplings to the $\PZ$ boson and the $\PW\PQt\PQb$ vertex, respectively, one also observes a complementarity between EW and Top observables, with the former controlling the bounds on the left-handed interactions (note that, e.g.\ a combination of $c_{\varphi Q}^{-}$ and $c_{\varphi Q}^{(3)}$ modifies the $Zbb$ vertex that enters in  EWPO) and the latter the right-handed ones.
In general, one observes a trend where the bounds on each operator are basically controlled by one of the two datasets. The exception are the coefficients $c_{\varphi \PW}$ and $c_{\varphi B}$. Looking in particular at this last one, which does not enter directly in top observables, the addition of top data in the fit results in a dramatic improvement on sensitivity. This occurs as a consequence of an approximate blind direction in Higgs observables between the contributions of such bosonic interactions and those from EW top dipoles. These two types of operators receive strong bounds from $\PSh \to \PGg\PGg$, but only via a linear combination. The dipole interactions are tested separately with the addition of $\epem\to\ttbar$ data, solving this blind direction and resulting in a much stronger bound for the bosonic operators. This was originally observed in Ref.~\cite{Durieux:2018ggn}.


\begin{figure}[!h]
    \centering
\includegraphics[width=0.85\linewidth]{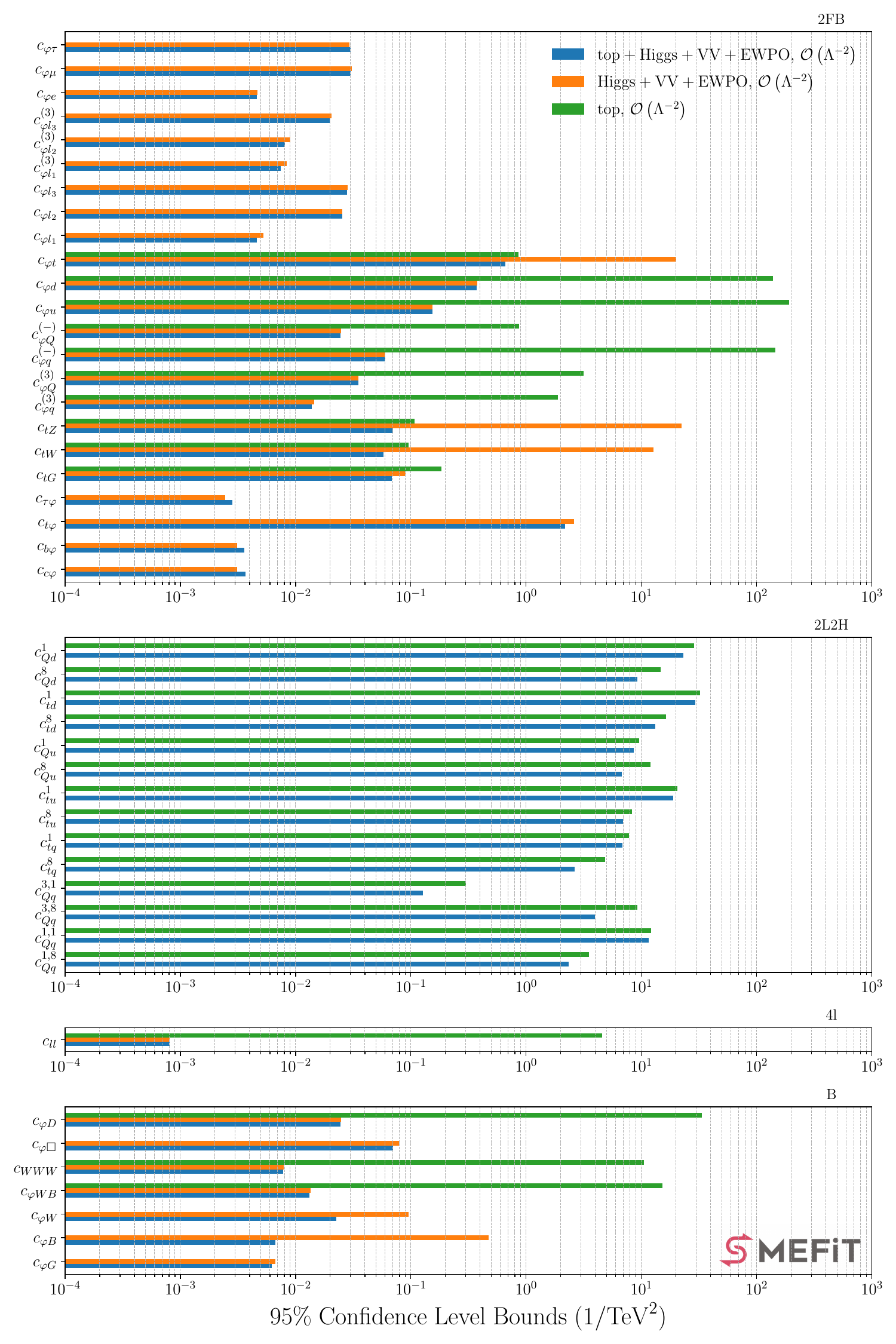}
    \caption{Impact of the top measurements on the bounds of the different Wilson coefficients in the global fit of Ref.~\cite{Celada:2024mcf}. We compare the bounds from the global fit (top+Higgs+VV+EWPO) at HL-LHC+FCC-ee, with those removing the top data set, as well as those obtained exclusively using top observables. 
    %
    }
\label{fig:Ops_DataSets}
\end{figure}

\subsection{BSM interpretation of SMEFT results}\label{sec:BSM_SMEFT}

The results of the SMEFT fit can be easily used to obtain information for the large class of models consistent with the SMEFT assumptions. This simply requires the matching between the EFT and the UV model of choice to be performed to obtain the values of the Wilson coefficients as function of the masses and couplings of the new particles of the model of choice. The SMEFT likelihood can then be projected in terms of these model parameters, to obtain bounds on the model of interest. This process of matching is not only fully known at the tree level to dimension six~\cite{deBlas:2017xtg}, but is also completely automatized up to 1-loop thanks to
tools like \textsc{MatchMakerEFT}~\cite{Carmona:2021xtq} or \textsc{Matchete}~\cite{Fuentes-Martin:2022jrf}.

The interpretation of the SMEFT results in terms of specific scenarios may also be illuminating for characterising properly the landscape of new physics scenarios explored at future colliders.  
Indeed, it is important to note that the flow of a specific new physics model to the SMEFT at low energies is not bidirectional.  It is true that any heavy new decoupling physics model will map into the SMEFT.  It is, however, not true that any SMEFT parameter point will necessarily map into a plausible heavy new physics model (i.e.\ models that, while possible, involve fine tunings not justified by any compelling theory argument).  As a result, agnostic SMEFT parameter fits may not accurately reflect the indirect power of a future collider.  This can arise in a number of ways.  There may be SMEFT flat directions which are difficult to constrain in a global fit which are otherwise impossible or implausible to realise within specific new physics models.  There may also be strongly constraining observables that correspond to one or more operators that arise frequently in specific models, rendering a collider a very powerful general probe, which otherwise from the SMEFT perspective seem to correspond to a small subset of a much larger whole of SMEFT possibilities.

For most of the results presented in this section, the interpretations will be based on 
the catalogue of models in Ref.~\cite{deBlas:2017xtg}. This comprises every scenario in which a dimension-6 SMEFT operator can be generated at the tree level, including the correlations between the various operators generated.  As such, while each individual model is not on its own particularly interesting or motivated, the collection of scenarios gives rise to a correlated set of SMEFT operators which can be considered representative of what can arise in specific well-motivated scenarios, whether weakly or strongly interacting.  The upshot is that while it might appear legitimate to consider, for example, the projected reach for a specific SMEFT operator, or direction in SMEFT parameter space, it may be that it is impossible or implausible to generate that isolated operator or pattern of operators in actual UV scenarios.  If the list were extended to scenarios generating SMEFT operators at the loop level then this point would only be strengthened, as typically a larger set of operators are generated at loop level than at tree level.

\subsubsection{Global constraints on single-particle SM extensions}\label{sec:SMEFfitUV}
%
The results in \cref{sec:SMEFT_EW_H_Top} can be recast in terms of the parameters of UV models for a broad class of UV scenarios matched to the SMEFT either at tree-level or at one-loop, thanks to the interface between \textsc{MatchMakerEFT}~\cite{Carmona:2021xtq} and \textsc{SMEFiT}~\cite{terHoeve:2023pvs}.
In this procedure, we restrict the possible UV couplings to ensure consistency with the \textsc{SMEFiT} flavour assumption after tree-level matching.
To illustrate the reach in the mass scale of new heavy particles achievable at future \epem circular colliders,
Fig.~\ref{fig2} displays results for the 95\% CL lower bounds on $M_{UV}$ achievable for different one-particle extensions of the SM matched at tree-level; see Ref.~\cite{Celada:2024mcf} for details.
Two possible values of the UV coupling, $g_{UV}=1$ and $4\pi$, are assumed.
For many scenarios, one finds that the FCC-ee measurements are sensitive to new physics mass scales up to a factor 10 larger than the EFT-based searches at the HL-LHC, reaching up to 100 TeV in models with $\mathcal{O}(1)$ couplings.
In Fig.~\ref{fig3b} we display the FCC-ee constraints on the couplings of one-particle UV extensions matched at one-loop  level and on a three-particle model matched at tree-level.

\begin{figure}[h]
    \centering
\includegraphics[width=0.90\linewidth]{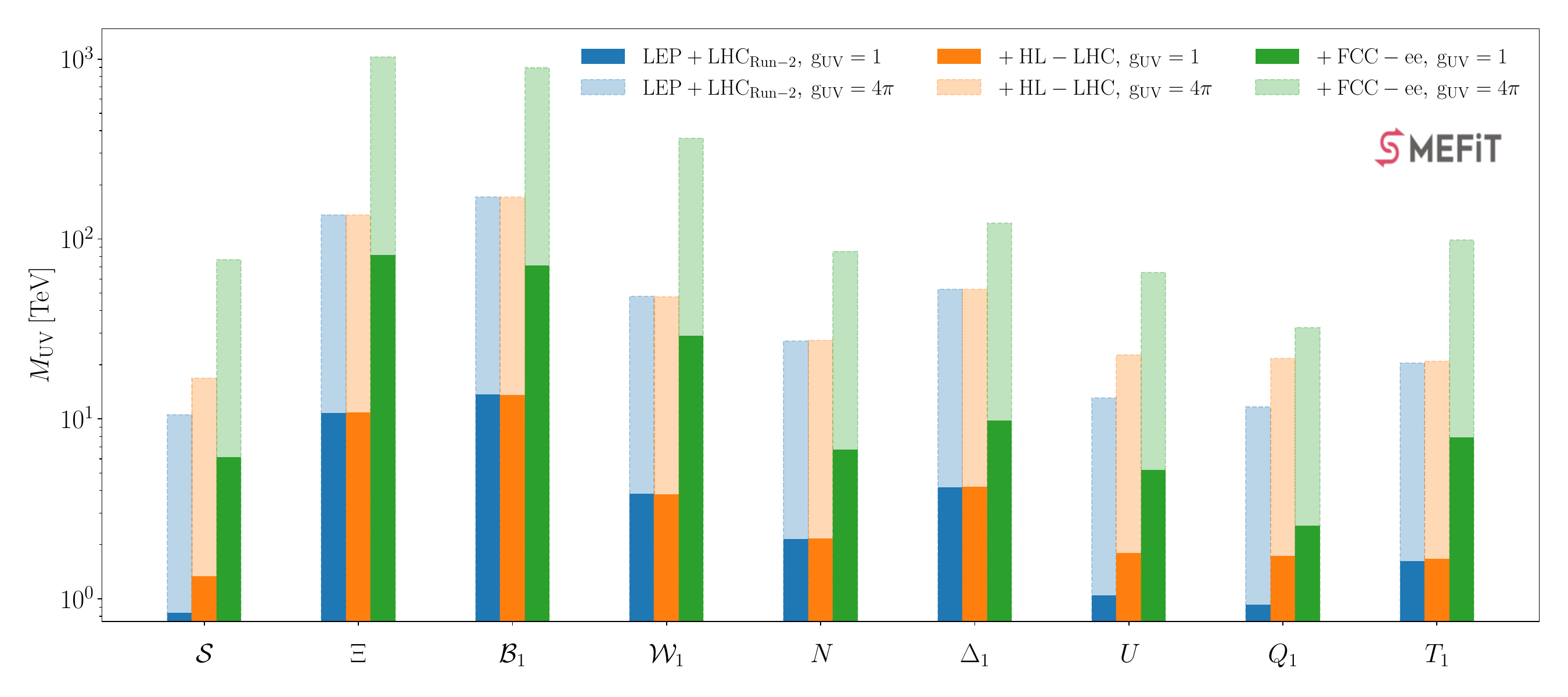}
    \caption{The reach in mass scale for new heavy particles in global SMEFT fits from current data, compared to that of the HL-LHC and the FCC-ee projections, for one-particle extensions of the SM matched at tree-level. Two different results are shown, assuming the new physics coupling $g_{UV}$ is either one (dark shaded bars) or $4\pi$ (light shaded bars).
    }
\label{fig2}
\end{figure}

%

\begin{figure}[h]
    \centering
\includegraphics[width=0.6\linewidth]{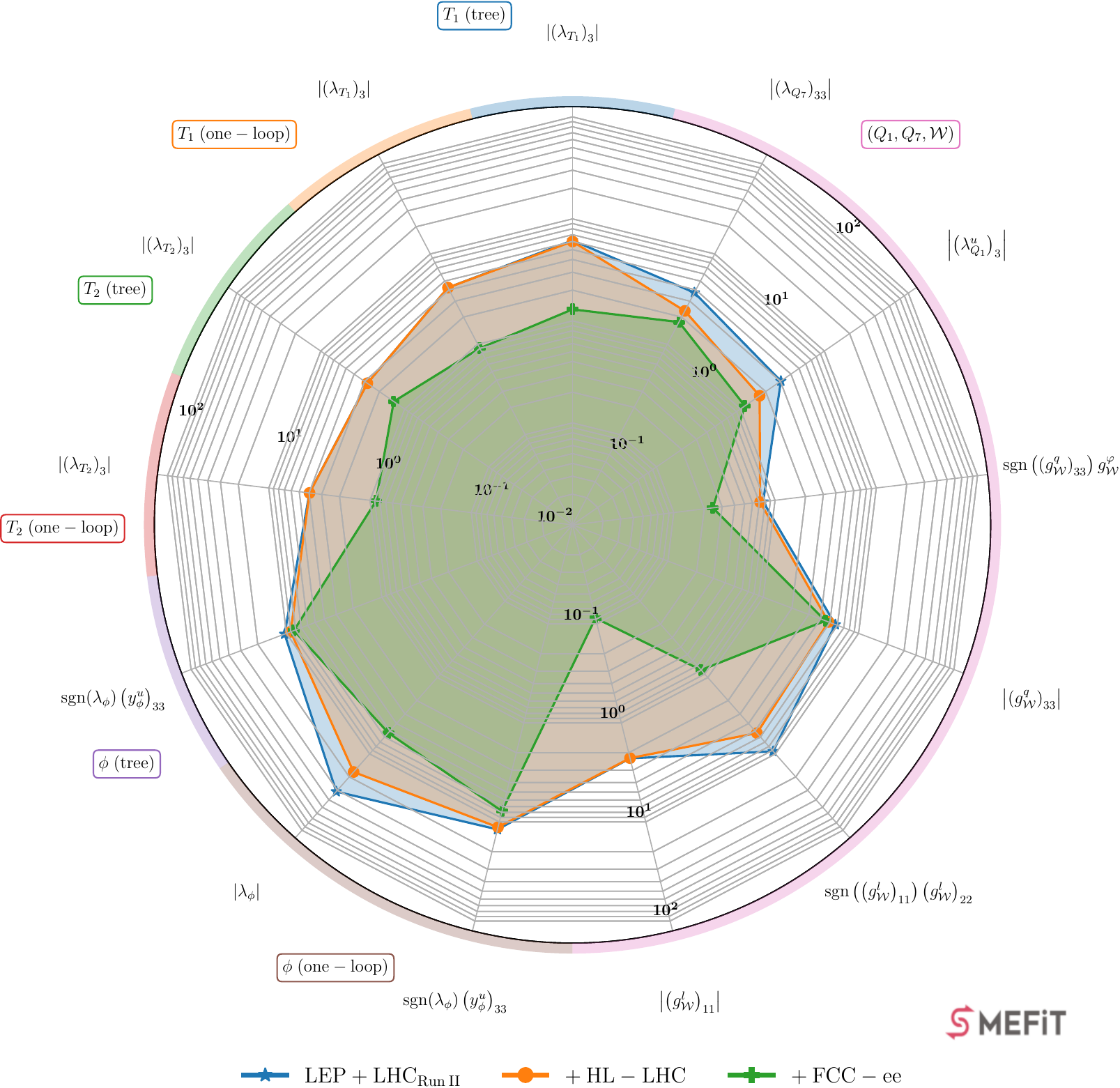}
    \caption{
    %
    FCC-ee constraints on one-particle UV models matched at one loop, and on a three-particle model matched at tree-level.
    }
\label{fig3b}
\end{figure}

In what follows, and to illustrate a bit more in detail how the previous picture can change once one considers NLO effects in the study, we discuss one particular example:
the case of EW-hypercharged scalar quadruplets. There are two types of such scalars that can contribute to dimension-six~\cite{deBlas:2017xtg} and we discuss the cases either by adding only one of these heavy scalars or by adding two of them with hypercharges $\frac{1}{2}$ and $\frac{3}{2}$ in a custodial-symmetric setting. 
By matching these UV models at the one-loop level to the SMEFT and accounting for Renormalisation Group Evolution (RGE) effects, we determine the reach in the UV parameter space from a global fit first at the HL-LHC and then at the FCC-ee. 
While naive considerations may suggest that the sensitivity is driven by the direct measurement of the Higgs self-coupling at the HL-LHC, we find that FCC-ee measurements, interpreted in the SMEFT framework, can improve the HL-LHC reach by at least a factor 2.

We consider an extension of the SM consisting of two scalar EW quadruplets of hypercharge $\frac{1}{2}$ and $\frac{3}{2}$, denoted by $\Phi$ and $\widetilde{\Phi}$ respectively, either together or separately. 
The relevant interactions of these heavy EW quadruplets with the SM fields are given by
\begin{equation}
\label{eq:lagrangian}
    \mathcal{L}_{\text{UV}} \supset - \lambda_{\Phi}\,H^* H^*\left(\varepsilon H\right)\Phi -\lambda_{\widetilde\Phi}  H^* H^* H^* \widetilde{\Phi} /\sqrt{3}+\text{h.c.}\, ,
\end{equation}
where $H$ indicates the SM Higgs field.
In the limit of $\lambda_{\widetilde\Phi} = \lambda_{\Phi} $ and $M_{\widetilde\Phi}=M_{\Phi}$, this model respects custodial symmetry and acquires an interesting phenomenology as studied in Refs.~\cite{Durieux:2022hbu,Allwicher:2024sso}.
Here we assume that the scalar masses are heavy enough to lie outside the reach of direct searches at the HL-LHC. 
The Lagrangian Eq.~(\ref{eq:lagrangian}) can be matched to the SMEFT both at tree and 1-loop level. 
We perform this matching using the \textsc{MatchMakerEFT} program~\cite{Carmona:2021xtq} and cross-check the results with those of Ref.~\cite{Durieux:2022hbu}.
Integrating out each quadruplet field generates only the $\mathcal{O}_{H}$ operator at tree level, whose main effect is to modify the Higgs self-coupling. At the 1-loop level, among others, the custodial-violating operator $\mathcal{O}_{H D}$ is generated. However, combining both quadruplets into a custodial-symmetric setup removes the contribution to $\mathcal{O}_{H D}$~\cite{Durieux:2022hbu}.

For this particular example, the global SMEFT analysis in \cref{sec:SMEFT_EW_H_Top} 
is extended with HL-LHC projections for double Higgs projections~\cite{Cepeda:2019klc}, sensitive to the Higgs self-coupling.
Additionally, we have included the dependence of $\PZ\PH$ production at FCC-ee on the $\mathcal{O}_{\PH}$ operator via EW loops by implementing the results in Ref.~\cite{Asteriadis:2024xts}.
The dynamical RGE effects are implemented through an interface with the \textsf{\emph{wilson}} package~\cite{Aebischer:2018bkb}.
%
%
The Wilson coefficients (WCs)  are run from a user-defined high scale $\Lambda$ (say multi-TeV) down to the scale of each observable, and then the fit results for the WCs (and UV parameters) are determined at $\Lambda$. 
Full details about this RGE implementation will be given in an upcoming publication~\cite{SMEFiT:RGE}.

We show in \cref{fig:bounds_quadruplet} the projected 95$\%$ CL on the absolute value of the coupling of the quadruplets to the Higgs bosons, $|\lambda_{\Phi}|$, at HL-LHC and the FCC-ee. 
On the left panel, we display the results for the custodial model with both $\Phi$ and $\widetilde{\Phi}$, meanwhile on the right panel, we show the case of considering only $\Phi$, which leads to a loop-suppressed custodial violation. In both cases, we consider masses of $M_{\Phi}=$ \SI{4}{\tera \electronvolt}, which we also take as the high scale $\Lambda$ for the RGE running.
We present results for different theory settings: for tree-level vs one-loop matching, and with and without RGE.
We also show results separately with and without the HL-LHC bounds on the di-Higgs cross-section and with and without NLO EW corrections to $\PZ\PH$ production at FCC-ee, which bring loop-induced sensitivity to $\mathcal{O}_{\PH}$.

When matched at tree level, both scenarios can be probed only via their contribution to $\mathcal{O}_{\PH}$, hence the sensitivity stems from only two processes: double Higgs production at HL-LHC, and \PZ{}\PH production at NLO EW at FCC-ee.
We observe how the loop-induced sensitivity at FCC-ee can improve the bound by a factor $\sim 2$ with respect to HL-LHC.
The RGE running leads to a slightly worse bound since the $\mathcal{O}_{H}$ operator coefficient runs down to smaller values as we lower the scale.

\begin{figure}[t]
    \centering
    %
    %
\includegraphics[width=\linewidth]{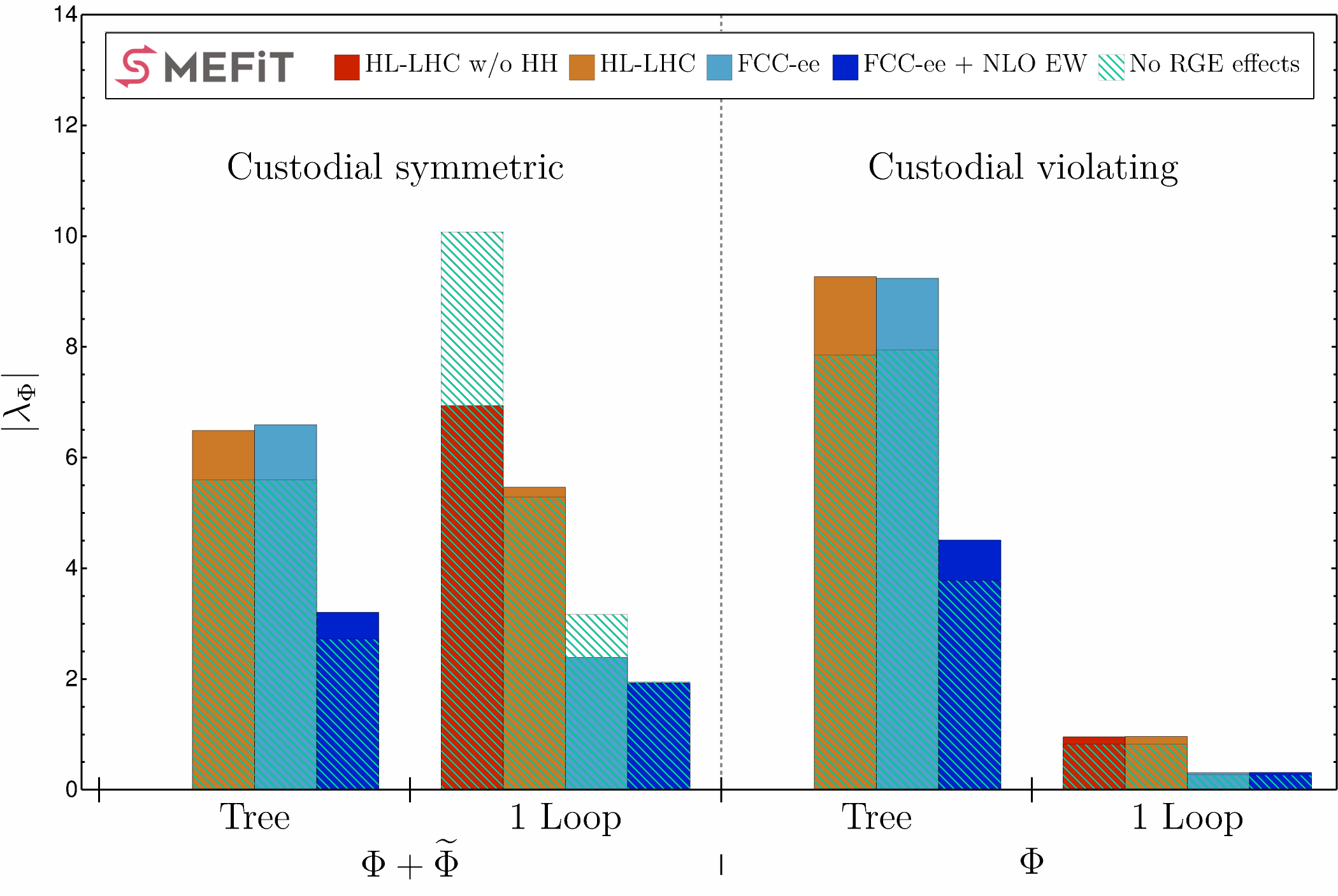}
    \caption{
The 95$\%$ CL upper bounds on the coupling $|\lambda_{\Phi}|$ between the heavy EW quadruplets and the Higgs boson, see Eq.~(\ref{eq:lagrangian}), obtained from a global SMEFT analysis. We consider the datasets corresponding to HL-LHC without HH projections (dark red), HL-LHC with HH projections (orange), HL-LHC+FCC-ee (light blue), and HL-LHC+FCC-ee with NLO EW corrections in the $\PZ\PH$ process. The full bars indicate the results with the RGE running from the matching scale $M_{UV}$ down to the scale of each observable and the hatched bars show the result of not including the RGE running. The mass of the quadruplets is set to $M_{UV}=4$~TeV. In the left half of the plot, we show the results of the custodial symmetric model that include both quadruplets, $\Phi$ and $\widetilde{\Phi}$. The right half shows the non-custodial case of considering only the quadruplet $\Phi$. For each model, we show the results of considering either tree-level or 1-loop-level matching.
}    \label{fig:bounds_quadruplet}
\end{figure}

The results with 1-loop matching clearly distinguish between the non-custodial and custodial scenarios. 
Starting with the latter, FCC-ee offers a sensitivity $\sim3$ times greater than HL-LHC when excluding double-Higgs measurements, regardless of the RGE effects.
Double Higgs production improves the HL-LHC sensitivity by a factor $\sim 2$. 
When including both RGE effects and HL-LHC double Higgs production, the FCC-ee sensitivity to this model is two times better than HL-LHC. 
%
When including 1-loop matching results, the impact of the loop-induced FCC-ee sensitivity to the $\mathcal{O}_{\PH}$ operator is reduced, in particular when including RGE effects. This shows the power of 1-loop matching and running to account partially for higher-order corrections.

The non-custodial scenario presents a rather different picture. 
There the sensitivity is driven by the 1-loop contribution to the custodial-violating operator $\mathcal{O}_{\PH D}$ at both HL-LHC and FCC-ee. Thus, the inclusion or not of RGE effects, double Higgs production, and loop-induced sensitivity to $\mathcal{O}_{\PH}$ has overall little effect.
The highly precise measurement of EW parameters and Higgs couplings expected at the FCC-ee implies a significant increase in sensitivity against HL-LHC for this non-custodial model. In particular, the impact associated to the power of EW measurements will be the focus of the next section.

In the custodial scenario, the FCC-ee advantage always arises from a 1-loop computation: either matching or loop-induced sensitivity to a tree-level generated operator~\cite{McCullough:2013rea,Carmona:2021xtq,DiVita:2017vrr,Asteriadis:2024xts}.
This stresses the need for progress in higher-order computations to match the future \epem precision and explore thoroughly the New Physics landscape.

\subsection{New Physics beyond leading order at Tera-Z}\label{sec:SMEFT_TeraZ}

Building on the last comment of the previous section, regarding the need and importance of higher order corrections for new physics interpretations, here we focus entirely on these type of effects, in particular on EWPO. Though expected to be suppressed by loop factors, we will see that the extreme precision of the measurements possible at a future Tera-Z run could bring significant indirect sensitivity to new physics, even when it does not contribute to such observables at the tree level.

\subsubsection{New physics in the third family}

New physics in the third generation remains highly motivated, as NP dominantly coupled to the Higgs and the top quark is minimally required to address the EW hierarchy problem. A classical example of this type of new physics includes composite Higgs models. 
Furthermore, as long as NP couples universally with the light families, having a different coupling to the third generation is compatible with experimental observations, namely that there are no deviations from the Standard Model (SM) in flavour-changing neutral currents and NP does not couple strongly to valence quarks at nearby energy scales. 
This scenario is described by an $U(2)^5$ flavour symmetry, which is a good approximate symmetry of SM Yukawa couplings. They provide an excellent starting point for reducing the number of SMEFT parameters needed to describe NP at nearby energy scales. 

 While bounds from proton colliders can be evaded by NP that couples mostly to the third family, the same is not true for electroweak (EW) precision tests at the \PZ-pole. The reason is the universality of the SM gauge interactions -- the produced \PZ bosons decay almost universally, meaning that NP coupled to any generation is probed nearly identically. While only 23 SMEFT operators affect \PZ-pole observables at leading order, hundreds more appear when the observables are computed at next-to-leading order (NLO)~\cite{Bellafronte:2023amz}. Intriguingly, this set includes almost all of the operators respecting $U(2)^5$ flavour symmetries~\cite{Allwicher:2023shc}. This makes the \PZ-pole a powerful indirect probe of NP that we could hope to directly produce at future colliders, a fact that has not been widely appreciated thus far in the literature.

Since the broad sensitivity of \PZ-pole observables to NP relies on loop effects, it is crucial to have a machine capable of providing a large amount of statistics. A ``\TeraZ'' program producing over $10^{12}$ \PZ bosons at a future circular $\epem$ colliders is required in order to fully take advantage of this opportunity. Generally speaking, a \TeraZ machine can probe SMEFT operators entering \PZ-pole observables at leading order up to 100 TeV, while there is sensitivity to operators entering at 1 or 2 loops of 10 or 1 TeV, respectively. This can be seen in~\cref{fig:EWplotvertical}, where the $\PZ/\PW$-pole (tree-level) and $\PZ/\PW$-pole (RGE) categories show the leading and NLO sensitivities of the \PZ-pole to various SMEFT operators. Furthermore, in the next section it will be demonstrated that, at the level of new particles, there is sensitivity at one-loop level to almost all perturbative UV models which generate a tree-level dimension-6 contribution to the SMEFT~\cite{Allwicher:2024sso}. The key point is that RG evolution brings nearly all models within \TeraZ reach, even if they do not contribute to \PZ-pole observables at tree level. Importantly, these RG contributions are IR-calculable and difficult to tune away, since they are proportional to $\log(\Lambda_{\text{NP}}^2/ m_Z^2)$.

The $\PZ/\PW$-pole (RGE) category of~\cref{fig:EWplotvertical} is not exhaustive but focuses for brevity on operators with top quarks, which have large RG mixings into \PZ-pole operators. These mixings are realized by closing top-quark loops and attaching Higgses with the top Yukawa. Though not shown, RG mixing via the SM gauge couplings is also important~\cite{Allwicher:2023aql,Allwicher:2023shc}. These RG effects in the first-leading-logarithmic approximation can also be seen via an NLO computation of the \PZ-pole observables. However, exactly solving the 1-loop RG equations also captures some higher-order effects, such as a 2-loop contribution of 4-top operators to the Peskin-Takeuchi $T$ parameter~\cite{Allwicher:2023aql,Stefanek:2024kds,Haisch:2024wnw}. It is remarkable that even with current data, this effect allows the \PZ-pole to provide better constraints on 4-top operators than high-energy probes from top production at the LHC. 

\begin{figure}[p]
    \centering
    \includegraphics[scale=0.9]{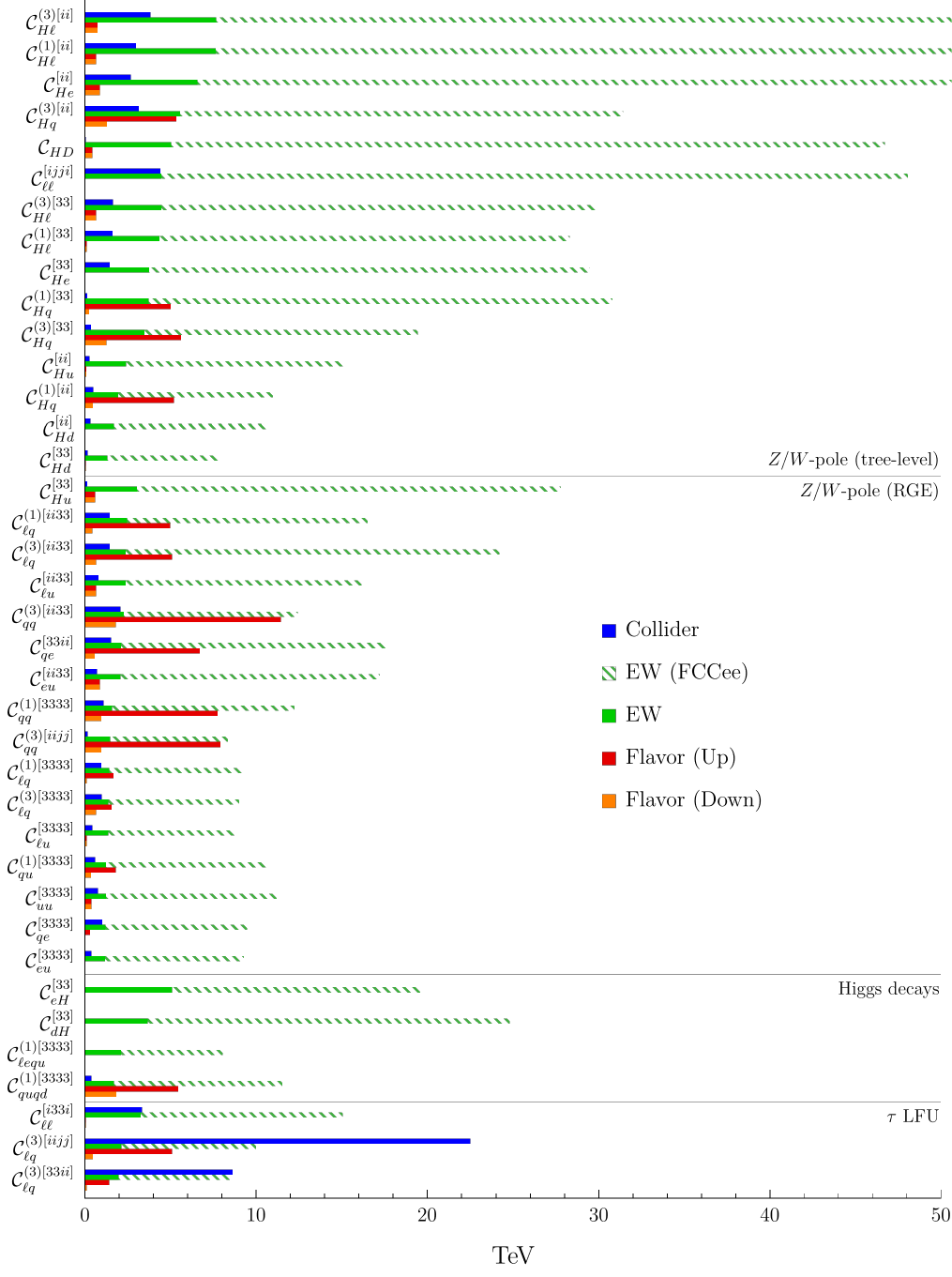}
    \caption{ FCC-ee projection for operators with the strongest current EW constraints, sorted by observable categories. The $\PZ/\PW$-pole (RGE) category contains operators which enter $\PZ/\PW$-pole observables only via RG mixing into $\PZ/\PW$-pole (tree-level) operators. See Ref.~\cite{Allwicher:2023shc} for more details.} 
    \label{fig:EWplotvertical}
\end{figure}

The fact that corrections to EW precision observables are important in the context of third-generation NP was first noticed in Ref.~\cite{Feruglio:2017rjo} and found to be even more comprehensive in Ref.~\cite{Allwicher:2023aql}. Connecting with the implications for specific new physics scenarios, like the case of composite Higgs models, the loop-level phenomenology was recently studied in the context of a \TeraZ program~\cite{Stefanek:2024kds}, taking into account the large RG mixings of operators unavoidably generated by a partially-composite top quark. It was shown that these effects, not considered in previous flavour-universal only analyses without RGE, allow a future \TeraZ machine to probe composite Higgs models up to a scale of $>25$ TeV. This would allow a \TeraZ program to test the naturalness of the EW scale at the $10^{-4}$ level.
In \cref{fig:CHright} we compare the current bounds on the $g_\star-m_\star$ plane, with $g_\star$ the typical interaction strength of the strong sector and $m_\star$ its mass scale, with the ones that would be possible at the FCC-ee for the case of right compositeness.

Trillions of \PZ bosons recorded in a clean $\epem$ environment would offer unprecedented microscopic resolution, necessarily revealing the interplay between the EW, Higgs, QCD and flavour sectors of the Standard Model that occurs beyond tree level. To make the most of the NP reach of a \TeraZ machine and demonstrate this interplay, NLO and even NNLO computations of \PZ-pole observables will be needed. Luckily, recent progress in the field means that semi-analytic results at NLO are already known~\cite{Bellafronte:2023amz} while the most important NNLO effects have begun to be accounted for~\cite{Haisch:2024wnw}.

\begin{figure}[t]
    \centering
    \includegraphics[scale=0.9]{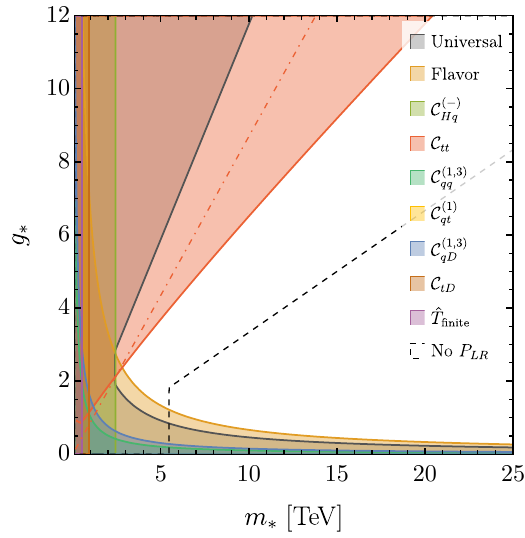}
    \includegraphics[scale=0.9]{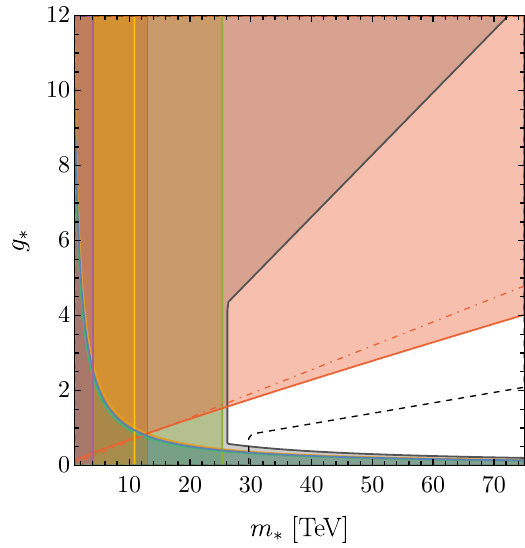}
    \caption{Comparison of current (left) and Tera-Z (right) bounds on the $g_\star-m_\star$ plane for a composite Higgs scenario (for the case of right compositeness). The assumed value of $\Lambda$ for current and future data is 2.5 and 25 TeV, respectively. See Ref.~\cite{stefanek2025nonuniversalprobescompositehiggs} for details.} 
    \label{fig:CHright}
\end{figure}

\subsubsection{Single-particle SM extensions}

This section, based on Ref.~\cite{Allwicher:2024sso}, builds on the observations about the exploratory power of the Tera-Z component of the FCC-ee physics programme highlighted in the previous section and puts it, again, in the context of sensitivity to the catalogue of models in Ref.~\cite{deBlas:2017xtg}, also used in \cref{sec:SMEFfitUV}. In that section, RGE effects were considered for the case of the scalar quadruplet, while here the analysis is extended to the whole set of models in the catalogue, but from the point of view of Tera-Z constraints only. Although this section focuses then on a more restricted set of observables than ~\cref{sec:SMEFfitUV}, the point of this study is to show how, even when some of these models do not contribute at leading-order, the exquisite precision of future Z-pole measurement will still bring great sensitivity to the NLO effects of these extensions, illustrated here by their leading-logarithimic contributions.

\begin{figure}[t!]
    \centering
    \includegraphics[width=0.48\textwidth]{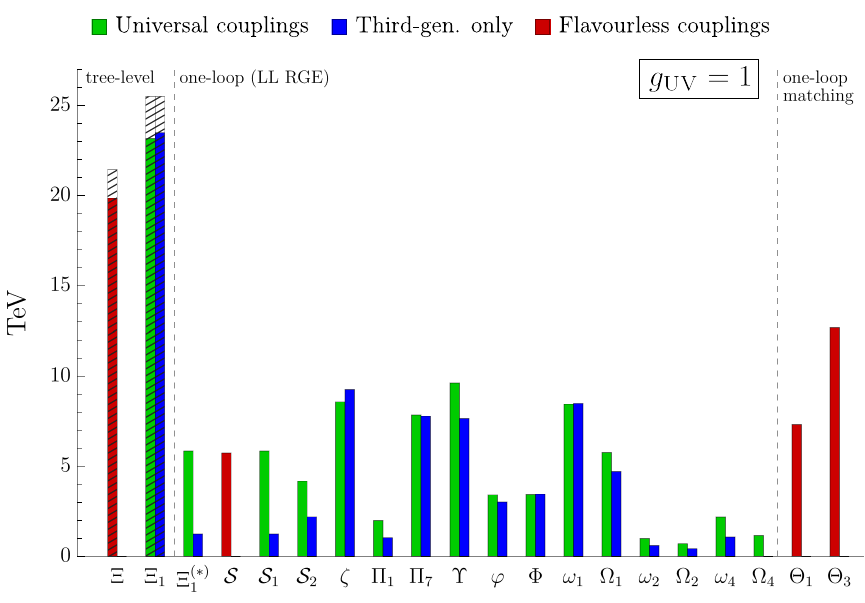}  \includegraphics[width=0.48\textwidth]{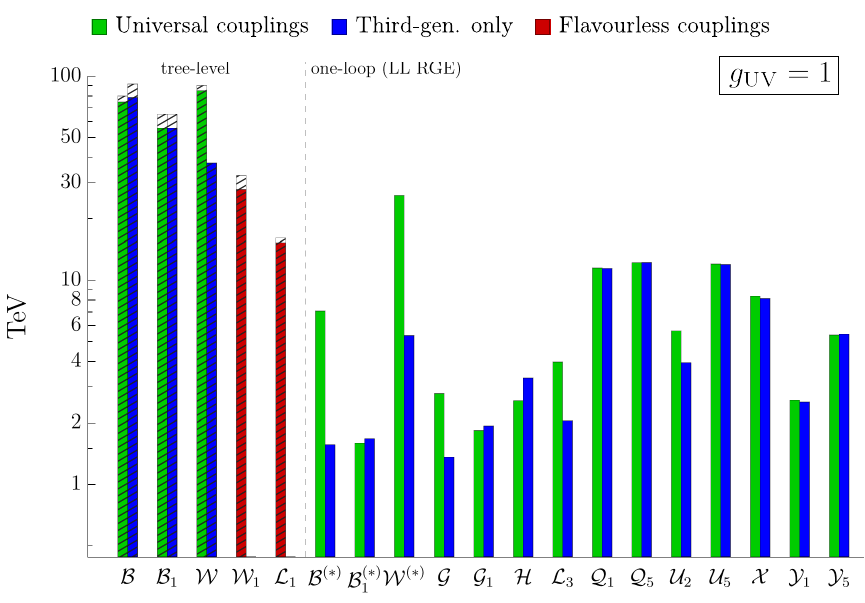} \newline
        \includegraphics[width=0.48\textwidth]{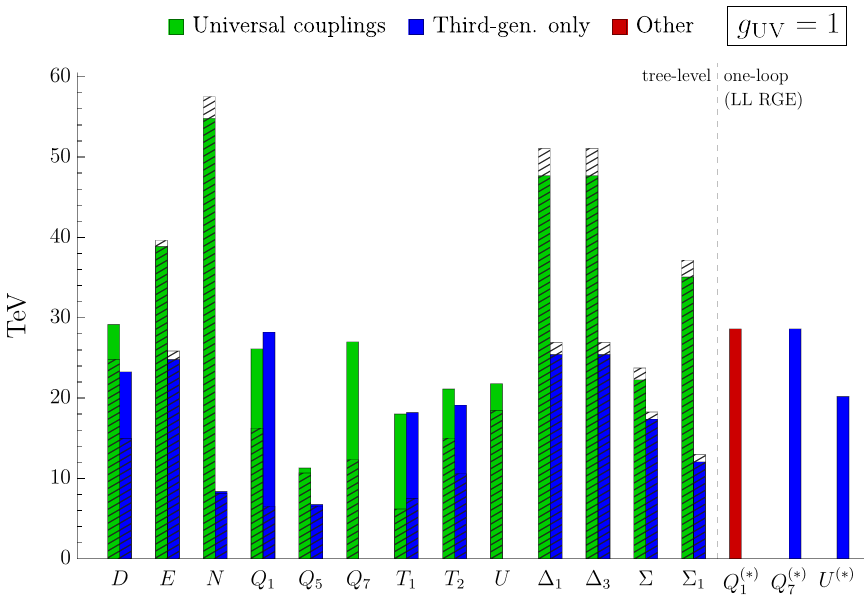}
    \caption{Projected bounds (95\% CL) on the masses of new scalar (top left), vector (top right) and fermion (bottom) fields, assuming unit couplings for the new particles, $g_{\rm UV}=1$. The vertical dashed lines separate fields which contribute to EWPOs at tree level, via one-loop RG evolution, and via one-loop matching. The green and blue bars correspond to different assumptions for the coupling to SM fermions, while red bars are used when the state does not have couplings to fermions.  For the fermions the red bar for $Q_1^{(*)}$ corresponds to the exceptional case where $Q_1$ couples only to right-handed top quarks. Fields indicated with a $^{(*)}$ correspond to cases where the tree-level contribution has been set to zero by forbidding a specific coupling. Hatched bars correspond to pure tree-level limits, without RG running.\label{fig:news}}
\end{figure}

Unit couplings are assumed throughout and no fine-tuning of UV parameters is undertaken.  For the most part only the RG evolution of SMEFT coefficients from the matching scale to the EW scale is performed at one loop, hence finite one-loop terms are overlooked, although including such terms would not change any of the qualitative conclusions.  Running is calculated at the first leading-log level and run from a scale of $2$ TeV down to the EW scale using DsixTools \cite{Celis:2017hod,Fuentes-Martin:2020zaz}.  The mapping from SMEFT coefficients to EW observables is described in Ref.~\cite{Allwicher:2024sso}.

Projected limits are shown in \cref{fig:news}.  It is evident that Tera-Z can indirectly probe the vast majority of scenarios with a sensitivity that, assuming order one couplings, can reach scales from the TeV up to above 50 TeV in some instances, offering the next major frontier in precision EW physics and a powerful and necessary complement to a Higgs factory programme.

It is notable that almost any model that can generate a non-vanishing dimension-6 SMEFT coefficient at tree-level will give contributions to precision Tera-Z observables either at tree-level or at one-loop.  The impact of this is that in specific models which give contributions to complementary components of a future collider programme, such as in Higgs physics, one also expects a complementary, or in some cases stronger, probe of that scenario to be realised at Tera-Z.  Moreover, if indirect evidence of new physics did begin to emerge at future colliders, Tera-Z would provide invaluable additional information in mapping out the nature of that new physics in advance of a next generation programme to explore new physics directly at higher energies.

\clearpage

\newcommand\srchinc[1]{\input{sections/searches/#1/#1.tex}}


\newcommand{\eqn}{equation}
\newcommand{\al}{\alpha}
\newcommand{\be}{\beta}
\newcommand{\lb}{\left(}
\newcommand{\rb}{\right)}

\section{Direct Searches for New Particles}
\label{sec:searches}
\editors{Rebeca Gonzalez Suarez, Roberto Franceschini, Aleksander Filip Zarnecki}

This chapter reviews the discovery potential of a future \epem collider in direct searches for new physics. The material is organised in sections covering a variety of exotic models (exotic scalars, new gauge bosons, heavy neutral leptons, SUSY, Dark Matter), with the exception of long-lived particles, which is a signature-driven search and for which the development of dedicated experimental techniques is usually required.


\subsection{General motivation for BSM searches at a Higgs/electroweak/top factory}

Higgs/electroweak/top factories have an enormous potential for probing the dynamics of the SM via precision measurements, as detailed in previous chapters. This potential largely relies on the high intensity that a Higgs/electroweak/top factory can achieve in colliding $\Pep \Pem$ beams. The $\sqrt{s}$ energy directly accessible at $\Pep \Pem$ is not much higher than already achieved for instance at LEP~2, thus it is important to explore the improvements brought compared with previous searches at a similar  $\Pep \Pem$ machine like LEP~2, but also the results of high-energy colliders such as the (HL-)LHC, and current high intensity facilities such as SuperKEK-B and the Belle~II experiment operating at lower energies. 

A Higgs/electroweak/top factory (HETF) clearly has the potential to discover new physics not accessible to current high-intensity machines, because it might be too heavy to be observed by a \SI{10}{\giga\electronvolt} machine like SuperKEK-B, or with the relatively low luminosity of previous higher energy machines such as LEP~2. The new physics in direct reach of a HETF typically has non-zero cross section at the HL-LHC, but suffers from backgrounds that are too large at a hadron machine. The experience gained in the search for the SM Higgs boson makes it evident that the discovery of even a relatively simple object such as the Higgs boson, would have been significantly challenging at the LHC if it were much lighter than \SI{125}{\giga\electronvolt}. In the following sections we propose a survey of models 
that can be difficult to observe at the LHC, but that have a great discovery potential at a HETF.

The list of possible new physics signals that are too faint at the LHC is extensive. In general the elusive new physics states have couplings to gluons that are very small --- or even zero --- and can be produced at the LHC only via electroweak interactions. This implies relatively small cross sections at the LHC, posing a significant challenge for their pursuit in a hadronic machine. For this reason an $\epem$ machine offers a discovery path that is complementary to that at the LHC for the direct search of new physics. New electroweak charged states from new physics have sizable production cross sections at a HETF, comparable to those of their background processes. Therefore, the general situation with a very small signal-to-noise ratio at the LHC gets reversed, with a typical signal-to-noise expected at a HETF of around 1 for simple $ 2 \to 2$ new physics production. 

New physics with ultra-weak couplings to the SM can be probed very efficiently at a HETF by exploiting the high intensity of collision rates. Depending on the type of signal, couplings that are of orders of magnitudes smaller than the sensitivity of previous or present machines can be probed, looking for peculiar events from new physics. In this category, it is natural to expect particles that could be long-lived, as weak couplings lead to long lifetimes, whose search is discussed in a mainly signature-driven approach in \cref{sec:LLP-focus-topic}. In addition, the search for long-lived particles can be sensitive to ``hidden sectors'' of new physics, which have a heavy mediator state that can interact with both the SM and the new physics sector. In this case, the long lifetime is expected to arise from the heaviness of the mediator state through which the hidden sector particles are produced and through which at least some of them will decay back into an observable states of the SM. 

For hidden new-physics sectors one can also expect signals involving an apparent imbalance of energy and momentum in the detectors of a HETF. This is a typical hallmark of states that can be candidates for (a fraction of) the Dark Matter of the universe. Therefore, a HETF has the potential to cover Dark Matter models and regions of parameter space that are inaccessible at the HL-LHC. We discuss these searches in \cref{sec:SRCH-darkmatter}. Hidden sectors are expected to have a complex structure, and not just be simply a single particle or a single-interaction ``portal''. A generic feature of the hidden sector is the existence of new gauge interactions, characterising the dynamics of that sector. In \cref{sec:SRCH-new-dark-gauge-bosons}, we present a survey of results relevant for the search and for the characterisation of the hidden sector that put a HETF at the forefront of this type of search for new physics. In addition, new gauge interactions can also be introduced to extend the SM to address open questions without necessarily advocating a whole new sector. Example studies of the discovery potential for new gauge bosons are presented in \cref{sec:SRCH-flavor-gauge-bosons}.

Assuming a more simplified point of view, a HETF can also probe the so-called ``portals'' to new physics. In these scenarios, a single (or few) interactions are assumed to be responsible for the SM interactions with the new physics sector. In this sense, a ``portal'' scenario is a very well-defined model, which can be explored in great detail. It may lack the complexity of a full model of new physics that addresses one or more of the open issues in the SM, but can be used as a very effective metric to evaluate the performance of a collider, allowing also to put the model in relation to other experimental approaches sensitive to the same new physics. In this category, in \cref{sec:SRCH-NHL} we show results for a new heavy neutral lepton, which could be involved in the mystery of neutrino masses, and can also serve as a benchmark to understand the detector capabilities at a HETF. This is also investigated in cases of long-lived signatures in \cref{sec:LLP-focus-topic}. Results from \cref{sec:exotic-scalars-searches} can also be applied to the so-called ``Higgs portal'', and results on new gauge boson mixing with the SM photon from \cref{sec:SRCH-new-dark-gauge-bosons} apply to the ``photon portal''. 

On top of these signals, there is a number of possible models to which a HETF is sensitive and that have to do with blind-spots of the searches for new physics at the LHC. Well-known examples of these blind-spots are the compressed mass spectrum scenarios in supersymmetry. These issues have been plaguing the reach of the LHC experiments for a long time. For colour-charged superpartners the combined reach of the many searches at the LHC might have covered all possible blind-spots already~\cite{Bagnaschi:2023cxg,Franceschini:2023nlp}, but this issue remains under study. Indeed, the shortcoming of the LHC searches is particularly severe for electroweak superpartners~\cite{Agashe:2023itp,Agashe:2024owh}. In these most challenging scenarios for the LHC, the  HETF will help enormously in removing any blind-spot that could remain unresolved until the end of the HL-LHC. 

In \cref{sec:SRCH-susy}, we present results on electroweak production of supersymmetric particles. We show that a HETF can access the regime of compressed spectra. In addition, we pick up a recent, still relatively small deviation from the Standard Model in searches for electroweak fermionic superpartners \cite{ATLAS:2019lng,ATLAS:2021moa,CMS:2021edw} and explore the possibility that these excesses will persist in the forthcoming LHC runs. These SUSY signatures are ``textbook'' manifestation of supersymmetry at colliders~\cite{Martin:1997ns,Canepa:2020ntc,Altakach:2023tsd,Agin:2024yfs}, thus they lend themselves for an assessment of highly detailed questions on the new physics model. In the following we discuss how a HETF is an ideal machine to characterise more precisely what new physics may be discovered at the LHC (or at a HETF itself). For this purpose, we show a study aimed at mapping  the signals of invisible particles to a univocal Dark Matter candidate of the Minimal Supersymmetric Standard Model.

Any excesses in the search for electroweak superpartners, or other excesses that appeared in the LHC data~\cite{cms-higgshunting24,Kundu:2022bpy,LeYaouanc:2023zvi,Yaouanc:2024xqg}, are still too weak to serve as convincing evidence of new physics. More data from the forthcoming runs of the LHC will help to clarify their origin.  The  exact improvement in the reach for light and electroweak states that can be achieved at future LHC runs remains to be seen. In any case, a HETF will have unique capabilities in the detailed study of any new physics discovered at the LHC.

 



\subsection{\focustopic Exotic scalar searches \label{sec:exotic-scalars-searches}}
\editor{Filip}

\subsubsection{Overview of scalar extensions of the Standard Model}
In this section, we briefly discuss typical new physics scenarios that still allow for additional neutral scalar states with masses that can be produced on-shell in a collider environment with a centre of mass energy of 240--250 \GeV. We do not claim to be inclusive but list straightforward popular extensions.

New physics scenarios with additional scalar content typically stem from extending the SM scalar sector by additional singlets/ doublets/ triplets, where the nomenclature decribes the transformation under the SM gauge group. Depending on the specific extension considered, the new physics scenarios contain several \CP-even or -odd additional neutral, additional charged, or multi-charged scalars. 

In general, new physics scenarios are subject to a long list of theoretical and experimental constraints, such as conditions on vacuum stability, unitarity, perturbativity of couplings (up to certain scales), as well as experimental findings: current and past collider search results, electroweak precision observables, agreement with flavour data, as well as dark matter findings in case the model contains a dark matter candidate. Another important ingredient is the fact that the new physics scenario has to comply with the properties of the 125 
~\GeV resonance discovered by the LHC experiments \cite{ATLAS:2022vkf,CMS:2022dwd}.
For the models presented below, we refer the reader to the corresponding publications for further details.

The main production mode for most models will be scalar strahlung involving the $\PGf \PZ\PZ$ vertex, however, we have to mention an important sum rule \cite{Gunion:1990kf} that relates the couplings of additional neutral scalars:
\begin{equation}
\sum_i g^2_{\PSh_i V V} \,=\,g^2_{\PSh_\text{SM} V V} \; ,
\end{equation}
where the sum runs over all neutral \CP-even scalars, $\PSh_\text{SM}$ denotes the SM Higgs, and $VV\,\in\,\left[\PZ\PZ; \PWp\PWm\right]$. As one of the scalars has to comply with the measurements of the 125 \GeV resonance at the LHC experiments, we see that from the signal strength alone we already have strong constraints on the new physics couplings. The above rule is amended in the presence of doubly charged scalars.

Reference \cite{Robens:2022zgk} has additional details, with updates presented in Ref.~\cite{Robens:2024wbw}.
\paragraph{Singlet extensions}
In singlet extensions (see e.g.\ Refs.~\cite{Pruna:2013bma,Robens:2015gla,Robens:2016xkb,Ilnicka:2018def,Robens:2019kga}), the SM scalar potential is enhanced by additional scalar states that are singlets under the SM gauge group. In such scenarios, the coupling of the novel scalar(s) to SM particles is typically inherited via mixing, i.e.\ mass-eigenstates are related to gauge eigenstates via a unitary mixing matrix. The corresponding couplings and interactions are mediated via a simple mixing angle. In general, in such models scalars have decay modes that connect them both to the SM as well as to the novel scalar sector. If the additional states are singlets, all interactions to SM final states are rescaled by a mixing angle, such that
$g_{\PSh_i,\text{SM}\text{SM}}\,=\, \sin\al_i\,\times\,g^\text{SM}_{\PSh_i,\text{SM}\text{SM}}$
for general $\text{SM}\text{SM}$ final states, where $\PSh_i$ and $\al_i$ denote the respective scalar and mixing angle. Depending on the specific kinematics, novel triple and quartic interactions are also possible, where the couplings depend on the specific values of the potential parameters. This allows in general for decays $\PSh_i\,\rightarrow\,\PSh_j\,\PSh_k$. Naturally, one of the \CP-even neutral scalars of the model has to have properties that concur with the SM-like scalar of mass 125\,GeV measured by the LHC experiments.

The status of the current searches \cite{Cepeda:2021rql} for the process
\begin{equation}\label{eq:lims}
\Pp\Pp\,\rightarrow\,\PSh_{125}\,\rightarrow\,\PSh_i\,\PSh_i\,\rightarrow\,X\,X\,Y\,Y \; ,
\end{equation}
which for such models can be read as a bound on 
$\sin^2\,\alpha\,\times\,\text{BR}_{\PSh_{125}\,\rightarrow\,\PSh_i\,\PSh_i\,\rightarrow\,X\,X\,Y\,Y}$,
can be seen in \cref{fig:current_stat}. Current bounds on the mixing angle for the 125 \GeV-like state are around $|\sin\,\al|\,\lesssim\,0.25$ \cite{Robens:2024wbw}, which in turn means that branching ratios $\text{BR}_{\PSh_{125}\,\rightarrow\,\PSh_i\,\PSh_i\,\rightarrow\,X\,X\,Y\,Y}$ down to $\mathcal{O}\lb 10^{-5}\rb$ can be tested. In particular, the $\PGm\PGm\PGm\PGm$ final states in the low mass region give interesting constraints on the $\PSh_{125}\,\rightarrow\,\PSh_i\,\PSh_i$ branching ratio down to $\,\sim\,0.03$. A related discussion can be found in Ref.~\cite{Carena:2022yvx}.
\begin{center}
\begin{figure}
\begin{center}
\includegraphics[width=0.95\textwidth]{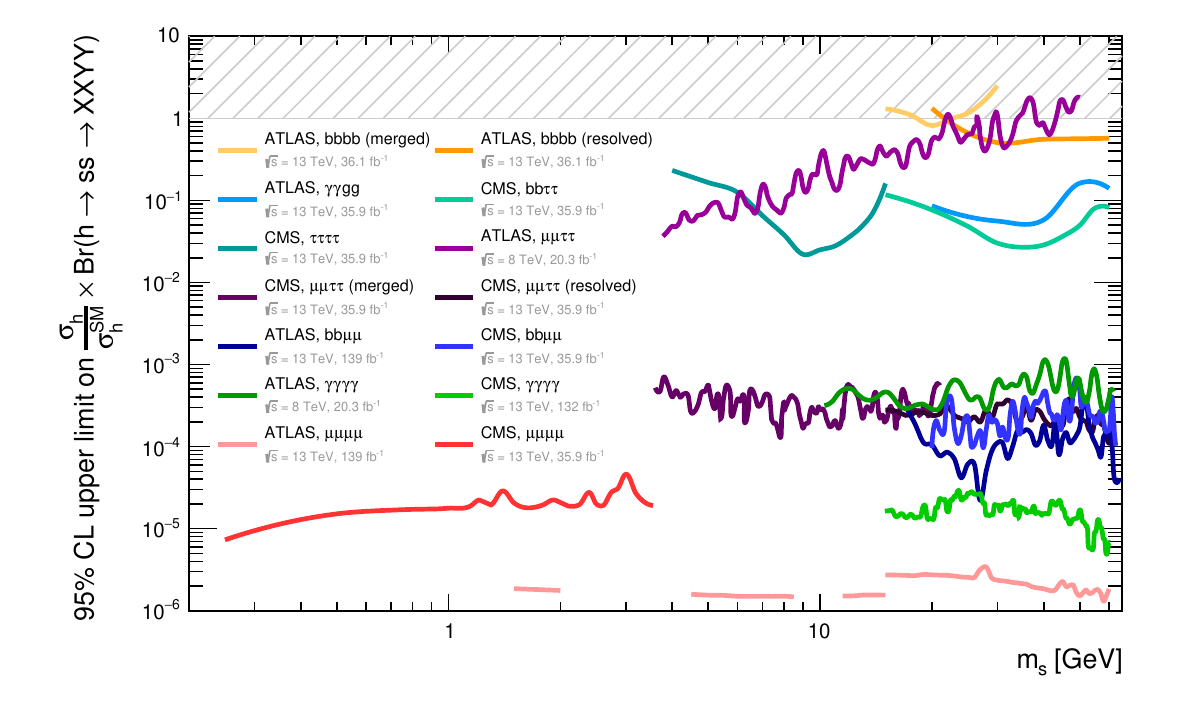}
\caption{\label{fig:current_stat} Current constraints on the process in Eq. (\ref{eq:lims})  ~\cite{Cepeda:2021rql}, which can be easily reinterpreted in extended scalar sector models, in particular models where couplings are inherited via a simple mixing angle. In this figure, the lighter scalar is denoted by $s$, which corresponds to $\PSh_i$ in the notation used in this report. 
}
\end{center}
\end{figure}
\end{center}

We proceed with an example of the allowed parameter space in a model with two additional singlets, the two real scalar extension studied in Refs.~\cite{Robens:2019kga, Robens:2022nnw}. In this model, three \CP-even neutral scalars exist that relate the gauge and mass eigenstates $\PSh_{1,2,3}$ via mixing. One of these states has to have couplings and mass complying with current measurements of the SM-like scalar; the other two can have higher or lower masses \cite{Robens:2019kga}. In \cref{fig:trsm}, we display two cases where either one (high-low) or two (low-low) scalar masses are smaller than $125\,\GeV$. 
The points were generated using ScannerS \cite{Coimbra:2013qq,Muhlleitner:2020wwk}, interfaced to HiggsTools~\cite{Bahl:2022igd} that builds on HiggsBounds \cite{Bechtle:2008jh,Bechtle:2011sb,Bechtle:2013wla,Bechtle:2020pkv} and HiggsSignals \cite{Bechtle:2013xfa,Bechtle:2020uwn}. Dominant decays are inherited from SM-like scalars with the respective masses, rendering the $\PQb\PAQb\PQb\PAQb$ final state dominant for large regions of this parameter space. Note that in the case where both additional scalars have masses $\lesssim\,125\,\GeV$, $\PSh_1\,\PSh_1\,\PSh_1$ final states would also be allowed in certain regions of parameter space. 
\begin{center}
\begin{figure}
\begin{center}
\includegraphics[width=0.45\textwidth]{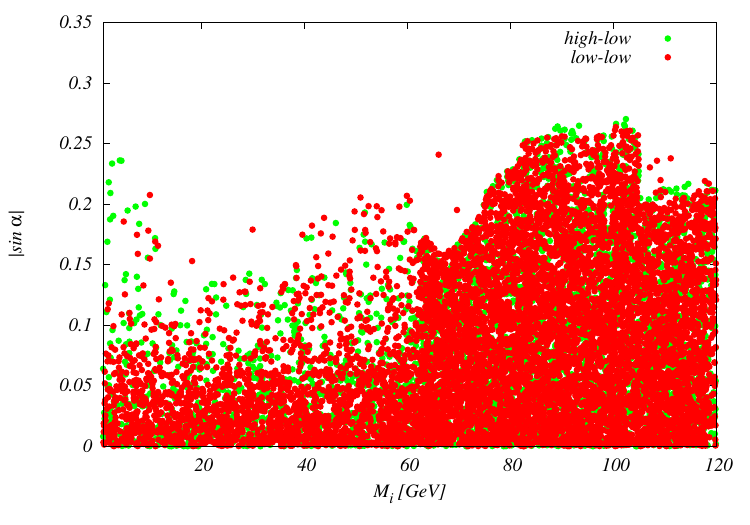}
\includegraphics[width=0.45\textwidth]{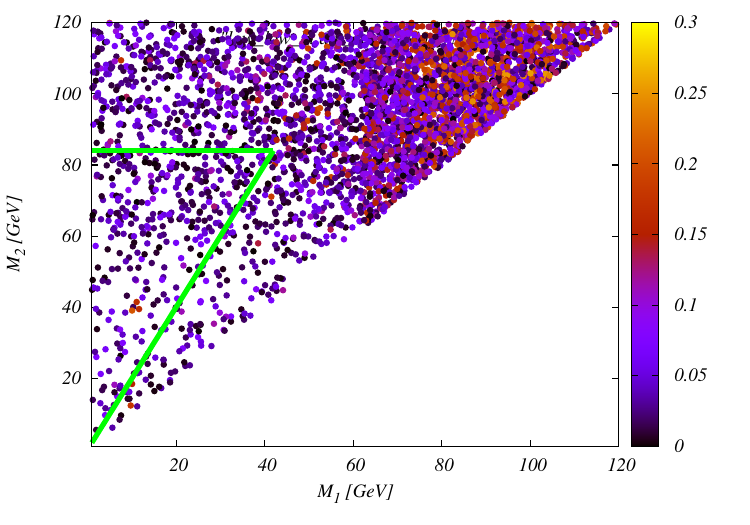}
\caption{\label{fig:trsm} Available parameter space in the Two-Real-Singlet-Model (TRSM), with one (high-low) or two (low-low) masses lighter than 125 \GeV. { Left}: light scalar mass and mixing angle, with $\sin\al\,=\,0$ corresponding to complete decoupling. { Right:} available parameter space in the $\lb m_{\PSh_1},\,m_{\PSh_2}\rb$ plane, with colour coding denoting the rescaling parameter $\sin\al$ for the lighter scalar $\PSh_1$. Within the green triangle, $\PSh_{125}\,\rightarrow\,\PSh_2 \PSh_1\,\rightarrow\,\PSh_1\,\PSh_1\,\PSh_1$ decays are kinematically allowed. Figures taken from \cite{Robens:2024wbw}.}
\end{center}
\end{figure}
\end{center}

\paragraph{Two Higgs Doublet Models}


Another interesting scenario is given by two Higgs doublet models (2HDM), where the SM scalar sector is augmented by a second doublet. Such models have been studied by many authors; see e.g.\ Ref.~\cite{Branco:2011iw} for an overview on the associated phenomenology. Here, we focus on the available parameter space in 2HDMs of type I, where all leptons couple to one doublet only. In such models, flavour constraints that are severely constraining the parameter space for other Yukawa structures (see e.g.\ Ref.~\cite{Misiak:2017bgg} for a discussion of type II models) are lessened.

Apart from the masses of the additional particles, important parameters in 2HDMs are given by a combination of the different mixing angles from gauge to mass eigenstates, $\cos\lb \be-\al \rb$, as well as $\tan\be$, which describes the ratio of the vacuum expectation values in a chosen basis\footnote{See e.g.\ Ref.~\cite{Davidson:2005cw} for a discussion of the meaning of physical parameters in a basis independent way.}. In general, if the lighter neutral \CP-even scalar $\Ph$ is identified with the 125 \GeV~ resonance (while $\PH$ is the heavier neutral \CP-even scalar and $\PSA$ is the \CP-odd state), the decoupling limit $\cos\lb \be - \al\rb\,=\,0$ denotes the configuration where the couplings to vector bosons become SM-like. The Higgs signal strength (i.e.\ the ratio of the measured signal yield to the corresponding SM prediction) poses important constraints on this quantity. In \cref{fig:atldec}, we display the current constraints from the ATLAS collaboration using the 125 \GeV scalar signal strength for various types of Yukawa structures \cite{ATLAS:2024lyh}.

\begin{center}
\begin{figure}
\begin{center}
\includegraphics[width=0.35\textwidth]{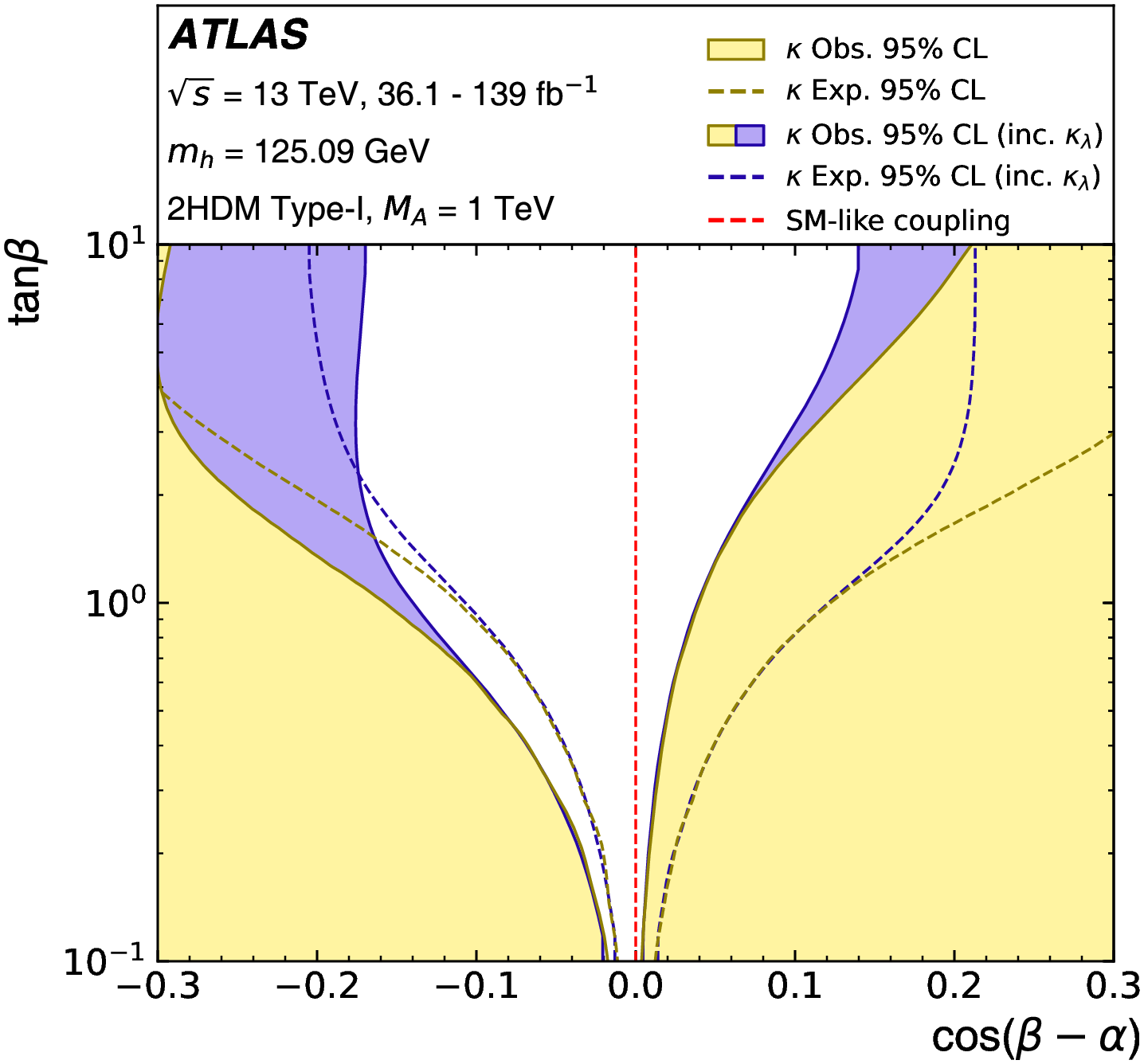}
\includegraphics[width=0.35\textwidth]{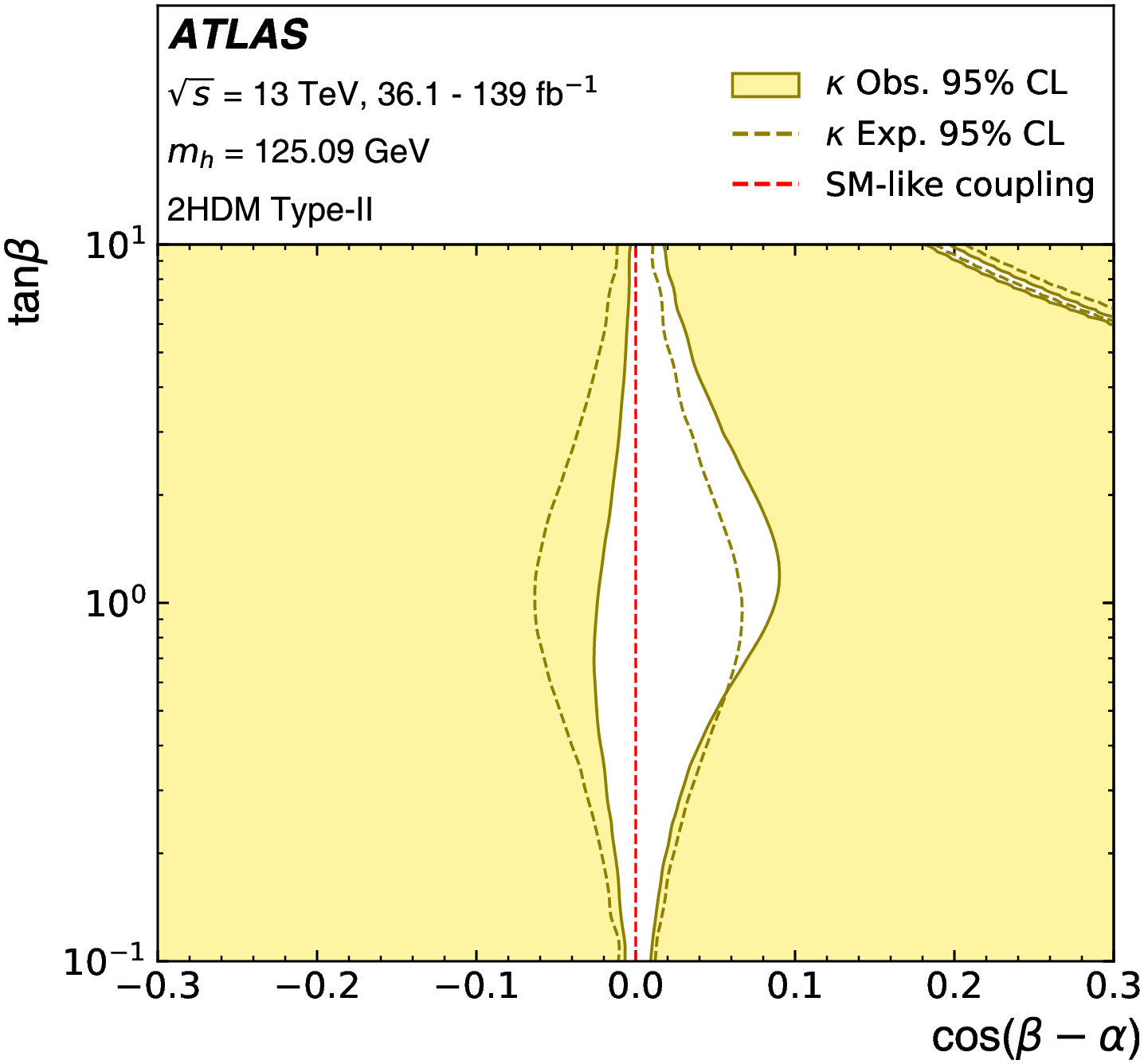}\\
\vspace{-0.7mm}
\includegraphics[width=0.35\textwidth]{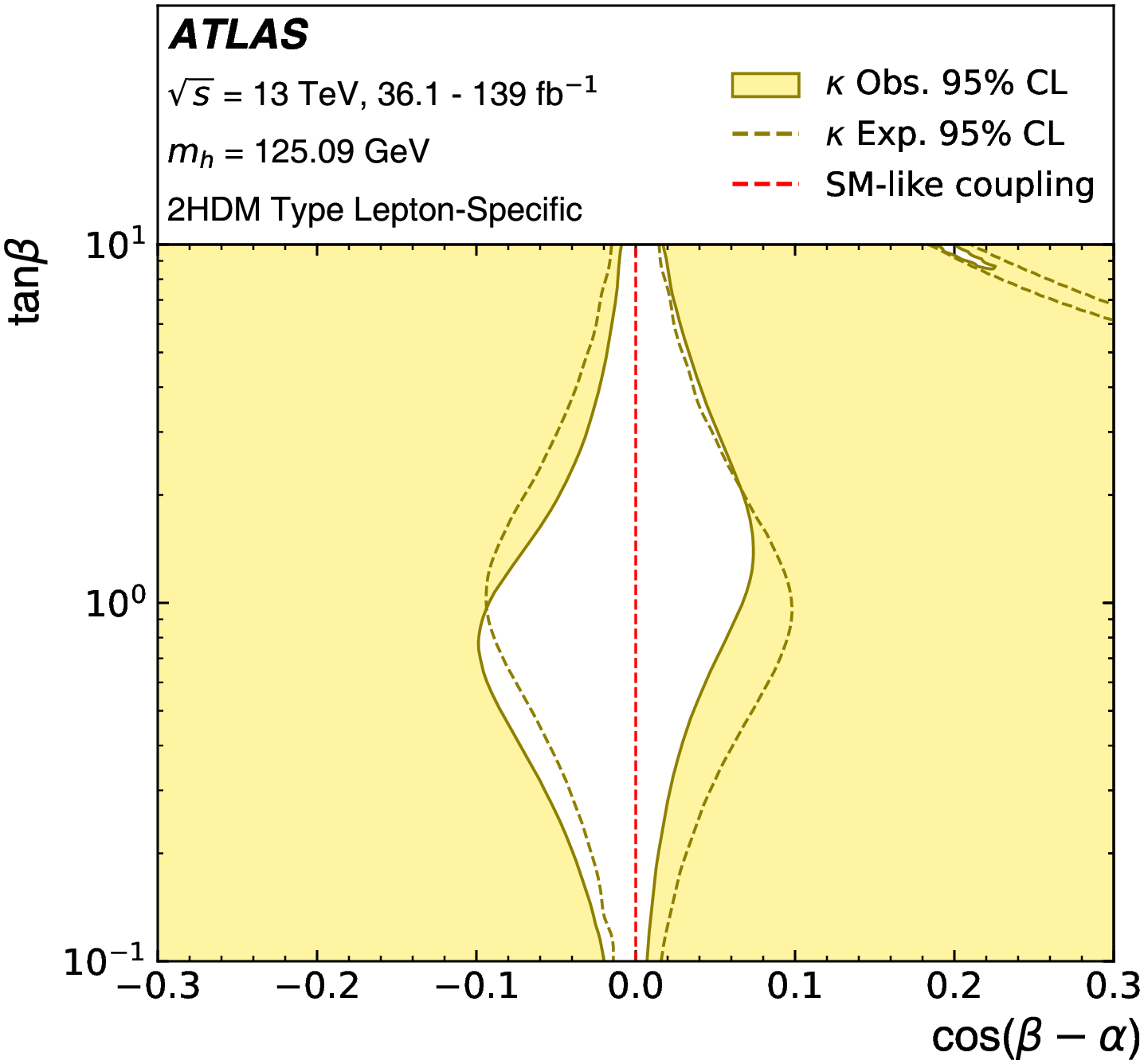}
\includegraphics[width=0.35\textwidth]{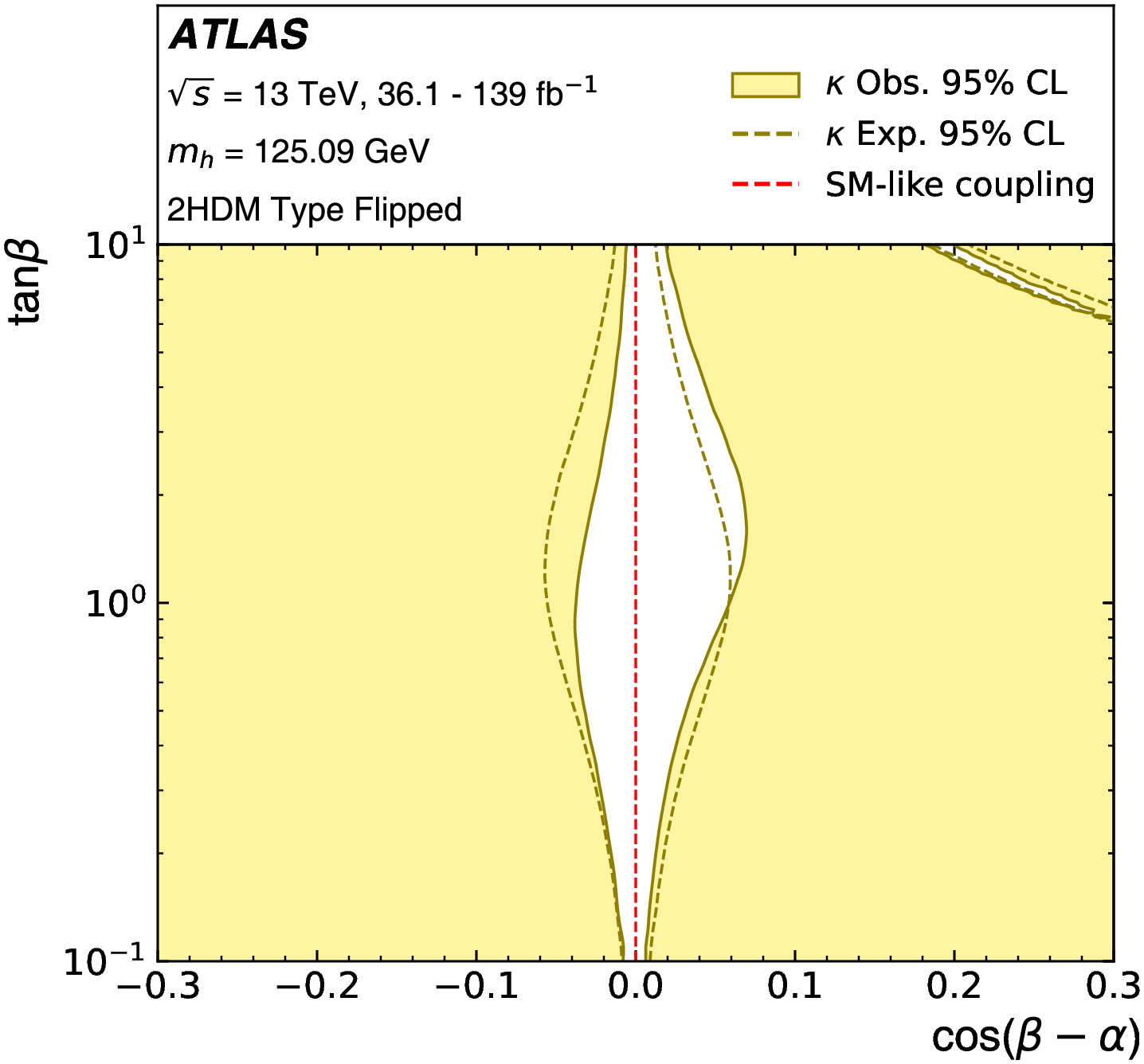}
\end{center}
\caption{\label{fig:atldec} Allowed and excluded regions in the $\lb \cos\lb \be-\al\rb,\tan\be \rb$ parameter space from Higgs signal strength from the ATLAS collaboration \cite{ATLAS:2024lyh}, for various Yukawa structures.}
\end{figure}
\end{center}

Another question of interest is which parameter space is still allowed in such models after all other theoretical and experimental constraints are taken into account. For this, we make use of the publicly available tool thdmTools  \cite{Biekotter:2023eil}, which allows for efficient scans of the 2HDM parameter space and contains an internal interface to HiggsTools for comparison with null-results from collider searches.

As the minimal version of the 2HDM contains in total 6 free parameters after one of the \CP-even neutral scalar masses has been fixed to guarantee agreement with current LHC findings, it is customary to reduce the number of these parameters in typical scan plots to two-dimensional planes. 
In \cref{fig:2hdm}, we show as an example the exclusion bounds from flavour physics (via an internal interface to SuperIso \cite{Mahmoudi:2007vz,Mahmoudi:2008tp}), Higgs signal strength, and currently implemented collider searches, for a type-I 2HDM for concrete scenarios of scalar masses and mixing angles.
%
Note that we chose to require a set of initial conditions, such as vacuum stability, perturbative unitarity, and agreement with electroweak precision data to be fulfilled. The relevant searches in this case are searches looking for a heavy resonance decaying into four-lepton final states via electroweak gauge bosons \cite{ATLAS:2020tlo}, heavy resonances decaying into di-tau final states \cite{CMS:2022goy}, searches for charged scalars into leptonic decay modes \cite{ATLAS:2018gfm}, resonance-enhanced discalar searches \cite{CMS:2018amk}, and the process $\PSA\,\rightarrow\,\Ph\,\PZ$ \cite{CMS:2019qcx}. 

\begin{center}
\begin{figure}
\begin{center}
\includegraphics[width=0.49\textwidth]{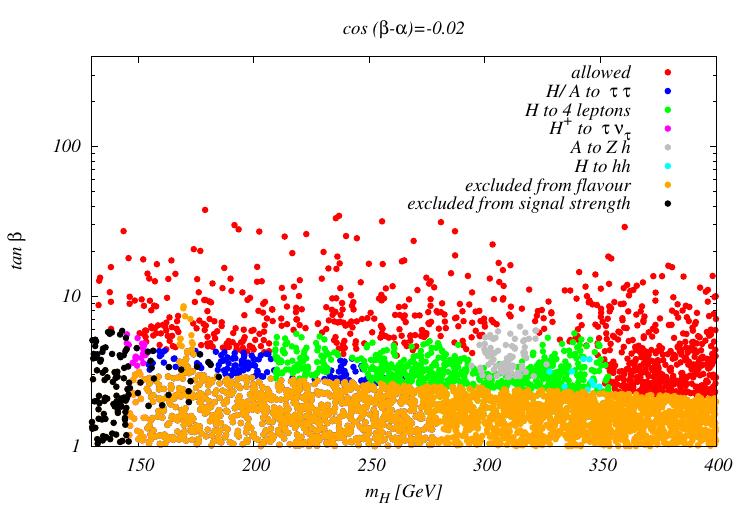}
\includegraphics[width=0.49\textwidth]{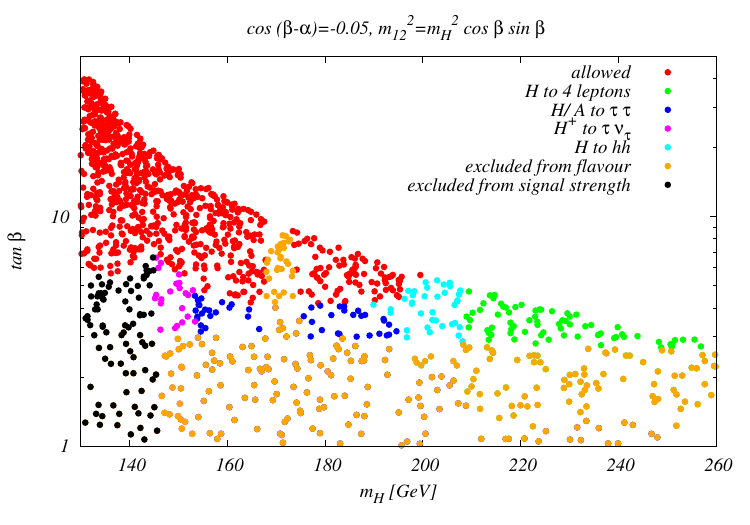}
\caption{\label{fig:2hdm} Allowed regions in the $\lb m_{\PH},\,\tan\be\rb$ plane for a type-I 2HDM after various constraints have been taken into account, including experimental searches at the LHC as currently available using thdmTools. { Left:} Here, $\cos\lb \be-\al\rb\,=\,-0.02$, and $m_{12}^2$ is floating freely. { Right:} We here fixed $\cos\lb \be-\al\rb\,=\,-0.05,\,m_{12}^2\,=\,m_{\PH}^2\,\sin\be\,\cos\be$. See text for further details. Taken from Ref.~\cite{Robens:2024wbw}. 
}
\end{center}
\end{figure}
\end{center}
In general, for the two scenarios shown in \cref{fig:2hdm}, the available parameter space is already severely constrained from current searches at the LHC. For the scenario where $\cos\lb \be -\al\rb=-0.02$, current searches cease to be sensitive for masses $\gtrsim\,400\,\GeV$, but we can find allowed parameter points in the whole mass range, where a lower scale for $\tan\be$ is set by flavour constraints. On the other hand, if we additionally fix the potential parameter $m_{12}^2$, we see that there is also an upper limit on $\tan\be$ from the remaining constraints, basically leading to a maximal mass scale of around 200 \GeV~ for the heavy scalars. Note that these bounds only apply when the other free parameters are fixed as discussed.


Other examples for 2HDMs including possibilities for low-mass scalars, in particular for the \CP-odd candidate $A$, can be found in Ref.~\cite{Wang:2022yhm}.

\paragraph{Other extensions}
The scalar sector of the SM can be extended by an arbitrary number of additional scalar fields, such as singlets, doublets, etc. One option that is also often considered is the extension of this sector by both singlets and doublets. Here, we will discuss a couple of examples of such extensions which are able to provide regions where some of the additional scalars are in the low-mass range of interest (typically below 125 \GeV). 

\begin{itemize}
\item{\bf IDM:}
The Inert Doublet Model \cite{Deshpande:1977rw, Barbieri:2006dq,Cao:2007rm} is an intriguing two Higgs doublet model that allows for a dark matter candidate. In this model, low-mass scalars that are within the reach of a lepton collider of 240 - 250 \GeV centre-of-mass energy are still allowed. This is because the additional new scalars do not couple to fermions, therefore evading otherwise strong constraints from flavour physics. In \cref{sec:exotics}, particular final states for this scenarios are investigated in a realistic collider scenario. The allowed points have been determined following Refs. \cite{Ilnicka:2015jba, Kalinowski:2018ylg} for the general parameter scan, including recent experimental updates on the bounds.
\item{\bf N2HDM:}
%
In Ref.~\cite{Abouabid:2021yvw}, a model is considered where the SM scalar sector is extended by an additional doublet as well as a real singlet. This model has 3 \CP-even neutral scalar particles, out of which one needs to be consistent with the measured properties of the 125 \GeV~ scalar. 
We show an example of the allowed parameter space in \cref{fig:n2hdm} that still allow for scalar masses below 125 \GeV. We see that in the \CP-even sector there are regions within this model that still allow for low mass scalars.
\begin{center}
\begin{figure}
\begin{center}
\includegraphics[width=0.65\textwidth]{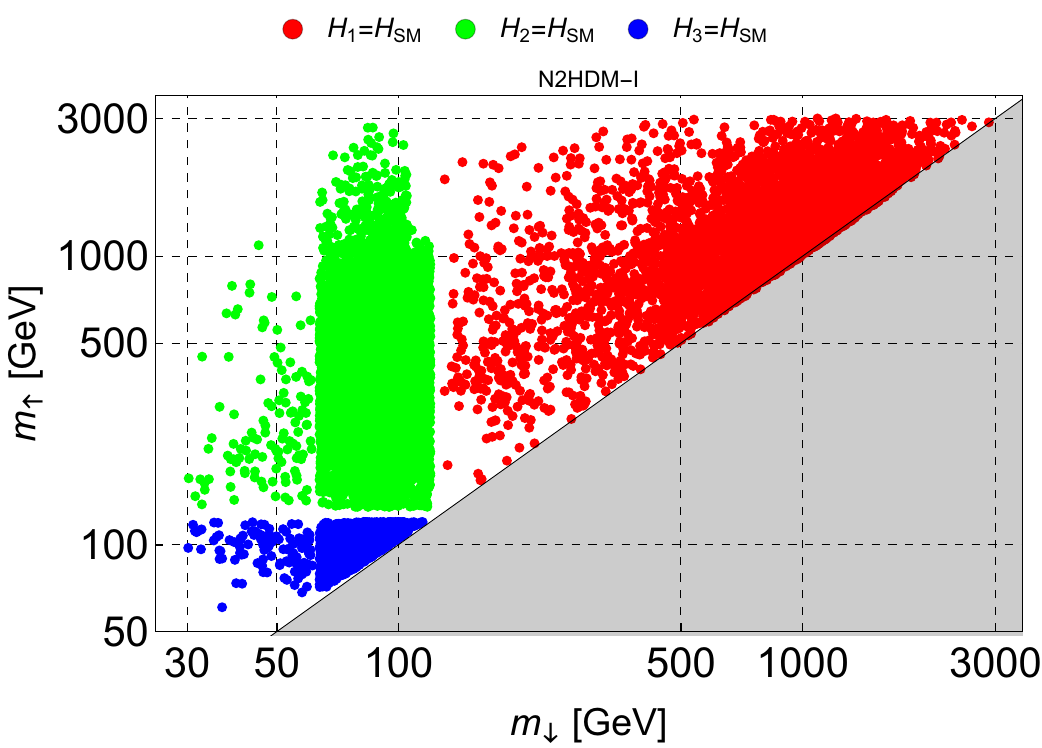}
\caption{\label{fig:n2hdm} Scan results in the N2HDM, showing the mass distributions of the non-SM-like neutral Higgs bosons, where $m_{\,\uparrow}$ is the larger of the two and $m_{\downarrow}$ the smaller, taken from Ref.~\cite{Abouabid:2021yvw}. There are regions in the model's parameter space where either one or two of the additional scalars have masses $\lesssim\,125\,\GeV$.}
\end{center}
\end{figure}
\end{center}
\item{\bf Lepton-specific Inert Doublet Model:}
In Ref.~\cite{Han:2021gfu}, a model is considered where the SM scalar sector is augmented by an additional doublet, imposing an exact $\mathbb{Z}_2$ symmetry which is broken by a specific coupling to the fermionic sector. Figure \ref{fig:idmv} shows regions in the model parameter space that agree with current searches as well as anomalous magnetic momenta of electron and muon, including regions where the second \CP-even scalar can have a mass $\lesssim\,30\,\GeV$. 
\begin{center}
\begin{figure}
\begin{center}
\includegraphics[width=0.85\textwidth]{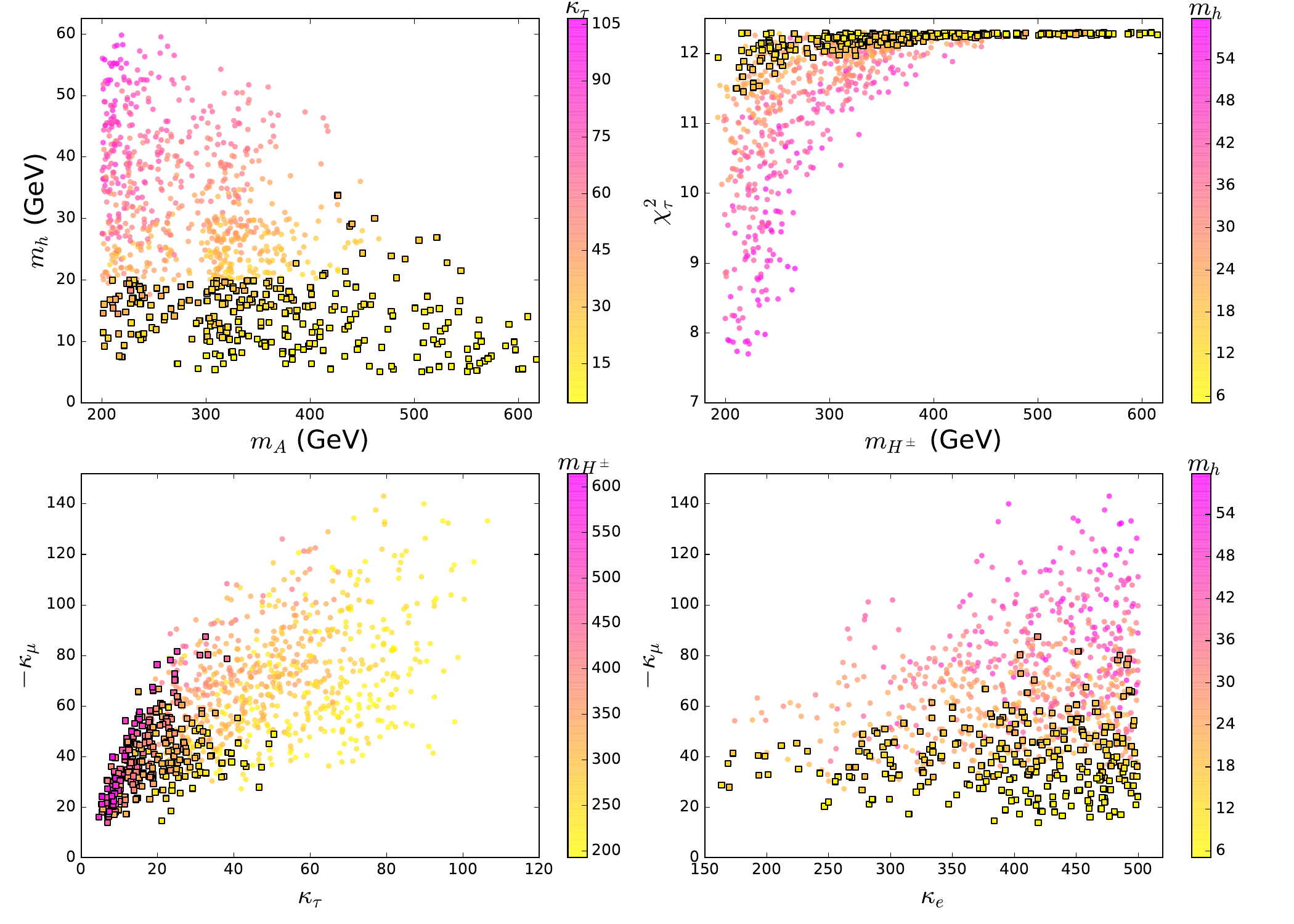}
\caption{\label{fig:idmv} Allowed regions in the model parameter space discussed in Ref.~\cite{Han:2021gfu}, 
where squares denote allowed and bullets excluded regions. \CP-even neutral scalars with low masses are viable within this model.}
\end{center}
\end{figure}
\end{center}
\item{\bf Scalar triplet model:}
A model containing scalar triplets, leading to a rich particle content as well as the possibility of \CP violating terms, has been studied in Ref.~\cite{Ferreira:2021bdj}. This model contains 5 neutral, 3 charged, and 2 doubly charged mass eigenstates. Figure \ref{fig:tripl} shows regions in parameter space where masses for some of these can be $\lesssim\,125\,\GeV$. A future HTE factory could investigate both neutral and charged scalars with low mass realizations, with a rescaling of the $\PSh\PZ\PZ$ coupling below LEP sensitivity. 
\begin{center}
\begin{figure}
\begin{center}
\includegraphics[width=0.5\textwidth]{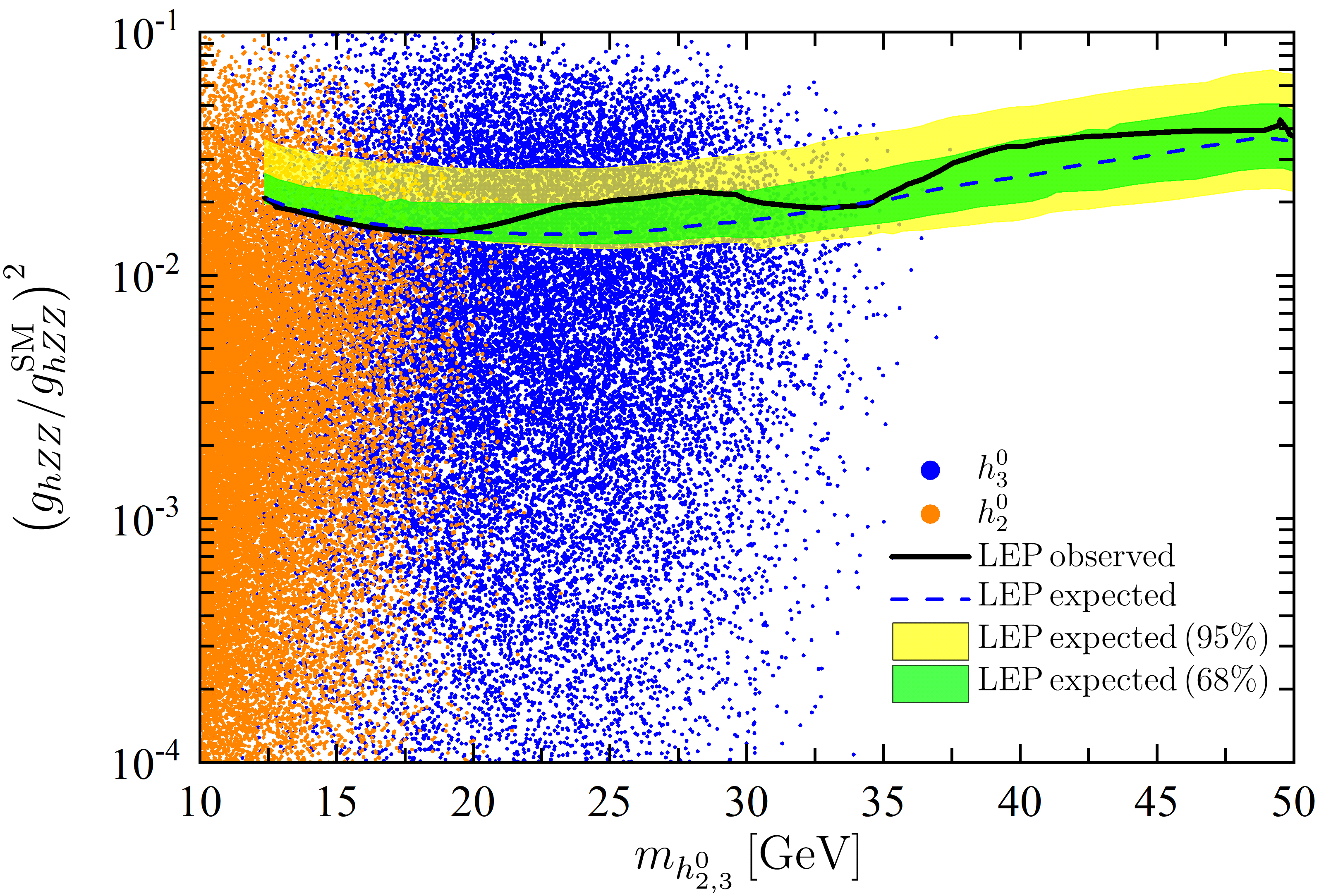}
\includegraphics[width=0.45\textwidth]{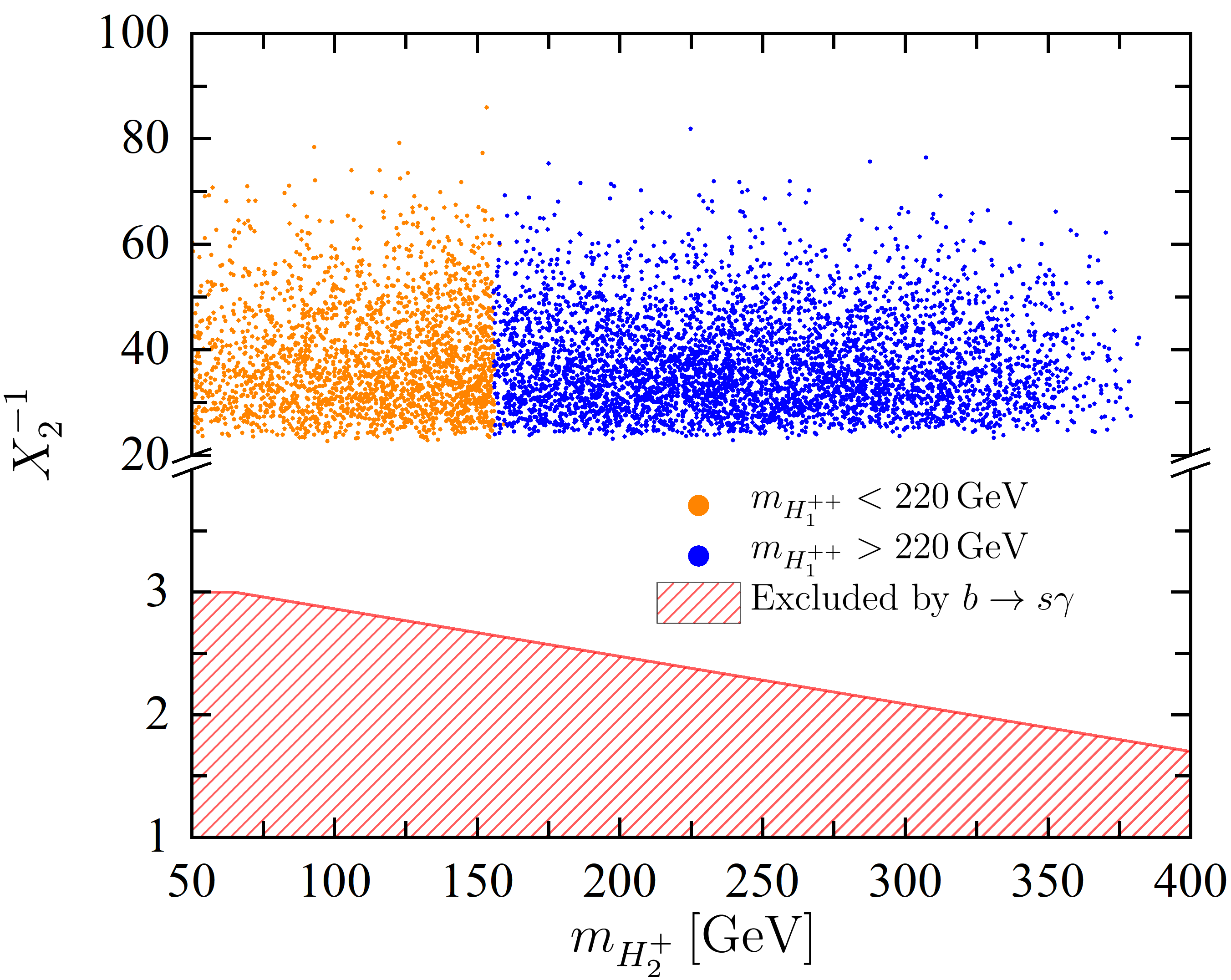}
\caption{\label{fig:tripl} Left: Upper bound on the ratio $(g_{\PSh\PZ\PZ}/g^{\rm SM}_{\PSh\PZ\PZ})^2$ at 95\% C.L, for $m_{\PSh^0_{2,3}}
\ge 10 \GeV$, from combined LEP data. 
In orange and blue the $m_{\PSh^0_{2}}$ and $m_{\PSh^0_{3}}$ vs. $(g_{\PSh\PZ\PZ}/g^{\rm SM}_{\PSh\PZ\PZ})^2$
points are shown, for the considered
mass range. Right: Allowed points in the ($X_2^{-1}$, $m_{\PH^+_2}$) plane. The hatched region corresponds to the fraction of the parameter space excluded by $\PQb \to \PQs \PGg$ constraints. The points for which the lightest doubly-charged scalar mass $m_{\PH^{++}_1}$ is above (below) the ATLAS lower bound of 220 \GeV extracted from $\PWpm \PWpm$ searches are shown in blue (orange). 
For neutral and charged new scalars, masses $\lesssim\,125\,\GeV$ are achievable. From Ref.~\cite{Ferreira:2021bdj}.}
\end{center}
\end{figure}
\end{center}
\end{itemize}

Other models with light scalar realizations have also been presented in Refs.~\cite{deLima:2022yvn,Kuncinas:2023ycz,Cao:2024axg,Yang:2024kfs,deBoer:2024jne}. Interesting connections between new physics scenarios with light new scalars and electroweak phase transitions have e.g.\ been discussed in Refs.~\cite{Kozaczuk:2019pet,Wang:2022dkz}. 

  

\DeclareRobustCommand{\PSS}{\HepParticle{S}{}{}\Xspace} 

\subsubsection{Focus topic targets}


Higgs factories can be sensitive to exotic scalar production even for very light scalars, thanks to the clean environment, as well as the precision and hermeticity of the detectors.
Different production and decay channels, including invisible decays, can be considered.
%
%
%
In Ref.~\cite{deBlas:2024bmz} two production modes for extra scalars were suggested, which deserve particular attention and more detailed studies.

\begin{itemize}
\item The first target selected for this focus topic was the production of a new scalar \PSS, in the scalar-strahlung process:
$\epem \to \PZ  \PSS$.
Higgs factories are best-suited to search for light exotic scalars in this process, similar to the Higgs-strahlung production process considered for the 125  \GeV Higgs boson, as new scalars can be tagged, independent of their decay, based on the recoil mass technique \cite{Yan:2016xyx}.  
Similar analysis methods can be used, looking for corresponding light scalar decay channels (e.g.\ $\bb$, $\PW^{+(*)}\PW^{+(*)}$, $\PGtp\PGtm$ or invisible decays), by relaxing the constraint imposed by the SM Higgs boson mass. 
%
%
%
As discussed in the previous section, additional light scalar particles, with masses of the order of or below the mass of the 125 GeV Higgs boson, are well motivated theoretically and are by far not excluded.
%
%
Shown in \cref{fig:exscalar_zs_cs} are benchmark points resulting from the parameter scan of the Two-Real-Singlet Model \cite{Robens:2022nnw,Robens:2023pax}, Two Higgs-Doublet Model \cite{Biekotter:2023eil} and Minimal R-symmetric Supersymmetric SM \cite{Diessner:2015iln}.
\begin{figure}[htbp]
\centering
\includegraphics[width=0.7\textwidth]{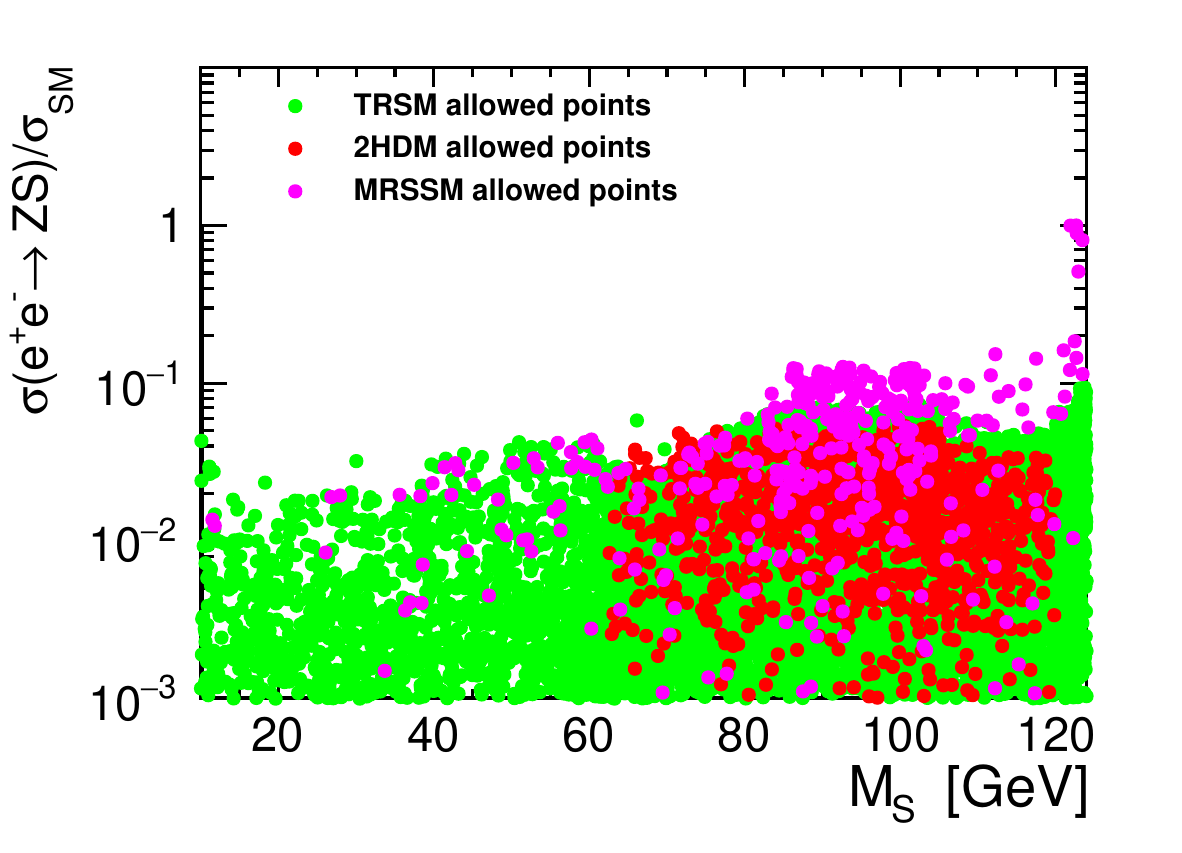}
\caption{Allowed cross section values for the scalar-strahlung process at 250 \GeV Higgs factory, relative to the SM predictions for the Higgs boson production at given mass, as a function of the new scalar mass.  Shown are benchmark points consistent with current experimental and theoretical constraints for three selected models.}
\label{fig:exscalar_zs_cs}
\end{figure}
For a wide range of scalar masses, cross sections up to the order of 1\% to 10\% of the SM production cross section are still allowed.
Taking into account different decay channels possible, this opens a wide range of search possibilities for future Higgs factory experiments.


%
\item As a second target for the exotic scalars focus topic, 
light scalar pair-production in 125\,GeV Higgs boson decays was proposed: 
$\epem \to \PZ  \PH \to \PZ  \Pa \Pa$.
Here again, different decay channels of the new scalar state; both SM-like decays (e.g.\ $\bb$, $\PGtp\PGtm$), and exotic ones (e.g.\ invisible decays), should be considered.
\end{itemize}

%
While new scalar states could in general be long-lived, only results referring to scenarios with prompt decays are included in the following section. Scenarios with long-lived are considered as a part of a separate focus topic in \cref{sec:LLP-focus-topic}. 

\subsubsection{Search for scalar-strahlung production}

\subsubsection*{Decay-mode-independent search}



The most model-independent approach for searches of new exotic scalars is to use their recoil against the \PZ boson. This approach has been explored before with a detailed, \geant 4-based simulation of the ILD detector concept~\cite{ILDConceptGroup:2020sfq} at the ILC with a centre-of-mass energy of 250~GeV. 
The new study presented below follows the same analysis approach, focusing on \PZ boson decays into two muons.

Full simulation background samples were generated with \whizard 2.8.5~\cite{Kilian:2007gr}, simulated and reconstructed using the ILD\_l5\_o2\_v02 detector model and  a modular \textsc{Marlin} framework \cite{Gaede:2006pj}, a part of \textsc{ILCSoft} package \cite{Ete:2021ljr} v02-02-01.
Beam luminosity spectra for ILC running at 250 \GeV as well as polarisation of electron and positron beams were taken into account according to the assumed H-20 running scenario \cite{Barklow:2015tja}.
Signal samples were generated with the same \whizard version using the
SGV fast detector simulation~\cite{Berggren:2012ar}, adapted to ILD,
for detector simulation and high-level reconstruction.
For the decay-mode-independent search, the decay branching ratios of the extra scalar were fixed to those of the 125~GeV Higgs boson.


The study was to search for $\epem\to \PSS\PZ$ production,
with the Z boson decaying to a pair of muons.
Two preselection cuts were applied. Events with a high energetic ISR photon were filtered out by using the energy and the angle of isolated photons. 
Secondly, events with two isolated muon candidates were selected, requesting  the di-muon invariant mass not to differ from the \PZ mass by more
than 40~\GeV and the recoil mass to be inside the kinematic range considered.
For the selected events, after applying the FSR correction for further calculation of the di-muon mass and momentum, two BDTGs classifiers based on TMVA \cite{hoecker:tmva} were trained separately against
two-fermion and four-fermion backgrounds.

In both cases, the same six input variables were used for the training:
the dimuon invariant mass, the polar angle of each muon,
the polar angle of the muon pair, the opening angle of the muon pair,
and the difference of the two muon azimuthal angles.
An additional selection based on the responses of these two BDTGs was
applied, with selection cuts depending on the scalar mass considered.
The expected exclusion limits were then computed applying the method of
fractional event counting to the final recoil mass distribution.

The limits for the 250 GeV ILC are shown in \cref{nunez:fig:limits}.

\begin{figure}[htbp]
  \centering
  \hspace{1cm}
       \includegraphics [width=0.72\textwidth]{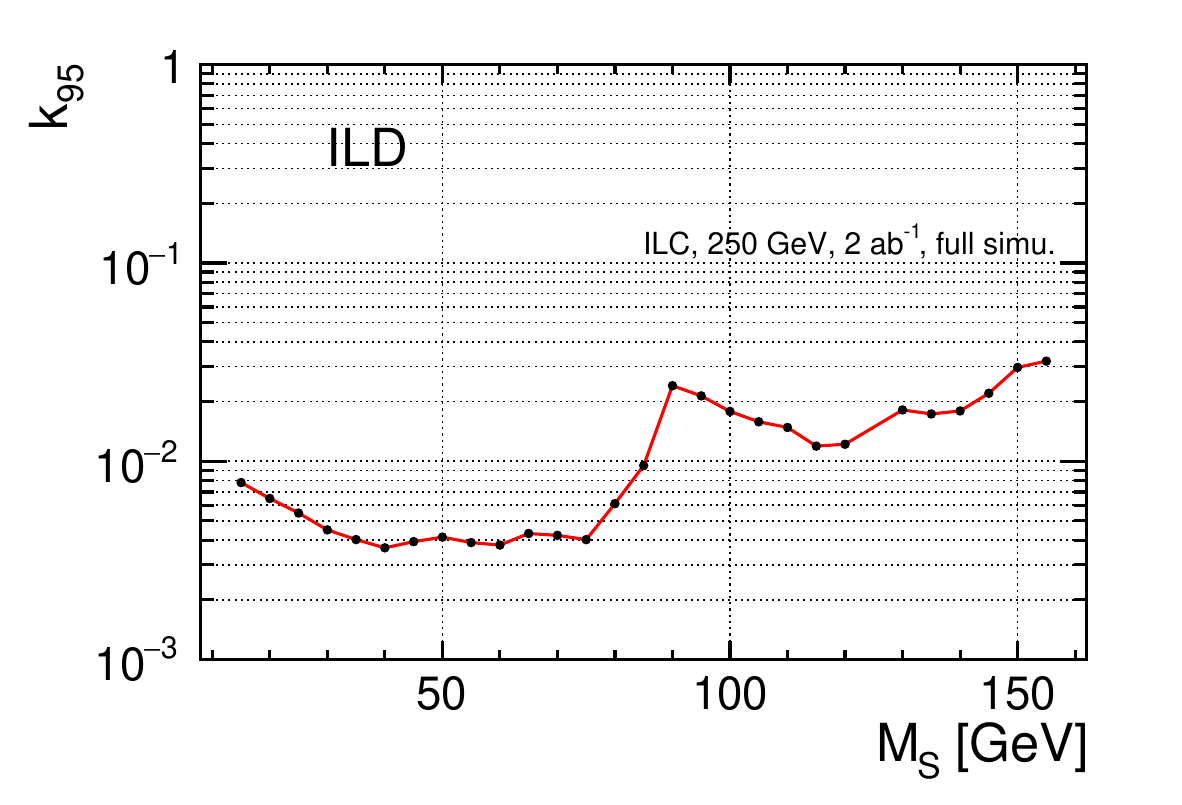}
       \caption{Exclusion limits for the ratio $k_{95}$ of the light scalar production cross section to the SM Higgs boson production cross section as a function of the scalar mass $M_S$, as expected from the decay independent analysis of 250 \GeV ILC data.
      }
    \label{nunez:fig:limits}
\end{figure}
2$\sigma$ limits on the scale factor between the scalar production
cross section and the cross section of the SM Higgs particle with
the same mass are shown for scalar masses from 10 to 155~GeV.
The results improve the LEP sensitivity by one to two orders of
magnitude, covering substantial new regions in the parameter space.

\subsubsection*{Search in \bb channel}



For many BSM scenarios the coupling structure of the new exotic scalars is similar to that of the SM Higgs boson, resulting in decays to \bb being the dominant decay channel for light scalars (below 125 \GeV). 
A study addressing prospects for observing these decays at 250 \GeV ILC was based on the same simulation framework as described above for the decay-mode-independent study (with full-simulation background samples and signal samples  produced with the SGV fast simulation framework \cite{Berggren:2012ar}).
Jet clustering and flavour tagging were performed within the \textsc{LCFIPlus} package. 
To avoid huge background from hadronic $\PWp\PWm$ events, only leptonic decays of the \PZ boson produced in association with the new scalar were considered. 
At the pre-selection stage, events with exactly two isolated leptons and two reconstructed jets, consistent with the expected event kinematics, were selected. Variables describing the two jets expected from the scalar decay and the \PZ boson decay products, including object 4-momenta, numbers of particles in jets and jet flavour tagging results, as well as the invariant mass of the scalar particle reconstructed using the recoil mass technique, were used for event classification.
BDT classifiers were trained separately for each beam polarisation setting and each scalar particle mass considered.
An example of the BDT response distribution, combined from the four beam polarisations, is shown in \cref{fig:brudnowski_combined_bdt_hist}. 
\begin{figure}[htbp]
    \centering
    \includegraphics[width=0.65\linewidth,trim=0 0 0 8mm,clip]{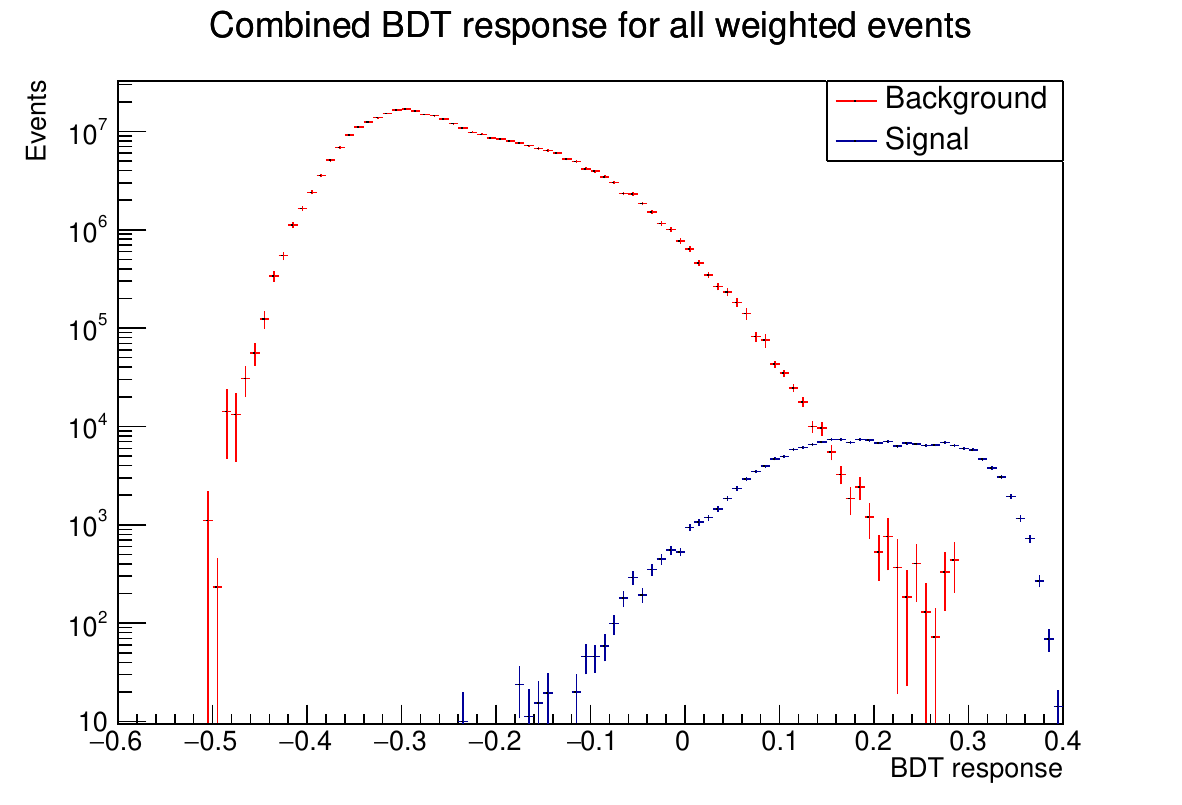}
    \caption{Combined distribution of the BDT response for the expected Standard Model backgrounds and the scalar signal production with a mass of 50 \GeV for 2 ab$^{-1}$ of data combined from the four beam polarisations. The scalar production cross section is normalised to 1\% of the SM cross section for this mass.}
    \label{fig:brudnowski_combined_bdt_hist}
\end{figure}


Expected 95\% C.L. exclusion limits on the scalar production cross section were calculated from the template fit to the BDT response distributions. 
For each scalar mass considered, limits were calculated for each beam polarisation setting as well as for the combined data (from the combined BDT response distribution). The results are shown in \cref{fig:brudnowski_limit_plot}.
\begin{figure}
    \centering
    \includegraphics[width=0.65\linewidth]{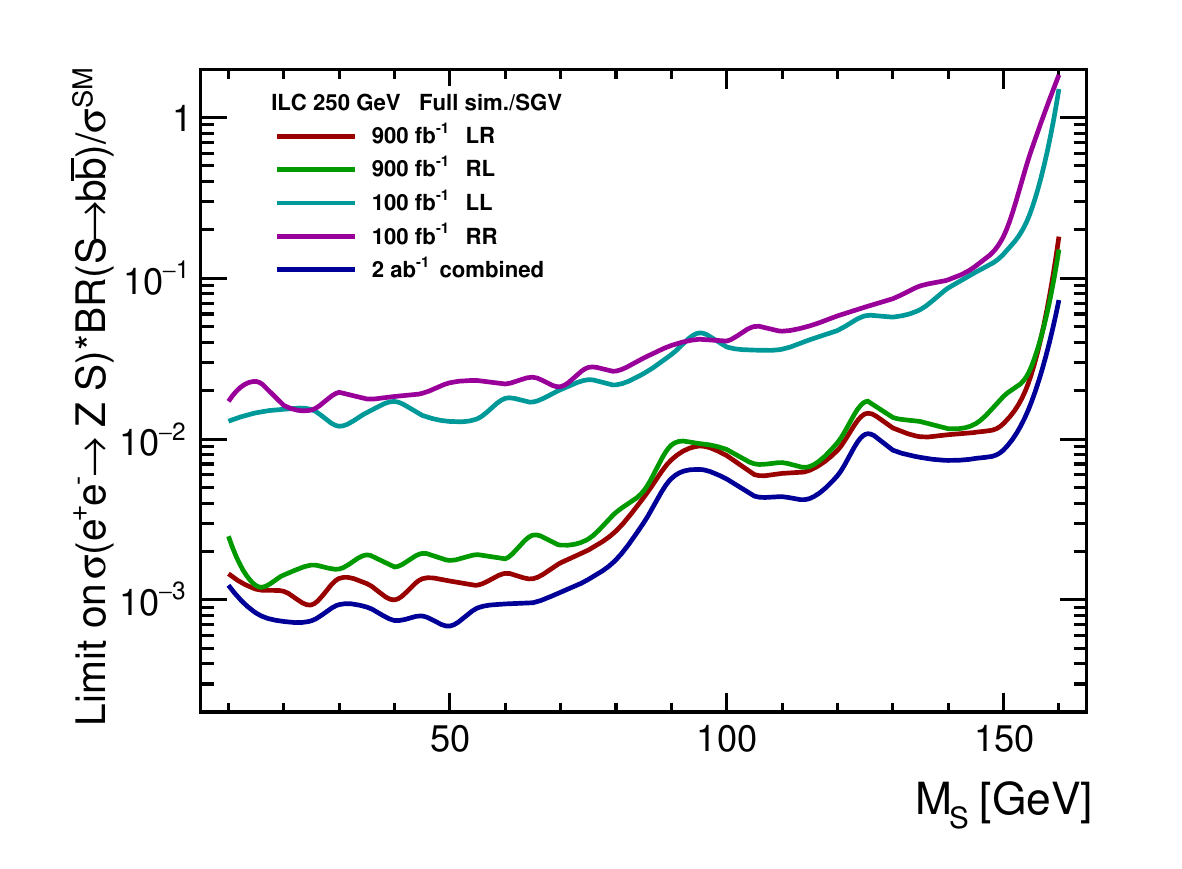}
    \caption{Expected exclusion limits on the ratio of the new scalar production cross section times branching ratio to \bb, to the SM Higgs boson production cross section at a given mass. Limits obtained for different beam polarisation settings and for the combined data sample are shown.}
    \label{fig:brudnowski_limit_plot}
\end{figure}
Exclusion limits obtained for LL and RR polarisation running are almost an order of magnitude weaker than for LR and RL. This is the result of much smaller cross sections and integrated luminosity expected in the H-20 running scenario for these beam polarisation settings \cite{Barklow:2015tja}. For all polarisations, the limits become weaker above 80 \GeV owing to the electroweak boson production becoming more difficult to separate from the scalar particle signal. At 125 \GeV, one can see a peak caused by the Higgs boson having a signature very similar to that of the new scalar particle at that mass. The limits become weaker also for masses approaching the kinematic limit of $\sqrt{s} - M_{\PZ}$.
Results correspond to about an order of magnitude increase in sensitivity with respect to the model-independent study presented above (see also Ref.~\cite{Wang:2018awp}), assuming that decays of the new scalar to \bb dominate. 

The sensitivity in this channel is clearly limited by considering only the leptonic decays of the \PZ boson produced in association with the new scalar. Estimates based on the fast simulation framework developed for the $\PGtp\PGtm$ channel analysis (see below) indicate that it can be improved by up to a factor of two, if hadronic \PZ boson decays are also considered.

\subsubsection*{Search in $\PGtp\PGtm$ channel}


The analysis of this channel was triggered by results presented in Ref.~\cite{Biekotter:2022jyr} indicating that some of the discrepancies from the SM predictions observed in LEP and LHC data could be explained by a new scalar with mass of about 95 GeV and enhanced
branching ratio to the $\PGtp\PGtm$ final state.
The study \cite{Brudnowski:2024iiu} is based on the samples of background and signal events generated with \whizard 3.1.2 \cite{Kilian:2007gr} using the built-in SM\_CKM model, 
and fast simulation of the detector response with the \delphes\ ILCgen model \cite{deFavereau:2013fsa, ilc_delphes}.
Beam polarisation and luminosity spectra for ILC running at 250 \GeV were taken into account assuming the H-20 running scenario \cite{Barklow:2015tja,Bambade:2019fyw}. 
Signal events were generated by varying the mass of the SM Higgs boson and forcing it to decay into a $\PGtp\PGtm$ pair. 

Three decay channels can be considered for the signal events: hadronic (with both taus decaying into hadrons), semi-leptonic (with one leptonic tau decay) and leptonic (with leptonic decays of both taus).
As a tight selection, we require each tau lepton to be identified either as an isolated lepton (with associated missing \pt) or hadronic jet with $\PGt$-tag.
However, as the efficiency of tau jet tagging implemented in \delphes\
is relatively poor (at most 70\%), we also consider loose selection of
hadronic and semi-leptonic events, when we require only one identified
tau candidate (isolated lepton or $\PGt$-tagged jet) and three
untagged hadronic jets, and take the jet with smallest invariant mass
as the second tau candidate. 
While the number of reconstructed jets was enforced by running the clustering algorithm in exclusive mode, it was required that there are only two untagged jets and two tau candidates in an event and no additional isolated leptons or photons. 

Separate BDT classifiers \cite{xgboost} were trained for each event category and beam polarisation combination, resulting in 20 independent classifiers.
In addition to the variables describing the reconstructed $\PSS$ and $\PZ$ candidates, as well as the total reconstructed energy and the recoil mass, the $y_{23}$ and $y_{34}$ variables from the jet clustering algorithm were included in the variable set used for event classification.

The expected exclusion limits on the exotic scalar production cross section times
di-tau branching ratio, relative to the total scalar production cross section predicted by the SM for a given mass,  
are presented in \cref{fig:zembaczynski2_limits_1}. Comparison of limits obtained for different event classes show that the best results are obtained for the tight semi-leptonic selection, due to high statistics and the relatively low background level. Including the loose selecion categories improves the limits by a further 20--30\%. 
Shown in  \cref{fig:zembaczynski2_limits_2} is the comparison of limits for the two polarisation settings with high integrated luminosity, $\HepParticle{\Pe}{L}{-}\Xspace \HepParticle{\Pe}{R}{+}\Xspace$ and $\HepParticle{\Pe}{R}{-}\Xspace \HepParticle{\Pe}{L}{+}\Xspace$.
Similar limits are obtained for the two configurations and limits corresponding to the combined analysis of the four polarisation configurations, according to the H-20 running scenario, correspond to about to a $\sim$ 10\%  improvement with respect to results obtained with the same integrated luminosity for unpolarised beams.
The impact of the uncertainties in theory predictions and data sample luminosities on the presented limits was found to be negligible.
\begin{figure}[htbp]
\centering
\begin{subfigure}{0.4\textwidth}
    \includegraphics[width=\textwidth]{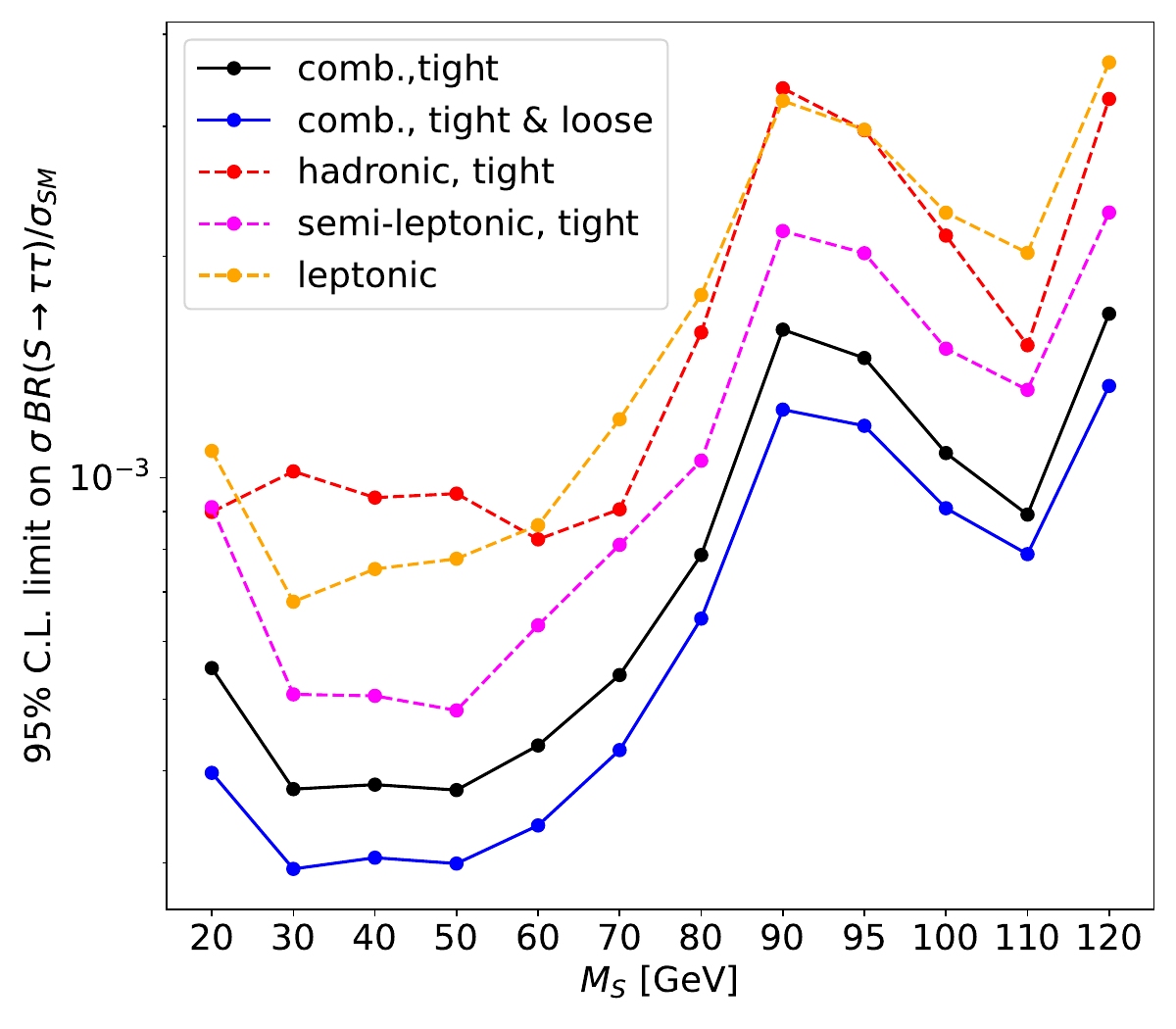}
    \caption{}
    \label{fig:zembaczynski2_limits_1}
\end{subfigure}
\hspace{0.5cm}
\begin{subfigure}{0.4\textwidth}
    \includegraphics[width=\textwidth]{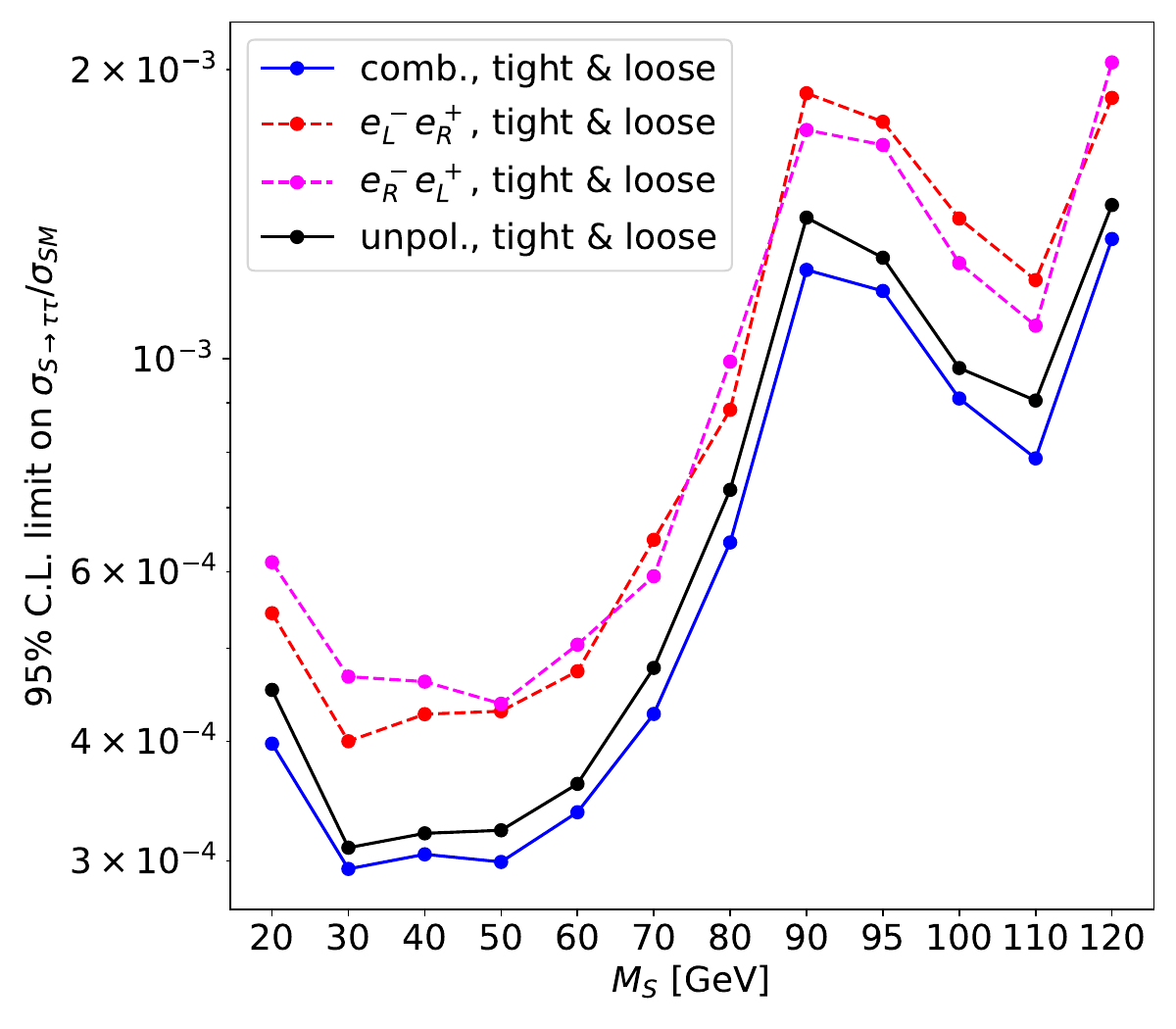}
    \caption{}
    \label{fig:zembaczynski2_limits_2}
\end{subfigure}
\caption{Expected 95\% C.L. cross section limits on the light scalar production cross section times
di-tau branching ratio as a function of the scalar mass for ILC running at 250 \GeV, assuming the H-20 running scenario.  (a) Comparison of combined limits for different event categories. 
(b) Comparison of combined limits for different beam polarisation configurations.}
\label{fig:zembaczynski2_limits}
\end{figure}
Assuming scalar decays to $\PGtp\PGtm$ dominate, the presented results correspond to over an order of magnitude improvement in sensitivity with respect to the decay-mode-independent study. They are also stronger than the limits expected for the \bb decay channel. However, taking into account that the branching ratio is included in the limit definition, this channel can be competitive only if the scalar branching ratio is of the order of 10\% or more. 
In \cref{fig:zembaczynski2_limits_bsm}, expected 95\% C.L. limits on the cross section times di-tau branching ratio are compared with the predictions of the benchmark scenarios \cite{Robens:2022nnw,Robens:2023pax,Biekotter:2023eil,Diessner:2015iln} discussed previously (\cref{fig:exscalar_zs_cs}).
\begin{figure}[htbp]
\centering
\includegraphics[width=0.7\textwidth]{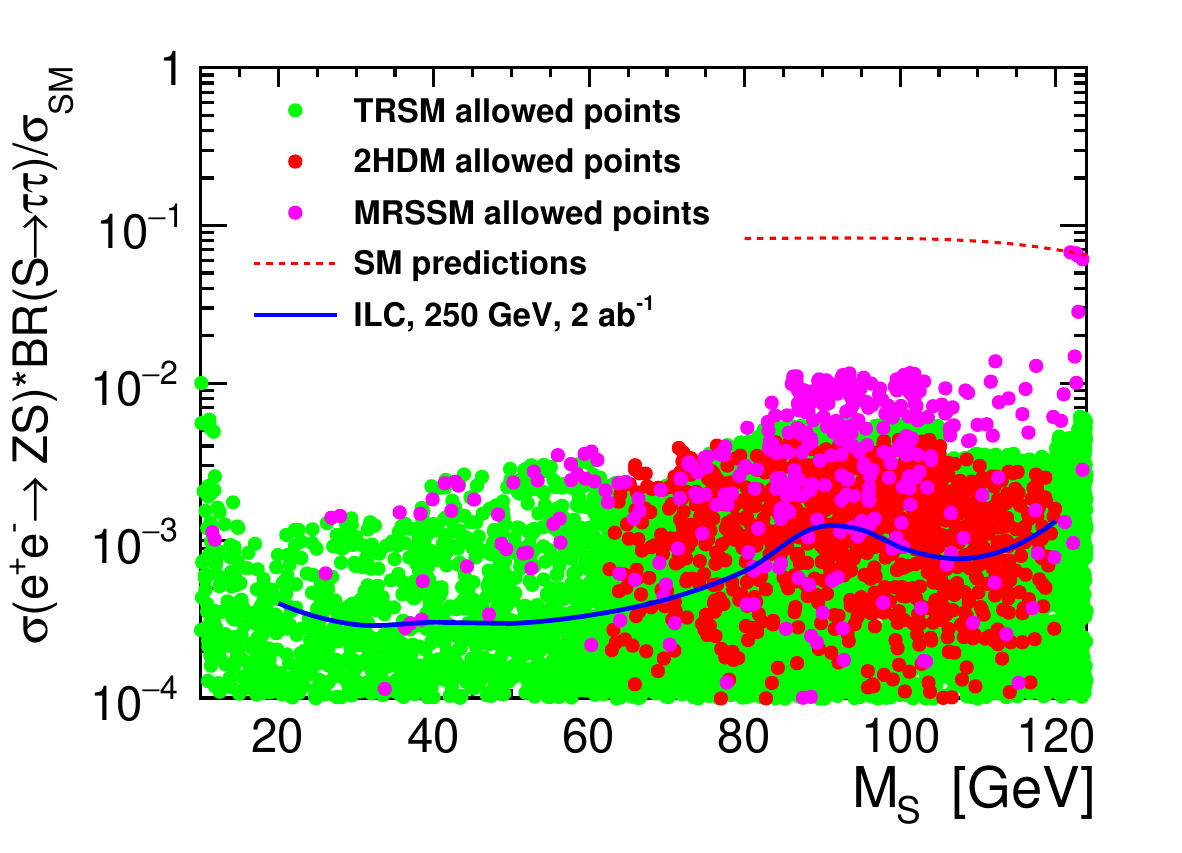}
\caption{Expected 95\% C.L. cross section exclusion limits on the light scalar production cross section times di-tau branching ratio as a function of the scalar mass for ILC running at 250 \GeV, assuming the \mbox{H-20} running scenario, compared with the predictions of three sets of benchmark models consistent with current experimental and theoretical constraints.}
\label{fig:zembaczynski2_limits_bsm}
\end{figure}
Measurements at the 250 \GeV Higgs factory will improve current limits for this channel by about an order of magnitude.

\subsubsection*{Search in $\PWp\PWm$ channel}


The presented study is based on the Two-Real-Singlet-Model (TRSM),
a model that enhances the SM scalar sector by two additional  real singlets
that obey a $\mathbb{Z}_2 \otimes \mathbb{Z}'_2$ symmetry \cite{Robens:2019kga}.
In this model, all scalar fields acquire a vacuum expectation value, such that the model
contains in total 3 CP-even neutral scalars that can interact with each other.  
Although measurements of the 125 \GeV Higgs boson at the LHC have already constrained the parameter space of the model, it remains far from being excluded, even for very light additional scalars with masses as low as 10 \GeV~\cite{Robens:2022nnw, Robens:2023oyz}.
Sizable production cross sections at a future Higgs factory are still not ruled out.
As two-fermion final states are expected to dominate light TRSM scalar decays and are covered in other contributions to this study, this contribution focuses on exploring the feasibility of observing the $\PSS \to \PWp\PWm$ decay at higher scalar masses.


At low masses the cross section for exotic scalar production with decay into $\PWp\PWm$ is highly suppressed due to the negligible branching fraction to \PW boson pairs. In the Standard Model, observing this decay channel only becomes feasible for scalar masses above 100 GeV. 
However, new scalar couplings to the gauge bosons need to be much smaller than those predicted by the SM, to be consistent with the measurements of the 125~\GeV Higgs boson at the LHC.
When the existing constraints are taken into account, possible signal cross sections are about an order of magnitude smaller than those predicted by the SM.
One can estimate that for the \epem Higgs factory running at 250 \GeV, observation of the new scalar production in the $\PWp\PWm$ decay channel could only be possible in the narrow mass window between 125 and 150 \GeV.


New scalar production in the scalar-strahlung process with subsequent scalar decay to two \PW bosons
results in the final state with three gauge bosons (\PZ, \PWp and \PWm). 
However, on-shell production of the three bosons is not possible at a 250 \GeV collider so that at least
one of them has to be off-shell.
Also, the decay widths of these bosons shall be properly treated and this can only be achieved by simulating the full process, with six-fermion final state, $\epem \to 6\;\Pf$.
To better understand the kinematics of the signal, the process
$\epem \to \mpmm \Pe \PAGne \PQq\PAQq$ was simulated in the TRSM model with an additional 140 GeV scalar.
Simulation results presented below were obtained using \whizard 3.1.2 \cite{Kilian:2007gr}, and the model UFO file is available via Ref.~\cite{TRSMUFO}. 
Contributing to the considered process are the new scalar production, the SM-like 125 \GeV Higgs boson production
and all the other SM diagrams with the same final state.
By the choice of this particular final state, matching of the final state fermions to \PZ, \PWp and \PWm is unique, simplifying the analysis.
Shown in \cref{fig:zarnecki2_mass_1d} are the invariant mass distributions for the \PZ candidate (muon pair)
and the \PSS candidate (remaining fermions giving a $\PWp\PWm$ pair).
The scalar-strahlung events can be selected by the very distinct peak in the di-muon invariant mass distribution (left plot).
Contributions coming from the scalar production (with 125 \GeV and 140 \GeV masses) are also clearly visible in the $\PWp\PWm$ invariant mass distribution (right plot).
\begin{figure}[tbp]
\centerline{
\includegraphics[width=0.9\textwidth]{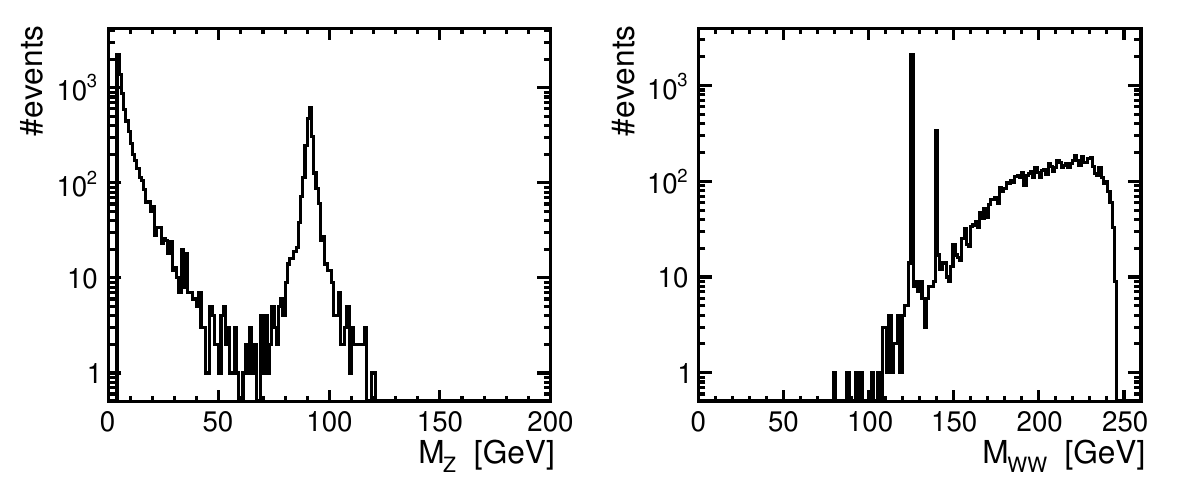}
}
\caption{Results of the \whizard simulation of the process
  $\epem \to \mpmm \Pe \PAGne \PQq\PAQq$ in the TRSM model with an additional 140 GeV scalar,
  at \roots=250 \GeV. Left: invariant mass distribution for the muon pair.
  Right: invariant mass distribution for the four-fermion final state corresponding to $\PWp\PWm$.}  
  \label{fig:zarnecki2_mass_1d}
\end{figure}
The separation between scalar-strahlung processes and other SM background contributions is also shown in
\cref{fig:zarnecki2_mass_2d}.
For the scalar-strahlung process, the \PZ boson is produced on-shell while at least one of the \PW bosons is off-shell.
The dominant background is the on-shell $\PWp\PWm$ pair production associated with a muon pair that comes from a virtual $\PZ$ or photon exchange.
While the detector resolution will clearly blur this picture, the separation of scalar-strahlung process
in the  $\PWp\PWm$ decay channel should still be possible with high efficiency.
\begin{figure}[tbp]
\centerline{
\includegraphics[width=0.45\textwidth]{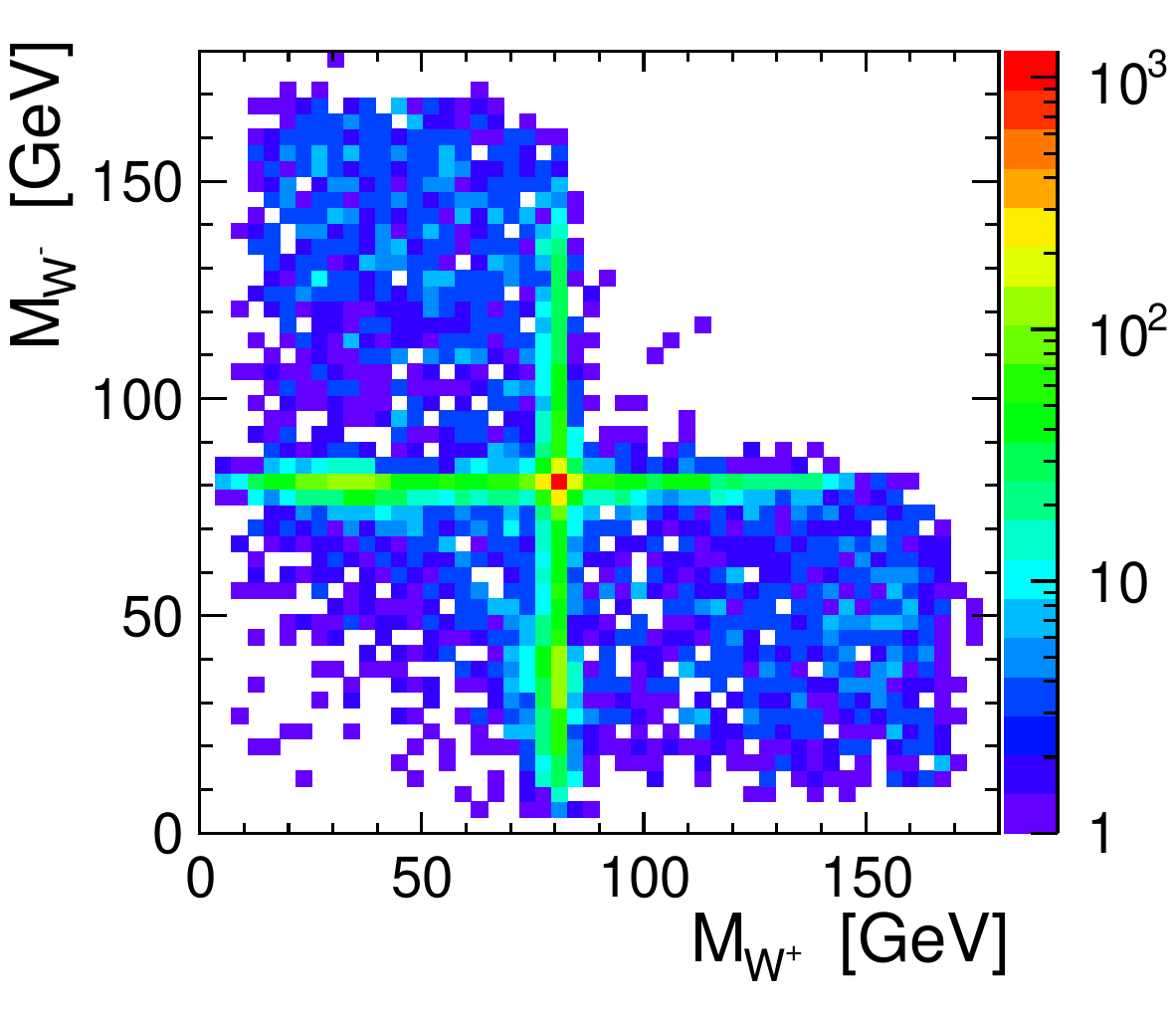}
\includegraphics[width=0.45\textwidth]{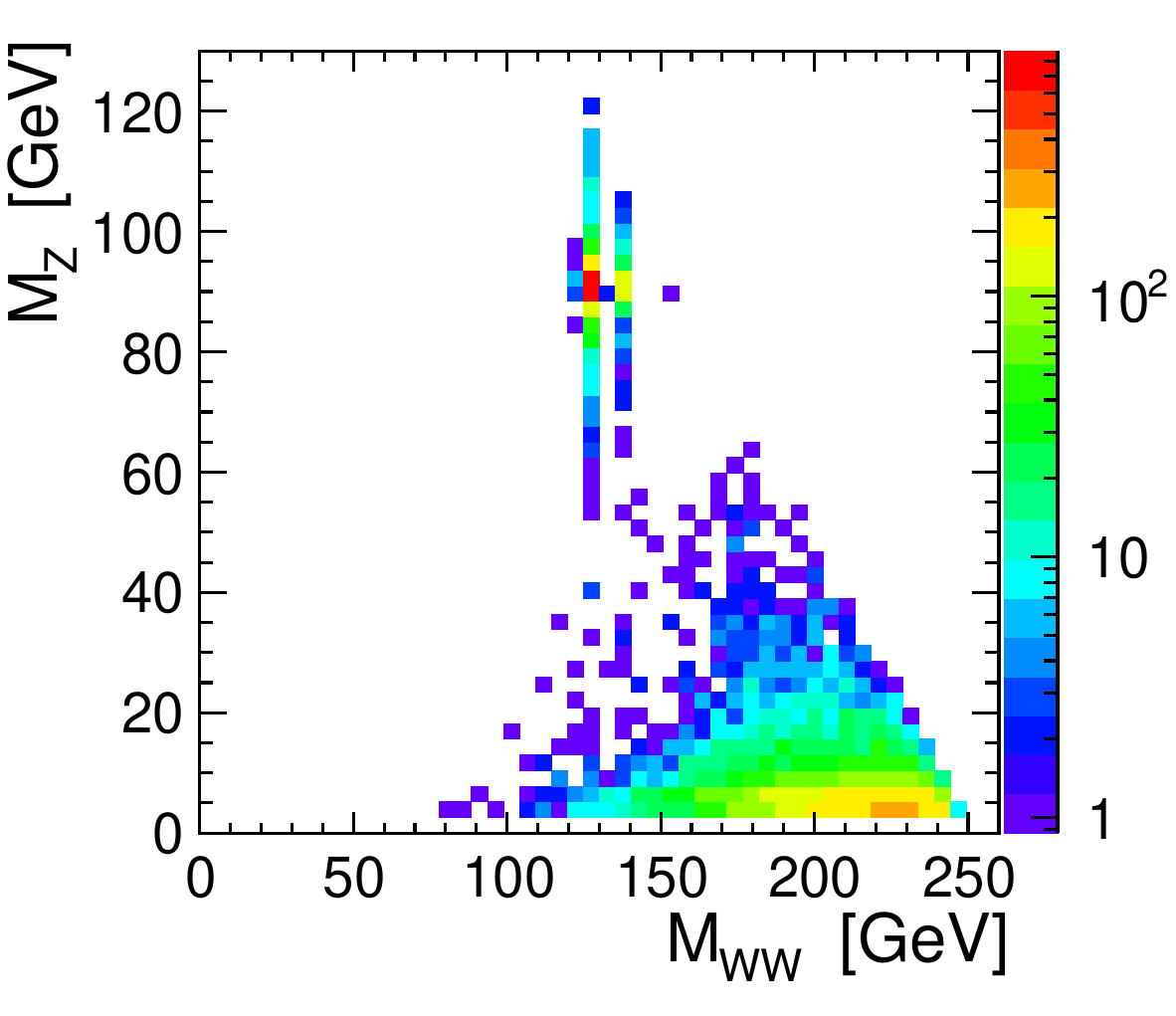}
}
\caption{Results of the \whizard simulation of the process
  $\epem \to \mpmm \Pe \PAGne \PQq\PAQq$ in the TRSM model with an additional 140 GeV scalar,
    at \roots=250 \GeV. Left: invariant mass distribution for the two $\PW$ boson candidates.
  Right: invariant mass of the \PZ boson candidate vs the invariant mass of the $\PWp\PWm$ pair. }  
  \label{fig:zarnecki2_mass_2d}
\end{figure}


Preliminary results from the Monte Carlo study indicate that the observation of the new scalar production
through the scalar-strahlung process with decay to two \PW boson should be feasible at a 250 \GeV Higgs factory
for scalar masses between 125 \GeV and 150 \GeV.
By exploiting the leptonic decay channel of the \PZ boson, the expected limits on the production cross section times 
BR$(\PSS\to\PWp\PWm)$ should be comparable to those obtained in the $\PSS\to\bb$ channel.
Furthermore, decays to $\PWp\PWm$ and $\PZ\PZ$ are highly promising when considering a future Higgs factory running at higher energy stages.

\subsubsection*{Searches in invisible decay channel}




The study is based on the framework developed and background samples generated for the search of light exotic scalar production with decays into two taus \cite{Brudnowski:2024iiu} presented above.
The scalar production in the scalar-strahlung process, $\epem \to \PZ \: \PSS$ is considered with hadronic \PZ decays (for highest sensitivity) and \PSS decays into invisible final state (e.g.\ dark sector).
A two-stage pre-selection was used to filter events consistent with the expected signal signature. First, all events with isolated leptons or photons reconstructed in the detector (including also forward calorimeters, LumiCal and BeamCal) were rejected. All reconstructed objects were then clustered into two jets (using the Durham algorithm in exclusive mode). Events were selected with reconstructed di-jet invariant mass in the $\pm$20 \GeV window around the \PZ mass and the missing transverse momentum greater than 10 \GeV. 
The following input variables were used for event classification: di-jet (\PZ candidate) invariant mass and energy, missing transverse momentum, cosine of the \PZ polar angle, angle between two jets, recoil mass (corresponding to the \PSS mass for signal events), and $y_{23}$ and $y_{34}$ variables from the clustering algorithm. Separate BDT classifiers \cite{xgboost} were trained for each beam polarisation configuration for each considered scalar mass. The dominant background source was found to be the $\PQq \PAQq \Pl \PGn$ process.


The expected exclusion limits on exotic scalar production cross section, calculated from the template fit to the BDT response distributions, are presented in \cref{fig:zembaczynski_limits_1}.
One should note that $\HepParticle{\Pe}{R}{-}\Xspace \HepParticle{\Pe}{L}{+}\Xspace$ running is expected to produce significantly better results than $\HepParticle{\Pe}{L}{-}\Xspace \HepParticle{\Pe}{R}{+}\Xspace$ running with the same luminosity. This is expected from the suppression of the $\PWp\PWm$ production, which contributes to the main background channel. 
Similar limits are obtained with 900 \fbinv collected with preferred polarisation configuration as for 2 \abinv with unpolarised beams. 
The combined analysis of the four polarisation configurations, according to the H-20 running scenario, results in about a 20\% improvement with respect to the same integrated luminosity with unpolarised beams.

\begin{figure}[htbp]
\centering
\begin{subfigure}{0.4\textwidth}
    \includegraphics[width=\textwidth]{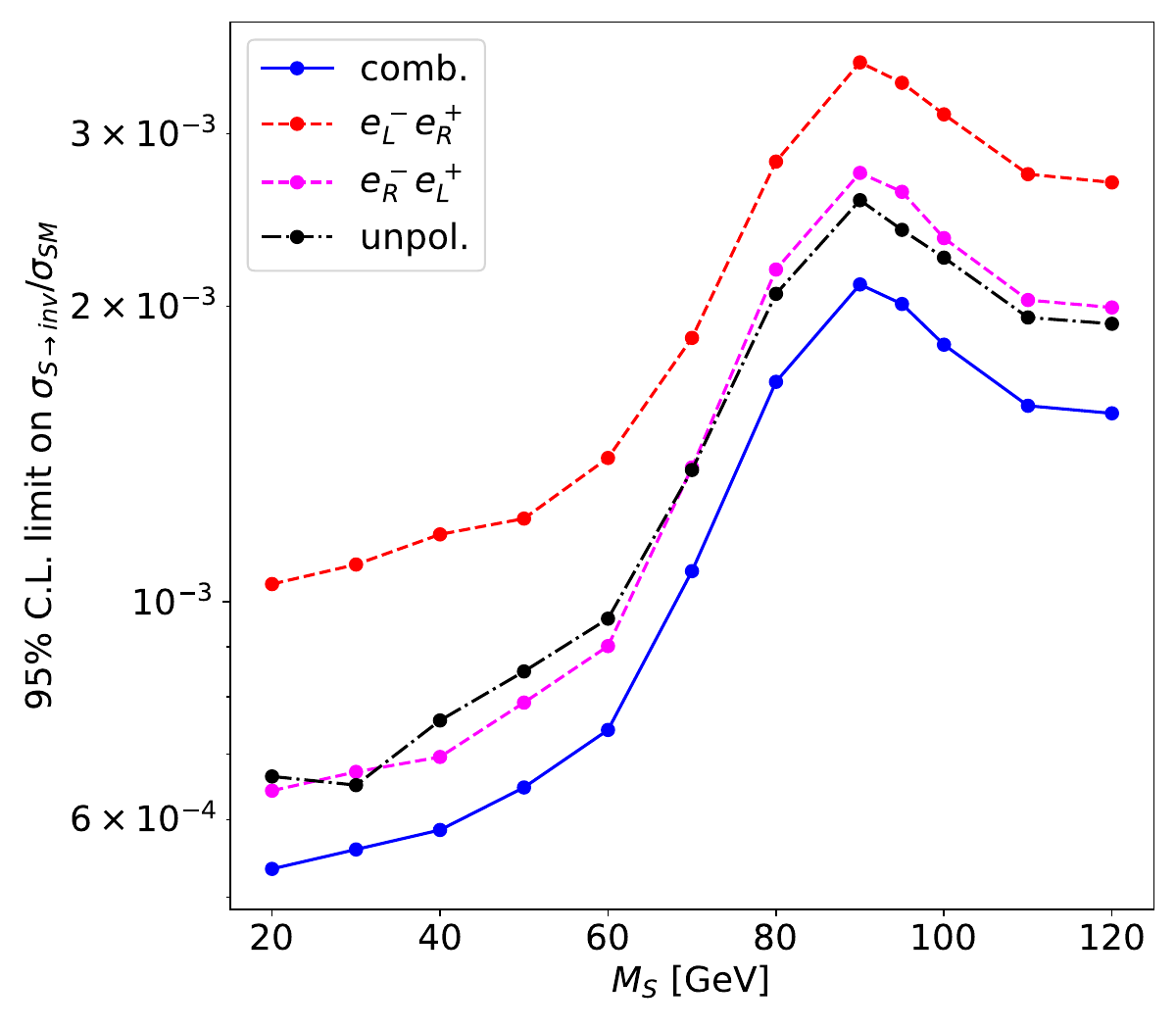}
    \caption{}
    \label{fig:zembaczynski_limits_1}
\end{subfigure}
\hspace{0.5cm}
\begin{subfigure}{0.4\textwidth}
    \includegraphics[width=\textwidth]{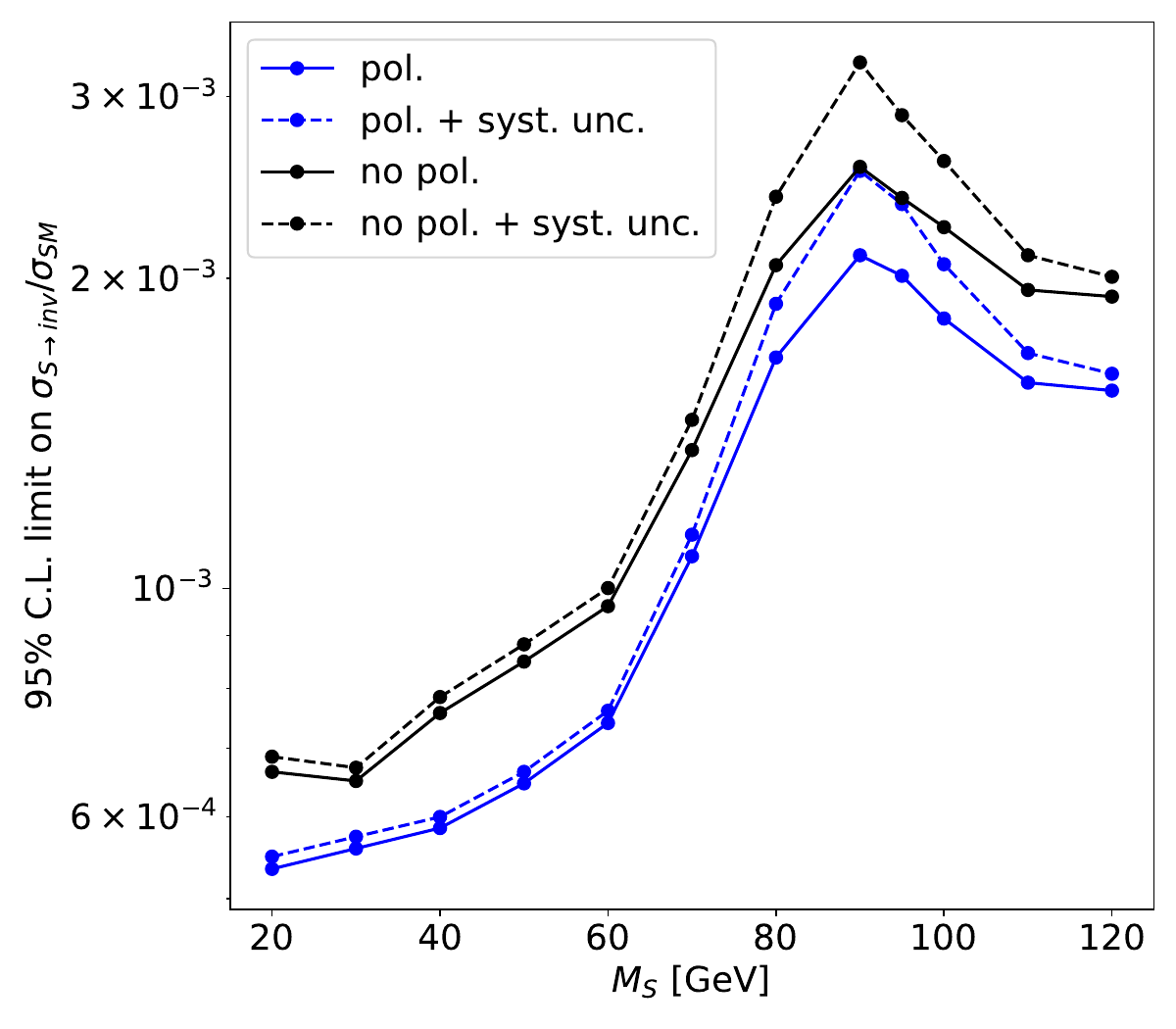}
    \caption{}
    \label{fig:zembaczynski_limits_2}
\end{subfigure}
\caption{Expected 95\% C.L. cross section exclusion limits on the light scalar production cross section times
invisible branching ratio as a function of the scalar mass for ILC running at 250 \GeV, assuming the H-20 running scenario.  (a) Comparison of combined limits obtained for different polarisation configurations. 
(b) Comparison of combined limits with and without polarisation and with and without systematic effects.}
\label{fig:zembaczynski_limits}
\end{figure}

To test the impact of systematic effects on the extracted cross section limits, nine additional nuisance parameters were introduced in the template fit procedure: four describing luminosity uncertainties of the data samples (for four polarisation settings) and five for the uncertainties in the theoretical predictions (for $\PAQq \PQq$, $\PWm \PWp$, $\PZ \PZ$, $\Pepm \PGg$ and $\PGg \PGg$ processes). The impact of systematic uncertainties is significant, as indicated in \cref{fig:zembaczynski_limits_2}, especially for the region of scalar masses around $\PWpm$ and $\PZ$ boson masses. The impact of systematic effects is also slightly larger for unpolarised beams, as expected.
%
%

The prospects for observing invisible decays of the new light scalar were also considered for FCC-ee running at 240 \GeV. 
Previously, this study was carried out for the CLIC accelerator at \roots = \SI{380} {\giga\electronvolt} \cite{Mekala:2020zys}. 
For event generation,  the following software chain is used: \textsc{MadGraph5\_aMC@NLO} \cite{Alwall:2014hca} to simulate the hard scattering processes, \pythiaeight\ \cite{Bierlich:2022pfr} for showering and hadronisation,  \delphes~\cite{Selvaggi:2014mya} to simulate the response of the IDEA detector \cite{IDEA1} and \fastjet \cite{Cacciari:2011ma} to cluster particles into two jets using the Durham \kT algorithm in exclusive mode. 
The above chain was handled within  the \keyhep framework \cite{Key4hep:2023nmr}.
We have used the SM model as implemented in \textsc{MadGraph5\_aMC@NLO} to simulate both the signal and background processes (ISR effects are not considered).

The events were reconstructed using information from two jets and missing momentum. In this case, we have used the ResonanceBuilder functionality within the FCCAnalyses software \cite{helsens_2024_13871482} that allows one to reconstruct the Z boson using jets whose invariant mass is close to the Z boson.  The recoil mass of the new scalar is defined as $M_{\mathrm{Recoil}} = s + m_Z^2 - 2 E_z \sqrt{s}$. 
Machine Learning techniques from  the TMVA framework \cite{Voss:2007jxm}  were applied to discriminate signal from background.
For training, the BDT classifiers were trained separately for each signal sample (i.e.\ scalar mass) against the combined SM background sample. 
No cuts were applied a-priori to BDT training. 
For the input variables, the four-momenta of the two jets, \PZ candidate mass and energy, missing momentum, $d_{23}$ and $d_{34}$ parameters of the jets, and the recoil mass were used.

\begin{figure}[htbp]
    \centering
    \includegraphics[width=0.5\textwidth]{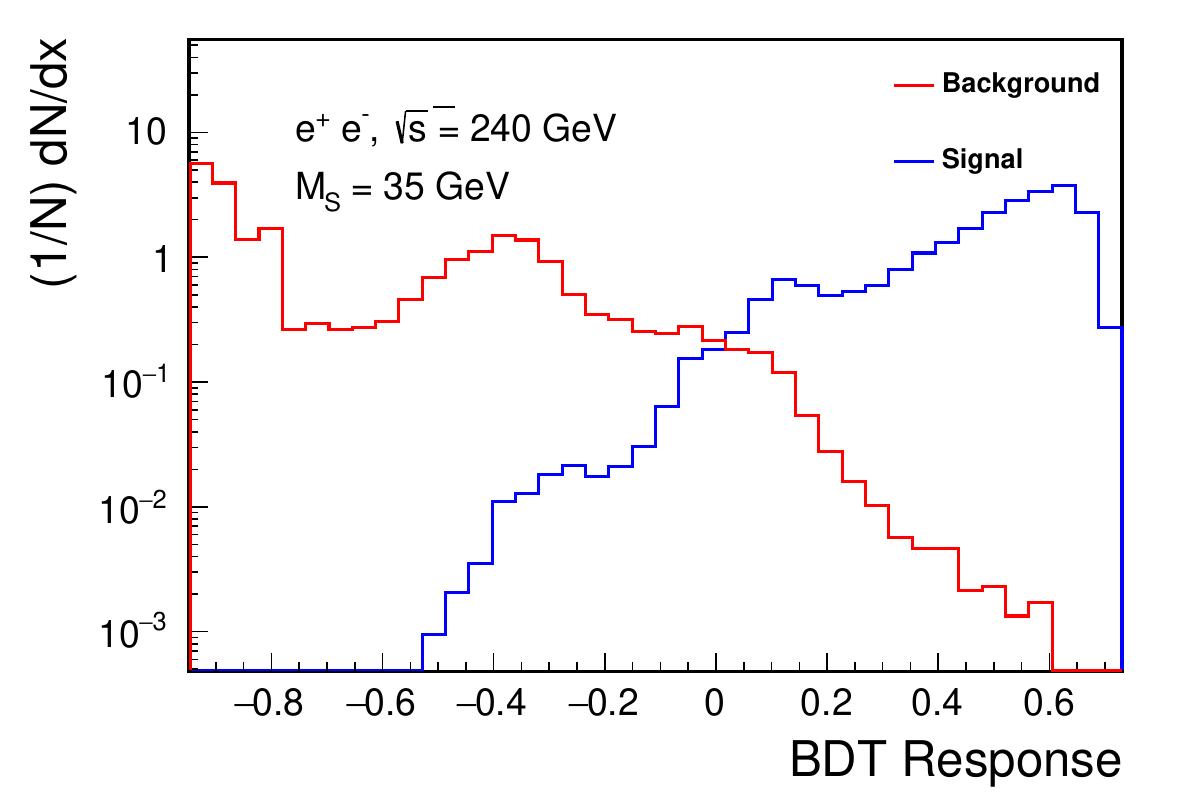}
    \includegraphics[width=0.35\textwidth]{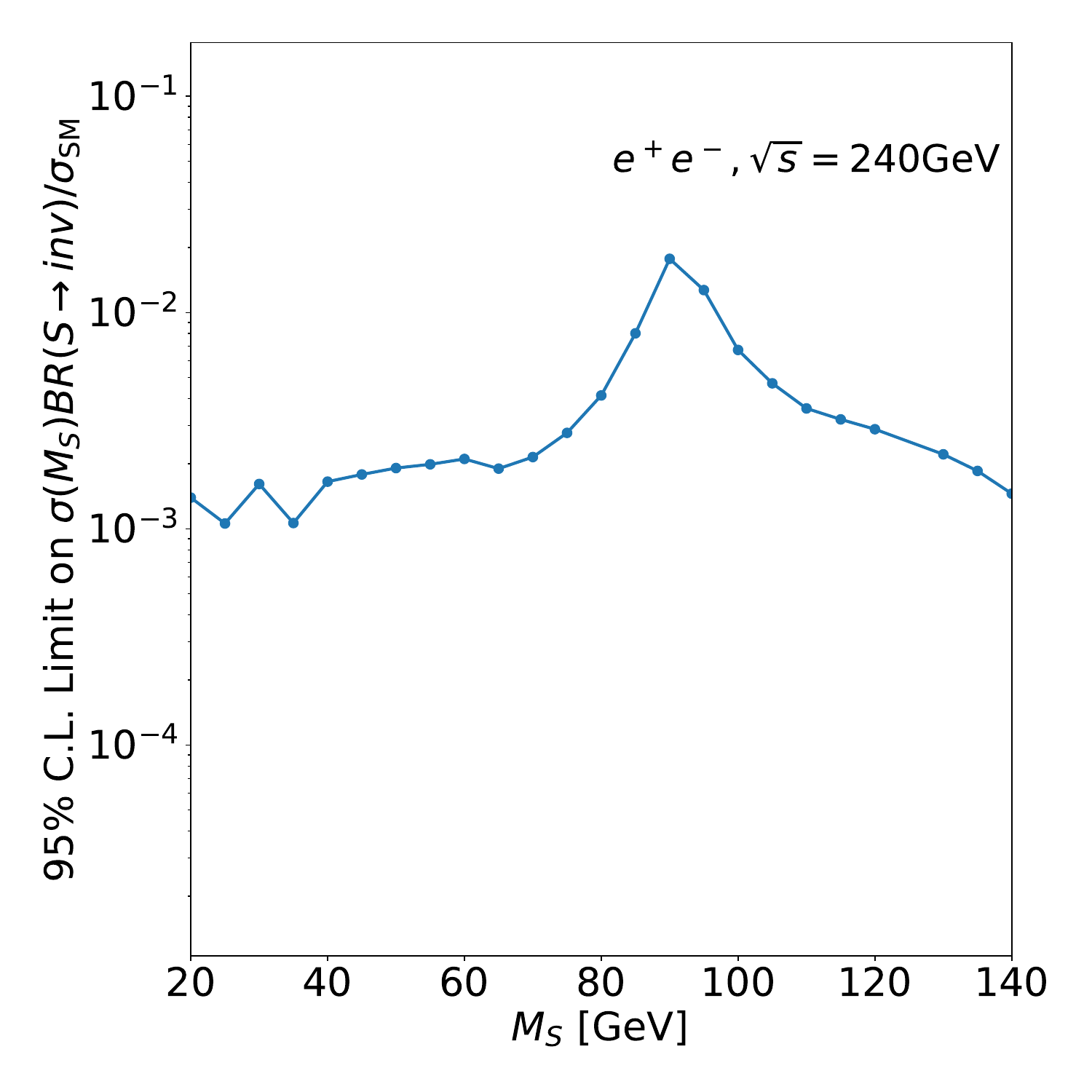}
    \caption{Left: BDT response for a scalar mass of 35 \GeV. Right: expected exclusion limits on scalar production cross section times the invisible branching ratio, relative to the SM Higgs production cross section as a function of the scalar mass.
    Results obtained with fast simulation for the IDEA detector at 240 \GeV FCC-ee. }
\label{fig:bdt_response_desai}
\label{fig:result_limit_desai}
\end{figure}

The BDT response distribution for the example signal scenario and the SM background is shown in \cref{fig:bdt_response_desai} (left), demonstrating that a very efficient signal-background separation is achievable. Exclusion limits on the scalar production cross section times the invisible branching ratio, relative to the SM Higgs production cross section are shown in \cref{fig:result_limit_desai} (right). 
%
\subsubsection*{Summary of scalar-strahlung searches}
BSM scenarios involving light scalars, with masses accessible at \epem 
Higgs factories, are still not excluded with the latest experimental data. 
Sizable production cross sections for new scalars can also coincide
with non-standard decay patterns, so a range of decay channels should
be considered.
Many new results on light exotic scalar searches in the scalar-strahlung process have been submitted as input to this report and their results are summarized in \cref{fig:exscalar_sz_summary}.
Strong limits are already expected from decay-mode-independent searches,
based on the recoil mass reconstruction technique.
If a light scalar preferentially decays to tau pairs, an improvement in search sensitivity by more than an order of magnitude is expected, making this channel the most sensitive one in this type of searches.
%
%
Expected exclusion limits in the invisible scalar channel, as well as limit estimates for the scalar decay to \bb with hadronic \PZ boson decays, are slightly weaker due to much higher backgrounds; while constraints on scalar decays to \bb with a leptonic signature are limited by the leptonic branching ratio of the \PZ boson.  
\begin{figure}[htbp]
    \centering
    \includegraphics[width=0.9\linewidth]{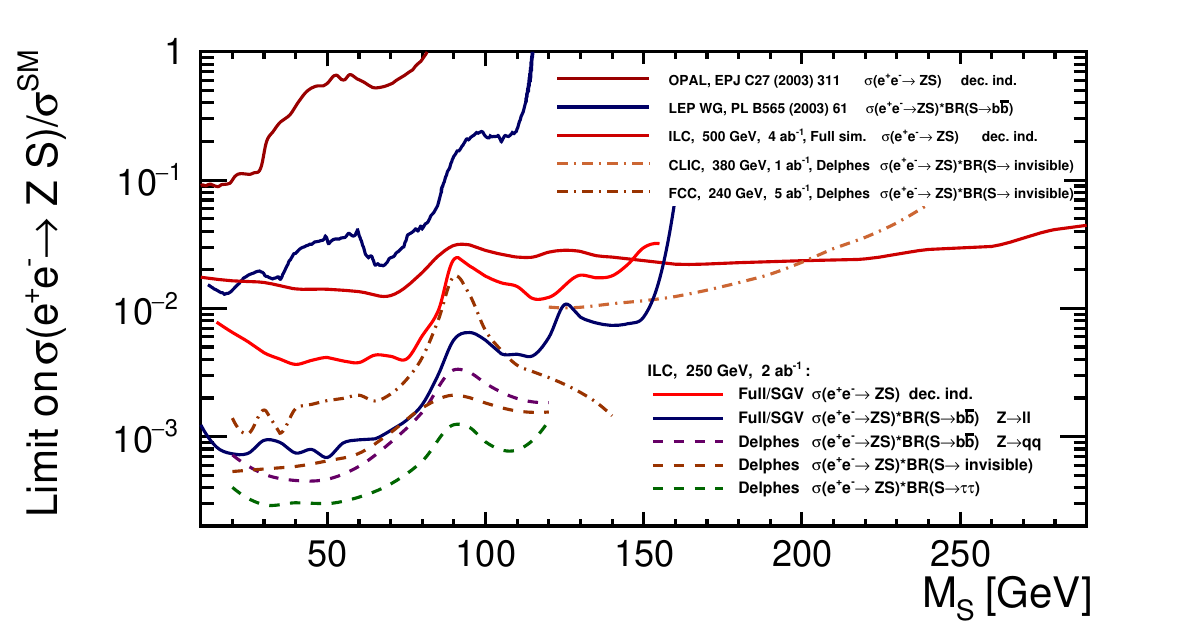}
    \caption{Comparison of expected exclusion limits for exotic scalar production searches in the scalar-strahlung processes,
      for different channels. }
    \label{fig:exscalar_sz_summary}
\end{figure}

\subsubsection{Production of exotic scalars in Higgs boson decays}\label{sec:exotics}





Future Higgs factories will be particularly suitable for searches for exotic decays of the 125 GeV Higgs boson  \cite{Liu:2016zki}.
Production of light (pseudo-)scalar bosons in 125 \GeV Higgs boson decays was studied for the Cool Copper Collider ($C^3$) \cite{Vernieri:2022fae} running at 250 \GeV. 
For efficient background suppression, only leptonic decays are considered for the \PZ boson produced in association with the Higgs boson, and with the two produced scalars (\Pa \Pa) decaying to \bb and $\PGtp\PGtm$ pairs. 
The final state considered is thus $\Pep\Pem \to \PZ\PH \to \PZ\Pa\Pa \to (\Plp\Plm)(\PQb\PAQb)(\PGtp\PGtm)$.
Two processes are expected to give leading contributions to the background: Higgs-strahlung with leptonic \PZ decays and all possible SM decays of the Higgs boson, and pair production of \PZ bosons, with one of them decaying to pair of electrons or muons.
Signal and background events were generated with \mgfive, followed by hadronisation using \pythiaeight, and detector simulation using \delphes\ with ILCgen detector cards.
The number of signal and background events was normalised to a luminosity of 2 \abinv at 250 \GeV. 





The reconstruction of the $\PZ$ boson is a critical step, as it serves as a tag for the associated production of the Higgs boson. Events were required to contain a pair of oppositely charged leptons ($\Plp\Plm$), consistent with the mass of the $\PZ$ boson. 
Tau leptons were reconstructed from energy flow (EFlow) objects, such as tracks, photons, and neutral hadrons. To identify taus, a combination of track isolation and clustering within a dynamic signal cone was employed. 
The reconstruction achieved a high matching efficiency and purity in the low \pt region for the hadronic $\PGt$, which is particularly significant for detecting the lighter scalars.
The identification of $\PQb$-jets is also an essential aspect of this analysis. To improve the efficiency and accuracy of $\PQb$-jet reconstruction, the ``Rest of Events'' (ROE) approach was employed,
constructing the $\PQb$-jets by clustering energy flow objects that are not attributed to leptons ($\Plp\Plm$) or taus ($\PGtp\PGtm$). This approach improves signal efficiency and achieves higher purity of $\PQb$-jets than the traditional jet reconstruction algorithm in \delphes.
For the $\PH$ boson reconstruction the recoil method was used, based on the precisely measured leptonic decays of the $\PZ$ boson. 
Compared to the reconstruction based on the measured $\PGt$-pairs, $\PQb$-jets and missing transverse energy (MET), it offers significantly improved reconstruction precision. 


The model-independent analysis yielded a 95\% confidence level (CL) upper limit on the cross section for the $\PZ\PH \to \PZ\Pa\Pa 	\to (\Pl\Pl)(\PQb\PAQb)(\PGtp\PGtm)$ process, as well as on the branching ratio BR$(\PH \to \Pa\Pa \to \PQb\PAQb\PGtp\PGtm)$, as illustrated in \cref{fig:Nee-Results_limits_1-aboson}. The limits were obtained using a maximum likelihood fit, with systematic uncertainties set at 0.2\% for signal and background, 1\% for leptons, and 2\% for taus. For instance, at $m_{\Pa} = 10~\GeV$, the branching ratio limit for $\PH \to \Pa\Pa \to (\PQb\PAQb)(\PGt^+\PGt^-)$ was found to be 1.2\%; a significant improvement compared to the 4.3\% reported by CMS \cite{CMS:2024uru}. The model-dependent analysis used the Two-Higgs-Doublet plus a Singlet (2HDM+S) framework to derive a limit on BR$(\PH \to \Pa\Pa)$ \cite{Curtin:2013fra}. For example, in 2HDM+S Type II with $\tan \beta = 5$ (\cref{fig:Nee-Results_limits_2-aboson}), the $\PH \to \Pa\Pa$ search shows improved sensitivity compared to CMS results \cite{CMS:2024uru}, especially in the lower $m_{\Pa}$ region. 

The $C^3$ analysis focuses on light or soft (pseudo-)scalar bosons, where the CMS trigger efficiency at HL-LHC may degrade due to increased background and pile-up. Thus, expected sensitivity improvements at HL-LHC might not fully materialize, particularly for low-mass $\Pa$ bosons. In contrast, $C^3$ can enhance the sensitivity by incorporating additional $\PZ$ decay channels, such as neutrino ($Z \to \nu\bar{\nu}$) and hadronic decays ($Z \to q\bar{q}$), building on its existing advantages over CMS at LHC.

\begin{figure}[htbp]
\centering
    \begin{subfigure}[t]{0.48\textwidth}
    \centering
    \includegraphics[width=0.88\textwidth,clip]{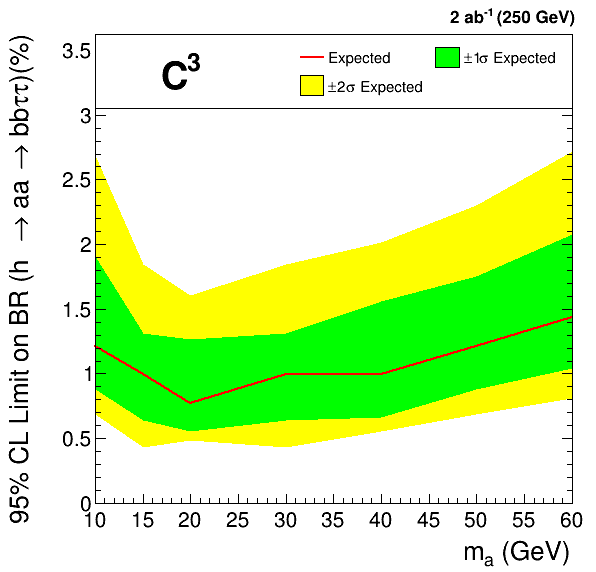}
    \caption{95\% CL upper limits on BR$( \PH \to \Pa\Pa \to \PQb\PAQb\PGtp\PGtm)$.}
    \label{fig:Nee-Results_limits_1-aboson} 
    \end{subfigure}
    \hfill
    \begin{subfigure}[t]{0.48\textwidth}
    \centering
    \includegraphics[width=0.88\textwidth,clip]{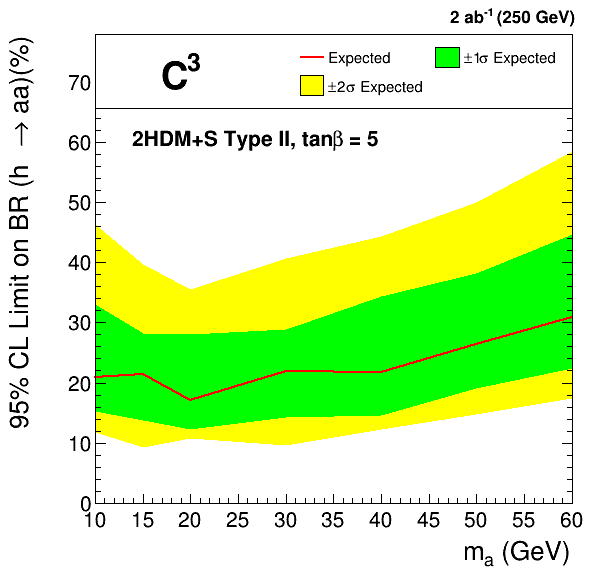}    
    \caption{95\% CL upper limits on BR$( \PH \to \Pa\Pa)$ in 2HDM+S models Type II with \(\tan\beta = 5\).}
    \label{fig:Nee-Results_limits_2-aboson} 
    \end{subfigure}
    \caption{Sensitivities to light (pseudo)-scalar bosons in 125~GeV Higgs boson decays, studied for the Cool Copper Collider running at 250~GeV.}
\label{fig:Nee-Results_combined-aboson}
\end{figure}



\subsubsection{\texorpdfstring{BSM triple Higgs couplings in exotic Higgs production}{BSM Triple Higgs Couplings in exotic Higgs production}}\label{sec:BSMHhh}


In this section we investigate the impact of one-loop corrections to the triple Higgs couplings (THCs)
on di-Higgs production in models with 
extended Higgs sectors at high-energy $\epem$ colliders, and explore the experimental sensitivity to such (BSM) THCs.
We work in the framework of the CP-conserving 2-Higgs doublet model (2HDM)~\cite{Branco:2011iw} and 
focus on the double Higgs-strahlung process, \ie\ $\epem\to \PSh\PSh\PZ$, with $\PSh$ being the 
Higgs boson discovered at the LHC.  
The process involves two THCs, which are the $\PSh$ self-coupling $\lambda_{\PSh\PSh\PSh}$ (or alternatively its ratio to
the SM prediction,  $\kappa_\lambda$), and the coupling $\lambda_{\PSh\PSh\PH}$, which enters through a resonant diagram mediated by 
the heavy Higgs boson $\PH$. 

Our calculation captures the leading EW corrections by including the predictions of the involved THCs at the one-loop
level, which can become very large in the 2HDM as has been shown in Ref.~\cite{Kanemura:2002vm}.
We compute the one-loop corrections to $\lambda_{\PSh\PSh\PSh}$ and $\lambda_{\PSh\PSh\PH}$ using the Coleman-Weinberg effective potential, as implemented in the code \texttt{BSMPT}~\cite{Basler:2018cwe,Basler:2020nrq,Basler:2024aaf}.
We furthermore compare our results to those of a fully diagrammatic computation of the one-loop corrections to $\lambda_{\PSh\PSh\PSh}$ using the code \texttt{anyH3}~\cite{Bahl:2023eau}.
This allows us to estimate the importance of the finite-momentum effects in the one-loop corrections to $\lambda_{\PSh\PSh\PH}$.

To summarize the (preliminary) results of our work, we focus on a particular point in the 2HDM 
with the following input parameters:
$
    m_{\PH}=\bar m = 300\ \GeV,\  m_{\PSA} = m_{\PSHpm}=650\ \GeV,\  \tan\beta=12,\  c_{\beta-\alpha}=0.12,
$
where we follow the same notation for the 2HDM input parameters as in Ref.~\cite{Arco:2020ucn}.
This point is allowed by the present constraints as described in Ref.~\cite{Muhlleitner:2020wwk}.
This leads to the following tree-level and one-loop THCs:
$
    \kappa_{\lambda}^{(0)} = 0.95,\ \kappa_{\lambda}^{(1)} = 4.69, \ \lambda_{\PSh\PSh\PH}^{(0)} = 0.02, \ \lambda_{\PSh\PSh\PH}^{(1)} = 0.21,
$
as given by the effective potential computation.

\Cref{fig:BP1} shows the differential distribution of the cross section w.r.t.\ the 
invariant di-Higgs mass $m_{\PSh\PSh}$ for the above input parameters.
The blue lines include the one-loop values of $\kappa_{\lambda}$ and $\lambda_{\PSh\PSh\PH}$ from the effective potential, the dashed red lines include the fully diagrammatic prediction for $\kappa_{\lambda}^{(1)}$, the yellow lines show the 2HDM tree-level prediction and the dotted black lines show the SM tree-level prediction.
Our cross section predictions are obtained with the tree-level formulas of Refs.~\cite{Djouadi:1999gv,Muhlleitner:2000jj} (adapted to the 2HDM), including the one-loop corrected THCs.
The right axis of the plots shows the expected final 4$\PQb$-jet events $\bar N_{4\PQb\PZ}$ after applying selection requirements as described in Ref.~\cite{Arco:2021bvf}.
We assume the projected ILC operating scenario with a centre-of-mass energy of $\sqrt{s}=500\ \GeV$
~\cite{Barklow:2015tja} with an integrated luminosity of 1.6\ \abinv and a beam polarisation of 
$P_{e^-} = -80\%$ and $P_{e^+} = +30\%$.
We furthermore include an experimental uncertainty due to the detector resolution by implementing a Gaussian smearing in the cross section distributions, as described in Ref.~\cite{Arco:2022lai}. 
The upper left, upper right, lower left and lower right plot assumes no uncertainty and a 2\%, 5\% and 10\% uncertainty, respectively.

First we comment on the effects of $\kappa_{\lambda}$ at one-loop.
The tree-level prediction is very close to the SM value, since we are very close to the alignment limit ($c_{\beta-\alpha}=0$).
On the contrary, the large one-loop corrections for $\kappa_{\lambda}$ lead to an enhancement of the non-resonant contributions to the cross section, due to the constructive interference of the $\lambda_{\PSh\PSh\PSh}$ diagram with the other non-resonant contributions.
These large values of $\lambda_{\PSh\PSh\PSh}$ at the one-loop level can be realized in the 2HDM (and in other models with extended Higgs sectors) with heavy Higgs bosons with a mass splitting between them.
We can also conclude that the finite-momentum effects in the one-loop corrections to $\kappa_{\lambda}$, see the red dashed line,
are minor, since they imply only a 5\% deviation in the total cross section prediction.
Therefore, the use of one-loop corrected THCs obtained from the effective potential provides a good approximation for the one-loop scalar corrections to the di-Higgs production.

We now turn to the effect of the one-loop corrections of $\lambda_{\PSh\PSh\PH}$, which enters the cross section 
through a resonant diagram mediated by the heavy Higgs boson $\PH$.
For our benchmark point, the resonance peak is expected at $m_{\PSh\PSh} = m_{\PH} = 300\ \GeV$, as it can be seen in all plots in \cref{fig:BP1}.
As an estimate of the relevance of the $\PH$ resonance peak, and therefore the possible access to the coupling $\lambda_{\PSh\PSh\PH}$, we compute the statistical significance $Z$ of the resonance peak.
We perform a profile likelihood ratio test between the no-resonance hypothesis (\ie\ with $\lambda_{\PSh\PSh\PH}=0$, corresponding to the dot-dashed lines in \cref{fig:BP1}) and the complete prediction with the $\PH$ peak, as described in Ref.~\cite{Cowan:2010js}.
For our considered point, the statistical significances containing the one-loop corrected THCs, $Z^{(1)}$, are larger than the ones from the tree-level distributions, $Z^{(0)}$. Both values are given in each plot.
This is due to the fact that for our benchmark point the tree-level prediction for $\lambda_{\PSh\PSh\PH}$ is very close to zero, while the one-loop correction increases the value of this coupling to a much larger value of 0.2, which makes the $\PH$ resonance peak more prominent.

We also find that the smearing of the cross section distributions has a relevant effect on the significance of the $\PH$ resonance peak.
A large experimental uncertainty in the cross section distributions (or in other words, large smearing values) makes the $\PH$ resonance peak less prominent and sharp, leading to a smaller significance, as can be inferred from the given $Z$ values.
We therefore conclude that the experimental resolution of the $m_{\PSh\PSh}$ distributions will be crucial for the determination of $\lambda_{\PSh\PSh\PH}$ from the $\PH$ resonance peak.
 
In conclusion, our study shows that one-loop corrections to THCs in models with extended Higgs sectors significantly affect the di-Higgs production cross section. 
These corrections can lead to a substantial enhancement of the cross section, even in the alignment limit. Therefore, it is crucial to include them in any accurate phenomenological analysis.

Finally, we provide a list of some open questions that should be addressed in future work:
1) How relevant can the remaining one-loop corrections be for the $\PSh\PSh$ production, especially from the $\PH\PZ\PZ$ vertex?
2) What is the expected experimental resolution on the $m_{\PSh\PSh}$ distributions?
3) What would be the effect of the experimental backgrounds and how would this affect the access to the BSM THCs?
4) How relevant could the finite momentum effects be for the loop-corrected $\lambda_{\PSh\PSh\PH}$ coupling?

\begin{figure}[t!]
    \centering
    \includegraphics[width=\textwidth]{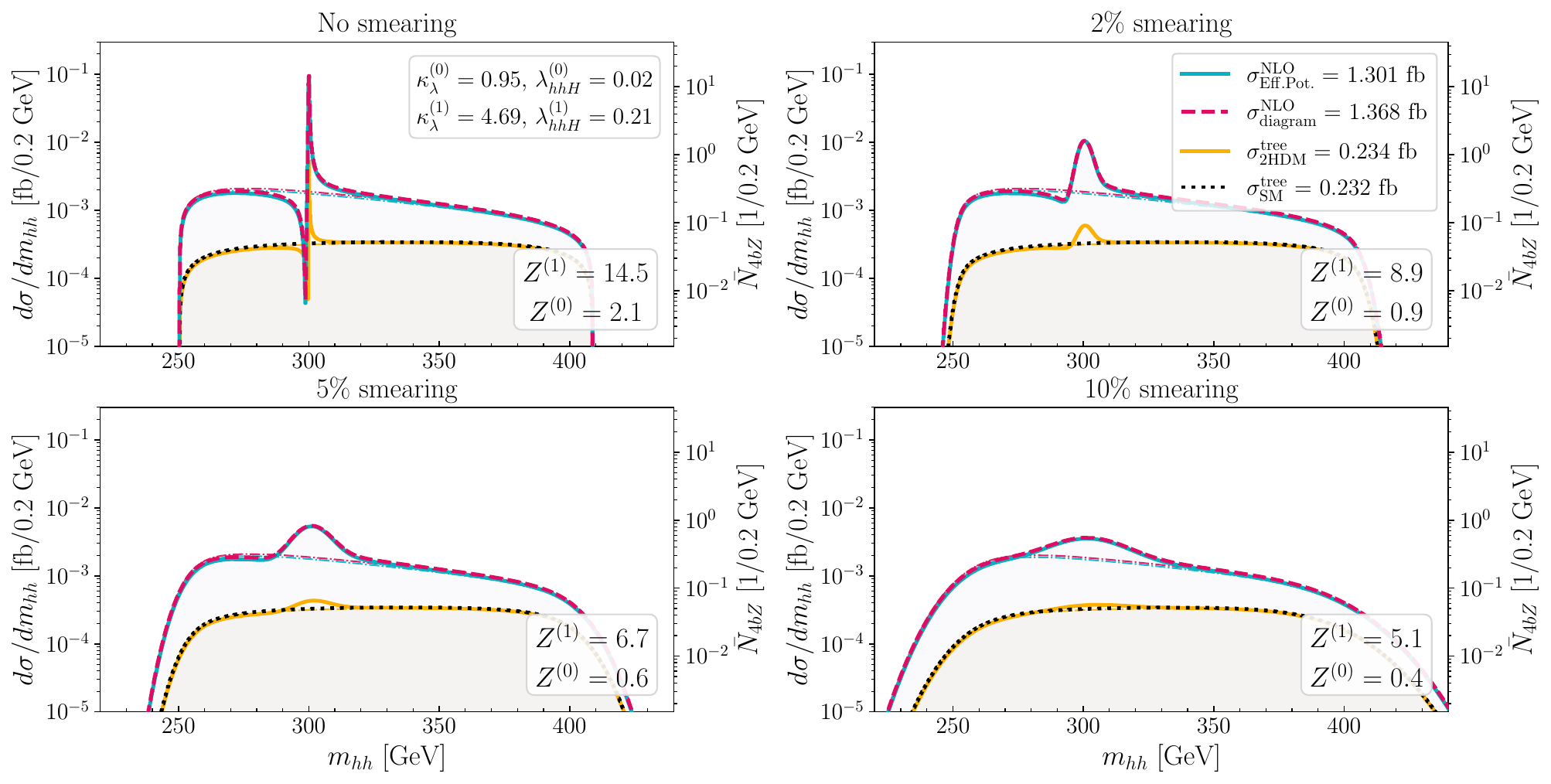}
        \caption{Differential distribution of the cross section as a function of $m_{\PSh\PSh}$ for the considered 2HDM input parameters given in the text, at $\sqrt{s}=500\ \GeV$ for different smearing assumptions. }
    \label{fig:BP1}
\end{figure}

\subsubsection{Searches in other production channels}

This study considers a search for pair production of the Inert Doublet Model (IDM) neutral scalars, dominated by the $\epem \to \PH \PSA$ process. Here we choose $\PH$ to be a lightest inert scalar, a dark matter candidate.
The IDM signal samples were generated using \textsc{Madgraph5\_aMC@NLO} v2.8.1~\cite{Alwall:2014hca}, using the UFO model available at Ref.~\cite{ufoidm},\footnote{Previous versions of this model file contained a non-unitary CKM matrix. We advise the readers to make sure this is remedied prior to using the model file.} interfaced with \pythia v8.2~\cite{Sjostrand:2014zea}, using $\epem$ collisions at centre-of-mass energies $\roots=240$ and 365~\GeV. Instead of targeting individual pair production modes, two final states are generated directly, in order to have all contributing diagrams and the proper interferences taken into account. We explore the signature of same-flavour dilepton and missing energy. For this study, we consider the  final states $\Plp\Plm \PH\PH$ and $\Plp\Plm\PGn\PAGn \PH\PH$, with $\Pl=\Pe$ or $\PGm$, and H the stable neutral scalar. 
The parameter $\lambda_2$ corresponds to couplings within the dark sector, while $\lambda_{345}$ determines the Higgs-DM coupling. 
The signal was generated fixing the coupling $\lambda_2$ to 0.1, $M_{H\P}$ between 70 and 115 (180)\,\GeV\ in steps of 5\,\GeV\ for $\roots=~240\,(365)$\,\GeV, and $M_{\PSA}-M_{\PH}$ between 2\,\GeV\ and the kinematic limit for on-shell production in steps of 2 to 5\,\GeV. Three scenarios are considered by setting $M_{\PH^{\pm}}$ and $\lambda_{345}$: (S1) $M_{\PH^{\pm}}=M_{\PSA}$ and $\lambda_{345}=0$; (S2) $M_{\PH^{\pm}}=M_{\PSA}$ and maximised $\lambda_{345}$; (S3) maximised $M_{\PH^{\pm}}$ and maximised $\lambda_{345}$.
The maximised values for $\lambda_{345}$ and $M_{\PH^{\pm}}$ are found taking into account various constraints, including SM precision measurements and direct searches. For a better estimate of the contribution stemming from the additional channel that depends on the coupling $\lambda_{345}$, we chose relatively large values that are currently in conflict with direct detection bounds.
About 500,000 events were generated per mass point. The cross sections calculated by \textsc{Madgraph5\_aMC@NLO} extend 
from 10\,fb at low $M_{\PH}$ and $M_{\PSA}-M_{\PH}$ down to $< 0.1$ ($<10^{-4}$)\,fb, 
close to the kinematic limit for the $\Plp\Plm \PH\PH$ ($\Plp\Plm\PGn\PAGn \PH\PH$) production in the S1 scenario.


Background processes include diboson and top pair production, $\epem\rightarrow \epem$ (generated with $M_{\epem} \in [30;150]$~\GeV), inclusive $\epem\rightarrow\mpmm$ and $\epem\rightarrow\tptm$ production, and SM Higgs production in association with a \PZ boson, or in VBF production. Samples were produced centrally with high statistics, within the so-called Winter2023 campaign, using \pythia v8.306~\cite{Bierlich:2022pfr} for the diboson and \ttbar samples, and \whizard v3.0.3~\cite{Kilian:2007gr} interfaced with \pythia v6.4.28~\cite{Sjostrand:2006za} for the other processes. The total integrated luminosity of the data sample is taken to be 10.8 (2.7) \abinv for the FCC-ee run with $\roots=240$\,(365) \GeV. 
Both signal and background samples are processed through \delphes\ v3.5.1pre05~\cite{deFavereau:2013fsa} with the IDEA~\cite{IDEA1} detector configuration as per the Winter2023 campaign. Events are produced using the \keyhep \cite{Ganis:2021vgv} framework. 

As preselection, a minimum momentum of 5 \GeV is imposed on the \delphes\ reconstructed electrons, muons and photons. Events with exactly two such electrons or exactly two such muons are selected. Additional selection criteria include requirements on the invariant mass and longitudinal momentum of the lepton pair, and lepton transverse momentum. Rejection factors of 98\% or above were obtained for all background channels, for a signal efficiency of around 50\%.
%
%
%
To further separate signal from backgrounds, a multivariate analysis is performed with input variables similar to those proposed in Refs.~\cite{Kalinowski:2018kdn,Zarnecki:2020swm}.

To take into account the dependence of the analysis and signal features on the dominant theoretical parameters (namely $M_{\PH}$ and $M_{\PSA}-M_{\PH}$), a parametric Neural Network (pNN) \cite{pNN} is chosen. With a pNN, only a single network has to be trained for the entire parameter space with optimal performance. With a smooth output, a continuous interpolation is done for points in the parameter space that were not trained on. The parametric neural network is implemented using PyTorch \cite{pytorch}.
A maximum likelihood fit of the pNN output distribution above 0.9 is performed using the CMS Combine package~\cite{CMS:2024onh}. A 95\% confidence-level (CL) upper limit is derived using the asymptotic approximation~\cite{Cowan:2010js}.
Only the MC statistical uncertainties are included in the fit~\cite{Conway:2011in}.

%
%
The 95\% CL exclusion region obtained is shown in \cref{fig:IDM-lim240} (left) for $\roots=240$~\GeV, for the total integrated luminosity of 10.8\,\abinv. The three different scenarios are shown as separate lines, well within the $\pm 1\sigma$ contour shown in green for scenario S1, including only the statistical uncertainties. The two final states $\epem$ and $\mpmm$ are combined for $(M_{\PSA}-M_{\PH})>30$ \GeV, below which only the $\mpmm$ result is shown given that backgrounds were not generated at lower masses for the $\epem$ channel. The kinematic limit is highlighted by the dashed line. The regions excluded by dark-matter relic density constraints and LEP SUSY recast are overlaid. With the full luminosity of $10.8\,\abinv$, almost the entire parameter space can be excluded at 95\% CL, reaching $M_{\PH} = 110$\,\GeV. The $>5$-sigma discovery region is shown in \cref{fig:IDM-lim240} (right) for $\roots=240$~\GeV, in the scenario S1, for the total integrated luminosity of $10.8$ and $2\,\abinv$.  Similarly, the 95\% CL limit (left) and $>5$-sigma discovery (right) regions are shown in \cref{fig:IDM-lim365} for $\roots=365$~\GeV, for the total integrated luminosity of $2.7\,\abinv$. The 95\% CL excluded region reaches $M_{\PH}=165$\,\GeV. The discovery reach extends to  $M_{\PH} = 157$\,\GeV\ for $M_{\PSA}-M_{\PH}=15$\,\GeV\ for the nominal FCC luminosity scenario, and is also shown for a scenario with about 10 times less luminosity.

\begin{figure}[htbp]
    \includegraphics[width=0.45\textwidth]{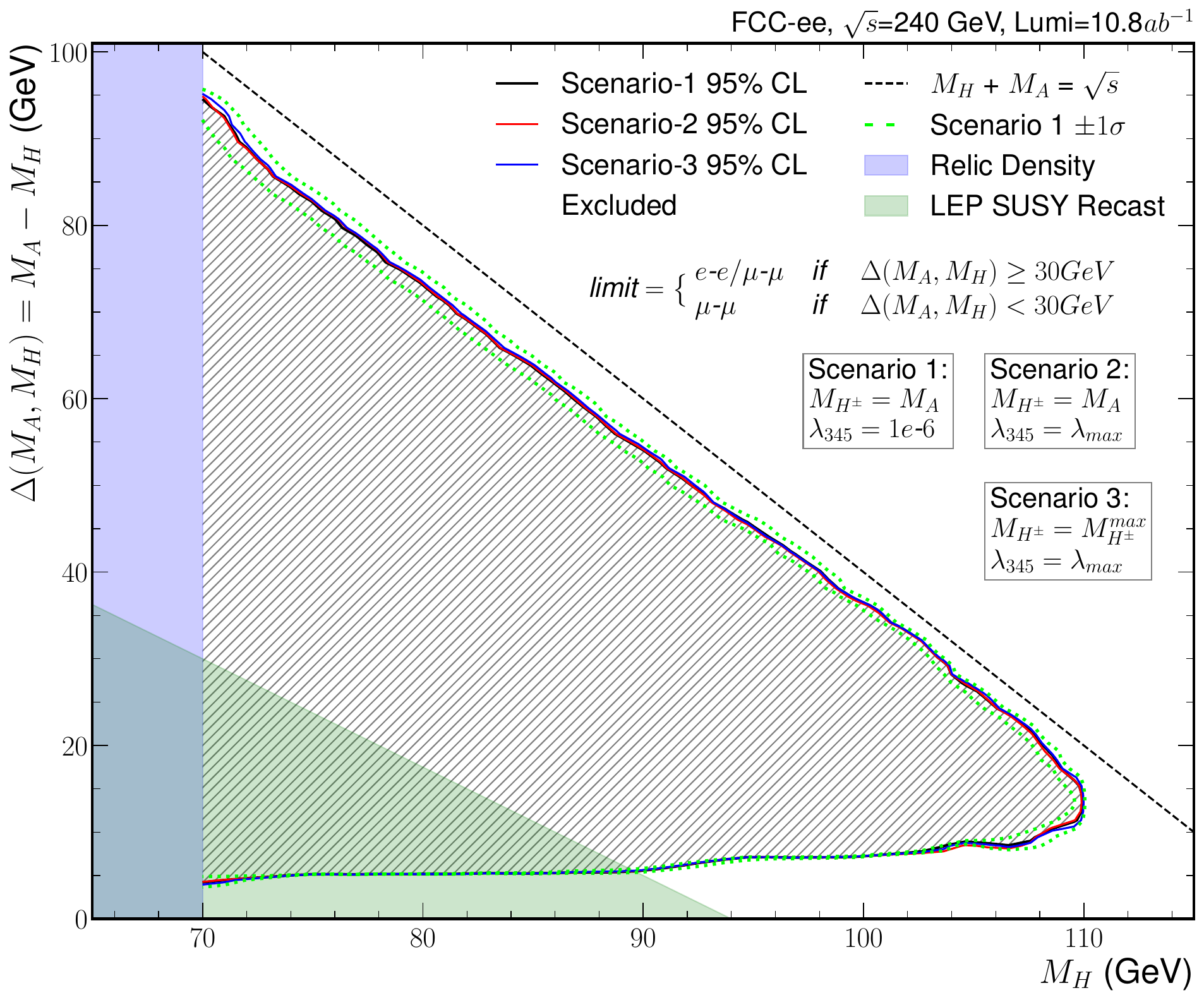} \hfill    \includegraphics[width=0.45\textwidth]{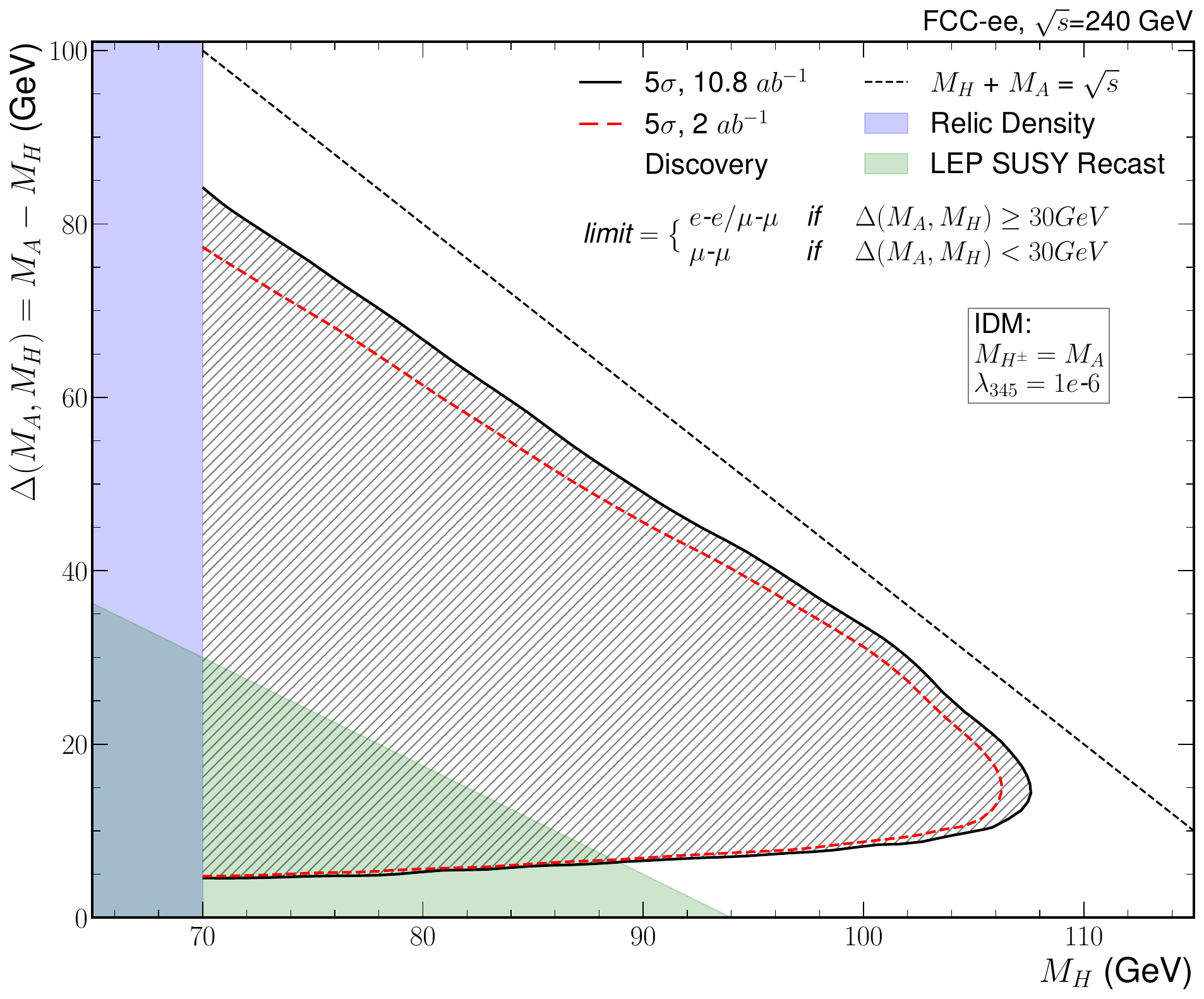}\\
    \caption{95\% CL expected exclusion (left) and $>5$-sigma discovery contour (right) in the $M_{\PSA}-M_{\PH}$ vs $M_{\PH}$ plane in an Inert Doublet Model, for $\roots=240$~\GeV with total integrated luminosities of 10.8 and 2\,\abinv.}
    \label{fig:IDM-lim240}
\end{figure}

\begin{figure}[htbp]
    \includegraphics[width=0.45\textwidth]{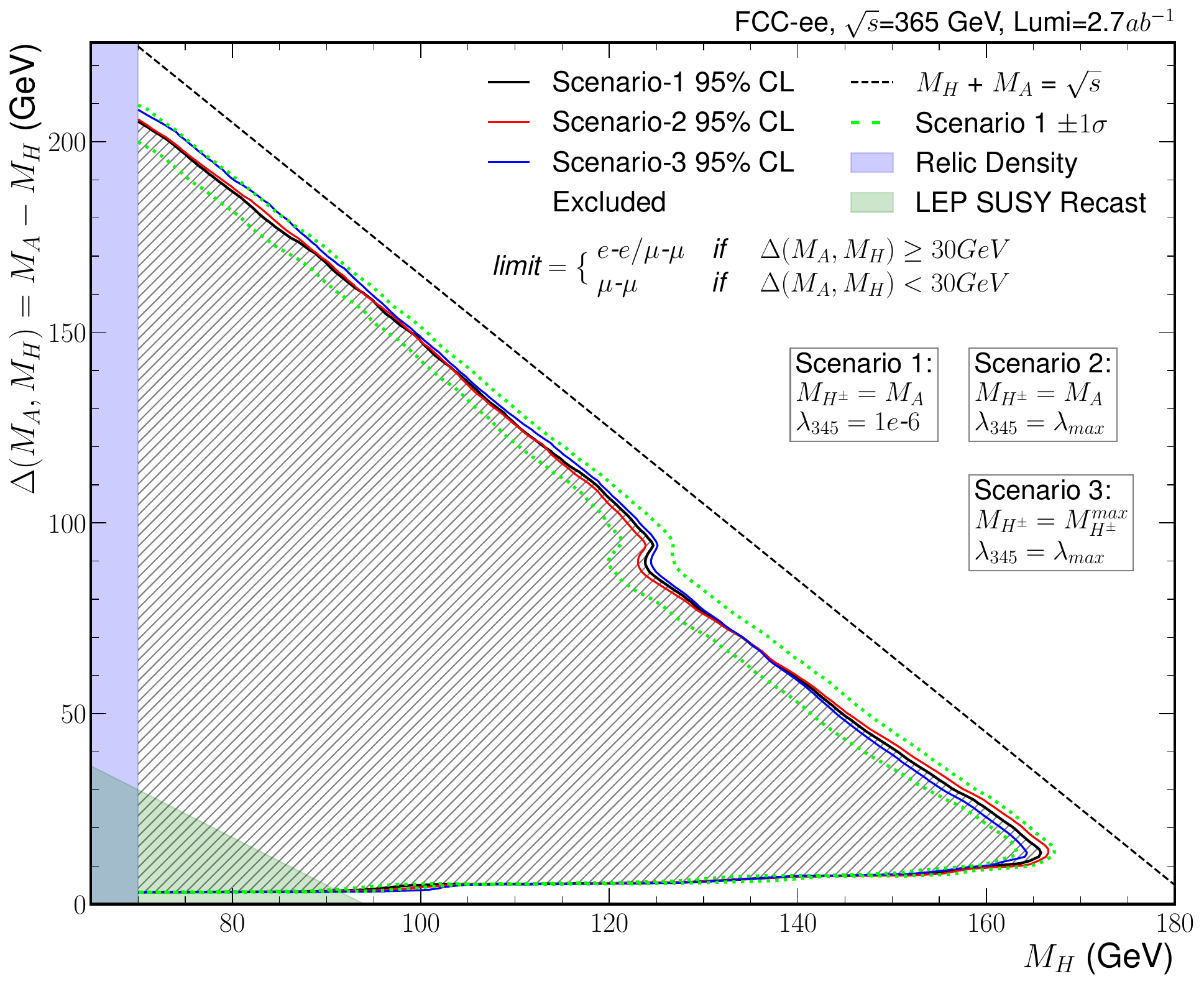} \hfill  \includegraphics[width=0.45\textwidth]{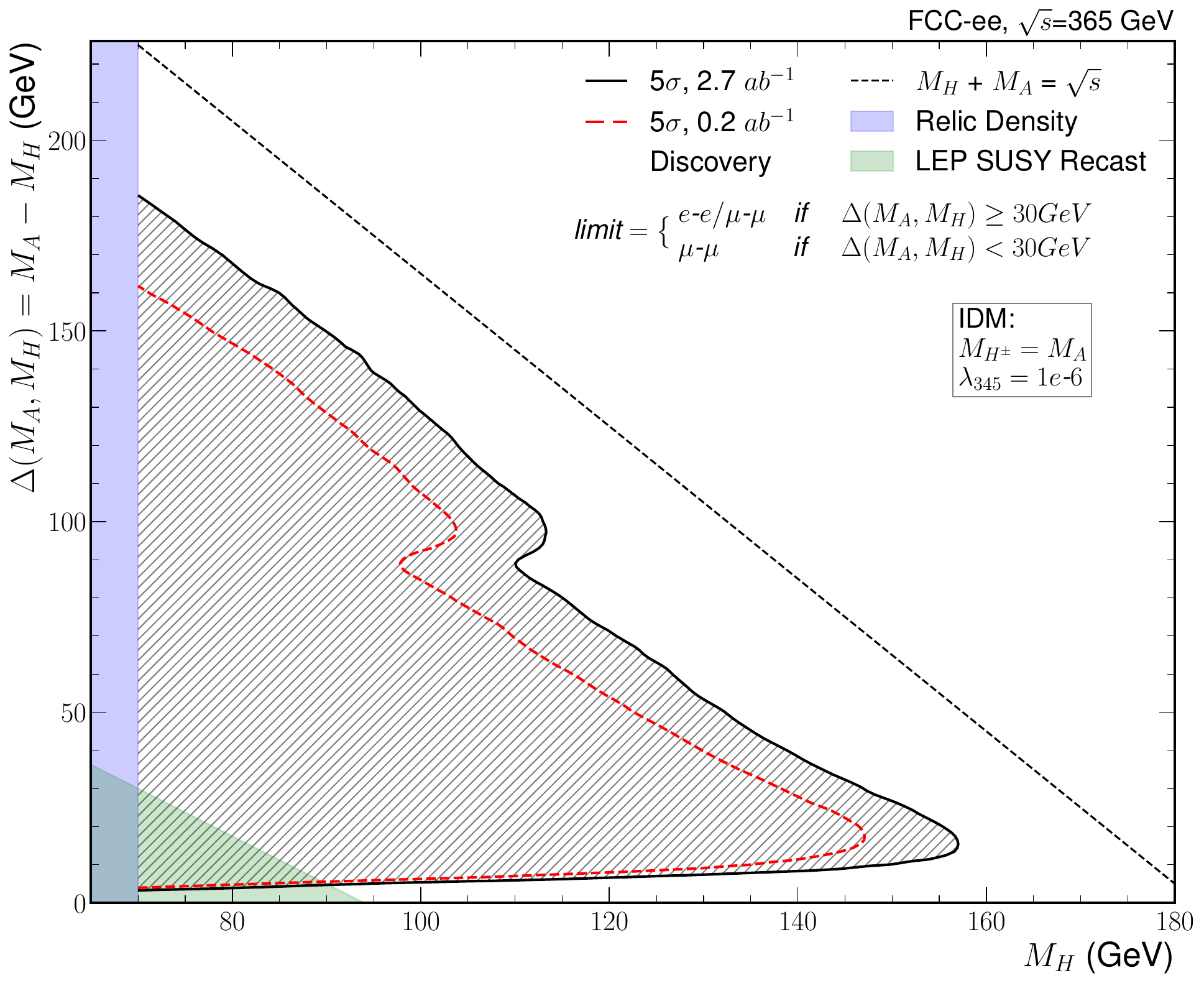}
    \caption{95\%CL limit (left)  and $>5$-sigma discovery contour (right) in the $M_{\PSA}-M_{\PH}$ vs $M_{\PH}$ plane in an Inert Doublet Model, for $\roots=365$ \GeV with total integrated luminosities of 2.7 and 0.2\,\abinv.}
    \label{fig:IDM-lim365}
\end{figure}

\subsection{\focustopic Long-lived particles \label{sec:LLP-focus-topic}}
\editor{Rebeca}
New BSM particles can have varied lifetimes in the same way SM particles do, depending on factors such as their mass and couplings. As colliders reach higher energies, heavier and shorter-lived particles become accessible, often leading to a focus on prompt decays. Searches for long-lived particles (LLPs) are however gaining momentum as an exciting alternative and complement to traditional searches~\cite{Curtin:2018mvb,Zurita:2019eug,Alimena:2019zri,Lee:2018pag,Agrawal:2021dbo}. This kind of search is signature-driven and can be linked to many BSM models in $\epem$ colliders~\cite{Blondel:2022qqo, Chrzaszcz:2021nuk}, such as Heavy Neutral Leptons (HNLs), Axion-Like Particles (ALPs), and exotic Higgs boson decays. 

Experimentally, LLPs have distinctive signatures like displaced vertices, tracks, and jets, as well as more complicated ones such as broken tracks, high energy loss, or delayed/stopped behaviour. Observing any clear LLP could be a sign of new physics independently of the type of search. However, while LLPs usually have low backgrounds, standard collider techniques typically struggle to identify and reconstruct them accurately. Triggering, one of the central issues at the LHC, may not be an issue in future $\epem$ colliders, but most other challenges will still exist across collider geometries and energies. By anticipating these challenges, it is possible to optimize future collider experiments to explore the full potential of LLP searches~\cite{Bose:2022obr}.

Besides the different studies proposed in the following, additional experiments following those running at the LHC or proposed for HL-LHC should also be considered, from additional detectors on- and off-axis to beam dump experiments~\cite{Sakaki:2020mqb,Asai:2021ehn}, depending on the considered facility. Mimicking the successful proposals and approvals of additional detectors at the LHC, similar additional experiments could be envisioned at the FCC and CEPC~\cite{Wang:2019xvx,Tian:2022rsi, Lu:2024fxs}. The civil engineering of the FCC and future FCC-hh space requirements will provide FCC-ee with significantly larger detector caverns than required for a lepton collider. This could enable the installation of instrumentation on the cavern walls to search for new long-lived particles~\cite{Chrzaszcz:2020emg}.

\subsubsection{Heavy Neutral Leptons}~\label{llps:hnls}
Sterile neutrinos under the SM gauge group can have both Dirac and Majorana mass terms. The heavy mass eigenstates that emerge from the small mixing with active SM neutrinos are commonly referred to as HNLs~\cite{Abdullahi:2022jlv}.
In addition to explaining the origin of neutrino masses, such models can address other unresolved questions in the SM, such as the baryon asymmetry of the universe (BAU) and the particle nature of dark matter. For SM neutrinos to remain light, collider-detectable HNLs must appear as pseudo-Dirac pairs of nearly degenerate Majorana fermions. In these models, the mass splitting of the pseudo-Dirac pair determines the light neutrino mass scale and the extent of lepton number violation (LNV) observable in processes involving HNLs.

Leptogenesis offers an attractive solution to the origin of matter by linking the observed BAU with the origin of light neutrino masses. In this mechanism, the same HNLs responsible for the light neutrino masses can generate the matter-antimatter asymmetry through their CP-violating decays (or, depending on the model, through their CP-violating oscillations) in the early universe.
Moreover, HNLs can be connected to dark sectors, such as dark photons and dark scalars, potentially explaining the observed dark matter. HNLs have been considered in dark sector models, particularly in fermionic extensions, and can interact with the SM through mixing between sterile neutrinos, dark fermions charged under new interactions, and standard neutrinos.
In minimal models, the production and decay of HNLs are governed by SM interactions and the mixing between HNLs and active neutrinos, typically resulting in relatively long lifetimes if their masses are in the MeV--GeV range. 

Figure~\ref{fig:sumHNL} shows the current constrains on the muon neutrino-sterile neutrino mixing $|U_{\mu N}|^2$ as a function of the sterile neutrino mass $m_N$~\cite{HNLlim}. Current collider searches cumulate in areas with relatively strong couplings and high masses. In the uncovered areas, approaching the see-saw line, couplings are very low, and so LLP signatures are more likely to occur. Only LLP signatures of HNLs will be discussed in this section, but HNLs as a more general topic will be expanded in Section~\ref{sec:SRCH-NHL}.

\begin{figure}[h!]
    \centering
        \includegraphics[width=0.85\textwidth]{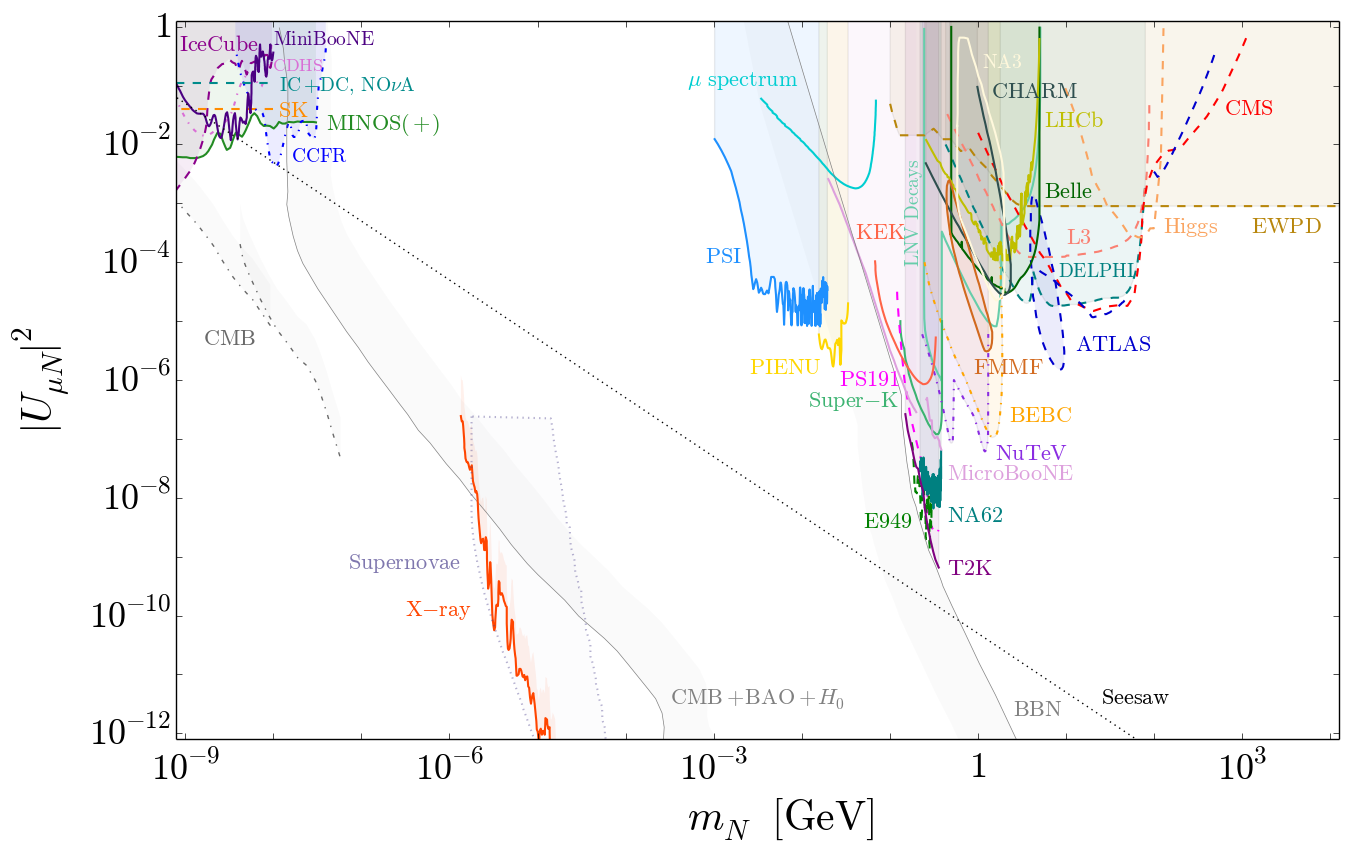}
    \caption{Current constraints on the muon neutrino-sterile neutrino mixing $|U_{\mu N}|^2$ as a function of the sterile neutrino mass $m_N$. From Ref.~\cite{HNLlim}.}
    \label{fig:sumHNL}    
\end{figure}
\FloatBarrier

In the last years, the community has been extremely active in the exploration of the potential of long-lived HNLs at future colliders~\cite{SisselMsC,RohiniMsC,LovisaMsC,TanishqMsC,SofiaMsC,DimitriMsC,TomMsC} across all design studies~\cite{Mekala:2022cmm} and a variety of final states. The potential for HNL searches is especially strong at FCC-ee, where the discovery potential is possible for HNL mass ranges from a few GeV up to the Z boson mass.

\subsubsection*{FCC-ee study (ee$\nu$ decay channel)}
This section documents one study performed as part of the FCC feasibility study \cite{FCC-FSR-Vol1}, which focused on a signal benchmark of a single HNL species of mass $m_{\text{N}}$ with a non-zero mixing with electron neutrinos $U_{\text{eN}}$ and all other mixings assumed negligible.
The signal is $\Pe\Pe\PGn$, and the selection included events with exactly two electrons, a veto on muons, photons, and jets, and missing momentum $p>10$\,GeV. 
The FCC feasibility study produced two central MC campaigns for physics and performance studies. For HNL studies, background samples were simulated using \pythia\  v8.303~\cite{Bierlich:2022pfr}.  For the signal HNL samples, the \mgfive\ v3.2.0 event generator~\cite{Alwall:2014hca} is used to simulate at leading order parton-level $\epem$ collisions, with \pythia\ used to simulate the parton showering and hadronisation. For the signal and background processes, the detector response was simulated using \delphes\ v3.4.2 ~\cite{deFavereau:2013fsa} with a version of the Innovative Detector for Electron-positron Accelerators (IDEA) FCC-ee detector card~\cite{FCC:2018evy}. 

This work uses an updated set of MC samples for the background modelling. The main changes relative to the first central MC sample production include an updated version of the \delphes\ detector card, with a smaller beam-pipe radius (R=1.0 cm compared to previously used $R$ = 1.5 cm) and improved resolution in the electromagnetic calorimeter (ECAL) through a crystal calorimeter. The machine parameters (beam energy spread and luminous region) are also updated to reflect developments in accelerator studies. For the leptonic $\PZ$-boson decays the statistics are increased by a factor of 10, whereas the increase is a factor of 5 and 4.39 respectively for $\PZ\rightarrow \PQc\PAQc$ and $\PZ\rightarrow \PQb\PAQb$. A standalone $\PZ\rightarrow \PQs\PAQs$ sample is also available.

Jets are reconstructed using \texttt{Fastjet}~\cite{Cacciari:2011ma} with the anti-\kT algorithm with a cone size of 0.4, and the minimum \pT of constituents set to be 1 GeV. In  previous studies, the default reconstructed objects returned by \delphes\ (electrons, muons, jets) were used. For experiments that use independent algorithms to reconstruct different physics objects double counting can occur due to the signals from a single particle being reconstructed as more than one different objects. To remove such ambiguities ``overlap'' removal procedures are typically applied that discard objects based on a pre-defined hierarchy for reconstructed objects. In this study, overlap removal is emulated by re-running the jet clustering algorithms with the reconstructed electrons and muons removed from the list of the reconstructed particles used for clustering. Secondary vertices are reconstructed using \texttt{LCFIPlus}~\cite{Suehara:2015ura} framework. The chosen configuration for secondary vertex (SV) reconstruction considered tracks with $\pT>1$ GeV and transverse impact parameter $|d_{0}|>2$ mm. This avoids reconstruction of secondary vertices from prompt particle decays and has a sufficiently high requirement on the track $d_{0}$ such that, when combined with vetoes on muons, photons and jets, the topology can be assumed to be background-free. After reconstruction, events are stored in the \edmhep\ event data format~\cite{Gaede:2869508} which is a common data format used for simulation in future colliders.
\begin{figure}[h!]
    \centering
        \includegraphics[width=0.7\textwidth]{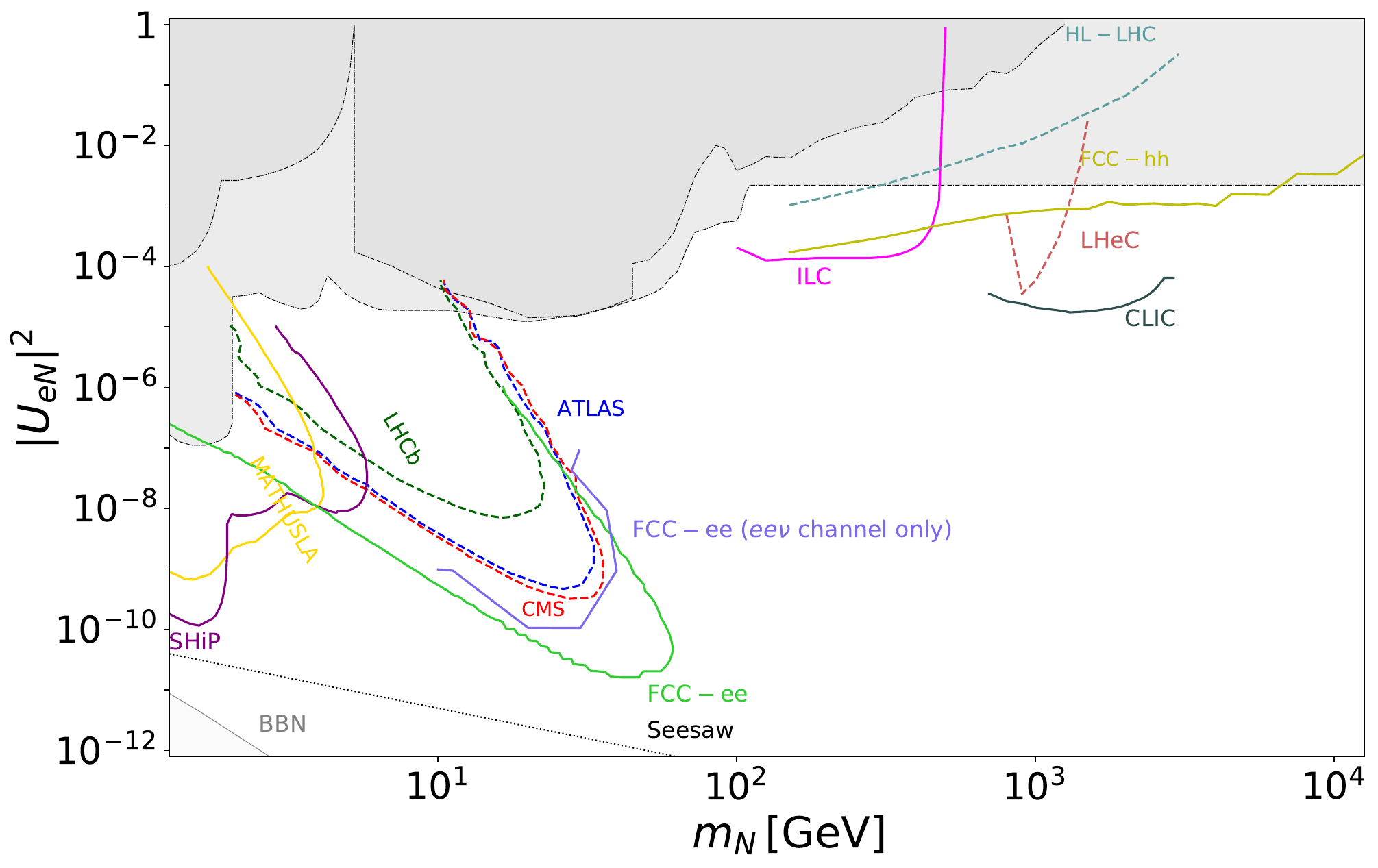}
    \caption{Comparison of the 3-event contour (solid purple line) for this study, compared to a selection of collider (and other) constraints compiled through the work of Ref.~\cite{Bolton:2019pcu},  as a function of the HNL mass and the electron-neutrino coupling.  For this study, signal points are normalised to $6\times 10^{12}$ \PZ bosons. } 
    \label{fig:contourHNL}    
\end{figure}
\FloatBarrier

Following studies of kinematic distributions for the signal and background after  an event selection requiring exactly two electrons, no electrons or muons and exactly one reconstructed secondary vertex, it was observed that a veto on hadronic jets would remove the remaining backgrounds. 
\Cref{fig:contourHNL} shows the 3-event contour for this initial selection, compared to a selection of collider (and other) constraints compiled through the work of Ref.~\cite{Bolton:2019pcu}. The FCC-ee line presented in this plot refers to the results of Ref.~\cite{Blondel:2014bra} which argued that it should be possible to perform background-free searches for HNLs in the 20--90 GeV mass range with decay lengths in the 10--100 cm range. The FCC-ee projections presented here should be seen as conservative for a number of reasons. Firstly, only leptonic HNL decays are considered; including the semi-leptonic decays would enable a more direct comparison with previous limits. Secondly, the limited MC statistics mean that the event selection criteria are not heavily optimised, which has to be done in a future study. Despite this, there are higher-mass and larger-coupling  regions where the constraints from this study extend beyond previous FCC-ee results, which motivates a more detailed study combining all HNL decay channels. This result further motivates additional studies using full simulation to further investigate the detector performance requirements for these benchmarks.

\subsubsection{Axion-like particles}
Axions were proposed in the 1980s to address the strong CP problem. More broadly, axion-like particles (ALPs)~\cite{Bauer:2018uxu} emerge in any theory with a spontaneously broken global symmetry, with a variety of possible masses and couplings to SM particles. At large symmetry-breaking scales, ALPs can signal a new physics sector at a scale ($\Lambda$) that would otherwise be experimentally unreachable. The leading couplings of ALPs to SM particles scale as ($\Lambda^{-1}$), making them weakly coupled at large new-physics scales. 
ALPs naturally implement spontaneous electroweak baryogenesis through a cosmic evolution that provides CPT violation~\cite{Im:2021xoy}. In this scenario, the ALP feebly couples to the Higgs field and gives a small contribution to the Higgs mass.

The search for ALPs in high-luminosity electron--positron colliders is one of the most interesting avenues for the discovery of physics beyond the SM. Probing the smallest possible couplings, and therefore longer lifetimes, is essential to uncovering information about this possible new physics sector.
A recent study explores the options for ALPs coupling to photons at FCC-ee and ILC~\cite{RebelloTeles:2023uig}. The study focuses on photon-fusion production of ALPs decaying into two photons over the light-by-light continuum background, for the design FCC-ee and ILC centre-of-mass energies and integrated luminosities. The study makes an analysis of the feasibility of measurements using parametrised simulations for two types of detectors. Upper limits at 95\% confidence level (CL) on the cross section for ALP production, $\sigma(\PGg \PGg \rightarrow \PXXa \rightarrow \PGg\PGg)$, and on the ALP-photon coupling are obtained over the $m_{\PXXa} \approx 0.1$--1000~GeV ALP mass range, and compared to current and future collider searches. Production cross sections down to $\sigma(\PGg \PGg \rightarrow \PXXa \rightarrow \PGg\PGg) \approx 1$~fb~(1~ab) will be probed at $m_{\PXXa}\approx 1$~(300)~GeV, corresponding to constraints on the axion-photon coupling as low as $g_{\PXXa\PGg\PGg} \approx 2\times 10^{-3}$~TeV$^{-1}$.
\begin{figure}[!htbp]
\centering
\includegraphics[scale=0.40]
{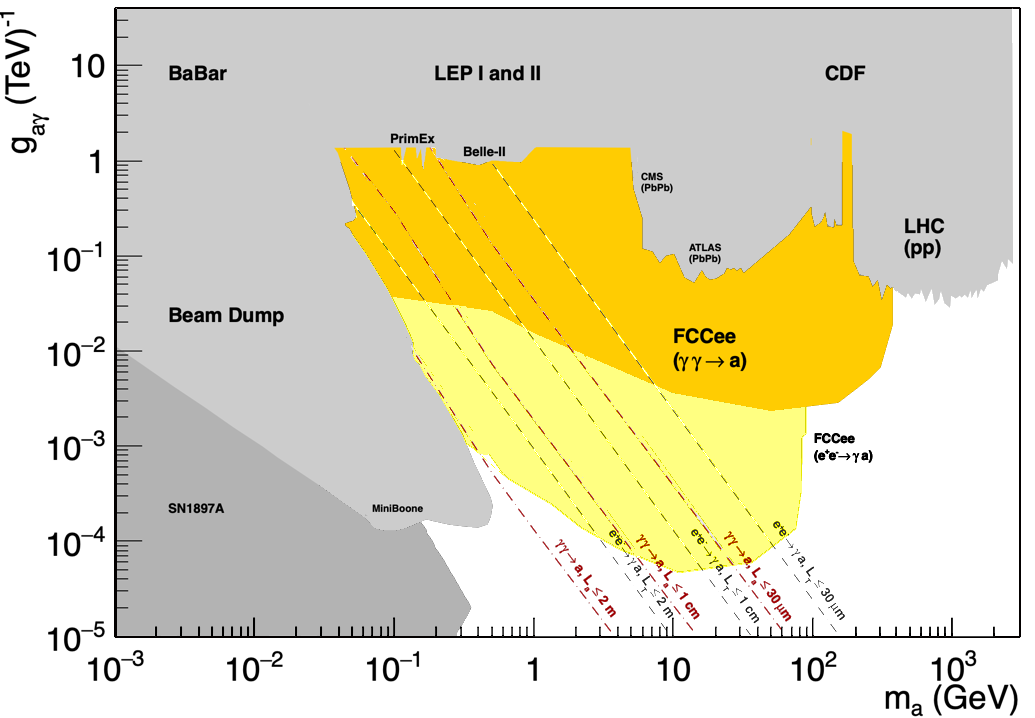}
\caption{Expected exclusion limits at 95\% CL on the ALP-photon coupling as a function of the ALP mass in $\epem$ collisions at FCC-ee for the combined $\PGg \PGg \to \PXXa$  (orange) and the $\epem\to\PXXa\PGg$ (yellow) processes, compared to current bounds (grey areas). Three reference average ALP transverse decay lengths are indicated with dashed diagonal lines for both ALP production processes. From Ref.~\cite{RebelloTeles:2023uig}.
\label{fig:FCCCombined}}
\end{figure}
\FloatBarrier

Figure~\ref{fig:FCCCombined} ~\cite{RebelloTeles:2023uig} compares the $\epem\to \PXXa\PGg$ (beige) and $\PGg \PGg \to \PXXa$ (orange)  expected limits at FCC-ee, with their three corresponding average ALP transverse decay length ranges, indicated with dashed diagonal lines. In the region of limits reachable at FCC-ee, one can see that most of the ALPs from the $\PGg\PGg$ production mode will have an average decay length below 1~cm and, thus, will be indistinguishable from the primary vertex, whereas a significant fraction of those coming from the $\epem \to \PZ \to \PGg \PXXa$ decay for the lowest $g_{\PXXa\PGg\PGg}$ couplings will feature secondary vertices within $\langle L_T \rangle \gtrsim 1.0$~cm~(1.0~m) for $m_{\PXXa}\lesssim 10$ (1.0)~GeV.

Finally, Fig.~\ref{fig:ILCCombined} compares the ALPs limits in the $(m_{\PXXa,g_{\PXXa\PGg\PGg}})$ plane expected via $\PGg\PGg\to\PXXa$ from all runs combined at the ILC (beige) and at FCC-ee (orange). Three reference average ALP transverse decay lengths $\langle L_{T} \rangle \approx 30~\mu$m,~1~cm,~2~m, are indicated with dashed diagonal lines. For most of the phase space covered, the ALPs will appear as coming from the primary vertex.

\begin{figure}[!htbp]
\centering
\includegraphics[scale=0.40]
{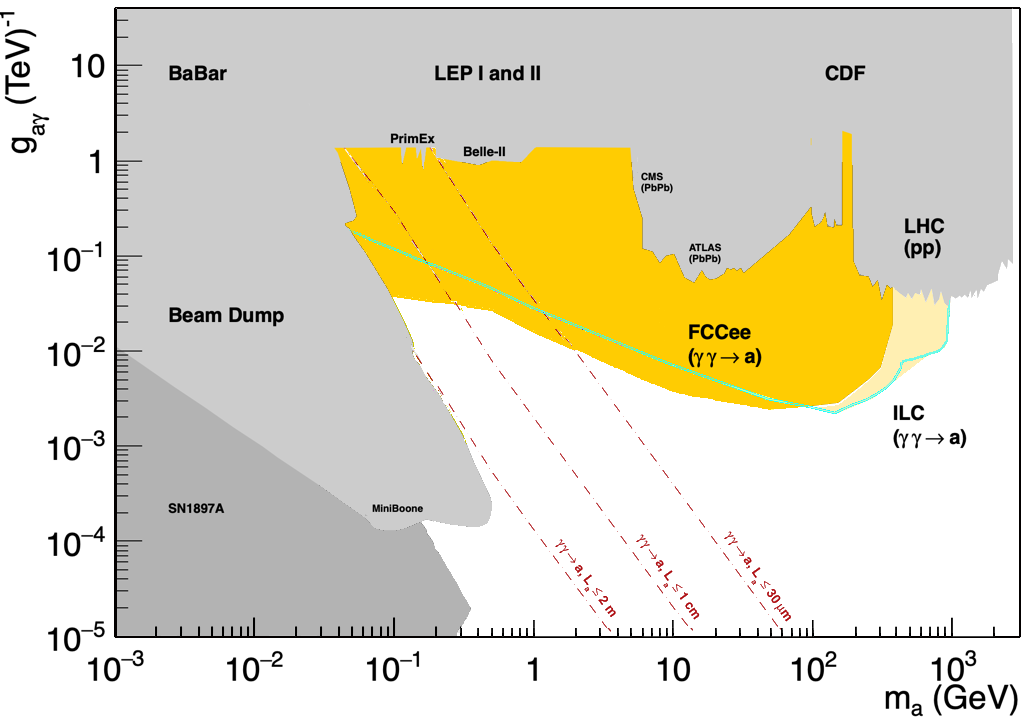}
\caption{Expected exclusion limits at 95\% CL on the ALP-photon coupling as a function of the ALP mass in all $\epem$ runs combined at ILC (beige) and at FCC-ee (orange) from the $\PGg \PGg \to a$ process, compared to current bounds (grey areas). Three reference average ALP transverse decay lengths, corresponding to $\langle L_{T} \rangle \approx 30~\mu$m,~1~cm,~2~m are indicated with dashed diagonal lines. From Ref.~\cite{RebelloTeles:2023uig}.
\label{fig:ILCCombined}}
\end{figure}
\FloatBarrier

\subsubsection{Exotic Higgs boson decays}~\label{exohiggsllp}
A third scenario to consider in future Higgs factories is exotic Higgs boson decays to LLPs~\cite{Alipour-Fard:2018lsf,Curtin:2013fra,Curtin:2017bxr}.
Such signals were first explored in the context of Hidden Valley models~\cite{Kucharczyk:2625054}, and subsequently found to arise in a variety of well-motivated scenarios, including ones involving the electroweak hierarchy problem, models of baryogenesis, and models of neutral naturalness.

\subsubsection*{FCC-ee study}
Previous studies suggest that the projected sensitivity of exotic Higgs decays to LLPs at FCC-ee is broadly competitive with that at the LHC, and potentially superior at lower LLP masses~\cite{Alipour-Fard:2018lsf}. The first such analysis for exotic Higgs boson decays at FCC-ee employng official FCC tools and simulations was performed recently~\cite{ripellino2024searchinglongliveddarkscalars}. 
The study targets a minimal scalar extension of the SM, where a new scalar field S interacts with the Higgs field doublet H via the portal term $\frac{1}{2}\kappa \text{S}^2 |\PH|^2$, where $\kappa$ is the coupling constant~\cite{Cepeda:2021rql}.
The new dark scalar S inherits the coupling structure of the SM Higgs boson, with the size of the couplings reduced by the sine of a common small mixing angle $\theta$ and the decay widths being those of a SM Higgs boson with the mass of the dark scalar, re-scaled by $\sin^2{\theta}$. At leading order, the partial decay width of the dark scalar into fermions is given by~\cite{Curtin:2013fra}
\begin{equation} \label{eq:scalarWidth}
    \Gamma(\text{S} \to \text{f} \bar{\text{f}}) \approx \sin^2 \theta \frac{N_c}{8 \pi} \frac{\ensuremath{m_\text{S}} m_\text{f}^2}{v_h^2} \Bigg( 1 - \frac{4 m_\text{f}^2}{\ensuremath{m_\text{S}}^2} \Bigg)^{3/2},
\end{equation}
where $m_\text{f}$ is the mass of the fermion and $N_c$ is the number of colours. For dark scalar masses above \SI{10}{\giga\electronvolt}, $\HepParticle{\PQb}{}{}\HepAntiParticle{\PQb}{}{}\Xspace$ is the dominant decay channel, with a branching ratio of $0.9$~\cite{Curtin:2013fra}. The lifetime of the new scalar is controlled by the mixing angle, with values of $\sin\theta \lesssim 10^{-5}$ resulting in lifetimes $c\tau \gtrsim O(\SI{1}{\milli\meter})$.

This study considers Higgs boson production through Higgsstrahlung in $\epem$ collisions at the ZH threshold at the FCC-ee, corresponding to a centre-of-mass energy of $\sqrt{s} = \SI{240}{\giga\electronvolt}$. The long-lived scalars are produced in pairs through decays of the Higgs boson and then subsequently decay exclusively into $\HepParticle{\PQb}{}{}\HepAntiParticle{\PQb}{}{}\Xspace$. 
The search strategy targets long-lived scalar decays within the inner-detector volume of the proposed detector structures for FCC-ee, resulting in the presence of at least two DVs that may be reconstructed from inner-detector tracks. The associated Higgs boson production topology is also probed, exploiting the clean signature from $\PZ\to \Plp\Plm$ (\Pl = \Pe,\PGm) decays. Events selected for the analysis are thus required to contain at least two DVs with high mass and track multiplicity relative to SM processes, as well as an $\epem$ or $\PGmp \PGmm$ pair with an invariant mass consistent with a $\PZ$ boson, effectively suppressing backgrounds.

The signal process is simulated for $\epem$ collisions at \SI{240}{\giga\electronvolt} in the Hidden Abelian Higgs Model (HAHM) using the HAHM\_MG5Model\_v3 model in \mgfive\ v.3.4.1 \cite{Stelzer:1994ta,Alwall:2014hca}. The HAHM model contains a dark photon in addition to the dark scalar, which for the purpose of this study is effectively decoupled by setting its mixing with the SM to zero. Two free parameters then remain in the model: the mass of the dark scalar, $m_\mathrm{S}$, and the coupling strength between the dark scalar and the Higgs boson, $\kappa$. Parton-level events are passed to \pythia\ v.8310 \cite{PYTHIA8} for parton showering and hadronisation. The detector response is simulated in \delphes \cite{deFavereau:2013fsa}, using the latest Innovative Detector for Electron-positron Accelerators (IDEA) FCC-ee detector concept~\cite{Antonello:2020tzq} card. Signal points are generated with masses between $m_\mathrm{S}=\SI{20}{\giga\electronvolt}$ and $m_\mathrm{S}=\SI{60}{\giga\electronvolt}$ and coupling values of $\sin\theta$ between $10^{-5}$ and $10^{-7}$.
The simulated background samples are centrally produced and correspond to the Winter2023 campaign. All backgrounds from ZH with H$\to$SM, as well as ZZ and WW processes at \SI{240}{\giga\electronvolt} are considered. All simulated samples are normalised to the expected luminosity of the ZH stage, corresponding to \SI{10.8}{\per\atto\barn}. 

This study is based on tracks and isolated electrons and muons reconstructed in \delphes. In addition, the analysis relies heavily on the secondary vertex finder of the analysis framework, which is based on the LCFIPlus framework \cite{Suehara:2015ura}. Within this algorithm, a custom track selection is developed to efficiently reconstruct the larger displacements of the long-lived scalar decays. The tracks are required to have a transverse momentum $p_\text{T} > \SI{1}{\giga\electronvolt}$ and to have a transverse impact parameter $|d_0| > \SI{2}{\milli\meter}$. 
Events selected for the analysis are required to contain exactly one $\epem$ or $\PGmp\PGmm$ pair, with an invariant mass in the range $70 < m_{ll} < 110~{\rm GeV}$. Additionally, events must have at least two reconstructed DVs, where each DV is required to have at least three associated tracks, a radius $r_{DV}$ between \SI{4}{\milli\meter} and \SI{2}{\meter}, and an invariant mass reconstructed from its tracks greater than \SI{2}{\giga\electronvolt}.
\begin{figure}[h!]
    \centering
        \includegraphics[width=0.5\textwidth]{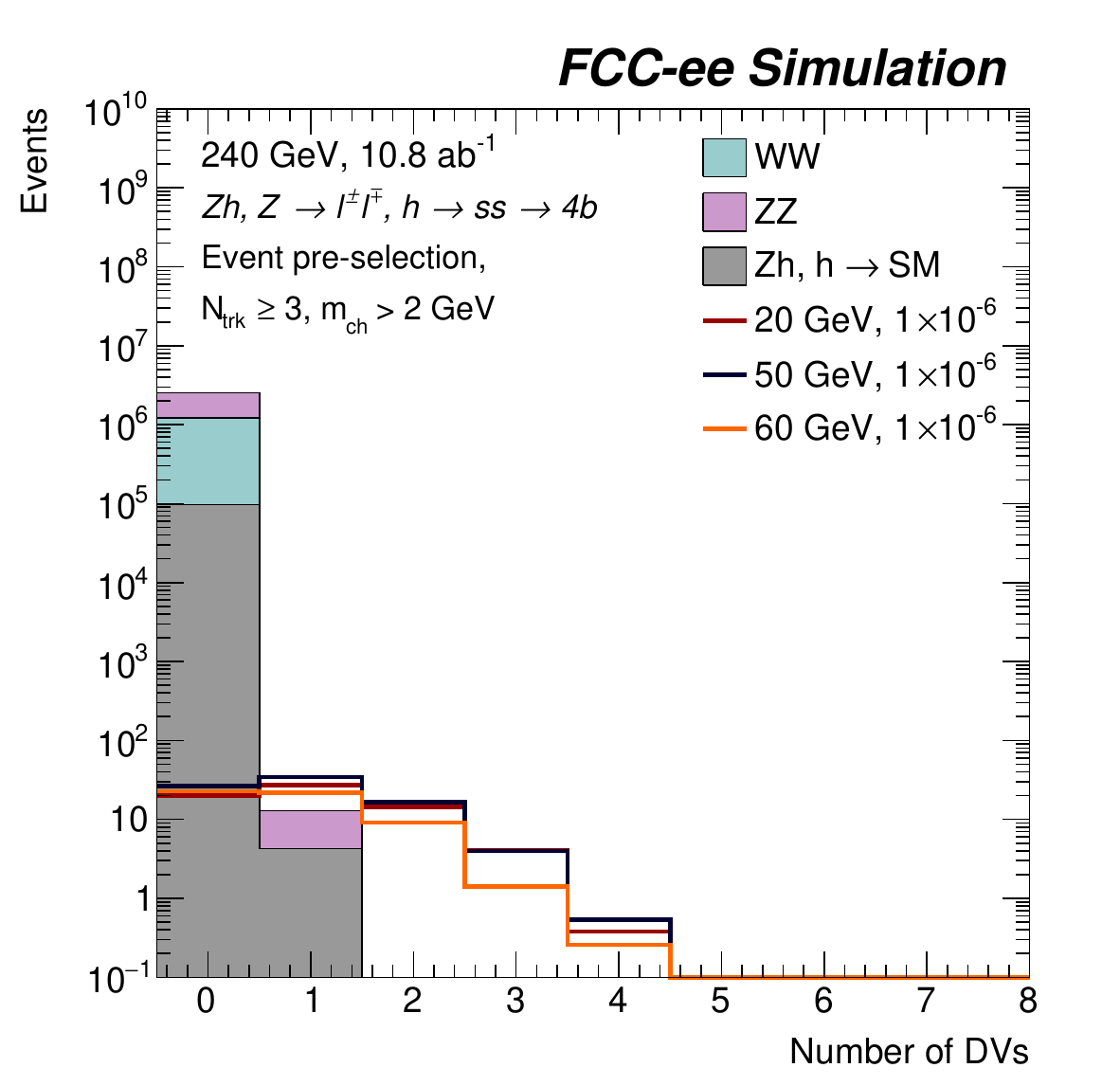}
    \caption{Number of DVs passing the full DV-level selection in events with exactly one $e^+ e^-$ or $\mu^+ \mu^-$ pair, with an invariant mass in the range $70 < m_{ll} < 110~{\rm GeV}$.}
    \label{fig:nDVs}
\end{figure}
%
\Cref{fig:nDVs} shows the number of DVs reconstructed in background and in three example signal points. Requiring each event to contain at least two DVs efficiently removes all background. Sensitivity can then be reached for all signal model points with at least three events surviving the full selection. This is demonstrated  in figure~\Cref{fig:limits} for all signal samples for the parameter space spanned by the dark scalar mass and lifetime $c\tau$. 
A shaded region is drawn around signal points with at least three events to indicate an approximate region of sensitivity to the signal. The results show that the search provides sensitivity to dark scalar masses between $m_\mathrm{s}=20$ GeV and $m_\mathrm{s}=60$ GeV and mean proper lifetimes $c\tau$ between approximately 10\,mm and 10\,m for the simulated Higgs to dark scalar branching ratios around 0.1\%.
%
The results can be rescaled to lower Higgs to dark scalar branching ratios. Such an exercise suggests that the search strategy has the potential to probe Higgs to dark scalar branching ratios as low as $10^{−4}$ for a mean proper lifetime $c\tau\approx 1$\,m.

\begin{figure}[h!]
    \centering
        \includegraphics[width=0.7\textwidth]{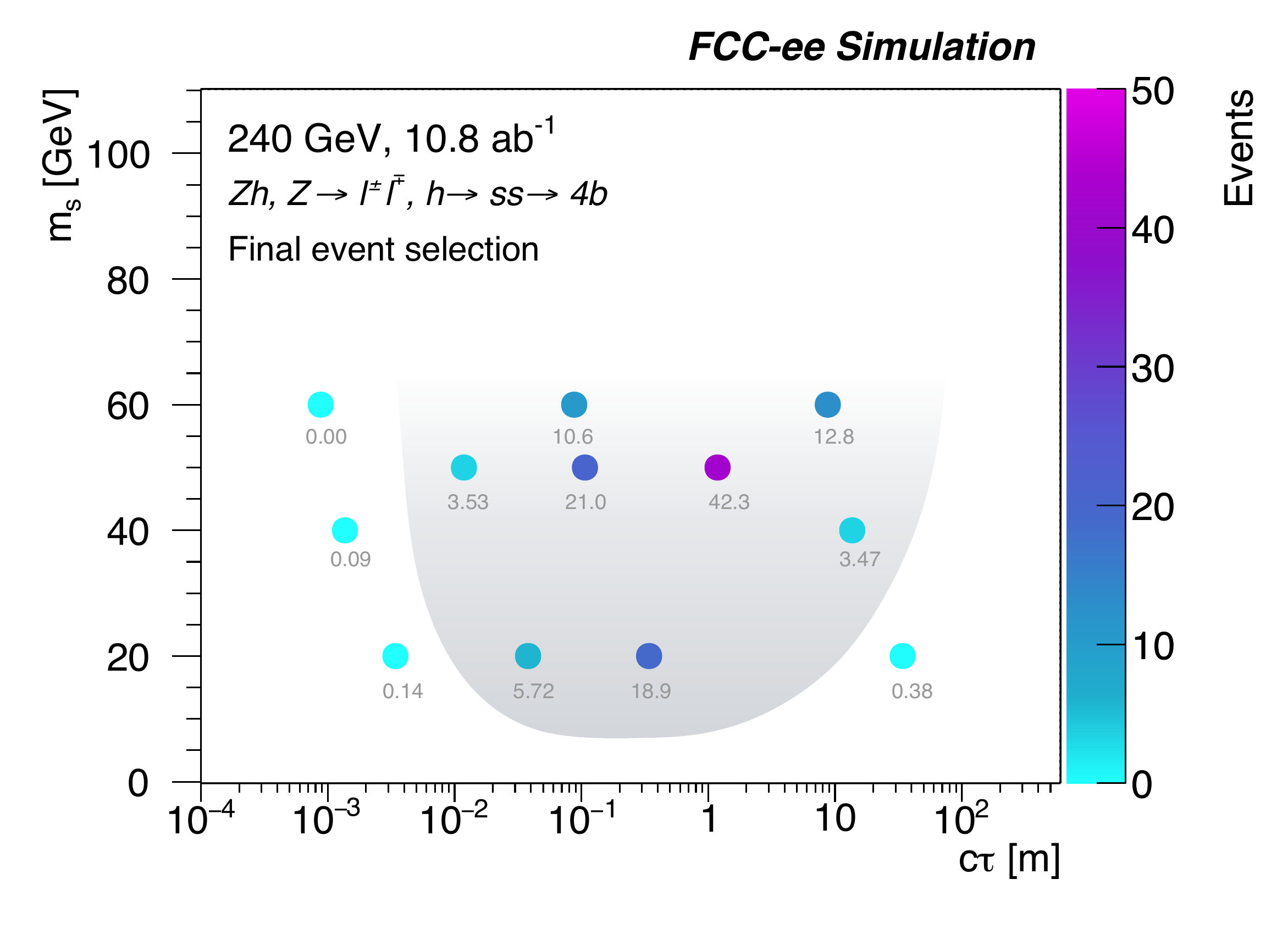}
    \caption{Number of signal events surviving the analysis selection as a function of dark scalar mass $m_\mathrm{s}$ and lifetime $c\tau$. The shaded region is drawn around signal points with at least three events.} 
    \label{fig:limits}
\end{figure}
%
%
Separately, analysis strategies incorporating hadronic and invisible $\PZ$ decays have also been presented, extending upon the leptonic $\PZ$ decays exclusively considered in the original analysis. 
For \textit{invisible \PZ decays}, a missing energy selection is applied to identify the presence of neutrinos from the \PZ boson decay. Preliminary results indicate that adding invisible \PZ decays significantly increases the signal yield, though further refinement is needed to improve background rejection.
For \textit{hadronic \PZ decays}, the total jet energy is used to tag hadronic \PZ boson decays. This mode offers a substantial increase in signal yield due to the larger branching ratio but poses challenges in distinguishing signal from background because of the high jet activity in the event. Future refinements, such as more stringent jet energy and vertex-related requirements, may improve background rejection.
\subsubsection*{ILC study}
This search is similarly possible at linear colliders like the ILC. A set of benchmarks with SM-like Higgs boson decays to long-lived scalars is considered in a recent full simulation analysis. The Higgs production channel used is $\epem\to\PSh\PZ$ at ILC250, with $\PZ\to\PGn\PAGn$ and $\PSh\to$ SS, where S is long-lived. Four benchmarks are chosen: two low-mass ($m_\mathrm{S} = \qty{0.4}{\giga\electronvolt} \text{ and } \qty{2}{\giga\electronvolt}$, with $c\tau=\SI{10}{\milli\metre}$) and two high-mass ($m_\mathrm{S} = \qty{50}{\giga\electronvolt} \text{ and } \qty{60}{\giga\electronvolt}$, with $c\tau=\SI{1}{\metre}$) scenarios.
Displaced vertex reconstruction follows the approach presented in Ref.~\cite{Klamka:2024gvd}. 

Because the $\PZ$ boson decays to neutrinos, and one of the LLPs can escape the detector acceptance, the expected signature is just at least one displaced vertex. Therefore, in addition to the selection described in Ref.~\cite{Klamka:2024gvd}, events containing prompt tracks with $\pT>\qty{2}{\giga\electronvolt}$ are rejected. Now, also the $\pT^{vtx}>\qty{10}{\giga\electronvolt}$ requirement was applied to allow us to fully neglect low-$\pT$ beam-induced backgrounds.

The expected number of background events is approximately 0.64 after the standard and the additional selection.
The corresponding 95\% C.L. expected limits on the signal production cross section (left) and the branching ratio BR$(\PSh\to\text{SS})=\sigma_{\text{95\%\,C.L.}}/\sigma_{h\nu\nu}$ (right) are presented in \cref{fig:klamka_higgs_llp_limits}, conservatively assuming that the observed number of events is one. 
ILD could improve the current limits~\cite{CMS:2024zqs} in the most extreme scenarios by an order of magnitude, or investigate longer lifetimes. Smaller decay lengths could be tested by a dedicated search using a vertex detector, while more data collected at higher energy stages of ILC could further enhance the reach.

\begin{figure}[bt]
	    \centering
	 	 \begin{subfigure}{0.49\textwidth}
	 	 	\centering
	 	 	\includegraphics[width=\textwidth]{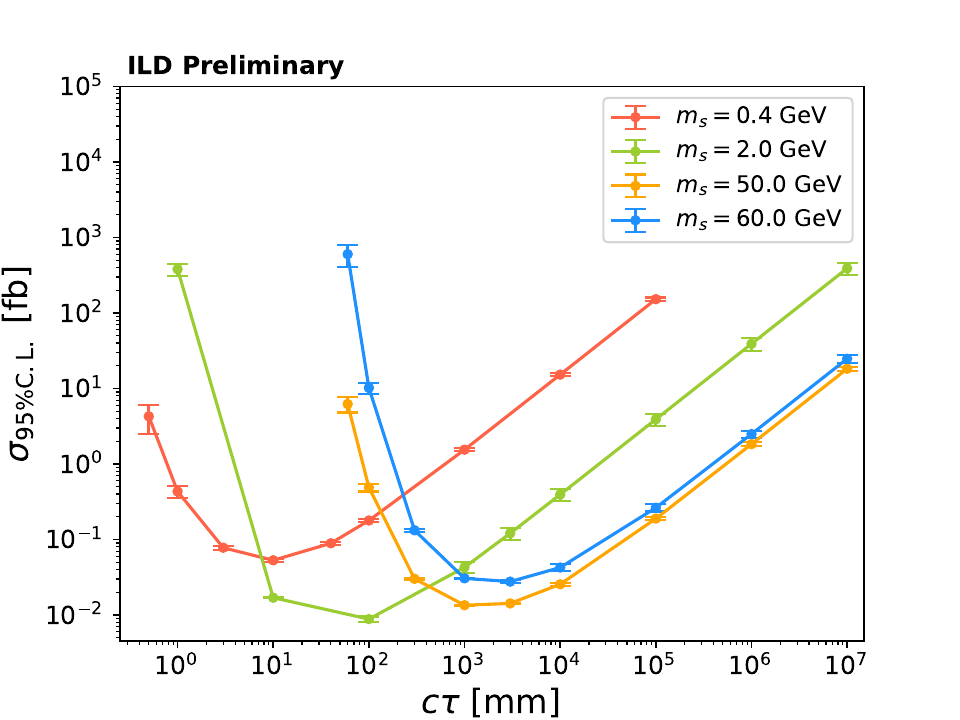}
	 	 \end{subfigure}%
	 	 \begin{subfigure}{0.49\textwidth}
	 	 	\centering
	 	 	\includegraphics[width=\textwidth]{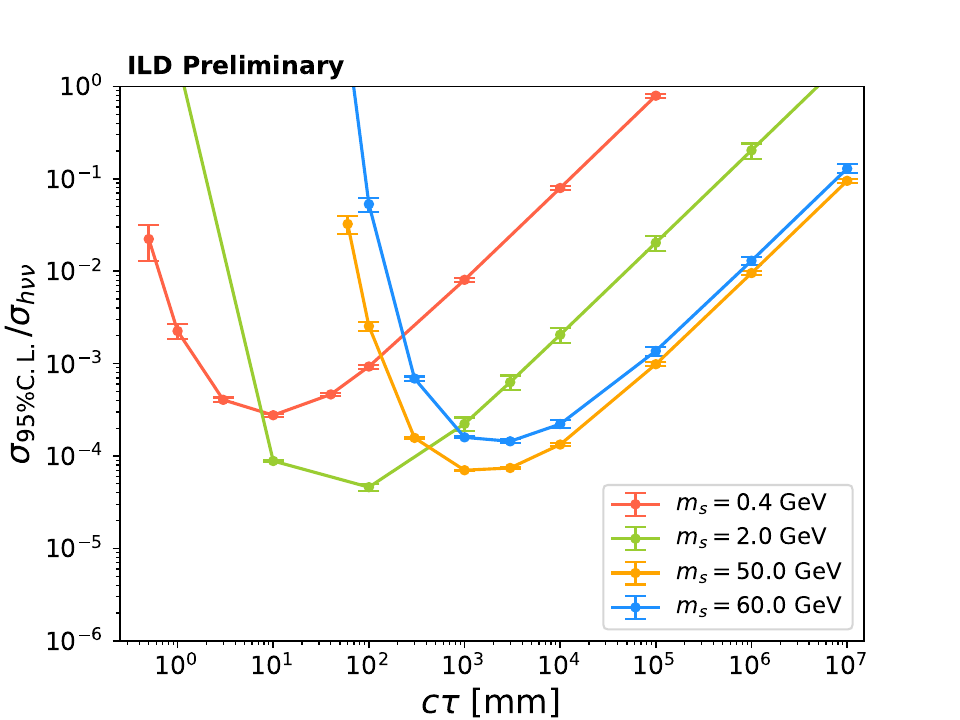}
	 	 \end{subfigure}
	 	 \caption{Expected 95\% C.L. upper limits on the signal production cross-section (left) and the branching ratio (right) for the considered benchmarks and different LLP mean decay lengths, for Higgs decays to long-lived scalars at $\sqrt{s}=\SI{250}{\giga\electronvolt}$. Only statistical uncertainties are included. 
         }
	 	 \label{fig:klamka_higgs_llp_limits}
\end{figure}
\FloatBarrier

\subsubsection{Model-independent searches for LLPs \label{sec:LLP-SRCH}}

Independently of the BSM model considered, LLPs searches are essentially signature-driven. That is, if future Higgs factories have methods in place to reconstruct e.g.\ displaced vertices, and a signal is observed, it would then have to be interpreted in the context of possible BSM models that can give rise to them. In particular, in the case of a discovery, an important question is what would be needed in order to determine the correct underlying model and its parameters.

A recent study~\cite{Klamka:2024gvd} presents prospects for the detection of neutral LLPs with the ILD detector at ILC operating at $\sqrt{s}=250$ GeV (ILC250) as a reference collider. 
A generic case of LLPs decaying to any two (or more) charged particles forming the displaced vertex is considered.
%
The benchmark scenarios considered in this study are not selected based on existing constraints on BSM models, but on challenging signatures from an experimental perspective. Therefore, two opposite classes of benchmarks were chosen. The first case considered is pair-production of heavy neutral scalars, $\PSA$ and $\PH$, where the former is the LLP and the latter is stable and escapes the detector undetected. The LLP decay channel is $\PSA$ $\to$ $\PZ^{*}\PH$, and its mass and proper decay length were fixed to $m_{\PSA} = \SI{75}{\giga\electronvolt}$ and $c\tau=\SI{1}{\metre}$. Four mass splitting values between $\PSA$ and $\PH$ were considered: $m_{\PSA} - m_{\PH} = 1,2,3,5\,\si{\giga\electronvolt}$. 
The second class features the production of a very light and highly boosted LLP with strongly collimated final-state tracks. It is generated using the associated production of a pseudoscalar LLP, a, with a hard photon. LLPs of four different masses are considered, $m_{\PXXa} = 0.3,1,3,10\,\si{\giga\electronvolt}$, with $c\tau= m_{\PXXa}\cdot \SI{10}{\milli\metre/\giga\electronvolt}$. 

The analysis relies on a vertex-finding algorithm designed for this study and is performed in a model-independent way, considering only the displaced vertex signature in the TPC and ignoring any other activity in the detector. The study was carried out using the full detector response simulation.
As no requirements related to charged particle identity are imposed, decays of $\PZ^{*}$ and a to muons are selected for simulation to minimize processing time.
Two types of background are taken into account: soft, beam-induced (low-$\pT$) processes; and hard (high-$\pT$) processes. The beam-induced processes occurring in each bunch-crossing constitute a significant standalone background if one wants to consider soft signals.
To reject fake vertices, a set of quality cuts is applied on the variables describing kinematic properties of tracks. The main background sources that remain include V$^0$ particles (long-lived neutral hadron decays and photon conversions) and secondary particle interactions with the detector material. In addition to rejection based on a dedicated ILD software for V$^0$ identification, selections corresponding to masses of different V$^0$s are applied. Further selection on the sum of $\pT^{vtx}$ of tracks forming the vertex, and on variables describing track-pair geometry, provide an overall reduction factor of $1.26\times 10^{-10}$ for beam-induced backgrounds.  Two- and four-fermion production with hadronic jets in the final state was considered as the high-$\pT$ background. To improve the high-$\pT$ background rejection, in addition to the \textit{standard} selection described above, we also consider the \textit{ tight} selection, where track pairs with an invariant mass below $\SI{700}{\mega\electronvolt}$ are rejected. An isolation criterion is also included in the tight selection.
Vertex finding rates for the signal and the background were used to obtain the expected 95\%\,C.L. upper limits on the signal production cross section, $\sigma_{95\% \mathrm{C.L.}}$, assuming an integrated luminosity of $\SI{2}{\per\atto\barn}$. 
An event re-weighting technique was employed to obtain the limits for a range of LLP lifetimes without generating and processing a large number of event samples. 

The results of the study are presented in \cref{fig:klamka_llp_limits} as a function of LLP proper decay length $c\tau$. 
The tight selection allows to strongly enhance the reach, but results in a loss of sensitivity for the $m_{\PXXa}=\SI{300}{\mega\electronvolt}$ scenario.
\begin{figure}[bt]
	    \centering
	 	 \begin{subfigure}{0.49\textwidth}
	 	 	\centering
	 	 	\includegraphics[width=\textwidth]{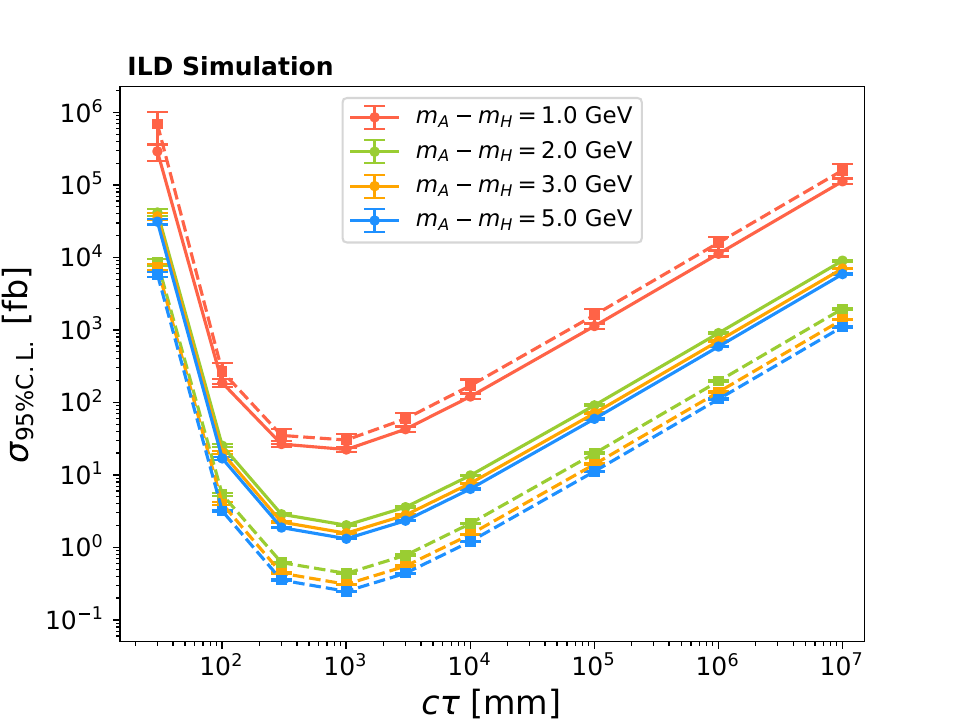}
	 	 \end{subfigure}%
	 	 \begin{subfigure}{0.49\textwidth}
	 	 	\centering
	 	 	\includegraphics[width=\textwidth]{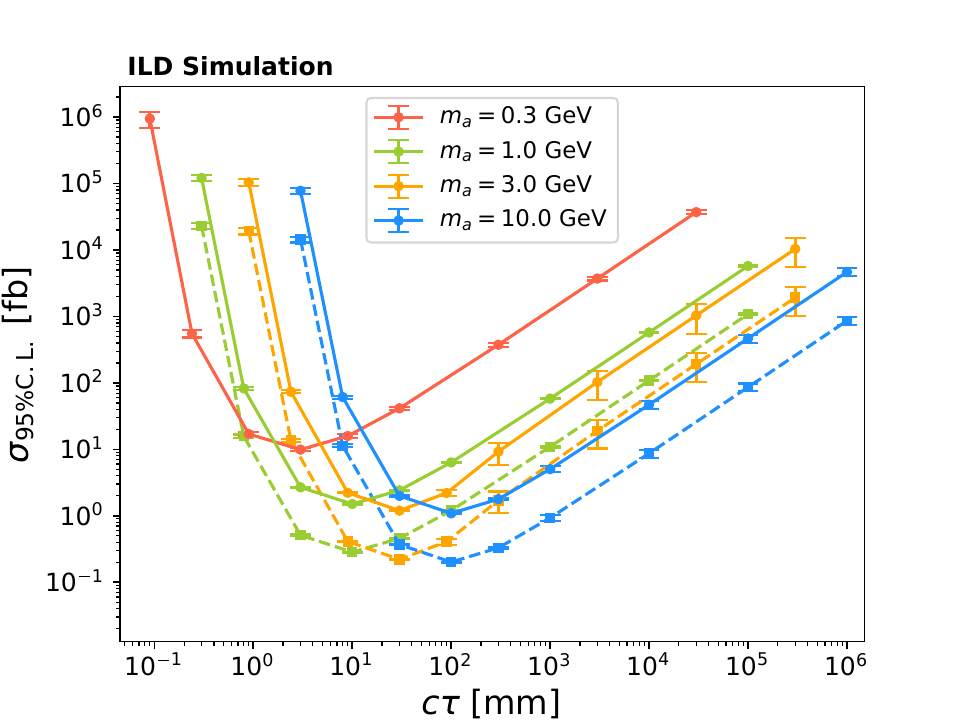}
	 	 \end{subfigure}
	 	 \caption{Expected 95\% C.L. upper limits on the signal production cross-section for the considered benchmarks and different LLP mean decay lengths, for the pair-production of neutral heavy scalars $\PSA$ and $\PH$ where the former is the LLP and the latter is stable and escapes the detector undetected(left); and the production of a light pseudoscalar with a hard photon (right) at $\sqrt{s}=\SI{250}{\giga\electronvolt}$. Solid lines correspond to the standard selection and dashed lines to the tight selection (see text for details). Only statistical uncertainties are included. From Ref.~\cite{Klamka:2024gvd}. }
	 	 \label{fig:klamka_llp_limits}
\end{figure}
\FloatBarrier
In addition, interesting opportunities for including projected LLP sensitivities of future colliders in general model-checking tools can be proposed, for instance the LLP-version of \textsc{ CheckMATE}~\cite{Desai:2021jsa,Dercks:2016npn}.

\subsubsection{Machine learning methods for LLP searches}~\label{ml}
Since LLPs exhibit unconventional experimental signatures, they provide a clear use case for machine learning/AI techniques. The potential of these methods in advancing the search for LLPs can be seen in e.g.\ a recent study for HNLs at FCC-ee. In traditional HNL searches, a cut-and-count approach is often employed to discriminate signal from background events. While effective, these methods are limited in terms of the complexity of the events they can analyze. In contrast, machine learning techniques, specifically Deep Neural Networks (DNNs), can handle multi-dimensional data more effectively, providing improved sensitivity. 

One unique feature of HNL production at the FCC-ee is the ability to distinguish between prompt and long-lived HNLs \cite{CMS2022}. HNLs with small mixing angles and low masses can exhibit extended decay lengths, creating a distinct long-lived signature that is rare in the context of most Standard Model particles. By training multivariate models utilizing also the transverse impact parameter and its significance, the analysis demonstrates the effective discrimination between prompt and long-lived HNLs across the studied parameter space.

|n this study a binary classification was used for  DNN models, where the algorithms are trained to distinguish HNL signals from background events. To achieve maximum sensitivity, training was conducted independently for each individual signal mass point. For DNN training, the following variables were used: $E$, $\phi$, $d_0$, $\sigma_{d_0}$, $\Delta R_{ejj}$ for the leading electron, $E_{\text{miss}}$, $\theta$ for the neutrino, $\Delta R_{jj}$, $\phi$ for the di-jet system, and $n_{\text{tracks}}$, $n_{\text{primary tracks}}$, $\chi^2_{\text{vertex}}$ for the vertex and the tracks.

The enhanced sensitivity of machine learning models is depicted in \cref{fig:llp:e_channel_limits}, which shows the Z-significance  as a function of the HNL mass and the mixing angle. The DNN method demonstrates considerably improved performance over the traditional cut-and-count approach across the entire mass range. In particular, the DNN model offers an improvement of approximately one order of magnitude in mixing angle reach. With more sophisticated architectures and better hyperparameter optimization, the DNN could significantly enhance our analysis sensitivity in future studies.  In the same plot, the DNN performance is shown after normalising to the full expected luminosity of FCC-ee of $L$ = 205 ab$^{-1}$.  However, the MC datasets are currently small and using MC datasets representative of the 
205~\abinv\ dataset will be essential to fully explore the analysis sensitivity and power of ML methods.
\begin{figure}[h!]
    \centering
        \includegraphics[width=0.7\textwidth]{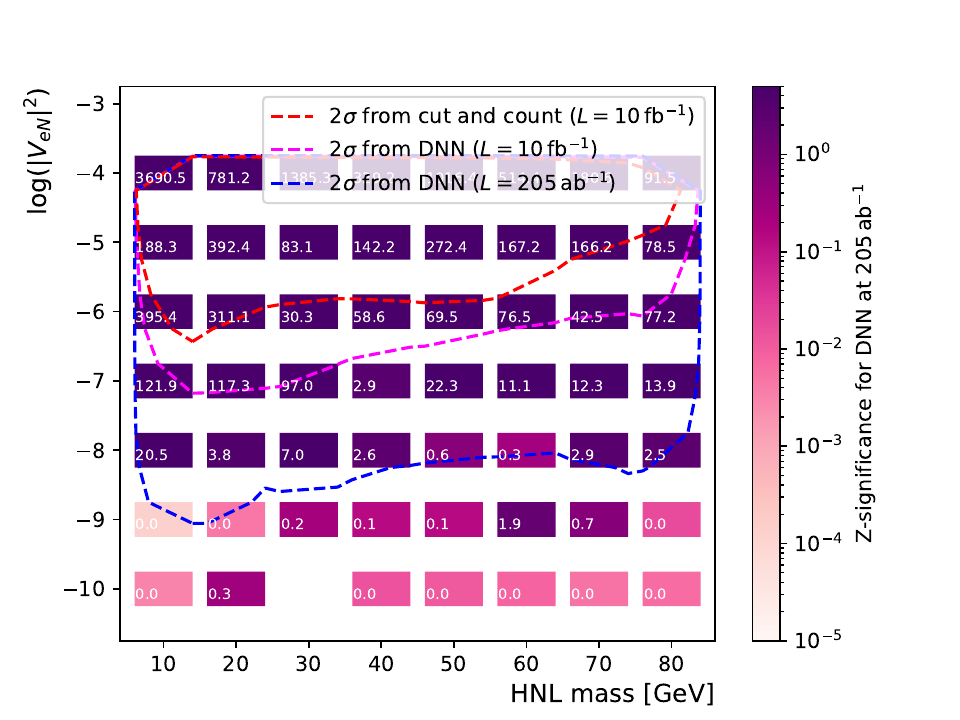}
    \caption{ Comparison of Z-significance obtained with machine learning and the traditional cut-and-count method in searches for HNL with masses ranging from 10 to 80 GeV, and mixing angles between $|U_{eN}|^2 = 10^{-4}$ and $10^{-10}$. The magenta line marks the 2$\sigma$ contour for the DNN analysis, while the red line represents the 2$\sigma$ contour from the cut-and-count study normalised to an integrated luminosity of $L$ = 10 fb$^{-1}$. The blue line marks the 2$\sigma$ contour for the DNN analysis normalised to an integrated luminosity of $L$ = 205 ab$^{-1}$. From Ref.~\cite{TomMsC}.} 
    \label{fig:llp:e_channel_limits}
\end{figure}
\FloatBarrier 
\subsubsection{Target detector performance aspects}
The LLP focus topic sets requirements on all subsystems of collider detectors, with an emphasis on tracking, timing, and calorimetry. In particular, gaseous main trackers  have advantages in specific energy loss measurements and  pattern recognition which might be difficult to compensate with all-silicon trackers. Similarly, the timing capabilities and the granularity of the chosen ECAL technology might have a significant impact on the ability to reconstruct displaced photons and out-of-time signatures. 
\subsubsection*{Calorimetric configurations for long-lived ALPs}
Searches for ALPs can inform detector design in future colliders. For example, light ALPs naturally couple to photons, and their photonic final states constitute an excellent benchmark for the photon performance of FCC detectors. It is possible to study the parameter space coverage of a simplified ALP model  for two  calorimetric configurations of the IDEA detector: monolithic Dual Readout (DR) fibre calorimeter, and Crystal DR electromagnetic  (EM) calorimeter.

This is a preliminary study based on available parametrisations of the performance of the two detector configurations that offers a first idea of  the impact of the different elements of the electromagnetic calorimeter performance on the analysis reach.
The study is based on the simplified model described in Ref.~\cite{Bauer:2018uxu}, and the couplings are set such that the ALP $a$ only couples to hypercharge and not to SU2, and it decays 100\% into two photons, yielding a three-photon final state from the process $\epem\rightarrow \PZ\rightarrow\PGg \PXXa$.
Under these assumptions the experimental reach can be represented in the  ($m_{\PXXa}$,$C_{\PGg\PGg}$) plane where $C_{\PGg\PGg}$ is the strength  of the $\PXXa\PGg\PGg$ vertex.
In different regions of this parameter space different experimental  signatures are accessible. For masses above $\sim10$~GeV, only ALPs with non-measurable flight path in the detector will be observable as 3 photons in the detector (``prompt'' signature). For masses below $\sim1$~GeV and low couplings, most of the ALPs decay outside
the detector, yielding a single detected photon and nothing else. In the intermediate mass region, it might be possible to experimentally measure the decay path of the ALP in LLP signatures.
The experimental  separation of the $3\PGg$ prompt signature and the LLP signature depends on how well one can detect an ALP decay away from the vertex. In the following,  a 3$\PGg$ analysis will be described making no assumptions on vertex detection. In addition, the reach of an analysis searching for a very long-lived ALP resulting in a single photon recoiling against missing energy (ME) from the undetected ALP will be assessed.

The signal samples were generated with the package \madgraphee \cite{Alwall:2014hca}, based on the UFO  \textsc{ALP\_NLO\_UFO}~\cite{Bauer:2018uxu}. The LHE files were hadronised with \pythiaeight\ \cite{Sjostrand:2014zea} and then fed into the \delphes \cite{deFavereau:2013fsa} fast simulation of the IDEA Detector \cite{Antonello:2020tzq}, based on the official datacards for the FCC-ee PED studies.
For the $3\PGg$  analysis, the dominant irreducible $\epem\rightarrow \PGg \PGg \PGg$ background was produced at LO with \madgraphee. For the single photon analysis, two background samples were likewise produced with \madgraphee:  $\epem\rightarrow \PGg\PGn\PGn $ and $\epem\rightarrow \PGg \epem$ . The background samples were passed through the same simulation chain as the signal.
The measured energy and impact point of the photons were simulated using a Gaussian smearing based on parametrisations for the resolution obtained with a detailed \geant 4 simulation of the response of the two calorimeter options under study, the main difference being a stochastic term of 11\% for the energy resolution of the EM fibre calorimeter, and of 3\% for the crystal EM calorimeter. It is found that the position resolution is the dominant contribution to ALP mass resolution up to $m_{\PXXa}\sim1$ GeV.

For the 3$\PGg$ analysis, the basic selection requires 3 photons within detector acceptance and energy larger than 0.1 GeV. Test masses $m_{\PXXa}$  between 0.1 and 85~GeV are addressed. For each $m_{\PXXa}$, the photon recoiling against the ALP is monochromatic, and its energy is determined by the recoil formula.
The three photons are labeled as produced in the ALP decay ($\PGg_1$,$\PGg_2$)  or recoiling against it ($\PGg_3$) by minimizing a compatibility function with the given mass hypothesis, which incorporates knowledge of the energy and mass resolution of the calorimeter system.
For a fixed mass, once the azimuthal symmetry is accounted for, the signal is fully defined by three variables. The following variables are chosen, as they provide a very good signal/background separation and are approximately independent from the value of $m_{\PXXa}$: the polar angle of the ALP in the lab system, the polar angle of $\PGg_1$  in the ALP rest system, and  the azimuthal angle of $\PGg_1$. In order to achieve optimal classification performance, for each value of  $m_{\PXXa}$ a BDT (XGB) is trained on 6 variables, the above three, $m(\PGg_1\PGg_2)$, $E_{\PGg_3}$, and the angular separation of $\PGg_1$ and $\PGg_2$ in the lab frame.

At low ALP masses ($<5$~GeV), the analysis is affected by two experimental issues: the fact that the ALP can decay far from the interaction point, producing a degradation in the mass measurement resolution; and the fact that the two photons from the ALP  decay are very near in the calorimeter, and may be observed as a single photon, and thus leading to a rejected event. Both these effects need to be studied with a detailed simulation of each calorimeter option. For the purpose of the present analysis, events where the ALP decays  at a distance larger than 5~cm from the interaction point are rejected, and events where the angular distance of the two ALP photons is  smaller than 0.01 (0.02) radians for the fibre (crystal) calorimeter option are rejected.

For each signal and background sample, the events after selection are normalised to  the expected integrated luminosity for the $\PZ$-pole run at FCC-ee , corresponding to  $6\times10^{12}$ $\PZ$ bosons. The cut on the XGB probability is applied by maximizing the significance defined according to the prescriptions of Ref. \cite{ATLASstat}, and the line corresponding to a significance of 2$\sigma$ is plotted in the ($m_{\PXXa}$,$C_{\PGg\PGg}$) plane.
In the  high mass range, the better energy resolution of the crystal option yields a significantly better reach. At low masses, the better granularity of the fibre calorimeter should allow a better separation of close-by photons.

For the 1$\PGg$+ME  analysis, $m_{\PXXa}<2$~GeV  is the relevant mass range and the signature is a monochromatic photon of energy $\sim45.5$~GeV with nothing else  in the detector. Two variables characterise the event, the energy and the polar angle of the photon. They  are combined through XGB, and the reach in the ($m_{\PXXa}$,$C_{\PGg\PGg}$) analysis is obtained with the same procedure as for the 3$\PGg$ analysis.
 
The reach of the two analyses for the crystal calorimeter option is shown in Figure~\ref{fig:ALP_results_total}, compared to existing ALP limits, and to an independent analysis addressing ALP production at FCC-ee via photon-photon fusion~\cite{RebelloTeles:2023uig}. In summary, the FCC-ee runs will allow the coverage of the ALP mass range 0.1--350~GeV down to couplings $C_{\PGg\PGg}$ of a few $10^{-3}$.
\begin{figure}
\centering
\includegraphics[width=0.7\textwidth]{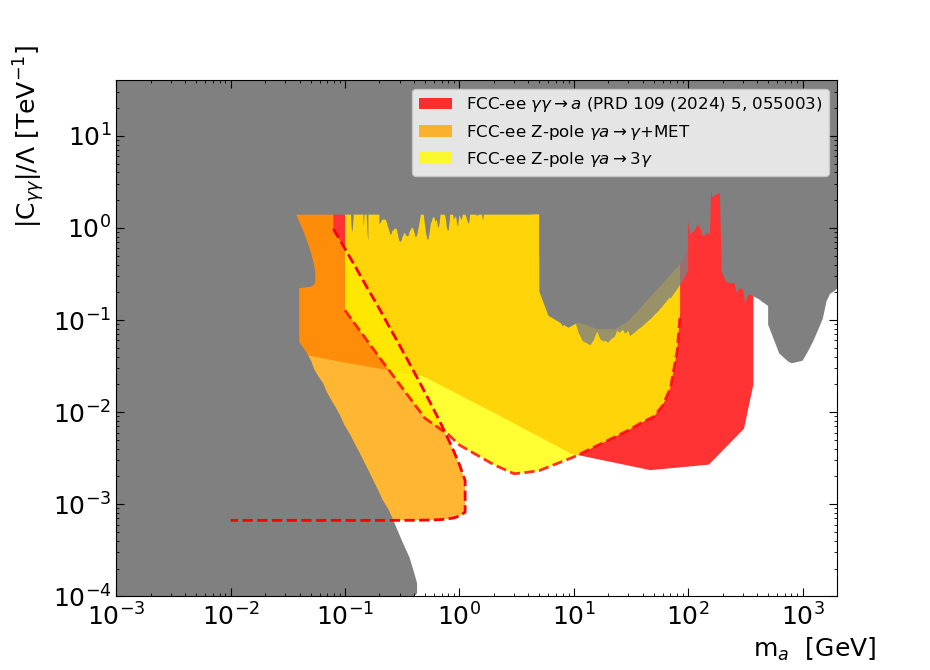}
        \caption{Reach at the FCC-ee for the discovery of
the ALP in the ($m_{\PXXa}$, $C_{\PGg\PGg}$ plane). The 
$3\PGg$ analysis is shown as a yellow area  and the $\PGg$+missing energy  analysis as an orange area. The red area corresponds to Ref.~\cite{RebelloTeles:2023uig}. Existing exclusion limits are shown as a grey area.  \label{fig:ALP_results_total}} 
\end{figure}
\FloatBarrier
%
%
%
Envisaged future improvements/refinements to this study include: the consideration of reducible backgrounds from SM Z decays and an improved treatment of long flight paths for the prompt analysis; the study of resolution effects on the impact angle of photons for the LLP case; the development of a dedicated CNN-based mass-reconstruction algorithm to study coalescing photons in the fibre calorimeter case; and the implementation of full detailed simulation for both calorimeter options.  The analysis reach can be reinterpreted  in the parameter space of more complete models featuring the same signature, such as the composite Higgs model of Ref.~\cite{Cacciapaglia:2021agf}.

\subsubsection*{Detector design comparison for long-lived HNLs}
Recent work on a sensitivity analysis for long-lived scalar particles from exotic Higgs boson decays at FCC-ee extends on the work presented in~\ref{exohiggsllp} to specifically compare results obtained using the IDEA and CLD detector designs, providing an assessment of displaced tracking performance with each tracker design. 
FCC-ee provides an ideal environment for searching for LLPs due to its high luminosity, fewer trigger constraints, and clean experimental conditions. However, the sensitivity to LLP signatures heavily depends on the tracking and vertexing capabilities. This study explores and compares two detector concepts, IDEA and CLD, with regards to identifying LLPs. Additionally, we investigate whether incorporating additional \PZ boson decay modes (invisible and hadronic decays) improve sensitivity to these LLP signatures.

The first part of the study compares the performance of two proposed FCC-ee detector designs: the Innovative Detector for Electron-Positron Accelerators (IDEA)~\cite{IDEA} and the CLIC-Like Detector (CLD)~\cite{CLD}. The goal is to evaluate how well these detectors can reconstruct the displaced vertices (DVs) that arise from LLP decays.
The IDEA tracker is optimized for low-momentum track reconstruction, featuring a lightweight drift chamber with 112 layers and a silicon pixel vertex detector. The design extends coverage from \SI{0.35}{\metre} to \SI{2}{\metre} from the interaction point (IP), enhancing the detection of displaced tracks from long-lived particles. 
The CLD design consists of inner and outer silicon trackers, with more limited coverage than the IDEA detector. The vertex detector covers a region up to \SI{13}{\centi\metre} from the IP, and the outer silicon layers provide coverage up to \SI{2.1}{\metre}. This detector has fewer layers than IDEA, limiting its ability to reconstruct displaced vertices, especially for long-lived particles that decay outside the innermost tracking layers.

The \textit{Hidden Abelian Higgs Model (HAHM)} within \mgfive\ is used to simulate signal events. The Higgs boson is assumed to decay into two long-lived scalar particles ($\PH \rightarrow \mathrm{SS}$), which in turn decay into bottom quarks ($\mathrm{S} \rightarrow \PQb\PAQb$). Six signal points are considered, varying scalar mass ($m_{\mathrm{S}} = 20, \SI{60}{\giga\electronvolt}$) and mixing angle ($\sin(\theta) = 10^{-5}, 10^{-6}, 10^{-7}$), corresponding to mean proper lifetimes ($c\tau$) ranging from millimetres to several metres. 
Each signal sample is processed through the IDEA and CLD detector simulations using \delphes. SM background processes considered include $\PW\PW$, $\PZ\PZ$, and $\PZ\PH$ production. 

The IDEA detector significantly outperforms CLD for long-lived particles, particularly for scalar particles with long decay lengths (e.g., greater than \SI{0.7}{\metre}), thanks to its increased number of layers and broader acceptance range for displaced tracks. Sensitivity loss in CLD is especially noticeable for the longer lifetime signal points, such as the $m_{\textrm{S}} = \SI{20}{\giga\electronvolt}$, $\sin(\theta) = 10^{-6} (c\tau = \SI{341.7}{\milli\metre})$ and $m_{\textrm{S}} = \SI{60}{\giga\electronvolt}$, $\sin(\theta) = 10^{-7} (c\tau = \SI{8769}{\milli\metre})$ cases, as shown in \cref{fig:ideacldtracks}. To mitigate the sensitivity loss, the minimum number of hits required to form a track is reduced for CLD. This improves sensitivity for longer lifetimes but does not fully recover the performance seen with IDEA. For shorter lifetimes, the two trackers show similar performance.
\begin{figure}[hbtp]
    \centering
    \includegraphics[width=0.5\linewidth]{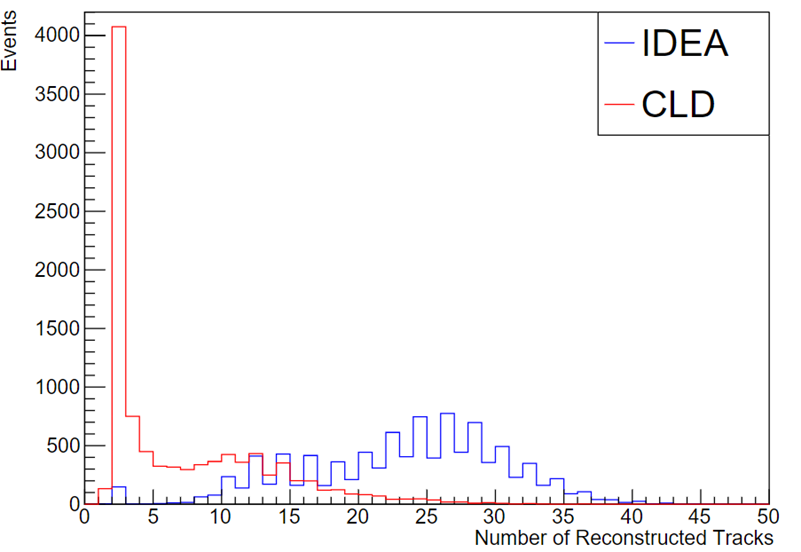}
    \caption{Number of reconstructed tracks per event for exotic decays of a Higgs into two long-lived scalar particles S, with $m_{\textrm{S}} = \SI{20}{\giga\electronvolt}$ and $\sin(\theta) = 10^{-6} (c\tau = \SI{341.7}{\milli\metre})$; a significant loss in sensitivity was found when using the CLD detector.}
    \label{fig:ideacldtracks}
\end{figure}
\FloatBarrier
The study provides valuable insights into the potential for long-lived particle searches at FCC-ee. The comparison between the IDEA and CLD detector concepts shows that IDEA offers superior performance for longer lifetime particles due to its extensive tracking system. Incorporating additional \PZ decay modes, particularly invisible and hadronic decays, could significantly increase sensitivity to LLP signals, though further work is required to optimize background rejection.



\subsection{New  gauge bosons }
\subsubsection{Flavoured gauge bosons\label{sec:SRCH-flavor-gauge-bosons}}
%
\begin{figure}
    \centering
    \includegraphics[width=0.49\textwidth]{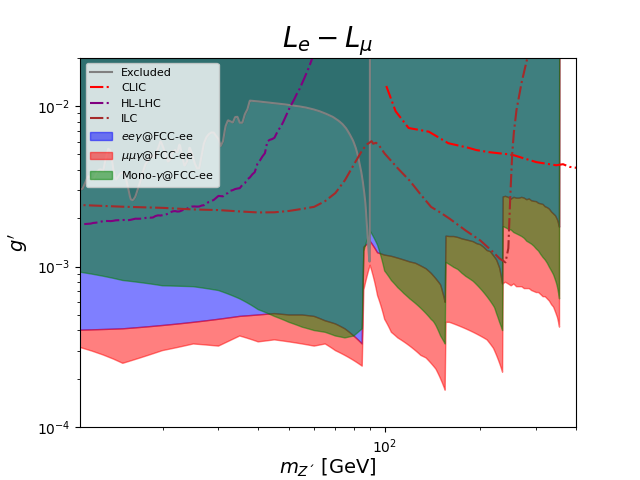}
    \includegraphics[width=0.49\textwidth]{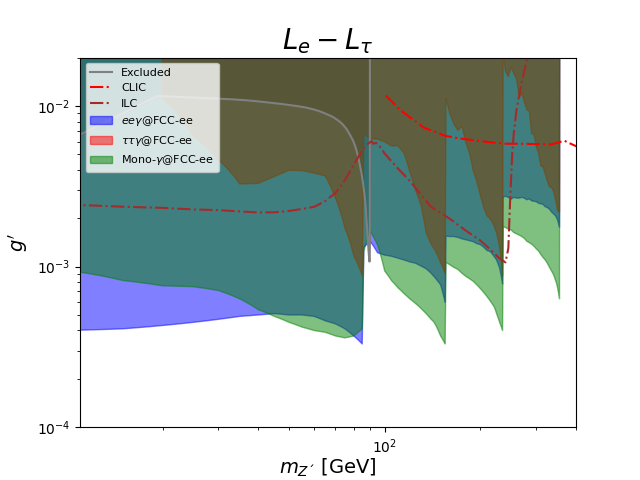}
    \caption{Projected {sensitivity} of FCC-ee to the $L_{\Pe} - L_{\PGm}$  and $L_{\Pe} - L_{\PGt}$ models through $\epem\to\Plp\Plm\PGg$. The red, blue and green bands correspond to the $\PGm\PGm\PGg$ ($\PGt\PGt\PGg$), $\Pe\Pe\PGg$ and mono-$\PGg$ search channels respectively, and show the strongest bound out of many possible ones that can be obtained by varying both the di-lepton mass and the photon energy windows in the range \SI{0.5}{} to \SI{10}{\giga\electronvolt}. \label{fig:optimistic-zurita}}
    \end{figure}
    \begin{figure}
        \centering
        \includegraphics[width=0.49\textwidth]{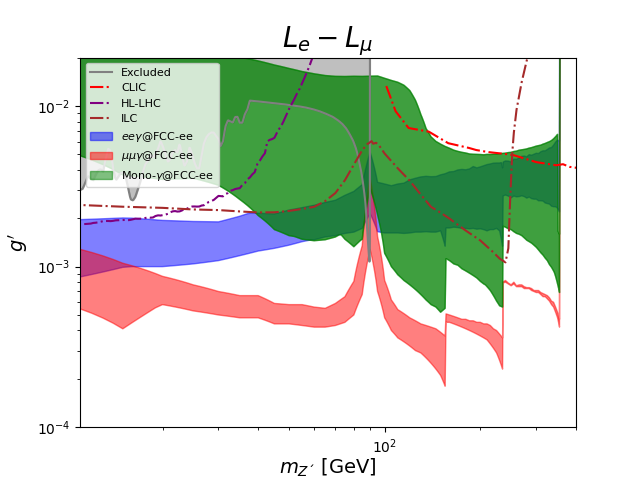}
        \includegraphics[width=0.49\textwidth]{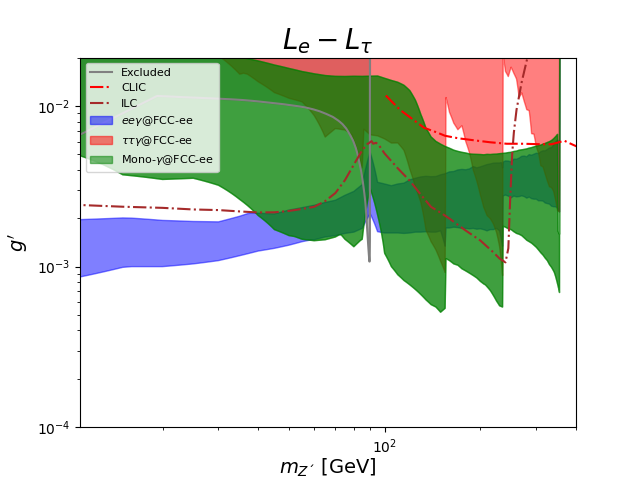}
        \caption{The impact of {systematic uncertainties} on the projected bounds shown in Figure \ref{fig:optimistic-zurita} from FCC-ee  on  the $L_e - L_\mu$ ($L_e - L_\tau$) models  through $\epem\to\Plp\Plm\PGg$. The red, blue and green bands correspond to the $\PGm\PGm\PGg$ ($\PGt\PGt\PGg$), $\Pe\Pe\PGg$ and mono-$\PGg$ search channels, respectively, and show the range of exclusions obtained by introducing a systematic uncertainty between 0.1\% and 1\% on the counts in each bin.}
        \label{fig:systematics}
        \end{figure}
It is known that gauging the groups $B-L$ and $L_x - L_y$, with  $x,y =\Pe, \PGm, \PGt$ and $x \neq y$, for the three SM matter families requires no additional particle content to cancel gauge anomalies~\cite{He:1991qd}. Since we are interested in probing this setup at FCC-ee, we focus here on the $L_{\Pe} - L_{\PGt}$ and $L_{\Pe} - L_{\PGm}$ models that can couple directly to the electrons and positrons in the beams. The Lagrangian for our study~\cite{GonzalezSuarez:2024dsp} is
\def\mzp{\ifmmode m_{\PZ^\prime} \else $m_{\PZ^{\prime}}$ \fi}
\def\gp{\ifmmode g^{\prime} \else $g^{\prime}$ \fi}
\begin{equation}
    \label{eq:lagzpl-Zurita}
{\cal L}
\supset - \gp  \left(\bar{l}_x  \slashed{Z}^{\prime} l_x + \bar{\nu}_x \slashed{Z}^{\prime}  \nu_x -  \bar{l}_y \slashed{Z}^{\prime} l_y - \bar{\nu}_y \slashed{Z}^{\prime}  \nu_y \right)  + \frac{1}{2} (m_{Z^{\prime}})^2 Z^{\prime \, \mu} Z^{\prime}_\mu \, .
\end{equation}
We have two free parameters in our study: $\gp$ and $\mzp$, which are the effective couplings of $\PZ^{\prime}$ to leptons and the mass of $\PZ^{\prime}$, respectively. 
We  focus on the mass regime $\mzp \gtrsim \SI{10}{\giga\electronvolt}$. Below that mass range there are strong constraints from BaBar~\cite{BaBar:2016sci}, $\gp \gtrsim 10^{-4}$ being excluded, while our results show that the FCC-ee sensitivity leads to constraints of about $10^{-3}$. In the mass range of our interest, LEP set strong constraints on searches for $\epem \to \Plp \Plm$ and also for $\epem \to \PGg \PGn \PGn$. In addition, IceCube searches for non-standard interactions set an important bound for masses in the 10--20 GeV range, for the $L_e - L_{\tau}$ model. For a more complete overview of the limits we refer the reader to Refs.~\cite{GonzalezSuarez:2024dsp,Dasgupta:2023zrh} and references therein. In \cref{fig:optimistic-zurita} we show the existing constraints, 
together with  projections for HL-LHC~\cite{GonzalezSuarez:2024dsp}, ILC~\cite{Kalinowski:2020lhp,limitsILC}, CLIC~\cite{Dasgupta:2023zrh}, and our FCC-ee sensitivity estimates.

We employ \textsc{MadGraph5\_aMC@NLO}~\cite{Alwall:2014hca} for parton level event generation, \pythiaeight~\cite{Bierlich:2022pfr} for showering and hadronisation and \delphes~\cite{deFavereau:2013fsa} for reconstruction, using as a baseline the IDEA detector card. Signal samples 
$\Pep \Pem  \to \Plp \Plm \PGg \,$,
with $\Pl= \Pe, \PGm, \PGt, \nu$ are generated with the \textsc{DMSimp}~\cite{Albert:2017onk} UFO model. We consider all $\Pep \Pem \to \PGg + X$ background processes, with $X$ being either two leptons or four leptons, considering both neutral and charged ones. 
We impose the following selection criteria on our events:
for electrons and muons, $\pT > \SI{0.5}{\giga\electronvolt}$ for $\Pe$ and $\SI{3}{\giga\electronvolt}$ for $\PGm$, $|\eta| \leq 2.56$, $\Delta R(l,X)$ > 0.5\,;
for photons, $E > \SI{2}{\giga\electronvolt}$, $\pT > \SI{1}{\giga\electronvolt}$,  $|\eta| < 3.0$,  $\Delta R(\PGg,X) >$ 0.5, reconstruction efficiency $\epsilon_{\PGg}$ = 0.99\,;
and for taus: 
$\pT > \SI{1}{\giga\electronvolt}$ ,  $|\eta| \lesssim 3.0$, $\Delta R(\PGt,X)$ > 0.5\,.

We consider as discrimination variables the energy of the photon (for all signals) and the invariant mass of the di-lepton for the final states with observable leptons. We consider bin widths varying   between \SI{0.5}{\giga\electronvolt} and \SI{10}{\giga\electronvolt} for the invariant mass distribution of charged leptons and for the energy of the photon. The 95\%~C.L. limit shown in \cref{fig:optimistic-zurita} is the best obtained varying the two bin sizes. Details on the impact of changing the bin-size can be found in Ref.~\cite{GonzalezSuarez:2024dsp}.  
To assess the impact of systematic uncertainties  we show in \cref{fig:systematics}  results including  systematic uncertainty between 0.1\% and 1\% on the counts in each bin. 
We find that FCC-ee can be sensitive to couplings $\gp \sim O(10^{-4})$ in the direct search for deviations from the SM in mono-$\PGg$ and the $\Plp \Plm \PGg$ channels, including  when systematic uncertainties are considered. 

We also remark that the HTE factory at the $\PZ$ pole will have sensitivity to this model by measuring with high precision the $\Pep \Pem  \to \Plp \Plm $ total rate and differential  quantities such as asymmetries, which are affected at one loop by the existence of the $\PZ^\prime$ of $L_e-L_\mu$ and $L_e-L_\tau$ that we are considering here, as well as by the $\PZ^\prime$ of $L_\mu-L_\tau$ that is more elusive for direct searches. Results from Ref.~\cite{Dasgupta:2023zrh} indicate that LEP1 already had sensitivity to $\gp\simeq O(10^{-2})$ for light $\PZ^\prime$ around \SI{10}{\giga\electronvolt}, thus promising bounds from the $\PZ$ pole run for even smaller $\gp$, potentially comparable with those probed by direct searches. 
In addition, the analysis of Ref.~\cite{Allwicher:2023shc} suggests sensitivity to heavier $\PZ^\prime$ bosons that would generate 4-lepton interactions, e.g.\ $\PGm\PGm\PGt\PGt$ or even $\PGt \PGt \PGt \PGt $ contact interactions, which have consequences for the electroweak observables measured at the $\PZ$ pole at the HTE factory.

\subsubsection{\texorpdfstring{New (dark) $U(1)$ gauge bosons}{New (dark) U(1) gauge bosons} \label{sec:SRCH-new-dark-gauge-bosons}}
We consider a model for a new $U(1)$ gauge boson,
called the dark photon  ($A_D$) in the literature, that couples  to the visible sector via kinetic mixing to the ordinary photon~\cite{Holdom:1985ag}
$\mathcal{L} \supset    \epsilon F^{\mu\nu} F_{\mu\nu}^D$,
where $F^{(D)}_{\mu\nu}=\partial_\mu A_\nu^{(D)} - \partial_\mu A_\nu^{(D)}$. 
The strength of the mixing, $\epsilon$, is a free parameter. Current bounds on this coupling imply it is rather small. 

The detector concept used to study $\Pep \Pem$ collisions for searches for a new $U(1)$ gauge boson coupled to the SM only via kinetic mixing of interest for this model is the ILD~\cite{ILDConceptGroup:2020sfq} at ILC ~\cite{ILCInternationalDevelopmentTeam:2022izu, Behnke:2013xla, ILC:2013jhg, Adolphsen:2013jya, Adolphsen:2013kya, Behnke:2013lya} operating at \SI{250}{\giga\electronvolt}. 
%
%
%
The signal process we consider in Refs.~\cite{HosseiniSenvan:614991,Berggren:2024zed} for $A_D$ production at \epem  colliders
is resulting in a visible signal for the decay of $A_D$ in SM fields. We focus on the decay to muons
\begin{equation}
\epem \rightarrow \PGg_{ISR} A_D \rightarrow \PGmp \PGmm   \PGg_{ISR} \, , \label{eq:signal-darkphoton-Berggren}
\end{equation}
where the energy of the ISR is such that the recoil-mass against it is peaked at $m_{A_D}$.
As shown in Ref.~\cite{Curtin:2014cca}, one can note that both production cross section  and the decay width scale with $\epsilon^2$. 
Considering a typical electroweak $2 \to 2$ scattering rate and applying the suppression by $\epsilon$ factors one finds that a generic Higgs factory should be sensitive to cross section of $\mathcal{O}$(\SI{1}{\femto\barn}) which implies a total width for $A_D$ corresponding to prompt decays, $c \tau < \SI{1}{\nano\meter}$, and a natural width too small to impact the measured peak in the di-muon mass distribution, which is then going to be determined by detector resolution.  
 
We generated events according to the model given in Ref.~\cite{Curtin:2014cca} with \whizard\ v3.0~\cite{Kilian:2007gr}.
   The generated events were then passed through the full \textsc{Geant4}-based simulation 
   and reconstruction 
   of ILD.     In the analysis the SM background included are $\epem \rightarrow \PGmp \PGmm   \PGg_{ISR}$ plus
 $t$-channel processes with beam-remnant electrons that go undetected, or mistaken for ISR photons. 
 To search for $A_D$ we   
  select events with two oppositely charged muons, and possibly an isolated photon and exclusively require nothing else being recorded in the event.
  In the selected di-muon sample, we search for an arbitrarily small narrow peak in the $m_{\PGm\PGm}$ distribution at which a peak from $A_D$ is expected for each $m_{A_D}$ hypothesis.

    \begin{figure}[h!]
    \begin{center}
      \subcaptionbox{}{
        \includegraphics [trim={0cm 0cm 0cm 0cm },clip,scale=0.8]{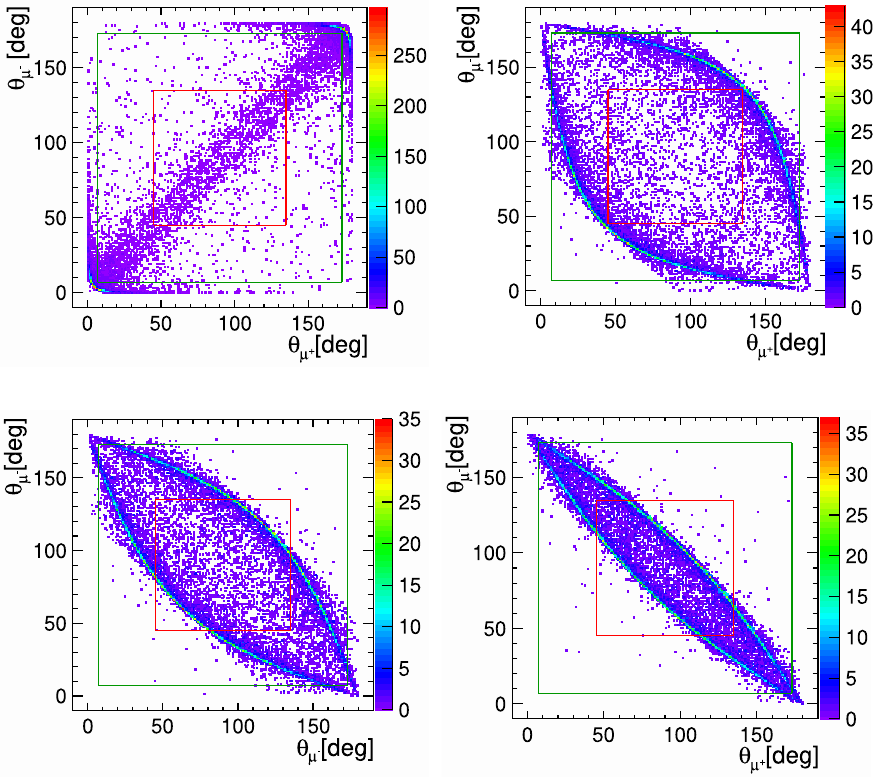}
      }
      \subcaptionbox{}{
        \includegraphics [scale=0.35]{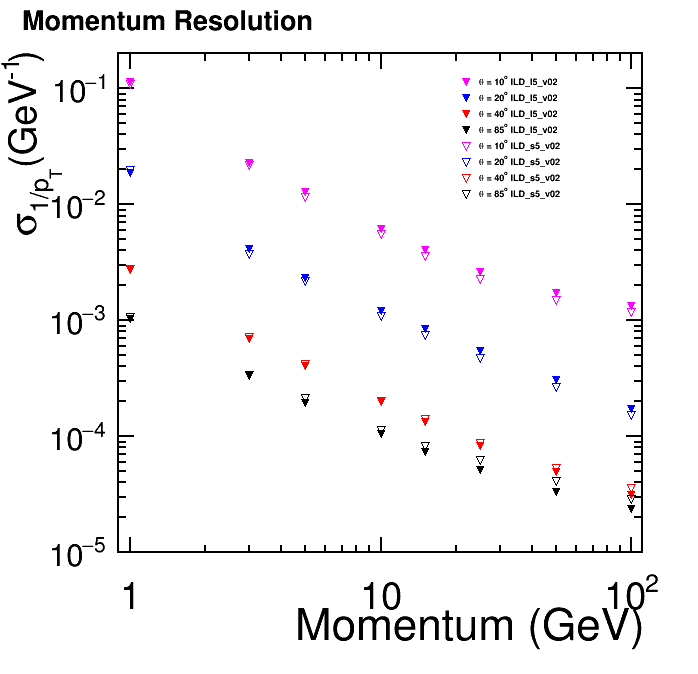}
      }
    \end{center}
    \caption{
      (a) The polar angle of the \PGmm  versus that of the \PGmp of
      the generated   \epem $\rightarrow \PGg_{ISR} A_D \rightarrow \PGmp \PGmm   \PGg_{ISR} $ events,
      for $m_{A_D}$ = 10, 100, 150 and 200 \GeV~(clock-wise, from upper-left). The green square indicates the acceptance
      of the ILD tracking system, and the red one indicates the coverage of the barrel tracking system;
      (b): Momentum resolution for charged particles in ILD from full detector simulation~\cite{ILDConceptGroup:2020sfq}.\label{fig:sigmapt_and_thetamu-Berggren}}
    \end{figure}

To understand the role of detector acceptance in this search we show in \cref{fig:sigmapt_and_thetamu-Berggren}(a) the angular distribution of the two muons at four different dark photon masses. The green square indicates the region of coverage of the tracking detectors, and, as can be seen in particular at the lower masses, a large fraction of the events will be lost for the reason that at least one of the muons is at angles below the acceptance of the tracking.
The efficiency to find both muons is therefore only $\sim$ 25\% for $m_{A_D}$=10 \GeV, but will approach 100\% as  $m_{A_D}$ becomes 100 \GeV~or more.
The expected mass resolution on each event can be inferred from the direction of the measured muons polar angle and momentum, using the knowledge of the momentum resolution from full detector simulation (shown in \cref{fig:sigmapt_and_thetamu-Berggren}(b)). Armed with this information we can optimize the width of the mass window around the sought $m_{A_D}$.  The result of this treatment of the resolution effects with respect to the more simple treatment of \cite{Karliner:2015tga}  can be appreciated in \cref{fig:exclusions-resoltuion-Berggren}(a) where the resolution on the mass from full simulation is compared to that in the simplified treatment. 
    \begin{figure}
    \begin{center}
      \subcaptionbox{}{
        \includegraphics [trim={0cm 0.0cm 0cm 0cm },clip,scale=0.90]{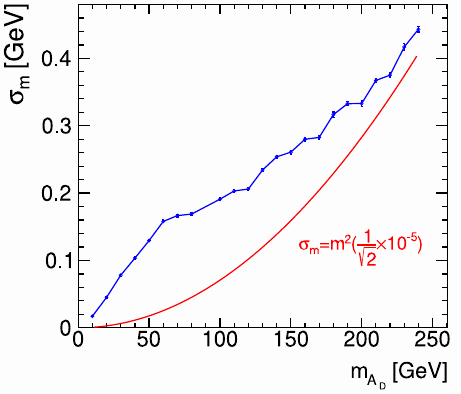}
      }
      \subcaptionbox{}{
        \includegraphics [trim={0cm 0.0cm 0cm 0cm },clip,scale=0.33]{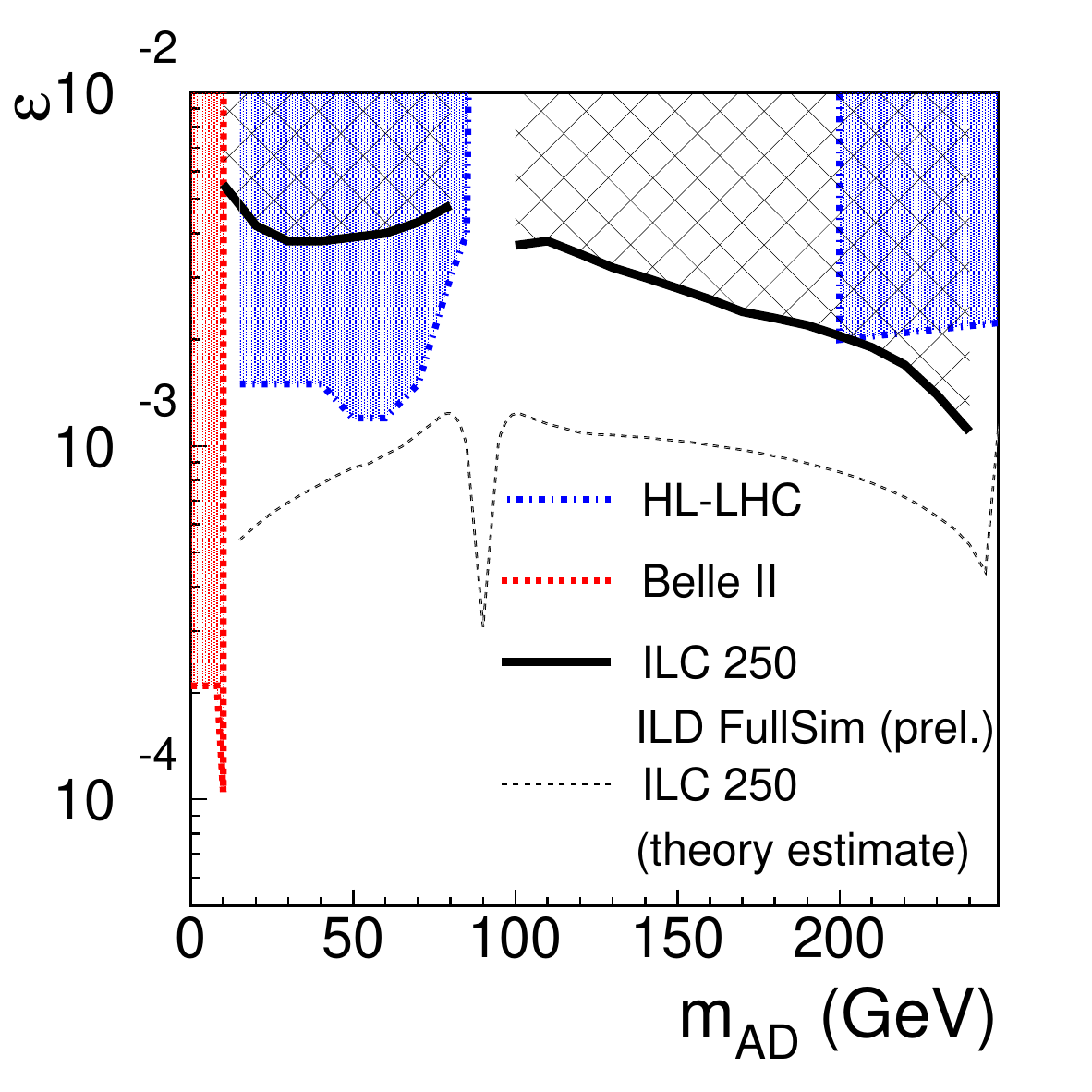}
      }
    \end{center}
    \caption{(a) The di-muon mass resolution versus dark-photon mass $m_{A_D}$. The blue curve is
      the full simulation results, the red one is the simplified one used in
      \cite{Karliner:2015tga,EuropeanStrategyforParticlePhysicsPreparatoryGroup:2019qin};
      (b) The exclusion reach of ILC 250 obtained from this full simulation study of ILD,
      and the expectations of
      Belle~II and HL-LHC (Fig. 8.16 of Ref.~\cite{EuropeanStrategyforParticlePhysicsPreparatoryGroup:2019qin})\label{fig:exclusions-resoltuion-Berggren}, recast to show $m_{A_D}$ on a linear scale). }
    \end{figure}
%

The expected sensitivity from full simulation is shown in  \cref{fig:exclusions-resoltuion-Berggren}(b). Compared with previous estimates with simplified detector effects treatment, the  sensitivity from full simulation is a factor between 2 and 4 worse for $m_{A_D}$ above $m_{\PZ}$. 
 Below $m_{\PZ}$, the impact of detector effects on the sensitivity  becomes even  larger, rendering   the HL-LHC limits~\cite{EuropeanStrategyforParticlePhysicsPreparatoryGroup:2019qin} stronger than the sensitivity of the HTE factory according to our study. This shows that it is crucial to employ a correct estimation of the momentum uncertainty for this kind of searches.
In addition, for light $m_{A_D}$ we find it is important to consider background from non-$\PZ \rightarrow \PGmp \PGmm $ processes,
and to consider the reduced probability of having both muons detected in the tracking system. 
Our result shows that, after including backgrounds previously not considered, and modelling the invariant mass resolution with a full detector simulation 
there can be significant losses of sensitivity, especially at low $m_{A_D}$. Extrapolating to higher centre-of-mass energies available at ILC and the Cool Copper Collider (C$^3$)
\cite{Vernieri:2022fae}, we expect to be able to cope rather well with heavier $m_{A_D}$, extending the reach of the HTE factory up to the kinematic limit of these higher energy machines. 
In addition, we expect that further sensitivity can be gained by exploiting the polarisation of the beams. 
Polarisation can help to fight the background that become relevant at low $m_{A_D}$. 

At the FCC-ee~\cite{Bernardi:2022hny, FCC:2018byv, FCC:2018evy} or
CepC~\cite{Gao:2022lew, CEPCPhysicsStudyGroup:2022uwl, CEPCStudyGroup:2018rmc, CEPCStudyGroup:2018ghi} circular machines,
the higher luminosity at lower centre-of-mass energy will be beneficial
for searches at lower $m_{A_D}$.
As shown above, the resolution on the  $A_D$ resonance  in $\PGmp \PGmm $ plays  a key role for this search. Thus  one should pay attention to the loss of sensitivity arising from lower   resolution due to the lower B-field employed in these machines.

\subsubsection{\texorpdfstring{Exotic Z boson decays into new $U(1)$ gauge bosons }{Exotic Z boson decays into new U(1) gauge bosons }}\label{sec:exoticZbosonDecays}
Models for new gauge $U(1)$ gauge bosons can in principle have further signatures, besides the production in $2 \to 2 $ scattering as in the signal \cref{eq:signal-darkphoton-Berggren}  explored above. The existence of matter then generates a kinetic mixing that may also give rise to a decay of the $\PZ$ boson into new gauge bosons:
 $ \PZ \to \PGg \PZpr$\,,
where we have denoted the new gauge boson as $\PZpr$ to highlight the fact that it may in general not coincide with the boson $A_D$ studied above. However, we stress that if the kinetic mixing is generated by electroweakly charged new matter, in general one expects also a trilinear coupling among $\PZ$, $\PGg$ and the new gauge boson.
The decay width for the $ \PZ \to \PGg \PZpr$  decay has been computed in Ref.~\cite{Michaels:2020fzj} for both anomaly-free choices of the new $U(1)$ quantum numbers, e.g.\ the combination of baryon and lepton number $B-L$, and for anomalous choices, e.g.\ the baryon number $B$. In the latter case, assuming anomaly cancellation from new matter introduced for this purpose has significant effects. Indeed, the anomaly cancelling fermions, which we call ``anomalons'', can become massive independently of the Higgs vacuum-expectation value, so their contribution to the $\PZ-\PZpr-\PGg$ vertex is non-decoupling.  

In~\cref{fig:FelixYu_ExoticZ_ZBconstraints}, we show the current constraints and projected ones from the HTE factory on the gauge coupling $g_X$ for the $U(1)_B$ gauge symmetry as a function of the \PZpr mass.  We show constraints compiled from Ref.~\cite{Michaels:2020fzj} that include searches for light dijet resonances in association with initial state radiation photons or jets, the constraint from the mixing with the $\Upsilon$ meson, and constraints on the excess contribution to the hadronic width of the \PZ boson. The current constraint  from LEP on the exotic decay   is also shown.
Extrapolating the  sensitivity by the expected increase in statistics, we draw a projection for a \TeraZ experimental program at the HTE factory. We remark that a \TeraZ run can be sensitive to coupling values not accessible at LHC. In addition, it can play a significant role in understanding the nature of a signal from new gauge bosons, as discussed at the beginning of \cref{sec:exoticZbosonDecays}.
%
\begin{figure}
\begin{center}
    \includegraphics[width=0.59\textwidth]{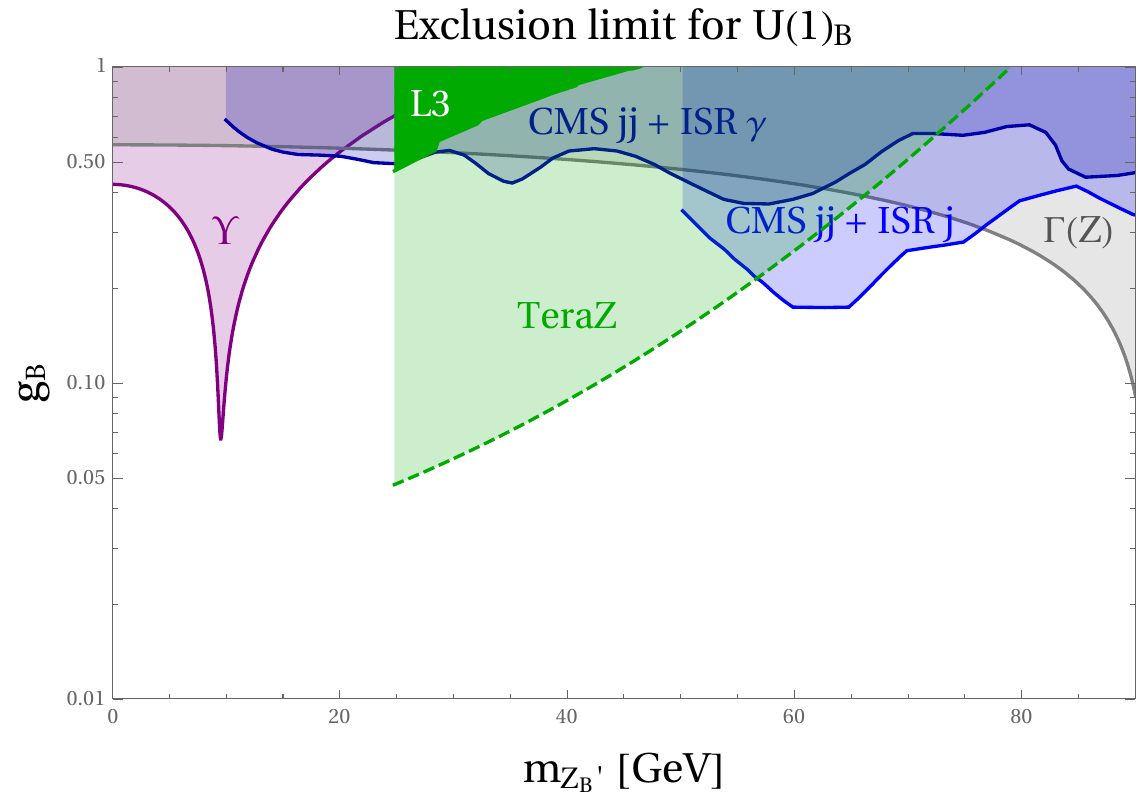}
\end{center}
    \caption{Constraints on the gauge coupling $g_B$ vs.~$m_{\PZpr}$ for a baryon-number coupled \PZpr boson in exotic $\PZ \to \PZpr\PGg$ decays, adapted from Ref.~\cite{Michaels:2020fzj}. Blue shaded regions are excluded by low-mass dijet resonance searches, $\Upsilon$ decays (purple), and the hadronic contribution to the \PZ width (grey). The search for the exotic decay $\PZ \to (jj)_\text{res} \PGg$ from the L3 experiment is shaded dark green.  The estimated improvement from a Tera-Z experiment is shaded in light green.
    \label{fig:FelixYu_ExoticZ_ZBconstraints}
    }
\end{figure}

\subsubsection{\texorpdfstring{A complete model for a new  $U(1)$ gauge interaction: (dark) Higgs and gauge bosons}{A complete model for a new  U(1) gauge interaction: (dark) Higgs and gauge bosons}}
Typical studies of gauge extensions of the SM concentrate on the observable effects of the new gauge bosons, as has been the case in the preceding parts of this section. However, there are important features of a realistic model that cannot be captured in the picture without discussing the origin of the mass of the new gauge boson. As discussed in Ref.~\cite{Liu:2017lpo},  there should be a Higgs boson related to the spontaneous symmetry breaking of the new $U(1)$ to explain the origin of the mass of the new gauge boson. The existence of such a Higgs boson for the new gauge sector has  important phenomenological consequences\footnote{We will ignore the issue of possible mass degeneracy between the new gauge boson and the SM $\PZ$. The interested reader can find a detailed in discussion in Ref.~\cite{LoChiatto:2024guj}.}. 
In general, one expects the new Higgs boson to have a quartic interaction with the Higgs boson of the SM, as in the so-called ``Higgs portal''
$\PH^2 \mathrm{S}^2$.
Such Higgs portal couplings lead to modifications in the interactions of the Higgs boson that can be parametrised via a mixing angle $\alpha$. 
The modification of   Higgs and gauge boson couplings induces deviations in the production cross section for Higgs and gauge bosons  at the HTE factory. 
For instance, the Higgs-strahlung cross section 
$\Pep \Pem \to \PZ \PH$ 
is modified by both the scalar mixing $\alpha$ and the kinetic mixing $\epsilon$. 
 In addition, a new ``Higgs-strahlung'' process 
$\epem \to K \PH$
allows to produce the new gauge boson, which we denote as $K$. This mechanism for the production of a new boson in association with a $\PH$ can be  probed directly if the visible decays of $K$ are tagged, or, if the $K$ decays invisibly, a recoil mass method using visible decays of \PH can be used instead.
A similar process is also possible in the model
$\epem \to \PZ \mathrm{S}$, 
for which the $\PZ$ boson can be used as tagger for the production of the new physics states S.
In addition, it is possible to have a fully BSM final state for the $\Pep \Pem$ scattering
    $\epem \to K \mathrm{S} \,$.

 In~\cref{sec:exotic-scalars-searches} we have a detailed discussion on the reach for production of exotic scalars. In a full model for a new gauge force, these results needs to be complemented by including searches for new gauge bosons discussed here. 
\begin{figure}[tbh]
    \begin{center}
        \includegraphics[width = 0.45\textwidth]{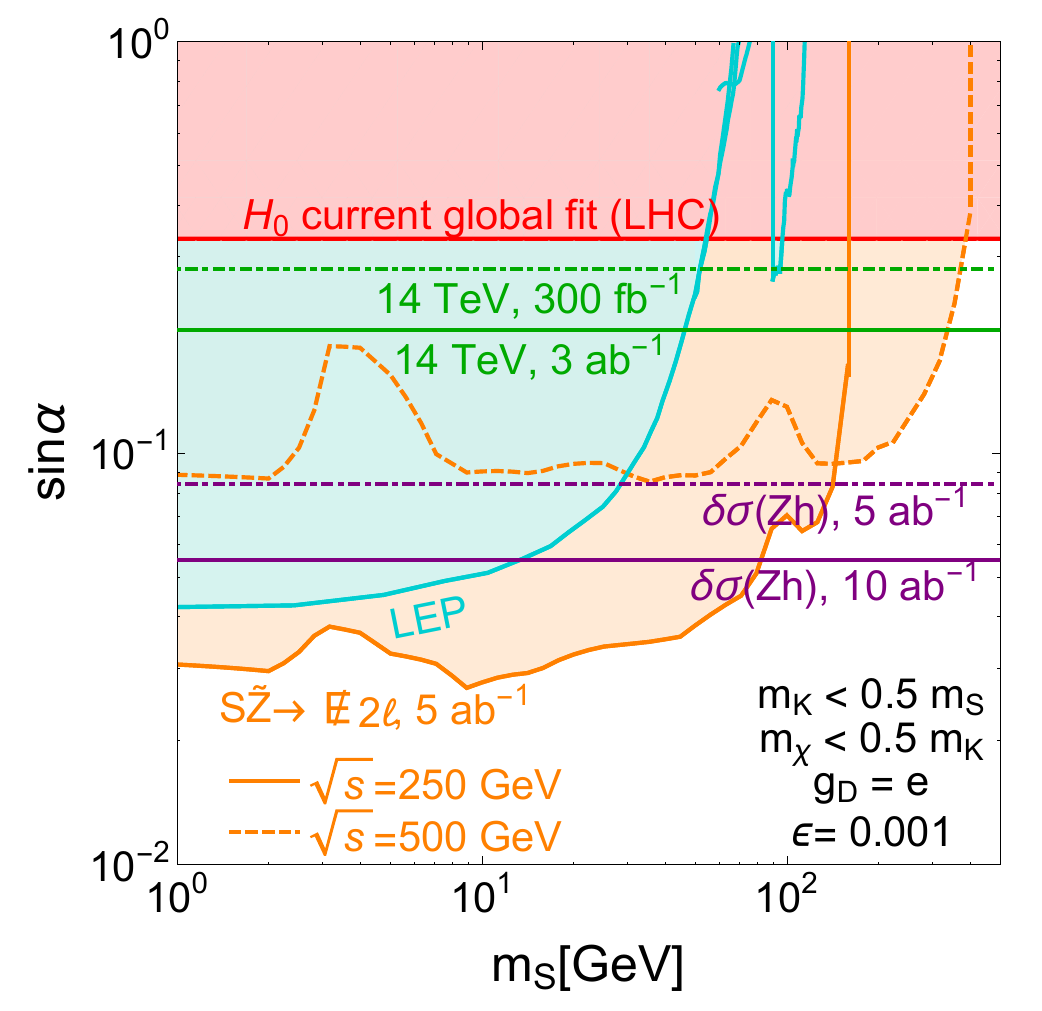}
        \includegraphics[width = 0.45\textwidth]{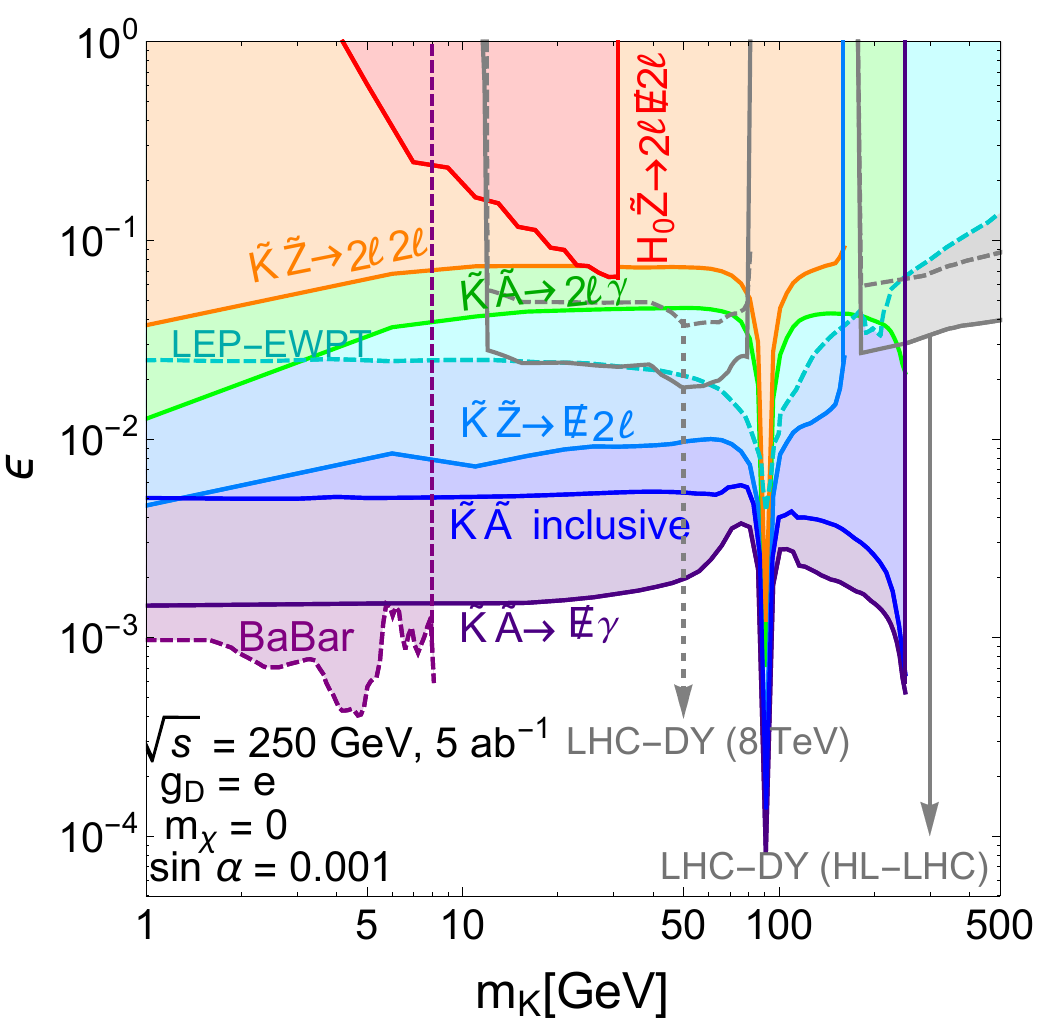}
    \end{center}
    \caption{Left: Constraints from LEP, LHC, and projected sensitivity from the HL-LHC and Higgs cross section measurements from FCC-ee on the scalar mixing angle $\sin \alpha$. Right: Sensitivity to the kinetic mixing parameter $\epsilon$ between the hypercharge field strength tensor and a new $U(1)$ field strength tensor as a function of the new gauge boson  mass $m_K$.  The  colours correspond to  different inclusive and exclusive searches at the HTE factory.  Constraints from BaBar and the Drell--Yan constraints from LHC and a HL-LHC projection are also shown. Results taken from Ref.~\cite{Liu:2017lpo}. 
    } \label{fig:Yu_DoubleDark_epsilon_vs_mK}
    \label{fig:Yu_DoubleDark_sinalpha_vs_mS}
\end{figure}
 In \cref{fig:Yu_DoubleDark_sinalpha_vs_mS} we offer an outlook on the possible combined study of $S$ and $K$. The sensitivity to a scalar mixing  $\sin\alpha$ as a function of the scalar mass $m_S$ for a fixed amount of kinetic mixing $\epsilon$ and fixed  coupling of the new gauge boson is illustrated. There is a significant degree of complementarity from the determination of Higgs boson couplings, in particular the coupling to the vector bosons, and the search for new scalars from their production in association with a SM $\PZ$ boson. For concreteness, we concentrated on the invisible decay mode for $S$ and estimated the reach from the $2\Pl$ +``invisible'' recoil mass~\cite{Liu:2017lpo}. Especially for larger centre-of-mass energies, the reach of the HTE factory can far surpass the sensitivity of LEP searches and that of current and future HL-LHC Higgs couplings measurements.

In addition to searching for the new scalar, one can search for the production of the new gauge boson $K$. This can be done in both visible and invisible decays channels of $K$ by producing the $K$ in association with a photon or a $\PZ$:
    $\Pep \Pem \to K + \PGg\;/\PZ\,$.
As discussed at the beginning of this section, the rate for this process depends  on $\epsilon$ and the presence of the new scalar should have no significant effects on this process. In \cref{fig:Yu_DoubleDark_sinalpha_vs_mS} we display the reach on $\epsilon$ that is expected for each mass of the new gauge boson state $K$ for several possible process, e.g.\ in association with a SM photon or with a $\PZ$ boson, for several possible final states, e.g.\ $\gamma$ + ``invisible'', $2\Pl$ + ``invisible'', $2\Pl +\PGg$, or with a untagged decay. From these results, we conclude that the HTE factory has the potential to be sensitive to both a new gauge vector boson and the new Higgs boson that is expected to be part of the symmetry-breaking sector that gives mass to the new gauge boson.
\subsubsection{Further characterisation of a new (dark confining) gauge interactions}
Once portals to a new sector ``charged'' 
under the new gauge interactions are established, e.g.\ the $\PGg$ or $\PZ$ gauge bosons or the Higgs boson, the HTE factory can further study the nature of the new sector. 
The new gauge boson is  expected to decay mainly into new physics particles, with minor branching fractions to SM states, as seen above. 
In the scenarios considered here, the new particles have no SM charge, so they appear as ``dark'', i.e.\ they leave no trace in our detectors, at least until they decay back into SM objects through the feeble portal coupling.  In general, we can expect that the new sector of ``dark'' states, while not being charged under the SM gauge interactions, might have other gauge interactions and an overall structure as complex as that of the SM; for example, possibly with many states coupled to the gauge bosons, and maybe even a similarly hierarchical structure of masses and other parameters, as well as familiar phenomena of bremsstrahlung and radiation of the new gauge bosons from the particles charged under the gauge interactions hidden to us ~\cite{Feng:2022aa,Feng:2011ab,Feng:2008aa,Hochberg:2014kqa,Strassler:2006im,Chacko:2005pe}.
If the  hidden sector has a confining non-abelian gauge interaction, the states charged under this non-abelian charge can be lighter than the confinement scale, similarly to the light quarks $\PQu$, $\PQd$, $\PQs$ for the usual QCD. In this case, one expects the formation of hidden sector hadrons, similarly to the hadronic  showers of ordinary QCD. Such ``dark showers'' can be a source of new distinctive signals from the new gauge interaction. Inspired by the SM QCD, we can imagine that  pseudo-Nambu--Goldstone bosons of chiral symmetry breaking exist in the hidden sector. By analogy with our QCD we call these states ``hidden pions'', $\hat{\pi}$. Due to the pseudo-Nambu--Goldstone bosons nature of these particles, they are likely to be the lightest states of the hidden sector, rendering them absolutely stable to decays through the very weak coupling between the hidden sector and ours. The hidden pions are expected to give rise to   long-lived signals, which we have explored in \cref{sec:LLP-SRCH}.
In the context of the study of new gauge interactions, these signals can provide an important handle to study the properties of the hidden sector or a smoking gun for discovery.

As discussed above, both the SM $\PZ$ and $\PH$ bosons can act as portals at the electroweak scale towards the hidden sector. An example of   portal interactions can be a contact interaction $ (i \PH^\dagger \overset{\leftrightarrow}{D}_\mu \PH  ) ( \bar\psi \gamma^\mu  \psi )+\text{h.c.}$ and $ (\PH^\dagger \PH )  ( \bar\psi \psi  )$ , where $\psi$ are quarks of the hidden sector. This contact interaction can arise from a heavy  mediator between the hidden sector and the Higgs and gauge sector. The most notable effect would be a BSM coupling of the $\PZ$ boson to hidden quarks that arises after substituting in the Higgs vacuum expectation value. Such coupling allows one to estimate the sensitivity of the high intensity run of the HTE factory at the $\PZ$ pole.  In the left panel of~\Cref{fig:TeraZ-DarkShowers} we show the sensitivity at the LHC and at the HTE factory from a Z pole run for \SI{150}{\per\atto\barn} for  a benchmark case with a hidden pion mass of $\SI{0.65}{\giga\electronvolt}$ for the displaced decay mode $\hat{\pi}\to \mpmm$. The sensitivity to branching ratio for this  exotic decay  of the $\PZ$ boson is displayed as a function of the life-time of the hidden pion.
The hidden sector can also have a peculiar structure of couplings to the SM that includes flavour-violating interactions, giving rise, among other possible processes, to the FCNC decay $\PB \to \PK \hat\pi (\to \mpmm)$. This could be detected by flavour experiments such as LHCb, as well as among the decay products of $\PQb \PAQb$ that are expected to be produced in large quantities at the $\PZ$ pole run of the HTE factory. The sensitivity to a long-lived decay of the hidden pion is shown in the right panel of \cref{fig:TeraZ-DarkShowers}. 
\begin{figure}[th!]
    \centering
    \includegraphics[width=8 cm]{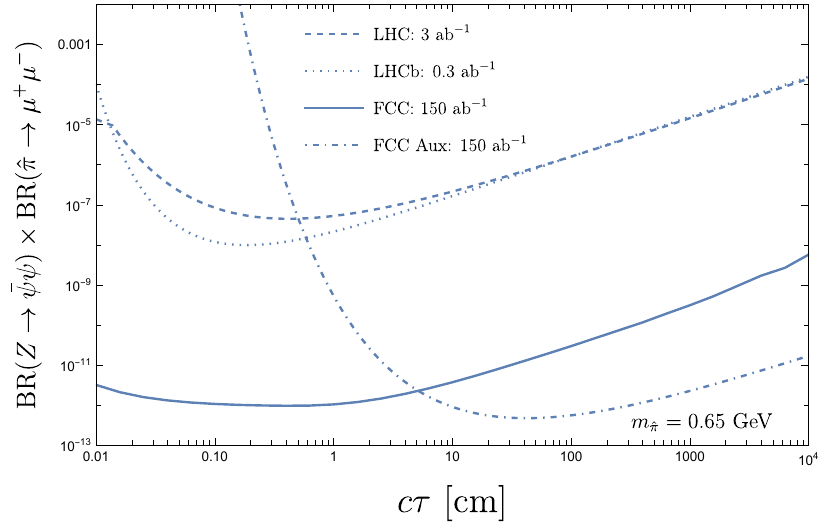}
    \includegraphics[width=8 cm]{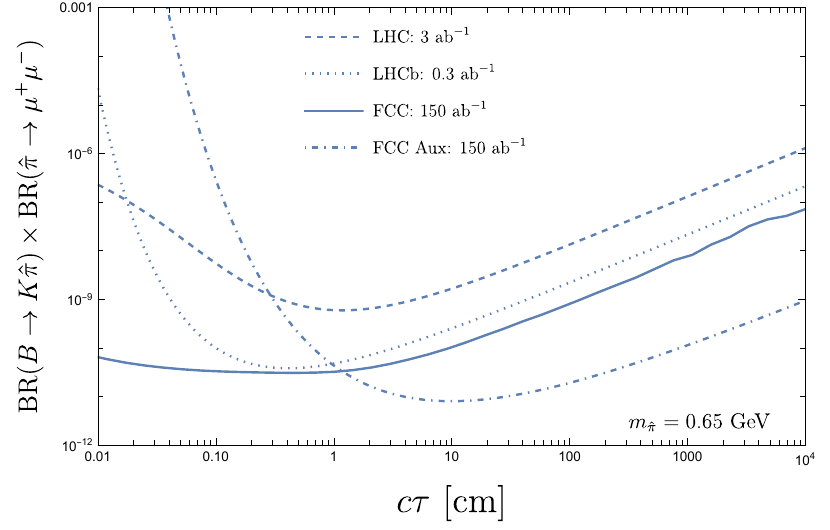}
    \caption{The projected 95\% C.L. exclusion limits at a $\PZ$ factory with luminosity $\SI{150}{\per\atto\barn}$ for a Dark Shower (left) and a $\PB$ meson FCNC decay (right) generating a displaced decay of a hidden pion ($\hat{\pi}$) in $\mpmm$.}
    \label{fig:TeraZ-DarkShowers}
    \end{figure}

The two examples of sensitivity in long-lived particle searches from high intensity studies of the $\PZ$ boson display how the flavour program and the signature-driven search for new particles can be used to search for new gauge bosons, adding important information on top of the traditional searches for new states.

\subsubsection{Two-particle angular correlations in the search for new physics}
Particle correlations in high-energy colliders can be a  complementary tool to other conventional searches for new phenomena including physics beyond the SM (BSM)~\cite{Sanchis-Lozano:2008zjj,Sanchis2009}. 
So far, they have revealed new phenomena in heavy-ion, proton--proton and proton--nucleus collisions~\cite{CMS_2015_ridge_pp,ALICE_ridge_pA,ATLAS_ridge_pA}. No clear ridge-like signal was found in the \epem ALEPH~\cite{Badea:2019vey} and Belle~\cite{Chen_2022} data, except for the recent ALEPH-archival-data~\cite{Chen:2023njr} analysis at high energies and high multiplicity. Such a study has been performed for Higgs factories~\cite{Musumeci:2023noi}.

The clean environment of \epem collisions is suitable for defining the \emph{thrust frame}, whose $z$-axis lies along 
the direction of the back-to-back jets. The azimuthal angle $\phi$ and the rapidity $y$ are defined event-by-event on the transverse plane to the thrust axis.
Rapidity and azimuthal differences of two final-state SM charged particles, $\Delta y \equiv y_1-y_2$, $\Delta \phi \equiv \phi_1-\phi_2$, are defined together with 
the two-particle correlation function:
\begin{equation}
    C^{(2)}(\Delta y, \Delta \phi) = \frac{S(\Delta y, \Delta \phi)}{B(\Delta y, \Delta \phi)},
    \qquad
    S(\Delta y, \Delta \phi) = \frac{1}{N_\text{pairs}} \frac{d^{2}N^\text{same}}{d \Delta y d \Delta \phi},
    \qquad
    B(\Delta y, \Delta \phi) = \frac{1}{N_\text{mix}} \frac{d^{2}N^\text{mix}}{d \Delta y d \Delta \phi},
\label{eq:mitsou_corr_function}
\end{equation}
where $S(\Delta y, \Delta \phi)$ and $B(\Delta y, \Delta \phi) $ are the particle-pair densities from the same and different (mixed) events, respectively.
The azimuthal yield, $Y(\Delta \phi)$, is of particular interest, calculated by integration over a given $\Delta y$ range defined by the lower/upper integration limit $y_\text{inf/sup}$ as:
\begin{equation}\label{eq:mitsou_yield}
    Y(\Delta \phi) = \frac{\int_{  y_\text{inf} \leq |\Delta y| \leq y_\text{sup} } S(\Delta y, \Delta \phi) dy}{\int_{y_\text{inf} \leq |\Delta y| \leq y_\text{sup}} B(\Delta y, \Delta \phi) dy}.
\end{equation}



The theoretical framework of \emph{Hidden Valley} (HV), which encompasses models with one or more BSM hidden sectors, serves as a test-bed here. 
In HV models \cite{Strassler:2006im}, the SM gauge group sector $G_\text{SM}$ is extended by a new gauge
group $G_\text{V}$ under which all SM particles are neutral. Hence, a new category of \textit{v-particles}, $q_\mathrm{v}$ emerges charged under $G_\text{V}$, but neutral under $G_\text{SM}$. ``Communicators'', charged under both $G_\text{SM}$ and $G_\text{V}$, are introduced allowing interactions between SM and HV particles. The communicator $D_v$ ($T_v$) considered here is the pair-produced hidden partner of the $\PQd$-($\PQt$-)quark, whereas $q_v$ are scalars.
The hidden and SM sectors may connect through, e.g., pair-produced communicators via the SM $\PGg^{\ast}/\PZ$  
coupling to a $D_v\bar{D}_v$ pair, yielding both visible and invisible cascades in the same event, seen in \cref{fig:mitsou_diagrams}(a), initiated by the decay $D_v \to d q_v$. The $q_v$ mass may strongly influence the kinematics of the visible cascade (leading to SM particles), thereby distinguishing it from the pure SM cascade seen in \cref{fig:mitsou_diagrams}(b). 




\begin{figure}[ht]
\centering
\subfloat[Hidden valley\label{fig:mitsou_hv}]{\includegraphics[width=4.cm]{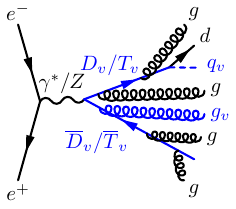} }
\hspace{2cm}
\subfloat[SM light quarks\label{fig:mitsou_sm}]{\includegraphics[width=4.cm]{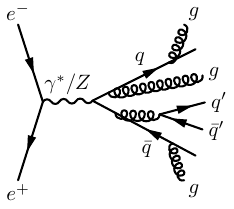}}
\caption{Leading diagrams in $\epem$ collisions for  (a) HV, and  (b) SM light quarks including bottom. \label{fig:mitsou_diagrams}} 
\end{figure}


The HV events were generated with \pythia~\cite{Sjostrand_2015}, whilst the  detector simulation was based on \textsc{SGV}~\cite{berggren2012sgv}, with the geometry and acceptance of  the ILD concept for  ILC~\cite{ILDConceptGroup:2020sfq}. 
The  reconstruction is based on Particle Flow Objects (PFOs), reconstructed by all individual particles of the final state via pattern recognition algorithms. 
%
%
The $\epem \to \PQq \PAQq$ background comes from the inclusive production of all the SM quark species except the (below threshold) top flavour. Unlike a previous preliminary study~\cite{Musumeci:2023rzf}, we also consider Initial State Radiation (ISR) and four-fermion production (dominated by $\epem\rightarrow \PW\PW$), but the contribution of the latter is negligible. 

In view of the small HV production cross sections, 
selection criteria are applied to suppress the background. Requirements on the number 
of neutral ($\leq22$) and charged ($\leq15$) PFOs are set~\cite{Irles:2024ipg}. Other selections include the emission angle ($|\cos{\theta_{\gamma_{\text{ISR}}}}|<0.5$) and energy ($E_{\gamma_\text{ISR}}< 40~\GeV$) of reconstructed ISR photons, as well as the di-jet invariant mass $m_{jj} < 130~\GeV$ and the leading-jet energy $E_j < 80~\GeV$. 
These selections achieve a drastic reduction of the SM background while largely maintaining the HV signal. 
%
In \cref{fig:mitsou_yield} the yield $Y(\Delta \phi)$, defined in Eq.~\eqref{eq:mitsou_yield} for short and long $y$ ranges, is shown. Various HV models are considered together with SM processes, while the SM background is also presented.
A near-side peak shows up at $\Delta \phi \simeq 0$, mainly from same-jet track pairs. An away-side correlation ridge around $\Delta \phi \simeq \pi$, and extending over a large rapidity range, results from back-to-back momentum balance. 
A clear difference becomes apparent for $0<|\Delta y|\leq 1.6$: a sizeable peak at $\Delta \phi \sim \pi$ characterises the HV scenario, unlike the pure SM case. This remarkable discrepancy of shapes would potentially serve as a valuable signature of a hidden sector.

\begin{figure*}[ht]
\centering
\includegraphics[width=7.00cm]{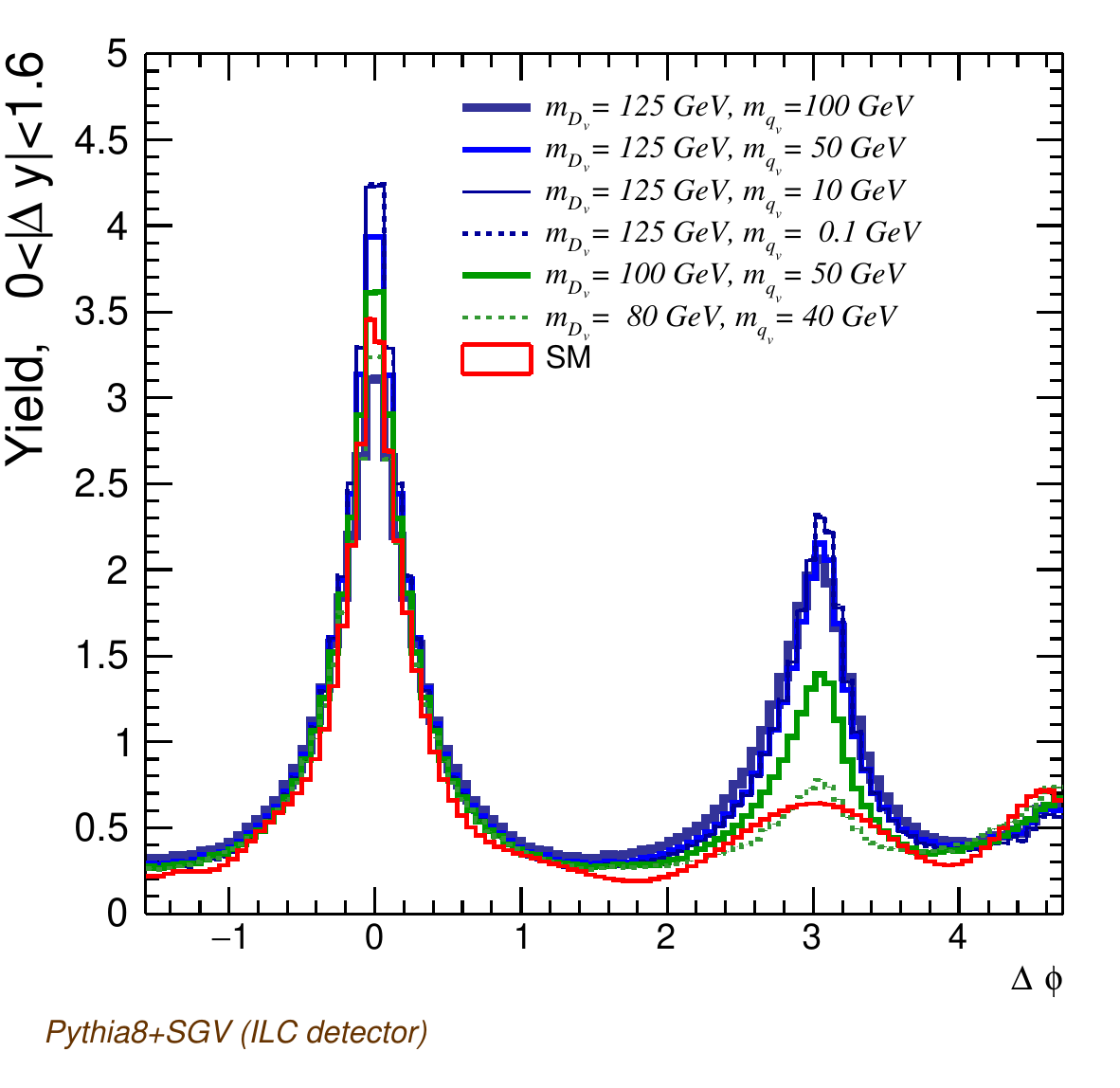}  
\hspace{1cm}
\includegraphics[width=7.00cm]{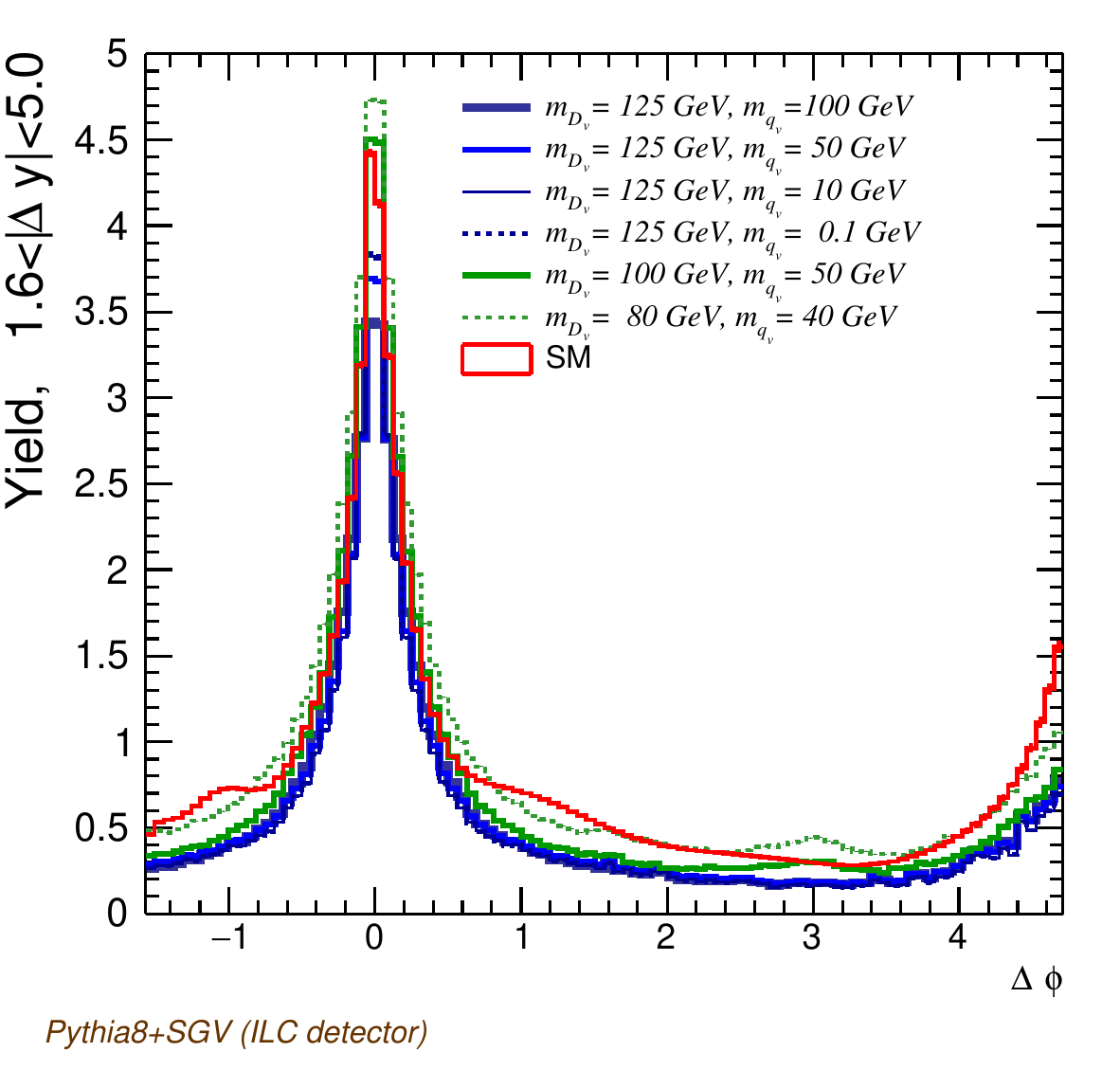}
  \caption{\label{fig:mitsou_yield} Detector-level yield $Y(\Delta \phi)$ for both HV signal in association with SM and for pure SM background (red line) for  $0<|\Delta y|<1.6$ (left) and $ 1.6<|\Delta y|<5$ (right), after selection cuts.}
\end{figure*}

A  search for BSM physics in high-energy collisions relying on angular correlations offers several advantages with respect to conventional direct methods. In particular, yield distributions  defined in Eq.~\eqref{eq:mitsou_yield} benefit from an almost total
cancellation of reconstruction efficiencies and detector acceptances, luminosity and cross-section dependence, etc. However, modelling uncertainties could be a limiting factor for these observables. 


The study is extended at particle level up to $\sqrt{s}=500~\GeV$ and 1~\TeV, also including $T_v\bar{T}_v$ pair production when kinematically possible. 
We consider the lightest communicator $D_v$, and the heaviest one $T_v$ which decays into $ q_v t$. 
Besides the $\PQq\PAQq$ and $\PW\PW\rightarrow 4\PQq$ backgrounds, we also take into account $\PQt\PAQt$, which becomes relevant at these energies. 
As expected, the SM backgrounds decrease with energy. For the HV signal, a reduction of the cross section by two orders of magnitude is obtained at $\sqrt{s}=1~\TeV$. Besides that, a study at $\sqrt{s}=240~\GeV$ for an ILD-for-FCC-ee detector showed similar results~\cite{fcc-note}.

Our results show that two-particle azimuthal correlations in an \epem Higgs factory could indeed become a useful tool to discover BSM, if kinematically accessible. Although a specific
HV model has been employed, other types of hidden sectors may yield similar signatures. 
In addition, collective effects stemming from a source other than BSM cannot be excluded.
Such searches, based on rather diffuse signals that spread over a large number of final-state particles, should be contemplated as complementary to other more conventional methods, thereby increasing the discovery potential of these machines.

\subsection{A portal scenario: heavy neutral leptons \label{sec:SRCH-NHL}}
In Section~\ref{llps:hnls} heavy neutral leptons (HNLs) were briefly introduced from a feebly-interacting point of view. In this chapter, the topic is further expanded beyond long-lived signatures and into prompt searches. 
\editors{Juergen Reuter}
\subsubsection{Theoretical introduction}
\label{sec:hnl_theo}
The existence of right-handed neutrinos is mandated by the observation of neutrino oscillations providing evidence that neutrinos have a non-zero mass. Right-handed neutrinos are not part of the SM, but they can be easily added with Yukawa interactions for up-type fermions to generate their masses through the Higgs mechanism. However, they do not have any SM gauge quantum numbers, and so a Majorana mass term is not forbidden by any SM symmetry. Through lepton-number violation and leptogenesis, several models do explain the baryon asymmetry of the universe via HNLs. Also, the missing CP violation needed for that could be hidden in the neutrino sector. HNLs could be connected to dark matter models if one of the sterile neutrinos were stable. There are several different mass regimes: ``light'' HNLs with $m_N \lesssim$ \SI{80}{\giga\electronvolt}, where they could show up in rare or invisible decays of the $\PZ, \PW, \PH$; then an intermediate regime, $80 \lesssim m_N \lesssim$ \SI{125}{\giga\electronvolt}; and the regime of ``heavy'' HNLs above the Higgs mass, where HNLs always decay promptly and could have decays into $\PW, \PZ, \PH$. For the light HNL category, the decays can still be prompt, feature displaced vertices or escape the detector as long-lived particles. This last category of signatures overlaps with~\cref{sec:LLP-focus-topic}. As with many other BSM signatures, also for HNLs there are (semi-)UV-complete models (e.g.\ type I seesaw) or simplified models in which there is one or several HNLs parametrised just by its mass(es) and width(s) and their couplings to the SM particles.

HNLs can be of Dirac nature with only lepton-number conserving (LNC) decays or of Majorana nature with lepton-number violating (LNV) production and decay processes. It is also possible that two quasi-generate Majorana neutrinos mix after electroweak symmetry and form so-called pseudo-Dirac neutrinos. In that case, neutrino--anti-neutrino oscillations are possible. Such a scenario is studied in~\cref{sec:hnl_osc,sec:HNL_Prop}. The study in~\cref{sec:HNL_Higgsdec} deals with the case of lighter neutrinos in exotic Higgs decays, while the studies in~\cref{sec:HNL_FCCee,sec:HNL_muX,sec:HNL_NPortal} look for ``light'' HNLs in $\PZ$ decays at the \PZ pole. Heavier neutrinos with masses above the Higgs boson and even into the multi-TeV range are considered in~\cref{sec:HNL_enjj}. The study in~\cref{sec:HNL_RHN} is the only one that investigates signatures where the HNL is pair-produced at a HTE factory. 

\subsubsection{Search for heavy neutral leptons through exotic Higgs decays}~
\label{sec:HNL_Higgsdec}
The theoretical basis of the study involves several models that predict HNLs. Depending on the model, the HNLs can explain the matter-antimatter asymmetry \cite{darkneutrino}, or the neutrino mass through the Seesaw mechanism \cite{hnl_snowmass, seesaw, inverse_seesaw, neutrino_mass, neutrino_mass2}. In many of these models, HNLs couple with SM particles and can be produced when the Higgs boson decays. This study focuses on an HNL $N_d$, with a mass between the \PZ and \PH bosons, making it detectable at the ILC. The free parameters related to this HNL are then the HNL mass $m_N$ and the mixing parameter $\varepsilon_{id}$ with the SM neutrinos, where $i\in\{\Pe, \PGm, \PGt\}$.

Full detector simulations of signal and background events are used. 
The \textsc{ILCSoft} v02-02 software package \cite{ilcsoftDoc} is used for MC simulation and reconstruction. \guineapig is used to simulate the beam parameters, beamstrahlung, and initial state radiation \cite{beamstrahlung, guinea-pig}. MC samples of the signal and SM background events are generated using \whizard 2.8.5 HNL masses, from 95 to \SI{120}{\giga\electronvolt}. \pythia 6.4 handles parton shower and hadronisation \cite{pythia6.4}, and the ILD detector response is simulated using \ddhep \cite{Frank_2014,Frank_2015} and \geant 4 \cite{geant4, geant4dev, geant4app}. Event reconstruction is performed with \marlin \cite{marlin}, employing the \pandora algorithm \cite{pfa} for particle flow analysis. For applying selections to the MC samples, the ROOT framework~\cite{root}, Jupyter notebooks \cite{jupyter}, and TMVA \cite{tmva} are used.
\begin{figure}[h]
    \centering
        \includegraphics[width=0.45\linewidth]{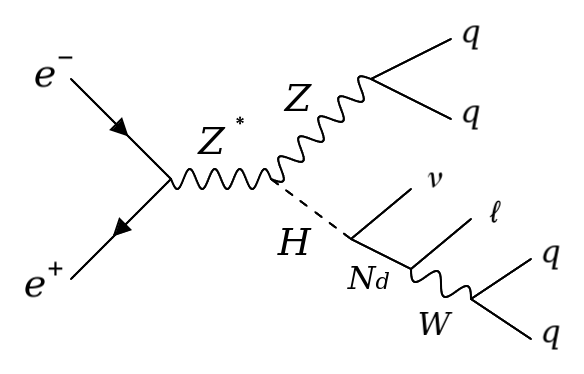}
        \includegraphics[width=0.45\linewidth]{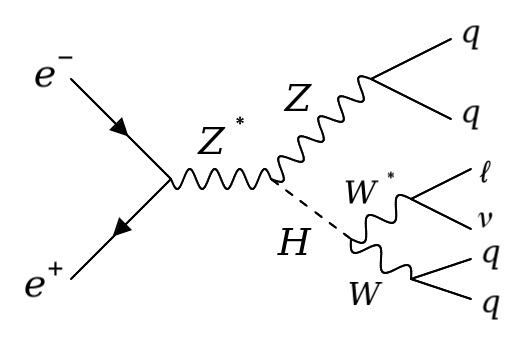}
    \caption{Feynman diagrams for the signal (left) and the main background (right) in the search for HNLs through exotic Higgs decays.}
    \label{fig:sig_bkg_feyn}
\end{figure}

In this study, we take advantage of the leading Higgs production channel at the ILC (Higgs-strahlung process) $\epem\to \PZ\PH$ and look for the Higgs exotic decay mode $\PH\to\PAGn N_d$. We concentrate on the dominant decay channel, where $\PZ\to \PQq \PAQq$ and $N_d\to \Plm \PWp\to \Pl\PQq\PAQq$\footnote{The charge conjugate channel is also targeted as our signal process and is implied throughout the section.}. The Feynman diagram of this signal process is shown in \cref{fig:sig_bkg_feyn} (left). The dominating background for this signal is shown in \cref{fig:sig_bkg_feyn} (right).
The observable will be the event rate of the signal process, which can be converted into a joint branching ratio of \PH and $N_d$ decays, BR$(\PH\to \PAGn N_d)\cdot BR(N_d\to \Plm \PWp)$, which can be computed as a function of the two free parameters $m_N$ and $\varepsilon_{id}$.

Events are filtered based on a one-dimensional selection, as well as a machine learning-based selection on various parameters, such as the reconstructed mass of the HNL, the invariant mass of the jets, the missing momentum, and the lepton momentum. The selections are applied separately for the different simulated HNL masses and the two beam polarisations $(-0.8, +0.3)$ and $(+0.8, -0.3)$ that have been proposed for the ILC. The integrated luminosity is set to \SI{1}{\per\atto\barn} per beam polarisation, totaling \SI{2}{\per\atto\barn}. The details of this analysis can be found in Ref.~\cite{Thor:2023nzu}. 
The signal significance, discovery and exclusion contours are shown as a function of the branching ratio and the HNL mass in \cref{fig:exclusion}. The exclusion limits present a factor of 25 improvement compared to the projected limit by HL-LHC \cite{hl-lhc-hn}. Reinterpreting the branching ratio as the mixing parameters $\varepsilon_{id}$ give the results in \cref{fig:overlay_exclusion}. Here, we see that the limit from ILC is a factor of 10 better than current limits.
\begin{figure}[h]
    \centering
    \includegraphics[width=0.6\linewidth]{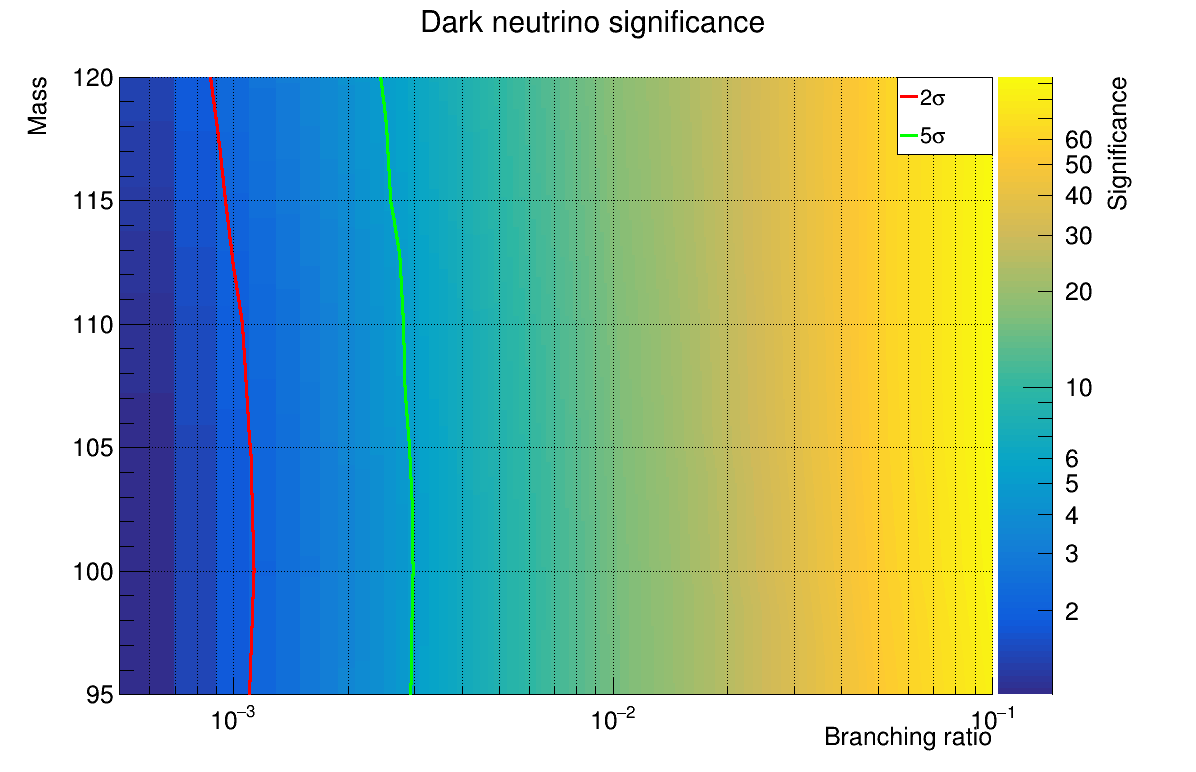}
    \caption{Search significance as a function of branching ratio BR$(H\to \nu N_d)BR(N_d\to lW)$ ($x$-axis) and HNL mass ($y$-axis), for the signal shown in \cref{fig:sig_bkg_feyn}. The red (green) curve indicates the $2\sigma$ exclusion (5$\sigma$ discovery) limit. Note that the significance is interpolated in between the tested mass points and branching ratios.}
    \label{fig:exclusion}
\end{figure}
%
\begin{figure}[htbp]
    \centering
        \includegraphics[width=0.48\linewidth]{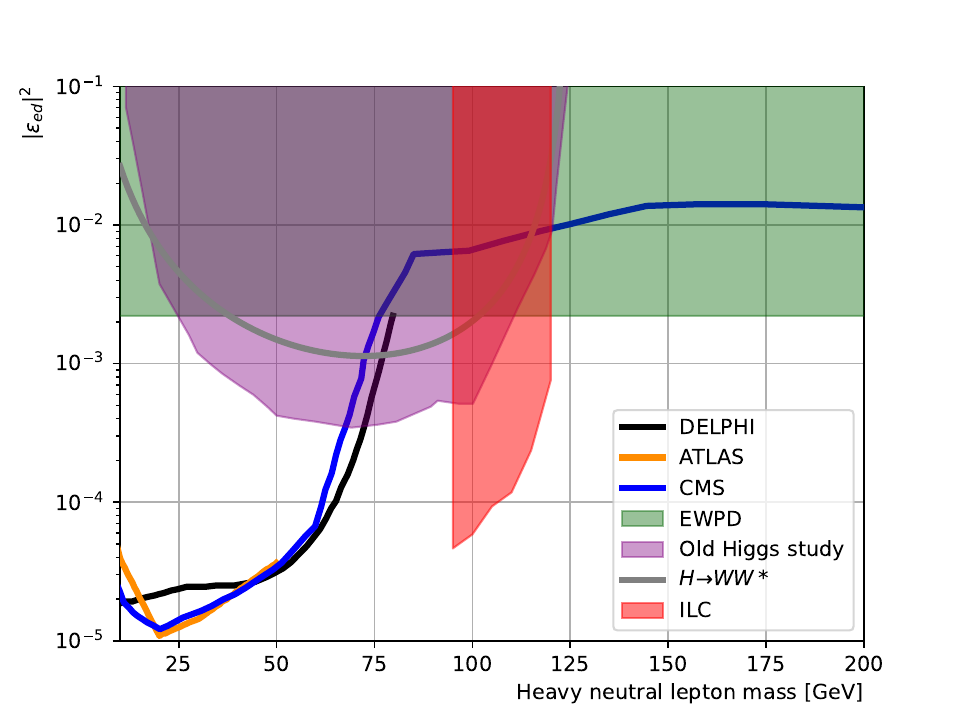}
        \includegraphics[width=0.48\linewidth]{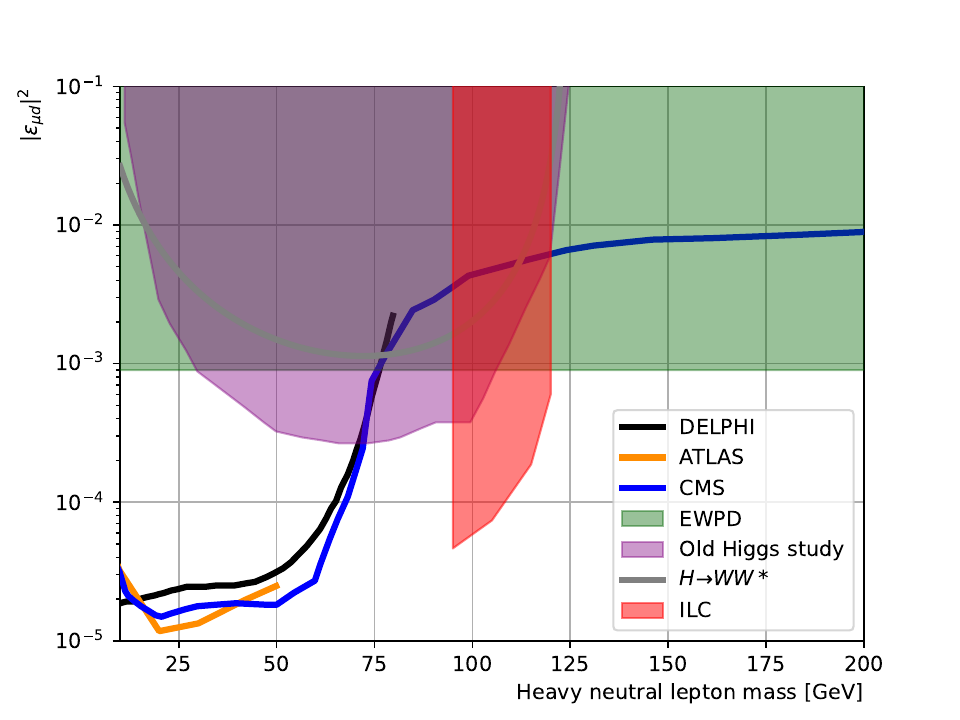}
    \caption{Exclusion curves as a function of the squared amplitude of the mixing matrix element ($y$-axis) and the HNL mass ($x$-axis) for the electron (left) and the muon neutrino (right). The red region shows the new results from the ILC study \cite{Thor:2023nzu}.
    }
    \label{fig:overlay_exclusion}
\end{figure}

In summary, 
six HNL masses and two beam polarisation schemes at ILC were analyzed. Both one-dimensional and machine learning selections were applied, which improved the signal-to-background ratio from $O(1/10^5)$ to $O(1)$. The study achieved a $2\sigma$ significance for a branching ratio of $0.1\%$, improving constraints on the HNL mixing matrix parameters by a factor of 10 compared to previous limits.
\subsubsection{Search for type I seesaw mechanism in a two heavy neutral leptons scenario}
\label{sec:HNL_FCCee}

The type I seesaw mechanism~\cite{Minkowski:1977sc} is a theoretical scenario of great interest in investigating the origin of neutrino masses beyond the SM via the introduction of HNLs.
One distinctive aspect of this study is the generalization of the underlying model's assumptions considering a more realistic scenario with two generations of quasi-degenerate HNLs coupling to all lepton flavours.  

This scenario is addressed with a simulation of the HNL mediated process in \madgraphee v.3.5.3 \cite{Alwall:2014hca} integrated with the model \textit{HeavyN} at leading order \cite{Alva:2014gxa,Degrande:2016aje}. Subsequently, \pythiaeight v.8.306 \cite{Bierlich:2022pfr} has been used to handle hadronisation as well as tau decays, and \delphes\ v.3.5.1pre05 \cite{deFavereau:2013fsa} to simulate the response of the IDEA detector~\cite{IDEA1} developed for FCC-ee. A few benchmark signals were chosen considering the ratios between mixing angles, $U^2_{\Pe}/U^2:U^2_{\PGm}/U^2:U^2_{\PGt}/U^2$, compatible with oscillation data \cite{Esteban:2020cvm}, leptogenesis production \cite{Granelli:2023tcj}, and flavour and electroweak measurements \cite{Blennow:2023mqx}. Furthermore, we choose $\Delta M=|M_{N_1}-M_{N_2}|=1\times10^{-5}$ GeV to account for the constraints given by the cosmological arguments \cite{Granelli:2023tcj,Canetti:2012kh} while 
enhancing the cross section. 
The main source of background comes from s-channel Z decays, in particular $\PZ\to\PGt\PGt$ with fully leptonic events and sizable missing energy. These processes have been centrally produced in the FCC-ee campaign, generated in \pythiaeight. Another background source is given by the SM processes producing the same final state as the HNL signal, $\Pl\Pl\PGn\PGn$. This background has been privately generated with \madgraphee using the UFO model \textit{sm-lepton\_masses}.

\begin{figure}[h!]
 \centering
    \begin{subfigure}[H]{0.49\textwidth}
       
        \includegraphics[width=\textwidth]{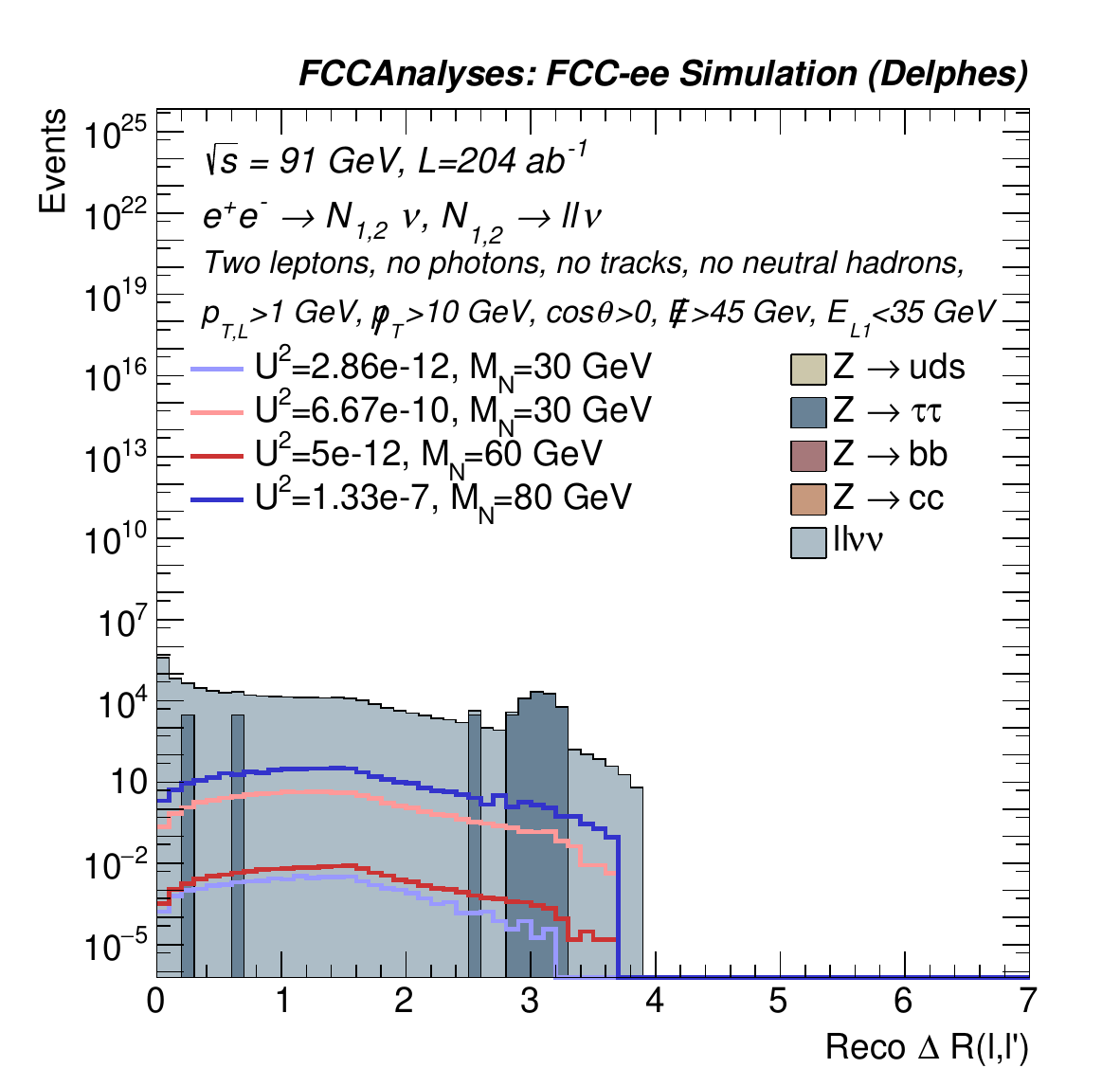}
    \end{subfigure}
    \begin{subfigure}[H]{0.49\textwidth}
        \includegraphics[width=\textwidth]{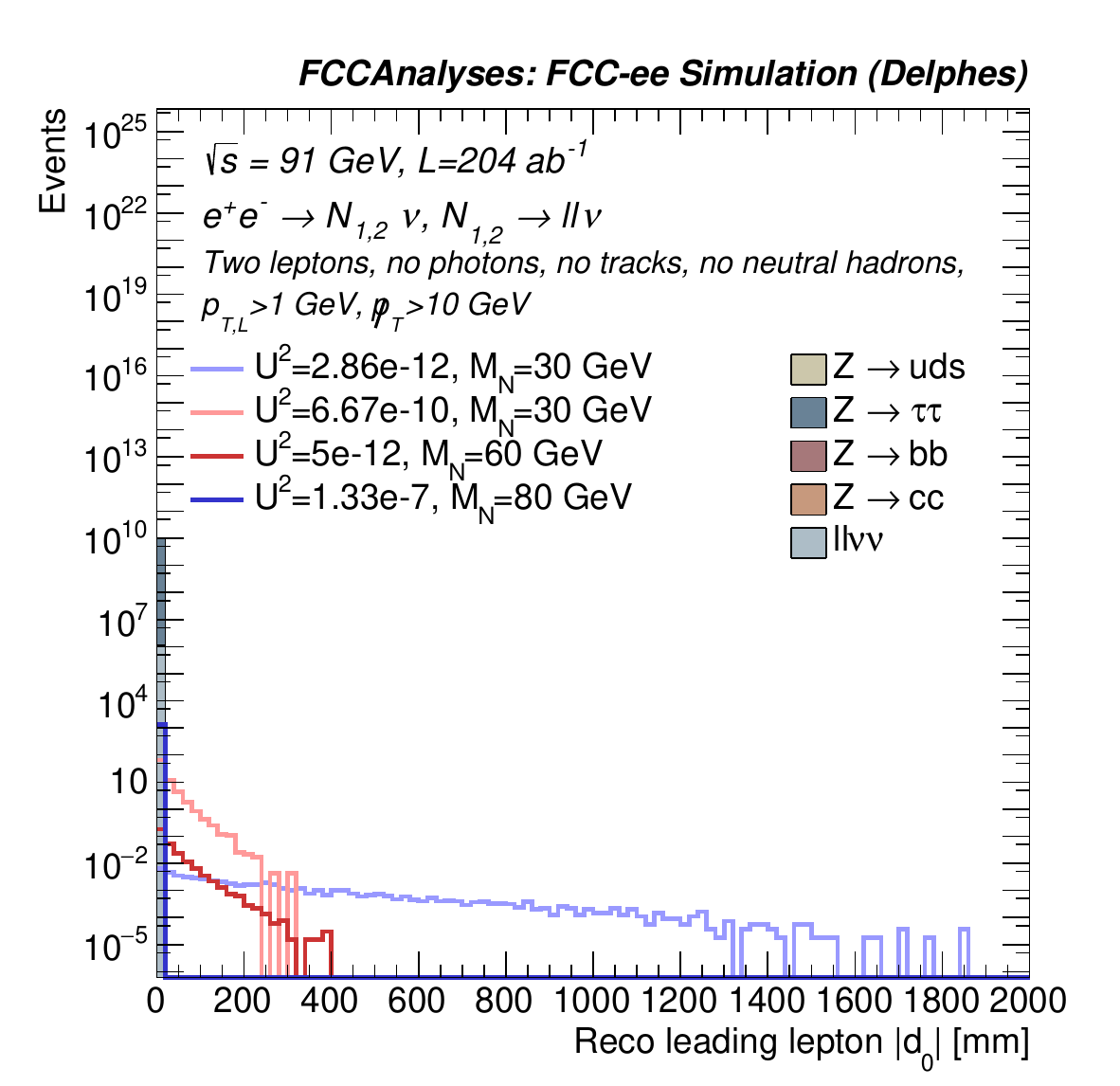}
    \end{subfigure}
    \caption{Left: Reconstructed angular distance between the final state leptons. Right: Lepton transverse impact parameter. Distributions are shown for selected signal benchmarks and backgrounds. 
    Only one mixing scenario, compatible with the inverted neutrino mass hierarchy, is shown.}
    \label{fig:KIT-HNL-results}
\end{figure}

The baseline selection includes two leptons with opposite charges with $p_{\mathrm{T}, {\Pl}}>1$ GeV, $E_{\Pl}>2$ GeV, ${p}^{miss}_{\mathrm{T}}>10$~GeV, and a veto on photons, additional tracks and neutral hadrons with $E>2$ GeV. An additional set of selections targetting specifically prompt decays includes an angular separation $\cos\theta>0$ between the two leptons, and requirements on the leading lepton energy $E_{leading\; \Pl}<35$ GeV, missing transverse energy ${E}^{miss}>45$ GeV and the invariant mass $M(\Pl, \Pl')<M_{HNL}$.
The angular distance between the two final state leptons models background processes better than other variables and is used for signal extraction. To evaluate the expected statistical significance, a binned maximum-likelihood fit is performed for each sample point chosen in the parameter space using the CMS Combine tool \cite{CMS:2024onh}. 
%
To specifically target long-lived HNLs, the baseline selection is augmented to also include requirements on the angular separation $\cos\theta>-0.8$ between the two leptons, their invariant mass $M(\Pl, \Pl')<80$ GeV, with additional requirements 
related to the vertex displacement, 
namely vertex $\chi^2<10$ and lepton transverse impact parameter $|d_0|>0.64$ mm.
Such selection yields a background-free analysis with current simulations. As the background modelling in this phase-space region is limited by the available sample sizes and the limitations of simulation, we simply provide the expected signal counts as the overall metric of the displaced analysis. Example distributions for the prompt and the displaced analyses for selected signal benchmarks and backgrounds are shown in Fig.\ref{fig:KIT-HNL-results}. 

\Cref{fig:KIT-HNL-results-final} shows the contours for the significance at 5$\sigma$ and the HNLs event counts in the case of the displaced analysis for one of the mixing patterns under study. Similar contours are obtained for all scenarios, consistent with the normal or inverted neutrino hierarchy. The results show that FCC-ee would be sensitive to HNLs in a realistic model containing two generations of HNLs in the type I seesaw mechanism in a large area of the available parameter space, using the data collected at the Z-pole run with an expected integrated luminosity of \SI{204}{\per\atto\barn}. 
Detailed information and additional figures can be found in Ref. \cite{Ajmal:2024kwi}.
\begin{figure}[h!]
\centering
    \includegraphics[width=0.65\textwidth]{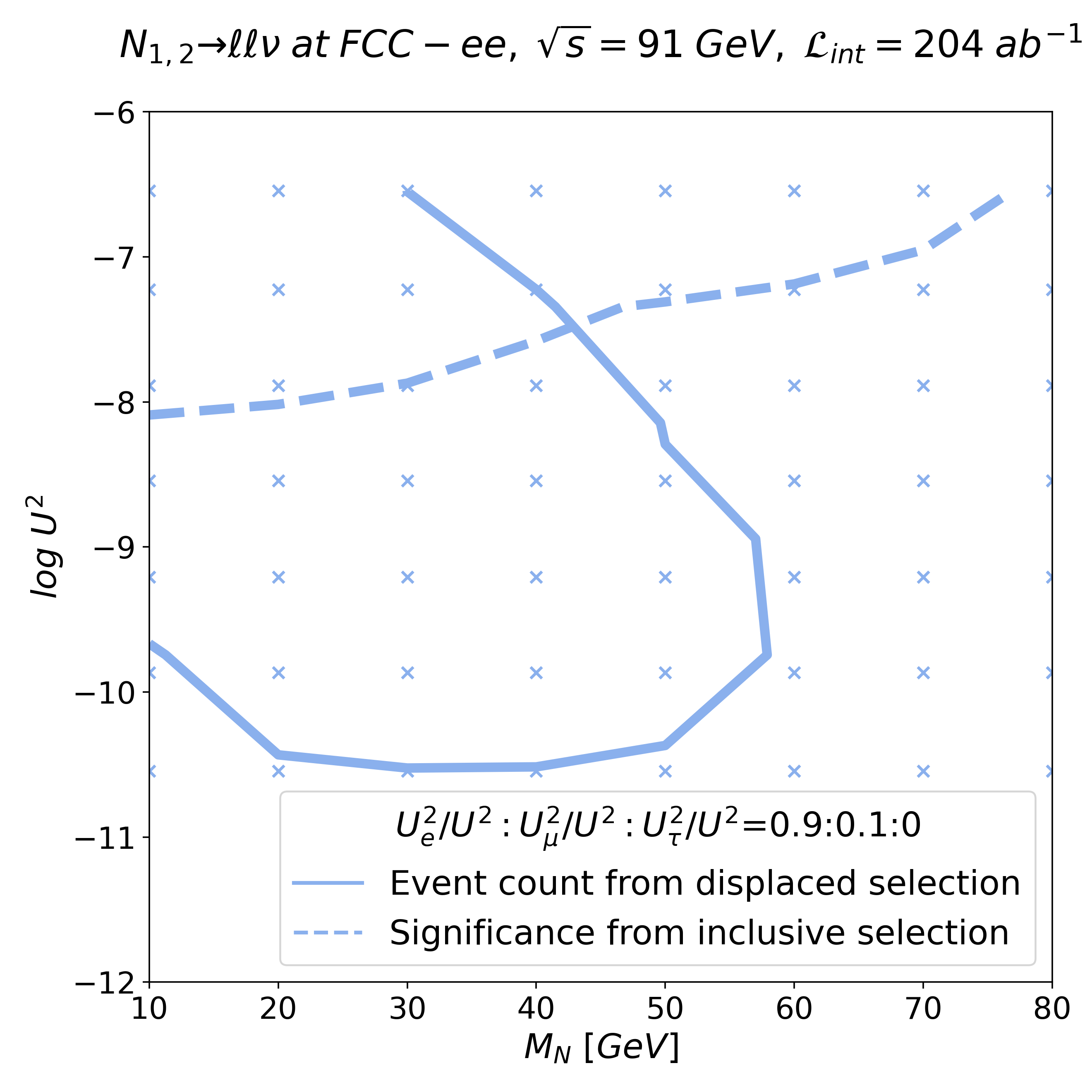}
    \caption{Expected 5$\sigma$ significance of HNLs extracted from the angular distance between the two final state leptons (dashed line) and at least 4 events of long-lived HNLs in a background-free scenario (solid line). Only one mixing scenario, compatible with the inverted hierarchy, is shown.}
    \label{fig:KIT-HNL-results-final}
\end{figure}

\subsubsection{Search for heavy neutrinos in prompt decays}
\label{sec:HNL_enjj}
%
Dirac and Majorana neutrinos with masses well above the electroweak scale have been proposed to solve certain issues of the SM. 
In our study~\cite{Mekala:2022cmm, Mekala:2023diu, Mekala:2023kzo}, we consider heavy Dirac and Majorana neutrinos decaying into two jets and a lepton at ILC running at 250\,GeV, 500\,GeV and 1\,TeV (with a total integrated luminosity of 2\,ab$^{-1}$, 4\,ab$^{-1}$ and 8\,ab$^{-1}$, respectively), CLIC at 3\,TeV (5\,ab$^{-1}$) and MuC at 3\,TeV and 10\,TeV (1\,ab$^{-1}$ and 10\,ab$^{-1}$, respectively). We assume that heavy neutral leptons mixing with the SM partners are the only relevant traces of New Physics, no other new phenomena occur and for simplicity, we study the case where only one heavy neutrino with a mass ranging from 100 GeV to 10 TeV couples to the SM particles.
The light-heavy neutrino pair production with their decay to two quarks and a lepton, $\Pl \Pl \to N \PGn \to \PQq \PQq \Pl \nu$, was considered. This channel offers the possibility of the full reconstruction of the heavy neutrino from two jets and a lepton measured in the detector. 

We generated signal and background events in \whizard v2.8.5~\cite{Moretti:2001zz,Kilian:2007gr} (ver. 3.0.0 was used for the Majorana signal generation). Parton shower and hadronisation were modelled  with \textsc{Pythia~6}~\cite{Sjostrand:2006za}. We generated reference samples with the mixing parameter $V_{\Pl N}$ set to the same value for all the leptons, and all the quark, electron and muon masses set to zero. To account for detector effects, the framework for fast detector simulation \delphes\ 3.5.0~\cite{deFavereau:2013fsa} was employed. The default cards for each collider project were used for detector parametrisation. Based on the expected signal topology consisting of one lepton and two reconstructed jets, we used the exclusive two-jet clustering mode.
In the next step, we trained a  BDT classifier, as implemented in the \textit{TMVA} package~\cite{Hocker:2007ht}. Eight input variables characterising the kinematics of the process were used. The BDT response distribution was then used to build a model describing the measurement within the \textit{RooStats} package~\cite{Moneta:2010pm}.
By scaling $V^2_{lN}$ with respect to the reference scenario, we extracted the expected 95\% C.L. limits on the mixing parameter using the CL$_s$ approach \cite{Read:2000ru}. 

The difference between Dirac and Majorana particles lies in their CP properties. This means that for Majorana particles for any specific decay channel also its CP-conjugated one exists, leading to lepton-number violation, while for Dirac neutrinos only one of them (lepton-number conserving). The chiral nature of weak decays together with the averaging over the decay process and its CP-conjugate for Majorana neutrinos leads to an experimental sensitivity, in particular, to the emission direction of a given final state particle (or anti-particle) in the rest frame of the decaying heavy neutrino.
Hence, for the Dirac vs. Majorana discrimination, we considered two additional variables: the cosine of the lepton and dijet emission angle multiplied by the lepton charge.
In this extended framework, the BDT algorithm was trained to distinguish between a signal sample of lepton-number-violating (LNV) heavy neutrino decays and a background sample contaminated with a lepton-number conserving (LNC) signal sample with some arbitrary weight. For the second training, the LNC decay sample was used as a signal and the LNV decay sample was used to contaminate the background. Then, 2-dimensional distributions of the sum and the difference of the two BDT responses were used for statistical analysis.
To find the minimal coupling allowing for model discrimination at 95\% C.L., which we will refer to as the discrimination limit, the signal normalisation was varied for each mass to obtain the value of the $\chi^2$-test statistic corresponding to the critical value of the $\chi^2$ distribution for probability $p=0.95$ and the considered number of degrees of freedom.

In Fig. \ref{mekala-fig:results}, the coupling limits obtained for Dirac neutrinos at future lepton colliders are compared with limits estimated for hadron machines. The CMS limits at LHC running at 13\,TeV (Fig. 2 in Ref.~\cite{Sirunyan:2018mtv}) were obtained assuming the Majorana nature of the neutrinos. The projections for HL-LHC and future possible successors of the LHC were adapted from Ref.~\cite{Pascoli:2018heg}.
\begin{figure}[t]
    \centering
    \includegraphics[width=0.65\textwidth]{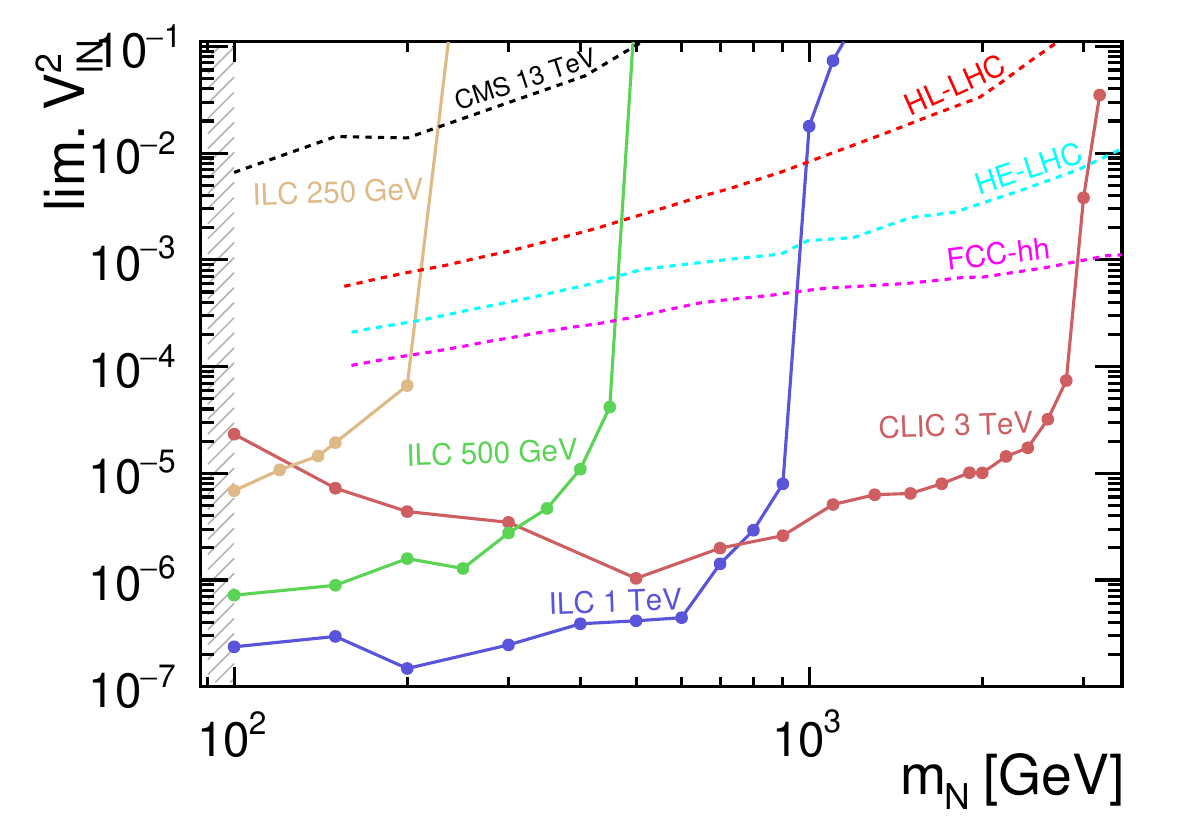}
    \caption{Expected 95\% CL exclusion limits on the mixing parameter $V^2_{lN}$ for future $\Pep \Pem $ and hadron colliders, along with current limits from LHC, as a function of the heavy neutrino mass, $m_N$.}
    \label{mekala-fig:results}
\end{figure}
\begin{figure}[t] 
    \centering
    \includegraphics[width=0.65\textwidth]{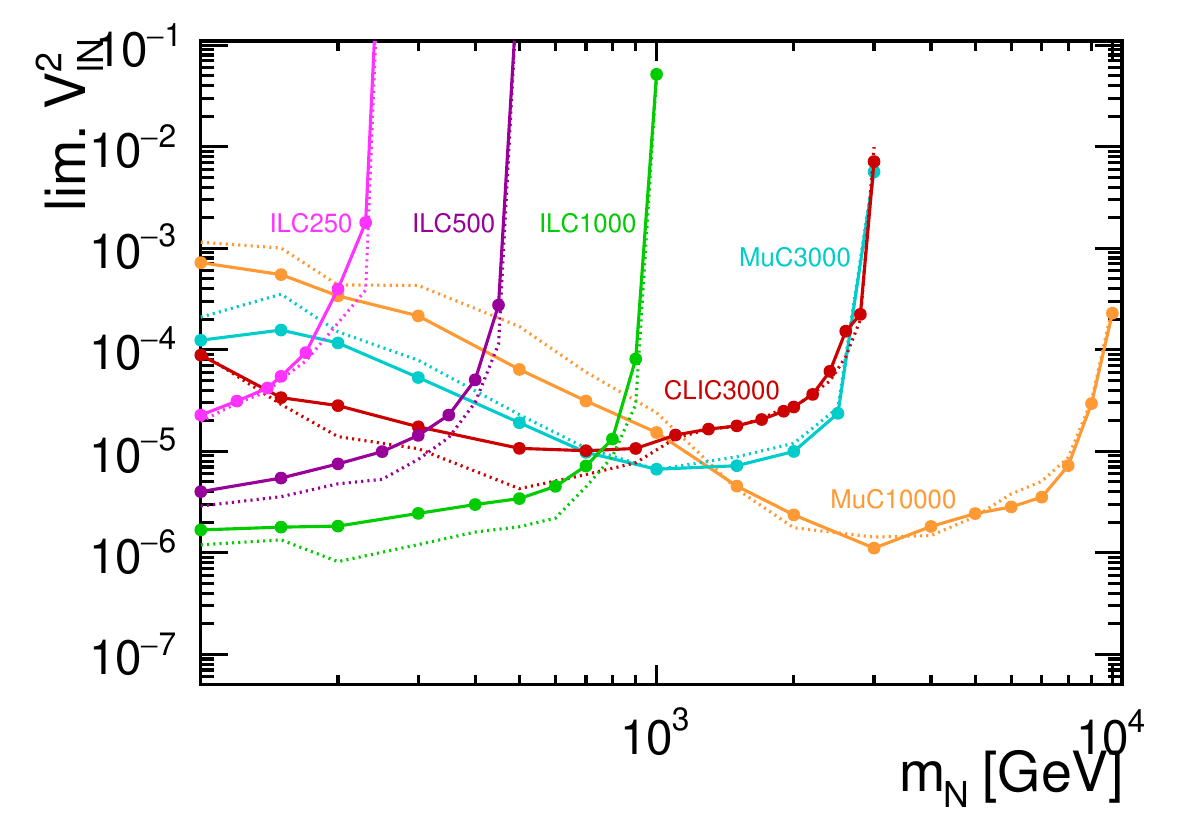}
    \caption{Comparison of expected 95\% C.L. discrimination limits (solid lines) and 5$\sigma$ discovery limits (dotted lines) between Majorana and Dirac neutrinos, as a function of the heavy neutrino mass, $m_N$, for different collider scenarios considered in the study.}
    \label{mekala-fig:results_DvM}
\end{figure}
The final results of our study for the Dirac vs. Majorana discrimination are shown in Figure~\ref{mekala-fig:results_DvM}. The 95\% C.L. discrimination limits are compared to the 5$\sigma$ discovery limits for the six collider scenarios considered in our work. The analysis confirms that once the heavy neutrinos are discovered at lepton colliders, it will be possible to determine their nature (real or complex Lorentz representation). 

The proposed analysis strategy resulted in estimating
limits on the $V^2_{lN}$ coupling which are much more stringent than any results for high-energy hadron colliders. 
Our analysis shows that, by employing variables encoding the chiral character of the particles, one may efficiently discriminate between complex and real Lorentz representations (i.e.\ Dirac or Majorana nature) of the heavy neutrinos simultaneously with their discovery at future lepton colliders.

\subsubsection{\texorpdfstring{Search for heavy neutral leptons in $\PGm+\text{X}$ channels}{Search for heavy neutral leptons in muon+X channels}}
\label{sec:HNL_muX}
Dirac or Majorana fermions with sterile neutrino quantum numbers are a very popular BSM extension, potentially addressing unresolved questions in neutrino physics, such as the origin of neutrino mass and oscillations through seesaw mechanism.
To fully exploit the discovery potential, the experimental setup must be designed to detect a wide range of signatures, from prompt to displaced decays, posing severe requirements on the detector performances. This study focuses on HNL production within a minimal scenario defined by two parameters: the HNL mass and its coupling to SM fermions. We present a study assessing the FCC-ee's ability to probe new regions of the HNL parameter space, providing new exclusion limits. 

The model was implemented using the \texttt{SM-HeavyN-LO} UFO \cite{Atre:2009rg} and the events were generated with\linebreak \madgraphee \cite{Alwall:2014hca}. Assuming a single HNL flavour production, all heavy neutrino mixing terms were set to zero except for the mixing parameter with the muon, $U_{\PGm}$. 
Signal samples were generated for masses ranging from 5 to 85 \SI{}{\giga\electronvolt} at the $\Pep\Pem$ centre-of-mass energy $\sqrt{s}=\SI{91.2}{\giga\electronvolt}$. Coupling values were scanned, from the minimum needed for decays within 2.5 metres from the interaction point, to a coupling squared excluded by current experiments ($U_{\PGm}^2 \simeq  \times 10^{-4}$). The generated partons were hadronised using \pythiaeight and processed with \delphes\ for a parametrised simulation of the IDEA detector \cite{IDEA1}, utilizing the official detector data cards. Background samples from FCC-ee productions were processed through the same \pythiaeight-\delphes\ \cite{winter2023samples}.

The decay channels investigated in this study consist of either a muon and a quark-antiquark pair ($\PGm\PQq\PQq'$) or two muons and a neutrino ($\PGm\PGm\PGnGm$), both mediated by an off-shell $\PW$ or $\PZ$ boson. The first channel results in a fully visible HNL final state with of one muon and two jets, with a branching fraction of approximately 50\% across the considered mass range. The second channel is a clean one, producing only two reconstructed muons and high missing energy.
At low HNL mass the highly boosted HNL produces collimated decay products. As the mass increases, the HNL is produced closer to rest, resulting in well-separated jets from the hadronic channel. 
The SM background includes hadronic and leptonic $\PZ$ decays, with reconstructed muons from meson decays or with jets in the final state, as well as the non-resonant and irreducible four-fermion process ($\Pep\Pem\to \PGm \PGn \PQq\PQq$ emulating the $\PGm \mathrm{jj}$ channel, and $\Pep\Pem \to \PGm\PGm \PGnGm \PGnGm$, emulating the $\PGm\PGm \PGnGm$ channel). 

For the $\PGm \mathrm{jj}$ channel, all of the visible particles in the signal are produced in the decay of a single resonance, therefore a precise measurement of the four-momentum of the HNL and of the neutrino recoiling against it is obtained by combining the four-momenta of all the reconstructed particle flow objects, with no need for jet reconstruction.
However, the number of jets in the event, the angular distance between them, and the energy sharing are useful variables for background rejection.
After requiring an isolated high-energy muon ($E_{\PGm} > \SI{3}{\giga\electronvolt}$), the reconstructed particles were clustered with \fastjet \cite{Cacciari:2011ma} requiring at most two (exclusive \kT) jets.
Kinematic selections were optimized to enhance signal detection by constraining the final state kinematics. Leptonic decays of the $\PZ$ boson were suppressed by nearly 99\% by imposing a requirement on the distance between the muon and the (single) reconstructed jet, along with requirements on compatibility with a single neutrino in the final state. At higher mass the hadronic decays are more relevant, and background was filtered by applying a requirement on the angle between jets and using an isolation selection for the muon relative to the jets.
The resolution achieved with the current \delphes\ simulation ensures that about 90\% of the signal is contained within a mass (or missing energy) window centred around the nominal visible mass (or nominal missing energy) with a width scaling approximately as $2\times 10\% \times \sqrt{M}$.
Selecting events in that window effectively suppresses the background to below the per-mille level.
The residual background is dominated by the four-fermions channel, and $\PZ \to \PGt\PGt$ at low mass, and $\PZ$ to charm and bottom at mass above $\sim \SI{60}{\giga\electronvolt}$. 

A distinguishing characteristic of those processes is the presence of secondary vertexes.
Therefore, the HNL decay vertex was isolated by requiring a single reconstructed vertex with good $\chi^2$ and by requiring that at most 5 tracks do not contribute to the vertex. This effectively suppresses heavy-flavoured Z decays. 
The vertex displacement on the transverse plane $d_{\textrm{vert}}$ was also used to distinguish prompt from displaced decays and to set up two statistically independent analyses. The separation point was set to  $d_{\textrm{vert}} = \SI{0.5}{\milli\meter}$, approximately a factor of 5 higher than the value of the extreme tails for the available background samples. In the case of no background the kinematic selections could be further relaxed, allowing for a selection efficiency of more than $60\%$ across the mass range.  
For the $\PGm\PGm \PGnGm$ channel, the selection required exactly two tracks reconstructed as muons with a momentum greater than $\SI{3}{\giga\electronvolt}$, along with a displacement requirement on the position of the vertex reconstructed out of these tracks, ensuring complete background rejection. The condition $d_{\textrm{vert}} > \SI{10}{\milli\meter}$ was applied, together with a selection on the angle between the reconstructed tracks at the decay vertex to address the presence of fake displaced vertexes due to poor vertex reconstruction in case of back-to-back topology.

Exclusion limits at the 95\% C.L. were obtained for $d_{\textrm{vert}}$ above and below 0.5 mm for the $\PGm \mathrm{jj}$ channel, and for $d_{\textrm{vert}}$ above $\SI{10}{\milli\meter}$ for the $\PGm\PGm \PGnGm$ channel. A total luminosity of  $\SI{205}{\per\atto\barn}$  corresponding to $\SI{6d12}{}$ $\PZ$ bosons was assumed, spread over three different $\Pep\Pem$ energies ($\SI{88}{}$, $\SI{91.2}{}$ and $\SI{94}{\giga\electronvolt}$) with different production cross sections. 
The expected number of events were calculated using the formulas provided in Ref.~\cite{Drewes:2022rsk}, and the MC  statistics were scaled accordingly. For the non-resonant four-lepton channels, the \madgraph cross section was used, assuming a mild dependence on the centre-of-mass energy.

\begin{figure}[h] \centering 
\includegraphics[width=0.7\textwidth]{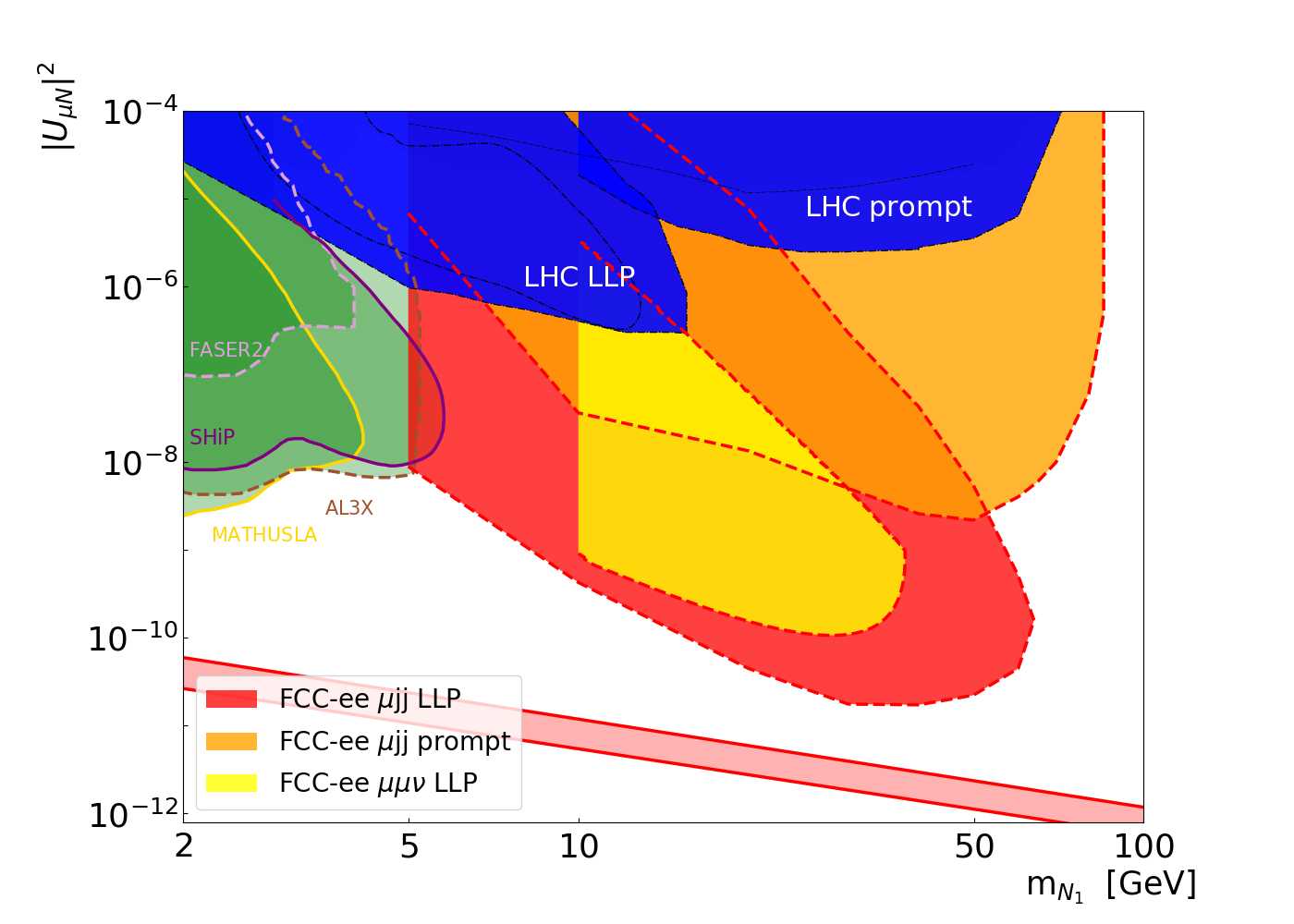} 
\caption{Expected exclusion limits at 95\% C.L. and 5$\sigma$ discovery potential in the $M_\mathrm{HNL}-U_{\PGm}^2$ plane. The FCC-ee results of this analysis are drawn as yellow, orange and red dashed lines \cite{Bolton:2019pcu}. The diagonal red band corresponds to the limit beyond which the neutrinos decay outside the detector region. \label{fig::HNLMU2::sensitivityres}}
\end{figure}
  
The results are presented and compared to existing experimental limits in \cref{fig::HNLMU2::sensitivityres}. The entirely suppressed background at large $d_{\textrm{vert}}$ greatly enhances sensitivity for HNLs  with masses below 50 GeV. Thanks to the large $\PZ$ event samples, couplings down to $U^2 \sim 10^{-11}$ and a range of HNL masses between 5 and $\SI{80}{\giga\electronvolt}$ could be covered by this analysis, far surpassing the theoretical projections for the coverage of HL-LHC across a similar mass range \cite{Drewes:2019fou}.

\subsubsection{Measuring heavy neutrino properties}
\label{sec:HNL_Prop}
%
In this study, we evaluate the potential for measuring different properties of HNLs at FCC-ee. First, we show how their mass can be precisely constrained through the measurement of the timing of the HNL decay vertex. Second, we analyze a realistic model where HNLs oscillate into their antiparticles, and study how the parameter space of the model can be experimentally constrained.

The samples used for the analysis are generated with \madgraphee  \cite{Alwall:2014hca}, hadronised with \pythiaeight and passed through a \delphes\ fast simulation of the IDEA detector \cite{winter2023samples}. A single HNL flavour is assumed, with mixing angle $U_{\PGm}\ll 1$ with the muon. The studied decay channel is $\mathrm{HNL} \to \PGm\PQq\PQq'$, having a branching fraction close to 50\%. The analysis is performed over a wide mass range and at values of $U_{\PGm}$ that remain unconstrained by existing experiments. A selection on the transverse position of the decay vertex ($d_{\textrm{vert}} > \SI{0.5}{\milli\meter}$) is applied to eliminate SM background while preserving sufficient statistics for the long lived HNL events. Minimal kinematic selections are applied to ensure high signal efficiency. Events are further selected requiring a single reconstructed HNL decay vertex with good $\chi^{2}$, and that most of the reconstructed tracks are attached to it \cite{HNLPavia::AN}.
The HNL decay channel into one muon and two jets is fully visible, allowing the kinematics of the HNL to be fully reconstructed on the basis of tracking and calorimetric measurements. The resolution on the HNL mass obtained in this way from current \delphes\ simulations is given in Ref.~\cite{HNLPavia::AN}. As a complementary measurement, in the process $\Pep\Pem \to \PGnGm \mathrm{HNL}\,$ the HNL mass can be expressed as a function of its velocity $\beta$ and the centre-of-mass energy of the collision. The velocity $\beta$ can be measured from the position of the HNL decay vertex and the time-of-flight of the decay products attached to it, thus providing an independent determination of the HNL mass.

The IDEA concept implements the central barrel with MAPS silicon sensors and ultralight drift chambers \cite{IDEA1}.
The intrinsic vertex resolution within the drift chambers is approximately $\SI{100}{\micro\meter}$, but the overall uncertainty is around \SI{250}{\micro\meter}, dominated by the limited knowledge of the coordinates of the $\Pep\Pem$ interaction vertex. The HNL decay time resolution depends on the resolution of the TOF detector, the number of tracks contributing to the vertex, and the precision in assigning their masses. This has been studied in detail in Ref.~\cite{Aleksan:2024hyq}. It has been demonstrated that, with TOF resolutions of a few tens of picoseconds, a mass resolution of a few percent can be achieved for long-lived particles  across a large part of the accessible parameter space.

A realistic HNL model is presented in Ref.~\cite{Antusch:2022ceb}. It introduces two nearly degenerate Majorana HNLs with identical mixing to SM particles and mass splitting $\Delta m$. As a consequence, the neutrino propagates as a pseudo-Dirac HNL oscillating into its antiparticle. 
This results in oscillations between lepton-number-conserving (LNC) and lepton-number-violating (LNV) processes. 
The parameter space includes the HNL mass ($M$), the mass splitting ($\Delta m$), which determines the oscillation length ($c\tau_\text{osc}$), and the HNL decay width, which determines the decay length ($c\tau_\text{dec}$).  For $c\tau_\text{osc} \gg c\tau_\text{dec}$, LNV processes are suppressed, and the HNL behaves effectively as a Dirac particle. Conversely for $c\tau_\text{osc} \ll c\tau_\text{dec}$, LNV is maximized and the model behaves effectively as a double Majorana model. Experimentally, the oscillation pattern can be observed as long as the oscillations are slow enough to be resolved by the detector;  otherwise, the HNL behaves as two independent Majorana states. Further details can be found in Ref.~\cite{Antusch:2024otj}.
The model described here is implemented in the software framework for FCC-ee PED studies including a detailed parametrisation of the IDEA detector. The aim of the present work is to study whether the striking signatures of this model can be actually measured in a realistic scenario, and whether the different regions of parameter space can be distinguished. To this purpose, samples have been simulated in the mass range between 15 and $\SI{35}{\giga\electronvolt}$, in a region accessible at FCC-ee with sufficient statistics, and varying $\Delta m$ from $\sim\SI{d1}{}$ to $\sim\SI{d3}{\micro\electronvolt}$ to cover the two opposite scenarios mentioned above.
\begin{figure}[hb!] \centering 
\includegraphics[width=0.4\textwidth]{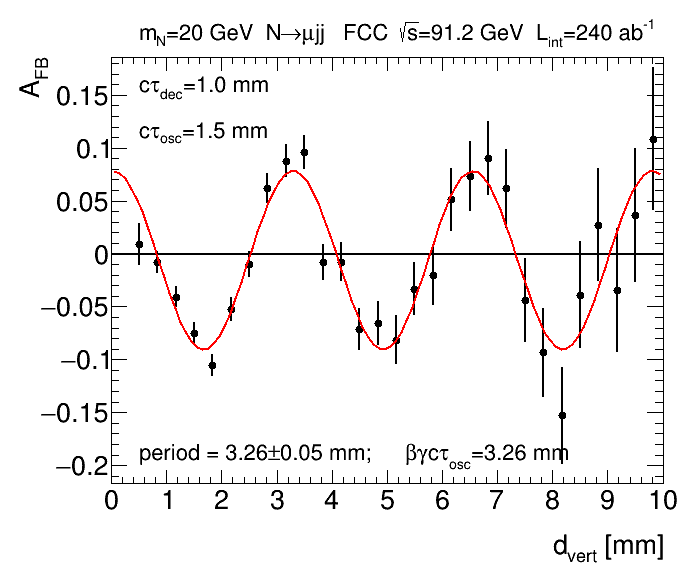} 
\includegraphics[width=0.4\textwidth]{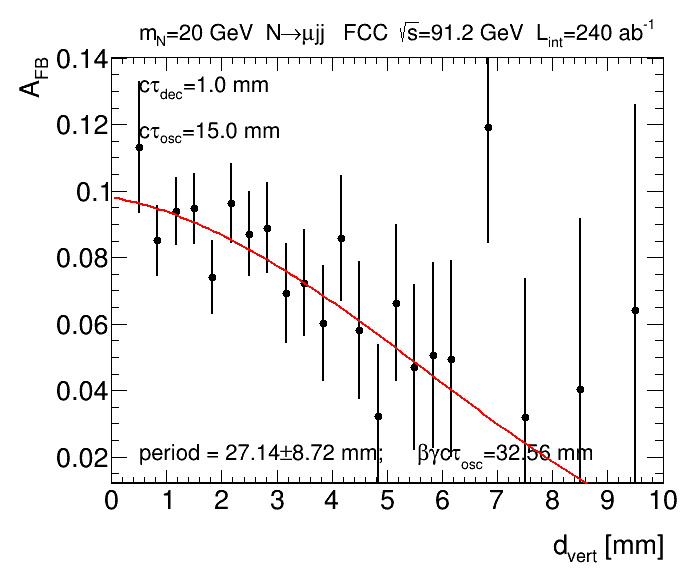} 

\includegraphics[width=0.4\textwidth]{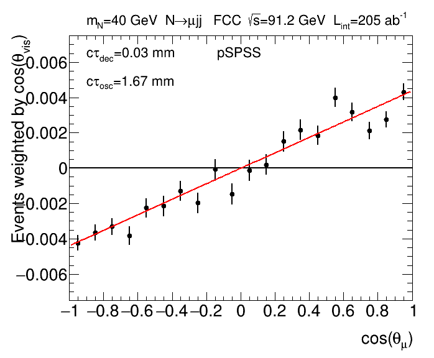} 
\includegraphics[width=0.4\textwidth]{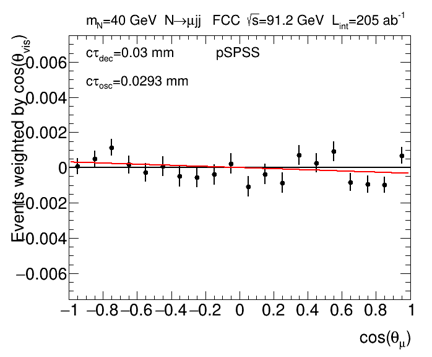}
\caption{Reconstructed observables for the pseudo-Dirac HNL model. \textit{Top:} Forward/backward asymmetry of the reconstructed muon  as a function of the vertex displacement, for $M_{\text{HNL}}=\SI{20}{\giga\electronvolt}$, $c\tau_\text{dec} = \SI{1}{\milli\meter}$ and two different oscillation lengths. In the left panel, where $\tau_\text{dec} \simeq \tau_\text{osc}$, the period can be measured with precision on the order of one percent. 
\textit{Bottom:} Distribution of $\cos \theta_{\PGm}$ in the HNL rest frame weighted by $\cos\theta_{HNL}$. The values of $(\tau_\text{dec}, \tau_\text{osc})$ represent complementary regions of the parameter space, showing either a pure Majorana (left) or a pure Dirac (right) behaviour. The error bar on the data points  correspond to the expected uncertainty for a real experiment with the $\PZ$-pole run statistics of $\SI{6d12}{}$ $\PZ$ bosons. } \label{fig::HNLPavia::oscillation}
\end{figure}

The experimental detection of the oscillation pattern is shown in the top panels of \cref{fig::HNLPavia::oscillation}. Although LNV and LNC processes cannot be directly distinguished because of the invisible $\nu$ produced together with the HNL, the polarisation of the intermediate $\PZ$ boson induces an angular asymmetry of the final-state particles. By correlating the charge of the muon (HNL daughter) with the hemisphere in which it is produced, a forward/backward asymmetry can be built. Its pattern as a function of the HNL decay path is shown for $c\tau_\text{osc} \simeq c\tau_\text{dec} = \SI{1}{\milli\meter}$ in the top-left panel of the figure. In this region of the parameter space, and with the given vertexing resolution, the mass splitting can be measured with percent-level precision. As $c\tau_\text{osc}$ becomes larger than $c\tau_\text{dec}$, a full oscillation period cannot be detected, as shown in the top-right panel. 
Similar or higher measuring power is obtained by measuring the dependence on the vertex displacement of variables sensitive to the polarisation of the HNL decay, such as the angle between the muon and missing momentum in the HNL rest frame, and of the total muon momentum.
Regions of parameter space with maximal or negligible LNV
fraction can be distinguished, even when the oscillations cannot be be measured, thanks to the Dirac-like or Majorana-like behaviour of the HNL mentioned above. A useful observable was found to be the polar angle distribution of the muon in the HNL rest frame, which is preferentially aligned with or against the HNL flight direction for Dirac HNLs.  If this variable is plotted weighted by the polar angle of the HNL in the lab frame, it  provides a clear distinction between the two cases. This is illustrated in the bottom panels of~\cref{fig::HNLPavia::oscillation}, where the  Dirac-like (left panel) and Majorana-like (right panel) cases are shown. Similar conclusions, and a comparable discriminating power, were obtained by analyzing the reconstructed HNL angle distribution in the laboratory frame.

The potential of FCC-ee to detect Heavy Neutral Leptons and measure their properties through long-lived decays and oscillations has been presented. A realistic simulation of the IDEA detector performance has been used to explore HNL mass constraints via decay timing and displacement, achieving percent-level precision. A realistic model with two nearly degenerate Majorana HNLs was implemented, and key observables were analyzed to distinguish between Dirac and Majorana behavior, evaluating how mass splitting and oscillation patterns can be measured experimentally.

\subsubsection{Search for heavy neutrino-antineutrino oscillations}
\label{sec:hnl_osc}
Low-scale type I seesaw models generically predict HNLs to be almost mass-degenerate pseudo-Dirac pairs of two Majorana fermions.
Despite lepton number (LN) being an approximate symmetry in this scenario, LN violation can nevertheless be observable at colliders due to neutrino-antineutrino oscillations. The observed neutrino oscillations can be explained by the type I seesaw mechanism after the introduction of at least two right-handed neutrinos \cite{Minkowski:1977sc}.
Measurements of final state distributions at future lepton colliders facilitate tests of this intriguing nature of HNLs.
We show a study scanning over the pseudo-Dirac HNL's parameter space that allows us to quantify the discovery potential of neutrino-oscillations at the \PZ-pole run of the FCC-\Pe{}\Pe.
%
In particular, displaced vertex searches for long-lived heavy neutrinos 
will be a powerful probe.

Models that do not introduce additional bosons or a high degree of fine-tuning must be symmetry protected in order to be detectable at FCC-\Pe{}\Pe.
This symmetry protection ensures that the two right-handed neutrinos form an almost mass-degenerate pseudo-Dirac HNL with a mass splitting $\Delta m$ that is correlated to the SM neutrino mass scale \cite{Antusch:2022ceb}.
Similar to the particle-antiparticle oscillations that have been observed in neutral meson decays, this mass splitting leads to heavy neutrino-antineutrino oscillations (NNOs).
The angular frequency of these oscillations is given by $\Delta m$ \cite{Antusch:2020pnn}.
The observation of such oscillations would constitute a remarkable measurement of the lepton number violation introduced by the seesaw model.

\begin{figure}[h]
\centering
\begin{subcaptionblock}{.25\linewidth}
\centering\includegraphics[width=\linewidth]{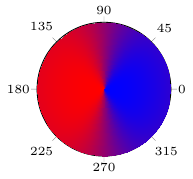}
\caption{Dirac limit} \label{fig:hnl-osc-BM1}
\end{subcaptionblock}\hfil
\begin{subcaptionblock}{.25\linewidth}
\centering\includegraphics[width=\linewidth]{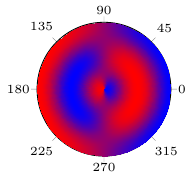}
\caption{Slow oscillations} \label{fig:hnl-osc-BM2}
\end{subcaptionblock}\hfil
\begin{subcaptionblock}{.25\linewidth}
\centering\includegraphics[width=\linewidth]{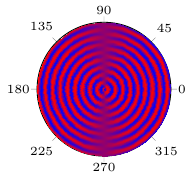}
\caption{Fast oscillations} \label{fig:hnl-osc-BM3}
\end{subcaptionblock}\hfil
\begin{subcaptionblock}{.25\linewidth}
\centering\includegraphics[width=\linewidth]{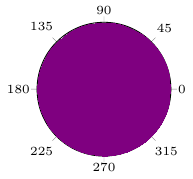}
\caption{Double-Majorana limit} \label{fig:hnl-osc-BM4}
\end{subcaptionblock}
\caption{
Illustrations of the asymmetric and oscillating distributions of the charges of the final state leptons (indicated by red and blue), taken from \cite{Antusch:2023jsa}.
The radial coordinate corresponds to the proper time of the HNL decay, while the angular direction corresponds to the polar angle of the position of the displaced vertex.
In the Dirac limit, the HNL exhibits a forward-backward asymmetry shown in panel (a).
With increased LN violation this asymmetry is modulated by oscillations shown for two different frequencies in panels (b) and (c).
Finally, in the double-Majorana limit, the oscillations wash out the asymmetry, leading to a uniform charge distribution indicated by purple in panel (d).
} \label{fig:hnl-osc}
\end{figure}
%
Since the interplay between decay and oscillations determines the amount of LN violation present in the model, the ratio $R$ between the number of LN-violating over the number of LN-conserving events is governed by the mass splitting $\Delta m$ and the decay width $\Gamma$.
On the one hand, when the decay happens much faster than the oscillation, this ratio approaches zero and the HNL behaves akin to a Dirac particle.
On the other hand, if the decay length is considerably longer than the oscillation length, this ratio approaches one and the HNL behaves similar to two Majorana particles whose mass peaks can not be distinguished from each other.
The generic case lies in between these two extreme scenarios.
However, the measured value of this ratio can differ from the naively predicted value due to the detector geometry \cite{Antusch:2022ceb} and decoherence of the oscillations~\cite{Antusch:2023nqd}.

\begin{figure}[h]
\centering\includegraphics[width=\linewidth]{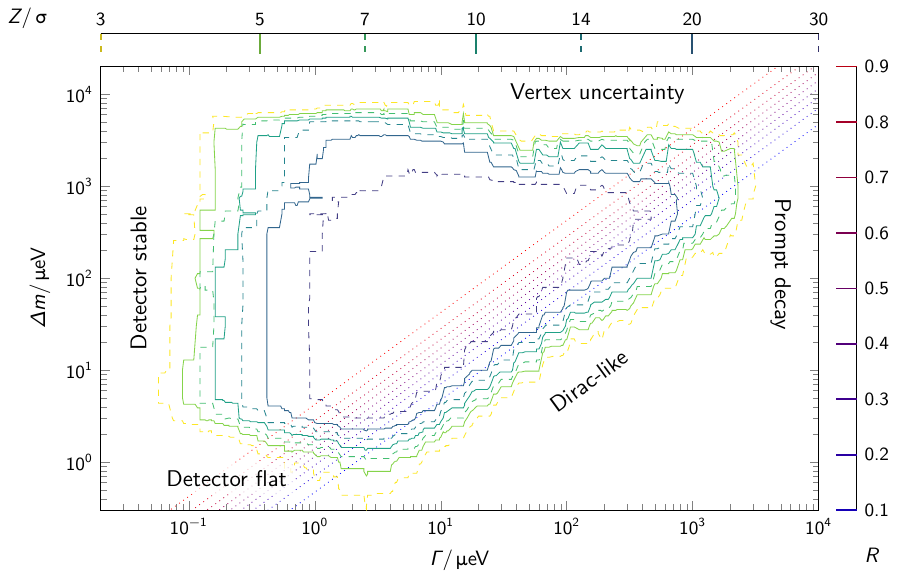}
\caption{
Discovery reach for heavy neutrino-antineutrino oscillations at FCC-\Pe{}\Pe with $6\times10^{12}$ \PZ bosons, taken from Ref.~\cite{Antusch:2024otj}.
The decay width $\Gamma$ corresponds to the point in the mass-coupling plane that has the largest significance $Z$ but has not yet been excluded by prior experiments.
The diagonal band corresponds to moderate values of the LN violation ratio $R$.
The discovery reach is constrained by five bounds.
If the decay width is too small or too large, the decays appear detector stable or fail to form a secondary vertex, respectively.
If the mass splitting $\Delta m$ is too small or too large, the oscillations are much larger than the detector geometry or smaller than the vertex uncertainty, respectively.
Finally, if the LN violation ratio is too small, the decay happens too fast for an oscillation to develop.
} \label{fig:hnl-osc-reach}
\end{figure}

While the measurement of LN violation at an hadron collider is straightforward \cite{Antusch:2022hhh}, the processes that are accessible at FCC-\Pe{}\Pe must be analysed using angular or momentum distributions \cite{Antusch:2024otj}.
Two of these distributions can be understood as the result of the polarisations of the \PZ-boson and the HNL, respectively.
The polarisation of the \PZ-boson leads to an angular distribution of the HNLs with respect to the beam axis in the lab frame.
This distribution is inherited by the final state leptons.
In the Dirac limit, it manifests itself as a forward-backward asymmetry shown in \cref{fig:hnl-osc-BM1}.
For oscillations whose frequency is comparable to the decay length, this forward-backward asymmetry gets superimposed with an oscillatory pattern shown in \cref{fig:hnl-osc-BM2,fig:hnl-osc-BM3}.
In the double-Majorana limit, the forward-backward asymmetry gets washed out by the oscillations and the distribution becomes flat.
The polarisation of the HNL leads to a further distribution that is either observable in the opening angle between the charged lepton and the reconstructed \PZ-boson in the HNL rest frame, or in the absolute value of the charged lepton momentum in the lab frame.
While the dependence of this distribution on the amount of LN violation mimics the behaviour demonstrated for the forward-backward asymmetry in \cref{fig:hnl-osc}, its analysis power is stronger.
Therefore, this observable yields a larger discovery potential.

We have performed a full MC study for the IDEA detector geometry at FCC-\Pe{}\Pe during the \PZ-pole run \cite{Antusch:2024otj}.
We find that at least 1,000 reconstructed HNL events are necessary to observe neutrino oscillations with a significance of 5$\sigma$.   
For discoverable mass splittings the contour lines of the discovery reach follow the usual shape in the mass coupling plane obtained by displaced vertex searches for HNLs.
Since the model contains as additional parameter the mass splitting $\Delta m$ that quantifies the amount of LN violation, we present in \cref{fig:hnl-osc-reach} our results in the plane spanned by the mass splitting and the decay width.
Here the decay width should be thought of as a function of the HNL mass and coupling.
While the discovery reach is limited by the interplay between detector geometry, decay width, and mass splitting, the overall reach of FCC-\Pe{}\Pe far exceeds the reach and potential of current experiments.

\subsubsection{Probing the neutrino portal at future lepton colliders}
\label{sec:HNL_NPortal}
%
In the most minimal scenarios, HNL mass eigenstates $N_i$ with mass $M_i$ can only be produced and decay through their mixing $\theta_{\alpha i}$ with SM neutrinos of flavour $\alpha = \Pe, \\PGm, \PGt$, while in extended scenarios they may have new scalar or gauge interactions that increase the production cross section. 
Future circular colliders would offer a unique opportunity to probe HNLs with masses below the electroweak scale, as these particles can be produced in large numbers at the Z-pole run. The unique opportunity here lies in the fact that FCC-ee or CEPC could not only discover HNLs, but potentially observe tens or even hundreds of thousands of long-lived HNLs in displaced vertex searches (\cref{antony}) within a phase-space region where they could simultaneously explain the light neutrino masses and the matter-antimatter asymmetry of the universe \cite{Drewes:2021nqr}. 
Acting as HNL factories, these machines hence permit the study of their properties in detail and could lead to the underlying neutrino mass mechanism. 
In view of the long time scales involved in planning and building each collider generation, having both a discovery and precision machine packed into one is a key asset. 

By exploiting complementarities in the broader landscape of activities in particle physics at all frontiers, this study covers several aspects of HNLs at future lepton colliders
relevant for the design of the machines and experiments.
\begin{itemize}
	\item \textbf{Model disambiguation:} Previous studies focus on the prospects for tests of the seesaw mechanism and leptogenesis in a minimal model with two HNL species \cite{Antusch:2017pkq}, in which in principle all model parameters can be constrained when collider data is combined with information from neutrinoless double $\beta$-decay ($0\nu\beta\beta$) and long-baseline neutrino experiments \cite{Drewes:2016jae}. Naively one would think that the considerably larger parameter space of models with three HNL species (18 instead of 11 new parameters in the mass and mixing matrices) makes it unfeasible to constrain individual model parameters. However, it turns out that this is partially compensated by the fact that more information about individual model parameters can be extracted from a given event number. This can be understood by comparing the quantities $U_{\alpha i}^2 = |\theta_{\alpha i}|^2$ that govern the HNL production cross section to the lower limit $U_0^2 =  \frac{\sum_i m_i}{M} $ for mass-degenerate HNLs with mass $M$, with $m_i$ the light neutrino masses (\emph{seesaw line}). Defining $U^2=\sum_{\alpha, i} U_{\alpha i}^2$ and $\epsilon = U_0^2/U^2$, one can expand the $U_{\alpha i}^2$ in powers of $\epsilon \ll 1$. In the model with two HNL flavours only integer powers appear, making it very difficult to access model parameters that do not affect the $U_{\alpha i}^2$ at the leading order $\mathcal{O}[1/\epsilon]$. With three HNL flavours, terms $\mathcal{O}[1/\sqrt{\epsilon}]$ appear \cite{Drewes:2024bla}. This enhancement allows to discriminate between the minimal and next-to-minimal models, and to constrain to more individual model parameters from a given number of events. The region where such terms can be accessed is indicated in \cref{antony}, which is based only on counting event numbers. Future studies will focus on more differential quantities and correlations between them, e.g.\ including searches for lepton number violation. 
	\item \textbf{Complementarity with Intensity and Cosmic Frontiers:}
 It is well-known that neutrino-oscillation experiments and $0\nu\beta\beta$ provide complementary probes of low scale seesaw models (see e.g.\ Ref.~\cite{Drewes:2016jae}). This connection was recently revisited in view of updated results for the $0\nu\beta\beta$ decay lifetime \cite{deVries:2024rfh}. 
 An observation of $0\nu\beta\beta$ decay would give access to a combination of CP-violating phases that can be relevant for leptogenesis and cannot be measured at colliders, hence complementing FCC-ee or CEPC.  
 Non-observation of $0\nu\beta\beta$ decay for inverted neutrino mass ordering can indicate that a precise cancellation between the HNL and light neutrino contributions is necessary, hence pointing to specific bands in the mass-mixing plane, cf. \cref{jordy}, pointing to regions where future lepton colliders could observe a large number of events. 
 The requirement of successful leptogenesis leads to additional constraints.
 This study was performed in the minimal model with two HNL species and no other new particles; generalising it to scenarios with three HNL species and new interactions (such as left-right symmetric models - LRSM) is an important future goal.
	\item  \textbf{Detector design and dedicated long-lived particle detectors:} Long-lived particle searches can give access to a very large number of HNL events, cf. \cref{antony}. This reach can be further increased by optimising the design of the main detectors and with dedicated long-lived particle detectors, cf. \cref{jordy} for an example. Studies on how to optimise the detector designs to maximise the discovery potential and efficiency of lepton colliders as HNL factories will continue.
\end{itemize}
\begin{figure}[h]
	\centering
 \includegraphics[width=0.53\textwidth]{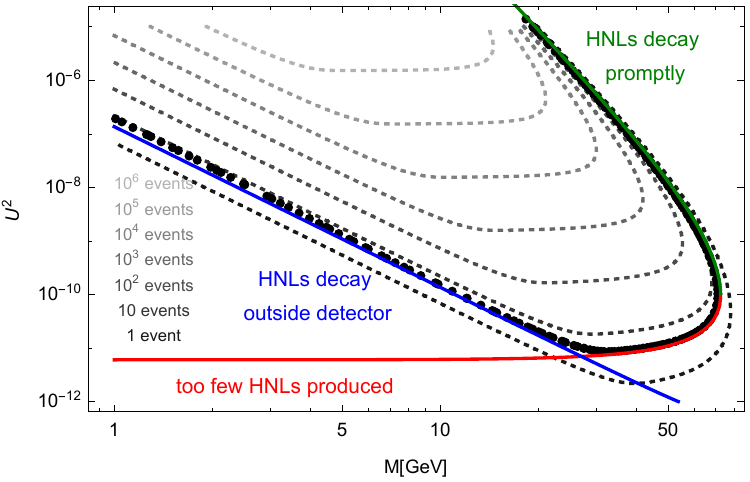}
	\includegraphics[width=0.35\textwidth]{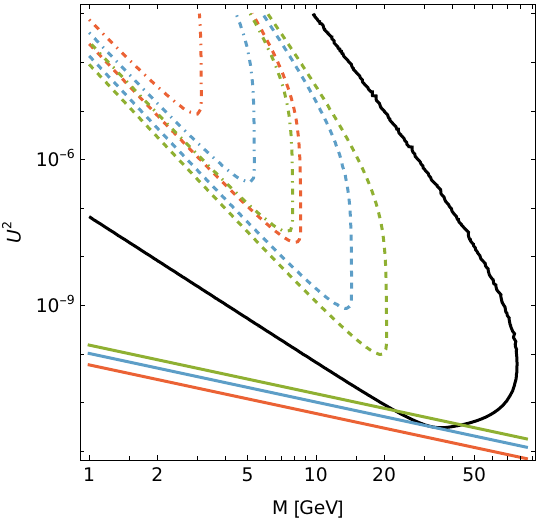}
	\caption{\label{antony}
  \emph{Left Panel (from Ref.~\cite{Drewes:2022rsk})}: Number of HNL decays inside an IDEA type detector at the Z-pole run at FCC-ee.
  \emph{Right Panel (from Ref.~\cite{Drewes:2024bla})}: Potential discovery range of FCC-ee (solid curve, corresponding to 4 or more events), 
  compared with the lower limit $U_0^2$ for different values of the lightest SM neutrino mass between zero (orange solid) and the cosmological upper bound (green solid). The region where FCC-ee or CEPC model discrimination power is enhanced lies inside the corresponding dashed and dotted-dashed lines   (3$\sigma$ and 5$\sigma$, respectively).
	}
\end{figure}

\begin{figure}
	\centering
	\includegraphics[width=0.53\textwidth]{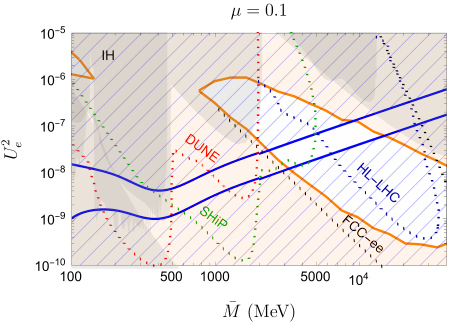}
 	\includegraphics[width=0.35\textwidth]{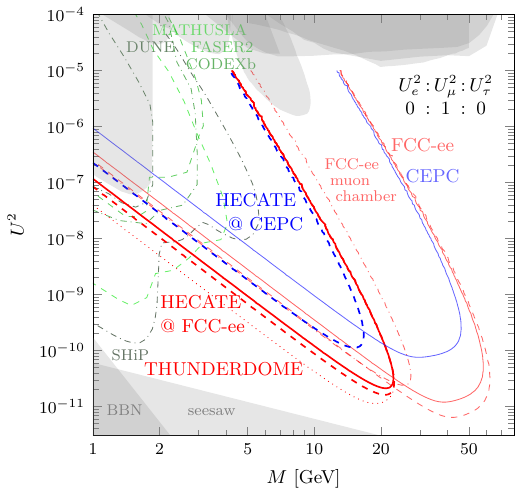}
	\caption{\label{jordy}%
\emph{Left Panel (from Ref.~\cite{deVries:2024rfh})}:
Non-observation of $0\nu\beta\beta$-decay with a limit of ${T_{1/2}^{0\nu}}(^{136}\text{Xe}) > 3.8\times 10^{28}$ y for inverted neutrino mass ordering would restrict the allowed range of HNL mass and mixing to the area between the blue curves for a given HNL mass splitting $\mu=|M_2-M_1|/(M_1+M_2)$ in the model with two HNL species, as a function of the seesaw scale, $\bar{M}$. The region within the orange contour produces correct BAU. In the region in which the LHC main detectors \cite{Drewes:2019fou} (or the planned SHiP experiment) can discover HNLs, FCC-ee could potentially see tens of thousands of HNL decays. \emph{Right panel (from Ref.~\cite{Chrzaszcz:2020emg})}: An example of the additional parameter range that becomes accessible at FCC-ee with a dedicated long-lived particle detector (HECATE), as a function of the Majorana mass, $M$.
	}
\end{figure}
%
\subsubsection{\texorpdfstring{Search for right-handed Majorana neutrinos}{Search for right-Handed Majorana neutrinos}}
\label{sec:HNL_RHN}
The type I seesaw mechanism, in which the SM is augmented by three SM-gauge-singlet Right Handed Neutrinos~(RHNs), is probably the simplest way to explain the origin of tiny neutrino masses.
RHNs are naturally introduced in the minimal $(B-L)$ (Baryon minus Lepton number) model in which the SM's accidental global
$U(1)_{(B-L)}$ symmetry is gauged~\cite{Das:2021esm}.
Breaking of the $U(1)_{(B-L)}$ gauge symmetry induces masses of the RHNs and the associated $\PZpr$ boson.
This study estimates the sensitivity to this class of models, using full simulation of the ILD concept at the ILC.
More details of the analysis can be found in Ref.~\cite{Nakajima:2022pkd}.

We focus on RHN pair production $\Pep \Pem \rightarrow \PN \PN$ at ILC-500.
If the RHN mass is sufficiently large, RHN can decay to $\Plpm \PWmp, \PZ \PGn$, and/or $\PH \PGn$.
In this study, we assume that the lightest RHN decays to the first lepton generation with the process $\PN \rightarrow \Pepm \PWmp$.
We consider benchmark points with the lightest RHN mass between 100 and 225~GeV,
$\PZpr$ mass $M_{\PZpr} = 7~\mathrm{TeV}$, coupling constant $g^\prime_{(B-L)} = 1$, and mixing angle $|V_{eN}| = 0.03$.
Model parameters are chosen to avoid experimental constraints.
We utilise a final state with two isolated same-sign leptons and hadronic jets, which has no irreducible backgrounds from
SM processes.
Some backgrounds do arise when a lepton originating from hadron decay is mis-identified as an isolated primary lepton,
and is paired with a charged lepton, for example from the decay of an associated \PW boson.
All SM processes leading to 4- and 6-fermion final states with at least one electron and two quarks are considered
in the background estimation.

Signal events were generated using the \whizard event generator v2.8.5~\cite{Kilian:2007gr}.
SM samples were prepared by the ILD software group using \whizard v1.9.5.
Both signal and background events included the effect of beam energy spread and Initial State Radiation (ISR).
Events were fully simulated and reconstructed using the standard tools developed for the ILD concept.
ILC is foreseen to have longitudinally polarised beams, 80\% for the \Pem and 30\% for the \Pep beams.
We assumed two data sets with opposite polarisations, each of $1,600~\mathrm{fb^{-1}}$, as envisioned in the current ILC run plan.

To select events with same-sign isolated electrons and hadronically decaying $\PW$ bosons, we used
the following filtering requirements: exactly two same-sign isolated \Pepm, and no isolated muons or photons; the energy of both isolated electrons should be $< 200~\mathrm{GeV}$; both isolated electrons should have polar angles $|\cos{\theta_{isoe}}| < 0.95$; the ``IsolatedLeptonTagging parameter'', i.e.\ the output of a multi-variate analysis designed to identify isolated leptons, of both electrons should be $> 0.9$; the jet clustering parameter with the Durham algorithm should be $\log_{10}{(y_{12})} > -1$; and, the magnitude of the missing momentum should satisfy the conditions $\mathrm{P}_\text{miss} < 100~\mathrm{GeV}$, and ($\mathrm{P}_\text{miss} < 40~\mathrm{GeV}$ or $|\cos{\theta_{\mathrm{P}_\text{miss}}}| > 0.95$).

In this study we consider RHN pair decays with two hadronically decaying $\PW$ bosons. After removing the isolated \Pepm, the remaining particles are clustered into 4 jets. To reconstruct the RHN mass, we search for the best jet-pairing. We assume that the jet-pair masses $M_{jj}$ should be consistent with the \PW mass, and choose the jet-pairing which minimizes the quantity
    $F_{1}=(M_{jj1}-M_{\PW})^{2}+(M_{jj2}-M_{\PW})^{2}$.
%
We also assume that the two reconstructed RHN masses $M_{jj\Pe 1}$, $M_{jj\Pe 2}$ should be equal,
so we choose the electron--jet-pair combination which minimizes the quantity
   $F_{2}=(M_{jj\Pe 1}-M_{jj\Pe 2})^{2}$.
The selected jet-lepton pairing is used to reconstruct the average RHN mass $M_{\PN}$.
A mass window from $-10~\mathrm{GeV}$ to $+15~\mathrm{GeV}$ around each tested RHN mass is then applied.
The background is consistent with a flat distribution in $M_{\PN}$.
Only 3 (20) SM background events remain in $1,600~\mathrm{fb^{-1}}$ with the $e_R^{80}p_L^{30}$ ($e_L^{80}p_R^{30}$) beam polarisation.

The signal significance ($N_{S}/\sqrt{N_{B}+N_{S}}$) and expected 95\% C.L. upper limits
on the partial cross section $\sigma( \Pep \Pem \rightarrow \PN \PN) \times BR(\PN \rightarrow \Pepm \PWmp)^{2}$
were calculated using the number of selected signal and background events $N_{S}, N_{B}$ and are shown in \cref{fig:exclusion_sigma}.
The expected cross section upper limits are up to 10 times smaller than those of the chosen benchmarks, depending on the RHN mass.
The results are significantly improved for the $e_R^{80}p_L^{30}$ polarisation,
a result of both the larger signal cross section and smaller SM background for this beam configuration. The SM backgrounds are expected to be significantly reduced in the di-muon final state, which probes different mixings between the lightest RHN and the SM leptons.
\begin{figure}[htb]
    \centering
    \includegraphics[scale=0.4]{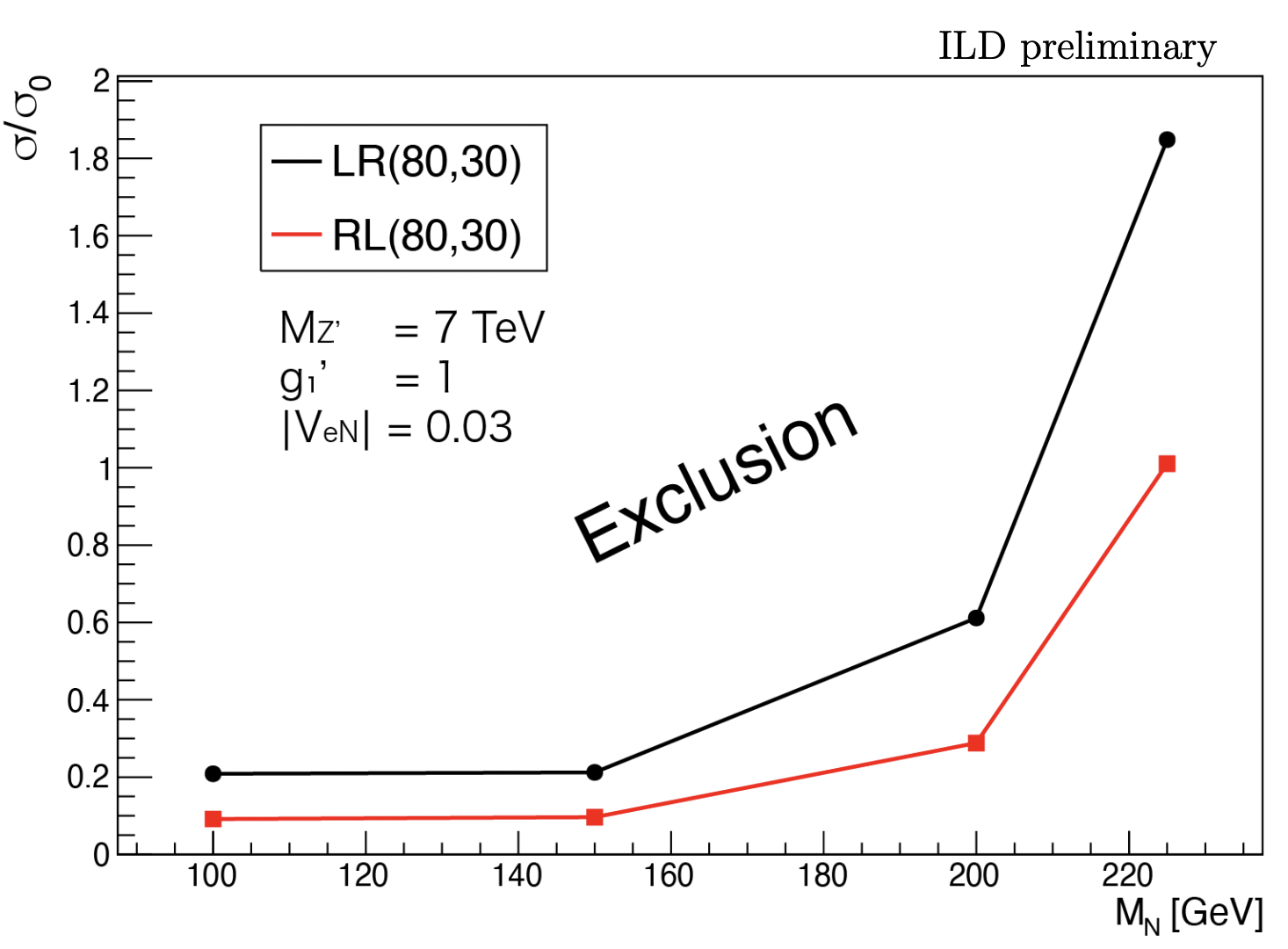}
    \caption{Obtained 95$\%$ C.L. upper limits on the partial cross section $\sigma(ee\rightarrow NN)\times (BR(N\rightarrow e^{\pm}W^{\mp}))^{2}$ normalised to the benchmark points' cross sections $\sigma_0$, as a function of $M_{N}$. $1,600~\mathrm{fb^{-1}}$ of data at ILC-500 in beam polarisations $e_L^{80}p_R^{30}$ (black) and $e_R^{80}p_L^{30}$ (red) are assumed. $g_1^\prime=1$ indicates SM-like couplings for the right-handed neutrinos.}
    \label{fig:exclusion_sigma}
\end{figure}

\subsection{SUSY searches\label{sec:SRCH-susy}}
%
Among the few internally consistent models for BSM, Super-Symmetry (SUSY)
\cite{Martin:1997ns,Wess:1974tw,Nilles:1983ge,Haber:1984rc,Barbieri:1982eh}, 
stands out as a prime candidate that offers solutions and/or hints to solutions to
several of the SM problems, including the naturalness and the hierarchy problems,
 the coupling constant unification at a unique GUT scale, and an explanation for the quantisation of charge. It can also provide a candidate for  Dark Matter,
and an explanation of the observed value of $g-2$ of the muon.
The neutralinos (charginos) are the supersymmetric partners of the neutral (charged) gauge and Higgs bosons and due to the SUSY breaking mechanism, 
the neutralinos are mixtures of the gaugino and higgsino
components, depending on the following fundamental MSSM parameters:
the U(1) gauge group parameter $M_1$ (``bino''),
the SU(2) gauge group parameter $M_2$ (``wino''), the higgsino mass parameter $\mu$ and 
the ratio of the vacuum
expectation values of both Higgs doublets $\tan \beta = v_2/v_1$. 
Chargino mixing depends only $M_2$, $\mu$ and $\tan \beta$.
The lightest supersymmetric particle (LSP) is, if R-parity is conserved, stable and 
a possible cold dark matter
candidate. 

No clear signal of SUSY has been seen in the data from the LHC so far, nor
did searches at LEP-II find any indications of SUSY. 
This has led to a sentiment in the community that SUSY is strongly
challenged.
In fact, what is strongly
challenged is the cMSSM (aka mSUGRA) paradigm that was popular pre-LHC.
This paradigm contains a minimal number of parameters,
and couples the electroweak and strong sectors of SUSY closely,
and thus predicted that coloured states (the squarks and the gluino) should
be in reach of the LHC. 
These have now been excluded up to masses well above 1 TeV.
But this coloured sector has little bearing on the issues mentioned above:
the issues only require rather light and close-together electroweak states to exist.
In fact, the precision electroweak measurements at LEP predicted that
the Higgs mass should be less than 140 \GeV if SUSY was assumed \cite{Djouadi:2005gj}, while
a much larger value of 285 \GeV would have been allowed by the SM alone~\cite{ALEPH:2005ab},
and indeed, a Higgs boson was observed below the SUSY-imposed limit.
Both LEP and LHC have observed an excess of Higgs-like events
at around 95 \GeV, which could be a sign of a second scalar Higgs,
required to exist in SUSY, but not in the SM.
Both ATLAS and CMS observe a persistent excess of events that can
be interpreted as Chargino/Neutralino production at a mass of around
200 \GeV and a mass-difference to the LSP of around 20 \GeV \cite{ATLAS:2021moa,ATLAS:2019lng,CMS:2021edw,CMS-PAS-SUS-23-003}.
While some specific models can be excluded by the LHC,
a full scan of the 18 parameters of R-parity and CP conserving SUSY
recently performed by ATLAS shows that hardly any points in the
parameter-plane beyond what was probed by LEP-II can be excluded \cite{ATLAS:2024qmx}.
The reason why LEP could conclusively exclude SUSY almost up to the
kinematic limit, while the LHC cannot, is that the blessing of the
high production cross section for strong processes becomes a curse
if the signal is colour-neutral: no increase of the signal from
strong production, only of the background.

 \begin{figure}[t]
    \begin{center}
\subcaptionbox{}{\includegraphics [scale=0.35]{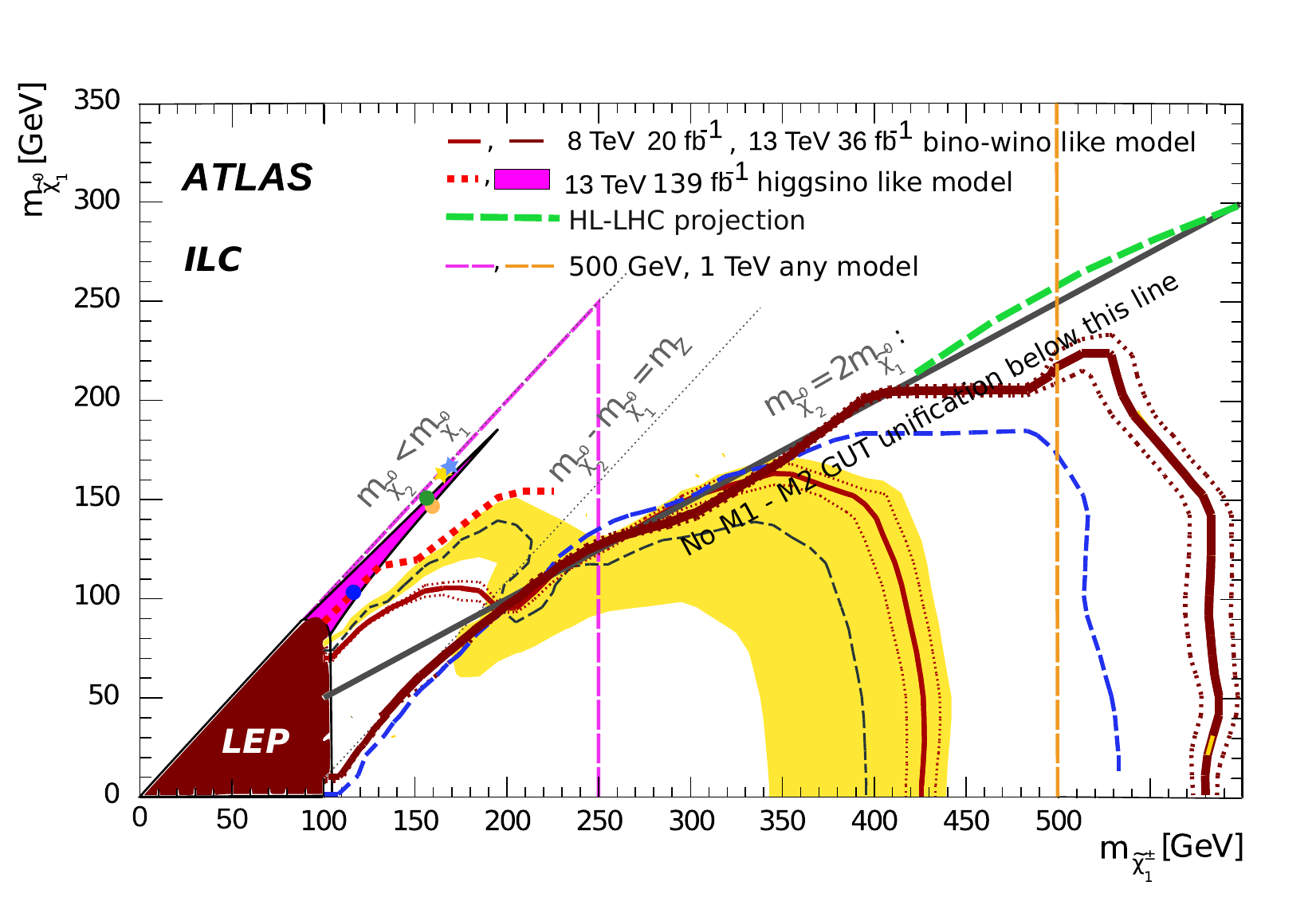}}
     \subcaptionbox{}{\includegraphics [scale=0.28]{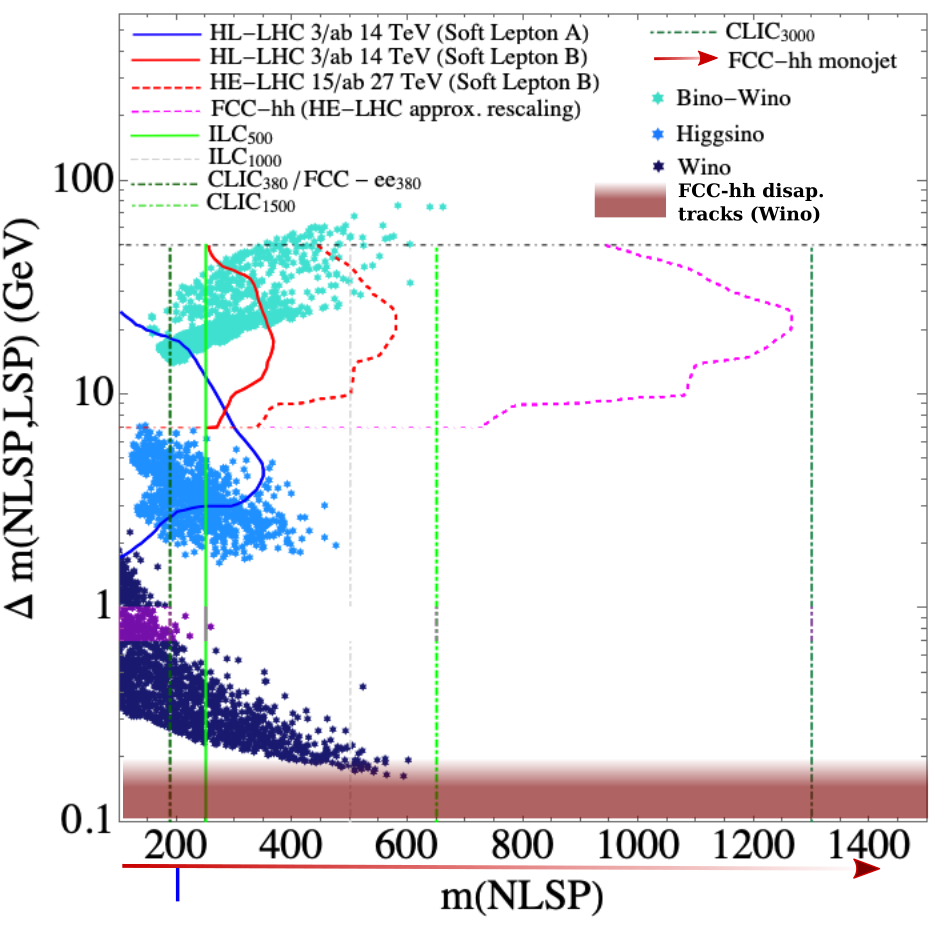}}
\end{center}
\caption{ Observed or projected exclusion regions for a \PSGcpmDo  NLSP for LEPII, LHC, HL-LHC and
  for ILC-500 and ILC-1000 are shown (a) in the $M_{ \PSGcpmDo }$--$M_{\PSGczDo}$ plane, and (b) in the  $M_{ \PSGcpmDo }$--$\Delta M$ plane. For the ILC curves, discovery and exclusion reach are both within the width of the lines. The points are the values allowed by other constraints, following  Ref.~\cite{Chakraborti:2021kkr}. \label{fig:X1summary}
}
 \end{figure}
Current and projected limits in the case that the next-to-lightest SUSY particle (NLSP) is a chargino is shown in \cref{fig:X1summary}. In the left panel (showing the $M_{ \PSGcpmDo }$--$M_{\PSGczDo}$ plane),
it should be noted that below the heavy black line, GUT unification of the
Bino and Wino mass-parameters  $M_1$ and $M_2$
is not possible: The difference between \PSGczDo and \PSGcpmDo cannot be larger than what the line indicates,
if such a unification is realised.
%
%

   \begin{figure}[htbp]
    \begin{center}
\includegraphics [scale=0.55]{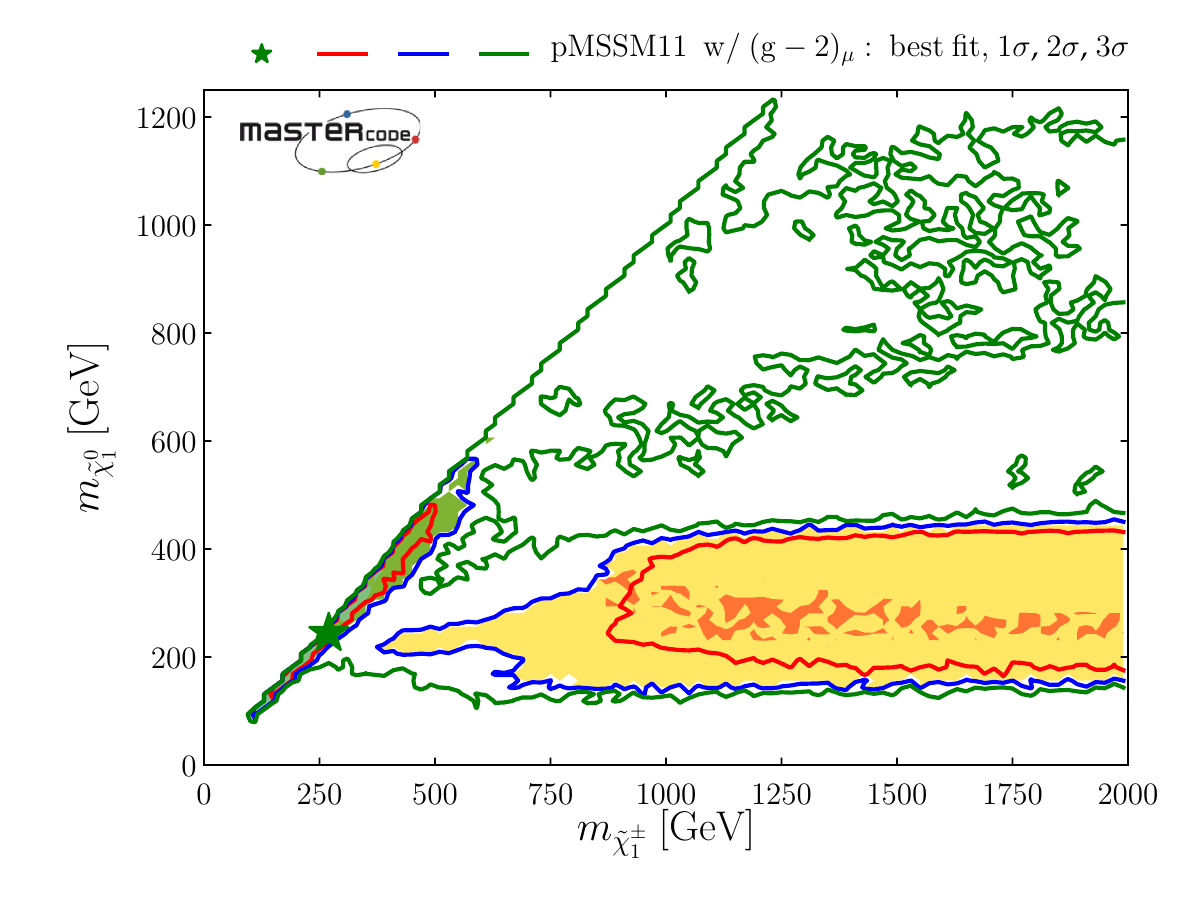} 
     \caption{ pMSSM11 fit by MasterCode to LHC13/LEP/g-2/DM(=100\%~LSP)/precision observables  in the \PSGcpmDo--\PSGczDo  plane. From Ref.~\cite{Bagnaschi:2017tru}. \label{fig:mastercode}}
     \end{center}
     \end{figure}
    Naturalness, the hierarchy problem, the nature of dark matter,
  or the observed value of the magnetic moment of the muon, are all reasons to prefer
  a light electroweak sector of SUSY.
  Moreover, many models
and the global set of constraints from observation~\cite{Bagnaschi:2017tru}
  point to a \textit{compressed spectrum}; see Figs. \ref{fig:X1summary}(b) and \ref{fig:mastercode}.
  If the Lightest SUSY Particle (the LSP) is a Higgsino or a Wino, there must be other
  bosinos  with a mass close to it, since the $\tilde{H}$ and $\tilde{W}$
  fields have several components, leading to a close connection between
  the physical states of the bosinos.
  Although the third possibility ---  a Bino-LSP --- has no such constraints,
  an overabundance of DM  is expected in this case \cite{Roy:2007ay}.
  To avoid such a situation,
  a balance between early universe LSP production and
    decay is needed.
  One compelling option is \PSGt~co-annihilation, and 
  for this process to contribute sufficiently, the density in the early universe of \PSGt and \PSGczDo 
  must be close to each other,
  which means that their masses must be quite similar.
Most sparticle-decays occur via cascades. In the case of compressed spectra, 
 the last decay in the cascade -- the one to SM particles and the LSP --
  has a small $\Delta M$, and hence low visible activity.
  For such decays, the current limits from LHC  are for specific models.
Therefore, a lepton collider with an energy well above the energy of LEP-II
will be paramount to be able to further exploit the SUSY parameter-space
in a model independent way. The proposed Higgs Factory can fill this role,
in particular if it is designed to be upgradable to reach energies up to the TeV range,
as the different proposals for linear colliders are. 

At an \epem Higgs factory, it is possible to perform a loophole free SUSY search,
since in SUSY, the properties of
the production and decay of NLSPs are fully predicted for given masses of the LSP and the NLSP.
All possible NLSP candidates can therefore be searched for in a systematic way
due to the
  SUSY-principle: ``sparticles couples as particles''.
Note that this does not depend on the (model dependent) SUSY breaking mechanism.
  By definition, there is only one NLSP, and it must have 100\% BR
  to its (on- or off-shell) SM-partner and the (stable or unstable) LSP.
  Also, there is only a handful of possible candidates to be the NLSP.
  Hence by performing searches for every possible NLSP, 
  model independent exclusion and  discovery reaches in the $M_{NLSP} - M_{LSP}$ plane,
separately for each NLSP candidate, or globally, by determining which NLSP gives the
weakest limit at any point. There will be no loopholes to the conclusion.
In particular,
one can aim at well motivated and maximally 
difficult NLSPs;
if one can find this, then one can find any other NLSP.
\subsubsection*{NLSP search at ILC: \PSGt and  \PSGcpmDo}\label{sect:stau}
 The \PSGt
  has two weak hypercharge eigenstates (\PSGtR , \PSGtL ),
  which are not mass degenerate.
  Mixing yields the physical states  (\PSGtDo , \PSGtDt ),
  the lightest one being
    likely to be
    the lightest sfermion, due to the stronger trilinear couplings
    expected for the third family SUSY particles.
   If R-parity is assumed to be conserved,
    the \PSGt will be pair-produced in the $s$-channel via
    $\PZ^0/\PGg$ exchange.
    The production cross section can be quite low,
    since  \PSGt-mixing can suppress the coupling
    to the $\PZ^0$ component of the neutral current, so that only $\PGg$
    exchange contributes.
    The \PSGt will decay to the LSP and a $\PGt$, implying a more
    difficult signal to identify  than that of other sfermions,
   owing to the invisible component of the $\tau$ decays.
   In addition,
  mixing can further reduce detectability.
 Furthermore, the presence of a \PSGt close in mass to the LSP
  can contribute to co-annihilation between the two in the early universe,
  and in this way avoid an over-abundance of SUSY WIMP dark matter~\cite{Ellis:1998}.
  Finally, the \PSGt is the SUSY particle least constrained from
  current data.
  We see that the \PSGt satisfies both conditions for being the prime target for
  the search: it is both a well motivated and maximally 
difficult NLSP candidate.
In Ref.~\cite{Berggren:2024ckz}, for the first time a study of this process has been performed in full detail,
both considering all SM background processes (fully simulated) as well as all
beam-induced backgrounds.
The result is shown in Fig.~\ref{fig:C1stauexcl}(a),
together with current limits from LEP and LHC, as well as projections to HL-LHC and recasts of
the ILC-500 result to ILC-250 and ILC-1000.
The study was done for the ILC at 500~\GeV,
but a recast of the results to FCC-ee at 240~\GeV is also given,
further discussed below.
The conclusion was that the often-made assumption that SUSY searches at \epem\ colliders can reach almost the kinematic limit, could for the first time be shown to be true, even in this worst case channel, and using fully realistic simulation.
Also shown in Fig.~\ref{fig:C1stauexcl}(b) is 
a conservative extrapolation from the LEP results in the case of a \PSGcpmDo NLSP \cite{PardodeVera:2020zlr},
once again showing the coverage up to quite close to the kinematic limit.
It could also be shown that the exclusion and discovery reaches are close to the same --- differing
only by a few \GeV\ 
--- which is quite different from the situation at hadron colliders, where the two
often are different by hundreds of \GeV.
%

\begin{figure}[t]
   \begin{center}
     \subcaptionbox{}{\includegraphics [scale=0.4]{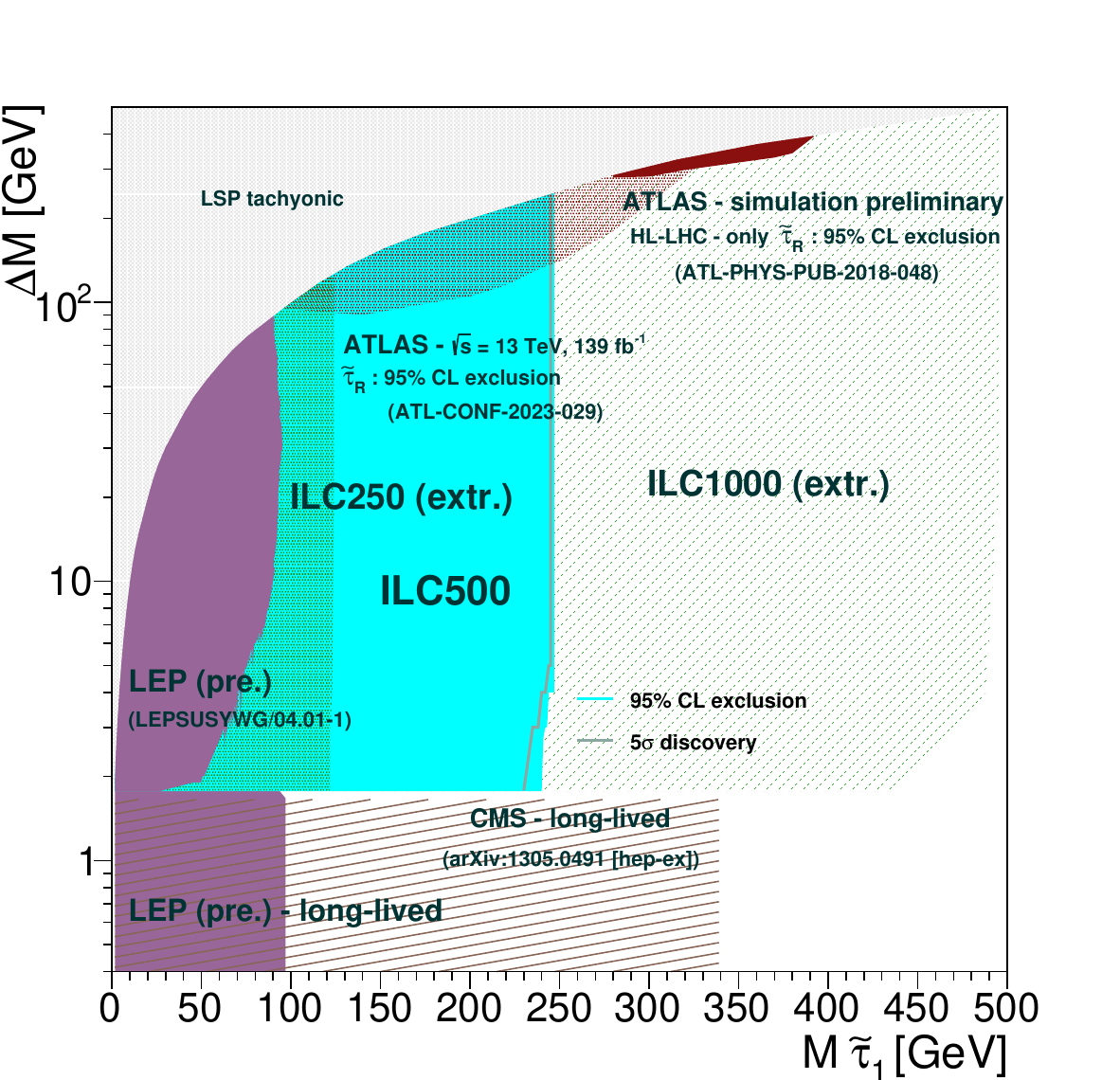}}
     \subcaptionbox{}{\includegraphics [scale=0.4]{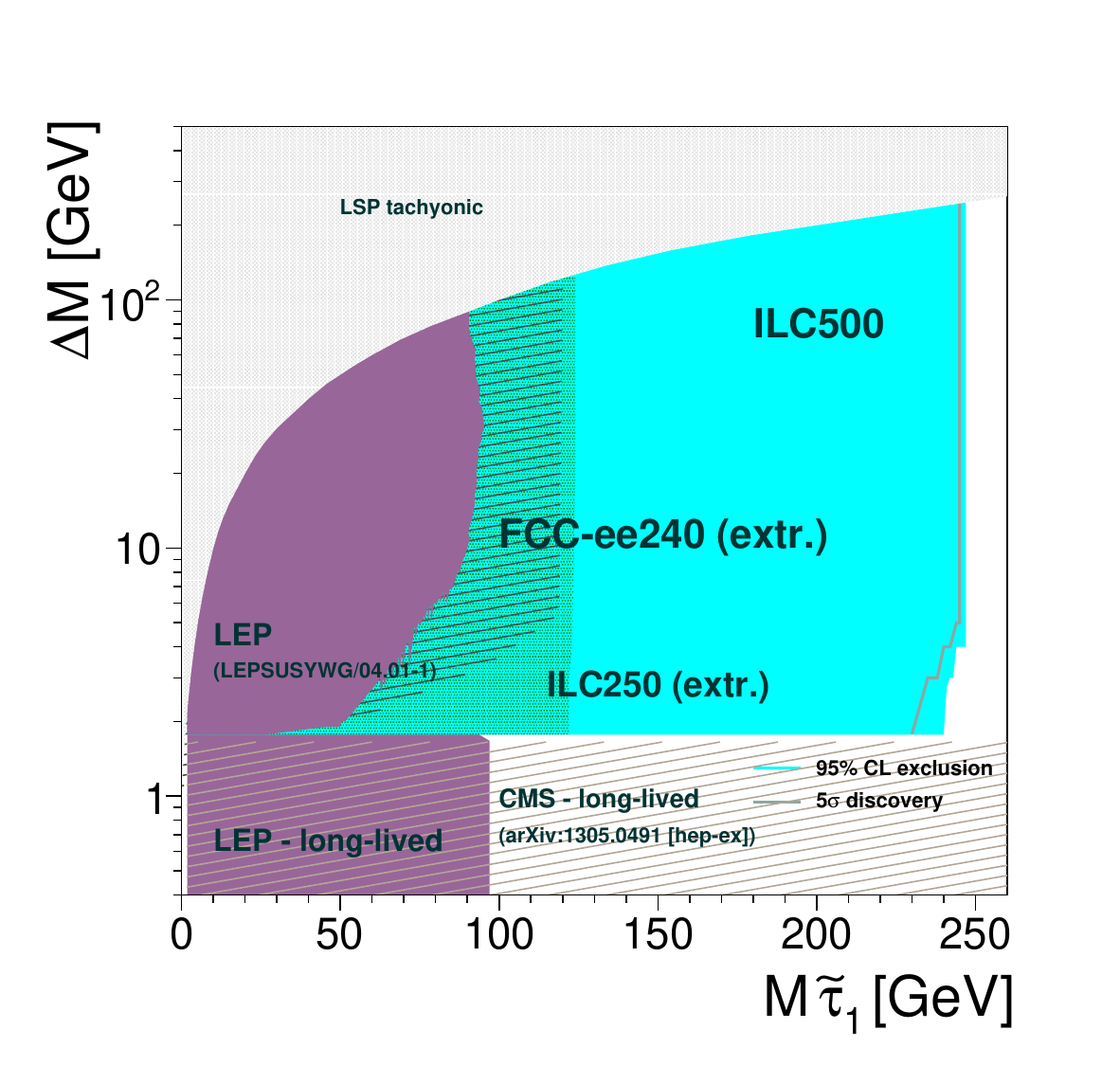}}
     \subcaptionbox{}{\includegraphics [scale=0.4]{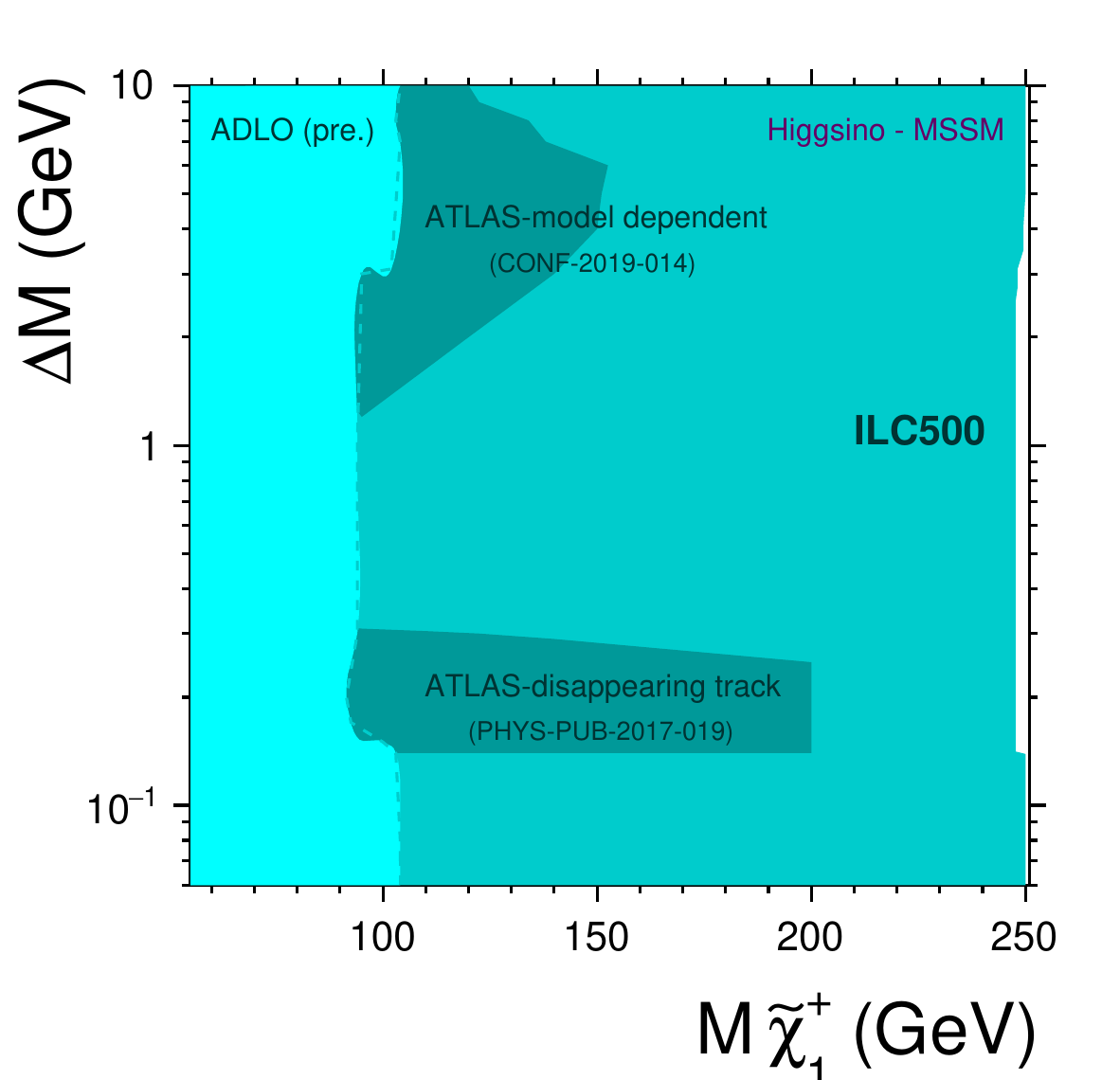}}
\end{center}
\caption{ Exclusion and discovery reaches for a \PSGtDo NLSP are shown in (a), as a function of the \PSGtDo mass and where $\Delta M$ is the difference between this and the LSP mass. The cyan area is the result of the full simulation
  study at ILC-500~\cite{Berggren:2024ckz}. Also shown are limits from LEP and ATLAS, as well as  extrapolations to ILC-250, ILC-1000 and HL-LHC.
  It should be noted that the limit and the HL-LHC projection from ATLAS are exclusion only,
and are for specific assumptions on the \PSGt properties, assumptions that are
not the most pessimistic.
  In (b) the same is shown zoomed
  in, where also the recast of the ILC-500 result to FCC-ee-240 conditions is shown.
  In (c), the exclusion reach for  a  \PSGcpmDo NLSP is
  shown~\cite{PardodeVera:2020zlr}.
  \label{fig:C1stauexcl}}
\end{figure}

\begin{figure}[tb] 
   \begin{center}
     \includegraphics [scale=0.25]{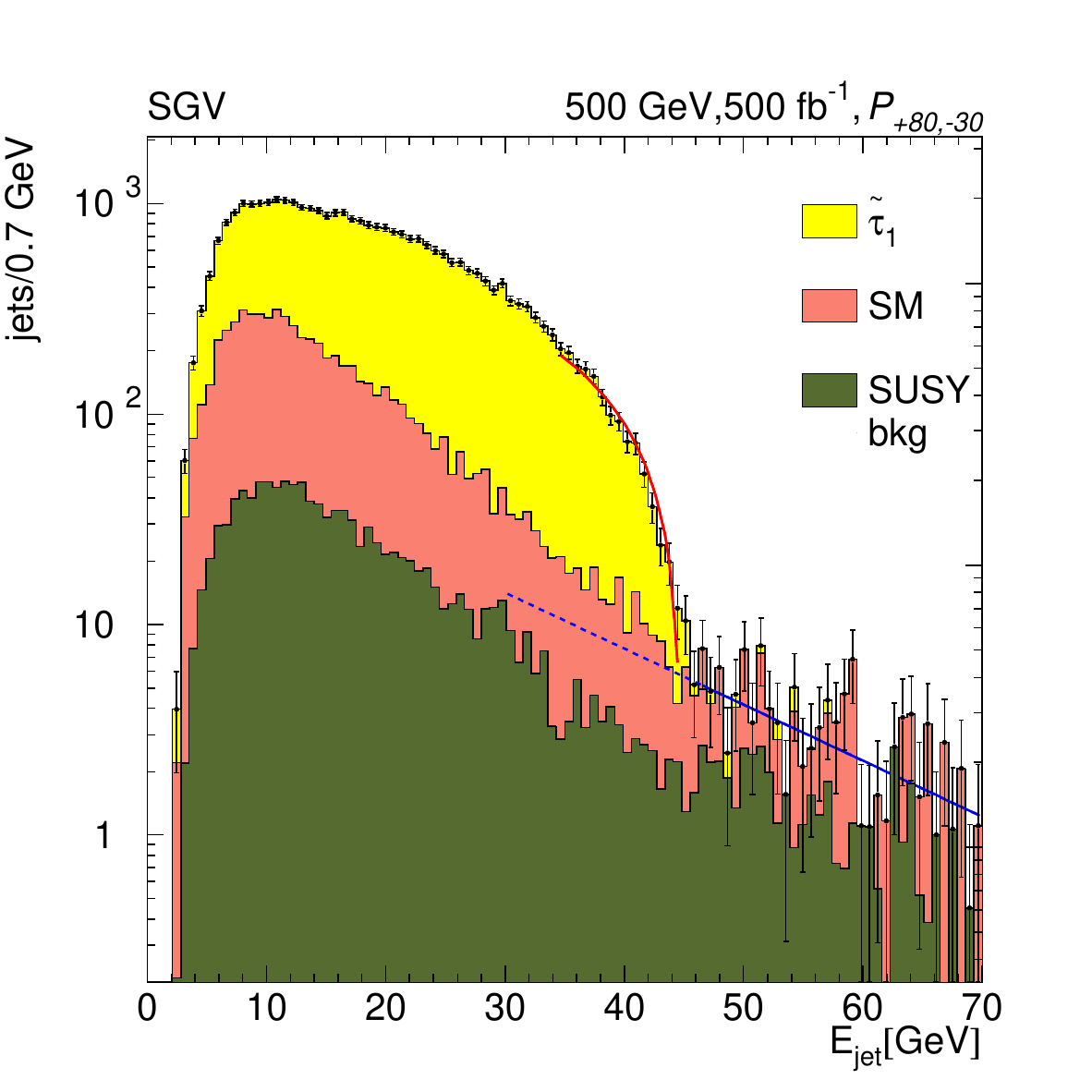}
     \includegraphics [scale=0.25]{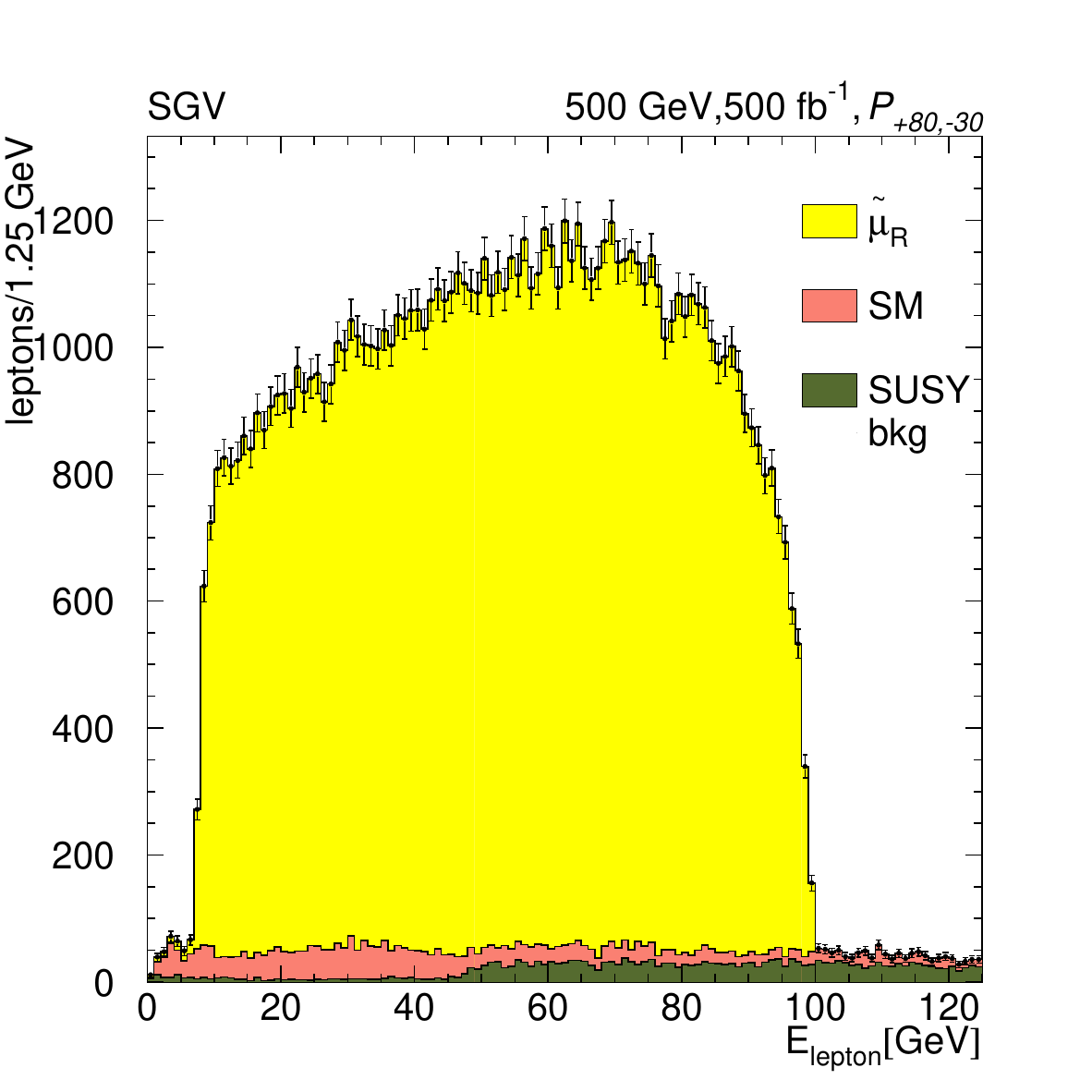}
     \includegraphics [scale=0.25]{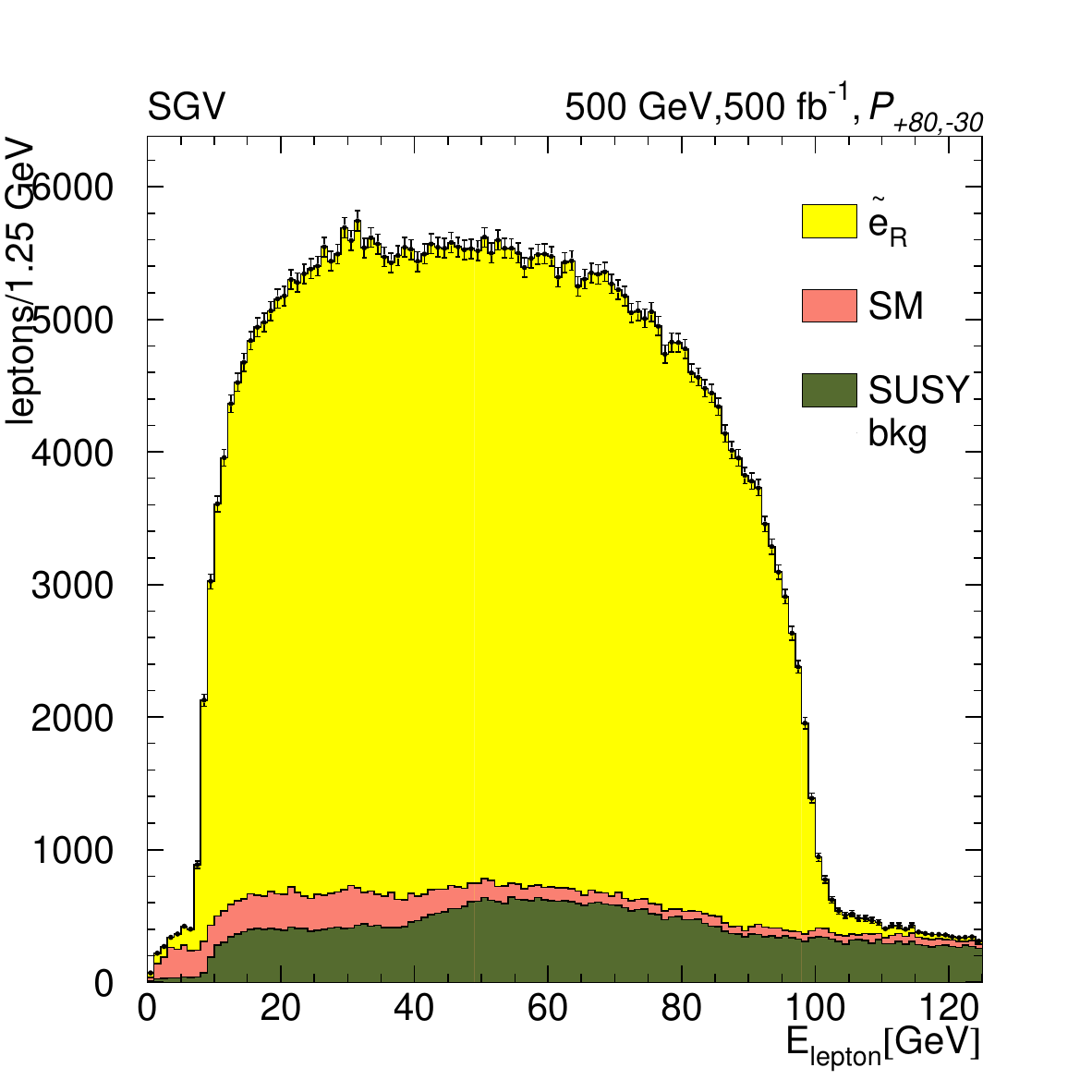}
     
      \includegraphics [scale=0.23]{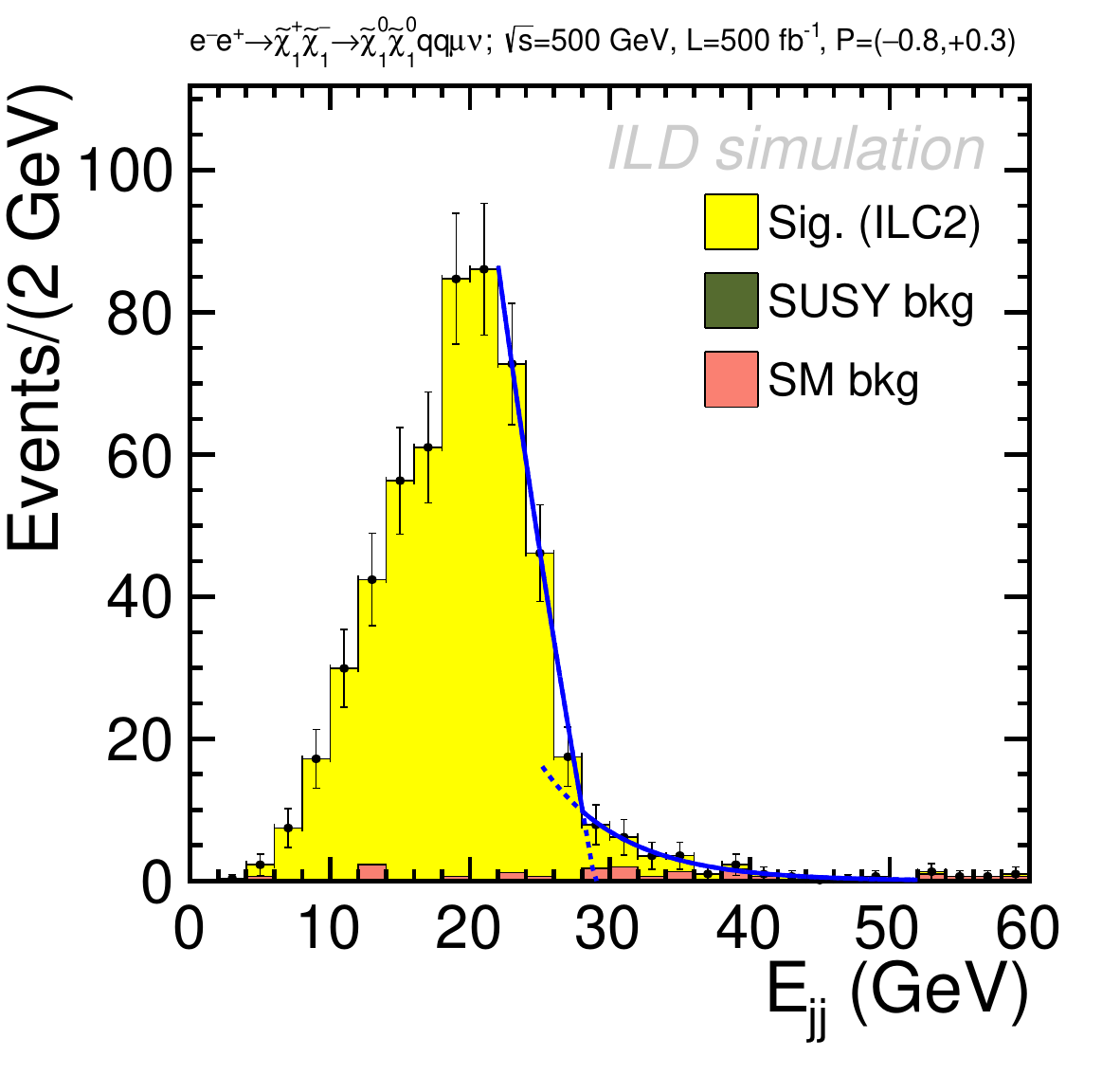}
      \includegraphics [scale=0.23]{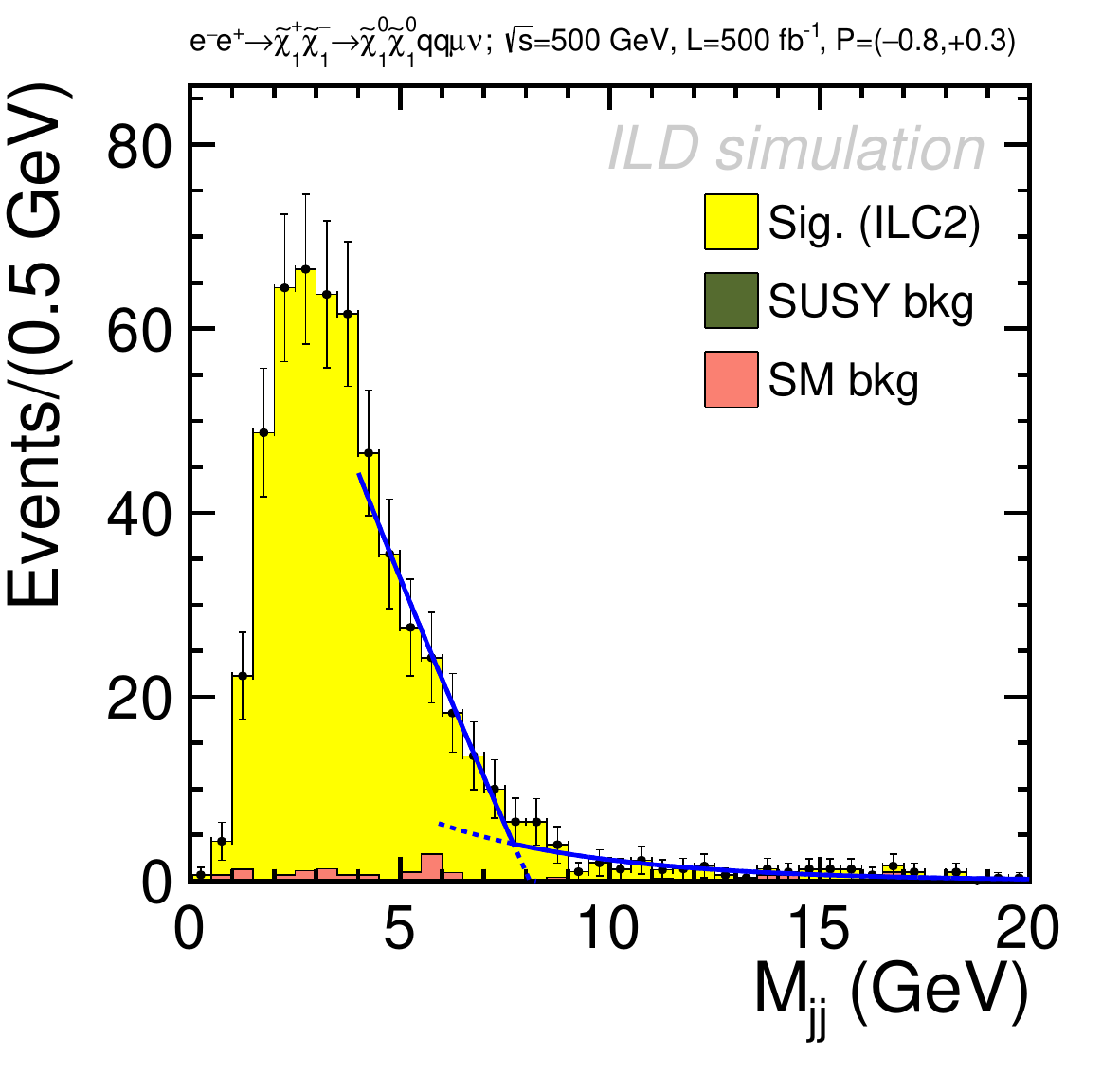}
      \includegraphics [scale=0.25]{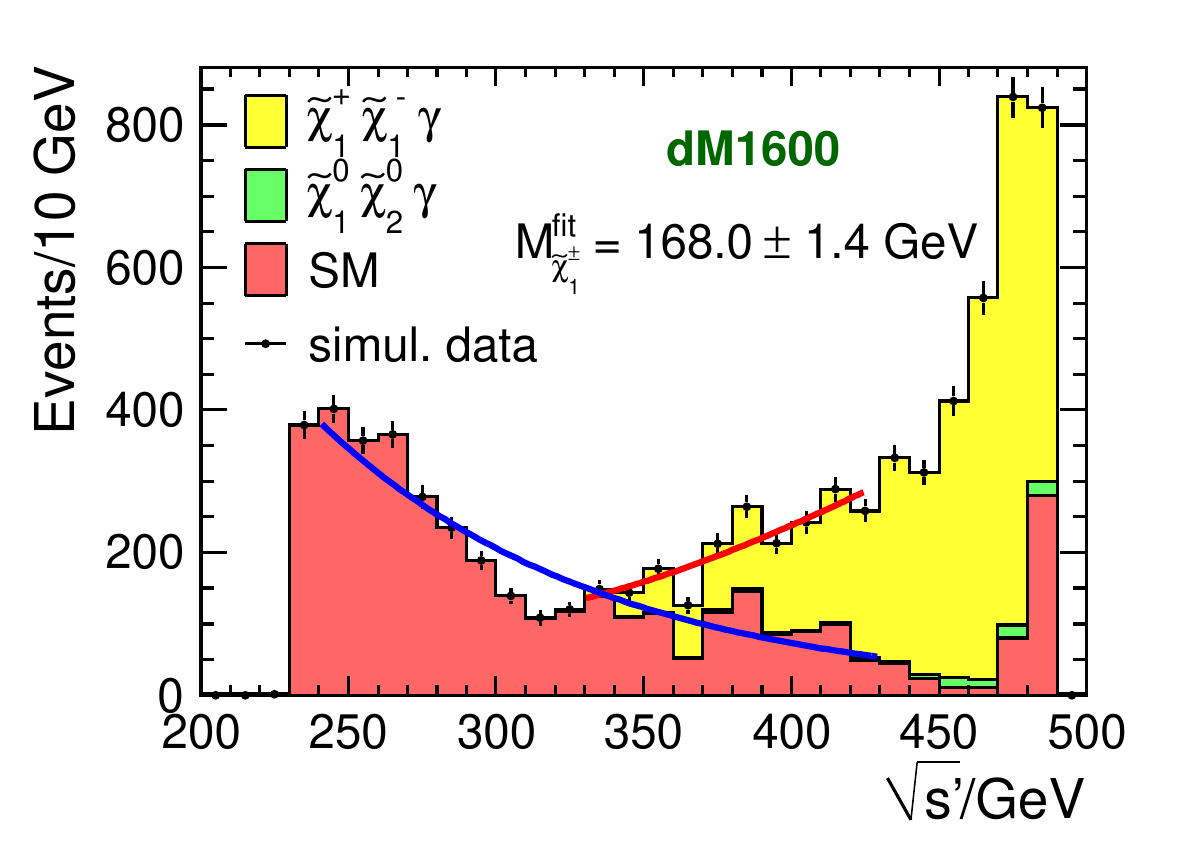}

      \includegraphics [scale=0.23]{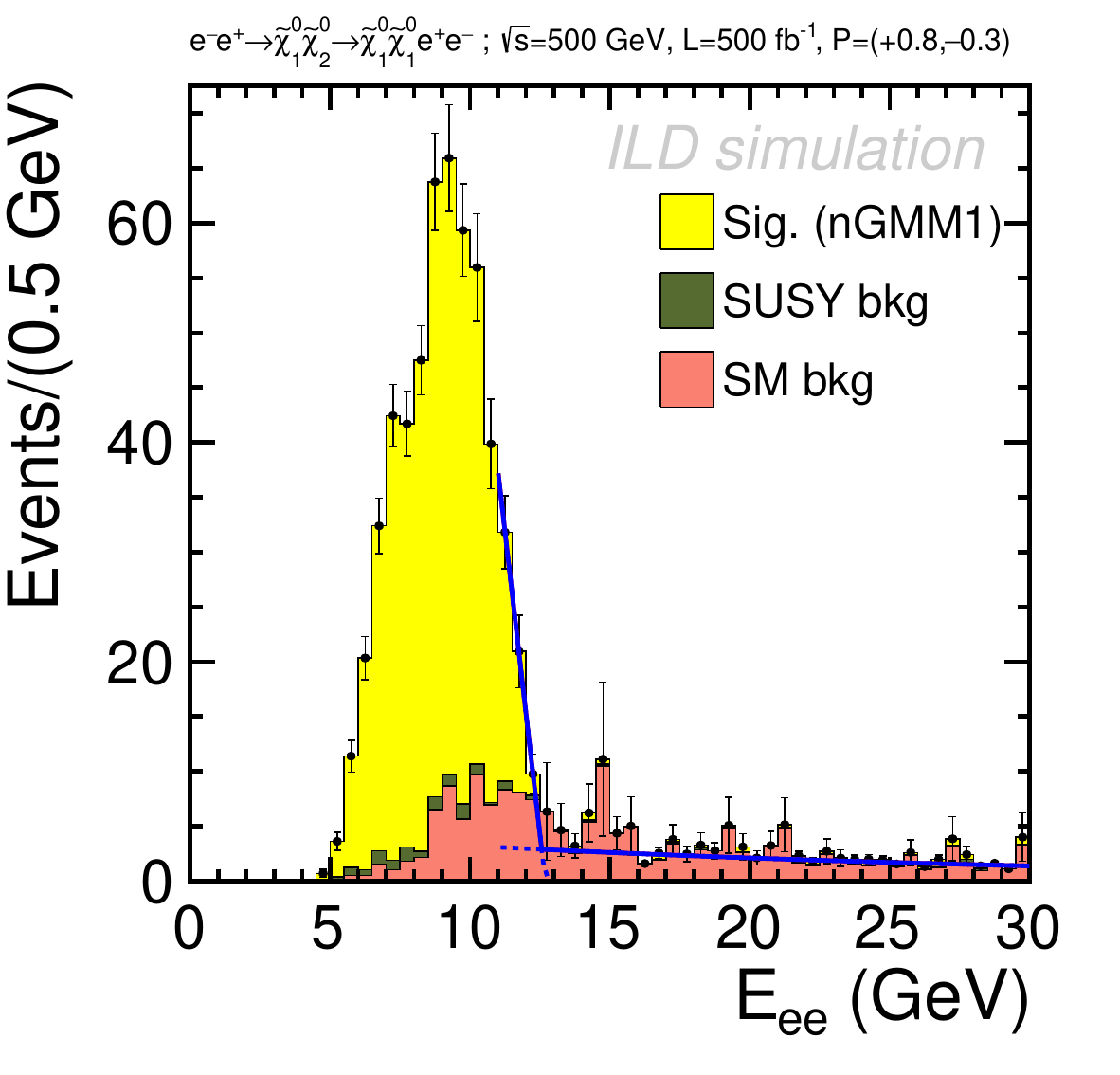}
      \includegraphics [scale=0.23]{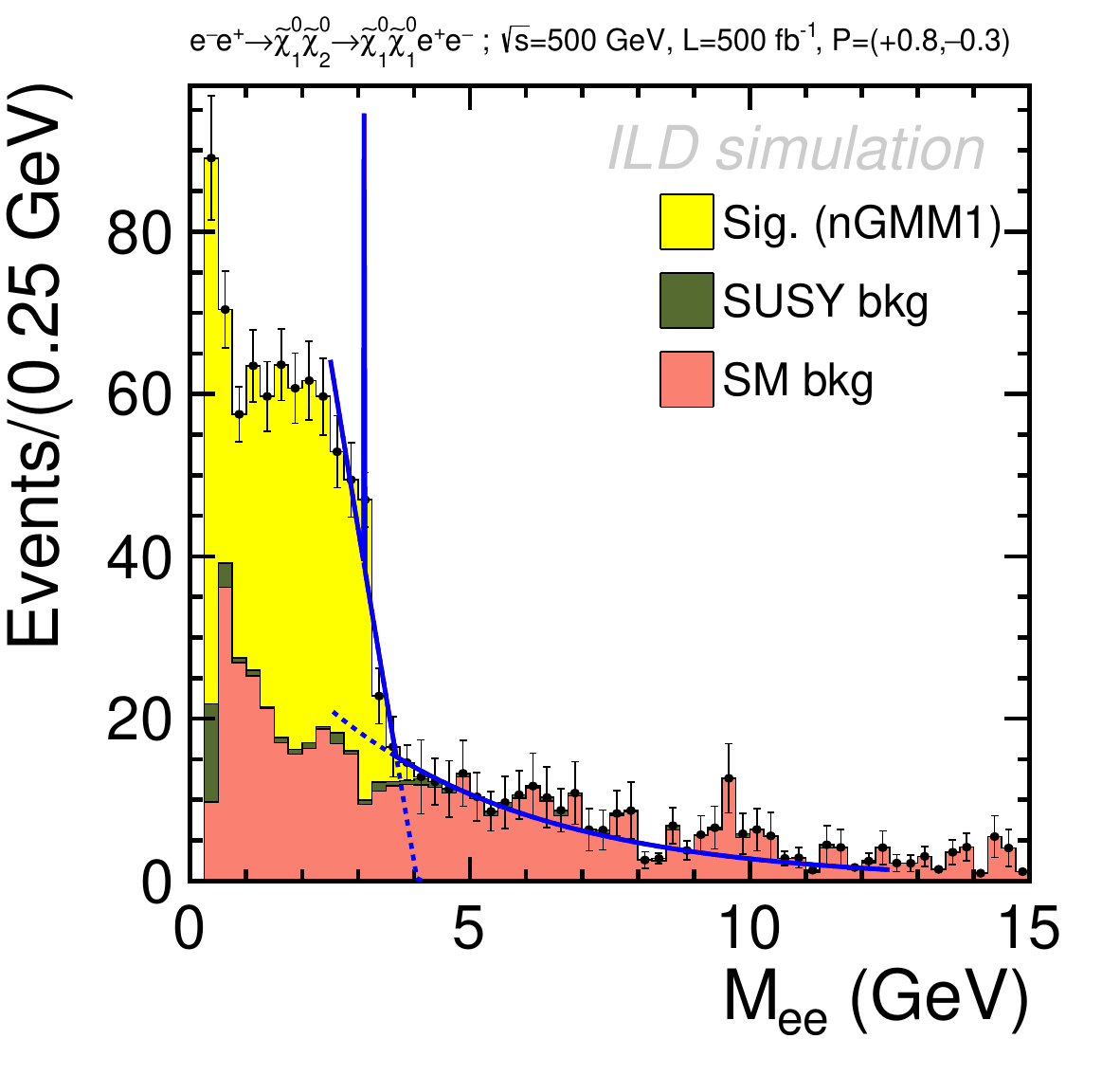}
      \includegraphics [scale=0.25]{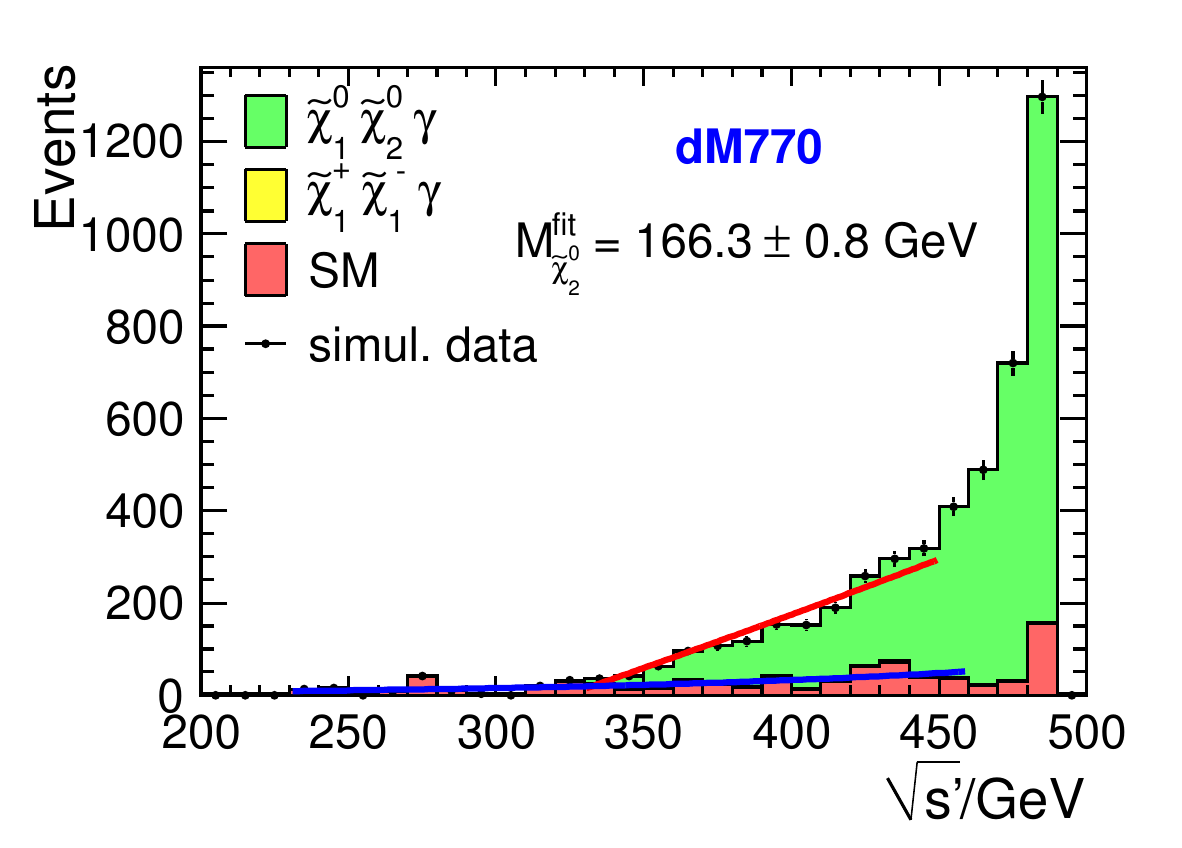}
    \end{center}
    \caption{Top row: \PSGt, \PSGm and \PSe energy spectra. Middle and bottom rows: Observables
      for three different Higgsino-LSP models. The middle row shows the case of \PSGcpmDo production, and the bottom row 
      that of \PSGczDt production.  \label{fig:sleptC1N2}}
\end{figure}  

\subsubsection*{NLSP search: ILC-500 to FCC-ee-240 comparison}\label{sect:extrap}

In Ref.~\cite{Berggren:2024ckz},
some well-founded conclusions were drawn by extrapolating the
ILC-500 results to FCC-ee-240 conditions.
This includes both re-scaling the results to a lower $E_{CM}$,
taking the different beam-conditions into account,
and evaluating the effect of the change in detector acceptance.
    The conclusion was that one expects S/B at one  $E_{CM}$ to be the same as that at another
    $E_{CM}$ if one scales the kinematic cuts and the SUSY masses with the
    ratio of the two
    $E_{CM}$, if all other conditions are identical.
    The authors then took into account the expected difference in cross sections both for signal and background
    when going from one $E_{CM}$ to another, the fact that ILC (unlike FCC-ee) foresees polarisation of the beams,
    that beam-beam effects are expected to be lower at FCC-ee, and that detector coverage at low angles is
    much compromised at FCC-ee due to the needs of the final focusing system. 
As an example,
the authors took the model-point with
$M_{\PSGt}$ = 245 \GeV and
$\Delta M$ = 10 \GeV.
Under ILC-500 conditions, this point can be excluded at the level of 2.35$\sigma$,
but the corresponding point at $E_{CM}$ = 240 \GeV, i.e.\  $M_{\PSGt}$ = 118 \GeV and
$\Delta M$ = 4.8 \GeV, cannot be excluded 
under FCC-ee conditions,
even if four experiments are assumed to collect data.
The main reason for this is the much worse detector coverage at low angles
mentioned above, but also the lack of polarisation.
In Fig.~\ref{fig:C1stauexcl}(b), a zoom of Fig.~\ref{fig:C1stauexcl}(a) is given,
where the extrapolation to FCC-ee-240 is overlaid.
\subsubsection*{SUSY at Higgs factories: not just exclusion or discovery}
  The fact that exclusion and discovery are so close under \epem conditions means that
a SUSY discovery would take place quite quickly.
The situation that an interesting SUSY signal is at the intermediate
level (neither excluded, nor discovered) for years will never occur: either the
process is not reachable and there is no sign of it, or it will be discovered immediately.
This means that SUSY studies at a Higgs factory would almost immediately enter the
realm of precision studies. The plots in \cref{fig:sleptC1N2} show a number of examples of the
type of signals that can be expected.
 Typical slepton signals (\PSGt, \PSGm and \PSe) are shown in the top row,
       in a  \PSGt co-annihilation model (FastSim) \cite{Berggren:2015qua}.
      Typical chargino and neutralino signals in different Higgsino LSP models are shown in the following rows.
       The two plots on the left are models  with moderate (a few to some tens \GeV) $\Delta{M}$ (FullSim) \cite{Baer:2019gvu},
       while those on the  right are  for a  model with very low (sub-\GeV) $\Delta{M}$ (Fast/FullSim) \cite{Berggren:2013vfa}.
      In all the cases illustrated,
 SUSY masses could be determined at the sub-percent level, and 
 the polarised production cross sections at a level of a few percent.
 Many other properties could also be obtained from the same data, such as
 decay branching fractions, mixing angles, and the spin of the sparticle~\cite{Berggren:2015qua,Baer:2019gvu,Berggren:2013vfa}.

\subsection{Dark matter \label{sec:SRCH-darkmatter}}
\editors{J.Kalinowski}


Weakly interacting massive particles (WIMPs) are among the primary candidates for dark matter (DM) and are being searched for via many different experimental approaches. While dark matter particles do not interact with the detector material and escape detection, visible particles recoiling against WIMPs can be used to identify their  signature. High energy $\epem$ colliders are considered perfect tools for the most
general search for pair-production of DM particles with a mono-photon signature, when only a single hard photon
radiated from the initial state is observed in the detector. So far, the only lepton collider bounds on WIMPs using the mono-photon signature have been derived at LEP~\cite{DELPHI:2003dlq}.
For the proper simulation of BSM  and SM background processes with mono-photon signature, a procedure that allows matching the soft ISR
radiation with the matrix element (ME) level calculation of detectable hard photons in \whizard\ \cite{Kilian:2007gr}, has been proposed in Ref.~\cite{Kalinowski:2020lhp}. 
\subsubsection{Simplified dark matter model}
We consider a simplified DM model which covers most popular scenarios of
dark matter pair-production at $\epem$ colliders.
In this model the dark matter particles, $\PGc_i$, couple to the
SM particles via the mediator, $Y_j$.
Each simplified scenario is characterised by one mediator 
and one dark matter candidate. 
The interaction between DM and the electrons can be mediated by a real
scalar $Y_R$ or a real vector $Y_V$, with the Lagrangian describing mediator
coupling to electrons given by
\begin{eqnarray*}
  \mathcal{L}_{eeY} & \ni & 
     \bar{e} (g_{\Pe Y_R}^{(1)} + \imath \PGg_5 g_{\Pe Y_r}^{(5)}) \Pe Y_R +
   \bar{\Pe} \PGg_\mu (g_{\Pe Y_V}^{(1)} + \PGg_5 g_{\Pe Y_V}^{(5)}) \Pe Y_V^\mu
   \, .
  \end{eqnarray*}
Only real values of couplings $g_{\Pe Y_R}^{(1)}$, $g_{\Pe Y_r}^{(5)}$, $g_{\Pe Y_V}^{(1)}$ and
$g_{\Pe Y_V}^{(5)}$ are taken into account, and, depending on the coupling choice, different scenarios can be considered. 

For DM particle masses significantly below half of
  the mediator mass, the pair-production cross section can be given in
  terms of the mediator partial widths (to electrons and DM particle),
  its total width and mass. 
  Since we assume that the total width is dominated by the DM partial width,
  the cross section dependence on the DM particle couplings is
  absorbed in the total mediator width and the limits extracted
  for fixed mediator mass and width hardly depend on the DM
  particle type or coupling structure. 
Therefore, only the Dirac fermion, $\PGc_D$, is considered as the DM particle,
and its interactions with possible mediators are described by
\begin{eqnarray*}
  \mathcal{L}_{\PGc\PGc Y} & \ni &
  \bar{\PGc}_D (g_{\PGc_D Y_R}^{(1)} + \imath \gamma_5 g_{\PGc_D Y_R}^{(5)}) \PGc_D Y_R +
   \bar{\PGc}_D \PGg_\mu (g_{\PGc_D Y_V}^{(1)} + \gamma_5 g_{\PGc_D Y_V}^{(5)}) \PGc_D Y_V^\mu
\, ,
\end{eqnarray*}
where, for simplicity,  the structure of mediator couplings to DM
fermions is assumed to be the same as to SM fermions.
(The DM particle type and its coupling structure become
  relevant only when the cross section limits are translated to the limits
  on the product of mediator couplings.)

For calculating the DM pair-production cross section and generating
signal event samples this simplified model \cite{SimpDMwiki} was encoded into \textsc{FeynRules}
\cite{Alloul:2013bka}  and exported in UFO format used as input to \whizard. 
The fast detector simulation framework
\textsc{Delphes} \cite{delph} with generic ILC detector model 
that included the 
description of the calorimeter systems in the very forward region 
was used to derive results presented in this contribution.
\subsubsection{Light mediator scenarios}
Most of the generated photons travel along the beam line and only a small fraction is reconstructed as mono-photon events in the detector.  The fraction of ``tagged'' events depends significantly on the mediator mass and width, as shown for the 250 GeV and 500 ILC and CLIC, respectively,  in \cref{fig:zarnecki_mono-cros}. 
The limits on the DM pair-production cross section and couplings are derived as a function of the mediator mass and width from two-dimensional distributions of the reconstructed mono-photon events in pseudorapidity and transverse momentum for background and signal events. In all cases,  DM particles are assumed to be light fermions: 1  \GeV for the 250 \GeV ILC case, and 50 \GeV for the 500 \GeV ILC and 3 TeV CLIC cases. The mass of the DM particle has a marginal effect as long as it is significantly below half of the mediator mass.
\begin{figure}[h!]
\begin{center}
\includegraphics[width=0.31\textwidth]{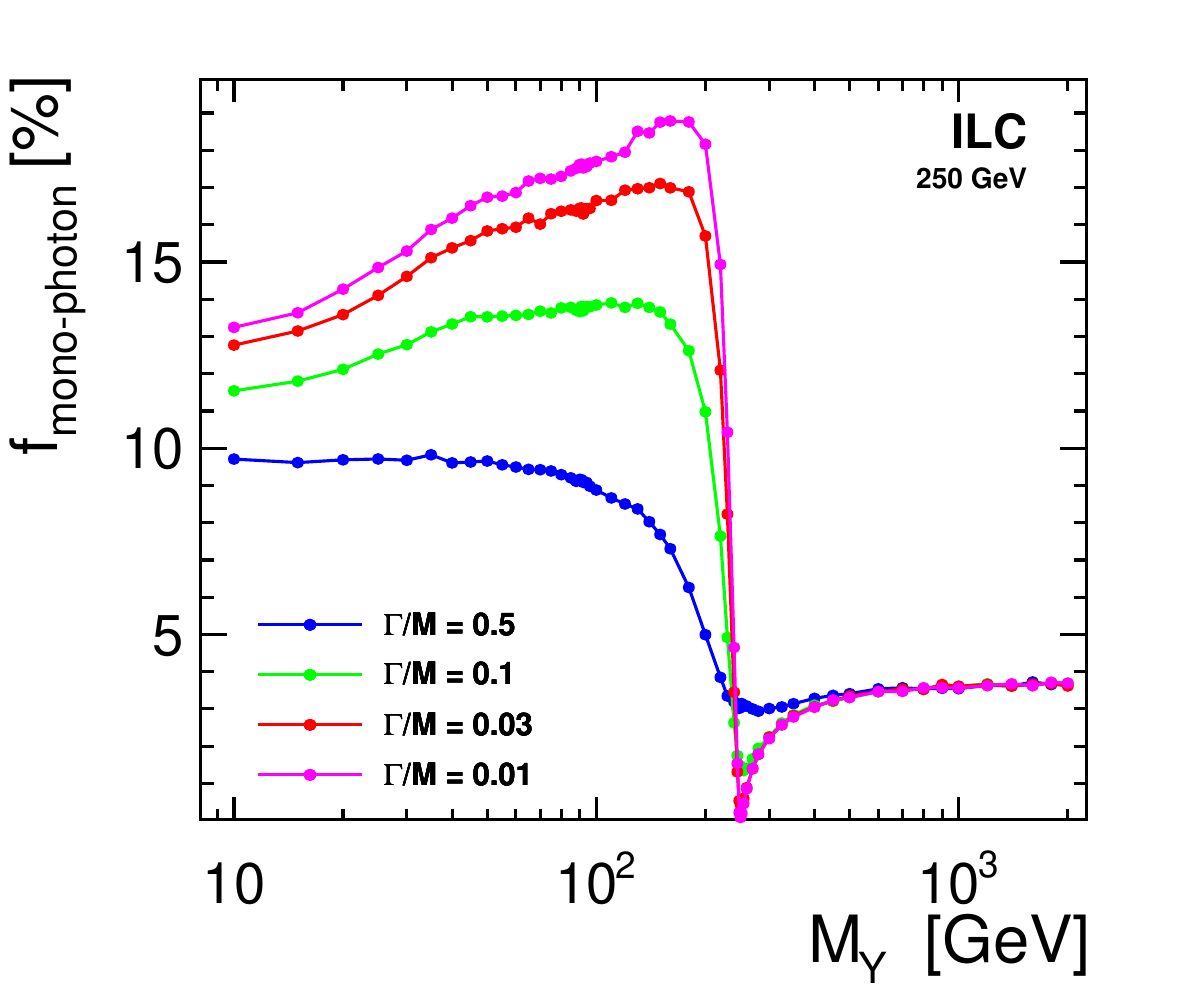}
\includegraphics[width=0.31\textwidth]{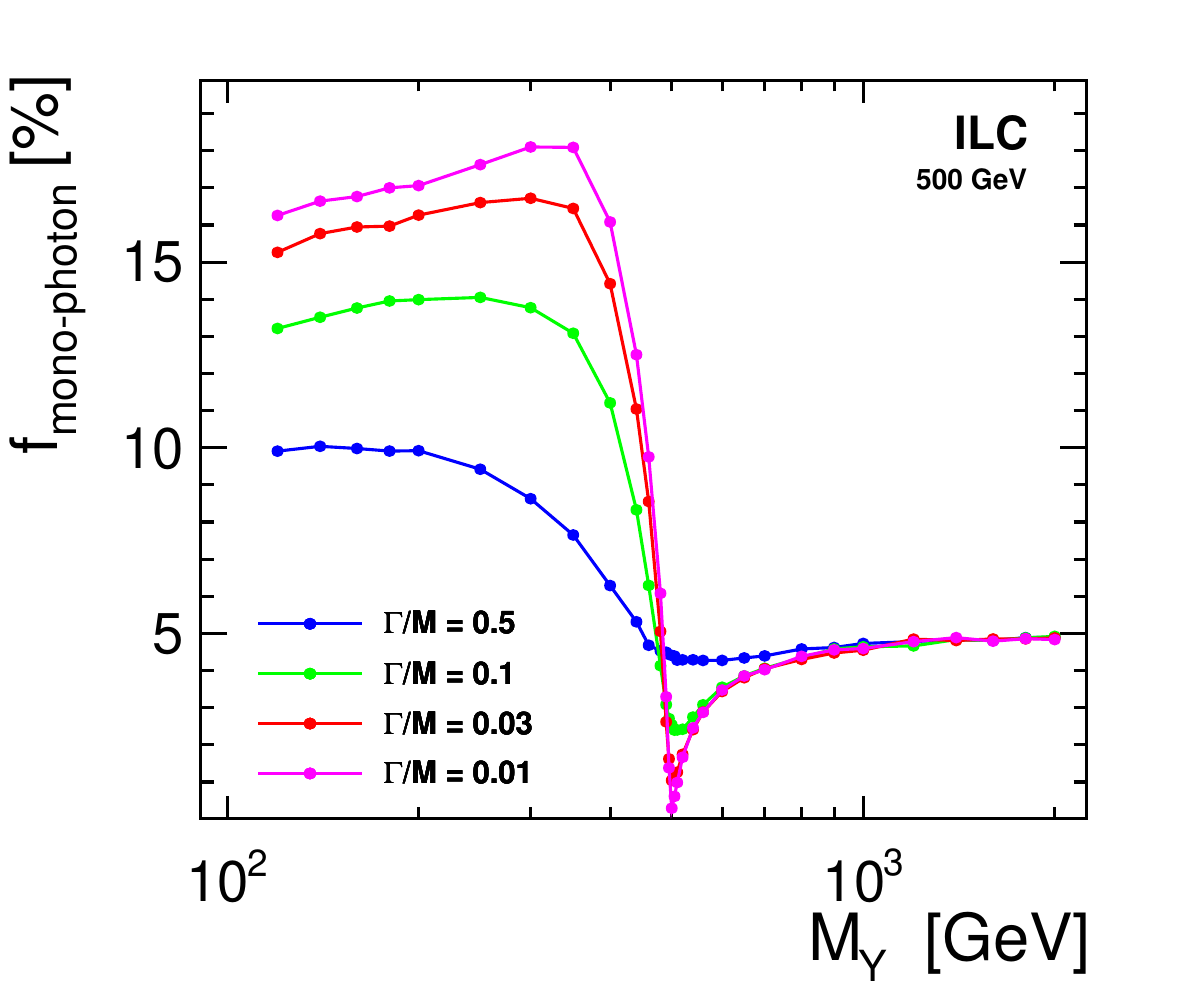}
\includegraphics[width=0.31\textwidth]{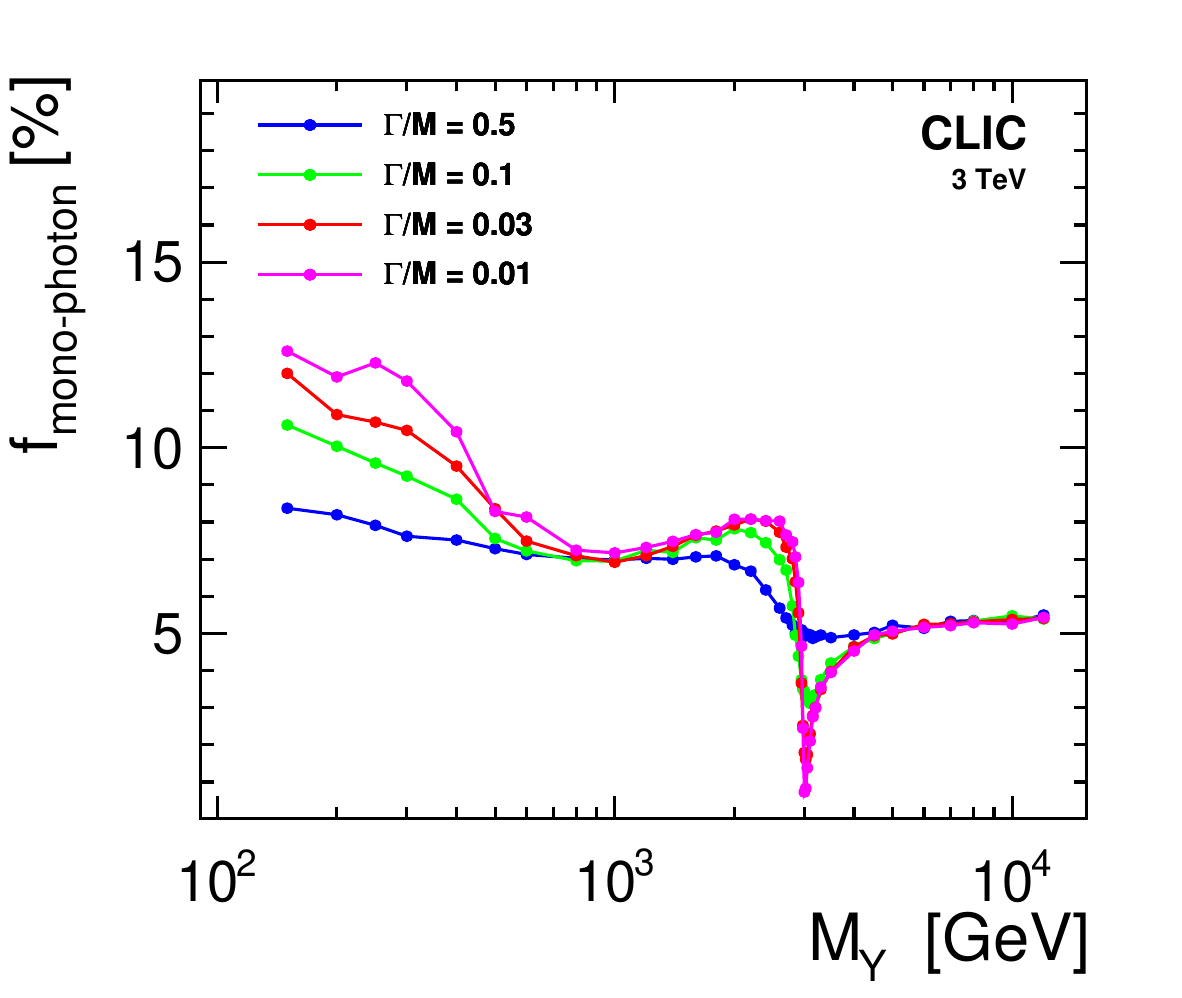}
\end{center}
\caption{Fraction of dark matter pair-production events which are reconstructed as mono-photon events in the detector,
presented as a function of mediator mass for different fractional mediator widths for ILC 250 (left), ILC 500 (middle) and CLIC (right).}
\label{fig:zarnecki_mono-cros}
\end{figure}

Cross section limits for  DM pair-production are extracted from the
combined analysis of data taken with different beam polarisations,
resulting in the strongest limits, also reducing the impact of
systematic uncertainties. In \cref{fig:zarnecki_mono-cros2},  cross section limits expected from the
combined analysis of 250 \GeV ILC data  are shown for vector mediator exchange and different fractional mediator widths.
The strongest limits are obtained for processes with light mediator
exchange and for narrow mediator widths, whereas for heavy mediator
exchange (M$_\text{Y} \gg \sqrt{s}$), cross section limits no longer
depend on the mediator width.

\begin{figure}[h!]
\begin{center}
\includegraphics[width=0.31\textwidth]{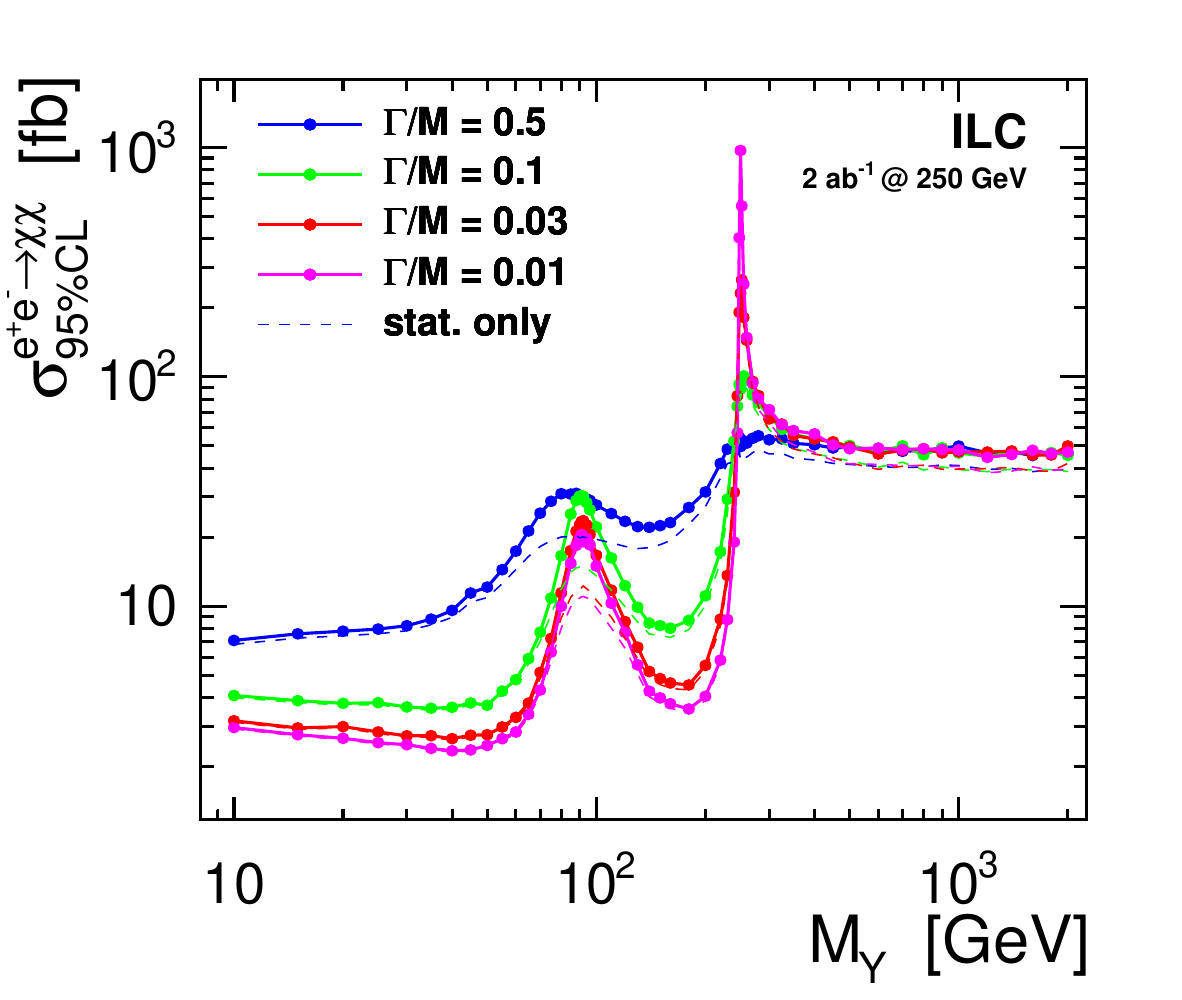}
\includegraphics[width=0.31\textwidth]{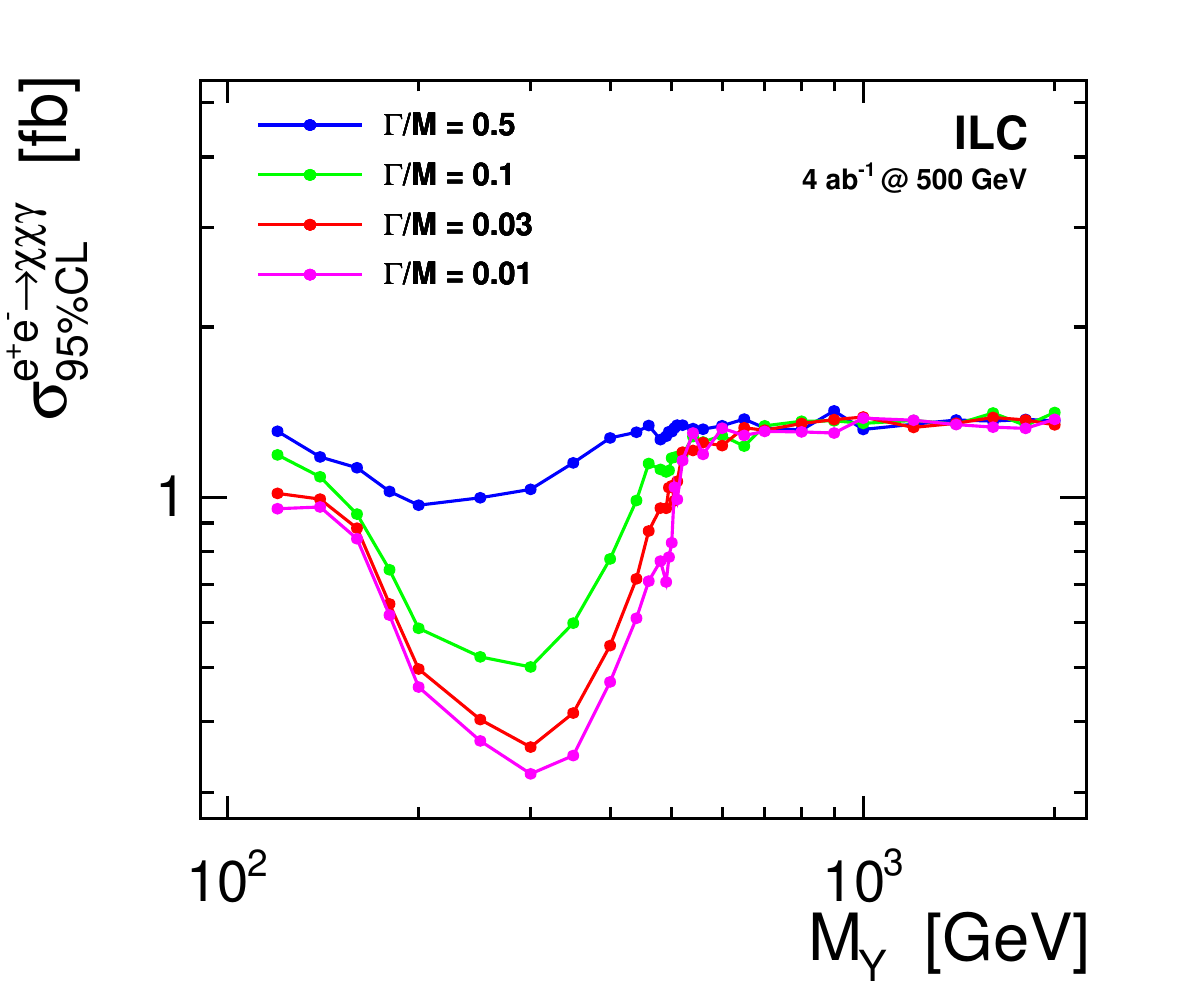}
\includegraphics[width=0.31\textwidth]{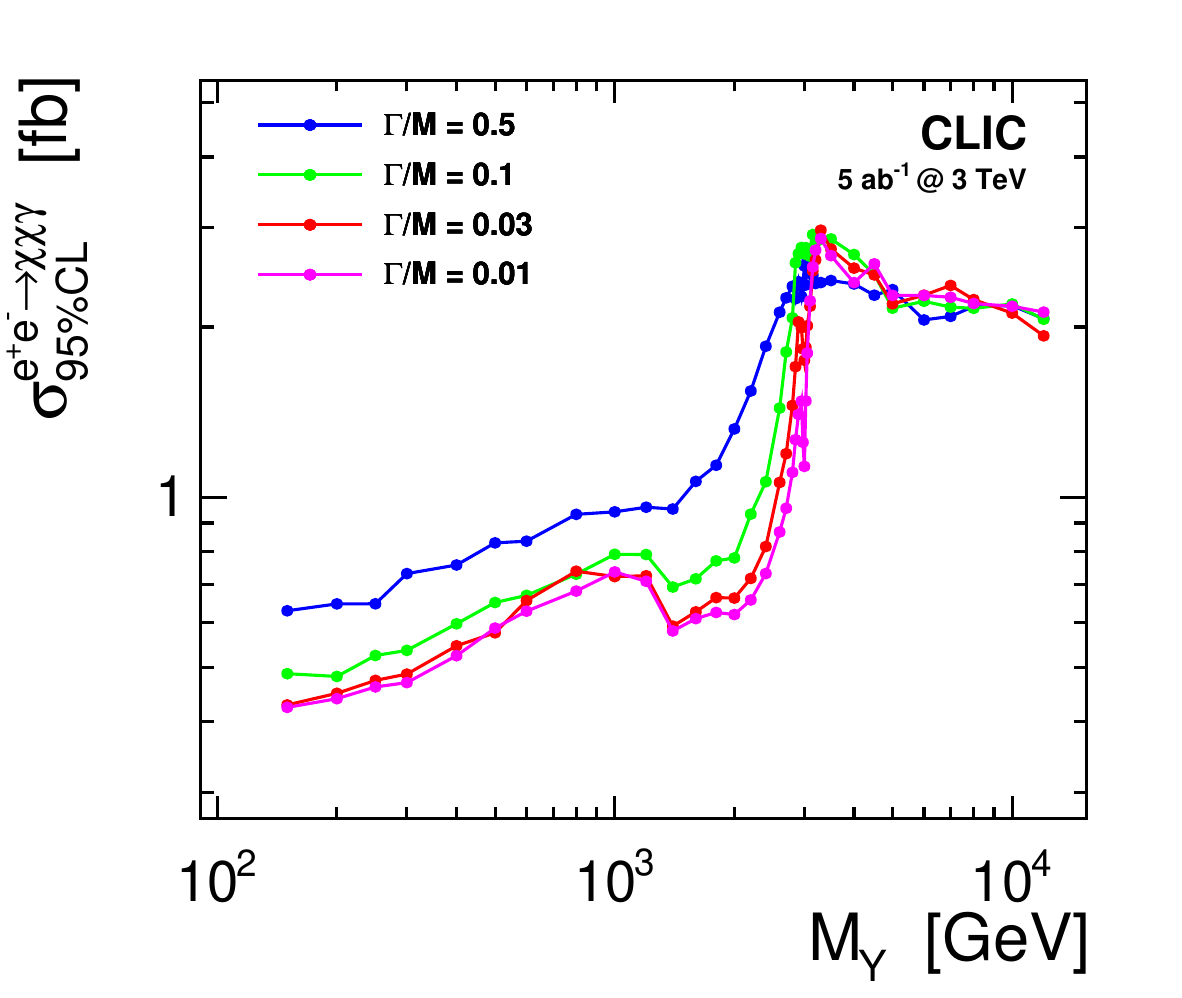}
\end{center}
\caption{Exclusion limits on the cross section for the radiative DM
pair-production processes with $s$-channel  vector mediator exchange
for the ILC running at 250 GeV (left) and 500\,GeV (middle), and CLIC running at 3\,TeV (right) and different fractional  mediator widths, as indicated in the plot. Combined limits corresponding to the assumed running scenarios
are presented with systematic uncertainties taken into account.}
\label{fig:zarnecki_mono-cros2}
\end{figure}
For heavy mediator scenarios, $M_{Y} \gg \sqrt{s}$, expected cross
section limits do not depend on the assumed mediator width and weakly depend
on the assumed mediator mass.
For lower masses, the cross section limits improve and the expected
sensitivity is highest, as expected, for narrow mediator scenarios.
For the 250 GeV ILC, limits are significantly weaker for narrow mediators with M$_\text{Y} \approx \sqrt{s}$, when photon radiation is significantly suppressed, and at M$_\text{Y} \approx \text{M}_{\PZ}$, when signal events are similar to radiative \PZ return events. 
Surprisingly, the shape of the mass dependence below the resonant
production threshold is very different for ILC and CLIC.
This can be understood as an effect of the much wider luminosity
spectra at CLIC significantly enhancing the resonant production cross
section for $M_{Y} < \sqrt{s}$.
For ILC running at 500\,GeV, the best limits are obtained for resonances with 
$M_{Y} \approx \frac{1}{2}\sqrt{s}$, while lowest mediator masses are
constrained best for CLIC.
Still, one should note that the presented 95\% C.L. limits on the
radiative DM pair-production depend rather weakly on the mediator mass
and width over a range of more than an order of magnitude.
%

As already discussed, we assume that the
mediator coupling to SM particles is small and its decays to SM particles
can be neglected.
The total mediator width is dominated by the invisible width resulting
from its coupling to DM particle, $g_{\PGc\PGc Y}$. 
This coupling is thus fixed when considering a given
mediator scenario by specifying mediator mass and width.
This allows us to translate the extracted cross section limits to the
limits on the mediator coupling to electrons, $g_{\Pe\Pe Y}$, 
the only free model parameter after fixing the {DM type,} coupling structure, masses
and mediator width. 
Limits on the mediator coupling to electrons, expected from combined
analysis of ILC and CLIC data,
are shown in Fig.~\ref{fig:coup_wid}
for different mediator widths.
\begin{figure}[tbp]
\includegraphics[width=0.31\textwidth]{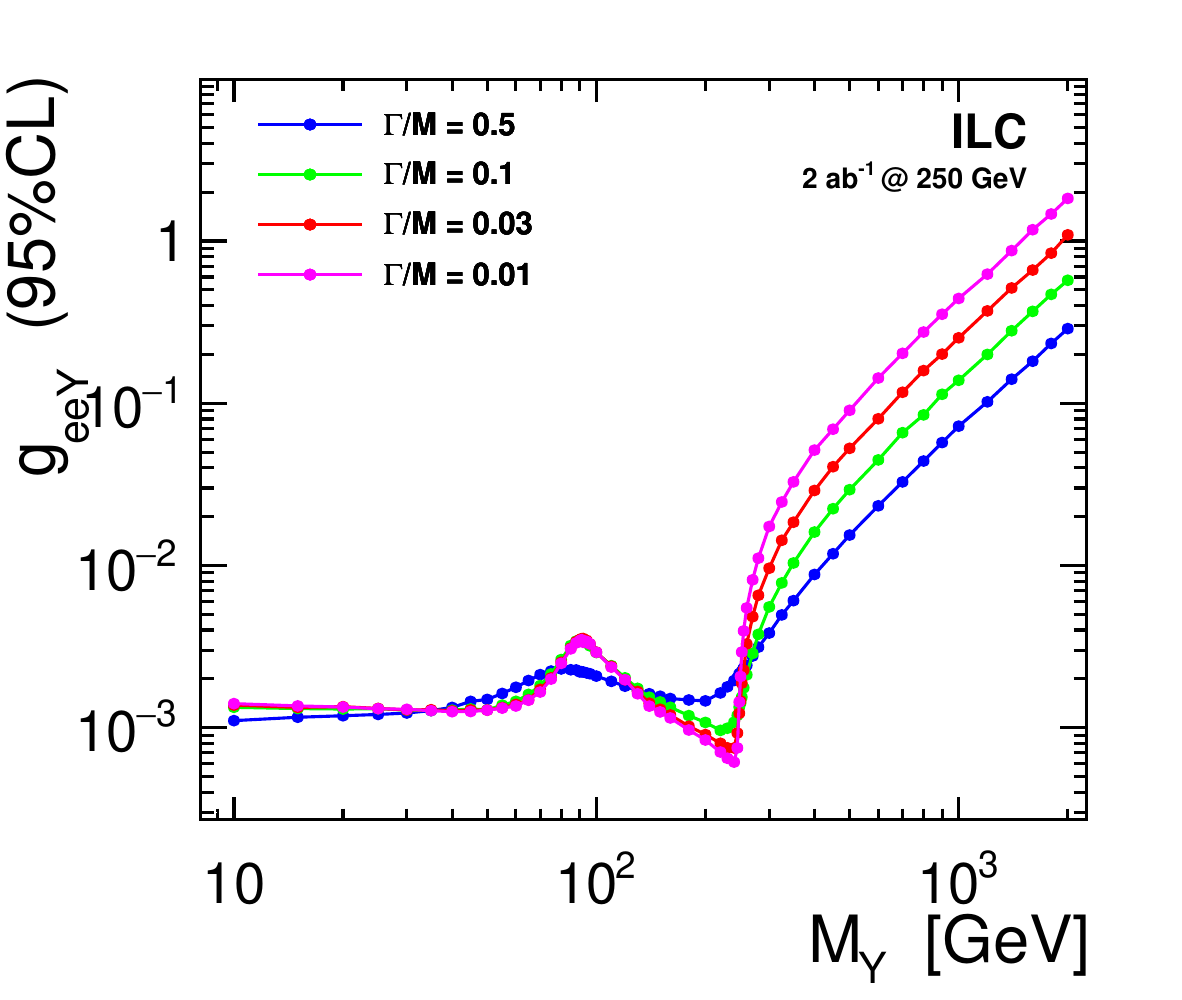}
 \includegraphics[width=0.31\textwidth]{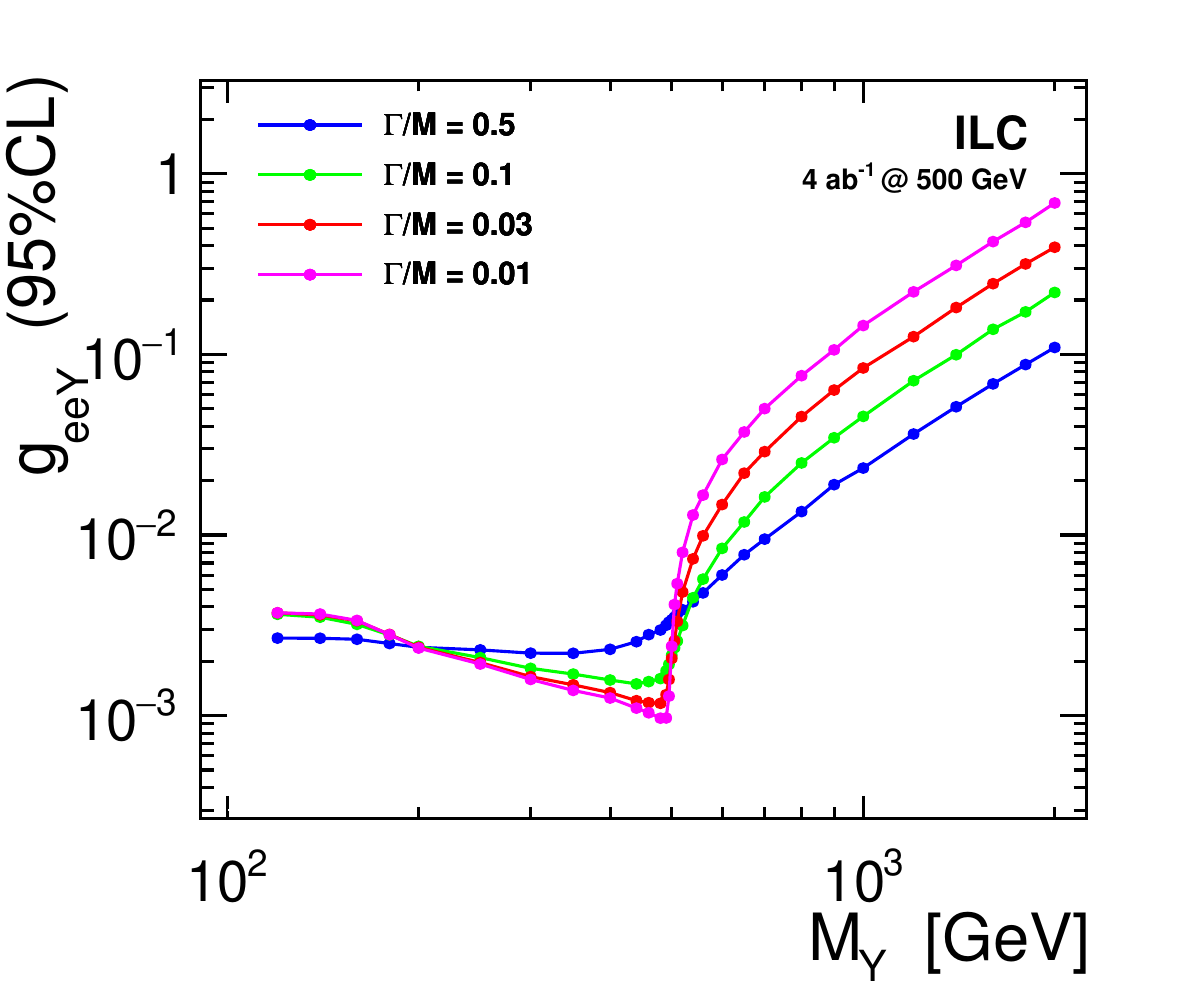}
 \includegraphics[width=0.31\textwidth]{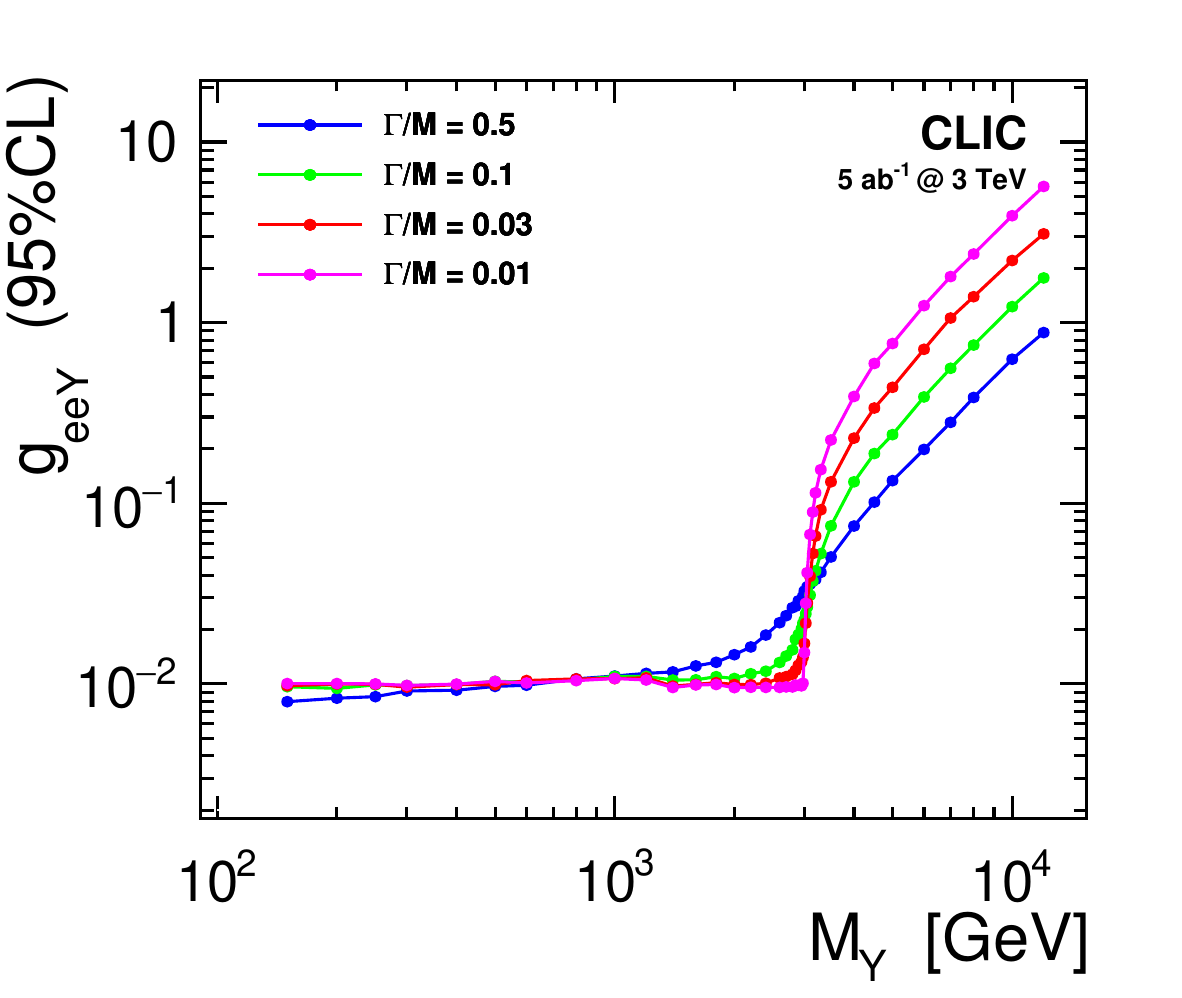}
  \caption{Exclusion limits on the vector mediator coupling to electrons in radiative DM pair-production for the ILC running at  250 GeV (left) and  500\,GeV (middle), and CLIC running at 3\,TeV (right) and different fractional mediator widths, as indicated in the plot. Combined limits corresponding to the assumed running
    scenarios are presented with systematic uncertainties taken into
    account.
  } 
  \label{fig:coup_wid}
\end{figure}
For scenarios with light mediator exchange, SM
coupling limits hardly depend on the mediator width. For the Vector
mediator scenario, limits of $(1--4) \times 10^{-3}$ are expected at
250 and 500\,GeV ILC, improving with mediator mass (except for the largest
mediator width), while a limit of about $10^{-2}$ is expected at 3\,TeV
CLIC, almost independent of the mediator mass (up to
kinematic limit) and width.
For a width of $\Gamma/M=0.03$, these coupling limits correspond to the
upper limits on the mediator branching ratio to electrons of about
$10^{-5}$ at 500\,GeV ILC and $10^{-4}$ at 3\,TeV CLIC.
This corresponds to single expected events and indicates that the
analysis of mono-photon spectra gives higher sensitivity to processes
with light mediator exchange than their direct searches in SM decay channels.
\subsubsection{Heavy mediator scenarios}
For heavy mediator exchange, the coupling limits increase with the
mediator mass squared, $\text{g}_\text{eeY} \sim
\text{M}_\text{Y}^2$, as expected in the
effective field theory (EFT) approach.
\begin{figure}[tbp]
\includegraphics[width=0.31\textwidth]{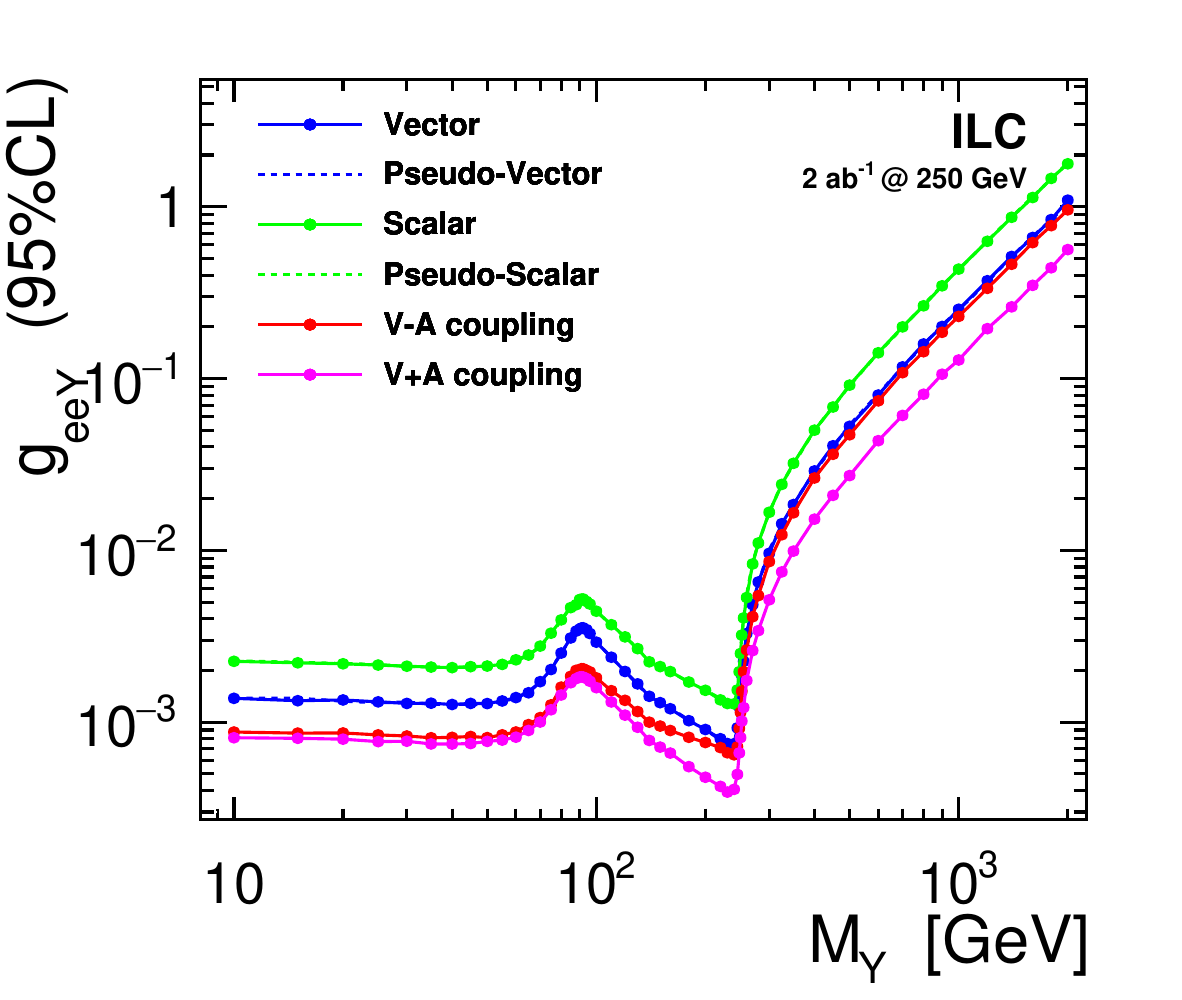}
 \includegraphics[width=0.31\textwidth]{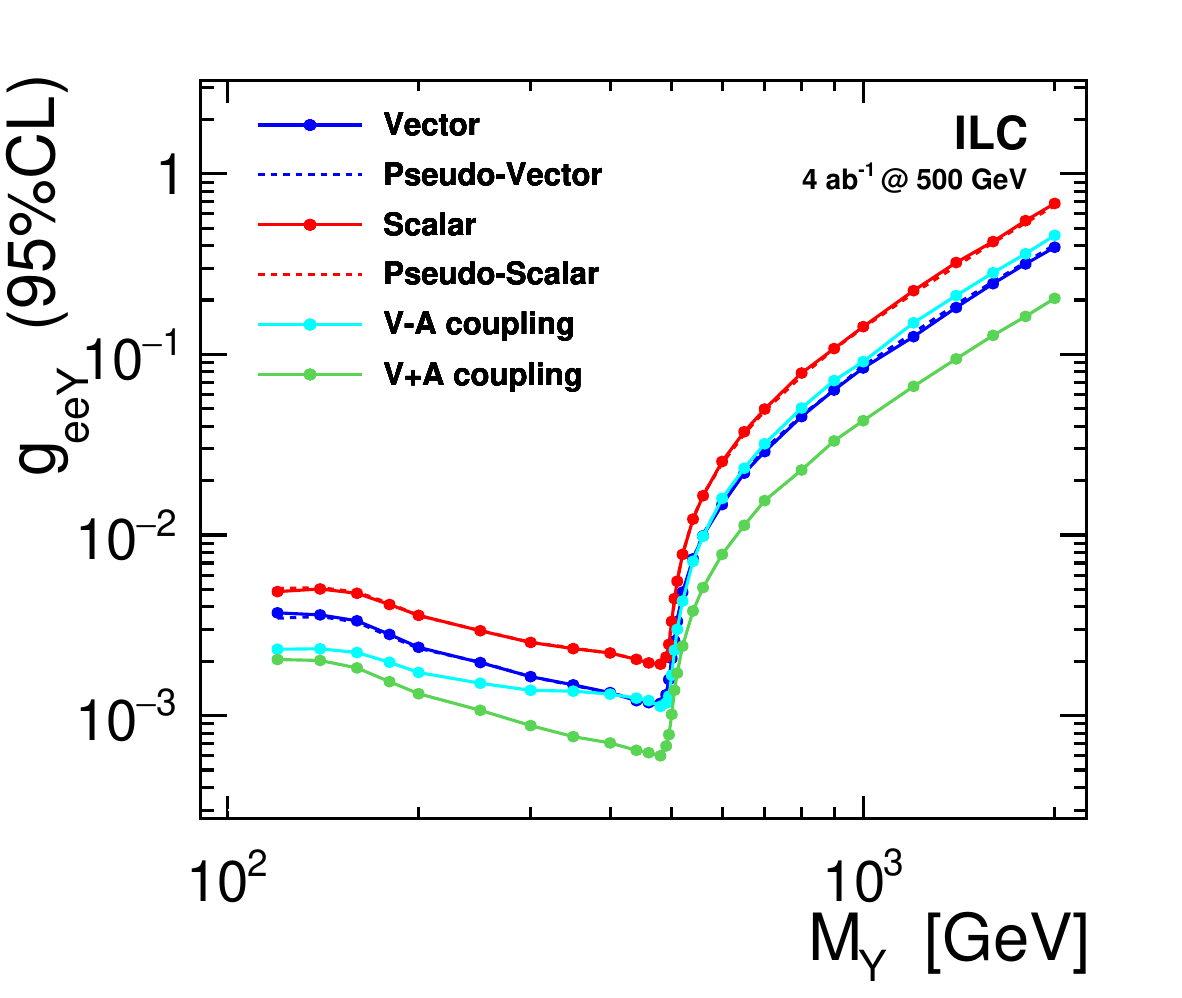}
 \includegraphics[width=0.31\textwidth]{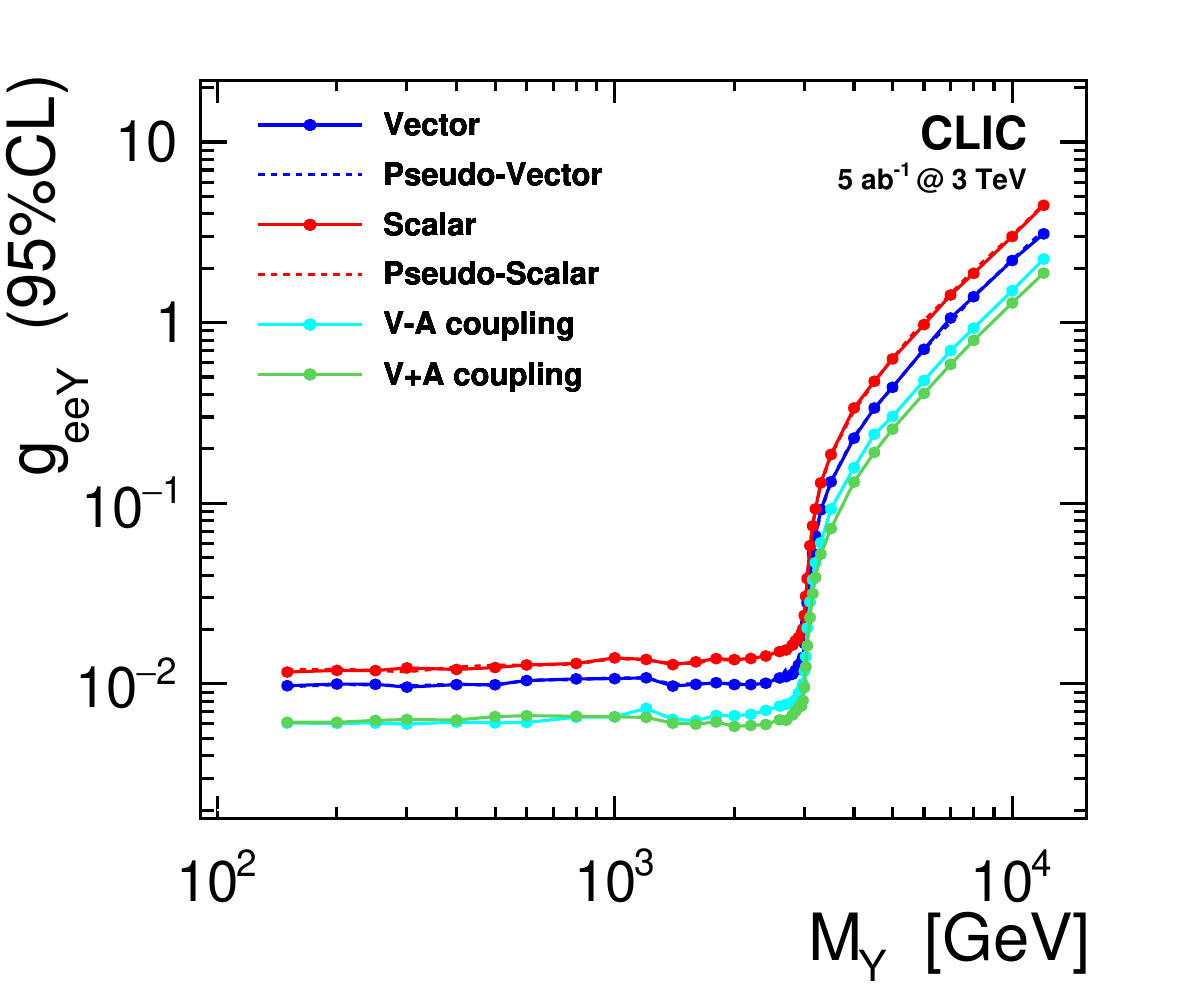}
  \caption{Limits on mediator coupling to electrons in radiative DM pair-production for the ILC
    running at 250 and 500\,GeV (left and centre) and CLIC running at 3\,TeV (right), for
    relative mediator width, $\Gamma/M = 0.03$, and different mediator
    coupling scenarios, as indicated in the plot. Combined limits
corresponding to the assumed running scenarios are presented with
systematic uncertainties taken into account.  
  } 
  \label{fig:coup_model}
\end{figure}
As shown in Fig.~\ref{fig:coup_model},  coupling limits expected from
the combined analysis of all data vary by up to a factor of 3 for ILC
and up to a factor of 2 for CLIC, depending on the assumed coupling
structure. 
The strongest limits are obtained for the V+A mediator coupling
structure.
For this scenario, the signal cross section is largest for polarisation
combinations corresponding to the lowest background levels.
The weakest limits are expected for Scalar and Pseudo-scalar coupling
scenarios.


As shown in Ref.~\cite{mttd2021},
results of the presented approach \cite{Kalinowski:2021tyr} are in very good agreement
with the limits from the ILD analysis \cite{Habermehl:2020njb} based on the full detector simulation and EFT  formalism. In the EFT framework, the sensitivity depends on the type of operator describing the WIMP production, the mass and spin of the WIMP, and on the parameter $\Lambda$, which defines the energy scale at which the new physics becomes important. Three different operators with vector, axial-vector and scalar tensor structure as presented in Table~\ref{tab:operators}  are used. The energy scale $\Lambda$ is related to the cross section as $\sigma\propto 1 / \Lambda^4$.
At lepton colliders, the probed energy scales are typically much higher than
the centre-of-mass energy, so that the validity of the EFT is ensured.

\begin{table}[h!]
 \begin{center}
  \begin{tabular}[h]{ l c c c }
\hline
   & four-fermion operator & $\sigma(\Pe^-_L,\Pe^+_R)=\sigma(\Pe^-_R,\Pe^+_L)$ & $\sigma(\Pe^-_L,\Pe^+_L)=\sigma(\Pe^-_R,\Pe^+_R) $ \\ \hline
  vector & $(\overline{f}\gamma^\mu f)(\overline{\PGc}\gamma_\mu\PGc)$ & $\sigma\propto 1 / \Lambda^4$ & 0 \\
  axial-vector & $(\overline{f}\gamma^\mu\gamma_5 f)(\overline{\PGc}\gamma_\mu\gamma_5\PGc)$ & 0 & $\sigma\propto 1 / \Lambda^4$ \\
  scalar & $(\overline{\PGc}\PGc)(\overline{f}f)$ & 0 & $\sigma\propto 1 / \Lambda^4$ \\
\hline
  \end{tabular}
  \caption[Effective operators used in this analysis.]{Effective operators and their chiral properties \cite{Habermehl:2020njb}.}
  \label{tab:operators}
 \end{center}
\end{table}

The sensitivity of the ILC to WIMP production was solely based on the full detector simulation study at a \com energy of 500\,GeV. The sensitivity will be presented in the plane of the new physics scale $\Lambda$ versus the WIMP mass $M_{\PGc}$, in which  all $\Lambda$ values below the curve can be discovered or excluded, depending on the tested hypothesis. The testable energy scales are always well above the \com energy and hence effective operators are a suitable approach. 
At a \com energy of 500\,GeV, the baseline running scenario for the ILC comprises 4\,ab$^{-1}$ of data shared between the different polarisation sign configurations of 40\% for each of $(-,+)$ and $(+,-)$, and 10\% for each of the equal-sign configurations.
\Cref{fig:H20_500} shows the $95$\% C.L. exclusion reach for this data set. WIMP masses up to almost half the \com energy can be tested. The sensitivity decreases for higher WIMP masses. In the case of the vector operator, the plateau with constant testable energy scales continues to significantly higher WIMP masses than for the other operators.  A grey area indicates a parameter space which cannot be tested using effective operators, taken here as~$\Lambda\leq\sqrt{s}$. The normalisation and shape-dependent systematic uncertainties are taken into account in the results. 
WIMP production could be discovered for new physics energy scales At 95\% C.L., energy scales up to $3$, $2.8$ and $2.6$\,TeV can be probed for the vector, axial-vector and scalar operators, respectively.

\begin{figure}[htb]
\includegraphics[width=0.33\textwidth]{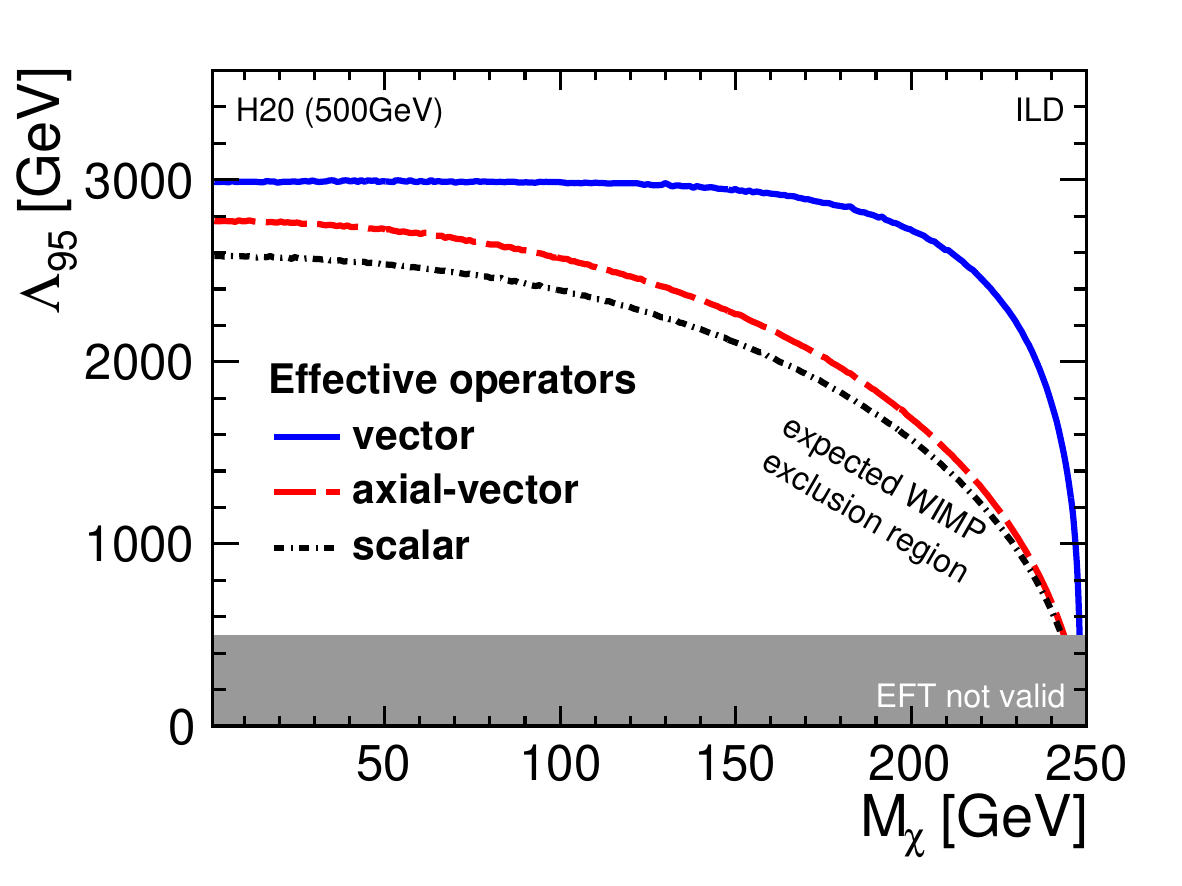} 
\includegraphics[width=0.33\textwidth]{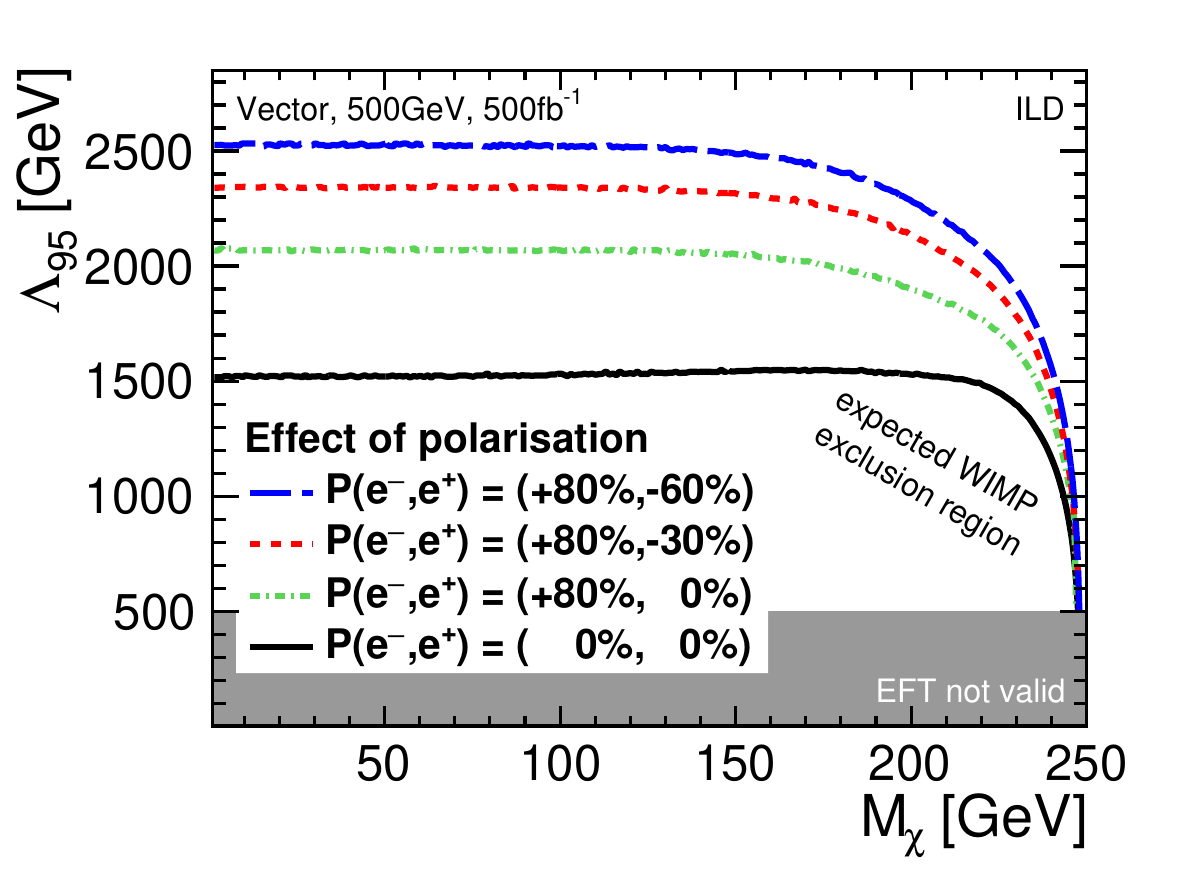} 
 \includegraphics[width=0.33\textwidth]{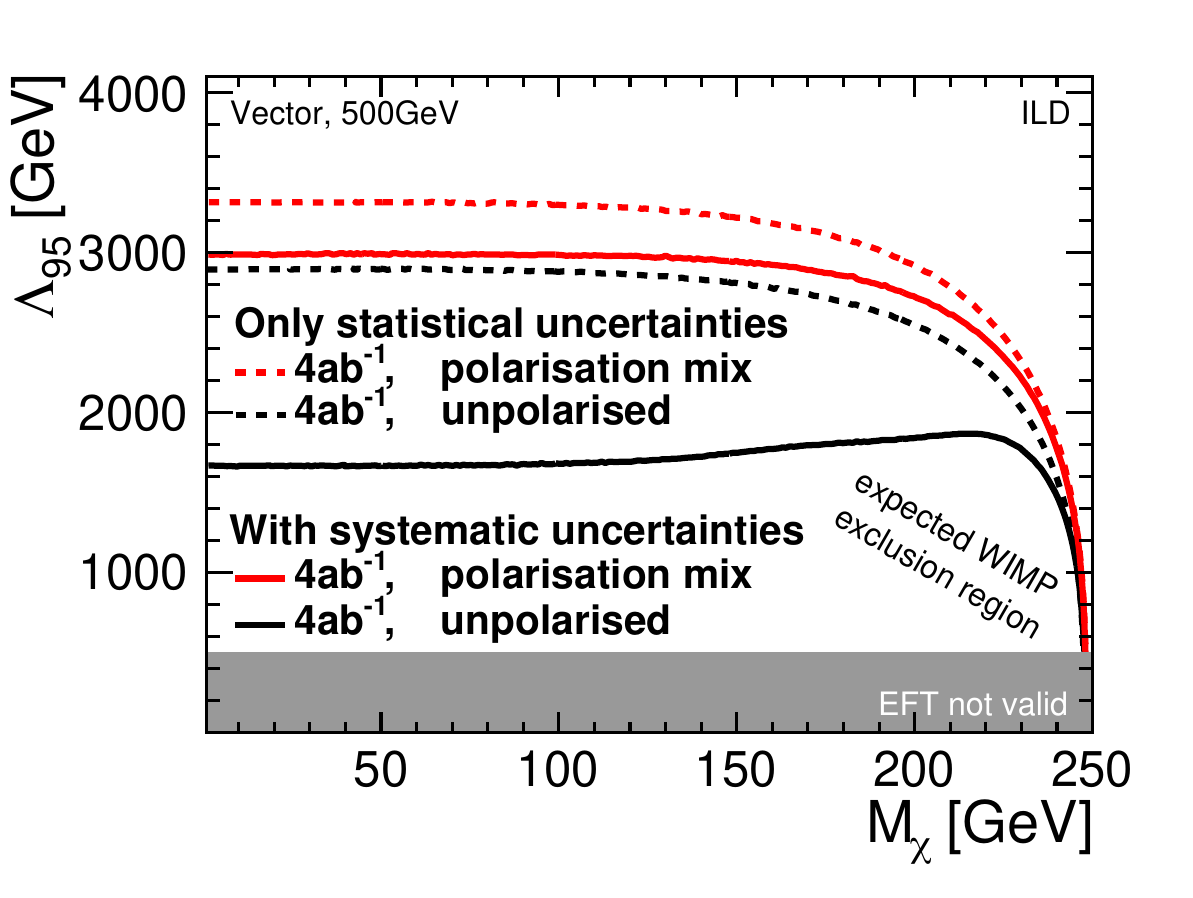}
  \caption{Expected sensitivity for the different effective operators  (left), effect of the beam polarisation for the example of the vector operator (middle), and impact of the systematic uncertainties on the expected exclusion limits (right). Results assume $4$\,ab$^{-1}$ at 500\,GeV.   }
  \label{fig:H20_500}
 \end{figure} 

The sensitivity to processes of radiative DM production is mainly limited by the ``irreducible''
background from radiative neutrino pair-production events, $\epem \to \PGn \,\PAGn  \PGg$.
As the structure of mediator couplings is unknown, data taken
with different polarisation combinations needs to be combined to obtain
the best sensitivity in all possible scenarios.
Moreover, by combining four independent data sets the impact of
systematic uncertainties is significantly reduced.
Exclusion limits from the combined
  analysis of data collected with different beam polarisations (red)
  are compared with limits expected for unpolarised data set with the
  same integrated luminosity in Fig.~\ref{fig:H20_500} (middle). 
The effect of systematic uncertainties is shown in Fig.~\ref{fig:H20_500} (right), where one can notice a significant impact. In particular, for unpolarised case the reach in $\Lambda$ shrinks by nearly a factor 2. The largest reach is provided by the data set with the polarisation mix, which loses only $\sim$10\% in reach.

The importance of the electron beam polarisation for the dark matter search sensitivity with the mono-photon signature was also demonstrated in the CLIC study \cite{Blaising:2021vhh}. When systematic uncertainties were taken into account, the best limits on the DM pair-production cross section were obtained when considering the ratio of the photon energy distributions for the left-handed and right-handed electron beam polarisation. 
The additional contribution from the BSM mediator exchange is expected to significantly change the dependence of the polarisation ratio on the photon energy, as seen in \cref{fig:clic_wimp_polar}. 
Cross section exclusion limits expected for 1 \abinv of data collected at 3 \TeV with polarised electron beam are about a factor of 3 stronger than those expected for 5 \abinv of data collected without polarisation.
\begin{figure}[htb]
\centerline{\includegraphics[width=0.4\textwidth]{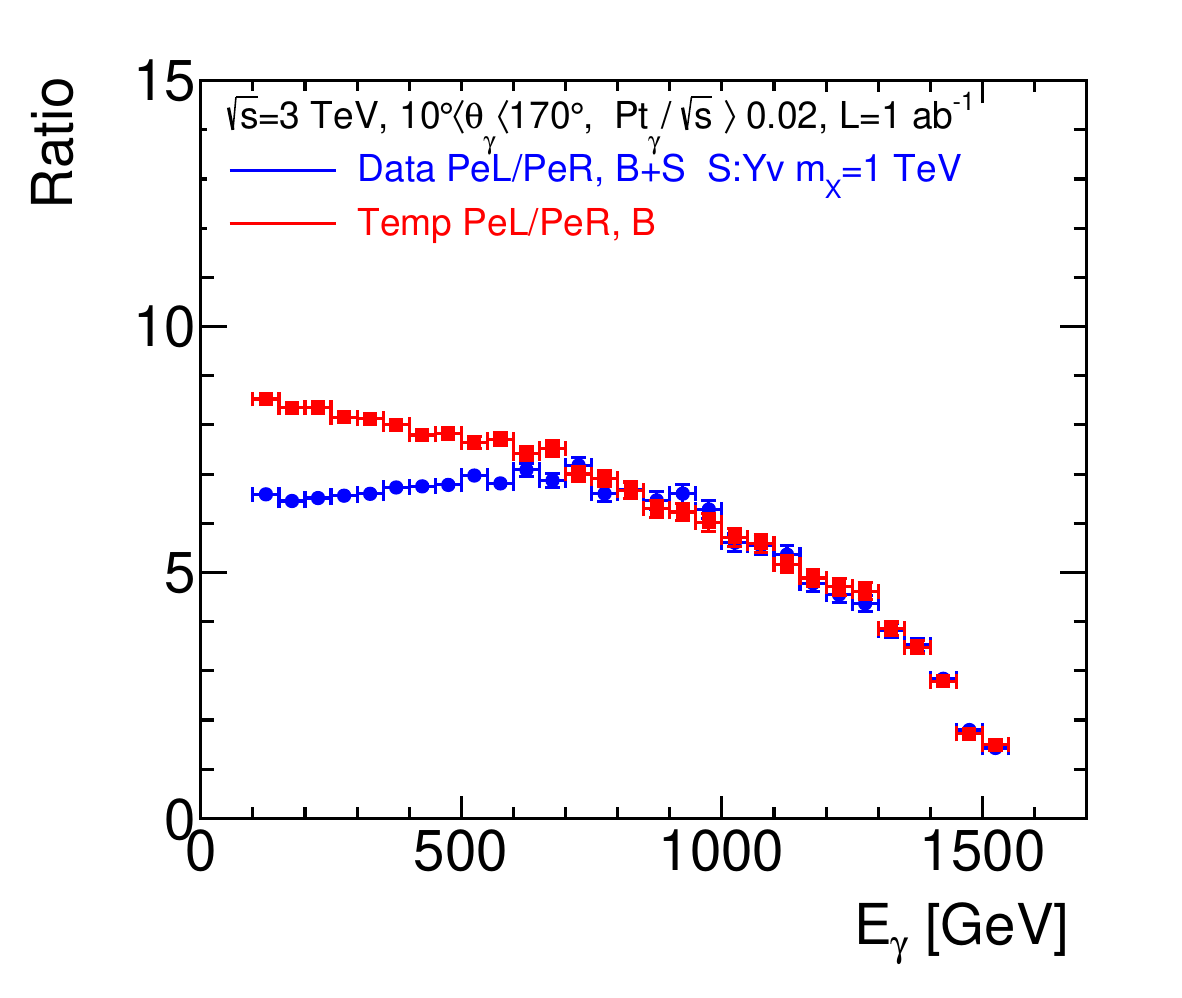} 
\includegraphics[width=0.4\textwidth]{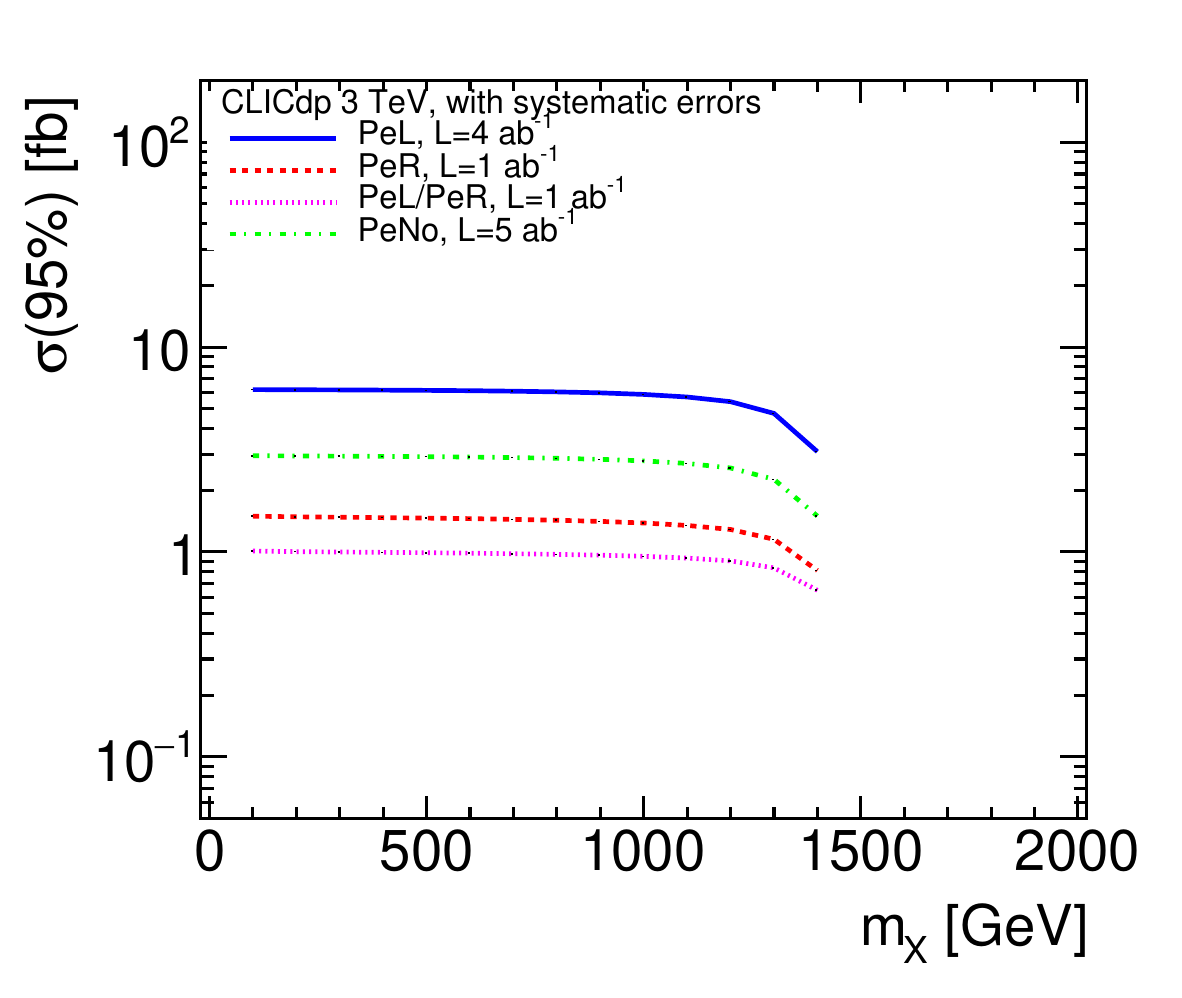}}
  \caption{Results of the CLIC Dark Matter search study using mono-photon signature \cite{Blaising:2021vhh}. Left: ratio of the photon energy distributions for two electron beam polarisations, left-handed over righ-handed, for the SM background expectations (red) and 1 TeV dark matter pair-production scenario (blue). Right: 95\% C.L. upper limit on the dark matter pair-production cross section as a function of the dark matter mass, for different polarisation and luminosity conditions. }
  \label{fig:clic_wimp_polar}
 \end{figure} 

A comparison of the sensitivity of the mono-photon approach in the case of a vector mediator and low mass WIMP for the proposed future lepton colliders with different luminosities, energies and polarisation settings is presented in Fig.~\ref{fig:compare} ~\cite{Habermehl:2020njb}. The polarisation is not only essential to reduce SM backgrounds, but also  to control systematic uncertainties, which becomes more important with increasing dataset size. One has to keep in mind that circular colliders accumulate the datasets necessary for the target precision in a much shorter time.  
%
\begin{figure}[htb]
\centerline{\includegraphics[width=0.8\textwidth]{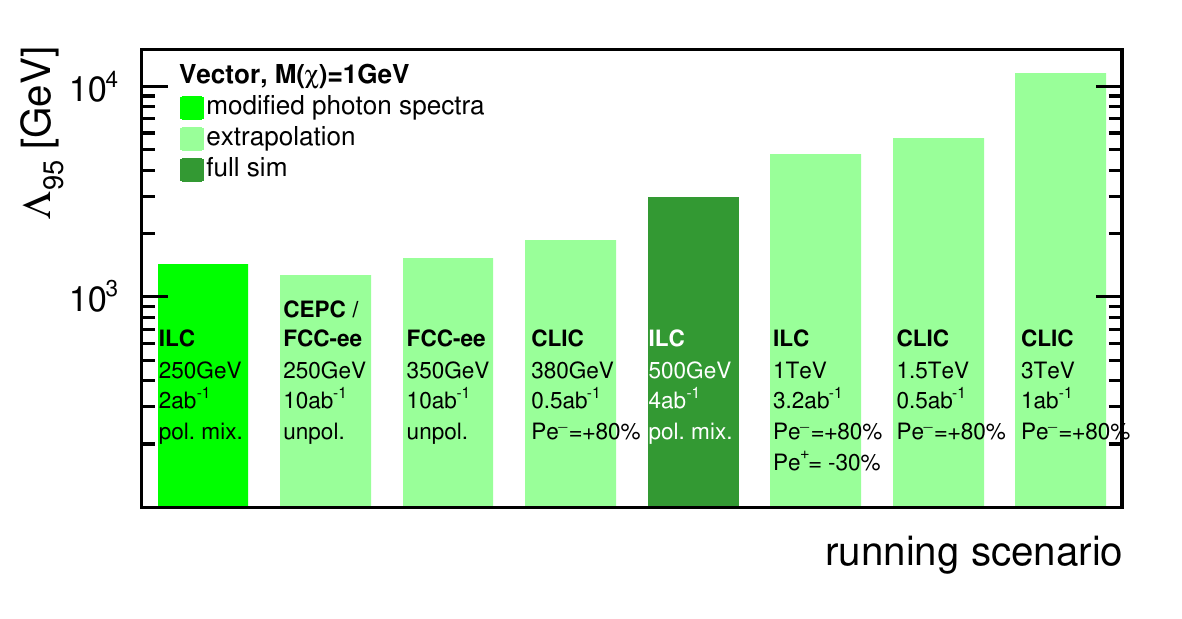}}
\caption{Comparison of sensitivity in the case of a vector mediator and low mass WIMP for different running scenarios. For the ILC setup at 1\,TeV and the CLIC configurations, only the data sets for $P(\Pem)=(+80\%)$ (and in the case of the ILC $P(\Pep)=(-20\%)$) are taken into account, as they dominate the sensitivity.} 
\label{fig:compare}
\end{figure}
%


\subsubsection{Scalar Rayleigh dark matter}
Although DM is expected to be neutral under electromagnetism, it can interact with SM photons and electroweak gauge bosons via higher-dimensional operators once heavy BSM particles are integrated out. Specifically, Rayleigh operators (see e.g.\ Ref.~\cite{Weiner:2012cb}) allow these
interactions at leading order in the EFT framework. However, because DM does not
couple directly to lighter SM particles such as quarks and gluons, the Rayleigh scattering cross section on
SM states is loop-suppressed, making direct detection highly challenging. Despite this, indirect detection of
DM annihilation into photons remains promising and FERMI data can be used to probe this benchmark for
DM masses up to 500 GeV.  The theoretical framework
relies on the presence of decoupled new physics at an arbitrary high energy scale $\Lambda$  plus a scalar DM candidate with mass at the EW scale. The following dimension-6 operators
\begin{equation}
{\cal{L}} = C_B \PGf^{2} B_{\mu\nu}B^{\mu\nu}  +C_W \PGf^{2} W_{\mu\nu}W^{\mu\nu} ,
\end{equation}
where $B,\, W$ are  SU(2)$_L$ and U(1)$_Y$ field strength tensors with $C_{B,W}=c_{B,W} \Lambda^{-2}$, have been considered in Ref.~\cite{Barducci:2025twe}. 
After the electroweak symmetry breaking the operators project onto the EW gauge bosons as
\begin{equation}
{\cal{L}} = \PGf^{2}( C_{\PGg\PGg} A_{\mu\nu}A^{\mu\nu} +
C_{\PGg \PZ} A_{\mu\nu}Z^{\mu\nu}  +C_{\PZ\PZ}Z_{\mu\nu}Z^{\mu\nu} 
+C_{\PW} W_{\mu\nu}W^{\mu\nu} )
\end{equation}
It is important to emphasize that this model can be UV-completed either at loop level or at tree level. The
constraints shown in \cref{fig:Barducci} pertain to the tree-level scale $\Lambda$. 

\begin{figure}[htb]
\begin{center}
    \includegraphics[width=0.46\textwidth]{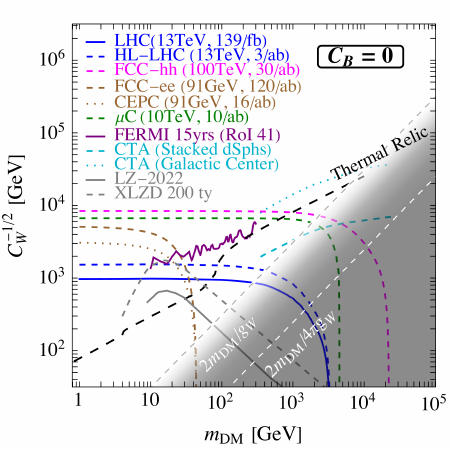}
    \includegraphics[width=0.46\textwidth]{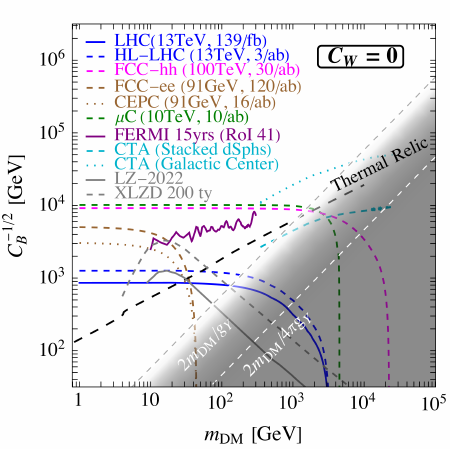}
\caption{Bounds for $\Lambda$ for a scalar Rayleigh dark matter candidate in the $C_B = 0$ (left) and $C_{\PW} = 0$
(right) Wilson Coefficient assumptions. Solid lines represent current experimental constraints, while
dashed lines illustrate future projections.
\label{fig:Barducci}
}
\end{center}
\end{figure}
Current and future colliders can constrain the Wilson coefficients of the Rayleigh operators defined above. For
all the experiments the Drell--Yan signal process with $\PGg\PGf\PGf$ final state has been considered, where the DM pairs are produced in association with a high-energy
photon in the final state. For the LHC running at
13 TeV, current and future projections for the
high-luminosity phase (HL-LHC) with $\sim 3$ ab$^{-1}$ of integrated luminosity are considered. The ATLAS search for DM
produced with a photon has been recast, imposing the same event selections and requiring $E^\text{miss}_{\mathrm{T}}=\pT^\gamma < \Lambda$ to ensure the EFT validity. 
Additionally, signals from the vector boson scattering (VBS)
process have been combined in the  bounds in Fig.~\ref{fig:Barducci}. 
Beyond the LHC, an estimate for a future
hadron collider (FCC-hh) with a centre-of-mass energy of 100 TeV and an integrated luminosity of
30~ab$^{-1}$ is included by rescaling the ATLAS search result to a higher energy for both the DY and VBS searches. 
Future $\epem$ facilities, such as CEPC and
FCC-ee, running at the
$\PZ$ pole with an integrated luminosity of ${\cal O}(10-100 ~\abinv)$, are also considered. The sensitivity to Rayleigh  focusing on a direct
signal from mono-photon searches in Drell--Yan production  leads to the strongest limits on
the scale $\Lambda$ for light DM masses. 
In the case of a muon collider ($\mu$C) operating at
$\sqrt{s} = 10$ TeV with 10 ab$^{-1}$
of integrated luminosity, 
both Drell--Yan production and the VBS process $\PGm \PGm \to \PGm\PGm \phi \phi$ were considered, requiring the detection of forward muons
with high pseudo-rapidity up
to $\eta$ = 7. The sensitivity is maximized by focusing on the invariant mass of the ``missing''
four-momentum calculated from the difference between the momentum of the initial state and that of the observables final
states in acceptance.

In addition to collider tests, the role of direct and indirect DM detection experiments in this scenario has been also considered.
Direct detection provides the most stringent limits on elastic spin-independent (SI) WIMP-nucleon cross sections
for dark matter masses ranging from a few GeV to tens of TeV. The LUX-ZEPLIN-2022 (LZ-2022)” experiment is the
leading direct detection experiment, but the Rayleigh DM benchmark remains elusive due to loop-suppressed
interactions with quarks and gluons. Looking at the SI limits and predictions for the Rayleigh differential
cross section  bounds are given  using data from LZ-2022  and projections from XLZD. Indirect
detection is also promising as DM can annihilate into gauge boson pairs,
which decay into stable SM particles. Focusing on the photon flux from DM annihilation  bounds are set using
FERMI-LAT measurements of galactic diffuse emission and future projections for CTA. 
More detailed results are presented in Ref.~\cite{Barducci:2025twe}.
\subsubsection{Leptophilic dark matter}
Thermal freeze-out is the only dark matter (DM) production mechanism that is insensitive to unknown high-energy physics above the scale of the DM mass. 
Over the past several decades, this mechanism has primarily been associated with the WIMP paradigm within a 10 GeV -- 10 TeV  mass window. 
However, the absence of a verified WIMP signal  motivates broadening the search effort outside this mass window and to a wider range of interactions.
In many studies, the most promising DM discovery channel is missing energy in association with a single SM object $X$ (typically a single jet or gauge boson). 
However, if DM couples preferentially to heavy flavour, there may be additional signals that improve upon the mono-$X$ strategy, offering more handles for both signal detection and background rejection.
For example, DM that couples mainly to $\PQb$-quarks, motivated by the galactic-centre excess, can be probed using $\Pp\Pp \to \PQb\PAQb \slashed{E}$ production.
This strategy can readily be adapted to leptophilic DM that couples primarily to muons and taus.
Production of DM is therefore accompanied by additional charged leptons, whose presence may enhance signal sensitivity.
The future collider discovery potential for models of leptophilic dark matter with mass-proportional
couplings to Standard Model leptons was investigated in Ref.~\cite{Cesarotti:2024rbh}. 
The representative benchmark scenario consists of a Majorana fermion DM candidate $\PGc$ with mass $m_{\PGc}$ 
coupled to a \CP-even scalar mediator 
$\varphi$ with mass $m_\varphi$ via 
\begin{equation}
\label{eq:lag} 
{\cal L}_{int} 
= -\frac{g_{\PGc}}{2} 
\varphi \PGc \PGc -
 \varphi \!\sum_{\Pl = \Pe, \PGm,\PGt} \! g_{\Pl} \, \PAl \Pl ,
 ~~~~ g_{\Pl} = g_{\Pe} m_{\Pl} /m_{\Pe},
\end{equation} 
where $g_{\PGc}$ and $g_{\Pl}$ are respectively the Yukawa couplings to DM and charged leptons $\Pl$. The mass
proportionality in $g_{\Pl}$ can naturally arise in a Type-III two-Higgs doublet model supplemented with the singlet
scalar $\varphi$ if the latter mass-mixes primarily with the heavy \CP-even eigenstate H and has suppressed mixing
with the SM-like 125 GeV eigenstate h.       
For thermal freeze-out to be predictive and testable with laboratory measurements, the cross section
for DM annihilation must depend on the $\varphi$ coupling to the SM, i.e.\ annihilation cannot occur strictly in the
dark sector. This requirement is readily achieved by demanding
$m_\varphi >  m_{\PGc}$, so that the
$\varphi\varphi\to \Plp\Plm$ topology
is kinematically allowed for all $\Pl$ satisfying $m_{\PGc} > m_{\Pl}$. 
During freeze-out, the process
$\PGc\PGc\to \varphi^*\to \Plp\Plm$ 
directly annihilates DM into SM leptons to yield the observed relic density $\Omega h^2 =0.120 \pm 0.001$.
Since the cross section for this process depends on the coupling between $\varphi$ and charged leptons, there is a
minimum value required for viable freeze-out. Thus, a thermal origin through the direct annihilation  predicts a minimum signal yield for DM production at $e_+e^-$ colliders via
$\epem \to \Plp\Plm \varphi$. 
For $m_\varphi>2 m_{\PGc}$ the mediator decays predominantly invisibly to dark matter, so the process has
a signature of $\epem \to \Plp\Plm  E^\text{miss}$.

Since the $\varphi$ coupling to the $\tau$ is
enhanced by a factor of $(m_\tau/m_e)^2 \sim 10^6$,  the signature
$\epem \to \PGtp \PGtm \varphi ~~,~~ (\varphi \to \PGc \PGc)$,
is considered, where the invisibly decaying mediator is produced as final state radiation and predicts deviations within well-measured SM uncertainties.
From the uncertainty of  LEP measurement of the di-tau partial width 
    $\Gamma({Z\to \PGtp\PGtm}) =   84.08 \pm 0.22 \, \mathrm{MeV}$
the value of $g_\tau$ at each value of $m_\varphi$ that satisfies 
$\Gamma(Z \to \PGtp \PGtm \varphi)  =  2 (0.22 \, \mathrm{MeV})$ 
 is extracted and translated  into a constraint on 
$     y \equiv  (g_{\Pe}/\Pe)^2 \alpha_{\PGc} (m_{\PGc}/m_\varphi)^4 $ 
for a given choice of $\alpha_{\PGc} \equiv g_{\PGc}^2/(4\pi)$ and $m_{\PGc}/m_\varphi$ ratio. The parameter $y$ is useful because it has a one-to-one relationship with the total annihilation cross section, independently of the $m_\phi/m_\chi$ or $g_\chi/g_e$ ratios. In Fig.~\ref{fig:leptophilic}, this constraint bounds the area shaded in grey and labeled as ``LEP''. 
\begin{figure}[htb]
\begin{center}
 \includegraphics[width=0.75\textwidth]{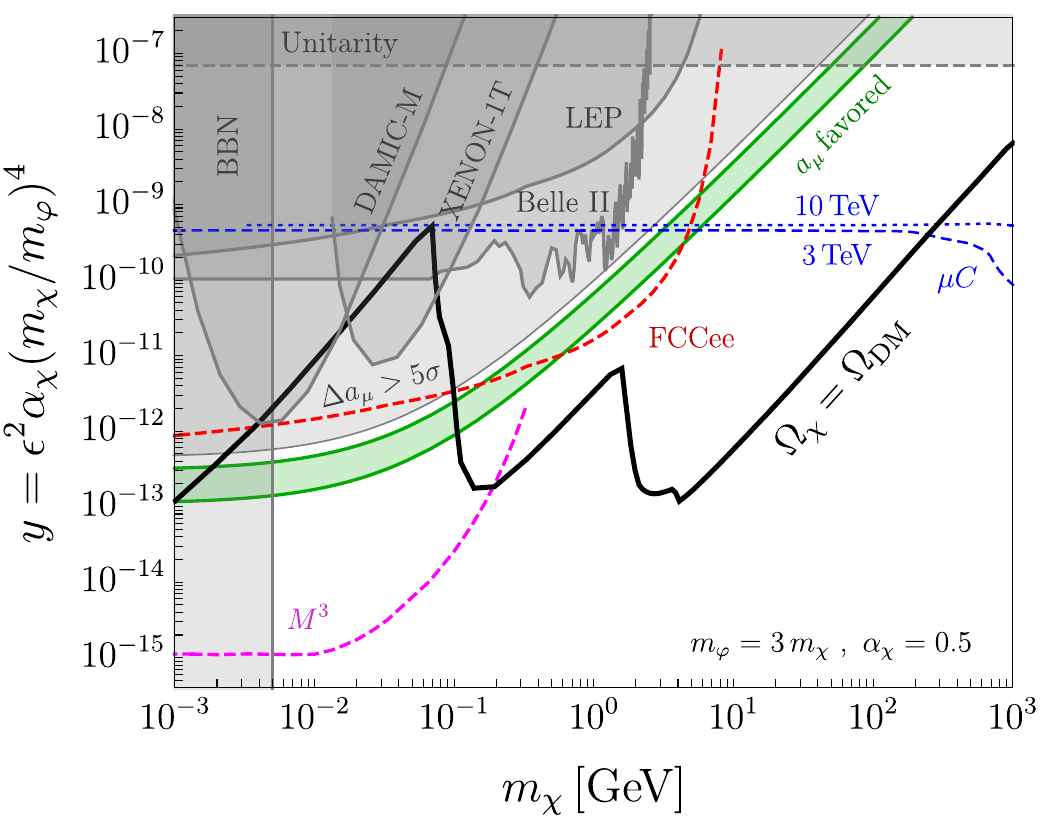}
\caption{Experimental milestone for $\PGc \PGc \to \Plp \Plm$ thermal freeze-out (black curve) plotted alongside 
grey-shaded constraints and dashed-line future projections. $M^3$ refers to the proposed $(g-2)_\mu$ experiment \cite{Cesarotti:2024rbh}. } 
\label{fig:leptophilic}
\end{center}
\end{figure}
Future accelerators will play a key role in probing the remaining thermal relic parameter space for the
leptophilic DM scenario, as also indicated in Fig.~\ref{fig:leptophilic}. 
At moderate masses, the proposed FCC-ee 
 $\epem$ collider can probe large regions of
viable parameter space with a Tera-Z run at the \PZ-pole, which can improve sensitivity to $\PZ \to \PGtp\PGtm$ decays,
currently limited by the luminosity of the legacy LEP dataset.

              



              
              

\clearpage
\section{Flavour Physics}
\label{sec:flavour}

\subsection{Introduction}
\label{sec:flav-intro}
\editors{Stephane Monteil, David Marzocca, Patrick Koppenburg}

Measurements in flavour physics have paved the way to the Standard Model (SM) as we know it to date. The GIM mechanism was inspired by studies in the kaon sector, namely the observation of the rarity of the Flavour Changing Neutral Current (FCNC) process $\PKzL \to \PGmp\PGmm$; the GIM mechanism yielded in turn to the prediction of the charm quark~\cite{Glashow:1970gm}; the observation of $C\!P$ violation in 1964 \cite{Christenson:1964fg} suggested the existence of a third generation~\cite{Kobayashi:1973fv}; and the first measurements of $\PBz-\PABz$  oscillations in 1986 immediately indicated the top quark to be massive ($m_{\PQt} > 50$~GeV) ~\cite{ARGUS:1987xtv}, well before its discovery. 
The discoveries of the top quark and the Higgs boson provided a complete formulation of the SM. Yet, the mass matrices and the mass mixing matrices do not receive a dynamical explanation in the SM and pose questions that are intimately connected to flavour physics and acknowledged as the Flavour puzzles: Why three generations? What explains the observed hierarchy in the fermion masses? What explains the distinctive structure of the Cabibbo-Kobayashi-Maskawa matrix? Is the SM $C\!P$ violation enough to describe the matter-antimatter asymmetry of the universe?  

In a bottom-up approach, flavour physics can be seen as a powerful tool of discovery of physics beyond the SM.
The constrained structure dictated by gauge invariance, combined with the peculiar values observed for the Yukawa couplings, endows the SM with a set of accidental properties and approximate symmetries.
A consequence is that some processes are very suppressed or entirely forbidden, such as FCNC transitions and lepton-flavour violating (LFV) decays. Those same accidental properties and symmetries are in general not respected by extensions of the SM and therefore one can expect new physics to induce large contributions to those rare or forbidden processes. 
In this context, measuring them indirectly probes the effects of new physics up to very large mass scales, which in some cases exceed by many orders of magnitude any scale that could be reached with present or planned future colliders (e.g.\ meson mixing, rare kaon or $\PB$-meson decays, LFV decays of muons and taus, electric dipole moments, etc.).

The current flagship experiments for quark flavour physics are LHCb at the Large Hadron Collider and Belle~II operating in the $\Pep\Pem$ environment at the $\PGU(4S)$ resonance. These experimental programs are planning upgrades that span over the next two decades, which will keep the field of beauty, charm, and $\PGt$ physics vibrant. 

We start in \cref{sec:flav-theory} by setting the context of the experimental study of rare/forbidden processes from the persective of their potential in probing new physics.
In \cref{sec:flav-explandscape} we address the anticipated potential of these experimental programmes and compare the attributes of the different experimental environment at $\Pp \Pp $ and $\Pep\Pem$ colliders for flavour physics, distinguishing the physics interest of an electroweak factory ($\PZ$ pole and $\PW \PW$ threshold).
The prospects for improvement in the knowledge of the CKM profile(s) are examined in \cref{sec:flav-CKM}.
The potential to study rare decays of heavy-flavoured particles is considered in \cref{sec:flav-rare}.
The $\PGt$ physics prospects at a future $\Pep\Pem$ electroweak factory are described in \cref{sec:flav-tau}.
The detector requirements necessary to achieve the physics objectives will be spelled out when available in these sections. Eventually, \cref{sec:flav-outlook} the possible interplays between a flavour physics programme and the rest of programme of a Higgs, top and electroweak factory, be they related to the experimental tools or the interpretation of the future results.   

Some of the studies described here require a high-luminosity run at the \PZ pole, providing several $10^{12}$ \PZ bosons, like proposed at FCC-ee or CEPC. We call such runs ``\TeraZ'', unless the studies are specifically done in the context of one accelerator or proposed detector design.

\subsection{The new physics perspective and UV models}
\label{sec:flav-theory}
\editors{David Marzocca}

While the hierarchy problem remains the best motivation for new physics at scales $M_{\text{NP}} \sim 1~\SI{}{\tera\electronvolt}$, precision flavour measurements push the effective scale of new physics contributing to FCNC and $C\!P$-violating processes to much higher scales.
As mentioned in the Introduction, the reason why rare SM processes are such powerful probes of new physics is their suppression in the SM due to loop, CKM, and other suppression factors.

Let us assume that one is able to measure the Wilson coefficient describing the short-distance SM contribution, $C_\text{SM}^\text{FCNC} \sim G_F \lambda_\text{SM}$ (here $\lambda_\text{SM}$ represents the product of all suppression factors while $G_F$ is the Fermi constant), with a relative precision $\delta$.
Heavy new physics contributing to the same FCNC process can also be parametrised in terms of a Wilson coefficient corresponding to a short-distance operator, $C_\text{NP}^\text{FCNC} \sim c^\text{FCNC} / M_{\text{NP}}^2$ (where $M_{\text{NP}}$ represents the new physics mass scale and $c^\text{FCNC}$ depends on the dynamics by which it mediates the rare process).
In many cases the reach can be estimated simply as
\begin{equation}
    C_\text{NP}^\text{FCNC} \lesssim C_\text{SM}^\text{FCNC} \times \delta~, \quad \rightarrow \quad
    M_{\text{NP}}/\sqrt{c^\text{FCNC}} \gtrsim (\delta \times G_F \lambda_\text{SM})^{-1/2}~.
\end{equation}
From this expression it is clear that, for a given precision $\delta$ on the SM contribution, the rarer the SM process is (i.e.\ the smaller $\lambda_\text{SM}$) the larger will be the new physics scale $M_{\text{NP}}$ that one is sensitive to.

The tension between the requirements of \emph{(i)} a low new physics scale $M_{\text{NP}}\sim 1~\SI{}{\tera\electronvolt}$ that addresses the Higgs hierarchy problem and \emph{(ii)} strong constraints from flavour physics that push the corresponding scale to large values, $M_{\text{NP}}/\sqrt{c^\text{FCNC}} \gg 10 \text{--} 100~\SI{}{\tera\electronvolt}$, is known as the new physics flavour problem.
It dictates that any TeV-scale solution to the hierarchy problem has a special flavour structure, i.e.\ that the new physics affecting FCNC processes should be suppressed, or in other words that $c^\text{FCNC} \ll 1$.

``Minimal flavour violation'' (MFV) is one such structure, whereby the BSM couples flavour-universally at leading order, $c_{ij} \sim \delta_{ij}$, and the SM Yukawas are the only sources of flavour violation~\cite{DAmbrosio:2002vsn}. But because MFV new physics couples strongly to valence quarks, such states are now subject to strong LHC bounds, typically $M_{\text{MFV}} \gtrsim \SI{10}{\tera\electronvolt}$.
Alternative flavour hypotheses in which the third generation couplings $c_{33}$ are disentangled from universal light generation couplings can give just as good flavour suppression~\cite{Barbieri:2011ci}. This ``$U(2)$-like'' scenario has two notable differences to MFV: \emph{(i)} if $|c_{11}|,|c_{22}|\ll |c_{33}| = \mathcal{O}(1)$, LHC bounds can be as weak as $M_{U(2)} \gtrsim \SI{1.5}{\tera\electronvolt}$~\cite{Allwicher:2023shc}, and \emph{(ii)} the flavour non-universal new physics might simultaneously explain the SM flavour puzzle.

\subsubsection{Flavour deconstruction models}
\label{sec:flav-UVdeconstruction}

In this section we focus on a class of theories that realise the $U(2)$-like hypothesis, wherein part of the SM gauge symmetry is ``flavour-deconstructed''. The FCC-ee or CEPC proposals would bring spectacular improvements for precision tests of this scenario, through the complementarity and reach of its electroweak and flavour programs. The present study was done in the context of FCC.

The basic hypothesis is that a SM gauge interaction $G$ is resolved into three copies $G_i$, one for each generation and with the Higgs coupled only to $G_3$. This symmetry, which 
arises e.g.\ from gauge-flavour unification ~\cite{Davighi:2022fer}, is broken to the SM in two steps, with $G_1 \times G_2 \to G_{12}$ at $v_{12}\sim 100\text{--} 1000~\SI{}{\tera\electronvolt}$, then $G_{12} \times G_3 \to G$ at $v_{23}\sim 1\text{--} 10~\SI{}{\tera\electronvolt}$; each step delivers a multiplet of heavy gauge bosons in the adjoint of $G$, denoted $X_{12}$ and $X_3$, with flavour non-universal couplings. 
While the third generation Yukawa couplings are unsuppressed, the light Yukawa couplings arise from higher-dimension operators suppressed by heavy mass scales in the model. 
\begin{table}[h]
\begin{center}
\begin{tabular}{|c|c|c|c|c|c|c|}
\hline
& Deconstructed force & $SU(3)$ & $SU(2)_L$ & $SU(2)_R$ & $U(1)_Y$ & $U(1)_{B-L}$ \\
\hline
Flavour & $|V_{cb}| \ll 1$ & $\checkmark$ & $\checkmark$ & $\times$ & $\checkmark$ & $\checkmark$ \\
 & $y_i\ll y_3$ & $\times$ & $\checkmark$ & $\checkmark$ & $\checkmark$ & ${\times}$ \\
\hline
EW & Natural upper limit of $|\tan\theta| M$ & 90 TeV & 20 TeV & 40 TeV & 40 TeV & 500 TeV \\
 & EWPOs order & 1-loop & Tree & Tree & Tree & 1-loop \\
\hline
\end{tabular}
\caption{Qualitative comparison of flavour deconstruction models.
The ``natural upper limits'' estimate the finite part of the leading radiative corrections to $m_{\PSh}^2$~\cite{Davighi:2023iks}, and require these to be less than $\SI{1}{\tera\electronvolt}^2$.
} \label{tab:options}
\end{center}
\end{table}

\noindent
Flavour deconstructing different SM forces can explain different aspects of the flavour puzzle~\cite{Davighi:2023iks}; see~\cref{tab:options}. Deconstructing $SU(2)_L$ or $U(1)_Y$, under which the Higgs is charged, can explain both  $y_{1,2}\ll y_3$ and the smallness of CKM matrix elements $V_{\PQc\PQb,\PQu\PQb}$. We focus on these options in the following. 
But the direct Higgs coupling also means these theories (i) give tree-level shifts in electroweak precision observables (EWPOs) that will be measured with unprecedented precision at a Tera-$\PZ$ machine, and (ii) give one-loop $\delta m_{\PSh}^2$ corrections, meaning they cannot be decoupled by taking $M^2 \gg m_{\PSh}^2$ without creating a hierarchy problem.
In~\cref{tab:options} we also consider deconstructing $SU(2)_R$ and $U(1)_{B-L}$~\cite{Barbieri:2023qpf,Barbieri:2024zkh,Covone:2024elw}, which gives more control over the Yukawa textures.
In all cases, the gauge coupling matching means we can define $g_{\text{SM}} = g_3 \cos\theta=g_{12}\sin\theta$. We focus on the phenomenology of $X_{23}$, which can be light thanks to its ``$U(2)$-like'' couplings. Values $\theta\approx\pi/2$ correspond to $g_3 \gg g_{12}$.
To illustrate the key prospects for HL-LHC and FCC-ee, in the following we focus on deconstructed $SU(2)_L$ or $U(1)_Y$~\cite{Davighi:2023xqn,Davighi:2023evx}.
\medskip

\noindent
\textbf{Electroweak precision.} Deconstructing an EW force yields bosons $X$ coupled to the Higgs and thus tree-level shifts in EWPOs. The bounds on $M$ in~\cref{tab:pheno}, which are saturated when the mixing angle $\theta \approx \pi/4$, come from a global fit to $\PZ$- and $\PW$-pole data from LEP, SLD, Tevatron, and the LHC~\cite{Davighi:2023xqn}.
For $SU(2)_L$, $m_W$ receives a strictly {\em negative} shift,
away from current $m_W$ measurements.
The $\PZ$-pole asymmetry observables $\delta A_{\Pe}$, $\delta A_{\PQb}$, and $\delta A_{\text{FB}}^{PQb}$ also receive negative corrections~\cite{Davighi:2023xqn}. 
These EWPOs will reach spectacular precision at the $\PZ$-pole run of FCC-ee. The FCC-ee projection in~\cref{tab:pheno} assumes SM-like central values and uses the projected sensitivities presented in~\cite{deBlas:2022ofj}, giving a bound $M \gtrsim \SI{30}{\tera\electronvolt}$ that almost covers the natural region. 
In the event one finds significant deviations,
being sensitive to the {\em signs} of the observables mentioned can clearly (dis)favour the deconstructed $SU(2)_L$ scenario. 
For deconstructed $U(1)_Y$, large contributions to $\PZ\to \PlR \PlR$ drive the current EWPO bound.
This time there is a strictly {\em positive} shift in $m_{\PW}$. This sign is a key reason why the bound is weaker for deconstructed $U(1)_Y$ than $SU(2)_L$; another reason is that $g_Y \approx g_L/2$, translating to roughly half as strong bounds on $M$. The FCC-ee projection in~\cref{tab:pheno} for deconstructed $U(1)_Y$ was computed slightly differently to the $SU(2)_L$ case, assuming a conservative improvement by a factor 10 for all EWPOs.

\begin{table}[h]
\begin{center}
\begin{tabular}{|c|c|c|c|}
\hline
& Deconstructed $SU(2)_L$ & Deconstructed $U(1)_Y$ \\
\hline
Electroweak: $\PZ$ pole \& $\PW$ pole & 9 TeV (5 TeV if exc. $m_{\PW}$) & 2 TeV  \\
Flavour: $\PBzs\to\PGm\PGm$ (up-alignment) & 7.5 TeV & 2 TeV \\
High $\pt$: Drell--Yan $\Pp\Pp\to \Pe\Pe,\PGm\PGm,\PGtp\PGtm$ & 4.5 TeV & 3.5 TeV \\
\hline
EW projection FCC-ee: on and off $\PZ$ pole \& $\PW$ pole & 30 TeV & 7 TeV  \\
\hline
\end{tabular}
\caption{ 
Constraints from flavour, high $\pt$, and EWPOs on the mass $M$ of the gauge bosons $X_{23}$.
} \label{tab:pheno}
\end{center}
\end{table}

\noindent
\textbf{Quark and Lepton Flavour.} The intrinsic flavour non-universality leads to tree-level effects in rare $\PB$- and $\PGt$-decays, giving multi-TeV constraints.
There are excellent future prospects, with FCC-ee or CEPC acting as a flavour factory in its \TeraZ run: of the $\mathcal{O}(10^{12})$ $\PZ$ bosons, 15\% decay to $\bb$, 12\% to $\cc$, and 3\% to $\tptm$. 

The observable $\text{BR}(\PBzs \to \PGmp\PGmm)$ currently provides the best flavour test (see~\cref{tab:pheno}), given the large $C_{10}^{\mu}$ Wilson coefficient predicted by both models.
The constraints from $\PBs$-mixing, while important, are weaker.
With HL-LHC data samples, the $\text{BR}(\PBzs \to \PGmp\PGmm)$ bounds should improve by $\times 1.5$ or so. Note that because this is such a rare decay, with a BR $\sim 10^{-9}$, even the production of $8\times 10^{10}$ $\PBzs$ mesons at FCC-ee is unlikely to deliver precision to rival HL-LHC.
Effects in $\PQb\PQs\PGtp\PGtm$ 
are, perhaps surprisingly, suppressed in both models. The tree-level prediction $C_{10}^\tau=0$ rules out testing these models via $\PBzs \to \PGtp\PGtm$ at FCC-ee. Even though $C_9^\tau\neq 0$, the shift in $\text{BR}(\PB \to \PK^{(\ast)} \PGtp\PGtm)$, which can be impressively well measured at FCC-ee~\cite{Li:2020bvr}, is too small~\cite{Davighi:2023xqn}.
Turning to $\PQb\PQs\PGn\PAGn$, 
both models predict $C_{L}^{\nu_\tau}=0$ at tree-level, meaning the contribution to $\PB\to \PK^{(\ast)}\PGn\PAGn$ comes only from $\PGne$, $\PGnGm$ and is thus directly correlated with $\text{BR} (\PBzs \to \PGmp\PGmm)$. Even though $\PBzs \to \PGmp\PGmm$ at LHC is currently much more precise than $\PB\to \PK\PGn\PAGn$ at B-factories, FCC-ee promises a tremendous improvement in $\PB\to \PK\PGn\PAGn$ perhaps down to $1\%$  precision~\cite{Amhis:2023mpj}. The complementary information from $\PBzs \to \PGmp\PGmm$ at HL-LHC plus $\PB\to \PK\PGn\PAGn$ at FCC-ee would be a crucial diagnostic for these models, due to the direct correlation.
A similar correlation occurs for other models, so this is an important synergy between HL-LHC and FCC-ee in flavour. If other gauge groups are deconstructed e.g.\ a quark-lepton unifying $SU(4)$ or $SU(2)_R$, even richer correlations are expected.

Turning to leptons, there are tree-level shifts in tau LFUV ratios. For deconstructed $SU(2)_L$ this gives the strong bound $M\geq \SI{4.5}{\tera\electronvolt}$~\cite{Davighi:2023xqn}. FCC-ee will significantly reduce the uncertainty on tau LFUV measurements, currently pioneered by Belle (II), by an estimated factor $\times 13$~\cite{Dam:2018rfz,Blondel:2021ema}; if measurements were SM-like at FCC-ee, the bound would jump to $M\geq \SI{11}{\tera\electronvolt}$. These tests, evidencing FCC-ee's capabilities as a  ``tau factory'', further complement the significant opportunities in quark flavour and electroweak precision.

\medskip

\noindent
\textbf{High-$\pt$.} The $X_{23}$ bosons are produced from $\PQq\PAQq$ in the $s$-channel, and so constrained by $\Pp\Pp\to \Plp\Plm$ Drell--Yan. 
The bounds in~\cref{tab:pheno} roughly correspond to the value of $\theta$ for which the $\Pl=\PGt$ and $\Pl=\Pe,\PGm$ constraints crossover.
For $SU(2)_L$, limits were computed using the \texttt{HighPT} tool~\cite{Allwicher:2022mcg} using an EFT-matched calculation, while those for $U(1)_Y$ were obtained using the full model prediction for $\sigma(\Pp\Pp\to\Plp\Plm)$.
HL-LHC would bring substantial increase in reach. For example, assuming an integrated luminosity of $3~\abinv$ and that a SM-like distribution is measured, the bounds on the $SU(2)_L$ triplet mass would improve from 4.5 TeV to 8 TeV.

\medskip

\noindent
\textbf{Summary.}
Flavour deconstructing an electroweak force offers a path to solving the flavour puzzle with a rich phenomenology whose effects cannot be decoupled from electroweak physics. FCC-ee is a powerful machine for probing these theories by achieving comparable sensitivity across diverse measurements traditionally separated into ``electroweak'' and ``flavour'' categories, but whose effects in such flavour models are tied together. 
Viewing the electroweak and flavour programmes of FCC-ee together, and in combination with high-$\pt$ and low-$\pt$  HL-LHC data, is therefore crucial to understanding its full power.

\subsection{Anticipated theoretical and experimental landscape at the dawn of future colliders}
\label{sec:flav-explandscape}
\editors{Stephane Monteil, Pablo Goldenzweig}

Flavour physics (in particular beauty, charm, and of $\tau$ leptons studies) is a vibrant field of study, with the current flagship experiments being LHCb at the Large Hadron Collider,  and Belle II at the SuperKEKB collider, featuring $\Pem\Pep$ collisions at the $\PGU(4S)$. Acknowledging the current LHC planning and plans for upgrade, the second part of the 2030 decade will see the operation of the second upgrade of LHCb, which aims at accumulating an integrated luminosity of  300\,fb$^{-1}$.  This experiment shall collect a dataset 50 times larger than those so far analysed~\cite{lhcbupgrade2}. The ATLAS and CMS experiments will continue to contribute strongly to the LHC flavour programme in selected areas, for example the measurement of the weak mixing phase $\phi_s$ and the studies of $\PBz \to \PGmp\PGmm$ and  $\PBzs \to \PGmp\PGmm$. The Belle~II experiment aims at integrating around 50\,ab$^{-1}$, with prospects of further running under discussion. In parallel to the great steps forward in experimental knowledge these projects will bring, one expects corresponding advances in lattice QCD calculations, and the development of other theoretical tools.
\Cref{fig:lhcbup2} shows a possible anticipated status of the Unitarity Triangle after LHCb Upgrade~II operation, based on results from that experiment alone and assuming that all results will be in agreement for the sake of an illustrative reference. It should be noted that this analysis assumed in addition plausible lattice QCD improvements~\cite{lhcbupgrade2}.   High precision will be similarly achieved in the studies of ``rare decays'', i.e.\ suppressed flavour-changing neutral-current (FCNC) processes, featuring either semileptonic or radiative decays. 

 \begin{figure}[h]
    \centering
    \resizebox{0.8\textwidth}{!}{\includegraphics{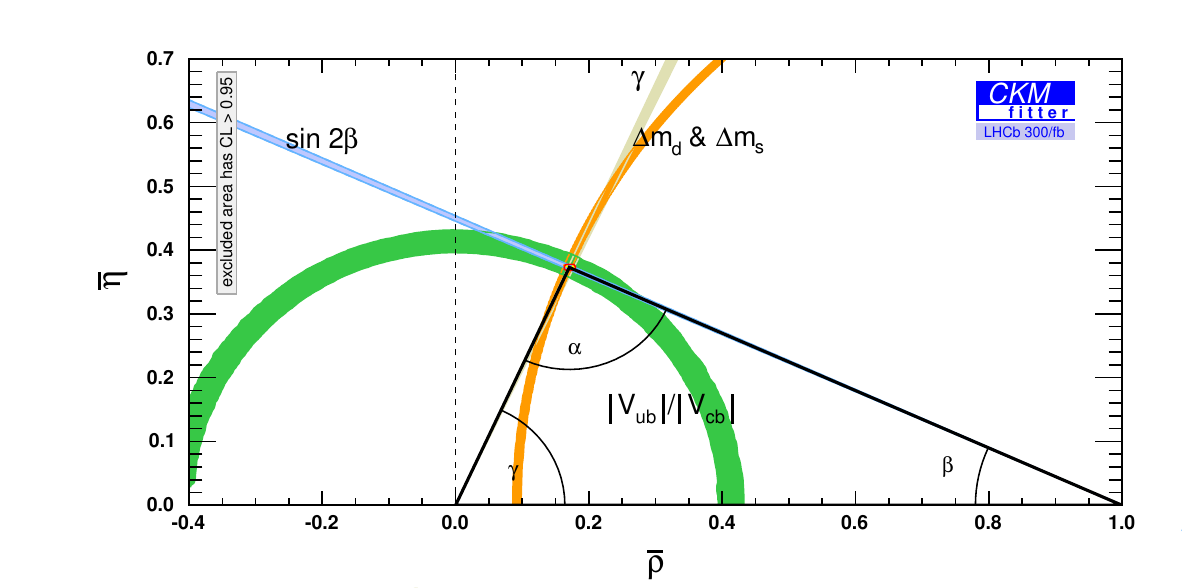}}
    \caption{Possible status of the Unitarity Triangle in the late 2030s, assuming LHCb measurements alone and improvements in lattice QCD~\cite{lhcbupgrade2}.}
    \label{fig:lhcbup2}
 \end{figure}

As introduced in Ref.~\cite{Monteil:2021ith}, \cref{tab:attributes} compares the advantages for flavour-physics studies at an $\Pep\Pem \to \PGU(4S) \to \PQb\PAQb$ experiment, such as Belle~II, a $\Pp\Pp \to \PQb\PAQb \PX$ experiment, such as LHCb, and an experiment that relies on $\Pep\Pem \to \PZ \to \PQb\PAQb$ production, such as \TeraZ. 
The $\PZ$ environment combines most of the advantages of Belle~II and LHCb: for the former these are the high signal-to-noise and fully efficient trigger, as well as a very high geometrical acceptance; for the latter they are the production of the full spectrum of hadrons, and the high boost. 
The momenta of $\PQb$ and $\PQc$ hadrons produced at the $\PZ$ are not known a priori, in contrast to the $\PGU(4S)$, although their distribution is well understood.  The momentum of the produced tau leptons at the $Z$ pole is exactly known in both $\Pep\Pem$ environments. The advantageous heavy flavour production cross-section at the LHC is only partially mitigated by the exquisite luminosity foreseen at a Tera-$\PZ$ factory.  

\begin{table}[h]
    \centering
    \begin{tabular}{lccc} \hline
    Attribute & $\PGU(4S)$ & $\Pp\Pp$ & $\PZ$ \\ \hline
All hadron species &  & \checkmark & \checkmark \\
High boost         &  & \checkmark & \checkmark \\
Enormous production cross-section & & \checkmark &  \\
Negligible trigger losses & \checkmark &  &  \checkmark \\
Low backgrounds & \checkmark &  &  \checkmark \\
Initial energy constraint & \checkmark &  & (\checkmark)  \\ \hline
    \end{tabular}
    \label{tab:attributes}
    \caption{Advantageous attributes for flavour-physics studies at Belle II ($\PGU(4S)$), LHC ($\Pp\Pp$) and \TeraZ runs~\cite{Monteil:2021ith}.}
\end{table}

The observables contributing to Unitarity Triangle tests are generally theoretically clean, and deserve therefore a continued experimental attention throughout the coming decades.  As an example, the angle $\gamma$ can be measured in $\PBm \to \PD\PKm$ decays with a relative theoretical uncertainty of $10^{-6}$ or better~\cite{Zupan:2011mn}. The second upgrade of LHCb can already bring the precision on this observable well below one degree. Other measurements, that may become limited by systematic uncertainties at the LHC, can benefit from the cleaner experimental environment of a high-luminosity $\PZ$ factory and a large dataset of hadronic $\PW$ decays. Examples include studies of semileptonic \CP-violating asymmetries or determinations of the matrix elements  $|V_{\PQu\PQb}|$ and $|V_{\PQc\PQb}|$,  be they measured in semileptonic $\PQb$-flavoured particle decays ($\PBzs$ mesons and $\PGLzb$ baryons are accessible in contrast to a $\PB$-factory) or in $\PW$ decays. Last but not least, it is likely that much will remain to be learned from semi-tauonic or tauonic $\PQb$-hadron decays, experimentally challenging at LHCb, and for which the $\PB$-factory studies will be limited by the sample size.

\subsubsection{Expected precision from lattice QCD }

The lattice regularisation of QCD allows for  
nonperturbative-physics contributions to SM processes to be computed from first principles, i.e., starting from the QCD Lagrangian without any model assumptions.
The lattice-regularised path integral is integrated by means of Monte-Carlo sampling, and the UV and IR regulators given by the lattice spacing and finite volume, respectively, are removed by taking the continuum and infinite-volume limits. Lattice QCD has come a long way since the first numerical simulations in the 1980s within the quenched approximation (only gluonic vacuum-polarisation effects), to simulations of the full QCD path integral including $\PQu,\PQd,\PQs,\PQc$ valence- and sea-quark, as well as $\PQb$ valence-quark effects. 
At the level of the effective electroweak theory, nonperturbative QCD effects in flavour physics can be computed as matrix elements of electroweak effective operators between initial and/or final hadronic states. 
The summary tables at the beginning of the Flavour Lattice Averaging Group  (FLAG) review~\cite{Colangelo:2010et,Aoki:2013ldr,Aoki:2016frl,FlavourLatticeAveragingGroup:2019iem,FlavourLatticeAveragingGroupFLAG:2021npn} provide an overview over results with reliable control over uncertainties, that are of direct relevance for the phenomenology of the SM.
The FLAG review provides results for meson and baryon form factors,  for the strong coupling constant and quark masses, which can only be computed with systematically improvable errors within lattice QCD.

\subsubsection*{Recent developments and opportunities}

\textbf{QCD+QED+strong isospin:} The majority of results in the FLAG review are based on isospin-symmetric QCD, neglecting  QED effects --- an approximation valid to the level of about 1\%. However, as the precision for some quantities within this approximation has now reached below 1\%, QED and strong isospin-breaking effects can no longer be ignored. Thanks to recent progress on formal aspects~\cite{deDivitiis:2011eh,deDivitiis:2013xla,Lucini:2015hfa,Fodor:2016bgu,Patella:2017fgk,Giusti:2017dmp,Basak:2018yzz,DiCarlo:2019thl,Boyle:2022lsi,DiCarlo:2024lue}, algorithms and computing infrastructure~\cite{Finkenrath:2023sjg,Kanwar:2024ujc}, we can now make predictions for SM parameters, hadron spectra and leptonic meson decay within QCD+QED. Work is underway to include QED effects also in other observables such as semileptonic decays~\cite{Christ:2023lcc} or scattering~\cite{Christ:2021guf}. \\
\textbf{2nd-order weak processes:} The past years have seen new developments~\cite{Isidori:2005tv,Bai:2014cva,Bai:2023lkr,Christ:2015aha,Christ:2016eae,Christ:2016mmq,Bai:2017fkh,Bai:2018hqu,Christ:2019dxu,RBC:2022ddw} towards computing long-distance contributions to rare decays such as $\PK\to\PGp\Plp\Plm(\PGn\PAGn)$, $\PKL \to \PGmp \PGmm$~\cite{Chao:2024vvl}, or $\PBzs \to \PGmp\PGmm\PGg$~\cite{Frezzotti:2024kqk}. There remain open questions, such as controlling the contributions from multi-particle intermediate states, but we expect these to be tackled in the coming years. \\
\textbf{Reconstruction of the spectral density:} Finding ways to reconstruct the spectral density from Euclidean correlation functions  has recently become a vibrant research topic~\cite{Hansen:2017mnd,Hansen:2019idp,Bruno:2020kyl,Bailas:2020qmv,Bulava:2021fre,Bergamaschi:2023xzx,Frezzotti:2023nun}. For instance, a first prediction of the ratio $R$ related to the hadronic vacuum polarisation has been made~\cite{ExtendedTwistedMassCollaborationETMC:2022sta}. Spectral reconstruction methods appear particularly attractive for
processes with multi-hadron in or out states, such as hadronic inclusive decays of the tau lepton~\cite{Evangelista:2023fmt,ExtendedTwistedMass:2024myu} and of mesons~\cite{Bailas:2020qmv,Gambino:2020crt,Gambino:2022dvu,Barone:2023tbl}. The latter development, for the first time, provides a clear path towards a lattice-computation for inclusive $\PB_{(\PQs)}$ and $\PD_{(\PQs)}$ decays on the lattice. This, in turn, allows one to be optimistic that resolving the persistent $V_{\PQc\PQb}$ and $V_{\PQu\PQb}$ puzzles can be achieved in the next decade. \\
\textbf{\CP-violation searches:} A first full computation of the indirect \CP-violation parameter $\epsilon_{\PK}$ was published last year \cite{Bai:2023lkr}. Entering this result are novel techniques for determining the long-distance contribution to the amplitude of $\PKz-\PAKz$ mixing, as well as the $\PK \to (\PGp \PGp)_I$ amplitude in the $I=0$ and $I=2$ isospin channels. Progress on calculating the $\PK \to \PGp \PGp$ amplitudes was presented at the most recent lattice conference \cite{Lat24:Tomii, Lat24:Kelly}. Initial steps towards computing hadronic $\PD$ decays have also been reported on \cite{Joswig:2022ctr,Lat23:Hansen}, providing a pathway for lattice QCD to contribute to the search of $C\!P$ violation in the charm sector.\\
\textbf{Heavy quarks on the lattice:} Simulating both light quarks (which requires a large volume to suppress finite-volume effects) and heavy quarks (which requires a fine lattice spacing to suppress discretization effects) presents significant challenges in lattice QCD. For the heavy $\PQb$ quark in particular, effective actions have to be employed, such as NRQCD \cite{Lepage:1992tx}, RHQ \cite{Christ:2006us} or the Fermilab action \cite{El-Khadra:1996wdx}. Alternatively, one must rely on a heavy-quark-mass extrapolation when simulating multiple heavy quark masses relativistically, an often more costly but systematically improvable approach that is starting to be employed by some precision calculations \cite{Boyle:2021kqn,Parrott:2022rgu,Aoki:2023qpa,Colquhoun:2022atw,Harrison:2023dzh,Harrison:2020gvo,Harrison:2021tol}. Recent computations \cite{Sommer:2023gap, Conigli:2023rod} have pioneered an interpolation between relativistic heavy quarks and static quarks to the physical $m_{\PQb}$ mass, showing promising reductions in systematic errors compared to previous methods. \\
\textbf{Multi-hadron weak decays:} Multi-hadron final states in weak decay transitions can be addressed in Lattice QCD using a dedicated finite-volume formalism \cite{Luscher:1986pf, Rummukainen:1995vs, Kim:2005gf, Hansen:2012tf, Briceno:2021xlc}. Initial applications have focused on the computations of radiative decays \cite{Alexandrou:2018jbt, Briceno:2016kkp, Radhakrishnan:2022ubg}, and progress reports have appeared on the $\PB \to \PGp \PGp \Pl \PGn$ decay \cite{Leskovec:2024wrn}. Advances are also being made in the study of weak decays involving $\PN \pi$ final states \cite{Barca:2024sub, Lat24:Gao}

\subsubsection*{Computing}

For most of the time since the first simulations of lattice QCD in the 1980s, the continued push to ever increasing computing resources has roughly followed Moore's law. Extreme-scaling is now pushing the boundaries of computing power. Lattice QCD is ideally suited for this development. At the core of the numerical problem is the discretised Dirac operator, which is a sparse matrix, ideally suited for data parallelism. The field is prolific in adapting to new computing paradigms in terms of software and algorithm development~\cite{Finkenrath:2023sjg,Kanwar:2024ujc}. As a result, the quality of simulations will continue to improve, allowing to further reduce statistical errors and better control systematic uncertainties. 
The continuum limit of lattice QCD is defined at a Gaussian fixed point of the theory. As a result, simulation algorithms cease to be ergodic, an effect known as critical slowing down~\cite{DelDebbio:2004xh,Schaefer:2009xx,Luscher:2011kk}. 
Ideas have been developed to mitigate the problem, however critical
slowing down is still limiting fully ergodic simulations well below a
lattice spacing of 0.04\,fm.
Novel directions like normalising-flow-based sampling~\cite{Luscher:2009eq,Albergo:2019eim} are attracting attention. Making progress towards finer lattice spacings is particularly important for the study of QCD observables involving the heavier $\PQc$ and $\PQb$ quarks. 
Substantial formal and numerical efforts are invested also into quantum computing~\cite{DiMeglio:2023nsa}. While current activities mainly concentrate on understanding how QFTs including QCD can be implemented on a quantum computer, this novel computing resource could in the future, when larger systems become available, provide for a step change in lattice-QCD predictions.

\subsubsection*{Outlook}
The last decade has seen enormous progress in the scope and quality of lattice-QCD computations with relevance for testing the SM. The increasingly precise predictions for hadronic matrix elements are crucial input to the interpretation and full exploitation of results from current experiments (e.g.\ LHCb, CMS, ATLAS, Belle~II, BES-III) and ones from future large facilities. The tremendous progress in lattice-QCD computations of the hadronic vacuum-polarisation contribution to the muon $g-2$~\cite{Aoyama:2020ynm,Kuberski:2023qgx,DaviesLattice:2024,Boccaletti:2024guq,Borsanyi:2020mff,Djukanovic:2024cmq,Kuberski:2024bcj,Ce:2022uix,RBC:2023pvn,RBC:2024fic,RBC:2018dos,FermilabLattice:2024yho,Bazavov:2024eou,FermilabLatticeHPQCD:2023jof,ExtendedTwistedMassCollaborationETMC:2024xdf,Wang:2022lkq,Aubin:2022hgm,Lehner:2020crt}, and the impact this is making on how we look at this decade-old puzzle (and potentially resolve it), and how this is reshaping the field, is an excellent example show-casing the importance and relevance of lattice QCD now, and going forward. The field has reached an impressive maturity, is diverse and vibrant, and continues developing clever new ideas to extend its reach. 

\subsection{CKM profile prospects} 
\label{sec:flav-CKM}
\editors{David Marzocca, Stephane Monteil, Andreas Juettner}

\subsubsection{Global analyses:  new physics in neutral meson mixings}
\label{section:NPinmixing}

Experiments at a  high-luminosity \PZ factory can perform sensitive studies of $C\!P$-violating phenomena by making use of the abundant production of boosted $\PQb$-flavoured hadrons (${\cal O}(10^{12})$), complemented by a flavour-tagging efficiency close to the one obtained at $\PB$-factories. This will allow for improved knowledge of the Unitarity Triangle angles $\alpha$, $\beta$ and $\gamma$, as well as the phase $\phi_{\PQs}$ between mixing and decay in the $\PBzs-\PABzs$ system.  Considerations about the sample sizes suggests that it will be possible to measure the relevant observables with similar or better precision than previous experiments, as reported in Ref.~\cite{Abada:2019lih}. This is particularly prominent with the studies of modes containing neutrals, with much larger sample sizes than will be available at Belle II, and better precision and efficiencies than at LHCb.  As a case in point, a wide range of charm-meson decay modes in measurements of $C\!P$ asymmetries in $\PBm \to \PD\PKm$ (where $\PD$ stands for  a superposition of $\PDz$ and $\PADz$) and $\PBzs \to \PD_{\PQs}^{(*)\pm} \PKmp$ are therefore accessible and can provide an enhanced knowledge of the CKM angle $\gamma$. Similar considerations apply to time-dependent measurements, where exciting possibilities include modes relevant for the angle $\alpha$ (time-dependent $C\!P$ asymmetries in $\PBz \to \PGpz\PGpz$).  

It is however quite possible that the most significant impact of a future  $\Pep\Pem$ collider  in the study of the CKM profile(s)  will come from $C\!P$-conserving observable measurements. Let's consider the results of a model-independent study of possible BSM contributions to the $\PBz-\PABz$ and $\PBzs-\PABzs$ mixing amplitudes, reported in Ref.~\cite{Charles:2020dfl}.  This class of phenomena is governed in the SM by box diagrams carrying $C\!P$-violating phases,  in which heavy particles (\PW and \PQt) are contributing. It constitutes therefore a natural entry point for any heavy degrees-of-freedoms from BSM physics and can serve as a figure of merit to assess the potential bottlenecks in precision for future machines. In Ref.~\cite{Charles:2020dfl}, the mixing amplitudes are modelled with complex numbers of modulus $h_{\PQq}$ ($\PQq=\PQd,\PQs$), each multiplying the SM $\PB^0_{\PQq}$  mixing Hamiltonian matrix element. Figure~\ref{fig:hdhs} shows the constraints on these BSM parameters (assuming the SM value for the sake of an unambiguous interpretation) for four scenarios: current measurements, the anticipated precision from the first LHCb upgrade and Belle II experiments' measurements, the anticipated precision measurements from  the second LHCb Upgrade (together with the more speculative Belle III programme) and finally including those additional measurements from a $\Pep\Pem$ Higgs/top/electroweak factory including the magnitude of the CKM element $V_{\PQc\PQb}$ from \PW decays. The projected performance at Belle~II and LHCb are based on the numbers reported in Refs.~\cite{lhcbupgrade2,Belle-II:2018jsg}, complemented with reasonable assumptions concerning progress in lattice QCD~\cite{Cerri:2018ypt}. Their values are kept the same in the fourth scenario.  

\begin{figure}
\centering
    \begin{subfigure}[b]{.35\textwidth}
    \resizebox{\textwidth}{!}{\includegraphics{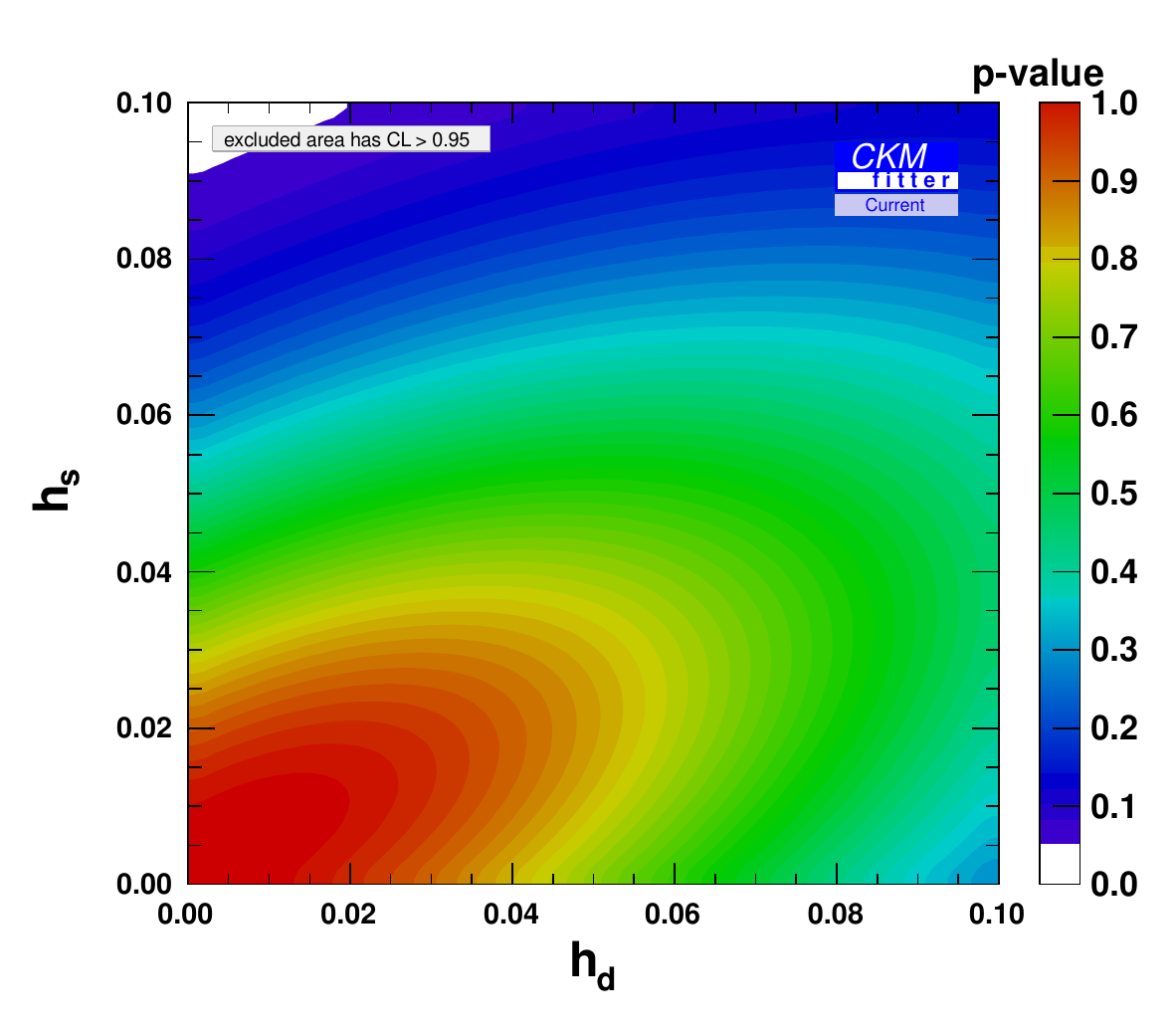}}
    \end{subfigure}
    \begin{subfigure}[b]{.35\textwidth}
    \resizebox{\textwidth}{!}{\includegraphics{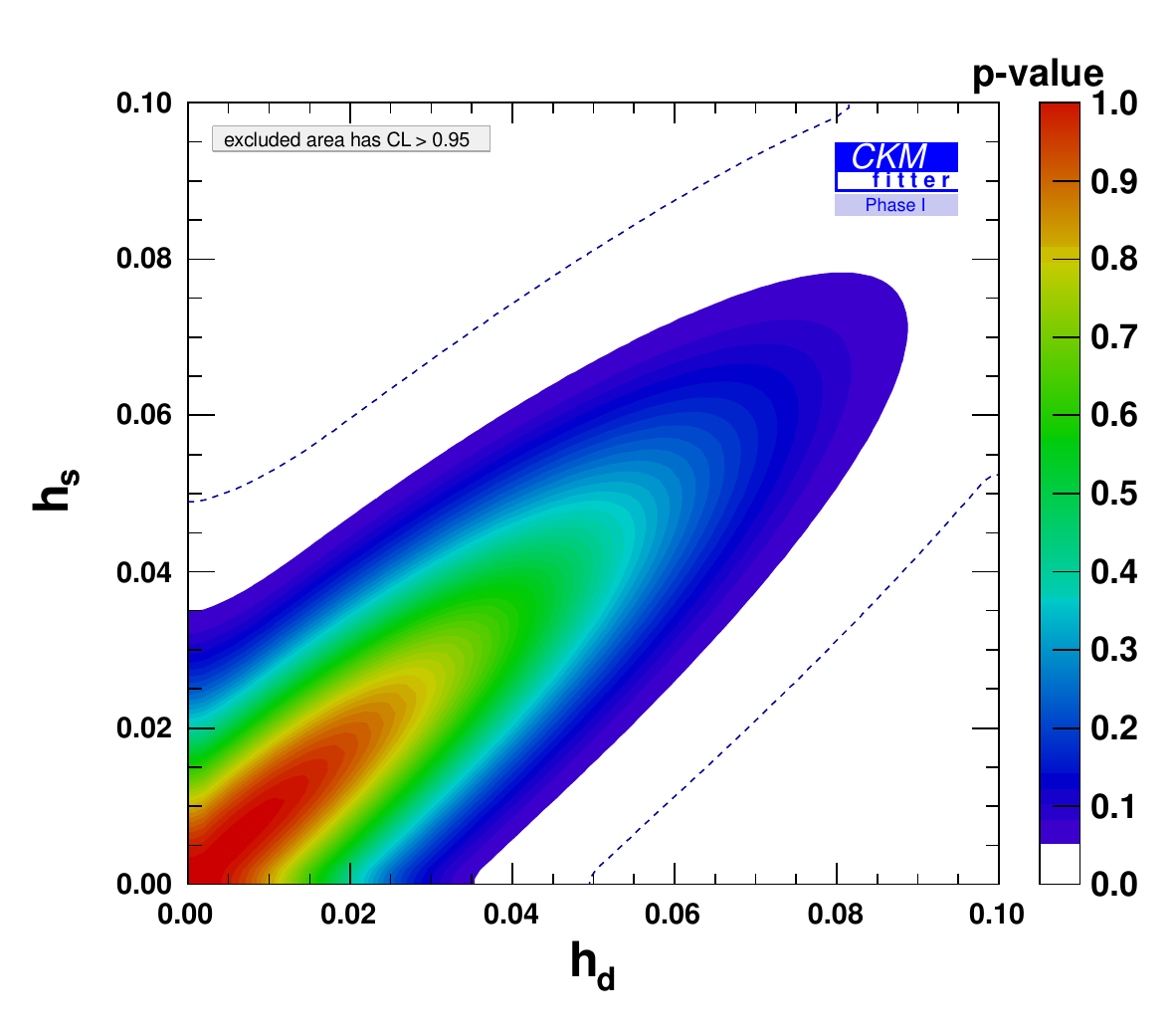}}
    \end{subfigure}
    \vskip\baselineskip
    \begin{subfigure}[b]{.35\textwidth}
    \resizebox{\textwidth}{!}{\includegraphics{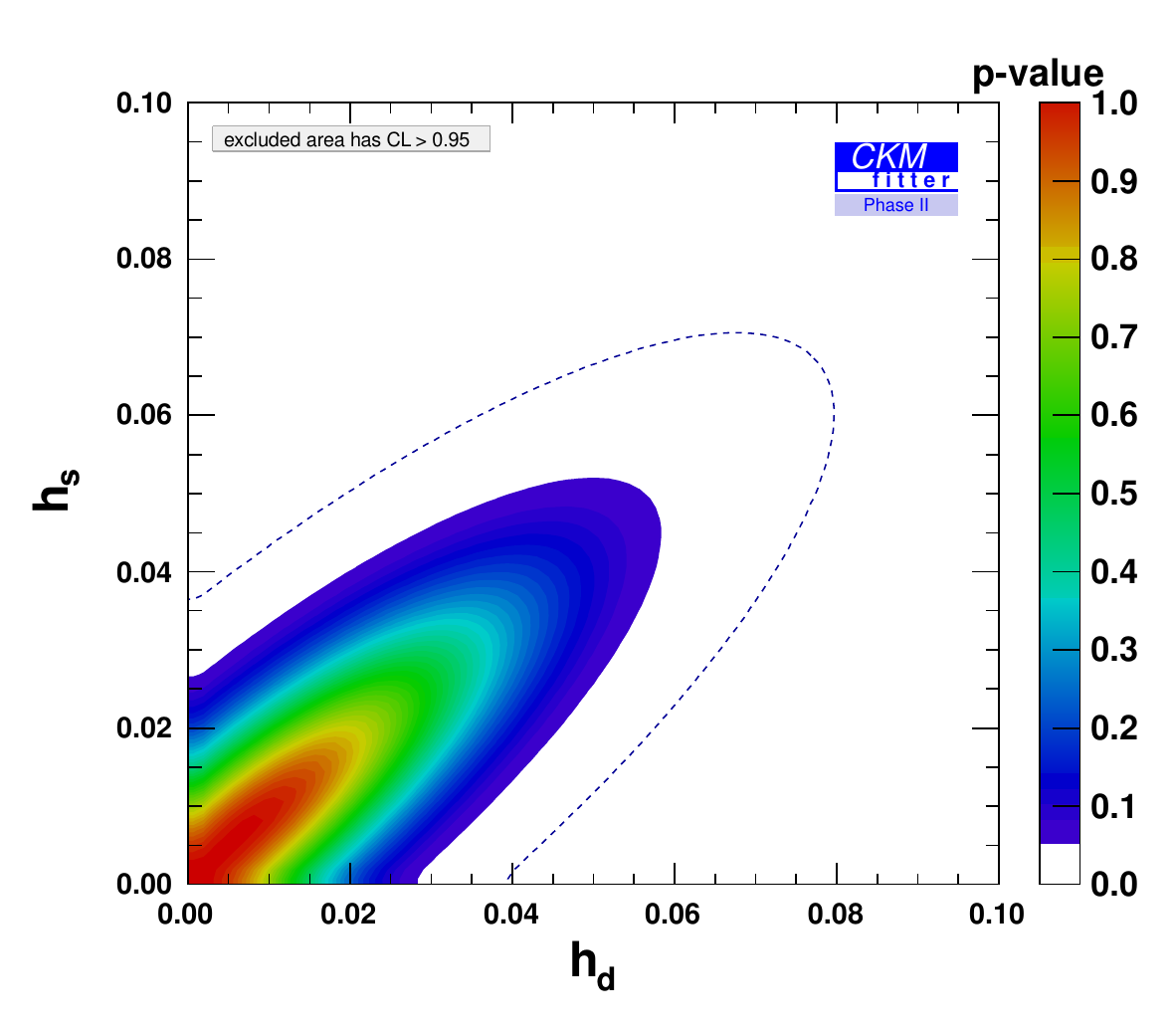}}
    \end{subfigure}
    \begin{subfigure}[b]{.35\textwidth}
    \resizebox{\textwidth}{!}{\includegraphics{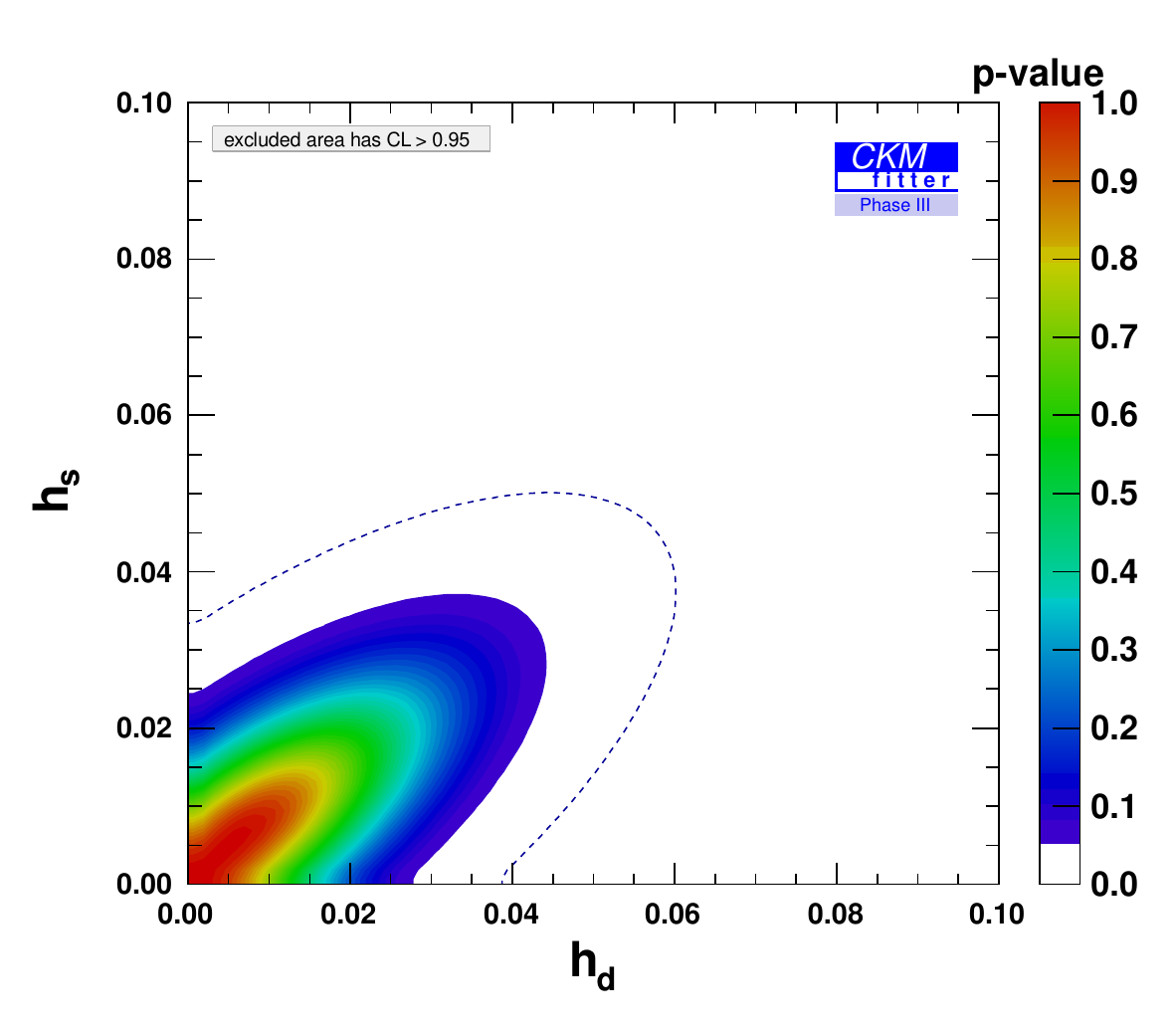}}
    \end{subfigure}

    \hfill
\caption{
Sensitivity to $h_{\PQd}-h_{\PQs}$ parameters in $\PBz$ and $\PBzs$ mixing
with  (top left) current sensitivity appropriately rescaled to SM values, (top right) the anticipated constraints after LHCb Upgrade I and Belle II operation,  (bottom) with  the second LHCb Upgrade (together with the more speculative Belle III program), and  (bottom right) with improved $V_{\PQc\PQb}$ precision from \PW decays. The dotted curves show the 99.7\%~C.L. ($3\sigma$) contours.  All plots are made with the SM inputs $h_{\PQd}=h_{\PQs}=0.$   Taken from Ref.~\cite{Charles:2020dfl}.
}
\label{fig:hdhs}
\end{figure}

Despite the expected large sample sizes of LHCb and Belle II and the precision of the $C\!P$-violating observables entering in this global analysis, two factors are limiting the sensitivity to the BSM parameters in Fig.~\ref{fig:hdhs}.  
\begin{itemize}
    \item  the precision of the CKM element $V_{\PQc\PQb}$, normalising the sides of the Unitarity Triangle, is limiting the sensitivity to the BSM parameters in Fig.~\ref{fig:hdhs}.  The CKM element $V_{\PQc\PQb}$ precision comes from measurements of semileptonic $\PB$ decays, either with inclusive or exclusive decays. In both cases, a hadronic form factor input, e.g.\ from lattice QCD,  is required for $|V_{\PQc\PQb}|$ to be determined.   Another approach  that has no such systematic limitation will open up with the several $10^8$ $\PW$-boson decays that can be produced at a future $\Pep\Pem$ collider.   This avenue will be explored in greater details in the next section. 
  \item Lattice QCD computations are required as well (and without an alternate determination in that case) to interpret the oscillation frequencies. These concerns the decay constants and bag factors hadronic mixing parameters. Given the importance of these inputs in order to maximally use the precision of the $C\!P$-violating observables in the global CKM profile test, a section of this report specifically discusses the challenges that  lattice QCD precision must meet to match the necessary precision.     

\end{itemize}


\subsubsection{\texorpdfstring{$\PB_{(\PQu,\PQc)}^+ \to \PGt \PGn$ leptonic decays as probes of $|V_{\PQu\PQb}|$, $|V_{\PQc\PQb}|$ and new physics.}{Bc -> tau nu leptonic decays as probes of |Vub|, |Vcb|  and new physics.}}
\label{sec:flav-leptonicdecays}

Lepton flavour universality in quark transitions is an important test of the SM. Measurements of tree-dominated observables as the ratio of the semileptonic branching fraction ${\cal R}(\PD^{(\ast)})$ are reported by experiments at the LHC and \PB-factories~\cite{BaBar:2013mob,Belle:2019rba,Aaij_2023}, yielding a longstanding $3\sigma$ deviation from SM predictions. 
The large \PQb-hadron samples provided by high-luminosity \PZ runs combined with exquisite vertex resolution gives access to permille-level precision in the determination of ${\cal R}(\PX)$ ratios of $\PB\to\PX\PGt\PGn$ to $\PB\to\PX\Pl\PGn$ branching fractions~\cite{Ai:2024nmn}, improving with respect to the potential of LHCb~\cite{LHCb:2018roe} and Belle~II~\cite{BelleII}. 

As counterpart of the same $\PQb \to \PQq\PGt\nu$ quark transition, the measurement of the branching fraction BR$(\PB_{\PQq}^+ \to \PGt^+ \PGn_{\PGt})$ ($\PQq=\PQc,\PQu$) can provide an independent way to determine BSM couplings. 
This has been studied in the context of low-energy EFT~\cite{Jenkins_2018}, providing constraints on the axial and pseudoscalar Wilson coefficients of the transition $\PQb \to \PQq \PGt \PGn$. 
The branching fractions of these leptonic decays BR$(\PB_{\PQq}^+ \to \PGt^+ \PGn_{\PGt})$ are furthermore sensitive to the magnitude of the CKM elements $|V_{\PQc\PQb}|$  and $|V_{\PQu\PQb}|$. Since these observables are only dependent on the decay constants, which are quite accurately determined by lattice QCD computations, they can contribute to understand and solve the longstanding inclusive vs exclusive puzzle~\cite{UTfit:2022hsi,CKMfitter}, with a tension of $3.3\sigma$~\cite{HFLAV:2022pwe}.   

Prospective studies of these leptonic decays~\cite{Amhis_2021,Zuo_2024} are conducted at FCC-ee in its \PZ-pole operation, assuming a total of $6\times10^{12}$ \PZ bosons.  The branching fraction precision, from idealistic to pessimistic estimates, are ranging from 1.8\% to 3.6\% for the $\PB^+\to \PGt^+\nu_{\PGt}$ decay channel and from 1.6\% to 2.3\% for the $\PB_{\PQc}^+\to \PGt^+\nu_{\PGt}$ decay channel. A similar sensitivity 
is achieved in studies at CEPC~\cite{Ai:2024nmn}.

The $|V_{\PQu\PQb}|$ determination corresponding to the $\PB^+\to \PGt^+\PGn_{\PGt}$ result is reported in the left plot of~\cref{fig:KIT_BuBc_interpretation}, illustrating the potential for a resolution of the inclusive vs exclusive puzzle. The $|V_{\PQc\PQb}|$ estimate is not provided as the extraction depends on the production fraction of the $\PB_{\PQc}^+$ meson, where no measurement is currently available. 

As far as the BSM interpretation is concerned, both $\PB^+\to \PGt^+\PGn_{\PGt}$ and $\PB_c^+\to \PGt^+\PGn_{\PGt}$ are used to set constraints on Wilson coefficients
of the effective Hamiltonian
\begin{align}
\label{eq:Bctaunu}
\mathrm{BR}(B_q^+ \to \tau^+ \nu_\tau)= \mathrm{BR}(B_q^+ \to \tau^+ \nu_\tau)^\mathrm{SM}\times\left|1-\left(C^q_{V_R} - C^q_{V_L}\right)+\left(C^q_{S_R} - C^q_{S_L}\right)\dfrac{m_{B_q}^2}{m_\tau(m_b+m_q)}\right|^2\,,
\end{align}
which can also be expressed in terms of common combinations $C^q_{A} \equiv C^q_{V_R}- C^q_{V_L}$ and $C^q_{P} \equiv C^q_{S_R}- C^q_{S_L}$.
The right plot of~\cref{fig:KIT_BuBc_interpretation} shows an example of the constraints on the $C^c_{S_L}$ coefficient from the $\PB_c^+\to \PGt^+\PGn_{\PGt}$ measurement.

\begin{figure}[h]
    \centering
    \includegraphics[height=0.26\textheight]{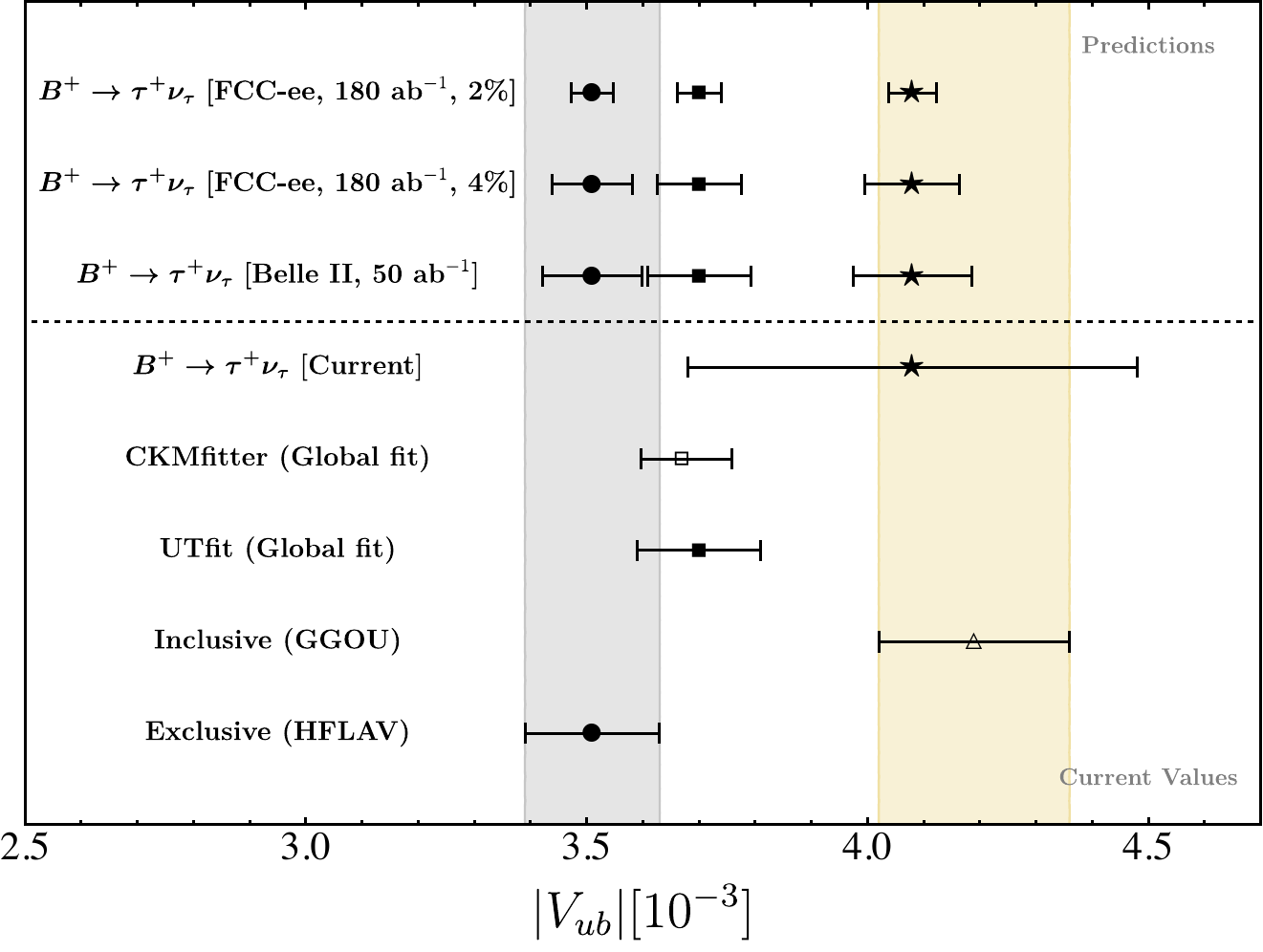}\hspace{0.05\textwidth}
    \includegraphics[height=0.26\textheight]{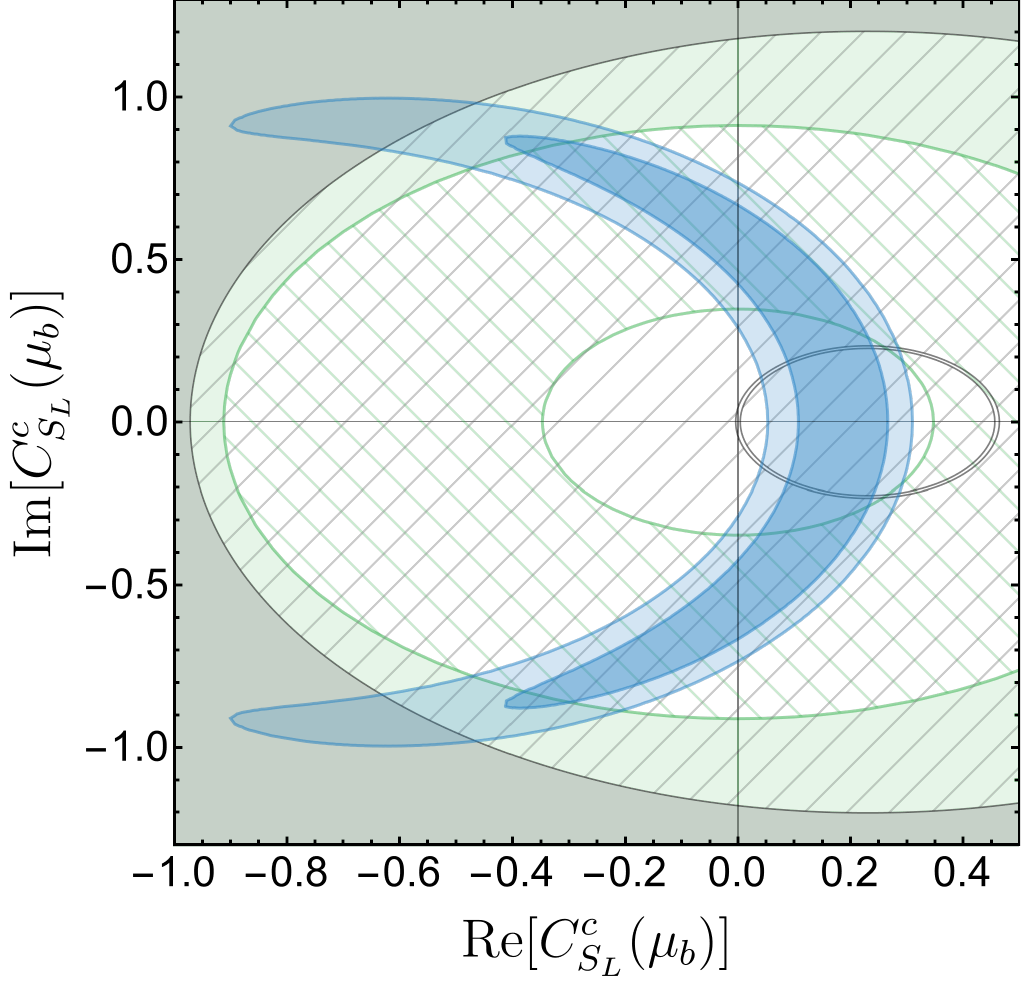}

    \caption{Left: current $|V_{\PQu\PQb}|$ determinations from various sources~\cite{HFLAV:2022pwe,UTfit:2022hsi,CKMfitter} compared with future projections at FCC-ee~\cite{Zuo_2024} and Belle II~\cite{BelleII}. Predictions are given for three values of $|V_{\PQu\PQb}|$~\cite{HFLAV:2022esi,UTfit:2022hsi,CKMfitter} and assumed precisions of 2\% or 4\% on BR$(\PB^+\to \PGt^+\PGn_{\PGt})$. The grey band presents the current exclusive determination and the yellow band the current inclusive determination. Right: constraints on the $C^c_{S_L}$ coefficient. The grey shaded and hashed regions represent current and FCC-ee exclusions from BR$(\PB_{\PQc}^+\to \PGt^+\PGn_{\PGt})$. The green shaded and hashed regions represent current and HL-LHC exclusions from leptoquark searches. The blue regions are the $2\sigma$ and $3\sigma$ bands from current $\PQb\to\PQc$ anomalies. Plots taken from Ref.~\cite{Zuo_2024}. }
    \label{fig:KIT_BuBc_interpretation}
\end{figure}



\subsubsection{\focustopic \texorpdfstring{Measuring $|V_{\PQc\PQb}|$ and $|V_{\PQc\PQs}|$ from $\PW$ decays}{Measuring Vcb and Vcs from W decays}}
\label{sec:flav-Vcb}
Future $\Pep\Pem$ Higgs factories, such as FCC-ee, CEPC, and ILC, present a unique opportunity to significantly enhance the precision in measuring the absolute values of specific CKM matrix elements, particularly $|V_{\PQc\PQb}|$ and $|V_{\PQc\PQs}|$ \cite{Charles:2020dfl,Monteil:VcbWW2024,Marzocca:2024dsz,Liang:2024hox}. These advancements are possible due to the production of large datasets of $\PW\PW$ boson pairs, expected to reach the order of a few $\times 10^8$, combined with state-of-the-art jet-flavour tagging techniques.

The decay of $\PW$ bosons into quark pairs, $\PW \to \PQu_i \PQd_j$, offers a direct method to measure CKM elements from the branching ratios of these processes. The decay width for these channels can be expressed as
\begin{equation}
\Gamma(\PWp \to \PQu_i \PAQd_j) = 3 |V_{ij}|^2 \Gamma_0 \equiv \Gamma^+_{ij}\,,
\end{equation}
where $\Gamma_0 \approx \frac{g_2^2 m_W}{48\pi}$ at leading order, with $g_2$ being the $\text{SU}(2)_L$ coupling constant and $m_{\PW} = 80.4$ GeV the mass of the \PW boson. The factor of 3 accounts for the number of quark colours. This relation holds for both $\PWp$ and $\PWm$ decays, allowing for a precise extraction of CKM elements by comparing the observed branching ratios to theoretical predictions.
Neglecting small quark-mass effects, the branching ratio of a specific channel can be expressed as~\cite{dEnterria:2020cpv,Marzocca:2024dsz}
\begin{equation}\label{eq:Bij}
\mathcal{B}_{ij} = \frac{\Gamma_{ij}^\pm}{\Gamma_{\text{tot}}} = \frac{|V_{ij}|^2}{\sum_{l=\PQu,\PQc;~ m=\PQd,\PQs,\PQb}|V_{lm}|^2} \mathcal{B}_{\text{had}} \, ,
\end{equation}
where $\mathcal{B}_{\text{had}}$ represents the well-measured total hadronic branching fraction of the $\PW$ boson. A precise measurement of  $|V_{\PQc\PQs}|$ is particularly important to reduce the parametric uncertainties in the extraction of the QCD coupling $\alpha_S$ via hadronic \PW boson decays, and reach a $\mathcal{O}(0.1\%)$ uncertainty on this SM parameter~\cite{dEnterria:2020cpv}.

\subsubsection*{Projected Sensitivity at FCC-ee and Systematic Uncertainties}

Early studies at FCC-ee suggested that the relative precision on $|V_{\PQc\PQb}|$ could reach approximately 0.4\% using previous BDT-based ILD jet-tagging performances as a reference \cite{Charles:2020dfl}. More recent evaluations have improved these estimates to about 0.15\% with optimized GNN-based IDEA performance data \cite{Monteil:VcbWW2024}. However, achieving these precision levels will depend critically on controlling systematic uncertainties, particularly those related to jet-flavour tagging.

A detailed analysis of the impact of systematic uncertainties on the tagging efficiencies, denoted as $\delta_\epsilon = \delta \epsilon / \epsilon$, indicates that these uncertainties play a crucial role in limiting the achievable precision on CKM elements \cite{Marzocca:2024dsz}. 
The calibration of the flavour tagging is discussed in more detail in \cref{Sec:Flavour:tagging}. Figure~\ref{fig:syst_scan}(left) illustrates the sensitivity reach for $|V_{\PQc\PQs}|$ and $|V_{\PQc\PQb}|$, in the $\Pep\Pem \to \PW\PW \to 4 j$ channel at FCC-ee, as a function of the systematic uncertainty in tagging efficiency. It highlights the importance of maintaining systematic uncertainties near the 0.1\% level to fully exploit the statistical power of future data. With this level of systematic uncertainties the projected precision could reach 0.15\% for $|V_{\PQc\PQb}|$ and 0.05\% for $|V_{\PQc\PQs}|$, to be compared with current precisions of 3.4\% and 0.6\%, respectively. 

\begin{figure}[t]
\begin{center}
\includegraphics[width=0.45\linewidth]{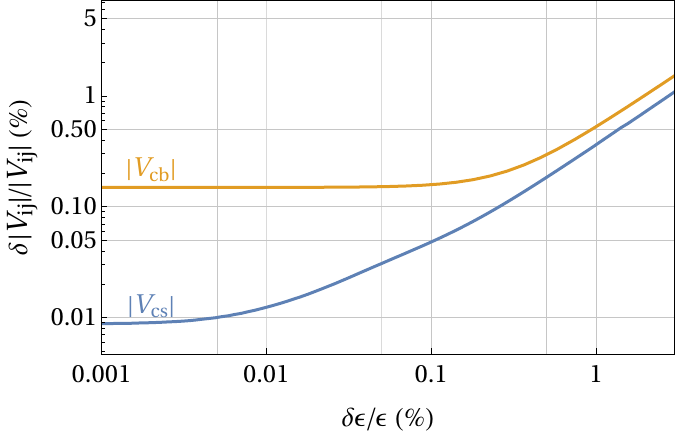}
\includegraphics[width=0.54\linewidth]{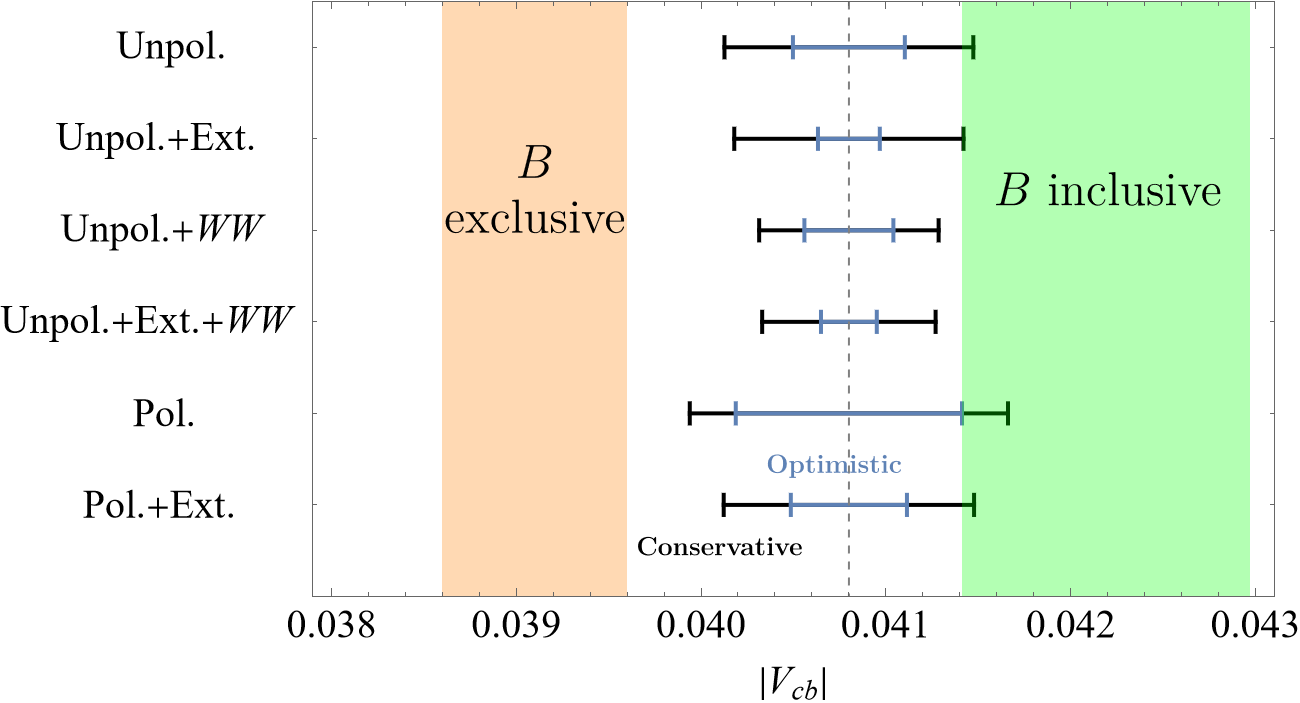}
\end{center}
\caption{Left: Sensitivity reach on $|V_{\PQc\PQs}|$ (blue) and $|V_{\PQc\PQb}|$ (orange) from $\Pep\Pem \to \PW\PW \to 4j$ as a function of the systematic uncertainty on the tagging parameters \cite{Marzocca:2024dsz}.
Right: Expected uncertainty on $|V_{\PQc\PQb}|$ from the study of $\Pep\Pem \to \PW\PW \to 2j \Pl \PGn$ for different collider running scenarios (unpolarised 5 and $20~\abinv$, with and without a \PW\PW-threshold run; and polarised 0.5 and $2~\abinv$), compared with present measurements from inclusive and exclusive $\PB$ decays \cite{Liang:2024hox}.}
\label{fig:syst_scan}
\end{figure}

\subsubsection*{\texorpdfstring{Precision and calibration for the determination of CKM matrix elements from \PW decays}{Precision and calibration for the determination of CKM matrix elements from W decays}}\label{Sec:Flavour:tagging}


Flavour tagging is discussed extensively in \cref{sec:com:flavourtagging}.
The understanding of the flavour tagger has been identified as a dominating systematic uncertainty in the determination of CKM matrix elements from \PW decays, hence precise calibration is key. 
Both circular and linear machines benefit from dedicated Z-pole running that allows determination of the flavour tagging efficiencies to a sub-permille precision, with the ILC projected to produce some $10^9$ Z bosons, and FCC-ee and CEPC reaching $10^{12}$ Z bosons.
This, however, requires the extrapolation of the calibration results from the Z-pole to the W threshold or the Higgs production energy of 250/240 GeV, whose kinematics are slightly and considerably different, respectively.
It remains subject of further studies to show that this is feasible at the required level of precision.

A summary of the jet-flavour tagging efficiencies expected at the IDEA detector (with Delphes) is provided in \cref{tab:JetTaggers} \cite{Bedeschi:2022rnj, Gouskos:taggers, Selvaggi:taggers}, with similar performance anticipated for the CEPC-baseline detector \cite{CEPCPhysicsStudyGroup:2022uwl, Liang:2023yyi} and ILD \cite{Suehara:taggers} (both with full-simulation).


\begin{table}[t]
\centering
\begin{tabular}{ccccccc}
\hline
 & $b$ & $s$ & $c$ & $u$ & $d$ & $g$ \\ \hline 
$\epsilon_\beta^b$ & 0.8 & 0.0001 & 0.003 & 0.0005 & 0.0005 & 0.007 \\ 
$\epsilon_\beta^c$ & 0.02 & 0.008 & 0.8 & 0.01 & 0.01 & 0.01 \\ 
$\epsilon_\beta^s$ & 0.01 & 0.9 & 0.1 & 0.3 & 0.3 & 0.2 \\ \hline
\end{tabular}
\caption{Jet-flavour taggers working points, indicating the probabilities $\epsilon_\beta^q$ to tag a $\beta$-jet as a $q$-jet. \label{tab:JetTaggers}}
\end{table}  

\subsubsection*{ILC/ILD Studies and Complementary Approaches}

A study utilizing the large ILD 250 GeV full-detector simulation sample and an ILC beam spectrum is currently still ongoing.
The effort focusses on the capabilities of new machine learning-based tagging methods like ParticleNet and Particle Transformer, which enhance flavour identification efficiencies.
While the ILC is expected to produce similar samples of $\PW$ bosons as FCC-ee and CEPC, the systematic uncertainty from jet-flavour tagging may be slightly larger due to the absence of a \TeraZ run for calibration. Nevertheless, the ILC remains a significant player in the precise measurement of CKM elements and can still provide competitive sensitivity for $|V_{\PQc\PQb}|$ and $|V_{\PQc\PQs}|$.

\subsubsection*{CEPC Studies and Semileptonic Channels}

In a complementary approach, recent studies with the CEPC baseline detector have focused on the semilep\-tonic channel, $\Pep\Pem \to \PW\PW \to \Pl\PGn \PQc\PQb$, at $\sqrt{s} = 240$ GeV \cite{Liang:2024hox}. Using full detector simulations, these studies have projected a relative statistical sensitivity of 0.34\% for $|V_{\PQc\PQb}|$ in an unpolarised scenario with 20 ab$^{-1}$ of integrated luminosity.
The study considers a conservative and an optimistic scenario for the combined systematic uncertainties, at 1.5\% and 0.2\%, respectively. This is summarised in \cref{fig:syst_scan}~(right), where results are also extrapolated to different collider running scenarios. While the conservative scenarios would typically not be able to resolve the $|V_{\PQc\PQb}|$ tension between inclusive and exclusive measurements, the optimistic scenarios would. The comparison also highlights to sensitivity of the WW production to beam polarisation: The polarised scenario with 2 ab$^{-1}$ provides about the same statistical precision as the unpolarised one with 5 ab$^{-1}$.

\subsubsection*{Conclusion}

Future $\Pep\Pem$ colliders hold great promise for advancing our understanding of the CKM matrix, particularly for $|V_{\PQc\PQs}|$ and $|V_{\PQc\PQb}|$, with potential precision improvements by an order of magnitude over current measurements. Achieving these goals will require controlling systematic uncertainties, particularly those related to jet-flavour tagging, and continuing to develop advanced machine learning techniques for event reconstruction.

\subsubsection{\texorpdfstring{Measuring $|V_{\PQt\PQs}|$ from $\PQt\to \PW \PQs$}{Measuring Vts from t -> W s}} \label{Sec:CKMWW}

\newcommand{\Vts}{\ensuremath{|V_{\PQt\PQs}|}\xspace} 

The current best determination of \Vts comes from $\PBs$--$\PABzs$ mixing, which is mediated by box diagrams with top quarks and W bosons.
The PDG~\cite{PDG2022} value is
\begin{align}
    \Vts=(41.5\pm0.9)\times 10^{-3},
\end{align}
which is based on the lattice QCD calculations of the bag parameters and decay constants~\cite{FLAG2021}. 
The theory uncertainty from lattice QCD is the dominating uncertainty in this measurement, outweighing the experimental uncertainty, from the $\Delta m_{\PQs}$ measurement~\cite{PDG2022}, by about 50~times.
Furthermore, the calculation on the loop-induced process assumes no new physics (NP) in the loop and is therefore model-dependent.

A direct measurement of \Vts would therefore be extremely valuable to reduce the model-dependence of the indirect one and to cross-check lattice QCD results.
The proposed \ttbar operation of FCC-ee at 365 GeV is expected to yield about 2M \ttbar events, in which the decay chain of top quarks can be fully reconstructed and the final state jet flavour can be identified with excellent efficiency.
Such a dataset presents the opportunity to directly measure \Vts via the $\PQt\to \PW\PQs$ decay for the first time.


Jet reconstruction and jet flavour identification is the core of this work.
For jet reconstruction, two clustering algorithms are tested for this analysis: the exclusive jet clustering ($\Pe\Pe$-$\kT$, or Durham, algorithm) and the inclusive jet clustering ($\Pe\Pe$ generalized $\kT$ algorithm), both provided by the \fastjet package~\cite{fastjet}.
The inclusive jet algorithm with radius parameter $R=0.5$ is found to provide a good profiling of jet energy and well-defined true jet flavour, and is consider as the nominal jet definition for this analysis.


The default flavour tagging algorithm at FCC-ee~\cite{Bedeschi_2022} is based on ParticleNet~\cite{Qu_2020}. 
The training is performed on 240 GeV $\Pep\Pem \to \PZ\PH$, $\PZ\to \nu\nu, \PH\to \Pg\Pg$ or $\PQq\PQq$ events, where $\PQq=(\PQu, \PQd), \PQs, \PQc, \PQb$, with $\Pe\Pe$-gen-$\kT$ jets.
As the jets in this analysis feature a different momentum range and potentially a different clustering algorithm from the jets used for the training, the tagging performance 
is sub-optimal compared to the reference.
In the current analysis, we apply the sub-optimal tagger as-is. 
Retraining the taggers with \SI{365}{\giga\electronvolt} samples can potentially improve the tagging performance. 

The \ttbar events can decay to several distinct final states.
This analysis targets 4 event categories: the ``dileptonic" category, ``semileptonic heavy" category, where the hadronic W decay is $\PW\to \PQc\PQs$, the ``semileptonic light" category, where the hadronic W decay is $\PW\to \PQu\PQd$, and the fully hadronic category, here also referred to as ``dihadronic" category.
The event categories are defined by the number of leptons and jets in the event. Electrons and muons with energy $>$ \SI{20}{\giga\electronvolt} and relative isolation $<$ 0.25 are considered as (hard) leptons in this analyses.
The inclusive jet algorithm with R5 clustering is chosen for the nominal jets, with additional selections of $m_j <$ \SI{50}{\giga\electronvolt} and $E_j >$ \SI{15}{\giga\electronvolt}. All electrons and muons with energy $>$ \SI{10}{\giga\electronvolt} are masked from the jet clustering.


In each event category, further selections are applied to refine signal purity. The jet with the highest s-tagging score is chosen as the signal jet candidate. The signal extraction is performed using two approaches: fitting the s-tagging score of the signal jet candidate (in a cut-based approach) and fitting an MVA discriminator trained on the full event kinematics. The MVA approach shows better results in the dileptonic and semileptonic-heavy categories and is chosen as the final fit approach. In the semileptonic-light and fully hadronic categories, no significant improvement is seen for the MVA approach and the cut-based analysis is used for the final result.

The binned maximum-likelihood fit is performed with the CMS \textsc{Combine} tool~\cite{combine}. The fit results are summarized in~\cref{tab:KIT_Vts_fit_result}. 
As the total $\PQt\PAQt$ cross section can be measured very precisely at FCC-ee, the measurement on the $\PQt\to \PW\PQs$ signal yield can be converted to a measurement on $\mathcal{B}(\PQt\to \PW\PQs)$ and therefore a determination of \Vts.
With $\sigma(\PQt\to \PW\PQs) \approx 15\%$, the precision on \Vts is expected to be about 7.5\%.

\begin{table}[h]
    \centering
    \begin{tabular}{l|c|c|c|c|c}
        category     &  dilep & semilep\_light & semilep\_heavy & dihad & combined\\
        \hline
        significance &  8.83 & 4.78 & 2.69 & 1.49 & 10.5 \\
        \hline
        uncertainty  & ${}^{+20\%}_{-18\%}$ & ${}^{+42\%}_{-33\%}$ & ${}^{+50\%}_{-41\%}$ & ${}^{+177\%}_{-99\%}$ & ${}^{+16\%}_{-14\%}$\\
    \end{tabular}
    \caption{Fit results in each category and combined.}
    \label{tab:KIT_Vts_fit_result}
\end{table}

\subsection{\texorpdfstring{\boldmath Rare decays of $\PQb$- and $\PQc$-flavoured particles}  {Rare decays of b- and c-flavoured particles}}  
\label{sec:flav-rare}
\editors{David Marzocca, Stephane Monteil, Pablo Goldenzweig}

As described in \cref{sec:flav-theory}, rare decays offer unique opportunities to test the SM and probe possible new physics contributions up to effective scales much higher than those explored directly at high-energy colliders.

The experimental study of rare FCNC decays requires a very large sample of meson decays (due to the rarity of the processes), very good detector capabilities to disentangle the small signal from possibly large backgrounds, and good control over both experimental systematics and theoretical uncertainties to minimise the relative uncertainty $\delta$.
The large dataset expected at FCC-$\Pe\Pe$ or at CEPC (expecting $\SI{6d12}{}$ $\PZ$ bosons), coupled with excellent vertexing and clean experimental conditions, provides a robust setting for probing rare decays with high precision.

In the following we describe some studies of expected sensitivities for a representative sample of rare $\PB$-meson decays. A study of semileptonic transitions $\PQb \to \PQs \Plp\Plm$ and $\PQb \to \PQd \Plp\Plm$ with light lepton flavours is provided first to illustrate the typical sensitivity for these modes. This section will then focus on the potential reach of the study of third generation couplings transitions that are unique to these facilities.

\subsubsection{\texorpdfstring{Time-dependent precision measurement of the $\PBzs \to \PGf \PGm \PGm$ decay}{Time-dependent precision measurement of the Bs0 -> phi mu mu decay}}
\label{sec:flav-bsmumu}
\newcommand{\Br}{\text{BR}}
\newcommand{\qsq}{q^2}
\newcommand{\ODf}{D_\text{f}}
\newcommand{\OACP}{A_{\CP}}
\newcommand{\OCf}{C_\text{f}}
\newcommand{\OSf}{S_\text{f}}

In this section we investigate the time-dependent $C\!P$ violation in the rare decay ${\PBzs \to \PGf(\to \PKp\PKm)\PGmp\PGmm}$ at the $\PZ$-pole run of the FCC-ee.
Such processes offer a unique opportunity to probe NP, especially through precise $C\!P$ violation measurements~\cite{Descotes-Genon:2022qce, Bobeth:2008ij, Fleischer:2024fkm, Fleischer:2022klb, Descotes-Genon:2015hea, Bobeth:2012vn}.
This analysis focuses on measuring the time-dependent $C\!P$ asymmetry with key observables such as $\ODf$, $\OSf$ and $\OCf$, which are crucial for determining $C\!P$ violation. These measurements offer insight into both the SM predictions and possible deviations due to NP contributions, which can be parametrised through Wilson coefficients in the Weak Effective Theory (WET) framework~\cite{Aebischer:2017gaw}. More details can be found in Ref.~\cite{Kwok:2025fza}.

\subsubsection*{Analysis}

To model the decay ${\PBzs \to \PGf\PGmp\PGmm}$, we use a combination of Monte Carlo event generators and detector simulations. The signal and background events were simulated using the \pythiaeight~\cite{Bierlich:2022pfr} event generator at the $\PZ$ pole, and the detector effects were accounted for using \delphes~\cite{deFavereau:2013fsa} with the IDEA detector concept~\cite{IDEA1, Ilg:2024+2}. The IDEA detector features a highly precise vertex tracker and an advanced particle identification system, which are crucial for reconstructing the final states and distinguishing signal from background. The signal sample is generated with $\Pep\Pem \to \PZ \to \PQb\PAQb$ with focus on ${\PBzs \to \PGf\PGmp\PGmm}$ decays, while the background includes dominant processes such as $\PZ \to \PQb\PAQb$ and $\PZ \to \PQc\PAQc$. The generated data was then passed through the cut-based analysis, based on information from reconstructed mass, momentum, etc. The selected ${\PBzs \to \PGf\PGmp\PGmm}$ signal and total background yields are expected to be around $\SI{3.9d4}{}$~\cite{HeavyFlavorAveragingGroupHFLAV:2024ctg} and $\SI{1.6d3}{}$, respectively.

At FCC-ee, several measurements can be performed with the ${\PBzs \to \PGf\PGmp\PGmm}$ decay. The branching ratio measurement and the untagged time-dependent decay rate, described by the observable $\ODf$, can be measured without the need of flavour tagging. Furthermore, with flavour tagging one can measure the time-integrated $\CP$ asymmetry, $\langle\OACP\rangle$ and the time-dependent $\CP$ asymmetry, mainly described by the observables $\OCf$ and $\OSf$. 

The large dataset collected at FCC-ee enables a significant reduction in statistical uncertainties. Table~\ref{tab:results_summary} summarizes the expected precision for the key observables, demonstrating how FCC-ee will vastly improve our understanding of rare decays and $\CP$ violation. Such a level of precision is unattainable at Belle-II due to the limited availability of $\PBzs$ data. Additionally, the tagged measurements are not feasible at LHCb because of insufficient tagging power, making FCC-ee uniquely suited for these studies.

\begin{table}[h!]
    \centering
    \begin{tabular}{ccccccc}
        \hline \hline\\[-12pt]
        $\frac{\delta {\Br}}{\Br}$ & $\frac{\delta {\Br}^{\qsq\in [1.1, 6]}}{{\Br}^{\qsq\in [1.1, 6]}}$ & $\frac{\delta {\Br}^{\qsq \geq 15}}{{\Br}^{\qsq \geq 15}}$ & $\delta \ODf$ & $\delta \langle\OACP\rangle$ & $\delta \OCf$ & $\delta \OSf$\\
        [7pt]\hline
        $0.515\%$ & $0.810\%$ & $1.577\%$ & $0.103$ & $9.21\times 10^{-3}$ & $0.0210$ & $0.0214$\\
        \hline \hline
    \end{tabular}
    \caption{Summary of the experimental precision of the observables. }
    \label{tab:results_summary}
    
\end{table}

\subsubsection*{Effective Field Theory Interpretation}

The study interprets the estimated precision of the observables within the framework of WET, focusing on the Wilson coefficients (WCs) $C_7$, $C_9$ and $C_{10}$ that describe the FCNC transitions responsible for ${\PBzs \to \PGf\PGmp\PGmm}$. These WCs play a crucial role in connecting the experimental results to potential NP effects.

In the SM, $\CP$ violation arises from a single complex phase in the Cabibbo-Kobayashi-Maskawa matrix. However, NP scenarios can introduce additional complex phases, affecting the imaginary components of the WCs. By precisely measuring the BR along with the $\CP$ observables such as $\OSf$, we can constrain both the real and imaginary parts of these WCs, offering insights into the possible contributions of NP.

Figure~\ref{fig:wc_constraints} shows the constraints on the real and imaginary parts of $C_7$, $C_9$ and $C_{10}$ from the BR and $\OSf$ measurements at FCC-ee.\footnote{Other measurements like $\ODf$ and $\OCf$ are not shown here, as they provide relatively weaker constraints.} The diagonal subplots display the 1D likelihood for each individual WC, while the 2D plots illustrate the constraints on a pair of WCs. In the 2D plots, the shaded regions and dashed lines represent the $1\sigma$ and $2\sigma$ boundaries, respectively. The black lines indicate the combined constraints from the measurements. 
\begin{figure}[t]
    \centering
    \includegraphics[width=0.7\textwidth]{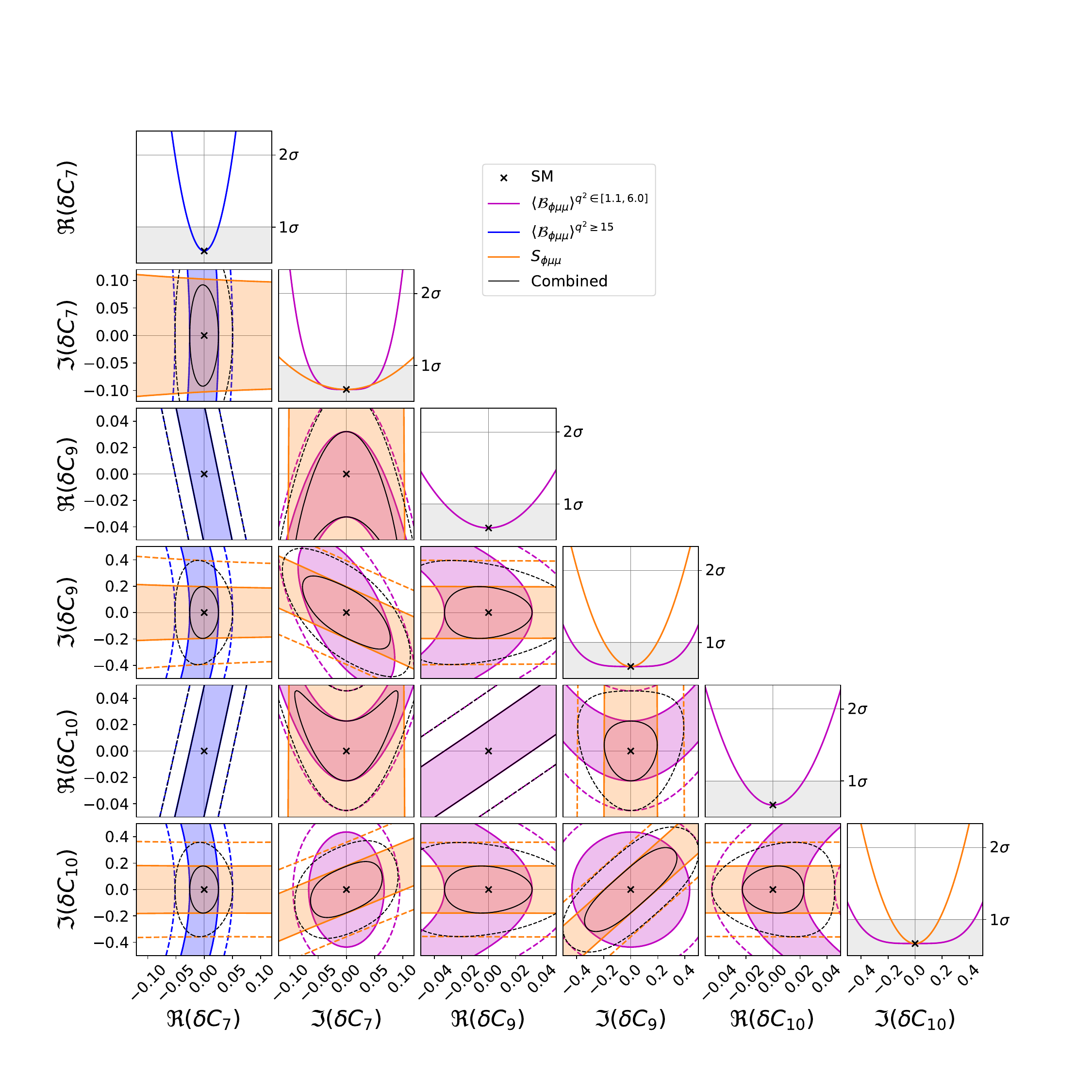}
    \caption{Constraints on the real and imaginary parts of the WCs $C_7$, $C_9$, and $C_{10}$ from the measurements of branching fraction and $\CP$ asymmetry at FCC-ee.}
    \label{fig:wc_constraints}
\end{figure}

The results show that the BR measurement is more sensitive to the real part of the Wilson coefficients, while $\OSf$ is particularly sensitive to the imaginary part, making these measurements complementary. By combining them, the flat directions present in individual measurements are reduced, significantly narrowing the allowed parameter space for NP contributions. FCC-ee is expected to achieve unprecedented sensitivity to deviations from SM predictions, probing NP at energy scales up to ${\cal O}(\SI{10}{\tera\electronvolt})$.

By constraining both the real and imaginary parts of the WCs, this study uniquely explores potential NP contributions, particularly in the imaginary components, which are less explored. The imaginary parts of the WCs carry critical information about $\CP$-violating phases introduced by NP, making them essential for a comprehensive understanding of $\CP$ violation. The precise measurements at FCC-ee allow for unprecedented sensitivity, offering a level of precision that is currently unattainable at other facilities. This makes FCC-ee one of the few experiments capable of probing these $\CP$-violating effects in such detail, reinforcing its importance in the search for new sources of $\CP$ violation and potential deviations from SM predictions.

\subsubsection{\focustopic \texorpdfstring{$\PQb \to \PQs \PGtp\PGtm$ and $\PQb \to \PQs \PGn \bar{\PGn}$}{b -> s tau+ tau- and b -> svv}}

\label{sec:flav-bstautau_bsnunu}
\subsubsection*{Theoretical and phenomenological motivations for $\PQb \to \PQs \PGtp \PGtm$ and $\PQb \to \PQs \PGn \overline{\PGn}$}

The process $\PBz \to \PK^*(892)^0\PGtp \PGtm$ tests the partonic FCNC transition $\PQb \to \PQs \PGtp \PGtm$. Present limits on this are weak, $\text{BR}(\PBp \to \PK^*(892)^0 \PGtp \PGtm) < 3.1 \times 10^{-3}$ at 90\% CL \cite{Belle:2021ecr}\footnote{The similar decay $\PBp \to \PKp \PGtp \PGtm$ has a slightly more stringent limit, $\text{BR}(\PBp \to \PKp \PGtp \PGtm) < 2.25 \times 10^{-3}$  at 90\% CL \cite{BaBar:2016wgb}.} and $\text{BR}(\PBzs \to \PGtp \PGtm) < 6.8 \times 10^{-3}$ at 95\% CL \cite{LHCb:2017myy}, while Belle II and LHCb will push these limits to the $10^{-4}$ ($10^{-5}$) level for the leptonic (semi-leptonic) mode. These should be compared with the SM prediction, which is at the $\sim 10^{-7}$ level for $\text{BR}(\PBz \to \PK^*(892)^0 \PGtp \PGtm)$ \cite{Kamenik:2017ghi}. The large number of boosted $\PB$ mesons produced from \PZ decays at a Tera-Z factory \cite{FCC:2018byv} could allow to improve substantially on the expected Belle-II limits and possibly reach a sensitivity to test the SM prediction.

The theoretical interest in this decay mode is wide, since it would allow to test a FCNC process involving third-generation leptons, complementing the present precise measurements of $\PQb \to \PQs \Plp \Plm$ with light leptons.
The general class of theories based on the $U(2)$-like scenarios discussed in \cref{sec:flav-UVdeconstruction}, for instance, predict the largest New Physics effects to appear in processes with third-generation fermions.
As a specific example, large effects due to New Physics can be expected in connection with the $R(\PD^{(*)})$ anomalies \cite{HFLAV:2022esi}. These involve the $\PQb \to \PQc \PGtm \PAGnGt$ transition, which in turn can be related to $\PQb \to \PQs \PGtp \PGtm$ via $SU(2)_L$ invariance and flavour symmetries. 
The same symmetries put these processes also in relation with $\PQb \to \PQs \PGnGt \PAGnGt$ transitions, which can be tested, for instance, with $\PB \to \PK^{(*)} \PGn \PAGn$ decays. Such possible correlation motivates including in this focus topic the so-called golden channels $\PB \to \PK^{(*)} \PGn \PAGn$. These are not affected by long-distance charm loop effects, allowing for lower theoretical uncertainties.

Belle II has recently searched for the $\PBp \to \PKp \PGn \PAGn$ decay and determined the branching fraction to be $[2.3 \pm 0.5 ^{+0.5}_{-0.4}] \times 10^{-5}$ \cite{Belle-II:2023esi}. This measurement has a significance of 3.5$\sigma$ and is the first evidence for the decay, resulting in a 2.7$\sigma$ tension with the SM prediction of $(5.58 \pm 0.37) \times 10^{-6}$. A recent study has been performed to give a baseline estimate for the sensitivity achievable at FCC-ee~\cite{Amhis:2023mpj}, including studies on the particle identification and vertex resolution requirements. This enables in addition measurements of $\PBzs$ and $\PGLzb$ decays which are impossible at Belle II. Experiments at a high-luminsoity \PZ-factory will be able to improve substantially upon Belle II measurements of $\PQb \to \PQs \PGn \PAGn$ decays by obtaining branching fraction measurements with $\mathcal{O}(1\%)$ sensitivity. This gives potential to make differential branching fraction measurements in these modes and will enable measurements of $\PBzs$ and $\PGLzb$ decays. 

\subsubsection*{Precision on the branching fraction of the transition $\PBz \to \PKst(892)^0 \PGtp\PGtm$ }

The experimental search for the transition $\PBz \to\PKst(892)^0  \PGtp\PGtm$ is challenging because of the two undetected neutrinos emitted from \PGt decays. The knowledge of all the vertices,  considering the fully charged $\PKst(892)^0 \to \PKp \PGpm$ decay and the three-prong hadronic $\PGtp \to  \PGpp \PGpm \PGpp$ decays, allows to close the kinematics of the process and opens the possibility to make an exclusive reconstruction of the decay \cite{Kamenik:2017ghi}. Events simulated with the FCC-ee simulation software featuring the IDEA tracking detector are used to evaluate the physics reach depending on the vertices measurement performances. A large number of physics backgrounds are considered in the study and a selection based on the topology of the decays is performed (see Fig.~\ref{fig:b2stautau}).  The vertexing performances are emulated to provide several working points for which the topological reconstruction is performed and the physics sensitivity is evaluated and reported in Fig.~\ref{fig:b2stautau}. Several characteristics of vertex detector were then simulated and placed on this curve. The improvements of single hit resolution and various reductions of the budget material of the IDEA vertex detector were considered. A remarkable conclusion of this study is that this physics case is limited by multiple coulombian scattering and therefore by the material budget of the beam pipe, triggering the interest and the thinking about novel vertex detector designs.           

\begin{figure}
    \centering
     \includegraphics[width=0.45\textwidth]{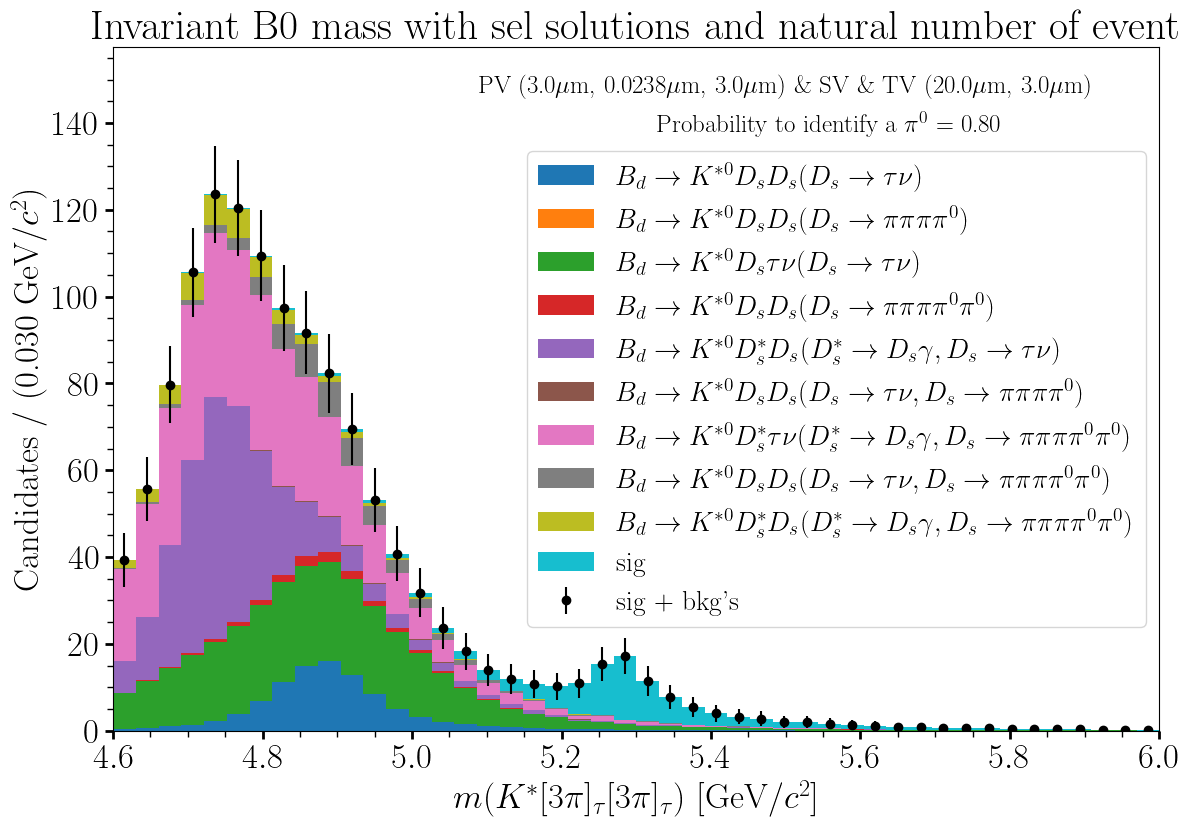}
    \includegraphics[width=0.45\textwidth]{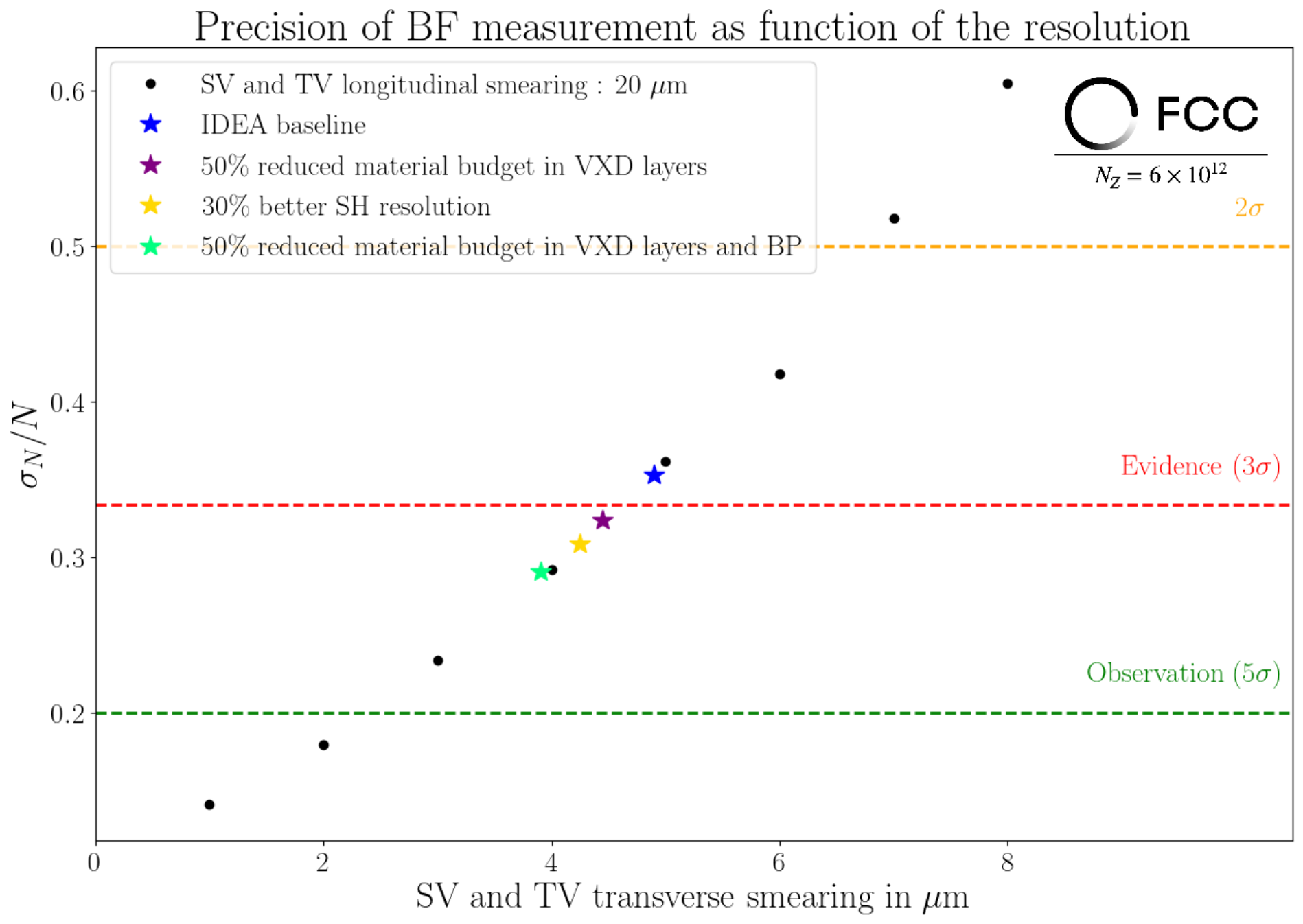}\\
    \caption{(Left) Invariant-mass distribution of $\PKst(892)^0\PGtp\PGtm$ candidates after all selection applied. (Right) Precision of the branching fraction measurement as function of the transverse vertexing resolution with a longitudinal resolution set at $\SI{20}{\micro\meter}$ for the emulated working points in addition of the IDEA baseline and regular improved detector working points.}
    \label{fig:b2stautau}
\end{figure}


\subsubsection*{Results of the $\PQb \to \PQs \PGn \PAGn$ study}

To study the $\PQb \to \PQs \PGn \PAGn$ transition, branching fraction measurements of the $\PBz \to \PKzS \PGn \PAGn$, $\PBzs \to \PGf \PGn \PAGn$, $\PBz \to \PKst(892)^0 \PGn \PAGn$ and $\PGLzb \to \PGL \PGn \PAGn$ decays have been considered.
In Ref.~\cite{Amhis:2023mpj} the statistical uncertainty on the BFs of $\PBz \to \PKst(892)^0\PGn \PAGn$, $\PBzs \to \PGf \PGn \PAGn$, $\PBz \to \PKzS \PGn \PAGn$ and $\PGLzb \to \PGL \PGn \PAGn$ decays are determined to be $0.53\%$, $1.20\%$, $3.37\%$ and $9.86\%$. For $\PBz \to \PKst(892)^0\PGn \PAGn$ and $\PBzs \to \PGf \PGn \PAGn$ decays this is determined by performing a simple two-stage BDT selection with cuts on the BDT responses optimised simultaneously, as well as a selection of rectangular cuts. The BDTs are trained on signal MC and inclusive background MC to model the various backgrounds originating from $\PZ \to \PQq\PAQq$ production, where $\PQq \in \{\PQu/\PQd/\PQs, \PQc, \PQb\}$. Due to the low number of background events that pass simultaneous cuts on the BDT responses, a bi-cubic spline is used to build a map of the efficiency in signal/background as a function of the BDT responses. For $\PBz \to \PKzS \PGn \PAGn$ and $\PGLzb \to \PGL\PGn \PAGn$ decay branching fraction, the appropriate reconstruction for the $\PKzS$ and $\PGL$ was not available and their precision was extrapolated assuming similar background as the $\PBz \to \PKst(892)^0\PGn \PAGn$ and $\PBzs \to \PGf \PGn \PAGn$ decays and a neutral reconstruction efficiency of $80\%$, suggested by companion studies. 
The dependence of the statistical sensitivity to the branching fraction of the $\PBzs \to \PGf \PGn \PAGn$ decay as a function of its value is shown in Fig.~\ref{fig:b2snunu}. This initial estimate is obtained assuming perfect particle identification (PID) and vertex seeding. To determine the dependence on PID and vertexing performance naive estimates are obtained. For the PID the signal efficiencies are recomputed after making random mass hypothesis swaps, kaon $\to$ pion and pion $\to$ kaon over a range of mis-identification rates. 
The resulting plots for these studies are shown in Fig.~\ref{fig:b2snunu}.

\begin{figure}
    \centering
    \includegraphics[width=0.48\textwidth]{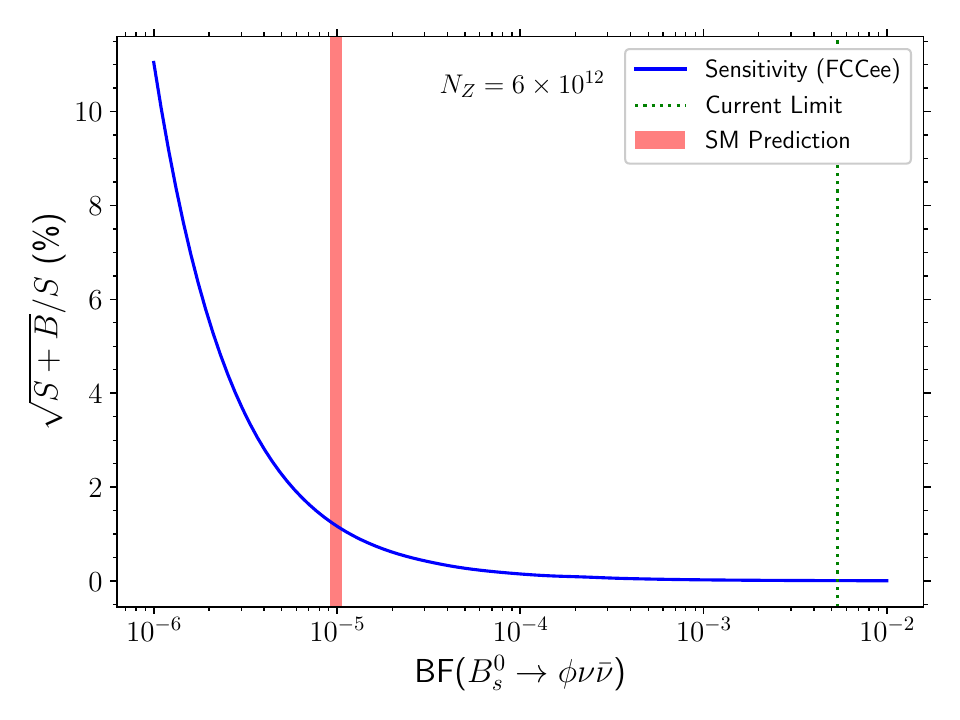}
    \includegraphics[width=0.48\textwidth]{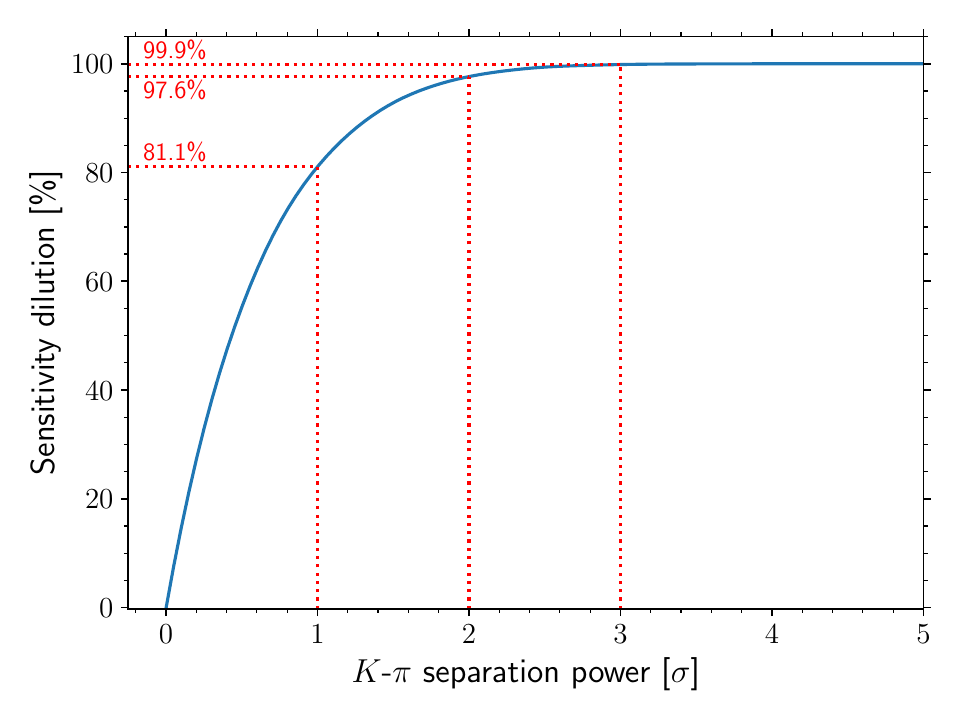}
    \caption{(Left) The statistical sensitivity to the $\PBzs \to \PGf \PGn \PAGn$ branching fraction  as a function of its value. (Right) Degradation of the sensitivity to the branching fraction, with respect to the nominal sensitivity assuming perfect PID, as a function of the kaon-pion separation power for the $\PBzs \to \PGf \PGn \PAGn$ decay. A similar sensitivity study is performed in Ref.~\cite{Amhis:2023mpj} for the correct secondary vertex association rate as a function of the expected vertex resolution.}
    \label{fig:b2snunu}
\end{figure}


\subsection{Tau physics}
\label{sec:flav-tau}


A high-luminosity  \PZ factory with $N_{\PZ} = 6\times 10^{12}$ \PZ decays~\cite{FCC:2018byv}  such as  FCC-ee or CEPC will see an abundant production of $2\times 10^{11}$ tau pairs allowing to deepen the studies of \PGt physics. In particular, the knowledge of the tau mass and lifetime and that of its leptonic branching fractions $\PGt \to \PGm \PGn \PAGn$, $\PGt \to \Pe \PGn \PAGn$ allows to test lepton universality in \PGt decays. This section reports the expected precisions on the tau mass and lifetime yielding a canonical lepton universality test as well as the sensitivity in the search for lepton-flavour violating \PGt  decays $\PGt \to 3 \Pl$,  $\PGt \to \Pl \PGg$.  



\subsubsection{\texorpdfstring{Tau lepton lifetime measurement at \TeraZ}{Tau lepton lifetime measurement at Tera-Z}}
\label{sec:tau-lifetime-fcc}


With a sample of $6 \times 10^{12}$ \PZ decays, it is convenient to measure the tau lifetime on the relatively small sub-sample of tau pairs where both tau leptons decay into a 3-prong topology, like Belle did~\cite{Belle:2013teo}. For these events, the two 3-prong vertices and the constraint of the very small luminous region precisely define the tau leptons' flight directions, significantly reducing systematics from Monte Carlo simulation of the effects of the undetected neutrinos on the reconstruction of the tau flight directions.

The DELPHI tau lifetime measure\-ment~\cite{McNulty:2001jt}, which includes a measurement done on tau pairs both decaying to 3 charged tracks (3--3-prongs topology), is extrapolated to the FCC sample. The DELPHI measurement is performed on the 1991--1995 sample, corresponding to about $4.0 \times 10^{6}$ hadronic \PZ decays~\cite{DELPHI:1995dsm}, hence about $N_{\PZ}^{\text{DELPHI 2004}} =4.0 \times  10^{6} / 70\% = 5.7\times 10^{6}$ \PZ decays.
The relative statistical uncertainty of the measurement restricted to the 3-3 prongs events is
$
\sigma(\tau_{\PGt}, \text{3--3})/\tau_{\PGt} \times 10^{6}\,\text{ppm}
\simeq 18000\,\text{ppm}
$,
where $\tau_{\PGt}$ is the tau lifetime~\cite{ParticleDataGroup:2024cfk}. 

The statistical and several sources of systematic uncertainties are scaled to the case of a \TeraZ run at FCC-ee as described in Ref.~\cite{Lusiani:2025}. 
Scaling the number of \PZ decays from DELPHI to FCC-ee, we estimate that the relative statistical uncertainty on the tau lifetime will be
$
\sigma(\tau_{\PGt}, \text{3--3}, \text{FCC}) / \tau_{\PGt}
\simeq
15.0\,\text{ppm}
$.

The largest sources of systematic uncertainty in the DELPHI analysis can be optimistically expected to scale down with the square root of the number of events: background subtraction (for which the simulation can be tuned with data control samples), reconstruction bias (which can be studied with data prompt events), and vertex alignment (done with data events). 
The tau lifetime measurement also requires the estimation of the average radiated energy in the initial state before the tau pair production. The DELPHI measurement relies on a $\Pep\Pem \to \PZ \to \PGtp\PGtm$ Monte Carlo simulation, whose estimated uncertainty contributes a 350\,ppm systematic uncertainty. We optimistically speculate that an improvement of a factor 30 may be achieved, reducing the related uncertainty on the tau lifetime to 11.5\,ppm at FCC. 
Additionally the uncertainty on the tau mass contributes 9 ppm, assuming the measurements described in Section~\ref{sec:tau-mass-fcc}.
In total the systematic uncertainties are expected to sum up to 21.5\,ppm.

\subsubsection{\texorpdfstring{Prospects for tau mass measurement at \TeraZ}{Prospects for tau mass measurement at Tera-Z}}
\label{sec:tau-mass-fcc}

Recently, the Belle\,II collaboration reported the most precise measurement of the tau mass, $\SI{1777.09}{} \pm \SI{0.08}{} \pm \SI{0.11}{\mega\eV\per c\squared}$~\cite{Belle-II:2023izd}. The
systematic uncertainties have been substantially reduced with respect to the previous $\PB$-factories' measurements, and the total uncertainty is smaller than the uncertainties of the measurements performed at the tau pair production threshold~\cite{BESIII:2014srs, Anashin:2023sch}. The Belle\,II statistical uncertainty of \SI{0.08}{\mega\eV\per c\squared} (45\,ppm) has been obtained with $175 \times 10^{6}$ tau pairs (\SI{190}{\per\femto\barn} of integrated luminosity), and could be improved to 1.3\,ppm with $2.0 \times 10^{11}$ tau pairs at FCC-ee, without taking into account the larger efficiency that can be expected at FCC-ee from the comparison of LEP versus $\PB$-factories tau measurements. Rescaling the statistical uncertainty of the OPAL tau mass measurement~\cite{OPAL:2000svx} to number of \PZ decays expected at FCC-ee gives an estimated statistical precision of 0.9\,ppm. The Belle\,II leading systematic uncertainty of \SI{0.07}{\mega\eV\per c\squared} (39\,ppm) is related to the knowledge of the beam energy, and is expected to be significantly smaller at FCC-ee, where the beam energy can be known with 1\,ppm precision. The other Belle\,II leading systematic uncertainty of \SI{0.06}{\mega\eV\per c\squared} (34\,ppm) is related to the understanding of the charged tracks reconstructed momentum scale, which can probably be calibrated with 2\,ppm precision at FCC-ee by matching the measured $\PJGy$ mass to its world average, presently known to 2\,ppm. Belle\,II reports systematic uncertainties related to the estimator bias (\SI{0.03}{\mega\eV\per c\squared}), to the choice of the fit function of the pseudo-mass distribution (\SI{0.02}{\mega\eV\per c\squared}), to the detector material (\SI{0.03}{\mega\eV\per c\squared}), and to the modeling of ISR, FSR and tau decay (\SI{0.02}{\mega\eV\per c\squared}), for a total of 29\,ppm.  We expect that these systematic uncertainties may be reduced by a factor 3 to 10\,ppm at FCC-ee, which we take as the estimated precision of the measurement of the tau mass at FCC-ee.

\subsubsection{\texorpdfstring{Lepton universality test in \PGt decays}{Lepton universality test in tau decays}}
\label{sec:tau-decays-fcc}

 The tau leptonic branching fractions were measured by the LEP experiments. As a case in point,   ALEPH measured the tau leptonic branching fractions BR$(\PGtm \to
  \PGmm \PAGnGm \PGnGt)$ and BR$(\PGtm \to \Pem \PAGnGm  \PGnGt)$ with a precision of about 0.44\% (0.40\% statistical and 0.19\% systematic), using $5.9 \times 10^{6}$ \PZ decays. The extrapolated statistical precision at a high-luminosity \PZ-factory with $6 \times 10^{12}$ \PZ decays amounts to 4.0\,ppm. The lepton universality test reported in Fig.~\ref{fig:tau-lept-univ-test} assumes that the ALEPH systematic uncertainty on the tau leptonic branching fractions can be reduced by a factor 10 to 0.019\%, 100\% correlated between the muon and the electron tau decay modes. The quantity $B^\prime_e$  denotes the average between the measured branching fraction BR$(\PGtm \to \Pem \PAGnGm \PGnGt)$ and its Standard Model prediction inferred from the measured branching fraction BR$(\PGtm \to \PGmm \PAGnGm \PGnGt)$. This lepton-universality test precision is improved w.r.t. the current knowledge by more than an order of magnitude and  would unravel the existence of new amplitudes if the central experimental values remain the same as to date.  This breakthrough is mostly brought by the improved \PGt lifetime measurement, unique to a $\PZ$-factory.    

  \begin{figure}[tb]\centering
    \includegraphics[width=0.60\linewidth]{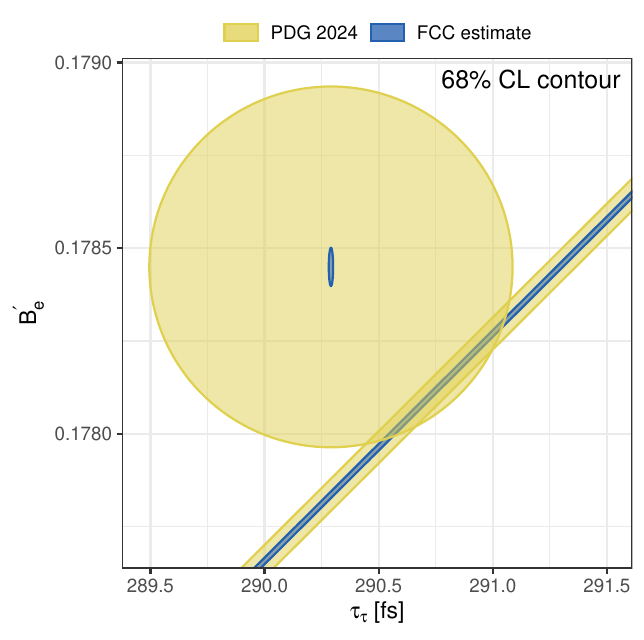}
    \caption{\label{fig:tau-lept-univ-test}%
      Lepton universality test using the tau mass, lifetime and leptonic branching fractions measurements. The test using the measurements reported in PDG 2024 is reported in yellow (lighter), and the estimated test at FCC-ee is reported in blue (darker). $B^\prime_e$ denotes the average between the measured branching fraction
      BR$(\PGtm \to \Pem \PAGnGm \PGnGt)$ and its Standard Model
      prediction using the measured branching fraction
      BR$(\PGtm \to \PGmm \PAGnGm \PGnGt)$.}
  \end{figure}

\subsubsection{\texorpdfstring{LFV from $\PGt$ decays}{LFV from tau decays}}
The large samples of $2.0 \times 10^{11}$ tau pairs produced at a \PZ-factory allows for high-precision searches for lepton-flavour violation. In particular, the prospects of the search for $\PGt\to\PGm\PGg$ and $\PGt \to \PGm\PGm\PGm$ are examined in the context of FCC-ee and compared to the current and foreseeable prospects at present or envisaged  facilities.  

\subsubsection*{\texorpdfstring{Search for $\PGt\to\PGm\PGg$}{Search for tau -> mu gamma}}

A simulation corresponding to $7 \times 10^{10}$ visible $\PZ$ decays has been used to estimate how many $\PGt\to\PGm\PGg$ decay candidates from background sources are to be expected for $3 \times 10^{12}$ \PZ decays at an FCC-ee experiment~\cite{ Dam:2018rfz,Lusiani:2025}. By assuming a reasonable reconstruction and selection efficiency, a sensitivity of $2 \times 10^{-9}$ has been estimated, corresponding to a signal equal to a double-sided $2\sigma$ fluctuation of the large number of expected background events, in the Gaussian approximation~\cite{Dam:2018rfz}. The sensitivity in terms of the expected upper limit at 90\% confidence level (CL) for a search of $\PGt\to\PGm\PGg$ on a sample of $6 \times 10^{12}$ $\PZ$ decays at FCC-ee is extrapolated to be 
 BR$(\PGt\to\PGm\PGg) \lessapprox 1.2 \times 10^{-9}\ \text{at 90\%\,CL}$.   \Cref{fig:tau-lfv-fcc} reports the present upper limits and the expected upper limits for BR$(\PGt \to \PGm\PGg)$. 

\subsubsection*{\texorpdfstring{Search for $\PGt\to\PGm\PGm\PGm$}{Search for tau -> mu mu mu}}

The Belle\,II collaboration reported a 90\% CL upper limit of $1.9 \times 10^{-8}$ for the lepton-flavour-violating branching fraction BR$(\PGt \to \PGm\PGm\PGm)$~\cite{Belle-II:2024sce}, using
$390\EE{6}$ tau pairs (corresponding to \SI{424}{\per\femto\barn} of integrated luminosity). The estimated selection efficiency is about 20.4\%, significantly larger than the one attained by the previous Belle search~\cite{Hayasaka:2010np}, 7.6\%, and is now comparable to the efficiency that has been reported at LEP\,1 for the DELPHI $\PGt\to\PGm\PGg$ search~\cite{DELPHI:1995mws}, 24.5\%, which is about 4 times the efficiency reported by the same $\PGt\to\PGm\PGg$ search by BaBar~\cite{BaBar:2010axs}. When assuming that for the $\PGt \to \PGm\PGm\PGm$ search an efficiency of 35.0\% may be obtained at FCC-ee, exploiting the highly efficient and pure muon selection at the \PZ peak energies, the expected upper limit at a high-luminosity \PZ-factory featuring  $2.0 \times 10^{11}$ tau pairs reads as BR$(\PGt\to\PGm\PGm\PGm) \lessapprox  2. \times 10^{-11}\  \text{at 90\%\,CL}$. \Cref{fig:tau-lfv-fcc} reports the present upper limits and the expected upper limits for BR$(\PGt \to \PGm\PGm\PGm)$.
\begin{figure}[tb]\centering
  \begin{subfigure}[t]{0.48\textwidth}
    \centering
    \includegraphics[width=\linewidth]{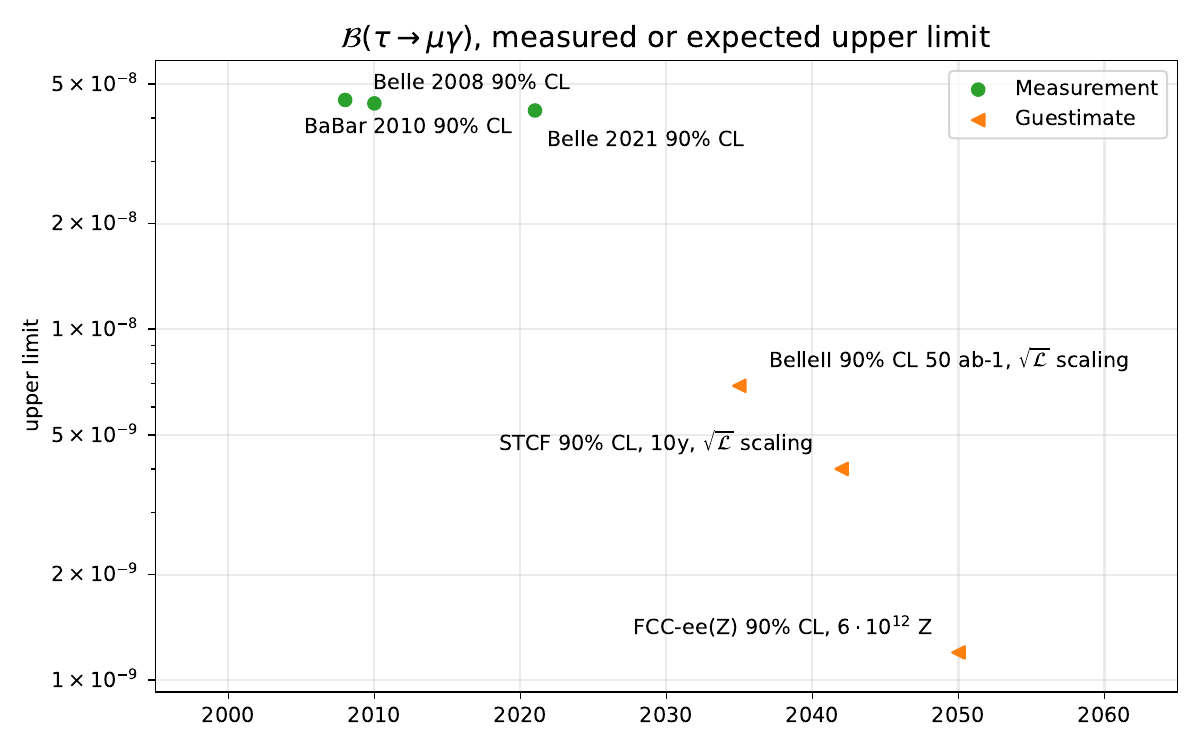}
  \end{subfigure}%
  \hfill
  \begin{subfigure}[t]{0.48\textwidth}
    \centering
    \includegraphics[width=\linewidth]{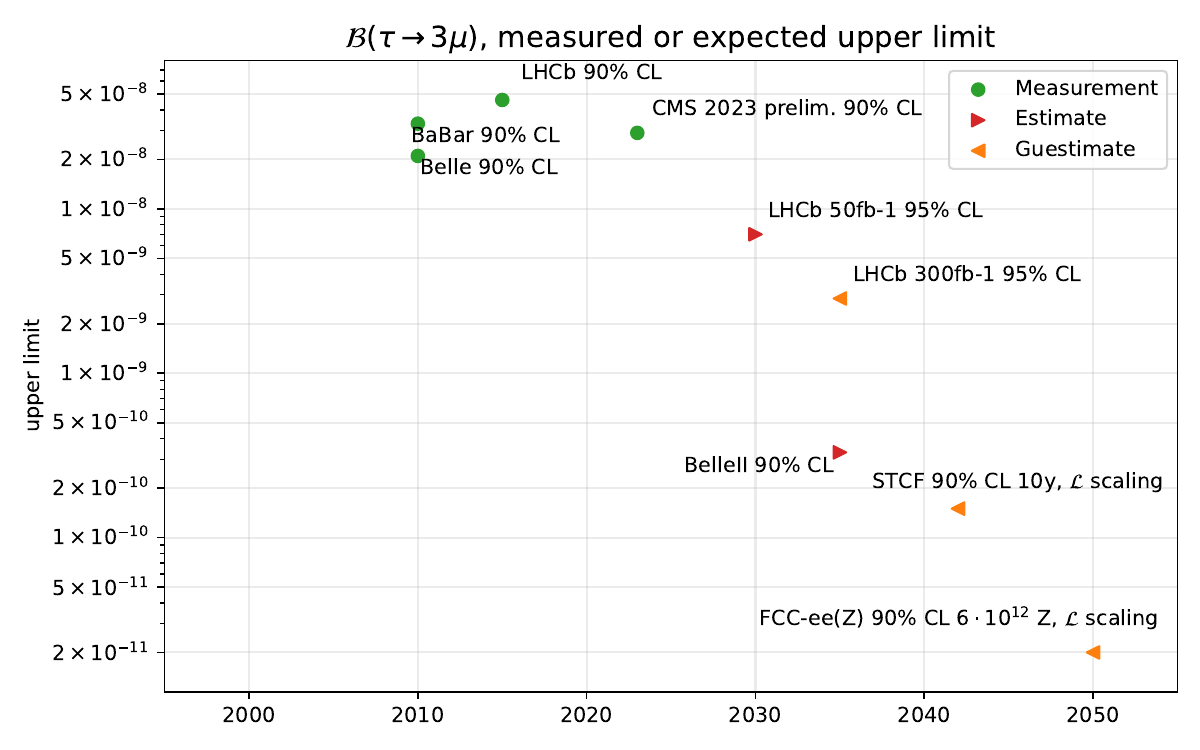}
  \end{subfigure}
\label{fig:exp-ul-tau-3mu}
  \caption{\label{fig:tau-lfv-fcc}%
    Present and future experimental reach on tau lepton-flavour-violating decays. Left: upper limits for ${\cal  B}(\PGt\to\PGm\PGg)$. Right:  upper limits for ${\cal B}(\PGt\to\PGm\PGm\PGm)$. The dates of the future measurements are speculative and mainly chosen for plotting purposes.}
\end{figure}




\subsection{Outlook on other avenues}
\label{sec:flav-outlook}


There was a surge of new hadrons observed at the LHC and the Belle~II and BESIII experiments. Many of those are exotic in nature, not fitting expectations from $\PQq\PQq\PQq$ or $\PQq\PAQq$ quark contents. So far 23 exotic particles containing at least one charm quark have been observed at the LHC, many in amplitude analyses of \PQb hadrons~\cite{Koppenburg:2024,Husken:2024rdk}. It is not clear whether they are compact objects of four or five quarks, or rather molecule-like bounds states of conventional hadrons, or even both. Many more such states are expected to be found LHCb and Belle~II. A Tera-\PZ run will add valuable information. As LHCb has reduced precision for neutral final states, and Belle~II lower data yields. The high precision of a \PZ run permits amplitude analyses involving neutral pions and kaons with similar precision as the corresponding mainly charged-only LHCb analyses. Such studies are essential to find isospin partners of known states, and find new states, thereby shedding light on their nature.


\clearpage
\section{References}
\printbibliography[heading=none]

\end{document}